%% file: hdr.tex
\pdfoutput=1
\documentclass[a4paper,11pt,twoside]{report}
\usepackage{epsfig,amsmath,amssymb,graphicx,fancyheadings,indentfirst,%
  undertilde,color,multirow,slashed}
\usepackage[english]{babel}
\usepackage[colorlinks=true,urlcolor=black]{hyperref}

\hypersetup{backref = true,pagebackref = true,hyperindex = true, colorlinks = true,
  breaklinks = true, urlcolor = blue, linkcolor = blue, bookmarks = true,
  bookmarksopen = true, citecolor=red}

\graphicspath{{figures/}}
\renewcommand{\chaptermark}[1]{\markboth{Chapter \thechapter\ - #1}{}}

\input{commands}

\begin{document}
\pagestyle{empty} \input{title}\cleardoublepage 

\pagestyle{fancy}\pagenumbering{roman}

\chapter*{Acknowledgements}\input{thanks} \cleardoublepage

\chapter*{Abstract}\input{abstract}
\tableofcontents \cleardoublepage

\chapter{Introduction} \pagenumbering{arabic} \input{intro}

\chapter{Supersymmetric quantum field theories} \input{susy}

\chapter{Supersymmetry in \feynrules} \input{superFR}

\chapter{Supersymmetry breaking} \input{brk}

\chapter{The Minimal Supersymmetric Standard Model} \input{mssm}

\chapter{Searching for non-minimal supersymmetry at hadron colliders} \input{nmsusy}

\chapter{From non-minimal supersymmetry to effective field theories} 
\input{effective}

\chapter{When theory meets experiment}
\input{cdf}

\chapter{Conclusions}
\input{conclusions}

\appendix
\renewcommand{\chaptermark}[1]{\markboth{Appendix \thechapter\ - #1}{}}

\chapter{Conventions} \input{conventions}

\cleardoublepage
\phantomsection
\addcontentsline{toc}{chapter}{Bibliography}
\bibliographystyle{utphys}
\footnotesize{\bibliography{biblio}}

\end{document}

%% file: commands.tex
\newcommand{\mysection}[1]{\setcounter{equation}{0}\section{#1}}

\definecolor{lightgrey}{gray}{0.9}

\def\be{\begin{equation}}
\def\ee{\end{equation}}
\def\bea{\begin{eqnarray}}
\def\eea{\end{eqnarray}}
\def\bsp#1\esp{\begin{split}#1\end{split}}
\def\bpm{\begin{pmatrix}} 
\def\epm{\end{pmatrix}} 
\def\bcen{\begin{center}}
\def\ecen{\end{center}}
\def\nn{\nonumber}
\def\bv{\begin{verbatim}}
\def\ev{\end{verbatim}}

\newcommand{\etc}{{\it etc.}}
\newcommand{\ie}{{\it i.e.}}
\newcommand{\eg}{{\it e.g.}}

\newcommand{\del}{\partial}
\renewcommand{\d}{{\rm d}} 
\newcommand{\g}{\mathfrak{g}}
\newcommand{\e}{\varepsilon} 
\def\lpp{{\lambda^{\prime\prime}}}
\def\hlpp{{\hat\lambda^{\prime\prime}}}
\newcommand{\gmu}{\gamma^\mu}

\newcommand{\tW}{{\widetilde W}}
\newcommand{\tF}{{\widetilde F}}

\newcommand{\delbar}{{\bar \del}}
\newcommand{\ebar}{{\bar \e}}
\newcommand{\Qbar}{{\bar Q}}
\newcommand{\Dbar}{{\bar D}}
\newcommand{\Wbar}{{\overline W}}
\newcommand{\sibar}{{\bar\sigma}} 
\newcommand{\chibar}{{\bar\chi}} 
\newcommand{\xibar}{{\bar\xi}} 
\newcommand{\psibar}{{\bar\psi}} 
\newcommand{\Phibar}{{\bar\Phi}} 
\newcommand{\Psibar}{{\bar\Psi}} 
\newcommand{\phibar}{{\bar\phi}} 
\newcommand{\lambar}{{\bar\lambda}} 
\newcommand{\thetabar}{{\bar\theta}} 
\newcommand{\Thetabar}{{\bar\Theta}} 
\newcommand{\betabar}{{\bar\beta}} 
\newcommand{\zetabar}{{\bar\zeta}} 
\newcommand{\etabar}{{\bar\eta}} 
\newcommand{\rhobar}{{\bar\rho}} 
\newcommand{\qbar}{{\bar q}}
\newcommand{\jbar}{{\bar j}}
\newcommand{\Jbar}{{\bar J}}
\newcommand{\Kbar}{{\bar K}}
\newcommand{\Fbar}{{\bar F}}
\newcommand{\Tbar}{{\bar T}}

\newcommand{\sbar}{{\bar s}}
\newcommand{\bbar}{{\bar b}}
\newcommand{\Bbar}{{\bar B}}
\newcommand{\ellbar}{{\bar \ell}}

\newcommand{\alphadot}{{\dot\alpha}} 
\newcommand{\betadot}{{\dot\beta}} 
\newcommand{\gammadot}{{\dot\gamma}} 

\newcommand{\is}{{i^\ast}}
\newcommand{\js}{{j^\ast}}
\newcommand{\ks}{{k^\ast}}
\newcommand{\ls}{{\ell^\ast}}
\newcommand{\ms}{{m^\ast}}

\newcommand{\D}{{\cal D}} 
\newcommand{\K}{{\cal K}} 
\newcommand{\G}{{\cal G}} 
 
\newcommand{\R}{{\cal R}} 
\newcommand{\E}{{\cal E}} 
\newcommand{\lag}{{\cal L}} 

\newcommand{\ip}{{i'}}
\newcommand{\jp}{{j'}}

\newcommand{\feynrules}{{\sc FeynRules}}
\newcommand{\feynarts}{{\sc FeynArts}}
\newcommand{\mathematica}{{\sc Mathematica}}
\newcommand{\calchep}{{\sc Calc\-Hep}}
\newcommand{\comphep}{{\sc CompHep}}
\newcommand{\mgme}{{\sc MadGraph/MadEvent}}
\newcommand{\sherpa}{{\sc Sherpa}}
\newcommand{\whizard}{{\sc Whi\-zard}}
\newcommand{\formcalc}{{\sc Form\-Calc}}
\newcommand{\gosam}{{\sc GoSam}}
\newcommand{\python}{{\sc Python}}
\newcommand{\madgraph}{{\sc MadGraph}}
\newcommand{\madgolem}{{\sc MadGolem}}
\newcommand{\madevent}{{\sc MadEvent}}
\newcommand{\madanalysis}{{\sc MadAnalysis}}
\newcommand{\alpgen}{{\sc AlpGen}}
\newcommand{\aloha}{{\sc Aloha}}
\newcommand{\herwig}{{\sc Herwig}}
\newcommand{\lanhep}{{\sc LanHep}}
\newcommand{\sarah}{{\sc Sarah}}
\newcommand{\spheno}{{\sc SPheno}}
\newcommand{\suspect}{{\sc SuSpect}}
\newcommand{\softsusy}{{\sc SoftSusy}}
\newcommand{\darksusy}{{\sc DarkSusy}}
\newcommand{\micromegas}{{\sc MicrOMEGAs}}
\newcommand{\pythia}{{\sc Pythia}}
\newcommand{\helas}{{\sc Helas}}

\newcommand{\delphes}{{\sc Delphes}}
\newcommand{\fastjet}{{\sc FastJet}}
\newcommand{\tauola}{{\sc Tauola}}
\newcommand{\fewz}{{\sc Fewz}}
\newcommand{\hathor}{{\sc Hathor}}

\newcommand{\hc}{{\rm h.c.}}

%% file: title.tex
\thispagestyle{empty} 
\begin{center}
  Universit\'e de Strasbourg\\
  \'Ecole doctorale de Physique et de Chimie Physique de Strasbourg\\

  \vspace{2cm}

  \large \textbf{Habilitation \`a diriger des recherches}\\
  Sp\'ecialit\'e: Physique des particules\\

  \vspace{1cm}

  pr\'esent\'ee par\\
  \textbf{Benjamin Fuks}\\ \vspace{1cm}

  \vspace{1.5cm}

  \LARGE \textbf{Supersymmetry\\
  When Theory Inspires Experimental Searches}\\

\vspace{4cm} \normalsize

Soutenue le 22 novembre 2013 devant le jury compos\'e de:\\
\vspace{1cm}

\renewcommand{\arraystretch}{.9}
\begin{tabular}{l c l}
Prof.\ Daniel Bloch & & Examinateur\\
Prof.\ Sabine Kraml & \hspace{1cm} & Pr\'esidente du jury\\
Prof.\ Eric Laenen & & Rapporteur\\
Prof.\ Fabio Maltoni & & Examinateur\\
Prof.\ Michelangelo Mangano & & Examinateur\\
Prof.\ Jean Orloff & & Rapporteur\\
Prof.\ Tilman Plehn & & Rapporteur\\
Prof.\ Michel Rausch de Traubenberg & & Garant\\
\end{tabular}\end{center}

%% file: thanks.tex
I would like to express my most sincere gratefulness to my
referees and the members of my habilitation committee. They
are all super-busy people who have taken up the challenge of
dedicating some time
for reading and commenting a 250+ pages manuscript.
I hope that they have got as much pleasure to do so
as me in writing this work. In order to avoid too long sentences
(although I love (very) long sentences), I will simply say to Sabine,
Daniel, Eric, Fabio, Michelangelo, Jean, Tilman and Michel:
Thank you!\\[0.05cm]

I am particularly grateful to Daniel for his support,
from his already-six-years-old warm welcome of a red-bearded theorist
in the CMS group of the IPHC laboratory up to now. Being accepted
as a theorist by CMS has allowed me to learn one very
important lesson. Even if it seems that theorists and experimentalists
speak different languages, we may easily discuss on common grounds with very
little effort from both sides. It is sufficient to try. \\[0.05cm]

I would also like
to thank Michel for having accepted to be the
advisor of this habilitation. In this world of phenomenology and
experimental physics, he has succeeded in bringing me back (at least once
in a while) to more formal and fundamental stuff,
linking me in this way to the Theory group of the lab.
I will even thank him twice as I have forced him to read
the entire manuscript word by word a second time, a task that he has seemed
happy to accept.\\[0.05cm]

Of course, I cannot forget Christelle Roy and Marc Rousseau for
their unconditional support to my activities, as well as Abdel-Mjid
Nourreddine for his welcome at the physics department of the University of
Strasbourg.\\[0.05cm]

Listing all my collaborators, colleagues and friends is 
a very difficult task as the probability to forget someone is definitely
equal to
one. Therefore, I will simply thank all of those with whom I have interacted
during my career. I know, this is cheating and I could have tried
a tentative list of names. I have started it, really. I have however stopped
after the 47$^{\rm th}$ name as too many letters in the alphabet were
remaining...\\[0.05cm]

And the last but not the least, I dedicate this work to my beloved wife
and son, my parents, grand-parents, godmother and
godfather as well as to my two brothers.

%% file: abstract.tex
We review, in the first part of this work, many pioneering works on supersymmetry
and organize these results to show
how supersymmetric quantum field theories arise from
spin-statistics, N\oe ther and a series of no-go theorems. We then introduce
the so-called superspace formalism dedicated to
the natural construction of supersymmetric Lagrangians and detail
the most popular mechanisms leading to soft supersymmetry breaking.

As an application, we describe the building of
the Minimal Supersymmetric Standard Model
and investigate current experimental limits on the parameter space
of its most constrained versions. To this aim, we use various flavor,
electroweak precision, cosmology and collider data. We then perform
several phenomenological excursions beyond this minimal setup and
probe effects due to non-minimal flavor violation in the squark sector, revisiting
various constraints arising from indirect searches for superpartners.

Next, we use several interfaced high-energy
physics tools, including the \feynrules\ package and its UFO interface
that we describe in detail, to study the phenomenology
of two non-minimal supersymmetric models at the Large Hadron
Collider. We estimate the sensitivity of this machine to
monotop production in $R$-parity violating supersymmetry and
sgluon-induced multitop production in $R$-symmetric supersymmetry.
We then generalize the results to new physics scenarios
designed from a bottom-up strategy and
finally depict, from a theorist point of view, a search
for monotops at the Tevatron motivated by these findings.

\cleardoublepage

%% file: intro.tex
After almost fifty years, the Standard Model of particle physics~\cite{Glashow:1961tr,
Salam:1964ry,Weinberg:1967tq, salamsm,
Glashow:1970gm, Weinberg:1971nd, Gross:1973ju, Kobayashi:1973fv, Gross:1974cs,
Politzer:1974fr} has been proved
to be a successful theory to describe all experimental
high-energy physics data.
It however leaves, despite its success,
many important questions open without providing any satisfactory answer.
Among those, one finds
the unexplained large hierarchy between the electroweak and the Planck scales,
the absence of a mechanism leading to neutrino oscillations, the unknown origins
of dark matter and of the cosmological constant as well as the strong $CP$-problem.
Consequently, the Standard Model is widely acknowledged as the low-energy
limit of a more fundamental theory. The recent discovery of a Higgs
boson~\cite{Aad:2012gk,Chatrchyan:2012gu} that seems to feature
properties as expected from the Standard Model reinforces this picture.
This first observation of a particle intrinsically unstable with
respect to quantum corrections indeed implies either a non-natural extreme fine-tuning
or a stabilization arising from a new physics sector which will emerge at scales that we will
probe soon.

As a result, model building activities in a beyond the Standard Model framework have
been very intense during the last decades. Among the leading candidates for
new physics, one finds extensions of the Standard Model where its
$SU(3)_c\times SU(2)_L\times U(1)_Y$ gauge group is embedded into a larger
structure, such as, \eg, $SU(5)$, $SO(10)$ or $E_6$~\cite{Georgi:1974sy,
Georgi:1974yf, Fritzsch:1974nn,Gursey:1975ki,Chang:1984uy}. In those contexts,
the Standard Model quark and lepton fields are encompassed into one or several representations
of the extended gauge group, together with possible additional matter content, and the
three gauge coupling constants have their strength unified at high energies
due to new effects in their renormalization group running. Those
models have also interesting additional properties, such that some of them could easily include
an explanation for neutrino masses or provide a mechanism leading to the quantization
of the electric charge. However, Grand Unified Theories have often difficulties to get in
agreement with the measured value for the electroweak mixing angle or even
to forbid the proton to decay in the case of the simplest extended gauge groups.

Another popular way to extend the Standard Model and solve at the same time the hierarchy problem
is to modify the structure of spacetime and include additional dimensions~\cite{Witten:1981me,
ArkaniHamed:1998rs,Randall:1999ee,Appelquist:2000nn}.
In this case, the Minkowski spacetime is extended by either a compact manifold, as in
pioneering extra-dimensional models, or by an orbifold, as in more modern approaches.
The large value of the Planck scale is then a consequence of the
presence of the extra dimensions. Moreover, each
field living in the extra-dimensions can be seen as an usual
four-dimensional field coming together with a series of more massive
excitations that can be possibly detected at collider experiments.

In this work, we choose to focus on another type of symmetry, dubbed supersymmetry, which naturally
extends the Poincar\'e algebra and links the fermionic and bosonic degrees of freedom of the
theory~\cite{Golfand:1971iw, Volkov:1973ix, Wess:1973kz, Wess:1974tw,Wess:1974jb,Salam:1974yz,
Salam:1974jj,Ferrara:1974ac,Ferrara:1974pu}. In particular, the minimal phenomenologically viable
supersymmetric model resulting from the direct supersymmetrization of the Standard Model,
the so-called Minimal Supersymmetric Standard Model (MSSM)~\cite{Nilles:1983ge,Haber:1984rc},
is one of the most studied options for new physics, both
at the theoretical and experimental levels. In addition of
associating with each fermion of the theory one bosonic superpartner, and \textit{vice versa},
weak scale supersymmetry also allows to solve several of the conceptual problems of the
Standard Model. Accounting for the supersymmetric degrees of freedom leads to a natural
unification of the three gauge couplings when run to higher energies~\cite{Ibanez:1981yh,
Dimopoulos:1981yj, Ellis:1990wk,
Amaldi:1991cn, Langacker:1991an, Giunti:1991ta} and stabilizes all scalar masses
with respect to quantum corrections, solving hence the hierarchy problem~\cite{Witten:1981nf}.
Furthermore, many supersymmetric models include a particle candidate for explaining the presence of dark
matter in the Universe~\cite{Goldberg:1983nd, Ellis:1983ew}. However, the superpartners
of the Standard Model particles have not been observed,
so that supersymmetry must be broken at low-energy.
In order not to reintroduce quadratically divergent quantum corrections in the theory,
this breaking must be soft and is therefore expected
to shift the supersymmetric particle masses around the TeV scale.

Consequently, the quest for supersymmetric particles is one of the main
topics of the experimental program at the Large Hadron Collider (LHC) at CERN.
However, there is no sign for a single superpartner so far and the latest
results of the general purpose experiments ATLAS and CMS are currently
pushing the bounds on the masses of the superpartners to higher and higher
scales~\cite{atlassusy,cmssusy}. In other words,
the supersymmetric parameter space turns out to be more and more constrained.
However, most analyses are only valid
in the context of the so-called constrained MSSM (cMSSM) framework, where the 105 free
parameters of the minimal supersymmetric model are reduced to a set of four
parameters and a sign, or for very specific simplified models inspired
by the cMSSM. In contrast, there are
much broader classes of supersymmetric theories valuable to be
studied both from a theoretical point of view and from an experimental one.
The results obtained from such studies could be further employed to design new search strategies
for new physics models, even possibly not supersymmetric when one accounts for a possible
recasting of the experimental analyses.

Phenomenological studies of such non-minimal supersymmetric models
in the context of hadron-collider experiments are often based on
the use of Monte Carlo event generators. In this framework,
a proper modeling
of the strong interactions, including parton showering,
fragmentation and hadronization, is essential for achieving a realistic
description of the hadronic collisions. The latter is efficiently provided
by packages such as
\pythia~\cite{Sjostrand:2000wi,
Sjostrand:2006za,Sjostrand:2007gs}, \sherpa~\cite{Gleisberg:2003xi,Gleisberg:2008ta} or
\herwig~\cite{Corcella:2000bw,Corcella:2002jc,Bahr:2008pv,Arnold:2012fq}.
However, any new physics signal is expected
to occur at the level of the underlying hard interaction. As a consequence, lots
of effort have been put into the development of matrix-element
generators such as \alpgen~\cite{Mangano:2002ea},
\comphep\ and \calchep~\cite{Pukhov:1999gg,Boos:2004kh,Pukhov:2004ca,Belyaev:2012qa},
{\sc Helac}~\cite{Kanaki:2000ey,Cafarella:2007pc},
\madgraph\ and \madevent~\cite{Stelzer:1994ta,Maltoni:2002qb,Alwall:2007st,Alwall:2008pm,Alwall:2011uj},
\sherpa~\cite{Gleisberg:2003xi,Gleisberg:2008ta} and
\whizard~\cite{Moretti:2001zz,Kilian:2007gr}, that allow for the generation of parton-level
events of large classes of beyond the Standard Model theories.

Historically, these packages have only supported the Standard Model and
a restricted subset of new physics theories, the reasons lying in the
complexity of the implementation of validated and ready-to-be-used model
files. This task indeed requires, first, a precise knowledge of
the Monte Carlo program itself, second, the implementation of thousands of lines
of code associated with the Feynman rules of the model and third,
a long and tedious process of debugging.
Implementing new models into these simulation packages has however recently drastically improved.
Parton-level matrix element generators have firstly begun to establish more general model
formats so that a less intimate knowledge of their internal code is now
necessary~\cite{Degrande:2011ua}. Secondly, several external programs,
such as \lanhep~\cite{Semenov:1996es,Semenov:1998eb,Semenov:2002jw,Semenov:2008jy,Semenov:2010qt},
\feynrules~\cite{Christensen:2008py,Christensen:2009jx,Christensen:2010wz,Duhr:2011se,Fuks:2012im,
Alloul:2013fw,Christensen:2013aua,Alloul:2013bka}
and \sarah~\cite{Staub:2008uz,Staub:2009bi,Staub:2010jh,Staub:2012pb},
have been developed
in order to allow the user to define a model via its Lagrangian rather than via the set of
its individual Feynman rules.

Thanks to the above-mentioned packages, a systematic investigation of the phenomenology of
any new physics model has been rendered possible and straightforward
following the path of a top-down approach. In this context, the theory is first
defined by its particle content, gauge symmetries, free parameters and Lagrangian.
Next, relevant benchmark
scenarios that are both theoretically motivated and not experimentally excluded
are constructed and finally
employed for predicting the model signatures at high-energy
experiments.
The design of such benchmarks is however not an easy task,
as many model parameters enter and cannot be fixed by the present constraints.
The conception of benchmarks is thus in general driven by simplicity, which
introduces at the same time some bias in the definition of signatures called typical
for a given model. Furthermore, a given signature is neither related to a single benchmark nor
to a specific model itself. Universal extra dimensions and supersymmetry share, for instance,
very similar signatures starting from the pair production of new states followed
by their cascade decays into an invisible state, jets and charged leptons.

For these reasons, it is useful to perform, in parallel to top-down phenomenological investigations,
alternative studies starting from a final state signature. In order to model
the mechanisms leading
to the production of a specific signature, a Lagrangian with a minimal set
of effective operators is usually supplemented to the Standard Model one. Results
obtained in this framework can then be
reinterpreted, in a second step, in the context of several beyond the Standard Model
theories simultaneously.

In this work, we have adopted a more pragmatic choice and rely both on the top-down
and bottom-up approaches for probing new physics at colliders.
We have first followed a top-down path and started by studying well defined non-minimal
supersymmetric theories. We have analyzed specific signatures
predicted by the models under consideration
and have designed several search strategies allowing for a possible
observation of the associated signals at the LHC. To this aim, we have performed
simulations of proton-proton collisions that have occurred at both past
LHC runs, at respective center-of-mass energies of 7~TeV and 8~TeV, and analyzed the generated
events within the \madanalysis~5\ framework~\cite{Conte:2012fm}. More into details,
events have been simulated
by means of the automated Monte Carlo program \madgraph~5,
whose necessary UFO model libraries have been produced
directly from the respective Lagrangians
by making use of the \feynrules\ package.
Accurate descriptions of both the new physics signals and
the different contributions to the Standard Model background
have been obtained by relying, on the one hand, on
multiparton matrix-element merging of leading-order event samples
with different final state multiplicities~\cite{Mangano:2006rw,Alwall:2008qv},
and, on the other hand, on results for total rates computed at the
next-to-leading order and next-to-next-to-leading order
accuracies in QCD. Moreover,
advanced simulations of the ATLAS and CMS
detector responses have been performed with the \delphes\
program~\cite{Ovyn:2009tx}.

In a second stage, we employ the investigated signatures as a starting
point and study them in a more general bottom-up context, well beyond the initial
non-minimal supersymmetric frameworks.
We construct
effective Lagrangians with a set of new interactions leading to the production
of the final states under consideration, possibly including mediation by additional new
states. Using the search strategies developed in the supersymmetric cases as guidelines,
we make use of Monte Carlo simulations as above to extract
the parameter space regions of the bottom-up inspired
theoretical models that can be reached by the LHC.

Finally, as the last step of this work, we describe (public)
experimental searches that have been motivated by our
phenomenological investigations. This allows this manuscript to illustrate
a full chain linking theory to experiment via both bottom-up and top-down phenomenological
excursions beyond the Standard Model.

In the next chapter (Chapter~\ref{chap:susy}), we describe in full generality
the building of a supersymmetric theory from very basic principles, namely the spin-statistics
theorem~\cite{Pauli:1940zz}, the N\oe ther theorem~\cite{Noether:1918zz}, as well as a series
of no-go theorems~\cite{Coleman:1967ad,Haag:1974qh}. We show how the only
knowledge of these theorems leads unavoidably to the Poincar\'e superalgebra underlying
phenomenologically relevant
supersymmetric models. We then move to a detailed
description of the superspace formalism~\cite{Salam:1974yz, Salam:1974jj, Ferrara:1974ac},
the natural approach for the building of supersymmetric Lagrangians. The content of this
chapter is based on the supersymmetry lectures given by the author at 
the University of Louvain-la-Neuve (Belgium) in January-March 2011 as well as on the book
(in French) of Ref.~\cite{livre},

{\small \begin{quote}
\textbf{B.~Fuks and M.~Rausch de Traubenberg},\\
\textit{Supersym\'etrie~: exercices avec solutions},\\
Ellipses Editions, 2011 (ISBN 978-2-729-86318-0).
\end{quote}}

Chapter~\ref{chap:FR} is dedicated to the implementation of supersymmetric models
in \feynrules, such a tool allowing for the study of the associated phenomenology
by means of Monte Carlo simulations thanks to dedicated interfaces to several automated
event generators. We first
describe the \feynrules\ package itself, together with the Universal \feynrules\ format
(UFO)
allowing to pass the information from \feynrules\ to any other program in a very generic way.
Emphasis is put on all the tasks that can be automated, reducing in this way
the risk of error by
the user. Examples of calculations that can be performed using the superspace module
of \feynrules\ are finally provided. The material
of this chapter is based on the manual of the version 2.0 of \feynrules~\cite{Alloul:2013bka},
on a series of specific papers that have recently appeared~\cite{Christensen:2009jx,
Christensen:2010wz, Duhr:2011se, Christensen:2013aua}, as well
as on the definition of the UFO conventions~\cite{Degrande:2011ua}, 

{\small\begin{quote}
  \textbf{A.~Alloul, N.~D.~Christensen, C.~Degrande, C.~Duhr and B.~Fuks},\\
    \textit{\feynrules~2.0, a complete toolbox for tree-level phenomenology},\\
    arXiv:1310.1921 [hep-ph] (submitted to Comput.\ Phys.\ Commun.).\\[.2cm]
  \textbf{N.~D.~Christensen, P.~de Aquino, N.~Deutschmann, C.~Duhr, B.~Fuks, C.~Garcia-Cely,
    O.~Mattelaer, K.~Mawatari, B.~Oexl and Y.~Takaesu},\\
    \textit{Simulating spin-3/2 particles at hadron colliders},\\
    Eur.\ Phys.\ J.\ C {\bf 73} (2013) 2580.\\[.2cm]
  \textbf{C.~Degrande, C.~Duhr, B.~Fuks, D.~Grellscheid, O.~Mattelaer and
      T.~Reiter},\\
      \textit{UFO - The Universal \feynrules\ Output},\\
      Comput.\ Phys.\ Commun.\  {\bf 183} (2012) 1201-1214.\\[.2cm]
  \textbf{N.~D.~Christensen, C.~Duhr, B.~Fuks, J.~Reuter and C.~Speckner},\\
    \textit{Introducing an interface between \whizard\ and \feynrules},\\
    Eur.\ Phys.\ J.\ C {\bf 72} (2012) 1990.\\[.2cm]
  \textbf{C.~Duhr and B.~Fuks},\\
    \textit{A superspace module for the \feynrules\ package},\\
    Comput.\ Phys.\ Commun.\  {\bf 182 } (2011)  2404-2426.\\[.2cm]
  \textbf{N.~D.~Christensen, P.~de Aquino, C.~Degrande, C.~Duhr, B.~Fuks,
    M.~Herquet, F.~Maltoni and S.~Schumann},\\
    \textit{A comprehensive approach to new physics simulations},\\
    Eur. Phys. J. {\bf C 71} (2011) 1541.
\end{quote}}

Realistic supersymmetric models must encompass supersymmetry breaking.
In Chapter~\ref{sec:susybrk}, we describe some general
features associated with any supersymmetry-breaking model~\cite{Witten:1981nf,Fayet:1974jb,
Wess:1978ns,Iliopoulos:1974zv,Fayet:1977vd, Fayet:1979yb,Ferrara:1979wa} and then turns
to a short review of the most popular mechanisms employed to achieve soft supersymmetry-breaking,
namely gravity-mediated supersymmetry breaking
\cite{Witten:1982hu, Deser:1976eh, Freedman:1976xh,
Freedman:1976py, Ferrara:1976um, Cremmer:1978hn, Cremmer:1978iv, Cremmer:1982en,
Chamseddine:1982jx, Barbieri:1982eh, Ibanez:1982ee, Ohta:1982wn, Ellis:1982wr,
AlvarezGaume:1983gj, Binetruy:2000zx}, gauge-mediated supersymmetry 
breaking \cite{Dimopoulos:1981au, Dine:1981za, Derendinger:1982tq, Fayet:1978qc,
Dine:1981gu, Nappi:1982hm, AlvarezGaume:1981wy, Dine:1993yw, Dine:1994vc,
Dine:1995ag, Giudice:1998bp} and anomaly-mediated supersymmetry breaking
\cite{Randall:1998uk, ArkaniHamed:1998kj, Bagger:1999rd, Derendinger:1991kr,
Derendinger:1991hq, LopesCardoso:1991zt, LopesCardoso:1992yd,
Kaplunovsky:1994fg}. This chapter is based on the above-mentioned book~\cite{livre}, on the
lectures given at the University of Louvain-la-Neuve, and on
the forthcoming publication~\cite{sugra},

{\small \begin{quote}
  \textbf{B.~Fuks and M.~Rausch de Traubenberg},\\
    \textit{A supergravity primer},\\
    In preparation.
\end{quote}}

The theoretical framework developed in Chapter~\ref{chap:susy} and Chapter~\ref{sec:susybrk}
is applied, in Chapter~\ref{chap:mssm}, to the building of the simplest supersymmetric
model, the Minimal Supersymmetric Standard Model~\cite{Nilles:1983ge,
Haber:1984rc}. We provide first detailed information
on the construction of the model itself. In a second step,
we present the most general features of the MSSM, addressing
the hierarchy problem, gauge coupling unification, the dark matter problematics and introducing
the most common properties of the MSSM
concerning the production of supersymmetric particles at colliders.
We then
establish, in the framework of three minimal MSSM scenarios with a small number of free parameters,
the parameter space regions compatible with up-to-date data from low-energy,
electroweak precision and flavor physics. A first step towards
non-minimal supersymmetric models is next achieved by studying the effects of
non-minimal flavor violation
in the squark sector. Finally, cosmological aspects are addressed, as well as constraints
originating from direct searches for supersymmetric particles at colliders
and in particular at the LHC. The results of this chapter are based, on the one hand,
on the above-mentioned Ref.~\cite{Duhr:2011se}, Ref.~\cite{livre}, and on the lectures given at
the University of Louvain-la-Neuve, as well as on the papers~\cite{Bozzi:2007me,
Fuks:2008ab, Fuks:2011dg},

{\small \begin{quote}
  \textbf{B.~Fuks, B.~Herrmann and M.~Klasen},\\
    \textit{Phenomenology of anomaly-mediated supersymmetry breaking scenarios with non-minimal
      flavor violation},\\
    Phys.\ Rev.\ D \textbf{86} (2012) 015002.\\[.2cm]
  \textbf{B.~Fuks, B.~Herrmann and M.~Klasen},\\
    \textit{Flavor violation in gauge-mediated supersymmetry breaking models: experimental constraints
     and phenomenology at the LHC},\\
    Nucl.\ Phys.\  B {\bf 810} (2009) 266-299.\\[.2cm]
  \textbf{G.~Bozzi, B.~Fuks, B.~Herrmann and M.~Klasen},\\
  \textit{Squarks and gaugino hadroproduction and decays in
    non-minimal flavor violating supersymmetry},\\
  Nucl.\ Phys.\  B {\bf 787} (2007) 1-54.
\end{quote}}

In the next chapter (Chapter~\ref{chap:nonmin}), we describe two non-minimal
supersymmetric theories, the MSSM with $R$-parity violation~\cite{Barbier:2004ez}
and the minimal version of a supersymmetric theory with an unbroken
$R$-symmetry~\cite{Fayet:1974pd, Salam:1974xa, Kribs:2007ac}. After implementing
both theories into \feynrules,
dedicated phenomenological analyses are performed for each of the models by means of Monte Carlo
simulations, employing the chain \feynrules-UFO-\madgraph~5-\pythia-\delphes-\madanalysis\ 5
for generating
and analyzing events including detector response effects. This allows us to investigate
one collider signature for each of the two models above, monotop production in
$R$-parity violating supersymmetry and multitop production induced by
the decay of a pair of sgluon fields as predicted in $R$-symmetric supersymmetric models. We show
that one can expect
visible signals at the LHC, running at a center-of-mass energy
of 7~TeV, in the case of specific benchmark scenarios. The material included in this
chapter is based on the work of Ref.~\cite{Fuks:2012im} for which it
also provides extra details,

{\small \begin{quote}
  \textbf{B.~Fuks},\\
    \textit{Beyond the Minimal Supersymmetric Standard Model: from theory to phenomenology},\\
    Int.\ J.\ Mod.\ Phys.\ A {\bf 27} (2012) 1230007.
\end{quote}}

Generalizing the supersymmetric picture,
we construct in Chapter~\ref{chap:effective} two frameworks based on an effective
field theory aiming to describe the production of the two new physics
signals under consideration. We explore in this way several
beyond the Standard Model theories at the same time, although
the reinterpretation process in the context of a specific model
goes beyond the scope of this work.
We hence address, using the chain of tools above,
the production of
a monotop state and the one of a multitop signature
arising from the decay of a pair of sgluons in an effective field theory context.
Detailed phenomenological investigations
are performed in order to estimate the regions of the parameter
spaces of both models covered by
the LHC, with 20~fb$^{-1}$ of collisions at a center-of-mass
energy of 8~TeV. We also provide, in this chapter, extensive details
about the simulation of the Standard model background.
We review the works of
the two published papers of Ref.~\cite{Andrea:2011ws} and Ref.~\cite{Calvet:2012rk}
and present new results that have been recently submitted~\cite{Agram:2013wda},

{\small \begin{quote}
  \textbf{J.~L.~Agram, J.~Andrea, M.~Buttignol, E.~Conte and B.~Fuks},\\
    \textit{Monotop phenomenology at the Large Hadron Collider},\\
    arXiv:1311.6478 [hep-ph] (accepted by Phys.\ Rev.\ D).\\[.2cm]
\textbf{S.~Calvet, P.~Gris, B.~Fuks and L.~Val\'ery},\\
      \textit{Searching for sgluons in multitop events at a center-of-mass energy of 8 TeV},\\
      JHEP {\bf 1304} (2012) 043.\\[.2cm]
  \textbf{J.~Andrea, B.~Fuks and F.~Maltoni},\\
      \textit{Monotops at the LHC},\\
      Phys.\ Rev.\ D \textbf{84} (2011) 074025.
\end{quote}
}

Motivated by our phenomenological results, several experimental searches at hadron
colliders have either been achieved or are currently on-going~\cite{Aaltonen:2012ek,
ATLAS-CONF-2013-051,monoATLAS, monoCMS}. We dedicate the last chapter of this document,
Chapter~\ref{chap:monotopsCDF}, to the presentation of a vision of a theorist for
one of these experimental
analyses\footnote{The author of this work has been enrolled in the
CDF collaboration for the considered analysis that is summarized in Ref.~\cite{Aaltonen:2012ek}.},

{\small \begin{quote}
  \textbf{CDF Collaboration}\\
    \textit{Search for a dark matter candidate produced in association with a
      single top quark in $p\bar{p}$ collisions at $\sqrt{s} = 1.96$ TeV},\\
      Phys.\ Rev.\ Lett.\  {\bf 108} (2012) 201802.
\end{quote}
}

Finally, we summarize our results in Chapter~\ref{chap:conclusions}
and collect, in Appendix~\ref{app:conv}, our conventions on indices, Pauli and Dirac matrices
and on relations among the Grassmann variables necessary for the superspace formalism.

\cleardoublepage

%% file: susy.tex
\label{chap:susy}
In this chapter, we review many pioneering works on supersymmetry
and organize the results to illustrate how
supersymmetric quantum field theories naturally arise from
spin-statistics theorem, N\oe ther theorem
and a series of no-go theorems. We then provide details on
the superspace formalism, a suitable mean
to construct supersymmetric Lagrangians, and build,
for the sake of the example, the most general
(non-renormalizable) supersymmetric Lagrangian.

\mysection{The Poincar\'e superalgebra}
\subsection{Quantum field theories and symmetries}

Particle physics model building relies on the framework of quantum field
theories which unifies two basic building
blocks, quantum mechanics and special relativity. Together with simple
principles of symmetry, this allows to classify and describe elementary
particles and their interactions by means of relativistic quantum fields and
their properties. Among those, we can emphasize two key features, the mass of
the particles and their spin, this last observable being 
associated to the famous spin-statistics theorem \cite{Pauli:1940zz}. This
theorem
proves that particles of half-odd-integer spin, \ie, fermions, obey Fermi-Dirac
statistics and are represented by anticommuting fields while particles of
integer spin, \ie, bosons, obey Bose-Einstein statistics and are described by
commuting fields.

We can define two classes of symmetries according to the
way they act on a
quantum field. Spacetime (often called external)
symmetries
explicitly modify spacetime coordinates $x^\mu$,
\be
  x^\mu \to x^{\prime\mu} = \xi^\mu(x)\ , 
\ee 
where $\xi$ is a Poincar\'e transformation of the spacetime variables and is thus by
definition invertible and differentiable. In contrast,
internal symmetries, such as gauge symmetries, act on the fields themselves, 
\be
  \varphi^a(x) \to \varphi^{\prime a}(x') = S^a{}_b\ \varphi^b(x) \ ,
\ee
where we have introduced a collection of generic fields $\{\varphi^a\}$ and
denote by $S^a{}_b$ the generators associated with a generic internal symmetry
operation. Moreover, in the notations above,
the field $\varphi$ can either be fermionic or bosonic.
As for particles and fields, generators of symmetries
can also be classified with respect to their bosonic or fermionic nature. In the
first case, particle spins are left unchanged by a symmetry operation,
while in the second case, particles of different spins could be related. 

\subsection{The Coleman-Mandula theorem}\label{sec:coleman}
We first focus on the construction of theories such as the Standard Model of
particle physics where the generators of the symmetry group are all bosonic. In
this case, combining spin-statistics \cite{Pauli:1940zz} and N\oe ther theorems
\cite{Noether:1918zz} leads naturally to a Lie algebra structure spanned by the
symmetry generators. This can be
shown by building a toy theory describing the dynamics of a set of
bosonic and fermionic fields $\phi^a$ and $\psi^i$ through a Lagrangian ${\cal
L}(\phi^a, \psi^i)$. We then assume that this Lagrangian is left invariant by
a symmetry operation to which we associate the continuous transformation of the
fields
\be\label{eq:varfield}
  \phi^a \to \phi^a + \delta_A \phi^a = \phi^a + (B^1_A)^a{}_b\ \phi^b 
\qquad\text{and}\qquad
  \psi^i \to \psi^i + \delta_A \psi^i = \psi^i + (B^2_A)^i{}_j\ \psi^j \ .
\ee
In the two equations above, the dependence on the spacetime coordinates is
understood for clarity and we have introduced the symmetry generators $B_A^1$
and $B_A^2$ acting on the bosonic and fermionic sectors of our toy theory,
respectively. From these transformation laws, we can deduce the corresponding
variation of the Lagrangian,
\be\bsp
  \delta_A {\cal L} =&\ 
     {\del {\cal L} \over \del \phi^a} \delta_A \phi^a + 
     {\del {\cal L} \over \del \psi^i} \delta_A \psi^i + 
     {\del {\cal L} \over \del \big(\del_\mu\phi^a\big)} 
       \delta_A \big(\del_\mu\phi^a \big) + 
     {\del {\cal L} \over \del \big(\del_\mu\psi^i\big)} 
       \delta_A \big(\del_\mu\psi^i \big) \\
   =&\  \del_\mu \bigg[
     {\del {\cal L} \over \del \big(\del_\mu\phi^a\big)} \delta_A \phi^a + 
     {\del {\cal L} \over \del \big(\del_\mu\psi^i\big)} \delta_A \psi^i
     \bigg]  \equiv \del_\mu j_A^\mu \ ,
\esp\label{eq:varlag}\ee
where the second equality is obtained after an integration by parts and using
Euler-Lagrange equations. By assumption, this Lagrangian ${\cal L}$ is
invariant
under the symmetry operation under consideration. Therefore, this implies the
conservation of the current\footnote{One can always redefine the current as
$J^\mu \to J^\mu + \kappa^\mu$ with $\del_\mu \kappa^\mu = 0$. This property
will be used in Chapter~\ref{sec:susybrk} when computing the supercurrent
yielding goldstino and
gravitino interactions.} 
\be
  J^\mu = j_A^\mu - K_A^\mu = 
    {\del {\cal L} \over \del \big(\del_\mu\phi^a\big)} \delta_A \phi^a + 
    {\del {\cal L} \over \del \big(\del_\mu\psi^i\big)} \delta_A \psi^i  - K_A^\mu
\ .
\label{eq:generalcurrent}\ee 
The quantity $K^\mu$ is obtained from a direct computation of the
variation of the Lagrangian, after applying Eq.\ \eqref{eq:varfield}
to ${\cal L}$ and then extracting $K^\mu_A$ from the relation
$\delta_A {\cal L} =\del_\mu K_A^\mu$.

N\oe ther theorem implies the conservation in time of the charge $B_A$
defined as the temporal component of the current $j^\mu$ integrated over the
entire tridimensional Euclidean space,
\be\label{eq:Bcharge}
  B_A = -i \int \d^3 x \bigg[
   {\del {\cal L} \over \del \big(\del_0\phi^a\big)} \delta_A \phi^a +
   {\del {\cal L} \over \del \big(\del_0\psi^i\big)} \delta_A \psi^i 
  \bigg] 
   = -i \int \d^3 x 
     \Big[\Pi_a (B^1_A)^a{}_b\ \phi^b + \rho_i (B^2_A)^i{}_j\ \psi^j\Big] \ .
\ee
In this last expression, we have introduced the momentum densities $\Pi_a$ and
$\rho_i$ conjugate to the fields $\phi^a$ and $\psi^i$,
\be
  \Pi_a = \frac{\del {\cal L}}{\del (\del_0 \phi^a)} 
  \qquad \text{and}\qquad
  \rho_i = \frac{\del {\cal L}}{\del (\del_0 \psi^i)} \ ,  
\ee
and employed the expressions of the variation of the fields of Eq.\
\eqref{eq:varfield}. On the basis of equal time (anti)commutation relations
\be\bsp
  \big[\phi^a(t,\mathbf{x}),\phi^b(t,\mathbf{y})\big] = 
  \big[\Pi_a(t,\mathbf{x}),\Pi_b(t,\mathbf{y})\big]= 
  \big\{\psi^i(t,\mathbf{x}),\psi^j(t,\mathbf{y})\big\} = 
  \big\{\rho_i(t,\mathbf{x}),\rho_j(t,\mathbf{y})\big\}= 0 \ , \\ 
  \big[\phi^a(t,\mathbf{x}),\Pi_b(t,\mathbf{y})\big] = i \delta^3(\mathbf{x} -
    \mathbf{y}) \delta^a{}_b \qquad \text{and} \qquad
  \big\{\psi^i(t,\mathbf{x}),\rho_j(t,\mathbf{y})\big\} = i \delta^3(\mathbf{x}
   - \mathbf{y}) \delta^i{}_j \ ,
\esp \ee 
and canonical quantization which prescribes commutators for bosonic
operators and anticommutators for fermionic operators, we now show that the
algebra spanned by the bosonic symmetry charges is a Lie algebra. The
combination of two symmetry operations, given by the commutator of the
associated charges, indeed reads
\be
  \big[B_A,B_B\big] = -i\int \d^3 x \Big(
    \Pi_a \big[B^1_A, B^1_B\big]^a{}_b \phi^b + 
    \rho_i\, \big[B^2_A, B^2_B\big]^i{}_j \psi^j\Big)\ .
\ee
Imposing the algebra to close enforces the relations 
\be
  \big[B_A^1, B_B^1 \big] = i f_{AB} {}^C B_C^1 \quad\text{and}\quad
  \big[B_A^2, B_B^2 \big] = i f_{AB} {}^C B_C^2\ ,
\label{eq:salgBBold}\ee 
where we have introduced the real constants $f_{AB}{}^C$, identical
for both the bosonic and fermionic sectors. The explicit
factors of $i$ are conventional, those normalizations being the ones
traditionally employed in particle physics. Eq.~\eqref{eq:salgBBold}
consequently leads to
\be
   \big[B_A, B_B \big] = i f_{AB} {}^C B_C \ .
\label{eq:salgBB}\ee
Since in addition, the Jacobi identities are verified due to the associativity
of the matrix product applied to the $B_A^1$ and $B_A^2$ matrices,
\be\label{eq:J1}
  \Big[B_A,\big[B_B,B_C\big]\Big] + 
  \Big[B_B,\big[B_C,B_A\big]\Big] + 
  \Big[B_C,\big[B_A,B_B\big]\Big] = 0 \ ,
\ee
this achieves to prove that the symmetry charges $B$ span a Lie algebra.

In 1967, Coleman and Mandula have proved that the structure of the symmetry 
group of the theory can only be expressed under the form of a direct product of
the Poincar\'e group and an internal symmetry group, $G\equiv ISO(1,3) \times
G_{\rm int}$ where $G_{\rm int}$ is a compact Lie group\footnote{For
massless theories, the Poincar\'e group can be enlarged by the conformal
group.}~\cite{Coleman:1967ad}.
In their proof, they have considered a relativistic quantum
field theory with a discrete spectrum of massive one-particle states where all
the symmetry generators are Lorentz-scalar quantities. In addition, the
$S$-matrix is
assumed non-trivial and the group $G$ contains, by definition, a subgroup
isomorphic to the Poincar\'e group. Applying this theorem to our toy theory, 
the generic $B$-charges introduced above can be split into two categories,
the generators of the Poincar\'e algebra (the four-momentum operator $P^\mu$ and
the Lorentz generators $M^{\mu\nu}$) and a set of generators for the internal
symmetry group which we denote generically by $T_a$. They fulfill a Lie algebra
which reads,
\be\bsp
  \big[M^{\mu\nu}, M^{\rho\sigma}\big] =&\ -i \big( \eta^{\nu\sigma} M^{\rho\mu} 
     - \eta^{\mu\sigma} M^{\rho\nu} + \eta^{\nu\rho} M^{\mu\sigma} - 
     \eta^{\mu\rho} M^{\nu\sigma}\big) \ ,  \\
  \big[M^{\mu \nu }, P^\rho \big] =&\ -i \big(\eta^{\nu\rho} P^\mu - 
     \eta^{\mu\rho} P^\nu\big)\ ,\\
  \big[T_a, T_b \big] =&\ i f_{ab}{}^c T_c \ , \\
  \big[P^\mu,P^\nu\big] =&\ \big[T_a,P^\mu\big] = \big[T_a,M^{\mu\nu}\big]=0 \ .
\esp\label{eq:PoincareAlg}\ee
In this set of equations, the Minkowski metric $\eta^{\mu\nu}$ is given by
$\text{diag} (1,-1,-1,-1)$ and $f_{ab}{}^c$ denote the structure constants of
the Lie algebra associated with the internal symmetry generators.
The last two vanishing commutators directly illustrate the Coleman-Mandula
theorem, since they show that the internal and external symmetry groups are
decoupled, \ie, the related symmetry operations commute with each other.

\subsection{Lie superalgebra}\label{sec:liesuperalg}
In the setup of Section \ref{sec:coleman}, it is assumed that all symmetry
generators are invariant under Lorentz transformations. Therefore, the spin of
the particles cannot be
modified by a symmetry operation. A way to bypass the Coleman-Mandula
theorem is to relax this constrain and allow for both fermionic and bosonic
symmetry generators. One extends the symmetry group of the toy theory built
in Section \ref{sec:coleman} by supplementing to the bosonic generators $B_A^1$
and $B_A^2$ the fermionic generators $F_I^1$ and $F_I^2$. The latter act on
bosonic and fermionic fields, respectively,
\be
  \phi^a \to \phi^a + \delta_I \phi^a = \phi^a + (F^1_I)^a{}_i\ \psi^i 
\qquad\text{and}\qquad
  \psi^i \to \psi^i + \delta_I \psi^i = \psi^i + (F^2_I)^i{}_a\ \phi^a \ ,
\ee
so that the fermionic or bosonic nature of the fields is now modified
by symmetry operations. From N\oe ther theorem,
one can express a fermionic charge $F_I$ in terms of the momentum densities and
the fields, as for the bosonic case in Eq.\ \eqref{eq:Bcharge}, 
\be
  F_I = -i \int \d^3 x 
        \Big[\Pi_a (F^1_I)^a{}_i\ \psi^i + \rho_i (F^2_I)^i{}_a\ \phi^a\Big] \ .
\ee

In order to derive the algebra spanned by the fermionic charges, we recall that
the 
combination of two fermionic operators through an anticommutation relation, as
prescribed by canonical quantization, leads to a bosonic operation. Therefore, 
one naturally asks the algebra of the fermionic charges to close in terms of the
bosonic ones. Since the anticommutator $\{F_I,F_J\}$ can be written as
\be
  \big\{F_I,F_J\big\}  = -i \int \d^3 x \Big(
     \Pi_a (F^1_I F^2_J + F^1_J F^2_I)^a{}_b\ \phi^b +
     \rho_j (F^2_I F^1_J + F^2_J F^1_I)^j{}_i\ \psi^i\Big) \ ,
\ee 
when employing the relations of Eq.\ \eqref{eq:varfield},
imposing the closure of the algebra implies that
\be
   F^1_I F^2_J + F^1_J F^2_I  = Q_{IJ} {}^C B_C^1 
   \quad\text{and}\quad 
   F^2_I F^1_J + F^2_J F^1_I = Q_{IJ} {}^C B_C^2 \ .
\ee
We have introduced a second set of real constants $Q_{IJ}{}^C$ which must again
identical for the bosonic and fermionic sectors of the theory so that we
eventually get
\be    
    \big\{F_I, F_J\big\} = Q_{IJ} {}^A B_A \ .
\label{eq:salgFF}\ee

We now turn to the combination of fermionic and bosonic symmetry operations.
Since the composition of a bosonic and a fermionic operator leads to an
operation of a fermionic nature, one computes the commutator 
\be
  \big[B_A,F_I\big] = -i \int \d^3 x \Big(
    \Pi_a (B^1_A F^1_I - F^1_I B^2_A)^a{}_k\ \psi^k 
     + \rho_k (B^2_A F^2_I - F^2_I B^1_A)^k{}_b\ \phi^b\Big) \ ,
\ee
using Eq.\ \eqref{eq:varfield}, and requires the algebra to close on the
fermionic charges. This enforces the properties 
\be
   B^1_A F^1_I - F^1_I B^2_A  = i R_{AI} {}^J\ F_J^1 
   \quad\text{and}\quad
    B^2_A F^2_I - F^2_I B^1_A = i R_{AI} {}^J\ F_J^2 \ ,
\ee 
where we have introduced a last set of real constants $R_{AI}{}^J$ that is once
again
identical for the two sectors of the theory, so that
\be
  \big[B_A,F_I\big] = i R_{AI} {}^J F_J \ .
\label{eq:salgFB}\ee

The three relations of Eq.\ \eqref{eq:salgBB}, Eq.\ \eqref{eq:salgFF} and Eq.\
\eqref{eq:salgFB} show, together with the Jacobi identities of Eq.\
\eqref{eq:J1} and
\be\label{eq:J2}\bsp
  \big[B_i,[B_j,F_a]\big] + \big[B_j,[F_a,B_i]\big] + \big[F_a,[B_i,B_j]\big]=0 \ ,\\
  \big[B_i,\{F_a,F_b\}\big] - \big\{[B_i,F_a],F_b\big\} -
    \big\{F_a,[B_i,F_b]\big\} = 0 \ ,\\
  \big[F_a,\{F_b,F_c\}\big] - \big[\{F_a,F_b\},F_c\big] + 
    \big[F_b,\{F_a,F_c\}\big]=0\ ,
\esp\ee
that the $F$-charges and $B$-charges fulfill a structure of Lie superalgebra. As
for Eq.\ \eqref{eq:J1}, the three additional Jacobi identities of Eq.\
\eqref{eq:J2} are naturally verified due to the associativity of the matrix
product applied to the matrices $B_A^1$, $B_A^2$, $F_I^1$ and $F_I^2$.

\subsection{The $N=1$ Poincar\'e superalgebra}
Fermionic symmetries such as those described by the $F$-charges in Section
\ref{sec:liesuperalg} allow to bypass the Coleman-Mandula theorem since some of
the generators of the symmetry group are not invariant under Lorentz
transformations~\cite{Golfand:1971iw}. In this case, the
most general superalgebra admissible for an interacting quantum field theory is
the $N-$extended Poincar\'e superalgebra with $N\leq 8$, as shown by
Haag, Lopuszanski and
Sohnius when they have extended the results of Coleman and Mandula to the
supersymmetric case \cite{Haag:1974qh}.  

The Poincar\'e superalgebra consists of a $\mathbb{Z}_2$-graded vectorial space
$\g= \g_0 \oplus \g_1$. The operators of $\g_0$ are all bosonic
and those in $\g_1$ are all fermionic. In Section
\ref{sec:coleman}, and in particular in Eq.\ \eqref{eq:salgBB}, we have proved
that the bosonic operators span a Lie algebra. The Coleman-Mandula theorem
further indicates that this Lie algebra is the direct product of the Poincar\'e
algebra and an internal algebra denoted by $\g_{\rm int}$,
\be
  \g_0 = \mathfrak{iso}(1,3) \times \g_{\rm int} \ ,
\ee
as given by Eq.\ \eqref{eq:PoincareAlg}.

The fermionic sector of the superalgebra $\g_1$ contains operators lying in a
non-trivial representation of $\g_0$ that are hence 
non-scalar with respect to the Lorentz group. Furthermore, we construct the fermionic
sector of the superalgebra $\g_1$ in a minimal way, using a set of $N$
Majorana spinors dubbed supercharges,
\be
  \g_1 = \big\{Q_\alpha^I, \alpha=1,2 \big\} \oplus 
   \big\{\Qbar^\alphadot_I; \alphadot=\dot{1},\dot{2}\big\} 
    \qquad\text{with}\qquad I=1,2,\ldots,N \ ,
\ee
referring to Appendix \ref{app:conv} for our conventions on spinor
indices. In this work, we focus on the simplest supersymmetric theories and
therefore restrict ourselves to the case $N=1$,
although in the general case, one can have up to 
$N=8$ supercharges.

We first derive the commutator obtained when combining the supercharge
$Q_\alpha$ with the
generators of the Lorentz group. Since $Q_\alpha$ is a left-handed Weyl spinor,
it transforms under the action of the generator of the Lorentz group as
\be
  Q_\alpha \to \exp\Big[ - \frac{i}{2} \omega_{\mu\nu} \sigma^{\mu\nu}
    \Big]_\alpha{}^\beta Q_\beta \ ,
\ee
where $\omega_{\mu\nu}$ are the transformation parameters and $\sigma^{\mu\nu}$ 
the generators of
the Lorentz algebra in the left-handed spinorial representation defined as
in Eq. \eqref{eq:simunu}. Considering the supercharge $Q_\alpha$ as an operator,
one also has the transformation law
\be\label{eq:QM}
  Q_\alpha \to \exp\Big[ \frac{i}{2} \omega_{\mu\nu} M^{\mu\nu}
    \Big]\ Q_\alpha\ \exp\Big[ -\frac{i}{2} \omega_{\rho\sigma} M^{\rho\sigma}
    \Big] \ ,
\ee
where $M^{\mu\nu}$ are Lorentz transformation operators. 
Combining these last two equations and performing an expansion to the first
order in the $\omega_{\mu\nu}$ parameters yield
\be
  \big[Q_\alpha, M^{\mu\nu}\big] = \sigma^{\mu \nu}{}_\alpha{}^\beta\ Q_\beta \ .
\ee
The right-handed supercharge $\Qbar^\alphadot$ being in the right-handed
spinorial representation of the Lorentz algebra, one similarly derives
\be\label{eq:QbarM}
  \big[\Qbar^\alphadot, M^{\mu\nu}\big] =
    \sibar^{\mu\nu}{}^\alphadot{}_\betadot\ \Qbar^\betadot \ .
\ee

In order to calculate the commutators of the four-momentum operator $P^\mu$ with
the supercharges $Q_\alpha$ and $\Qbar_\alphadot$, we recall that the latter
are respectively in the $(\utilde{\bf 2},\utilde{\bf 1})$ and $(\utilde{\bf
1},\utilde{\bf 2})$ representations of the Lorentz group\footnote{Investigating
the representations of the Lorentz algebra is equivalent to studying the
finite-dimensional representations of
$\mathfrak{sl}(2,\mathbb{C}) \oplus \mathfrak{sl}(2,\mathbb{C})$, the latter
being in one-to-one correspondance with the finite-dimensional representations of
$\mathfrak{so}(3) \oplus \mathfrak{so}(3)$.
Therefore, we denote, in our notations, the representation of any object under
the Lorentz group as $(\utilde{S},\utilde{S}')$ where $\utilde{S}$ and
$\utilde{S}'$ stand for the representations of the considered object under each
of the $\mathfrak{so}(3)$ algebra. We also employ the conventions of $S=2s+1$
and $S'=2s'+1$ so that the quantity $s+s'$ equals to the spin of the object
under consideration.}, as shown by Eq.\ \eqref{eq:QM} and Eq.\ \eqref{eq:QbarM}.
Moreover, the operator $P^\mu$ lies in the vectorial representation $(\utilde{\bf
2},\utilde{\bf 2})$, so that
\be
  (\utilde{\bf 2},\utilde{\bf 2}) \otimes (\utilde{\bf 2},\utilde{\bf 1}) =
    (\utilde{\bf 1},\utilde{\bf 2}) \oplus (\utilde{\bf 3},\utilde{\bf 2})
  \quad\text{and}\quad
  (\utilde{\bf 2},\utilde{\bf 2}) \otimes (\utilde{\bf 1},\utilde{\bf 2}) =
    (\utilde{\bf 2},\utilde{\bf 1}) \oplus (\utilde{\bf 2},\utilde{\bf 3}) \ .
\ee
Since there is no generator of the complete symmetry group 
in the $(\utilde{\bf 2},\utilde{\bf 3})$ and $(\utilde{\bf 3},\utilde{\bf 2})$
representations of the Lorentz algebra, the only natural expressions that can be
written for the commutators $[P_\mu, Q_\alpha]$ and $[P_\mu, \Qbar^\alphadot]$
read, {\it a priori},
\be\label{eq:PQone}
  \big[P_\mu, Q_\alpha\big] = a\ \sigma_\mu{}_{\alpha\alphadot} \Qbar^\alphadot
  \quad\text{and}\quad 
  \big[P_\mu, \Qbar^\alphadot]= b\ \sibar_\mu{}^{\alphadot\alpha} Q_\alpha \ ,
\ee
after introducing the appropriate index structure by means of the Pauli
matrices and where $a$ and $b $ are constants to be determined. From the first
Jacobi identity of Eq.\ \eqref{eq:J2}, one gets
\be
  0= \Big[P_\mu,\big[P_\nu,Q_\alpha\big]\Big] + 
     \Big[P_\nu,\big[Q_\alpha,P_\mu\big]\Big] +
     \Big[Q_\alpha,\big[P_\mu,P_\nu\big]\Big] 
   = 4 ab\ \sigma_{\nu \mu}{}_\alpha{}^\beta Q_\beta \ ,
\ee
after employing Eq.\ \eqref{eq:PoincareAlg}, Eq.\ \eqref{eq:PQone} and the
definition
of the $\sigma^{\mu\nu}$ matrices given in Eq.~\eqref{eq:simunu}. Since the two
supercharges are adjoint operators, fixing $a=0$ ensures $b=0$. Therefore, one
gets
\be\label{eq:PQ}
  \big[P_\mu, Q_\alpha\big] = 
  \big[P_\mu, \Qbar^\alphadot]= 0 \ . 
\ee

We now turn to the fermionic sector of the superalgebra and compute the
anticommutator of two supercharges. The structure of an operator resulting
from the direct combination of two Weyl spinorial
operators can be again deduced from group theory arguments, since%
\renewcommand{\arraystretch}{1.2}%
\be\begin{array}{c}
  \quad
  (\utilde{\bf 2},\utilde{\bf 1}) \otimes (\utilde{\bf 1},\utilde{\bf 2})=
    (\utilde{\bf 2},\utilde{\bf 2}) \ , \\
  (\utilde{\bf 2},\utilde{\bf 1}) \otimes (\utilde{\bf 2},\utilde{\bf 1}) =
    (\utilde{\bf 3},\utilde{\bf 1}) \oplus (\utilde{\bf 1},\utilde{\bf 1})\ ,
  \qquad 
  (\utilde{\bf 1},\utilde{\bf 2}) \otimes (\utilde{\bf 1},\utilde{\bf 2}) =
    (\utilde{\bf 1},\utilde{\bf 3}) \oplus (\utilde{\bf 1},\utilde{\bf 1}) \ .
\end{array}\ee
\renewcommand{\arraystretch}{1}%
On the basis of the self-duality properties of the Pauli matrices of Eq.\
\eqref{eq:dual},
one observes that $\sigma^{\mu \nu} M_{\mu \nu}$ and $\sibar^{\mu \nu} M_{\mu
\nu}$ lie in the $(\utilde{\bf 3},\utilde{\bf 1})$ and $(\utilde{\bf
1},\utilde{\bf 3})$ representations of the Lorentz group, respectively. 
Therefore, the only possible structure for the three
anticommutators of the supercharges is given by
\be
  \big\{Q_\alpha,\Qbar_\alphadot\big\} = c\ \sigma^\mu{}_{\alpha\alphadot}
    P_\mu \ ,  \quad
  \big\{Q_\alpha, Q_\beta\big\} = d\ \sigma^{\mu \nu}{}_{\alpha\beta} M_{\mu
    \nu} \ ,  \quad
  \big\{\Qbar_\alphadot, \Qbar_\betadot\big\} = e\ \sibar^{\mu
    \nu}{}_{\alphadot\betadot} M_{\mu \nu} \ , 
\label{eq:QQ}\ee
where $c$, $d$ and $e$ are constants that can be obtained from Jacobi
identities. Applying the second relation of Eq.\ \eqref{eq:J2} to the
operators $Q_\alpha$, $Q_\beta$ and $P^\mu$, one finds $d=0$. Similarly, it can
be shown that $e=0$. Furthermore, $\big\{Q_1, \Qbar_{\dot 1} \big\} +
\big\{Q_2, \Qbar_{\dot 2} \big\}=\big\{Q_1,  Q^\dag_1 \big\} +
\big\{Q_2,  Q^\dag_2 \big\}$ is a unitary and positively defined operator.
Therefore, one
finds that the constant $c$  has to be positive, since the energy operator $P^0$
is positively defined. By conventions, we set $c=2$. 

In addition to all the relations derived so far, we impose that the supercharges
are singlet under the internal symmetry group,
\be\label{eq:TQ}
  \big[Q_\alpha,T_a] =  \big[\Qbar_\alphadot,T_a\big]= 0 \ .
\ee
This choice is however not the most general one. Inspecting the relation of Eq.\
\eqref{eq:QQ} (with $c=2$ and $d=e=0$), one observes that it admits a $U(1)$
group as an automorphism group. The two supercharges being Hermitian conjugate
of each other, they have consequently opposite $U(1)$ quantum numbers and
Eq.\ \eqref{eq:TQ} can therefore be generalized to
\be\label{eq:RQ}
  \big[Q_\alpha,R] =  Q_\alpha \quad\text{and}\quad 
   \big[\Qbar_\alphadot,R\big]= -\Qbar_\alphadot \ ,
\ee
where the operator $R$ stands for the generator of the automorphism group of the 
Poincar\'e superalgebra. Such a symmetry is commonly known as the $R$-symmetry
embedded in the $N=1$ Poincar\'e superalgebra.

Collecting the results of Eq.\ \eqref{eq:PoincareAlg}, Eq.\
\eqref{eq:QM}, Eq.\ \eqref{eq:QbarM}, Eq.\ \eqref{eq:PQ}, Eq.\ \eqref{eq:QQ},
Eq.\ \eqref{eq:TQ} and Eq.\ \eqref{eq:RQ}, 
the $N=1$ Poincar\'e superalgebra is finally given by %
\renewcommand{\arraystretch}{1.3}%
\be\boxed{\bsp
  \big[M^{\mu\nu}, M^{\rho\sigma}\big] =&\ -i \big( \eta^{\nu\sigma} M^{\rho\mu} 
     - \eta^{\mu\sigma} M^{\rho\nu} + \eta^{\nu\rho} M^{\mu\sigma} - 
     \eta^{\mu\rho} M^{\nu\sigma}\big) \ , \\
  \big[M^{\mu \nu }, P^\rho \big] =&\ -i \big(\eta^{\nu\rho} P^\mu - 
     \eta^{\mu\rho} P^\nu\big)\ , \\
  \big[T_a, T_b \big] =&\ i f_{ab}{}^c T_c \ , \\
  \big[Q_\alpha, M^{\mu\nu}\big] =&\ \sigma^{\mu \nu}{}_\alpha{}^\beta\ Q_\beta \
     , \qquad \big[\Qbar^\alphadot, M^{\mu\nu}\big] =
    \sibar^{\mu\nu}{}^\alphadot{}_\betadot\ \Qbar^\betadot \ , \\
  \big\{Q_\alpha,\Qbar_\alphadot\big\} =&\ 2 \sigma^\mu{}_{\alpha\alphadot}
    P_\mu \ , \\
  \big[Q_\alpha,R] =&\  Q_\alpha \ , \qquad 
   \big[\Qbar_\alphadot,R\big]= -\Qbar_\alphadot \ , \\
  \big[P^\mu,P^\nu\big] =&\ \big[P_\mu, Q_\alpha\big] = \big[P_\mu,
    \Qbar^\alphadot] = \big\{Q_\alpha, Q_\beta\big\} = \big\{\Qbar_\alphadot,
    \Qbar_\betadot\big\} =0 \ , \\
  \big[T_a,P^\mu\big] =&\ \big[T_a,M^{\mu\nu}\big]= 
  \big[Q_\alpha,T_a] =  \big[\Qbar_\alphadot, T_a\big]= 0 \ .
\esp}\label{eq:poincaresalg}  
\ee%
\renewcommand{\arraystretch}{1.}%

\mysection{Representations of the $N=1$ Poincar\'e superalgebra}
\subsection{Representations of the Poincar\'e algebra}
\label{sec:palgrep}
In order to build any ($N=1$) supersymmetric quantum field theory, it is
necessary to use representations of the Poincar\'e superalgebra
summarized in Eq.\ \eqref{eq:poincaresalg}.
In the following subsections, we address the derivation of these
representations, first in the massless case, then in the massive one. Such
representations have been originally derived in the works of
Refs.~\cite{Salam:1976ib, Nahm:1977tg, Ferrara:1980ra}. 
Before moving on, let us however recall some basic properties of the Poincar\'e algebra
employed for building non-supersymmetric quantum field theories.

We define multiplet states representing the particle content of a theory by
the eigenvalues of the Casimir operators associated with the algebra
under consideration. In the
case of the Poincar\'e algebra, one has one quadratic Casimir operator ${\cal
C}_2$ and one quartic operator ${\cal C}_4$,
\be
  {\cal C}_2 = P^\mu P_\mu \qquad\text{and}\qquad 
  {\cal C}_4 = W^\mu W_\mu \ , 
\label{eq:Casimir}\ee
where $W_\mu$ is the Pauli-Lubanski operator,
\be
  W_\mu = \frac12 \e_{\mu\nu\rho\sigma}  P^\nu M^{\rho\sigma} \  .
\ee
This can be proved as follows. Since the four-momentum operator $P^\mu$ commutes
with itself, as presented in the relations of Eq.\ \eqref{eq:PoincareAlg}, and since 
\be
  \big[W_\mu,P^{\tilde \mu}\big] = 
    \frac12\e_{\mu\nu\rho\sigma} P^\nu [M^{\rho\sigma}, P^{\tilde\mu}] =
    \frac12\e_{\mu\nu\rho\sigma} P^\nu (\eta^{\sigma \tilde\mu} P^\rho
      -\eta^{\rho \tilde\mu} P^\sigma)=0 \ , 
\ee
the operators ${\cal C}_2$ and ${\cal C}_4$ commute with the four-momentum
operator. In
the derivation of this last equation, we have used the fact that
$\e_{\mu\nu\rho\sigma}P^\nu P^\rho$ vanishes due to symmetry properties under
the exchange of the Lorentz indices. Moreover, both $P_\mu P^\mu$ and $W_\mu
W^\mu$ are Lorentz scalar quantities and they therefore commute with all the
generators of
the Lorentz group. Consequently, ${\cal C}_2$ and ${\cal C}_4$ are indeed
Casimir operators of the Poincar\'e algebra since they commute with
all its generators.

Since in addition, both Casimir operators commute with the generators of the
internal symmetry algebra $\g_{\rm int}$, all the members of an irreducible
multiplet of $\g_{\rm int}$ have the same mass and spin. They can subsequently
be labeled with at minimum two quantum numbers $|m,\omega,\ldots\rangle$ where
$m^2$ and $m^2 \omega(\omega+1)$ are the eigenvalues of the operators ${\cal
C}_2$ and ${\cal C}_4$. In the massless case, both
Casimir eigenvalues vanish and cannot thus be employed for characterizing the states
(see below).
In these notations, the dots stand for extra quantum numbers
related to operators that commute with both Casimir operators, such as 
the eigenvalue $p^\mu$ of the four-momentum operator $P^\mu$ or the eigenvalues of the
generators of the little algebra associated with the Lorentz algebra, \ie, the
subalgebra of the Lorentz algebra whose the generators fix $p^\mu$.

\subsubsection{Massless particles}

In the so-called standard frame, the four-momentum of a massless state can be
written as
\be\label{eq:pmassless}
  p^\mu = \bpm E \\0\\0\\E\epm \ ,
\ee 
where $E$ is an arbitrary positive real number. In order to derive the
representations of the Poincar\'e algebra for massless
particles, one must first work out the structure of the associated little algebra. Under a
finite transformation, $p^\mu$ transforms as
\be
  p^\mu \to \Big(\exp
    \Big[-\frac{i}{2}\omega_{\nu\rho} M^{\nu\rho}\Big]\Big)^\mu\Big._\sigma
   p^\sigma \ ,
\ee
which must be read as $p^\mu\to p^\mu$ in the case of the little group. At the
operator level, this becomes
\be
  P^{\tilde\mu} \to \exp\Big[ \frac{i}{2} \omega_{\mu\nu} M^{\mu\nu}\Big]\
    P^{\tilde\mu} \exp\Big[ -\frac{i}{2} \omega_{\rho\sigma} M^{\rho\sigma} \Big] =
    P^{\tilde \mu} \ .
\ee
Applying this relation to a state which the eigenvalue of the
four-momentum operator is given by Eq.\ \eqref{eq:pmassless}, one deduces that
the generators of the little algebra are
\be
  M \equiv M^{12} \ , \qquad T_1 \equiv M^{10} - M^{13} \qquad\text{and}\qquad
  T_2 \equiv M^{20} - M^{23} \ .
\ee
These consist of the generators of the algebra of the rotations and
translations in two
dimensions, $\mathfrak{iso}(2)$. To avoid the continuous degrees of freedom
yielded by the translation operators $T_1$ and $T_2$, the related eigenvalues
are set to zero so that we are left with one single generator $M$ whose eigenvalue
$\lambda$ is called the helicity. Moreover, it can be shown that the quantum numbers
associated with the two Casimir operators are vanishing, so that massless states are
labeled by $|0,0; p^\mu, \lambda\rangle$, the four-momentum being given by Eq.\
\eqref{eq:pmassless}.

The study of the representations of the Lorentz algebra
ensures that the allowed values for $\lambda$ are either integer or
half-odd-integer (see, \eg, Ref.\ \cite{weinbergqft}). Moreover, the $CPT$
theorem implies that a state with a non-vanishing helicity 
$-\lambda$ is always supplemented by a state with an helicity
$\lambda$.

\subsubsection{Massive particles}
The standard frame for massive particle consists of its rest frame, so that the
eigenvalue of the four-momentum operator reads
\be\label{eq:pmassive}
  p^\mu = \bpm m \\0\\0\\0\epm \ .
\ee 
The constant $m$ is a real positive number related to the eigenvalue of the
Casimir operator ${\cal C}_2$, which is in this case 
$m^2$. From Eq.\ \eqref{eq:pmassive}, one deduces that the little algebra is
$\mathfrak{so}(3)$, the algebra of the rotations in three dimensions whose the
generators leave $p^\mu$ invariant. Since the quartic 
Casimir operator can be expressed
as ${\cal C}_4 = m^2 \mathbf M^2$ where $\mathbf M^2$ is the quadratic Casimir
operator of the rotation algebra, the massive representations of the Poincar\'e
algebra are labeled as $|m,j; p^\mu, j_3\rangle$, where the
four-momentum is given as in Eq.\ \eqref{eq:pmassive}, the eigenvalues of
$\mathbf M^2$ are $j(j+1)$ and $j_3$ denotes the eigenvalue of the third
rotation generator $M_3$.

\subsection{Representations of the Poincar\'e superalgebra: general features}
\label{sec:repr_general}
Turning now to the representations of the $N=1$ Poincar\'e superalgebra, one
first observes that among the two Casimir operators of Eq.\
\eqref{eq:Casimir}, ${\cal C}_2$ is still a good Casimir operator. Therefore,
all the states of a specific irreducible multiplet of the Poincar\'e
superalgebra, also called a supermultiplet, have the same mass. In contrast,
the quartic Casimir operator ${\cal C}_4$ is not commuting with the
supercharges so that the different components of a supermultiplet can therefore have
different spins. In order to label
the irreducible representations of the Poincar\'e superalgebra,
the second Casimir operator must be generalized to a new
operator commuting with the supercharges,
\be
  {\cal C}_4 = \Big(\tW_\mu P_\nu - \tW_\nu P_\mu\Big)
  \Big(\tW^\mu P^\nu - \tW^\nu P^\mu\Big)
  \qquad\text{with}\qquad
  \tW_\mu = W_\mu - \frac14 \Qbar\sibar_\mu Q \ ,
\ee
the operator $\tW_\mu$ being the supersymmetric counterpart of the
Pauli-Lubanski operator.

As a general feature of any supermultiplet, the
number of fermionic degrees of freedom equals the number of bosonic ones. This
is shown by introducing the fermion number operator $(-)^N$ which returns an
eigenvalue of $+1$ when acting on a bosonic state and $-1$ when acting on a
fermionic state. We have
\be
  (-1)^N \ Q_\alpha= - Q_\alpha(-1)^N \ , 
\ee
so that
\be
  \text{Tr} \Big[(-1)^N \big\{Q_\alpha, \Qbar_\alphadot \big\}\Big] =
  \text{Tr} \Big[-Q_\alpha (-1)^N \Qbar_\alphadot + (-1)^N \Qbar_\alphadot Q_\alpha
    \Big] = 0 \ ,
\ee
by cyclicity of the trace. However, using the superalgebra relations presented
in Eq.\ \eqref{eq:poincaresalg}, we also have
\be
  \text{Tr} \Big[(-1)^N \big\{Q_\alpha, \Qbar_\alphadot \big\}\Big] =
   2 \sigma^\mu{}_{\alpha\alphadot} \text{Tr} \Big[(-1)^N P_\mu \Big] \ .
\ee
Consequently, the trace of the fermionic operator $(-1)^N$ vanishes, which implies
an equal number of bosonic and fermionic degrees of freedom in each
supermultiplet.

\subsection{Massless representations of the Poincar\'e superalgebra}
\label{sec:masslesssupermul}
For massless representations of
the Poincar\'e superalgebra, the four-momentum in the standard frame is
given by Eq.\ \eqref{eq:pmassless}. One can show that both Casimir operators
vanish, as for the Poincar\'e algebra. In order
to derive all the quantum numbers labeling an irreducible supermultiplet, one
must derive the structure of the little algebra. In addition to $M=M^{12}$ (see
Section \ref{sec:palgrep}), the generators of the little algebra now also include
the two supercharges $Q_\alpha$ and $\Qbar_\alphadot$. The little algebra is
then deduced from Eq.\ \eqref{eq:poincaresalg} and reads 
\be\label{eq:littlemassless}\bsp
  \big[M, Q_1\big] = \frac12 Q_1\ , \quad
  \big[M, \Qbar_{\dot 1}\big] = & -\frac12 \Qbar_{\dot 1}\ ,\quad
  \big[M, Q_2\big] = -\frac12 Q_2\ ,\quad
  \big[M, \Qbar_{\dot 2}\big]= \frac12 \Qbar_{\dot 2}\ ,\\
  \big\{Q_\alpha, \Qbar_\alphadot\big\} =&\ 2 E
    (\sigma^0+\sigma^3)_{\alpha\alphadot} = 4 E \bpm 1& 0\\ 0 & 0 \epm\ .
\esp\ee
From the last relation, it can be seen that if we consider a vacuum state
$|\Omega\rangle$ annihilated by $Q_2$, one gets
\be
  0= \langle \Omega| \{ Q_2, \Qbar_{\dot 2}\}|\Omega\rangle =
     \langle \Omega| Q_2 \Qbar_{\dot 2} |\Omega\rangle = 
     \big|\big| \Qbar_{\dot 2} |\Omega\rangle \big|\big|^2 \ .
\ee
As a consequence, the two operators $Q_2$ and $\Qbar_{\dot 2}$ are vanishing by
unitarity and we
only have two active supercharges $Q_1$ and $\Qbar_{\dot 1}$ from which
we can construct the creation and annihilation operators
\be
  a^\dag = \frac{1}{2\sqrt{E}} \Qbar_{\dot 1} \qquad\text{and}\qquad
  a = \frac{1}{2\sqrt{E}} Q_1
\ee
fulfilling standard anticommutation relations
\be
  \big\{a,a\big\} = \big\{a^\dag,a^\dag\big\} = 0 \qquad\text{and}\qquad
  \big\{a,a^\dag\big\} = 1 \ .
\ee
Considering a state $|0,0; p^\mu,\lambda\rangle$, one observes that
\be\bsp
  M a |0,0; p^\mu,\lambda\rangle =&\ \big( a M + [M,a] \big) |0,0;
    p^\mu,\lambda\rangle = \big( \lambda + \frac12 \big) a |0,0;
    p^\mu,\lambda\rangle \ , \\
  M a^\dag |0,0; p^\mu,\lambda\rangle =&\ \big( a^\dag M + [M,a^\dag] \big) |0,0;
    p^\mu,\lambda\rangle = \big( \lambda - \frac12 \big) a^\dag |0,0;
    p^\mu,\lambda\rangle  \ ,
\esp\ee
by using the little algebra relations of Eq.\ \eqref{eq:littlemassless}. As a
consequence,
the creation operator $a^\dag$ raises the helicity of the state by half a unit
and the annihilation operator $a$ reduces it by half a unit.

The field content of a specific supermultiplet is obtained by starting from a
vacuum state
$|\Omega_\lambda \rangle$ corresponding to the state with the highest helicity
$\lambda$. This state is annihilated both by the operators $a$ and
$a^\dag a^\dag$, since $a^\dag$ is a fermion (which implies $a^\dag a^\dag = 0$). 
The degrees of freedom included in
a given supermultiplet consist thus of one single state of helicity $\lambda$
related to the vacuum state and another state of helicity $\lambda-1/2$ obtained
after applying the creation operator
$a^\dag$ to the vacuum state. 

In addition, each representation is required to be $CPT$-conjugate. Therefore,
special care must be taken when a state with a non-vanishing helicity
$\lambda_1$ is present within a supermultiplet. In the case
the $CPT$-conjugate state of helicity $-\lambda_1$ is absent, one needs to
include the degrees of freedom that are derived when repeating the above procedure
but starting from the conjugate
vacuum state $|\Omega_{-\lambda}\rangle$. This leads to the doubling
of the degrees of freedom so that the field content of a
$N=1$ supermultiplet now consists of states of helicities equal to $\pm
\lambda$ and $\pm\lambda\mp1/2$. 

Assuming the highest possible helicity being 1/2, 1 and 2, one defines the
so-called matter, gauge and gravity supermultiplets, respectively. Their
field content in terms of helicity states is given by
\renewcommand{\arraystretch}{1.1}%
\bcen\begin{tabular}{c|cc|cc|cc}
\multirow{2}{*}{Helicity} & 
  \multicolumn{2}{c}{Matter} &
  \multicolumn{2}{|c}{Gauge} & 
  \multicolumn{2}{|c}{Gravitation}\\
& $|\Omega_{1/2}\rangle$ & $|\Omega_{-1/2}\rangle$ & 
  $|\Omega_1\rangle$ & $|\Omega_{-1} \rangle$ &
  $|\Omega_2\rangle$ & $|\Omega_{-2} \rangle$\\\hline
$\phantom{-}2$       &   &   &   &    & 1 &   \\
$\phantom{-}\frac32$ &   &   &   &    & 1 &   \\
$\phantom{-}1$       &   &   & 1 &    &   &   \\
$\phantom{-}\frac12$ & 1 &   & 1 &    &   &   \\
$\phantom{-}0$       & 1 & 1 &   &    &   &   \\
$-\frac12$           &   & 1 &   & 1  &   &   \\
$-1$                 &   &   &   & 1  &   &   \\
$-\frac32$           &   &   &   &    &   & 1 \\
$-2$                 &   &   &   &    &   & 1 \\
\end{tabular}\ecen
\renewcommand{\arraystretch}{1.0}%
One observes that a $N=1$ massless matter supermultiplet contains two real
scalar degrees of freedom, \ie, one complex scalar field, and one Weyl fermion.
In contrast, the field content of a gauge supermultiplet consists of one
massless real
vector boson (with two degrees of freedom) and one Majorana spinor. Finally, a
gravity supermultiplet contains one massless
spin-two field and one massless two-component Rarita-Schwinger field. In each
case, the numbers of bosonic and fermionic degrees of freedom are equal, as
proved in Section \ref{sec:repr_general}.

\subsection{Massive representations of the Poincar\'e superalgebra}
\label{sec:massivesupermul}
In this section, we focus on massive $N=1$ representations of the Poincar\'e
algebra for which the standard frame is the rest frame and the four-momentum is given by
Eq.\ \eqref{eq:pmassive}. We deduce from the results of Section
\eqref{sec:palgrep} and from the Poincar\'e superalgebra of Eq.\ 
\eqref{eq:poincaresalg} that the little algebra takes the form
\be\bsp
  \big[Q_\alpha, M^{ij}\big] = \sigma^{ij}{}_\alpha{}^\beta Q_\beta\ , \qquad
  &\  \big[\Qbar^\alphadot, M^{ij}\big] = \sibar^{ij}{}^\alphadot{}_\betadot
     \Qbar^\betadot\ ,\\
  \big\{Q_\alpha, \Qbar_\alphadot\big\} = 2 m&\ \sigma^0_{\alpha\alphadot} = 2 m 
    \bpm 1& 0\\ 0 & 1 \epm\ ,
\esp\ee
where the Latin indices are defined by $i,j=1,2,3$. The two Casimir operators
are, in contrast
to the massless case, non-vanishing and read
\be
  {\cal C}_1 = m^2 \qquad\text{and}\qquad {\cal C}_2 = -2 m^4 \mathbf Y^2
\qquad\text{with}\qquad Y_i = M_i - \frac{1}{4m} \Qbar\sibar_i Q \ ,
\ee
denoting the rotations by $M_i\equiv M_{jk}$ where $(i,j,k)$ is a cyclic
permutation of $(1,2,3)$.
The quantum number $y(y+1)$ associated to the squared operator $\mathbf Y^2$
is called the superspin. Therefore, a massive representation of the $N=1$
Poincar\'e superalgebra is specified by the label $|m,y\rangle$. 

All the components of the representation $|m,y\rangle$ have the same superspin, although
they have different spins. In order to work out the spin structure,
one defines the creation and annihilation operators
\be
  a_{1,2}^\dag = \frac{1}{\sqrt{2m}} \Qbar_{\dot 1,\dot 2} \qquad\text{and}\qquad
  a_{1,2} = \frac{1}{\sqrt{2m}} Q_{1,2}
\ee
satisfying the anticommutation relations
\be
  \big\{a_i,a_j\big\} = \big\{a_i^\dag,a_j^\dag\big\} = 0 \qquad\text{and}\qquad
  \big\{a_i,a_j^\dag\big\} = \delta_{ij} \ .
\ee
The degrees of freedom embedded into a supermultiplet are obtained by starting 
from a vacuum state $|\Omega\rangle =|m,j;p^\mu,j_3 \rangle$ and iteratively
acting on it with the creation operators above. Since we are
interested in the spin of the states, the labels $j$ and $j_3$ refer to the
spin and its projection on the third axis and are therefore not related to the
superspin. This vacuum state is also assumed to be annihilated by the
annihilation operators $a_1$ and $a_2$. One observes that the action of the
creation operators allows to get
the spin structure of the massive
representations of the Poincar\'e superalgebra,
\be\bsp
  M_3 a_1^\dag |\Omega\rangle = \big(j_3-1/2\big) a_1^\dag|\Omega\rangle \ , \qquad
  &\ M_3 a_2^\dag |\Omega\rangle = \big(j_3-1/2\big) a_2^\dag |\Omega\rangle \ ,
\\
  M_3 a_2^\dag a_1^\dag |\Omega\rangle =&\ \big(j_3 - 1\big) a_2^\dag a_1^\dag 
    |\Omega\rangle \ . 
\esp\ee

Massive matter supermultiplets are built from a vacuum state of spin 1/2 and
therefore contain one massive Majorana fermion and one
massive complex scalar field as degrees of freedom. As another example, 
massive gauge
supermultiplets are derived from a scalar vacuum state and contain, after
doubling the spectrum due to the $CPT$ theorem, two states of spin zero, two
pairs of 
states of spin $\pm1/2$ and two states of spin $\pm 1$. These are the degrees of
freedom of one massive vector field, one real scalar field and one massive 
Dirac fermion.

Coming back to the superspin, it can be seen as the linear
combination of a spin $j$ and a spin 1/2 since the creation operators
$a_1$ and $a_2$ are
fermionic. Moreover, from the form of the operator $\mathbf Y$, one observes
that for the vacuum state, $\mathbf Y = \mathbf M$, its superspin being thus
equal to its spin.

\mysection{The superspace formalism}
\label{sec:superspace}
\subsection{Supercharges and superderivatives in the
superspace}\label{sec:schsder}
In order to built supersymmetric theories in a way where supersymmetry is
manifest, it is conventional to employ the superspace formalism
\cite{Salam:1974yz, Salam:1974jj, Ferrara:1974ac}. This offers the possibility
to combine the
different components of the supermultiplets derived in Section
\ref{sec:masslesssupermul} and Section \ref{sec:massivesupermul} into a single
object dubbed superfield. Superfields are then used for simplifying
and writing in a compact form most of the objects related to supersymmetric
model building.  

The superspace is constructed as an extension of the ordinary spacetime
by adjoining a Majorana spinor $(\theta_\alpha,
\thetabar^\alphadot)$ to the usual spacetime coordinates $x^\mu$. The
anticommuting parameters $\theta_\alpha$ and $\thetabar^\alphadot$ are
Grassmannian two-component Weyl fermions, satisfying the Grassmann algebra
relations of Eq.\ \eqref{eq:grassmannalg}.
One can interpret the superspace coordinates $(x^\mu, \theta_\alpha,
\thetabar^\alphadot)$ as a representation of the Poincar\'e superalgebra in the
same way as the spacetime coordinates $x^\mu$ are interpreted as a
representation of the Poincar\'e algebra.
This results from Eq.\ \eqref{eq:poincaresalg} which implies that
\be
  \big[ \theta\!\cdot\! Q, \Qbar \!\cdot\! \thetabar \big] = 2 
     \theta\sigma^\mu\thetabar P_\mu  \qquad\text{and}\qquad
  \big[ \theta \!\cdot\! Q, \theta \!\cdot\! Q \big] = 
  \big[ \Qbar \!\cdot\! \thetabar, \Qbar\!\cdot\!\thetabar \big] =  0  \ ,
\ee
after imposing that $\{Q_\alpha,\thetabar^\alphadot\} =
\{\Qbar_\alphadot,\theta^\alpha\} = 0$ and where we again refer to
Appendix \ref{app:conv} for our conventions on spinors and for
the construction of invariant products of spinorial fields. From these
considerations, one defines finite supersymmetry
transformations as elements of the coset space $G/H$, $G$ being the
Poincar\'e supergroup and $H$ the Lorentz group. In this context, a superspace
point is parametrized as a translation in superspace,
\be
  G(x,\theta,\thetabar) = \exp \Big[ i \big(x^\mu P_\mu +  \theta\!\cdot\! Q + 
   \Qbar \!\cdot\! \thetabar\big) \Big] \ .
\ee
The elements $G(a,0,0)$ are pure translations of
parameter $a$ in Minkowski space, whilst $G(0,\e,\ebar)$ are pure supersymmetric
transformations of parameters $(\e,\ebar)$. 

Multiplying two group elements allows to compute the variations of the
superspace coordinates under a pure supersymmetric transformation of spinorial 
parameters $(\e,\ebar)$, the latter being imposed to be anticommuting with the
Grassmann variables and the supercharges. One gets, for an action from
the left,
\be
  G(0,\e,\ebar) G(x,\theta,\thetabar)  = \exp \Big[ i\big(
   x^\mu + i \e \sigma^\mu \thetabar - i \theta \sigma^\mu \ebar\big) P_\mu + 
   i \big(\theta + \e) \!\cdot\! Q + i \Qbar \!\cdot\! 
   \big(\thetabar + \ebar\big)\Big]
   \ ,
\ee
after employing the Baker-Campbell-Hausdorff identity. The variations of the
coordinates can be further rewritten as
\be
  \delta x^\mu =\big[i \big(\e \sigma^\nu \thetabar - \theta \sigma^\mu
    \ebar\big)\del_\nu, x^\mu \big]\ ,\qquad
  \delta \theta^\alpha = \big[\e\!\cdot\!\del,\theta^\alpha\big]
  \qquad\text{and}\qquad
  \delta \thetabar^\alphadot = -\big[\delbar\!\cdot\!\ebar,
    \thetabar^\alphadot\big] \  ,
\ee
where we have introduced the variables $(\del_\mu, \del_\alpha,
\delbar^\alphadot)$, conjugate to the superspace coordinates, that are
defined in Eq.\ \eqref{eq:spaceconju}.
Comparing to a direct application of the supersymmetric transformation
generators on the superspace coordinates, 
\be
   \delta x^\mu = \big[i\e \!\cdot\! Q  + i\Qbar \!\cdot\! \ebar, x^\mu]\ ,\quad
   \delta \theta^\alpha = \big[ i\e \!\cdot\! Q  + i\Qbar \!\cdot\! \ebar,
   \theta^\alpha  \big] \quad\text{and}\quad
   \delta\thetabar^\alphadot = \big[i\e \!\cdot\! Q  + i\Qbar \!\cdot\! \ebar,
    \thetabar^\alphadot\big] \ , 
\ee
one can derive the form of the supercharges as differential operators acting on
functions on superspace. Similarly, starting from a multiplication from the
right, $G(x,\theta,\thetabar) G(0,\e,\ebar)$, one can express the
superderivatives $D_\alpha$ and $\Dbar_\alphadot$ in terms of the conjugate
variables $(\del_\mu, \del_\alpha, \delbar^\alphadot)$, after introducing
appropriate normalization factors. The results for both the supercharges
and the superderivatives read %
\renewcommand{\arraystretch}{1.2}%
\be\boxed{\begin{array}{l l}
     Q_\alpha = -i\big(\del_\alpha + i \sigma^\mu{}_{\alpha\alphadot}
    \thetabar^\alphadot \del_\mu\big) \  , 
  &\qquad \Qbar_\alphadot = i \big(\delbar_\alphadot + i \theta^\alpha
    \sigma^\mu{}_{\alpha \alphadot} \del_\mu\big) \ , \\
     D_\alpha = \del_\alpha - i \sigma^\mu{}_{\alpha\alphadot} \thetabar^\alphadot 
     \del_\mu \ , 
  &\qquad \Dbar_\alphadot =  \delbar_\alphadot - i\theta^\alpha
     \sigma^\mu{}_{\alpha\alphadot}\del_\mu \ .
\end{array}} \label{eq:schsder} 
\ee%
\renewcommand{\arraystretch}{1}%

Since the actions from the left and from the right commute, the supercharges and
the superderivatives anticommute. One gets, in addition,
\be
  \big\{Q_\alpha, \Qbar_\alphadot \big\} = 2 i \sigma^\mu{}_{\alpha\alphadot}
     \del_\mu = -2 \sigma^\mu{}_{\alpha\alphadot} P_\mu \quad\text{and}\quad
  \big\{D_\alpha, \Dbar_\alphadot \big\} = -2 i \sigma^\mu{}_{\alpha\alphadot}
     \del_\mu =  2 \sigma^\mu{}_{\alpha\alphadot} P_\mu\ .
\ee
Since the derivative form of the four-momentum operator reads
$P_\mu\equiv -i \del_\mu$, the form of the supercharges that we have
derived is consistent with the Poincar\'e superalgebra of Eq.\ 
\eqref{eq:poincaresalg}, up to a global sign. This is not surprising since
we have chosen to derive a differential representation of the supercharges
starting from an action from the left of the group elements. Conversely, the
superderivatives, associated to an action from the right, yield an
anticommutation relation with the correct sign.

\subsection{General superfields}

Any function $\Phi(x,\theta,\thetabar)$ defined on the $N=1$ superspace is called a
superfield and can be expanded as a Taylor series with respect to the
coordinates $\theta$ and $\thetabar$. Since the square of an anticommuting
object vanishes and due to the relations of Eq.\ \eqref{eq:thetarelations},
this series has a finite number of terms and its most general expression can be
written as 
\be\label{eq:genSF}\bsp
  \Phi(x,\theta,\thetabar) = &\ z(x) 
    +\theta \!\cdot\! \xi(x) + \thetabar\!\cdot\!\zetabar(x)
    + \theta \!\cdot\! \theta f(x) + \thetabar \!\cdot\! \thetabar g(x) 
  + \theta \sigma^\mu \thetabar v_\mu(x) 
    + \thetabar \!\cdot\! \thetabar\ \theta \!\cdot\! \omega(x)\\ &\  
    + \theta \!\cdot\! \theta\ \thetabar \!\cdot\! \bar \rho(x)
    + \theta \!\cdot\! \theta\ \thetabar \!\cdot\! \thetabar d(x) \ . 
\esp \ee
In the equation above, we have assumed that the superfield $\Phi$ is a scalar
superfield, \ie, it does not carry any Lorentz or spin index. However, 
extensions to non-scalar superfields are immediate. Specific examples can be found
in the rest of this chapter, with, \eg,
the computation of the superfield strength tensors (see
Eq.\ \eqref{eq:WandWbar}) where we expand in terms of Grassmann variables 
superfields carrying a spin index. The coefficients of the expansion in
Eq.\
\eqref{eq:genSF} form a supermultiplet and are referred to as the component
fields of the superfield. They correspond to the usual scalar, fermionic and
vector fields employed in particle physics. The fields $z$, $f$, $g$ and $d$
are hence complex scalar whilst $\xi$, $\zeta$, $\omega$ and $\rho$
denote complex Weyl fermions. Finally, $v_\mu$ is a complex vector field. This
leaves an equal number of 16 bosonic and 16 fermionic degrees of freedom.

\subsection{Chiral superfields}\label{sec:CSF}

The superfield $\Phi$ of Eq.\ \eqref{eq:genSF} contains too many degrees of
freedom compared to the field content of the supermultiplets derived in Section
\ref{sec:masslesssupermul} and Section \ref{sec:massivesupermul}. It has
therefore to be reduced by imposing constraining relations compatible with
supersymmetry transformations. We first consider the case of the left and
right-handed chiral superfields \cite{Salam:1974yz, Salam:1974jj,
Ferrara:1974ac} employed to embed matter supermultiplets whose degrees of
freedom consist of one complex scalar field and one Weyl fermion.

The chiral and antichiral superfields $\Phi_L$ and $\Phi_R$ are defined to satisfy the
constraints
\be\label{eq:chiralconst}
  \Dbar_\alphadot \Phi_L (x,\theta,\thetabar) = 0 
  \quad \text{and} \quad
  D_\alpha \Phi_R (x,\theta,\thetabar) = 0 \ ,
\ee 
where the superderivatives have been introduced in Eq.\ \eqref{eq:schsder}.
Since the supercharges and the superderivatives anticommute (see Section
\ref{sec:schsder}), these two relations are preserved by supersymmetry
transformations,
\be\bsp
  \Dbar_\alphadot\delta_\e\Phi_L = 
   i\Dbar_\alphadot \big(\e \!\cdot\! Q + \Qbar \!\cdot\! \ebar\big) \Phi_L  =
   i\big(\e \!\cdot\! Q + \Qbar \!\cdot\! \ebar\big) \Dbar_\alphadot \Phi_L = 0
   \ , \\
  D_\alpha\delta_\e\Phi_R = 
   iD_\alpha \big(\e \!\cdot\! Q + \Qbar \!\cdot\! \ebar\big) \Phi_R  =
   i\big(\e \!\cdot\! Q + \Qbar \!\cdot\! \ebar\big) D_\alpha \Phi_R = 0 \ ,
\esp\ee
where the notation $\delta_\e \Phi$ has been introduced to indicate the variation of a
generic superfield $\Phi$
under a supersymmetry transformation of parameters $(\e,\ebar)$. In order to
work out the component field structure of both left-handed and right-handed chiral
superfields in a straightforward fashion, it is important to note that
\be
  \Dbar_\alphadot x^\mu = -i \big(\theta\sigma^\mu\big)_\alphadot
  \qquad\text{and}\qquad
  D_\alpha x^\mu = -i \big(\sigma^\mu\thetabar\big)_\alpha \ .
\ee
Equivalently, these equations can be rewritten as
\be
  \Dbar_\alphadot \big(x^\mu - i \theta\sigma^\mu\thetabar\big) = 0
  \qquad\text{and}\qquad
  D_\alpha \big(x^\mu  +i \theta\sigma^\mu\thetabar\big) = 0 \ .
\ee
This motivates a change of spacetime variables
\be
  x^\mu \to y^\mu = x^\mu - i \theta\sigma^\mu\thetabar\ .
\label{eq:xtoy}\ee
A left-handed chiral superfield therefore consists of a quantity depending only
on $y$ and $\theta$, since
\be
  0 = \Dbar_\alphadot \Phi_L 
    = \delbar_\alphadot \Phi_L - i\big(\theta\sigma^\mu\big)_\alphadot \del_\mu
       \Phi_L
    = \del_{y^\mu} \Phi_L \delbar_\alphadot y^\mu  + \delbar_\betadot \Phi_L
       \delbar_\alphadot \thetabar^\betadot - i \big( \theta \sigma^\mu
       \big)_\alphadot \del_{y^\mu} \Phi_L 
    = \delbar_\alphadot \Phi_L \ .
\ee
For the first equality, we have employed the definition of the superderivatives
presented in Eq.~\eqref{eq:schsder}. For the second equality, we have performed
the change of variables of Eq.\ \eqref{eq:xtoy}, recalling that $\del_\mu\Phi_L=
\del_{y^\mu}\Phi_L$,  the operator $\del_{y^\mu}$ indicating a derivation
with respect to the $y$-variable, in contrast to $\del_\mu$ where we derive with
respect to the spacetime coordinates $x^\mu$. In the expression above,
we have also removed the arguments
of the superfield $\Phi_L$ for clarity. In a similar
fashion, a right-handed chiral superfield $\Phi_R$ only depends on $y^\dag$ and
$\thetabar$. Consequently, the most general solutions to the constraints of Eq.\
\eqref{eq:chiralconst} and their expansion in terms of their
scalar components $\phi$ and $\phi^\dag$, fermionic components $\psi$ and
$\psibar$ and auxiliary components $F$ and $F^\dag$ can be written as%
\renewcommand{\arraystretch}{1.2}%
\be\boxed{\bsp
  \Phi_L =&\ \phi(y) + \sqrt{2} \theta \!\cdot\! \psi(y) - \theta
    \!\cdot\! \theta F(y)  \\
  =&\ \phi(x) +\sqrt{2} \theta \!\cdot\! \psi(x) - \theta \!\cdot\! \theta F(x) 
    - i \theta \sigma^\mu \thetabar \del_\mu \phi(x) 
    +\frac{i}{\sqrt{2}} \theta \!\cdot\! \theta \del_\mu \psi(x) \sigma^\mu
      \thetabar
   -\frac14  \theta \!\cdot\! \theta \thetabar \!\cdot\! \thetabar \Box
      \phi(x) \ , \\
  \Phi_R  =&\ \phi(y^\dag) + \sqrt{2} \thetabar \!\cdot\!
    \psibar(y^\dag) - \thetabar \!\cdot\! \thetabar F(y^\dag) \\
  =&\ \phi^\dag(x) +\sqrt{2} \thetabar \!\cdot\! \psibar(x)  - \thetabar
   \!\cdot\!
    \thetabar F^\dag(x) + i \theta \sigma^\mu \thetabar \del_\mu  \phi^\dag(x)
   - \frac{i}{\sqrt{2}} \thetabar \!\cdot\! \thetabar \theta \sigma^\mu \del_\mu
     \psibar (x) 
   - \frac14 \theta \!\cdot\! \theta \thetabar \!\cdot\! \thetabar \Box
    \phi^\dag (x)  \ .
\esp}\label{eq:chiralSF}
\ee%
\renewcommand{\arraystretch}{1.}%
For the second equality of each equation, a Taylor expansion of the $y$-variable
around the spacetime coordinates $x^\mu$ has been performed.
Concerning left-handed (right-handed) chiral superfields, the
normalizations of the components in $\theta$ ($\thetabar$) and
$\theta\!\cdot\!\theta$ ($\thetabar\!\cdot\!\thetabar$) are conventional.

As stated above, chiral and antichiral superfields are appropriate to describe
matter supermultiplets having as degrees of freedom
one two-component Weyl fermion and one complex scalar field, as can
be seen from the results of Section \ref{sec:masslesssupermul} and Section
\ref{sec:massivesupermul}.
When inspecting the physical degrees of freedom included in the
expansions of $\Phi_L$ and $\Phi_R$ in terms of their component fields, one
indeed observes the presence of a complex scalar field $\phi$ and a two-component
fermion $\psi$. However, chiral superfields contain an extra complex scalar
field $F$
that does not correspond to any physical degree of freedom of the matter
supermultiplets. This field is nevertheless
mandatory to restore the equality between the numbers of fermionic and bosonic
degrees of freedom when the component fields are off-shell, since an off-shell
Weyl fermion contains four degrees of freedom instead of two in the on-shell
case. The two additional bosonic degrees of freedom carried by $F$ are then
compensating that lack and the equations of motion of this non-physical field are fixed 
appropriately so that it vanishes on-shell.

\subsection{Vector superfields}\label{sec:VSF}

Chiral superfields that have just been introduced do not contain any
vectorial component. Consequently, they cannot be used for dealing with 
gauge supermultiplets whose the field content has been given in Section
\ref{sec:masslesssupermul}. This leads us to the
introduction of a new type of superfields, the
vector superfield \cite{Salam:1974jj, Ferrara:1974ac,
Fayet:1976cr}. A gauge boson
being a real vector field, one naturally demands a vector superfield
$V(x,\theta, \thetabar)$ to satisfy the reality condition 
\be \label{eq:Vreal}
  V = V^\dag \ .
\ee
Under this condition, the expansion of $V$ in terms of its component fields can
be written as 
\be\bsp
  V(x,\theta, \thetabar) =&\ C(x)  + i\theta \!\cdot\! \chi(x) - i\thetabar
    \!\cdot\!
    \chibar(x) + \frac{i}{2} \theta \!\cdot\! \theta f - \frac{i}{2} \thetabar
    \!\cdot\!  \thetabar f^\dag
    + \theta \sigma^\mu \thetabar v_\mu(x)  \\
    &\ + i \theta \!\cdot\! \theta \thetabar \!\cdot\! \Big(\lambar(x) - \frac{i}{2} 
       \sibar^\mu \del_\mu \chi(x) \Big)
    -i \thetabar\!\cdot\!\thetabar \theta\!\cdot\! \Big( \lambda(x) - \frac{i}{2} 
       \sigma^\mu \del_\mu \chibar(x)\Big)\\
    &\ + \frac12 \theta\!\cdot\!\theta \thetabar\!\cdot\!\thetabar \Big(D(x) - 
        \frac12 \Box  C(x) \Big)\ .
\esp\label{eq:genVSF}\ee 
The set of component fields includes, among others, the degrees of freedom of
the $N=1$
gauge supermultiplets, \ie, a massless vector boson $v_\mu$ and a Majorana
fermion dubbed gaugino $(\lambda_\alpha, \lambar^\alphadot)$. As for
chiral superfields,
although the number of fermionic degrees of freedom equal the number of bosonic
ones for on-shell fields, one bosonic degree of freedom is missing when going
off-shell. Consequently, one auxiliary non-propagating field must be added,
similarly to the $F$-field in the chiral case. This field is then eliminated
when going on-shell through its equations of motion.

Inspecting Eq.\
\eqref{eq:genVSF}, we observe that the expansion of $V$ contains more than one
additional auxiliary field, since we have two real scalar fields $C$ and
$D$, one complex scalar field $f$ as well as one Majorana fermion $(\chi_\alpha,
\chibar^\alphadot)$. The way the expansion has been performed is however not
obvious and will be justified below when addressing gauge
transformations of vector superfields, the latter allowing in fact
to eliminate all the non-necessary auxiliary fields.

When working out the transformation laws of the vector superfield under a
supersymmetry transformation (see Section \ref{sec:deltasusy}), one observes
that the component fields $v$, $(\lambda,\lambar)$ and $D$ transform into each
other when eliminating the $C$,
$M$, $N$ and $(\chi,\chibar)$ fields which are then enforced to vanish.
Being left with only the real scalar field $D$ as an auxiliary
component of $V$, one is thus able to ensure the equality between the number of
bosonic and fermionic degrees of freedom when going off-shell and map the
remaining degrees of freedom to the field content of gauge supermultiplets.

Let us now consider a left-handed chiral superfield $\Lambda$. Since $i(\Lambda-
\Lambda^\dag)$ is real, the transformation $V \to V - i (\Lambda - \Lambda^\dag)$
preserves the reality condition of $V$ and we get
\be\bsp 
  C    \to C -i (\phi - \phi^\dag) \ ,  \qquad
  &\ \chi \to \chi - \sqrt{2} \psi \ , \qquad
  f  \to f + 2F \ ,  \\
  v_\mu \to v_\mu - \del_\mu(\phi + \phi^\dag) \ ,\qquad
  &\ \lambda \to \lambda \ , \qquad\qquad\quad
   \ D \to D \ ,
\esp \label{eq:Vgauge}\ee
where $\phi$, $\psi$ and $F$ stand for the component fields of the chiral
superfield $\Lambda$.
The transformation laws of the vectorial field $v_\mu$ correspond exactly to an
abelian gauge transformation. This strongly suggests to interpret $V \to V -i 
(\Lambda - \Lambda^\dag)$ as a generalized (abelian) gauge transformation at the
superfield level. We adopt a specific gauge fulfilling $i(\phi-\phi^\dag)=C$,
$\sqrt{2} \psi = \chi$ and $2F=-f$. Consequently, all the unphysical component
fields but the $D$-field are eliminated. Such a convenient gauge is called the
Wess-Zumino
gauge, in which a vector superfield can be expanded in terms of its component
fields as 
\be\boxed{
  V_{W.Z.}(x,\theta,\thetabar) = \theta \sigma^\mu \thetabar v_\mu(x) + 
    i \theta \!\cdot\! \theta \thetabar \!\cdot\! \lambar(x) - 
    i \thetabar \!\cdot\! \thetabar \theta \!\cdot\! \lambda(x) + 
    \frac12 \theta \!\cdot\! \theta \thetabar \!\cdot\! \thetabar  D(x) \ . }
\label{eq:VSF}\ee
Since we still have the freedom to
fix the real part of the scalar component of the gauge transformation
parameters, \ie, $(\phi+\phi^\dag)$, it is still possible to adopt
a specific gauge choice for the vector field $v_\mu$.

\mysection{From superfields to Lagrangians}\label{sec:susylag}
\subsection{Chiral Lagrangians}\label{sec:Lchiral}
The main advantage of writing down supersymmetric Lagrangians in terms of
(chiral and vector) superfields rather than in terms of their component fields
lies in
the size of the corresponding expressions. As it will be explicitly shown in
Section \ref{sec:deltasusygen}, the $\theta^2 \thetabar^2$-component of a
general
superfield, its $D$-term, transforms under a supersymmetric
transformation as a total derivative. Similarly, the $\theta^2$-component of a
chiral superfield, its $F$-term, also transforms as a total
derivative (see Section \ref{sec:deltasusychir}). This is the cornerstone of the
method for Lagrangian construction in supersymmetric quantum
field theories within the superspace formalism. Products and sums of superfields
being superfields, a
supersymmetric Lagrangian then consists of $F$-terms and $D$-terms of sums and
products of the elementary superfields representing the supermultiplets
included in the model under consideration.
In this section, we focus on the Wess and Zumino model \cite{Wess:1974tw,
Wess:1978ns} describing chiral supermultiplets in interaction and construct
the associated Lagrangian by employing the superspace formalism.

The kinetic Lagrangian terms describing the propagation of the degrees of
freedom included in a matter supermultiplet represented by a chiral superfield
$\Phi$ is obtained from the quantity $\Phi^\dag \Phi$. Its series expansion in
terms of the Grassmann variables $\theta$ and $\thetabar$ is given by
\bea
  \Phi^\dag \Phi &=&  
    \phi^\dag \phi +  
    \sqrt{2} \theta\!\cdot\!\phi^\dag\psi +  
    \sqrt{2} \thetabar\!\cdot\!\psibar \phi -  
    \theta\!\cdot\!\theta \ \phi^\dag F - 
    \thetabar\!\cdot\!\thetabar\  F^\dag \phi +
    \theta \sigma^\mu \thetabar \Big[  
        i \phi^\dag \del_\mu \phi 
      - i \del_\mu \phi^\dag \phi 
      - \psibar \sibar_\mu \psi\Big] 
\nn\\  && + 
    \theta\!\cdot\!\theta\ \thetabar\!\cdot\! \Big[ 
      \frac{i}{\sqrt{2}} \phi^\dag \sibar^\mu \del_\mu \psi
      - \sqrt{2} \psibar F  
      - \frac{i}{\sqrt{2}} \del_\mu \phi^\dag \sibar^\mu \psi\Big] 
\nn\\&& +
    \thetabar\!\cdot\!\thetabar\ \theta\!\cdot\! \Big[ 
        \frac{i}{\sqrt{2}} \sigma^\mu \del_\mu \psibar \phi 
      - \sqrt{2} F^\dag \psi  
      - \frac{i}{\sqrt{2}} \sigma^\mu \psibar \del_\mu \phi \Big] \\  
\label{eq:lagchiral}  &&\nn   +
    \theta\!\cdot\!\theta \thetabar\!\cdot\!\thetabar \Big[ 
      - \frac14 \phi^\dag \Box \phi  
      + \frac12 \del_\mu \phi^\dag \del^\mu \phi  
      - \frac14 \Box \phi^\dag \phi  
      - \frac{i}{2} \psibar \sibar^\mu \del_\mu \psi  
      + \frac{i}{2} \del_\mu \psibar \sibar^\mu \psi  
      + F^\dag F \Big] \ , \eea
where $\phi$, $\psi$ and $F$ are the scalar, fermionic and auxiliary components
of the chiral superfield $\Phi$. The $D$-term of the object $\Phi^\dag \Phi$,
which reads after an integration by parts and omitting the total derivative, 
\be\bsp
{\cal L}=\del_\mu \phi^\dag \del^\mu \phi + 
  \frac{i}{2}(\psi \sigma^\mu \del_\mu \bar \psi -
  \del_\mu \psi \sigma^\mu \bar \psi) + F^\dag F \ ,
\esp\ee
is therefore appropriate to describe standard kinetic terms
for the scalar and fermionic fields $\phi$ and $\psi$ since we recover the
Klein-Gordon and Weyl Lagrangian densities. The non-derivative
term associated to the auxiliary field $F^\dag F$ ensures that this field
is non-propagating and the associated 
equations of motion give $F=0$. Therefore, as required above, the
$F$-field is vanishing in the on-shell case.

We now fix the chiral content of a generic theory to a set of chiral and
antichiral superfields $\Phi^i$ and $\Phi^\dag_\is$. In our notations, we adopt
starry Latin letters, such as $\is$, for antichiral indices, and
normal Latin letters, such as $i$, for chiral indices. This choice allows to
underline the
K\"ahler manifold structure spanned by the matter superfields, as it will
be explicitly worked out below. 

We can build the chiral action ${\cal S}_{\rm K}$
from the Lagrangian of Eq.\ \eqref{eq:lagchiral} as an integral over the
eight-dimensional superspace
\be 
  {\cal S}_{\rm K} = \int \d^4 x\ \d^2 \theta\ \d^2 \thetabar\ K(\Phi,
    \Phi^\dag) \qquad\text{with}\qquad
  K(\Phi, \Phi^\dag) =  \delta^\is{}_i \Phi^\dag_\is \Phi^i \ ,
\label{eq:SK}\ee 
since integration upon the Grassmann coordinates fulfills the relations of Eq.\
\eqref{eq:intgrass}. In the equation above, we have explicitly introduced the
K\"ahler potential $K(\Phi, \Phi^\dag)$. This form for the action ${\cal S}_{\rm
K}$ allows us to immediately generalize it, although
the K\"ahler potential has been taken trivial in this case.

As sketched in Eq.\ \eqref{eq:SK}, in their most general and non-renormalizable
versions,
supersymmetric chiral actions are entirely expressed with the help of a single 
fundamental function of the chiral (and antichiral) superfield content of the theory. This
function is dubbed the K\"ahler
potential $K(\Phi,\Phi^\dag)$ \cite{Zumino:1979et, AlvarezGaume:1981hm}. In the
renormalizable case, the K\"ahler potential takes the very simple form given in
Eq.\
\eqref{eq:SK}, $K(\Phi, \Phi^\dag) =  \delta^\is{}_i \Phi^\dag_\is \Phi^i$. In 
the following, it will however be left unspecified and kept fully
generic. The K\"ahler potential founds its name from the fact that it satisfies
the properties, presented below, of a K\"ahler manifold \cite{Witten:1982hu,
Bagger:1983tt, Bagger:1982ab, Bordemann:1985xy, Itoh:1985ha}.

An action being real, $K(\Phi,\Phi^\dag)$ is an
arbitrary real superfield that can be expanded as 
\be 
  K(\Phi,\Phi^\dag) = W_I(\Phi^\dag) W^I(\Phi)\ , 
\ee 
where $W^I(\Phi)$ and $W_I(\Phi^\dag)$ are holomorphic and anti-holomorphic 
functions of the chiral and antichiral superfields $\Phi^i$ and 
$\Phi^\dag_\is$, respectively. In the rest of this subsection, we denote the
scalar, fermionic and auxiliary component fields
of $\Phi^i$ by $\phi^i$, $\psi^i$ and $F^i$, respectively, while those of the
antichiral superfield $\Phi^\dag_\is$ are denoted by $\phi^\dag_\is$,
$\psibar_\is$ and $F^\dag_\is$. Performing Taylor expansions around the scalar
components of two functions $W^I(\Phi)$ and $W_I(\Phi^\dag)$, one derives 
\be\label{eq:expW}\bsp
  W^I(\Phi) =  &\
      W^I +  
      \sqrt{2} \theta\!\cdot\!\psi^i \frac{\del W^I}{\del\phi^i} - 
      \theta\!\cdot\!\theta \Big( 
        F^i \frac{\del W^I}{\del\phi^i} + 
        \frac12 \psi^i\!\cdot\!\psi^j \frac{\del^2 W^I}{\del\phi^i \del\phi^j}
      \Big) \ ,\\
  W_I(\Phi^\dag) =  &\
    W_I +  
      \sqrt{2}\thetabar\!\cdot\!\psibar_\is \frac{\del W_I}{\del\phi^\dag_\is} - 
      \thetabar\!\cdot\!\thetabar \Big( 
         F^\dag_\is \frac{\del W_I}{\del \phi^\dag_\is} + 
         \frac12 \psibar_\is\!\cdot\!\psibar_\js \frac{\del^2 
           W_I}{\del\phi^\dag_\is \del\phi^\dag_\js}\Big) \ , 
\esp\ee 
where $W_I\equiv W_I(\phi^\dag_\is)$ and  $W^I\equiv W^I(\phi^i)$.
From these expressions, we can compute the expansion in terms of the Grassmann
variables of the most general expression of the K\"ahler potential. Following
standard superspace techniques as, \eg, presented in Ref.\ \cite{livre} and
Ref.\ \cite{sugra}, one obtains
\be\label{eq:K}\bsp 
  & K(\Phi,\Phi^\dag) =
    K +  
    \sqrt{2} \theta\!\cdot\! \Big[K_i \psi^i\Big] +  
    \sqrt{2} \thetabar\!\cdot\! \Big[K^\is \psi_\is \Big] - 
    \theta\!\cdot\!\theta \Big[ 
      K_i F^i + \frac12 K_{ij} \psi^i\!\cdot\!\psi^j \Big] 
\\ &\
- \thetabar\!\cdot\!\thetabar \Big[ 
      K^\is F^\dag_\is +  
      \frac12 K^{\is\js} \psibar_\is\!\cdot\!\psibar_\js\Big] + 
    \theta \sigma^\mu \thetabar \Big[ 
      -i K_i \del_\mu \phi^i +i K^\is \del_\mu \phi^\dag_\is + 
      K^\is{}_i \psi^i \sigma_\mu \psibar_\is \Big]
\\ &\ +
    \theta\!\cdot\!\theta\ \thetabar\!\cdot\! \Big[ 
      - \sqrt{2} K^\is{}_i F^i \psibar_\is  
      - \frac{1}{\sqrt{2}} K^\is{}_k \Gamma_i{}^k{}_j \psi^i\!\cdot\!\psi^j 
       \psibar_\is  
      + \frac{i}{\sqrt{2}} K^\is{}_i \sibar^\mu \psi^i \del_\mu \phi^\dag_\is 
\\&\ \qquad
      - 
       \frac{i}{\sqrt{2}} \sibar^\mu  
          \big(K_i \D_\mu + \D_i K_j \del_\mu\phi^j\big)\psi^i 
    \Big]
\\ &\ + 
    \thetabar\!\cdot\!\thetabar\ \theta\!\cdot\! \Big[ 
      - \sqrt{2} K^\is{}_i F^\dag_\is \psi^i 
      - \frac{1}{\sqrt{2}} K^\ks{}_i \Gamma^\is{}_\ks{}^\js  
       \psibar_\is\!\cdot\!\psibar_\js \psi^i 
      + \frac{i}{\sqrt{2}} K^\is{}_i  \sigma^\mu \psibar_\is \del_\mu 
        \phi^i 
\\ &\ \qquad -
      \frac{i}{\sqrt{2}} \sigma^\mu \big( 
          K^\is \D_\mu +  {\cal D}^\is K^\js \del_\mu \phi^\dag_\js\big) \psibar_\is 
     \Big]
\\ &\ + 
    \theta\!\cdot\!\theta\ \thetabar\!\cdot\!\thetabar \Big[ 
      - \frac14 \del_\mu\big(K_i \del^\mu \phi^i + K^\is \del^\mu
        \phi^\dag_\is\big)  
      + K^\js{}_i \del_\mu \phi^i \del^\mu \phi^\dag_\js  
      + K^\js{}_i F^i F^\dag_\js
\\&\ \qquad + 
    \frac14 K^{\ks \ls}{}_{ij} \psi^i\!\cdot\!\psi^j
      \psibar_\ks\!\cdot\!\psibar_\ls +  
    \frac{i}{2} \big(  
        K^\js{}_i \psi^i \sigma^\mu {\cal D}_\mu \psibar_\js  
      - K^\js{}_i {\cal D}_\mu\psi^i \sigma^\mu \psibar_\js\big) 
\\&\ \qquad + 
        \frac12 K^\js{}_i \Gamma^\ks{}_\js{}^\ls F^i \psibar_\ks\!\cdot\!\psibar_\ls 
      + \frac12 K^\is{}_\ell \Gamma_i{}^\ell{}_j F^\dag_\is \psi^i\!\cdot\!\psi^j  
   \Big] \ . 
\esp\ee 
The symbol $K\equiv K(\phi,\phi^\dag)$ stands for the K\"ahler potential
expressed in terms of the scalar components of the chiral and antichiral
superfields $\Phi^i$ and $\Phi^\dag_\is$. The derivatives of $K$ are indicated
by the shorthand notations 
\be \bsp  
  &\ K_i    = \frac{\del K(\phi,\phi^\dag)}{\del \phi^i} \ , \qquad  
     K^\is  = \frac{\del K(\phi,\phi^\dag)}{\del \phi^\dag_\is} \ , \\ 
  &\ K_{ij} = \frac{\del^2 K(\phi,\phi^\dag)}{\del \phi^i\del \phi^j} \ , \qquad
     K^{\is\js} = \frac{\del^2 K(\phi,\phi^\dag)}{\del \phi^\dag_\is 
       \del\phi^\dag_\js} \ , \qquad 
  K^\is{}_{i} = \frac{\del^2 K(\phi,\phi^\dag)}{\del \phi^i \del 
       \phi^\dag_\is} \ , \\  
  &\ K_i{}^\ks{}_{j} = \frac{\del^3 K(\phi,\phi^\dag)}{\del \phi^i \del \phi^j 
       \del \phi^\dag_\ks} = K^\ks{}_\ell \Gamma_i{}^\ell{}_j\ , \qquad 
     K^\is{}_k{}^\js = \frac{\del^3 K(\phi,\phi^\dag)}{\del \phi^\dag_\is \del 
       \phi^\dag_\js \del \phi^k} = K^\ls{}_k \Gamma^\is{}_\ls{}^\js\ , \\ 
  &\ K^{\is\js}{}_{k \ell} = \frac{\del^4 K(\phi,\phi^\dag)}{\del \phi^\dag_\is 
       \del \phi^\dag_\js \del \phi^k \del \phi^\ell}\ .
\esp\label{eq:derK}\ee 
We have introduced here the natural tensors of the K\"ahler manifold 
spanned by the scalar components of the chiral and antichiral superfields, 
the K\"ahler metric $K^\is{}_i$ and the elements of the connection $\Gamma$.
 
A K\"ahler manifold is an analytic Riemann manifold with specific properties. As
for any analytic Riemann manifold, it can  be
parametrized in terms of two sets of complex coordinates. In our case,
these sets of coordinates consist of the scalar fields $\{\phi^i\}$ and
$\{\phi^\dag_\is\}$ and two different types of indices (starry and
non-starry letters from the middle of the Latin alphabet) are
attached to the two series of coordinates. In contrast to standard
analytic Riemann manifolds, K\"ahler manifolds are endowed with an Hermitian
metric,
positively defined and invertible. Moreover, this metric is derived from a
scalar function of the coordinates, the K\"ahler potential. In the example
studied in this section, the metric is denoted by $K^\is{}_i$ and the 
potential, which depends on the coordinates, by $K\equiv K(\phi,\phi^\dag)$.
By definition, the metric and its inverse allow to raise, lower and change the
nature of the indices,
\be\label{eq:kahlermetric}
  \phi^\dag_j = K^\is{}_j \phi^\dag_\is \ ,\quad
  \phi^{\dag j} = (K^{-1})^j{}_\is \phi^{\dag\is} \ ,\quad
  \phi^\is = K^\is{}_j \phi^j \quad\text{and}\quad
  \phi_\is = (K^{-1})^j{}_\is \phi_j \ .
\ee
Furthermore, we impose that derivation preserves the analytical nature of the
coordinates, \ie, that the derivatives are covariant with respect to the
transformations
\be
  \phi^i \to \phi^{\prime i} = \varphi^i(\phi) \qquad\text{and}\qquad 
  \phi^\dag_\is \to \phi^{\prime\dag}_\is = \varphi^\dag_\is(\phi^\dag) \ .
\ee
where $\varphi$ and $\varphi^\dag$ are analytical functions of $\phi^i$ and
$\phi^\dag_\is$, respectively. The derivatives are hence 
made covariant by introducing
appropriate elements of the connection $\Gamma$. Working out the structure of
the connection, it is found out that its only non-vanishing components 
are $\Gamma_i{}^j{}_k$ and $\Gamma^\is{}_\js{}^\ks$. Consequently, 
the covariant derivatives appearing in Eq.\ \eqref{eq:K} have the structure
\be \bsp 
  {\cal D}_\mu \psibar_\is =\del_\mu \psibar_\is + \Gamma^\js{}_\is{}^\ks 
    \del_\mu \phi^\dag_\js \psibar_\ks \ , &\qquad 
  {\cal D}_\mu \psi^i = \del_\mu \psi^i + \Gamma_j{}^i{}_k \del_\mu \phi^j 
    \psi^k  \ , \\  
  \D^\is K^\js = K^{\is \js} -\Gamma^\is{}_\ks{}^\js K^\ks \quad\ , &\qquad 
  \D_i K_j = K_{ij} -\Gamma_i{}^k{}_j K_k \ ,
\esp\label{eq:covderK}\ee
where elements of the connection such as $\Gamma^{\is j}{}_k$ do not appear. 
In addition, fourth-order derivatives of the K\"ahler potential are related to
the components of the curvature tensor derived from commutators of
covariant derivatives. It can be shown that its elements of the form
$R_i{}^\js{}_k{}^\ls$, relevant for our purposes, obey to the relation
\be 
  R_i{}^\js{}_k{}^\ls = K^{\js \ls}{}_{ik} - K^\ms{}_{n} \Gamma^\js{}_\ms{}^\ls 
   \Gamma_i{}^n{}_k \ . 
\ee

Inspecting more into details Eq.\ \eqref{eq:K}, several terms such as, \eg,
$K_i F^i + \frac12 K_{ij} \psi^i\!\cdot\!\psi^j$, are not covariant with
respect to the K\"ahler manifold. However, solving the equations of motion for
the auxiliary fields and inserting the solution in the expansion of the K\"ahler
potential of Eq.~\eqref{eq:K} render them
fully covariant, and lead, \eg, to four fermions couplings to the curvature
tensor.

Mass and interaction terms among chiral superfields can be added through the
$F$-term of a quantity dubbed the superpotential. This object is, in its
most general form, an arbitrary holomorphic function $W(\Phi)$ depending on the
chiral superfield content. It is hence itself a chiral superfield and its
$F$-term is therefore a good supersymmetric Lagrangian candidate, as mentioned
above. The expansion of the
superpotential in terms of the Grassmann variables as well as the one of the
conjugate function
$W^\star(\Phi^\dag)$ is derived in a similar way as the computations leading to
Eq.\ \eqref{eq:expW}, 
\be \label{eq:W}\bsp
  W(\Phi) =&\  W +   \sqrt{2} \theta\!\cdot\! W_i \psi^i 
    - \theta\!\cdot\!\theta \Big[ F^i W_i + \frac12 W_{ij} \psi^i\!\cdot\!\psi^j 
    \Big] \ ,  \\
  W^\star(\Phi^\dag) =&\  W^\star + \sqrt{2} \thetabar\!\cdot\! W^{\star\is}
    \psibar_\is - \thetabar\!\cdot\!\thetabar \Big[ F^\dag_\is W^{\star\is} + 
     \frac12 W^{\star\is\js} \psibar_\is\!\cdot\!\psibar_\js 
    \Big] \ ,
\esp\ee
where the symbols $W$ and $W^\star$ stand for $W\equiv W(\phi)$ and
$W^\star\equiv W^\star(\phi^\dag)$. The shorthand
notations for the derivatives of the
superpotential follow the same structure as those of the K\"ahler potential,
\ie,
\be 
  W_i   = \frac{\del   W(\phi)}{\del \phi^i} \ , \quad 
  W_{ij}= \frac{\del^2 W(\phi)}{\del \phi^i \del \phi^j} \ , \quad 
  W^{\star\is} = \frac{\del W^\star(\phi^\dag)}{\del \phi^\dag_\is} 
  \quad \text{and} \quad
  W^{\star\is\js}= \frac{\del^2 W^\star(\phi^\dag)}{\del \phi^\dag_\is \del
    \phi^\dag_\js} \ . 
\label{eq:derW}\ee
The corresponding action ${\cal S}_{\rm int}$ is directly built from a
six-dimensional integration upon the (chiral and antichiral) superspace
\be 
  {\cal S}_{\rm int} = \int \d^4 x\ \d^2 \theta \ W(\Phi)  + 
    \int \d^4 x\ \d^2 \thetabar \ W^\star(\Phi^\dag) \ .
\label{eq:Sint}\ee 
Finally, in the renormalizable case, the
superpotential is reduced to functions at most trilinear in the chiral
superfields to avoid higher-dimensional non-renormalizable operators appearing
in its expansion.

Collecting the results of Eq.\ \eqref{eq:SK}, Eq.\ \eqref{eq:K}, Eq.\
\eqref{eq:W} and Eq.\ \eqref{eq:Sint}, the most general action describing the
dynamics of chiral and antichiral superfields in
interaction is built from two fundamental functions, the K\"ahler
potential $K$ and the superpotential $W$ (coming with its Hermitian conjugate 
counterpart $W^\star$), 
\be\boxed{
  {\cal S} = \int \d^4 x\ \d^2 \theta\ \d^2 \thetabar\ K(\Phi, \Phi^\dag)  + 
    \int \d^4 x\ \d^2 \theta \ W(\Phi)  + 
    \int \d^4 x\ \d^2 \thetabar \ W^\star(\Phi^\dag) \ . 
}\label{eq:chiralaction}\ee
The corresponding Lagrangian ${\cal L}$ is derived from Eq.\ \eqref{eq:K}, and
reads, after omitting total derivatives,
\be\boxed{\bsp
  {\cal L} =&\
      K^\js{}_i \del_\mu \phi^i \del^\mu \phi^\dag_\js  
      + K^\js{}_i F^i F^\dag_\js
  + \frac{i}{2} \big(  
      K^\js{}_i \psi^i \sigma^\mu {\cal D}_\mu \psibar_\js  
    - K^\js{}_i {\cal D}_\mu\psi^i \sigma^\mu \psibar_\js\big) 
\\&\
  - F^i W_i 
  - F^\dag_\is W^{\star\is} 
  - \frac12 W_{ij} \psi^i\!\cdot\!\psi^j
  - \frac12 W^{\star\is\js} \psibar_\is\!\cdot\!\psibar_\js \ .
\\&\  
  + \frac12 K^\js{}_i \Gamma^\ks{}_\js{}^\ls F^i \psibar_\ks\!\cdot\!\psibar_\ls 
  + \frac12 K^\is{}_\ell \Gamma_i{}^\ell{}_j F^\dag_\is
    \psi^i\!\cdot\!\psi^j
  + \frac14 K^{\ks \ls}{}_{ij} \psi^i\!\cdot\!\psi^j
    \psibar_\ks\!\cdot\!\psibar_\ls \ .
\esp}\label{eq:chirallag}\ee
where the derivatives of the K\"ahler potential and those of
the superpotential are defined as in Eq.\ \eqref{eq:derK} and Eq.\
\eqref{eq:derW} and where the covariant derivatives acting on the component fields are
defined in Eq.\ \eqref{eq:covderK}. 
As already introduced above,
the equations of motion for the auxiliary fields $F^i$ and $F^\dag_\is$ are
subsequently solved and the solutions are inserted in the Lagrangian ${\cal L}$,
which renders it explicitly covariant with respect to the K\"ahler manifold.

\subsection{Vector Lagrangians}\label{sec:Lvector}
When introducing gauge interactions, the Lagrangian of
Eq.\ \eqref{eq:chirallag} must be made covariant with respect to
gauge transformations. This is addressed in details in Section
\ref{sec:mattergauge} and yields, through the supersymmetric version of the N\oe
ther procedure, the introduction of a set of vector superfields in the adjoint
representation of the gauge group.
Kinetic and gauge interaction terms must be added for
the component fields of these vector superfields, which is the scope of this
section. Abelian and non-abelian supersymmetric gauge theories have been
first constructed in the pioneering works of Refs.\ \cite{Salam:1974jj,
Fayet:1976cr, Salam:1974ig, Wess:1974jb, Ferrara:1974pu}, both in terms of
component fields and within the superspace formalism.

We first address the abelian case, adopting the Wess-Zumino gauge.
From the expansion in terms of component fields of a vector superfield $V$ as
presented in Eq.\ \eqref{eq:VSF}, one can calculate the first powers of $V$,
\be\label{eq:VSFpow}
  V^2 = \frac12 \theta \!\cdot\! \theta \ \thetabar \!\cdot\! \thetabar\ 
    v^\mu v_\mu \qquad\text{and}\qquad V^3=0 \ ,
\ee
using the relations of Eq.\ \eqref{eq:thetarelations} and where we recall that $v^\mu$ stands
for the vector component field of $V$. It is therefore clear
that neither $V$ nor any of its powers are appropriate to generate kinetic
terms for the degrees of freedom included in gauge supermultiplets and new
quantities have to be constructed. To this aim, we
employ the superderivatives presented in Eq.~\eqref{eq:schsder} and we
account, in addition to supersymmetry invariance, for invariance 
under (abelian) gauge transformations (see Section \ref{sec:VSF}),
\be
  V \to V -i (\Lambda - \Lambda^\dag) \ ,
\label{eq:abgaugetr}\ee
denoting the transformation parameters by a pair of Hermitian
conjugate chiral and antichiral superfields $\Lambda$ and $\Lambda^\dag$. 
The lowest-order derivatives of the vector superfield $V$ satisfying 
gauge invariance are of the third-order, 
\be\label{eq:Wabelian}
  W_\alpha = -\frac14 \Dbar \!\cdot\! \Dbar D_\alpha V 
  \qquad\text{and}\qquad
  \Wbar_\alphadot = \frac14  D \!\cdot\! D \Dbar_\alphadot V\ ,
\ee
where the normalization factors are conventional.
Those (Hermitian conjugate) spinorial superfields are chiral and antichiral,
respectively, since 
\be
  \Dbar_\alphadot W_\alpha =  D_\alpha \Wbar_\alphadot = 0 \ .
\ee
From the
expressions of the superderivatives given in Eq.\ \eqref{eq:schsder} and from
the expansion of $V$ in terms of the Grassmann variables,
one can compute the expansion of the $W_\alpha$ and $\Wbar_\alphadot$ quantities
in terms of the component fields of $V$. In the
Wess-Zumino gauge presented in Eq.~\eqref{eq:VSF}, the latter consist of
the gaugino field
$(\lambda,\lambar)$, the vector field $v$ and the auxiliary field $D$. Following
those notations, the
expansions of $W_\alpha$ and $\Wbar_\alphadot$ are given by
\be\bsp 
  W_\alpha =&\ 
    - i \lambda_\alpha
    + \Big[ 
        - \frac{i}{2} (\sigma^\mu\sibar^\nu\theta)_\alpha F_{\mu\nu}
        + \theta_\alpha D
      \Big] 
    - \theta\!\cdot\!\theta \big(\sigma^\mu \del_\mu \lambar\big)_\alpha  \ , \\ 
  \Wbar_\alphadot =&\ 
    i \lambar_\alphadot
    + \Big[ 
        - \frac{i}{2} (\thetabar\sibar^\mu\sigma^\nu)_\alphadot F_{\mu\nu}
        + \thetabar_\alphadot D
      \Big] 
    - \thetabar\!\cdot\!\thetabar
        \big(\del_\mu\lambda\sigma^\mu\big)_\alphadot \ ,  
\esp\label{eq:WandWbar}\ee 
where we have introduced the abelian field strength tensor
\be
  F_{\mu\nu} = \del_\mu v_\nu - \del_\nu v_\mu  \ .
\ee
This suggests that the $W_\alpha$ and $\Wbar_\alphadot$ objects are
the supersymmetric counterparts of the gauge field strength tensor and are
therefore 
often called superfield strength tensors. We remind that in Eq.\
\eqref{eq:WandWbar}, all the fields depend on the (omitted for clarity)
$y$-variables and not on the usual
spacetime coordinates. As for non-supersymmetric quantum field theories, the
gauge action is built from squaring the superfield strength tensors, 
\be\label{eq:WW}\bsp
  W^\alpha W_\alpha = &\  
     - \lambda\!\cdot\!\lambda  
     + \theta\!\cdot\!
       \Big[\sigma^\mu \sibar^\nu \lambda\ F_{\mu\nu} - 2 i \lambda D \Big] 
     + \theta\!\cdot\!\theta \Big[ 
         D^2 - \frac12 \big(F_{\mu\nu} F^{\mu\nu}  + i F_{\mu \nu} \tF^{\mu
          \nu}\big) + 2 i \lambda \sigma^\mu \del_\mu \lambar 
       \Big] \ , \\
  \Wbar_\alphadot \Wbar^\alphadot = &\ 
     - \lambar\!\cdot\!\lambar 
     + \thetabar\!\cdot\!
        \Big[2 i \lambar D - \sibar^\mu \sigma^\nu \lambar\ F_{\mu \nu} \Big] 
     + \thetabar\!\cdot\!\thetabar \Big[ 
         D^2 - \frac12 \big(F_{\mu\nu} F^{\mu\nu} - i F_{\mu\nu} \tF^{\mu\nu}
         \big)  - 2 i \del_\mu\lambda \sigma^\mu \lambar 
       \Big] \  .
\esp\ee
To derive those expressions, we have employed the relations of Eq.\
\eqref{eq:idsigma} and introduced the dual gauge field strength tensor
\be\label{eq:tfmunu}
  \tF^{\mu \nu} = \frac12 \e^{\mu\nu\rho\sigma} F_{\rho\sigma} \ . 
\ee
Inspecting Eq.\ \eqref{eq:WW}, one observes that the sum of the highest order
component fields of the two computed quantities
gives the
expected kinetic Lagrangian terms for the component fields of the vector
superfield $V$, after introducing an additional factor of $1/4$ to
get a correct normalization, 
\be
  {\cal L} = -\frac14 F^{\mu\nu} F_{\mu\nu} 
     + \frac{i}{2}(\lambda \sigma^\mu \del_\mu\lambar - \del_\mu
       \lambda\sigma^\mu \lambar) + \frac12 D^2  \ .
\label{eq:lVab}\ee
As for the chiral case, the kinetic term associated to the auxiliary $D$-field
ensures that it is non-propagating and its
equations of motion are $D=0$, so that it vanishes when going on-shell, like
the $F$-field in the case of chiral actions (see Section \ref{sec:Lchiral}). 

In order to generalize the results presented above to the non-abelian case, we
introduce a non-abelian Lie algebra $\g$ and a representation 
specified by the generators $T_a$ fulfilling standard commutation relations 
\be 
  \big[ T_a, T_b \big] = i f_{ab}{}^c T_c \ , 
\ee 
where $f_{ab}{}^c$ are the antisymmetric structure constants of the algebra.
We associate a vector superfield $V^a$ with each representation matrix and define
$V=V^a T_a$, so that the superfield $V$ is now naturally
endowed with a gauge
invariance structure. Gauge transformation laws for vector superfields are
derived from the N\oe ther procedure leading to the covariantization of the
chiral action of Eq.\ \eqref{eq:chiralaction}. They are given, at the finite
level, by
\be
  e^{2 g V} \to e^{-2 i g \Lambda} e^{2 g V} e^{2 i g \Lambda^\dag}
  \qquad\text{and}\qquad
  e^{-2 g V} \to e^{-2 i g \Lambda^\dag} e^{-2 g V} e^{2 i g \Lambda}
\label{eq:gaugeV}\ee
where the conjugate chiral and antichiral superfields $\Lambda =
\Lambda^a T_a$ and $\Lambda^\dag=\Lambda^{a\dag} T_a$ are
the transformation
parameters. In our conventions, we have explicitly introduced the associated
gauge coupling constant $g$ and factors of two. Expanding these expressions at
the first order in the parameters $\Lambda$ and $\Lambda^\dag$ and employing the
Baker-Campbell-Hausdorff identity, the variation of $V$ under a gauge
transformation is computed as
\be
  \delta V = i\ \text{ad}(gV) \!\cdot\! (\Lambda + \Lambda^\dag) 
    - i\ \text{ad}(gV)\ \text{coth} [\text{ad}(gV)] \!\cdot\! (\Lambda-
      \Lambda^\dag) \ ,
\ee
where the operator ad is defined by  $\text{ad}(X) \!\cdot\! Y \equiv [X,Y]$.
Imposing that the scalar component of $\Lambda-\Lambda^\dag$ vanishes (or
equivalently, that the scalar component of $\Lambda$ is a real scalar field) and
that 
there exists a Wess-Zumino gauge where the expansion of the superfield $V$ is
expressed as in Eq.\ \eqref{eq:VSF}, the variation of $V$ can be simplified to 
\be
  \delta V = -i (\Lambda-\Lambda^\dag) + i g \big[V,\Lambda+\Lambda^\dag\big] \
   ,
\label{eq:nagaugetra}\ee
which generalizes the abelian limit of Eq.\ \eqref{eq:abgaugetr}.

In order to build super-Yang-Mills Lagrangians describing the dynamics of the
components of non-abelian vector superfields, one must generalize the
superfield strength tensors of Eq.~\eqref{eq:WandWbar} to the non-abelian case. 
This is achieved by computing (third-order) superderivatives of the quantity
$\exp[-2 g V]$,
\be\bsp
  W_\alpha =&\ -\frac14 \Dbar\!\cdot\!\Dbar e^{2gV} D_\alpha e^{-2gV}
   \ ,\\
  \Wbar_\alphadot =&\ -\frac14 D\!\cdot\!D e^{-2gV} \Dbar_\alphadot e^{2gV} \ ,
\esp\label{eq:defWWbar}\ee
that transform, under a gauge transformation, as
\be\bsp
  W_\alpha \to&\ e^{-2ig\Lambda} W_\alpha e^{2ig\Lambda} \ , \\
  \Wbar_\alphadot \to&\ e^{-2ig\Lambda^\dag} \Wbar_\alphadot e^{2ig\Lambda^\dag}\
  .
\esp \ee
This generalizes the usual transformation laws of the field
strength tensors $F_{\mu\nu}^a T_a$ under a gauge symmetry operation. 
As for non-supersymmetric gauge theories, the trace of the squared
superfield strength tensors is gauge invariant and thus a good candidate for
constructing Lagrangians for the gauge sector. The computation of the components 
of the superfield strength tensors relies on the properties of the
Wess-Zumino gauge of Eq.\ \eqref{eq:VSFpow}, so that
\be 
  e^{-2gV} = 1 - 2gV + 2 g^2 V^2 \qquad\text{and}\qquad
  e^{ 2gV} = 1 + 2gV + 2 g^2 V^2 \ ,
\label{eq:expV}\ee
which allows to get 
\be\bsp
    W_\alpha^a =&\ -2g \bigg[
    -i \lambda_\alpha^a 
    - \frac{i}{2} (\sigma^\mu \bar \sigma^\nu \theta)_\alpha F_{\mu \nu}^a 
    + \theta_\alpha D^a
    - \theta \!\cdot\! \theta (\sigma^\mu D_\mu \bar \lambda^a)_\alpha \bigg] \ , \\
  \Wbar^a_\alphadot =&\ -2g \bigg[
    i \lambar_\alphadot^a
    - \frac{i}{2} (\thetabar\sibar^\mu\sigma^\nu)_\alphadot F_{\mu\nu}^a
    + \thetabar_\alphadot D^a
    - \thetabar\!\cdot\!\thetabar
        \big(D_\mu\lambda^a\sigma^\mu\big)_\alphadot \bigg] \ .  
\esp\label{eq:WandWbarna}\ee
We have introduced in the two relations of Eq.\ \eqref{eq:WandWbarna}, the
non-abelian field strength tensors and the covariant derivative in the
adjoint representation%
\renewcommand{\arraystretch}{1.2}%
\be\begin{array}{c}
  D_\mu \lambda^a = \del_\mu \lambda^a  + g f_{bc}{}^a v_\mu^b \lambda^c \ ,
  \qquad
  D_\mu \lambar^a = \del_\mu \lambar^a  + g f_{bc}{}^a v_\mu^b \lambar^c \ , \\
  F_{\mu\nu}^a = \del_\mu v_\nu^a - \del_\nu v_\mu^a + g f_{bc}{}^a v_\mu^b
    v^c_\nu \ .
\end{array}\label{eq:gaugecov}\ee%
\renewcommand{\arraystretch}{1.}%
Squaring the superfield strength tensors, one subsequently obtains
\be\label{eq:wwwbwb}\bsp
  \frac{1}{4 g^2} W_a^\alpha W^a_\alpha = &\  
     - \lambda_a\!\cdot\!\lambda^a  
     + \theta\!\cdot\!
       \Big[\sigma^\mu \sibar^\nu \lambda_a\ F^a_{\mu\nu} - 2 i \lambda_a D^a
        \Big]
\\&\qquad
     + \theta\!\cdot\!\theta \Big[ 
         D^a D_a - \frac12 \big(F^a_{\mu\nu} F_a^{\mu\nu}  + i F^a_{\mu \nu}
         \tF_a^{\mu\nu}\big) + 2 i \lambda_a \sigma^\mu D_\mu \lambar^a 
       \Big] \ , \\
  \frac{1}{4 g^2} \Wbar^a_\alphadot \Wbar_a^\alphadot = &\ 
     - \lambar_a\!\cdot\!\lambar^a 
     + \thetabar\!\cdot\!
        \Big[2 i \lambar_a D^a - \sibar^\mu \sigma^\nu \lambar_a\ F^a_{\mu \nu}
        \Big]
\\&\qquad
     + \thetabar\!\cdot\!\thetabar \Big[ 
         D^a D_a - \frac12 \big(F^a_{\mu\nu} F_a^{\mu\nu} - i F^a_{\mu\nu}
         \tF_a^{\mu\nu} \big)  - 2 i D_\mu\lambda^a \sigma^\mu \lambar_a 
       \Big] \  ,
\esp\ee
where the non-abelian dual field strength tensor $\tF_a^{\mu\nu}$ is given as in
Eq.\ \eqref{eq:tfmunu},
\be\label{eq:tfmununa}
  \tF_a^{\mu \nu} = \frac12 \e^{\mu\nu\rho\sigma} F_{\rho\sigma a} \ . 
\ee
Summing the highest order components of the superfield products calculated in
Eq.\ \eqref{eq:wwwbwb} allows to 
generalize the Lagrangian of Eq.\ \eqref{eq:lVab} to the non-abelian case,
\be
  {\cal L} = -\frac14 F_a^{\mu\nu} F^a_{\mu\nu} 
     + \frac{i}{2}(\lambda_a \sigma^\mu D_\mu\lambar^a - D_\mu
       \lambda_a\sigma^\mu \lambar^a) + \frac12 D^a D_a  \ ,
\ee
after including, as in the abelian case, an extra factor of $1/4$ in order to
get standard 
normalizations for kinetic and gauge interaction terms. Yang-Mills actions,
written in terms of six-dimensional integrals over the chiral (and antichiral)
superspace coordinates, read, under their most general form,
\be\boxed{
  {\cal S}_{\rm SYM} = 
    \int\d^4 x\ \d^2 \theta \ 
        \frac{h_{ab}(\Phi) }{16 g^2} W^{\alpha a} W_\alpha^b  +
    \int\d^4 x\ \d^2 \thetabar \ \frac{h^\star_{ab}(\Phi^\dag)}{16 g^2}
        \Wbar_\alphadot^a \Wbar^{\alphadot b} \ ,
}\label{eq:SYM}\ee
which reduces in the abelian case to 
\be\boxed{
  {\cal S}_{U(1)} = 
    \int\d^4 x\ \d^2 \theta \ \frac14 W^\alpha W_\alpha  +
    \int\d^4 x\ \d^2 \thetabar \ \frac14 \Wbar_\alphadot\Wbar^\alphadot \ .
}\label{eq:SU1}\ee
In the super-Yang-Mills case, we have
explicitly introduced the gauge kinetic function $h$ and its Hermitian
conjugate counterpart $h^\star$ which depend on the chiral content of the
theory \cite{Bagger:1982fn, Hull:1985pq, wb}. For renormalizable theories, the
gauge kinetic function takes a very simple form, $h_{ab} = \delta_{ab}$, but can
be more complicated in the case of non-renormalizable theories.

We now generalize the expressions of Eq.\ \eqref{eq:wwwbwb} by including the
gauge kinetic function and its component fields. Following standard superspace 
techniques \cite{livre, sugra}, we first expand the gauge kinetic function and
the Hermitian conjugate function as in Eq.\ \eqref{eq:expW},
\be \bsp
  h_{ab}(\Phi) =&\  h_{ab} +   \sqrt{2} \theta\!\cdot\! h_{abi} \psi^i 
    - \theta\!\cdot\!\theta \Big[ F^i h_{abi} + \frac12 h_{abij}
      \psi^i\!\cdot\!\psi^j \Big] \ ,  \\
  h_{ab}^\star(\Phi^\dag) =&\  h_{ab}^\star + \sqrt{2} \thetabar\!\cdot\!
      h_{ab}^{\star\is} \psibar_\is - \thetabar\!\cdot\!\thetabar \Big[
      F^\dag_\is h_{ab}^{\star\is} + 
     \frac12 h_{ab}^{\star\is\js} \psibar_\is\!\cdot\!\psibar_\js 
    \Big] \ ,  
\esp\ee
after introducing the notations $h_{ab}\equiv h_{ab}(\phi)$ and
$h^\star_{ab}\equiv h^\star_{ab}(\phi^\dag)$. As in Section \ref{sec:Lchiral},
we denote the scalar components of the chiral and antichiral superfields
$\Phi^i$ and $\Phi^\dag_\is$ by $\phi^i$ and $\phi^\dag_\is$, their fermionic
components by $\psi^i$ and $\psibar_\is$ and the auxiliary pieces by
$F^i$ and $F^\dag_\is$. In the two expressions above, we have also defined the
first-order and second-order derivatives of the gauge kinetic function as
\be \label{eq:derh}
  h_{abi}  = \frac{\del h_{ab}(\phi)}{\del \phi^i}\ , \quad  
  h_{abij} = \frac{\del^2 h_{ab}(\phi)}{\del \phi^i \del \phi^j} \ , \quad
  h^{\star\is}_{ab}  = \frac{\del h^\star_{ab}(\phi^\dag)}{\del \phi^\dag_\is}
  \quad\text{and}\quad 
  h^{\star\is\js}_{ab} = \frac{\del^2 h^\star_{ab}(\phi^\dag)}{\del
    \phi^\dag_\is \del \phi^\dag_\js} \ .
\ee 
Computing the superfield products, one gets
\be\bsp 
  & \frac{h_{ab}(\Phi)}{4 g^2} W^{a\alpha}  W^b_\alpha =
   -h_{ab} \lambda^a\!\cdot\!\lambda^b  
    + \theta\!\cdot\! \Big[ 
      h_{ab} \big(\sigma^\mu \sibar^\nu \lambda^a F^b_{\mu \nu}  
      -2 i \lambda^a D^b\big) 
      - \sqrt{2} h_{abi} \lambda^a\!\cdot\!\lambda^b \psi^i 
    \Big] \\ 
  &\qquad + \theta\!\cdot\!\theta \Big[ 
      h_{ab} D^a D^b
    - \frac12 h_{ab} \big(F^a_{\mu \nu} F^{b \mu \nu} 
    + i F^a_{\mu \nu} \tF^{b\mu \nu} \big)
    + 2 i h_{ab} \lambda^a \sigma^\mu D_\mu \lambar^b   
\\ &\qquad
    +   \big(F^i h_{abi} + \frac12 h_{abij} \psi^i\!\cdot\!\psi^j\big) 
        \lambda^a\!\cdot\!\lambda^b 
    +\sqrt{2} h_{abi} \big(-\frac12 \psi^i \sigma^\mu \sibar^\nu \lambda^a 
       F^b_{\mu\nu}  + i \psi^i\!\cdot\!\lambda^a D^b \big)
   \Big] \ ,   \\
  & \frac{h^\star_{ab}(\Phi^\dag)}{4 g^2} \Wbar_\alphadot^a \Wbar^{\alphadot b}=
   -h^\star_{ab} \lambar^a\!\cdot\!\lambar^b  
    + \thetabar\!\cdot\! \Big[ 
      h^\star_{ab} \big(2 i \lambar^a D^b - \sibar^\mu \sigma^\nu \lambar^a\
      F^b_{\mu \nu}\big) 
      - \sqrt{2} h^{\star\is}_{ab} \lambar^a\!\cdot\!\lambar^b \psibar_\is 
    \Big] \\ 
  &\qquad + \thetabar\!\cdot\!\thetabar \Big[ 
      h^\star_{ab} D^a D^b
    - \frac12 h^\star_{ab} \big(F^a_{\mu \nu} F^{b \mu \nu} 
    - i F^a_{\mu \nu} \tF^{b\mu \nu} \big)
    - 2 i h^\star_{ab} D_\mu\lambda^a \sigma^\mu \lambar^b   
\\ &\qquad
    +   \big(F^\dag_\is h^{\star\is}_{ab} + \frac12 h^{\star\is\js}_{ab}
       \psibar_\is\!\cdot\!\psibar_\js\big) \lambar^a\!\cdot\!\lambar^b 
    + \sqrt{2} h^{\star\is}_{ab} \big(\frac12 \psibar_\is \sibar^\mu \sigma^\nu
       \lambar^a F^b_{\mu\nu}  - i \psibar_\is\!\cdot\!\lambar^a D^b \big)
   \Big] \ .
\esp\label{eq:hWW}\ee
This allows to deduce the most general Lagrangian for supersymmetric
Yang-Mills theories built from the knowledge of one single fundamental function,
the gauge kinetic function $h$ and its conjugate counterpart $h^\star$, %
\renewcommand{\arraystretch}{1.2}%
\be\boxed{\bsp 
  {\cal L} =&\ \frac12 \Re\{h_{ab}\} D^a D^b
    \!-\! \frac14 \Re\{h_{ab}\} F^a_{\mu \nu} F^{b \mu \nu} 
    \!+\! \frac14 \Im\{h_{ab}\} F^a_{\mu \nu} \tF^{b\mu \nu} 
    \!-\! \frac12 \Im\{h_{ab}\} D_\mu\big(\lambda^a \sigma^\mu \lambar^b \big)
\\&\
    + \frac{i}{2} \Re\{h_{ab}\} \Big[
        \lambda^a\sigma^\mu D_\mu\lambar^b - D_\mu\lambda^a\sigma^\mu\lambar^b   
      \Big]
    +  \frac14 \big(F^i h_{abi} + \frac12 h_{abij} \psi^i\!\cdot\!\psi^j\big) 
        \lambda^a\!\cdot\!\lambda^b 
\\ &\
    +  \frac14 \big(F^\dag_\is h^{\star\is}_{ab} + \frac12 h^{\star\is\js}_{ab}
       \psibar_\is\!\cdot\!\psibar_\js\big) \lambar^a\!\cdot\!\lambar^b 
    +\frac{\sqrt{2}}{4} h_{abi} \big(i \psi^i\!\cdot\!\lambda^a D^b  
       \!-\! \frac12 \psi^i \sigma^\mu \sibar^\nu \lambda^a F^b_{\mu\nu} \big)
\\&\
    \!+\! \frac{\sqrt{2}}{4} h^{\star\is}_{ab} \big(\frac12 \psibar_\is \sibar^\mu
        \sigma^\nu \lambar^a F^b_{\mu\nu} \!-\! i \psibar_\is\!\cdot\!\lambar^a
        D^b \big) \ ,
\esp}\label{eq:SYMlag}\ee%
\renewcommand{\arraystretch}{1.}%
where the derivatives of the gauge kinetic functions are defined by Eq.\
\eqref{eq:derh} and where the gauge covariant derivatives, the field strength tensor and
its dual are given in Eq.~\eqref{eq:gaugecov} and Eq.~\eqref{eq:tfmununa}.
Moreover, the abelian limit is straightforward to obtain, as
well as the renormalizable version of this Lagrangian for which
the gauge kinetic function is given by
\be
  h_{ab}(\Phi) = h^\star_{ab}(\Phi^\dag) = \delta_{ab}  \ .
\ee

\subsection{Gauge interactions of chiral superfields} \label{sec:mattergauge}
In this section, we introduce the procedure allowing to compute the gauge-invariant version
of the chiral Lagrangian of Eq.\ \eqref{eq:chirallag}. 
Promoting field gauge transformation laws at the superfield level, the chiral
and antichiral superfields $\Phi^i$ and $\Phi^\dag_\is$ obey to 
\be 
  \Phi \to e^{-2 i g \Lambda} \Phi
  \qquad\text{and}\qquad
  \Phi^\dag \to \Phi^\dag e^{2 i g \Lambda^\dag} \ ,
\label{eq:gaugevarchir}\ee 
respectively, where the superfield indices are understood and where we recall that the
transformation parameters are defined by $\Lambda = \Lambda^a T_a$, the matrices
$T_a$
specifying the representation of the gauge group 
under consideration in which lies the chiral superfield $\Phi$. Moreover, 
as in the previous subsection, the parameter $g$ denotes the associated coupling constant. 
Consequently, the K\"ahler potential $K(\Phi,\Phi^\dag)$ is 
not a gauge-invariant object, since $\Lambda^\dag$ is in general
different from
$\Lambda$. One can however recover gauge invariance after following a
supersymmetric version of the N\oe ther procedure leading to the
covariantization of the chiral Lagrangian with respect
to gauge transformations,
\be 
 K(\Phi,\Phi^\dag) \to \K \equiv \frac12 \Big[K(\Phi, \Phi^\dag e^{-2gV}) +
   K(e^{-2gV} \Phi,\Phi^\dag) \Big] \ ,
\label{eq:KK}\ee 
where the vector superfield $V$ is defined as $V=V^a T_a$.
From the gauge transformation laws of vector superfields given in Eq.\
\eqref{eq:gaugeV}, this modified version of the K\"ahler potential
is thus a gauge invariant quantity. 

The computation of the expansion of $\K$ in terms of the component fields of
the vector and chiral superfields of the theory can be performed by considering
the K\"ahler potential $K$ as a polynomial in the chiral and antichiral
superfields $\Phi$ and $\Phi^\dag$,  
\be
  K(\Phi,\Phi^\dag) = \sum_{m,n} k^{j^*_1 \cdots j^*_m}{}_{i_1 \cdots 
    i_n} \Phi^\dag_{j_1^*} \cdots \Phi^\dag_{j_m^*} \Phi^{i_1} \cdots \Phi^{i_n}
    \ .
\label{eq:Kexpanded}\ee
As in Section \ref{sec:Lchiral}, we use two different sets of indices for
chiral and
antichiral superfields in order to make the K\"ahler manifold structure related
to the matter supermultiplets apparent. The same index structure is employed for
the component fields of the chiral and antichiral superfields, and we denote, as usual, by
$\phi^i$
and $\phi^\dag_\is$ their scalar components, by $\psi^i$ and $\psibar_\is$ their
fermionic components and by $F^i$ and $F^\dag_\is$ their auxiliary
components. As in Section \ref{sec:Lvector}, the component fields of the vector
superfield $V^a$, taken in the Wess-Zumino gauge, are denoted by $v_\mu^a$,
$(\lambda^a,\lambar^a)$ and $D^a$ for the associated gauge boson, gaugino and
auxiliary fields, respectively. 

To compute the first of the two terms contributing to
$\K$ of Eq.\ \eqref{eq:KK}, we consider
the expansion in terms of the Grassmann variables of one
monomial
term of the expansion of Eq.~\eqref{eq:Kexpanded}. We then perform the
summation~\cite{sugra}, focusing on the quantity 
\be
  \Omega_J \Omega^I  = \bigg[\Phi^\dag_{k_1^*} (e^{-2gV})^{k_1^*}{}_{j_1^*}
    \cdots 
    \Phi^\dag_{k_m^*} (e^{-2gV})^{k_m^*}{}_{j_m^*}\bigg] \bigg[\Phi^{i_1} \cdots
    \Phi^{i_n}\bigg] \ ,
\ee
where the chiral  superfields $\Omega^I$ and general superfield $\Omega_J$ are
defined by
\be
  \Omega_J = \Phi^\dag_{k_1^*} (e^{-2gV})^{k_1^*}{}_{j_1^*} \cdots 
    \Phi^\dag_{k_m^*} (e^{-2gV})^{k_m^*}{}_{j_m^*} \qquad\text{and}\qquad
  \Omega^I  = \Phi^{i_1} \cdots \Phi^{i_n}\ .
\ee
In a first step, we address the expansion in terms of the Grassmann variables of
both superfields $\Omega^I$ and $\Omega_J$ separately. From these results,
we  derive, in a second step, the expansion of the product of these
two superfields.

The computation of $\Omega^I$ is immediate and can be
deduced from Eq.\ \eqref{eq:expW}. After expanding the $y$-variable
in terms of the spacetime coordinates (as shown, \eg, in Eq.\
\eqref{eq:chiralSF}), one gets
\be\bsp 
  \Omega^I =&\ 
      X^I + \sqrt{2} X^I_i\ \theta\!\cdot\!\psi^i - 
      \theta\!\cdot\!\theta \Big[
        X^I_i F^i + \frac12 X^I_{ij} \psi^i\!\cdot\!\psi^j \Big]
      - i X^I_i\ \theta \sigma^\mu \thetabar\ \del_\mu \phi^i 
\\ &\
      - \frac{i}{\sqrt{2}} \theta\!\cdot\!\theta\ \thetabar \sibar^\mu \Big[
         X^I_i \del_\mu \psi^i + X^I_{ij} \psi^i \del_\mu \phi^j \Big] - 
       \frac14 \theta\!\cdot\!\theta\  \thetabar\!\cdot\!\thetabar \Big[
          X^I_{ij} \del_\mu \phi^i \del^\mu \phi^j + X^I_i \square \phi^i \Big]
\ , 
\esp \ee 
where we have introduced the object $X^I$ and its derivatives,
\be
  X^I = \Phi^{i_1} \cdots \Phi^{i_n} \ , \qquad
  X^I_i = \frac{\del X^I}{\del \phi^i} \qquad\text{and}\qquad
  X^I_{ij} = \frac{\del^2 X^I}{\del\phi^i\del\phi^j} \ .
\label{eq:XI}\ee
In contrast, the computation of the antichiral superfield $\Omega_J$ is more
complicated. We start by expanding the exponential factors, which gives,
after introducing explicitly the representation matrices of the gauge
group and employing the properties of the Wess-Zumino gauge of Eq.~\eqref{eq:expV}, 
\be\bsp
  \big(e^{-2gV}\big)^\ks{}_\js =  &\ 
    \delta^\ks{}_\js + g \Big[
       -2 \theta \sigma^\mu \thetabar \ v_\mu^a 
       - 2 i \theta\!\cdot\!\theta \ \thetabar\!\cdot\!\lambar^a  
       + 2 i \thetabar\!\cdot\!\thetabar \ \theta\!\cdot\!\lambda^a
       - \theta\!\cdot\!\theta \ \thetabar\!\cdot\!\thetabar D^a
     \Big] T_a{}^\ks{}_\js
 \\ &\quad
    + \theta\!\cdot\!\theta \ \thetabar\!\cdot\!\thetabar \
       g^2 v_\mu^a v^{\mu b} \big(T_a T_b\big)^\ks{}_\js \ .
\esp\label{eq:expgvbis}\ee
Introducing quantities conjugate to those of Eq.\ \eqref{eq:XI},
\be
  X_J = \Phi^\dag_{j_1^*} \cdots \Phi^\dag_{j_m^*} \ , \qquad
  X_J^\is = \frac{\del X_J}{\del \phi^\dag_\is} \qquad\text{and}\qquad
  X_J^{\is\js} = \frac{\del^2 X_J}{\del\phi^\dag_\is\del\phi^\dag_\js} \ ,
\ee
one gets, using in addition the results of Eq.\ \eqref{eq:chiralSF} and Eq.\
\eqref{eq:expW} and after performing the expansion of the $y^\dag$-variable in terms of the
spacetime coordinates,
\be\bsp 
 & \Omega_J = 
      X_J 
    +  \sqrt{2} X_J^\is\ \thetabar\!\cdot\!\psibar_\is  
    - \thetabar\!\cdot\!\thetabar \ \Big[ X_J^\is F^\dag_\is + \frac12
       X_J^{\is\js} \psibar_\is\!\cdot\!\psibar_\js\Big] 
    + X_J^\is  \theta \sigma^\mu \thetabar 
       \Big[ i \del_\mu \phi^\dag_\is - 2 g (\phi^\dag T_a)_\is v_\mu^a \Big] 
\\&\
   -  2 i g X_J^\is \theta\!\cdot\!\theta\  \thetabar\!\cdot\!\lambar^a 
     (\phi^\dag T_a)_\is 
   +  \thetabar\!\cdot\!\thetabar\ \theta\!\cdot\! \bigg[ 
     X_J^{\is\js} \Big( 
           \sqrt{2} g (\phi^\dag T_a)_\is v_\mu^a \sigma^\mu \psibar_\js
          - \frac{i}{\sqrt{2}} \sigma^\mu \psibar_\is \del_\mu \phi^\dag_\js 
          \Big)
\\&\ 
     +  X_J^\is \Big( 
             \sqrt{2} g \sigma^\mu (\psibar T_a)_\is v_\mu^a  
           - \frac{i}{\sqrt{2}} \sigma^\mu \del_\mu\psibar_\is  
           + 2 i g  (\phi^\dag T_a)_\is \lambda^a\Big) 
      \bigg] 
  + \thetabar\!\cdot\!\thetabar \ \theta\!\cdot\!\theta \bigg[ 
      X_J^\is \Big( 
        - \frac14  \square \phi^\dag_\is 
\\&\
        - i g v^{\mu a} (\del_\mu\phi^\dag T_a)_\is 
        - g D^a (\phi^\dag T_a)_\is 
        + \sqrt{2} i g \lambar^a\!\cdot\! (\psibar T_a)_\is  
        + g^2 v_\mu^a v^{\mu b} (\phi^\dag T_a T_b)_\is \Big) 
     + X_J^{\is\js}
\\&\ \times\Big( 
           \sqrt{2} i g \psibar_\is\!\cdot\!\lambar^a (\phi^\dag T_a)_\js
         - \frac14 \del_\mu\phi^\dag_\is\del^\mu\phi^\dag_\js 
         - i g \del_\mu \phi^\dag_\is (\phi^\dag T_a)_\js v^{\mu a} 
         + g^2 v_\mu^a v^{\mu b}  (\phi^\dag T_a)_\is (\phi^\dag T_b)_\js 
         \Big) 
    \bigg] \ ,  
\esp\ee
The product of the two superfields
$\Omega^I$ and $\Omega_J$ is thus given by 
\be\bsp 
& \Omega_J \Omega^I =  
   X_J X^I 
   + \sqrt{2} X_J X^I_i\ \theta\!\cdot\!\psi^i 
   + \sqrt{2} X_J^\is X^I\ \thetabar\!\cdot\!\psibar_\is  
   - \theta\!\cdot\!\theta  \Big[ 
       \frac12 X_J X^I_{ij}\ \psi^i \!\cdot\! \psi^j + X_J X^I_i\ F^i \Big] 
\\ &\
   - \thetabar\!\cdot\!\thetabar \Big[
       \frac12 X_J^{\is\js} X^I\ \psibar_\is\!\cdot\!\psibar_\js 
       + X_J^\is X^I\ F^\dag_\is \Big]
  +  \theta \sigma^\mu \thetabar \Big[  
       i X_J^\is X^I \big(\del_\mu \phi^\dag_\is
         \!+\! 2 i g (\phi^\dag T_a)_\is v_\mu^a\big)
       - i X_J X^I_i \del_\mu \phi^i 
\\ &\
     +  X_J^\is X^I_i \psi^i \sigma^\mu \psibar_\is \Big]  
     + \theta\!\cdot\!\theta\ \thetabar\!\cdot\! \Big[ 
       \frac{i}{\sqrt{2}} X_J^\is X^I_i \sibar^\mu \psi^i \big(
         \del_\mu\phi^\dag_\is  + 2 i g (\phi^\dag T_a)_\is v_\mu^a \big)
      - \sqrt{2} X_J^\is X^I_i F^i \psibar_\is 
\\ &\
      - \frac{1}{\sqrt{2}} X_J^\is X^I_{ij} \psi^i \!\cdot\! \psi^j \psibar_\is 
      - \frac{i}{\sqrt{2}} \sibar^\mu \big(X_J X^I_i \del_\mu\psi^i 
        +  X_J X^I_{ij} \psi^i \del_\mu \phi^j \big)
      - 2 i g X_J^\is X^I (\phi^\dag T_a)_\is \lambar^a \Big]  
\\ &\
     + \thetabar\!\cdot\!\thetabar\ \theta\!\cdot\! \Big[ 
         \frac{i}{\sqrt{2}} X_J^\is X^I_i \sigma^\mu \psibar_\is \del_\mu\phi^i 
       - \frac{i}{\sqrt{2}} X_J^{\is\js} X^I \sigma^\mu \psibar_\is \big(
          \del_\mu \phi^\dag_\js + 2 i g (\phi^\dag T_a)_\is v_\mu^a \big)
       - \sqrt{2} X_J^\is X^I_i \psi^i F^\dag_\is 
\\&\
       - \frac{1}{\sqrt{2}} X_J^{\is\js}  X^I_i \psibar_\is \!\cdot\!
         \psibar_\js \psi^i 
       - \frac{i}{\sqrt{2}} X_J^\is X^I \sigma^\mu \big( 
          \del_\mu\psibar_\is + 2 i g (\psibar T_a)_\is v_\mu^a \big)
       +  2 i g  X_J^\is X^I (\phi^\dag T_a)_\is \lambda^a 
    \Big] 
\\  &\ 
 + \thetabar\!\cdot\!\thetabar \ \theta\!\cdot\!\theta \Big[ 
         g^2 X_J^\is X^I v_\mu^a v^{\mu b} (\phi^\dag T_a T_b)_\is  
       \!-\! \frac14 \del_\mu \big( X_J X^I_i \del^\mu \phi^i \!+\!
              X_J^\is X^I \del^\mu \phi^\dag_\is \big)
       \!-\! i g X_J^\is X^I v^{\mu a} (\del_\mu \phi^\dag T_a)_\is 
\\ &\
       - i g X_J^{\is\js} X^I \del_\mu \phi^\dag_\is (\phi^\dag T_a)_\js v^{\mu a} 
       + g^2 X_J^{\is\js} X^I v_\mu^a v^{\mu b} (\phi^\dag T_a)_\is 
        (\phi^\dag T_b)_\js 
       + X_J^\is X^I_i F^i F^\dag_\is 
\\ &\
       + X_J^\is X^I_i \del_\mu\phi^i \big(\del^\mu\phi^\dag_\is + i g  
           (\phi^\dag T_a)_\is v^{\mu a} \big)
       - \frac{i}{2} X_J^\is X^I_i \Big(\del_\mu\psi^i \sigma^\mu \psibar_\is 
         - \psi^i \sigma^\mu \big(\del_\mu\psibar_\is + 2 i g  (\psibar
           T_a)_\is v_\mu^a\big) \Big)
\\&\
       + \frac{i}{2} X_J^{\is\js} X^I_i \psi^i \sigma^\mu \psibar_\is 
           \big(\del_\mu\phi^\dag_\js \!+\! 2 i g (\phi^\dag T_a)_\js 
         v_\mu^a \big)
       \!-\! \frac{i}{2} X_J^\is X^I_{ij} \psi^i \sigma^\mu \psibar_\is
         \del_\mu\phi^j 
       \!+\! \frac14 X_J^{\is\js} X^I_{ij} \psibar_\is \!\cdot\! \psibar_\js
         \psi^i\!\cdot\! \psi^j 
\\ &\
       + \frac12 X_J^{\is\js} X^I_i \psibar_\is \!\cdot\! \psibar_\js F^i 
       + \frac12 X_J^\is X^I_{ij} F^\dag_\is \psi^i\!\cdot\! \psi^j 
       - g X_J^\is X^I D^a (\phi^\dag T_a)_\is 
       - \sqrt{2} i g X_J^\is X^I_i \psi^i \!\cdot\! \lambda^a (\phi^\dag T_a)_\is 
\\&\  
       + \sqrt{2} i g X_J^{\is\js} X^I \psibar_\is \!\cdot\! \lambar^a 
           (\phi^\dag T_a)_\js
       + \sqrt{2} i g X_J^\is X^I (\psibar T_a)_{\is}\!\cdot\!\lambar^a 
   \Big] \ . 
\esp\label{eq:OO}\ee 
Summing over all the monomial terms of Eq.\ \eqref{eq:Kexpanded} and adding the
second contribution to ${\cal K}$, \ie, the second term in Eq.\ \eqref{eq:KK}
conjugate to the first one,
we work out the expansion of the gauge-invariant version of the K\"ahler
potential in terms of $\theta$ and $\thetabar$, 
\bea
 {\cal K} &=&
   K 
    + \sqrt{2} K_i \theta\!\cdot\!\psi^i 
    + \sqrt{2} K^\is \thetabar\!\cdot\!\psibar_\is 
    - \theta\!\cdot\!\theta 
     \Big[K_i F^i + \frac12 K_{ij} \psi^i\!\cdot\!\psi^j \Big]
\nn\\&&
    - \thetabar\!\cdot\!\thetabar \Big[
      K^\is F^\dag_\is + \frac12K^{\is\js}\psibar_\is\!\cdot\!\psibar_\js\Big]
    + \theta \sigma^\mu \thetabar \Big[ 
      i K^\is D_\mu \phi^\dag_\is - i K_i D_\mu \phi^i + 
      K^\is{}_i \psi^i \sigma_\mu \psibar_\is \Big]
\nn\\&&
   + \theta\!\cdot\!\theta\ \thetabar\!\cdot\! \Big[ 
      - \frac{i}{\sqrt{2}} \sibar^\mu \big(K_i {\cal D}_\mu \!+\! {\cal D}_{i}
         K_j D_\mu\phi^j\big)\psi^i 
      - \sqrt{2} K^\is{}_i F^i \psibar_\is  
      + \frac{i}{\sqrt{2}} K^\is{}_i \sibar^\mu \psi^i D_\mu \phi^\dag_\is 
\nn\\&&\qquad
      - \frac{1}{\sqrt{2}} K^\is{}_k \Gamma_i{}^k{}_j\ \psi^i\!\cdot\!\psi^j 
          \psibar_\is 
      - i g K^\is \lambar^a (\phi^\dag T_a)_\is 
      - i g K_i \lambar^a (T_a \phi)^i 
    \Big] 
\nn\\&&
   + \thetabar\!\cdot\!\thetabar\ \theta\!\cdot\! \Big[ 
      - \frac{i}{\sqrt{2}} \sigma^\mu \big( K^\is {\cal D}_\mu \!+\! {\cal
           D}^\is K^\js D_\mu \phi^\dag_\js\big) \psibar_\is
      - \sqrt{2} K^\is{}_i F^\dag_\is \psi^i 
      + \frac{i}{\sqrt{2}} K^\is{}_i  \sigma^\mu \psibar_\is D_\mu \phi^i
\nn\\&&\qquad
      - \frac{1}{\sqrt{2}} K^\ks{}_i \Gamma^\is{}_\ks{}^\js  
         \psibar_\is\!\cdot\!\psibar_\js \psi^i 
      + i g K^\is \lambda^a (\phi^\dag T_a)_\is  
      + i g K_i   \lambda^a (T_a \phi)^i  \Big] 
\label{eq:gaugeinvariantK}\\ && 
    + \theta\!\cdot\!\theta \ \thetabar\!\cdot\!\thetabar \Big[ 
      - \frac14 \del_\mu\big(K_i D^\mu \phi^i + K^\is D^\mu \phi^\dag_\is\big)  
      + K^\is{}_i D_\mu \phi^i D^\mu \phi^\dag_\is 
      + K^\is{}_i F^i F^\dag_\is 
\nn\\&&\qquad
      + \frac14 K^{\is \js}{}_{ij} \psi^i\!\cdot\!\psi^j 
        \psibar_\is\!\cdot\!\psibar_\js 
      + \frac{i}{2} \big(  
          K^\is{}_i \psi^i \sigma^\mu {\cal D}_\mu \psibar_\is  
        - K^\is{}_i {\cal D}_\mu\psi^i \sigma^\mu \psibar_\is\big)  
\nn\\&&\qquad
      + \frac12 K^\is{}_i \Big(\Gamma^\js{}_\is{}^\ks F^i
         \psibar_\js\!\cdot\!\psibar_\ks 
      +  \Gamma_j{}^i{}_k F^\dag_\is \psi^j\!\cdot\!\psi^k \Big)
      - \frac{g}{2}D^a\Big( K^\is \big(\phi^\dag T_a\big)_\is 
          + K_i \big(T_a\phi\big)^i\Big) 
\nn\\&&\qquad 
      -\sqrt{2} i g (\phi^\dag T_a)_\is K^\is{}_i \psi^i\!\cdot\!\lambda^a 
      +\sqrt{2} i g \lambar^a\!\cdot\!\psibar_\is K^\is{}_i (T_a \phi)^i 
  \Big] \ .
\nn\eea
To obtain the expression above, we have performed simplifications by means of the relations
\be\bsp 
0 = \delta_a K = &\ K_i (T_a \Phi)^i - K^\is (\Phi^\dag T_a)_\is\ , \\  
0 = \delta_a \delta_b K = &\  
   \Big[ 
     K_{ij} (T_a \Phi)^i (T_b \Phi)^j +  
     K_i (T_a T_b \Phi)^i - 
     K^\is{}_i (T_a\Phi)^i (\Phi^\dag T_b)_\is  
   \Big] + \\ &\ 
   \Big[ 
     K^{\is\js} (\Phi^\dag T_a)_\is (\Phi^\dag T_a)_\js + 
     K^\is (\Phi^\dag T_a T_b)_\is - 
     K^\is{}_i(T_a\Phi)^i (\Phi^\dag T_b)_\is 
   \Big] \ , 
\esp\ee 
which only express the gauge invariance of the K\"ahler potential.
Moreover, we have gathered terms forming covariant derivatives and introduced
the K\"ahler potential $K\equiv K(\phi,\phi^\dag)$ as in Eq.~\eqref{eq:K},
together with its derivatives as defined in Eq.~\eqref{eq:derK}.
However, in contrast to Eq.\ \eqref{eq:K}, the
derivatives of the component
fields of the chiral and antichiral superfields $\Phi^i$ and $\Phi^\dag_\is$ are
now covariant both with respect to the gauge group,
\be\label{eq:covderKg}\begin{array}{ll}
  D_\mu \phi^i = \del_\mu\phi^i - i g v_\mu^a (T_a \phi)^i \ , 
  &D_\mu \psi^i = \del_\mu\psi^i - i g v_\mu^a (T_a \psi)^i \ , \\ 
  D_\mu \phi^\dag_\is = \del_\mu\phi^\dag_\is + igv^a_\mu (\phi^\dag T_a)_\is \ , 
  &D_\mu \psibar_\is  = \del_\mu \psibar_\is + i g v^a_\mu (\psibar T_a)_\is \ ,
\end{array}\ee 
and with respect to the K\"ahler manifold,
\be\label{eq:covderKg2}\bsp
  {\cal D}_\mu \psi^{i} = D_\mu \psi^i + \Gamma_j{}^i{}_k D_\mu\phi^j \psi^k \ , 
  &\qquad 
  {\cal D}_\mu \psibar_\is = D_\mu \psibar_\is + \Gamma^\js{}_\is{}^\ks 
    D_\mu \phi^\dag_\js \psibar_\ks\ ,\\
  \D_i K_j = K_{ij} -\Gamma_i{}^k{}_j K_k\ ,  &\qquad
  \D^\is K^\js = K^{\is \js} -\Gamma^\is{}_\ks{}^\is K^\ks \ . 
\esp\ee

Collecting all the results derived in this section, the gauge-invariant version
of the first term of the supersymmetric chiral action of Eq.~\eqref{eq:chiralaction} reads
\be\boxed{
  {\cal S} = \frac12 \int \d^4 x\ \d^2 \theta\ \d^2 \thetabar\ 
     \Big[K(\Phi, \Phi^\dag e^{-2gV}) + K(e^{-2gV} \Phi,\Phi^\dag) \Big] \ . 
}\ee
Introducing the component fields, the corresponding Lagrangian ${\cal L}$ is
derived from Eq.\ \eqref{eq:gaugeinvariantK},%
\renewcommand{\arraystretch}{1.2}%
\be\boxed{\bsp
  {\cal L}  =&\
    - \frac14 \del_\mu\big(K_i D^\mu \phi^i + K^\is D^\mu \phi^\dag_\is\big)  
      + K^\is{}_i D_\mu \phi^i D^\mu \phi^\dag_\is 
      + K^\is{}_i F^i F^\dag_\is 
\\&\
      + \frac14 K^{\is \js}{}_{ij} \psi^i\!\cdot\!\psi^j 
        \psibar_\is\!\cdot\!\psibar_\js 
      + \frac{i}{2} \big(  
          K^\is{}_i \psi^i \sigma^\mu {\cal D}_\mu \psibar_\is  
        - K^\is{}_i {\cal D}_\mu\psi^i \sigma^\mu \psibar_\is\big)  
\\&\
      + \frac12 K^\is{}_i \Big(\Gamma^\js{}_\is{}^\ks F^i
         \psibar_\js\!\cdot\!\psibar_\ks 
      \!+\!  \Gamma_j{}^i{}_k F^\dag_\is \psi^j\!\cdot\!\psi^k \Big)
      \!-\! \frac{g}{2}D^a\Big( K^\is \big(\phi^\dag T_a\big)_\is 
          \!+\! K_i \big(T_a\phi\big)^i\Big) 
\\&\
      -\sqrt{2} i g (\phi^\dag T_a)_\is K^\is{}_i \psi^i\!\cdot\!\lambda^a 
      +\sqrt{2} i g \lambar^a\!\cdot\!\psibar_\is K^\is{}_i (T_a \phi)^i  \ .
\esp}\label{eq:gaugechiral}\ee%
\renewcommand{\arraystretch}{1.2}%
In the equation above, the derivatives of the K\"ahler potential
are defined as in Eq.\ \eqref{eq:derK}, 
while the covariant derivatives acting on the component fields
are defined as in Eq.\ \eqref{eq:covderKg} and Eq.\ \eqref{eq:covderKg2}.

Mass and interaction terms among the chiral superfields are
still described by the parts of the action derived from the superpotential given
in Section \ref{sec:Lchiral}, with the extra condition that the superpotential
must now be a gauge-invariant quantity. All the previous results for the
superpotential Lagrangian therefore still hold (see Section~\ref{sec:Lchiral}).

\subsection{Non-renormalizable and renormalizable supersymmetric gauge theories}
We collect in this section all the results of Section \ref{sec:Lchiral},
Section \ref{sec:Lvector} and Section \ref{sec:mattergauge} to
construct a generic supersymmetric theory. We firstly fix a gauge
group $G$ and one of its representation $\R$   
spanned by the Hermitian matrices $T_a$. We then 
associate to this gauge group a vector superfield $V =V^a T_a$. Secondly,
we set the chiral sector of the theory which consists of a collection of chiral 
(antichiral) superfields $\Phi^i$ ($\Phi^\dag_\is$), lying in the 
representation ${\cal R}$ (the complex conjugate representation $\bar{\cal R}$) 
of $G$. 

The most general action describing the dynamics of our theory is given
by
\be\bsp 
  {\cal S} =&\  
    \int \d^4 x\ \d^2 \theta\ \d^2 \thetabar\ \frac12 \Big[ K(\Phi, \Phi^\dag 
      e^{-2gV}) + K(e^{-2gV} \Phi, \Phi^\dag) \Big] 
\\ &\quad 
   + \int \d^4 x\ \d^2 \theta \ W(\Phi)  
   +   \int \d^4 x\ \d^2 \thetabar \ W^\star(\Phi^\dag)
\\ &\quad 
 + \frac{1}{16 g^2}\int \d^4 x\ \d^2 \theta\ h_{ab}(\Phi) W^{a\alpha} W^b_\alpha 
 + \frac{1}{16 g^2}\int \d^4 x\ \d^2 \thetabar\ h_{ab}^\star(\Phi^\dag) 
      \Wbar^a_\alphadot \Wbar^{b\alphadot} \ ,
\esp\label{eq:generalsusyaction}\ee 
where we have introduced three fundamental functions of the chiral content of
the theory, \ie, the K\"ahler potential $K(\Phi, \Phi^\dag)$, the gauge kinetic
function $h(\Phi)$ and the superpotential $W(\Phi)$, $K$ being
real and $W$ gauge-invariant. The superfield strength tensors $W_\alpha^a$ and
$\Wbar^a_\alphadot$ are related to the vector superfields $V^a$ 
and are defined as in Eq.\ \eqref{eq:defWWbar}.

To extract the Lagrangian, we introduce the component fields $\phi^i$
($\phi^\dag_\is$), $\psi^i$ ($\psibar_\is$) and $F^i$ ($F^\dag_\is$) as the
scalar, fermionic and auxiliary components of the superfields $\Phi^i$
($\Phi^\dag_\is$), respectively, as well as the vector, the gaugino and the
auxiliary components of the vector superfields $V^a$ which we denote by
$v_\mu^a$, $(\lambda^a,\lambar^a)$ and $D^a$. The most general
non-renormalizable supersymmetric Lagrangian is then written, after omitting
total derivatives, by
\be\boxed{\bsp
  {\cal L} =&\
   K^\is{}_i D_\mu \phi^i D^\mu \phi^\dag_\is 
   + K^\is{}_i F^i F^\dag_\is 
   + \frac{i}{2}K^\is{}_i  \Big[  
       \psi^i \sigma^\mu {\cal D}_\mu \psibar_\is  
   - {\cal D}_\mu\psi^i \sigma^\mu \psibar_\is\Big]  
   - \frac14 \Re\{h_{ab}\} F^a_{\mu \nu} F^{b \mu \nu} 
\\&\
   + \frac14 \Im\{h_{ab}\} F^a_{\mu \nu} \tF^{b\mu \nu} 
   + \frac{i}{2} \Re\{h_{ab}\} \Big[
       \lambda^a\sigma^\mu D_\mu\lambar^b - D_\mu\lambda^a\sigma^\mu\lambar^b   
     \Big]
   - \frac12 \Im\{h_{ab}\} D_\mu\big(\lambda^a \sigma^\mu \lambar^b \big)
\\ &\
   + \frac12 \Re\{h_{ab}\} D^a D^b
   + \frac14 K^{\is \js}{}_{ij} \psi^i\!\cdot\!\psi^j 
     \psibar_\is\!\cdot\!\psibar_\js 
   + \frac12 K^\is{}_i \Big[\Gamma^\js{}_\is{}^\ks F^i
      \psibar_\js\!\cdot\!\psibar_\ks 
   +  \Gamma_j{}^i{}_k F^\dag_\is \psi^j\!\cdot\!\psi^k \Big]
\\&\
      - \frac{g}{2}D^a\Big[ K^\is \big(\phi^\dag T_a\big)_\is 
          + K_i \big(T_a\phi\big)^i\Big] 
      -\sqrt{2} i g K^\is{}_i  \Big[ (\phi^\dag T_a)_\is \psi^i\!\cdot\!\lambda^a 
       - \lambar^a\!\cdot\!\psibar_\is (T_a \phi)^i \Big] 
\\&\
    +  \frac14 \big(F^i h_{abi} + \frac12 h_{abij} \psi^i\!\cdot\!\psi^j\big) 
        \lambda^a\!\cdot\!\lambda^b 
    +  \frac14 \big(F^\dag_\is h^{\star\is}_{ab} + \frac12 h^{\star\is\js}_{ab}
       \psibar_\is\!\cdot\!\psibar_\js\big) \lambar^a\!\cdot\!\lambar^b 
\\&\
    +\frac{\sqrt{2}}{4} h_{abi} \big(i \psi^i\!\cdot\!\lambda^a D^b  
       \!-\! \frac12 \psi^i \sigma^\mu \sibar^\nu \lambda^a F^b_{\mu\nu} \big)
    \!+\! \frac{\sqrt{2}}{4} h^{\star\is}_{ab} \big(\frac12 \psibar_\is \sibar^\mu
        \sigma^\nu \lambar^a F^b_{\mu\nu} \!-\! i \psibar_\is\!\cdot\!\lambar^a
        D^b \big)
\\&\
   - F^i W_i - \frac12 W_{ij} \psi^i\!\cdot\!\psi^j
   - F^\dag_\is W^{\star\is} - \frac12 W^{\star\is\js} \ .
\esp}\label{eq:Lgeneral}\ee
The shorthand notations for the derivatives of the three fundamental functions
are given in Eq.~\eqref{eq:derK}, Eq.~\eqref{eq:derW} and Eq.\ \eqref{eq:derh}. Derivatives
covariant with respect to the gauge group are presented in Eq.\
\eqref{eq:gaugecov} and Eq.\  \eqref{eq:covderKg}, 
\be\begin{array}{ll}
  D_\mu \lambda^a = \del_\mu \lambda^a  + g f_{bc}{}^a v_\mu^b \lambda^c \ ,
  &D_\mu \lambar^a = \del_\mu \lambar^a  + g f_{bc}{}^a v_\mu^b \lambar^c \ , \\
  D_\mu \phi^i = \del_\mu\phi^i - i g v_\mu^a (T_a \phi)^i \ , 
  &D_\mu \phi^\dag_\is = \del_\mu\phi^\dag_\is + igv^a_\mu (\phi^\dag T_a)_\is \
  , \\
  D_\mu \psi^i = \del_\mu\psi^i - i g v_\mu^a (T_a \psi)^i \ , 
  &D_\mu \psibar_\is  = \del_\mu \psibar_\is + i g v^a_\mu (\psibar T_a)_\is \ ,
\end{array}\label{eq:gaugecovbis}\ee
while the derivatives covariant with respect to both the K\"ahler manifold and
the gauge group are presented in Eq.\ \eqref{eq:covderKg2},
\be\bsp
  {\cal D}_\mu \psi^{i} = D_\mu \psi^i + \Gamma_j{}^i{}_k D_\mu\phi^j \psi^k \ , 
  &\qquad 
  {\cal D}_\mu \psibar_\is = D_\mu \psibar_\is + \Gamma^\js{}_\is{}^\ks 
    D_\mu \phi^\dag_\js \psibar_\ks\ ,\\
  \D_i K_j = K_{ij} -\Gamma_i{}^k{}_j K_k\ ,  &\qquad
  \D^\is K^\js = K^{\is \js} -\Gamma^\is{}_\ks{}^\is K^\ks \ . 
\esp\ee 
Finally, Eq.\ \eqref{eq:gaugecov} and Eq.\ \eqref{eq:tfmununa} contain our
conventions for the gauge field strength tensor and its dual, 
\be\label{eq:fmunubis}
   F_{\mu\nu}^a = \del_\mu v_\nu^a - \del_\nu v_\mu^a + g f_{bc}{}^a v_\mu^b
    v^c_\nu \qquad\text{and}\qquad
  \tF_a^{\mu \nu} = \frac12 \e^{\mu\nu\rho\sigma} F_{\rho\sigma a} \ . 
\ee
Solving the equations of motion for the auxiliary fields leads to 
\be\bsp
  &\ F^i = (K^{-1})^i{}_\is W^{\star\is}  - \frac12 \Gamma_j{}^i{}_k \psi^j
    \!\cdot\! \psi^k  \ , \qquad
  F^\dag_\is=(K^{-1})^i{}_\is W_i - \frac12 \Gamma^\js{}_\is{}^\ks \psibar_\js
    \!\cdot\! \psibar_\ks\ ,  \\
  &\ D^a =\big(\Re\big\{h^{-1}\}\big)^{ab} \Big[ 
    \frac12 g \big(K_i (T_b \phi)^i + K^\is (\phi^\dag T_b)_\is \big)
    - \frac{\sqrt{2}i}{4} \big( h_{bci} \psi^i \!\cdot\! \lambda_c 
      - h_{bc}^{\star\is} \psibar_\is \!\cdot\! \lambar_c\big)\Big]\ ,
\esp\label{eq:auxsol}\ee
which gives, after inserting those solutions in the Lagrangian of Eq.\
\eqref{eq:Lgeneral}, additional interactions among the fermions, the scalar
fields and the gauginos. 

In most relevant phenomenological supersymmetric models, it is enough to
consider a renormalizable version of the Lagrangian of Eq.\
\eqref{eq:Lgeneral}. In this case, the K\"ahler potential and the gauge kinetic
function take a simple form, 
\be
  K(\Phi, \Phi^\dag) =  \delta^\is{}_i \Phi^\dag_\is \Phi^i
  \qquad\text{and}\qquad
  h_{ab}(\Phi) = h^\star_{ab}(\Phi^\dag) = \delta_{ab} \ .
\label{eq:normthcond}\ee
In addition, the superpotential is a gauge-invariant function at most trilinear
in the chiral content of the theory,
\be
   W(\Phi) = \frac16 \lambda_{ijk} \Phi^i \Phi^j \Phi^k + \frac12 \mu_{ij}
     \Phi^i \Phi^j + \xi_i \Phi^i \ ,
\label{eq:renosuperW}\ee
where $\lambda$, $\mu$ and $\xi$ are free parameters of the model. The
corresponding action reads thus
\be\boxed{\bsp
  {\cal S} =&\
    \int \d^4 x\ \d^2 \theta\ \d^2 \thetabar\ \Phi^\dag e^{-2gV} \Phi 
   + \int \d^4 x\ \d^2 \theta \ W(\Phi)  
   +   \int \d^4 x\ \d^2 \thetabar \ W^\star(\Phi^\dag)
\\ &\
 + \frac{1}{16 g^2}\int \d^4 x\ \d^2 \theta\ W^{a\alpha} W_{a\alpha }
 + \frac{1}{16 g^2}\int \d^4 x\ \d^2 \thetabar\  
      \Wbar^a_\alphadot \Wbar_a^\alphadot  \ ,
\esp}\label{eq:gensusyaction}\ee
and the associated
Lagrangian is obtained after expanding the superfields in terms of the Grassmann
variables, as shown in Eq.\ \eqref{eq:Lgeneral}. Integrating over $\theta$ and
$\thetabar$ leads to
\be\boxed{\bsp
  {\cal L} =&\
   D^\mu \phi^\dag_i D_\mu \phi^i 
   + \frac{i}{2} \Big[  
       \psi^i \sigma^\mu D_\mu \psibar_i
      \!-\! D_\mu\psi^i \sigma^\mu \psibar_i
      \!+\! \lambda^a\sigma^\mu D_\mu\lambar_a
      \!-\! D_\mu\lambda^a\sigma^\mu\lambar_a   
  \Big]  
   - \frac14 F^a_{\mu \nu} F_a^{\mu \nu} 
\\ &\
   + F^i F^\dag_i
   + \frac12 D^a D_a
      - g D^a \phi_i^\dag T_a{}^i{}_j \phi^j
      -\sqrt{2} i g \Big[ \phi^\dag_i T_a{}^i{}_j \psi^j\!\cdot\!\lambda^a 
       - \lambar^a\!\cdot\!\psibar_i T_a{}^i{}_j \phi^j \Big] 
\\&\
   - F^i W_i - \frac12 W_{ij} \psi^i\!\cdot\!\psi^j
   - F^\dag_i W^{\star i} - \frac12 W^{\star ij} \psibar_i\!\cdot\!\psibar_j
   \ .
\esp}\label{eq:gensusylag}\ee
We remind that the gauge covariant derivatives are given by Eq.\ 
\eqref{eq:gaugecovbis} and the gauge field strength tensor by Eq.\
\eqref{eq:fmunubis}. 
Inserting the solution of the equations of motion for $F^i$ and $D^a$
simplifies the Lagrangian ${\cal L}$ to
\be\boxed{\bsp
  {\cal L} =&\
   D^\mu \phi^\dag_i D_\mu \phi^i 
   + \frac{i}{2} \Big[  
       \psi^i \sigma^\mu D_\mu \psibar_i
      \!-\! D_\mu\psi^i \sigma^\mu \psibar_i
      \!+\! \lambda^a\sigma^\mu D_\mu\lambar_a
      \!-\! D_\mu\lambda^a\sigma^\mu\lambar_a   
  \Big]  
   - \frac14 F^a_{\mu \nu} F_a^{\mu \nu} 
\\ &\
   - \frac12 g^2 \phi_i^\dag T_a{}^i{}_j \phi^j\ \phi_k^\dag T_a{}^k{}_\ell
     \phi^\ell
   -\sqrt{2} i g \Big[ \phi^\dag_i T_a{}^i{}_j \psi^j\!\cdot\!\lambda^a 
       - \lambar^a\!\cdot\!\psibar_i T_a{}^i{}_j \phi^j \Big] 
\\ &\
    - W^{\star i} W_i - \frac12 W_{ij} \psi^i\!\cdot\!\psi^j 
    - \frac12 W^{\star ij}  \psibar_i\!\cdot\!\psibar_j
     \ ,
\esp}\label{eq:gensusylag2}\ee
the abelian limit being again trivially recovered.

\cleardoublepage

%% file: superFR.tex
\label{chap:FR}
The program \feynrules\ has been developed to facilitate the implementation of
new physics theories into high-energy physics tools. Starting from
the model gauge symmetries, particle content,
parameters and Lagrangian, \feynrules\ provides, in the context
of a supersymmetric theory, all necessary routines to derive semi-automatically
the Lagrangian, extract the associated Feynman rules and pass the information
to other tools allowing for phenomenological investigations of the model. We dedicate
this chapter to the description of \feynrules\ and the Universal
\feynrules\ Output format, as widely used in this work to obtain
the main results of this work.

\mysection{The \feynrules\ package}
\subsection{Basic features of the \feynrules\ package}
The program \feynrules\ \cite{Christensen:2008py, Christensen:2009jx,
Christensen:2010wz, Duhr:2011se, Fuks:2012im, Alloul:2013fw, Christensen:2013aua,
Alloul:2013bka} is a \mathematica
\footnote{{\sc
Mathematica}\ is a registered trademark of Wolfram Research, Inc.} package
allowing for the extraction of the Feynman rules from any perturbative quantum
field theory-based Lagrangian in an automated fashion. In a second step, the
Feynman rules can be exported automatically to several matrix element
generators. This procedure allows for phenomenological investigations
of a large class of models whose a hand-made implementation in a Monte
Carlo event generator can be considered as too involved. Currently,
interfaces exist to the \comphep/\calchep\ \cite{Pukhov:1999gg, Boos:2004kh,
Pukhov:2004ca, Belyaev:2012qa}, \feynarts/\formcalc\ \cite{Hahn:1998yk,
Hahn:2000kx, Hahn:2009bf, Agrawal:2011tm}, \mgme\ \cite{Stelzer:1994ta,
Maltoni:2002qb, Alwall:2007st, Alwall:2008pm, Alwall:2011uj}, \sherpa\
\cite{Gleisberg:2003xi, Gleisberg:2008ta} and \whizard\ programs
\cite{Moretti:2001zz, Kilian:2007gr}. In addition, any model can also be
exported as a set of \python\ classes and objects representing particles,
parameters and vertices under the so-called Universal \feynrules\ Output
(UFO)
format \cite{Degrande:2011ua}. The produced \python\ library contains the full
model information, without any restriction on the allowed Lorentz and/or color
structures appearing in the Lagrangian, in contrast to the other interfaces. The
latter indeed reject a vertex which would not be compliant with the structures
supported by the related program. Presently, the UFO format is used by the
\madgraph\ 5 and the \gosam\ \cite{Cullen:2011ac, Cullen:2011xs} generators as
well as by the \madanalysis\ 5 analysis package \cite{Conte:2012fm} and the
\aloha\ program \cite{deAquino:2011ub}. Its use by \herwig++~\cite{Bahr:2008pv,Arnold:2012fq}
is currently being validated.

In order to implement a particle physics model in \feynrules, the user needs to
provide, on the one hand, the particle content and the free parameters of the
model, and on the other hand, the Lagrangian describing the interactions among
the different particles. In the rest of this section, we describe the
basic features of the package and how to implement model files in general,
considering both supersymmetric and non-supersymmetric theories
\cite{Christensen:2008py, Butterworth:2010ym}\footnote{By the time this work
has been completed, additional modules have been supplemented to \feynrules. They
allow for automated spectrum generation~\cite{Alloul:2013fw} and decay width
computations~\cite{Alwall:xxx}. These are not described in this chapter and we refer
to the appropriate references for more information.}
Concerning the implementation of a supersymmetric model and
computations to be performed within the superspace, dedicated functions have
been implemented and are described in Section \ref{sec:spacemod} and in
Refs.~\cite{Duhr:2011se, Fuks:2012im}. We also present how to run the code in
order to derive the interaction vertices and export them to Monte Carlo
event generators \cite{Christensen:2009jx, Christensen:2010wz,
Degrande:2011ua}. For more details on the \feynrules\ package as well as on its
interfaces, we refer the reader to the \feynrules\ webpage \cite{FRwebpage} and
manual~\cite{Christensen:2008py,Alloul:2013bka}.

The \feynrules\ model format is an extension of the model file structure of
\feynarts~\cite{Hahn:2000kx} so that the definitions of particles,
parameters and gauge groups characterizing the model consist of lists of
\mathematica\ replacement rules. The model file itself is a text file
with a \texttt{.fr} extension. This file starts with a preamble containing 
model and author information as well as definitions for the indices 
carried by the fields. The
declaration of the gauge groups, particles and parameters follows,
and the Lagrangian describing the interactions among the particles is eventually
given. The way to
implement the different parts of the model file is described in the next
subsections.

\subsection{Preamble of the model file: information and indices}
\label{sec:FRidx}
The preamble of a \feynrules\ model implementation contains
the two optional variables \texttt{M\$ModelName} and
\texttt{M\$Information} as well as the mandatory declaration of all 
the indices carried by the fields and parameters of the model. 

The variable \texttt{M\$ModelName} is a string with the name of the model.
If not included in the model file, \feynrules\ is using as a default value 
the name of the file containing the model implementation. Even if optional, the
second variable \texttt{M\$Information} is 
crucial in the sense that it acts as the electronic signature of the author of
the implementation. It consists of a \mathematica\ replacement list
providing information about its name, institution and email.
In addition, the date on which the model has been implemented as well
as the version of the implementation can be provided, together with a list of
references and a link to a webpage. This set of
information ensures a good traceability so that all information about the
physics content of the model, the choices for the free parameters and
contact information about the model author can be recovered.

The content of the \texttt{M\$Information} variable can be accessed once the
model has been loaded by issuing the command
\begin{verbatim}
  ModelInformation[] 
\end{verbatim}
As an example, the two variables \texttt{M\$ModelName} and
\texttt{M\$Information} included in the implementation of the Minimal
Supersymmetric Standard Model (MSSM) \cite{Duhr:2011se} read
\begin{verbatim}
  M$ModelName = "MSSM";

  M$Information = { 
    Authors      -> {"Benjamin Fuks"}, 
    Emails       -> {"benjamin.fuks@iphc.cnrs.fr"}, 
    Institutions -> {"IPHC Strasbourg / University of Strasbourg"},
    Date         -> "21.08.12", 
    Version      -> "1.3.12",
    References   -> {"C. Duhr, B. Fuks, CPC 182 (2011) 2404-2426"},
    URLs         -> {"http://feynrules.irmp.ucl.ac.be/view/Main/MSSM"} 
  };
\end{verbatim}

In general, fields and parameters carry several indices. For instance, a gluon
field $g_\mu^a$ carries a Lorentz index $\mu$ ranging from 1 to 4 and
an adjoint gauge index $a$ ranging from 1 to 8. Similarly, the
Cabibbo-Kobayashi-Maskawa (CKM) matrix $V^{\rm CKM}_{ij}$ is an object with two
generation indices $i$ and $j$ ranging from 1 to 3.

\feynrules\ treats fields and parameters as objects of the form
\texttt{head[index1, index2,...]} or \texttt{head}. In both cases, the name of
the object is given by \texttt{head}, whilst only the first case refers to a
quantity carrying indices. In order to have \feynrules\ properly running, all
the index types have to be declared in the preamble of the model file, together
with the allowed range of values they can take. This is done with the use of the
\texttt{IndexRange} command as, \eg, in
\begin{verbatim}
  IndexRange[Index[SU2D]]   = Unfold[Range[2]]; 
  IndexRange[Index[Colour]] = NoUnfold[Range[3]]; 
  IndexRange[Index[SU2W]]   = Unfold[Range[3]]; 
  IndexRange[Index[Gluon]]  = NoUnfold[Range[8]]; 
  IndexRange[Index[NEU]]    = Range[4];
\end{verbatim}
In the set of commands above, we declare fundamental indices of $SU(2)_L$ and
$SU(3)_c$ labeled as \texttt{SU2D} and \texttt{Colour} and ranging from 1 to 2 and
1 to 3,
respectively. Adjoint indices for the same groups are also declared
as \texttt{SU2W} and \texttt{Gluon} and range from 1 to 3
and 1 to 8, respectively, Finally, a neutralino index, ranging from 1 to 4, is
defined by the label \texttt{NEU}. Whilst the choice for the index names is in principle
left freely to the user, the names of
the $SU(3)_c$ indices employed above
are driven by the fact that the strong gauge group has a special
significance in many Feynman diagram calculators. Consequently, the symbols for
the indices related to the fundamental and adjoint representations of $SU(3)_c$
are reserved names denoted by \texttt{Colour} and \texttt{Gluon},
respectively.

The \texttt{NoUnfold} function appearing in the index
declarations serves when passing the model information to \feynarts. It teaches
\feynarts\ that the
corresponding indices have not to be unfolded at the particle level. In other
words, it means that one desires to consider, \eg, one single quark with a
generic color index instead of three different quarks with a well-defined color
index. In contrast, the presence of the \texttt{Unfold} command indicates
\feynrules\ that the related indices have to be explicitly replaced by their
numerical values when an expansion over the flavor indices is performed (see
below). If these indices are repeated, the sum is thus expanded to several
terms.

To make the screen output more readable, the user can also specify how
\feynrules\ should print an index through the \texttt{IndexStyle} command.
For instance, including in a model file 
\begin{verbatim}
  IndexStyle[ Gauge, a ]
\end{verbatim}
tells \feynrules\ to print all the indices of type \texttt{Gauge} with a symbol starting by the
letter \texttt{a}, followed by a unique number to avoid name clashes. 

Finally, four-vector indices ranging from 1 to 4
(\texttt{Lorentz}), Dirac indices ranging to 1 to 4 (\texttt{Spin}) as well as
left-handed and right-handed Weyl indices ranging from 1 to 2
(\texttt{Spin1} and \texttt{Spin2}) are predefined and do not need to be
declared.

\subsection{Defining gauge groups} \label{sec:FRgroups}

The structure of the interactions described by Lagrangians is in
general governed by gauge symmetries. In order to facilitate the implementation
of these interactions, several dedicated functions allowing,
\eg, for an
automated treatment of the covariant derivatives or of the (super)field strength
tensors have been implemented into \feynrules. To run properly, these functions
rely on gauge group classes that must be declared in the model file.

Gauge interactions are embedded within either simple or semi-simple groups.
Equivalently, the gauge group consists either of a single simple group or of
a product of several simple groups. These are collected into the list 
\texttt{M\$GaugeGroups} included in the \feynrules\ model file,
\begin{verbatim}
  M$GaugeGroups = { Group1=={options},  Group2=={options},  ...  };
\end{verbatim}
In this series of \mathematica\ equalities, \texttt{\{options\}} stands
for a set of rules defining the properties of the groups \texttt{Group1},
\texttt{Group2}, \etc. 

These options are illustrated below by a concrete example as the 
implementation of the three factors of the Minimal Supersymmetric Standard Model
gauge group \cite{Duhr:2011se}, which could be implemented as
\begin{verbatim}
  U1Y  == {                         
    Abelian          -> True,      
    CouplingConstant -> gp,        
    Superfield       -> BSF,       
    GaugeBoson       -> B,       
    Charge           -> Y          
  }
\end{verbatim}
for the abelian group describing the hypercharge interactions and
\begin{verbatim}
  SU3C ==  {                          SU2L == { 
    Abelian           -> False,         Abelian           -> False, 
    CouplingConstant  -> gs,            CouplingConstant  -> gw,
    Superfield        -> GSF,           Superfield        -> WSF, 
    StructureConstant -> f,             StructureConstant -> ep,
    Representations   -> {              Representations   -> {{Ta,SU2D}}, 
      {T,Colour},                       Definitions       -> {
      {Tb,Colourb}},                        Ta[a__]        -> PauliSigma[a]/2,
    SymmetricTensor   -> dSUN               ep             -> Eps}
      }                               }
\end{verbatim}
for the two non-abelian factors, \ie, the strong and weak interactions. The
respective \feynrules\ labels are denoted by \texttt{U1Y}, 
\texttt{SU3C} and \texttt{SU2L}.

Gauge group classes are divided into abelian and non-abelian groups,
distinguished by the option \texttt{Abelian} which takes the value
\texttt{True} or \texttt{False}. 

In the case of non-abelian gauge groups, the user can
specify the symmetric and antisymmetric structure constants of the gauge group
through the options \texttt{SymmetricTensor} and \texttt{StructureConstant}, the
right-hand side of these rules being the related \mathematica\ symbols.
Representation matrices of the group can be
specified by the user as a list through the option \texttt{Representations}. 
Each element of this list consists of another
list whose the first element is the \mathematica\ symbol standing for the 
matrices themselves and the second element is the index type on
which they act, assumed to be properly implemented as presented in
Section \ref{sec:FRidx}. Taking the matrices \texttt{Ta} introduced above as an
example, the gauge group declaration teaches \feynrules\ that these matrices are 
tensors
of the form \texttt{Ta[SU2W,SU2D,SU2D]}, the index \texttt{SU2W} being the
adjoint index of the group. The latter is indirectly defined through the
options \texttt{GaugeBoson} and/or \texttt{Superfield} of the gauge group
class, which refer to the symbols of the gauge boson and/or vector superfield
associated to the gauge group. The index they carry is indeed the index related
to the adjoint representation of the group. In the case both options are
specified by the user (as in the \texttt{U1Y} example above), they
must be consistent, the gauge boson being the vector component of the
vector superfield.  Finally, the option \texttt{Definitions} allows the user to
define representation
matrices and structure constants in terms of the model parameters (see Section
\ref{sec:FRprm}) or standard \mathematica\ variables. As for
the indices related to QCD, predefined symbols are implemented for the QCD
structure constants and representation matrices (see the \feynrules\
manual~\cite{Christensen:2008py, Alloul:2013bka}).

In the case of abelian groups, the user has the possibility to define the $U(1)$
charge associated with the gauge group through the option \texttt{Charge}. This 
allows to further check charge conservation at the Lagrangian or at the
Feynman rules level.

The last option introduced in the examples above consists of the model parameter
to be used as
the gauge coupling constant, which is defined by using the option
\texttt{CouplingConstant}.

\subsection{Declaring the model parameters}\label{sec:FRprm}

In a \feynrules\ implementation, all the model parameters (coupling constants,
mixing angles and matrices, masses, \etc) are collected in the list
\texttt{M\$Parameters}, 
\begin{verbatim}
  M$Parameters = { param1=={options},  param2=={options},  ...  };
\end{verbatim}
where \texttt{param1}, \texttt{param2}, \etc, are user-defined names for the
parameters and where the replacement rules provided as \texttt{options} contain
optional information defining each parameter. 

The model parameters are split into two
categories according to the fact they carry indices or not. As examples of
scalar parameters, the strong coupling constants $g_s$ and $\alpha_s$ could be
implemented as
\begin{verbatim}
  aS == {                                gs == {
    TeX              -> Subscript[a,s],     TeX              -> Subscript[g,s],
    ParameterType    -> External,           ParameterType    -> Internal,
    ComplexParameter -> False,              ComplexParameter -> False,
    InteractionOrder -> {QCD, 2},           InteractionOrder -> {QCD, 1},
    Value            -> 0.1184,             Value            -> Sqrt[4 Pi aS],
    BlockName        -> SMINPUTS,           ParameterName    -> G,
    OrderBlock       -> 3,
    Description      -> "QCD coupling"      Description      -> "QCD coupling"
  }                                      } 
\end{verbatim}
The symbol \texttt{aS} stands for an external (or independent) parameter of the
model to which we associate a numerical value through the option \texttt{Value}
of the parameter class. All the declared external parameters are organized
following a structure inspired by the Supersymmetry Les Houches Accord
\cite{Skands:2003cj, Allanach:2008qq} which is 
effectively used by several Monte Carlo tools. The
Les Houches block name associated to the parameter and the position in the
block are specified through the options \texttt{BlockName} and
\texttt{OrderBlock} of the parameter class. If left unspecified, \feynrules\
assigns automatically a
different Les Houches block for each parameter, the order in the block being set
to one. In contrast to \texttt{aS}, \texttt{gs} is an internal
parameter and depends on other parameters that have to be
declared previously in the model file. The formula defining the internal
parameter is provided via the option \texttt{Value} of the parameter class.
As illustrated in the example above, the external or internal nature of the
parameters is specified through the option \texttt{ParameterType} which takes
the value \texttt{External} or \texttt{Internal}. Instead of
using the option \texttt{Value}, the user can employ the option \texttt{Definitions}
which refers to a list of \mathematica\ replacement rules. Taking the example of
the parameter $g_s$ introduced above, one would have
\begin{verbatim}
  Definitions -> { gs -> Sqrt[4 Pi aS] }
\end{verbatim}
In this case, the parameter is replaced by its
value before the derivation of the interaction vertices by \feynrules, in
contrast to the option \texttt{Value} where the symbol referring to the
parameter is kept.

The option \texttt{ComplexParameter}, which takes the values \texttt{True} or
\texttt{False} (default), determines whether FeynRules treats the parameter as
complex, and the option \texttt{Description} allows the user to enter a string
describing
the physical meaning of the parameter. 

Besides these options which are directly used by \feynrules, there is an
additional set of options needed by some of the interfaces to Feynman diagram
calculators. Whilst \mathematica\ symbols (such as Greek letters) could be used
for parameter names, they are inappropriate at the level of the Monte Carlo
model files using programming language different from \mathematica. The option
\texttt{ParameterName} specifies what to replace the symbol by before writing
out the Monte Carlo model files. By default, it is equal to the parameter
symbol used in \mathematica. 

On different footings, some Monte Carlo programs such as \madgraph\
require the knowledge of the order of a coupling. For instance, $g_s$ is a
coupling 
of order one in QCD while $\alpha_s$ is a coupling of order two. Using this
information allows to speed up event generation by selecting only specific
diagrams, \eg, strong production modes with respect to weak production modes. The
information on the coupling order of a parameter can be passed into \feynrules\ via the
\texttt{InteractionOrder} option of the \texttt{Parameter} class. Furthermore,
the hierarchy among the different coupling orders to be used within a model
implementation are included in the lists \texttt{M\$InteractionOrderHierarchy}.
Each element of these list has two components, the tag of a coupling order and
an integer number standing for its relative strength compared to all the coupling
orders of the model. In addition, the list \texttt{M\$InteractionOrderLimit}
specifies the maximum number of occurrences that a coupling order can reach in a
single diagram, the 
value 99 indicating that no restriction holds. It consists as well of a list of
two-component lists. Taking the example of the MSSM, we have
\begin{verbatim}
  M$InteractionOrderHierarchy = { {QCD, 1}, {QED, 2} }
  M$InteractionOrderLimit = { {QCD, 99}, {QED, 99}   }
\end{verbatim}
which reflects the fact that $g_s^4$ is of the same order of magnitude as $e^2$.

The second category of parameters that can be implemented in \feynrules\ model
files consists of tensorial parameters, carrying indices. While
the index structure can be specified through the option
\texttt{Indices}, all the
attributes described in the case of scalar parameters can still be employed for
declaring tensorial parameters, with one exception. The {\tt OrderBlock}
option is not allowed since one complete Les Houches block must be associated with each single
tensorial parameter, the Les Houches counters referring by definition to the
different possible numerical values
for the indices. In many cases, tensors correspond to unitary, Hermitian or
orthogonal matrices, which can be indicated by turning the \texttt{Unitary},
\texttt{Hermitian} or \texttt{Orthogonal} options to \texttt{True}, the default
value being \texttt{False}. In contrast to scalar parameters, the 
option \texttt{Value} refers this time to a list of values for each possible
choice for the indices. Moreover, by default, tensors are complex
quantities. A complete tensor declaration could read, taking the example of 
the CKM matrix, 
\begin{verbatim}
  CKM == {
    ParameterType    -> Internal,
    Indices          -> {Index[Generation], Index[Generation]},
    Unitary          -> True,
    ComplexParameter -> True,
    Definitions      -> {CKM[3,3]->1, CKM[i_,3]:>0/;i!=3, CKM[3,i_]:>0/;i!=3},
    Value            -> {CKM[1,2]-> Sin[cabi], CKM[1,1]->Cos[cabi],
                         CKM[2,1]->-Sin[cabi], CKM[2,2]->Cos[cabi]},
    Description      -> "CKM-Matrix"
  } 
\end{verbatim}
where the indices \texttt{Generation} have been declared as in Section
\ref{sec:FRidx} and \texttt{cabi} stands for another parameter. 
In the replacement rules above, we have simultaneous used the \texttt{Value} and the
\texttt{Definitions} options for the sake of the example. As a result, zero
vertices are removed from the Lagrangian before computing the Feynman rules and
they are not exported to Monte Carlo generators.

Finally, it is important to keep in mind that many Feynman diagram calculators
have the strong and electromagnetic interactions built in. Therefore, it
is necessary to use special names for the parameters associated with these
interactions (see the \feynrules\ manual).

\subsection{Implementing fields and particles}\label{sec:FRfields}
Following the original \feynarts\ conventions, particles are gathered into
classes. Each class consists of a multiplet whose the members share the same
quantum numbers and possibly different masses. As for 
parameters and gauge groups, each particle class is defined in terms of a
set of properties given by \mathematica\ replacement rules. Collecting the
particle content of the model into classes also
allows the user to write compact expressions for Lagrangians. For the
sake of the example, we consider QCD interactions among massless quarks and
gluons. The associated Lagrangian can be written as
\be
  {\cal L}_{\rm QCD} = -\frac14 g_{\mu\nu}^a g_a^{\mu\nu} + i \qbar_f
   \slashed{\del} q^f + g_s\ \qbar_f \gamma^\mu T^a q^f\ g_\mu^a\ .
\label{eq:lagqcd}\ee
In the equation above, $q^f$ denotes the class symbol representing massless
quarks, $T^a$ the fundamental representation matrices of the QCD gauge group,
$g_s$ its gauge
coupling constant and $g_\mu^a$ and $g_{\mu\nu}^a$ the gluon field and the
associated field strength tensor. Having a class containing all quarks avoids
writing out explicitly one Lagrangian term for each quark flavor.

All the declared instances of the particle class are collected into the list
\texttt{M\$Classes\-Description}, 
\begin{verbatim}
  M$ClassesDescription = { particle1=={options}, particle2=={options}, ...  };
\end{verbatim}
where \texttt{particle1}, \texttt{particle2}, \etc, are user-defined names
for the particle classes of the model and \texttt{options} contains the 
properties of each class. The \mathematica\ name to be used for a
particle class has to
obey strict rules. It consists of one single letter among \texttt{F}
(spin-1/2 Dirac or Majorana fermion), \texttt{S} (scalar field),
\texttt{T} (spin-two field), \texttt{R} (four-component
spin-3/2 field), \texttt{U} (ghost field), \texttt{V} (vector
field), \texttt{W} (two-component spin-1/2 fermionic field) and
\texttt{RW} (two-component spin-3/2 fermionic field) followed by a number
chosen by the user and put between squared brackets. To illustrate this last
statement, together with most of the possible options for the particle class,
we consider the declarations, 
\begin{verbatim}
                                        F[1] == { 
                                          ClassName        -> x,
                                          SelfConjugate    -> False,
                                          Indices          -> {Index[CHA]},
                                          FlavorIndex      -> CHA,
 W[1] == {                                WeylComponents   -> {xp,xmbar}, 
   ClassName      -> chi,                 ParticleName     -> {"x1+","x2+"}, 
   SelfConjugate  -> False,               AntiParticleName -> {"x1-","x2-"},
   Unphysical     -> True,                QuantumNumbers   -> {Q->1},
   Chirality      -> Left,                ClassMembers     -> {x1,x2},
   Indices        -> {Index[Colour]},     Mass             -> {Mx,Mx1,Mx2},
   Definitions    -> {chi[c_]->...}       Width            -> {Wx,Wx1,Wx2},
 }                                        PDG              -> {124,125},
                                          PropagatorLabel  -> {"x","x1","x2"},
                                          PropagatorType   -> Straight,
                                          PropagatorArrow  -> Forward 
                                        } 
\end{verbatim}
This declares a left-handed Weyl fermion $\chi$ and a class of
four-component fermions $x$ containing two members, $x_1$ and $x_2$. The
labels \texttt{W[1]} and \texttt{F[1]} indicate that a Weyl fermion and a
four-component fermions are respectively declared, the left-handed
(\texttt{Left}, default choice) or right-handed (\texttt{Right}) chirality of
the Weyl fermion being specified through the option \texttt{Chirality}. The
particle class has two
mandatory attributes, the \texttt{ClassName} option assigning a \mathematica\
symbol to the class that can be further used when constructing the 
Lagrangian and the \texttt{SelfConjugate} option taking the value
\texttt{True} or \texttt{False}.

In addition to these two features, particle classes have several optional
properties that can be divided into
two categories according to the fact that they are used directly by \feynrules\
or only serve at the level of the interfaces to Feynman diagram
calculators. We start by describing the properties directly related to
\feynrules.

Quantum fields carry in general a collection of indices either related to
symmetry groups, as the index \texttt{Colour} in the declaration of
the field \texttt{chi}, or to labels, as the index \texttt{CHA} specifying the
generation number for the fermion $x$.
While Lorentz (\texttt{Index[Lorentz]}) and spin (\texttt{Index[Spin]},
\texttt{Index[Spin1]} and \texttt{Index[Spin2]}) indices are automatically
handled by \feynrules, the user has to declare, as a list, the rest of
the carried indices. They are specified via the option
\texttt{Indices} which by default refers to an empty list. 
The ordering in which the indices are declared defines the one 
in which they must be employed in the construction of the model
Lagrangian. 

In addition to indices, a field may also carry charges related to
abelian groups. They are passed through
the \texttt{QuantumNumbers} option. The value of this option 
consists of a list of rules mapping each abelian charge to its value. 
Inspecting the declaration of the field $x$ above, one can observe that $x_1$
and $x_2$ are positively charged fields (\texttt{Q->1}). Setting the quantum
numbers of the fields properly in the model file
allows to check their conservation when \feynrules\ computes Feynman rules.

When a particle class contains several members, the option
\texttt{ClassMembers}, referring by default to an empty list, has to refer to a
list of \mathematica\ symbols associated with the different class members.
Moreover, the
index playing the role of the generation index must be specified through the
option
\texttt{FlavorIndex}, as shown in the declaration of the field $x$ above.
The declaration of the \texttt{FlavorIndex} attribute allows
\feynrules\ to perform the expansion of a Lagrangian in terms of the class
members (see Section \ref{sec:FRlag}) or to derive Feynman rules with
flavor expansion performed.

It is often simpler to write down Lagrangians in the gauge basis or using Weyl
fermions instead of employing four-component fermions or mass eigenstates.
However,
interfaces to Monte Carlo generators require the Lagrangian to be entirely
expressed in terms of the physical
eigenstates. Fields can therefore be tagged as unphysical through the option
\texttt{Unphysical} of the particle class, taking the value \texttt{True} or
\texttt{False}. The relations between the unphysical and physical fields can be 
provided in two ways. The user can firstly specify the option
\texttt{Definitions}, which refers to
a list of replacement rules as in the declaration of the Weyl fermion $\chi$
above. In the case we have several class members, this list contains one
element for each of the class members. Secondly, Weyl fermions can be
associated to the left-handed or right-handed components of a Dirac fermion
through the option \texttt{WeylComponents}, as in the declaration of the field
$x$ above. In the case of a Majorana fermion, the option
\texttt{WeylComponents} refers to a single Weyl field instead of a list with a
left-handed and right-handed Weyl spinor.

Finally, masses and widths can be assigned to the different class members 
using the \texttt{Mass} and \texttt{Width} attributes of the particle class.
Their value consists of a list containing, as a first element, a generic mass (or
width), followed by the masses (or widths) of all the class members. Masses and
widths are automatically declared as model parameters by \feynrules, which
employs scalar parameters for the masses of class members and tensorial
parameters for generic masses, the latter carrying a flavor index.
In the case no numerical value is given by the user, a default value of 1 is
assigned by \feynrules. Otherwise, the values can be specified as, \eg,
\begin{verbatim}
   Mass -> {MW, Internal} 
   Mass -> {MZ, 91.188}
   Mass -> {Mu, {MU, 0}, {MC, 0}, {MT, 174.3}}
\end{verbatim}
In the first example, \texttt{MW} is given the value \texttt{Internal} which
teaches \feynrules\ that this mass is an internal parameter defined by the
user in \texttt{M\$Parameters}. This is the only case in which a user needs to
define a mass in the parameter list. The two other examples above being
intuitive, we omit their description. 

The last two optional attributes which serve at the \feynrules-level
are related to the phase possibly carried by Majorana
fermions and to the ghost and Goldstone particles which are connected to a gauge
boson. The Majorana phase can be specified through the \texttt{MajoranaPhase}
property of the particle class and the attributes \texttt{Ghost} and
\texttt{Goldstone} indicate to \feynrules\ the name of the connected gauge
boson.

We now turn to the options that are used by the interfaces to Feynman diagram
calculators. As in the case of parameters, the \mathematica\ symbols
standing for the particle names (\texttt{ClassName})
may not be appropriate for Feynman diagram calculators. The 
options \texttt{ParticleName} and \texttt{AntiParticleName} allow the
user to specify the string (or list of strings) that should be used instead of
the name of the class or the name of the class members (see the example of the
$x$ fermion above). 

According to the Particle Data Group (PDG) 
\cite{Beringer:2012zz}, each particle is represented by a numerical code, 
the PDG code of the particle. Such codes can be specified in the model
description via the attribute \texttt{PDG}, whose value is the PDG code of
the particle or a list with the PDG numbers of all the class members, as also
shown with the declaration of the $x$ field above. If absent,
PDG codes are automatically assigned by \feynrules. 

To achieve this section, we remind that many Feynman diagram calculators draw
the Feynman diagrams that they generate.  Some of these programs allow the user
to specify how to draw and label the lines associated with the fields. 
This information can be provided in the model declaration through the attributes
\texttt{PropagatorLabel}, \texttt{PropagatorArrow} and \texttt{PropagatorType}.
The first of these options takes a string or a list of strings as value, the
default one corresponding to the name of the particle or to a list with the
names of all the class members.
The option \texttt{PropagatorArrow} determines whether an arrow
should be put on the propagator, the allowed value being \texttt{None}
or \texttt{Forward}. Finally, the option \texttt{PropagatorType} can
take any of the values \texttt{ScalarDash} (a straight dashed line),
\texttt{Sine} (a sinusoidal line), \texttt{Straight} (a straight solid line),
\texttt{GhostDash} (a dashed line) and \texttt{Curly} (a curly line).

We refer to the \feynrules\ manual \cite{Christensen:2008py, Alloul:2013bka} as
well as to Ref.\
\cite{Duhr:2011se}, Ref.~\cite{Christensen:2013aua} and
Ref.\ \cite{Butterworth:2010ym} for more information. 

\subsection{Implementing Lagrangians}\label{sec:FRlag}
Model Lagrangians are entered using ordinary \mathematica\ commands and objects, 
augmented by some new symbols such as 
\texttt{Ga}, \texttt{si} or \texttt{sibar} representing the Dirac matrices
$\gamma^\mu$ and the Pauli matrices $\sigma^\mu$ and $\sibar^\mu$, the
left-handed and right-handed chirality projectors 
\texttt{ProjM} and \texttt{ProjP} or the Minkowski metric and the fully
antisymmetric tensors \texttt{ME} and \texttt{Eps}.
All these \mathematica\ symbols are objects of the form \texttt{symbol[a,b,...]}
with an appropriate collection of indices. This syntax is also the one to be
used for quantum fields, when building Lagrangians.
The indices \texttt{a}, \texttt{b},
\etc, denote thus the indices carried either by the fields or by the
parameters in the same order in which they have been declared in the model
file. Moreover, Lorentz and spin indices always appear in the first
position and a class member does not carry the flavor index linking it
to the generic class symbol.

The symbols associated to antiparticles are automatically created by \feynrules\
by suffixing the ending \texttt{bar} to
the name of the particle. For instance, if \texttt{e} denotes the electron field, 
then \texttt{ebar} stands for the positron field. There are two additional ways
to get the names of the antiparticles, employing one of the commands \texttt{HC}
and \texttt{anti}. The function \texttt{HC} can however also be used to
compute the
Hermitian conjugate of any object or expression (even a Lagrangian). In this
sense, \texttt{HC[e]} returns the field $\bar e$ defined by $\bar e =
e^\dagger\gamma^0$, which is equivalent to the symbol \texttt{ebar}, in contrast
to \texttt{anti[e]} which corresponds to $e^\dag$. In the same
way, the charge conjugate of a field or of an expression can be obtained by
means of the \texttt{CC} function.

For anticommuting objects (fields and parameters), the \mathematica\
\texttt{Dot} command has to be employed. As a consequence, \mathematica\ is
prevented  from changing the relative ordering of these objects, which ensures
correct anticommutation relations to be accounted for in the calculation of the
Feynman rules.

There are several ways of dealing with field and parameter indices when entering
Lagrangians in the \feynrules\ model file. If there is no ambiguity, the user
has the option to suppress all indices and \feynrules\ is capable of restoring
them. For the sake of the example, there are two equivalent ways of implementing
the QCD interactions included in Eq.~\eqref{eq:lagqcd}. The most compact
implementation reads
\begin{verbatim}
  gs qbar.Ga[mu].T[a].q G[mu, a]
\end{verbatim}
where \texttt{gs} stands for the symbol denoting the strong coupling constant,
\texttt{q} and \texttt{qbar} for the quark field and its Hermitian conjugate
counterpart, \texttt{T} for the fundamental representation matrices of the QCD
gauge group and \texttt{G} for the gluon field. The spin, color and generation
indices of the quark fields
are here all implicit. They could have been explicitly implemented as
\begin{verbatim}
  gs Ga[mu, s, r] T[a, m, n] qbar[s, f, m].q[r, f, n] G[mu, a]
\end{verbatim}
This interaction Lagrangian, expressed in terms of the generic class name
\texttt{q} for the quark field can be expanded in terms of the six class members
corresponding to the six flavors of quarks. The 
\texttt{Ex\-pand\-Indices} function has been implemented for this purpose, 
\begin{verbatim}
  ExpandIndices[gs Ga[mu, s, r] T[a, m, n] qbar[s, f, m].q[r, f, n] G[mu, a],
    FlavorExpand->FLA];
\end{verbatim}
where \texttt{FLA} is the symbol standing for quark flavors. Replacing
\texttt{FLA} by \texttt{True} leads to an expansion over all the indices
that have been declared as flavor indices at the time of the particle
declarations. In the case the indices have been suppressed by
the user
in the Lagrangian implementation, \texttt{ExpandIndices} first restores all
indices before performing the expansion.

The implementation of the kinetic terms included in Eq.\ \eqref{eq:lagqcd}
follows the same rules with respect to indices and can be equivalently
implemented using one of the two expressions
\begin{verbatim}
  I Ga[mu, s, r] uqbar[s, f, m].del[uq[r, f, m], mu]
  I uqbar.Ga[mu].del[uq, mu]
\end{verbatim}
where the symbol \texttt{del} stands for the spacetime derivative operator.
In the case the QCD gauge group has been properly declared as presented in
Section \ref{sec:FRgroups}, the interaction terms can be automatically handled
together with the kinetic terms by \feynrules. To this aim, it is sufficient to
use the built-in covariant derivative function \texttt{DC},
\begin{verbatim}
  I Ga[mu, s, r] uqbar[s, f, m].DC[uq[r, f, m], mu]
  I uqbar.Ga[mu].DC[uq, mu]
\end{verbatim}
This feature allows for a very compact implementation of Lagrangians since the
user does not have to worry about gauge interaction terms at all. Along the same
footings, the implementation of the gauge boson kinetic terms is also automatic
thanks to the
function \texttt{FS} standing for the field strength tensor. The first term of
the Lagrangian of Eq.\ \eqref{eq:lagqcd} can hence be implemented as 
\begin{verbatim}
  -1/4 FS[G, mu, nu, a] FS[G, mu, nu, a]
\end{verbatim}

Lagrangians, and in particular supersymmetric Lagrangians, are sometimes more
easily written in terms of Weyl fermions. If the
\texttt{WeylComponents} attributes of the particle class have been properly set,
the Lagrangian can be automatically translated in terms of four-component
fermions by means of the function \texttt{WeylToDirac}.

The \feynrules\ package comes with a set of additional functions dedicated to
Lagrangian manipulations. The kinetic terms,
defined as the quadratic terms of a Lagrangian including a spacetime derivative,
can be extracted by issuing, in a \mathematica\ session, the command
\begin{verbatim}
  GetKineticTerms[ Lag, options ]
\end{verbatim}
where \texttt{Lag} is the symbol standing for the Lagrangian. As for any of the
functions described in the rest of this section, \texttt{GetKineticTerms} shares
the same options as the \texttt{FeynmanRules} routine (see below). Similarly, the
functions \texttt{GetMassTerms}, \texttt{GetQuadraticTerms} and
\texttt{GetInteraction\-Terms} can be used to select the
mass terms, quadratic terms and interaction terms included in the Lagrangian,
respectively. The \texttt{GetMassSpectrum} routine offers the possibility to
calculate the values of the masses that appear in the Lagrangian, both in
numeric and symbolic ways, provided that the Lagrangian is implemented in a mass
diagonal form.
Finally, it is also possible to filter out terms related
to specific interactions by employing the routine \texttt{SelectFieldContent}, 
\begin{verbatim}
  SelectFieldContent[ Lag, { fields1, fields2, ... } ] 
\end{verbatim}
where \texttt{fields1}, \texttt{fields2} denote lists of fields which the user is 
interested in their interactions. 

In general, a quantum field theory Lagrangian has to fulfill a set of basic
requirements, such as hermiticity, gauge invariance, \etc. The 
associated checks can be performed by \feynrules. First of all, hermiticity 
can be checked by using the command \texttt{CheckHermiticity}. 
Secondly, for \feynrules\ to work properly, all the
quadratic terms of the Lagrangian must be diagonal. This can be checked by
means of the three intuitive commands \texttt{CheckDiagonalQuadraticTerms},
\texttt{CheckDiagonalKineticTerms} and \texttt{Check\-Diagonal\-MassTerms}. In the
case the kinetic terms are diagonal, the user has the possibility to verify their
normalization by employing the \texttt{Check\-KineticTerm\-Normalisation} function.
In the same way, if the mass terms are diagonal, the routine
\texttt{CheckMassSpectrum} compares the masses extracted from the Lagrangian to
those provided in the field declarations.

\subsection{Running \feynrules}\label{sec:FRrun}
The \feynrules\ package can be loaded as any other
\mathematica\ package, with the only difference that the path where
\feynrules\ has been downloaded must be specified in the
\texttt{\$FeynRulesPath} variable,
\begin{verbatim}
  $FeynRulesPath = SetDirectory[ <the address of the package> ];
  << FeynRules`
\end{verbatim}
A model can then be loaded as 
\begin{verbatim}
  SetDirectory[ <path to the model file(s)> ];
  LoadModel[ < file1.fr >, < file2.fr>, ... ];
\end{verbatim}
The model can be equivalently contained in
one single model file or split among several files whose extension is
\texttt{.fr}. The Feynman rules can subsequently be extracted by using the
command \texttt{FeynmanRules},
\begin{verbatim}
  vertices = FeynmanRules[ Lag ];
\end{verbatim}
where \texttt{Lag} is the \mathematica\ symbol containing the expression of the
Lagrangian. The latter must be entirely written in four-dimensional spacetime
and in terms of four-component spinors (in contrast to two-component 
fermions). As a result of the command above,
the vertices derived by \feynrules\ are written on the screen and
stored in the variable \texttt{vertices}. In the case the user is not interested
in printing the Feynman rules to the screen, the function \texttt{FeynmanRules}
has to be called with the option \texttt{ScreenOutput} set to \texttt{False},
\begin{verbatim}
  vertices = FeynmanRules[ Lag, ScreenOutput -> False];
\end{verbatim}

The extracted Feynman rules consist of a list of vertices. Each
vertex is written as a list of two elements, the first one being the list of
particles
incoming to the vertex and the second one the analytical expression for the
vertex.
Each particle is also written as a two-component list. The first element is
the name of the particle while the second one is an integer number. This number
is employed in the analytical expression of the vertex to distinguish the
indices belonging to the different incoming particles. For instance, the
output obtained for the interaction vertex
between quarks and gluons included in the Lagrangian of Eq.~\eqref{eq:lagqcd}
is
\be\nn
  \hspace*{-4.8truecm}
  \texttt{\{  \{\{G, 1\}, \{qbar, 2\}, \{q, 3\}\},} 
     \qquad i g_s\, \delta_{f_2f_3}\, \gamma^{\mu_1}_{s_2s_3}\, T^{a_1}_{m_2m_3}
\texttt{\}}
\ee
where spin indices start with the letter \texttt{s}, Lorentz indices with the
Greek letter $\mu$, generation indices with the letter \texttt{f}, color indices
with the letter \texttt{m} and gluon indices with the letter \texttt{a}. Each
letter is followed by the integer number related to the relevant field.

If the flavor expansion has not been performed at the Lagrangian level, the
Feynman rules returned by \feynrules\ are expressed in terms of generic class
names for the fields instead of the class members. Flavor expansion
can be
achieved by typing in \mathematica 
\begin{verbatim}
  vertices = FeynmanRules[ Lag, FlavorExpand -> True];
\end{verbatim}
If several flavor indices are present in the model, such as in the MSSM where we
have generation indices, squark indices, neutralino indices, \etc, and if one is
interested in only expanding given indices, the symbol \texttt{True} above can
be replaced with the list of indices to be expanded. 
 
The list of Feynman rules can be quite long and it may sometimes be desirable to
extract one or a few vertices. In order to limit the number of constructed
vertices, several options have been implemented.  Setting \texttt{MaxParticles
-> n} (\texttt{MinParticles -> n}) teaches \feynrules\ to derive vertices with
at most (least) \texttt{n} external legs. Similarly, setting
\texttt{MaxCanonicalDimension -> n} (\texttt{MinCanonicalDimension -> n})
reduces the output to vertices whose the canonical dimension does not exceed (is
at least) \texttt{n}.

If the user is interested in only specific interaction vertices, one can call
the function 
\texttt{FeynmanRules} with the option \texttt{SelectParticles} pointing towards a
list containing the names of the particles allowed to enter a vertex.
Similarly, the options
\texttt{Contains} and \texttt{Free}, both referring to a list of particles,
instruct
\feynrules\ to only derive vertices that contain or not the particles
indicated in the list. The \texttt{SelectVertices} routine has been designed in
the same purpose and allows to perform a vertex selection among a list of
already derived vertices. It accepts the options \texttt{MaxParticles},
\texttt{MinParticles}, \texttt{SelectParticles}, \texttt{Contains} and
\texttt{Free} which have to be used in the same way as for
\texttt{FeynmanRules},
\begin{verbatim}
  SelectVertices[vertices, options]
\end{verbatim}

By default, checks whether the quantum numbers that have been defined in the
model file are conserved at each vertex are performed and warnings are returned
in the relevant cases. Checks can be turned off by setting 
\texttt{ConservedQuantumNumbers} to \texttt{False} when calling the 
\texttt{FeynmanRules} function. Alternatively, the value 
of this option could be set to a list with the quantum numbers that \feynrules\
has to check for their conservation.

It is also possible to compute Feynman rules for
several sub-Lagrangians and merge them in a second stage by means of the
\texttt{MergeVertices} function,
\begin{verbatim}
  vertices = MergeVertices[ vertices1, vertices2, ... ];
\end{verbatim}
where the lists of vertices labeled \texttt{vertices1}, \texttt{vertices2}, \etc,
have been previously extracted by employing the function \texttt{FeynmanRules}. 

The \feynrules\ program includes two special functions for manipulating vertex
lists. In the case a vertex depends explicitly on the particle four-momenta, it
can be sometimes useful to simplify its analytical expression by
replacing one of the four-momentum by the opposite of the sum of the others.
This is achieved by issuing in \mathematica,
\begin{verbatim}
  MomentumReplace[vertex, n ]
\end{verbatim}
where \texttt{vertex} represents a single vertex, \ie, an element of a vertex
list. The effect of this function is to replace the four-momentum of the
\texttt{n}$^{\rm th}$ particle in terms of the four-momentum of the other
particles incoming to the vertex. In the case an entire set of vertices has to
be simplified using momentum conservation, the user can employ the routine
\texttt{ApplyMomentumConservation}. For each vertex of the list, \feynrules\
uses the function {\tt MomentumReplace}, cycling through each momentum, and
compares the size of each expression. The shortest one is kept. 

The parameters also play an important role in the model file. Their numerical
values can be obtained by employing the \texttt{NumericalValue} routine,
\begin{verbatim}
  NumericalValue[ function ]
\end{verbatim}
where \texttt{function} is an analytical function of one or several of the model
parameters. The numerical value of external parameters can be modified by means
of the \texttt{UpdateParameters} command, 
\begin{verbatim}
  UpdateParameters[ param1 -> value1, param2 -> value2, ... ]
\end{verbatim}
which sets the value of the external parameters \texttt{param1}, \texttt{param2}, \etc,
to \texttt{value1}, \texttt{value2}, \etc, respectively. Any
change tried to be made to an internal parameter is ignored. 

In the case a whole spectrum has to be loaded, the user has the more efficient
possibility to import an entire Les Houches card following the Les Houches
block structure of the \feynrules\ model
file. This is performed by typing in a \mathematica\ session the command 
\begin{verbatim}
  ReadLHAFile[Input -> "file.dat"];
\end{verbatim}
where \texttt{file.dat} is the Les Houches file containing the numerical values
of all the model parameters. Conversely, a Les Houches file can be written by
\feynrules\ by issuing the command
\begin{verbatim}
  WriteLHAFile[Output -> "file.dat"];
\end{verbatim}
which yields the creation of the file \texttt{file.dat} containing the current
values of all the external parameters, the latter being ordered according to the
Les Houches block structure of the \feynrules\ model file.

The user has also the possibility to implement his own set of functions for
manipulating vertex lists or even Lagrangians. To facilitate this task,
\feynrules\ comes with a set of Boolean functions that allow to probe the
properties attached to a symbol. 
The (self-explained) functions \texttt{FieldQ}, \texttt{FermionQ}, 
\texttt{BosonQ}, \texttt{SelfConjugateQ}, \texttt{ScalarFieldQ},
\texttt{WeylFieldQ},
\texttt{DiracFieldQ}, \texttt{Majorana\-FieldQ}, \texttt{VectorFieldQ},
\texttt{Spin2FieldQ}, \texttt{SuperfieldQ}, \texttt{ChiralSuperfieldQ},
\texttt{Vector\-SuperfieldQ} and \texttt{Ghost\-FieldQ} hence test the nature of a
(super)field. 
The two functions \texttt{numQ} and \texttt{CnumQ} return true or
false whether a parameter is real or complex, while
\texttt{TensQ}, \texttt{CompTensQ}, \texttt{UnitaryQ}, \texttt{HermitianQ} and
\texttt{OrthogonalQ} are dedicated to tensorial parameters. 
In addition, \texttt{MR\$Quantum\-Numbers[part]} returns a list
with the quantum numbers related to the particle \texttt{part} and
\texttt{\$IndList[fld]} gives a list with the indices declared for the field
\texttt{fld}. 

\subsection{Interfaces to Monte Carlo event generators}\label{sec:FRint}

Once Feynman rules have been obtained, \feynrules\ can export them, together
with the model definition, to various matrix-element generators by means 
of dedicated interfaces. Currently, 
interfaces exist to the \comphep/\calchep, \feynarts/\formcalc,
\madgraph/\ \madevent, \sherpa\ and \whizard\ programs.
The output of an interface consists of a set of files organized in a single
directory which can be copy-pasted into the model directory of the
relevant matrix-element generator. \feynrules\ models can then be directly used
as any other built-in models. However, many other
Feynman diagram calculators exist and are not directly
interfaced to \feynrules. To this aim, a universal output of the \feynrules\
model is available as a \python\ shared library which can be further loaded by
any external program. Section \ref{sec:ufo} is dedicated to its description,
while we focus on the generic properties of the other interfaces below. For more
information, we also refer to the \feynrules\ manual and to
Ref.~\cite{Christensen:2009jx} and Ref.~\cite{Christensen:2010wz}.

All the interfaces are invoked with commands of the form
\begin{verbatim}
  Write__Output[lag1, lag2, ..., options]
\end{verbatim}
where the underscores are replaced by a sequence of 
letters according to the interface under consideration. Hence,
\texttt{WriteCHOutput}, \texttt{WriteFeynArtsOutput}, \texttt{WriteMGOutput},
\texttt{WriteSHOutput} and \texttt{WriteWOOutput} are the commands related to 
\comphep/\calchep, \feynarts/\formcalc, \mgme, \sherpa\ and \whizard,
respectively. The arguments of these functions consist of a sequence of
Lagrangians \texttt{lag1}, \texttt{lag2}, \etc, as well as options specific to
each interface (see Ref.\ \cite{Christensen:2009jx} and Ref.\
\cite{Christensen:2010wz}). Moreover, the model can also be exported as a \TeX-
file by means of the command \texttt{WriteLatexOutput}.

Matrix-element generators have very often strong constraints on the allowed names
for the model particles and parameters and on the type of supported fields.
Moreover, most of the time, the treatment of the color and Lorentz structures is
hard-coded. Therefore, the different interfaces check whether all the particles,
parameters and vertices
are compliant with the matrix-element generator requirements and discard them if
necessary.

Furthermore, several diagram calculators employ Feynman gauge while others use
the
unitary gauge. Since these gauges involve different particles and Lagrangian
terms, \feynrules\ model files have to include the flag 
\texttt{FeynmanGauge} set to {\tt True} or {\tt False}. For instance, setting it
to false automatically removes the Goldstone bosons and ghost fields at the
interface level.

As already mentioned above, several parameters and particles have a special
significance and must be clearly identified by Feynman diagram calculators.
Their names are therefore fixed at the \feynrules\ level, which ensures a proper
running of the interfaces. Hence, the indices for the fundamental and adjoint
representations of the QCD gauge group have to be called \texttt{Colour}
and \texttt{Gluon}, respectively. Furthermore, the names of the gluon field,
strong coupling constant, $SU(3)$ totally antisymmetric and symmetric structure
constants, as well as the fundamental representation matrices of the group must
be \texttt{G}, \texttt{gs}, \texttt{f}, \texttt{dSUN} and \texttt{T}. As an
example, we refer to the QCD gauge group implementation shown in Section
\ref{sec:FRgroups}. In addition, the strong
coupling constant and its square over $4\pi$ must be declared as presented in
Section \ref{sec:FRprm}, where $\alpha_s$ is given as the external parameter and
included in the Les Houches block \texttt{SMINPUTS}. In contrast, $g_s$ is an
internal parameter.

Electromagnetic interactions also have a special role in many Monte
Carlo generators. Consequently, a valid model implementation in \feynrules\ has to
include a parameter for the electromagnetic coupling constant labeled as
\texttt{ee}, and the electric charge has to be called \texttt{Q}. Following
the Les Houches accord conventions \cite{Skands:2003cj, Allanach:2008qq}, the
external parameter is the inverse of the square of the electromagnetic coupling
constant over $4 \pi$. Together with the Fermi constant and the $Z$-boson pole
mass, these parameters have to be organized within the Les Houches block {\tt
SMINPUTS}, following Les Houches accord standards.

\mysection{The Universal \feynrules\ output - the UFO}\label{sec:ufo}
\subsection{Basic features}
The universal output format of \feynrules, dubbed the UFO~\cite{Degrande:2011ua},
has been designed to
overcome all restrictions related to model formats commonly used by Monte Carlo programs.
These restrictions are in general related to the form of the interactions
compliant with a specific tool, such as,
\eg, the allowed Lorentz and color structures in the vertices.
The key features of the UFO are flexibility and modularity through the translation of the
model in an abstract way, employing \python\ classes and objects to represent
particles, parameters and vertices. Consequently, it is not tied to any specific
matrix element generator and is universal in the sense that the entire \python\
library can be directly used for further interfacing, without any modification,
by any code.
This format is presently
employed by four programs, \aloha\ \cite{deAquino:2011ub}, \gosam\
\cite{Cullen:2011ac, Cullen:2011xs},  \madanalysis\ \cite{Conte:2012fm} and
\madgraph\ 5 \cite{Alwall:2011uj} and is expected to be used in the future
by \herwig++~\cite{Bahr:2008pv,Arnold:2012fq}.
In order to extract the UFO files from a \feynrules\ model, a dedicated
interface has been implemented and can be invoked by issuing 
\begin{verbatim}
   WriteUFO[ Lag ] ; 
\end{verbatim}
where the symbol {\tt Lag} stands for the model Lagrangian. As results of the
command above, a set of \python\ files are generated. These files can be
split into two categories, model-independent files, identical for all models,
and model-dependent files containing among others the definitions of the
particles, parameters, \etc. All those files are provided as self-contained
\python\ modules and are described in the next sections\footnote{By the time
this work has been completed, the UFO conventions have been generalized
to allow for non-standard
propagators~\cite{Christensen:2013aua} and $1\to 2$ partial widths~\cite{Alwall:xxx}.
These features being irrelevant for this work, we do not describe them in this chapter
and refer to the appropriate references.}
have been. They could also be
possibly generated by the programs \lanhep\ 
\cite{Semenov:1996es,Semenov:1998eb,Semenov:2002jw,Semenov:2008jy,Semenov:2010qt} and
\sarah\ \cite{Staub:2008uz,Staub:2009bi, Staub:2010jh, Staub:2012pb}.

\subsection{Initialization and structure of the objects and functions}

A UFO module comes with an initialization file
\texttt{\textunderscore\textunderscore init\textunderscore\textunderscore .py}.
This file is standard in the \python\ language and corresponds to a tag for
importing the complete module. This task is achieved by issuing, in a \python\
interpreter or within another \python\ program,
\begin{verbatim}
  import Directory
\end{verbatim}
where \texttt{Directory} refers to the name of the directory containing the
UFO files. This \texttt{\textunderscore\textunderscore
init\textunderscore\textunderscore .py} file also contains links to
the different lists gathering the set of objects implemented in a UFO module,
\texttt{all\textunderscore particles},
\texttt{all\textunderscore vertices}, \texttt{all\textunderscore pa\-rameters},
\texttt{all\textunderscore couplings},
\texttt{all\textunderscore lorentz}, 
\texttt{all\textunderscore coupling\textunderscore orders} and
\texttt{all\textunderscore functions}.
The names of these lists are intuitive (more information being provided below)
and they allow, \eg, to access the full particle content of a model or the list
of all parameters. Moreover, each time an
instance of a given class is created, it is 
automatically added to the corresponding list.

There are only six basic classes necessary for the implementation of a UFO
model. They are denoted \texttt {Particle}, \texttt{Parameter}, \texttt{Vertex},
\texttt{Coupling}, \texttt{Lorentz} and \texttt{CouplingOrder}. These classes
are derived from a mother class
\texttt{UFOBaseClass} that contains general methods and attributes accessible by
each of its daughters. Hence, the method \texttt{get\textunderscore all} allows
to list all the attributes of an object while \texttt{nice\textunderscore
string} returns a string 
with a representation of an object including the values of all its attributes.
In addition, the usual functions \texttt{get} and \texttt{set} allow to read
and modify the value of an attribute.  The structure of the mother class as well
as the one of its children are defined in the model-independent
file \texttt{object\textunderscore library.py}.

For some model implementations, the user might need to define his own set of
routines. The latter can be included in the file \texttt{function\textunderscore
library.py} which
contains functions based on the class {\tt Function}. These allow for the
translation of functions that can be defined within a single \python\ line to
other programming languages such as {\sc Fortran} or {\sc C++}. An instance of
the class {\tt Function} comes with three mandatory attributes, called
\texttt{name}, \texttt{arguments} and \texttt{expression}, which respectively
refer to a string representing the name of the function, a list of strings with
the arguments of the function and the analytical expression
defining the
function itself provided as a string. Common mathematical functions for which the
standard \python\ module {\tt cmath} is insufficient are
predefined in the function library, so that 
\texttt{complexconjugate} (complex
conjugation), \texttt{csc} (cosecant), \texttt{acsc} (arccosecant), \texttt{im}
(the imaginary part of a complex number), \texttt{re} (the real part of a
complex number), \texttt{sec} (secant) and \texttt{asec} (arcsecant) are
available form the start and can be further employed in parameter declarations. 
For the sake of the example, we show below the implementation of the secant function
that is given by
\begin{verbatim}
  sec=Function(name='sec', arguments=('z',), expression='1./cmath.cos(z)')
\end{verbatim}

\subsection{Implementing the model particle content} 
The particle content of a UFO model is implemented as a set of instances of the
class \texttt{Particle} which are collected in the file \texttt{particles.py}.
They describe the mass-eigenstates of the model under
consideration, possibly together with some auxiliary non-propagating fields that
can be helpful to, \eg, model higher-dimensional operators as in some Monte Carlo
generators. As it will be shown below, the definition of a particle in
the UFO is almost a direct translation in \python\ of its declaration within
\feynrules, after expanding all particle classes with respect to the
flavor indices.

A particle and the associated antiparticle are uniquely defined by their names,
which are provided as strings under the values of the attributes \texttt{name}
and \texttt{antiname} of the \texttt{Particle} class. In addition, those names
are also provided under their \TeX-form as the values of the
attributes \texttt{texname} and
\texttt{antitexname}. At the matrix-element generator level, particles are
rather identified, in general, through their PDG code \cite{Beringer:2012zz}.
The latter is therefore stored as the
value of the attribute \texttt{pdg\textunderscore code} that can be set to any
integer value. Among the basic features defining a particle, one also finds its
mass and width that are encoded in the
\texttt{mass} and \texttt{width} attributes of the \texttt{Particle} class, which
refers to instances of the \texttt{Parameter} class declared in the
file \texttt{parameters.py}\footnote{The \texttt{Parameter} objects are 
imported by including at the beginning of the \texttt{particles.py} file
the command \texttt{import parameters as Param}, \texttt{Param} being a
user-chosen name.} (see Section
\ref{sec:UFOprm}). 

The transformation properties of the particle under the Lorentz group as well as
under the
QCD and electromagnetic gauge groups are specified through the \texttt{spin},
\texttt{color} and \texttt{charge} attributes of the \texttt{Particle} class.
Concerning
the attribute \texttt{spin}, its value has to be an integer number
given under the form $2s+1$, $s$ being the particle spin assumed to be smaller
than or equal to two. For ghost fields, it takes by convention the value $-1$.
With respect to color, the only supported representations are singlets ($1$),
triplets and 
antitriplets ($\pm3$), sextets and antisextets ($\pm6$) as well as octets ($8$).
Finally, the electric charge is provided as
a rational number. 

All the attributes of the particle class introduced so far are mandatory.
It is however also possible to include three predefined optional
attributes
\texttt{goldstone}, \texttt{propagating} and \texttt{line}. The first
two attributes take a Boolean value tagging the corresponding particle as a
Goldstone boson (the default value is \texttt{false}) and as a propagating
particle (the default value is \texttt{true}), respectively. The attribute
\texttt{line} returns a string representing how the propagator of the particle
should be drawn in a Feynman diagram, the possible values being
\texttt{'dashed'}, \texttt{'dotted'}, \texttt{'straight'}, \texttt{'wavy'},
\texttt{'curly'}, \texttt{'scurly'}, \texttt{'swavy'} and \texttt{'double'}. The
default value is deduced from the spin and color representations of the
particle.

In addition, an instance of
the \texttt{Particle} class can be declared with an arbitrary number of optional
attributes. Every attribute different from those presented above hence
refers to an integer number representing an extra additive quantum
number associated with the model under consideration.

As an example, an instance of the UFO \texttt{Particle} class describing the top
quark could read,
\begin{verbatim}
  t = Particle(
    name         = 't',
    antiname     = 't~',
    texname      = 't',
    antitexname  = '\\bar{t}',
    pdg_code     = 6,
    mass         = Param.MT,
    width        = Param.WT,
    spin         = 2,
    color        = 3,
    charge       = 2/3,
    line         = 'straight',
    LeptonNumber = 0
  )

  t__tilde__ = t.anti()
\end{verbatim}
The second command allows for the instantiation of the top antiquark in an
automated fashion, from the knowledge of the top quark properties. In the
\python\ commands above, we have introduced the parameters standing for the top
mass and width \texttt{MT} and \texttt{WT} as well as an additional quantum
number defined by the attribute \texttt{LeptonNumber}.

\subsection{Implementing the model parameters}
\label{sec:UFOprm}
All the parameters in a UFO model are collected as instances of the
\texttt{Parameter} class in the file \texttt{parameters.py}. As for the model
description in \feynrules\ (see Section \ref{sec:FRprm}), 
parameters are either external or internal, and, following
the \feynrules\ conventions, the user has to provide numerical values for 
external parameters and algebraic functions of the other parameters for
internal parameters.

Information on the name and on the external or internal nature of a parameter is
provided through the attributes \texttt{name}, \texttt{texname} and
\texttt{nature} of the \texttt{Parameter} class. The first two attributes take
strings as values referring to the name of the parameter under a \python\ form and
a \TeX\ form, respectively. The last attribute receives either the value
\texttt{'external'} or the value \texttt{'internal'} according to the nature of
the parameter.
The numerical values of external parameters and the analytical formulas defining 
internal parameters are eventually provided through the
attribute \texttt{value} of the \texttt{Parameter} class.

As in Section \ref{sec:FRprm}, external parameters are organized  in
Les Houches blocks and counters \cite{Skands:2003cj,Allanach:2008qq}. The
structure adopted for the model parameters 
is encoded via the two attributes \texttt{lhablock} and
\texttt{lhacode} of the \texttt{Parameter} class. The value of the attribute
\texttt{lhablock} is a string with the name of the block in which the
parameter under consideration is stored, whilst \texttt{lhacode} refers to
a list of integer numbers related to the counter specifying the position of the
parameter within the block.

Finally, parameters can either be real or complex, which is specified by the
attribute \texttt{type} of the {\tt Parameter} class, under the constraint that 
external parameters are always real numbers. In contrast,
internal parameters can be complex, \texttt{type} being set to the value
\texttt{'complex'} in this case. 

All the attributes of the \texttt{Parameter} class described above are mandatory.
We illustrate the description of the parameter implementations with the same
examples as in Section
\ref{sec:FRprm} and present a possible UFO implementation for the strong
coupling constant and its square over $4 \pi$, 
\begin{verbatim}
  aS = Parameter(
    name     = 'aS',
    texname  = '\\alpha_s',
    nature   = 'external',
    type     = 'real',
    lhablock = 'SMINPUTS',
    lhacode  = [3],
    value    = 0.118
  )

  G  = Parameter(
    name    = 'G',
    texname = 'G',
    nature  = 'internal',
    type    = 'real',
    value   = 'cmath.sqrt(4 * cmath.pi * aS)'
  )
\end{verbatim}
Let us note that it is mandatory that every internal
parameter depends only on parameters which have been previously declared. 

In order to print out the numerical values of all the external parameters in an
efficient way, the model-independent file {\tt write\textunderscore
param\textunderscore card.py} contains the implementation of the 
class {\tt ParamCardWriter}. 
Calling from another \python\ module or from a \python\ interpreter
\begin{verbatim}
  ParamCardWriter('./param_card.dat', qnumbers=True)
\end{verbatim}
outputs a parameter file named {\tt param\textunderscore card.dat} which
follows the Les Houches block structure encoded in the declaration of the
external parameters of the model.  The second argument specifies whether 
the \texttt{QNUMBERS} blocks with the quantum numbers of all the particles
implemented in the model \cite{Alwall:2007mw} have to be included in the
output. If set to \texttt{True}, the full set of masses and widths, even if they
are dependent parameters, are included.

Since most matrix-element generators have information on the Standard Model
input parameters hard-coded, the latter must be correctly
identified. We refer to Section \ref{sec:FRprm} and Section \ref{sec:FRint} for
the \feynrules\ conventions that are also those followed by UFO models.

\subsection{Implementing the interactions of the model} 

The vertices corresponding to the interactions included in a model are defined
in the file {\tt vertices.py} using instances of the {\tt Vertex} class. Their
declaration relies on the expansion of each vertex in a color
$\otimes$ spin basis, 
\be\label{eq:ufovert}
 \begin{cal}V\end{cal}^{a_1\ldots a_n, \ell_1\ldots\ell_n}(p_1,\ldots,p_n) =
    \sum_{i,j}C_i^{a_1\ldots a_n}\,G_{ij}\,L_j^{\ell_1\ldots\ell_n}(p_1,\ldots,p_n)
     \ ,
\ee
for the generic example of a $n$-point interaction. 
In the equation above, the variables $p_i$ denote the four-momenta of the
particles incoming to the vertex and $G_{ij}$ the coupling
strengths. The quantities $C_i^{a_1\ldots a_n}$ and
$L_j^{\ell_1\ldots\ell_n}(p_1,\ldots,p_n)$ are tensors in color and spin
space\footnote{The terminology {\it spin indices} refers here to both Lorentz and
Dirac indices.}, respectively, and together defines the mentioned color $\otimes$ spin
basis. They can be shared by several vertices, which
reduces possible redundancies in a model implementation. This
explains the term \textit{basis} that has been employed above as well as the one of
\textit{coordinates} for the coupling strengths.
As an example, the QCD four-gluon vertex could be written as 
\be\hspace{-.25truecm}
   \Big(f^{a_1a_2b} f_b{}^{a_3a_4}\     f^{a_1a_3b}f_b{}^{a_2a_4}\      
     f^{a_1a_4b}f_b{}^{a_2a_3}\Big)
   \bpm 
     ig_s^2 & 0 & 0\\
     0 & ig_s^2 & 0\\
     0 & 0 & ig_s^2 
  \epm
  \bpm
    \eta^{\mu_1\mu_4}\eta^{\mu_2\mu_3} - \eta^{\mu_1\mu_3}\eta^{\mu_2\mu_4} \\
    \eta^{\mu_1\mu_4}\eta^{\mu_2\mu_3} - \eta^{\mu_1\mu_2}\eta^{\mu_3\mu_4}\\
    \eta^{\mu_1\mu_3}\eta^{\mu_2\mu_4} - \eta^{\mu_1\mu_2}\eta^{\mu_3\mu_4}\
  \epm \ . 
\label{eq:4gluon}\ee

The UFO format for vertex implementation mimics this structure by using 
different \python\ objects to represent the vertex itself,
the relevant Lorentz and color structures, as well as the
coupling strengths. Going on with the example of Eq.\ \eqref{eq:4gluon}, its
implementation read
\begin{verbatim}
 V_1 = Vertex(
    name      = 'V_1',
    particles = [P.G,P.G,P.G,P.G],
    color     = ['f(1,2,-1)*f(-1,3,4)','f(1,3,-1)*f(-1,2,4)',
                 'f(1,4,-1)*f(-1,2,3)'],
    lorentz   = [L.VVVV1,L.VVVV2,L.VVVV3],
    couplings   = {(0,0):C.GC_1,(1,1):C.GC_1,(2,2):C.GC_1}
 )
\end{verbatim}
where the particles, Lorentz tensors and coupling objects have been imported in
\texttt{vertices.py} as the objects \texttt{P},
\texttt{L} and \texttt{C}, respectively\footnote{This is achieved by including
at the beginning of the \texttt{vertices.py} file the commands \texttt{import
particles as P}, \texttt{import lorentz as L} and \texttt{import couplings as
C}.}.

The {\tt Vertex} class comes with the five mandatory attributes shown in the
example. First, each vertex is identified by a string stored in 
the attribute \texttt{name}. Next, the attribute \texttt{particles}
contains a list with the particles incoming to the vertex, represented by the
corresponding {\tt Particle} objects. 
The attributes \texttt{color} and \texttt{lorentz} respectively refer to lists
with the color and Lorentz tensor bases associated to the vertex under
consideration, \ie, the
quantities $C_i^{a_1\ldots a_n}$ and $L_j^{\ell_1\ldots\ell_n}(p_1,\ldots,p_n)$
of Eq.\ \eqref{eq:ufovert}. The coordinates in the spin $\otimes$ color basis,  
\ie, the $G_{ij}$ quantities of Eq.\ \eqref{eq:ufovert}, are also implemented as
a list which is stored as the value of the attribute \texttt{couplings}.
This list is given as a
\python\ dictionary relating the coordinate $(i,j)$ to a particular
\texttt{Coupling} object.

As it can be observed in the example, each color tensor is given as a string
representing a polynomial combination of elementary color tensors. The UFO
conventions\footnote{In this paragraph, $i_n$ ($\alpha_n$) stands for triplet
(sextet) color indices while $\jbar_n$ ($\betabar_n$) denotes antitriplet
(antisextet) color indices; adjoint color indices are
written as $a_n$. The symbol $n$ is an integer number related to
the $n^{\rm th}$ particle incoming to the vertex.} for the latter imply to use
\texttt{1} for the trivial
tensor, \texttt{Identity(1,2)} for the Kronecker delta
$\delta^{\jbar_2}{}_{i_1}$,
\texttt{T(1,2,3)} for a fundamental representation matrix
$\big(T^{a_1}\big)^{\bar\jmath_3}{}_{i_2}$, \texttt{f(1,2,3)} for an
antisymmetric structure constant $f^{a_1a_2a_3}$, \texttt{d(1,2,3)} for a
symmetric structure constant $d^{a_1a_2a_3}$, \texttt{Epsilon(1,2,3)} for a 
fundamental Levi-Civita tensor $\e_{i_1i_2i_3}$, \texttt{EpsilonBar(1,2,3)} for
an antifundamental Levi-Civita tensor $\e^{\jbar_1\jbar_2\jbar_3}$ and
\texttt{T6(1,2,3)} for a sextet representation matrix
$\big(T_6^{a_1}\big)^{\betabar_3}{}_{\alpha_2}$. In addition, one is allowed to
employ sextet Clebsch-Gordan coefficients
$\big(K_6\big)_{\alpha_1}{}^{\jbar_2\jbar_3}$ arising from the combination of
two generators of $SU(3)$ lying in the fundamental representation  as well as
their conjugate counterparts $\big(\Kbar_6\big)^{\betabar_1}{}_{i_2j_3}$. These
coefficients are implemented as the objects \texttt{K6(1,2,3)} and
\texttt{K6Bar(1,2,3)}, the conventions for sextet and
antisextet representations of $SU(3)$ following those presented in 
Ref.\ \cite{Han:2009ya}. All the UFO objects above take integer numbers as
arguments which refer to the position of the relevant particles in the list
provided as the value of the 
\texttt{particle} attribute of the \texttt{Vertex} class. Repeated negative
numbers also appear and are related to contracted indices.

As illustrated in the four-gluon vertex implementation above, the
\texttt{lorentz} attribute of the {\tt Vertex} class contains a list with the
 spin structures relevant for the vertex. The latter are implemented as instances
of the class \texttt{Lorentz} and are all declared in the \texttt{lorentz.py}
file. The three mandatory attributes of a
\texttt{Lorentz} object are its name (stored as a string in the attribute
\texttt{name}), the list of spin states interacting at the vertex given under the
$2s+1$ form (stored in the attribute \texttt{spins}) and the Lorentz
structure itself given as an analytical formula (stored as a string in the
attribute \texttt{structure}). For instance, the \texttt{VVVV1} structure
appearing in the four-gluon vertex is implemented as 
\begin{verbatim}
  VVVV1 = Lorentz(
     name      = 'VVVV1',
     spins     = [3,3,3,3],
     structure = 'Metric(1,4)*Metric(2,3) - Metric(1,3)*Metric(2,4)'
  )
\end{verbatim} 
In order to write down the structure, one can employ several tensors which are
represented in the UFO conventions\footnote{In this paragraph, we denote by
$s_n$ and $\mu_n$ spin
and Lorentz indices, $n$ being an integer number related to the $n^{\rm th}$
particle incoming to the vertex.} by
\texttt{C(1,2)} for the charge conjugation matrix $C_{s_1 s_2}$, 
\texttt{Epsilon(1,2,3,4)} for the rank-four totally antisymmetric tensor
$\epsilon^{\mu_1 \mu_2 \mu_3\mu_4}$, \texttt{Gamma(1, 2, 3)} and
\texttt{Gamma5(1,2)}  for the Dirac matrices $(\gamma^{\mu_1})_{s_2 s_3}$ and
$(\gamma^5)_{s_1 s_2}$, \texttt{Identity(1,2)} for the Kronecker delta
$\delta_{s_1 s_2}$, \texttt{Metric(1,2)} for the Minkowski metric $\eta_{\mu_1
\mu_2}$, \texttt{P(1,N)} for the momentum of the $N^{\text{th}}$ particle
$p^{\mu_1}_N$, \texttt{ProjP(1,2)} and \texttt{ProjM(1,2)} for the right-handed
and left-handed chirality projectors $[1\pm\gamma_5)/2]_{s_1 s_2}$ and
\texttt{Sigma(1,2,3,4)} for the $\gamma^{\mu_1 \mu_2}_{s_3 s_4}$ matrices. 
The arguments of these objects follow the same conventions as those employed for 
the color tensors, negative indices standing thus for contracted indices and any
positive integer number $i$ referring to the $i^{\rm th}$ particle incoming to the
vertex.

Each coordinate in the spin $\otimes$ color basis in which the vertex is
expanded is implemented as a \texttt{Coupling} object, all these objects being
all declared in the file \texttt{couplings.py} as internal
parameters. For instance, the coupling \texttt{GC\textunderscore 1} appearing in
the four-gluon vertex is implemented as 
\begin{verbatim}
  GC_1 = Coupling(
    name  = 'GC_1',
    value = 'complex(0,1)*G**2',
    order = {'QCD':2}
  )
\end{verbatim}
The attribute \texttt{name} is a string with the name of the \texttt{Coupling}
object and \texttt{value} contains, as a string, the algebraic expression of the
coupling in terms of the model parameters. 
The \texttt{order} attribute is related to the coupling orders (See Section
\ref{sec:FRprm}). The value of this attribute consists of a \python\ dictionary
where
the key of each entry is a string and its value a non-negative integer. In the
example above, this means that the \texttt{GC\textunderscore 1} coupling is
proportional
to two powers of the strong coupling. 

Coupling orders are specified in the UFO by instances of
the \texttt{Coupling\-Order} class. All the coupling orders that have to be
implemented for a specific model are collected into the file
\texttt{coupling\textunderscore orders.py}. A \texttt{CouplingOrder} object has
three attributes, its name given as a string (\texttt{name}), the maximum
number of occurrences of the coupling order for a specific Feynman diagram given
as an
integer number (\texttt{expansion\textunderscore order}), and its relative
strength, compared to the other coupling orders of the model given as an
integer (\texttt{hierarchy}). For instance, the declarations of the coupling
orders {\tt QCD} and {\tt QED} read
\begin{verbatim}
  QCD = CouplingOrder(name=`QCD', expansion_order=99, hierarchy=1)
  QED = CouplingOrder(name=`QED', expansion_order=99, hierarchy=2)
\end{verbatim}
where the value 99 of the attribute \texttt{expansion\textunderscore
order} indicates that any number of QCD and QED couplings are allowed in a
Feynman diagram.

\mysection{Implementing supersymmetric models in \feynrules} \label{sec:spacemod}
\subsection{Calculations in superspace}\label{sec:FRsuperspace}
As presented in Section \ref{sec:superspace}, the $N=1$ superspace is an
extension of the ordinary spacetime defined by adjoining a Majorana spinor
$(\theta_\alpha, \thetabar^\alphadot)$ to the usual spacetime coordinates. These
Weyl fermions can be used within \feynrules\ through the symbols \texttt{theta}
and \texttt{thetabar} 
\be
  \texttt{theta[alpha]} \leftrightarrow \theta_\alpha 
  \quad \text{and} \quad
  \texttt{thetabar[alphadot]} \leftrightarrow \bar\theta_{\dot\alpha} 
\nn \ee
where, by conventions, both spin indices are assumed to be lower indices.
The position of the spin indices can be modified by employing the
rank-two antisymmetric tensors as introduced in Eq.\ \eqref{eq:rank2eps}.
Levi-Civita tensors with lower and upper indices are implemented in \feynrules\
as the objects \texttt{Deps} and \texttt{Ueps}, respectively, which can
equivalently take left-handed or right-handed indices.

In order not to loose minus signs in superspace computations, it is mandatory to
keep track of the position of the spin indices as well as of the fermion
ordering. To this aim, \feynrules\ always assumes that an explicit spin index is a
lower index. Moreover, the environment \texttt{nc[chain]} has to be used,
where {\tt chain} stands for any ordered sequence of fermions (with lower spin
indices).
As simple examples, scalar products such as those introduced in
Eq.~\eqref{eq:scalprod} could be implemented as
\be\bsp
 &\  \lambda \!\cdot\! \lambda^\prime \leftrightarrow 
     \texttt{nc[lambda[a], lambdaprime[b]] Ueps[b,a]}  \\
 &\  \chibar \!\cdot\! \chibar^\prime \leftrightarrow 
     \texttt{nc[chibar[bd], chibarprime[ad]] Ueps[bd,ad]} 
\esp\nn\ee
where $(\lambda, \lambda')$ and $(\chibar, \chibar')$ are two pairs of
left-handed and right-handed Weyl fermions, respectively.
Even though the form of the scalar products above is the canonical form used
inside the code and not too complicated to implement, the situation becomes 
tricky when dealing with longer, more complex, expressions. For this
reason, the package contains an environment \texttt{ncc} which has the 
same effect as the \texttt{nc} environment. However, all spin indices and
$\e$-tensors can be omitted,
\be
   \lambda\cdot\lambda^\prime \leftrightarrow 
    \texttt{ncc[lambda, lambdaprime]} 
\qquad\text{and}\qquad
   \chibar \!\cdot\! \chibar^\prime \leftrightarrow 
    \texttt{ncc[chibar, chibarprime]}\vspace*{-.1cm}
\nn\ee
The code is assuming that all the
suppressed indices are contracted according to the conventions of Appendix
\ref{app:conv} and outputs the results into their canonical form, employing
rank-two
antisymmetric tensors and two-component spinors with lowered indices ordered
within a \texttt{nc} environment where relevant.

As a consequence of this canonical form, the \mathematica\ output for superspace
expressions could be difficult to read, especially when the expressions are
long. To bypass this issue and facilitate the readability, invariant products of
spinors can be formed. In particular, products of the Grassmann variables
$\theta$ and $\thetabar$ can always be simplified using the relations of
Eq.~\eqref{eq:grassmannalg}. This is achieved in \feynrules\ with the help of
the \texttt{ToGrassmannBasis} command,
\begin{verbatim}
  ToGrassmannBasis[expression] 
\end{verbatim}
where {\tt expression} stands for any function of the superspace coordinates.
This method allows to rewrite any expression in terms of a restricted set of
scalar products involving
Grassmann variables and Pauli matrices. In addition, the index naming scheme
employed for the results has been optimized, so that \mathematica\ expressions
that are equal
up to the names of contracted indices, such as 
\begin{verbatim}
  Dot[theta[al], theta[al]] -  Dot[theta[be], theta[be]] 
\end{verbatim}
are collected and summed\footnote{As the two terms represent different patterns in
\mathematica, the cancellation does not take place.}.

For instance, applying the \texttt{ToGrassmannBasis} function on the scalar
product $\lambda\!\cdot\!\lambda'$,
\begin{verbatim}
  ToGrassmannBasis[ nc[lambda[sp1],lambdaprime[sp2]] Ueps[sp2,sp1] ]
\end{verbatim}
leads to a \mathematica\ output very close to the original form. This 
simplification method also works on spinorial and tensorial
expressions containing non-contracted spin indices. In this case, the results can 
contain upper or lower free indices. Each free index is either attached to a
single fermion or to a chain
containing one fermion and a given number of Pauli matrices, such as 
$(\sigma^\mu\sibar^\nu\lambda)_\alpha$. Applying the \texttt{ToGrassmannBasis}
function on such an expression, 
\begin{verbatim}
  ToGrassmannBasis[ nc[lambda[b]] * si[mu,a,ad] * sibar[nu,ad,b] ]
\end{verbatim}
one obtains 
\begin{verbatim}
  nc[ TensDot2[si[mu,a,ad], sibar[nu,ad,b], lambda[b]][down,Left,a] ] 
\end{verbatim}
A chain with two Pauli matrices and the fermion has been formed
and stored in a \texttt{TensDot2} environment. This new environment follows the
pattern 
\begin{verbatim}
  TensDot2[ chain ][pos, chir, name]
\end{verbatim}
where \texttt{chain} is a
sequence of one Weyl fermion and possibly one or several Pauli matrices,
\texttt{pos} is the \texttt{up} or \texttt{down} position of the free spin
index, \texttt{chir} its dotted (\texttt{Left}) or undotted (\texttt{Right}) 
nature and \texttt{name} the name of the free index. 

The optimization of the index naming scheme can also be
performed without forming scalar products involving Grassmann variables. The
standalone version of this method consistently renames the
indices of an expression and is called in \feynrules\ by issuing
\begin{verbatim}
  OptimizeIndex[expression, list] 
\end{verbatim}
where \texttt{list} is an optional list of variables carrying indices to be
included in the index renaming procedure and that are neither fields nor model
parameters.

The basis corresponding to the output of the
\texttt{To\-Grassmann\-Basis} function can also be used to input superspace
expressions. The rules for this format are the following. First, invariant
products of spinors, connected or not by one or several Pauli matrices, are
written as
\begin{verbatim}
  ferm1[sp1].ferm2[sp2] chain[sp1,sp2] 
\end{verbatim}
where the symbols \texttt{ferm1} and \texttt{ferm2} denote two Weyl fermions of
the model. The quantity \texttt{chain} stands for a series of Pauli matrices
linking the
spin indices \texttt{sp1} and \texttt{sp2}, such as, \eg, \texttt{si[mu,sp1,spd]
sibar[nu,spd,sp2]}. Trivially, in the case the two spin indices are equal, no
Pauli matrix is included.
Next, the implementation of any fermionic expression carrying a free spin index
must employ both the \texttt{nc} environment and the \texttt{TensDot2}
structure, following the
syntax detailed above. 
In a second step, the implemented expressions can be converted to their
canonical form by employing the \texttt{Tonc} function, 
\begin{verbatim}
  Tonc[ expression ]
\end{verbatim}
We remind that converting expressions to their canonical form is mandatory to
use any method, described below, included in the superspace module of \feynrules.

\subsection{Supercharges, superderivatives and supersymmetric transformations}
\label{sec:FRsch}
The generators $Q_\alpha$ and $\Qbar_\alphadot$ of the supersymmetric
transformations and the superderivatives $D_\alpha$ and $\Dbar_\alphadot$ that
have been computed in Section \ref{sec:schsder} and collected in Eq.\
\eqref{eq:schsder} can be called in \feynrules\ via the commands
\be \bsp
 &\ \texttt{QSUSY[expression,alpha]} \leftrightarrow Q_\alpha
    (\text{expression})  \ , \\ 
 &\ \texttt{QSUSYBar[expression,alphadot]} \leftrightarrow \bar Q_\alphadot
    (\text{expression})  \ , \\ 
 &\ \texttt{DSUSY[expression,alpha]} \leftrightarrow D_\alpha
    (\text{expression})  \ , \\ 
 &\ \texttt{DSUSYBar[expression,alphadot]} \leftrightarrow \bar D_\alphadot
    (\text{expression})  \ .
\esp \nn \ee
For a proper treatment of the spin indices, all the indices appearing in the
quantity
\texttt{expression} must be explicit and the environment {\tt nc} has to be used.
The only exception consists of the single fermion case where the
\texttt{nc} environment can be omitted since fermion ordering is trivially
irrelevant.

As presented in Chapter \ref{chap:susy}, the variation of a superfield under a 
supersymmetric transformation can be computed using the supercharges, 
\be
  \delta_\e \Phi = i \big(\e\!\cdot\! Q + \Qbar \!\cdot\! \ebar\big) \Phi \ ,
\ee
where the Majorana spinor $(\e,\ebar)$ stands for the transformation parameters.
After replacing the supercharges by their derivative representation and 
expanding the superfield equation above in terms of the component fields, 
we can immediately read off the transformations laws the component fields of
$\Phi$. 

The operator $\delta_\e$ is implemented via the {\tt DeltaSUSY} function,
\begin{verbatim}
  DeltaSUSY [ expression , epsilon ] 
\end{verbatim}
In this expression, the symbol \texttt{expression} is any polynomial function of
superfields and/or component fields while \texttt{epsilon} refers to the
left-handed piece of the supersymmetric transformation parameters, to be given
without any spin index. There are ten of such parameters predefined in the
superfield module, labeled by \texttt{eps\it{x}} with \texttt{\it{x}} being an
integer between zero and nine. The output of the
\texttt{DeltaSUSY} method is given as the full series expansion in terms of the 
Grassmann variables.

\subsection{Implementing and manipulating superfields in \feynrules}
\label{sec:FRSF}
Since most of the phenomenologically relevant supersymmetric theories can be
built with only chiral and vector superfields (in the Wess-Zumino gauge),
\feynrules\ allows for a very
efficient way of implementing these two types of superfield. Superfields are
declared as instances of the superfield class which are collected, in a model
file, in the list \texttt{M\$Superfields},
\begin{verbatim}
  M$Superfields = { superfield1=={options}, superfield2=={options}, ...  };
\end{verbatim}
where \texttt{superfield1}, \texttt{superfield2}, \etc, are user-defined
names for the declared superfield classes and \texttt{options} contains 
\mathematica\ replacement rules with the properties defining each superfield. As for
particles, the syntax for the names is constrained. They start
with the letters \texttt{CSF} or \texttt{VSF} for chiral and vector
superfields, respectively, followed by a number chosen by the user and put
between squared brackets. 

Superfields can be declared in \feynrules\ model files in a way similar to the
one shown in Section \ref{sec:FRfields} for the ordinary fields. In order to
illustrate the features associated with superfield declaration, we take the examples
of a left-handed chiral superfield $\Phi_L$, a right-handed chiral superfield
$\Phi_R$ and a vector superfield $\Phi_V$. Their implementation in a \feynrules\
model file reads
\begin{verbatim}
 CSF[1] == {            CSF[2] == {              VSF[1] == { 
   ClassName -> PHIL,     ClassName -> PHIR,       ClassName  -> PHIV,
   Chirality -> Left,     Chirality -> Right,      GaugeBoson -> V,
   Weyl      -> psi,      Weyl      -> psibar,     Gaugino    -> lambda,
   Scalar    -> z,        Scalar    -> zbar        Indices    -> {Index[SU2W]}
   Auxiliary -> FF         
 }                      }                          }
\end{verbatim}
which results in the declaration of two chiral superfields labeled by the
tags \texttt{CSF[1]} and \texttt{CSF[2]} and of one
vector superfield labeled by \texttt{VSF[1]}. The link to the
symbol to be used when employing such superfields in Lagrangian building or
superspace computations is provided as the value of the option
\texttt{ClassName}, similarly to the declaration of normal fields. Moreover, the
chirality of chiral superfields is assigned through the value of the option 
\texttt{Chirality} of the superfield class (being set to \texttt{Left} or
\texttt{Right}). 

Each superfield must also be related to its component fields. 
The fermionic and scalar components of a chiral superfield are referred to as the
value of the options \texttt{Weyl} and \texttt{Scalar}, whilst the bosonic and
gaugino components of a vector superfield are defined by setting the options
{\tt GaugeBoson} and {\tt Gaugino} to the name of the corresponding fields,
all component fields having been properly declared in the \feynrules\
model file (see Section \ref{sec:FRfields}). In contrast, the declaration of
the
auxiliary fields is optional. If absent from the model implementation,
\feynrules\ takes care of it internally (as for the $\Phi_R$ and $\Phi_V$
superfields above), otherwise, the user must set the value of the
\texttt{Auxiliary} attribute of the superfield class referring to the name of a 
non-physical scalar field (as for the $\Phi_L$ superfield above).

In addition, all the other attributes available to all particle classes and
reviewed in Section~\ref{sec:FRfields}, such as \texttt{QuantumNumbers} and
\texttt{Indices}, can also be employed for superfield declarations. 

As already mentioned in Section \ref{sec:FRgroups}, a vector
superfield
can be linked to a gauge group through the option \texttt{Superfield} of the
gauge group class which points towards the name of the corresponding vector
superfield, \ie, the symbol referred to by the \texttt{ClassName}
attribute. 

Superfield strength tensors could be implemented by hands by means
of the superderivatives, following their definitions of Eq.\
\eqref{eq:Wabelian} and Eq.\ \eqref{eq:defWWbar}. However, \feynrules\ comes
with built-in functions dedicated to this task, that can be called in a
\mathematica\ session as
\begin{verbatim}
  SuperfieldStrengthL[ V, sp    ]        SuperfieldStrengthL[ VV, sp,    ga ]
  SuperfieldStrengthR[ V, spdot ]        SuperfieldStrengthR[ VV, spdot, ga ]
\end{verbatim}
The two methods of the first line allow to derive the left-handed superfield
strength
tensors $W_\alpha$ and $W_\alpha^a$ associated with abelian and non-abelian
vector superfields denoted by \texttt{V} and \texttt{VV}, respectively. In
contrast, the last 
two methods are dedicated to the abelian and non-abelian right-handed superfield
strength tensors $\Wbar_\alphadot$ and $\Wbar_\alphadot^a$. In the expressions
above, the symbols \texttt{sp} and \texttt{spdot} denote respectively an
undotted and dotted spin index, while \texttt{ga} is an adjoint gauge index
relevant for non-abelian gauge groups. It is important to emphasize that these
spinorial superfields are not hard-coded in \feynrules\ and are recalculated
each time it is necessary. However, the results of the
\texttt{SuperfieldStrengthL} and
\texttt{SuperfieldStrengthR} functions are only evaluated at the time an
expansion in terms of the component fields is performed.

In Section \ref{sec:CSF} and Section \ref{sec:VSF}, we have worked out the
expansion of chiral and vector superfields as a series in the Grassmann
variables. This can be automatically performed in \feynrules\ via the
\texttt{SF2\-Components} function,
\begin{verbatim}
  SF2Components [ expression ] 
\end{verbatim}
This expands in a first step all the chiral and vector superfields appearing in
the quantity \texttt{expression} in terms of their component fields and the
usual spacetime coordinates (in contrast to the $y$-variable related to chiral
superfields). Secondly, scalar products of Grassmann variables are simplified
and the expression is reduced to a human-readable form by internally calling the 
\texttt{ToGrassmannBasis} function. During this procedure,
representation matrices of the gauge Lie algebra could appear, as for instance,
when expanding vector superfields of the type $V=V^a T_a$, $T_a$ being such
representation matrices and $V^a$ a set of vector superfields. If necessary, the
commutation relations between the generators are internally employed for
simplifications. 

The output of the \texttt{SF2Components} function consists of a two-component
list of the form 
\begin{verbatim}
  { Full series , List of the nine coefficients }
\end{verbatim}
The first element of this list (\texttt{Full series}) consists of the full
series expansion in terms of the Grassmann variables. This could equivalently
be obtained with the \texttt{GrassmannExpand} function,
\begin{verbatim}
  GrassmannExpand [ expression ] 
\end{verbatim}
The second element of the list above is itself a list containing the nine
coefficients of the series, \ie, the scalar piece independent of the Grassmann
variables, followed by the coefficients of the $\theta_\alpha$,
$\thetabar_\alphadot$, $\theta \sigma^\mu\thetabar$, $\theta\!\cdot\!\theta$,
$\thetabar\!\cdot\!\thetabar$, $\theta\!\cdot\!\theta \thetabar_\alphadot$,
$\thetabar\!\cdot\!\thetabar \theta_\alpha$ and $\theta\!\cdot\!\theta
\thetabar\!\cdot\!\thetabar$ terms. Each of these could also be obtained using
the dedicated functions \texttt{ScalarComponent}, \texttt{ThetaComponent},
\texttt{ThetabarComponent}, \texttt{Theta\-Thetabar\-Component},
\texttt{Theta2Component}, \texttt{Thetabar2Component},
\texttt{Theta2ThetabarComponent}, \texttt{Thetabar2\-Theta\-Component} and
\texttt{Theta2Thetabar2Component}. A spin index can also be specified in the arguments of
the functions related to fermionic coefficients and a Lorentz index for
\texttt{ThetaThetabarComponent}.

\subsection{Automatic generation of supersymmetric Lagrangians}
\label{sec:frsusylag}
Several built-in functions are available in \feynrules\ for generating
the Lagrangians associated with renormalizable supersymmetric theories given 
by Eq.\ \eqref{eq:gensusylag} and Eq.\ \eqref{eq:gensusylag2}.

The kinetic part of the Lagrangian describing the dynamics of a chiral 
superfield $\Phi$ (neglecting
for the moment its gauge interactions) can be implemented by means of the
function \texttt{Theta2Thetabar2} introduced in Section \ref{sec:FRSF}, 
\begin{verbatim}
  Theta2Thetabar2Component[PHIbar PHI]
\end{verbatim} 
In this expression, \texttt{PHI} is the symbol representing the superfield
$\Phi$, assuming that it has been correctly declared in the \feynrules\ model
file. This Lagrangian is a direct translation in the \feynrules\ language of the 
superfield Lagrangian given in Eq.\ \eqref{eq:SK}. However, this Lagrangian can
also be automatically derived by employing the \texttt{CSFKineticTerms}
function,
\begin{verbatim} 
  Theta2Thetabar2Component[ CSFKineticTerms[PHI] ]
\end{verbatim}
which automatically accounts, in addition, for the possible gauge interactions
of the superfield
$\Phi$. Since the \texttt{CSFKineticTerms} method returns a non-expanded
superfield expression, the relevant component field must be selected by applying
the \texttt{Theta2Thetabar2Component} function to its result. The full chiral
Lagrangian is obtained by summing explicitly over all the chiral content of the
model under consideration, or by issuing the command
\begin{verbatim} 
  Theta2Thetabar2Component[ CSFKineticTerms[] ]
\end{verbatim}
where \texttt{CSFKineticTerms} is called without any argument. 

As stated in Chapter \ref{chap:susy}, the interactions among the chiral
superfields are driven by the superpotential.
Implementing it in the \feynrules\ model under a variable that we label by
\texttt{SuperPot} in our example,
we can derive the associated interaction Lagrangian shown in 
Eq.~\eqref{eq:Sint} by employing the functions allowing to
extract the $\theta\cdot\theta$ and $\thetabar\cdot\thetabar$ components
of a superfield object, 
\begin{verbatim}
  Theta2Component[SuperPot] + Thetabar2Component[HC[SuperPot]] 
\end{verbatim}

We now turn to the gauge sector of the supersymmetric Lagrangians. 
From the superfield strength tensor implementation described in Section
\ref{sec:FRSF}, we can easily build kinetic terms for vector superfields.
However, this can also be done in an automated way by issuing 
\begin{verbatim}
 Theta2Component[VSFKineticTerms[V]] + Thetabar2Component[VSFKineticTerms[V]]
\end{verbatim}
where \texttt{V} stands for a vector superfield properly declared in the model
file. As for the \texttt{CSF\-Kinetic\-Terms} function, \texttt{VSFKineticTerms}
does not perform any expansion in terms of the Grassmann variables. Therefore,
the \texttt{Theta2Component} and \texttt{Thetabar2Component} routines have
to be employed to get the super-Yang-Mills Lagrangian of Eq.\ \eqref{eq:SYM} or
the supersymmetric abelian vector Lagrangian of Eq.\ \eqref{eq:SU1}.
Similarly to the automatic generation of the complete chiral Lagrangian, 
issuing \texttt{VSFKineticTerms} without any argument leads
to the derivation of kinetic and gauge interaction terms for all the vector
superfields defined in the model.

To summarize, generating a Lagrangian density for a supersymmetric model in
\feynrules\ is reduced to the task of defining
the superpotential \texttt{SuperPot} in terms of
the superfield content. The full (supersymmetric) Lagrangian can then be
calculated as 
\begin{verbatim}
 LC=Theta2Thetabar2Component[CSFKineticTerms[]]; 
 LV=Theta2Component[VSFKineticTerms[]]+Thetabar2Component[VSFKineticTerms[]];
 LW=Theta2Component[SuperPot]+Thetabar2Component[HC[SuperPot]];
 Lag = LC + LV + LW;
\end{verbatim}

The Lagrangian density obtained in this way however still depends on the
auxiliary $F$-fields and $D$-fields that have to eliminated by inserting in the
Lagrangian the solution of their equations of motion. This can be automatically
performed via the functions \texttt{SolveEqMotionD}
and \texttt{SolveEqMotionF}, 
\begin{verbatim}
  Lag = SolveEqMotionF[SolveEqMotionD[Lag]]; 
\end{verbatim}
where \texttt{Lag} is the Lagrangian calculated above. 

Finally, in order to pass the Lagrangian to the \texttt{FeynmanRules} function
presented in Section~\ref{sec:FRrun} or to use the interfaces to Monte Carlo
event generators (see Section~\ref{sec:FRint}), the Lagrangian has still to be
re-expressed in terms of four-component Dirac and Majorana fermions rather than
in terms of two-component fermions. As already mentioned in Section
\ref{sec:FRlag}, this step is automated and it is sufficient to type
\begin{verbatim}
  Lag = WeylToDirac[ Lag ];
\end{verbatim}
However, a subtlety occurs for the QCD gauge group because of the reserved names
of the color indices. We refer to a specific example as the implementation
of the MSSM which is detailed in Section \ref{sec:FRMSSM} or in Ref.\
\cite{Duhr:2011se}. Other examples of non-minimal supersymmetric model
implementations can be found in Ref.\ \cite{Fuks:2012im}.

\mysection{Supersymmetric transformations}\label{sec:deltasusy}

\subsection{Variation of chiral superfields}\label{sec:deltasusychir}
In this section, we present a first example of the usage of \feynrules\ for
calculations in superspace. We address a detailed computation of the variation
$\delta_\e \Phi$ of a left-handed
chiral superfield $\Phi$ under a supersymmetry transformation of parameters
$(\e,\ebar)$. This superfield is declared in the \feynrules\ model file as 
\begin{verbatim}
  CSF[1] == {
    ClassName -> PHI,
    Chirality -> Left,
    Weyl      -> psi,
    Scalar    -> z,
    Auxiliary -> F
  }
\end{verbatim}
according to the rules presented in Section \ref{sec:FRSF}. Moreover, we
remind that this declaration must be included in the list
\texttt{M\$Superfields} containing all the superfields of the model. The scalar,
fermionic and auxiliary
components of $\Phi$, denoted by $z$ (\texttt{z}), $\psi$ (\texttt{psi}) and $F$
(\texttt{F}),  are declared within the list
\texttt{M\$Classes\-Description}, as explained in Section
\ref{sec:FRfields}. Following the standard syntax for declaring fields, we have
\begin{verbatim}
  S[1] == {
    ClassName     -> F,
    SelfConjugate -> False,
    Unphysical    -> True
  }

  S[2] == {
    ClassName     -> z,
    SelfConjugate -> False
  }

  W[1] == {
    ClassName     -> psi,
    SelfConjugate -> False,
    Chirality     -> Left
  }
\end{verbatim}

\begin{figure}[t!]
 \centering
 \includegraphics[width=.66\columnwidth]{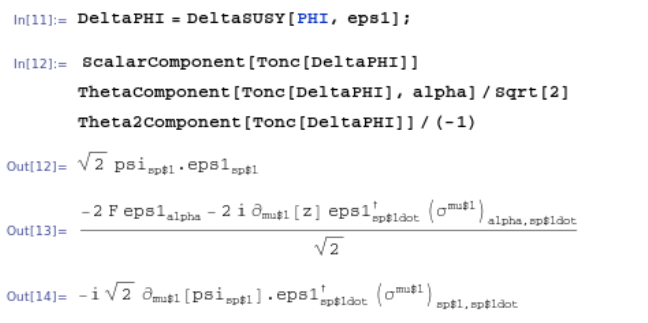}
 \caption{\label{fig:susytr}
 Screenshot of a \mathematica\ session where the variations under a
 supersymmetric transformation of parameters $(\e_1,\ebar_1)$ of the component
 fields of a chiral superfield  are computed. We refer to the text for more
 details.}
\end{figure}

The easiest way to proceed with the computation of the variations of the
components of the superfield $\Phi$ under a supersymmetric transformation is to
employ the \texttt{DeltaSUSY} function introduced in Section
\ref{sec:FRsch}. As stated above, this function admits two types of arguments,
either a polynomial function of the model superfields or an expression depending
on the Grassmann variables and the component fields. Therefore, the variation
$\delta_\e\Phi$ can be calculated by issuing equivalently one of the two
commands,
\begin{verbatim}
  DeltaPHI = DeltaSUSY[ PHI , eps1 ]
  DeltaPHI = DeltaSUSY[ Tonc[GrassmannExpand[PHI]] , eps1 ]
\end{verbatim}
where we have employed one of the ten predefined Weyl fermions of the form
\texttt{eps\it{x}} for the transformation parameters.
The variation of the scalar, fermionic and auxiliary component fields are
deduced from the lower order coefficients of the series in the Grassmann
variables, 
\begin{verbatim}
  ScalarComponent[ Tonc[DeltaPHI] ]
  ThetaComponent[ Tonc[DeltaPHI], alpha]/Sqrt[2]
  Theta2Component[ Tonc[DeltaPHI] ]/(-1)
\end{verbatim}
where the numerical denominators are related to the normalization conventions
for chiral superfields (see Eq.\ \eqref{eq:chiralSF}). 
The function \texttt{Tonc} is necessary as the \texttt{XXXXComponent} methods 
require their arguments to be expressed in the canonical form. Furthermore, the
explicit \texttt{alpha} index included in the second command ensures that the
free spin index of the
$\theta$-component of $\delta_\e \Phi$ is denoted by \texttt{alpha} in the
\mathematica\ output. We recover the well-known textbook expressions,
\be\bsp
  \delta_\e z =&\ \sqrt{2} \e \!\cdot\! \psi \ ,\\
  \delta_\e \psi_\alpha =&\ -\sqrt{2} \e_\alpha F -i \sqrt{2} \big(\sigma^\mu
    \ebar\big)_\alpha \del_\mu z \ , \\
  \delta_\e F =&\ -i \sqrt{2} \del_\mu \psi \sigma^\mu \ebar 
    = -i \sqrt{2} \del_\mu\big[ \psi \sigma^\mu \ebar \big ]\ ,
\esp\label{eq:susyvarchircomp}\ee
where for the last equality, we remind that the transformation parameters are
constant. These formulas can be compared to the expressions obtained with
\feynrules\ that are illustrated in Figure \ref{fig:susytr}.
In Section \ref{sec:susylag}, we have mentioned that the $\theta\cdot\theta$
term of a chiral superfield can be used as a Lagrangian candidate since it is 
invariant, up to a total derivative, under supersymmetric transformations. This
statement is proved above, since $\delta_\e F$ is exactly a total derivative.
%
%

\subsection{Variation of general superfields}\label{sec:deltasusygen}
As a second example, we perform the same exercise as in Section
\ref{sec:deltasusychir} but in the case of general superfields. There is no dedicated
function in \feynrules\ to get the expansion of such superfields in terms of the
Grassmann variables so that we have to implement it directly in the
\mathematica\ session.  In our example, we consider a general
superfield represented by the symbol \texttt{Phi} and use the \texttt{ncc}
environment presented in Section \ref{sec:FRsuperspace} to facilitate its
implementation,
\begin{verbatim}
  Phi = z + ncc[theta, xi] + ncc[thetabar, zetabar] + ncc[theta, theta] F + 
    ncc[thetabar, thetabar] G + ncc[theta, si[mu], thetabar]  v[mu] + 
     ncc[thetabar, thetabar] ncc[theta, omega] + 
     ncc[theta, theta] ncc[thetabar, rhobar] +
     ncc[theta, theta] ncc[thetabar, thetabar] d
\end{verbatim}
The names and symbols related to the component fields follow those of Eq.\
\eqref{eq:genSF}, and the list \texttt{M\$\-Classes\-Declaration} accordingly
contains 
\begin{verbatim}
  V[1] == {ClassName -> v, SelfConjugate->True}
  W[1] == {ClassName->zeta,  Chirality->Left, SelfConjugate->False}
  W[2] == {ClassName->xi,    Chirality->Left, SelfConjugate->False}
  W[3] == {ClassName->omega, Chirality->Left, SelfConjugate->False}
  W[4] == {ClassName->rho,   Chirality->Left, SelfConjugate->False}
  S[1] == {ClassName->z, SelfConjugate->False}
  S[2] == {ClassName->F, SelfConjugate->False}
  S[3] == {ClassName->G, SelfConjugate->False}
  S[4] == {ClassName->d, SelfConjugate->False}
\end{verbatim}

The variation of the general superfield under a supersymmetric transformation of
parameters $(\e,\ebar)$ is computed as in the previous section,
\begin{verbatim}
  DeltaPhi = DeltaSUSY[ Phi , eps1 ]
\end{verbatim}
Extracting the nine component fields by means of the
\texttt{XXXXComponent[DeltaPhi]} commands, one gets the textbook expressions
\be\bsp
  \delta z   = &\ \e \!\cdot\!\xi + \ebar \!\cdot\! \zetabar \ , \\
  \delta \xi_\alpha =&\ 2 f \e_\alpha + \big(\sigma^\mu \ebar\big)_\alpha
     \big(v_\mu - i\del_\mu z\big) \
    , \\ 
  \delta \zetabar^\alphadot = &\ 2 g \ebar^\alphadot  - \big(\sibar^\mu
     \e\big)^\alphadot \big(v_\mu + i\del_\mu z\big) \ , \\
  \delta v_\mu =&\ -\frac{i}{2} \e\!\cdot\!\del_\mu \xi - \e \sigma_{\nu\mu}
    \del^\nu \xi + \frac{i}{2} \ebar \!\cdot\!\del_\mu \zetabar - \ebar
    \sibar_{\nu \mu} \del^\nu \zetabar + \omega\sigma_\mu\ebar +
    \e\sigma_\mu\rhobar \ ,\\
  \delta f=&\ \frac{i}{2} \del_\mu \xi \sigma^\mu \ebar + \ebar \!\cdot\!
    \rhobar\ ,\\ 
  \delta g=&\ -\frac{i}{2} \e \sigma^\mu \del_\mu \zetabar + \e \!\cdot\!
    \omega\ ,  \\
  \delta \omega_\alpha =&\ -i  \big(\sigma^\mu \ebar\big)_\alpha  \del_\mu g 
     + \frac{i}{2} \e_\alpha \del_\mu v^\mu - \frac{i}{2} \big(\sigma^{\mu \nu}
     \e\big)_\alpha \ F^0_{\mu \nu} + 2 \e_\alpha d \ ,  \\
  \delta \rhobar_\alphadot =&\ i\big(\e\sigma^\mu\big)_\alphadot \del_\mu f
     - \frac{i}{2} \ebar_\alphadot \del_\mu v^\mu  - \frac{i}{2} \big(\ebar
       \sigma^{\mu\nu}\big)_\alphadot F^0_{\mu \nu} + 2 \ebar_\alphadot d \ , \\
  \delta d=&\ \frac{i}{2} \del_\mu \omega \sigma^\mu \ebar -
     \frac{i}{2} \e\sigma^\mu \del_\mu \rhobar \ ,
  = \frac{i}{2} \del_\mu \big[\omega \sigma^\mu \ebar -
     \frac{i}{2} \e\sigma^\mu  \rhobar \big]\ ,
\esp\label{eq:susyvar}\ee
where $F^0_{\mu\nu}$ is the derivative part of the field strength tensor,
\be
   F^0_{\mu\nu} = \del_\mu v_\nu - \del_\nu v_\mu \ .
\ee
The variation of the $D$-term proves the statement that the highest-order
component field of a general superfield is a good candidate for a supersymmetric 
Lagrangian (see Section~\ref{sec:susylag}) since
it transforms as a total derivative under supersymmetric transformations.

\subsection{Vector superfields in the Wess-Zumino gauge}

In this section, we apply the results of the previous section to the
general vector superfield of Eq.~\eqref{eq:genVSF}, considering only the
non-abelian case since the abelian limit can be easily derived. 
Switching to notations and normalization conventions of Section
\ref{sec:VSF}, the variations of the component fields are directly read from Eq.
\eqref{eq:susyvar}. Adopting the Wess-Zumino gauge where the component fields $C^a$,
$(\chi^a,\chibar^a)$, $f^a$ and $g^a$\footnote{We make the gauge indices
explicit.} vanish, one observes that these five fields transform
as
\be\bsp
  C^a=0 \to&\ C^{a\prime} =  C^a + i \e \!\cdot\!\chi^a + i \ebar \!\cdot\!
    \chibar^a = 0 \ , \\
  \chi^a_\alpha = 0 \to&\ \chi^{a\prime}_\alpha = \chi^a_\alpha + f^a \e_\alpha - i
     \big(\sigma^\mu \ebar\big)_\alpha  \big(v^a_\mu - i\del_\mu C^a\big) = - i
     \big(\sigma^\mu \ebar\big)_\alpha v_\mu^a
      \ ,  \\
  \chibar^{a\alphadot} = 0 \to &\ \chibar^{a\prime\alphadot}  =
     \chibar^{a\alphadot} + 
     f{a^\dag} \ebar^\alphadot  - i\big(\sibar^\mu \e\big)^\alphadot \big(v^a_\mu +
     i\del_\mu C^a\big)  = -i \big(\sibar^\mu \e\big)^\alphadot v^a_\mu \ ,  \\
  f^a = 0 \to&\ f^{a\prime} = f^a + i \del_\mu \chi^a \sigma^\mu \ebar + 2 \ebar
    \!\cdot\!
    \big(\lambar^a - \frac{i}{2} \sibar^\mu \del_\mu \chi^a \big) = 2 \ebar
    \!\cdot\! \lambar^a \ , \\
  f^{a\dag} = 0 \to &\ f^{a\prime\dag}  = f^{a\dag} -i \e \sigma^\mu \del_\mu
    \chibar^a +
    2 \e \!\cdot\! \big(\lambda^a - \frac{i}{2} \sigma^\mu \del_\mu \chibar^a\big)
    = 2 \e\!\cdot\!\lambda^a \ .
\esp\label{eq:VSFvar1}\ee 
The Wess-Zumino gauge is thus not supersymmetric as not
preserved by supersymmetric transformations. In Section \ref{sec:VSF}, we have
also motivated the use of vector superfields in the
Wess-Zumino gauge instead of general vector superfield
by the argument that the
component fields $(\lambda,\lambar)$, $v$ and $D$ are transforming into each
other under supersymmetric transformations. This is true according to
Eq.~\eqref{eq:susyvar} since in the notations of Eq.~\eqref{eq:genVSF}, we have
\be\bsp
  \lambda^a_\alpha \to &\ \lambda_\alpha^{a\prime} = \lambda^a_\alpha  -i 
    (\sigma^{\mu \nu} \e)_\alpha \ F^{0a}_{\mu \nu} + i \e_\alpha D^a \ , \\
  \lambar^a_\alphadot \to &\ \lambar^{a\prime}_\alphadot = \lambar^a_\alphadot +
    i (\ebar
    \sibar^{\mu\nu})_\alphadot F^{0a}_{\mu \nu} - i \ebar_\alphadot D^a \ , \\
  v_\mu^a \to &\ v^{a\prime}_\mu = v^a_\mu  -i \lambda^a\sigma_\mu\ebar + i \e
     \sigma_\mu\lambar^a \ , \\
  D^a \to &\ D^{a\prime} = D^a + \del_\mu \lambda^a \sigma^\mu \ebar +
    \e\sigma^\mu \del_\mu \lambar^a = D^a + \del_\mu\big[ \lambda^a \sigma^\mu
    \ebar + \e\sigma^\mu\lambar^a \big] \ ,
\esp\label{eq:VSFvar2}\ee 
where $F^{0a}_{\mu \nu}$ denotes the derivative terms of the non-abelian field
strength tensor. However, vector superfields are closely related to gauge
transformations, and in order to preserve the
Wess-Zumino gauge, it is necessary to accompany supersymmetry transformations by
gauge transformations. From Eq.\ \eqref{eq:nagaugetra}, the variation of a
vector superfield $\delta_g \Phi_V^a$ under a gauge transformation reads 
\be
  \delta_g \Phi_V^a =-i(\Lambda^a - \Lambda^{a\dag}) 
    - g f_{bc}{}^a \Phi_V^b (\Lambda^c + \Lambda^{c\dag}) \ , 
\label{eq:gaugetrV}\ee 
where $\Lambda$ is the (superfield) transformation parameter. We now use
\feynrules\ to show that fixing
\be
  \Lambda^a = -i \theta\sigma^\mu\ebar v^a_\mu + \theta\!\cdot\!\theta\
\ebar\!\cdot\! \lambar^a \ ,
\label{eq:deflam}\ee
allows to restore the Wess-Zumino gauge.

First, we assume that the non-abelian vector superfield $\Phi_V$ and its
component fields $\lambda$, $v$ and $D$ have been properly declared, as shown in
Section \ref{sec:FRfields} and Section \ref{sec:FRSF}. Moreover, the gauge group
associated to $\Phi_V$, as well as the related adjoint gauge index and the
structure constants $f_{bc}{}^a$, must also have been
declared following the syntax of Section~\ref{sec:FRidx} and
Section~\ref{sec:FRgroups}. Finally, the
coupling constant has to be added to the list of parameters, as presented in
Section~\ref{sec:FRprm}. 

Then, we associate the symbol \texttt{LAM} to the superfield $\Lambda$ of Eq.\
\eqref{eq:deflam}. Its implementation in \feynrules\ reads
\begin{verbatim}
  LAM[a_] := -I ncc[theta,si[mu],eps1bar] V[mu, a] + 
     ncc[theta,theta] ncc[eps1bar,lambdabar[a]];
\end{verbatim}
where as in the beginning of this section, we employ the predefined spinor
\texttt{eps1} for the transformation parameters and the \texttt{ncc} environment
to handle the canonical form of the superspace expressions. The variation under
a gauge transformation is finally computed as 
\begin{verbatim}
  deltag := -I (LAM[a]-HC[LAM[a]]) - g f[b,c,a] PHIV[b] (LAM[c]+HC[LAM[c]])
\end{verbatim}
assuming that the vector superfield, the structure constants and the
coupling strength related the gauge group are represented by the symbols
\texttt{PHIV}, \texttt{f[a,b,c]} and \texttt{g}. Expanding this expression
in terms of the Grassmann coordinates,
\begin{verbatim}
  GrassmannExpand[deltag]
\end{verbatim}
one gets
\be\bsp
  \delta_g \Phi_V^a =&\ -\theta \sigma^\mu\ebar v_\mu^a
  + \thetabar \sibar^\mu\e v_\mu^a 
  - i \theta\!\cdot\!\theta \lambar^a\!\cdot\!\ebar
  + i \thetabar\!\cdot\!\thetabar \lambda^a\!\cdot\!\e 
  - g f_{bc}{}^a \thetabar\!\cdot\!\thetabar 
     \theta\sigma^{\mu\nu} \e v_\mu^b v_\nu^c
\\ &\quad
  + i g f_{bc}{}^a \theta\!\cdot\!\theta
     \thetabar\sibar^{\mu\nu} \ebar v_\mu^b v_\nu^c
  + \frac12 g f_{bc}{}^a v_\mu^c \big[\e\sigma^\mu\lambar^b +
   \lambda^b\sigma^\mu\ebar\big]
\esp\ee
Merging this result with the normalization of the fields introduced in Eq.\
\eqref{eq:genVSF} and the variations of Eq.\ \eqref{eq:VSFvar1} and Eq.\
\eqref{eq:VSFvar2}, one obtains 
\be
  C^a= C^{a\prime} = \chi^a_\alpha = \chi^{a\prime}_\alpha = 
  \chibar^{a\alphadot} = \chibar^{a\prime\alphadot}  = f^a = f^{a\prime} = 
   f^{a\dag} = f^{a\prime\dag}  = 0 \ ,
\ee
which effectively restores the Wess-Zumino gauge. In addition, the variation of
the components of the vector supermultiplet of Eq.~\eqref{eq:VSFvar2}
are now covariant with respect to gauge transformations,
\be\bsp
  \lambda^a_\alpha \to &\ \lambda_\alpha^{a\prime} = \lambda^a_\alpha  -i (\sigma^{\mu
    \nu} \e)_\alpha \ F^a_{\mu \nu} + i \e_\alpha D^a \ , \\
  \lambar^a_\alphadot \to &\ \lambar^{a\prime}_\alphadot = \lambar^a_\alphadot +
    i (\ebar
    \sibar^{\mu\nu})_\alphadot F^a_{\mu \nu} - i \ebar_\alphadot D^a \ , \\
  v_\mu^a \to &\ v^{a\prime}_\mu = v^a_\mu  -i \lambda^a\sigma_\mu\ebar + i \e
     \sigma_\mu\lambar^a \ , \\
  D^a \to &\ D^{a\prime} =  D^a + D_\mu\big[ \lambda^a \sigma^\mu
    \ebar + \e\sigma^\mu\lambar^a \big] \ ,
\esp\ee 
where $F^a_{\mu\nu}$ is the standard non-abelian field strength tensor and
$D_\mu\lambda^a$ and $D_\mu\lambar^a$ the covariant derivatives in the
adjoint representation given in Eq.\ \eqref{eq:gaugecov}.

We achieve this section by considering the variations of the chiral
superfield of Section \ref{sec:deltasusychir} under a joint 
supersymmetric and gauge transformation. At the infinitesimal level, Eq.\
\eqref{eq:gaugevarchir} reads
\be 
  \delta_g \Phi =  -2 i g \Lambda^a T_a \Phi \ .
\ee 
This can be implemented in \feynrules\ as
\begin{verbatim}
  dgphi[i_] := -2 I g LAM[a] T[a,i,j] PHI[j]
\end{verbatim}
where we assume that the superfield $\Phi$ lies in a representation of the gauge
group spanned by the matrices $T_a$\footnote{The superfield $\Phi$ now carries
gauge indices. We omit all details about the corresponding modifications of the
model file for brevity. Information can be found
in the previous sections.}. Extracting the scalar, fermionic and
auxiliary coefficients of the chiral superfield $\delta_g\Phi$ by means of the
functions \texttt{XXXXComponent} and introducing appropriate normalization
factors, one gets
\be\bsp
  \delta_g z =  &\ 0\ , \\
  \delta_g \psi_\alpha =&\ -\sqrt{2} g \big(\sigma^\mu\ebar\big)_\alpha 
     v_\mu^a \big(T_a z\big)\ , \\
  \delta_g F = &\ 2 i g \big(T_a z\big) \lambar^a\!\cdot\!\ebar - \sqrt{2} g
   \big(T_a\psi\big)\sigma^\mu\ebar v_\mu^a \ .
\esp\label{eq:gaugesusyvar1}\ee
Combining these results with those of Eq.\ \eqref{eq:susyvarchircomp},
the variations of the component fields of a chiral superfield under the
composition of a gauge and a supersymmetric transformation are now
gauge covariant, 
\be\bsp
  \delta_\e z =&\ \sqrt{2} \e \!\cdot\! \psi \ ,\\
  \delta_\e \psi_\alpha =&\ -\sqrt{2} \e_\alpha F -i \sqrt{2} \big(\sigma^\mu
    \ebar\big)_\alpha D_\mu z \ , \\
  \delta_\e F =&\ -i \sqrt{2} D_\mu\big[ \psi \sigma^\mu \ebar \big ] + 
    2 i g \big(T_a z\big) \lambar^a\!\cdot\!\ebar\ ,
\esp\label{eq:gaugesusyvar2}\ee
which recovers well-known textbook results. We recall that the covariant derivatives of
the scalar and fermionic fields, automatically expanded in \feynrules,
are given by Eq.\ \eqref{eq:covderKg}.

\mysection{Non-renormalizable supersymmetric model building}
In this section, we give an additional example to illustrate the power of
\feynrules\ for superspace calculations by computing the
K\"ahler potential of Eq.\ \eqref{eq:K}. We start with the declaration of two
sets of indices, denoted by \texttt{II} and
\texttt{JS}, related to chiral and antichiral superfields, respectively,
\begin{verbatim}
  IndexRange[Index[II]] = Range[12];    IndexRange[Index[JS]] = Range[12];
\end{verbatim}
where the ranges of these indices are arbitrary since irrelevant
for the following. We then include in \texttt{M\$ClassesDescription} two sets of
component fields associated with these indices,
\begin{verbatim}
  W[1] == {ClassName->psiL, Chirality->Left, SelfConjugate->False, 
    Indices->{Index[II]}}
  W[2] == {ClassName->psiR, Chirality->Left, SelfConjugate->False, 
    Indices->{Index[JS]}}
  S[1] == {ClassName->phiL, SelfConjugate->False, Indices->{Index[II]}}
  S[2] == {ClassName->phiR, SelfConjugate->False, Indices->{Index[JS]}}
  S[3] == {ClassName->FL, SelfConjugate->False, Unphysical->True, 
    Indices->{Index[II]}}
  S[4] == {ClassName->FR, SelfConjugate->False, Unphysical->True, 
    Indices->{Index[JS]}}
\end{verbatim}
and we omit the declaration of the corresponding chiral and antichiral superfields
$\Phi^i$ and $\Phi^\dag_\is$ as not necessary for our scope. Since these fields
are all members of chiral (and not antichiral) supermultiplets, the Hermitian
conjugate of the fields carrying a starry index \texttt{JS} have to be
further used.

The K\"ahler potential is defined from the left-handed and right-handed chiral 
superfields $W^I(\Phi)$ and $W_I(\Phi^\dag)$ of Eq.\ \eqref{eq:expW}. These two
superfields are declared in the \texttt{M\$Superfields} list,
\begin{verbatim}
  CSF[1] == { ClassName->WISF, Chirality->Left, Weyl->psWL, Scalar->phWL, 
    Auxiliary->FWL}
  CSF[2] == { ClassName->WISFbar, Chirality->Right,  Weyl->psWR, Scalar->phWR, 
    Auxiliary->FWR}
\end{verbatim}
where their component fields have been
declared in \texttt{M\$ClassesDescription},
\begin{verbatim}
  W[3] == {ClassName->psWL, Chirality->Left,  SelfConjugate->False}
  W[4] == {ClassName->psWR, Chirality->Right, SelfConjugate->False}
  S[5] == {ClassName->phWL, SelfConjugate->False}
  S[6] == {ClassName->phWR, SelfConjugate->False}
  S[7] == {ClassName->FWL,  SelfConjugate->False, Unphysical->True}
  S[8] == {ClassName->FWR,  SelfConjugate->False, Unphysical->True}
\end{verbatim}
To get the correct expansion of the K\"ahler potential in terms of the Grassmann
variables, these fields have
to be further replaced by the relevant expressions. These replacement rules
require the declarations as parameters of the functions of the scalar fields
$W^I$ and $W_I$ and their derivatives. They are included
in the \texttt{M\$Parameters} list,
\begin{verbatim}
  WI   == { TeX->Superscript[W,"I"] }
  DWI  == { TeX->Superscript[W,"I"], Indices->{Index[II]} }
  DDWI == { TeX->Superscript[W,"I"], Indices->{Index[II], Index[II]}}
  WIbar   == { TeX->Subscript[W,"I"] }
  DWIbar  == { TeX->Subscript[W,"I"], Indices->{Index[JS]} }
  DDWIbar == { TeX->Subscript[W,"I"], Indices->{Index[JS], Index[JS]}}
\end{verbatim}
Consequently, the replacement rules for the component fields associated with the
chiral superfield $W^I$, being implemented following a standard \mathematica\
syntax, read
\begin{verbatim}
  rules := {
    del[del[phWL,mu_],mu_] :> Module[{ii,jj}, DDWI[ii,jj] del[phiL[ii],mu] * 
      del[phiL[jj], mu] + DWI[ii] del[del[phiL[ii],mu],mu]], 
    del[phWL,mu_] :> Module[{ii}, DWI[ii] del[ phiL[ii], mu]], 
    phWL -> WI,
    del[psWL[sp_], mu_] :> Module[{ii,jj}, del[psiL[sp,ii],mu] DWI[ii] + 
      psiL[sp,ii] del[phiL[jj],mu] DDWI[ii,jj]], 
    psWL[sp_] :> Module[{ii}, psiL[sp,ii] DWI[ii]],
    FWL :> Module[{ii,jj}, FL[ii] DWI[ii] + 1/2 ncc[psiL[ii],psiL[jj]] *
      DDWI[ii,jj]]
  }  
\end{verbatim}
The rules associated with the antichiral superfield $W_I$ being similar,
they are omitted from the present document for brevity\footnote{We however
remind, as stated above, to employ Hermitian conjugate fields in the
right-hand side of the rules.}.

We are now ready to compute the K\"ahler potential, by simply issuing in a
\mathematica\ session
\begin{verbatim}
  kahler = GrassmannExpand[WISFbar WISF];
  kahler = Tonc[kahler]/.rules;
  kahler = Expand[kahler];
\end{verbatim} 
The first command above gives a rather long expression not yet depending on the
K\"ahler potential and its derivatives. We choose to get back to the
\texttt{nc} environment by means of the \texttt{Tonc} function (see Section
\ref{sec:FRsuperspace}) before applying the replacement rules in the second
command. This allows to 
avoid the \mathematica\ \texttt{Dot} environment, explicitly removed by the
\texttt{Tonc} method, which could lead to expansion issues when replacing the
fermionic components of the $W_I$ and $W^I$ superfields by their correct expressions.

In order to map the products of derivatives of the (scalar)
function $W^I$ and $W_I$, a second set of standard \mathematica\ replacement
rules has to be applied on the result,
\begin{verbatim}
  {WI WIbar->K[{},{}], WIbar*DDWI[i__]->K[{i},{}], WIbar*DWI[i__]->K[{i},{}], 
  WI*DDWIbar[i__]->K[{},{i}], WI*DWIbar[i__]->K[{},{i}],
  DDWIbar[j__]*DDWI[i__]->K[{i},{j}], DWIbar[j__]*DDWI[i__]->K[{i},{j}], 
  DDWIbar[j__]*DWI[i__]->K[{i},{j}], DWIbar[j__]*DWI[i__]->K[{i},{j}]}
\end{verbatim}
In the commands above, we have introduced a function \texttt{K} taking two lists
as arguments, the first one being related to derivation operations with respect
to the chiral fields and the second one to those associated with the antichiral fields.
It can be checked that the results match the one presented in Eq.\ \eqref{eq:K},
after having expanded all the covariant derivatives included in this equation as
given by Eq.\ \eqref{eq:covderK} and employed the definitions of Eq.\
\eqref{eq:derK}. Moreover, the \mathematica\ output readability can be improved
by adding the formatting rule 
\begin{verbatim}
  Format[K[{ii___},{jj___}]]:=Subsuperscript[K, StringJoin@@(ToString/@{ii}), 
    StringJoin@@(ToString/@{jj})] 
\end{verbatim}

\cleardoublepage

%% file: brk.tex
\label{sec:susybrk}
Since not a single superpartner with the mass of its
Standard Model counterpart has been observed up to now,
supersymmetry must be broken. We present in this chapter the main features
of supersymmetry breaking and detail the most studied mechanisms.

\mysection{Supersymmetry breaking: general features}

\subsection{Motivations for softly broken supersymmetric theories}
For many years, the Standard Model of particle physics \cite{Glashow:1961tr,
Salam:1964ry,Weinberg:1967tq, salamsm,
Glashow:1970gm, Weinberg:1971nd, Gross:1973ju, Kobayashi:1973fv, Gross:1974cs,
Politzer:1974fr} has passed impressively all experimental tests. Only the
mechanism of electroweak symmetry breaking and the one of generation of mass
have for a long time remained unsolved questions. Those two issues are however
presently
addressed by the general-purpose experiments ATLAS and CMS at the LHC, since
both experiments have observed a new neutral
scalar particle that seems compatible with a
Standard-Model-like Higgs boson \cite{Aad:2012gk, Chatrchyan:2012gu}. However,
despite of the success of the Standard Model, many fundamental questions remain
unanswered.

One of the most infamous conceptual problem of the Standard Model lies in the 
non-explained large hierarchy between the electroweak scale and the Planck
scale. The latter lying orders of magnitude away from the weak scale, the mass
of any fundamental scalar field, and in particular, the one of the new 
state observed by the ATLAS and CMS experiments, is found to be drastically
affected by quantum corrections. 
Furthermore, many questions arise from the
complicated structure of the $SU(3)_c \times SU(2)_L \times U(1)_Y$ Standard
Model gauge group. For instance, there is no fundamental reason for the chosen
representation of the matter fields and this choice additionally implies
the non-unification of the gauge coupling constants
at high energies. Moreover, the Standard Model does not provide a
mechanism for neutrino oscillations or a candidate particle to account for the
presence of dark matter in the universe. As last examples, one can also state
that it also does not
motivate the non-vanishing cosmological constant, does not provide an
explanation for the strong $CP$-problem and
does not include gravity.

Over the last decades, large classes of theories have been proposed
in order to extend the Standard Model and address one or several of its open
issues. Among all these new physics theories, we focus in this work on
supersymmetry, a particularly appealing extension of the Standard Model as 
theoretically well-motivated (see Chapter \ref{chap:susy}).
Supersymmetry is known to solve the hierarchy problem
by the introduction of the superpartners of the Standard Model degrees of
freedom. Their presence also leads to a modification of
the renormalization group equations
driving the evolution of the gauge coupling constants between two energy scales,
allowing for supersymmetry to tackle the problem of 
the unification of the
gauge couplings at high energies. Many supersymmetric models contain, in
addition to gauge symmetries, extra discrete symmetries that typically render
the lightest supersymmetric particle stable, which can then be
seen as a viable dark matter candidate. Finally,
local supersymmetry, also known as supergravity, provides a
possible way to include gravity next to the other interactions.
 
Even if supersymmetry is very attractive both from the theoretical and
phenomenological points of view, we have shown in Section \ref{sec:repr_general}
that all the components of a given
supermultiplet have the same mass. However, not a single supersymmetric partner of the
Standard Model particles has been observed so far, especially at the LHC
\cite{atlassusy, cmssusy, Lowette:2012uh}. In
particular, no hint for a scalar electron with a 511 keV mass or for a very
light 
scalar quark of a few MeV has been found. Therefore, supersymmetry cannot be
an exact symmetry of nature and building phenomenologically viable
supersymmetric models requires to break supersymmetry at low energies.
In order to remain
a viable solution to the hierarchy problem, this breaking is required to be soft,
yielding supersymmetric masses around the TeV scale that are thus reachable at the LHC.

We dedicate the rest of this section to a presentation of the 
general properties of supersymmetry
breaking, while three of the most studied mechanisms implying
soft supersymmetry breaking are briefly reviewed in Section \ref{sec:popbrk}.
For all those mechanisms, supersymmetry breaking generally occurs at a higher
scale where some organizing relations among the model parameters hold. In contrast,
the current particle physics experimental program is designed to probe a
much lower scale, the
electroweak scale, so that associated phenomenological works
require to evolve those parameters down to the
electroweak scale. This evolution is controlled by highly coupled differential
equations, the supersymmetric renormalization group equations, that we
present in Section \ref{sec:rge}.

\subsection{The Goldstone theorem for supersymmetry}
As for any symmetry, supersymmetry is spontaneously broken under the condition
that the vacuum state $|\Omega\rangle$ is not invariant under supersymmetry
transformations. In other words, the action of the supercharges on the vacuum
state obeys to
\be
  Q_\alpha | \Omega\rangle \neq 0 \qquad\text{and}\qquad 
  \Qbar_\alphadot | \Omega\rangle \neq 0 \ .
\label{eq:qbr}\ee
From the energy operator $P^0$ and the superalgebra of Eq.\
\eqref{eq:poincaresalg}, one can derive a constraint on the energy $E$ of the
vacuum state, 
\be\bsp
  \langle\Omega| E |\Omega\rangle = &\ \langle \Omega | P^0 |\Omega\rangle 
   = \frac14 \Big\langle \Omega \Big| 
    \big\{Q_1,Q_1^\dag\big\} + \big\{Q_2,Q_2^\dag\big\} 
    \Big| \Omega\Big\rangle  \\ =&\ \frac14 \Big[ 
   \big|\big| Q_1 | \Omega\rangle\big|\big| + 
   \big|\big| Q_2 | \Omega\rangle\big|\big| + 
   \big|\big| Q_1^\dag | \Omega\rangle\big|\big| + 
   \big|\big| Q_2^\dag | \Omega\rangle\big|\big| \Big] \neq 0 \ ,
\esp\label{eq:Eomega}\ee
so that $E$ is non-vanishing on the basis of 
Eq.\ \eqref{eq:qbr}.
Moreover, a norm being positively defined, the energy, being equal to the sum of
four norms, is thus positive too.

Considering the vacuum state, the Hamiltonian $P^0$, as the Lagrangian ${\cal
L}$, is reduced to the scalar
potential $V$. Therefore, the vacuum expectation value of the potential is
positively defined, $\langle V \rangle > 0$, as shown in Eq.\ \eqref{eq:Eomega}
and the scalar
potential gets a non-trivial minimum. To investigate under which conditions
this minimum is reached, we start from the solutions of the equations of
motion for the auxiliary fields given in Eq.\ \eqref{eq:auxsol}. In the limit of
a renormalizable theory\footnote{From now on and in the rest of this document, we
only focus on renormalizable theories, unless stated otherwise.}, these 
solutions become, after accounting for Eq.\
\eqref{eq:normthcond} and the properties of the K\"ahler metric of Eq.\
\eqref{eq:kahlermetric}, 
\be
  F^i = W^{\star i}  \ , \qquad
  F^\dag_i=W_i \qquad\text{and}\qquad
  D^a =  g \phi^\dag_i \big(T^a \phi\big)^i  \ .
\label{eq:eqmota}\ee
In our theoretical setup, we construct a generic supersymmetric theory
describing the dynamics of
a set of chiral superfield $\{\Phi^i\}$ in interaction. The scalar, fermionic
and auxiliary component fields of these superfields are denoted by $\phi^i$,
$\psi^i$ and $F^i$, respectively, as in Chapter \ref{chap:susy}. Moreover, we
also recall our conventions for the shorthand notations employed for the
first-order derivatives of the
superpotential $W$ (and those of the Hermitian conjugate function $W^\star$) that
appear in Eq.\ \eqref{eq:eqmota}, 
\be
  W_i \equiv \frac{\del W(\phi)}{\del \phi^i} \qquad\text{and}\qquad
  W^{\star i} \equiv \frac{\del W^\star(\phi^\dag)}{\del \phi^\dag_i} \ . 
\ee
Finally, the matrices $T_a$, on which depend the solutions of the equations of
motion for the $D$-fields, are representation matrices of the gauge group associated
with the representation in which the scalar fields $\phi^i$ and
$\phi^\dag_i$ lie, and as usual, the coupling constant of the gauge group is denoted
by $g$. Inserting Eq.\ \eqref{eq:eqmota} in the Lagrangian of Eq.\
\eqref{eq:gensusylag}, the scalar
potential $V$ can be rewritten exclusively in terms of the auxiliary fields, 
\be
  V =  F^i F^\dag_i + \frac12 D^a D_a \ .
\label{eq:scalpot}\ee
As stated above, the vacuum expectation value of the scalar potential is
positive, $\langle V \rangle > 0$. From the results of Eq.\ \eqref{eq:scalpot}, 
supersymmetry is thus broken in
the case at least one of the auxiliary fields gets a non-vanishing
vacuum expectation value. 

The minimization
conditions for the scalar potential
imply that the first-order derivatives of $V$ with respect to the
scalar fields vanish at the minimum, 
\be
  0 = \left\langle \frac{\del V}{\del\phi_j}\right\rangle = 
  \left\langle  F^i \frac{\del F^\dag_i}{\del \phi^j} + D^a
    \frac{\del D^a}{\del \phi^j}\right\rangle = \left\langle  F^i W_{ij} + g D^a
    (\phi^\dag T_a)_j\right\rangle \qquad \forall j \ ,
\label{eq:susybrcond1}\ee
where the equalities are derived after inserting the solutions of the equations
of motion for the auxiliary fields of Eq.\ \eqref{eq:eqmota} in the expression
of the scalar potential of Eq.\ \eqref{eq:scalpot} .
Additional information can be obtained from the fact that 
superpotential is a gauge-invariant quantity. Consequently, its variation under
a gauge transformation vanishes, 
\be
  0 = \delta_\omega W^\star = W^{\star i} \delta_\omega \phi^\dag_i = 
    F^i \Big[ i g \omega^a (\phi^\dag T_a)_i \Big] \ , 
\label{eq:susybrcond2}\ee
where $\omega^a$ are the transformation parameters. In order to derive this last
result, we have used
the equation of motions for the auxiliary $F$-field and we recall that, at the
infinitesimal level, the scalar fields transform as
\be
  \phi^i \to \phi^i -i g \omega^a \big(T_a \phi\big)^i 
  \qquad\text{and}\qquad
  \phi^\dag_i \to \phi^\dag_i + i g \omega^a \big(\phi^\dag T_a\big)_i\ .
\ee
The two conditions of Eq.\ \eqref{eq:susybrcond1} and Eq.\
\eqref{eq:susybrcond2} can be collected into a single matrix equation
characterizing the vacuum state,
\be\boxed{
 \bpm 
    \langle W_{ij} \rangle & \langle g  \phi_k^\dag(T_a)^k{}_j\rangle  \\
    \langle g \phi_k^\dag (T_a)^k{}_i\rangle  & 0
  \epm \bpm \langle F^i\rangle \\ \langle D^a \rangle\epm  = \bpm 0 \\ 0 \epm
 \ .} 
\label{eq:susybrcondall}\ee

We now investigate the fermionic sector, and in particular the mass matrices
resulting from the spontaneous breaking of supersymmetry. The
interaction terms of the Lagrangian yielding fermionic mass terms,
that we denote by ${\cal L}_{\rm
int}^\psi$, can be extracted from Eq.~\eqref{eq:gensusylag},
\be
  {\cal L}_{\rm int}^\psi = - \frac12 W_{ij} \psi^i\!\cdot\!\psi^j
     - \sqrt{2} i g \phi^\dag_i (T_a)^i{}_j \psi^j\!\cdot\!\lambda^a 
     + \hc \ .
\ee
Shifting the scalar fields with respect to their vacuum expectation
values then generates the mass terms
\be
  {\cal L}_{\rm mass}^\psi = - \frac12
  \bpm \psi^j & \sqrt{2} i \lambda^a\epm^\alpha\bpm 
    \langle W_{ij} \rangle & \langle g  \phi_k^\dag(T_b)^k{}_j\rangle  \\
    \langle g \phi_k^\dag (T_a)^k{}_i\rangle  & 0
  \epm  \bpm \psi^i \\ \sqrt{2} i \lambda^b\epm_\alpha  + \hc \ .
\label{eq:fermmass}\ee
Using the minimization conditions of Eq.\ \eqref{eq:susybrcondall}, we observe
that the fermionic field defined by
\be\boxed{
  | \psi_G \rangle =  \frac{\sqrt{2}}{2} \langle F_i^\dag \rangle\ | \psi^i
    \rangle + \frac{i}{2} \langle D^a\rangle\ | \lambda_a \rangle 
\label{eq:defgoldstino}}\ee
is massless.

The prediction of the existence of such a massless field, dubbed a
goldstino, consists of the Goldstone theorem for supersymmetry. When
supersymmetry is spontaneously broken, the scalar
potential is minimum for a configuration in which at least one  of the
auxiliary fields gets a non-vanishing vacuum expectation value $\langle
F^i\rangle$ or
$\langle D^a\rangle$ and the particle spectrum consequently contains a massless
fermionic state $\psi_G$ defined by Eq.\ \eqref{eq:defgoldstino}
\cite{Witten:1981nf,Fayet:1974jb}.

\subsection{Properties of the goldstino field}\label{sec:propgold}
As explained in the previous section, spontaneous supersymmetry breaking
predicts the existence of a massless (Majorana) fermion, the goldstino. The
interactions of such a field
with the rest of the particle spectrum can be derived from the conservation of the
N\oe ther current. We recall that the current is defined from the variation of
the Lagrangian ${\cal L}$ under a supersymmetry transformation, as shown in Eq.\
\eqref{eq:generalcurrent},
\be
  \delta_\e {\cal L} = \del_\mu K^\mu = \del_\mu \bigg[
     {\del {\cal L} \over \del \big(\del_\mu X\big)} \delta_\e X 
     \bigg] \ ,
\label{eq:current1}\ee
where the quantity $K^\mu$ is obtained after directly varying the fields
appearing in the Lagrangian and where the Majorana fermion $(\e,\ebar)$ denotes the
transformation parameters. As in Chapter \ref{chap:susy}, the second 
equality is obtained on the basis of Euler-Lagrange equations, a sum 
over all the fields $X$ of the theory being understood.
Consequently, the supercurrent $(J^\mu,\Jbar^\mu)$, defined as
\be
  \e\!\cdot\!J^\mu + \ebar\!\cdot\!\Jbar^\mu =  {\del {\cal L} \over \del
\big(\del_\mu X\big)} \delta_\e X - K^\mu \ ,
\label{eq:current3}\ee
is a conserved quantity.

In Chapter \ref{chap:susy}, we have derived supersymmetric Lagrangians both in
terms of superfields and component fields and have shown that, in terms of
superfields, a 
supersymmetric Lagrangian can generically be written as a sum of five terms,
\be
  {\cal L}= \Big[\Phi_i^\dag e^{-2gV} \Phi^i\Big]_{\theta^2 \thetabar^2} + 
    \frac{1}{16 g^2} \Big[W^\alpha_a  W^a_\alpha\Big]_{\theta^2} +
    \frac{1}{16 g^2} \Big[\Wbar_\alphadot^a \Wbar_a^\alphadot\Big]_{\thetabar^2}
  + \Big[W(\Phi)\Big]_{\theta^2} + \Big[W^\star(\Phi^\dag)\Big]_{\thetabar^2} \
  ,
\label{eq:susylagsf}\ee
where the subscripts indicate which coefficients of the expansion in terms of
the Grassmann variables have to be selected. We recall that in this expression,
the (generic)
coupling constant of the gauge group is denoted by $g$ and we consider a theory
describing the dynamics of a set of chiral superfields $\{\Phi^i\}$. We have
also associated with the gauge group of the theory a
set of vector superfields $\{V^a\}$ lying in its adjoint
representation. The related Lagrangian terms,
\ie, the second and third terms of
Eq.\ \eqref{eq:susylagsf}, rely on the superfield strength
tensors built from the vector superfields $V^a$,
the spinorial superfields
$W_\alpha^a$ and $\Wbar_\alphadot^a$ defined by Eq.\ \eqref{eq:defWWbar}. 

From the variation laws of the different 
component fields of a general superfield, as collected in Eq.\
\eqref{eq:susyvar}, as well as from those of a chiral superfield shown in 
Eq.\ \eqref{eq:susyvarchircomp}, the variation of the Lagrangian above under a
supersymmetry transformation of parameters $(\e,\ebar)$ reads
\be\bsp
   \delta_\e {\cal L}=&\ -\frac{i}{2} \e \sigma^\mu \del_\mu \Big[\Phi_i^\dag
       e^{-2gV} \Phi^i\Big]_{\theta^2\thetabar} - \frac{i}{2} \ebar \sibar^\mu
        \del_\mu \Big[\Phi_i^\dag e^{-2gV} \Phi^i\Big]_{\theta\thetabar^2} 
\\ &\
   - i \e \sigma^\mu  \del_\mu \Big[W^\star(\Phi^\dag) + \frac{1}{16 g^2}
       \Wbar_\alphadot^a \Wbar_a^\alphadot\Big]_\thetabar
   - i \ebar \sibar^\mu  \del_\mu \Big[W(\Phi) + \frac{1}{16 g^2} W^\alpha_a
      W^a_\alpha\Big]_\theta \ .
\esp \ee
Expanding the superfields in terms of their component fields and selecting
the proper coefficients of the series in the Grassmann variables, one finds, after
inserting the solutions of the equations of motion for the auxiliary fields,
\be\bsp
  \delta_\e {\cal L} =& \
    \e \!\cdot\!\del_\mu \bigg[
      \frac{\sqrt{2}}{2} D_\nu\phi^\dag_i\sigma^\mu \sibar^\nu  \psi^i 
      - \frac{\sqrt{2}i}{2}\sigma^\mu  \psibar_i W^{\star i}
      - \frac12 g \phi_i^\dag \big(T_a\phi\big)^i \sigma^\mu\lambar^a 
      - \frac{i}{4}\sigma^\mu \sibar^\nu \sigma^\rho\lambar_a F^a_{\nu\rho}
\\&\
      - \frac{\sqrt{2}}{4} \del_\nu\big[\phi^\dag_i\sigma^\mu \sibar^\nu
          \psi^i\big] 
     \bigg]
      + \ebar \!\cdot\!\del_\mu \bigg[
        \frac{\sqrt{2}}{2} D_\nu\phi^i\sibar^\mu \sigma^\nu \psibar_i 
        - \frac{\sqrt{2}i}{2} \sibar^\mu \psi^i W_i
\\&\
        + \frac12 g \big(\phi^\dag T_a\big)_i  \phi^i \sibar^\mu\lambda^a
        - \frac{i}{4} \sibar^\mu  \sigma^\nu \sibar^\rho\lambda_a F^a_{\nu\rho} 
        - \frac{\sqrt{2}}{4} \del_\nu\big[\phi^i\sibar^\mu \sigma^\nu \psibar_i
          \big]
 \bigg] \ .
  \esp\ee
To compute this last result, we have performed two integrations by parts
(explicitly including all total derivatives)
and introduced the gauge covariant derivatives of
the scalar fields given in Eq.~\eqref{eq:gaugecovbis}. One derives from this expression the
quantity $K^\mu$ of Eq.\ \eqref{eq:current1} and Eq.\ \eqref{eq:current3},
\be\bsp
  K^\mu_\alpha =&\ \frac{\sqrt{2}}{2} D_\nu\phi^\dag_i \big(\sigma^\mu
    \sibar^\nu  \psi^i \big)_\alpha   
      - \frac{\sqrt{2}i}{2} \big(\sigma^\mu  \psibar_i\big)_\alpha W^{\star i}
      - \frac12 g \phi_i^\dag \big(T_a\phi\big)^i
          \big(\sigma^\mu\lambar^a\big)_\alpha
\\&\
      - \frac{i}{4} \big(\sigma^\mu \sibar^\nu \sigma^\rho\lambar_a\big)_\alpha
          F^a_{\nu\rho}
      - \frac{\sqrt{2}}{4} \del_\nu\Big[\phi^\dag_i \big(\sigma^\mu \sibar^\nu
          \psi^i\big)_\alpha\Big]  \ .
\esp\label{eq:scuretmpo1}\ee

To achieve the calculation of the supercurrent, it is also necessary to compute
the first term of the right-hand side of Eq.\ \eqref{eq:current3}. To this aim,
it is enough to select the derivative terms of the supersymmetric Lagrangian of
Eq.\ \eqref{eq:gensusylag2}, 
\be\bsp
  {\cal L} =&\
- \frac14 \del_\mu \big(\phi^\dag_i \del^\mu \phi^i + \del^\mu \phi^\dag_i
  \phi^i \big) + D^\mu \phi^\dag_i D_\mu \phi^i 
   + \frac{i}{2} \Big[  
       \psi^i \sigma^\mu D_\mu \psibar_i
      - D_\mu\psi^i \sigma^\mu \psibar_i\big]
\\ &\
   - \frac14 F^a_{\mu \nu} F_a^{\mu \nu}    
   + \frac{i}{2}\Big[
        \lambda^a\sigma^\mu D_\mu\lambar_a
      - D_\mu\lambda^a\sigma^\mu\lambar_a \Big]  \ ,
\esp\ee
and perform the derivation of this expression with respect to the first-order
derivatives of the fields. We have included in this Lagrangian 
the total derivative that was originally originally present (see, \eg,
Eq.\eqref{eq:gaugechiral}) and then omitted in the computations
of Chapter \ref{chap:susy}. From the variations
of the fields computed in Eq.\ \eqref{eq:gaugesusyvar1} and Eq.\
\eqref{eq:gaugesusyvar2}, one gets, after inserting again the solutions of the
equations of motion for the auxiliary fields,
\be\bsp
   {\del {\cal L} \over \del \big(\del_\mu X\big)} \delta X = &\ 
    \e \!\cdot\! \Big[ 
        \sqrt{2} D_\mu \phi_i^\dag \psi^i 
      - \frac{\sqrt{2}}{4} \del^\mu(\phi_i^\dag \psi^i)
      + \frac{\sqrt{2}i}{2} \sigma^\mu \psibar_i W^{\star i}  
      + \frac{\sqrt{2}}{2} D_\nu \phi_i^\dag  \sigma^\nu\sibar^\mu \psi^i
\\&\
      - i \sigma_\nu \lambar^a F_a^{\mu\nu} 
      + \frac12 g  \phi^\dag_i \big(T_a  \phi\big)^i \sigma^\mu \lambar^a
      - \frac{i}{4}  \sigma^\rho \sibar^\nu \sigma^\mu \lambar_a  F^a_{\nu\rho}    \Big] 
    + \ebar \!\cdot\! \Big[ 
        \sqrt{2} D_\mu \phi^i \psibar_i 
\\ &\ 
      - \frac{\sqrt{2}}{4} \del^\mu(\phi^i \psibar_i)
      + \frac{\sqrt{2}i}{2} \sibar^\mu \psi^i W_i  
      + \frac{\sqrt{2}}{2} D_\nu \phi^i \sibar^\nu\sigma^\mu \psibar_i
      - i \sibar_\nu \lambda^a F_a^{\mu\nu} 
\\&\
      - \frac12 g  \big(\phi^\dag T_a\big)_i  \phi^i \sibar^\mu\lambda^a
      - \frac{i}{4}  \sibar^\rho \sigma^\nu \sibar^\mu \lambda_a  F^a_{\nu\rho}
        \Big] \ .
\esp\label{eq:scuretmpo2}\ee
The supercurrent is finally given, collecting the two contributions of
Eq.~\eqref{eq:scuretmpo1} and Eq.~\eqref{eq:scuretmpo2} and employing the
definition of Eq.~\eqref{eq:current3}, by
\be\boxed{
  J^\mu_\alpha =
      \sqrt{2} D_\nu\phi^\dag_i \big(\sigma^\nu
         \sibar^\mu  \psi^i \big)_\alpha   
    + \sqrt{2}i \big(\sigma^\mu  \psibar_i\big)_\alpha W^{\star i}
    + g \phi^\dag T_a \phi \big(\sigma^\mu\lambar^a\big)_\alpha
    - \frac{i}{2} \big(\sigma^\rho \sibar^\nu \sigma^\mu \lambar_a\big)_\alpha
        F^a_{\nu\rho} \ ,}
\label{eq:supercurrent}\ee
where we have made use of the identities of Eq.\ \eqref{eq:sisi} and Eq.\ 
\eqref{eq:idsigma}. This result agrees with those of the pioneering works of
Ref.\ \cite{Wess:1978ns} and Ref.\ \cite{Iliopoulos:1974zv}.
As stated in Chapter \ref{chap:susy}, the conserved supercurrent is always
defined up to a quantity $\kappa^\mu$ that fulfills the relation 
$\del_\mu \kappa^\mu= 0$. From the
computations above, this quantity is found to be
\be
  \kappa^\mu_\alpha = 
      \frac{\sqrt{2}}{4} \del_\nu\Big[\phi^\dag_i \big(\sigma^\mu \sibar^\nu
          \psi^i\big)_\alpha\Big] 
      - \frac{\sqrt{2}}{4} \del^\mu \Big[\phi_i^\dag \psi^i_\alpha\Big] \ .
\ee

These calculations can also be performed automatically by employing the
superspace module of \feynrules. The routine allowing for
the extraction of the supercurrent in the case of any supersymmetric model
exactly follows the approach
described above \cite{Christensen:2013aua}. To compute the
supercurrent associated with a given model implementation, it is enough to type in
a \mathematica\ session
\begin{verbatim}
  SuperCurrent[lv,lc,lw,sp,lor]
\end{verbatim}
where the variables \texttt{lv}, \texttt{lc} and \texttt{lw} contain the
Lagrangian terms related to the gauge
sector (the second and third terms in Eq.\ \eqref{eq:susylagsf}), those 
associated with the chiral content of the theory (the
first term of Eq.\ \eqref{eq:susylagsf}) and those related
to the interaction terms driven from the superpotential (the last two terms of
Eq.\ \eqref{eq:susylagsf}). Following the syntax presented above, each of the
three quantities represented by the symbols
\texttt{lv}, \texttt{lc} and \texttt{lw} must be
given as a full series in the Grassmann variables. Finally, the symbols
\texttt{sp} and \texttt{lor} stand for the spin and Lorentz indices attached to
the supercurrent.

Once extracted, the supercurrent can be further employed in the
building of Lagrangians, deriving the supercurrent being the first step to
the construction of an effective action describing the interactions
of the goldstino field.

For the sake of the example, we assume that only the auxiliary
component of one single chiral supermultiplet acquires a vacuum
expectation value. In other words, one single chiral supermultiplet is
responsible for supersymmetry breaking. 
The general case can however be easily deduced from Eq.\
\eqref{eq:defgoldstino}. According to our simplification assumption, this
equation shows that the goldstino field is the fermionic component of the
supermultiplet that leads to the breaking of supersymmetry. We therefore denote
it by $(\phi_G,\psi_G,F_G)$. 
After shifting the auxiliary field with respect
to its vacuum expectation value, $F_G\to v_F/\sqrt{2} + F_G'$, the supercurrent
can be rewritten as
\be
  J^\mu = 
    i v_F\ \big(\sigma^\mu\psibar_G\big)_\alpha + {\cal J}^\mu_\alpha \
   .
\ee
However, current conservation enforces that $\del_\mu J^\mu = 0$, or
equivalently
\be
  i v_F\ \big(\sigma^\mu\del_\mu\psibar_G\big)_\alpha + 
    \del_\mu{\cal J}^\mu_\alpha  = 0\ .
\ee
This equation can be seen as the equations of motion for the goldstino field, 
so that the associated (effective) Lagrangian reads, after introducing standard
normalization \cite{Fayet:1977vd, Fayet:1979yb},
\be\boxed{
  {\cal L}_G = \frac{i}{2}(\psi_G \sigma^\mu \del_\mu\psibar_G -
     \del_\mu \psi_G \sigma^\mu \psibar_G) 
  + \frac{1}{2 v_F} \psi_G \!\cdot\! \del_\mu {\cal J}^\mu 
  + \frac{1}{2 v_F} \psibar_G \!\cdot\! \del_\mu \bar{\cal J}^\mu \ . 
}\ee

It is important to note that this goldstino Lagrangian does not depend on the
supersymmetry-breaking mechanism itself, but only
on supercurrent conservation. Therefore, the form of ${\cal L}_G$ given above
is a very general result.
One can check that similar Lagrangians are obtained in the case
supersymmetry is broken via the vacuum expectation value of a $D$-term, or
through a linear combination of several $F$-terms and $D$-terms.

From these considerations, pioneering mechanisms for supersymmetry breaking have
been proposed based on either a spontaneous supersymmetry breaking via a
$D$-term, or through a $F$-term. From the names of the authors of such
mechanisms, they are known as 
the Fayet-Iliopoulos \cite{Fayet:1974jb, Fayet:1974pd} and the O'Raifeartaigh
mechanisms \cite{O'Raifeartaigh:1975pr}, respectively. However, they have been
found to be not phenomenologically viable as they both predict superpartners
lighter than their Standard Model counterparts, which contradicts the
experimental (non-)observations. We therefore refer to the
literature for more details about these supersymmetry-breaking models and only
focus, in Section \ref{sec:popbrk}, on viable supersymmetry-breaking scenarios
commonly used for phenomenological and experimental studies.

\subsection{The supertrace constraint}\label{sec:str}
As briefly mentioned at the end of Section \ref{sec:propgold}, neither
$F$-term-induced
nor $D$-term induced supersymmetry-breaking mechanisms are satisfactory. In this
section, we focus on the derivation of an important constraint on viable
supersymmetry-breaking mechanisms which 
arises from the existence of a (tree-level) sum rule on the
model particle masses \cite{Ferrara:1979wa}. This rule is obtained from
inspecting the traces of the scalar, fermion and vector squared mass
matrices and strongly limits the possibilities for designing realistic
supersymmetry-breaking models.

The squared scalar mass matrix ${\cal M}_0^2$ can be deduced from the
second-order derivatives of the scalar potential. It reads, in the $(\phi^i,
\phi^\dag_j)$ basis, 
\be\bsp &
  {\cal M}_0^2 = \\ & \left\langle\!\bpm 
    W^{\star kj} W_{ki} \!+\! 
      g^2 \big[T^a{}^j{}_i \phi^\dag T_a \phi 
      \!+\!\! \big(\phi^\dag T^a\big)_i \big(T_a\phi\big)^j\Big]
&
    W^{\star kij} W_k + g^2\big(T^a\phi\big)^j \big(T_a\phi\big)^i\\
    W^{\star k} W_{kij} + g^2\big(\phi^\dag T^a\big)_i \big(\phi^\dag T_a\big)_j
&
    W^{\star ki} W_{kj}\!+\! g^2 \big[T^a{}^i{}_j \phi^\dag T_a\phi
      \!+\!\! \big(\phi^\dag T^a\big)_j \big(T_a \phi\big)^i \big]
  \epm \!\right\rangle  .
\esp\ee
The trace of this matrix is thus given by
\be
  {\rm Tr}\big[ {\cal M}_0^2 \big] = 
    2 \left\langle\frac{\del^2 V}{\del \phi^i \del \phi^\dag_i}\right\rangle =
    2 \langle W^{\star ki} W_{ki} \rangle 
       + 2 g^2 \big\langle \phi^\dag T^a T_a \phi \big\rangle
       + 2 g^2 \big\langle \phi^\dag T^a \phi \big\rangle  {\rm Tr}
         \big[T_a\big] \ .
\label{eq:tr1}\ee

We now turn to the vectorial field masses. The corresponding squared mass matrix arises
from Lagrangian terms including gauge covariant derivatives of the scalar fields.
After having shifted these scalar fields by their vacuum expectation values, one
gets terms bilinear in the vector fields, so that the associated squared mass
matrix reads
\be
  {\cal M}_1^2 = \left\langle\bpm 
    2 g^2 \big(\phi^\dag T_a\big)_i \big(T^b\phi\big)^i
  \epm \right\rangle \ ,
\ee
the trace of such a matrix being given by
\be
  {\rm Tr}\big[ {\cal M}_1^2 \big] = 2 g^2 \big\langle \phi^\dag T^a T_a \phi
    \big\rangle \ . 
\label{eq:tr2}\ee

Finally, considering the fermion sector, one has two contributions to their mass
matrix. The first one arises from the supersymmetric masses included in the
superpotential and the second one is related to the
gaugino-fermion-scalar interactions included in the K\"ahler
potential, as shown in Eq.\ \eqref{eq:fermmass}. The trace of the
square of this matrix can then be computed as
\be
   {\rm Tr}\big[ {\cal M}_{1/2}^2 \big]  = 
   {\rm Tr}\big[ {\cal M}_{1/2}^\dag {\cal M}_{1/2} \big] =
   \langle W^{\star ki} W_{ki} \rangle + 4 g^2
    \big\langle \phi^\dag T_a T^a \phi\big\rangle \ .
\label{eq:tr3}\ee 

Collecting the results of Eq.\ \eqref{eq:tr1}, Eq.\ \eqref{eq:tr2} and Eq.\
\eqref{eq:tr3}, one deduces the supertrace formula
\be\boxed{
  {\rm sTr} \big[ {\cal M}^2\big] = \sum_{\ell=0,1/2,1} (-)^{2\ell} (2 \ell + 1)
   {\rm Tr}\big[ {\cal M}_\ell^2 \big] =    
   2 g^2 \big\langle \phi^\dag T^a \phi \big\rangle  {\rm Tr}\big[T_a\big] \ .
\label{eq:supertrace}}\ee
This sum rule is very difficult to accommodate when building
phenomenologically viable models for supersymmetry breaking. Accounting for the
masses of the Standard
Model particles, it indeed imposes that some superparticles are
always unacceptably light. The strategy to evade this rule 
is to break supersymmetry either radiatively or via
non-renormalizable interactions. 

In general, the breaking of supersymmetry is assumed to occur in a 
hidden sector of particles that have no or reduced couplings to the visible
sector. The latter consists of the Standard Model particles, together with 
their superpartners. Among the most
popular mechanisms, one finds gravity-mediated supersymmetry breaking
\cite{Witten:1982hu, Deser:1976eh, Freedman:1976xh,
Freedman:1976py, Ferrara:1976um, Cremmer:1978hn, Cremmer:1978iv, Cremmer:1982en,
Chamseddine:1982jx, Barbieri:1982eh, Ibanez:1982ee, Ohta:1982wn, Ellis:1982wr,
AlvarezGaume:1983gj, Binetruy:2000zx}, gauge-mediated supersymmetry 
breaking \cite{Dimopoulos:1981au, Dine:1981za, Derendinger:1982tq, Fayet:1978qc,
Dine:1981gu, Nappi:1982hm, AlvarezGaume:1981wy, Dine:1993yw, Dine:1994vc,
Dine:1995ag, Giudice:1998bp} and anomaly-mediated supersymmetry breaking
\cite{Randall:1998uk, ArkaniHamed:1998kj, Bagger:1999rd, Derendinger:1991kr,
Derendinger:1991hq, LopesCardoso:1991zt, LopesCardoso:1992yd,
Kaplunovsky:1994fg}. We review
the main properties of those three mechanisms in the next sections.
Other mechanisms, such as, \eg, gaugino-mediated
supersymmetry breaking,  have however been proposed more recently and now receive a
sensible attention. They are not discussed in this work.

\mysection{Examples of soft supersymmetry-breaking mechanisms}
\label{sec:popbrk}

\subsection{Supergravity: general features}\label{sec:sugra}
In this section, we briefly review how gravity-mediated supersymmetry breaking
arises. We refer to Ref.\ \cite{sugra} for technical details and only describe,
in the following, the main features of this mechanism since a deep and
detailed study of supergravity theories is clearly going beyond the scope of
this work.

To construct supergravity theories, one must add the gravity effects which
matter and vector supermultiplets are sensitive to through a coupling to the
gravitation supermultiplet. In Section~\ref{sec:masslesssupermul},
we have shown that the degrees of freedom included
in the latter consist of the graviton field and the spin 3/2 gravitino field.
The derivation
of these gravity-related interactions relies on a local extension of
supersymmetry. Similarly to global
supersymmetry, Lagrangians are more easily constructed by employing the
superspace formalism. However, the superspace structure of supergravity slightly 
differs from the one used in  global supersymmetry
that has been presented in Section \ref{sec:superspace}.

In general relativity, the spacetime is curved and the
standard Minkowski spacetime is only recovered locally, when considering
reference frames where gravity effects are eliminated. Similarly, in
supergravity theories, the superspace is curved and at each superspace point,
we consider a (different) reference frame where gravity is eliminated. In
addition, care must be taken with the choice of this frame so that fields
with spins higher than two, naturally arising in the general case, are
eliminated~\cite{west}. This procedure
allows to recover locally a flat tangent superspace at each point.

These flat and curved superspaces are connected by the supervierbein, \ie, the
supersymmetric version of the vierbein of general relativity which allows to
convert flat quantities to their curved counterpart. Moreover, the
superconnection allows to define covariant superderivatives ${\cal D}_M$
accounting for the curvature of the space. These superderivatives
are related to the torsion and curvature (superfield)
tensors $T_{MN}{}^P$ and $R_{MNPQ}$, the indices $M$, $N$, $P$ and $Q$
generically denoting Lorentz ($\mu$) and spin ($\alpha$, $\alphadot$) indices.
They help to write (anti)commutation relations among superderivatives
\be\label{eq:DD} 
  \Big[ {\cal D}_M, {\cal D}_N \Big]_{|M||N|} = T_{MN}{}^P {\cal D}_P - \frac12
   R_{MN\underline{\alpha}\underline{\beta}}
    J^{\underline{\beta}\underline{\alpha}} \ ,
\ee
where we have introduced the graded commutator $[\cdot , \cdot]_{\rm grading}$
which consists of an anticommutator for two fermionic quantities and a
commutator otherwise.
More generally, the gradings of the spin and Lorentz indices are defined as
$|\mu|=0$ and $|\alpha| = |\alphadot| =1$.
Finally, the underlined index $\underline{\alpha}$ denotes a generic index being either
a left-handed spin index $\alpha$ or a right-handed spin index $\alphadot$, and
the operators $J_{\alpha\beta}$ and $J_{\alphadot\betadot}$
are the generators of the Lorentz algebra in the two two-component spinorial
representations.

Imposing well-chosen constraints on the elements of the torsion tensor allows to
recover,
when taking the flat limit of Eq.\ \eqref{eq:DD}, the superalgebra of Eq.\
\eqref{eq:poincaresalg} \cite{wb, Wess:1977fn}. Furthermore, by means of the 
(supersymmetric) Bianchi identities, 
\be \bsp  
  0=&\  
    (-)^{|M_1||M_3|} \Big[ \D_{M_1}, \big[\D_{M_2},\D_{M_3}\big]_{|M_2| 
     |M_3|}]\Big]_{|M_1|(|M_2|+|M_3|)}  \\ 
  &\  + (-)^{|M_2||M_1|} \Big[\D_{M_2},\big[\D_{M_3},\D_{M_1}\big]_{|M_3| 
     |M_1|}]\Big]_{|M_2|(|M_3|+|M_1|)}  \\ 
  &\  + (-)^{|M_3||M_2|} \Big[\D_{M_3},\big[\D_{M_1},\D_{M_2}\big]_{|M_1| 
     |M_2|}]\Big]_{|M_3|(|M_1|+|M_2|)} \ , 
\esp\ee 
one can show that all the elements of the torsion and the
curvature tensors can be entirely defined from three basic superfields, 
a scalar chiral superfield ${\cal R}$, a real
vectorial superfield $G_\mu$ and a chiral superfield with three left-handed spin
indices $W_{(\alpha\beta\gamma)}$ symmetric under the exchange of two indices,
together with its antichiral Hermitian-conjugate equivalent
$\Wbar_{(\alphadot\betadot\gammadot)}$.

The lowest order coefficients of the expansion of those superfields in
terms of the Grassmann variables, together with the lowest order components 
of the supervierbein,
are the key ingredients allowing to construct chiral and vector superfields in
curved superspace. In other words, gravity effects in local supersymmetry can be
entirely modeled by means of
a reduced set of component fields, identified to the graviton and the gravitino fields
as well as one supplementing complex scalar and one extra real vectorial
auxiliary fields. As for chiral and vector
superfields (see Section \ref{sec:CSF} and Section \ref{sec:VSF}), the auxiliary
fields of the gravity supermultiplet allow to recover a same
number of fermionic and bosonic degrees of freedom when considering off-shell
fields.

In supergravity, expanding chiral and vector superfields in terms of the
Grassmann variables consists of a non-trivial task since the Grassmann
variables are local and thus depend on the superspace point under consideration. 
However, the building of Lagrangians can be facilitated
by introducing a hybrid system of Grassmann variables, $\Theta$ and $\Thetabar$,
depending both on curved and flat indices \cite{wb}. The price to pay 
is the introduction
of a more complicated invariant measure, the capacity ${\cal E}$, a superfield
that the expansion however only depends on the components of the gravity
supermultiplet.

Skipping all technical details, the supergravity Lagrangian generalizing the
action presented in Eq.~\eqref{eq:generalsusyaction} can be written, under a fully chiral
form, as 
\be\bsp
  {\cal L} = &\
    \frac{3}{8 \kappa^2}
       \int \d^2 \Theta\  \E \big[ \bar{\cal D}\!\cdot\!\bar{\cal D} -8
      \R\big] e^{-\frac13\kappa^2  K(\Phi, \Phi^\dag e^{-2gV})} 
    + \int \d^2 \Theta \E W(\Phi) 
\\&\
    +  \frac{1}{16 g^2} \int \d^2 \Theta\ \E h_{ab}(\Phi) 
       W^{a \alpha} W_\alpha^b
    + \hc \ ,
\esp\label{eq:lagsugra}\ee
where the parameter $\kappa$ stands for the inverse of the Planck mass.  
In this expression, the functions $K$, $W$, $h$ are the curved versions of the
K\"ahler potential, the superpotential and the gauge kinetic function,
respectively. We also stress that the chiral superfields $\Phi$ and the
superfield strength tensors $W_\alpha^a$ are now curved quantities, in contrast
to their flat counterparts introduced in Chapter \ref{chap:susy} and used in
Eq.\ \eqref{eq:generalsusyaction}.

Extracting the expression of this Lagrangian after expanding the
superfields in terms of the Grassmann variables is rather tedious \cite{sugra,
wb}. Furthermore, at the end of this procedure, kinetic terms are obtained
in the unconventional Brans-Dicke form \cite{Brans:1961sx}. In order to recover
standard normalizations, 
some factors have to be absorbed in the fields. This normalization 
procedure relies on the symmetries of the theory. Since
the Weyl supergroup is the symmetry group of the (curved)
superalgebra \cite{Howe:1978km, Wess:1978bu, Siegel:1979wr}, a
super-Weyl transformation can be
employed to restore standard normalizations. This is similar to the Weyl rescaling
of the vierbein inferred by the Weyl group in general relativity~\cite{wb}.

Standard normalizations can also be recovered by introducing Weyl
compensators $\tilde\Phi$ which render the action of Eq.\ \eqref{eq:lagsugra} 
superconformal \cite{Kaplunovsky:1994fg, Gates:1983nr, Siegel:1978mj},
\be\bsp
  {\cal L} = &\
    \frac{3}{8\kappa^2} 
      \int \d^2 \Theta\  \E  \big[ \bar{\cal D}\!\cdot\!\bar{\cal D} - 8 \R\big]
        \big[\tilde\Phi \tilde\Phi^\dag  e^{-\frac13\kappa^2 K(\Phi, \Phi^\dag
        e^{-2gV})} \big] 
      + \int  \d^2 \Theta \E \tilde \Phi^3  W(\Phi)
\\&\ 
    +  \frac{1}{16 g^2} \int \d^2 \Theta\ \E h_{ab}(\Phi) W^{a \alpha} W_\alpha^b
    + \hc \ .
\esp\label{eq:sugracompe}\ee
Fixing appropriately the lowest-order coefficient of the compensator superfield
$\tilde \Phi$ allows to get standard normalizations for the kinetic and gauge
interaction terms.

Since we focus on the mediation of supersymmetry breaking through gravitational
interactions, we omit the complete expression of the Lagrangian ${\cal L}$,
irrelevant for our purposes, and
only consider the scalar
potential $V$. Its form generalizes the one that can be extracted from Eq.\
\eqref{eq:generalsusyaction} so that gravity effects are now incorporated. After
having eliminated the auxiliary fields, the scalar potential reads, skipping
again all technical details,
\be\label{eq:scalpotsugra}\bsp
  V= &\ \frac18 \big(\R\{h^{-1}\}\big)^{ab} \Big[K_i (T_a \phi)^i + (\phi^\dag
      T_a)^\is \phi^\dag_\is\Big] \Big[K_j (T_b \phi)^j + (\phi^\dag T_b)^\js
     \phi^\dag_\js\Big] \\ 
   &\  + \kappa^2 e^\G \Big[ \G_i (\G^{-1})^i{}_\is \G^\is - 3 \Big] \ ,
\esp\ee
where $\G$ is the generalized
K\"ahler potential that unifies the superpotential $W$ and the K\"ahler
potential $K$ as
\be
   \G = \kappa^2 K + \log \big|W\big|^2 \ .
\ee
The first and second order derivatives of the generalized K\"ahler
potential included in Eq.~\eqref{eq:scalpotsugra} follow the notation
conventions introduced in Chapter \ref{chap:susy} for the
K\"ahler potential (see Eq.~\eqref{eq:derK}) and the superpotential (see Eq.\
\eqref{eq:derW}),
\be\bsp
  \G_i  = \frac{\del \G}{\del\phi^i} = \kappa^2 K_i + \frac{W_i}{W}\ , \qquad
    \qquad & \G^\is = \frac{\del \G}{\del\phi^\dag_\is} =\kappa^2 K^\is +
    \frac{W^{\star\is}}{W^\star}\ , \\
  \G^\is{}_i = \frac{\del^2 \G}{\del\phi^i\del\phi^\dag_\is} =& \kappa^2
    K^\is{}_i \  .
\esp\ee

\subsection{Gravity-mediated supersymmetry breaking}\label{sec:grmsb}

In this section, we employ the supergravity framework presented in Section
\ref{sec:sugra} in order to break supersymmetry by means of gravity effects.
Although this could be achieved in several ways, we choose to focus
on a minimal approach, following the framework of Ref.~\cite{polo}.
As mentioned in Section \ref{sec:str}, supersymmetry breaking occurs in a
hidden sector. We consequently introduce a gauge singlet supermultiplet $Z\equiv
(z,\psi_z, F_z)$ that is the only
relevant part of the hidden sector. In contrast, the
set of chiral superfields $\{\Phi^i \equiv (\phi^i, \psi^i,
F^i)\}$ denotes the superfield content of the visible sector. In this setup,
supersymmetry is broken when the auxiliary component of $Z$ gets a vacuum
expectation value. In order to mediate
supersymmetry breaking to the visible sector, both sectors are coupled via 
superpotential and/or K\"ahler interactions. In the minimal version of
supergravity theories, the K\"ahler potential is
chosen as in the renormalizable case,
\be
   K= \phi^\dag_i \phi^i + z^\dag z \ ,
\ee
while the superpotential reads
\be\label{eq:superpotsugra}
  W= W_v(\phi) + \frac{\mu}{\kappa} (z + \beta) \ ,
\ee 
both quantities being expressed as polynomial functions of the scalar degrees of freedom.
In the expressions above, we have split the $K$ and $W$ functions 
into terms depending exclusively on the visible sector and terms related only 
to the hidden sector. In particular, the quantity $W_v$ denotes the
superpotential interactions of the visible 
sector and we assume that the (\textit{a priori}) free parameters of the
model related to the hidden sector, $\mu$ and $\beta$, are 
real, for simplicity.

After having eliminated the auxiliary $F_z$ field by inserting back into the
Lagrangian the solution of its equations of motion,
the vacuum state corresponds to a field configuration where only the scalar
component of the hidden superfield, \ie, the field $\phi_z$,
gets an expectation value $v_z/\sqrt{2}$. The vacuum expectation
value of the 
scalar potential is then given by 
\be
  \langle V \rangle = \mu^2 e^{\frac{v_z^2}{2 m_p^2}} \bigg(
     \Big[ m_p + \frac{v_z}{\sqrt{2} m_p} \big(\frac{v_z}{\sqrt{2}} + \beta\big)\Big]^2 
        - 3 \Big[ \frac{v_z}{\sqrt{2}}+\beta\Big]^2 \bigg)\ , 
\ee
after having reintroduced the Planck mass $m_p = 1/\kappa$ and after omitting
all terms arising from the gauge sector. The potential lies
at a minimum under two conditions. On the one hand, the first-order derivative
of $\langle V\rangle$ with respect to $v_z$ must vanish. On the other hand, 
its second-order derivative must be positive. Further imposing that  $\langle
V\rangle = 0$, or in other words asking for a vanishing cosmological constant, 
one derives the relations
\be\label{eq:vevsugra}
  v_z = \pm \Big( \sqrt{2}(\sqrt{3} - 1) m_p\Big) 
  \qquad\text{and}\qquad 
  \beta = \pm \Big((2 - \sqrt{3}) m_p\Big) \ . 
\ee

Considering the case where both $v_z$ and $\beta$ are positive real numbers, we
now study the particle spectrum resulting from supersymmetry breaking. 
We start from the scalar potential of Eq.\ \eqref{eq:scalpotsugra} and shift
the scalar component of the hidden sector supermultiplet by its vacuum
expectation value, 
\be
  z \to \frac{1}{\sqrt{2}} \Big[ v_z + \Re\{z\} + i \Im\{z\} \Big] \ ,
\label{eq:sugrazshift}\ee
where we have also split the shifted complex field into its scalar and
pseudoscalar components. In the limit of a large Planck mass, \ie, when 
$\kappa\to 0$ or $m_p\to\infty$, the terms quadratic in $\Re\{z\}$ and
$\Im\{z\}$ included in Eq.\ \eqref{eq:scalpotsugra} read 
\be
  V(z) \approx 
    e^{2 (2-\sqrt{3})} \mu^2 \Big[ \sqrt{3} \Re\{z\}^2 +
\big(2-\sqrt{3}\big) \Im\{z\}^2 \Big] \ ,
\ee
after simplifying the results by means of Eq.\ \eqref{eq:vevsugra}. The complex
scalar state of the supermultiplet $Z$, responsible for supersymmetry breaking, 
becomes thus massive, and one furthermore observes a mass splitting among its
scalar and pseudoscalar components.

From the form of the superpotential of Eq.\ \eqref{eq:superpotsugra},
it can be shown, starting from the complete supergravity Lagrangian, that 
the fermionic component of the hidden supermultiplet also becomes massive. This
could be surprizing as this field is expected to be the goldstino as predicted by
Eq.\ \eqref{eq:defgoldstino}. However, this last equation is only valid for the
case of global supersymmetry. In the context of local supersymmetry, the
goldstino field can be massive and will be identified with the
longitudinal polarizations of the (massive) gravitino field (see below).
The entire field content of the hidden sector consists
thus, after supersymmetry breaking, of 
one massive fermionic field $\psi_z$ identified with the goldstino field and
two massive real scalar fields, $\Re\{z\}$ and $\Im\{z\}$.

Next, we turn to the chiral content of the visible sector. Collecting the
leading terms of the expansion of the scalar
potential in terms of $m_p$, one gets, omitting all terms
independent of the scalar fields $\phi^i$ and $\phi^\dag_i$,
\be
   V(\phi,\phi^\dag) \approx 
     e^{2(2-\sqrt{3})} \Big[ \mu^2 \phi^\dag_i \phi^i + W_{v i} W_v^{\star i}
        \Big] + e^{2-\sqrt{3}} \mu \Big[\big(-\sqrt{3} W_v + 
        \phi^i W_{v i} + \hc \big) \Big] \ .
\label{eq:scalbrk}\ee
This last expression is again obtained after shifting the field $z$ by its vacuum
expectation value. We have also introduced the quantities $W_{vi}$ and
$W_v^{\star i}$ as the
first-order derivatives of the parts of the superpotential related to the visible
sector. The results of Eq.\ \eqref{eq:scalbrk} show that
supersymmetry breaking has lead to the generation of scalar mass terms as well
as of multiscalar
interactions. Moreover the form of the superpotential
is such that no new fermionic mass terms are generated additionally to the
supersymmetric masses possibly included in $W_v$.

Getting back to the globally supersymmetric Lagrangian of 
Eq.\ \eqref{eq:Lgeneral}, one observes that the gauge kinetic function also couples
to the gaugino fields when one inspects the terms of the fifth line, provided
that we account for the
solutions of the equations of motion for the auxiliary $F$-fields of Eq.\
\eqref{eq:auxsol}. Those Lagrangian terms are still present for local
supersymmetry and are not even modified by gravity effects, 
\be
 {\cal L}_{\rm ino} = 
    \frac14 F^i h_{abi}  \lambda^a\!\cdot\!\lambda^b 
  + \frac14 F^\dag_\is h^{\star\is}_{ab} \lambar^a\!\cdot\!\lambar^b  \ .
\label{eq:sugrall}\ee
In contrast, the solutions of the equations of motion for the $F$-fields, which
are originally given by Eq.\
\eqref{eq:auxsol} in global supersymmetry, are modified as soon as matter and
gauge fields are coupled to gravity, 
\be
  F^i=\kappa^2 (\G^{-1})^i{}_\is e^{\frac12 \G} \G^\is  + \ldots\ , 
\ee
where the dots stand for additional terms irrelevant for our purposes.
Gaugino mass terms
arise after inserting this last relation in Eq.\ \eqref{eq:sugrall} and 
shifting the scalar component of the supermultiplet $Z$
by its vacuum expectation value. For a model where the gauge kinetic function is
given by
\be
  h_{ab} = \alpha z \ \delta_{ab} \ , 
\ee
the parameter $\alpha$ being taken real for simplicity, supersymmetry breaking
subsequently generates gaugino mass terms of the form
\be
  {\cal L}_{\rm mass}^{(\lambda)} = 
    \frac{\sqrt{3}}{4} e^{2-\sqrt{3}}\ \mu m_p\ \alpha \big[
     \lambda^a\!\cdot\!\lambda^a + \lambar^a\!\cdot\!\lambar^a \big] \ .
\label{eq:inomass}\ee

In order to improve the readability of all the mass and interaction terms
generated by gravity-mediated supersymmetry breaking, it is useful to introduce
the gravitino mass $m_{3/2}$. Initially massless, the gravitino field also
acquires a mass after supersymmetry breaking. This can be seen by starting from 
the Lagrangian terms
coupling the gravitino field to the generalized K\"ahler potential,
\be
  {\cal L}_{\rm gravitino} = - \frac{i}{2} \kappa^2 e^{\frac12 \G}
\psi_\mu\sigma^{\mu\nu}\psi_\nu + \hc \ ,
\ee
where $\psi_\mu$ stands for the Rarita-Schwinger gravitino field. After shifting
the
scalar field $z$ with respect to its vacuum expectation value, a mass
term is generated,
\be
  {\cal L}_{\rm mass}^{(\psi)} = -\frac{i}{2} e^{2-\sqrt{3}} \mu
    \psi_\mu\sigma^{\mu\nu}\psi_\nu + \hc \ ,
\ee
the gravitino mass being thus
\be
  m_{3/2} =  e^{2-\sqrt{3}} \mu\ .
\label{eq:ginomass}\ee

To summarize the effects of gravity-mediated supersymmetry breaking on the
fields of the visible sector, we now assume
that the superpotential interactions of the visible sector $W_v$ are
renormalizable and thus given by Eq.\ \eqref{eq:renosuperW},
\be
   W_v = 
    \frac16 \lambda_{ijk} \phi^i \phi^j \phi^k + \frac12 \mu_{ij}\phi^i \phi^j +
    \xi_i \phi^i \ ,
\ee
$\lambda$, $\mu$ and $\xi$ being free parameters of the model. Collecting
the results
from Eq.\ \eqref{eq:scalbrk} and Eq.\ \eqref{eq:inomass}, the
supersymmetry-breaking Lagrangian terms that have been generated 
are all soft, \ie, the related coupling strengths have
strictly positive mass dimensions,
\be\boxed{\bsp
  {\cal L}_{\rm soft} = &\
    -\frac12 m_{1/2} 
      \big[ \lambda^a\!\cdot\!\lambda_a + \lambar^a\!\cdot\!\lambar_a \big] 
    - m_0^2  \phi^\dag_i \phi^i  \\ &\quad
    - \bigg[\frac16 A_0\ \lambda_{ijk} \phi^i \phi^j \phi^k 
    + \frac12 B_0\ \mu_{ij} \phi^i \phi^j 
    + C_0\ \xi_i \phi^i 
    + \hc \bigg] \ .
\esp}\label{eq:sugralsoft}\ee
In the equation above, we have introduced the universal gaugino and scalar
masses $m_{1/2}$ and $m_0$, as well as the universal trilinear coupling $A_0$.
These three universal parameters are however not
independent and can all be rewritten in terms of the gravitino 
mass given in Eq.\ \eqref{eq:ginomass},%
\renewcommand{\arraystretch}{1.4}%
\be\boxed{\begin{array}{c}
   m_{1/2} = \frac{\sqrt{3}}{2} \alpha m_{3/2} m_p \ , \qquad
   m_0 = m_{3/2} \ , \qquad\\
   C_0 = (1-\sqrt{3}) m_{3/2}  \ , \qquad
   B_0 = (2-\sqrt{3}) m_{3/2}  \ , \qquad
   A_0 = (3-\sqrt{3}) m_{3/2}  \ .
\end{array}}\label{eq:sugrarelat}\ee%
\renewcommand{\arraystretch}{1.}%
The universality of those parameters is driven by our choices for the K\"ahler
potential, the gauge kinetic function and the superpotential. The model
presented in this section consists of
the so-called Polonyi model \cite{polo} with the simplest parametrization of
the hidden sector. Among other rather popular choices, one finds 
dilaton-dominated models \cite{Kaplunovsky:1993rd, Barbieri:1993jk,
Brignole:1993dj} or no-scale models \cite{Brignole:1997dp}. They all lead to a
supersymmetry-breaking Lagrangian similar to the one presented in Eq.\
\eqref{eq:sugralsoft}, with however different relations
among the soft parameters.

Constructing phenomenologically models viable with respect to the current
experimental bounds on supersymmetric masses, \ie, with superpartners of
about 1 TeV, implies that the gravitino mass is of about 1 TeV too.
The gravitino, initially massless, has absorbed the goldstino field
after supersymmetry breaking
so that it gets two helicity $\pm
1/2$ components and becomes massive. This is called the
super-Brout-Englert-Higgs
mechanism \cite{Cremmer:1978iv, Deser:1977uq}, the supersymmetric analog of the
Brout-Englert-Higgs mechanism in
non-supersymmetric quantum field theories.

The universality feature of Eq.\
\eqref{eq:sugrarelat} has inspired the so-called 
constrained versions of supersymmetric models. In this case, the soft
supersymmetry-breaking Lagrangian is still given by Eq.\ \eqref{eq:sugralsoft}, 
the soft parameters are taken universal and independent.

\subsection{Gauge-mediated supersymmetry breaking}\label{sec:GMSB}
In contrast to gravity-mediated supersymmetry breaking which is inferred by 
non-re\-nor\-ma\-li\-za\-ble interactions, gauge-mediated supersymmetry breaking
is a mechanism that can be 
entirely expressed in terms of renormalizable loop effects and standard gauge
interactions \cite{Dine:1981gu,
Nappi:1982hm, AlvarezGaume:1981wy, Dine:1993yw, Dine:1994vc, Dine:1995ag}.  As
for supergravity, supersymmetry breaking still occurs in a hidden sector. The
latter contains, in its minimal version,
a gauge singlet chiral supermultiplet $Z\equiv (z,\psi_z, F_z)$ that both the
scalar and the auxiliary components acquire vacuum expectation values,
\be
  \langle Z \rangle = \frac{1}{\sqrt{2}} \Big[ v_z - \theta\!\cdot\!\theta\ v_F^2
    \Big] \ .
\label{eq:gmsbvev}\ee

In order to mediate supersymmetry breaking to the visible sector, one
introduces several messenger fields, organized in two sets of 
chiral supermultiplets ${\bf \Phi}^i$ and ${\bf \Phibar}_i$ lying in
(non-trivial) complex
conjugate representations of the gauge group. Contrary to the fields of the
visible sector that are not connected to the hidden sector by any mean, the
messenger fields are allowed to couple to the hidden sector through interactions
driven by the superpotential. For simplicity, we consider, in the
rest of this section, that the messenger sector consists of one single
pair of messenger superfields ${\bf \Phi}$ and ${\bf \Phibar}$. The
generalization to a fully generic setup goes along the same lines of what is
presented below and is omitted as minimal
gauge-mediated supersymmetry-breaking scenarios are sufficient for
depicting the main
features of a spontaneous breaking of supersymmetry by gauge interactions.
The messengers are coupled to the hidden sector through
superpotential interactions,
\be
   W_{\rm mes} = \lambda {\bf \Phibar} {\bf \Phi} Z  \ ,
\label{eq:gmsbwmess}\ee
where $\lambda$ is a free parameter of the model.

Since the messenger superfields lie in non-trivial conjugate representations
of the gauge group, they also communicate with the visible sector
by means of ordinary gauge interactions that are included in the
K\"ahler potential. Focusing on the hidden and messenger sectors, the
K\"ahler potential is given by 
\be
  K({\bf \Phi}, {\bf \Phibar}) = 
    {\bf \Phi}^\dag e^{-2 g V} {\bf \Phi} + {\bf \Phibar}^\dag e^{-2gV'} {\bf
    \Phibar} + Z^\dag Z \ ,
\label{eq:gmsbkahl}\ee
where we have adopted the simplest renormalizable form for the function $K$,
as given in Eq.~\eqref{eq:gensusyaction}.
In Eq.\ \eqref{eq:gmsbkahl}, we denote respectively by $V^{(')}$ and
$g$ the vector superfield and coupling constant associated with the gauge group of
the model (the prime denoting different representations of the gauge group).

Multiscalar interaction terms are included
in both the Lagrangian related to the superpotential of Eq.\
\eqref{eq:gmsbwmess} and in the one associated with the K\"ahler potential of Eq.\
\eqref{eq:gmsbkahl},
\be
  {\cal L}_{\rm scal.} = F^\dag F + \Fbar^\dag \Fbar + F_z^\dag F_z - 
    \lambda \Big[ F_z  \phibar \phi + z \Fbar \phi + z \phibar F \Big] + \hc\ .
\ee
In our notations, the fields $\phi$ and $\phibar$ are the scalar components of
the messenger supermultiplets ${\bf \Phi}$ and ${\bf \Phibar}$ while $F$ and
$\Fbar$ are their 
auxiliary components, respectively. After the components of the 
superfield $Z$ get their vacuum expectation values as in Eq.\
\eqref{eq:gmsbvev}, mass terms are generated from this
interaction Lagrangian ${\cal L}_{\rm scal.}$. 
Solving the equations of motion for the auxiliary fields and
inserting back the solutions into the Lagrangian, ${\cal L}_{\rm scal.}$ can
then be rewritten as 
\be
  {\cal L}_{\rm scal.} = 
    - \frac12 \big| \lambda v_z \big|^2  \phi^\dag \phi 
    - \frac12 \big| \lambda v_z \big|^2 \phibar\phibar^\dag  
    + \frac{1}{\sqrt{2}} \lambda v_F^2 \phibar \phi 
    + \frac{1}{\sqrt{2}} \lambda^\ast v_F^{2\ast} \phi^\dag \phibar^\dag +
     \ldots\ ,
\ee
where the dots stand for trilinear and quartic terms. In the $(\phi,
\phibar^\dag)$ basis, the squared mass matrix is thus given by 
\be
  {\cal M}^2_{\rm mes} = \bpm 
    \frac12 \big| \lambda v_z \big|^2 & - \frac{1}{\sqrt{2}} \lambda v_F^2\\
    - \frac{1}{\sqrt{2}} \lambda^\ast v_F^{2\ast} & \frac12 \big|\lambda
      v_z\big|^2
  \epm \ . 
\ee
Consequently, the scalar components of the pairs of messenger superfields
${\bf \Phi}$ and ${\bf \Phibar}$, initially massless, mix to two mass eigenstates
defined by
\be
  \phi_1= \frac{1}{\sqrt{2}}\bigg[ \frac{\lambda v_F^2}{\big|\lambda v_F^2\big|}
   \phi - \phibar^\dag\bigg]
  \qquad\text{and}\qquad
  \phi_2= \frac{1}{\sqrt{2}}\bigg[ \frac{\lambda v_F^2}{\big|\lambda v_F^2\big|}
   \phi + \phibar^\dag\bigg] \ ,
\label{eq:gmsbphyssca}\ee
the masses being given by
\be
  m_1^2 = \frac12 \big|\lambda v_z\big|^2 - \frac{1}{\sqrt{2}} \big|\lambda
    v_F^2 \big|
  \qquad\text{and}\qquad
  m_2^2 = \frac12 \big|\lambda v_z\big|^2 + \frac{1}{\sqrt{2}} \big|\lambda
    v_F^2 \big|\ .
\label{eq:gmsbmasss}\ee

We now turn to the fermionic fields of the hidden and messenger sectors. 
The interaction Lagrangian terms derived from 
the superpotential contain the Yukawa interactions
\be
  - \lambda \Big[ 
      z\ \psi_{\bf \Phibar}\!\cdot\!\psi_{\bf \Phi}
    + \phi\ \psi_z\!\cdot\!\psi_{\bf \Phibar} 
    + \phibar\ \psi_{\bf \Phi}\!\cdot\!\psi_z \Big] + \hc \ ,
\ee
where $\psi_{\bf \Phi}$ and $\psi_{\bf \Phibar}$ are the fermionic components of
the messenger superfields ${\bf \Phi}$ and ${\bf \Phibar}$, respectively, and we 
recall that the field $\psi_z$ denotes the fermionic component of the
superfield $Z$ of the hidden sector. Shifting the
scalar field $z$ by its vacuum expectation value as shown in Eq.
\eqref{eq:gmsbvev}, one observes that
supersymmetry breaking has rendered the messenger fermions massive,
\be
  m_\psi^2 = \frac12 \big|\lambda v_z\big|^2 \ ,
\label{eq:gmsbmassf}\ee
whilst the $\psi_z$ fermionic field stays massless and can be identified
with the goldstino field, in agreement with Eq.\ \eqref{eq:defgoldstino}.

From Eq.\ \eqref{eq:gmsbmasss} and Eq.\ \eqref{eq:gmsbmassf}, it can be seen
that the effects of supersymmetry breaking on the messenger sector is to split
its spectrum apart if $v_F\neq 0$. Since this condition
is also necessary to ensure supersymmetry breaking, it is however always fulfilled. 
One of the messenger scalar mass
eigenstate has thus become lighter than the fermionic messenger field while the
other ones is now heavier.

\begin{figure}
 \centering
   \vspace*{-8.2cm}\hspace*{2.5cm}
   \includegraphics[width=\columnwidth]{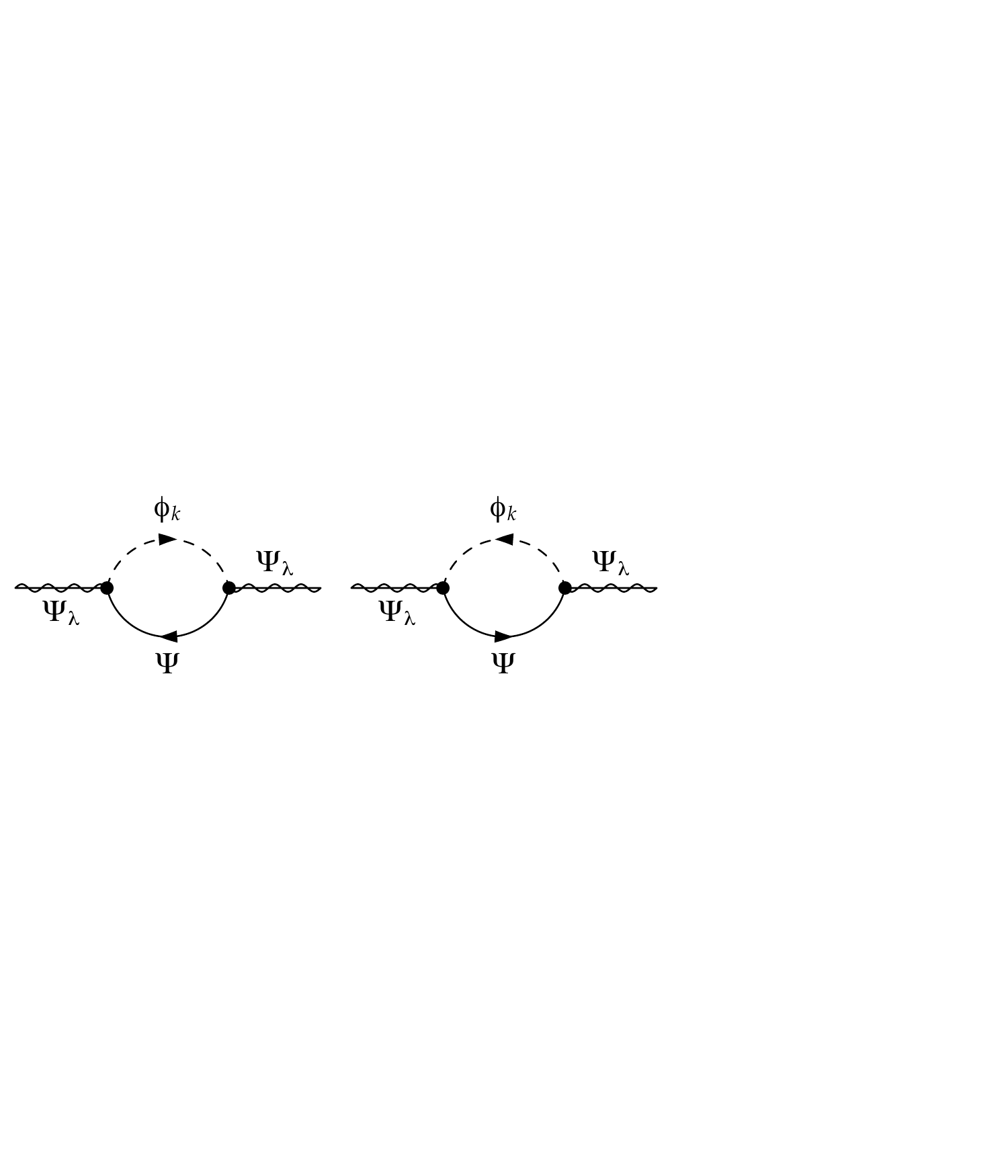}
   \vspace*{-8.4cm}
 \caption{\label{fig:gmsbino}Virtual contributions to gaugino self-energies
leading to radiatively-induced gaugino masses. In the notations employed in this
figure, $\Psi_\lambda$
stands for the Majorana gaugino field, $\Psi$ for the Dirac messenger field and
$\phi_k$ for the two scalar messenger fields, with $k=1,2$.}
\end{figure}

Supersymmetry breaking is subsequently communicated to the visible sector
through radiative corrections. Both the messengers and the fields of the
visible sector of the theory are sensitive to the gauge interactions. 
In particular, the Lagrangian of Eq.\ \eqref{eq:gensusylag}
includes terms coupling the scalar and fermionic components of a specific chiral
supermultiplet of the theory to the gaugino field $\lambda_a$. This feature 
holds both for the visible and the messenger sectors and is the key for
understanding the
radiative generation of mass terms for the visible sector fields. 
Focusing on the messenger superfields
lying in complex conjugate representations $\R$ and $\bar \R$ of the gauge
group that are specified by the
Hermitian matrices $T_a$ and $\Tbar_a = - T_a^t$, respectively, the relevant
interactions read
\be
   {\cal L}_2 = -i \sqrt{2} g \Big[ \phi^\dag T^a \psi_{\bf \Phi} \!\cdot\!
    \lambda_a - \lambda_a \!\cdot\! \psi_{\bf\Phibar} T^a \phibar^\dag \Big] +
    \hc\ .
\ee
In order to more easily compute the gaugino mass terms generated by such
interactions, we
rewrite this Lagrangian, currently expressed in terms of two-component Weyl
fermions, in terms of a four-component Majorana field and a four-component Dirac
field
\be
  \Psi_\lambda = \bpm \phantom{-}i \lambda \\ -i \lambar \epm 
  \qquad\text{and}\qquad
  \Psi = \bpm \psi_{\bf \Phi} \\ \psi_{\bf \Phibar} \epm \ .
\ee
Usual Feynman rules can therefore be extracted and standard
Feynman diagram techniques with four-component spinors employed in order to 
calculate radiatively-induced mass terms.
The Lagrangian ${\cal L}_2$ is rewritten, in
four-component notations and after the rotation of the scalar messenger fields
$\phi$ and $\phibar^\dag$ to the physical basis $(\phi_1,\phi_2)$, as
\be
  {\cal L}_4 = - \sqrt{2} g \sum_{k=1}^2 
    \Big[ \phi_k^\dag R_{k1} T_a \Psibar_\lambda^a P_L
    \Psi - \Psibar P_L \Psi_\lambda^a T_a R_{k2}^\ast \phi_k \Big]+ \hc\ ,
\ee
where the relation between mass eigenstates and gauge eigenstates, together with
the definition of the elements of the mixing matrix $R$, can be
obtained from Eq.\ \eqref{eq:gmsbphyssca}, %
\renewcommand{\arraystretch}{1.4}%
\be
  R = \frac{1}{\sqrt{2}} \bpm  \frac{\lambda v_F^2}{\big|\lambda v_F^2\big|} & -1 \\
    \frac{\lambda v_F^2}{\big|\lambda v_F^2\big|} & 1 \epm \ .
\label{eq:gmsbmessmix}\ee%
\renewcommand{\arraystretch}{1}%
In addition, we have also introduced the left-handed and right-handed
chirality projectors $P_L$ and $P_R$ acting on four-component spinors.

At the first order, the (unrenormalized) gaugino propagator $-i \Sigma$ receives
contributions from the one-loop diagrams presented in Figure \ref{fig:gmsbino}.
This propagator can always be rewritten by 
splitting its scalar and vectorial pieces $\Sigma_S$ and $\Sigma_V$, 
\be
 -i \Sigma_{ab}(p) = - i \delta_{ab} \Big[ \Sigma_V(p^2) \slashed{p} +
   \Sigma_S(p^2) P_L + \Sigma^\ast_S(p^2) P_R \Big]\ ,
\ee
where $p_\mu$ stands for the gaugino four-momentum and $a$ and $b$ for
adjoint gauge
indices attached to the external legs. Evaluated on-shell, the
quantities $\Sigma_S$ and $\Sigma_V$ allow to derive the quantum corrections to
the gaugino mass $\delta m_\lambda$,
\be
  \delta m_\lambda  = -m_\lambda  \Sigma_V(m_\lambda ^2)
      - \Re\big\{\Sigma_S(m_\lambda ^2)\big\}  = - \Re\big\{\Sigma_S(0)\big\} \ .
\ee
We recall that for the second equality, we have employed the fact that the
gaugino field is massless at tree-level. Computing the loop-diagrams of Figure
\ref{fig:gmsbino} using standard loop-computation
techniques, the self-energy corrections can be written in terms of two-point 
Passarino-Veltman functions $B_0(0; m_k^2, m_\psi^2)$
with $k=1,2$ \cite{Passarino:1978jh},
\be
  \delta m_\lambda  =   - \frac{g^2 m_\psi}{4 \pi^2} \tau_\R \sum_{k=1}^2\bigg[ 
       \Re\big\{R_{k2}^\ast R_{k1}\Big\} B_0(0; m_k^2, m_\psi^2)\bigg] \ .
\ee
In this expression, we account for the fact that both scalar messenger states
propagate in the quantum loops. In addition, 
products of representation matrices of the gauge group have been simplified by
means of the relation
\be
  {\rm Tr} \big[ T_a T_b\big] = \tau_\R \delta_{ab} \ ,
\ee 
where the group invariant $\tau_\R$ is the Dynkin index related to the
representation $\R$. 
Inserting the analytical expressions for the elements of
the mixing matrix of Eq.\ \eqref{eq:gmsbmessmix}, the values of the messenger
masses of Eq.\ \eqref{eq:gmsbmasss} and Eq.\ \eqref{eq:gmsbmassf} and computing
explicitly the Passarino-Veltman integrals, one obtains \cite{Dine:1993yw,
Dine:1994vc, Dine:1995ag}
\be\boxed{ 
  \delta m_\lambda  = \frac{g^2}{8 \pi^2} \tau_\R \Lambda \Big[ 
    \frac{M_{\rm mes}^2 + \Lambda M_{\rm mes}}{\Lambda^2} \log \frac{M_{\rm
     mes}+\Lambda}{M_{\rm mes}} + \frac{M_{\rm mes}^2 -
    \Lambda M_{\rm mes}}{\Lambda^2} \log \frac{M_{\rm mes}-\Lambda}{M_{\rm
    mes}} \Big] \ .}
\label{eq:gmsb_ino_mass}\ee
In this expression, we have introduced the supersymmetry-breaking scale
$\Lambda$ standing for
the ratio of the vacuum expectation values of the scalar and auxiliary
components of the 
superfield $Z$ and the messenger scale $M_{\rm mes}$ being equal to the mass of
the fermionic messenger field,
\be\boxed{
   \Lambda= \frac{v_F}{v_z} \qquad\text{and}\qquad
   M_{\rm mes}=m_\psi  = \frac12 \big|\lambda v_z\big|^2\ . }
\ee

To summarize, the splitting of the messenger spectrum has
lead to radiatively induced gaugino masses. In other words, supersymmetry
breaking has been successfully transferred to the visible sector by means of
standard gauge interactions.

\begin{figure}
 \centering
   \vspace*{-1.5cm}
    \includegraphics[width=.9\columnwidth]{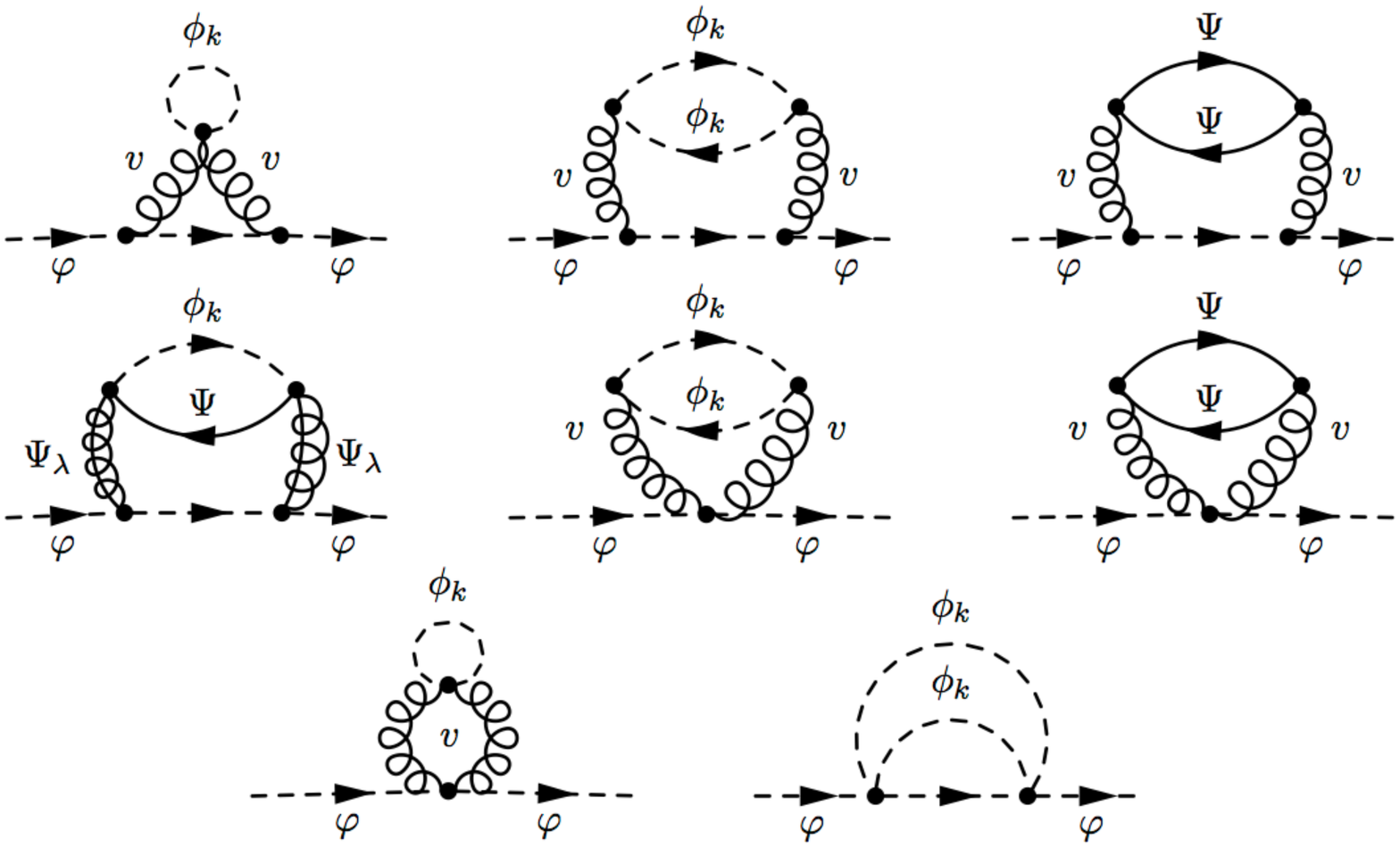}
   \vspace*{-1.5cm}
 \caption{\label{fig:gmsbsf}Virtual contributions to the self-energies of the 
scalar fields of the visible sector that lead to radiatively-induced masses.
In
the notation conventions of this figure, $\varphi$ stands for a generic scalar
field of the visible
sector, $v$ and $\Psi_\lambda$ for the gauge boson and gaugino field associated with the
model gauge group, and $\phi_k$ and $\Psi$ for the scalar and four-component
fermionic messenger fields, respectively.}
\end{figure}
Similarly, masses for the scalar fields of the visible sector can also be
generated by quantum corrections, but this time at the two-loop level, as shown
in Figure \ref{fig:gmsbsf}.
The relevant loop diagrams involve, as for the gaugino case,
standard gauge interactions depicted in the Lagrangian of Eq.\
\eqref{eq:gensusylag2} that apply both for the scalar fields of the visible
sector and the messenger fields. 
Quantum corrections to the squared mass of a massless
scalar field $\varphi$ of the visible sector are derived from the expression of
the scalar propagator $-i \Pi$ evaluated on-shell
\be
  -i \Pi_{mn}(p) = - i \Pi (p) \delta_{mn} \qquad\Rightarrow\qquad
  \delta m^2_\varphi = \Pi(0) \ ,
\ee
where $m$ and $n$ are gauge indices related to the representation of the
gauge group in which the field $\varphi$ lies. After an omitted (for brevity)
computation, these two-loop quantum corrections are evaluated as
\be\boxed{\bsp
  \delta m^2_\varphi =&\ \frac{g^4}{128 \pi^4} \tau_\R C_\R \bigg[
    \Big(M_{\rm mes}^2 \!+\! \Lambda M_{\rm mes}\Big) \Big(
      \log \frac{M_{\rm  mes}\!+\!\Lambda}{M_{\rm mes}} \!-\! 2 {\rm Li}_2
      \frac{\Lambda}{M_{\rm mes}\!+\!\Lambda} \!+\! \frac12 {\rm Li}_2\frac{2
      \Lambda}{M_{\rm mes}\!+\!\Lambda} \Big) \\
&\ +    
    \Big(M_{\rm mes}^2 \!-\! \Lambda M_{\rm mes}\Big)\Big(
      \log \frac{M_{\rm  mes}\!-\!\Lambda}{M_{\rm mes}} \!-\! 2 {\rm Li}_2
      \frac{\Lambda}{\Lambda \!-\! M_{\rm mes}} \!+\! \frac12 {\rm Li}_2\frac{2
      \Lambda}{\Lambda \!-\! M_{\rm mes}} \Big)
   \bigg] \ ,
\esp}\label{eq:gmsb_sc_mass}\ee
where $C_\R$ is the Casimir invariant associated to the representation $\R$,
\be
  \big(T^a T_a\big)^m{}_n = C_\R\ \delta^m{}_n \ .
\ee
The originally massless scalar fields have hence been rendered massive.

In gravity-mediated supersymmetry breaking, multiscalar soft interactions have
been generated as illustrated in the Lagrangian of Eq.\
\eqref{eq:sugralsoft}. The same kind 
of supersymmetry-breaking interactions are also induced by gauge-mediation.
However, they arise, as for the scalar masses, at the two-loop level but can,
in contrast to the mass terms, be approximately neglected at the messenger
scale. The
supersymmetry-breaking Lagrangian hence reads, collecting the terms related to
the visible sector,
 \be\label{eq:Lbrkgmsb}\boxed{\bsp
  {\cal L}_{\rm soft} = &\
    - \frac12 m_{\lambda^a}
      \big[ \lambda^a\!\cdot\!\lambda_a + \lambar^a\!\cdot\!\lambar_a \big] 
    - m_{\varphi^i}^2  \varphi^\dag_i \varphi^i  \ .
\esp}\ee
The gaugino and scalar masses $m_{\lambda^a}$ and $m_{\phi^i}$ are deduced
from 
Eq.~\eqref{eq:gmsb_ino_mass} and Eq.~\eqref{eq:gmsb_sc_mass}, respectively, and
multiscalar interactions are only generated when evolving the soft
supersymmetry-breaking parameters down to the electroweak scale.

\subsection{Anomaly-mediated supersymmetry breaking}\label{sec:AMSB}
In this section, we present a mechanism where supersymmetry breaking is mediated
to the visible sector through quantum loop effects related to anomalous
rescaling violations
\cite{Randall:1998uk, ArkaniHamed:1998kj, Bagger:1999rd}. This mechanism has
the virtue to operate as soon as we have a hidden sector in the theory by means
of several sources of anomalies and possibly competes with other
sources of supersymmetry breaking. We take here the example of
supergravity where supersymmetry breaking is also mediated by
Weyl anomalies, although their contribution is strongly suppressed with respect to gravity
effects.

In order to underline the effects of anomaly-mediated supersymmetry
breaking, we start from the supergravity Lagrangian of Eq.\
\eqref{eq:lagsugra} and replace the capacity $\E$ by the superfield product $\E
\langle{\bf \Phi}\rangle^3$, introducing a spurious
superfield ${\bf \Phi}$\footnote{It can be noted that a superfield defined as
the product of a capacity by a chiral superfield still verifies the properties
of a capacity \cite{sugra, wb} and can therefore safely be used as a superspace integral
invariant measure.}. The vacuum expectation value of the latter is
explicitly given by
\be
  \big\langle {\bf \Phi}\big\rangle  = 1 + M_{\rm aux}\ 
    \theta\!\cdot\!\theta \ ,
\label{eq:amsbvev}\ee
so that $\langle{\bf \Phi}\rangle$ can be employed as a tool to select the
lowest-order component of the supergravity multiplet included in the
supervierbein,
\be
  \E = e \big\langle {\bf \Phi}^3 \big\rangle + \ldots 
\label{eq:capacity}\ee
where higher spin states are contained in the dots. This computation
artifact allows us to factorize out the pure gravity Lagrangian and
only focus on the chiral and gauge content of the theory.
The spurious
superfield is hence used to rewrite the first term of the Lagrangian of
Eq.~\eqref{eq:lagsugra} under an eight-dimensional integral upon the superspace
coordinates. This uses a recasting of the kinetic and interaction
terms included in the K\"ahler potential, neglecting gravitational
contributions involving non-scalar states of the gravitation supermultiplet,
where the relevant terms are made proportional to $\langle{\bf
\Phi \Phi}^\dag\rangle$~\cite{Siegel:1978mj,Bagger:1995ay}.
Inserting the spurious superfield $\langle {\bf \Phi}
\rangle$ into the other terms of the supergravity
Lagrangian, the latter can be rewritten as
\be\bsp
  {\cal L} = &\
      \frac{3 e}{2 \kappa^2} \int \d^2 \Theta\  \d^2\Thetabar\ \big\langle {\bf
      \Phi \Phi}^\dag \big\rangle\ e^{-\frac13\kappa^2  K(\Phi, \Phi^\dag
       e^{-2gV})}  
    + e \int \d^2 \Theta\ \big\langle {\bf \Phi}^3 \big\rangle\ W(\Phi) 
\\&\
    +  \frac{e}{16 g^2} \int \d^2 \Theta\ h_{ab}(\Phi) W^{a \alpha} W_\alpha^b
    + \hc + \ldots \ ,
\esp\label{eq:amsblag}\ee
where the superfield $\Phi$ represents in a generic fashion the chiral
content of the theory, $W_\alpha$
the superfield strength tensor associated with the gauge group and the dots stand
for the omitted terms related to the gravity supermultiplet. We moreover assume 
that all the three
fundamental functions, \ie, the K\"ahler potential $K$, the superpotential $W$ and
the gauge kinetic function $h$, are independent of the spurious 
superfield ${\bf \Phi}$ and can be written as sums of two terms, one of them
depending exclusively on the superfields of the visible sector and the other
one being only related to the hidden sector where supersymmetry
breaking occurs. The same Lagrangian could also have been obtained
by employing the Weyl compensator formalism for the rescaling. In this case,
it is sufficient to replace the compensator superfield appearing in the
Lagrangian 
of Eq.\ \eqref{eq:sugracompe} by its vacuum expectation value 
given by Eq.\ \eqref{eq:amsbvev}.

Inspecting the Lagrangian of Eq.\ \eqref{eq:amsblag}, we observe that the
super-Yang-Mills part of the Lagrangian (\ie, the term of the second line of
this equation) does not couple to
the spurious superfield $\langle{\bf
\Phi}\rangle$.
This is not surprising as the gauge sector is invariant under super-Weyl
transformations. However, this symmetry is anomalous and broken via quantum loop
effects~\cite{Derendinger:1991kr, Derendinger:1991hq, LopesCardoso:1991zt,
LopesCardoso:1992yd, Kaplunovsky:1994fg}. Canceling this anomaly by appropriate
counterterms balancing the effects of the relevant loop-diagrams yields
a shift of the gauge kinetic function,
\be
  h_{ab}(\Phi) \to h_{ab}(\Phi) - 2\ b_0\ \delta_{ab}\ \langle\log({\bf
    \Phi})\rangle\ ,
\label{eq:amsbshift}\ee
where $b_0$ stands for the one-loop coefficient of the gauge beta function
$\beta_g$
\be
 \beta_g = -b_0 g^3  + \ldots \ .
\ee
After evaluating the integral upon $\Theta$ in the super-Yang-Mills Lagrangian, 
the shift of
Eq.~\eqref{eq:amsbshift} generates a non-vanishing gaugino mass $m_\lambda$, 
\be\boxed{
  m_\lambda = \frac{\beta_g}{g} M_{\rm aux} \ .}
\label{eq:amsbino}\ee
This consists of a strong prediction of anomaly-mediated supersymmetry-breaking
scenarios as this last equation means that the different gaugino masses appear as the
ratio of the corresponding gauge beta functions  when the gauge group is a
semi-simple group.

As briefly mentioned above, there are other ways to generate anomalous mass
terms for the gaugino fields, but they can all be considered as results of the
violation of the super-Weyl invariance. We can 
associate a cutoff scale $\Lambda_{\rm UV}$ to these effects, related to a
hidden sector embedding some mechanism for anomaly cancellation. In the case of
supergravity, this scale is
naturally the Planck scale. Renormalizable physics, \ie, physics related to
the visible sector, is however independent of the nature of the cutoff.
Therefore, the previous results of Eq.\ \eqref{eq:amsbino} can be safely
generalized. To illustrate this statement, we
introduce a renormalization scale $\mu_R$ and turn to the study of
anomaly-induced supersymmetry-breaking terms in the general case. 

After renormalization, the Wilsonian effective Lagrangian obtained after
integrating out all the modes lying above $\Lambda_{\rm UV}$ (and thus included
in the hidden sector) is given by
\be\bsp
  {\cal L} = &\ 
    \int \d^2 \theta\  \d^2\thetabar\ Z^i{}_j\bigg[\frac{\mu_R}{\Lambda_{\rm UV}
      \langle{\bf \Phi} \rangle},  \frac{\mu_R}{\Lambda_{\rm UV} \langle 
      {\bf \Phi^\dag} \rangle} \bigg] \Phi_i^\dag e^{-2gV} \Phi^j 
    + \int \d^2 \theta\ W(\Phi) 
\\ &\
    +  \frac{1}{16 g^2} \int \d^2 \theta\ \tau_{ab}\bigg[\frac{\mu_R}{\Lambda_{\rm
      UV}\langle{\bf \Phi}\rangle}\bigg] W^{a \alpha} W_\alpha^b
    + \hc \ .
\esp\label{eq:amsbrenlag}\ee
This last expression can be computed by starting from Eq.\ \eqref{eq:amsblag},
then performing a Taylor expansion of the exponential and finally evaluating the
results in the flat space limit, 
\be
  \Theta\to\theta \ , \qquad \Thetabar\to\thetabar \qquad\text{and}\qquad e\to 1
\ .
\ee
The tree-level dependence in the spurious superfield is shifted away by the
superfield redefinitions
\be
  \Phi \langle{\bf \Phi}\rangle \to \Phi \ ,
\label{eq:amsbphiresc}\ee
so that the dependence in $\langle{\bf \Phi}\rangle$ now only occurs through
loop-effects embedded into the functions $Z$ and $\tau$. In
other words, there is no tree-level communication of supersymmetry breaking.
In order to write such a form for the Lagrangian, we have assumed 
that the superpotential is trilinear and homogeneous in the chiral
superfields of the visible sector\footnote{Taking the superpotential as a
trilinear and homogeneous function of the chiral superfields of the visible
sector renders it $R$-symmetric. This symmetry, being anomalous, is employed in
the following to investigate the structure of the generated soft
supersymmetry-breaking terms.}, as it originally multiplies a third
power of
the spurious superfield and there is no more dependence on $\langle {\bf
\Phi}\rangle$ in the superpotential Lagrangian of Eq.\ \eqref{eq:amsbrenlag}.
The generation of superpotential mass and
linear terms is however still possible dynamically, once supersymmetry is broken 
\cite{Randall:1998uk, Giudice:1988yz, Giudice:1998xp, Pomarol:1999ie}.
Moreover,
the absence of renormalization effects in the superpotential term of
Eq.\ \eqref{eq:amsbrenlag} is justified by the supersymmetric
non-renormalization theorems~\cite{Salam:1974jj,Wess:1973kz,
Iliopoulos:1974zv,  Ferrara:1974fv, Zumino:1974bg, Ferrara:1975ye, 
Grisaru:1979wc}. To summarize, the spurious superfield, and in particular its
vacuum expectation value $\langle {\bf \Phi} \rangle$, are
the only quantities encompassing the effects of the hidden sector which left in
the effective Lagrangian.

The form of the functions $Z$ and $\tau$, as well as the one of their arguments
can be obtained by means of the classical $R$-symmetry of the Lagrangian.
Before the rescaling of Eq.\ \eqref{eq:amsbphiresc}, the tree-level Lagrangian
is
manifestly $R$-symmetric after assigning a vanishing $R$-charge to the chiral
superfields $\Phi$ and a $R$-charge of 2/3 for the spurious superfield, since
the superpotential has been chosen as a trilinear and homogeneous function of
the chiral superfields. At the
one-loop level, the situation is however highly different. Firstly, counterterms
that depend on a cutoff scale $\Lambda_{\rm UV}$
must be added to the theory so that the one-loop
ultraviolet divergences are compensated.
Secondly, the rescaling of the
matter
fields does not allow to fully eliminate the dependence in $\langle{\bf
\Phi}\rangle$ issued from the loop diagrams but only the tree-level one.
It has been found that the
latter always appears through a product with the cutoff
scale~\cite{Randall:1998uk, Kaplunovsky:1994fg, Gaillard:1998bf}, so that the
holomorphic function $\tau$ only depends on the product $\Lambda_{\rm UV}
\langle{\bf
\Phi}\rangle$, while the function $Z$ depends in addition on the conjugate
quantity $\Lambda_{\rm UV} \langle{\bf \Phi}^\dag\rangle$.
After the rescaling, all chiral superfields $\Phi$ get a $R$-charge of 2/3, as
shown by Eq.\ \eqref{eq:amsbphiresc}. Since the $R$-symmetry is (formally)
exact, the function $Z$ is enforced to depend on the spurious superfield only
through a $R$-invariant products of ${\bf \Phi}$ and ${\bf \Phi}^\dag$, 
\be
 Z^i{}_j\bigg[\frac{\mu_R}{\Lambda_{\rm UV}
   \langle{\bf \Phi} \rangle},  \frac{\mu_R}{\Lambda_{\rm UV} \langle 
   {\bf \Phi^\dag} \rangle} \bigg]  \to
 Z^i{}_j\bigg[\frac{\mu_R}{\Lambda_{\rm UV} \langle |{\bf \Phi} |\rangle}\bigg]
  \ .
\ee

Turning off the effects of the hidden sector, \ie, fixing $\langle{\bf
\Phi}\rangle$ to unity, the $R$-symmetry becomes explicitly broken in the gauge
sector, a
consequence of the well-known anomalies of the $R$-symmetry. Curing these
anomalies by means of appropriate counterterms therefore implies a shift of the
function $\tau$ that restores the $R$-symmetry. The anomalies impose in this way the
form of the
function $\tau$, which is found to be similar to Eq.\ \eqref{eq:amsbshift},
\be
  \tau_{ab}\bigg[\frac{\mu_R}{\Lambda_{\rm
    UV}\langle{\bf \Phi}\rangle}\bigg] = h_{ab} + 2 b_0 \delta_{ab}
    \log\frac{\mu_R}{\Lambda_{\rm UV}\langle{\bf \Phi}\rangle} \ .
\ee
This shows that gaugino mass term at a scale $\mu_R$ are obtained according to
renormalization group running from the cutoff scale $\Lambda_{\rm UV}
\langle{\bf \Phi}\rangle$ to $\mu_R$, the results of Eq.\
\eqref{eq:amsbshift} being retrieved after Taylor-expanding the logarithm of
the spurion superfield.

We have so far proved that super-Weyl anomalies are responsible for generating
gaugino mass
terms. Other soft supersymmetry-breaking terms can also be generated by
anomalies. To underline such effects, one proceeds with the computation of
the Taylor expansion of the 
wave function $Z$ associated with the chiral superfield $\Phi$ with respect to
$M_{\rm aux}$. Understanding superfield indices, one finds
\be\bsp
   \log Z\bigg[\frac{\mu_R}{\Lambda_{\rm UV} \langle |{\bf \Phi}| \rangle}\bigg] = &\
   \log\frac{\mu_R}{\Lambda_{\rm UV}} 
   - \frac12 \gamma(g,y) M_{\rm aux}\ \theta\!\cdot\!\theta
   - \frac12 \gamma(g,y) M_{\rm aux}^\ast\ \thetabar\!\cdot\!\thetabar
\\&\
   + \frac14 \big|M_{\rm aux}\big|^2 \Big[ \frac{\del\gamma(g,y)}{\del\beta} \beta_g
       + \frac{\del\gamma(g,y)}{\del y} \beta_y\Big]\ \theta\!\cdot\!\theta\
     \thetabar\!\cdot\!\thetabar \ ,
\esp \ee
where we have introduced the anomalous dimension of the superfield $\Phi$,
\be
  \gamma(g,y) = \frac{\del \log Z}{\del \log \mu_R} \ ,
\ee
and the beta functions
\be
  \beta_g = \frac{\del g}{\del\log\mu_R} \qquad\text{and}\qquad
  \beta_y = \frac{\del y}{\del\log\mu_R} \ .
\ee
In our notations, the quantity $y$ generically denotes any superpotential
coupling and $g$ stands for the gauge coupling constant. 
As far as mass terms are concerned, the lower-order
coefficients of the Grassmann expansion of $\log Z$ can be rescaled
away\footnote{This rescaling is also anomalous, but these new effects are of
higher-order and can thus be safely neglected.},
\be
  \exp\bigg[ \frac12 \log\frac{\mu_R}{\Lambda_{\rm UV}} 
   - \frac12 \gamma(g,y) M_{\rm aux}\ \theta\!\cdot\!\theta \bigg] \langle{\bf
     \Phi}\rangle \to \langle{\bf \Phi}\rangle \ ,
\label{eq:amsbfinalshift}\ee
so that after integration upon the Grassmann variables, the first term of the
Lagrangian of Eq.~\eqref{eq:amsbrenlag}
contains a scalar mass term, the related mass parameter
$m_\phi^2$ evaluated at
the renormalization scale $\mu_R$ reading
\be\boxed{
  m_\phi^2(\mu_R) = -\frac14 \big|M_{\rm aux}\big|^2 \bigg[
    \frac{\del\gamma}{\del\beta} \beta_g + \frac{\del\gamma}{\del y}
    \beta_y\bigg] \ .
}\label{eq:amsbscal}\ee
Scalar masses hence arise
at the two-loop level, in contrast to the gaugino masses which are
induced by one-loop effects shown in Eq.\ \eqref{eq:amsbino}, since the
leading contributions to the scalar masses are derived from the knowledge
of coefficients of the anomalous dimension and the beta
functions already computed at the one-loop order. Schematically, they can be written as
\be
  \gamma(g,y) = \gamma_g g^2 + \gamma_y y^2 \ , \qquad
  \beta_g = -b_0 g^3 \qquad\text{and}\qquad
  \beta_y = y\big(y_y y^2 + y_g g^2\big) \ ,
\ee
where $\gamma_g$ is always positive and real while the other parameters could be
positive or negative real numbers. In particular, $\gamma_y$ is negative for
infrared-free gauge
theories, so that some scalar masses can become tachyonic if no other source of
supersymmetry breaking allows for balancing the negative effects.

The shift of Eq.\ \eqref{eq:amsbfinalshift} also induces effects in the
superpotential Lagrangian, \ie, in the second term of Eq.\
\eqref{eq:amsbrenlag}. Recalling that
the superpotential only consists of trilinear interaction terms, this shift 
leads to supersymmetry-breaking trilinear scalar interactions. Denoting
generically by $y_{ijk}$ a
superpotential interaction that couples the superfields $\Phi^i$, $\Phi^j$ and
$\Phi^k$, the associated soft supersymmetry-breaking coupling strength
$A_{ijk}$ is hence given by
\be\boxed{
  A_{ijk} = \frac12 y_{ijk} \Big[\gamma_i + \gamma_j + \gamma_k\Big] M_{\rm aux}
\ ,}
\label{eq:amsbtri}\ee
where $\gamma_i$, $\gamma_j$ and $\gamma_k$ are the anomalous dimensions related
to the superfields $\Phi^i$, $\Phi^j$ and $\Phi^k$, respectively.

Collecting all the results derived above, the soft supersymmetry-breaking
Lagrangian induced by anomalies can be written as
\be\boxed{\bsp
  {\cal L}_{\rm soft} = &\
    - \frac12 m_\lambda 
      \big[ \lambda^a\!\cdot\!\lambda_a + \lambar^a\!\cdot\!\lambar_a \big] 
    - m_{\phi^i}^2\  \phi^\dag_i \phi^i  
    - \Big[\frac16 A_{ijk} \phi^i \phi^j \phi^k 
    + \hc \Big] \ ,
\esp}\ee
where the gaugino masses are given by Eq.\
\eqref{eq:amsbino}, the
scalar masses by Eq.\ \eqref{eq:amsbscal} and the trilinear couplings by
Eq.\ \eqref{eq:amsbtri}. In addition, the form of these
soft terms is renormalization-group invariant and therefore holds at any scale. This
consequently entails large predictivity for anomaly-mediated
supersymmetry-breaking scenarios, with fixed mass ratios and distinctive signatures
\cite{Gherghetta:1999sw, Feng:1999fu, Feng:1999hg, Rattazzi:1999qg,
Barr:2002ex}.

\mysection{Renormalization group equations for supersymmetry}\label{sec:rge}
In the three supersymmetry-breaking scenarios introduced in Section
\ref{sec:grmsb}, Section \ref{sec:GMSB} and Section \ref{sec:AMSB},
supersymmetry breaking occurs at a high energy-scale. In order to investigate
the related phenomenology, but at the electroweak scale, the generated soft
terms must be subsequently run down. Assuming a renormalizable
(softly-broken) supersymmetric theory and only focusing on the visible sector,
the supersymmetric pieces of the action are given as in Eq.\
\eqref{eq:gensusyaction},
\be\bsp
  {\cal S} =&\
    \int \d^4 x\ \d^2 \theta\ \d^2 \thetabar\ \Phi^\dag e^{-2gV} \Phi 
   + \int \d^4 x\ \d^2 \theta \ W(\Phi)  
   +   \int \d^4 x\ \d^2 \thetabar \ W^\star(\Phi^\dag)
\\ &\
 + \frac{1}{16 g^2}\int \d^4 x\ \d^2 \theta\ W^{a\alpha} W_{a\alpha }
 + \frac{1}{16 g^2}\int \d^4 x\ \d^2 \thetabar\  
      \Wbar^a_\alphadot \Wbar_a^\alphadot  \ ,
\esp\ee
where we have chosen the gauge kinetic function and the K\"ahler
potential in a minimal way, as given by Eq.\ \eqref{eq:normthcond}.
We recall that in our conventions, the Lagrangian above describes the dynamics
of a set of matter supermultiplets 
that are represented by chiral superfields $\Phi$. These
superfields lie in given representations of a gauge group $G$ with which we
associate the coupling constant $g$ and the vector superfield $V$. Finally, 
kinetic and gauge interaction terms for the components of the superfield $V$
are derived from the sum of the squares of the superfield strength tensors
$W_\alpha$ and $\Wbar_\alphadot$, as usual.

For a renormalizable theory, the superpotential $W$ is a gauge-invariant
function at most trilinear 
in the chiral content of the theory, as given in Eq.\ \eqref{eq:renosuperW},
\be
   W(\Phi) = \frac16 \lambda_{ijk} \Phi^i \Phi^j \Phi^k + \frac12 \mu_{ij}
     \Phi^i \Phi^j + \xi_i \Phi^i \ ,
\ee
where $\lambda$, $\mu$ and $\xi$ are free parameters of the model.

In addition to the supersymmetric Lagrangian, one
also needs to consider supersymmetry-breaking soft terms generically written as 
\be
  {\cal L}_{\rm soft} =  -\frac16 a_{ijk} \phi^i \phi^j\phi^k - \frac12 b_{ij}
    \phi^i \phi^j - c_i \phi^i - \frac12 M \lambda \!\cdot\! \lambda
    -\frac12 \phi^\dag_i (m^2)^i{}_j \phi^j+ \hc ,
\ee
where the parameters $a$, $b$, $c$ denote the interaction strengths of the
supersymmetry-breaking  
multiscalar interactions, $M$ the gaugino masses and $m^2$ the 
squared mass matrix associated with the scalar fields of the theory. In the
Lagrangian above, the scalar components of the matter
supermultiplets are denoted by $\phi$ while the gaugino component of the vector
supermultiplet $V$ is represented by $\lambda$.

These parameters are related to the supersymmetry-breaking mechanism and are in
principle known at the (high) scale where supersymmetry breaking occurs. The
derivation of the values of those parameters at the electroweak scale,
relevant for, \eg, collider phenomenology, is driven by supersymmetric 
renormalization
group equations. These equations link the low-energy parameters to their
high-energy counterparts. Although
the number of free parameters can be quite large, some organizing
principles in general hold at the supersymmetry-breaking scale so that we end up
with a few input parameters at the high scale.

The supersymmetric renormalization group equations have
been known at the one-loop level~\cite{Derendinger:1983bz, Falck:1985aa} and
two-loop
level for a long time~\cite{Martin:1993zk, Yamada:1993uh, Yamada:1993ga,
Yamada:1994id, Jack:1994rk} and are generically written as
\be
  \frac{{\rm d}}{{\rm d} t } x =  \beta_x \qquad\text{with}\qquad
    \beta_x = \frac{1}{16 \pi^2} \beta_x^{(1)}  + \frac{1}{(16 \pi^2)^2}
   \beta_x^{(2)} + \ldots \ ,
\ee
where $x$ stands for any parameter and $\beta_x$ for the associated beta
function. This function is thus calculated perturbatively,
$\beta_x^{(1)}$ and $\beta_x^{(2)}$ being the one-loop and two-loop
coefficients, respectively.

Starting with the gauge sector, the first coefficients of the
$\beta$-functions of the gauge coupling constant $g$ and of the
gaugino mass parameter $M$ read
\be\label{eq:betafuncgauge1}
  \beta_g^{(1)} =  g^3 \Big[ \tau - 3 C_G \Big] \qquad\text{and}\qquad
  \beta_M^{(1)} =  2 g^2 M \Big[ \tau - 3 C_G \Big] \ ,
\ee
whilst the two-loop coefficients are given, in the $\overline{DR}$
renormalization scheme, by
\be\bsp
 \beta_g^{(2)} = &\ g^5 \Big[ - 6 C_G^2 + 2 C_G \tau + 4 \tau_\R C_\R \Big] 
      - g^3 \lambda^{\ast ijk} \lambda_{ijk} C_k/d_G \ , \\
 \beta_M^{(2)} = &\ 4 g^4 M \Big[  - 6 C_G^2 + 2 C_G \tau + 4 \tau_\R C_\R \Big]
   + 2 g^2 \lambda_{ijk} \Big[ a^{\ast ijk}  - M \lambda^{\ast ijk}\Big] C_k/d_G \
  .
\esp\label{eq:betafuncgauge2}\ee
Considering a chiral superfield lying in a representation $\R$ of the gauge
group, we denote by $C_\R$ the quadratic Casimir invariant associated with this
representation. Similarly, $C_G$ is the quadratic Casimir invariant
relative to the adjoint representation and $C_k$ corresponds to the Casimir
invariant associated with the representation in which the chiral superfield
$\Phi^k$ lies. In addition, we have also introduced $\tau$ as the
total Dynkin index of the gauge group, \ie, the Dynkin index summed over
all the chiral content of the model,
accounting for the superfield multiplicity\footnote{In
the case of an abelian group, we define $\tau$ as the sum of the squared charges
over the whole chiral content of the theory, accounting for the multiplicity
of each superfield.}. Moreover,
the quantity $C_\R \tau_\R$ contains an implicit summation. 
It refers to a sum of the Dynkin
indices over all the chiral superfields of the model, weighted by the
corresponding quadratic Casimir invariants. Finally, the dimension of the gauge
group is denoted by $d_G$. 

On the basis of the supersymmetric non-renormalization theorems
\cite{Wess:1973kz, Salam:1974jj, Iliopoulos:1974zv,Ferrara:1974fv,
Zumino:1974bg, Ferrara:1975ye, Grisaru:1979wc}, the evolution of
the superpotential parameters is entirely driven by the anomalous dimensions of
the fields (see Section \ref{sec:AMSB}). The first and second coefficients of the
$\beta$-functions of the
linear, bilinear and trilinear interaction parameters $\xi$, $\mu$ and $\lambda$
are then given by
\be\bsp
  \beta_{\xi_i}^{(n)} =&\ \xi_p\ (\gamma^{(n)})^p{}_i \ , \\ 
  \beta_{\mu_{ij}}^{(n)} =&\ \mu_{ip}\ (\gamma^{(n)})^p{}_j + 
    \mu_{pj}\ (\gamma^{(n)})^p{}_i  \ , \\ 
  \beta_{\lambda_{ijk}}^{(n)} =&\ \lambda_{ijp}\ (\gamma^{(n)})^p{}_k + 
    \lambda_{ipk}\ (\gamma^{(n)})^p{}_j + \lambda_{pjk}\ (\gamma^{(n)})^p{}_i  \
   ,
\esp\ee
respectively, with $n=1$ or 2. The
first two coefficients of the anomalous dimensions $\gamma$ appearing in these
beta functions are given by
\be\bsp
  (\gamma^{(1)})^j{}_i =&\  \frac12 \lambda_{ipq} \lambda^{\ast jpq} 
    - 2 \delta^j{}_i g^2 C_i \ , \\
  (\gamma^{(2)})^j{}_i = &\ g^2 \lambda_{ipq} \lambda^{\ast jpq} \Big[2C_p \!-\!
    C_i\Big] \!-\! \frac12 \lambda_{imn} \lambda^{\ast npq} \lambda_{pqr}
    \lambda^{\ast mrj} \!+\! 2 \delta^j{}_i g^4 C_i \Big[ \tau \!+\! 2 C_i \!-\! 3
    C_G\Big] \ .
\esp\ee

We now turn to the evolution of the supersymmetry-breaking parameters. The
first coefficients of the beta functions related to the linear, bilinear and
trilinear scalar interactions read
\be\bsp
  \beta_{c_i}^{(1)} =&\ \frac12 \lambda_{ipq} \lambda^{\ast rpq} c_r \!+\! 
    \lambda^{\ast rpq} a_{ipq} \xi_r \!+\! \mu_{ir} \lambda^{\ast rpq} b_{pq}
    \!+\! 2 \lambda_{ipq} (m^2)^q{}_r \mu^{\ast pr} \!+\! a_{ipq} b^{\ast pq} \
    , \\
  \beta_{b_{ij}}^{(1)} =&\ \frac12 b_{ip} \lambda^{\ast pqr} \lambda_{qrj} \!+\! 
    \frac12 \lambda_{ijr} \lambda^{\ast rpq} b_{pq} \!+\! \mu_{ip} \lambda^{\ast
    pqr} a_{qrj} \!-\! 2 g^2 C_i \big[ b_{ij} \!-\! 2 M \mu_{ij} \big] \\ 
  &\ + \frac12 b_{jp} \lambda^{\ast pqr} \lambda_{qri} \!+\! 
    \frac12 \lambda_{jir} \lambda^{\ast rpq} b_{pq} \!+\! \mu_{jp} \lambda^{\ast
    pqr} a_{qri} \!-\! 2 g^2 C_j \big[ b_{ji} \!-\! 2 M \mu_{ji} \big]\ , \\ 
  \beta_{a_{ijk}}^{(1)} =&\ \frac12 a_{ijr} \lambda^{\ast rpq} \lambda_{pqk}
    \!+\! \lambda_{ijr} \lambda^{\ast rpq} a_{pqk} \!-\! 2 g^2 C_k 
    \big[ a_{ijk} \!-\! 2
    M \lambda_{ijk}\big] 
\\&\ + \frac12 a_{kjr} \lambda^{\ast rpq} \lambda_{pqi}
    \!+\! \lambda_{kjr} \lambda^{\ast rpq} a_{pqi} \!-\! 2 g^2 C_i 
    \big[ a_{kji} \!-\! 2
    M \lambda_{kji}\big]
\\ &\ + \frac12 a_{ikr} \lambda^{\ast rpq} \lambda_{pqj}
    \!+\! \lambda_{ikr} \lambda^{\ast rpq} a_{pqj} \!-\! 2 g^2 C_j 
    \big[ a_{ikj} \!-\! 2
    M \lambda_{ikj}\big]\ ,
\esp\ee
while the second coefficients of these beta functions are  given by
\be\bsp
  \beta_{c_i}^{(2)} = &\
   4 g^2 C_q\Big[ 
   \big(a_{ipq} \!-\! \lambda_{ipq} M\big)
     \big( \lambda^{\ast rpq} \xi_r \!+\! b^{\ast pq}\big)
   \!+\! \frac12 \lambda_{ipq} \lambda^{\ast rpq} c_r 
   \!-\! \lambda^{\ast rpq} \big(\mu_{pq} M \!-\! b_{pq}\big) \mu_{ir} 
\\ &\
   + \big( 
        \lambda_{ipq}(m^2)^p{}_r
      \!+\! \lambda_{irp}(m^2)^p{}_q
      \!+\! 2\lambda_{irq} |M|^2 
      \!-\! a_{irq} M^*
    \big)\mu^{\ast rq}
     \Big]
\\ &\
   - \Big[ \lambda_{ipq} a_{trs} + a_{ipq} \lambda_{trs}\Big] \lambda^{\ast qrs}
       \lambda^{\ast kpt}\xi_k
   - \frac12 \lambda_{ipq} \lambda^{\ast qrs} \lambda_{trs}\lambda^{\ast kpt}
      c_k
\\ &\
   - \Big[ a_{qst}\mu_{pk} + \lambda_{qst} b_{pk} \Big] \lambda^{\ast
       kst}\lambda^{\ast rpq}\mu_{ir}  
   - \Big[\lambda_{ipq} a_{rst} + a_{ipq} \lambda_{rst} \Big]
     \lambda^{\ast qst} b^{\ast pr}
\\&\
   -\! \Big[ 
       \big(\lambda_{ikq} (m^2)^k{}_p \lambda_{rst} 
     \!+\! \lambda_{ipq} (m^2)^k{}_r  \lambda_{kst}
     \!+\! 2\lambda_{ipq} (m^2)^t{}_k \lambda_{rst} 
     \!+\! \lambda_{ipk}(m^2)^k{}_q \lambda_{rst}\big) 
        \lambda^{\ast qst}
\\&\
     +\! \big(\lambda_{ipq} a_{rst} - a_{ipq} \lambda_{rst} \big) a^{\ast qst}
  \Big]\mu^{\ast pr} \ ,
\\
  \beta_{b_{ij}}^{(2)} =&\
    g^2 \lambda^{\ast rpq} \Big\{ 
    2 C_p \lambda_{ijr} \big(b_{pq} \!-\! \mu_{pq} M\big)
    \!+\! \big(2 C_p \!-\! C_i\big) 
      \Big[ b_{ir} \lambda_{pqj} \!+\! 2 \mu_{ir} \big(a_{pqj} \!-\! 
    \lambda_{pqj}M \big) \Big] \Big\}
\\&\
    + 2 g^4 \big(b_{ij} \!-\! 4 \mu_{ij} M\big) \big(C_i \tau \!+\! 2 C_i^2 \!-\! 
      3 C_G C_i \big)
    \!-\! \frac12 \Big[ b_{ip} \lambda_{qkj} \!+\! b_{qk} \lambda_{ijp} \Big] 
        \lambda_{str} \lambda^{\ast pqr} \lambda^{\ast stk}
\\&\
 -\! \Big[\frac12 \lambda_{ijs} \mu_{tr} a_{pqk}
    \!+\! \mu_{is} \big(a_{kpq} \lambda_{trj} \!+\! \lambda_{kpq} a_{trj}\big)
  \Big]\lambda^{\ast pqr}  \lambda^{\ast stk}
 + (i \leftrightarrow j)   \ ,
\\
  \beta_{a_{ijk}}^{(2)} =&\
    - \lambda^{\ast smn} \lambda^{\ast pqr} \Big[ 
        \frac12 a_{ijs} \lambda_{npq} \lambda_{mrk}  
        + \lambda_{ijs} \big(\lambda_{npq}a_{mrk} + a_{npq}\lambda_{mrk}\big) 
    \Big]
\\ &\ 
    +  g^2 \big(2 C_p - C_k \big) \lambda^{\ast rpq} \Big[ 
      a_{ijr} \lambda_{pqk} + 2 \lambda_{ijr} \big( a_{pqk} - M
      \lambda_{pqk}\big) \Big]
\\ &\ 
    + 2 g^4 \big(a_{ijk} - 4 M \lambda_{ijk} \big)
     \big( C_k \tau + 2 C_k^2  - 3 C_G C_k\big)
    + (k \leftrightarrow i) + (k \leftrightarrow j) \ .
\esp\ee
Finally, we achieve this section by providing the first two coefficients of the
beta function associated to the scalar mass parameters,
\be\bsp
  \beta^{(1)}_{ (m^2)^i{}_j} =&\ \frac12 \lambda^{\ast ipq} \lambda_{pqr} 
    (m^2)^r{}_j + \frac12 \lambda_{jpq} \lambda^{\ast pqr} (m^2)^i{}_r + 
    2 \lambda^{\ast ipq} \lambda_{jpr}
    (m^2)^r{}_q + a_{jpq} a^{\ast ipq}\\ &\ - 8 \delta^i{}_j M M^\dag g^2 C_i + 2
     g^2 (T_a)^i{}_j {\rm Tr}\big[ T^a m^2 \big] \ , 
\esp\ee
and
\be\bsp
  \beta^{(2)}_{ (m^2)^i{}_j}=  &\
    \delta^i{}_j g^4 \bigg[ 
        M M^\dag \Big(24 \tau C_i \!+\! 48 C_i^2 \!-\! 72 C_G C_i \Big)
       + 8 C_i \Big({\rm Tr}\big[\tau_\R m^2\big] \!-\! C_G M M^\dag \Big)
   \bigg] 
\\&\
   - 2 g^2 (T_a)^i{}_j \big( T^a m^2 \big)^r{}_s \lambda^{\ast spq} \lambda_{rpq}
   + 8 g^4 (T_a)^i{}_j {\rm Tr}\big[ T^a C_\R m^2 \big] 
\\ &\ 
 + g^2 \big(C_p \!+\! C_q \!-\! C_i \big) \Big[ 
    (m^2)^i{}_s \lambda^{\ast spq} \lambda_{jpq}
   \!+\! \lambda^{\ast ipq} \lambda_{spq} (m^2)^s{}_j 
   \!+\! 4 \lambda^{\ast ipq} \lambda_{jps} (m^2)^s{}_q
\\&\
   + 2 a^{\ast ipq} \big( a_{jpq} - \lambda_{jpq} M\big) 
   - 2 \lambda^{\ast ipq} \big( a_{jpq} M^\dag
   - 2 \lambda_{jpq} M M^\dag\big)
  \Big]
\\&\
    - \lambda^{\ast ism} \bigg[
          \lambda_{jnm}\Big( (m^2)^r{}_s \lambda^{\ast npq} \lambda_{rpq}
         \!+\! (m^2)^n{}_r \lambda^{\ast rpq} \lambda_{spq} \Big)
      \!+\! \lambda_{jnr} (m^2)^n{}_s \lambda^{\ast pqr} \lambda_{pqm}
\\&\
      + 2 \lambda_{jsn}  (m^2)^r{}_q \lambda^{\ast npq} \lambda_{mpr}
    \bigg]
   \!-\! \Big[ a^{\ast ism} \lambda^{\ast npq} \!+\! \lambda_{jsn} a_{mpq} \Big]
     \Big[ a_{jsn} \lambda_{mpq} \!+\! \lambda_{jsn} a_{mpq} \Big]
\\ &\
    -\frac12  \lambda_{pqn} \lambda^{\ast pqr} \Big[
        (m^2)^i{}_s \lambda^{\ast smn} \lambda_{mrj} 
      + (m^2)^s{}_j \lambda_{smr} \lambda^{\ast mni}\Big]  
  \  .
\esp\ee
The terms depending explicitly on the
representation matrices of the gauge group vanish identically for non-abelian
groups. We recall that in the abelian case, these matrices must be read
as squared abelian charges. Moreover, the Casimir invariant $C_\R$ and  Dynkin
index $\tau_\R$ appearing in
some of these terms is associated with the representation of the relevant fields
in the traces.

\cleardoublepage

%% file: mssm.tex
\label{chap:mssm}

In Chapter~\ref{chap:susy} and Chapter~\ref{sec:susybrk}, we have shown how
to construct any softly-broken supersymmetric field theory from very basic
principles. In this chapter, we apply all the concepts that have been introduced
so far in order to build the simplest phenomenologically relevant
supersymmetric model, namely
the Minimal Supersymmetric Standard Model or MSSM. We start with a detailed
description of the construction of the
model itself in Section~\ref{sec:mssmbuilding}, including
details on the most popular choices for breaking
supersymmetry in that framework and then detail
the main features of the model in Section~\ref{sec:mssmpheno}. One of the drawbacks of the
MSSM, similarly to large classes of beyond the Standard Model theories,
consists of its very large parameter space and the subsequent difficulties in
designing non-experimentally excluded benchmark scenarios which implies to account
for strong constraints from
many experimental data.
In this prospect, we study in Section~\ref{sec:mssm_indirect}
several low-energy electroweak and flavor observables in the context of the MSSM
and dedicate Section~\ref{sec:mssm_cosmo}
to the investigation of its cosmological aspects. We next address
the more recent direct constraints extracted from the Large Hadron Collider data
in Section~\ref{sec:direct}, including the observation
of a 125~GeV state compatible with the Standard Model Higgs boson~\cite{Aad:2012gk, Chatrchyan:2012gu}
and finally shortly motivate the needs to go beyond the
MSSM in Section~\ref{sec:eqbyd}.

\mysection{Construction of the model}
\label{sec:mssmbuilding}
\subsection{Field content}\label{sec:mssmfields}
The Minimal Supersymmetric Standard Model is the simplest supersymmetric model
extending the Standard Model of particle physics \cite{Nilles:1983ge,
Haber:1984rc}. It results from the straightforward supersymmetrization of the
Standard Model with the same gauge interactions, based on the semi-simple gauge
group $SU(3)_c \times SU(2)_L \times U(1)_Y$. Because of their quantum numbers,
the Standard Model particles cannot be gathered into $N=1$ representations of
the Poincar\'e superalgebra (see Section \ref{sec:masslesssupermul}). It is
consequently necessary to embed each of the Standard Model degrees of freedom
within a chiral or vector supermultiplet, supplementing it with one new state.

\renewcommand{\arraystretch}{1.2}
\begin{table}[!t]
  \begin{center}
  \begin{tabular}{|c||c|c|c|c|}
    \hline
    Supermultiplet & \begin{tabular}{c}Standard Model\\ fermion\end{tabular} & 
     Superpartner & Representation \\ \hline 
    \multirow{4}{*}{$Q_L^i$}&&&\\
       &$q_L^i = \bpm u_L^i \\ d_L^i \epm$ & 
       $\tilde q^i_L = \bpm \tilde u^i_L\\  \tilde d^i_L\epm$ &
       $({\utilde {\bf 3}}, {\utilde{\bf 2}}, \frac16)$ \\&&&\\
    $U^i_R$ & 
       $u_R^{ic}$ & 
       $\tilde u_R^{i\dag}$ & 
       $({\utilde{\bf \bar 3}},{\utilde{\bf 1}}, -\frac23)$\\
    $D_R^i$ &
       $d_R^{ic}$ & 
       $\tilde d_R^{i\dag}$ & 
       $({\utilde{\bf \bar 3}},{\utilde{\bf 1}}, \frac13)$\\&&&\\
    \hline
    \multirow{4}{*}{$L_L^i$}&&&\\
       &$\ell_L^i = \bpm \nu_L^i \\ e_L^i\epm$ & 
       $\tilde \ell_L^i = \bpm\tilde \nu^i_L\\  \tilde e^i_L\epm$ &
       $({\utilde{\bf 1}}, {\utilde{\bf 2}}, -\frac12)$\\&&&\\
    $E_R^i$&
       $e_R^{ic}$ & 
       $\tilde e_R^{i\dag}$ & 
       $({\utilde{\bf 1}},{\utilde{\bf 1}}, 1)$\\
    $N_R^i$&
       $\nu_R^{ic}$ & 
       $\tilde \nu_R^{i\dag}$ & 
       $({\utilde{ \bf 1}},{\utilde{\bf 1}}, 0)$\\&&&\\
    \hline 
  \end{tabular}
  \end{center} 
  \caption{\label{tab:matter}The MSSM matter sector resulting from
    the supersymmetrization of the Standard Model quark and lepton fields. The 
    representations under $SU(3)_c \times SU(2)_L \times U(1)_Y$ are 
    provided in the last column of the table and the superscript $c$ denotes
    charge conjugation. }
\end{table}
\renewcommand{\arraystretch}{1}

The matter sector of the theory\footnote{By the terminology \textit{matter 
sector}, we refer to
quark and lepton supermultiplets.} consists of  three
generations of six chiral 
supermultiplets containing the Standard Model quarks and leptons, together with
their squark and slepton superpartners. The corresponding superfields and 
their representations under the MSSM gauge group read 
\be\label{eq:SFnames}\bsp
 &\ Q_L^i = ({\utilde {\bf 3}}, {\utilde{\bf 2}}, \phantom{-}\frac16) \quad  ,\quad
    U_R^i = ({\utilde {\bf \bar 3}}, {\utilde{\bf 1}},-\frac23) \quad ,\quad
    D_R^i = ({\utilde {\bf \bar 3}}, {\utilde{\bf 1}}, \frac13) \quad , \quad \\
 &\ L_L^i = ({\utilde {\bf 1}}, {\utilde{\bf 2}},-\frac12) \quad , \quad
    E_R^i = ({\utilde {\bf 1}}, {\utilde{\bf 1}},\phantom{-} 1) \quad , \quad 
    N_R^i = ({\utilde {\bf 1}}, {\utilde{\bf 1}}, 0) \quad ,
\esp\ee
where $i=1,2,3$ stands for a generation index and where the
right-handed neutrino superfields $N_R$ are included for completeness. 
In our notations, we have
employed the hypercharge quantum numbers as `representations' for the
$U(1)_Y$ subgroup. The physical component
fields embedded into these superfields are detailed in Table
\ref{tab:matter}. Usually,
four-component spinorial representations of the Poincar\'e algebra are
employed to describe (massive) quarks and leptons. Denoting such a
four-component spinor $\Psi$ by
\be
  \Psi = \bpm \chi_\alpha\\ \xibar^\alphadot \epm \ ,
\ee
allows us to put an emphasis on its two-component fermionic content. This 
field combines
a left-handed Weyl fermion $\chi$ and a right-handed Weyl fermion $\xibar$.
However, only left-handed Weyl fermions 
can be employed to construct supersymmetric theories. 
This issue is cured by means of the
charge conjugate Dirac field $\Psi^c$,
\be
  \Psi^c = C \Psibar^t = \bpm \xi_\alpha\\ \chibar^\alphadot \epm \ ,
\ee
where $C$ is the charge-conjugation operator.
To supersymmetrize a non-supersymmetric
theory with Dirac fermions by employing left-handed chiral superfields,
the left-handed components of both the Dirac fields and
their charge-conjugate counterparts are hence employed. In the example of the MSSM, 
it means that the superfields lying in the fundamental representation
$\utilde{\bf 2}$ of $SU(2)_L$ ($Q_L$ and $L_L$) are built upon the left-handed
component of the Standard
Model quarks and leptons while those being singlet of $SU(2)_L$ ($U_R$, $D_R$, $E_R$
and $N_R$) are based on the left-handed component of the conjugate fields. As
shown in Table \ref{tab:matter}, the scalar components of the chiral
supermultiplets are introduced accordingly, employing
conjugate fields where relevant.

In contrast to the Standard Model, the Higgs sector of the MSSM contains two
chiral supermultiplets $H_D$  and $H_U$. Since the
superpotential is an holomorphic function of the chiral superfields (see
Section \ref{sec:Lchiral}), two $SU(2)_L$ doublets are necessary to give
mass to both up-type and down-type particles.
Moreover, two supermultiplets with opposite hypercharge quantum numbers are necessary 
to cancel the chiral anomalies resulting from the fermionic components
of $H_U$ and $H_D$, dubbed higgsinos. A good choice for their
representation under the MSSM gauge group is thus 
\be\bsp
  H_D = ({\utilde{\bf 1}}, {\utilde{\bf 2}}, -\frac12)\quad ,\quad
  H_U = ({\utilde{\bf 1}}, {\utilde{\bf 2}},  \frac12) \ ,
\esp\ee
where in this way, the superfield
$H_U$ couples to up-type particles whilst $H_D$ couples to down-type particles.
The component fields are collected in Table \ref{tab:higgs}.

\begin{table}[!t]
  \begin{center}
  \begin{tabular}{|c||c|c|c|c|}
    \hline
    Supermultiplets & Scalar fields & Higgsino fields & Representation \\ \hline 
    \hline 
    \multirow{4}{*}{$H_D$}&&&\\
       &$H_d = \bpm H_d^0 \\ H_d^- \epm$&
        $\widetilde H_d = \bpm \widetilde H_d^0 \\ \widetilde H_d^- \epm$&
        $({\utilde{\bf 1}}, {\utilde{\bf 2}}, -\frac12)$ \\&&&\\
    \multirow{4}{*}{$H_U$}&&&\\
       &$H_u = \bpm H_u^+ \\ H_u^0 \epm$ & 
        $\widetilde H_u = \bpm \widetilde H_u^+ \\ \widetilde H_u^0\epm$&
        $({\utilde{\bf 1}}, {\utilde{\bf 2}}, \frac12)$ \\&&&\\
\hline
\end{tabular}
  \end{center} 
  \caption{\label{tab:higgs}The Higgs sector of the Minimal
    Supersymmetric Standard Model. The representations under the MSSM gauge
    group $SU(3)_c \times SU(2)_L \times U(1)_Y$ are indicated, together with the
    components of the two Higgs supermultiplets.}
\end{table}

\renewcommand{\arraystretch}{1.2}
\begin{table}[!t]
  \begin{center}
  \begin{tabular}{|c||c|c|c|c|}
    \hline
    Supermultiplet & Gauge boson & Gaugino field & Representation \\ \hline 
    \hline 
    $V_B$ & $B_\mu$ & $\widetilde B$ & $({\utilde{\bf 1}}, {\utilde{\bf 1}}, 0)$ \\
    $V_W$ & $W_\mu$ & $\widetilde W$ & $({\utilde{\bf 1}}, {\utilde{\bf 3}}, 0)$ \\
    $V_G$ & $g_\mu$ & $\widetilde g$ & $({\utilde{\bf 8}}, {\utilde{\bf 1}}, 0)$ \\
    \hline
  \end{tabular}
  \end{center} 
  \caption{\label{tab:gauge}The gauge sector of the Minimal Supersymmetric
    Standard Model.  The representations under the MSSM gauge  group $SU(3)_c
    \times SU(2)_L \times U(1)_Y$ are indicated, together with the components
    fields of the gauge supermultiplets.}
\end{table}
\renewcommand{\arraystretch}{1}

We finally turn to the gauge sector of the model which contains one vector
superfield for each of the direct factors of the gauge group. These superfields
lie in the corresponding adjoint representation and are singlets under all
the other gauge symmetries,
\be \bsp
SU(3)_c \leftrightarrow &\ V_G = ({\utilde{\bf 8}}, {\utilde{\bf 1}}, 0)\ ,\\
SU(2)_L \leftrightarrow &\ V_W = ({\utilde{\bf 1}}, {\utilde{\bf 3}}, 0)\ ,\\
 U(1)_Y \leftrightarrow &\ V_B = ({\utilde{\bf 1}}, {\utilde{\bf 1}}, 0)\ .
\esp\ee
As shown in Table \ref{tab:gauge}, these supermultiplets contain, in addition to
the Standard Model gauge bosons, their fermionic partners dubbed gauginos.

\subsection{Supersymmetry-conserving Lagrangian}\label{sec:mssmsusylag}
As shown in Eq.\ \eqref{eq:gensusyaction}, kinetic and gauge interaction terms
for the chiral and vector superfields of the theory are entirely fixed by gauge
invariance and supersymmetry. For the MSSM vector superfield content of 
Table \ref{tab:gauge}, they read 
\be\bsp
  \lag_{\rm vector} = &\ \bigg[
    \frac14 W_B^\alpha W_{B\alpha} +
    \frac{1}{16 g_w^2} W_{Wk}^\alpha W^k_{W\alpha}  + 
    \frac{1}{16 g_s^2} W_{Ga}^\alpha W^a_{G\alpha} 
   \bigg]_{\theta\cdot\theta} + \hc \ ,
\esp\label{eq:lagmssmvec}\ee
where we recall that the notation $[\ .\ ]_{\theta\cdot\theta}$ indicates
that only the $\theta\cdot\theta$-component of the expansion of the
superfield lying inside the squared brackets must be kept. The Lagrangian
$\lag_{\rm vector}$ depends, as presented in Chapter \ref{chap:susy}, on the
superfield strength tensors defined by Eq.~\eqref{eq:Wabelian} and
Eq.~\eqref{eq:defWWbar}, 
\be\bsp
  W_{B\alpha} =&\ -\frac14 \Dbar \!\cdot\! \Dbar D_\alpha V_B \ , \\
  W_{W\alpha} =&\ -\frac14 \Dbar \!\cdot\! \Dbar e^{2 g_w V_W} D_\alpha e^{-2 g_w
    V_W}  \ , \\
  W_{G\alpha} =&\ -\frac14 \Dbar \!\cdot\!\Dbar e^{2 g_s V_G} D_\alpha e^{-2 g_s
    V_G} \ .
\esp\label{eq:mssmsfs}\ee
In these expressions, we have contracted the adjoint indices of the vector
superfields associated with $SU(2)_L$ and $SU(3)_c$. with the 
fundamental representation matrices $\frac12 \sigma_k$ and $T_a$ of these two
groups, so that we define $V_W = V_W^k \frac12 \sigma_k$ and 
$V_G = V_G^a T_a$. Moreover, we have also introduced the gauge coupling
constants $g_w$ and $g_s$ and for further references, we denote the hypercharge
coupling constant by $g_y$. In Eq.\ \eqref{eq:mssmsfs}, we hence employ the superfields
\be
   W_{W\alpha} = W_{W\alpha}^k \frac12\sigma_k \qquad\text{and}\qquad
   W_{G\alpha} = W_{G\alpha}^a T_a \ ,
\ee
that can be used to further extract the quantities 
included in the Lagrangian of Eq.\ \eqref{eq:lagmssmvec}.

Gauge interaction and kinetic terms for the chiral superfields of 
Table \ref{tab:matter} and Table \ref{tab:higgs} are given, according to Eq.\
\eqref{eq:gensusyaction}, by
\be\label{eq:lagmssmchir}\bsp
 \lag_{\rm chiral} = &\ \bigg[
    Q_L^\dag \Big(e^{-\frac13 g_y V_B} e^{-2 g_w V_W} e^{-2 g_s V_G}\Big)
      Q_L  + 
    U_R^\dag \Big(e^{ \frac43 g_y V_B} e^{-2 g_s V^\prime_G} \Big) U_R  +
\\ &\quad
    D_R^\dag \Big(e^{-\frac23 g_y V_B} e^{-2 g_s V^\prime_G} \Big) D_R  + 
    L_L^\dag \Big(e^{ g_y V_B} e^{-2 g_w V_W} \Big) L_L   +  
    E_R^\dag \Big(e^{-2 g_y V_B} \Big) E_R   +
\\&\quad
    N_R^\dag N_R  \ + 
    H_D^\dag \Big(e^{ g_y V_B} e^{-2 g_w V_W} \Big) H_D + 
    H_U^\dag \Big(e^{-g_y V_B} e^{-2 g_w V_W} \Big) H_U 
  \bigg]_{\theta\cdot\theta \bar\theta\cdot\bar\theta}  \ ,
\esp\ee
where all indices are understood and where the notation $[\ .\
]_{\theta\cdot\theta\bar\theta\cdot\bar\theta}$ indicates that 
only the $\theta\!\cdot\!\theta\bar\theta\!\cdot\!\bar\theta$-component of the
expansion of the superfield lying inside the squared brackets has to be kept.
Concerning the $SU(3)_c$ exponential factors, we have introduced the
antifundamental representation matrices $\bar T_a$ within $V_G^\prime =
V_G^a \bar T_a = - V_G^a T_a^t$ since the superfields $U_R$ and $D_R$ 
lie in the ${\utilde{\bf \bar 3}}$ representation of the QCD gauge group.

Superpotential interactions contain the
Yukawa couplings generating quark and lepton masses. According to
the quantum numbers of the MSSM chiral superfields (see Table
\ref{tab:matter} and Table \ref{tab:higgs}), additional terms can be added 
and the most general superpotential is written as 
\be \bsp
  W_{\rm MSSM} = &\ 
    ({\bf y^u})_{ij}  \  U_R^i\ Q_L^j \!\cdot\! H_U - 
    ({\bf y^d})_{ij}  \  D_R^i\ Q_L^j \!\cdot\! H_D +
    ({\bf y^\nu})_{ij}\  N_R^i\ L_L^j \!\cdot\! H_U  \ - 
\\ &\ 
    ({\bf y^e})_{ij}\   E_R^i\ L_L^j \!\cdot\! H_D +
    \mu\ H_U \!\cdot\! H_D + 
    ({\bf m^\nu})_{ij}\ N_R^i N_R^j + W_{RPV}\ , 
\esp \label{eq:wmssm}\ee
where ${\bf y^u}$, ${\bf y^d}$, ${\bf y^\nu}$  and ${\bf y^l}$ denote the
$3\times3$ Yukawa matrices in flavor space, $\mu$ the Higgs off-diagonal
mass-mixing parameter and ${\bf m^\nu}$ the $3\times 3$ (right-handed) neutrino
mass matrix. The dot products stand for $SU(2)$ invariant
products, defined, for two generic chiral superfields $\Phi$ and $\Phi'$, by
\be
   \Phi \!\cdot\! \Phi' = \e_{k\ell}\ \Phi^k \Phi^{'\ell} \ ,
\ee
where $k$ and $\ell$ are fundamental $SU(2)_L$ indices and $\e_{12} = 
-\e^{12} = 1$. In the last term of the superpotential of Eq.\
\eqref{eq:wmssm}, we have included the so-called $R$-parity violating interactions,
\be \bsp
  W_{RPV} = 
      \frac12 \lambda_{ijk} L_L^i \!\cdot\! L_L^j E_R^k 
    + \lambda^\prime_{ijk} L_L^i \!\cdot Q_L^j D_R^k 
    + \frac12 \lambda^{\prime\prime}_{ijk} U_R^i D^j_R D_R^k
    - \kappa_i L_L^i \!\cdot\! H_U \ .
\esp\label{eq:wrpv}\ee  
The lepton-Higgs mixing parameter $\kappa$ is a three-dimensional vector in
generation space, while the Yukawa-like parameters $\lambda$, $\lambda^\prime$
and $\lambda^{\prime\prime}$ are $3 \times 3 \times 3$ tensors of 
this space. These interactions explicitly violate either the lepton number
$L$ or the baryon number $B$ and lead to disastrous phenomenological
consequences. Therefore, it is desirable to forbid them. This is achieved 
by imposing a discrete symmetry,
dubbed $R$-parity \cite{Farrar:1978xj}, defined by 
\be 
   R =  (-1)^{3B+L} \qquad\text{and}\qquad 
   R =  (-1)^{3B+L+2S} \ ,
\label{eq:rpdef}\ee 
at the superfield and component field level, respectively, 
$S$ standing for the spin. Following standard
conventions for constructing the \textit{minimal} version of a supersymmetric
theory based upon the Standard Model, $R$-parity
conservation is imposed so that the interactions induced by $W_{RPV}$ are not allowed. 
We will come back to $R$-parity violation in
Section~\ref{sec:rpv}.

Still in the aim of constructing a minimal model, the right-handed neutrino
is assumed decoupled so that the MSSM superpotential finally reduces to 
\be \boxed{
  W_{\rm MSSM} = 
    ({\bf y^u})_{ij}   U_R^i\ Q_L^j \!\cdot\! H_U - 
    ({\bf y^d})_{ij}   D_R^i\ Q_L^j \!\cdot\! H_D +
    ({\bf y^e})_{ij}   E_R^i\ L_L^j \!\cdot\! H_D +
    \mu H_U \!\cdot\! H_D \ . 
\label{eq:wmssm2}}\ee
The corresponding interaction terms are obtained by extracting the
$\theta\!\cdot\!\theta$-component of this quantity, 
\be
  \lag_W = \Big[W_{MSSM}\Big]_{\theta\cdot\theta} + \hc \ .
\label{eq:wmssm2l}\ee

Collecting the results of Eq.\ \eqref{eq:lagmssmvec}, Eq.\
\eqref{eq:lagmssmchir}, Eq.\ \eqref{eq:wmssm2} and Eq.\
\eqref{eq:wmssm2l}, the supersymmetric part of the MSSM Lagrangian
is summarized by 
\be
  \lag_{\rm MSSM,susy}  =  \lag_{\rm vector} + \lag_{\rm chiral} + \lag_W \ .
\label{eq:mssmlag}\ee
The expansion of this Lagrangian in terms of the component fields being
straightforward, we refer to Eq.\ \eqref{eq:gensusylag2} without
providing any further detail.

\subsection{Supersymmetry-breaking Lagrangian}
Realistic supersymmetric theories are
theories where supersymmetry is spontaneously broken. The Lagrangian density
therefore respects supersymmetry invariance, but the vacuum state does not.
As mentioned in Chapter \ref{sec:susybrk}, supersymmetry breaking is assumed
to occur at some high-energy scale. It is then mediated to the visible
sector of the model by some mechanisms which generate soft mass and interaction
terms. The specific way in which those mechanisms work is however not addressed in this 
section (see Chapter \ref{sec:susybrk}), and we rather choose to adopt a
phenomenological point of view at low energy and supplement to the
Lagrangian of Eq.\ \eqref{eq:mssmlag} all
possible soft terms breaking supersymmetry explicitly \cite{Girardello:1981wz}, 
\be \boxed{\bsp 
   \lag_{\rm soft} =
   &\ \frac12 \Big[ 
     M_1 \widetilde B \!\cdot\! \widetilde B + 
     M_2 \widetilde W \!\cdot\! \widetilde W + 
     M_3 \widetilde g \!\cdot\! \widetilde g + 
     \hc \Big]
    - ({\bf m^2_{\tilde Q}})^i{}_j \tilde q_{Li}^\dag \tilde q_L^j 
      - ({\bf m^2_{\tilde U}})^i{}_j \tilde u_{Ri} \tilde u_R^{j\dag} 
\\&\
      - ({\bf m^2_{\tilde D}})^i{}_j \tilde d_{Ri} \tilde d_R^{j\dag} 
      - ({\bf m^2_{\tilde L}})^i{}_j \tilde \ell_{Li}^\dag \tilde \ell_L^j 
      - ({\bf m^2_{\tilde E}})^i{}_j \tilde e_{Ri} \tilde e_R^{j\dag} 
      - m_{H_u}^2 H_u^\dag H_u  
      - m_{H_d}^2 H_d^\dag H_d\\
   &\ - \Big[  ({\bf T^u})_{ij} \tilde u_R^{i\dag} \tilde q_L^j \!\cdot\! H_u 
           - ({\bf T^d})_{ij} \tilde d_R^{i\dag} \tilde q_L^j \!\cdot\! H_d 
           - ({\bf T^e})_{ij} \tilde e_R^{i\dag} \tilde \ell_L^j \!\cdot\! H_d
           + b H_u \!\cdot\! H_d  + \hc \Big]\ .
\esp \label{eq:lmssmbrk}}\ee
The first bracket of Eq. \eqref{eq:lmssmbrk} contains gaugino mass terms. The
sign of these terms may seem \textit{a priori} surprising. However, a phase
is further absorbed in field redefinitions (see Section~\ref{sec:mssmewsb})
so that one eventually gets mass terms with the correct sign. 
The next seven terms of $\lag_{\rm soft}$ consist of
scalar mass terms, where the parameters ${\bf m_{\tilde
Q}}$, ${\bf m_{\tilde L}}$, ${\bf m_{\tilde u}}$, ${\bf m_{\tilde d}}$ and 
${\bf m_{\tilde e}}$ are $3\times3$ Hermitian matrices in flavor space and 
$m_{H_u}$ and $m_{H_d}$ are Higgs mass parameters.
Finally, the remaining soft terms are bilinear and trilinear scalar
interactions, where ${\bf T_u}$, ${\bf T_d}$, and ${\bf T_e}$ are
$3\times 3$ matrices in generation space and $b$ is the strength of the
supersymmetry-breaking Higgs off-diagonal mixing.

The form of the 
Lagrangian given in Eq.\ \eqref{eq:lmssmbrk} also holds, in general, at high
energy. In realistic supersymmetry-breaking scenarios, the parameters are however
deduced from a reduced set of key parameters related to the
supersymmetry-breaking mechanism (see, \eg, the relations of 
Eq.\ \eqref{eq:sugrarelat} for gravity-mediated supersymmetry breaking). 
The parameters of the
low-energy Lagrangian are then recovered by means of the
supersymmetric renormalization group equations introduced in Section
\ref{sec:rge}.

If one allows for $R$-parity violation, additional soft
terms are permitted. Their form is very similar to the one of the superpotential of Eq.\
\eqref{eq:wrpv}, with extra slepton-Higgs off-diagonal mass terms.
We refer to Section \ref{sec:rpv} for more information. 
In addition, we
also recall that all possible contributions depending on the right-handed 
sneutrino fields $\tilde \nu_R$ have been omitted, those fields being assumed 
decoupled.

\subsection{Electroweak symmetry breaking and particle mixings}
\label{sec:mssmewsb}
The classical Higgs potential is extracted from the Lagrangian derived in
the previous section. In order to simplify its analysis, we first
employ an $SU(2)_L$ gauge
transformation to rotate away the possible vacuum expectation value of the
charged component of one of the two Higgs doublets. Then,
the minimization equations of the scalar potential imply that the
vacuum expectation of the charged component of the second doublet also
vanishes, in agreement with electromagnetism conservation.
We concentrate from now on on the terms depending exclusively on the
neutral components of the two Higgs doublets $H_u^0$ and  $H_d^0$,
\be \bsp
   V(H_d^0, H_u^0) = &\ 
    \frac{g_y^2+g_w^2}{8} \Big[ H_d^{0^\dag} H_d^0 - H_u^{0^\dag} H_u^0\Big]^2 
   - b \Big[H_d^0 H_u^0 + H_d^{0^\dag}H_u^{0^\dag} \Big]
\\&\
   +  \big(|\mu|^2 + m_{H_d}^2 \big) H_d^{0^\dag} H_d^0 
   +  \big(|\mu|^2 + m_{H_u}^2 \big) H_u^{0^\dag} H_u^0 
\ ,
\esp \label{eq:mssmHpot} \ee
where we have redefined the Higgs fields so that the soft parameter $b$ 
is real and positive. 
In order to break spontaneously the electroweak symmetry, the Higgs potential
must be bounded from below. For arbitrary large and different 
vacuum expectation values
$v_d/\sqrt{2}$ and $v_u/\sqrt{2}$ of the two fields $H_d^0$ and $H_u^0$,
the quartic term of Eq.\ \eqref{eq:mssmHpot} dominates so that the potential is
always stabilized. However, in the case the two vacuum expectation values are
equal, the quartic terms vanish at the minimum of the potential.
Since these terms are issued from the contributions of the auxiliary
$D$-fields to the scalar potential, these configurations are referred to as 
$D$-flat directions of the Higgs potential. In order to bound the potential 
from below along these directions, one asks for the condition
\be 
   2 |\mu|^2 + m_{H_d}^2 +m_{H_u}^2 - 2 b > 0 \ . 
\label{eq:ewsbmssmcond1} \ee

A second condition among the Higgs sector parameters arises by forbidding 
the trivial solution $v_u = v_d = 0$ to be a minimum of the
potential. This is achieved from the associated Hessian matrix, %
\renewcommand{\arraystretch}{1.2}%
\be
   M_H^2 = \bpm |\mu|^2 + m_{H_d}^2 &  -b\\ -b & |\mu|^2 +
    m_{H_u}^2 \epm \ ,
\ee%
\renewcommand{\arraystretch}{1.}%
evaluated at $v_u = v_d = 0$.
If its determinant is negative, then $v_u = v_d = 0$ is a saddle
point. Equivalently, this corresponds to 
\be
  \big(|\mu|^2 + m_{H_d}^2 \big) \big( |\mu|^2 + m_{H_u}^2  \big) < b^2
  \ .
\label{eq:ewsbmssmcond2} \ee
Contrary, if the determinant of the $M_H^2$ is positive,
\be
  \big(|\mu|^2 + m_{H_d}^2 \big) \big( |\mu|^2 + m_{H_u}^2  \big) - b^2 >
0 \ . 
\ee
From Eq.\ \eqref{eq:ewsbmssmcond1} and since
the Higgs soft mixing parameter $b$ is positive and real,
the sums $(|\mu|^2 + m_{H_d}^2)$ and  
$(|\mu|^2 + m_{H_u}^2)$ have the same
sign. Consequently, a positive determinant of
$M_H^2$ always leads to an unacceptable minimum in $v_u = v_d = 0$. 

Collecting the previous results, the electroweak symmetry is broken only
if the two conditions of Eq.\ \eqref{eq:ewsbmssmcond1} and Eq.\
\eqref{eq:ewsbmssmcond2} are satisfied,
\be\boxed{
   2 |\mu|^2 + m_{H_d}^2 +m_{H_u}^2 - 2 b > 0 \qquad\text{and}\qquad
  \big(|\mu|^2 + m_{H_d}^2 \big) \big( |\mu|^2 + m_{H_u}^2  \big) < b^2 \ .
}\label{eq:ewsbmssmcond}\ee
Supersymmetric renormalization group equations drive the evolution of the Higgs
squared mass parameters $m_{H_d}^2$ and $m_{H_u}^2$ between the high scale where
supersymmetry is broken down to the electroweak scale. Due to the strong Yukawa
coupling between the superfields $H_U$, $Q_L^3$ and $U_R^3$ in the
superpotential (related to the mass of the top quark), these equations naturally
push $m_{H_u}^2$ to be negative or
much smaller than $m_{H_d}^2$, which helps to satisfy the conditions of Eq.\
\eqref{eq:ewsbmssmcond}. In many viable supersymmetry-breaking scenarios, this
effect is even sufficient to guarantee the spontaneous breaking of the electroweak
symmetry therefore referred to as radiative electroweak symmetry
breaking.

Finally, all the parameters of the Higgs sector are not
independent. The minimization conditions of the Higgs potential impose its
first-order derivatives with respect to the neutral Higgs fields to vanish 
at the minimum,
\be\bsp
 0 =&\ \frac{g_w^2+g_y^2}{8} \big[ v_d^2-v_u^2\big] + |\mu|^2 + m_{H_d}^2 - b
   \frac{v_u}{v_d}  \ , \\
 0 =&\ \frac{g_w^2+g_y^2}{8} \big[ v_u^2-v_d^2\big] + |\mu|^2 + m_{H_u}^2 - b
   \frac{v_d}{v_u} \ .
\esp\label{eq:mssmminc}\ee
Two parameters can then be deduced from the knowledge of the others.

After electroweak symmetry breaking,  the
$SU(2)_L$ and $U(1)_Y$ vector bosons mix and get massive, as in the Standard
Model. 
Shifting the neutral scalar Higgs bosons by their vacuum expectation value, 
\be
  H_u^0 \to  \frac{v_u}{\sqrt{2}} + h_u^0 \quad \text{and} \quad  
  H_d^0 \to  \frac{v_d}{\sqrt{2}}  + h_d^0 \ ,
\ee 
where $h_u^0$ are $h_d^0$ are complex scalar fields, we extract the squared
mass matrix of the neutral electroweak gauge bosons
$B_\mu$ and $W_\mu^3$ from the Higgs kinetic and gauge interaction terms,
\be
  \lag_{\rm EW} = D^\mu H_u^\dag D_\mu H_u +  D^\mu H_d^\dag D_\mu H_d \ .
\ee
This matrix reads, in the $(B^\mu, W_\mu^3)$ basis, 
\be
  M^2_{V^0} = \frac14 \big[v_d^2+v_u^2\big] \bpm g_y^2 & - g_y g_w \\
   -g_y g_w & g_w^2 \epm \ , 
\ee
and is diagonalized by introducing the photon $A_\mu$ and neutral electroweak
boson $Z_\mu$ defined by the rotation
\be\boxed{
  \bpm  A_\mu \\ Z_\mu \epm = \bpm \cos\theta_w & \sin\theta_w\\ -\sin\theta_w
    & \cos\theta_w \epm \bpm B_\mu \\ W_\mu^3 \epm \ .
}\ee
The electroweak mixing angle $\theta_w$ and
the masses $M_A$ and $M_Z$ of the physical states are calculated as
\be\boxed{
   \cos^2\theta_w = \frac{g_w^2}{g_w^2+g_y^2} \ , \qquad M_A = 0 \qquad\text{and}\qquad
   M_Z = \frac{g_w \sqrt{v_u^2+v_d^2}}{2 \cos\theta_w}  \ .
}\label{eq:zparams}\ee
Similarly to the Standard Model, the physical charged weak boson states are
obtained after diagonalizing the third generator of $SU(2)_L$ in the adjoint
representation. The transformation rules relating the mass and interaction bases
and the physical $W$-boson mass $M_W$ are given by 
\be\boxed{
  W_\mu^\pm = \frac{1}{\sqrt{2}} (W_\mu^1 \mp i W^2_\mu) \qquad\text{and}\qquad 
    M_W = \frac{g_w\sqrt{v_u^2+v_d^2}}{2} \ .
}\label{eq:wparams}\ee

Before getting back to the Higgs sector, one can emphasize that the relations of
Eq.\ \eqref{eq:zparams} allows us to rewrite the minimization conditions of Eq.\
\eqref{eq:mssmminc} as 
\be\boxed{
 \sin 2\beta =  \frac{2 b}{2 |\mu|^2 + m_{H_d}^2  + m_{H_u}^2}
 \quad\text{and}\quad
  M_Z^2 = \frac{\big|m_{H_u}^2 \!-\! m_{H_d}^2\big|}{\big|\cos 2 \beta\big|}
   -\! 2 |\mu|^2 \!-\! m_{H_d}^2 \!-\! m_{H_u}^2 \ ,
}\label{eq:mssmminc2}\ee
where we have introduced the ratio of the vacuum expectation values of the
neutral Higgs fields, $\tan\beta = v_u/v_d$. These two equations enlighten the
so-called $\mu$-problem of the MSSM~\cite{Kim:1983dt}. Without fine-tuned
cancellations, all the
parameters $b$, $|\mu|$, $m_{H_d}$ and $m_{H_u}$ must be roughly of the order of
the $Z$-boson mass. However, $\mu$ is a superpotential parameter, not related
to the breaking of the electroweak symmetry, and there is therefore no
strong reason to impose its value to be of the order of the electroweak scale.
Furthermore, from a theoretical point of view, the $\mu$-parameter being the
only dimensionful quantity of the superpotential, its natural size can only be
either of the order of the supersymmetry-breaking scale, the only inherent mass
scale in the setup, or zero when forbidden by a symmetry. Several
solutions exist to address this issue, such as, \eg, the elegant proposal 
(however not considered in this work) of
generating the $\mu$ term dynamically from the vacuum expectation value of a new
singlet chiral superfield as in the so-called Next-to-Minimal
Supersymmetric Standard Model  \cite{Fayet:1976cr, Fayet:1974pd,
Derendinger:1983bz, Fayet:1977yc,  Ellis:1988er, Drees:1988fc, Ellwanger:1993xa,
Elliott:1994ht, Ellwanger:1995ru, Ellwanger:1996gw, Maniatis:2009re,
Ellwanger:2009dp}.

When becoming massive, the $W$ and $Z$ gauge bosons eat three out of the eight
real degrees of freedom included in the two Higgs doublets, \ie, the
Goldstone bosons $G^\pm$ and $G^0$ which become the longitudinal modes of
the weak bosons. The
five other degrees of freedom mix to the physical
Higgs fields, $h^0$, $H^0$, $A^0$ and $H^\pm$. The definition of these eight states, 
as well as their mass, can be obtained from the diagonalization of the scalar,
pseudoscalar and charged Higgs mass matrices $M_S^2$, $M_P^2$ and $M_\pm^2$, 
extracted from the full Higgs potential. They read,
in the $(\Re \{h_d^0\}, \Re\{h_u^0\})$, $(\Im \{h_d^0\}, \Im
\{h_u^0\})$ and $(H_d^{-\dag}, H_u^+)$ bases, %
\renewcommand{\arraystretch}{1.2}%
\be\bsp
 M_S^2 =&\ \bpm
    M_Z^2 c^2_\beta \!+\! \big(2 |\mu|^2 \!+\! m_{H_d}^2 \!+\!
    m_{H_u}^2\big) s^2_\beta &  - (2 |\mu|^2 \!+\! m_{H_d}^2  \!+\! m_{H_u}^2
    \!+\! M_Z^2)  s_\beta c_\beta\\
    - (2 |\mu|^2 \!+\! m_{H_d}^2  \!+\! m_{H_u}^2 \!+\! M_Z^2)  s_\beta c_\beta
      & M_Z^2 \sin^2 \beta \!+\! \big( 2 |\mu|^2 \!+\! m_{H_d}^2
      \!+\! m_{H_u}^2 \big)c^2_\beta
\epm \ , \\
 M_P^2 =&\ b \bpm
     t_\beta & 1\\
     1 & \frac{1}{t_\beta} 
     \epm  \qquad\text{and}\qquad
  M_\pm^2 = \bpm
     M_W^2 s_\beta^2 + b t_\beta & M_W^2 s_\beta c_\beta  +b \\
 M_W^2 s_\beta c_\beta + b & M_W^2 \cos^2 \beta + \frac{b}{t_\beta}\\
\epm \ , 
\esp\ee%
\renewcommand{\arraystretch}{1.}%
respectively, after employing Eq.\ \eqref{eq:mssmminc}, Eq.\
\eqref{eq:zparams} and Eq.\ \eqref{eq:wparams} for simplifications. We have also 
introduced the shorthand notations
$s_\beta = \sin\beta$, $c_\beta = \cos\beta$ and $t_\beta = \tan\beta$. After
diagonalizing those matrices, one rewrites the gauge eigenstates $H_u^+$,
$h_u^0$, $h_d^0$ and $H_d^-$ in terms of the mass eigenstates as
\be\boxed{\bsp
  h_u^0 =&\  \cos\alpha\ h^0 + \sin\alpha\ H^0 + 
    i \cos\beta\ A^0 + i \sin\beta\ G^0\ , \\ 
  h_d^0 =&\ -\sin\alpha\ h^0 + \cos\alpha\ H^0 + 
    i \sin\beta\ A^0 - i \cos\beta\ G^0\ , \\
  H_u^+ =&\  \cos\beta\ H^+ +\sin\beta\ G^+\ , \\  
  H_d^- =&\  \sin\beta\ H^- - \cos\beta\ G^-\ , 
\esp}\label{eq:HiggsMixing}\ee 
where the neutral Higgs mixing angle $\alpha$ is defined as
\be
  \tan 2 \alpha = \tan 2 \beta \frac{2 |\mu|^2 \!+\! m_{H_d}^2 \!+\!
    m_{H_u}^2 + M_Z^2}{2 |\mu|^2 \!+\! m_{H_d}^2 \!+\!
    m_{H_u}^2 - M_Z^2} \ .
\ee 
Furthermore, the physical squared masses are the eigenvalues of the three matrices
above,
\be\boxed{\bsp
  M_{G^0}^2  = &\ 0 \quad {\rm and} \quad M_{A^0}^2 = 2 |\mu|^2 + m_{H_d}^2  + m_{H_u}^2 \ , \\
  M_{G^\pm}^2= &\ 0 \quad {\rm and} \quad  M_{H^\pm}^2 = M_W^2 + 2 |\mu|^2 +
    m_{H_d}^2  + m_{H_u}^2 \ , \\
  M_{h^0}^2 = &\ \frac12 \Big[ M_{A^0}^2 + M_Z^2 - \sqrt{(M_{A^0}^2 - M_Z^2)^2 +
    4 M_{A^0}^2 M_Z^2 \sin^2 2\beta} \Big] \ , \\
  M_{H^0}^2 = &\ \frac12 \Big[ M_{A^0}^2 + M_Z^2 + \sqrt{(M_{A^0}^2 - M_Z^2)^2 +
    4 M_{A^0}^2 M_Z^2 \sin^2 2\beta} \Big]  \ .
\esp}\ee
One observes that the mass of the lightest Higgs field $h^0$ is bounded from
above since 
\be\label{eq:hcst}
  M_{h^0} < M_Z \big| \cos 2\beta \big| \ ,
\ee
which contradicts current observations. However, the particle masses are subject to
quantum corrections. The lightest Higgs mass is hence shifted to a higher
scale, in particular through quantum loops of top quarks and squarks.

In the fermionic sector, the mass matrix of the neutral partners of
the gauge and Higgs bosons reads, in the $(i\widetilde B, i \widetilde W^3, 
\widetilde H_d^0, \widetilde H_u^0)$ basis,
\be
  M_{\tilde{\chi}^0} = \bpm
    M_1&0& - M_W \tan \theta_w \cos \beta & M_W \tan \theta_w \sin \beta \\ 
    0&M_2& M_W \cos \beta & - M_W \sin \beta\\
    - M_W \tan \theta_w \cos \beta & M_W\cos\beta & 0 & -\mu\\
      M_W \tan \theta_w \sin \beta & -M_W\sin\beta & -\mu  & 0       
\epm  \ .
\ee
We have obtained this expression after employing Eq.\ \eqref{eq:zparams}
and Eq.\ \eqref{eq:wparams} for simplifications. Factors 
of $i$ have been
absorbed in gaugino field redefinitions so that the mass matrix $M_{\tilde
\chi^0}$ is now real. Moreover, since $M_{\tilde \chi^0}$ is in addition 
symmetric, it can be diagonalized by means of a unitary matrix $N$,
\be\boxed{
  N^* M_{\tilde{\chi}^0} N^{-1} = {\rm diag}\,
    (M_{\tilde{\chi}^0_1}, M_{\tilde{\chi}^0_2},
     M_{\tilde{\chi}^0_3}, M_{\tilde{\chi}^0_4}) \ .
}\ee
The masses $M_{\tilde{\chi}^0_1} < M_{\tilde{\chi}^0_2} <
M_{\tilde{\chi}^0_3} < M_{\tilde{\chi}^0_4}$ are the masses of the two-component
fields $\chi^0_i$, dubbed neutralinos, related to the
gaugino and higgsino interaction eigenstates by
\renewcommand{\arraystretch}{1.2}%
\be\boxed{ 
  \bpm \chi^0_1 \\ \chi^0_2 \\ \chi^0_3 \\ \chi^0_4 \epm = N 
  \bpm i \widetilde B \\ i \widetilde W^3 \\ \widetilde H_d^0 \\ 
     \widetilde H_u^0 \epm \ .
}\ee %
\renewcommand{\arraystretch}{1.}%
Following the second version of the
Supersymmetry Les Houches Accord conventions~\cite{Allanach:2008qq},
the eigenvalues of the $M_{\tilde{\chi}^0}$ matrix are
chosen non-negative and real so that the mixing matrix $N$ is generally
complex\footnote{This contrasts with the original agreement~\cite{Skands:2003cj}
where the mixing matrix $N$ is imposed to be real, so that one or several of the
mass eigenvalues can be negative.}.
Their analytical expressions can be written under
a relatively compact form by means of projection operators but we however refer to the
literature for further details \cite{ElKheishen:1992yv,
Gounaris:2001fx, Ibrahim:2007fb}.

Similarly, the mass matrix of the charged fermionic partners reads, in the $(i
\widetilde W^-, \widetilde H^-_d)$ and $(i \widetilde W^+, \widetilde H^+_u)$
bases, 
\be
 M_{\tilde{\chi}^\pm} = \bpm M_2 & \sqrt{2} M_W \sin\beta \\
 \sqrt{2} M_W \cos\beta &  \mu \epm \ , 
\ee
after employing Eq.\ \eqref{eq:wparams} for simplifications. The charged wino 
states $\widetilde W_\mu^\pm$ are defined
after diagonalizing the third generator of $SU(2)$ in the adjoint
representation, as for the $W$-boson states,
\be
  \widetilde W^\pm = \frac{1}{\sqrt{2}} (\widetilde W^1 \mp i \widetilde
    W^2) \ . 
\ee
The matrix $M_{\tilde{\chi}^\pm}$ is diagonalized by two unitary
matrices $U$ and $V$ so that
\be\boxed{
  U^* M_{\tilde{\chi}^\pm} V^{-1} = {\rm diag} (M_{\tilde{\chi}^\pm_1},
    M_{\tilde{\chi}^\pm_2}) \ ,
\label{eq:diaginos}}\ee
where $M_{\tilde{\chi}^\pm_1} < M_{\tilde{\chi}^\pm_2}$ are the masses of
the so-called chargino states. Those two rotations relate the
interaction eigenstates to the physical charginos states
$\chi^\pm_i$ by
\be\boxed{
  \bpm \chi^+_1 \\ \chi^+_2 \epm = V \bpm i \widetilde W^+ \\ \widetilde H^+_u \epm 
  \quad \text{and}\quad  
  \bpm \chi^-_1 \\ \chi^-_2 \epm = U \bpm i \widetilde W^- \\ \widetilde H^-_d
    \epm \ .
}\ee
The computation of $V$ stems from the diagonalization of the Hermitian matrix
$M_{\tilde{\chi}^\pm}^\dag M_{\tilde{\chi}^\pm}$, so that one has, 
from Eq.\ \eqref{eq:diaginos}, 
\be
  V M_{\tilde{\chi}^\pm}^\dag M_{\tilde{\chi}^\pm} V^{-1} = {\rm
    diag}\,(M_{\tilde{\chi}^\pm_1}^2, M_{\tilde{\chi}^\pm_2}^2) \ .
\ee
This allows to derive the squared chargino masses,
\be
 M_{\tilde{\chi}^\pm_{1,2}}^2 = \frac12\Big[|M_2|^2+|\mu|^2+
 2 M_W^2 \mp \sqrt{(|M_2|^2+|\mu|^2+ 2 M_W^2)^2 - 4 |\mu M_2- M_W^2
 s_{2\beta}|^2}\Big] \ , 
\ee
after having introduced the shorthand notation $s_{2\beta} =
\sin 2\beta$, and the rotation matrix $V$ reads
\be
  V = \bpm \cos\theta_+&\sin\theta_+\,e^{-i\phi_+}\\
         -\sin\theta_+\,e^{i\phi_+}&\cos\theta_+\epm \ . 
\ee
The phase of the off-diagonal element cannot be eliminated as it is needed to 
rotate away
the imaginary part of the off-diagonal matrix element in
$M_{\tilde{\chi}^\pm}^\dag M_{\tilde{\chi}^\pm}$, 
\be
 \Im \Big[ (M_2^* \sin\beta + \mu \cos\beta) e^{i\phi_+} \Big] = 0 \ .
\ee
Moreover, the rotation angle $\theta_+\in[0;\pi]$ is uniquely fixed by the two 
conditions
\be\bsp
  \tan2\theta_+ = &\ \frac{2\sqrt{2}M_W \big( M_2^* \sin\beta + \mu\cos\beta \big)
    e^{i\phi_+}}{|M_2|^2 -|\mu|^2 + 2 M_W^2 c_{2\beta}} \ , \\
  \sin2\theta_+ = &\ \frac{-2\sqrt{2} M_W \big( M_2^* \sin\beta + \mu
    \cos\beta\big) e^{i\phi_+}}{\sqrt{\Big[|M_2|^2 -|\mu|^2 + 2
    M_W^2c_{2\beta}\Big]^2
    +8 M_W^2\Big[(M_2^* \sin\beta + \mu\cos\beta) e^{i\phi_+}\Big]^2}} \ ,
\esp\ee
where $c_{2\beta} = \cos 2\beta$.
Similar relations can be obtained from the rotation matrix $U$ starting from 
\be
  U^\ast M_{\tilde{\chi}^\pm} M_{\tilde{\chi}^\pm}^\dag U^t = {\rm
    diag}\,(M_{\tilde{\chi}^\pm_1}^2, M_{\tilde{\chi}^\pm_2}^2) \ .
\ee
However, it is more convenient to derive it directly from the knowledge of
$V$ by employing 
\be
 U = {\rm diag}\,(\frac{1}{M_{\tilde{\chi}^\pm_1}},\frac{1}{M_{\tilde{\chi}^\pm_2}})
  V^* M_{\tilde{\chi}^\pm}^t \ .
\ee

As in the Standard Model, the diagonalization of the quark sector requires four
unitary matrices $V_u$, $V_d$, $U_u$ and $U_d$, so that
\be
  d_L^i \to \big(V_d d_L\big)^i \ , \quad 
  d_R^{ic} \to \big(d_R^c U_d^\dag\big)^i\ , \quad 
  u_L^i \to \big(V_u u_L\big)^i \quad \text{and} \quad
  u_R^{ic} \to  \big(u_R^c U_u^\dag\big)^i\ ,
\label{eq:fermrot}\ee
where the generation index $i$ is explicitly indicated and the
subscript $c$ stands for charge conjugation. 
In contrast, the diagonalization of the lepton sector proceeds only through 
three unitary rotations $V_e$, $V_\nu$ and $U_e$, 
\be
  e_L^i \to \big(V_e e_L\big)^i \ , \quad 
  e_R^{ic} \to \big(e_R^c U_e^\dag\big)^i\quad\text{and} \quad 
  \nu_L^i \to \big(V_\nu \nu_L\big)^i\ , 
\label{eq:fermrot2}\ee
since the right-handed neutrino fields have been decoupled.
Promoting these rotations to the superfield level, Eq.\ \eqref{eq:fermrot} and
Eq.\ \eqref{eq:fermrot2} lead to a redefinition of the
parameters of the superpotential, 
\be 
  W_{\rm MSSM} = 
    ({\bf \hat y^u})_{ij}   U_R^i\ Q_L^j \!\cdot\! H_U - 
    ({\bf \hat y^d} V_d^\dag V_u)_{ij}   D_R^i\ Q_L^j \!\cdot\! H_D +
    ({\bf \hat y^e})_{ij}   E_R^i\ L_L^j \!\cdot\! H_D +
    \mu H_U \!\cdot\! H_D \ ,
\label{eq:wmssm3}\ee 
where the general Yukawa matrices of Eq.\ \eqref{eq:wmssm2} have been replaced by 
\be\boxed{
  {\bf y^u} \to {\bf \hat y^u} = U_u^\dag {\bf y^u} V_u  \ ,\quad
  {\bf y^d} \to {\bf \hat y^d} V_d^\dag V_u = U_d^\dag {\bf y^d} V_u  
  \quad\text{and}\quad
  {\bf y^e} \to {\bf \hat y^e} = U_e^\dag {\bf y^e} V_e  \ .
}\ee
The matrices ${\bf \hat y^u}$, ${\bf \hat y^d}$ and ${\bf \hat y^e}$ are diagonal and
real $3 \times 3$ matrices in flavor space and the rotations of Eq.\ 
\eqref{eq:fermrot} have been absorbed into two new superfield redefinitions, 
\be\boxed{
  D_L^i \to \big(V_{\rm CKM} D_L\big)^i = \big(V_u^\dag V_d D_L\big)^i 
  \qquad\text{and}\qquad
  N_L^i \to \big(V_{\rm PMNS} N_L\big)^i = \big(V_l^\dag V_\nu N_L\big)^i \ ,
}\label{eq:rotsckm}\ee
which corresponds
to the supersymmetrization of the well-known Standard Model CKM and PMNS mixings. 
In these notations, the superfield $D_L$ stands for the down-type component of the 
$SU(2)_L$ doublet $Q_L$ while the superfield $N_L$ is the up-type component of
the doublet $L_L$. This defines the so-called super-CKM and
super-PMNS bases \cite{Hall:1985dx}.

To prevent the rotation matrices of Eq.\ \eqref{eq:fermrot} and Eq.\ 
\eqref{eq:fermrot2} to appear explicitly in the soft supersymmetry-breaking 
Lagrangian of Eq.\ \eqref{eq:lmssmbrk}, appropriate redefinitions of the soft 
parameters are also performed, 
\be \bsp 
   \lag_{\rm soft} =
   &\ \frac12 \Big[ 
     M_1 \widetilde B \!\cdot\! \widetilde B + 
     M_2 \widetilde W \!\cdot\! \widetilde W + 
     M_3 \widetilde g \!\cdot\! \widetilde g + 
     \hc \Big]
      - m_{H_u}^2 H_u^\dag H_u  
      - m_{H_d}^2 H_d^\dag H_d
\\&\
      - (V_{\rm CKM} {\bf \hat m^2_{\tilde Q}} V_{\rm CKM}^\dag)^i{}_j 
          \tilde q_{Li}^\dag \tilde q_L^j 
      - ({\bf \hat m^2_{\tilde U}})^i{}_j \tilde u_{Ri} \tilde u_R^{j\dag} 
      - ({\bf \hat m^2_{\tilde D}})^i{}_j \tilde d_{Ri} \tilde d_R^{j\dag} 
\\&\
      - ({\bf \hat m^2_{\tilde L}})^i{}_j \tilde \ell_{Li}^\dag \tilde \ell_L^j 
      - ({\bf \hat m^2_{\tilde E}})^i{}_j \tilde e_{Ri} \tilde e_R^{j\dag} 
\\&\ 
      - \Big[  ({\bf \hat T^u})_{ij} \tilde u_R^{i\dag} \tilde q_L^j \!\cdot\! H_u 
           - ({\bf \hat T^d} V_{\rm CKM}^\dag)_{ij} \tilde d_R^{i\dag} \tilde q_L^j \!\cdot\! H_d 
           - ({\bf \hat T^e})_{ij} \tilde e_R^{i\dag} \tilde \ell_L^j \!\cdot\! H_d
           + \hc \Big]\\
   &\ - \Big[  b H_u \!\cdot\! H_d  + \hc \Big]\ ,
\esp\label{eq:lmssmbrk2}\ee
so that only the rotations of Eq.\ \eqref{eq:rotsckm} 
have still to be applied to the full Lagrangian.
The trilinear interaction strengths have been redefined according to
\be \boxed{
  {\bf T^u} \to  {\bf \hat T^u} = U_u^\dag {\bf T^u} V_u\ , \quad
  {\bf T^d} \to  {\bf \hat T^d V_{\rm CKM}^\dag} = U_d^\dag {\bf T^d} V_u\quad\text{and}\quad
  {\bf T^e} \to  {\bf \hat T^e} = U_e^\dag {\bf T^e} V_e \ ,
}\ee
and the sfermion masses have been replaced by
\be\boxed{\bsp
  {\bf m^2_{\tilde Q}} \to V_{\rm CKM} {\bf \hat m^2_{\tilde Q}} V_{\rm CKM}^\dag 
    \ ,\quad 
 &\ {\bf m^2_{\tilde U}} \to {\bf \hat m^2_{\tilde U}} =
     U_u^\dag {\bf m^2_{\tilde U}} U_u \ , \quad
\\
  {\bf m^2_{\tilde D}} \to {\bf \hat m^2_{\tilde D}} =
     U_d^\dag {\bf m^2_{\tilde D}} U_d \ , 
  {\bf m^2_{\tilde L}} \to {\bf \hat m^2_{\tilde L}} =&\
     V_e^\dag {\bf m^2_{\tilde L}} V_e \ ,  \quad
  {\bf m^2_{\tilde E}} \to {\bf \hat m^2_{\tilde E}} =
     U_e^\dag {\bf m^2_{\tilde E}} U_e \ . 
\esp}\ee

It is important to emphasize that the `hatted' soft parameters are not necessarily
diagonal in flavor space so that  
in the the super-CKM and super-PMNS bases, fermion
and sfermion fields can be possibly misaligned. Keeping full generalities, 
the $3\times 3$ sneutrino mass matrix reads, in the 
$(\tilde\nu_e, \tilde\nu_\mu, \tilde \nu_\tau)$ basis,%
\renewcommand{\arraystretch}{1.4}%
\be
  M_{\tilde{\nu}}^2 = V_{\rm PMNS}^\dag {\bf m^2_{\tilde L}} V_{\rm PMNS} 
   +\frac12 \cos 2\beta M_Z^2 I_3 \ ,
\ee%
where $I_3$ is the three-dimensional identity matrix, 
while the $6\times 6$ squark and charged slepton mass matrices are respectively 
given, in the 
$(\tilde u_L, \tilde c_L, \tilde t_L, \tilde u_R, \tilde c_R, \tilde t_R)$, 
$(\tilde d_L, \tilde s_L, \tilde b_L, \tilde d_R, \tilde s_R, \tilde b_R)$
and 
$(\tilde e_L, \tilde \mu_L, \tilde \tau_L, \tilde e_R, \tilde \mu_R, \tilde \tau_R)$ 
bases, by
\be\bsp
  M_{\tilde{u}}^2 = &\ \bpm   
    V_{\rm CKM} {\bf \hat m^2_{\tilde Q}} V_{\rm CKM}^\dag + M_{q_u}^2 + 
      (\frac12 -\frac23 s_w^2) c_{2\beta} M_Z^2 I_3 & 
       \frac{v_u}{\sqrt{2}} {\bf \hat T^u}{}^\dag - \frac{1}{t_\beta} \mu M_{q_u}\\
    \frac{v_u}{\sqrt{2}} {\bf \hat T^u}  - \frac{1}{t_\beta} \mu^\ast M_{q_u} & 
    {\bf \hat m^2_{\tilde U}} + M_{q_u}^2 + \frac23
     s_w^2 c_{2\beta} M_Z^2 I_3
  \epm \ ,\\
  M_{\tilde{d}}^2 = &\ \bpm   
    {\bf \hat m^2_{\tilde Q}} + M_{q_d}^2 - (\frac12 - \frac13 s_w^2) c_{2\beta}
        M_Z^2 I_3 & 
       \frac{v_d}{\sqrt{2}} {\bf \hat T^d}{}^\dag - t_\beta \mu M_{q_d}\\
    \frac{v_d}{\sqrt{2}} {\bf \hat T^d}  - t_\beta \mu^\ast M_{q_d}& 
    {\bf \hat m^2_{\tilde D}} + M_{q_d}^2 - \frac13 s_w^2 c_{2\beta}
       M_Z^2 I_3
  \epm \ ,\\
  M_{\tilde{e}}^2 = &\ \bpm   
    {\bf \hat m^2_{\tilde L}} + M_\ell^2 - (\frac12 - s_w^2)  c_{2\beta}M_Z^2 I_3 & 
       \frac{v_d}{\sqrt{2}} {\bf \hat T^e}{}^\dag - t_\beta \mu M_\ell\\
    \frac{v_d}{\sqrt{2}} {\bf \hat T^e}  - t_\beta \mu^\ast M_\ell & 
    {\bf \hat m^2_{\tilde E}} + M_\ell^2 - s_w^2  c_{2\beta} M_Z^2 I_3
  \epm \ .
\esp\ee%
\renewcommand{\arraystretch}{1.}%
To derive those results, we have employed Eq.\ \eqref{eq:zparams}, introduced 
the shorthand notations $s_w =
\sin\theta_w$ and $c_{2\beta} = \cos 2\beta$ and defined the (diagonal and
real) fermion mass matrices $M_{q_u}$, $M_{q_d}$ and $M_\ell$ by
\be
  M_{q_u} = \frac{v_u {\bf \hat y^u}}{\sqrt{2}} \ , \qquad
  M_{q_d} = \frac{v_d {\bf \hat y^d}}{\sqrt{2}} \ , \qquad
  M_\ell  = \frac{v_d {\bf \hat y^e}}{\sqrt{2}} \ .
\ee
The four sfermion mass matrices can be diagonalized by means of four additional
rotations $R^u$, $R^d$, $R^e$ and $R^\nu$, 
\be\boxed{\bsp
 {\rm diag}(M_{\tilde u_1}^2, \ldots, M_{\tilde u_6}^2) = R^u
   M_{\tilde{u}}^2 R^{u\dag} \ , \qquad 
 & {\rm diag}(M_{\tilde d_1}^2, \ldots, M_{\tilde d_6}^2) = R^d
   M_{\tilde{d}}^2 R^{d\dag}\ , \\
 {\rm diag}(M_{\tilde e_1}^2, \ldots, M_{\tilde e_6}^2) = R^e
   M_{\tilde{e}}^2 R^{u\dag} \ , \qquad 
 &{\rm diag}(M_{\tilde \nu_1}^2, M_{\tilde \nu_2}^2, M_{\tilde \nu_3}^2) = R^\nu
   M_{\tilde{\nu}}^2 R^{\nu\dag} \ .
\esp}\ee
By convention, the states are mass-ordered from the lightest to the heaviest,
$M_{\tilde f_1} < \ldots <
M_{\tilde f_6}$ for $f=u$, $d$ and $e$ and $M_{\tilde \nu_1} < M_{\tilde \nu_2} <
M_{\tilde \nu_3}$ and the physical mass eigenstates are related to their gauge
counterparts (in the super-CKM and super-PMNS bases) by 
\be\boxed{\bsp 
  \bpm \tilde u_1\\ \tilde u_2\\ \tilde u_3\\ \tilde u_4\\ \tilde u_5 \\ \tilde
    u_6 \epm = R^u \bpm \tilde u_L\\ \tilde c_L\\ \tilde t_L\\ \tilde u_R \\
    \tilde c_R\\ \tilde t_R \epm \ ,\quad
  \bpm \tilde d_1\\ \tilde d_2\\ \tilde d_3\\ \tilde d_4\\ \tilde d_5 \\ \tilde
    d_6 \epm =&\ R^d \bpm \tilde d_L\\ \tilde s_L\\ \tilde b_L\\ \tilde d_R \\
    \tilde s_R\\ \tilde b_R \epm \ ,\quad
  \bpm \tilde e_1\\ \tilde e_2\\ \tilde e_3\\ \tilde e_4\\ \tilde e_5 \\ \tilde
    e_6 \epm = R^e \bpm \tilde e_L\\ \tilde \mu_L\\ \tilde \tau_L\\ \tilde e_R \\
    \tilde \mu_R\\ \tilde \tau_R \epm \ , \\
 \bpm \tilde \nu_1\\ \tilde \nu_2\\ \tilde \nu_3 \epm = &\ R^\nu \bpm
    \tilde\nu_e\\ \tilde\nu_\mu\\ \tilde\nu_\tau\epm \ .
\esp}\label{eq:sfmix}\ee

Finally, the four-component spinor representations $\psi$ of
the fermionic fields are related to the two-component ones by
\be\label{eq:weyltodirac} \boxed{\bsp
  &\
    \psi_u^i = \bpm u_L^i \\ \bar u_{Ri}^c\epm \ , \quad 
    \psi_d^i = \bpm u_L^i \\ \bar d_{Ri}^c\epm \ , \quad 
    \psi_e^i = \bpm e_L^i \\ \bar e_{Ri}^c\epm \ , \quad 
    \psi_\nu^i = \bpm \nu_L^i \\ \bar \nu_{Ri}^c\epm \ , \\
  &\qquad \qquad
    \psi_{\chi^0}^i = \bpm \chi^0_i \\ \bar \chi^{0i} \epm \ , \quad 
    \psi_{\chi^\pm}^i = \bpm \chi^\pm_i \\ \bar \chi^{\mp i} \epm \ , \quad 
    \psi_{\tilde g} = \bpm i \tilde g \\ -i \overline {\tilde g}\epm \ .
\esp}\ee 

We have now achieved the derivation of the MSSM particle spectrum, 
together with the associated
mass matrices, at tree-level.
Loop corrections are mandatory to compute an accurate enough particle spectrum.
One-loop contributions to the mass matrices are known for almost two decades and
the associated analytical results can be found in Ref.\
\cite{Chankowski:1992er}, Ref.\ \cite{Dabelstein:1994hb} and Ref.\
\cite{Pierce:1996zz}. We omit their complete expressions from the present 
manuscript for brevity but will use them when addressing the generation of realistic
supersymmetric spectra in the rest of this section.

\subsection{Supersymmetry-breaking models in the
MSSM framework}\label{sec:mssmbrkex}
The supersymmetry-breaking Lagrangian of Eq.\ \eqref{eq:lmssmbrk2} contain 105
masses, phases and mixing angles that cannot be rotated away
by field redefinitions \cite{Dimopoulos:1995ju}. Most of them are however
strongly constrained by the experiment as they imply new sources of flavor
violation and/or $CP$ violation. When non-vanishing or non-drastically
reduced, these parameters could enhance the rates of
several observables related to $K^0-\bar{K}^0$, $D^0-\bar{D}^0$ and
$B^0-\bar{B}^0$ oscillations, flavor-changing decays of kaons, $D$-mesons,
$B$-mesons, muons and taus, \etc.

This clearly restricts the class of phenomenologically viable supersymmetry-breaking
scenarios that can be constructed. 
We choose to focus, in this section, on the minimal version of the three 
classes of models presented in
Section \ref{sec:popbrk}, \ie, gravity-mediated, gauge-mediated and
anomaly-mediated supersymmetry-breaking theories. In those scenarios, 
organizing principles relate all the model free parameters to a reduced set of
quantities and impose no additional flavor and $CP$ violation with respect to
those already included in the CKM and PMNS mixings. 
Deviations from this assumption will be shortly
discussed in Section \ref{sec:mssm_indirect}, following
a phenomenological approach where some of the off-diagonal parameters of the sfermion
mass matrices are taken, at low energy, arbitrary \cite{Bozzi:2007me,
Fuks:2008ab, Fuks:2011dg}.

The first class of supersymmetry-breaking models investigated in this work are
inspired from gravity-mediated supersymmetry-breaking theories, as introduced in
Section \ref{sec:grmsb}, and are commonly called constrained MSSM (cMSSM)
scenarios. First, one
assumes that all the model parameters are real, with the exception of the
elements of the CKM matrix. Moreover, the PMNS matrix is taken equal to the
identity $3\times 3$ matrix $I_3$. Next, on the basis of Eq.~\eqref{eq:sugrarelat}, all the
free parameters included in the soft Lagrangian of Eq.\ \eqref{eq:lmssmbrk2} are
defined, at a high scale, in terms of three parameters, the universal scalar
mass $m_0$, the universal gaugino mass $m_{1/2}$ and the universal trilinear
couplings $A_0$,
\be\bsp
  M_1 = &\ M_2 = M_3 = m_{1/2} \ , \\
  {\bf \hat m^2_{\tilde Q}} =&\ {\bf \hat m^2_{\tilde U}} =  {\bf \hat
    m^2_{\tilde D}} = {\bf \hat m^2_{\tilde L}} = {\bf \hat m^2_{\tilde E}} =
    m_0^2\ I_3 \ , \qquad 
   m_{H_u}^2 =  m_{H_d}^2 = m_0^2 \ ,\\
   {\bf \hat T^u} =&\ {\bf \hat y^u}  A_0 \ , \qquad 
   {\bf \hat T^d} = {\bf \hat y^d}  A_0 \ , \qquad 
   {\bf \hat T^e} = {\bf \hat y^e}  A_0 \ . 
\esp\ee
These relations hold at a high-scale being usually taken as the scale where the three
gauge couplings of the model unify and the values of the Yukawa couplings at the 
high scale are derived,
using the supersymmetric renormalization group equations, from the knowledge of
the Standard Model fermion masses. Moreover, concerning the Higgs sector
parameters, the values of the off-diagonal Higgs mixing parameters $\mu$ and $b$
are indirectly fixed by the minimization conditions of Eq.
\eqref{eq:mssmminc} or Eq.\ \eqref{eq:mssmminc2} and the electroweak input
parameters. The electroweak
symmetry being broken at low-energy, the values of $\mu^2$ and $b$ 
at the high scale are deduced by
employing again the supersymmetric renormalization group equations.

It is however necessary to 
specify the ratio of the vacuum expectation values of the two neutral Higgs
fields (at low-energy) as well as the sign of $\mu$ since both these quantities 
cannot be extracted from the knowledge of the other parameters of the model. 
This leaves a total of
four free parameters and a sign, 
\be\boxed{
  m_0\ , \qquad m_{1/2}\ , \qquad  A_0 \ , \qquad
  \tan\beta = \frac{v_u}{v_d} \qquad\text{and}\qquad {\rm sign} \big(\mu\big) \ ,
}\label{eq:cmssmprm}\ee
supplementing the Standard Model input parameters.

The second class of supersymmetric scenarios for the MSSM 
which have been studied in this work are
scenarios where the spontaneous breaking of supersymmetry is induced by gauge
interactions. As presented in Section \ref{sec:GMSB}, the source of
supersymmetry breaking is parametrized through a gauge singlet superfield whose
both the scalar and auxiliary components acquire vacuum expectation values. This
singlet superfield communicates to the visible sector of the MSSM (described in
Section \ref{sec:mssmfields}) by means of superpotential couplings to messenger
fields. In realistic gauge-mediated supersymmetry-breaking models, the
messenger sector is taken more general than in Section \ref{sec:GMSB} and
consists of $N_q$ and $N_\ell$ pairs of
colored and non-colored messenger superfields, being respectively
charged and singlet under the QCD gauge group. This freedom in choosing
independently $N_q$ and $N_\ell$ allows to study various model
configurations. We adopt here minimal benchmark scenarios where
$N_q = N_\ell$ and where additionally,
the messenger sector is embedded in complete representations of
$SU(5)$. 

The soft Lagrangian terms at the messenger scale are derived from Eq.\
\eqref{eq:Lbrkgmsb} and read 
\be \bsp 
   \lag_{\rm soft} =
   &\ \frac12 \Big[ 
     M_1 \widetilde B \!\cdot\! \widetilde B + 
     M_2 \widetilde W \!\cdot\! \widetilde W + 
     M_3 \widetilde g \!\cdot\! \widetilde g + 
     \hc \Big]
      - ({\bf \hat m^2_{\tilde Q}})^i{}_j \tilde q_{Li}^\dag \tilde q_L^j 
      - ({\bf \hat m^2_{\tilde U}})^i{}_j \tilde u_{Ri} \tilde u_R^{j\dag} 
\\&\
      - ({\bf \hat m^2_{\tilde D}})^i{}_j \tilde d_{Ri} \tilde d_R^{j\dag} 
      - ({\bf \hat m^2_{\tilde L}})^i{}_j \tilde \ell_{Li}^\dag \tilde \ell_L^j 
      - ({\bf \hat m^2_{\tilde E}})^i{}_j \tilde e_{Ri} \tilde e_R^{j\dag} 
      - m_{H_u}^2 H_u^\dag H_u  
      - m_{H_d}^2 H_d^\dag H_d \ ,
\esp\ee
where all the mass parameters are given by Eq.\
\eqref{eq:gmsb_ino_mass} and Eq.\ \eqref{eq:gmsb_sc_mass}. Gauge interactions being
flavor-blind, new flavor and $CP$ violating effects are naturally 
suppressed so that
the soft scalar mass matrices are diagonal and real in flavor space with a good 
approximation. Moreover, as
for the considered cMSSM scenarios, we also assume $V_{\rm PMNS} = I_3$. 
Supersymmetric renormalization group equations are then
employed to derive the values of the model parameters at the electroweak scale.
Although absent at the high scale, trilinear couplings at low energy are
generated when run down so that the supersymmetry-breaking Lagrangian 
at the TeV scale is given by Eq. \eqref{eq:lmssmbrk}.

Comments on the superpotential and Higgs parameters
similar to those mentioned in the context of the CMSSM holding, the free
parameters of the MSSM with gauge-mediated supersymmetry breaking 
are 
\be\boxed{
  \Lambda\ , \qquad M_{\rm mes}\ , \qquad  N_q= N_\ell \ , \qquad
  \tan\beta = \frac{v_u}{v_d} \qquad\text{and}\qquad {\rm sign} \big(\mu\big) \
  ,
}\label{eq:gmsbprm}\ee
in addition to the Standard Model inputs. In contrast to
gravity-mediated scenarios where the gravitino mass is related to the other soft
parameters, it has to be included, in the setup above, as an extra input.
Moreover, its mass being given by the ratio of the
supersymmetry-breaking
scale and the Planck mass, the gravitino is expected to be the lightest
supersymmetric particle. 
The value of this mass could induce, in particular, a long 
lifetime for the next-to-lightest supersymmetric particle,
leading to collider signatures with
displaced vertices.

Finally, the last class of supersymmetry-breaking scenarios
investigated in this work focuses on supersymmetry breaking by anomalies.
According to the results of Section \ref{sec:AMSB}, the soft terms given by
Eq.\ \eqref{eq:lmssmbrk2} are derived from Eq.\ \eqref{eq:amsbino}, Eq.\
\eqref{eq:amsbscal} and Eq.\ \eqref{eq:amsbtri}. These equations being
renormalization group invariant, they hold at any scale. Again, we assume
that the PMNS matrix is equal to the identity matrix and that flavor and $CP$ 
violation
are only possible through CKM mixing. Therefore, the parameters appearing in 
Eq.\ \eqref{eq:lmssmbrk2} have to be read as flavor-diagonal and real. 
Moreover, the values of $\mu$ and $b$ are obtained after imposing, as above,
a correct electroweak symmetry breaking.  The input parameters of the model
consists thus in the auxiliary mass $M_{\rm aux}$, the ratio of the Higgs
vacuum expectation values $\tan\beta$ and the sign of the $\mu$-parameter.

This very simple setup however unfortunately leads to tachyonic slepton fields
because of the form of the related anomalous dimensions, QED being an
infrared-free theory. This problem must be cured in
order to have a phenomenologically viable model and several solutions have been
proposed
\cite{Randall:1998uk, Pomarol:1999ie, Chacko:1999am, Katz:1999uw, Jack:2000cd,
Jack:2003qg, Murakami:2003pb, Kitano:2004zd, Ibe:2004gh, Hodgson:2005en,
Jones:2006re, Hodgson:2007kq}. We adopt here the phenomenological
approach of assuming non-negligible contributions to the scalar
soft masses, induced, \eg, by gravity-mediated supersymmetry breaking. This
renders all squared masses positive at the weak scale and allows to construct
viable theories. The free parameters of the model consist thus of 
\be\boxed{
 M_{\rm aux} \ , \qquad m_0 \ , \qquad \tan\beta = \frac{v_u}{v_d}
   \qquad\text{and}\qquad {\rm sign} \big(\mu\big) \ ,
}\label{eq:amsbprm}\ee
in addition to the Standard Model inputs.

\mysection{Main phenomenological features of the MSSM}\label{sec:mssmpheno}
\subsection{The hierarchy or the fine-tuning problem}\label{sec:hierarchy}
Despite its success, it seems very likely that the Standard Model has to be
extended at the TeV scale. The main motivation lies in the so-called
\textit{hierarchy
problem}, or equivalently the \textit{fine-tuning problem}, related to the
relative order of magnitude between the Planck scale and
the electroweak scale \cite{Witten:1981nf}. In the Standard Model,
successful spontaneous electroweak symmetry
breaking implies the existence of a Higgs boson whose mass $M_h$ has
been found to be of about 125 GeV.
However,
scalar fields receive enormous contributions from quantum corrections so that
the Standard Model parameters must be fine-tuned up to the 16$^{\rm th}$ digit
in order to maintain the Higgs-boson mass within the ${\cal O}(100)$ GeV range.
Retaining only the dominant top quark, $Z$-boson, $W$-boson and Higgs-boson loop
diagrams and cutting off the loop-integral momentum integration at a scale $\Lambda_{\rm
UV}$, the quantum corrections to the Higgs-boson mass are calculated as
\be
  \delta M_h^2 = \frac{3 \Lambda^2_{\rm UV}}{8 \pi^2 v^2} \Big[ M_h^2 +
2 M_W^2 + M_Z^2 - 4 M_t^2 \Big] \ ,
\ee
where $v$ is the Higgs-boson vacuum expectation value and $M_Z$, $M_W$ and $M_t$
the masses of the $Z$-boson, $W$-boson and top quark. 
The cut-off scale $\Lambda_{\rm UV}$ can be seen as the scale
where new physics
appears and where the Standard Model is known not to be valid anymore and is
thus believed to be of the order of the Grand Unification or of the
Planck scale.

\begin{figure}
 \centering
    \vspace*{-7.2cm}
   \includegraphics{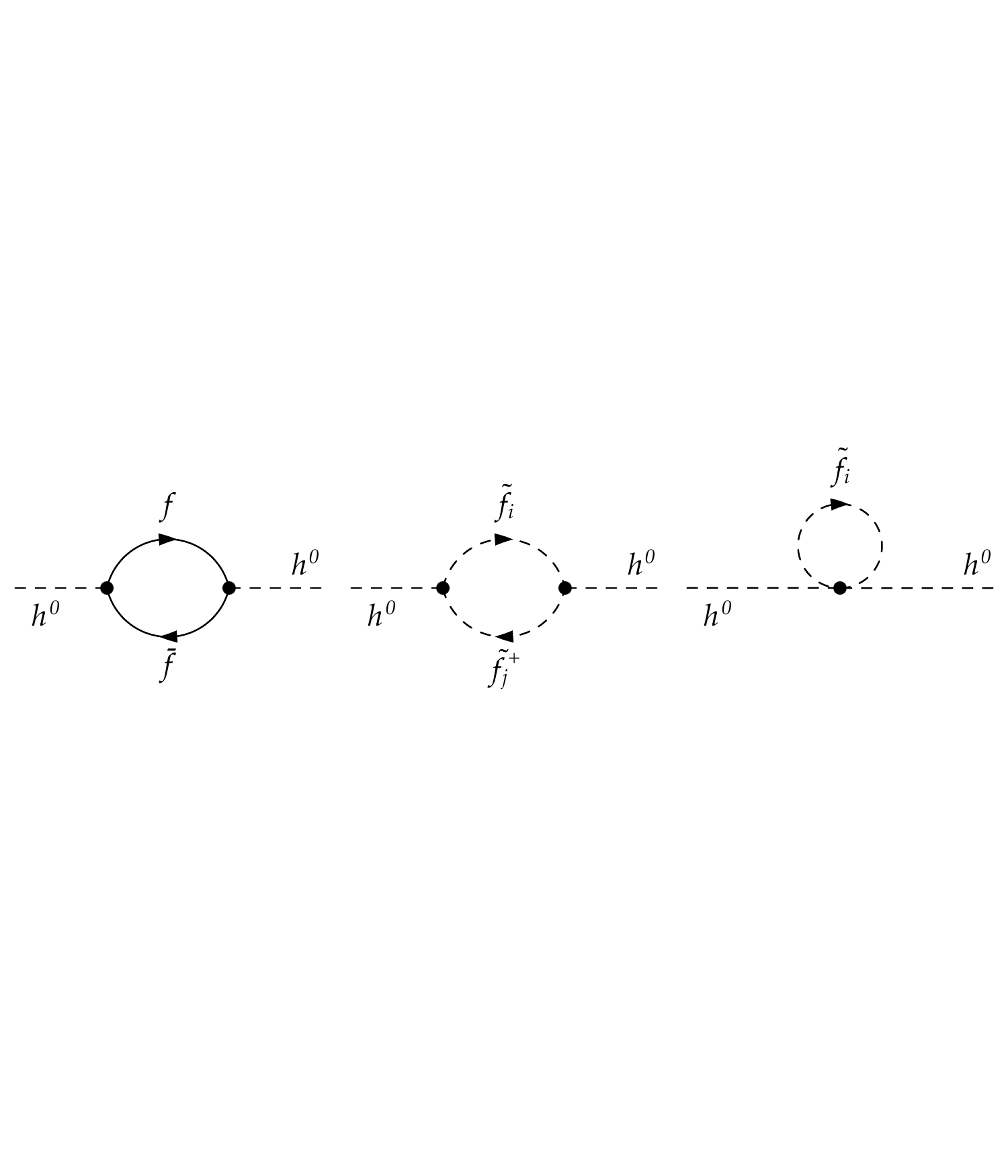}
    \vspace*{-7.5cm}
 \caption{\label{fig:hmass_ferm}Fermion and sfermion loop-contributions to the
lightest MSSM Higgs boson self-energies.}
\end{figure}

To illustrate how the hierarchy problem is cured in supersymmetry, we take the
example of the lightest Higgs boson of the MSSM. The 
results presented below can however be safely generalized to the case of any softly
broken supersymmetric theory and for any of its scalar degrees of freedom.
We start from the Lagrangian of Eq.\
\eqref{eq:gensusylag2}, adapt it to the MSSM field content (including the
diagonalization of the mass spectrum) and convert
the Weyl fermions into Dirac fermions as shown in Eq.\ \eqref{eq:weyltodirac}. 
The Feynman rule describing the interaction of the lightest Higgs boson
$h^0$ with a fermion-antifermion pair $f_k \bar f_k$\footnote{We recall that in
the super-CKM and super-PMNS bases, the Yukawa matrices are diagonal.} can then
be extracted and 
reads, considering all particles as incoming to the vertex,\\
\begin{center}\begin{tabular}{r l}
  \parbox{0.25\textwidth}{
    \vspace*{-1cm}
    \includegraphics[width=.25\columnwidth]{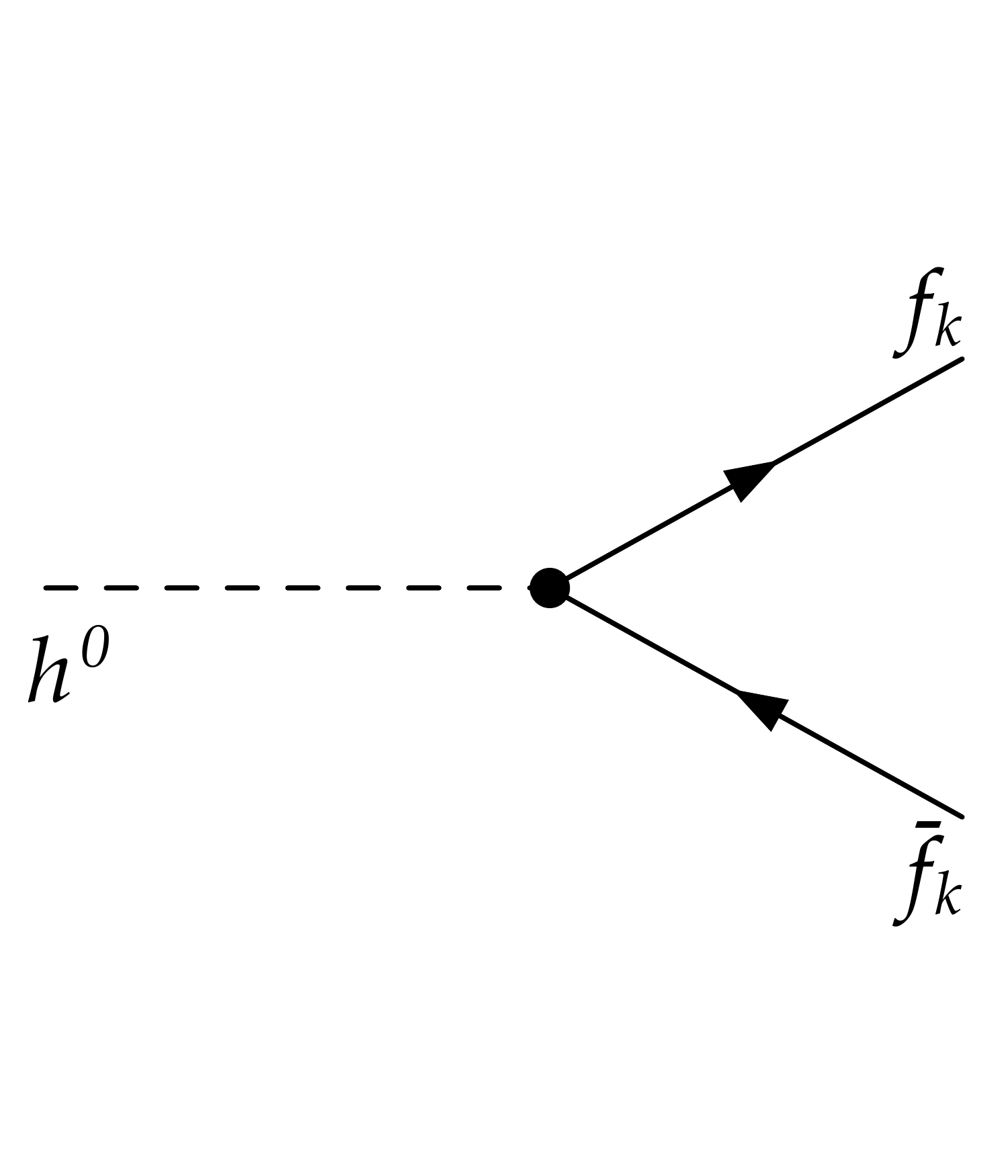}
    \vspace*{-1cm}
    }
  & \hspace{2truecm}
\parbox{0.3\textwidth}{$\displaystyle 
   -\frac{i}{\sqrt{2}} ({\bf \hat y^f})_{kk}X_\alpha$\     .}  
\end{tabular}\end{center}$~$\vspace{-0.5cm}\\
The color and spin structures being trivial, they are understood. Moreover, 
the index $k$ denote the (non-summed) generation index attached to the fermion field
$f_k$. We have also introduced the quantity
$X_\alpha = \cos\alpha$ for up-type quarks and $X_\alpha = -\sin\alpha$ for
down-type quarks and charged lepton. This
interaction induces quantum corrections to the lightest Higgs boson mass
illustrated by the Feynman diagram of 
left panel of Figure \ref{fig:hmass_ferm}, which gives, 
\be
   -i \Pi(p) = -\frac{i N_c ({\bf \hat y^f})_{kk}^2 X^2_\alpha }{16 \pi^2} \Big[
      A_0(M_f^2) - \big(\frac12 p^2 - 2 M_f^2\big) B_0(p; M_f^2,M_f^2)\Big] \ ,
\label{eq:hprop}\ee
after inserting the Feynman rule above and introducing standard
Passarino-Veltman integrals~\cite{Passarino:1978jh}.
We have denoted the mass of the fermion $f$ by $M_f$ and 
the factor $N_c$ accounts for its representation under the
QCD gauge group, being $N_c=3$ for quarks and $N_c=1$ for leptons. The
Passarino-Veltman functions appearing above diverge in four dimensions so that
we introduce an explicit cut-off scale $\Lambda_{\rm UV} \gg M_f$ to regulate
the integration over of the loop momentum. The computation of the two integrals
$A_0(M_f^2)$ and $B_0(p; M_f^2,M_f^2)$ is then straightforward and they read, in the
limit of a large scale $\Lambda_{\rm UV}$, 
\be\bsp
  A_0(M_f^2) = \int_{(q^2<\Lambda_{\rm UV}^2)} \frac{\d^4 q}{i \pi^2} \frac{1}{q^2
   - M_f^2} = -\Big[ \Lambda_{\rm UV}^2 + M_f^2
     \log\frac{M_f^2}{\Lambda_{\rm UV}^2}\Big] + {\cal
     O}\Big(\frac{1}{\Lambda_{\rm UV}}\Big)\ , \\
  B_0(p; M_f^2, M_f^2) = \int_{(q^2<\Lambda_{\rm UV}^2)} \frac{\d^4 q}{i \pi^2}
    \frac{1}{\Big[q^2 - M_f^2\Big]\Big[(q+p)^2 - M_f^2\Big]} =
    -\log\frac{M_f^2}{\Lambda^2_{\rm UV}} + {\cal O}(\frac{1}{\Lambda_{\rm UV}}) \
    .
\esp\label{eq:pavered}\ee
We deduce the corresponding
quantum corrections to the mass of the lightest MSSM Higgs boson 
from the evaluation of the propagator of Eq.\ \eqref{eq:hprop} for an on-shell
Higgs boson (see Section \ref{sec:GMSB}), \ie, for $p^2 = M_{h^0}^2$. One
gets, after summing over all fermion species, 
\be
  \delta_1 M_{h^0}^2 = \frac{1}{16 \pi^2} \sum_{f,k}\bigg[N_c ({\bf \hat
   y^f})_{kk}^2 X^2_\alpha
    \Big[ -2 \Lambda_{\rm UV}^2 + (2 M_{h^0}^2 - 6 M_f^2)
    \log\frac{M_f^2}{\Lambda^2_{\rm UV}} \Big] \bigg]+ {\cal
    O}(\frac{1}{\Lambda_{\rm  UV}})\ .
\label{eq:dmhferm}\ee
Consequently, these
corrections drive the lightest Higgs-boson squared mass by about 30 orders of
magnitude above the squared electroweak scale for a cut-off scale of the order
of the Planck mass or of the Grand Unification scale. Moreover,
the naive option of lowering $\Lambda_{\rm UV}$ to an appropriate scale ensuring
a Higgs-boson mass of about 100 GeV is excluded since this in general implies 
the failure of either unitarity or causality \cite{Eliezer:1989cr}.

In the MSSM, each of the Standard Model fermions has a scalar partner
which the couplings to the lightest Higgs boson $h^0$ are driven by
supersymmetry. These new states, absent in the Standard Model, also
contribute to $\delta M_{h^0}^2$ and 
regularize the quadratic dependence on $\Lambda_{\rm UV}$
of Eq.\ \eqref{eq:dmhferm}. From the Lagrangian of Eq.\
\eqref{eq:gensusylag2}, one can derive the Feynman rule associated with 
the three-point interaction of a $h^0$ particle with 
a sfermion pair $\tilde f_i \tilde f_j^\dag$,  \\
\begin{center}\begin{tabular}{r l}
  \parbox{0.20\textwidth}{
    \vspace*{-1cm}
    \includegraphics[width=.25\columnwidth]{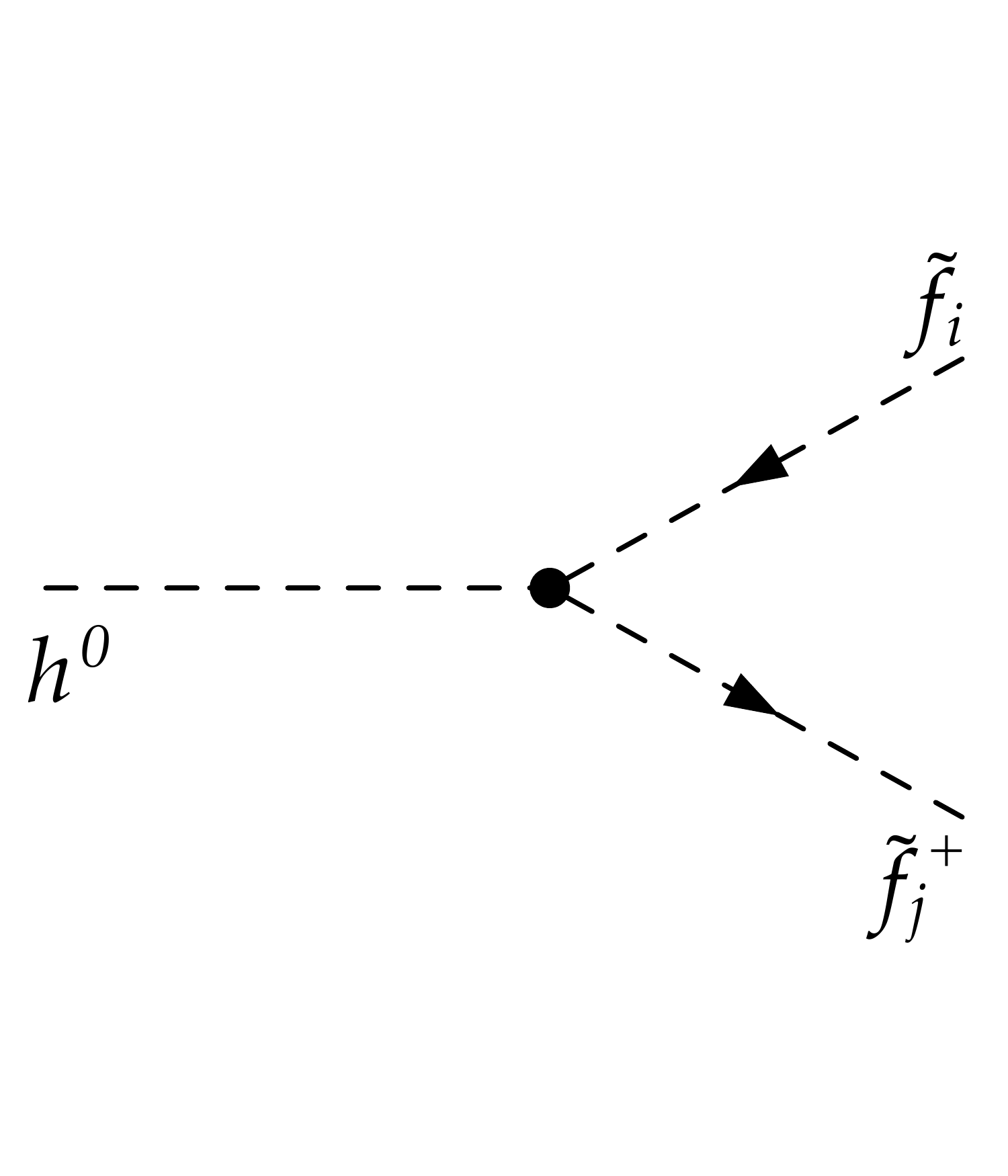}
    \vspace*{-1cm}
    }
  & \hspace{.6truecm}
\parbox{0.64\textwidth}{$\displaystyle  \frac{-i}{\sqrt{2}} C^f_{ij}
   \!\equiv\!  \frac{-i}{\sqrt{2}} \bigg\{\sum_k \Big[({\bf \hat
   y^f})^2_{kk} Y_\alpha \big( R^{f\ast}_{ik}R^f_{jk} \!+\!
   R^{f\ast}_{i(k+3)}R^f_{j(k+3)}\big) + ({\bf \hat
   y^f})_{kk} \big(Z_\alpha R^{f\ast}_{i(k+3)}R^f_{jk} + Z^*_\alpha
   R^{f\ast}_{ik} R^f_{j(k+3)}\big)\Big] + X_\alpha \sum_{k,l}\Big[\big( ({\bf
   \hat T^f})_{lk} R^{f\ast}_{ik} R^f_{j(l+3)} + ({\bf \hat
   T^f})^*_{lk} R^{f\ast}_{i(k+3)} R^f_{jl}\big) \Big]\bigg\}$\
    ,\\}  
\end{tabular}\end{center}$~$\\
all particles incoming to the vertex. 
The color structure is again understood and we have introduced the
shorthand notations $Y_\alpha = \sqrt{2}v_u X_\alpha$ ($\sqrt{2} v_d X_\alpha$)
and $Z_\alpha = \mu \sin\alpha$ ($- \mu \cos\alpha$) for up-type squarks 
(down-type squarks and charged sleptons). Moreover, in the expression above,
we do not follow the Einstein summation conventions for the generation
indices $k$ and $l$. This interaction contributes to the quantum corrections to
the lightest MSSM Higgs-boson mass, as illustrated by the loop diagram 
of the middle panel
of Figure \ref{fig:hmass_ferm}, the corresponding unrenormalized propagator
reading, after summing over all possible internal sfermion fields,
\be
   -i \Pi(p) =\frac{i N_c}{32 \pi^2} \sum_f \sum_{i,j=1}^6 \Big[ \big|
     C_{ij}^f\big|^2 B_0(p; M_{\tilde f_i}^2,M_{\tilde f_j}^2)\Big] \ .
\label{eq:hlooplogs}\ee
Introducing again a cut-off scale $\Lambda_{\rm  UV}$ for regularizing 
the integration over
the loop momentum in the Passarino-Veltman function $B_0(p; M_{\tilde
f_i}^2,M_{\tilde f_j}^2)$, one gets
\be\bsp
  B_0(p; M_{\tilde f_i}^2, M_{\tilde f_j}^2) =&\ \int_{(q^2<\Lambda_{\rm UV}^2)}
    \frac{\d^4 q}{i \pi^2}
    \frac{1}{\Big[q^2 - M_{\tilde f_i}^2\Big]\Big[(q+p)^2 - M_{\tilde
f_j}^2\Big]} \\ = &\
     \frac12\Big[ \log\frac{M_{\tilde f_i}^2}{\Lambda^2_{\rm UV}} +
    \log\frac{M_{\tilde f_j}^2}{\Lambda^2_{\rm UV}}\Big] + {\cal
    O}(\frac{1}{\Lambda_{\rm  UV}}) \ ,
\esp\ee
so that the associated
contributions to the one-loop corrections to the lightest Higgs mass are
\be
  \delta_2 M_{h^0}^2 =  -\frac{N_c}{64 \pi^2} \sum_f \sum_{i,j=1}^6 
    \Big[ \big| C_{ij}^f\big|^2 \big(\log\frac{M_{\tilde f_i}^2}{\Lambda^2_{\rm
    UV}} + \log\frac{M_{\tilde f_j}^2}{\Lambda^2_{\rm UV}}\big)\Big]
    + {\cal O}(\frac{1}{\Lambda_{\rm UV}})\ .
\ee
This result illustrates that soft supersymmetry-breaking terms ($\hat T^f$)
contribute at most logarithmically to the scalar field mass corrections and 
are then not dangerous with this respect.

Finally, the Lagrangian also contains
four-point interactions among the lightest Higgs boson and the scalar partners
of the Standard Model particles, the corresponding interaction strength being
again related to the Yukawa couplings by supersymmetry. The relevant Feynman rule
reads,\\ 
\begin{center}\begin{tabular}{r l}
  \parbox{0.25\textwidth}{
    \vspace*{-1.5cm}
    \includegraphics[width=.25\columnwidth]{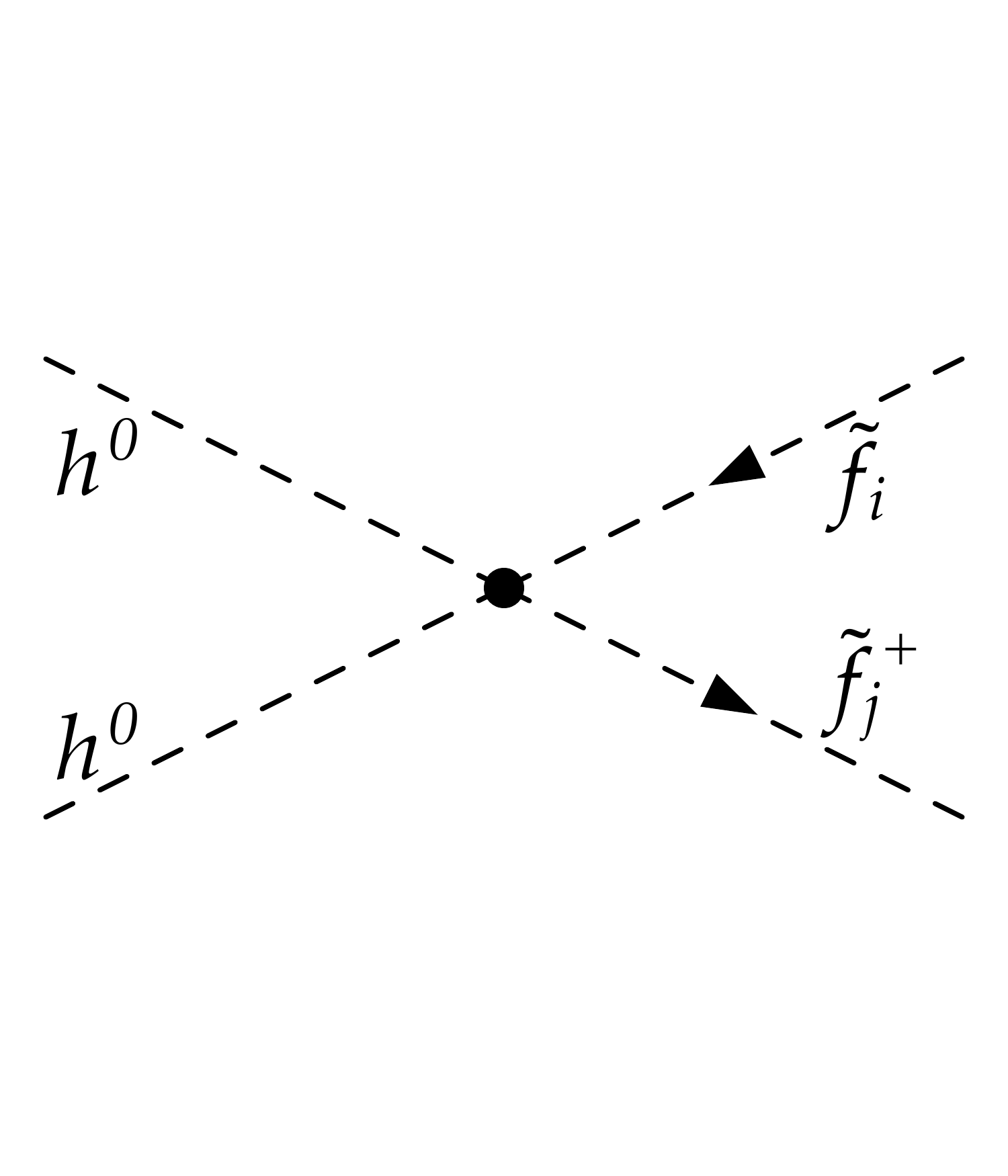}
    \vspace*{-1.7cm}
   }
  & \hspace{1.0truecm}
\parbox{0.5\textwidth}{$\displaystyle  -i X_\alpha^2 \sum_{k} 
   \Big[ ({\bf \hat y^f})^2_{kk}\big( R^{f\ast}_{ik}R^f_{jk} +
   R^{f\ast}_{i(k+3)}R^f_{j(k+3)}\big)\Big]$\
    ,\\}  
\end{tabular}\end{center}$~$\\
all particles being again incoming to the vertex and the color structure 
understood. As above, the Einstein summation conventions
do not apply to the generation index $k$, and the matrices $R^f$ denote the
$6\times 6$ sfermion mixing matrices introduced in Eq.\ \eqref{eq:sfmix}. This
interaction contributes
to the lightest Higgs boson mass quantum corrections through the diagram
presented in the right panel of Figure \ref{fig:hmass_ferm}. Evaluating this
diagram  after summing over all the scalar partners of the Standard
Model fermions, we obtain the unrenormalized propagator 
\be
   -i \Pi(p) =\frac{i N_c}{16 \pi^2} \sum_{f,k} \sum_{i=1}^6 \Big[X_\alpha^2 
    ({\bf \hat y^f})^2_{kk}\big( R^{f\ast}_{ik}R^f_{ik} +
   R^{f\ast}_{i(k+3)}R^f_{i(k+3)}\big) A_0(M_{{\tilde f}_i}^2) \Big]  \ .
\ee
The associated contribution to the lightest MSSM Higgs boson mass is then
deduced after employing the unitarity properties of the sfermion mixing
matrices and Eq.\ \eqref{eq:pavered} for the calculation of the
Passarino-Veltman function,
\be\bsp
  \delta_3 M_{h^0}^2 =&\ \sum_{f,k}\bigg\{\frac{N_c X_\alpha^2 
    ({\bf \hat y^f})^2_{kk}}{16 \pi^2} \Big[  2 \Lambda_{\rm UV}^2 + 
     \sum_{i=1}^6 \Big( R^{f\ast}_{ik}R^f_{ik} +
      R^{f\ast}_{i(k+3)}R^f_{i(k+3)}\Big) M_{\tilde f_i}^2
    \log\frac{M_{\tilde f_i}^2}{\Lambda^2_{\rm UV}} \Big] \bigg\} \\&\quad+ {\cal
   O}(\frac{1}{\Lambda_{\rm UV}})\ . 
\esp\label{eq:scaltad}\ee

Confronting this last equation to Eq.\ \eqref{eq:dmhferm}, the
quadratic divergences cancel out, which is a strong prediction for any
softly-broken supersymmetric theory. Supersymmetry associates with each fermionic
loop diagram contributing to scalar mass corrections (\eg, left panel
of Figure~\ref{fig:hmass_ferm}) 
a scalar loop (\eg, right panel of the figure). The coupling strengths being related
by supersymmetry, the sum of the two contributions makes all
quadratic divergences cancel. In addition, \textit{soft} supersymmetry breaking
leads to contributions that  
are at most logarithmically divergent and thus do not reintroduce 
the hierarchy problem. 

Another important aspect of Eq.\ \eqref{eq:scaltad} is that it 
predicts the scale of the masses of the scalar superpartners
strongly coupling to the Higgs bosons.
In order to have a Higgs mass lying in the ${\cal O}(100)$ GeV range,
the superpartners of the Standard Model top quark must hence lie around the TeV scale so
that the term in $M_{\tilde f_i}^2$ is not too large.

Similar results
hold when considering gauge boson and gaugino loops. Therefore, the
searches for the superpartners of the Standard Model
particles at the LHC, which is currently exploring the TeV scale, is one of the 
hottest topic of the present experimental program in
particle physics. These results can be generalized
for any scalar degree of freedom of a softly-broken supersymmetric theory 
and are still valid at any loop level since
Supersymmetry in fact implies a net cancellation of the quadratic divergences at all
orders.

\subsection{Gauge coupling unification}
Another good motivation for the existence of supersymmetry as a symmetry
of Nature lies in the unification of the gauge interactions at high energies, as
predicted by several supersymmetric models. The superpartners modify the
various beta functions of the model gauge group, as shown in Eq.\
\eqref{eq:betafuncgauge1} and Eq.\ \eqref{eq:betafuncgauge2}. Consequently,
it can be shown that this implies a consistent unification of the gauge coupling
at high energies \cite{Ibanez:1981yh, Dimopoulos:1981yj, Ellis:1990wk,
Amaldi:1991cn, Langacker:1991an, Giunti:1991ta}.

%
\begin{figure}[t!]
 \centering
 \includegraphics[width=.49\columnwidth]{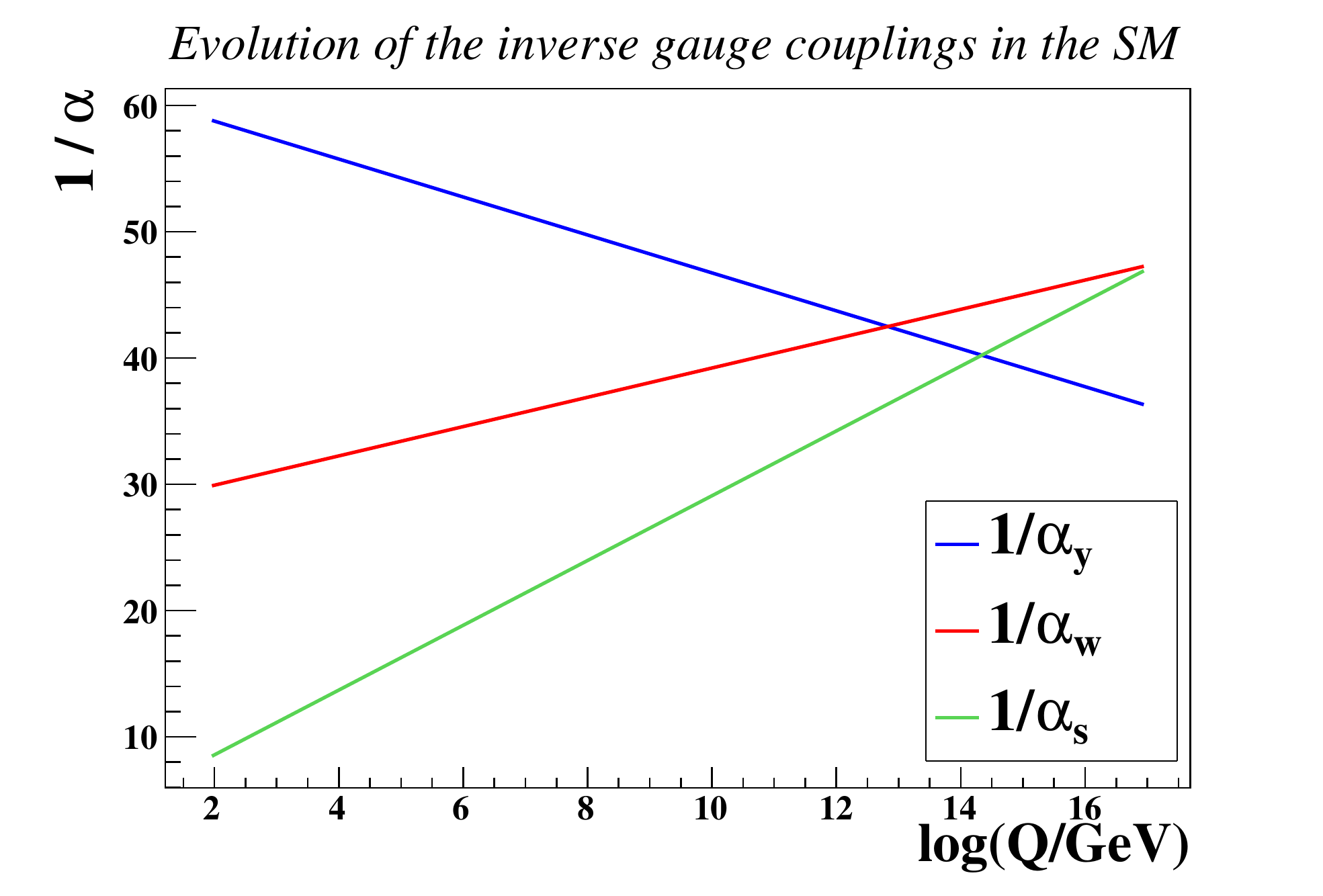}
 \includegraphics[width=.49\columnwidth]{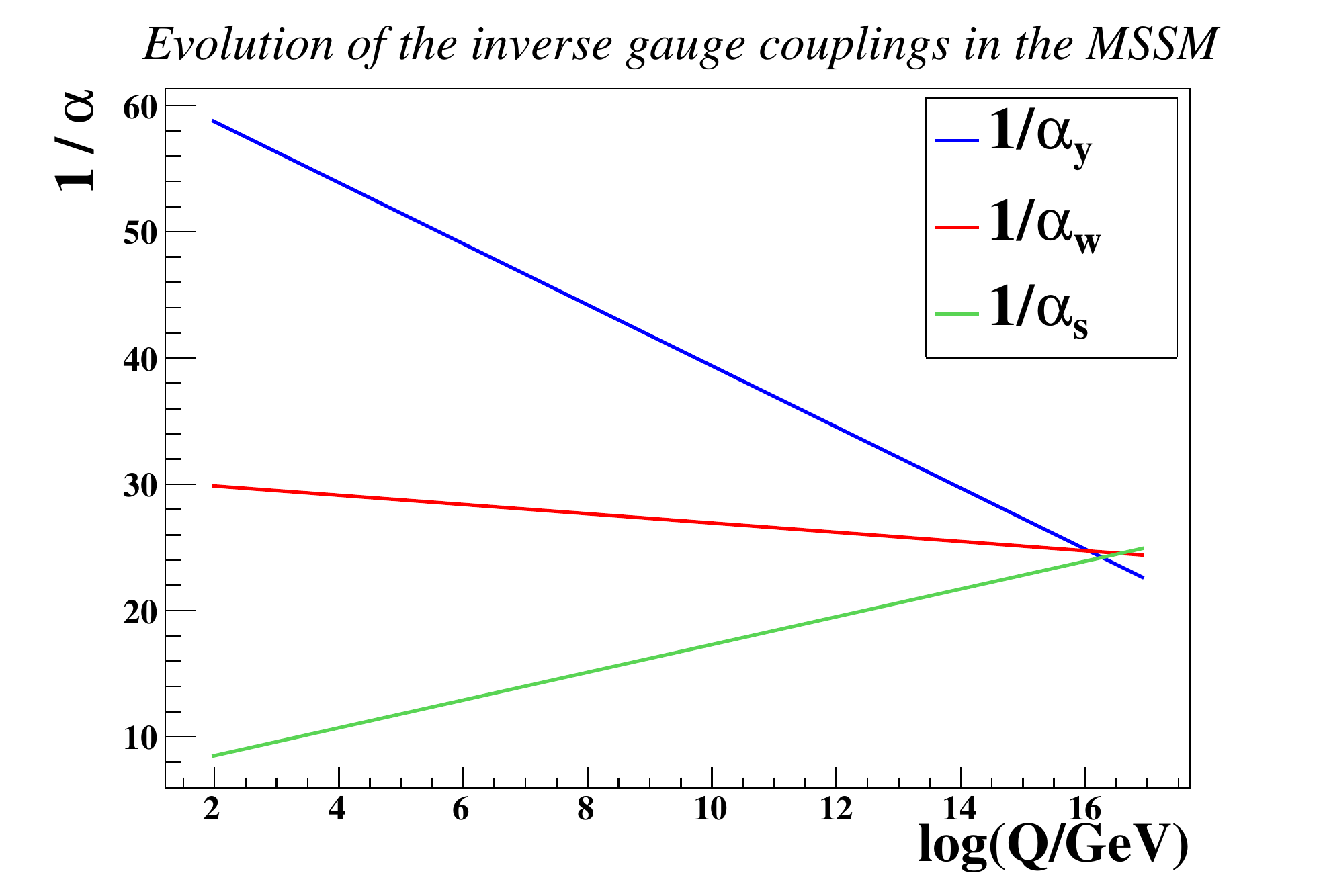}
 \caption{\label{fig:rge}
 One-loop renormalization group evolution, as a function of the energy scale $Q$, of the
  inverse hypercharge ($\alpha_y^{-1}$),
  weak ($\alpha_w^{-1}$) and strong ($\alpha_s^{-1}$) gauge couplings in the
  Standard Model (left) and in the MSSM (right).}
\end{figure}
%

Starting from Eq.\ \eqref{eq:betafuncgauge1},
the one-loop
renormalization group equations for the hypercharge, weak and strong gauge
coupling strengths $g_y$, $g_w$ and $g_s$ are given, when applied to the MSSM
case, by
\be\bsp
  \frac{{\rm d}}{{\rm d} t } g_y =  \frac{33}{80 \pi^2} g_y^3 \ , \qquad
  \frac{{\rm d}}{{\rm d} t } g_w =  \frac{1}{16 \pi^2}  g_w^3 \qquad\text{and}\qquad
  \frac{{\rm d}}{{\rm d} t } g_s =  -\frac{3}{16 \pi^2} g_s^3 \ . 
\esp\label{eq:rgemssm}\ee
The evolution variable is taken as $t = \log(Q/Q_0)$, where $Q_0$ is the scale
at which the associated boundary conditions are imposed.  
Since the experimental measurements at the $Z$-boson scale allows for 
extracting very precisely the gauge couplings, $Q_0$ is in general chosen 
equal to $M_Z$.
In contrast, in the Standard Model, the
gauge couplings evolve according to 
\be\bsp
  \frac{{\rm d}}{{\rm d} t } g_y =  \frac{41}{160 \pi^2}g_y^3  \ , \qquad
  \frac{{\rm d}}{{\rm d} t } g_w =  -\frac{19}{96 \pi^2}g_w^3   \qquad\text{and}\qquad
  \frac{{\rm d}}{{\rm d} t } g_s =  -\frac{7}{16 \pi^2} g_s^3 \ . \\
\esp\label{eq:rgesm}\ee
In the two series of equations above, the normalization of the hypercharge
coupling $g_y$ has been chosen as predicted by the grand unification of the MSSM
(or Standard Model)
gauge group into $SU(5)$ or $SO(10)$. In this setup, the electromagnetic coupling
constant $e$ is related to the hypercharge and weak coupling constants through
the sine and cosine of the weak mixing angle $\theta_w$, 
\be
  e = g_w \sin\theta_w = \sqrt{\frac35} g_y \cos\theta_w \ .
\ee
This specific choice for the normalization of $g_y$ is adequate to ensure that Tr$(T_a
T_b) = 1/2 \delta_{ab}$ for any pair $T_a$ and $T_b$ of generators of $SU(5)$ or
$SO(10)$ in the fundamental representation, including the 
hypercharge operator. Fixing the boundary conditions at the $Z$-pole as
\be
  M_Z = 91.1876 \text{ GeV }\ , \ \ \sin\theta_w^2 = 0.23361 \ , \ \
  \alpha^{-1} = \frac{4\pi}{e^2} = 127.934  \ \ \text{and}\ \
  \alpha_s = 0.118 \ ,
\ee
we show in Figure \ref{fig:rge} the evolution of the three gauge coupling
constants with
the energy. The Standard Model predictions of 
Eq.\ \eqref{eq:rgesm} are given in the
left panel of the figure,
while the MSSM results driven by the differential equations of Eq.\
\eqref{eq:rgemssm} are presented in its right panel.
In the MSSM, the three gauge
couplings appear to unify at a scale of the order of 
$2\cdot 10^{16}$ GeV, 
in contrast to the Standard Model where no unification occurs. 
This value for the unification scale can be possibly seen as a hint favoring 
grand unified
theories, which in general accommodate gauge coupling unification below the
Planck scale.

The results presented in Figure \ref{fig:rge} do not 
illustrate directly the evolution of the
gauge coupling strengths $g_i$ (with $i=y$, $w$ and $s$) but rather exploit the
one of the inverse of the quantities $\alpha_i = g_i^2/(4 \pi)$. This is motivated
by the fact that from
Eq. \eqref{eq:rgemssm} and Eq.\ \eqref{eq:rgesm}, the $\alpha_i$ coupling 
constants are found to have the convenient property to run linearly
with the renormalization group scale at the first order.

All these results can be extended to (and hold at) the two-loop level after also 
possibly including mass
thresholds for the superpartners \cite{Carena:1993ag}. In this section,
we however restrict ourselves to an illustration of the main
phenomenological features of the MSSM and rather refer to 
public packages such as \spheno\ \cite{Porod:2003um, Porod:2011nf},
\suspect\ \cite{Djouadi:2002ze} or \softsusy\ \cite{Allanach:2001kg,
Allanach:2009bv, Allanach:2011de} concerning results
beyond the one-loop level.

\subsection{Consequences of $R$-parity conservation}
Imposing the conservation of $R$-parity, whose associated quantum number is
obtained from Eq.\ \eqref{eq:rpdef}, has important phenomenological
consequences. First, the
superpartners (sfermions, gauginos and higgsinos) are $R$-parity odd particles,
\be\bsp
  & R(\widetilde H_u) = R(\widetilde H_d) = R(\widetilde B) = R(\widetilde W) = 
    R(\widetilde g) = -1 \ , \\
  & R(\tilde q_L) = R(\tilde u_R) = R(\tilde d_R) = R(\tilde \ell_L) = R(\tilde
    e_R) = -1\ ,
\esp\ee
flavor indices being understood, 
while the particles of the Standard Model sector (quarks and leptons, gauge
and Higgs bosons) are $R$-parity even states,
\be\bsp
  & R(H_u) = R(H_d) = R(B) = R(W) = R(g) = +1 \ , \\
  & R(q_L) = R(u_R) = R(d_R) = R(\ell_L) = R(e_R) = +1\ .
\esp\ee
Next, the exact
conservation of $R$-parity implies that no particle mixing occurs between the
Standard Model particles and their superpartners, as already shown in Section
\ref{sec:mssmewsb}. In contrast, $R$-parity violating
superpotential contributions 
such as the bilinear term of Eq.\ \eqref{eq:wrpv} leads to important
mixings
among the neutral particles of the spectrum (neutrinos and neutralinos; 
sneutrinos and neutral Higgs bosons) as well as among their charged counterparts
(charged leptons and charginos; charged Higgs bosons and charged sleptons). 

Furthermore, $R$-parity conservation imposes that each interaction vertex
always involves an even number of $R$-parity-odd particles. Consequently, each
superparticle always decays into an odd number of other, lighter,
superparticles. The lightest supersymmetric particle is thus stable since 
all possible decay channels are kinematically
closed. In addition, if this state is electrically and color neutral, it
can be seen as an attractive candidate to explain the presence of
non-baryonic dark matter in the universe \cite{Goldberg:1983nd, Ellis:1983ew}.

From the point of view of collider experiments, supersymmetric
particles are always produced in pairs and each 
of the produced particles further decays following a possibly complicated 
chain where the latest link consists of the lightest supersymmetric particle, forming
cascade decays \cite{Baer:1985yd,
Gamberini:1986eg, Baer:1986au, Barnett:1987kn}.
Moreover, once produced (either directly or from the decays
of heavier superparticles), the lightest superpartner escapes the detector invisibly, which
brings us to the typical supersymmetric signatures searched for
at colliders, consisting of final states containing a fair amount of missing
transverse energy produced in association with a possibly large number of jets and
charged leptons.

\mysection{Constraints from low-energy and electroweak precision measurements}
\label{sec:mssm_indirect}
Apart from direct searches at colliders, supersymmetric 
scenarios can be indirectly constrained by means of 
numerous low-energy and electroweak precision measurements restricting the
masses and mixings of the superpartners. We dedicate this section to the study
of the most stringent of these constraints and translate their effects in terms
of parameter space scans in the context of the three supersymmetry-breaking
scenarios presented in Section \eqref{sec:mssmbrkex}, \ie, the constrained
version of the MSSM with four parameters and a sign (see Eq.\
\eqref{eq:cmssmprm}), the MSSM with gauge-mediated supersymmetry breaking 
described by four parameters and a sign (see Eq.\
\eqref{eq:gmsbprm}) and the MSSM with anomaly-mediated supersymmetry-breaking 
modeled by three parameters and a sign (see Eq.\
\eqref{eq:amsbprm}). 

\subsection{Rare $B$-meson decays}\label{sec:raredec}
Flavor physics observables are in general very sensitive to new physics, the
associated measurements being one of the most promising ways to indirectly probe
beyond
the Standard Model effects. In particular, rare $B$-meson decays are already
loop-suppressed in the Standard Model so that new physics
diagrams are expected to contribute with the same order of
magnitude. 

The framework traditionally employed to compute
theoretical predictions for $B$-physics observables consists of a low-energy
heavy-quark effective field theory described by the Hamiltonian
\cite{Buchalla:1995vs}
\be
  \mathcal{H}_{\rm eff} = -2 \sqrt{2} G_F\ (V_{\rm CKM})_{ts}^\ast\ (V_{\rm
    CKM})_{tb}\  \sum_i {\cal C}_i(\mu_b) {\cal O}_i(\mu_b) \ ,
\label{eq:heffb}\ee 
where $G_F$ is the Fermi constant. The most relevant effective operators ${\cal O}_i$
in the context of calculations related to $B$-decays 
depend on the electromagnetic and strong coupling
constants $e$ and $g_s$, on the pole mass of the bottom quark $M_b$ and on 
the photon and gluon field strength
tensors $F^{\mu\nu}$ and $g^{\mu\nu}$. Introducing in addition the fundamental
representation matrices of the $SU(3)_c$ gauge group $T^a$, these operators read, 
employing four-component notations for the fermionic fields, 
\be\bsp
  {\cal O}_7 =&\ \frac{e M_b}{16\pi^2} \sbar_L \gamma_{\mu\nu} b_R F^{\mu\nu}\ ,
    \qquad
  \tilde{\cal O}_7 = \frac{e M_b}{16\pi^2} \sbar_R
    \gamma_{\mu\nu} b_L F^{\mu\nu} \\   
  {\cal O}_8 =&\ \frac{g_s M_b}{16\pi^2} \sbar_L \gamma_{\mu\nu} T^a b_R
    g^{\mu\nu}_a\ , \qquad
  \tilde{\cal O}_8 = \frac{g_s M_b}{16\pi^2} \sbar_R \gamma_{\mu\nu} T^a b_L
    g^{\mu\nu}_a \ , \\
  {\cal O}_9 =&\ \frac{e^2}{16\pi^2} \big[\sbar_L
    \gamma_\mu b_L\big] \big[ \ellbar \gamma^\mu\ell\big] \ , \qquad
  {\cal O}_{10} = \frac{e^2}{16\pi^2} \big[\sbar_L
    \gamma_\mu b_L\big] \big[ \ellbar \gamma^\mu \gamma^5\ell\big] \ . 
\esp \ee
In these expressions, we have explicitly
indicated, as a subscript, the left-handed ($L$) or
right-handed ($R$) chirality of the charm ($c$), strange ($s$) and bottom ($b$)
quark fields instead of employing the chirality projectors $P_L$ and $P_R$. 
Moreover, we generically note a charge lepton field as
$\ell$.
Getting back to Eq.\ \eqref{eq:heffb}, the renormalization scale $\mu_b$
is conveniently chosen of order of the mass of the bottom quark $M_b$, which 
ensures that
all large logarithmic terms which could appear in computations
based on the Hamiltonian $\mathcal{H}_{\rm eff}$ are embedded in the
Wilson coefficients ${\cal C}_i$.

In order to perform accurate and reliable predictions,  a calculation of the
Wilson coefficients beyond the leading order in QCD is mandatory. In the
framework of the MSSM, these quantities are known up
to two loops with respect to QCD radiative corrections \cite{Kagan:1998bh,
Kagan:1998ym} and up to one loop when including supersymmetric
particles~\cite{Hou:1992sy, Logan:2000iv, Baek:2001kh, Bobeth:2001jm,
Buras:2002vd, Huber:2005ig, Hahn:2005qi, Lunghi:2006hc}. Under that setup, 
the theoretical uncertainties
are under good control and found to be reduced to a level of about $10\%$
\cite{Buras:1997bk, Chetyrkin:1996vx}. The MSSM predictions obtained in that
way can be confronted to experimental data in order to extract constraints on the
squark, chargino, neutralino and gluino masses and couplings.
More generally, these bounds can be translated in terms of limits on the reduced
set of model parameters associated with the supersymmetry-breaking scenarios. 

The inclusive branching ratio
BR($b\to s\gamma$) has been determined from combined measurements of the BABAR,
BELLE, and CLEO experiments \cite{Amhis:2012bh} and reads
\be\label{eq:bsg}
  {\rm BR}(b\to s\gamma) = \big( 3.55 \pm 0.24_{\rm exp} \pm 0.23_{\rm theo}
    \big) \times 10^{-4}  \ ,
\ee
where the theoretical uncertainties have been chosen according to the
discussions presented in Ref.\ \cite{Hurth:2003dk} and Ref.\
\cite{Misiak:2006zs}. Possible contributions
arising from supersymmetric diagrams have however been ignored. For these reasons,
the ranges employed in the analysis of this section are estimated at the $2\sigma$ level,
both for this observable as well as for any other
of the observables presented below. In the context of a more general version of
the MSSM where non-minimal $6 \times 6$ squark mixings are allowed (see Eq.
\eqref{eq:sfmix}), this measurement represents
one of the most stringent constraints on the second and third generation mixing
parameters in many popular supersymmetry-breaking mechanisms \cite{Bozzi:2007me,
Fuks:2008ab, Fuks:2011dg}. 
A second important observable consists of the
$b\to s\mu^+\mu^-$ branching fraction, experimentally measured as \cite{Amhis:2012bh} 
\be\label{eq:bsmumu}
  {\rm BR}(b\to s\mu^+\mu^-) = \big( 2.23 \pm 0.98_{\rm exp} \pm
    0.11_{\rm theo} \big) \times 10^{-6} \ ,
\ee
where we have taken the theoretical uncertainty from the results presented in
Ref.\ \cite{Huber:2007vv}. Finally, due to very recent experimental data, 
the $B_s^0$-meson
decay rate to a muon-antimuon pair has also become a strong constraint on new physics,
with a branching ratio given by~\cite{Aaij:2012ct},
\be\label{eq:bs0}
  {\rm BR}(B_s^0 \to \mu^+\mu^-)  = \Big( 3.2^{+1.5}_{-1.2}\Big) \times 10^{-9}
    \ ,
\ee
where theoretical uncertainties are included in the errors. This measurement 
performed by the LHCb collaboration
consists of the first evidence for the $B_s^0 \to \mu^+\mu^-$ decay.
Data has been found compatible with the Standard Model
expectation, the order of magnitude of the branching fraction
being explained by both a loop-suppression and a helicity-suppression associated with
the small muon mass.

Other observables could also be employed to constrain new
physics, such as the branching ratios BR$(B_u \to \tau\nu_\tau)$ or
BR$(B\to D \tau\nu_\tau)$. However, either their predictions cannot be computed
in an accurate fashion due to the uncertainties on the
relevant model parameters, or the associated experimental analysis is rather
complex, which lead to large errors on the measurements. Therefore, we focus from now 
on the three $B$-meson decay constraints given by Eq.\ \eqref{eq:bsg}, Eq.\ \eqref{eq:bsmumu}
and Eq.\ \eqref{eq:bs0}.

To compute these observables in the context of the supersymmetry-breaking
mechanisms of Section \ref{sec:mssmbrkex}, our procedure starts at a
high-energy scale with the few input parameters given either in Eq.\
\eqref{eq:cmssmprm}, Eq.\ \eqref{eq:gmsbprm} or Eq.\ \eqref{eq:amsbprm}.
The soft
supersymmetry-breaking terms at the electroweak scale are then obtained through
renormalization group running using the \spheno\ package version 3.2.1
\cite{Porod:2011nf}, which solves the renormalization group equations
numerically to two-loop order (see Section \ref{sec:rge}). This program next
extracts the particle spectrum and mixings at the electroweak scale including 
one-loop corrections to the mass matrices of matter and gauge fields
\cite{Pierce:1996zz} and both one-loop and two-loop contributions to the  
Higgs mass matrices~\cite{Chankowski:1992er, Dabelstein:1994hb,
Degrassi:2001yf, Brignole:2001jy, Brignole:2002bz, Dedes:2002dy, Dedes:2003km}.
Finally, it computes several flavor physics observables, and in particular 
the branching ratio associated with the inclusive $b\to s
\gamma$ decay \cite{Baek:2001kh, Bobeth:2001jm, Lunghi:2006hc}, the one related
to the inclusive $b\to s\mu^+ \mu^-$ decay \cite{Baek:2001kh, Bobeth:2001jm,
Huber:2005ig} and the exclusive branching ratio BR$(B_s^0 \to \mu^+ \mu^-)$
\cite{Baek:2001kh, Logan:2000iv, Buras:2002vd}.  

For the numerical values of the
Standard Model parameters, we fix the top quark pole mass to $M_t = 173.5$ GeV, 
the bottom quark mass to $M_b(M_b) = 4.2$ GeV and the $Z$-boson mass to $M_Z =
91.1876$ GeV. The Fermi constant has been taken as $G_F = 1.16637 \!\cdot\!
10^{-5}$ GeV$^{-2}$, and the strong and electromagnetic coupling constants at the
$Z$-pole as $\alpha_s(M_Z) = 0.1176$ and $\alpha(M_Z)^{-1} = 127.934$
\cite{Beringer:2012zz}. Based on LHC searches with about 1 fb$^{-1}$ of data and
on various sources of indirect constraints, benchmark planes for future searches
on supersymmetry at the LHC have been defined from discussions among the
supersymmetry working groups of the ATLAS and CMS experiments, together the LHC
Physics Center at CERN \cite{AbdusSalam:2011fc}. For
cMSSM scenarios, two $(m_0, m_{1/2})$ planes have been
proposed, motivated by the 
constraints derived from the measurements of the anomalous magnetic moment of
the muon (see Section \ref{sec:gm2}) and the rare $b\to s\gamma$ decay. Only
models with a positive
off-diagonal Higgs mixing parameter $\mu>0$ and $\tan\beta=10$, $A_0=0$ GeV or
$\tan\beta=40$, $A_0=-500$ GeV have been adopted. While flavor bounds are
not likely to strongly constrain regions of the parameter space with a 
low $\tan\beta$ value,
loop-induced flavor-changing couplings of the Higgs bosons of the MSSM are
enhanced for large values of $\tan\beta$ so that important effects 
are expected in the $\tan\beta=40$ case.

%
\begin{figure}[t!]
 \centering
 \includegraphics[width=.32\columnwidth]{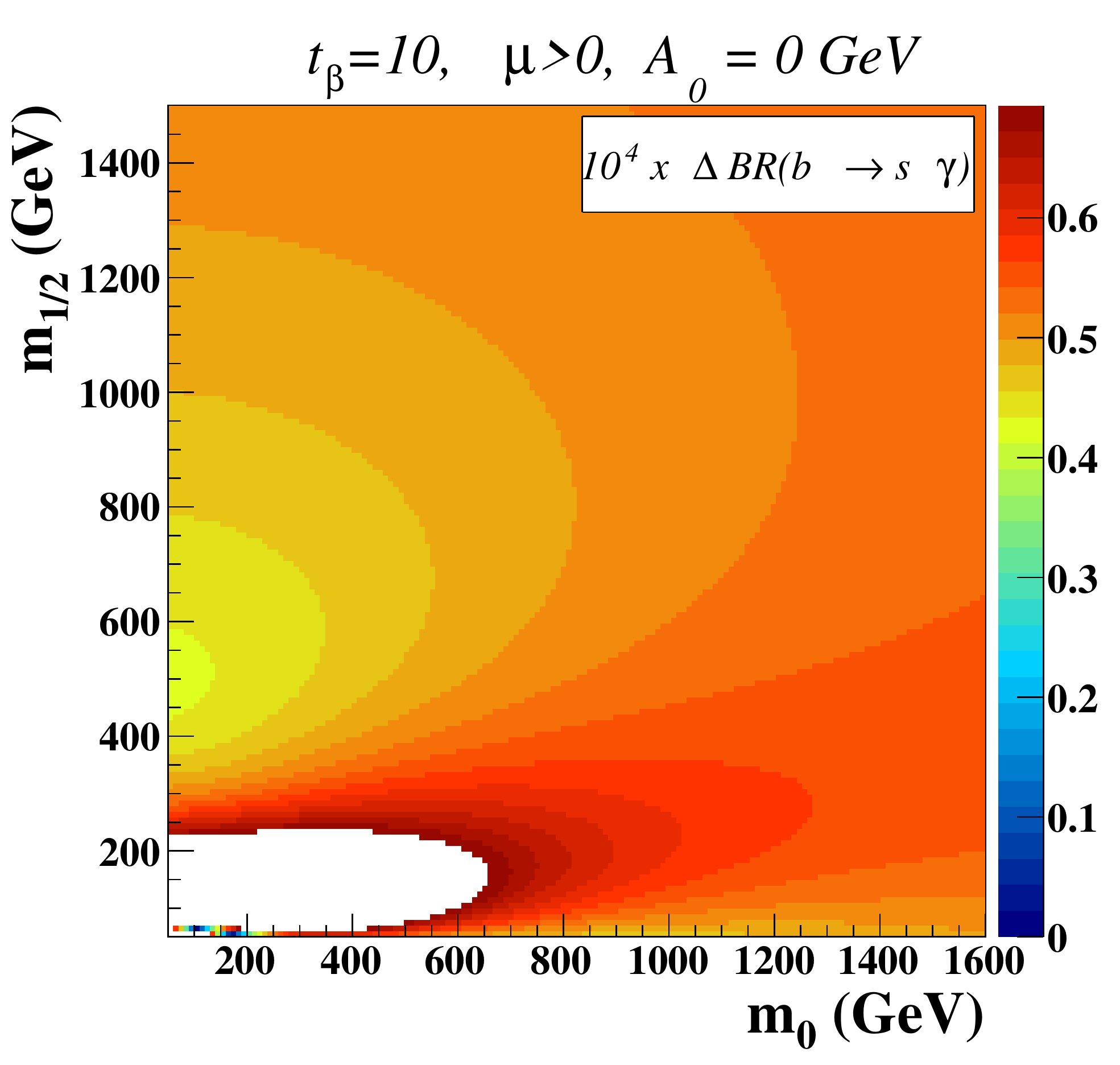}
 \includegraphics[width=.32\columnwidth]{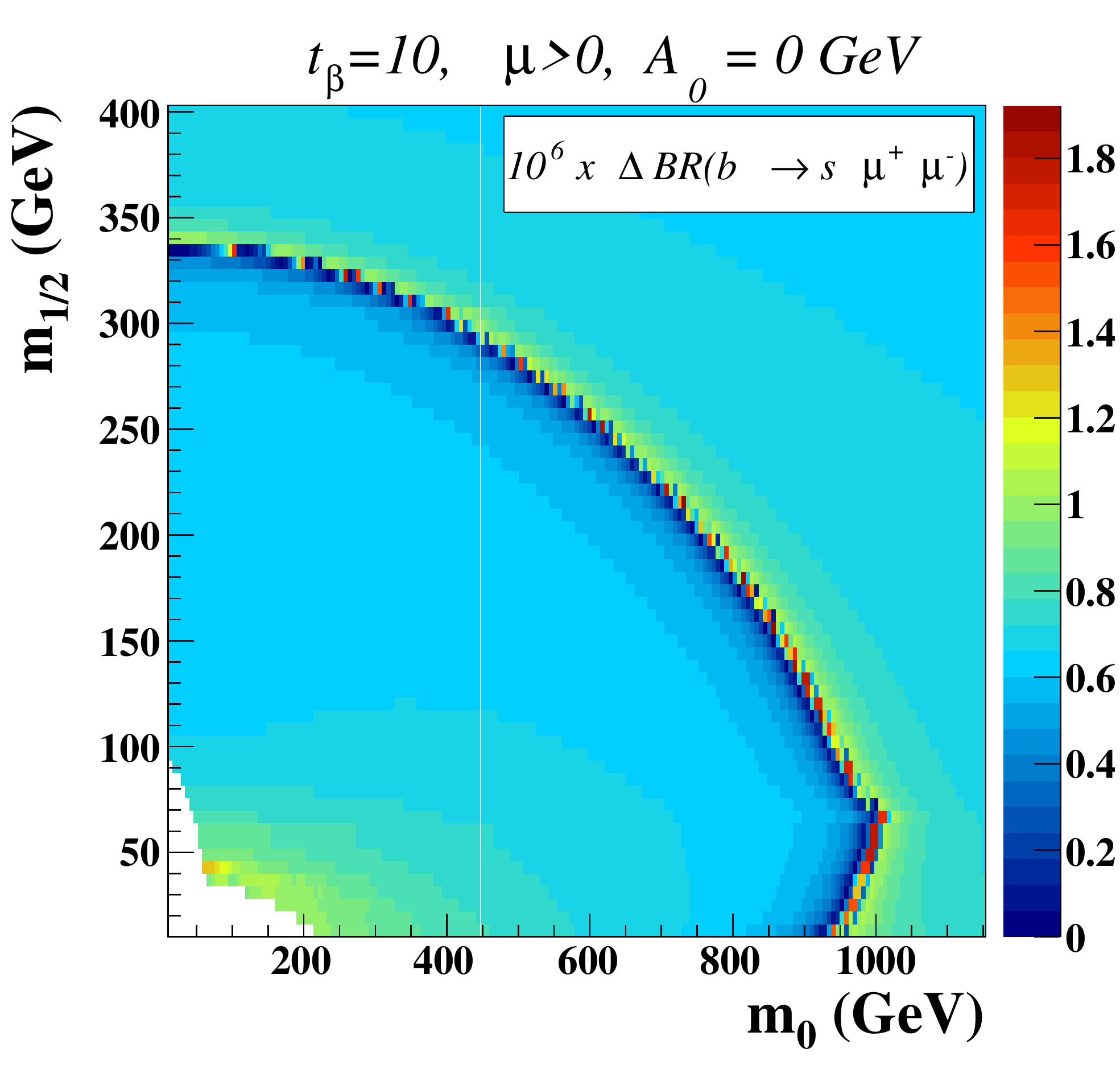}
 \includegraphics[width=.32\columnwidth]{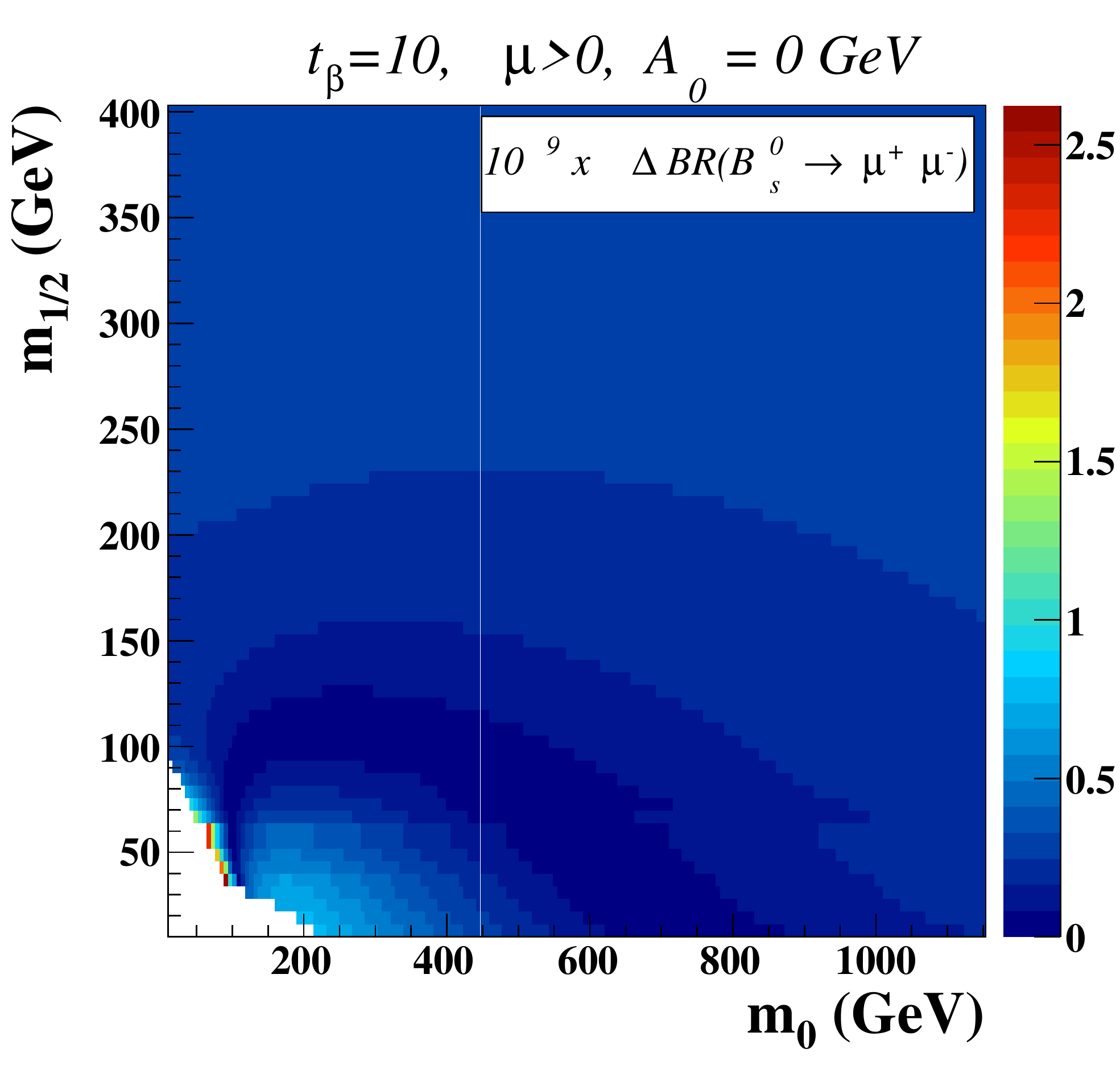}
 \caption{\label{fig:cmssm10_bdecays} Theoretical predictions for the branching
ratios associated with the inclusive $b\to s\gamma$ (left panel) and $b\to s
\mu^+ \mu^-$ (central panel) decays as well as with the exclusive $B_s^0 \to
\mu^+\mu^-$ decay given in terms of deviations from the central measured
values given in Eq.~\eqref{eq:bsg}, Eq.\ \eqref{eq:bsmumu} and Eq.\
\eqref{eq:bs0}. We present the results in $(m_0,m_{1/2})$-planes of the cMSSM 
for fixed values of $\tan\beta=10$, $A_0=0$ GeV and a positive Higgs mixing
parameter $\mu>0$. The regions depicted in white correspond to excluded regions
when applying the bounds of Eq.\ \eqref{eq:bsg}, Eq.\ \eqref{eq:bsmumu} and Eq.\
\eqref{eq:bs0} at the $2\sigma$-level, or to regions for which there is no solution
to the supersymmetric renormalization group equations.}
\vspace{.3cm}
 \centering
 \includegraphics[width=.32\columnwidth]{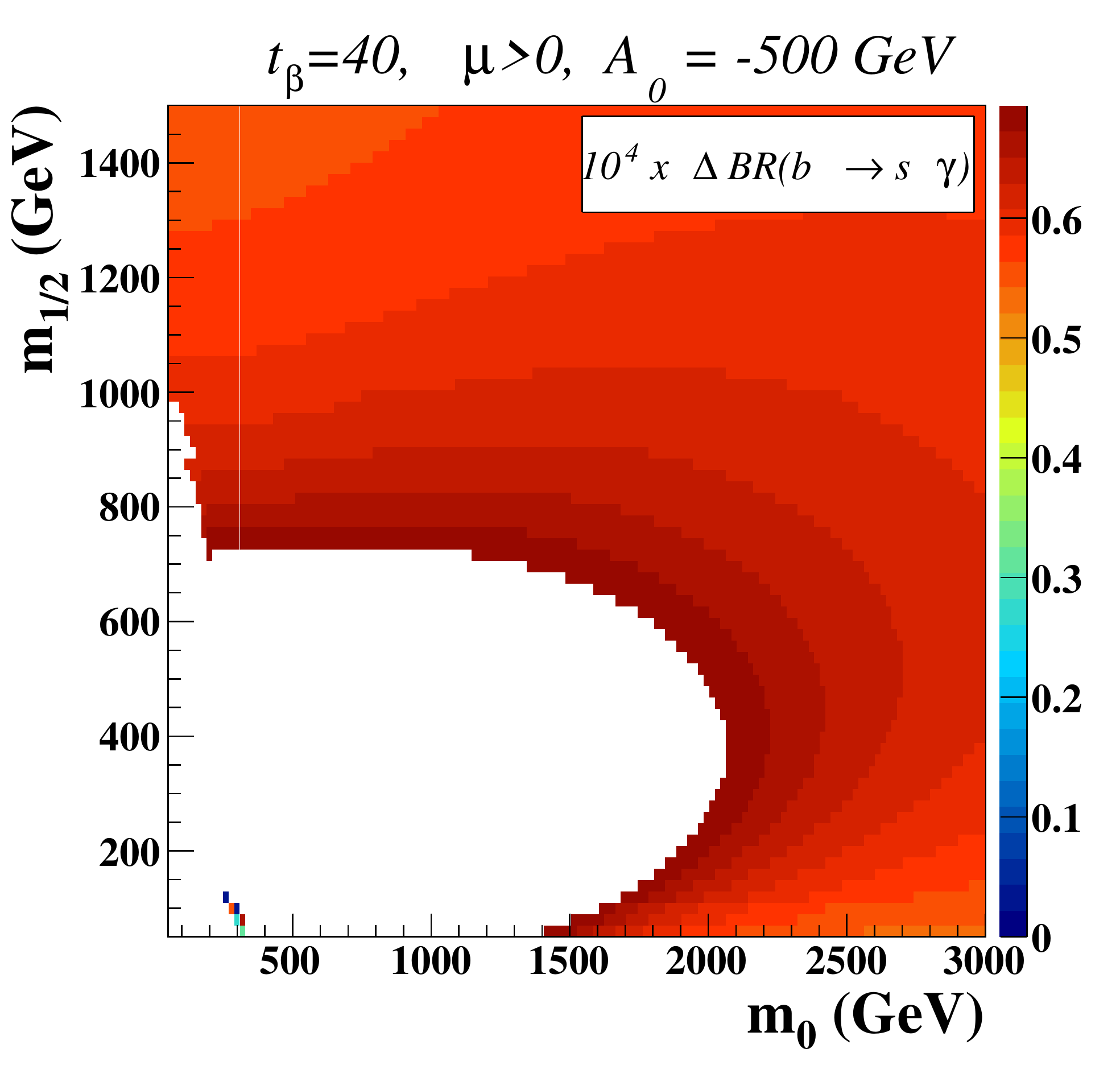}
 \includegraphics[width=.32\columnwidth]{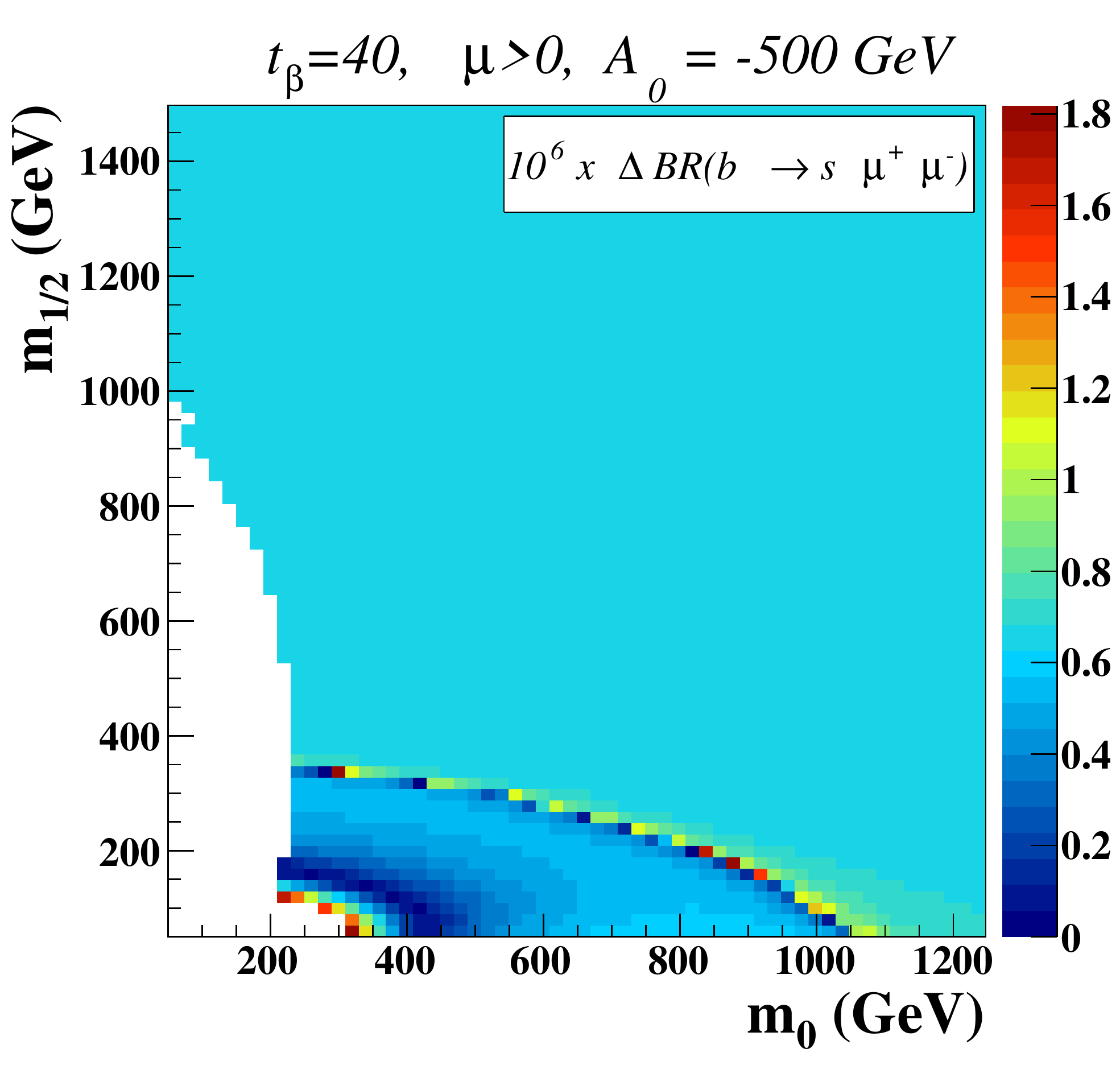}
 \includegraphics[width=.32\columnwidth]{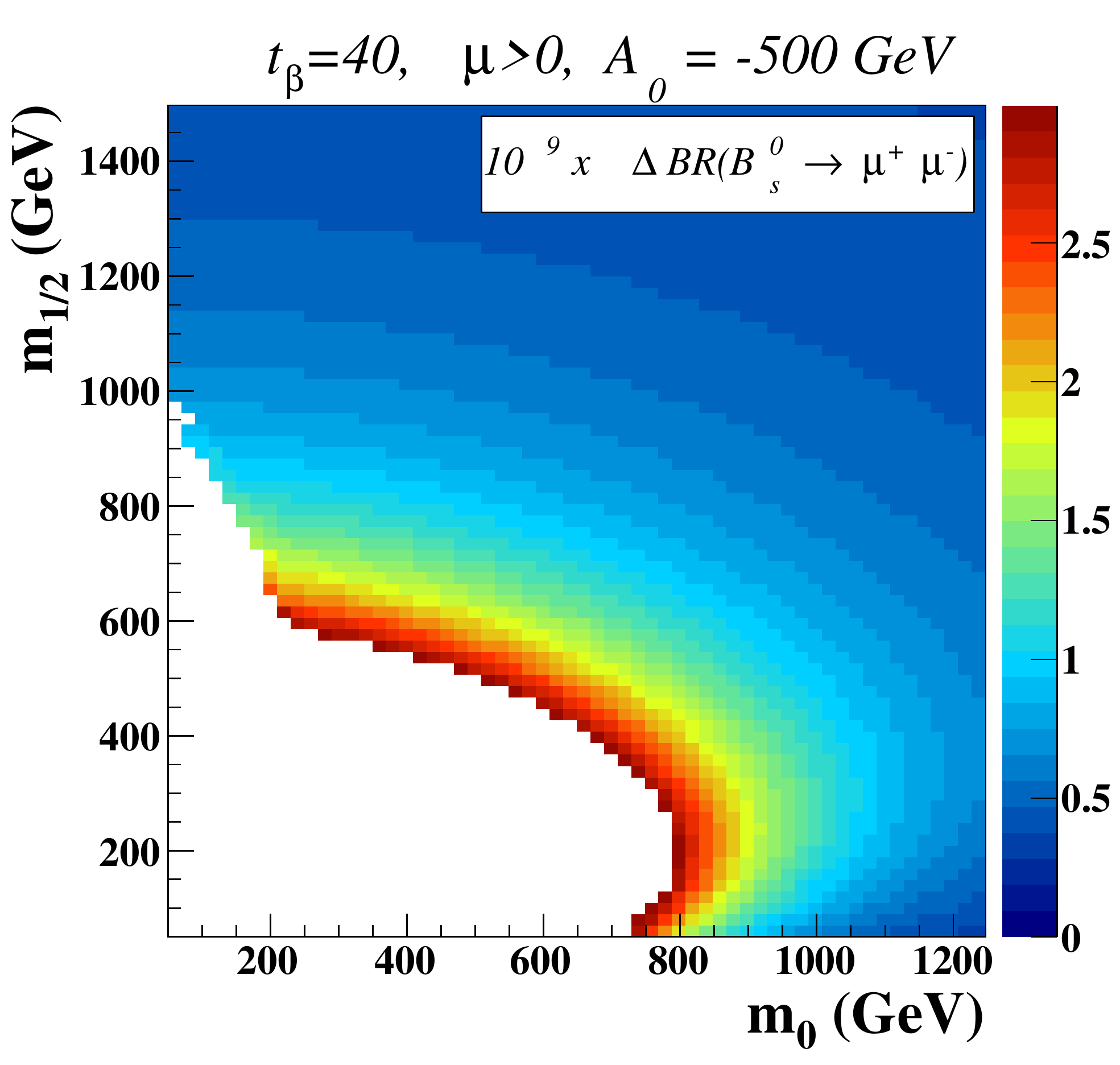}
 \caption{\label{fig:cmssm40_bdecays} Same as Figure \ref{fig:cmssm10_bdecays}
but for fixed values of
$\tan\beta=40$, $A_0=-500$ GeV and a positive Higgs mixing
parameter $\mu>0$.}
\end{figure}
%

The theoretical predictions associated with the three $B$-physics observables
discussed above are illustrated, for these two planes, on Figure
\ref{fig:cmssm10_bdecays} and Figure \ref{fig:cmssm40_bdecays}, respectively. In
the left panel of the figures, we present theoretical predictions for the   
inclusive $b\to s \gamma$ branching ratio as deviations from the central
value given in Eq.\ \eqref{eq:bsg}. In addition, we depict as white areas 
parameter space regions either excluded by the experimental limits of Eq.\
\eqref{eq:bsg}, imposed at the $2\sigma$-level, or for which there is no
low-energy phenomenologically viable solution to the supersymmetric
renormalization group equations. 

For the
two considered benchmark planes, the low-mass regions, even if attractive from a
collider point of view, are strongly disfavored by indirect constraints
derived from the measurement of the
$b\to s \gamma$ branching ratio that is compatible with the Standard Model
predictions. For relatively small values of $m_0$ and
$m_{1/2}$, theoretical predictions are never found included in the
$2 \sigma$-range deduced from Eq.\ \eqref{eq:bsg}. Moreover, comparing results
for $\tan\beta=10$ and $\tan\beta=40$, the constraints are found, as expected, 
enhanced for larger values of $\tan\beta$. The construction of
viable,
collider-friendly, benchmark scenarios with such a large $\tan\beta$ value is therefore
challenging. Acceptable choices can however be made
in two ways. The first option consists of keeping the
gauginos relatively light (a small universal gaugino mass
$m_{1/2}$) together with imposing the scalar particles of the model to be heavy
(large universal scalar mass $m_0$ of several TeV). This allows for the heavy scalar propagators
to tame the supersymmetric loop-diagram contributions to the $b\to s
\gamma$ predictions, so that they
lie well within the $2\sigma$-window derived from Eq.\
\eqref{eq:bsg}. The second option consists of fixing
$m_{1/2}$ above one TeV, implying heavy gauginos that then reduce the supersymmetric
effect on the $b\to s \gamma$ predictions for any value of $m_0$.

%
\begin{figure}[t!]
 \centering
 \includegraphics[width=.32\columnwidth]{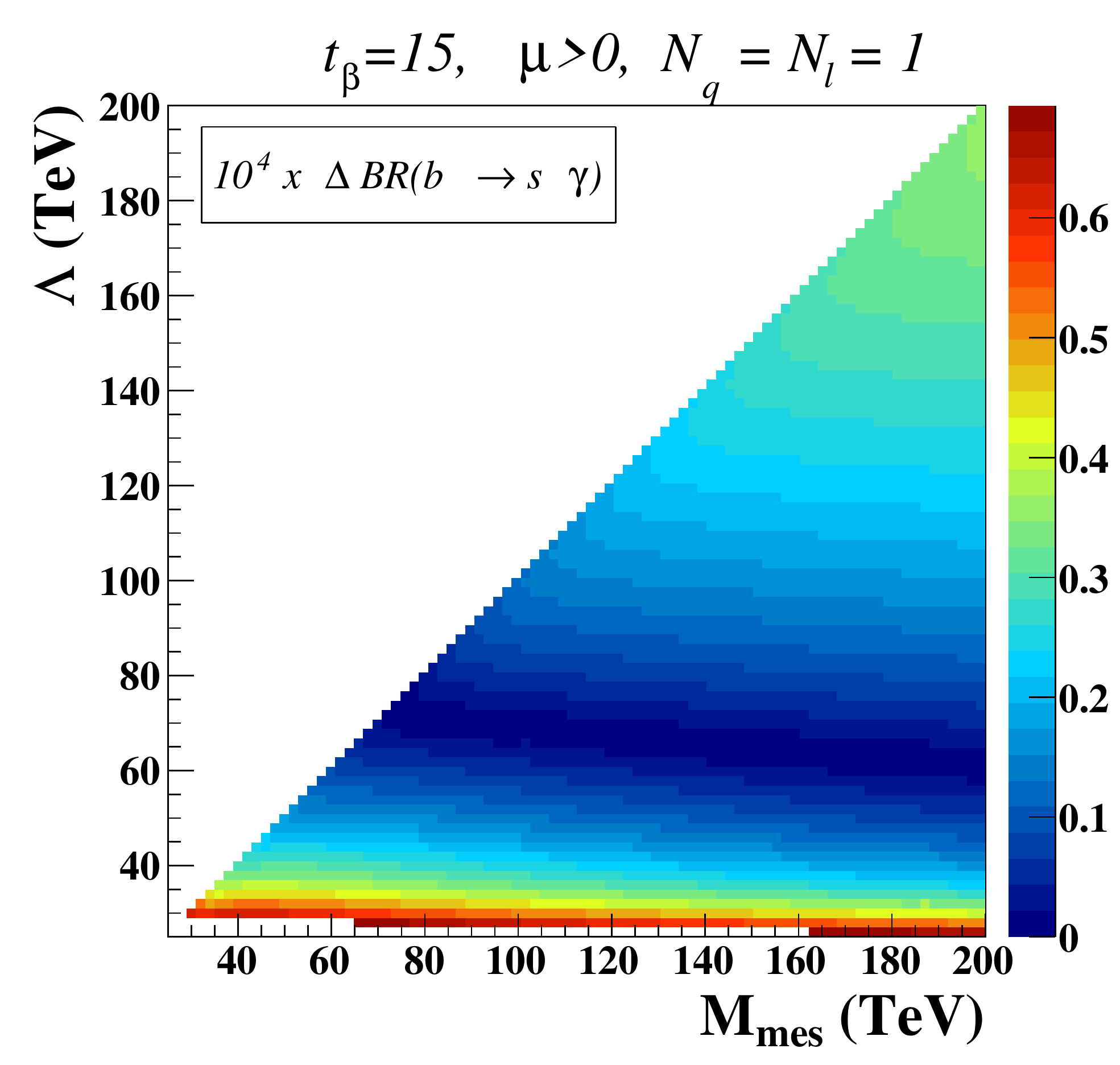}
 \includegraphics[width=.32\columnwidth]{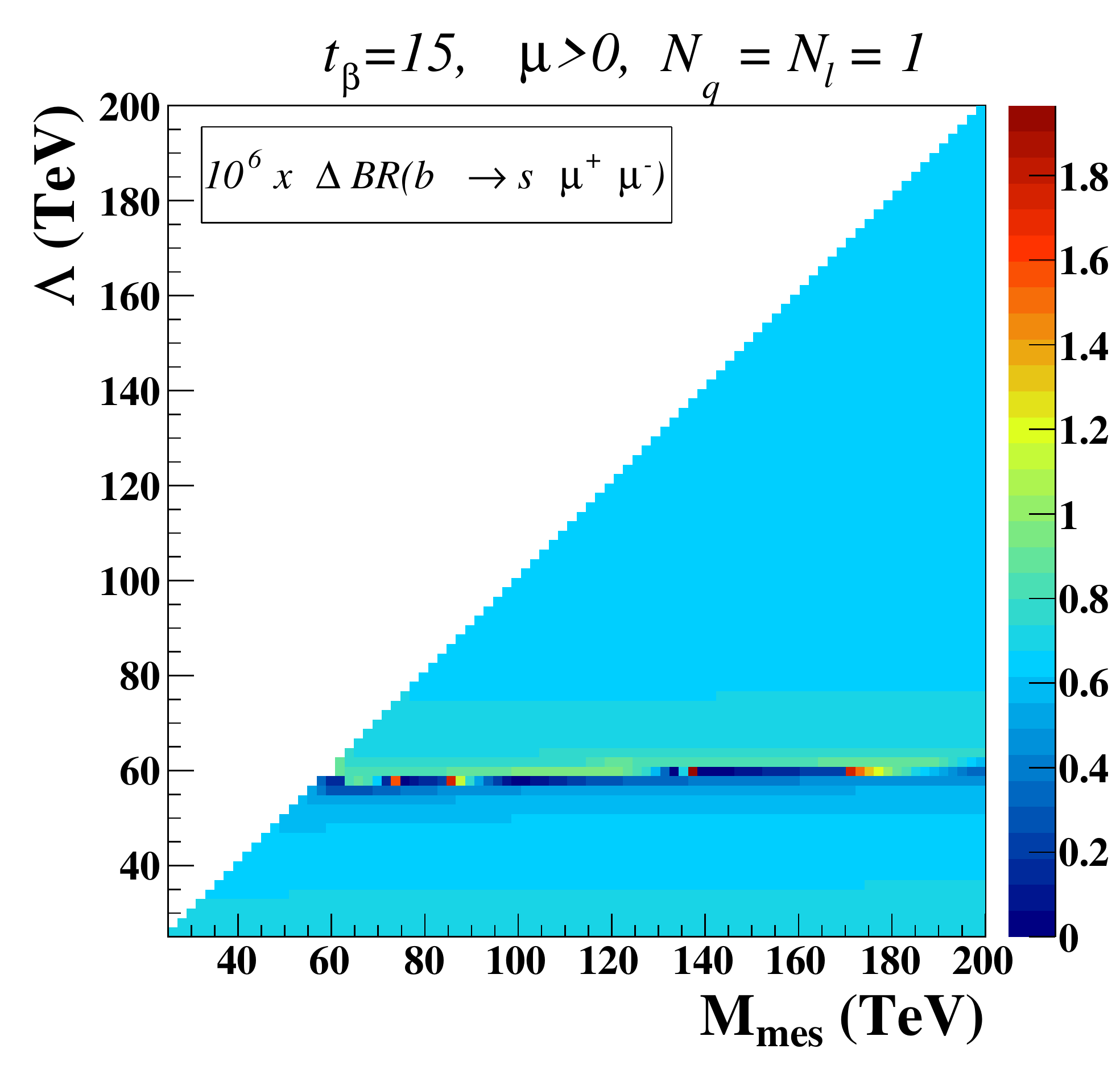}
 \includegraphics[width=.32\columnwidth]{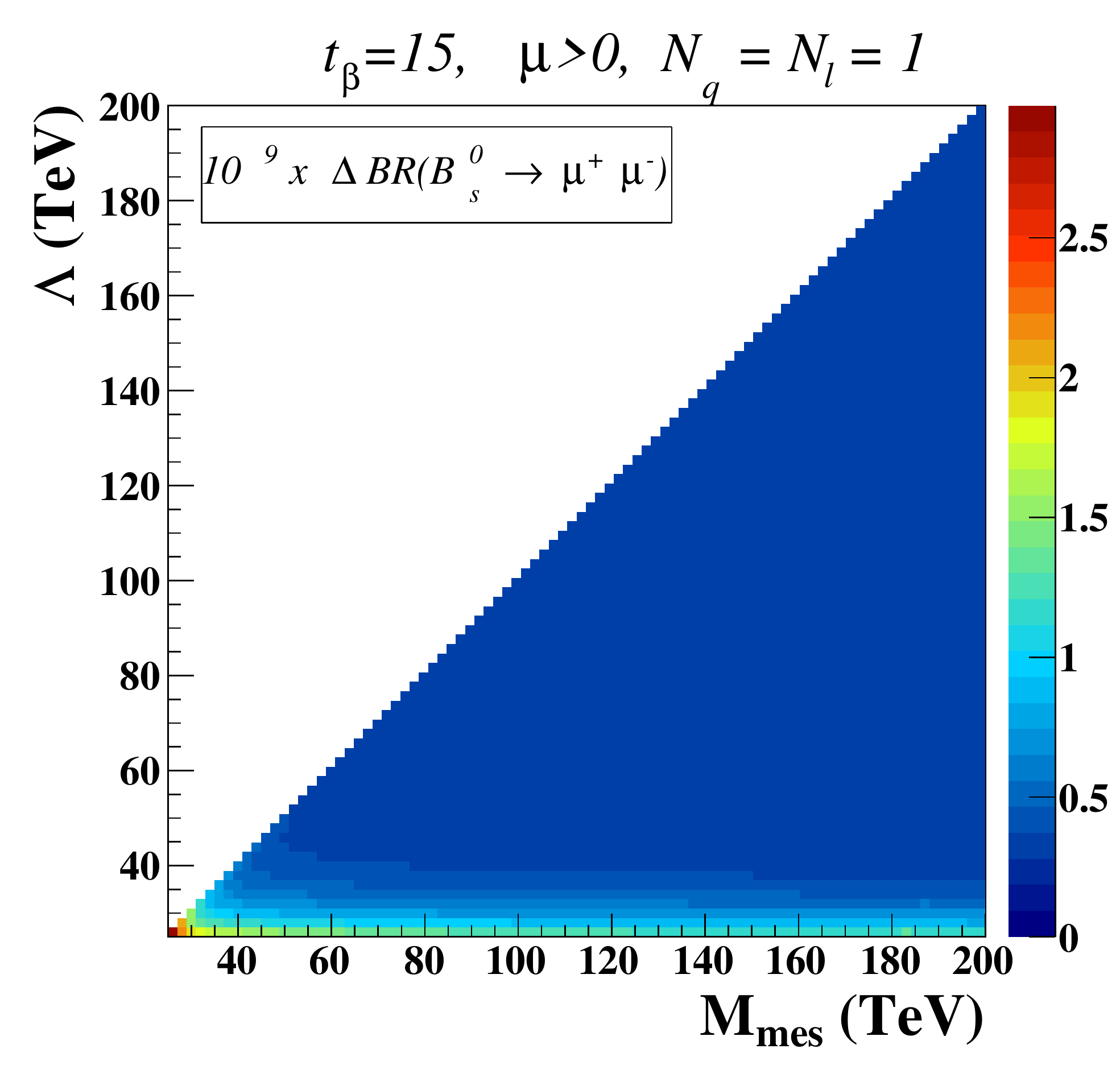}
 \caption{\label{fig:gmsb1_bdecays} Same as Figure \ref{fig:cmssm10_bdecays}
but for gauge-mediated supersymmetry-breaking MSSM scenarios.
We present $(M_{\rm mes},\Lambda)$ planes with
$\tan\beta=15$, $N_q = N_\ell =1$ and a positive Higgs mixing
parameter $\mu>0$.}
 \vspace{.3cm}
 \includegraphics[width=.32\columnwidth]{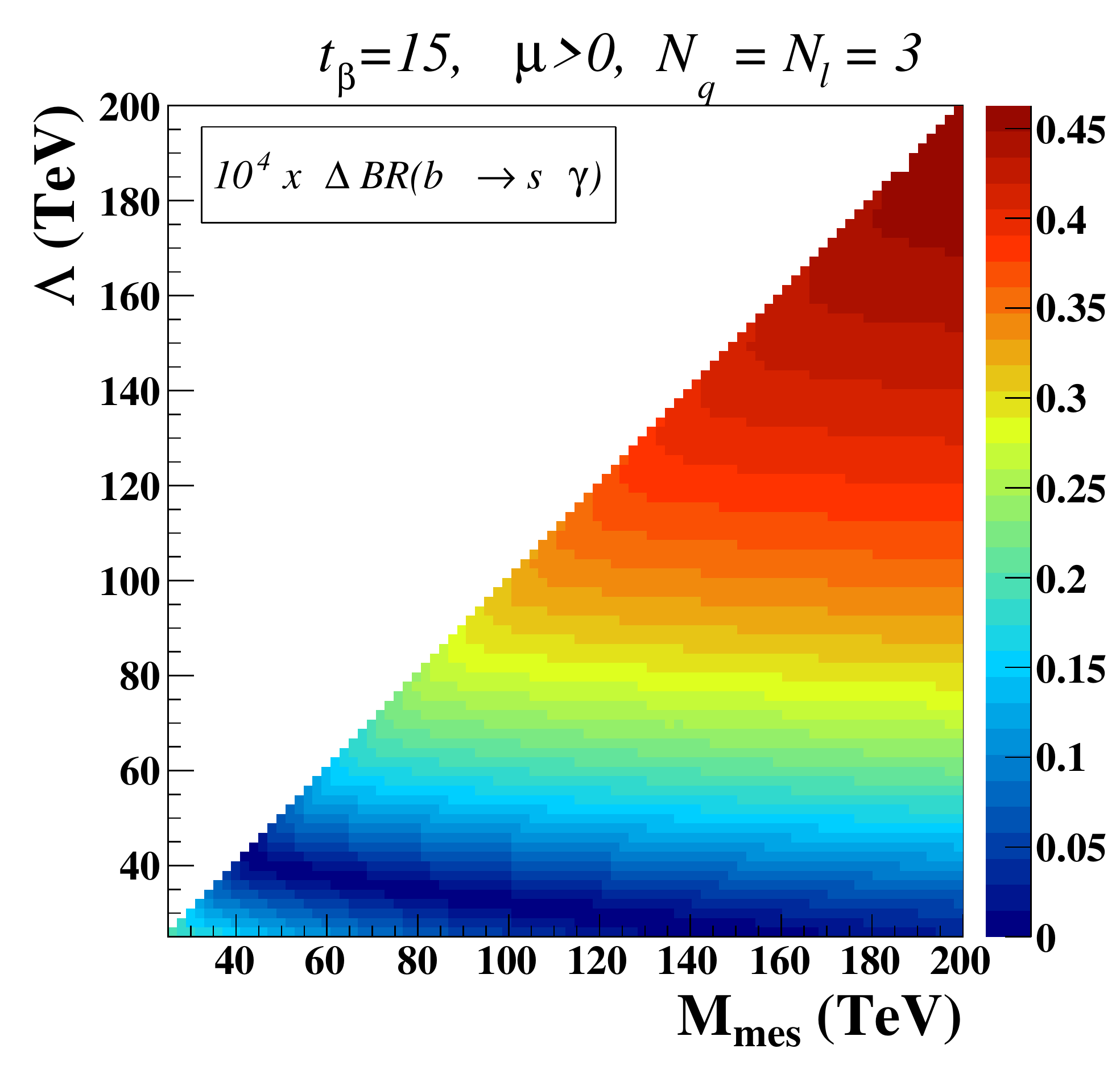}
 \includegraphics[width=.32\columnwidth]{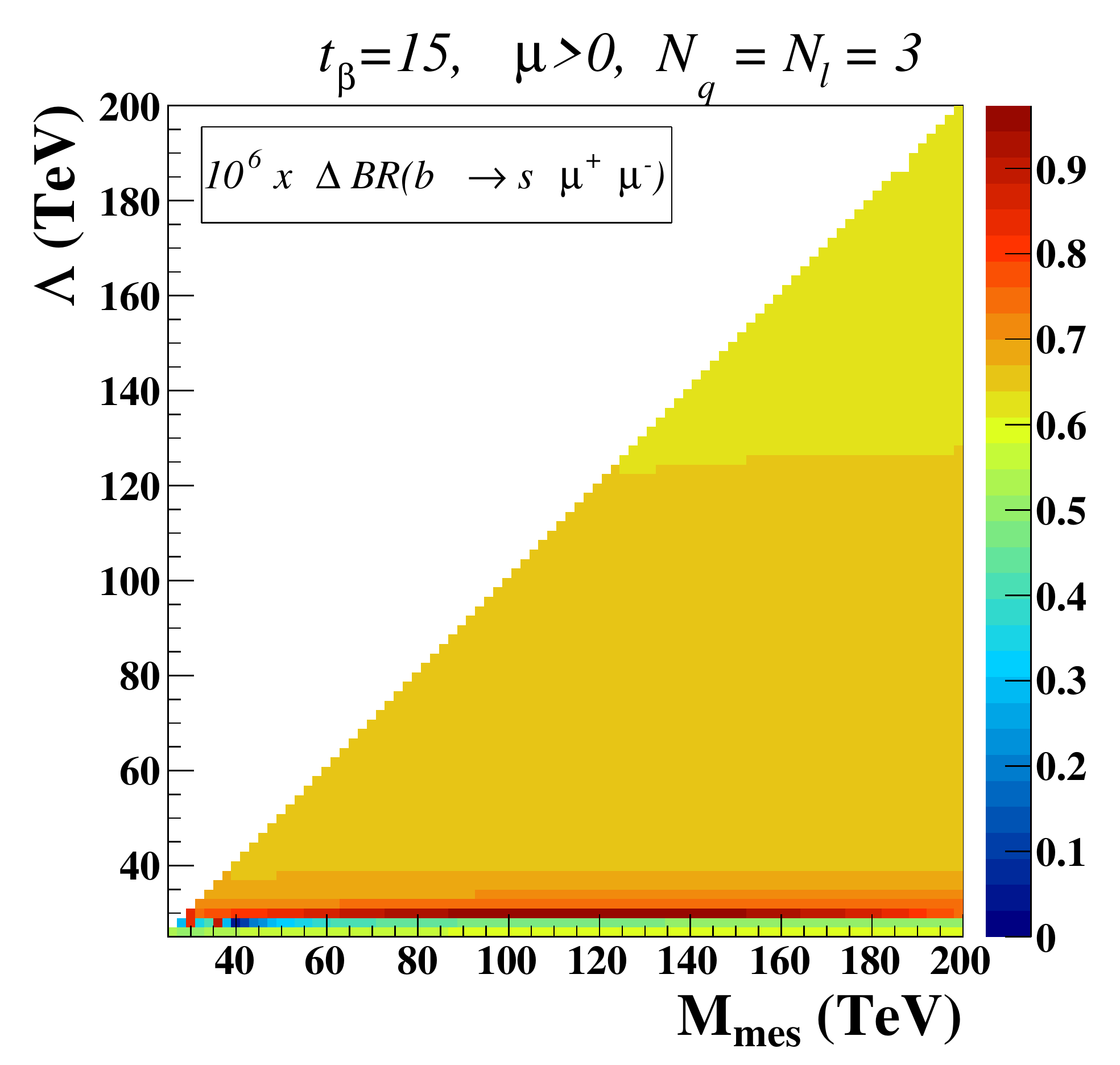}
 \includegraphics[width=.32\columnwidth]{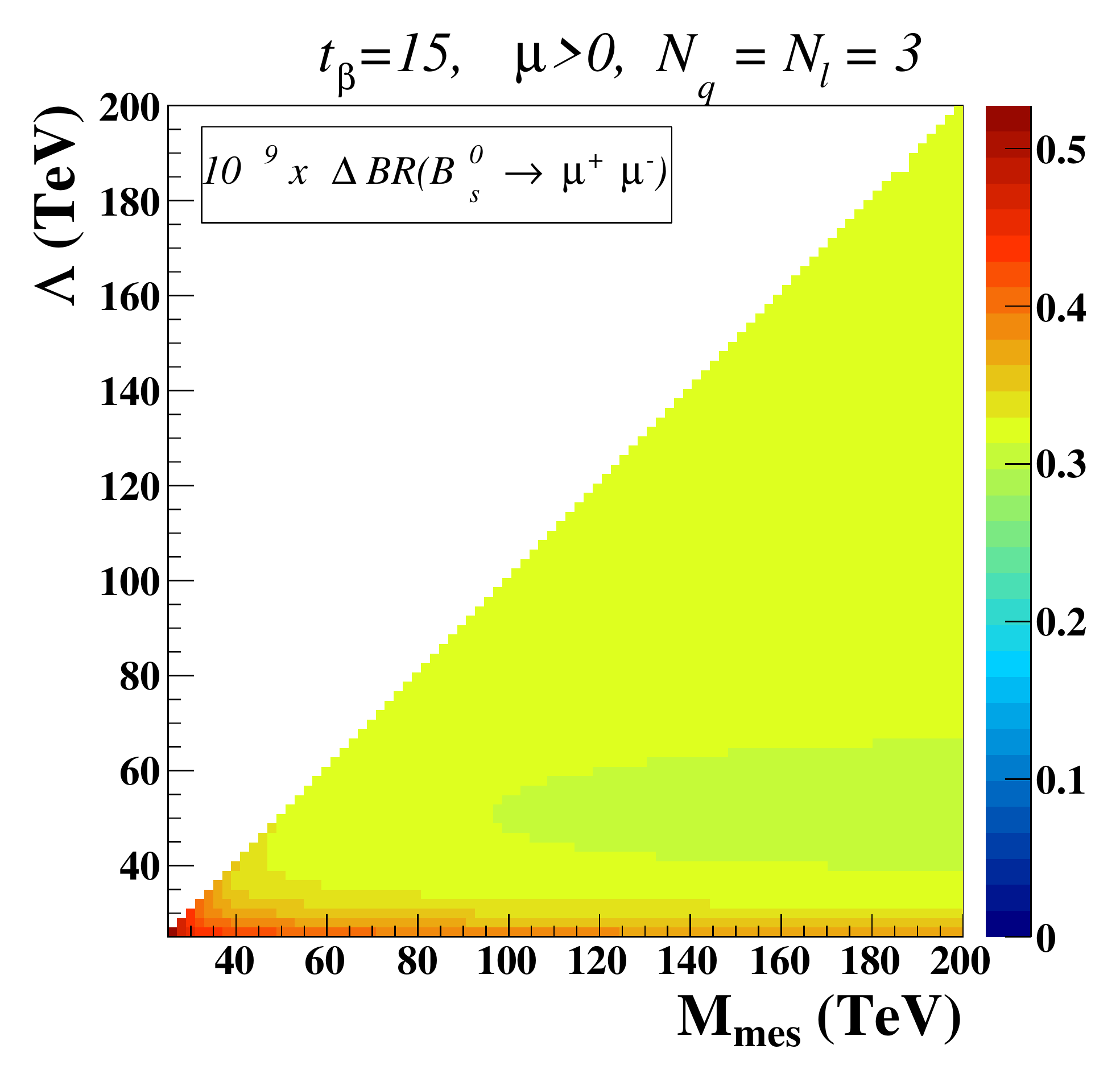}
 \caption{\label{fig:gmsb3_bdecays} Same as Figure \ref{fig:gmsb1_bdecays}
but for $N_q = N_\ell =3$.}
\end{figure}
%

The dependence of the $B$-physics observable on $\tan\beta$ is also illustrated
on Figure \ref{fig:cmssm10_bdecays} and Figure \ref{fig:cmssm40_bdecays} in the
context of the $b\to s \mu^+ \mu^-$ (central panel) and $B_s^0 \to \mu^+\mu^-$
decays (right panel), where
we present the two branching ratios as deviations with respect
to the central value of Eq.\ \eqref{eq:bsmumu} and Eq.\ \eqref{eq:bs0}, respectively. 
The regions excluded at the $2\sigma$-level,
as well as those for which there is no solution to the supersymmetric
renormalization group equations, are again shown as white areas. 
For the case $\tan\beta=10$ and $A_0
= 0$ GeV (Figure \ref{fig:cmssm10_bdecays}), the effects of the
supersymmetric contributions to these two observables are found to be
numerically reduced due to the low value
of $\tan\beta$. Additionally, both experimental measurements are suffering from
large uncertainties. Consequently, the constraints on the parameter space
are not competitive with those induced by the $b\to s\gamma$ observable.

In contrast, for larger values of $\tan\beta$, loop-corrections involving 
the bottom Yukawa coupling are enhanced so that the associated diagrams become
as important as those involving the top Yukawa coupling. 
Enhancements of several orders of magnitude
are in this way possible for the predictions of the BR$(B_s^0 \to
\mu^+ \mu^-)$ observable \cite{Choudhury:1998ze, Babu:1999hn}. The case
$\tan\beta=40$ is illustrated on Figure \ref{fig:cmssm40_bdecays}.
The constraints derived from the $b\to s \mu^+\mu^-$ and $B_s^0\to
\mu^+\mu^-$ branching ratios are found not as restrictive as those induced by the
inclusive $b\to s \gamma$ decay.

We now turn to the investigation of the MSSM with gauge-mediated supersymmetry
breaking. In  Ref.\ \cite{AbdusSalam:2011fc}, two benchmark $(M_{\rm
mes}, \Lambda)$ planes have been adopted, with a fixed value of $\tan\beta=15$
and a positive
$\mu$-parameter. The difference between the two planes lies in the number of
messenger fields. In the first case, the model is constructed on the basis of
$N_q = N_\ell =1$ messenger field while
in the second case, it incorporates three sets of messengers ($N_q = N_\ell
=3$). These two choices imply two different natures for the next-to-lightest
superpartner, the
lightest supersymmetric particle being always the gravitino.
The
next-to-lightest supersymmetric particle is hence the lightest stau for $N_q = N_\ell
=3$ and the lightest neutralino for $N_q = N_\ell =1$. 

The predictions for the three considered $B$-physics observables are presented
on Figure~\ref{fig:gmsb1_bdecays} and Figure~\ref{fig:gmsb3_bdecays} for the
$N_q = N_\ell =1$ and $N_q = N_\ell =3$ benchmark planes, respectively. For the
entire scanned regions, one gets agreement between theory and experiment
for all three observables, which is not surprising since 
gauge-mediated
supersymmetry breaking naturally solves the so-called
supersymmetric flavor problem. Supersymmetry is here usually
broken
within a few orders of magnitude from the weak scale, whereas the unrelated
flavor-breaking scale can be chosen much higher. Consequently, the important
flavor-violating terms included in the soft supersymmetry-breaking Lagrangian of
Eq.\ \eqref{eq:lmssmbrk2} are avoided and one obtains, after diagonalization of
the fermion sector, approximately flavor-conserving scalar mass matrices.
Good agreement with measurements of flavor-changing neutral current observables
is therefore foreseen, as shown in Figure \ref{fig:gmsb1_bdecays}
and Figure
\ref{fig:gmsb3_bdecays}. Finally, the regions with $\Lambda >
M_{\rm mes}$ are theoretically excluded as they do not allow for physical
solutions of the supersymmetric renormalization group equations.

%
\begin{figure}[t!]
 \centering
 \includegraphics[width=.32\columnwidth]{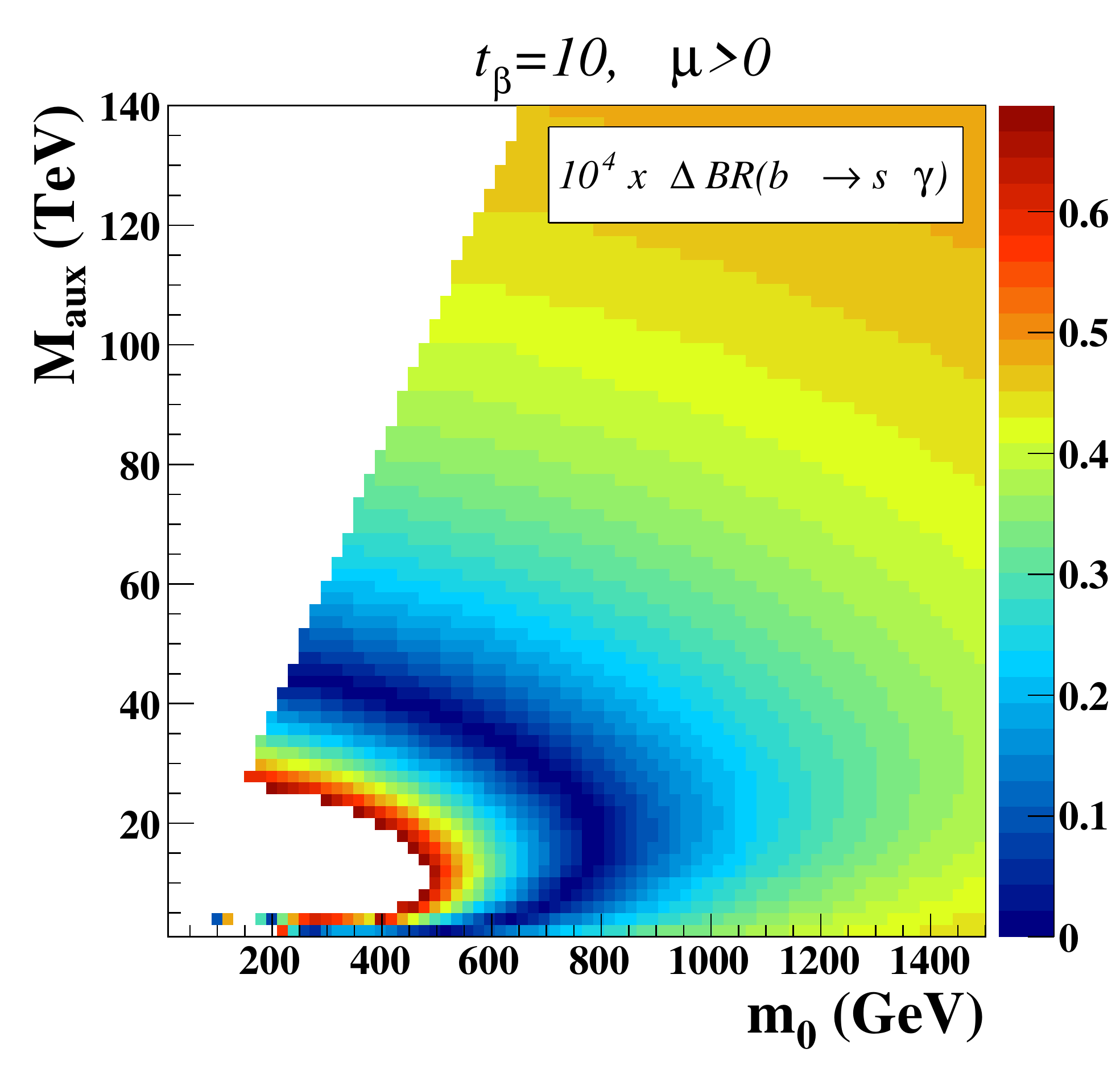}
 \includegraphics[width=.32\columnwidth]{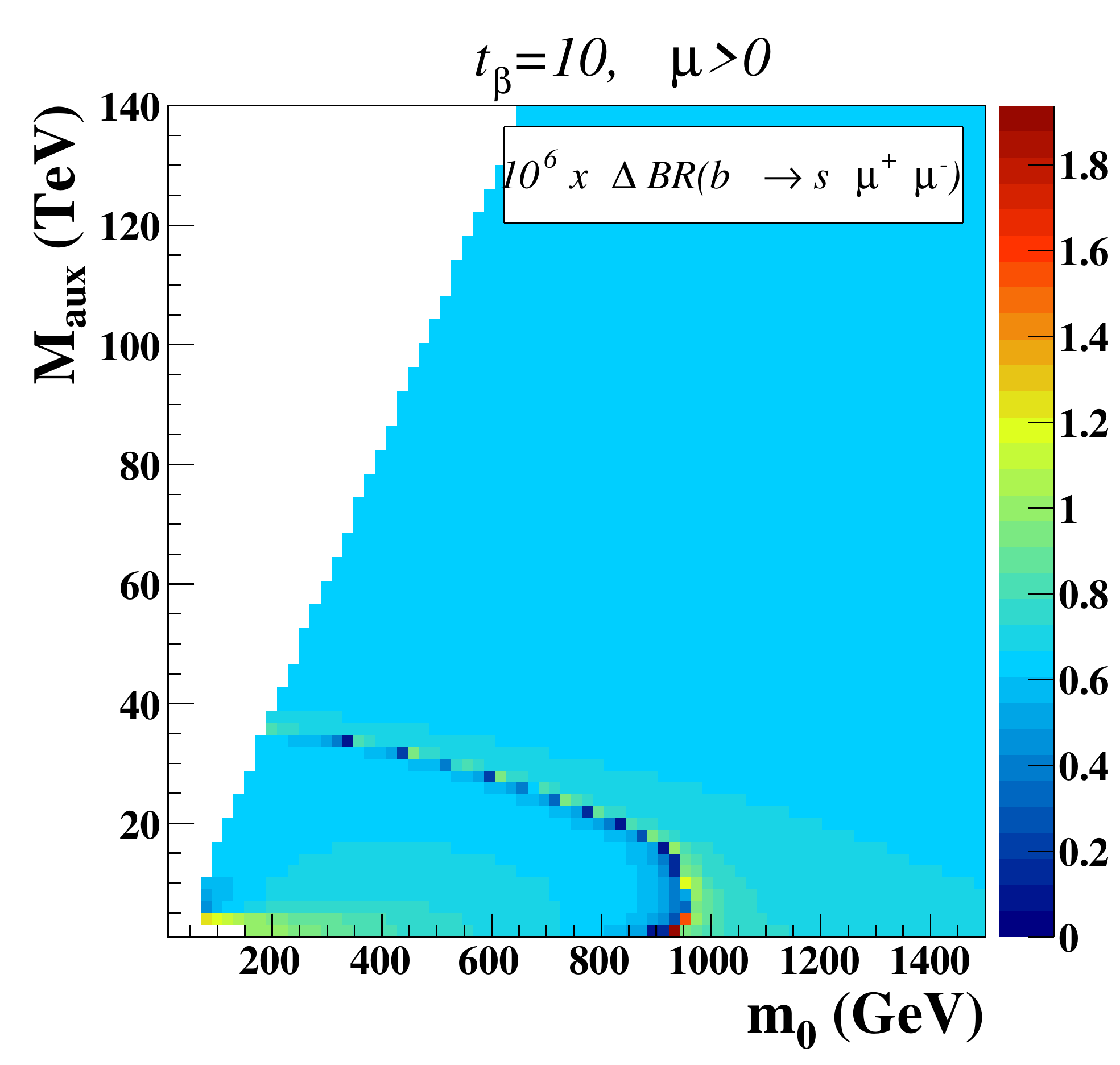}
 \includegraphics[width=.32\columnwidth]{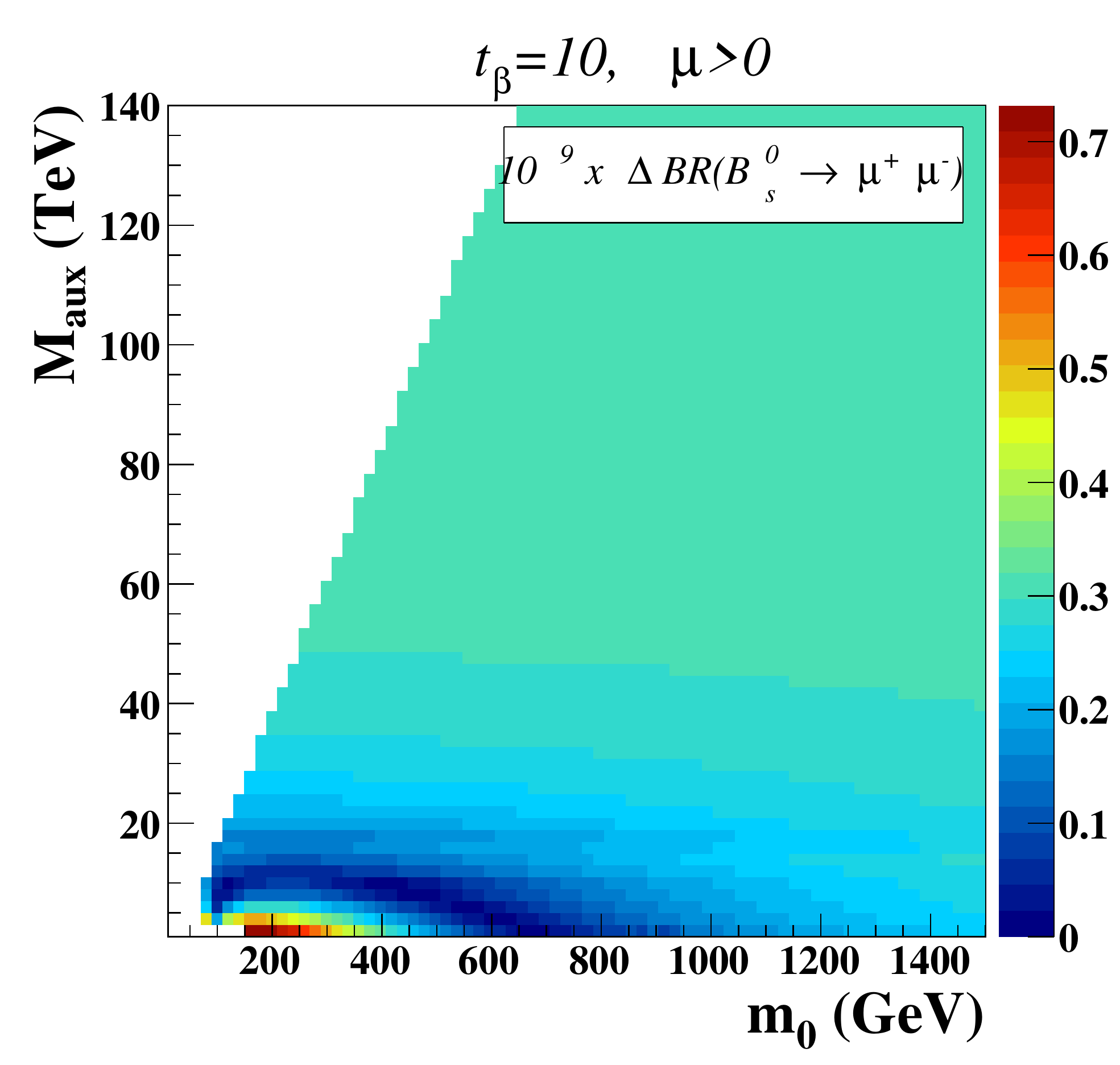}
 \caption{\label{fig:amsb_bdecays} Same as Figure \ref{fig:cmssm10_bdecays}
but for anomaly-mediated supersymmetry-breaking MSSM scenarios.
We present $(m_0, M_{\rm aux})$ planes with
$\tan\beta=10$ and a positive Higgs mixing parameter $\mu>0$.}
\end{figure}
%

In order to study the effect of the constraints derived
from rare $B$-decays on MSSM scenarios where supersymmetry is broken via
anomalies, we again follow the prescriptions of Ref.\ \cite{AbdusSalam:2011fc}.
We adopt a benchmark plane inspired by 
the anomaly-mediated supersymmetry breaking benchmark point 
\texttt{SPS9} \cite{Allanach:2002nj}. The ratio of the
expectation values of the two neutral Higgs boson is fixed 
to $\tan\beta = 10$ and 
a positive sign for the $\mu$-parameter is chosen. The predictions for the three
considered $B$-physics observables are presented in
$(m_0,M_{\rm aux})$ planes, in Figure \ref{fig:amsb_bdecays}, where we show, as above,
their deviations from the central experimental values of Eq.\
\eqref{eq:bsg}, Eq.\ \eqref{eq:bsmumu} and Eq.\ \eqref{eq:bs0}. As for 
the cMSSM,
the low-mass regions with a relatively small $m_0$, attractive from a
collider point of view, are strongly disfavored by data on the $b\to
s\gamma$ branching ratio, the two other observables not bringing any
additional restriction on the parameter space.

In the following, we turn to the investigation of the constraints induced by 
$B$-physics observables when more general squark mixings, like those given in Eq.\
\eqref{eq:sfmix}, are allowed. We adopt a phenomenological approach where we assume
non-negligible contributions to the off-diagonal terms of the squark soft mass
matrices that we rewrite, in the super-CKM basis, as%
\renewcommand{\arraystretch}{1.2}%
\be M_{\tilde{q}}^2 = \left(
  \begin{array}{ccc|ccc} 
    M^2_{L_{q_1}} & \Delta^{q_1 q_2}_{LL} & \Delta^{q_1 q_3}_{LL} & 
      X_{q_1} & \Delta^{q_1 q_2}_{LR} & \Delta^{q_1 q_3}_{LR} \\
    \Delta^{q_1 q_2\ast}_{LL} & M^2_{L_{q_2}} & \Delta^{q_2 q_3}_{LL} & 
      \Delta^{q_1 q_2\ast}_{RL} & X_{q_2} & \Delta^{q_2 q_3}_{LR} \\ 
    \Delta^{q_1 q_3\ast}_{LL} & \Delta^{q_2 q_3\ast}_{LL} & M^2_{L_{q_3}} & 
      \Delta^{q_1 q_3\ast}_{RL} & \Delta^{q_2 q_3\ast}_{RL} & X_{q_3} \\ 
    \hline 
    X_{q_1}^\ast & \Delta^{q_1 q_2}_{RL} & \Delta^{q_1 q_3}_{RL} &
      M^2_{R_{q_1}} & \Delta^{q_1 q_2}_{RR} & \Delta^{q_1 q_3}_{RR} \\
    \Delta^{q_1 q_2\ast}_{LR}& X_{q_2}^\ast &  \Delta^{q_2 q_3}_{RL} & 
      \Delta^{q_1 q_2\ast}_{RR} & M^2_{R_{q_2}} & \Delta^{q_2 q_3}_{RR} \\ 
    \Delta^{q_1 q_3\ast}_{LR}& \Delta^{q_2 q_3\ast}_{LR} & X_{q_3}^\ast&
      \Delta^{q_1 q_3\ast}_{RR} & \Delta^{q_2 q_3\ast}_{RR} & M^2_{R_{q_3}} 
  \end{array} \right) \ .
\ee \renewcommand{\arraystretch}{1.}%
We have explicitly separated the left-left, left-right, right-left and right-right chiral 
sectors by means of horizontal and vertical line and additionally introduced 
shorthand notations for the flavor-diagonal elements,
\be\bsp
   M_{L_{u_i}}^2 = (V_{\rm CKM} {\bf \hat m^2_{\tilde Q}} V_{\rm CKM}^\dag 
      \!+\! M_{q_u}^2)^i{}_i 
     \!+\! (\frac12 \!-\! \frac23 s_w^2)c_{2\beta} M_Z^2  \ , &\quad
   M_{R_{u_i}}^2 = ({\bf \hat m^2_{\tilde U}} \!+\! M_{q_u}^2)^i{}_i 
     \!+\! \frac23 s_w^2 c_{2\beta} M_Z^2 \ , \\
   M_{L_{d_i}}^2 = ({\bf \hat m^2_{\tilde Q}} \!+\! M_{q_d}^2)^i{}_i 
     \!+\! (-\frac12  \!+\!\frac13 s_w^2) c_{2\beta} M_Z^2 \ , &\quad
   M_{R_{d_i}}^2 = ({\bf \hat m^2_{\tilde D}} \!+\! M_{q_d}^2)^i{}_i 
     \!-\! \frac13  c_{2\beta} M_Z^2 s_w^2  \ , \\ 
   X_{u_i} = (\frac{v_u}{\sqrt{2}}{\bf \hat T^u}{}^\dag - \frac{\mu}{\tan\beta} M_{q_u})^i{}_i \ , 
     &\quad
   X_{d_i} = (\frac{v_d}{\sqrt{2}}{\bf \hat T^d}{}^\dag - \mu  \tan\beta M_{q_d})^i{}_i \ . 
\esp\label{eq:SU2L}\ee
In these expressions, Einstein summation
conventions are not applied on the generation index~$i$.
The off-diagonal parameters of the mass matrices that we denote by
$\Delta$ are arbitrary and in general
normalized to the diagonal entries of the soft supersymmetry-breaking
mass matrices~\cite{Gabbiani:1996hi},
\be
   \Delta_{LL}^{u_{i} u_{j}} = \lambda^{u_{i} u_{j}}_{LL} ({\bf \hat m^2_{\tilde Q}})^i{}_i 
     ({\bf \hat m^2_{\tilde Q}})^j{}_j  \ ,\ \ 
   \Delta_{LR}^{d_{i} d_{j}} = \lambda^{d_{i} d_{j}}_{LR} ({\bf \hat m^2_{\tilde Q}})^i{}_i 
     ({\bf \hat m^2_{\tilde D}})^j{}_j  \ , \textit{etc.} \ .
\ee
Additional sources of squark flavor violation are then parametrized through the 
21 dimensionless (possibly complex) new variables $\lambda^{q_{i} q_{j}}_{ab}$, recalling
that the $\lambda_{LL}$ quantities for both the up-type and down-type squark sectors are
identical. This also implies 
that both squark mass matrices cannot be simultaneously diagonal without
neglecting the CKM matrix which we
calculate using the Wolfenstein
parametrization. The corresponding four free parameters are set to
$\lambda_{\rm CKM}=0.2272$, $A_{\rm CKM}=0.818$, $\rhobar_{\rm CKM} = 0.221$ and
$\etabar_{\rm CKM} = 0.34$ \cite{Beringer:2012zz}.

Extensive studies of the kaon sector, $B$- and $D$-meson
oscillations, rare decays, and electric dipole moments suggest that only flavor
mixing involving the second and third generations of squarks can be substantial,
and this only in the left-left and right-right chiral sectors 
\cite{Gabbiani:1996hi, Hagelin:1992tc, Brax:1995up, Ciuchini:2007cw}. For this
reason, we restrict to scenarios where only mixing of
the second and third generation squarks is non-vanishing and
that squarks with different chiralities can only
mix in a flavor-conserving way. 
The only non-zero $\lambda$-parameters are thus given by
\be
	\lambda_{\rm L} ~\equiv~ \lambda_{LL}^{ct}, \qquad
	\lambda_R ~\equiv~ \lambda_{RR}^{ct}  = \lambda_{RR}^{sb} \ ,
\label{eq:lambda}\ee
in addition to $\lambda_{LL}^{sb}$ which is
connected to $\lambda_L$ through the CKM matrix
(see Eq.~\eqref{eq:SU2L}).
The equality $\lambda_R = \lambda_{RR}^{ct}  = \lambda_{RR}^{sb}$ has been enforced for
simplicity in order to avoid handling too many free parameters.

%
\begin{figure}[t!]
 \centering
 \includegraphics[width=.32\columnwidth]{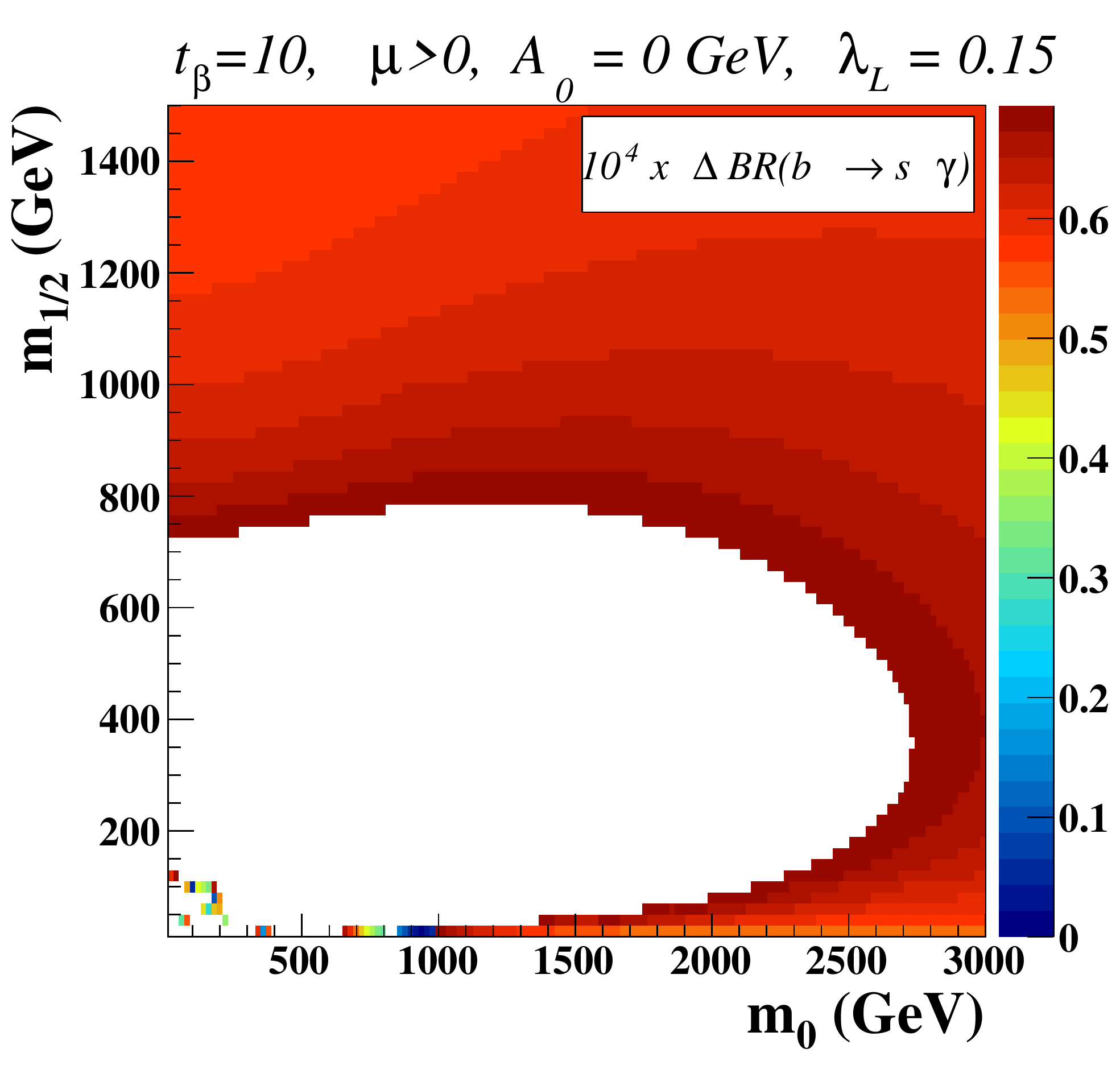}
 \includegraphics[width=.32\columnwidth]{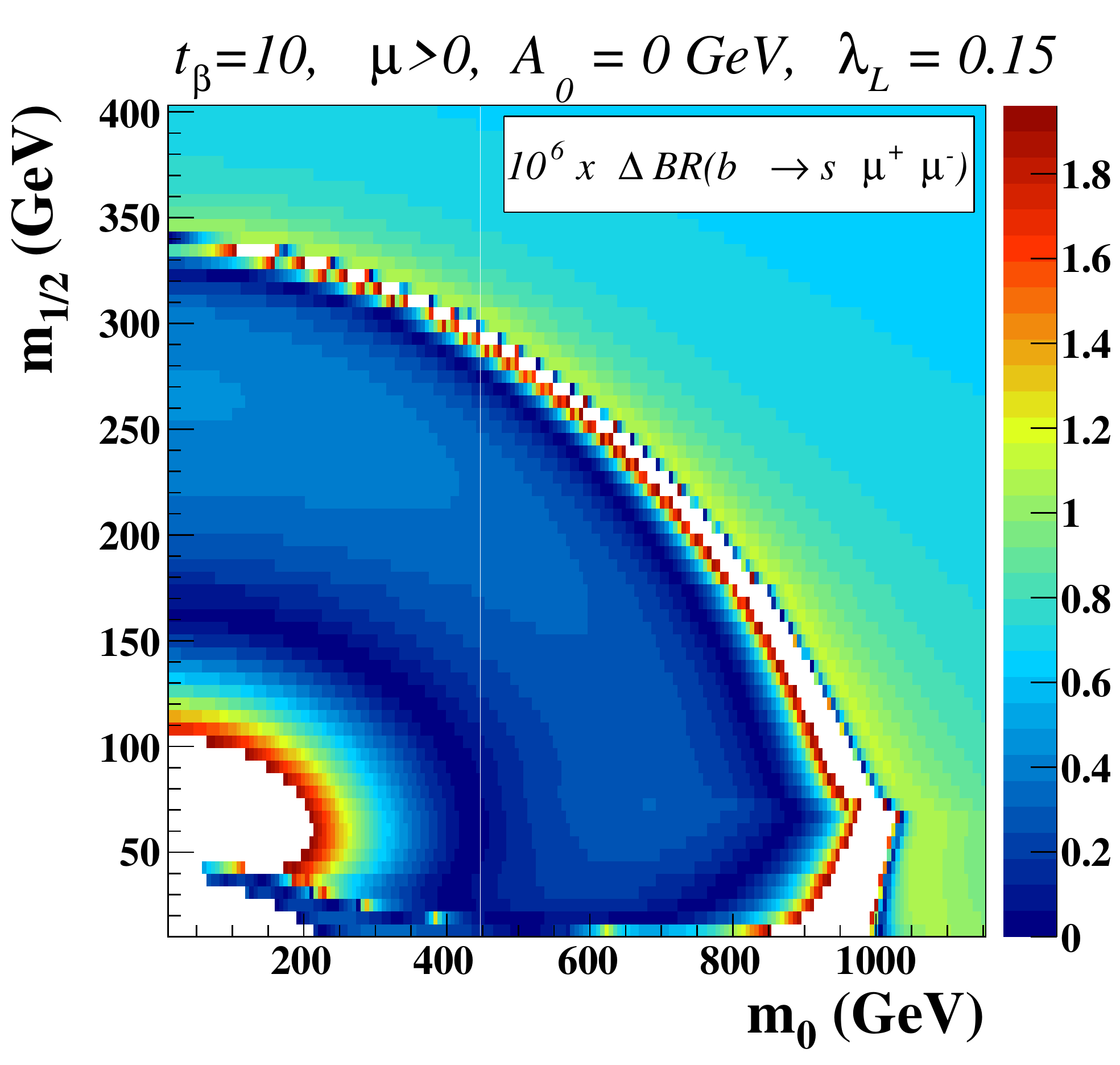}
 \includegraphics[width=.32\columnwidth]{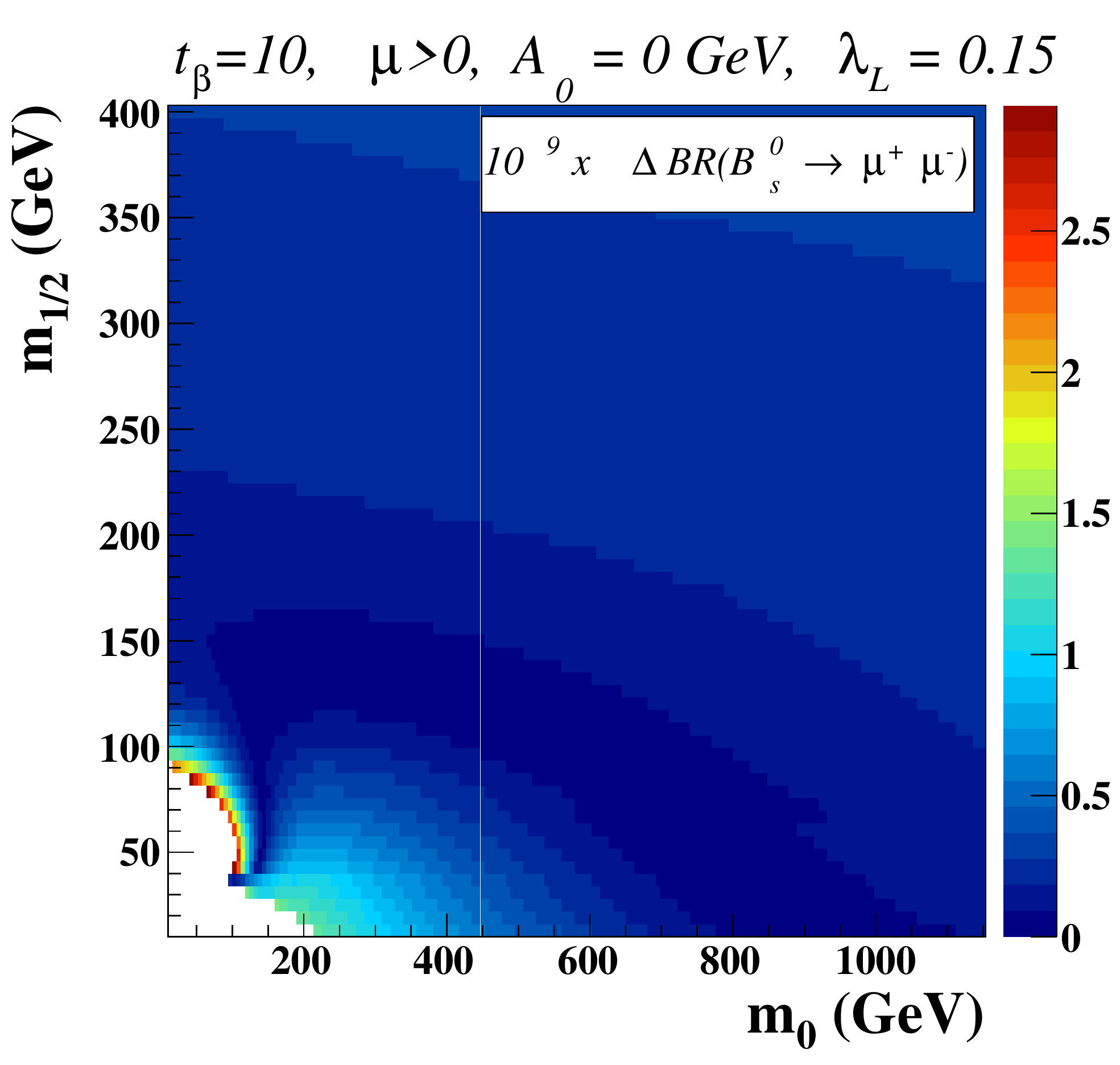}
 \caption{\label{fig:cmssm10_bdecaysL} Same as in Figure
\ref{fig:cmssm10_bdecays} (cMSSM, $\tan\beta=10$, $A_0=0$ GeV, $\mu>0$)
  with $\lambda_L = 0.15$.}
\vspace{.3cm}
 \includegraphics[width=.32\columnwidth]{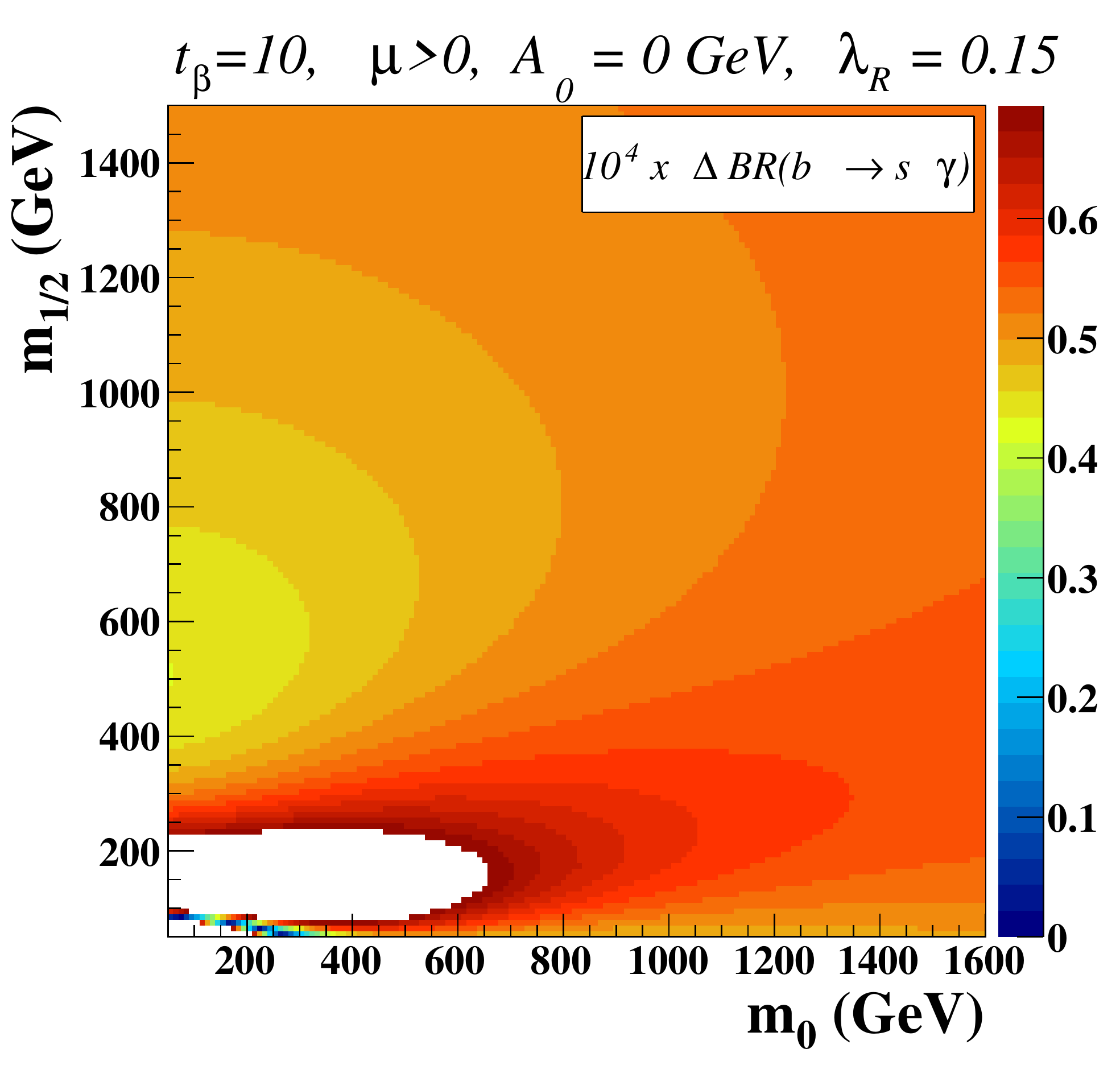}
 \includegraphics[width=.32\columnwidth]{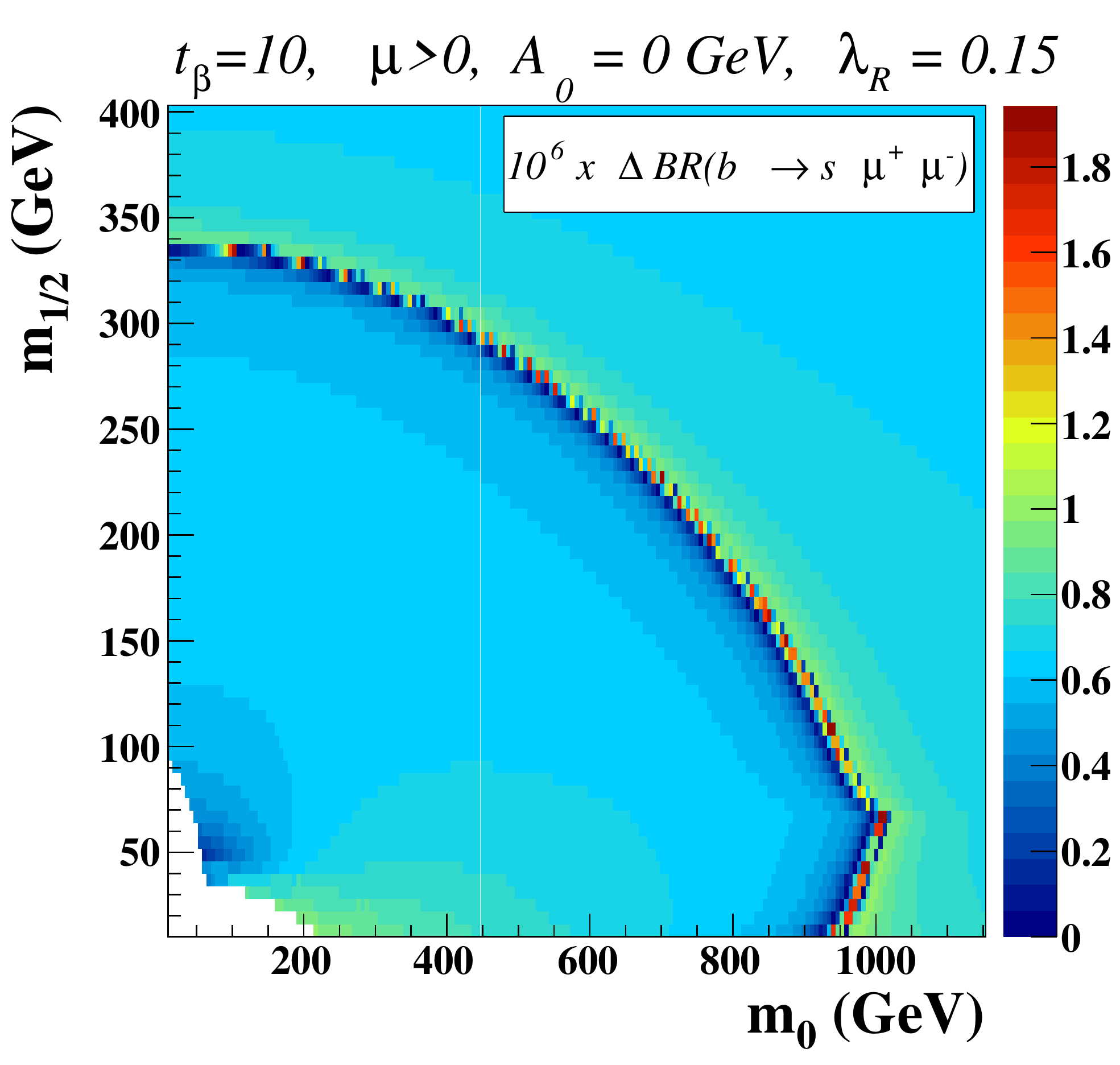}
 \includegraphics[width=.32\columnwidth]{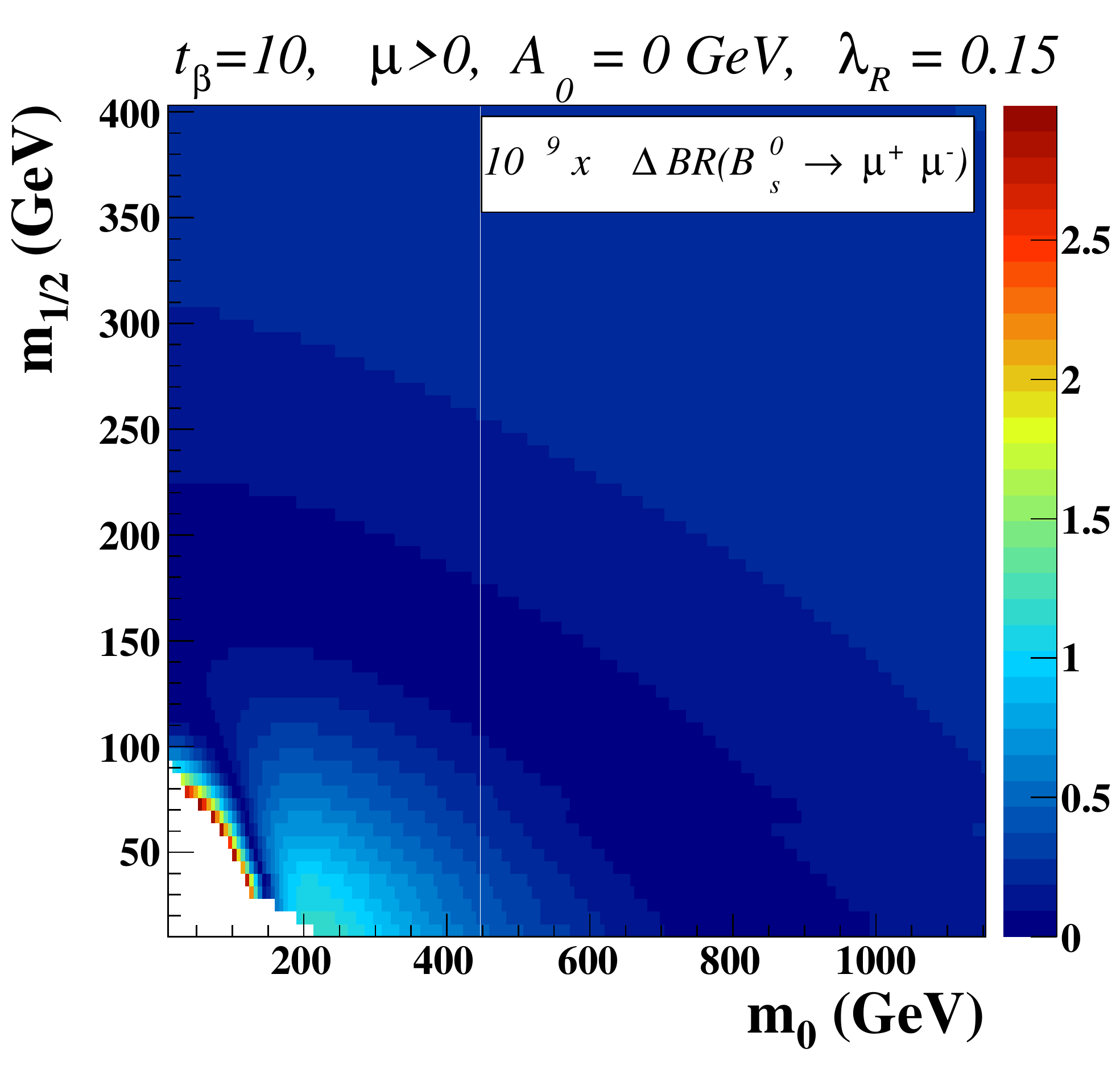}
 \caption{\label{fig:cmssm10_bdecaysR} Same as in Figure
\ref{fig:cmssm10_bdecays} (cMSSM, $\tan\beta=10$, $A_0=0$ GeV, $\mu>0$) with
 $\lambda_R = 0.15$.}
\vspace{.3cm}
 \includegraphics[width=.32\columnwidth]{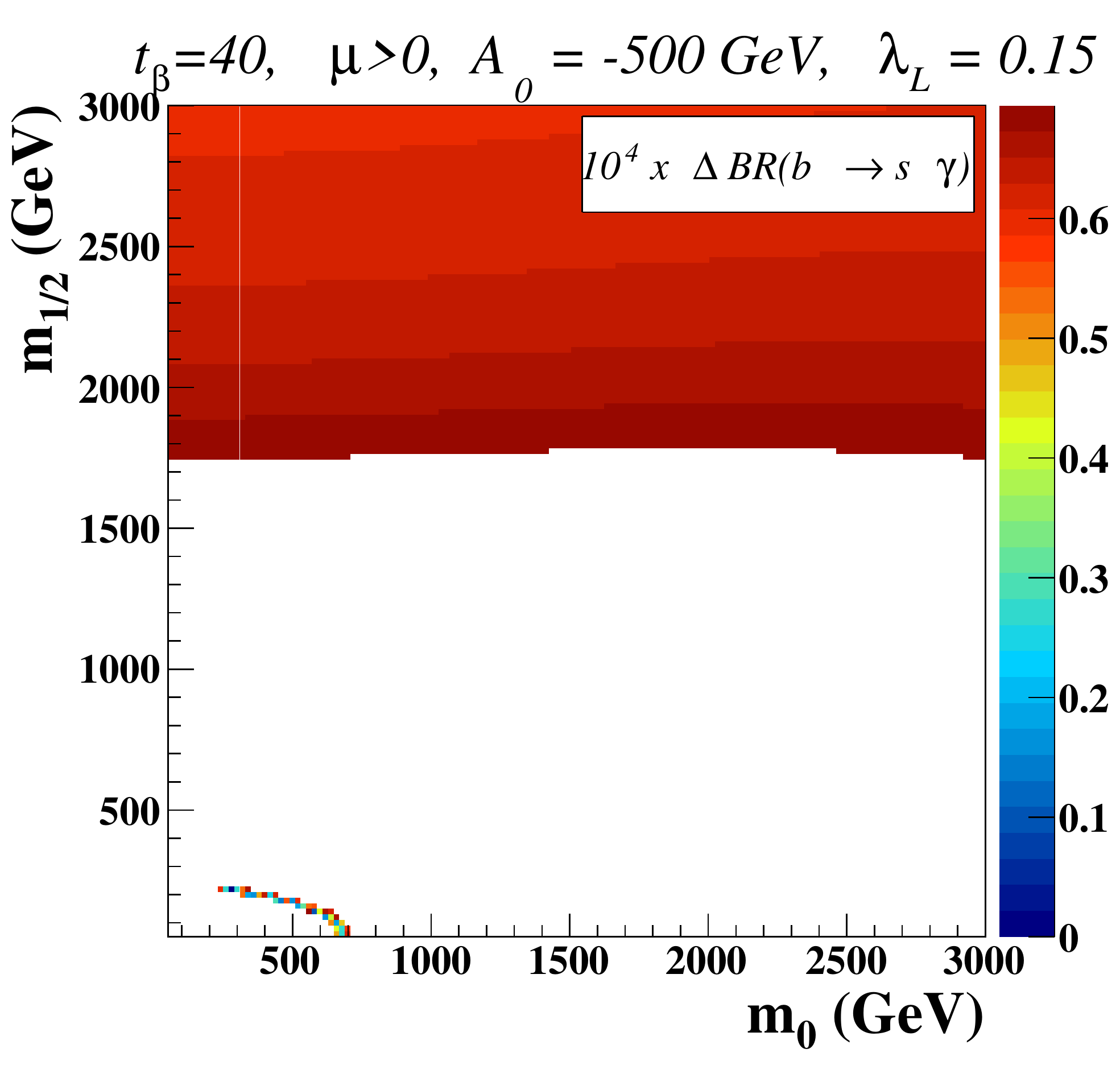}
 \includegraphics[width=.32\columnwidth]{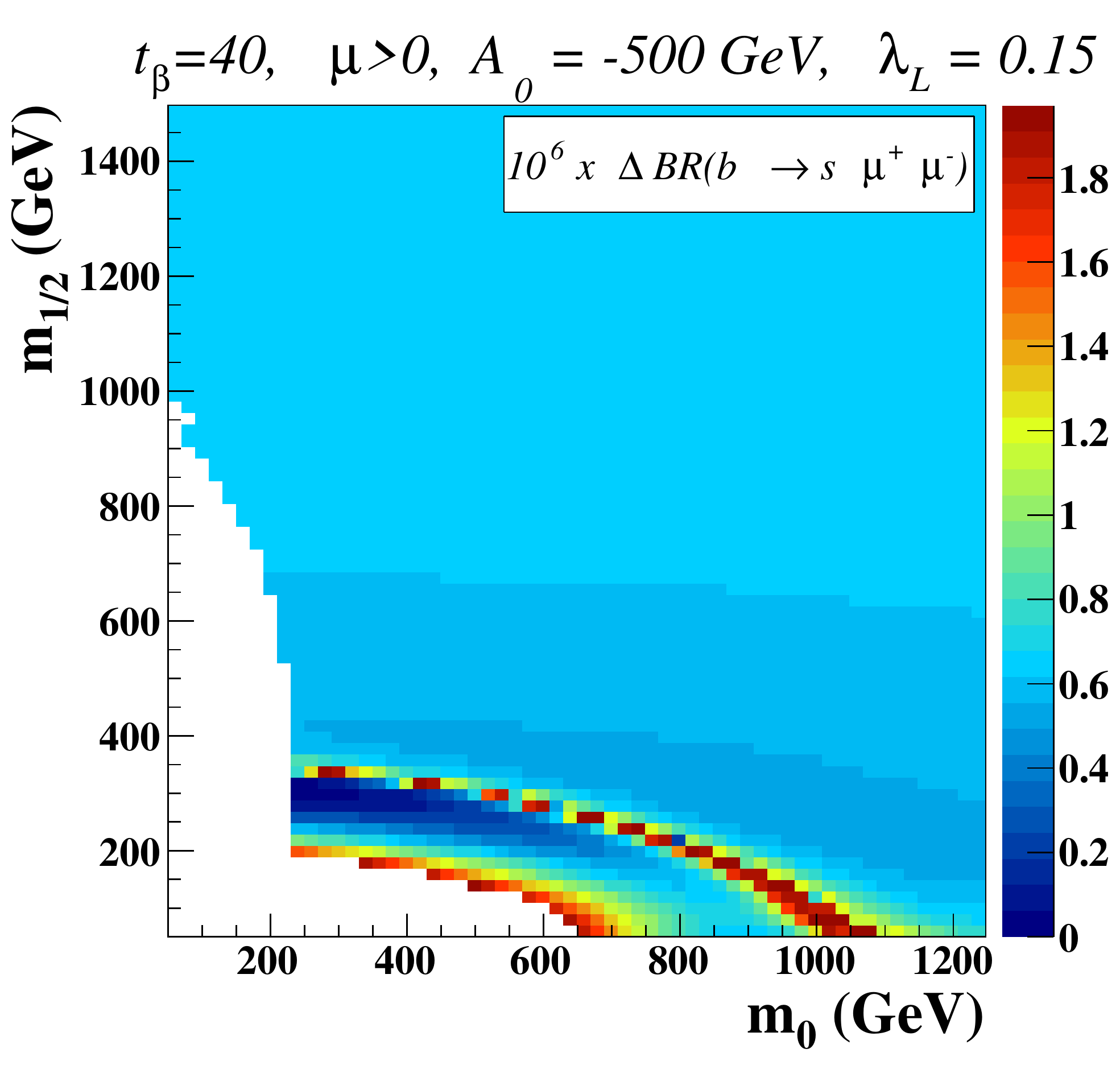}
 \includegraphics[width=.32\columnwidth]{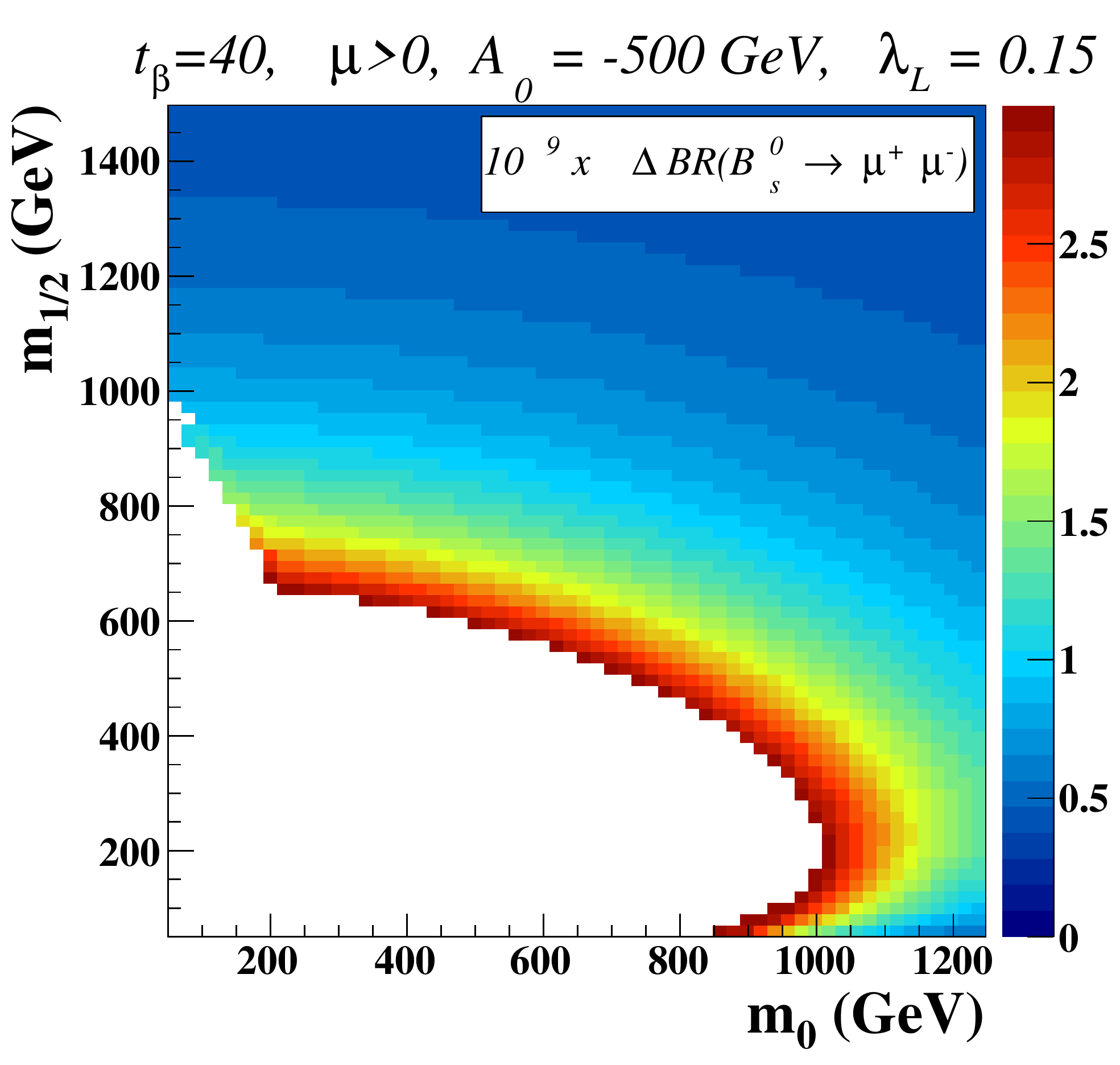}
 \caption{\label{fig:cmssm40_bdecaysL} Same as in Figure
\ref{fig:cmssm40_bdecays} (cMSSM, $\tan\beta=40$, $A_0=-500$ GeV, $\mu>0$)
  with $\lambda_L = 0.15$.}
\end{figure}
%

We now revisit the previous studied constraints and address
the phenomenology of MSSM scenarios including non-minimal
flavor violation when the squark mass
matrices are generalized by including the two
flavor-violating parameters presented in Eq.\ \eqref{eq:lambda} at
low-energy.
We however do not address the conception itself of fundamental
mechanisms leading to non-vanishing flavor-violating squark mixing terms but
provide instead several examples where 
such terms arise. In supergravity, they can be induced by non-trivial K\"ahler
interactions. Although gauge-mediated supersymmetry-breaking scenarios are
known to efficiently address the supersymmetric flavor problem, 
several possibilities for reintroducing flavor-violating terms have been pointed
out recently
\cite{Giudice:1998bp, Tobe:2003nx, Dubovsky:1998nr}. For example, mixing between
messenger and matter fields may lead to important flavor violation in the
squark and slepton sectors. In Section \ref{sec:AMSB}, we have shown that
in its simplest form, the
mechanism yielding supersymmetry-breaking via anomalies leads to
the problematics of tachyonic sleptons. In Section
\ref{sec:mssmbrkex}, we have presented a phenomenological approach to cure this
problem by introducing additional and universal soft supersymmetry-breaking
contributions to the scalar masses. This task could also be achieved in a
non-universal way so that non-minimal flavor violation can be 
introduced in the theory~\cite{Fuks:2011dg}.

%
\begin{figure}
\centering
 \includegraphics[width=.32\columnwidth]{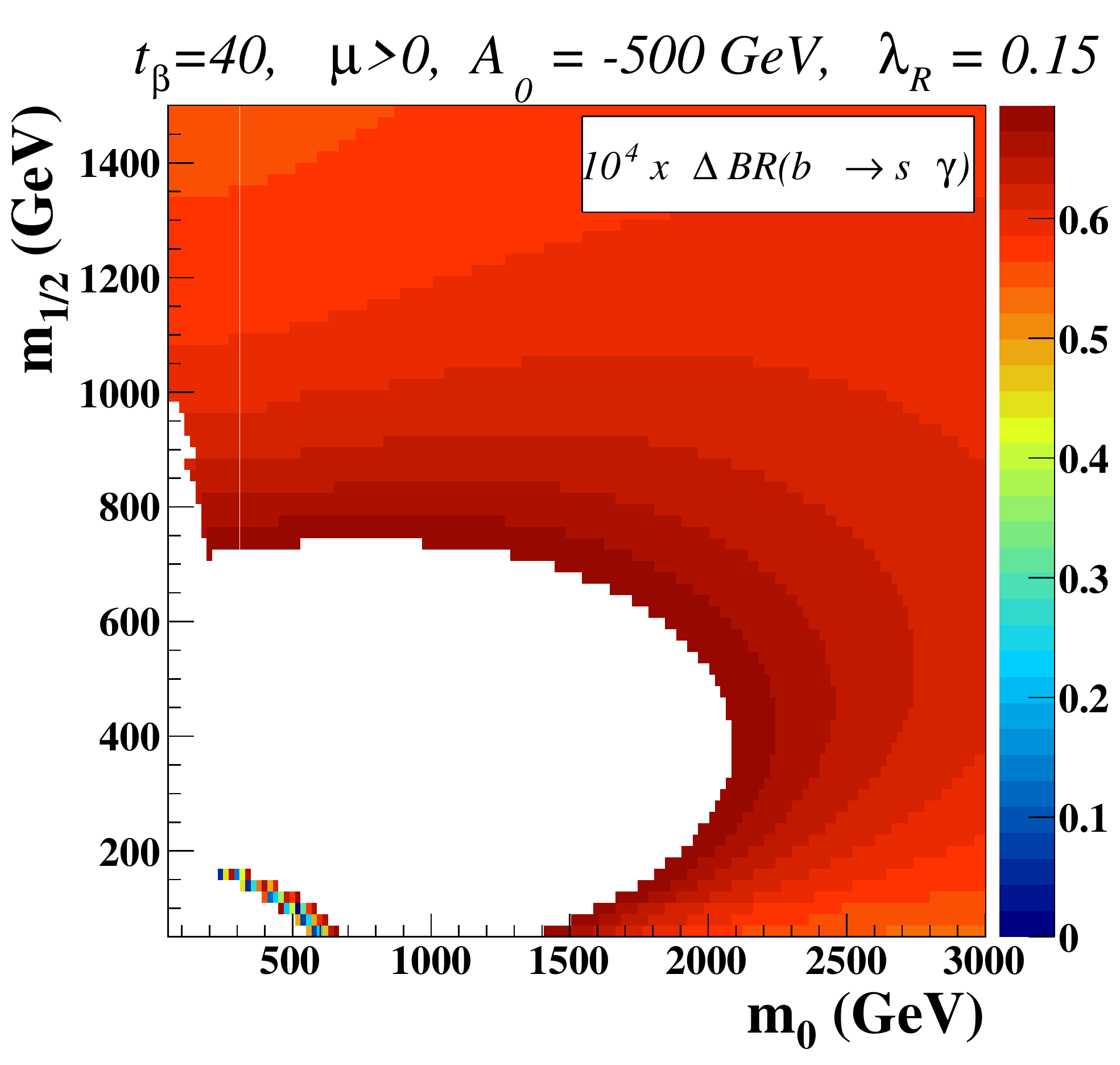}
 \includegraphics[width=.32\columnwidth]{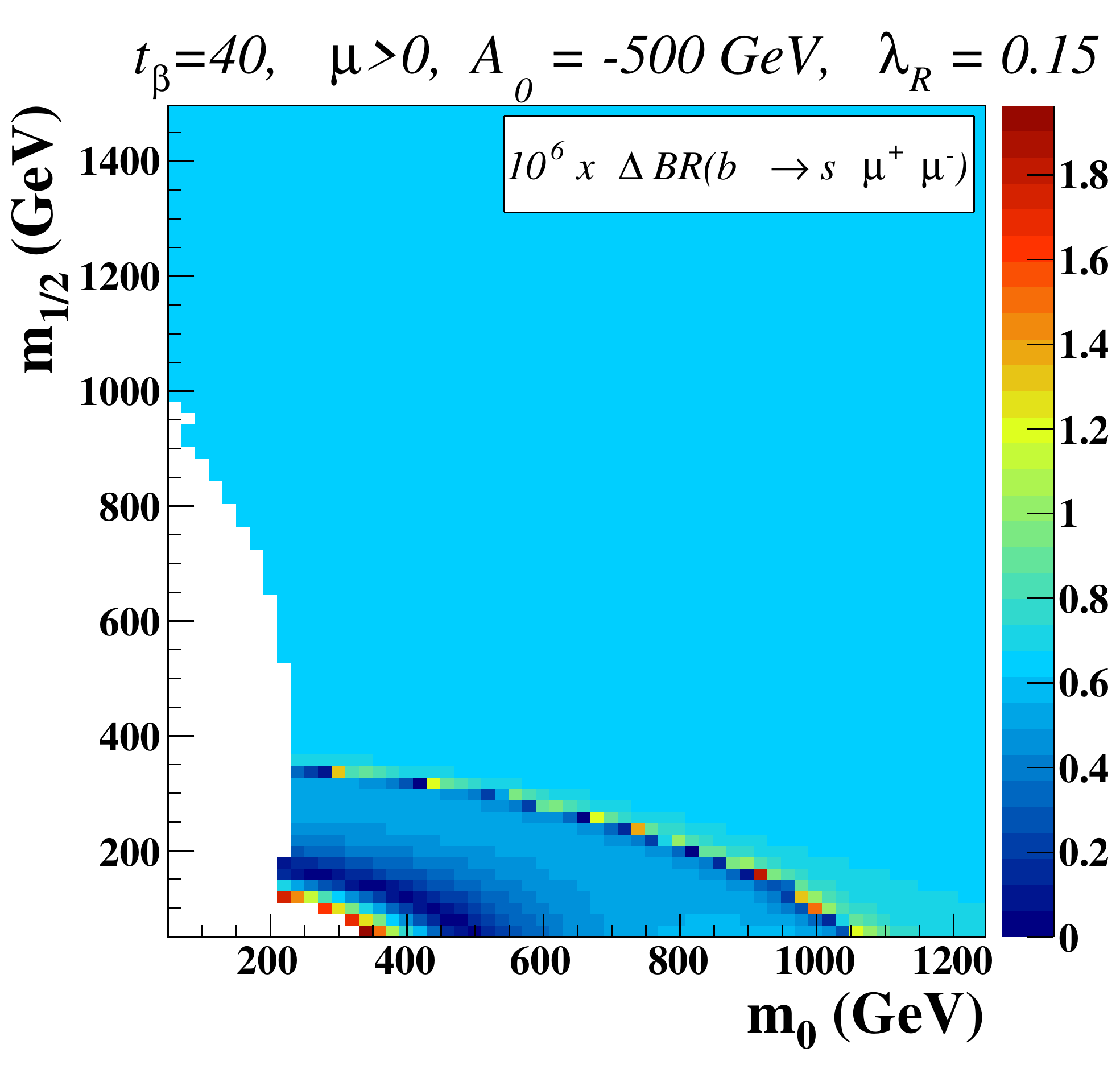}
 \includegraphics[width=.32\columnwidth]{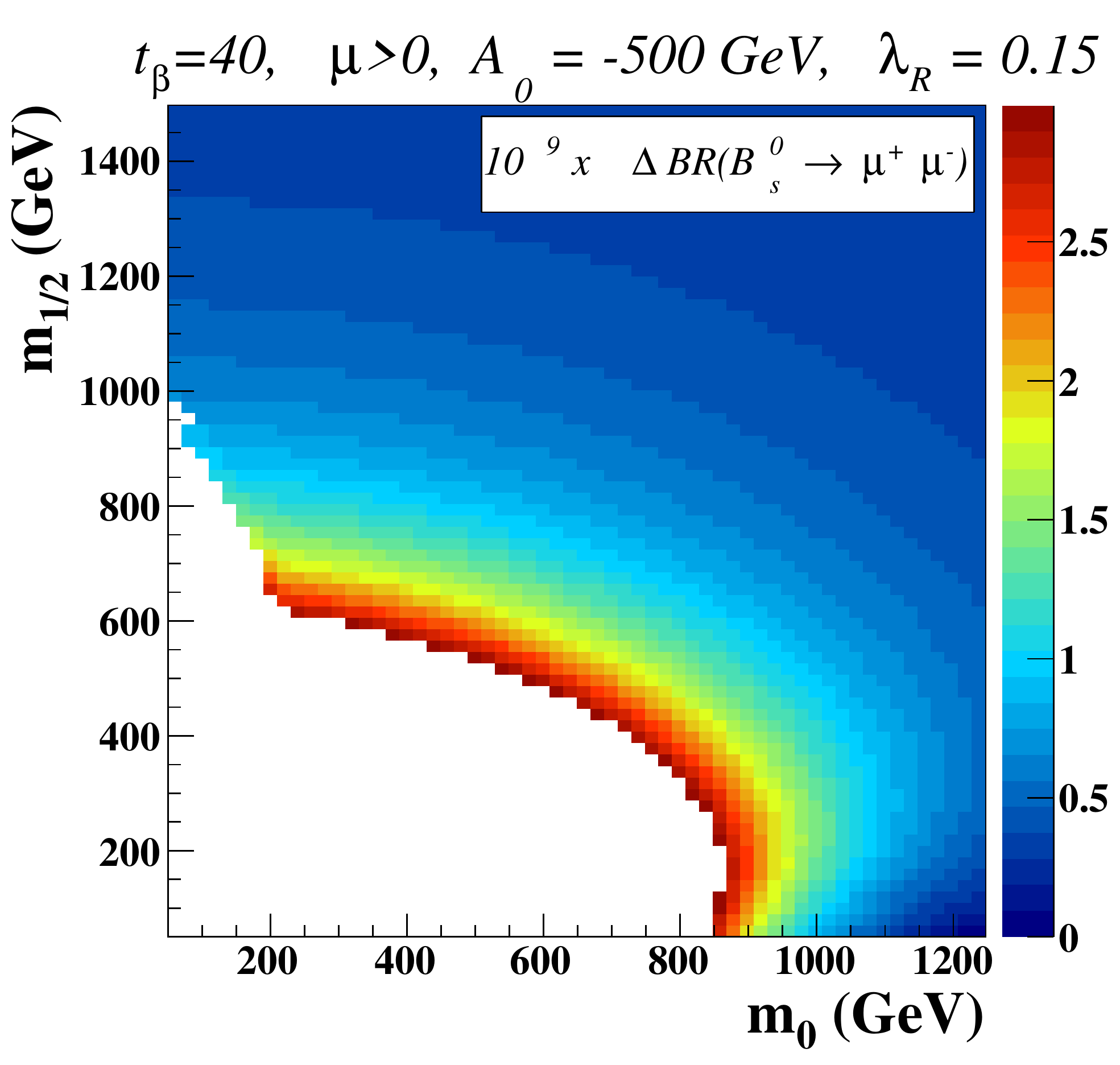}
 \caption{\label{fig:cmssm40_bdecaysR} Same as in Figure
\ref{fig:cmssm40_bdecays} (cMSSM, $\tan\beta=40$, $A_0=-500$ GeV, $\mu>0$) with
  $\lambda_R = 0.15$.}
\end{figure}
%

In Figure \ref{fig:cmssm10_bdecaysL}, Figure \ref{fig:cmssm10_bdecaysR}, Figure
\ref{fig:cmssm40_bdecaysL} and Figure \ref{fig:cmssm40_bdecaysR}, we depict the
impact of the $\lambda_L$ and $\lambda_R$ parameters of Eq.\ \eqref{eq:lambda}
on the theoretical predictions for the three considered $B$-physics observables
in the case of cMSSM scenarios with 
non-minimal flavor-violation. We adopt two
scenarios, one where second
and third generation squark mixing is only allowed in the left-left chiral sector
($\lambda_L = 0.15$, $\lambda_R = 0$) and one where it is only allowed in the
right-right chiral sector ($\lambda_R = 0.15$, $\lambda_L = 0$). 
The $b\to s\gamma$ observable (left panel of the figures)
allows us to probe non-minimally
flavor-violating squark mixings in the left-left chiral sector through the effects
of new flavor-violating loop-diagram contributions where lighter squark and
(the wino component of the) neutralino or chargino fields propagate
into the loops. Compared with the flavor-conserving
case, larger
fractions of the scanned $(m_0,m_{1/2})$ planes are found to be excluded,
so that phenomenologically viable scenarios imply heavier superpartners.
%
\begin{figure}
\centering
 \includegraphics[width=.32\columnwidth]{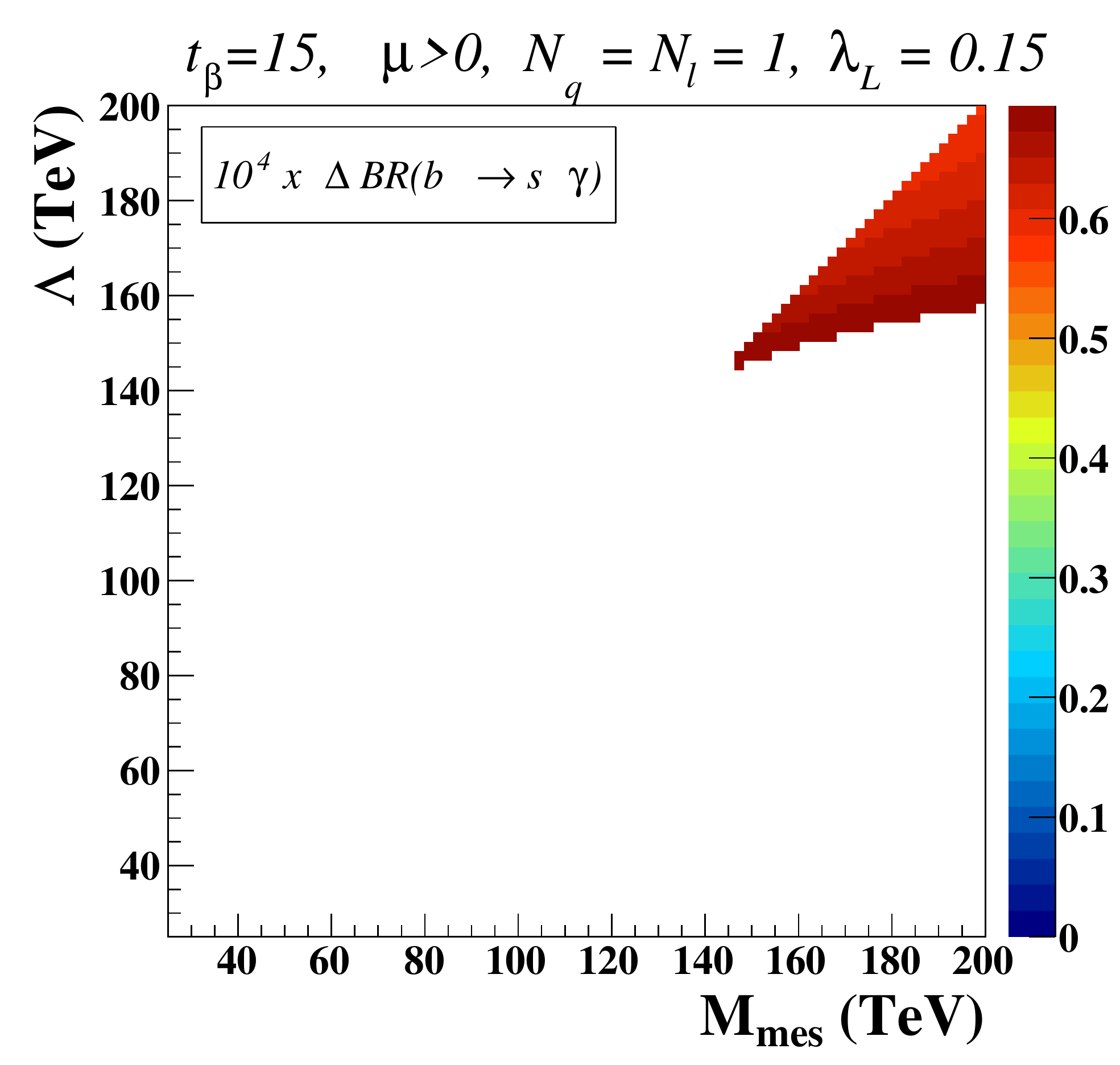}
 \includegraphics[width=.32\columnwidth]{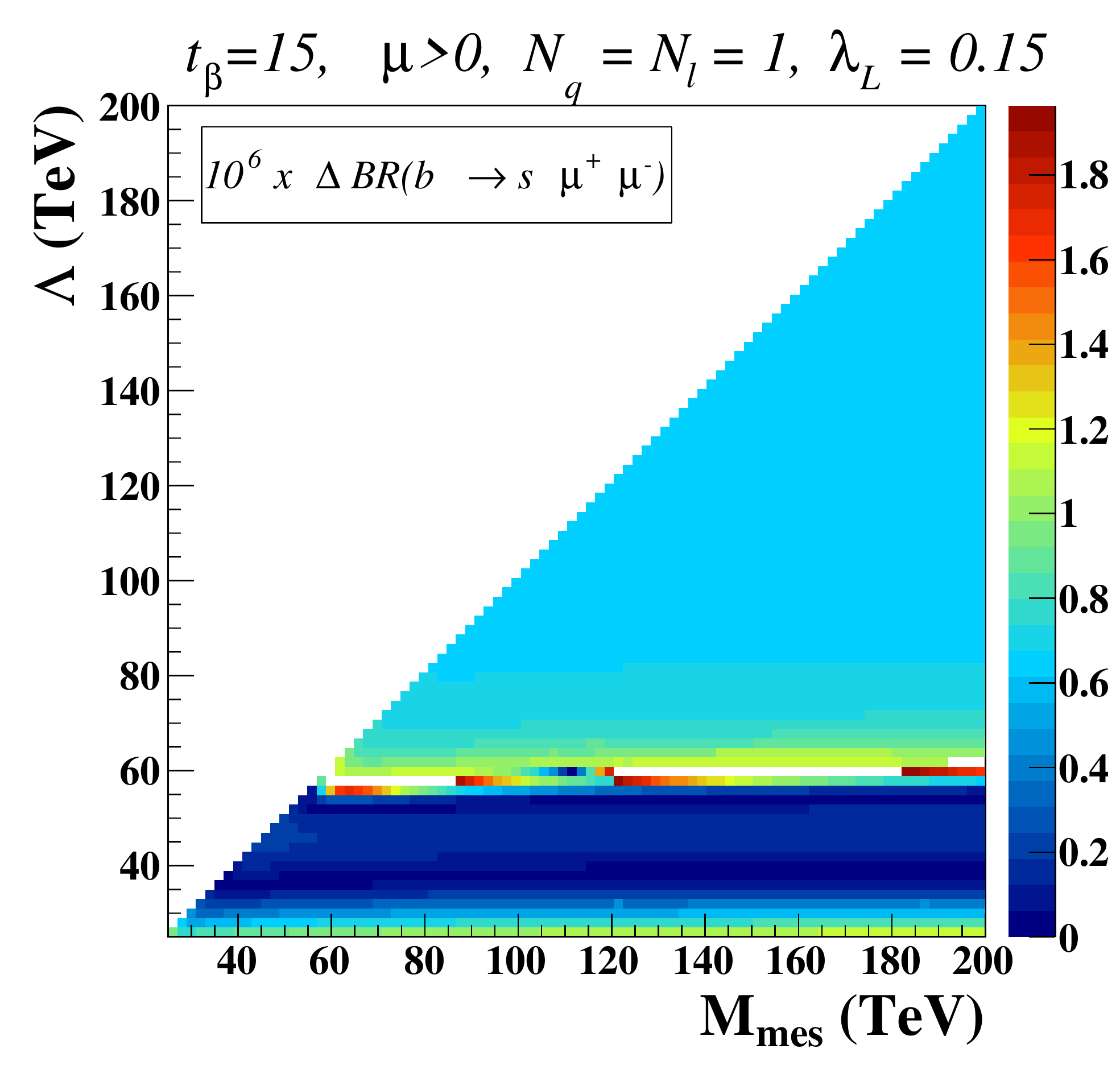}
 \includegraphics[width=.32\columnwidth]{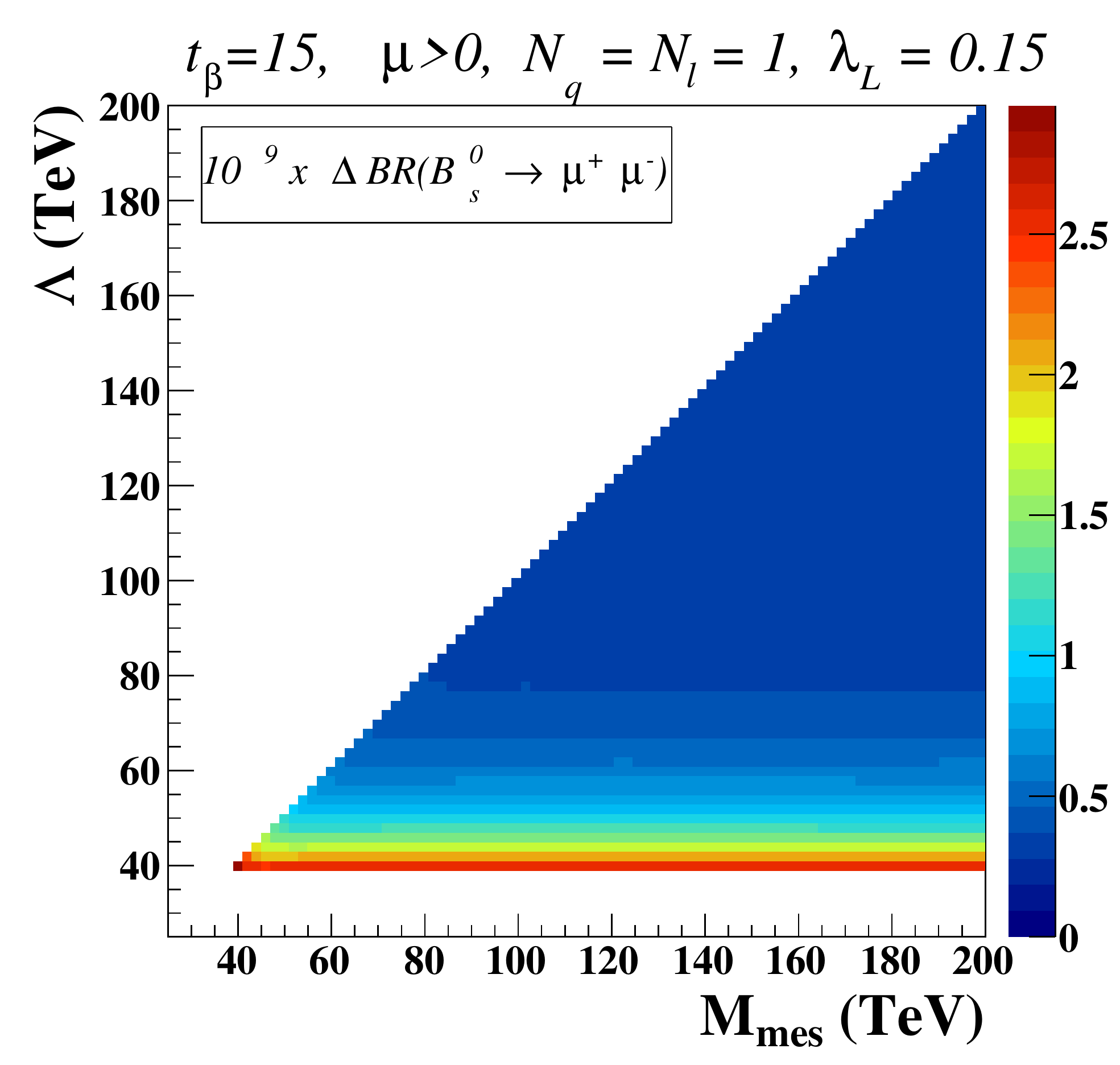}
 \caption{\label{fig:gmsb1_bdecaysL} Same as in Figure
\ref{fig:gmsb1_bdecays} (MSSM with gauge-mediated supersymmetry breaking,
  $\tan\beta=15$, $N_q = N_\ell =1$, $\mu>0$) with
  $\lambda_L = 0.15$.}
\vspace{.3cm}
 \includegraphics[width=.32\columnwidth]{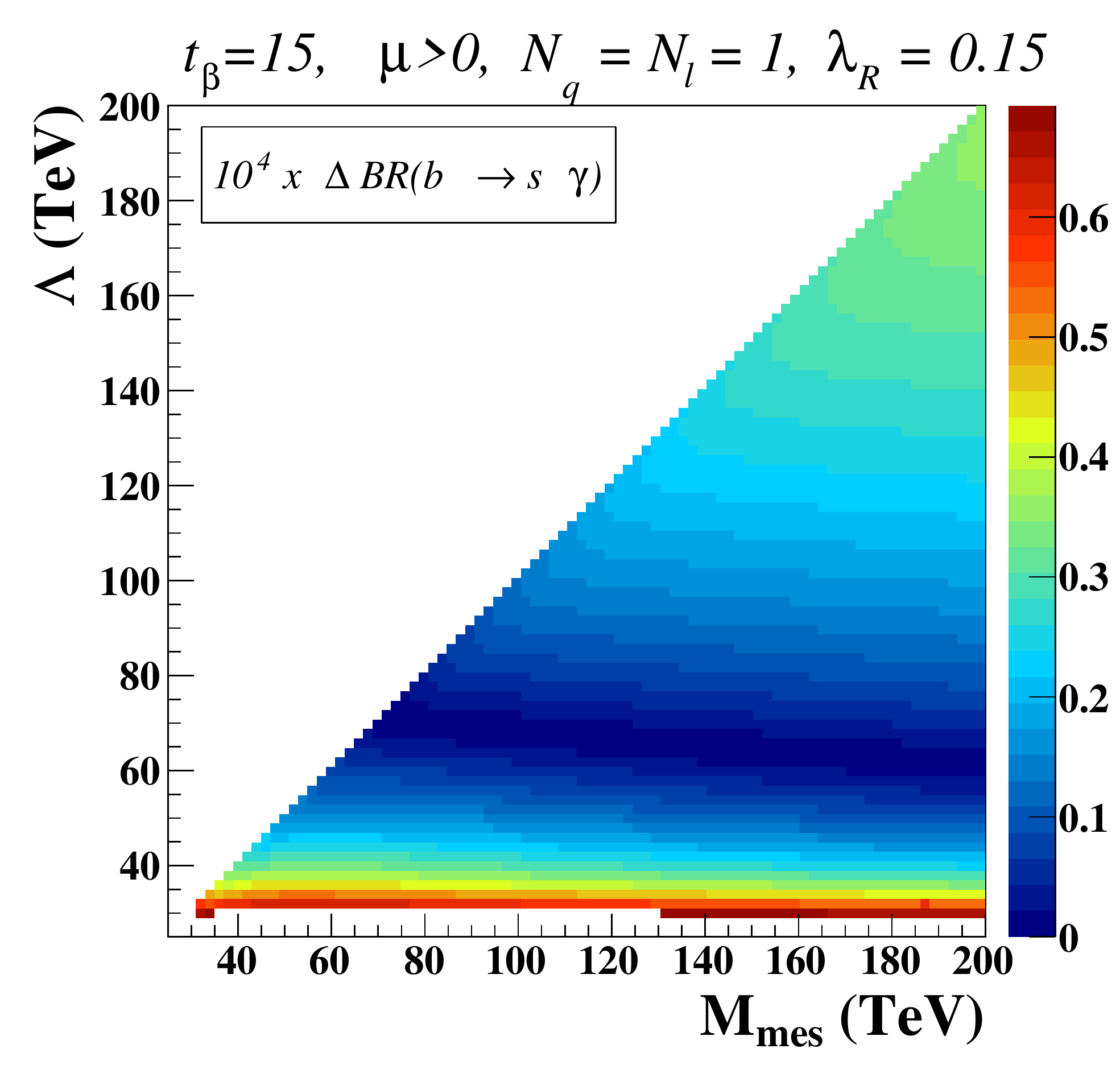}
 \includegraphics[width=.32\columnwidth]{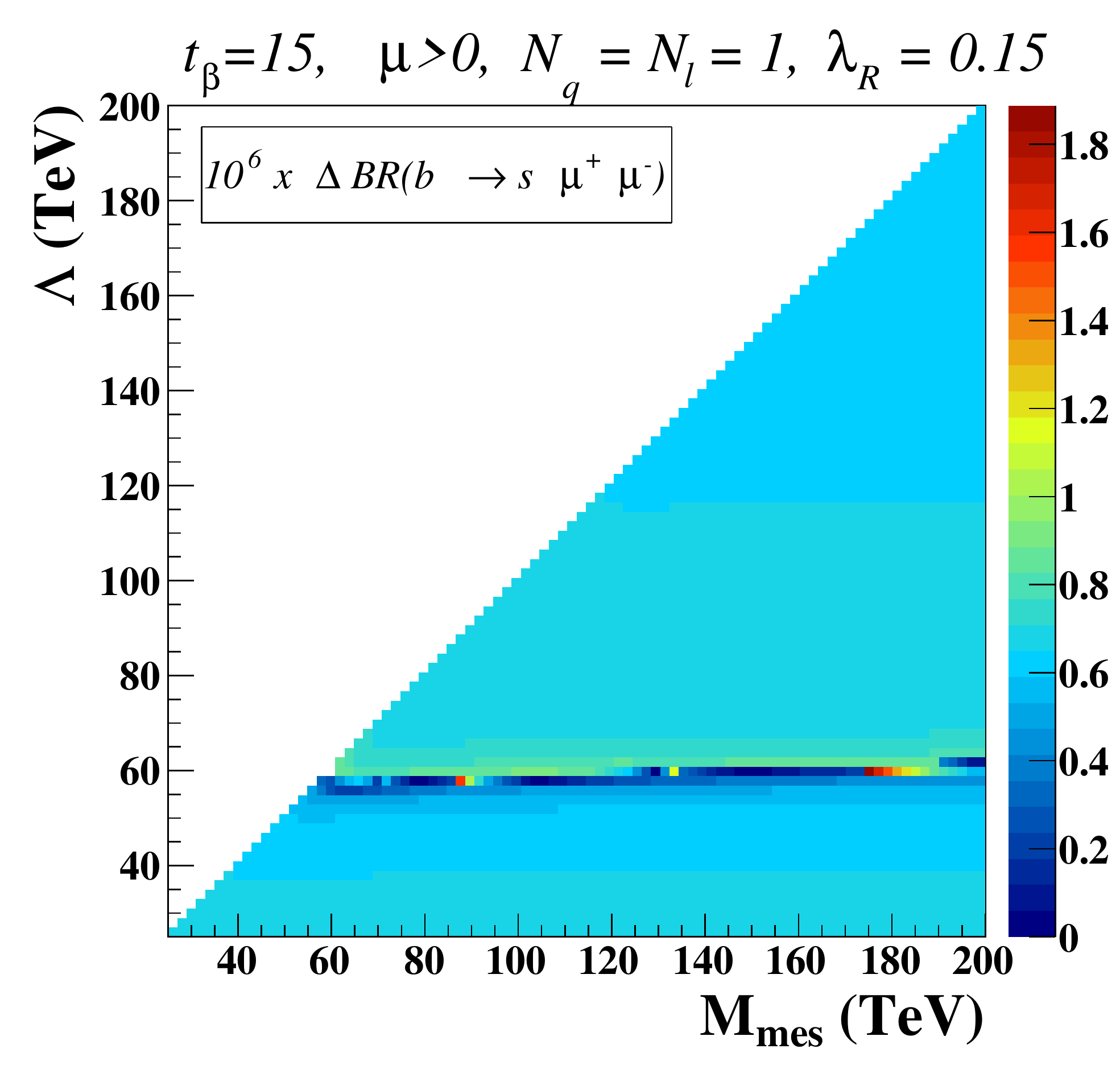}
 \includegraphics[width=.32\columnwidth]{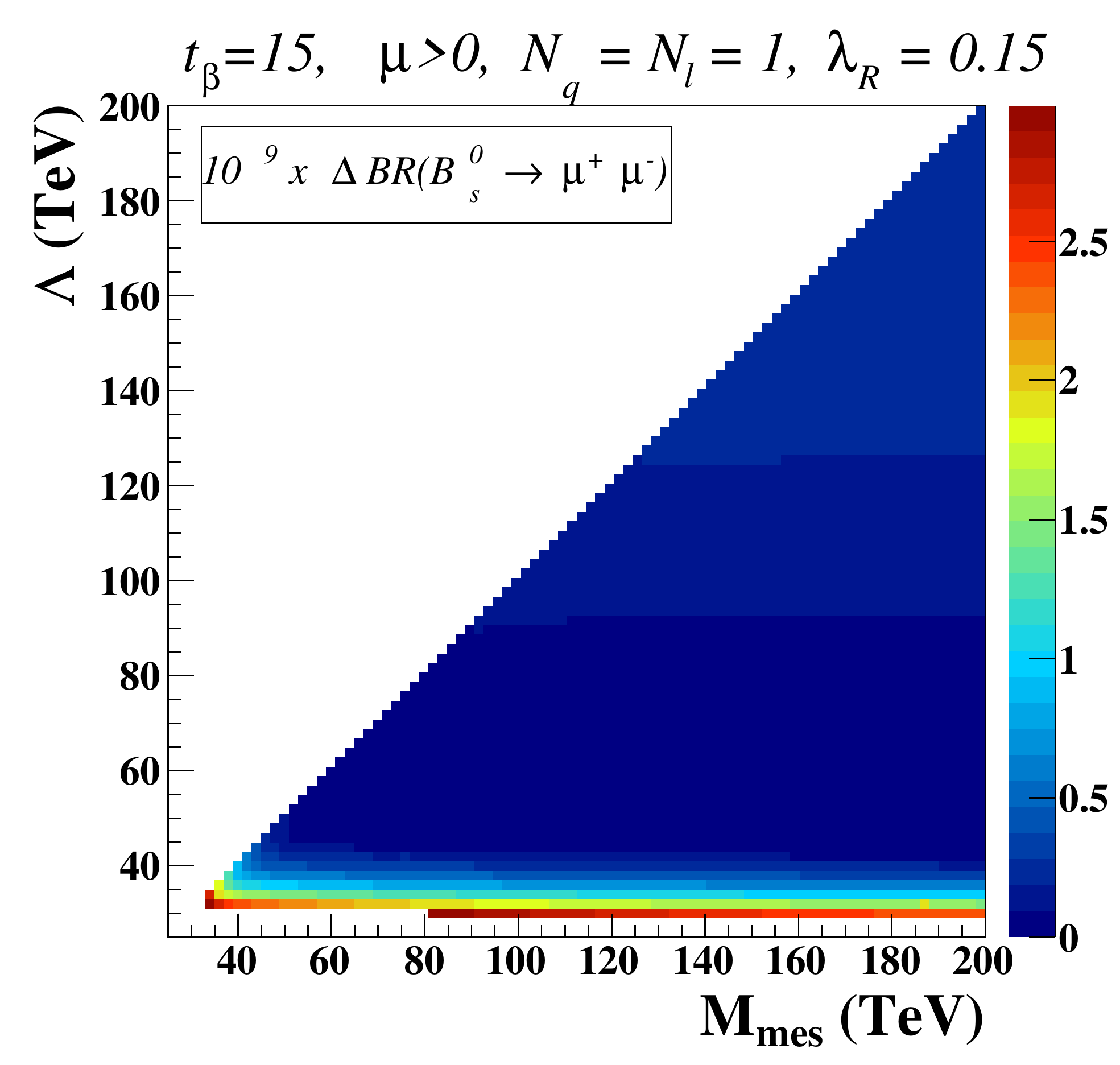}
 \caption{\label{fig:gmsb1_bdecaysR} Same as in Figure
\ref{fig:gmsb1_bdecays} (MSSM with gauge-mediated supersymmetry breaking,
  $\tan\beta=15$, $N_q = N_\ell =1$, $\mu>0$) with
  $\lambda_R = 0.15$.}
\vspace{.3cm}
 \includegraphics[width=.32\columnwidth]{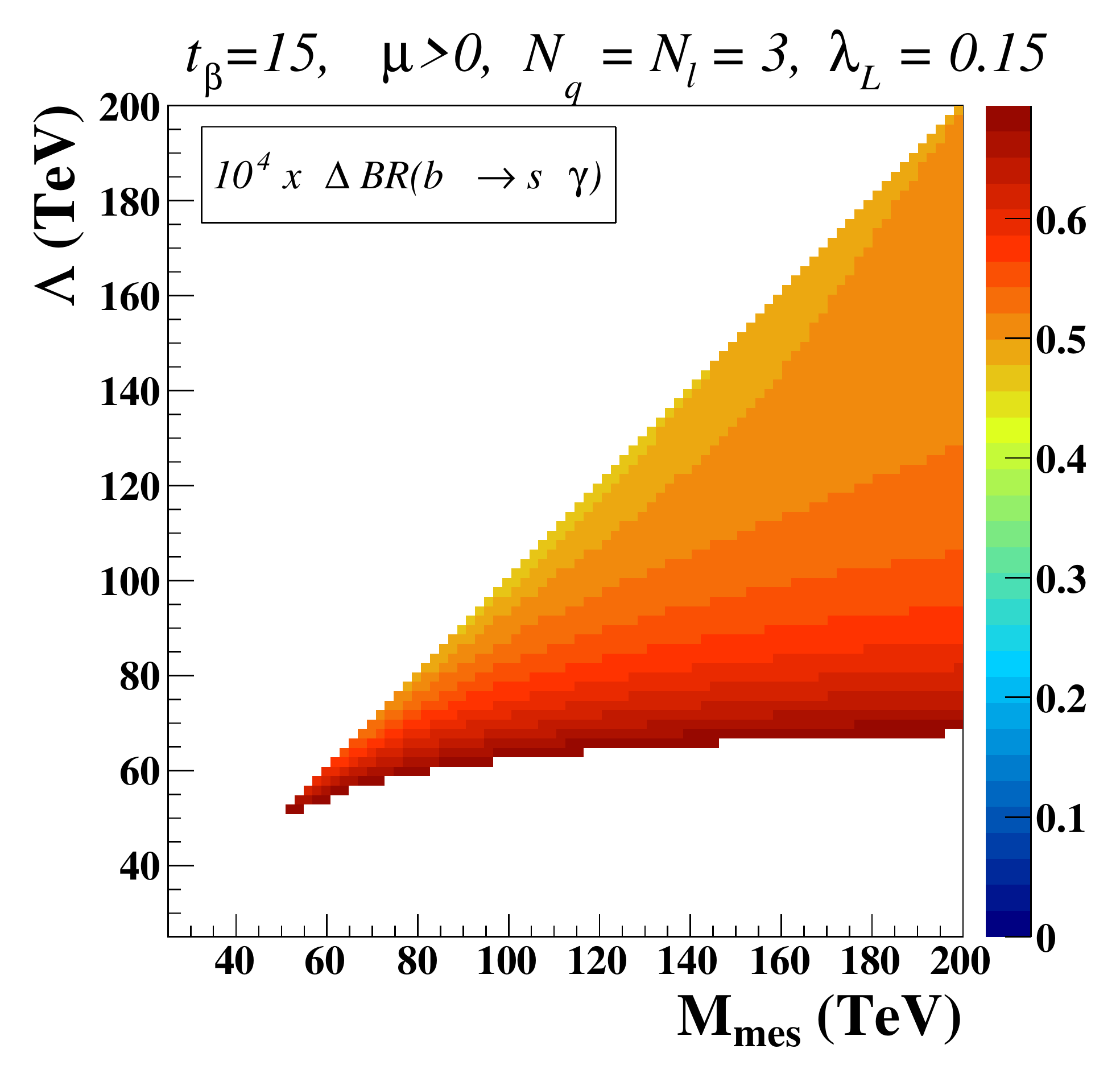}
 \includegraphics[width=.32\columnwidth]{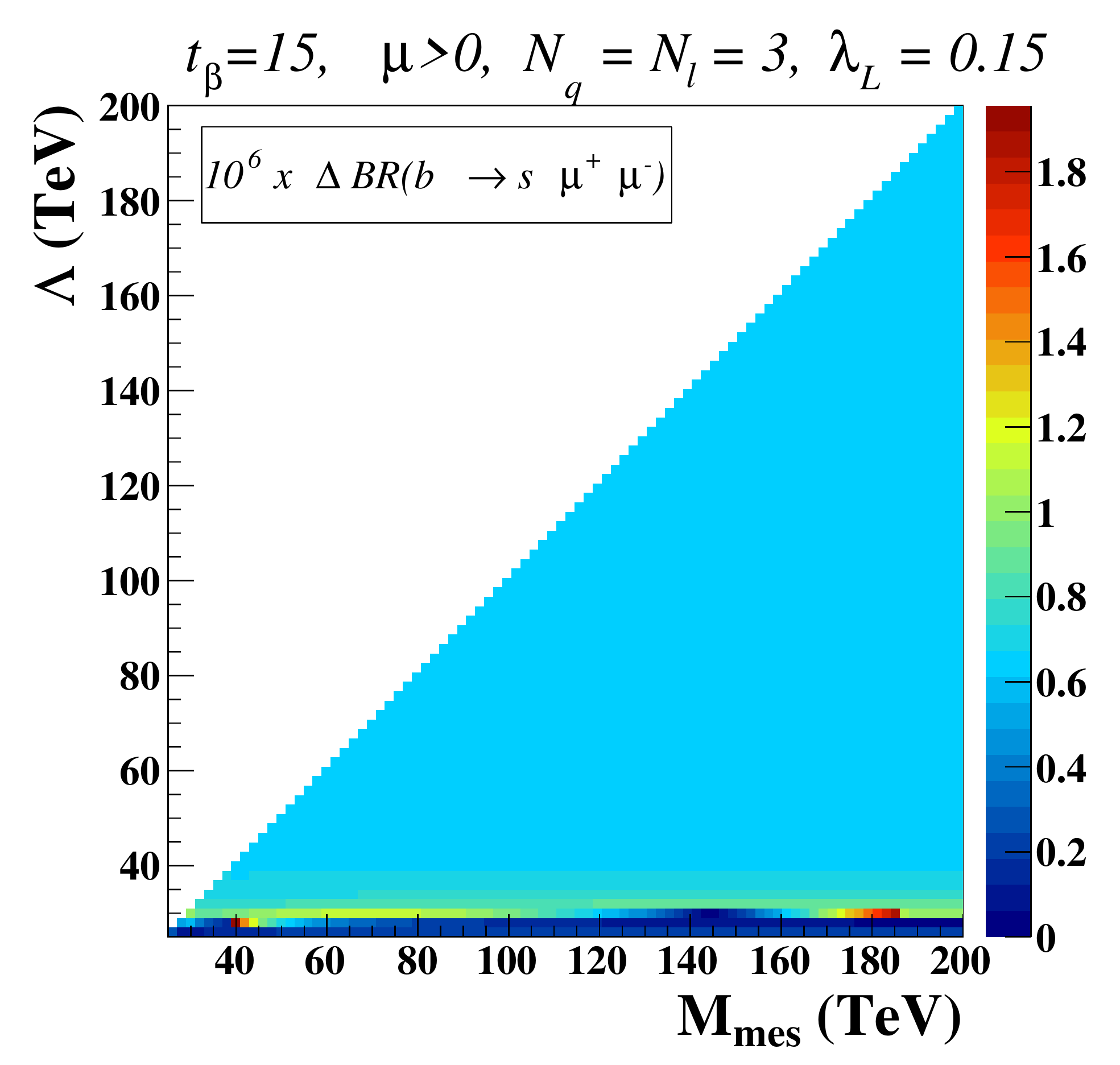}
 \includegraphics[width=.32\columnwidth]{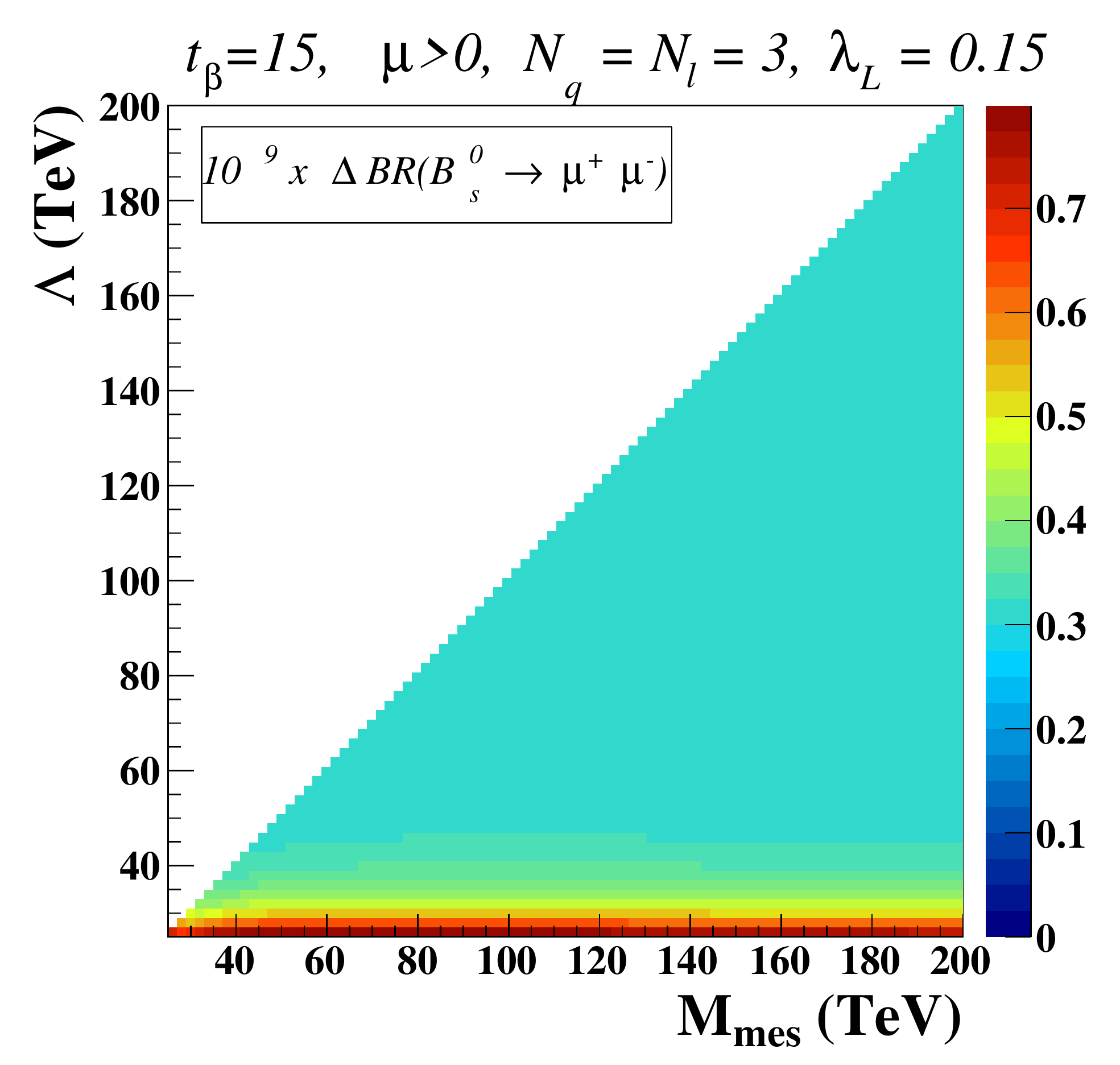}
 \caption{\label{fig:gmsb3_bdecaysL} Same as in Figure
\ref{fig:gmsb3_bdecays} (MSSM with gauge-mediated supersymmetry breaking,
  $\tan\beta=15$, $N_q = N_\ell =3$, $\mu>0$) with
  $\lambda_L = 0.15$.}
\end{figure}
%
In contrast, the $b\to s
\gamma$ branching ratio is insensitive to non-minimal 
right-right chiral squark mixings. The possible effects of a non-vanishing 
$\lambda_R$ parameter are related to the couplings of the higgsino component of the chargino and
neutralino states and cMSSM scenarios usually imply lighter winos and heavier 
higgsinos, the associated $\lambda_R$-dependent contributions to the $b\to s\gamma$ branching 
ratio being thus suppressed.

%
\begin{figure}
\centering
 \includegraphics[width=.32\columnwidth]{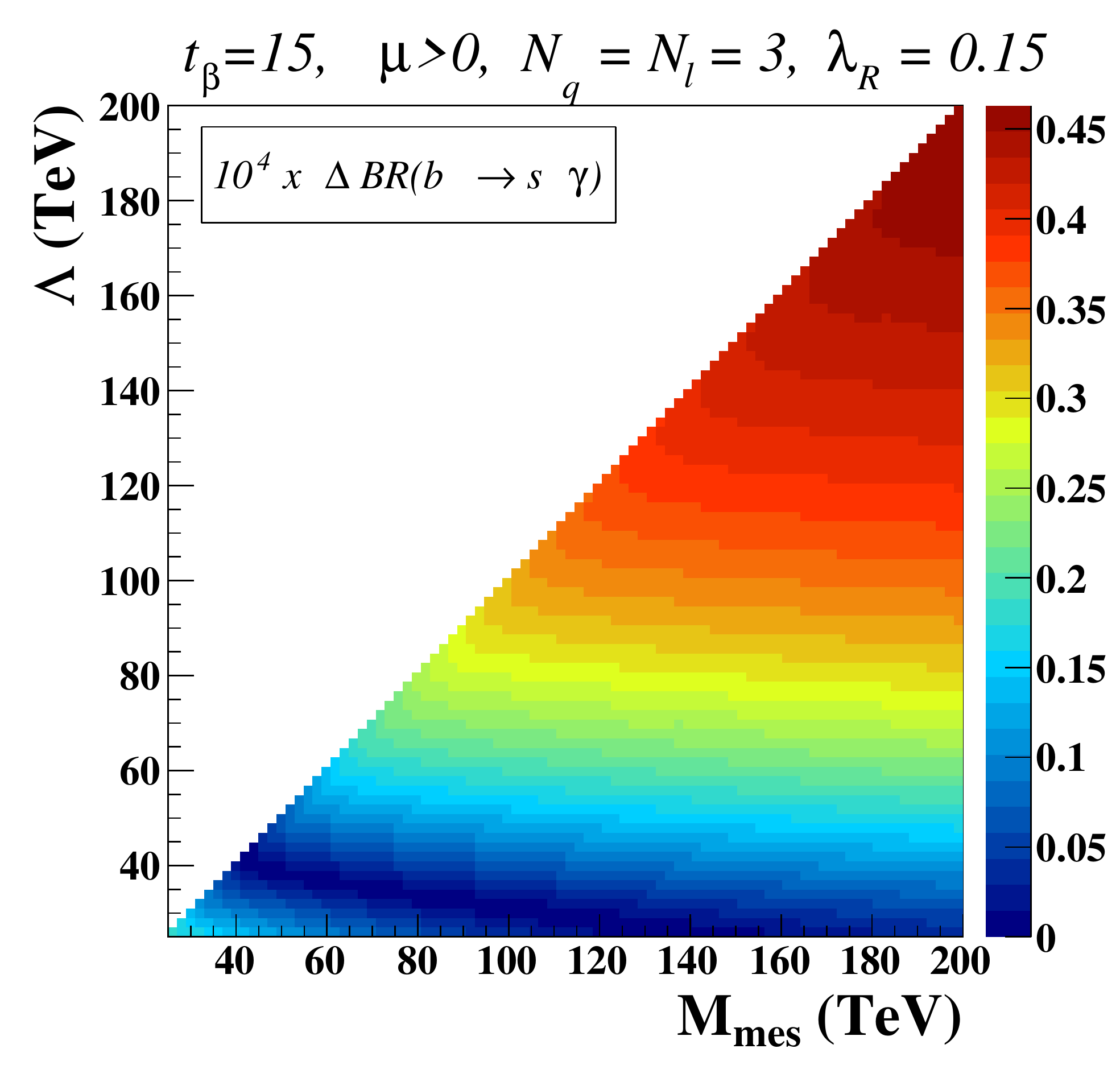}
 \includegraphics[width=.32\columnwidth]{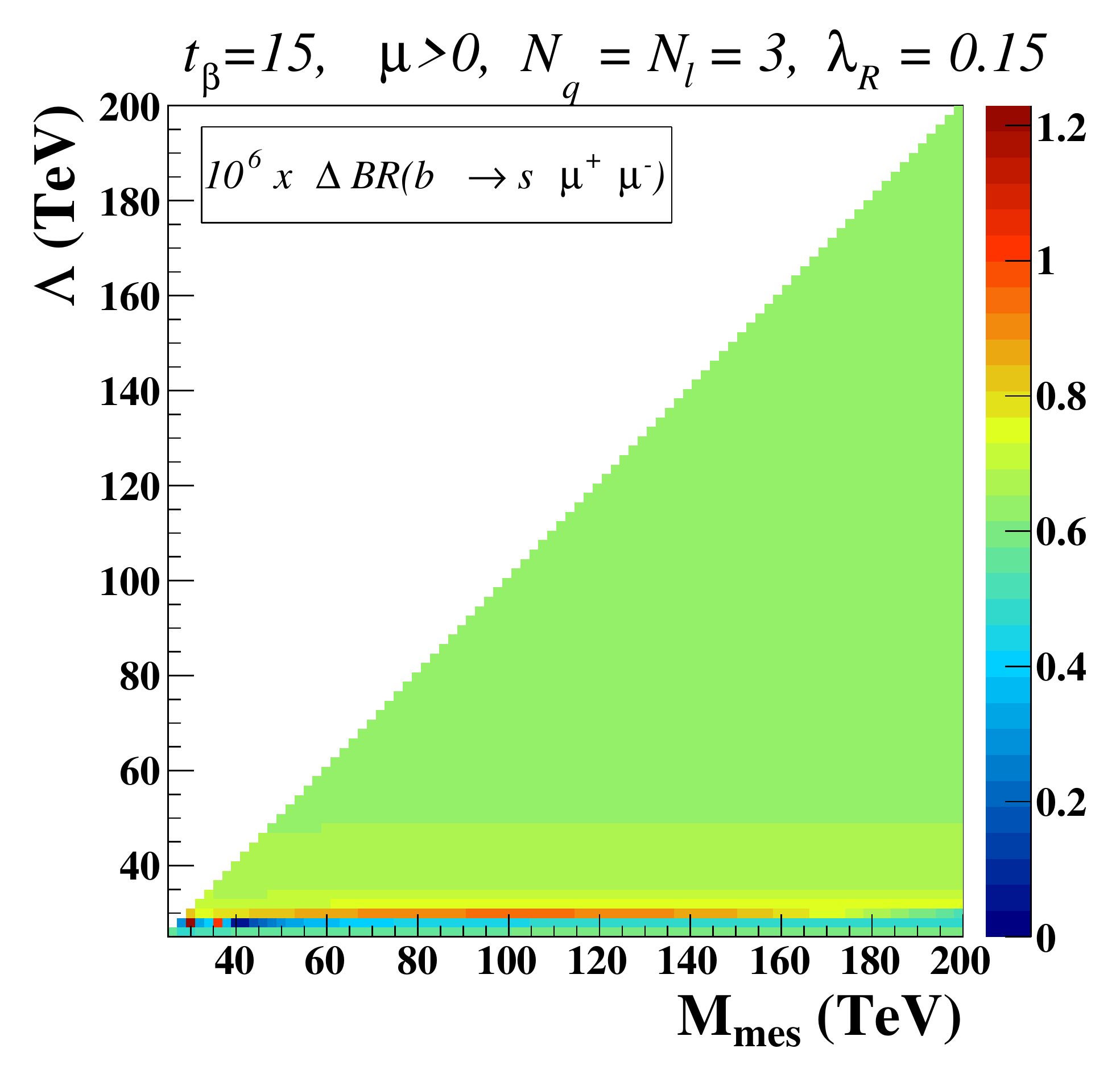}
 \includegraphics[width=.32\columnwidth]{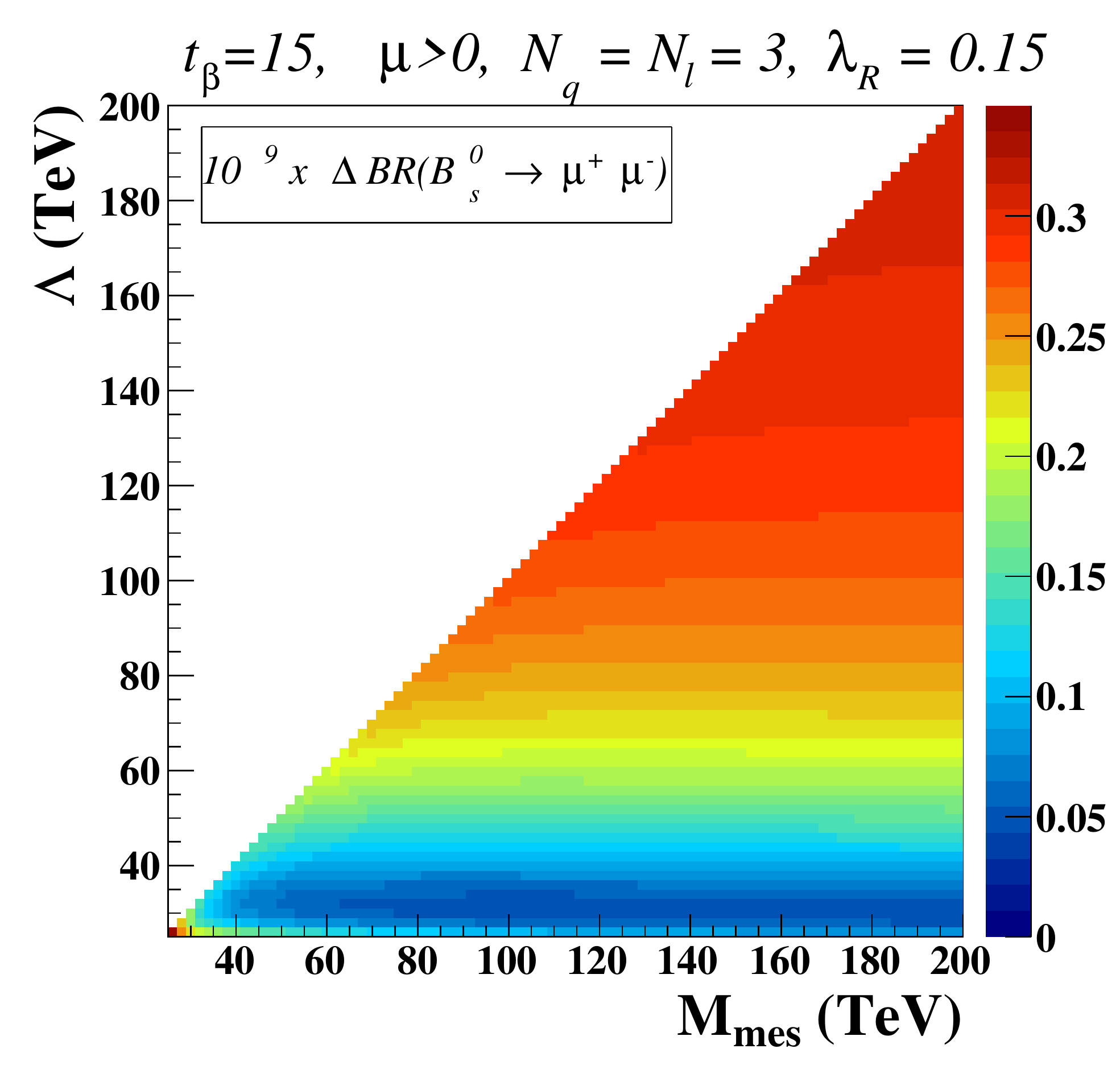}
 \caption{\label{fig:gmsb3_bdecaysR} Same as in Figure
\ref{fig:gmsb3_bdecays} (MSSM with gauge-mediated supersymmetry breaking,
  $\tan\beta=15$, $N_q = N_\ell =3$, $\mu>0$) with
  $\lambda_R = 0.15$.}
\vspace{.3cm}
 \includegraphics[width=.32\columnwidth]{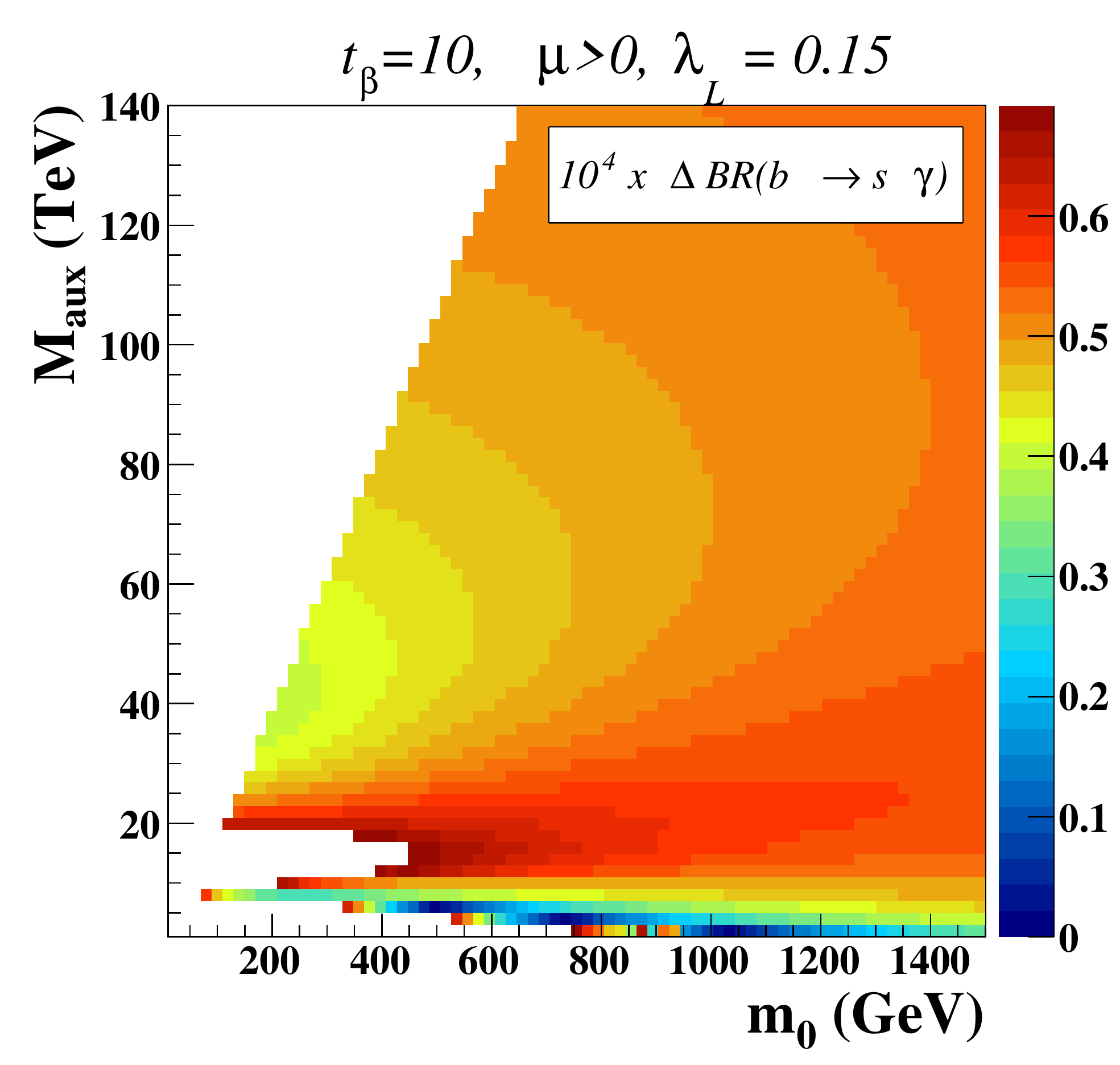}
 \includegraphics[width=.32\columnwidth]{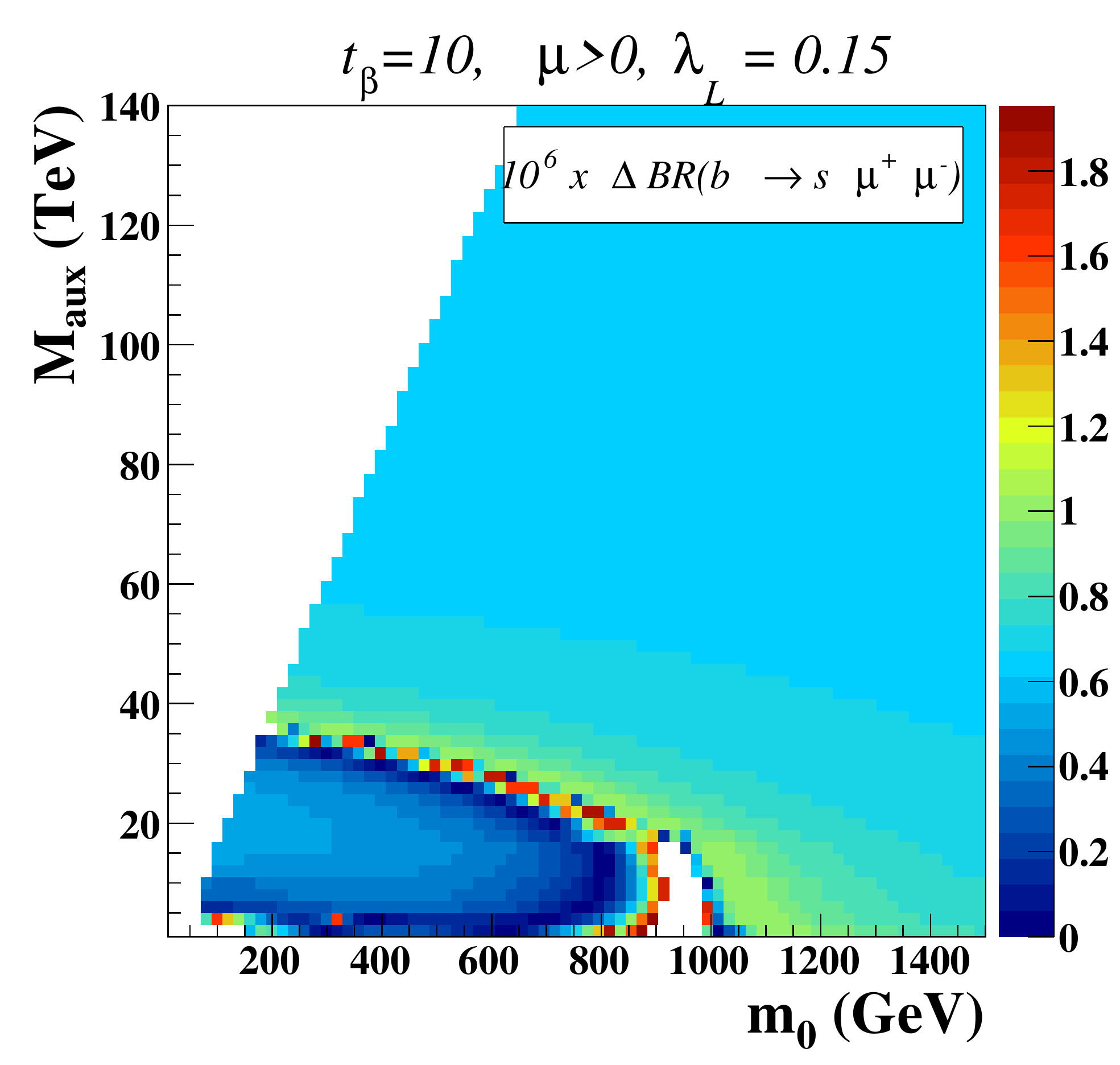}
 \includegraphics[width=.32\columnwidth]{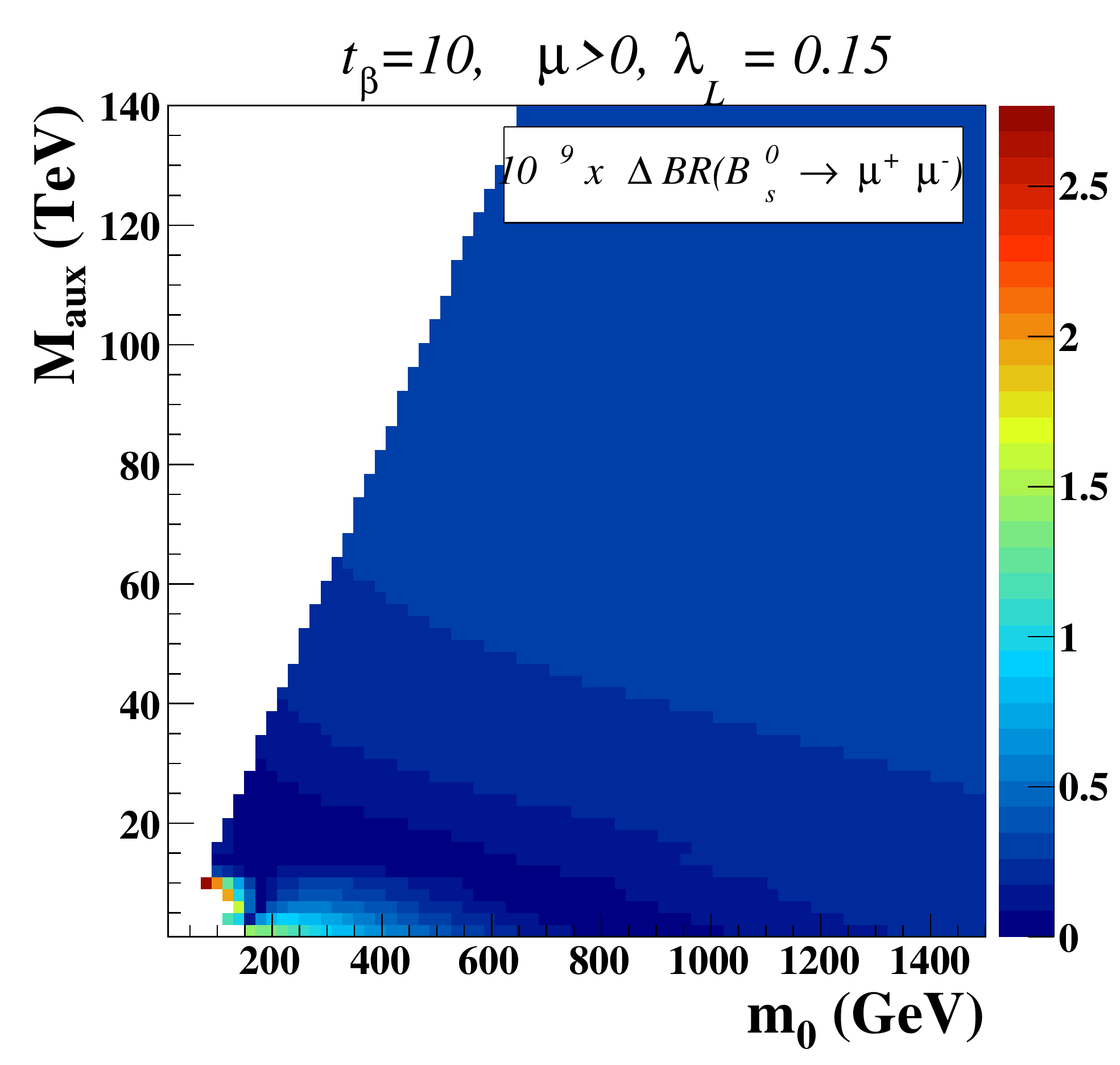}
 \caption{\label{fig:amsb_bdecaysL} Same as in Figure
\ref{fig:amsb_bdecays} (MSSM with anomaly-mediated supersymmetry breaking,
  $\tan\beta=10$, $\mu>0$) with $\lambda_L = 0.15$.}
\vspace{.3cm}
 \includegraphics[width=.32\columnwidth]{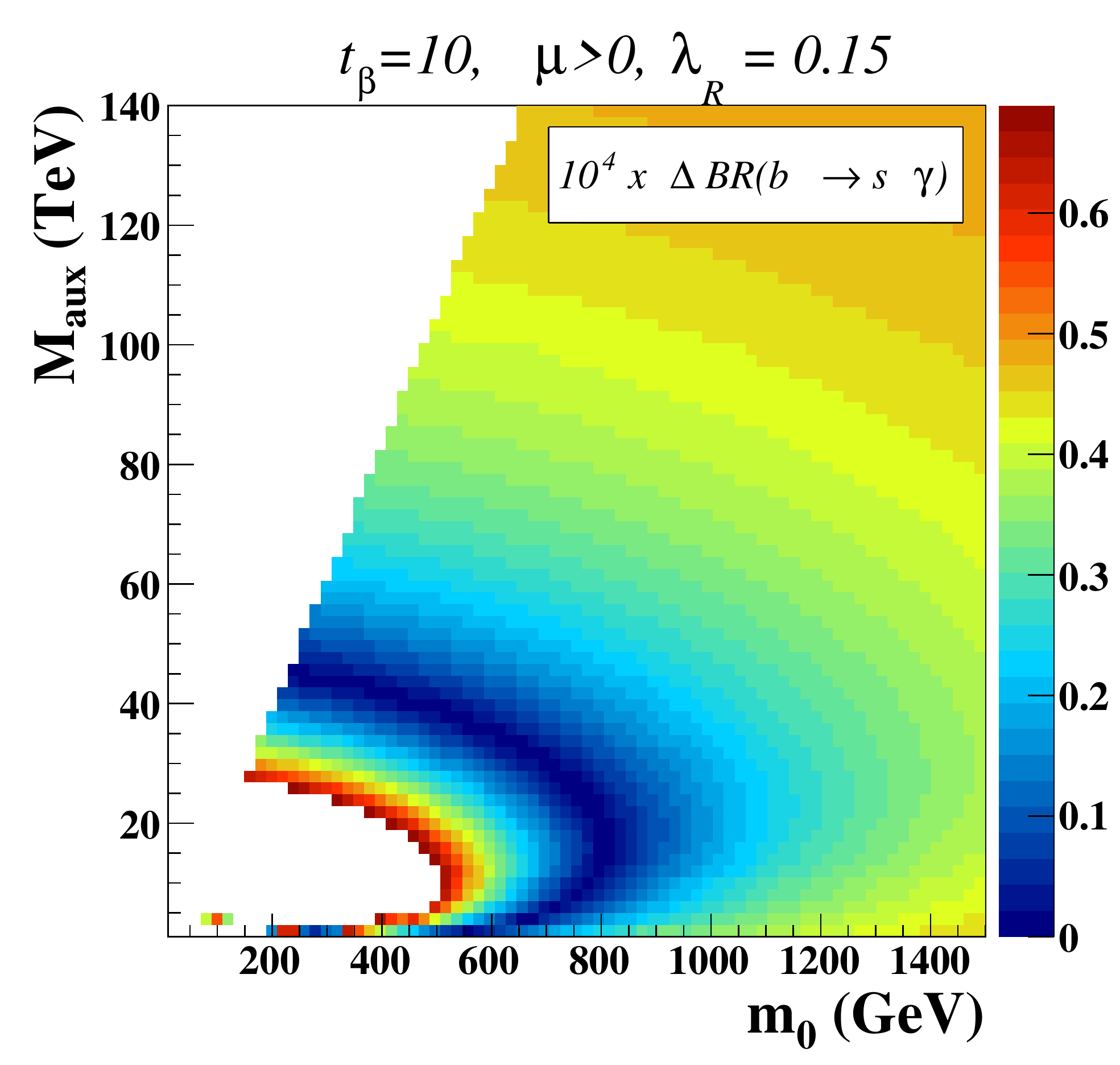}
 \includegraphics[width=.32\columnwidth]{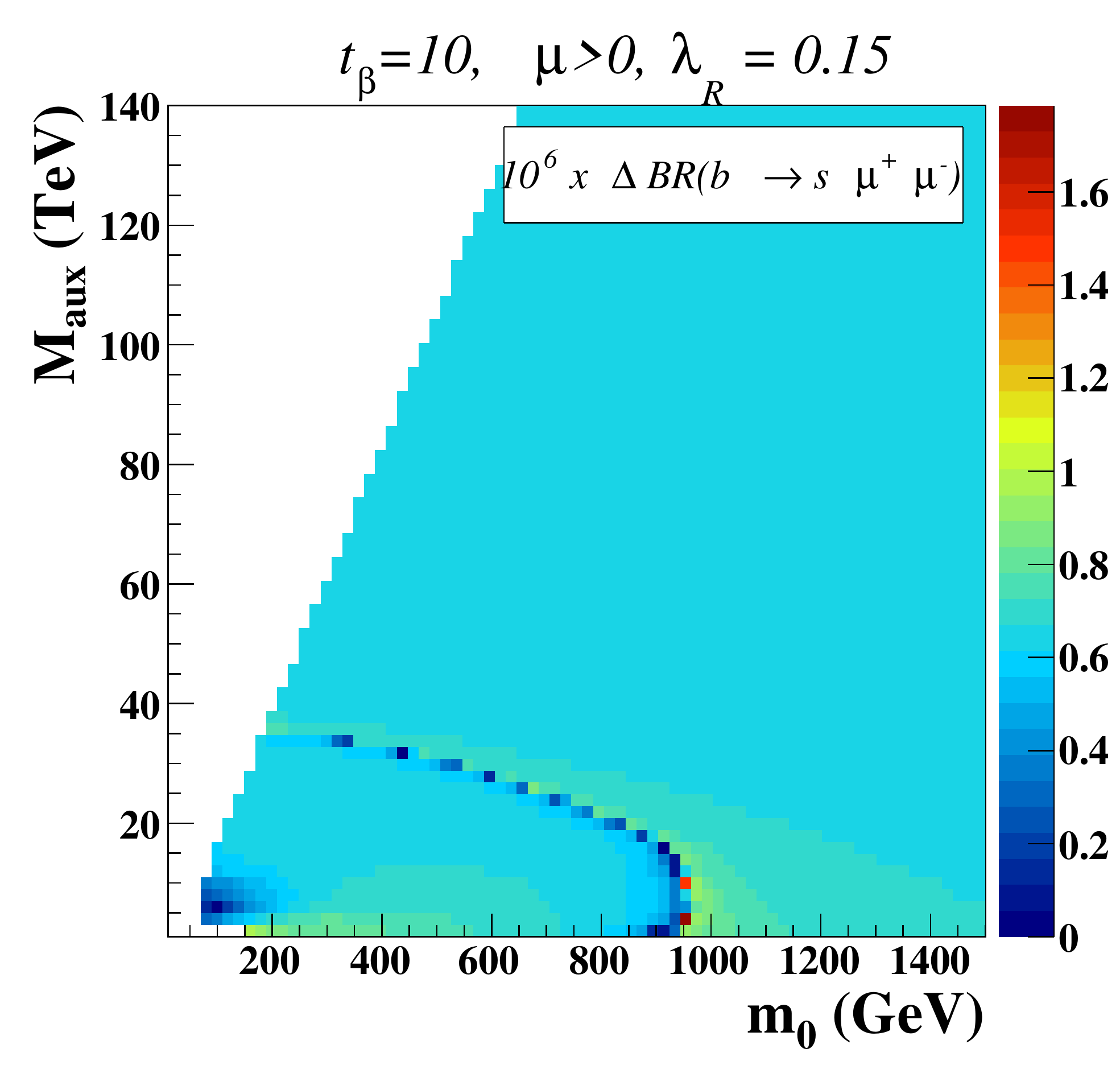}
 \includegraphics[width=.32\columnwidth]{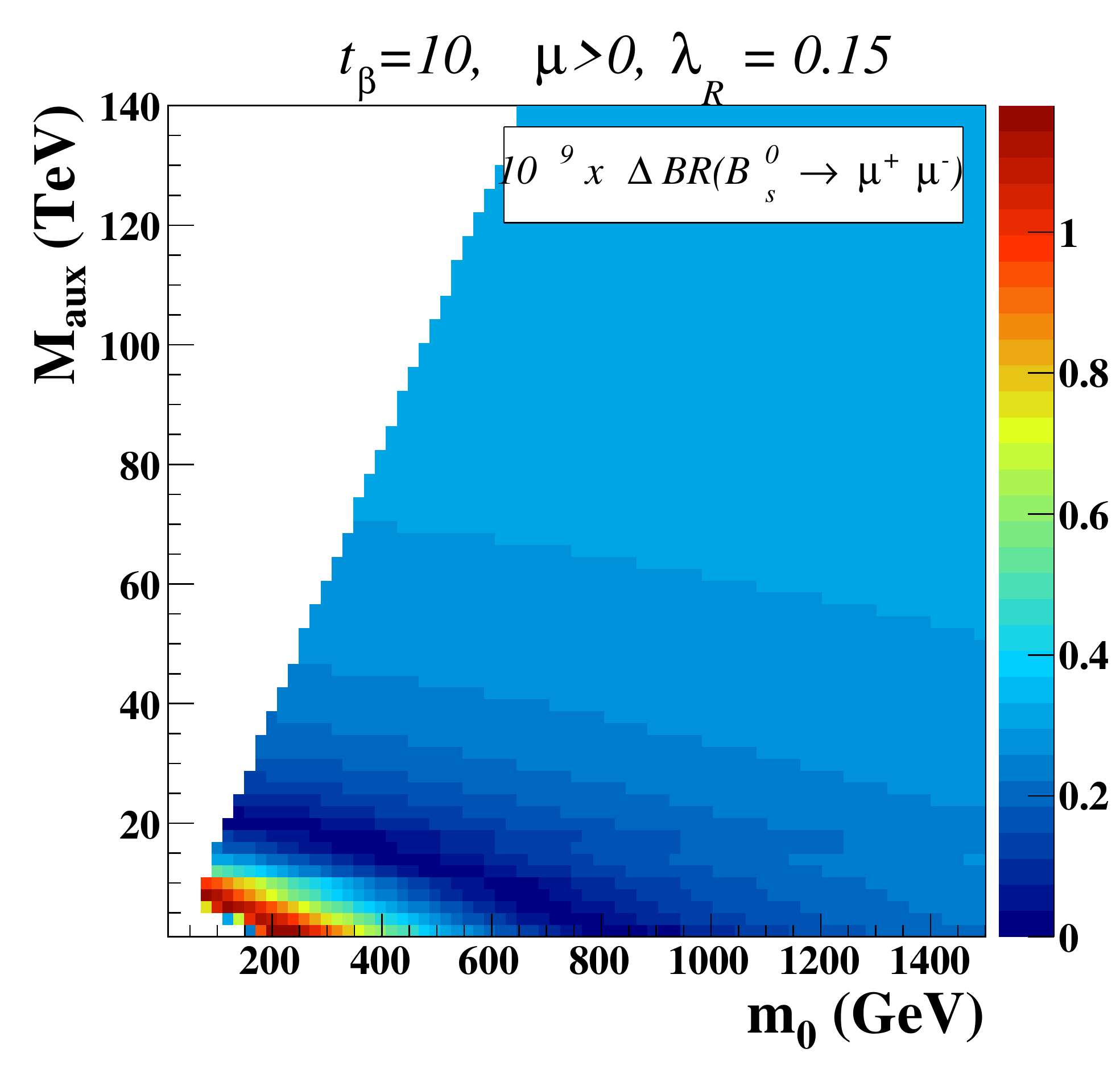}
 \caption{\label{fig:amsb_bdecaysR} Same as in Figure
\ref{fig:amsb_bdecays} (MSSM with anomaly-mediated supersymmetry breaking,
  $\tan\beta=10$, $\mu>0$) with $\lambda_R = 0.15$.}
\end{figure}
%

Non-minimal left-left squark chiral mixings also affect 
predictions for the $b\to s\mu^+ \mu^-$ 
branching ratio (middle panels of the figures). All cMSSM regions
regions excluded by $b\to s \mu^+ \mu^-$ branching ratio
measurements are however found to be also excluded by the $b\to s \gamma$ data. 
Contrary, right-right chiral mixings (right panels of the figures) cannot be probed,
for the same reason as in the
$b\to s\gamma$ case.

The large uncertainty on the $B_s^0\to\mu^+
\mu^-$ branching ratio measurement potentially reduces the relevance of 
this observable to constrain squark flavor-violating mixings, in particular
when confronting the constraining power of the $b\to s \gamma$ rare decay (see
Figure \ref{fig:cmssm10_bdecaysL}, Figure \ref{fig:cmssm10_bdecaysR},
Figure~\ref{fig:cmssm40_bdecaysL} and Figure \ref{fig:cmssm40_bdecaysR}).

The same effects can be observed for gauge-mediated
supersymmetry-breaking scenarios, as illustrated on Figure
\ref{fig:gmsb1_bdecaysL}, Figure \ref{fig:gmsb1_bdecaysR},  Figure
\ref{fig:gmsb3_bdecaysL} and Figure \ref{fig:gmsb3_bdecaysR}, and for
anomaly-mediated supersymmetry breaking MSSM scenarios as shown on
Figure \ref{fig:amsb_bdecaysL} and Figure \ref{fig:amsb_bdecaysR}.
The theoretical predictions are
insensitive to non-minimal flavor-violating 
right-right chiral mixings and the $b\to s \gamma$
branching ratio is the only observable capable to restrict the
possibilities of constructing viable scenarios with non-minimal left-left chiral 
squark mixing.

\subsection{$B$-meson oscillations}\label{sec:osc}

Weak interactions are responsible for providing different
masses to the flavor eigenstates $|B_q^0 \rangle = | \bbar q\rangle$ and
$|\Bbar_q^0 \rangle = | b \qbar \rangle$, with $q = s$ or $d$, as well as mass
mixing terms among those states. Consequently, the light ($L$) and heavy ($H$)
neutral $B_q$-meson
mass-eigenstates $|B_q^L\rangle$ and $|B_q^H\rangle$ differ in their
masses and in their decay widths. Neglecting $CP$-violation since it is
expected to be very small \cite{Beneke:1998sy, Lenz:2006hd, Lenz:2011ti}, the
mass eigenstates are also $CP$-eigenstates, the light $|B_q^L\rangle$
state being $CP$-even and the heavy $|B_q^H\rangle$ state $CP$-odd.
The evolution of a state prepared as a pure flavor eigenstate
$|B_q^0 \rangle$ or $|\Bbar_q^0 \rangle$ 
at a time $t=0$ is driven by the time-dependent probabilities ${\cal P}_{\rm
unmix}(t)$ and ${\cal P}_{\rm mix}(t)$ that the flavor remains unchanged or
oscillates, 
\be\bsp
   {\cal P}_{\rm mix}(t) =&\ \frac12 \Gamma_q e^{-\Gamma_q t} \Big[ 1 -
   \frac{\Delta\Gamma_q^2}{4 \Gamma^2_q}\Big] \Big[ \cosh \frac{\Delta\Gamma_q}{2}
   t - \cos\Delta M_q t \Big] \ , \\
   {\cal P}_{\rm unmix}(t) =&\ \frac12 \Gamma_q e^{-\Gamma_q t} \Big[ 1 -
   \frac{\Delta\Gamma_q^2}{4 \Gamma^2_q}\Big] \Big[ \cosh \frac{\Delta\Gamma_q}{2}
   t + \cos\Delta M_q t \Big] \ ,
\esp\label{eq:bmesosc}\ee
respectively.
In those expressions, $M_q$ and $\Gamma_q$ are the average mass and width of the
$B_q$-mesons while $\Delta M_q$ and $\Delta \Gamma_q$ are the mass and width
differences between the two eigenstates $|B_q^L\rangle$ and $|B_q^H\rangle$.

Among all the parameters introduced in Eq.\ \eqref{eq:bmesosc}, 
the mass difference $\Delta M_s$ is one of the key observables restricting the
design of
phenomenologically viable supersymmetric scenarios. The
$B_s^0-\Bbar_s^0$ oscillations have been observed for the first
time in 2006 by the CDF and D0 collaborations \cite{Abulencia:2006mq,
Abazov:2006dm} and the precision on the measurement
increases after
the LHCb collaboration publishes results derived from the latest LHC data
\cite{Aaij:2011qx}. The oscillation
frequency, given by the mass difference $\Delta M_s$, then reads
\cite{Amhis:2012bh}
\be
  \Delta M_s = \big( 17.719 \pm 0.043_{\rm exp} \pm 3.3_{\rm
   theo}\big) \text{ ps}^{-1} \ ,
\label{eq:dmbs}\ee
where the theoretical uncertainty of 3.3 ps$^{-1}$ has been calculated in Ref.\
\cite{Ball:2006xx}.

%
\begin{figure}[t!]
 \centering
 \includegraphics[width=.32\columnwidth]{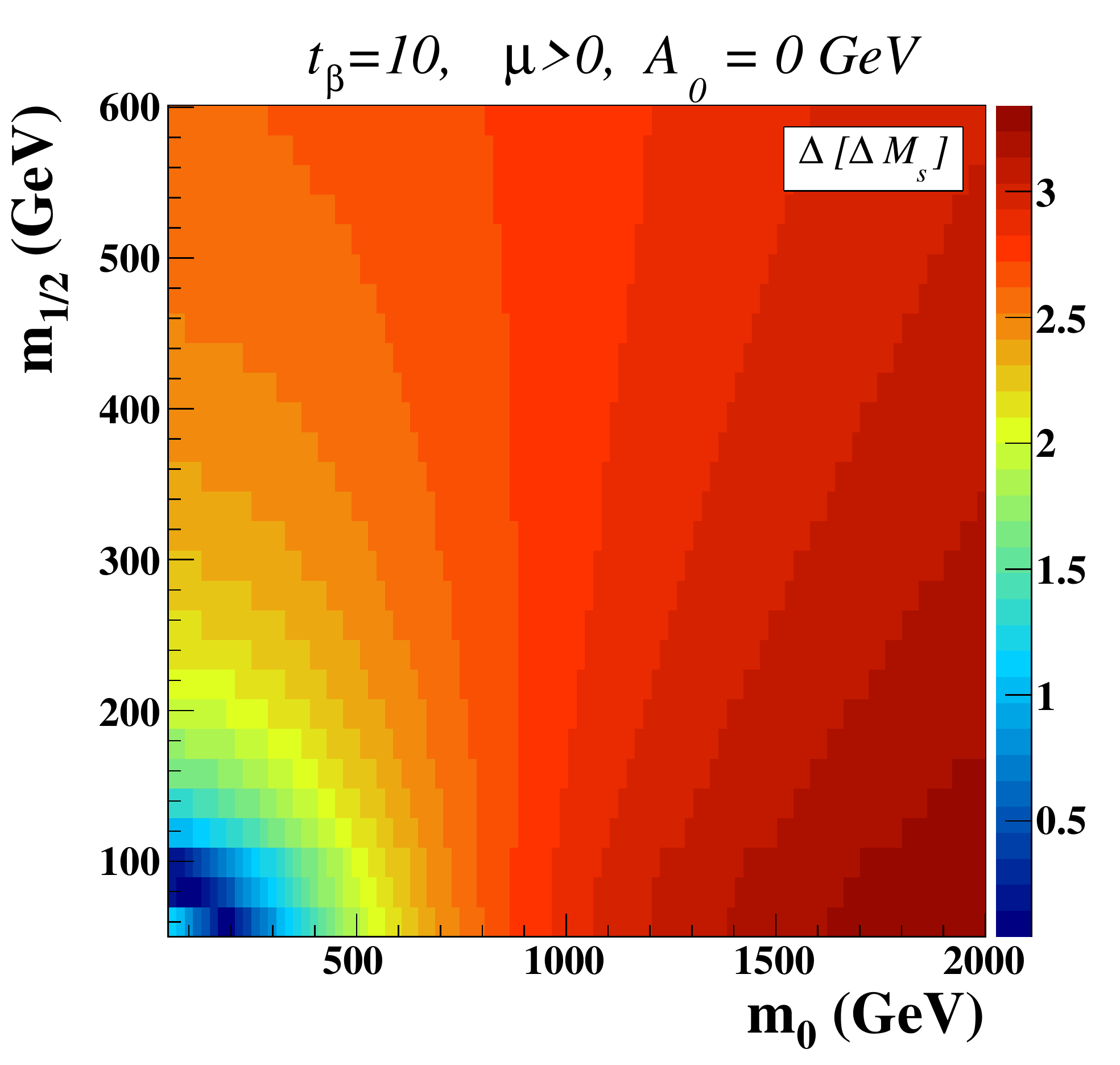}
 \includegraphics[width=.32\columnwidth]{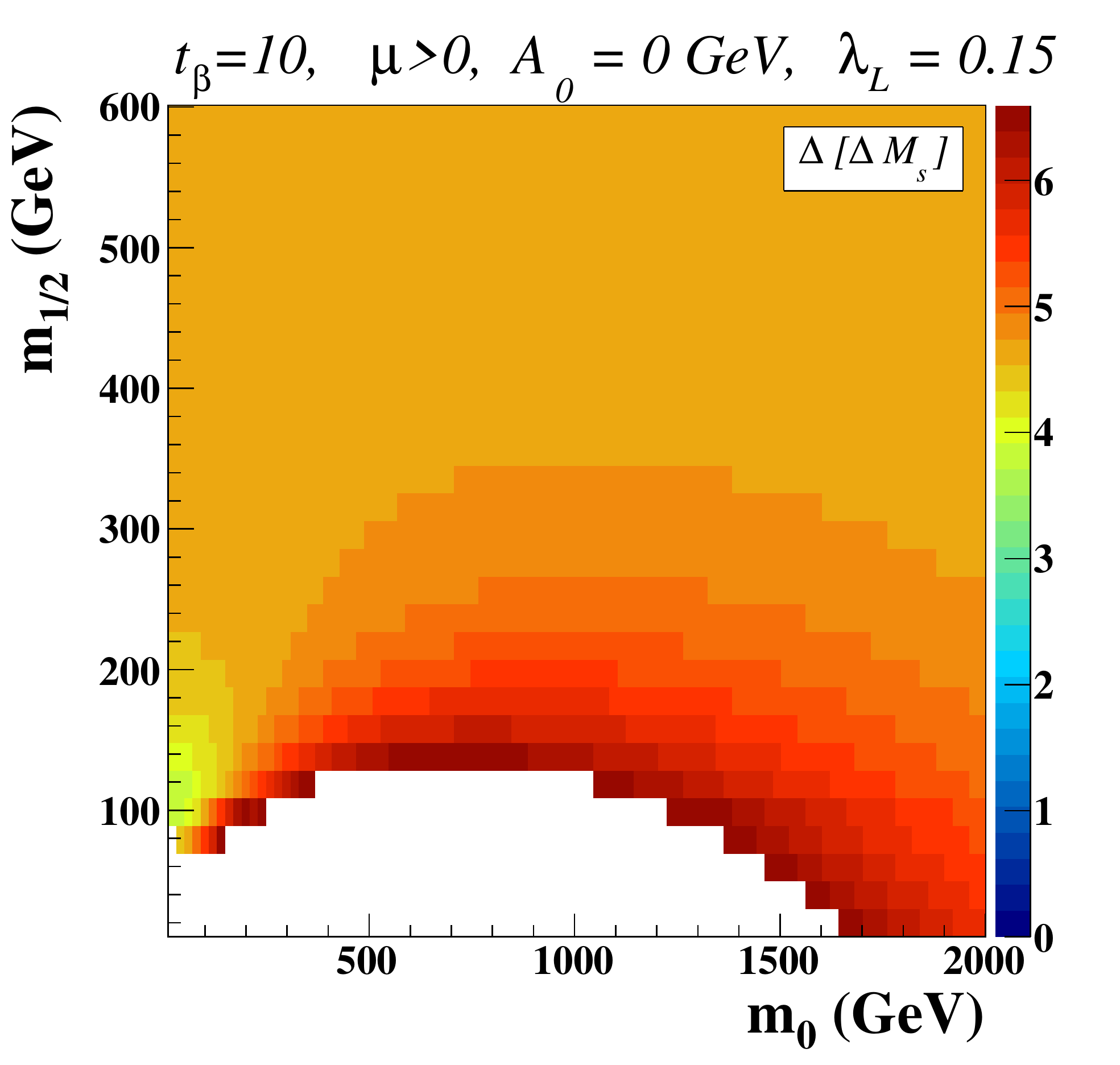}
 \includegraphics[width=.32\columnwidth]{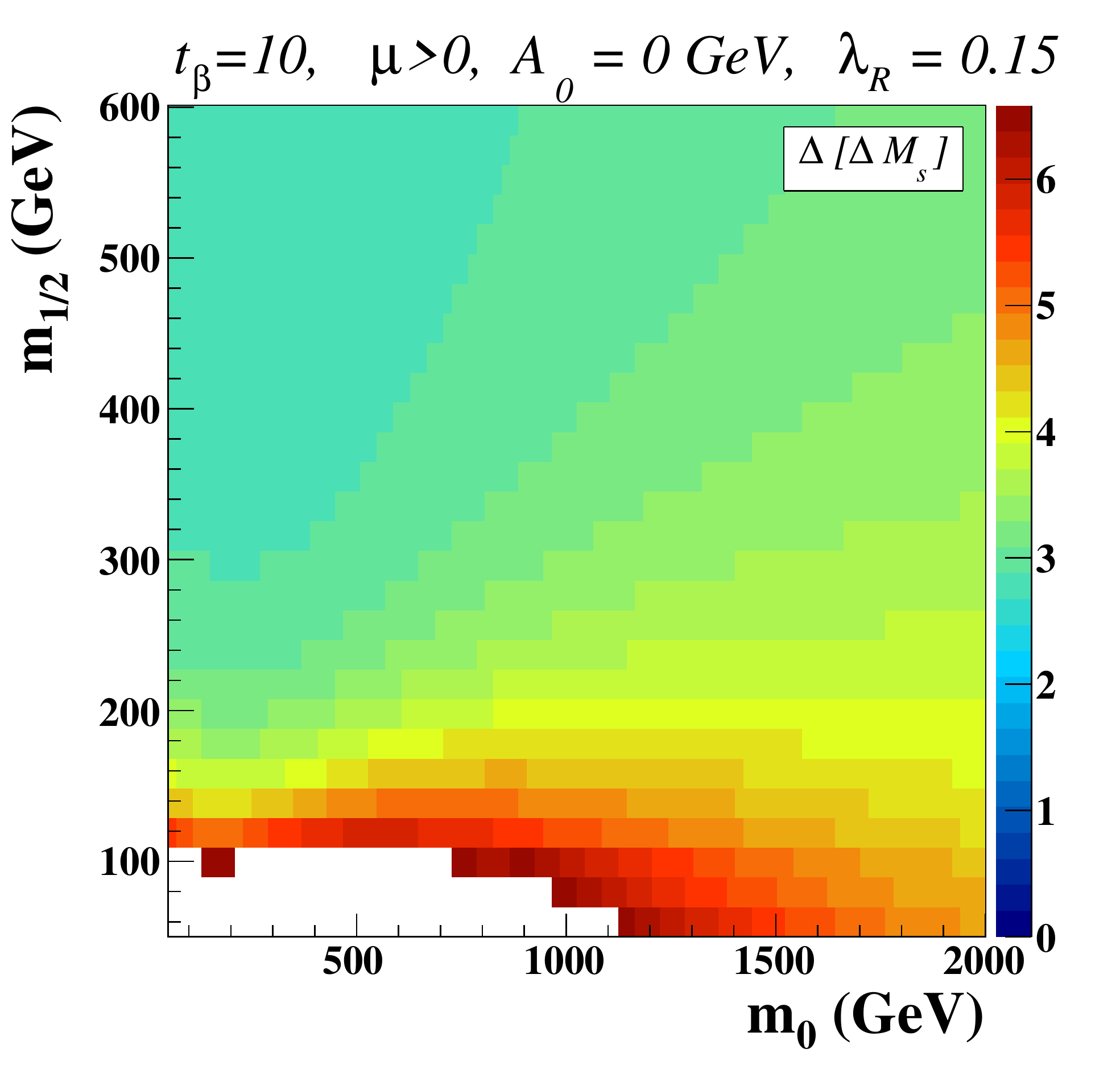}
 \caption{\label{fig:cmssm10_dMBs} Predicted deviations from the central
measured value for the mass difference between the two $B_s$-meson
mass-eigenstates, as given in Eq.\ \eqref{eq:dmbs}, in the framework of the 
cMSSM. We present the results in
$(m_0,m_{1/2})$-planes for fixed values of $\tan\beta=10$, $A_0=0$
GeV and a positive Higgs mixing parameter $\mu>0$. The regions depicted in white
correspond to excluded regions when applying the bounds of Eq.\ \eqref{eq:dmbs} 
at the $2\sigma$-level, or to regions for which there is no solution to the
supersymmetric renormalization group equations. Squark non-minimal flavor-violation is
not allowed in
the left panel of the figure, while we include 
non-vanishing flavor-violating squark mixing parameters $\lambda_L$ and
$\lambda_R$ in its middle and right panels, respectively (see
Eq.\ \eqref{eq:lambda}).}
\vspace{.1cm}
 \includegraphics[width=.32\columnwidth]{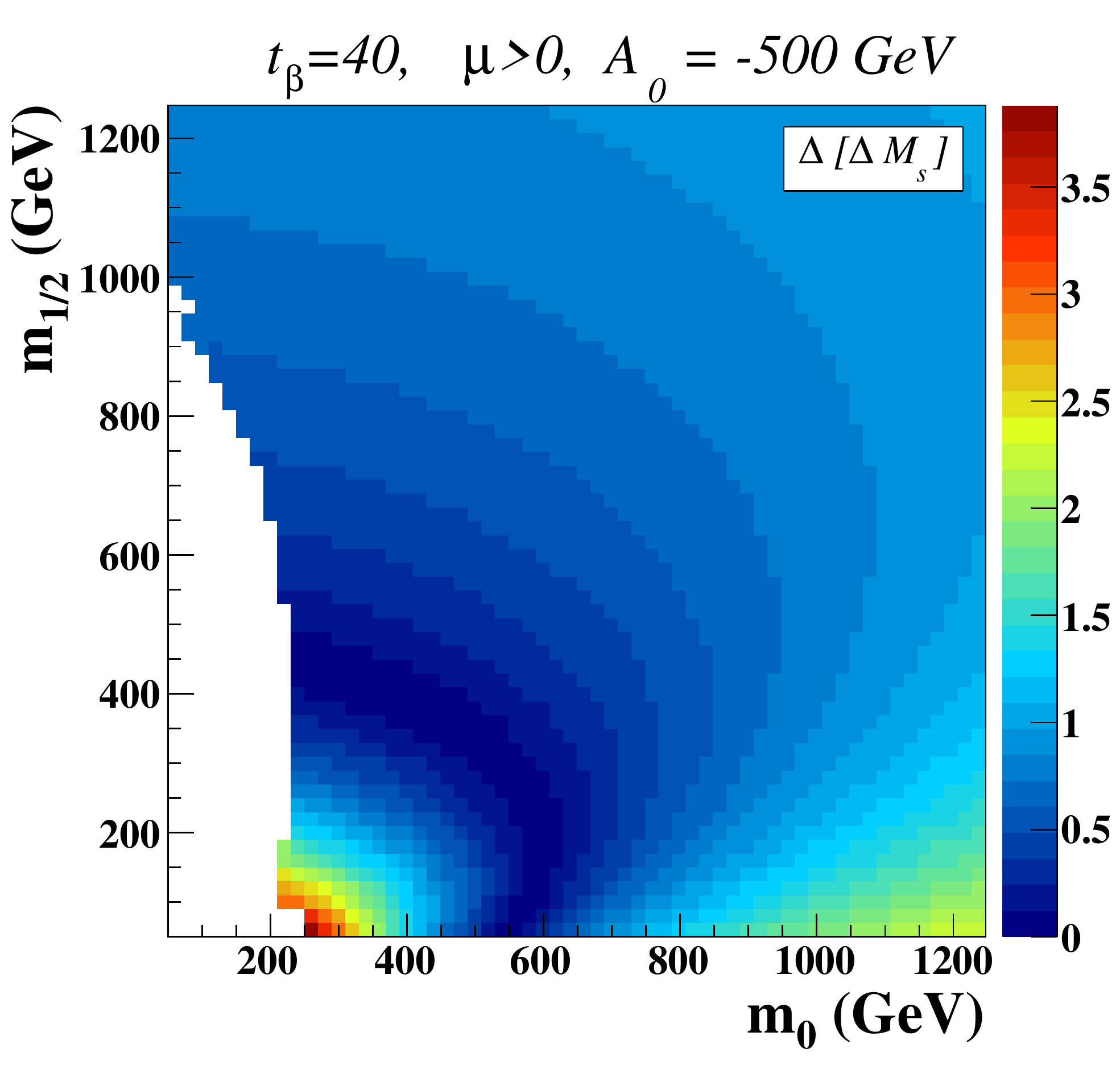}
 \includegraphics[width=.32\columnwidth]{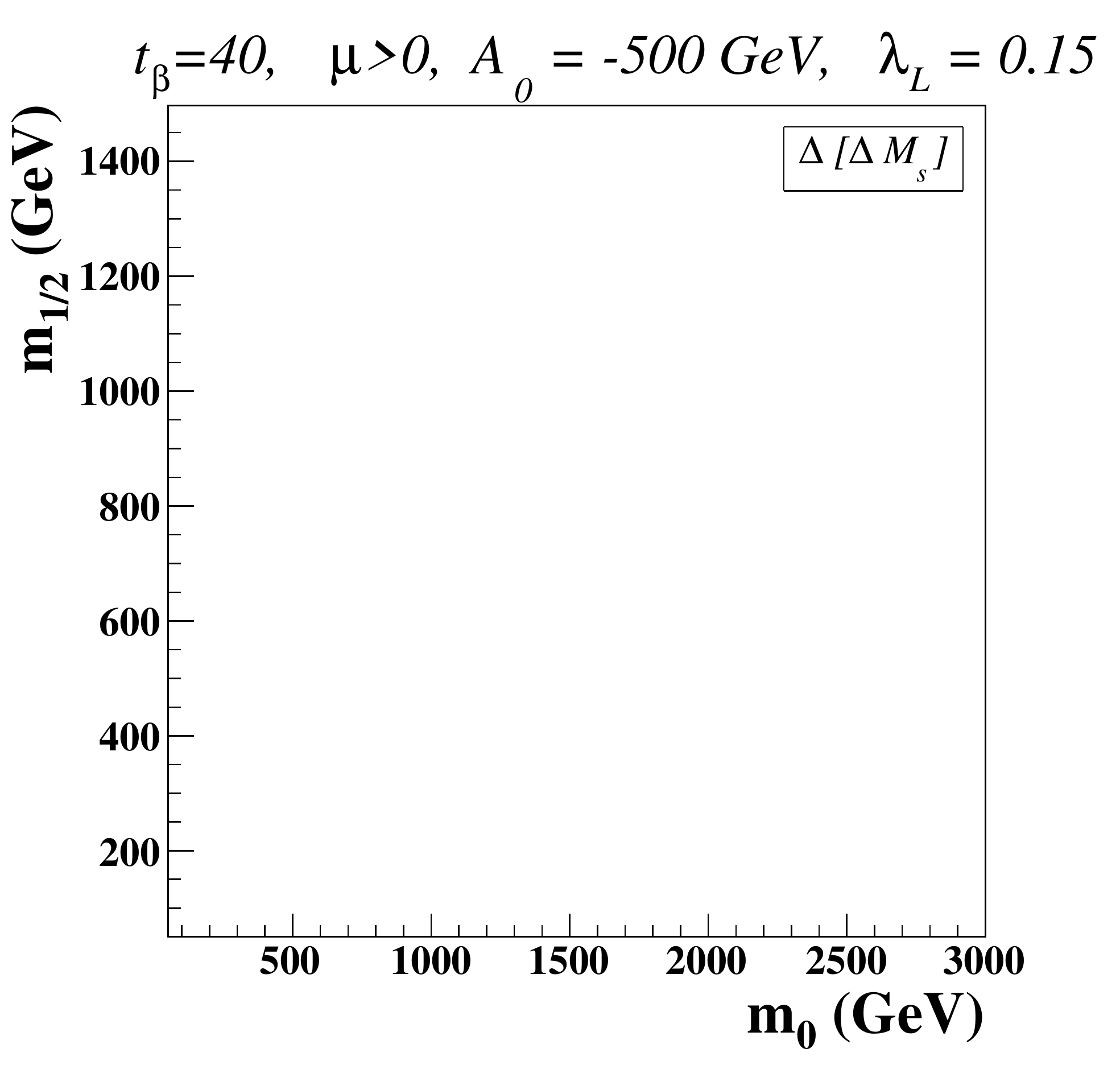}
 \includegraphics[width=.32\columnwidth]{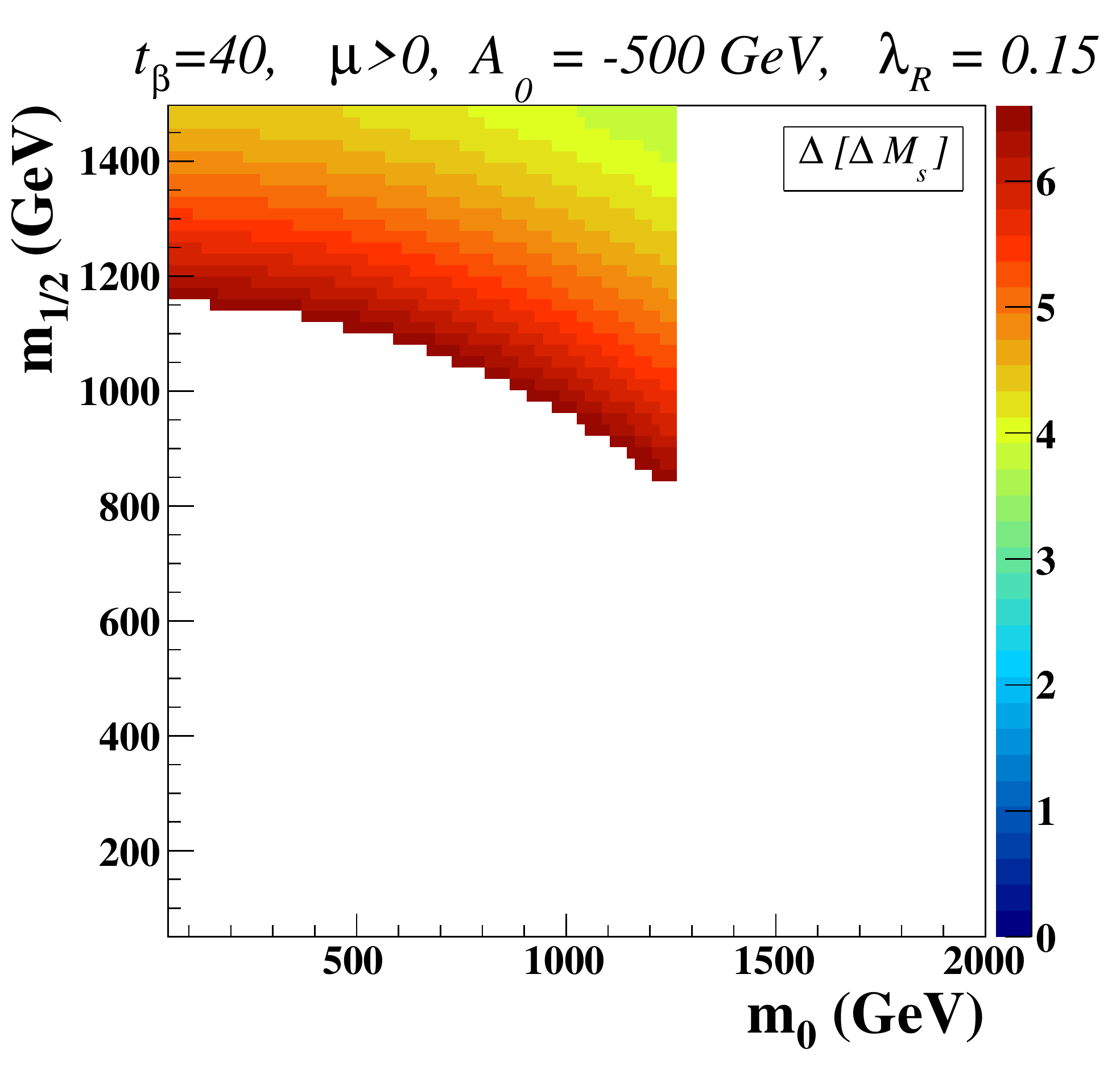}
 \caption{\label{fig:cmssm40_dMBs} Same as in Figure \ref{fig:cmssm10_dMBs} but
for $\tan\beta=40$, $A_0=-500$ GeV.}
\end{figure}
%

The computation of the associated theoretical predictions are based on the
effective Hamiltonian of Eq.\ 
\eqref{eq:heffb} and the four-fermion operators
\be\bsp
  {\cal O}_1^q = \big[\sbar_L\gamma_\mu T^a q_L\big]\big[\qbar_L\gamma^\mu T_a
    b_L \big]\ , \quad &\ 
  {\cal O}_4 = \big[\sbar_L\gamma_\mu T^a b_L\big] \sum_Q\big[\Qbar\gamma^\mu
      T_a Q\big] \ , \\
  {\cal O}_2^q = \big[\sbar_L\gamma_\mu q_L\big]\big[\qbar_L\gamma^\mu 
    b_L \big]\ , \quad &\ 
  {\cal O}_5 = \big[\sbar_L\gamma_\mu \gamma_\nu \gamma_\rho b_L\big]
    \sum_Q\big[\Qbar \gamma^\mu \gamma^\nu \gamma^\rho Q\big] \ , \\
  {\cal O}_3 = \big[\sbar_L\gamma_\mu b_L\big] \sum_Q\big[\Qbar\gamma^\mu Q\big]
    \ , \quad &\ 
  {\cal O}_6 = \big[\sbar_L\gamma_\mu  \gamma_\nu \gamma_\rho T^a b_L\big]
    \sum_Q\big[\Qbar\gamma^\mu \gamma^\nu \gamma^\rho  T_a Q\big] \ ,
\esp \ee 
with $q=u$ or $c$ dominantly contributing.
The MSSM contributions to these operators have been
calculated at the one-loop level in Ref.\ \cite{Baek:2001kh} and Ref.\
\cite{Buras:2002vd} and these
results have been implemented in the computer code \spheno\ \cite{Porod:2011nf},
which we employ to
confront theory to data and constrain the MSSM parameter space.

In the context of the cMSSM, these constraints are translated in terms of
scans of $(m_0, m_{1/2})$ planes at fixed values of $\tan\beta$ and $A_0$. We
refer to Section \ref{sec:raredec} for more
details on the benchmark plane choices, and show, on
Figure \ref{fig:cmssm10_dMBs} and Figure \ref{fig:cmssm40_dMBs}, theoretical
predictions for the mass difference $\Delta M_s$ associated with the $B_s$-meson
system. These predictions are presented as deviations from the central value
given in Eq.\ \eqref{eq:dmbs}.
Regions either excluded after imposing these constraints at the $2\sigma$-level,
or for which there is no
solution to the supersymmetric renormalization group equations linking high
scale to low-energy scale physics, are depicted as white areas. In the case no
squark non-minimal flavor
violation is allowed (left panel of the figures),
the large theoretical uncertainty of Eq.\
\eqref{eq:dmbs} strongly weakens the relevance of $\Delta M_s$ as
a constraining observable for the construction of 
phenomenologically viable scenarios.  

This contrasts with the case of scenarios featuring
non-minimal flavor-violating mixings among the second and third
generation squarks as defined in Eq.\ \eqref{eq:lambda} (see the middle and right panels
of Figure \ref{fig:cmssm10_dMBs} and Figure \ref{fig:cmssm40_dMBs}).
The $\Delta M_s$ observable is here highly sensitive
to both mixings in the left-left chiral sector and in the right-right chiral
sector as expected from the form of the dominant effective operators, contrary to
$B$-meson rare decays which mainly 
probe mixings in the left-left chiral sector.
Next, supersymmetric contributions
to the $B_s^0-\Bbar_s^0$ oscillations are strongly enhanced in parameter space
regions where the value of $\tan\beta$ is large.

%
\begin{figure}[t!]
 \centering
 \includegraphics[width=.32\columnwidth]{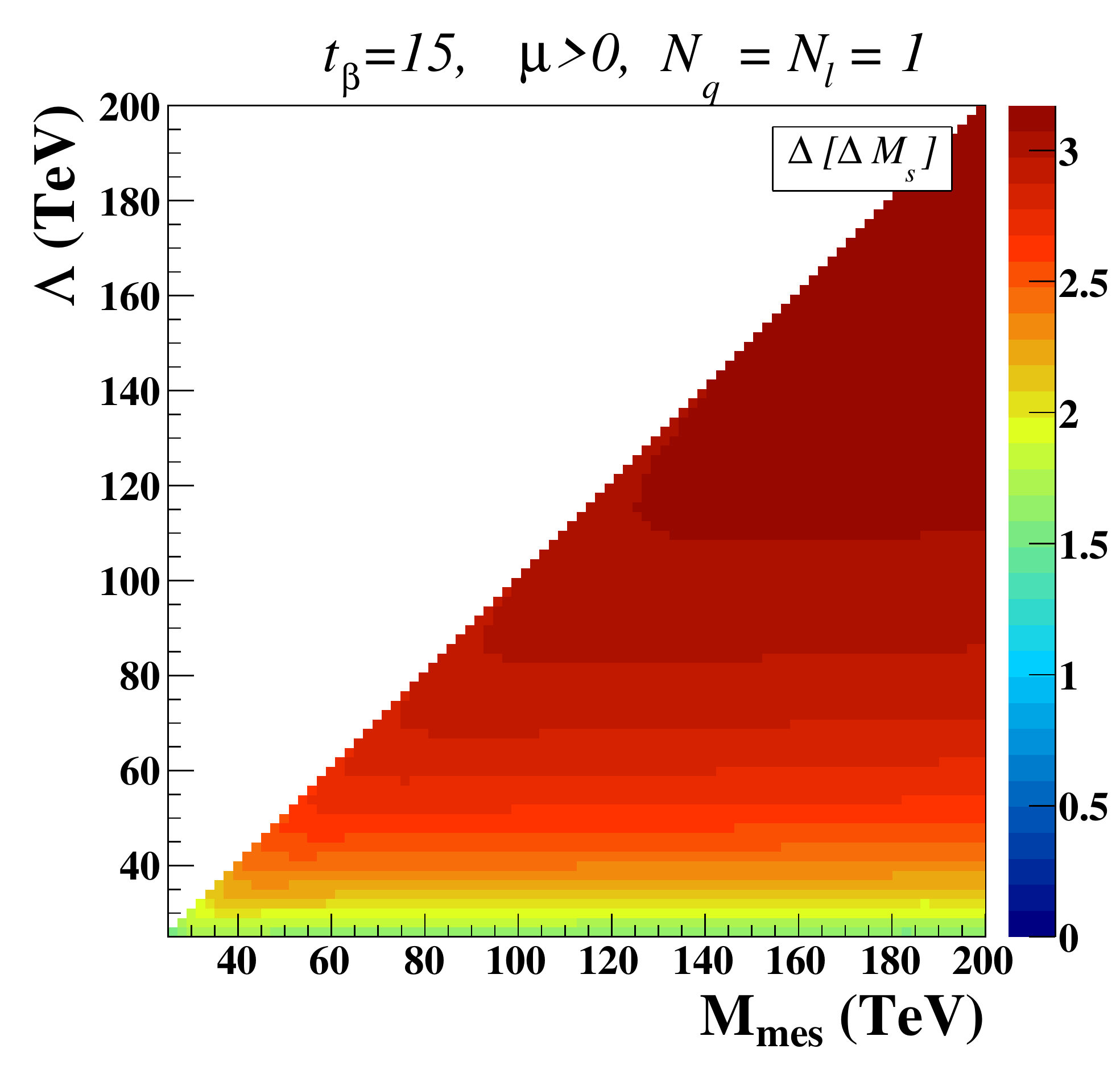}
 \includegraphics[width=.32\columnwidth]{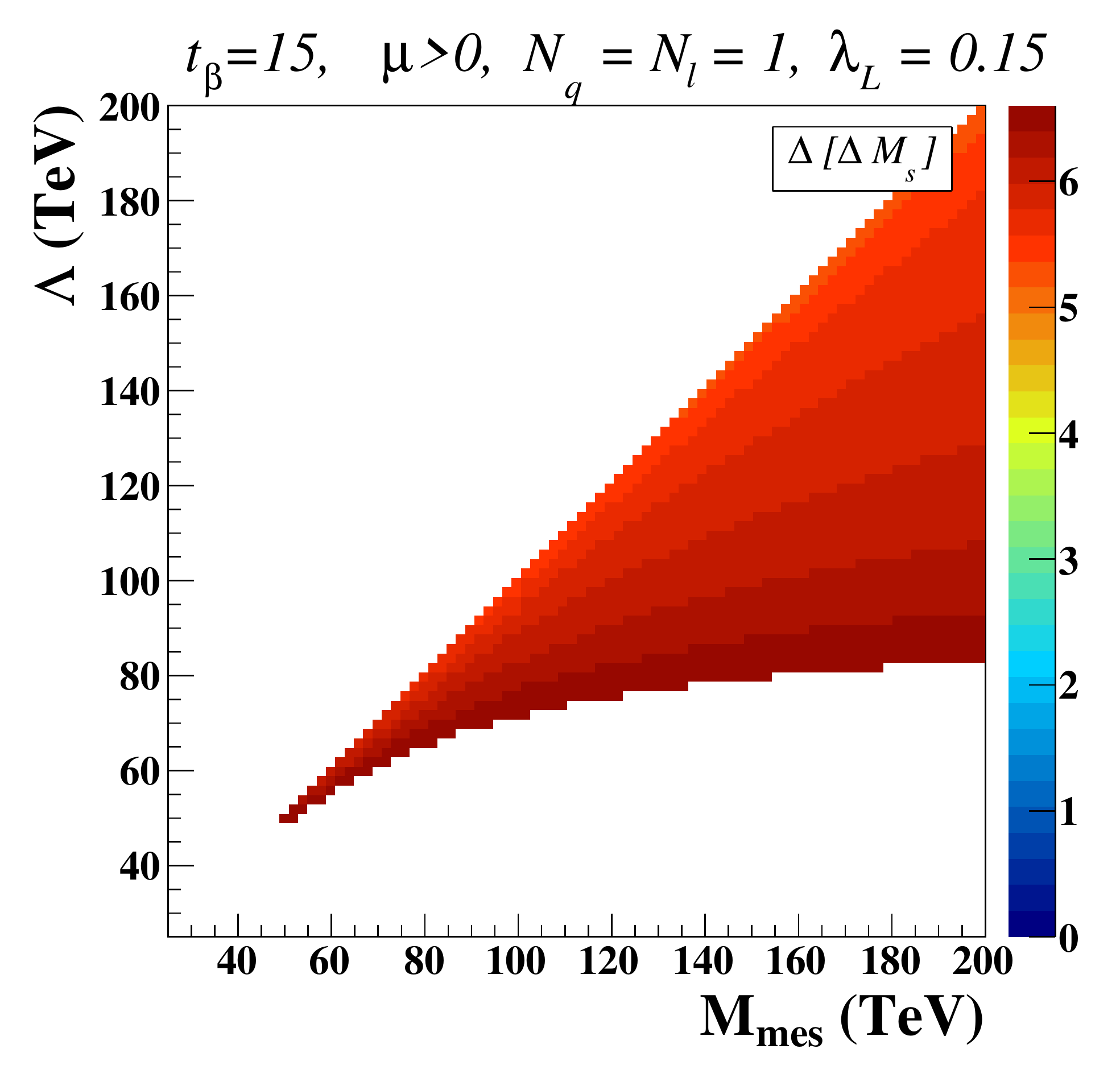}
 \includegraphics[width=.32\columnwidth]{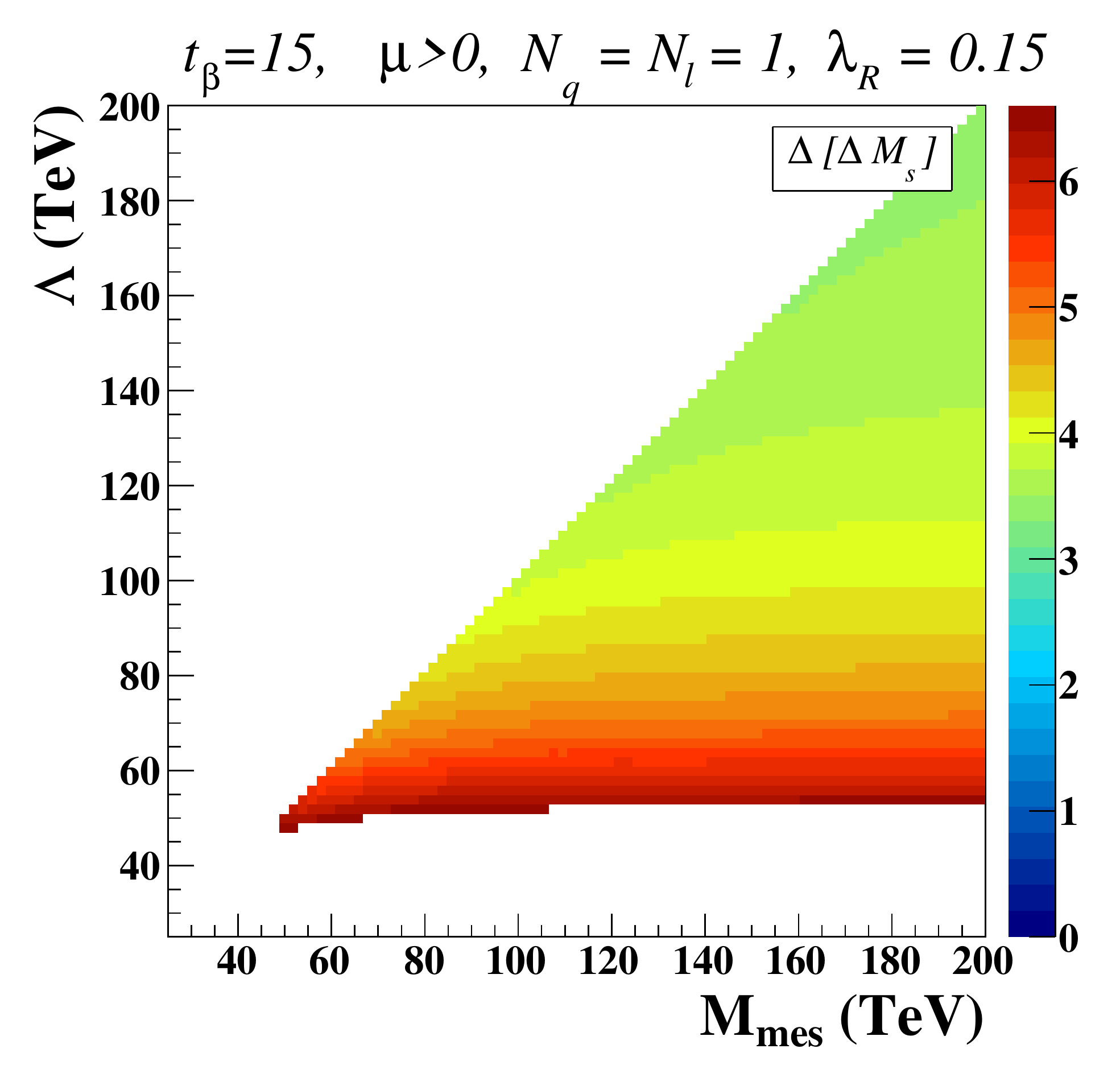}
 \caption{\label{fig:gmsb1_dMBs} Same as in Figure \ref{fig:cmssm10_dMBs} but
for the MSSM with gauge-mediated supersymmetry-breaking. We present $(M_{\rm
mes},\Lambda)$ planes with $\tan\beta=15$, one
single series of messenger fields $N_q = N_\ell =1$ and a positive Higgs
mixings parameter $\mu>0$.}
\vspace{.3cm}
 \includegraphics[width=.32\columnwidth]{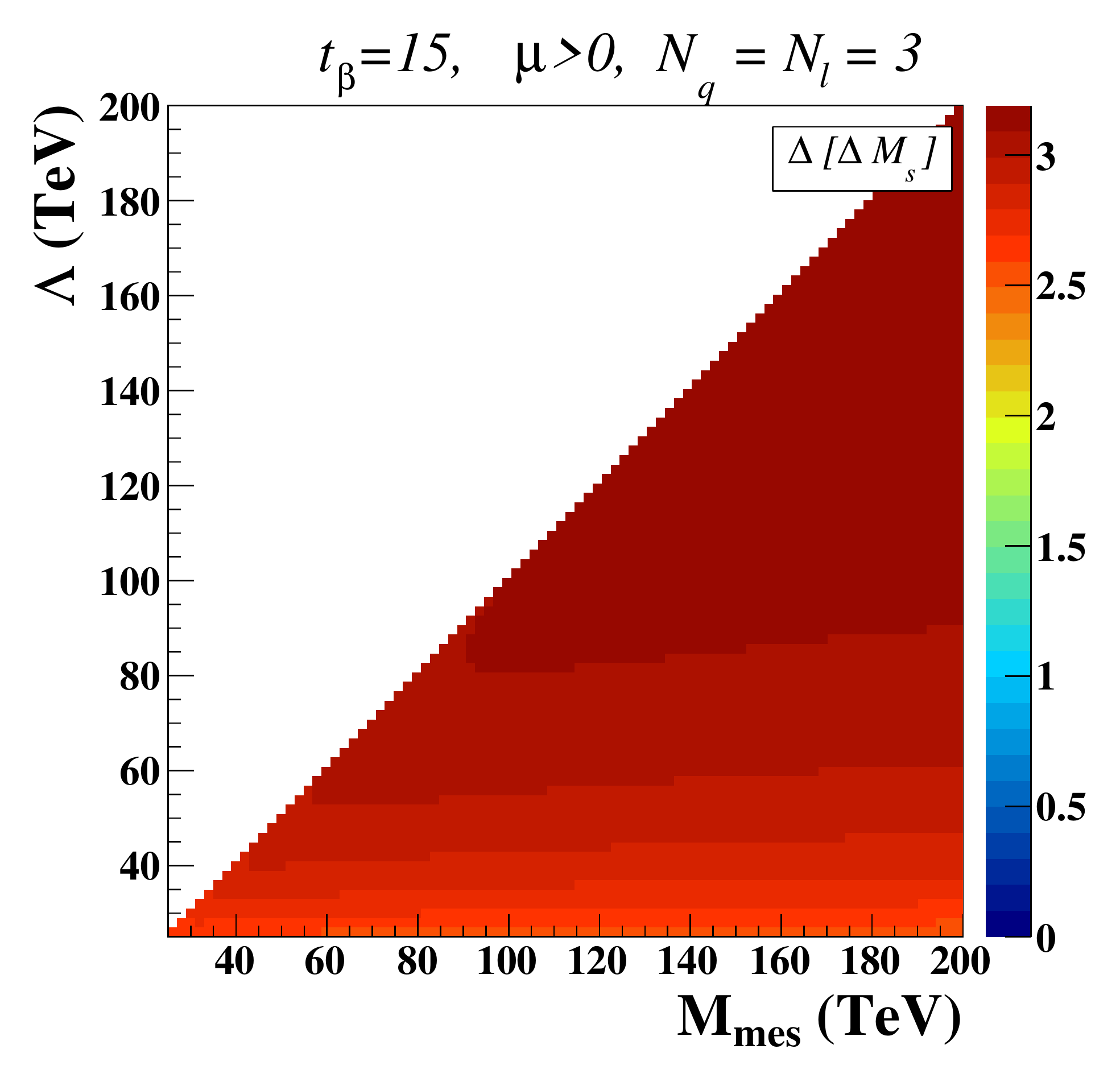}
 \includegraphics[width=.32\columnwidth]{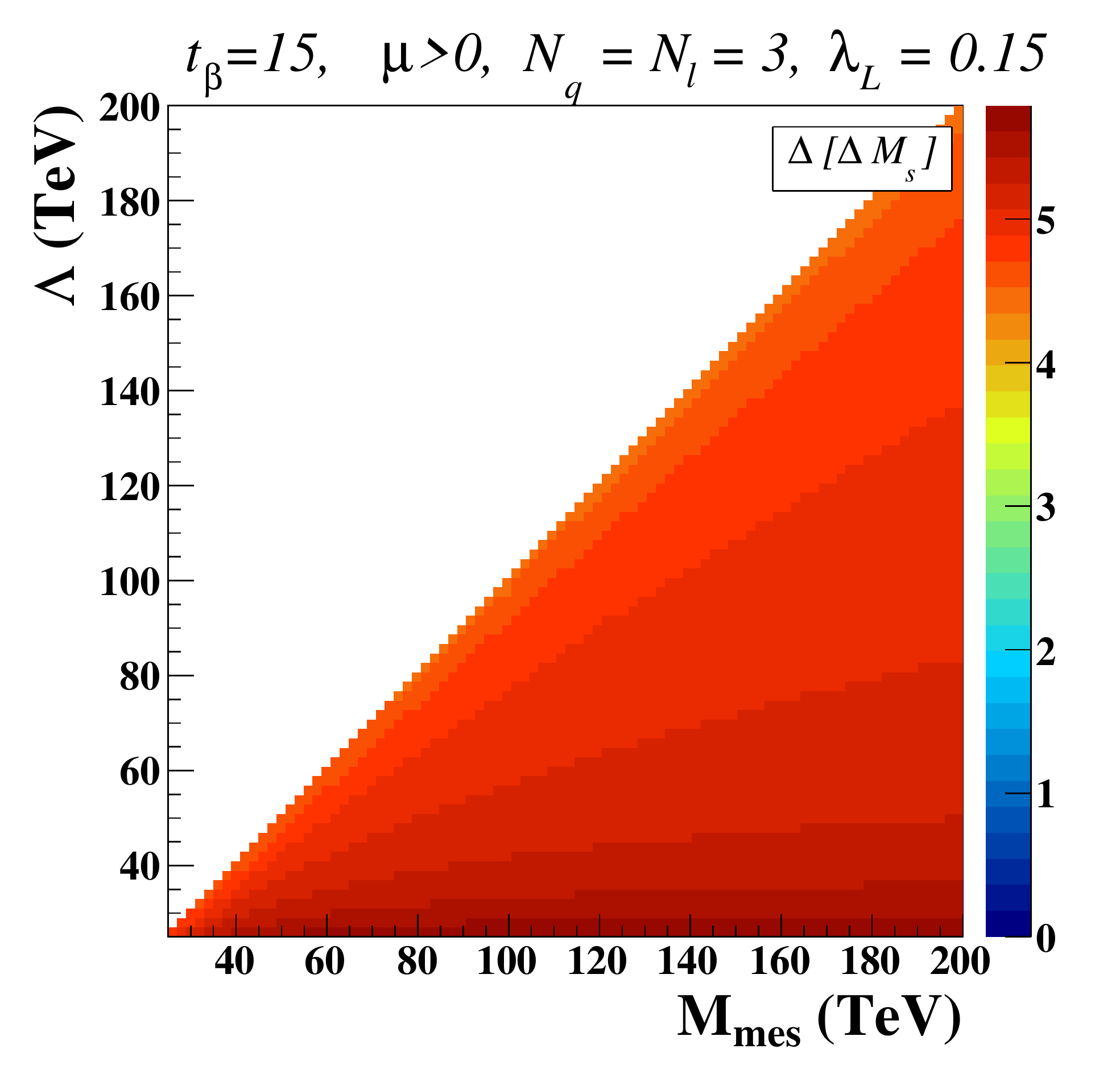}
 \includegraphics[width=.32\columnwidth]{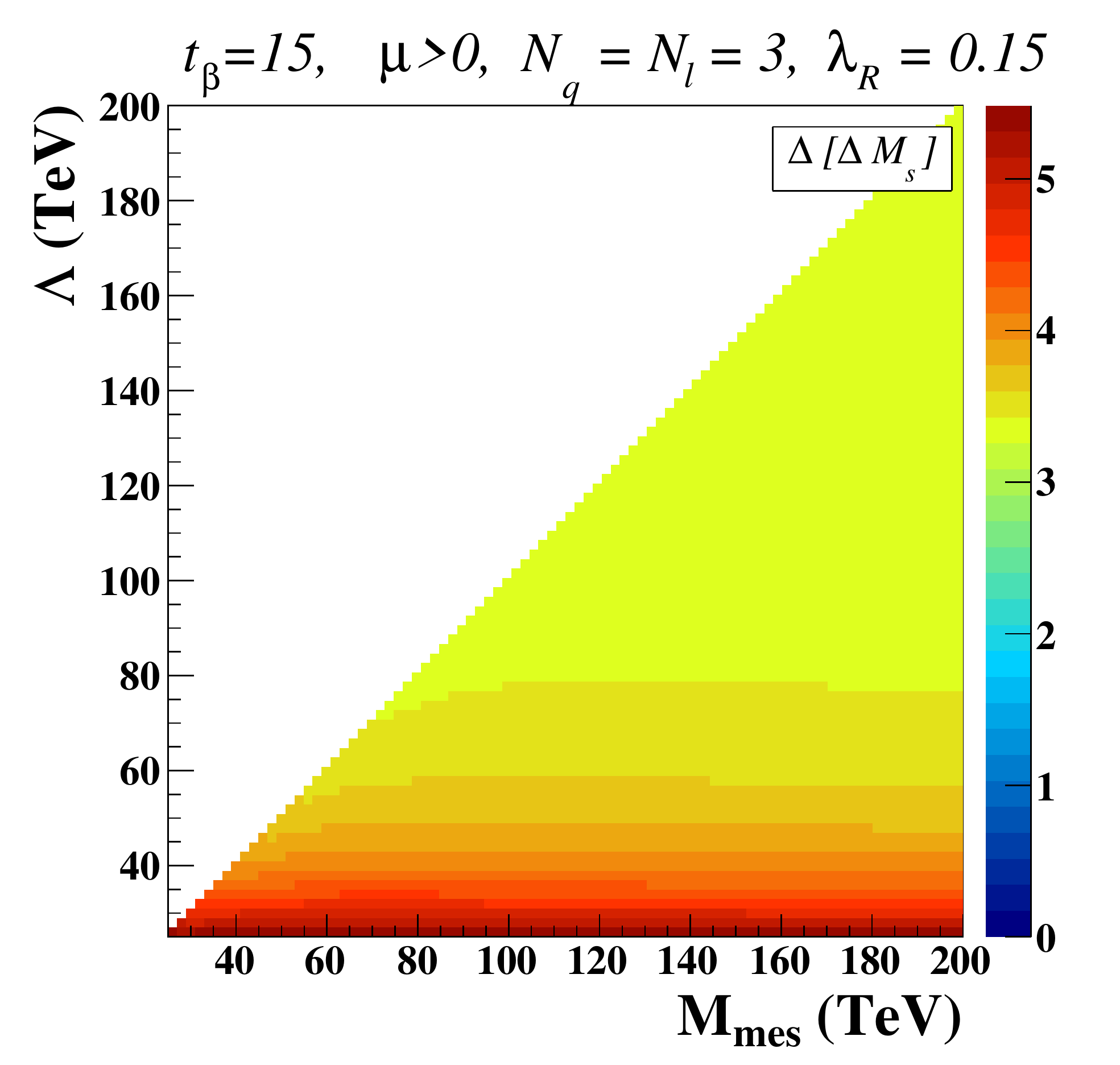}
 \caption{\label{fig:gmsb3_dMBs} Same as in Figure \ref{fig:cmssm10_dMBs} but
for the MSSM with gauge-mediated supersymmetry-breaking. We present $(M_{\rm
mes},\Lambda)$ planes with $\tan\beta=15$, three
series of messenger fields $N_q = N_\ell =3$ and a positive Higgs
mixings parameter $\mu>0$.}
\end{figure}
%

%
\begin{figure}[t!]
 \centering 
 \includegraphics[width=.32\columnwidth]{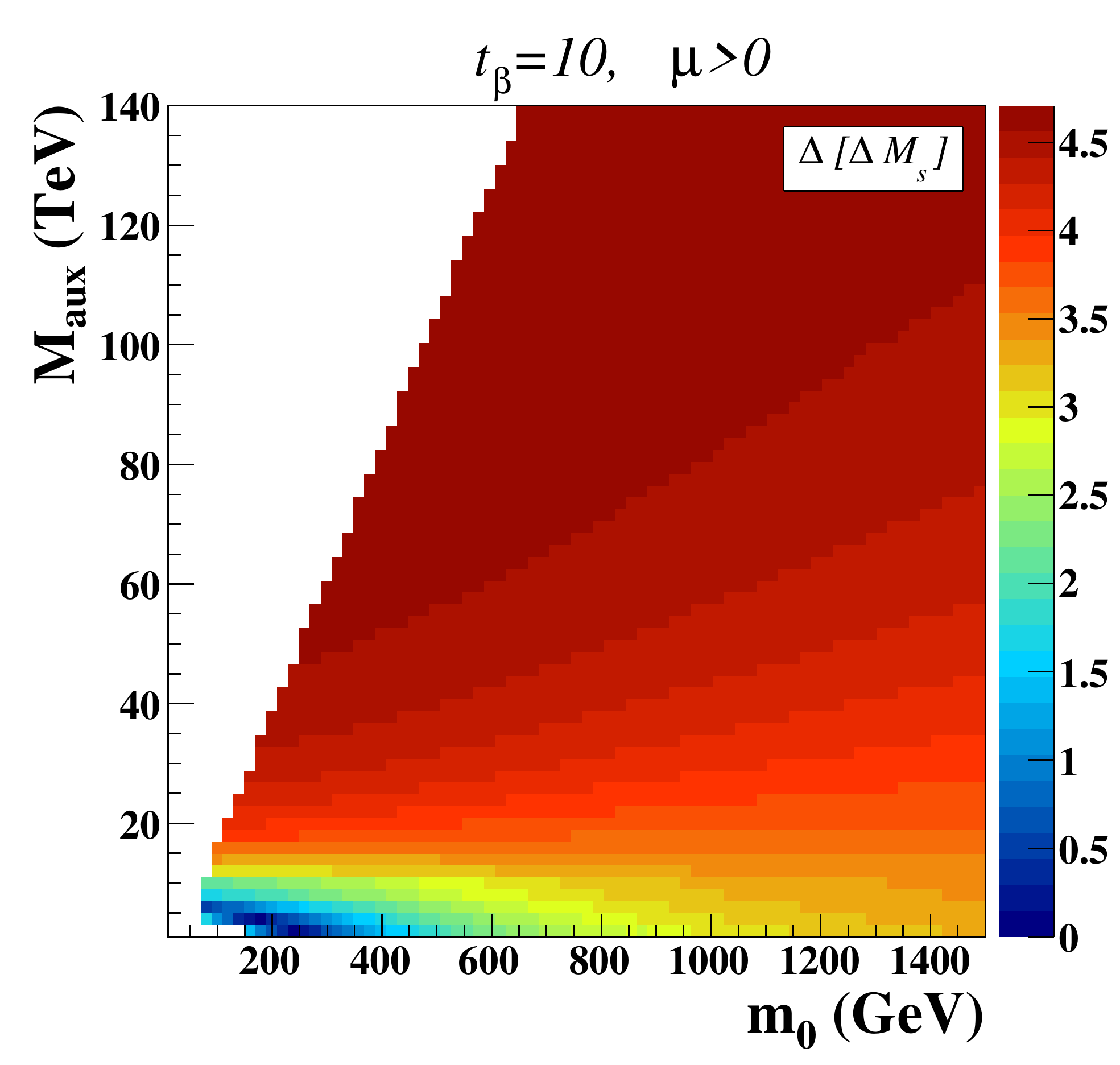}
 \includegraphics[width=.32\columnwidth]{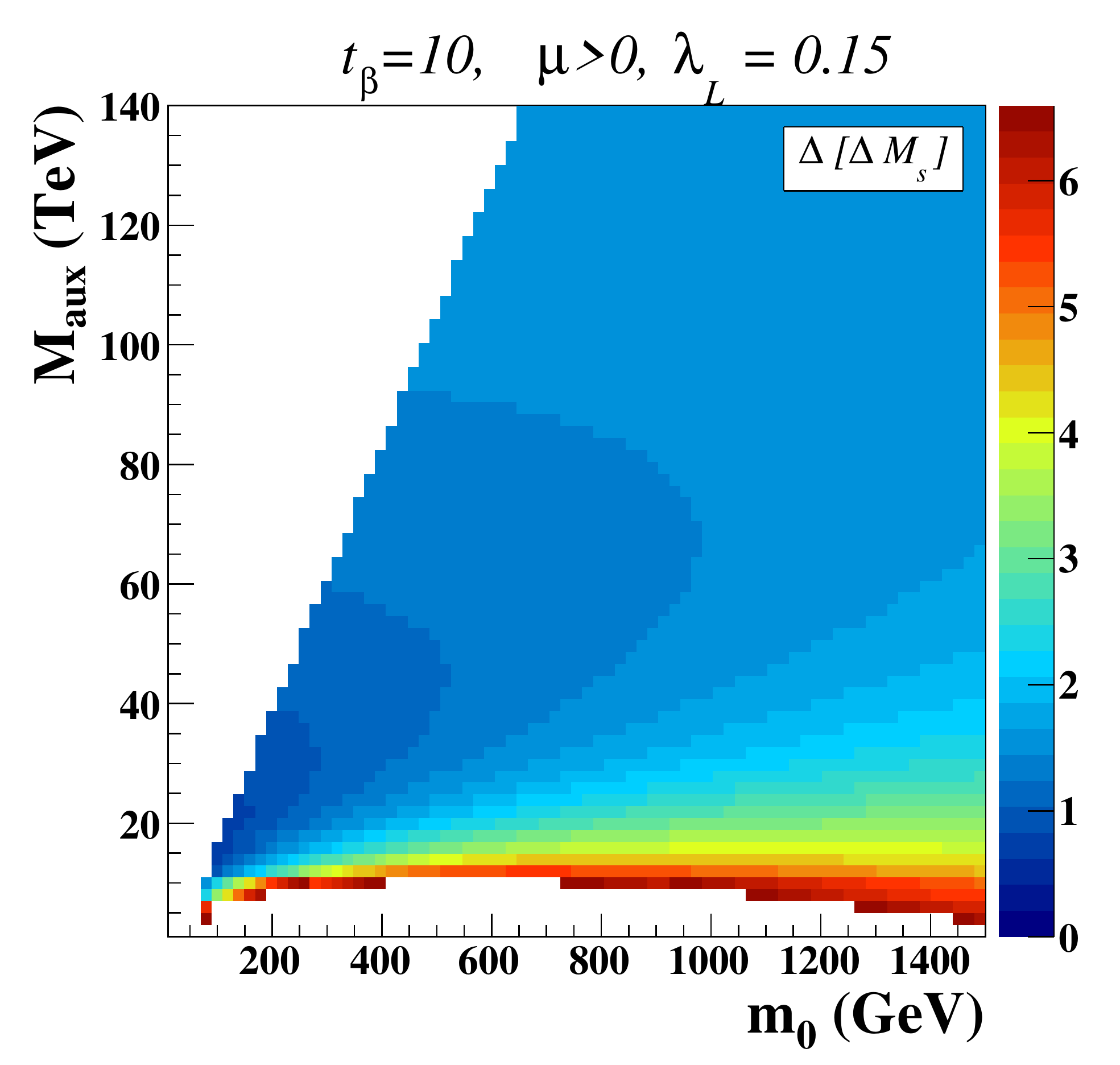}
 \includegraphics[width=.32\columnwidth]{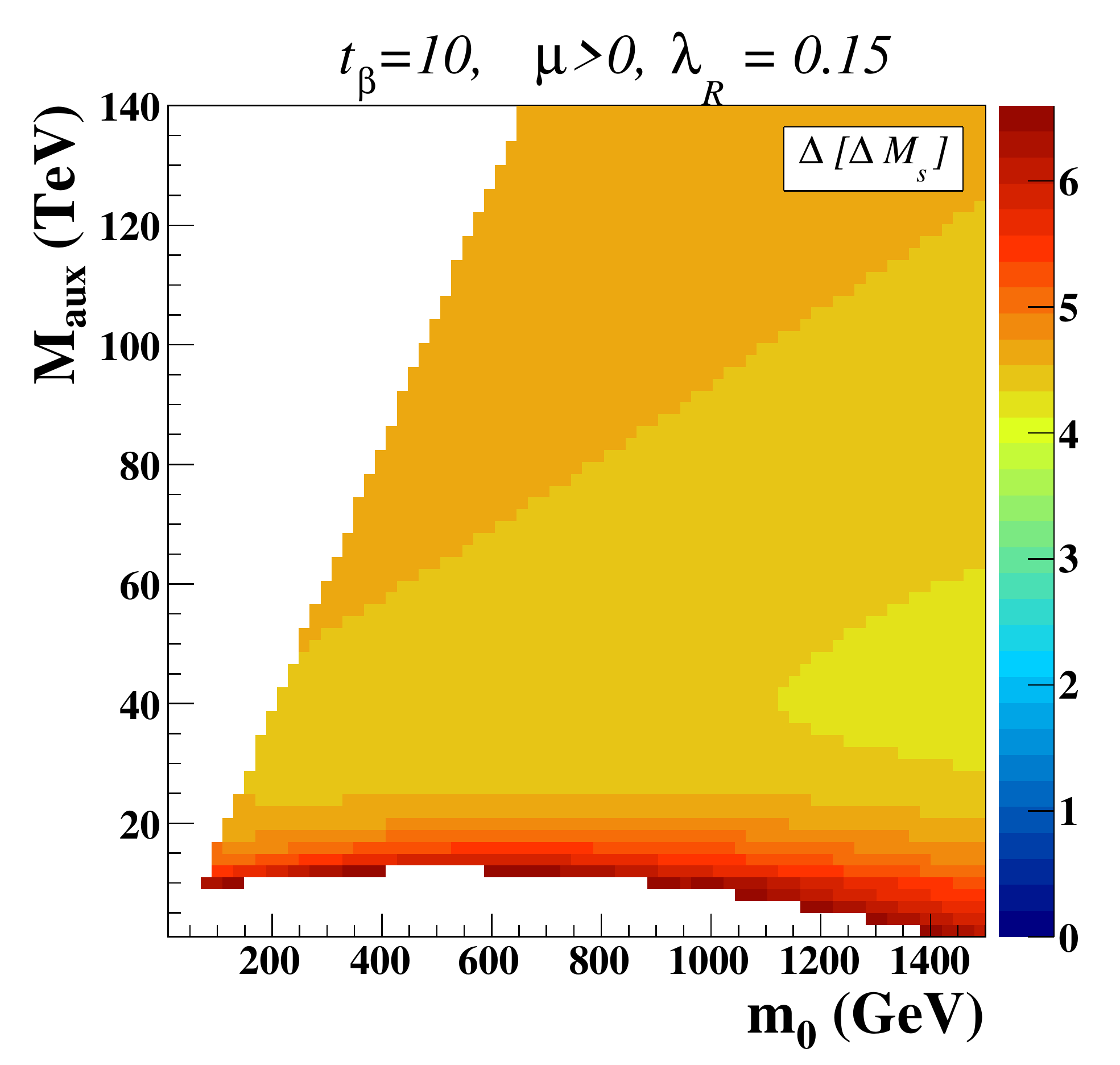}
 \caption{\label{fig:amsb_dMBs} Same as in Figure \ref{fig:cmssm10_dMBs} but
for the MSSM with anomaly-mediated supersymmetry-breaking. We present $(m_0,
M_{\rm aux})$ planes with
$\tan\beta=10$ and a positive Higgs mixing parameter $\mu>0$.}
\vspace{1cm}
\end{figure}
%

On Figure \ref{fig:gmsb1_dMBs} and Figure \ref{fig:gmsb3_dMBs}, we turn to the
case of the MSSM with gauge-mediated supersymmetry breaking and present $(N_{\rm
mes}, \Lambda)$ planes at fixed $\tan\beta$ and number of messenger fields. The
investigated benchmark planes are those already described in Section
\ref{sec:raredec} with $\tan\beta=15$ and 
$N_q=N_\ell=1$ or $N_q=N_\ell=3$. When
flavor is conserved, \ie, when  all $\lambda$-parameters of Eq.\
\eqref{eq:lambda} are vanishing, good agreement is found  between
data and theory for the entire scanned regions
after accounting for the large theoretical uncertainties.
When allowing for
non-minimal flavor violation, new constraints on the parameter space appear
but are
found to be less severe than those induced by rare $B$-meson decays
(see Section \ref{sec:raredec}).

Finally, we investigate
MSSM scenarios with anomaly-mediated supersymmetry breaking on Figure 
\ref{fig:amsb_dMBs}, where  $(m_0, M_{\rm
aux})$ planes at a fixed $\tan\beta=10$ value are presented.
$B$-meson oscillations are found
complementary to $B$-meson decays to constrain
the parameter space in cases with non-minimal flavor violation 
in the squark sector, contrary to the MSSM with a flavor-conserving setup.

\subsection{The anomalous magnetic moment of the muon}\label{sec:gm2}

The world average experimental value for the muon anomalous
magnetic moment is dominated by data collected by the E821 experiment at
Brookhaven \cite{Bennett:2004pv},
\be
  a_\mu^{\rm exp} = \big(11659208.0 \pm 6.3\big) \times 10^{-10} \ .
\ee
Theoretical predictions in the Standard Model
include QED contributions up
to four loops and are even analytical up to the three-loop level \cite{Li:1992xf,
Laporta:1992pa, Laporta:1996mq, Czarnecki:1998rc, Erler:2000nx,
Kinoshita:2004wi, Passera:2004bj, Kinoshita:2005ti}. In addition,
the leading logarithmic terms of the five-loop results are known~\cite{Hughes:1999fp,
Kinoshita:2001pn, Czarnecki:2001pv, Davier:2004gb, Kataev:2005av,
Kinoshita:2005sm, Miller:2007kk, Jegerlehner:2007xe}, as well as
both the electroweak~\cite{Brodsky:1966mv, Burnett:1967467, Jackiw:1972jz, Bars:1972pe,
Fujikawa:1972fe, Altarelli:1972nc, Bardeen:1972vi, Kukhto:1992qv, Peris:1995bb,
Czarnecki:1995wq, Czarnecki:1995sz, Degrassi:1998es} and hadronic contributions
\cite{Melnikov:2003xd, Erler:2006vu} up to three loops. Moreover, the
associated theoretical uncertainties are dominated by light-by-light scattering 
diagrams. The
theoretical predictions hence read \cite{Beringer:2012zz}
\be
  a_\mu^{\rm th} = \big(11659184.1 \pm 4.8\big) \times 10^{-10} \ ,
\ee
which leads to a discrepancy of about $3\sigma$ between
data and theory,  
\be
  \Delta a_{\mu} = \big( 23.9 \pm 7.92 \big) \times 10^{-10} \ ,
\ee
that we assume to be explained by new physics.
In the context of supersymmetry, the dominant contributions
$a_\mu^{\text{susy}, 1}$ to the anomalous magnetic moment of the muon consist 
of smuon, sneutrino, chargino and neutralino loops~\cite{Moroi:1995yh},
\be
 a_\mu^{\text{susy}, 1} \simeq 13 \times 10^{-10} \Big(
\frac{100~{\rm GeV}}{M_{\rm susy}} \Big)^2 \tan\beta \ {\rm sign}(\mu) \ ,
\ee
where $M_{\rm susy}$ is a representative supersymmetry mass scale.
This motivates
the choice of a positive off-diagonal mixing $\mu$ parameter for all the
considered benchmark
scenarios as negative $\mu$-values would increase
rather than decrease the discrepancy between data and theory.

%
\begin{figure}[t!]
 \centering
 \includegraphics[width=.32\columnwidth]{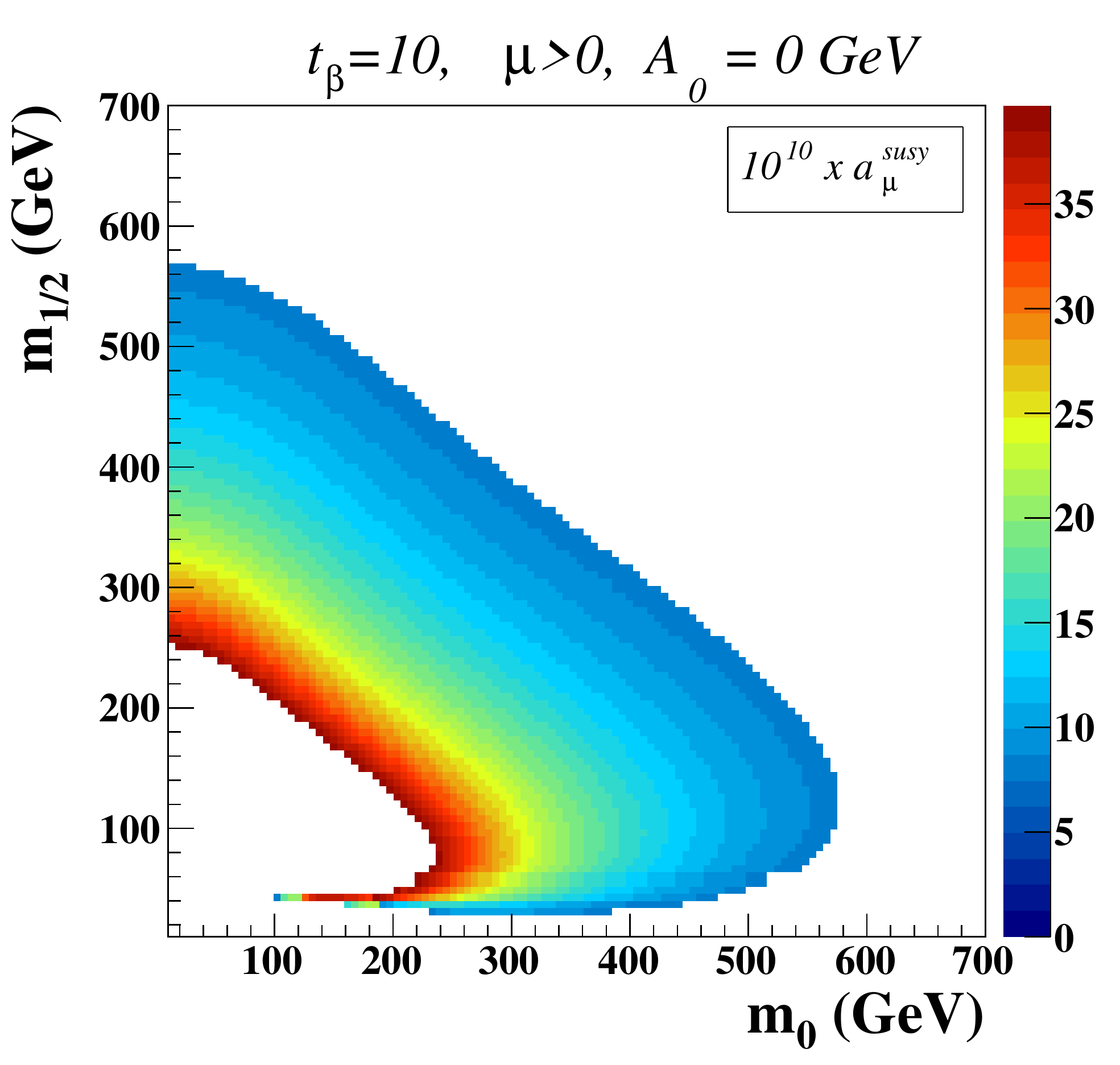}
 \includegraphics[width=.32\columnwidth]{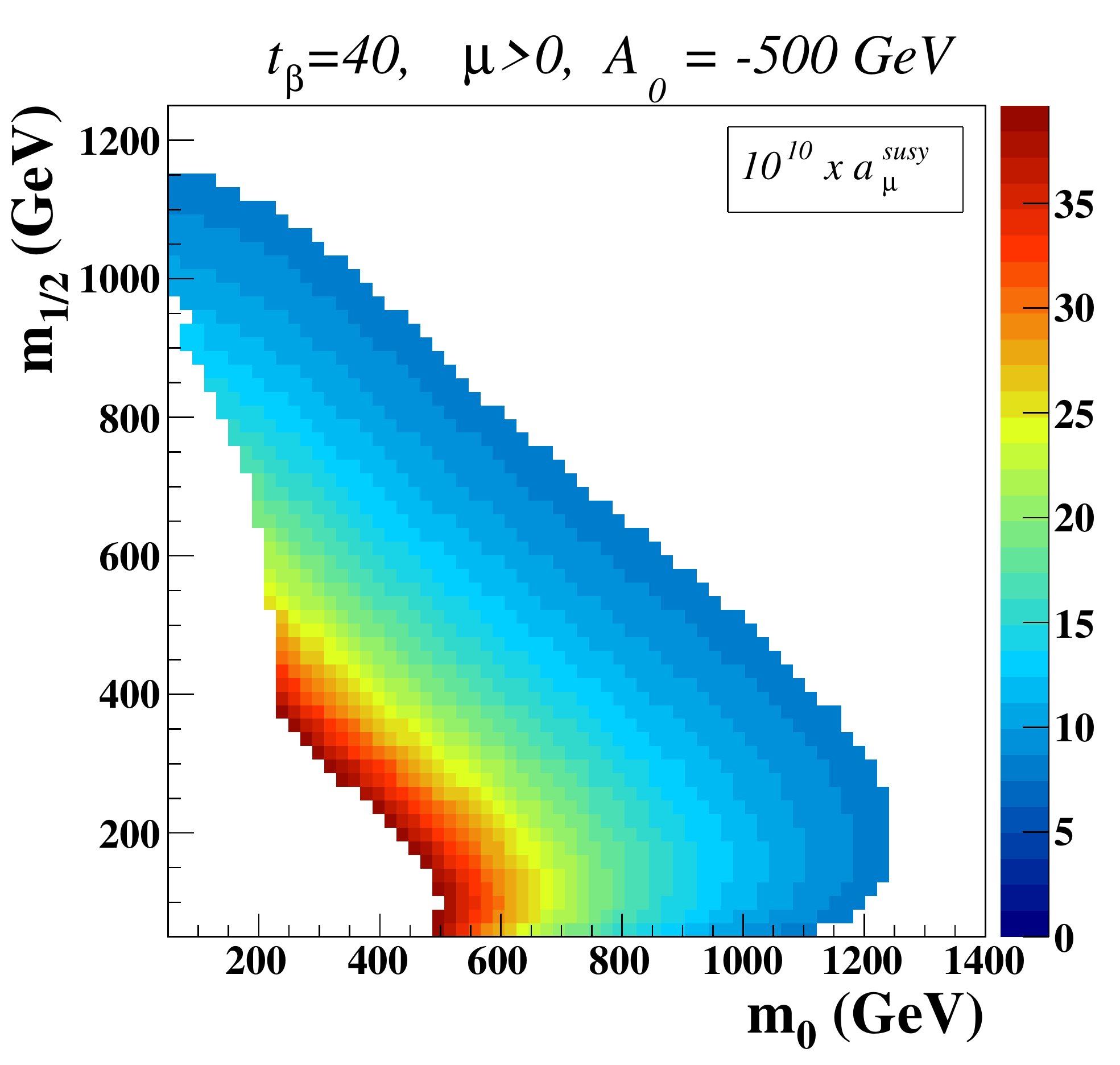}
 \caption{\label{fig:cmssm10_gm2} Supersymmetric contributions to the anomalous
magnetic moment of the muon. Only the regions of the parameter space 
for which an agreement between the theoretical predictions and the
measurements is found, at a $2\sigma$ level, are shown. We present the results in
$(m_0,m_{1/2})$-planes of the cMSSM for fixed values of $\tan\beta=10$ (40), 
$A_0=0$~GeV ($-500$~GeV)
and a positive Higgs mixing parameter $\mu>0$ in the left (right) panel
of the figure.}
\vspace{.2cm}
 \includegraphics[width=.32\columnwidth]{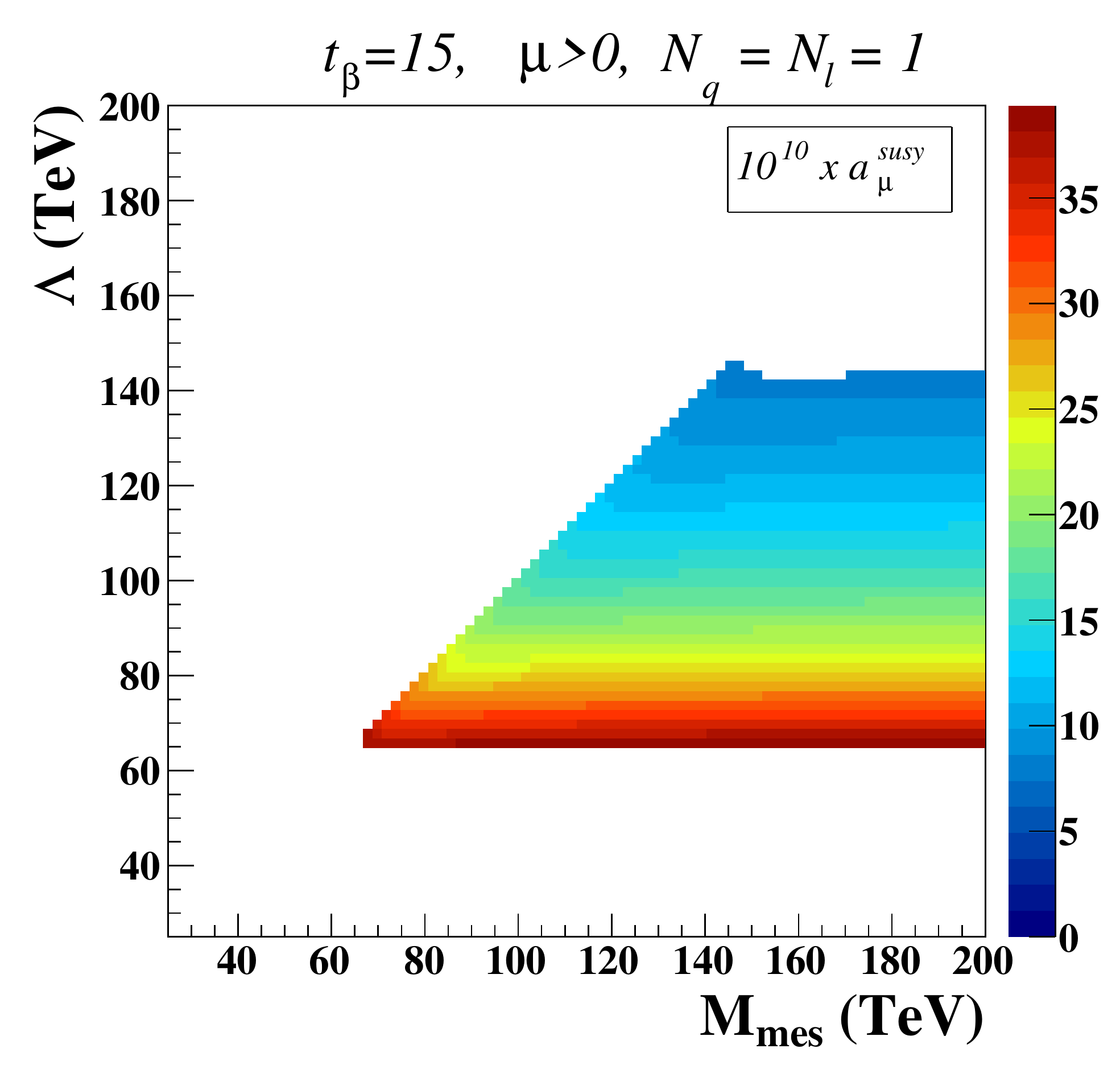}
 \includegraphics[width=.32\columnwidth]{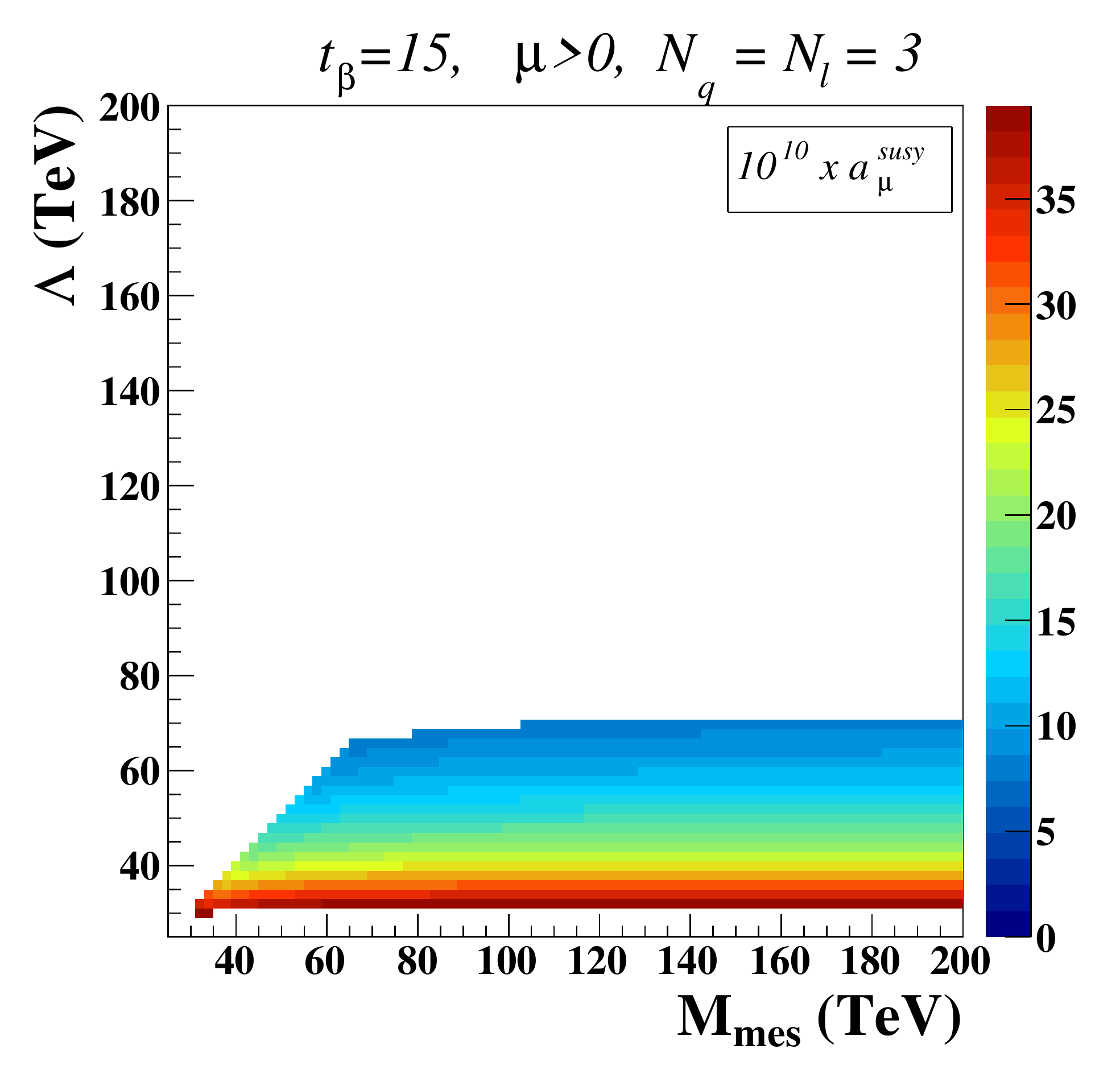}
 \caption{\label{fig:gmsb1_gm2} Same as in Figure \ref{fig:cmssm10_gm2} but
for the MSSM with gauge-mediated supersymmetry-breaking. We present, in the left
(right) panel of the figure, $(M_{\rm
mes},\Lambda)$ planes for $\tan\beta=15$, one (three)
single series of messenger fields $N_q = N_\ell =1$ (3) and a positive Higgs
mixings parameter $\mu>0$.}
\vspace{.2cm}
 \includegraphics[width=.32\columnwidth]{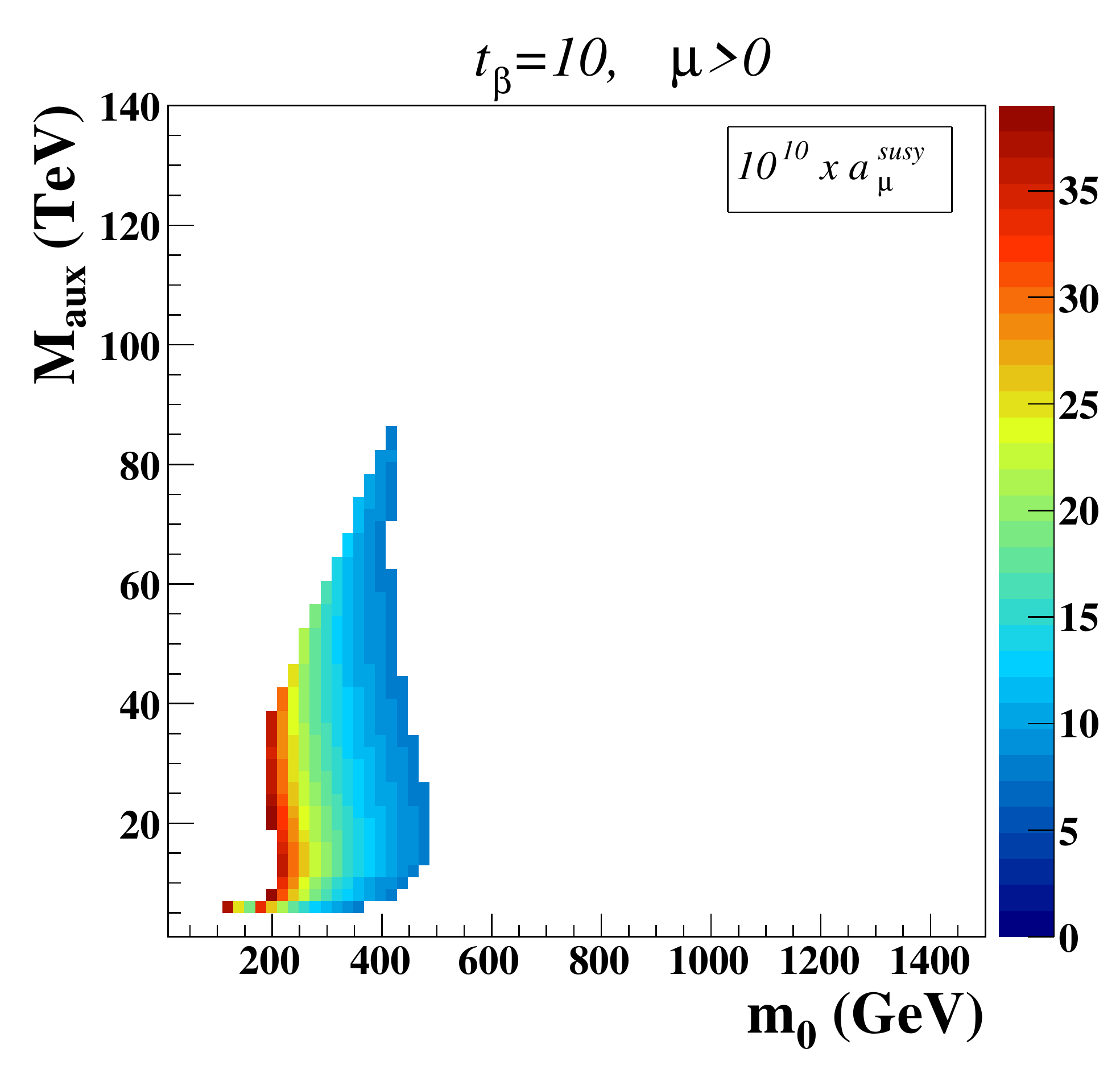}
 \caption{\label{fig:amsb_gm2} Same as in Figure \ref{fig:cmssm10_gm2} but
for the MSSM with anomaly-mediated supersymmetry-breaking. We present $(m_0,
M_{\rm aux})$ planes with $\tan\beta=10$ and a positive Higgs mixing parameter $\mu>0$.}
\end{figure}
%

We compute below the supersymmetric contributions to the
muon anomalous magnetic moment including up to two-loop diagrams \cite{Ibrahim:1999hh,
Heinemeyer:2003dq, Heinemeyer:2004yq} and scan the parameter spaces
associated with the three considered supersymmetry-breaking scenarios. In
Figure \ref{fig:cmssm10_gm2}, we show results
in the cMSSM and demand that supersymmetry restores the agreement
between theory and
data within two standard deviations. We hence present parameter space regions
compliant with this requirement for the two $(m_0,m_{1/2})$ benchmark planes
of the previous sections. Since squarks only contribute to 
the anomalous magnetic moment of the muon at the 
two-loop level, this 
considerably reduces the dependence of the $a_\mu$ on squark
non-minimal flavor violation so that the results are independent of
the considered $\lambda$-parameters.

Comparing with the results of Section
\ref{sec:raredec} and Section \ref{sec:osc}, a significant
fraction of the flavor-conserving cMSSM parameter space is found favored by both the
$B$-physics constraints and the anomalous magnetic moment of the muon
requirement, in particular for the small $\tan\beta$ region. Accommodating all constraints
when non-minimal
flavor violation is allowed becomes however challenging, as illustrated
in the case of moderate left-left and right-right chiral squark mixings.

The same conclusions hold for MSSM scenarios with 
gauge-mediated supersym\-me\-try-break\-ing, as shown
on Figure \ref{fig:gmsb1_gm2} in which we present
$(M_{\rm mes}, \Lambda)$ benchmark planes for $\tan\beta=15$
and $N_q=N_\ell=1$ (left panel) and  $N_q=N_\ell=3$ (right panel)
messenger fields.

In contrast, it is very difficult to ask predictions of the anomalous
magnetic moment of the muon to be compliant with data in the framework of MSSM
scenarios
with anomaly-mediated supersymmetry breaking, as illustrated on
Figure \ref{fig:amsb_gm2}. We show, by presenting theoretical results in
a $(m_0, M_{\rm aux})$ plane with $\tan\beta=10$,
that only a very small part of the parameter space exhibits
compatibility with experimental data. Imposing, in addition,
constraints from $B$-physics (see previous sections) renders the 
design of viable benchmark scenarios complicated,
as also found by other authors~\cite{Martin:2001st, Allanach:2009ne}. 
Predictions for the anomalous magnetic moment of the muon
are however strongly correlated to the slepton sector. The latter being 
linked to the way employed to solve the tachyonic slepton problem mentioned in Section
\ref{sec:mssmbrkex}, it is therefore
convenient to ignore constraints from the anomalous magnetic moment of the muon
when designing phenomenologically
viable scenarios with anomaly-mediated supersymmetry breaking.

\subsection{The electroweak $\rho$-parameter}\label{sec:drho}
%
\begin{figure}[t!]
 \centering
 \includegraphics[width=.32\columnwidth]{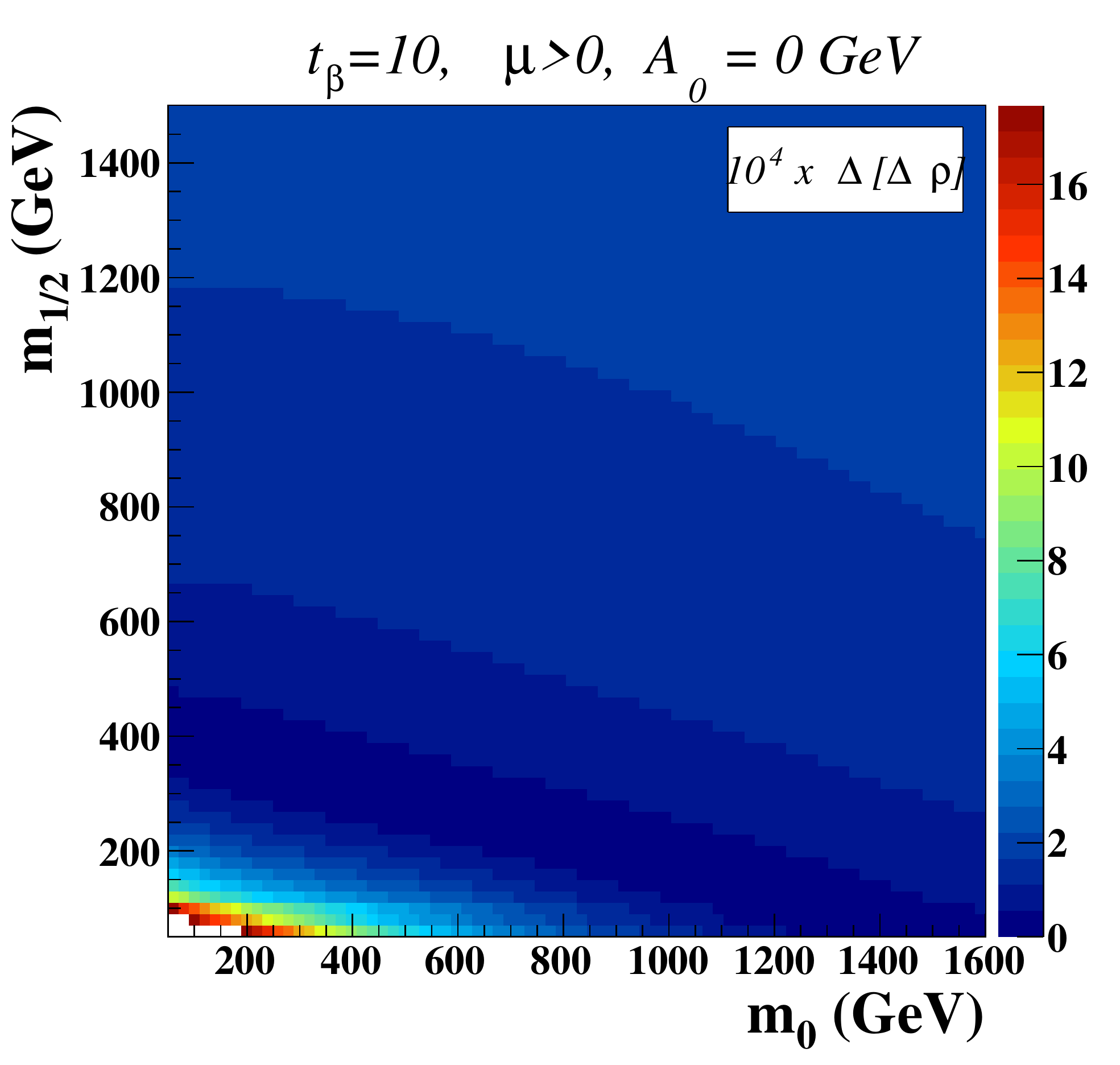}
 \includegraphics[width=.32\columnwidth]{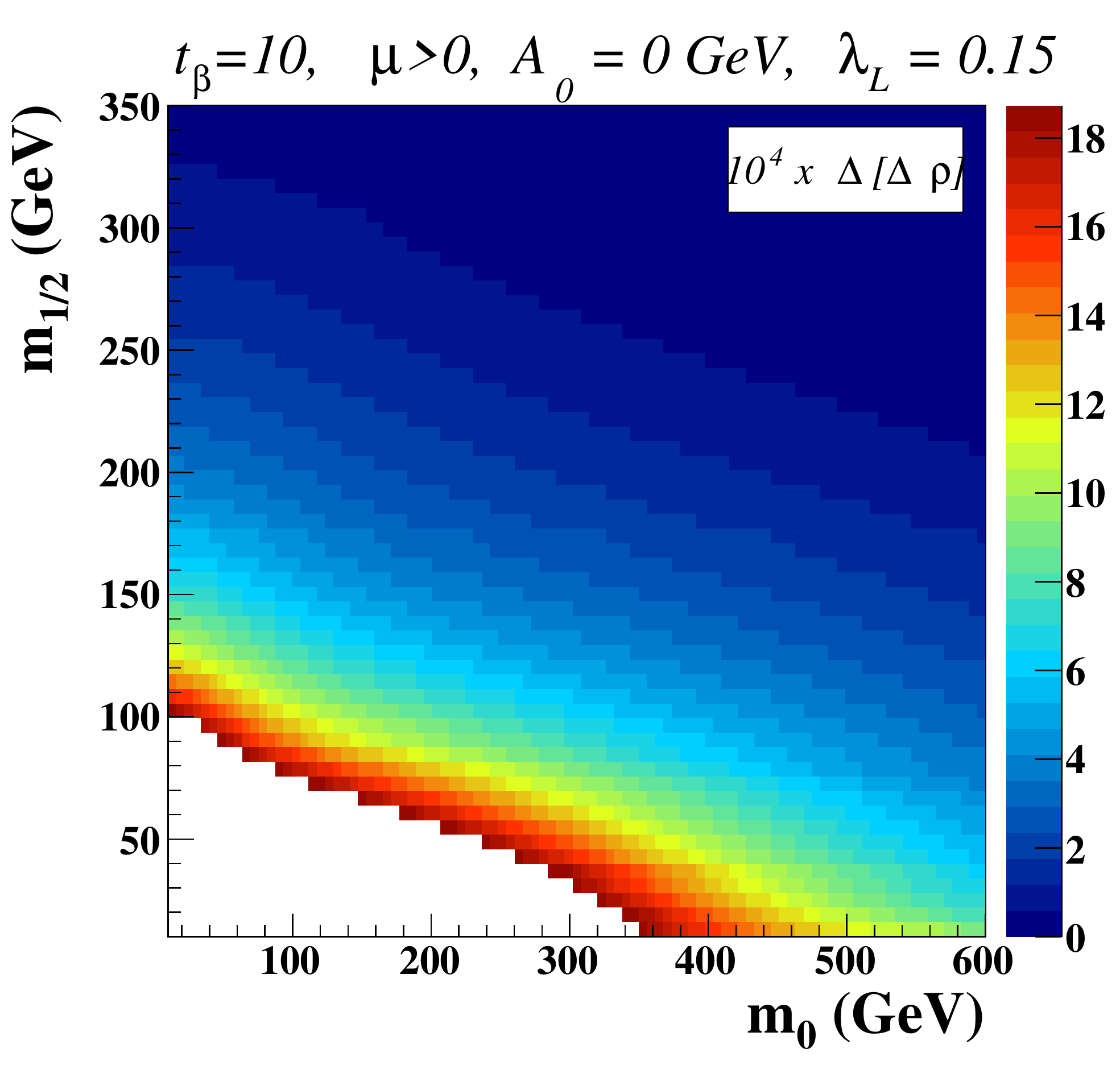}
 \includegraphics[width=.32\columnwidth]{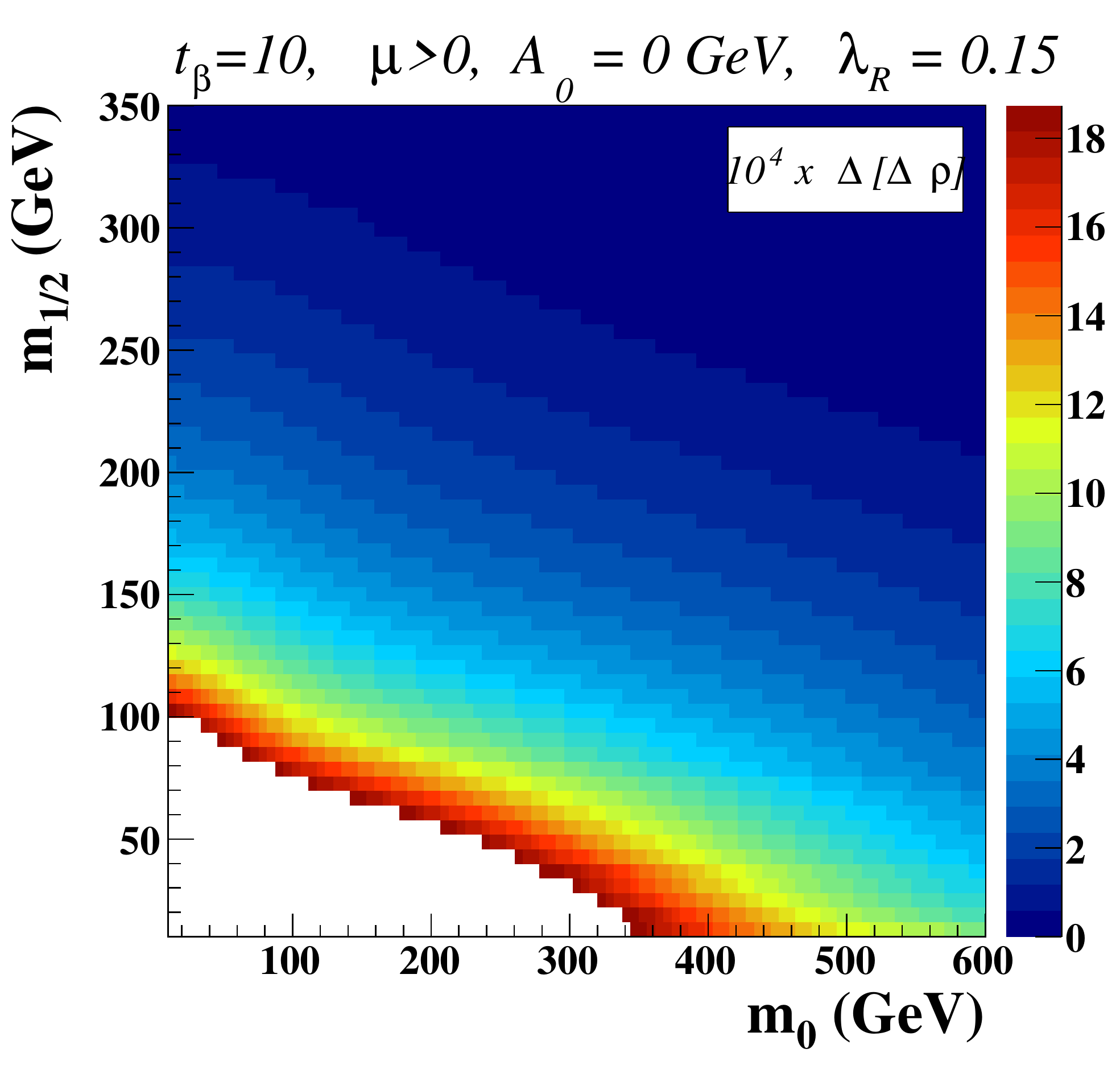}
 \caption{\label{fig:cmssm10_dr} Supersymmetric contributions to
electroweak precision observables given as deviations from the measured central
value of the $\Delta\rho$ quantity defined in
Eq.\ \eqref{eq:drho}. We present the results in
$(m_0,m_{1/2})$-planes of the cMSSM with $\tan\beta=10$, $A_0=0$
GeV and a positive Higgs mixing parameter $\mu>0$. The regions depicted in white
correspond to excluded regions when applying the bounds of Eq.\
\eqref{eq:drhoexp} 
at the $2\sigma$-level, or to regions for which there is no solution to the
supersymmetric renormalization group equations. Non-vanishing 
flavor-violating squark mixing parameters $\lambda_L$ and
$\lambda_R$ are permitted in the middle and right panels of the figure, respectively (see
Eq.\ \eqref{eq:lambda}), while flavor violating squark mixing is forbidden in the left
panel.}
\vspace{.3cm}
 \includegraphics[width=.32\columnwidth]{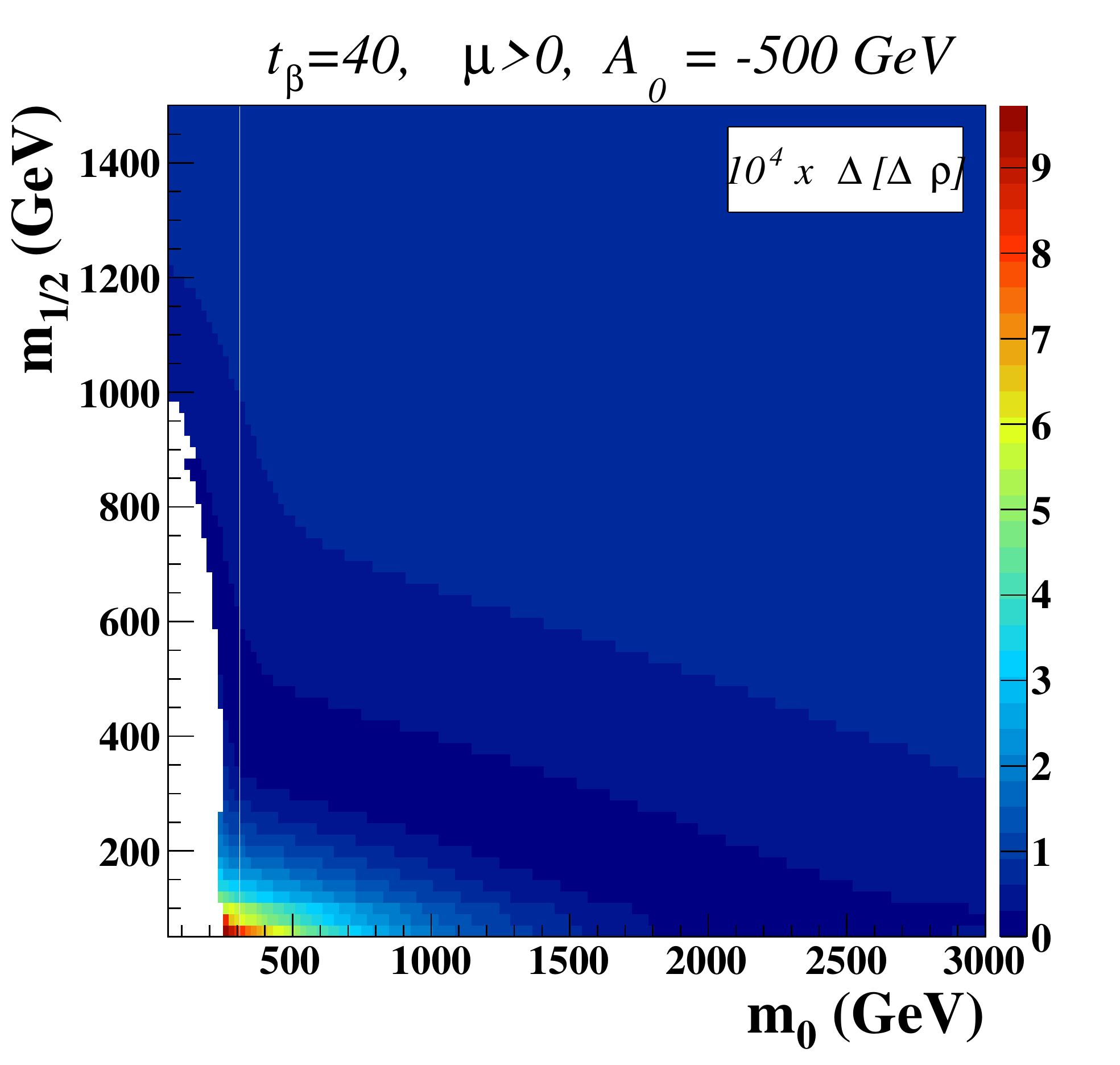}
 \includegraphics[width=.32\columnwidth]{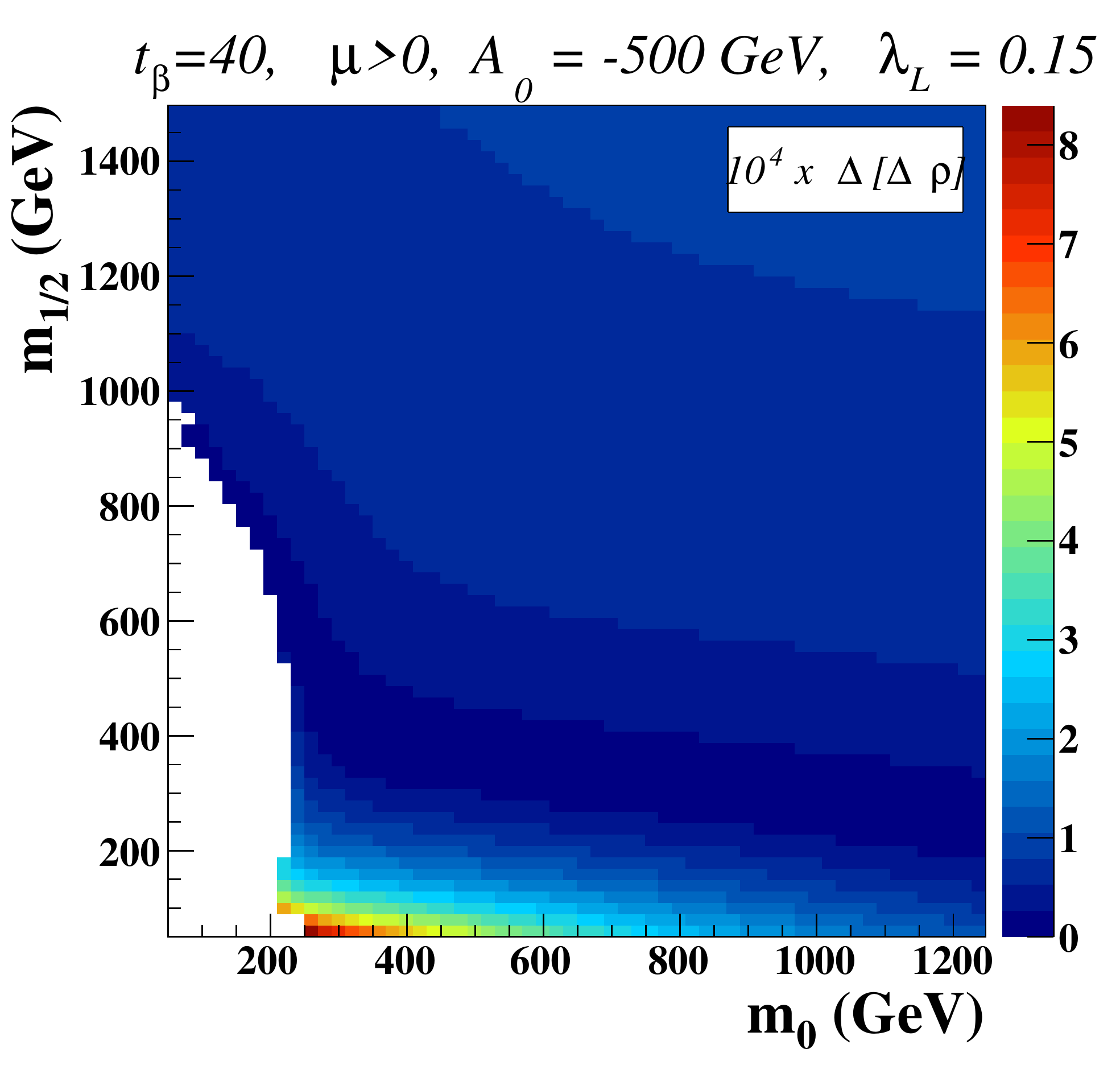}
 \includegraphics[width=.32\columnwidth]{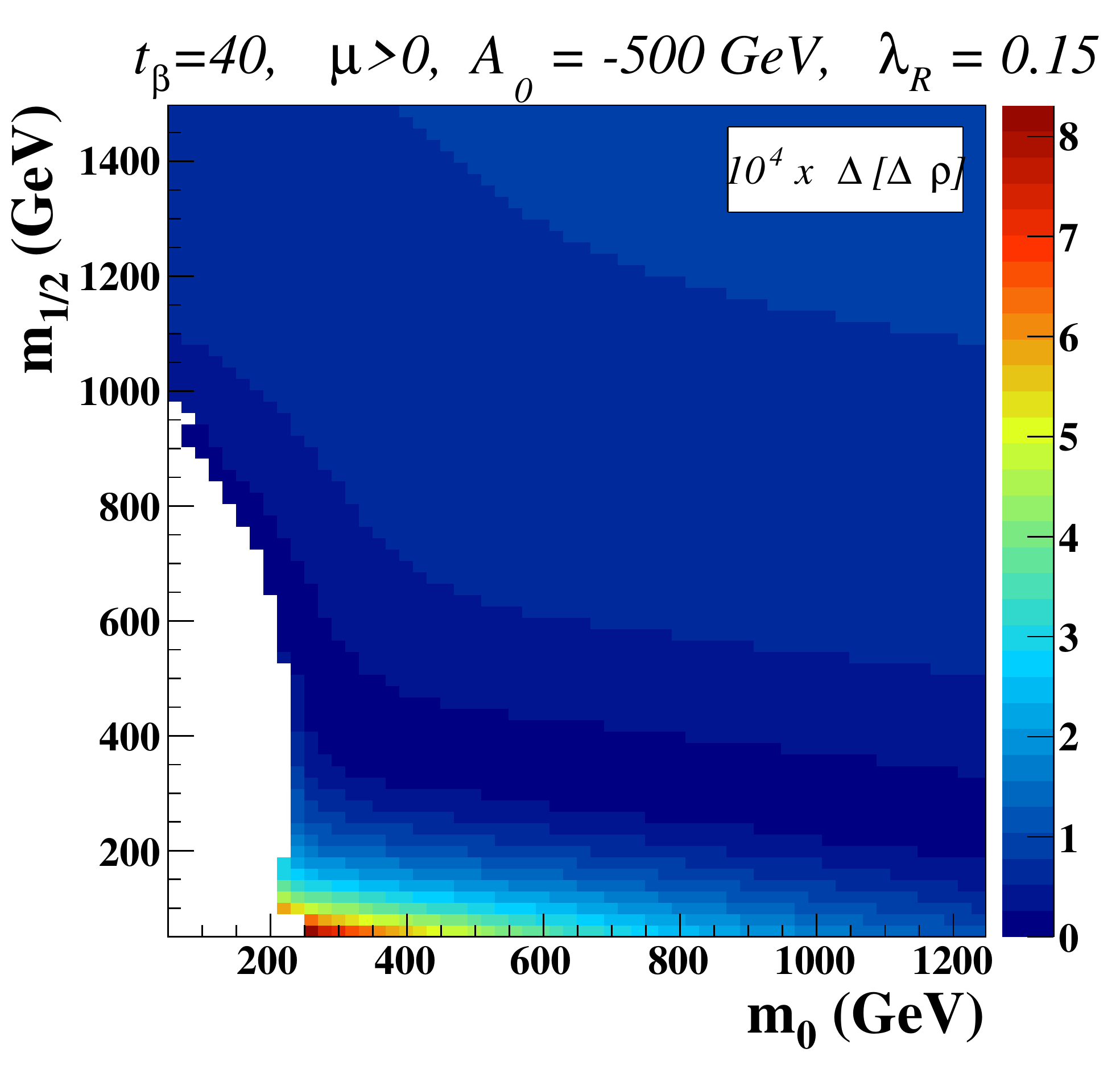}
 \caption{\label{fig:cmssm40_dr} Same as in Figure \ref{fig:cmssm10_dr} but
for $\tan\beta=40$, $A_0=-500$ GeV.}
\end{figure}
%

Supersymmetry, as many extensions of the Standard Model, can be probed by means
of electroweak precision observables such as the 
$\rho_0$-parameter. The latter is defined by
\be
  \rho_0 = \frac{M_W^2}{M_Z^2 \cos\theta_w^2 \rho} \ ,
\label{eq:ewrho0}\ee
where the electroweak mixing angle $\theta_w$ is evaluated at the $Z$-pole and
all loop effects are embedded into the $\rho$-parameter.
For $\rho=\rho_0=1$, one recovers the well-known tree-level relation among
the $Z$-boson and $W$-boson masses $M_Z$ and $M_W$. Eq.\
\eqref{eq:ewrho0} provides a way to generalize this tree-level
relation at higher orders
\cite{Veltman:1977kh,  Sirlin:1980nh, Kennedy:1988rt, Bardin:1989di,
Hollik:1988ii}. In the Standard Model, the $\rho_0$-parameter is
defined as equal to one, following the conventions of the Particle Data Group
\cite{Beringer:2012zz}. In contrast, for 
extensions of the Standard Model affecting the weak sector, $\rho_0$
usually differs
from unity, the associated new physics contributions being more
conveniently re-expressed as
\be
  \Delta \rho = \frac{\Sigma_Z(0)}{M_Z^2} - \frac{\Sigma_W(0)}{M_W^2} \ , 
\label{eq:drho}\ee 
where $\Delta\rho$ measures the deviation of the $\rho_0$ parameter from unity
in terms of the weak vector boson self-energies at zero-momentum $\Sigma_Z(0)$
and $\Sigma_W(0)$. The $\Delta\rho$ quantity is traditionally employed 
to evaluate new physics contributions to electroweak precision
observables such as the squared sine of the electroweak mixing angle
or the $W$-boson mass. The latest combined fits of the $Z$-boson
mass, width, pole asymmetry, $W$-boson and top-quark masses restrict $\Delta\rho$
to be in the range \cite{Beringer:2012zz}
\be
 \Delta\rho = \big(1.564 \pm 9.381\big) \times 10^{-4} \ .
\label{eq:drhoexp}\ee 

%
\begin{figure}[t!]
 \centering
 \includegraphics[width=.32\columnwidth]{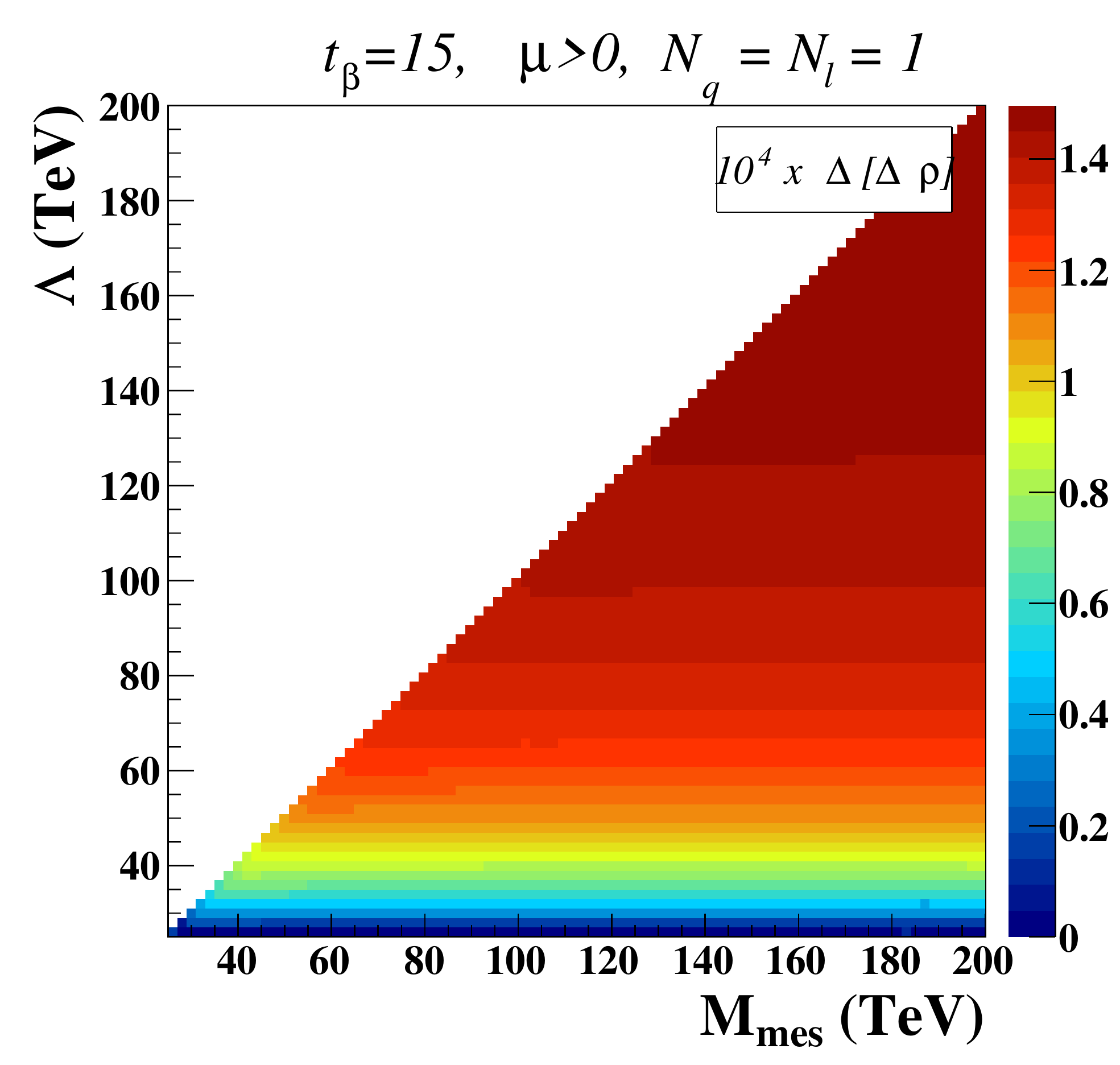}
 \includegraphics[width=.32\columnwidth]{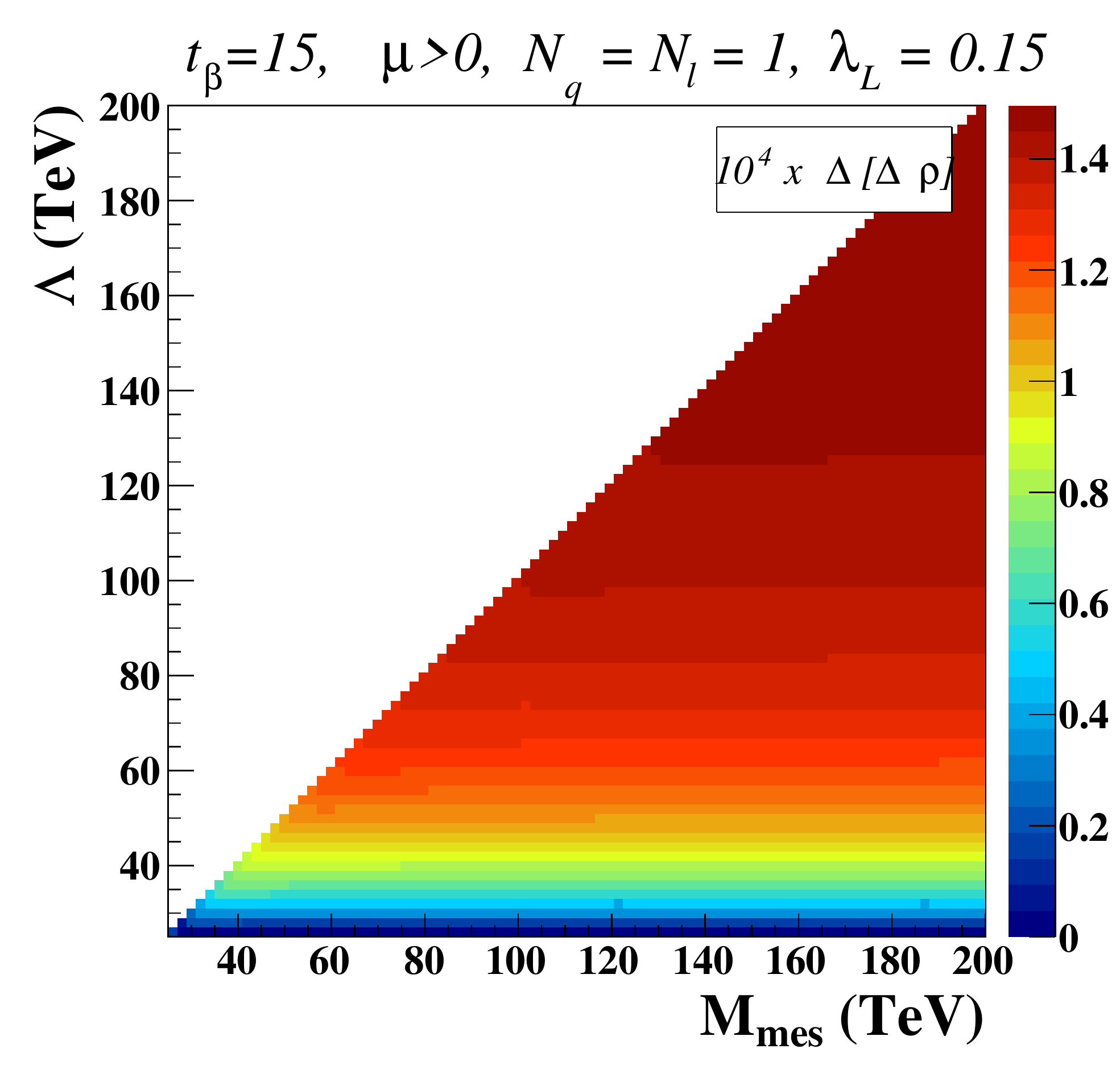}
 \includegraphics[width=.32\columnwidth]{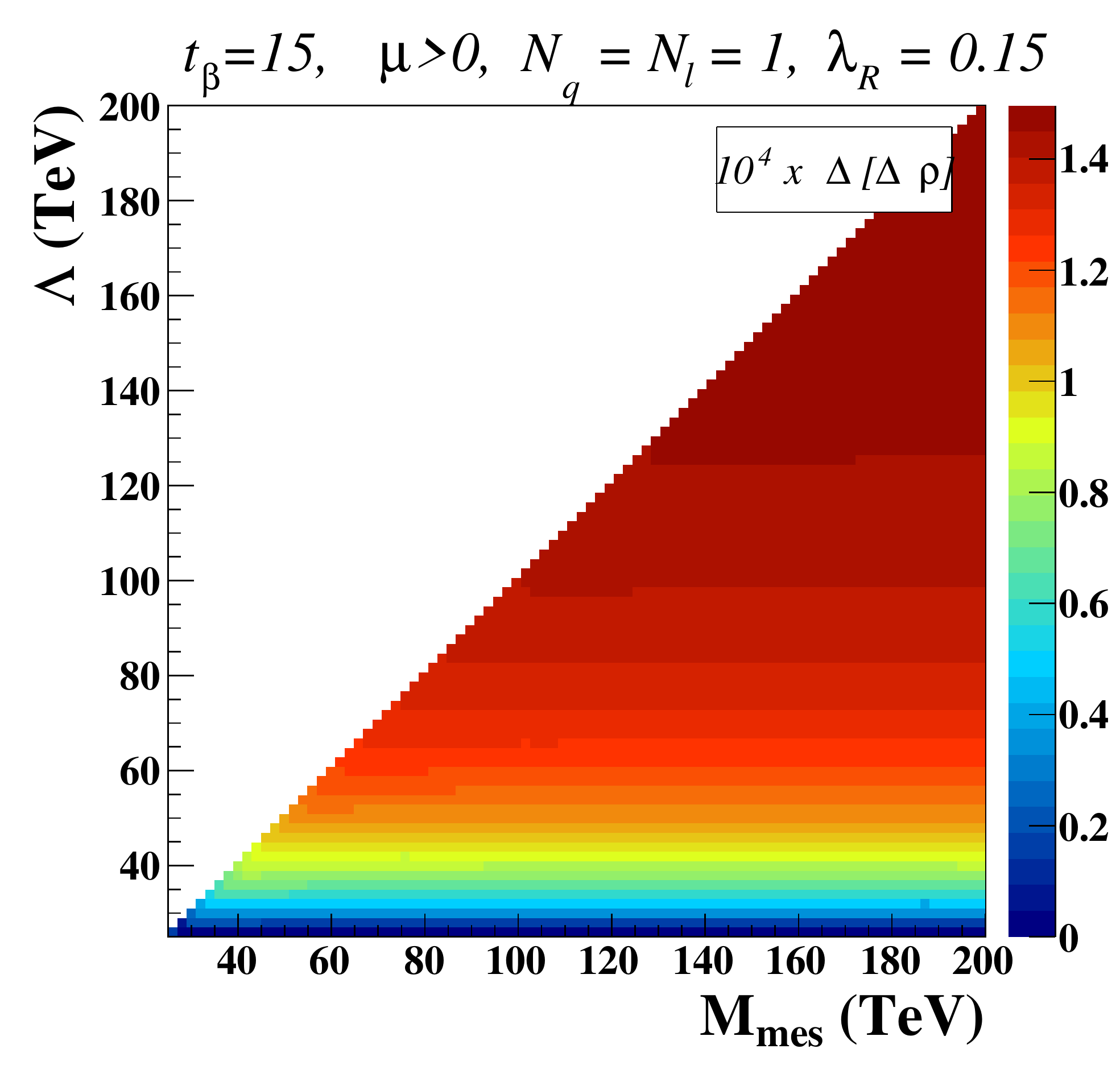}
 \caption{\label{fig:gmsb1_dr} Same as in Figure \ref{fig:cmssm10_dr} but
for the MSSM with gauge-mediated supersymmetry-breaking. We present $(M_{\rm
mes},\Lambda)$ planes with $\tan\beta=15$, one
single series of messenger fields $N_q = N_\ell =1$ and a positive Higgs
mixings parameter $\mu>0$.}
\vspace{.3cm}
 \includegraphics[width=.32\columnwidth]{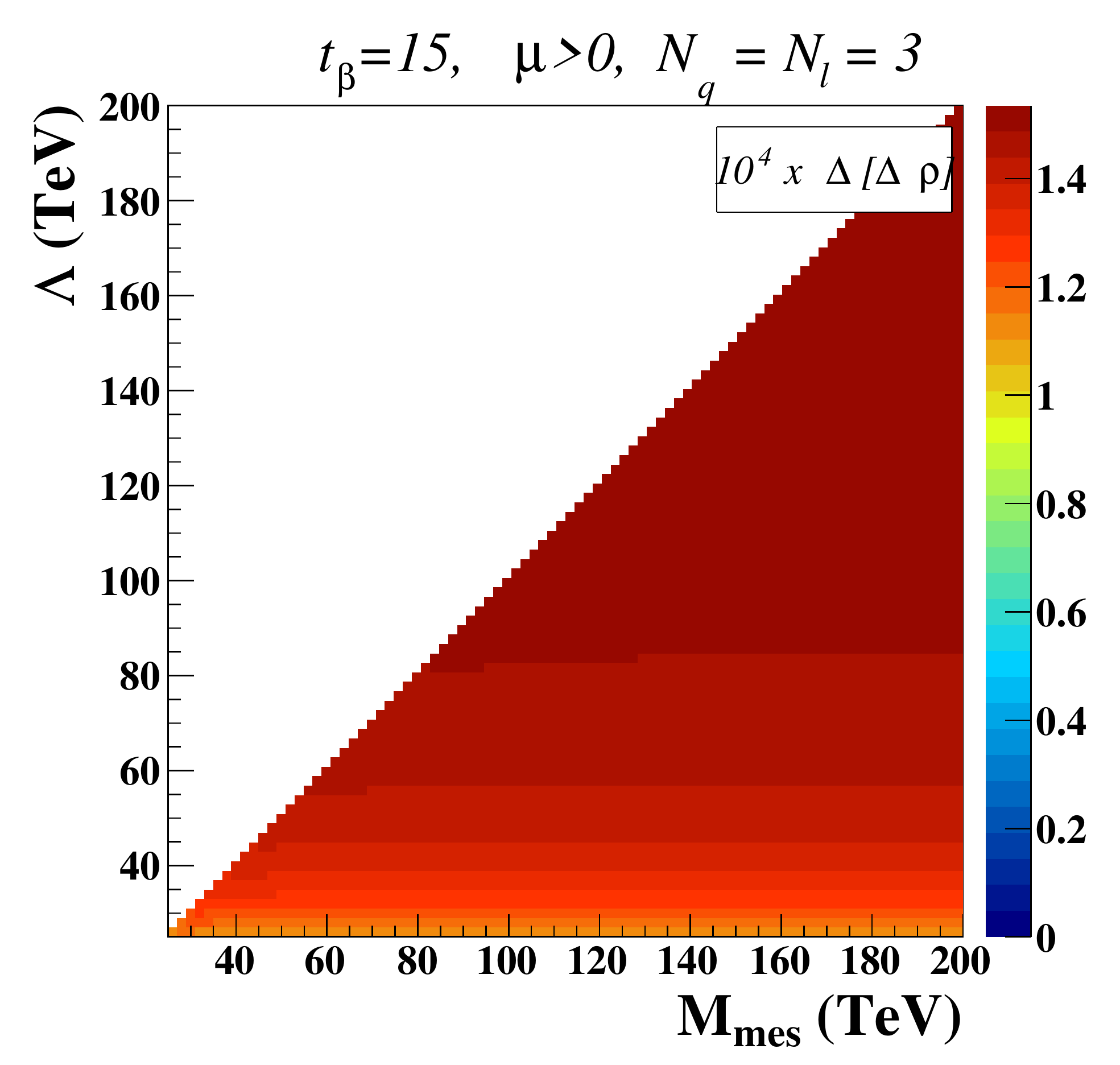}
 \includegraphics[width=.32\columnwidth]{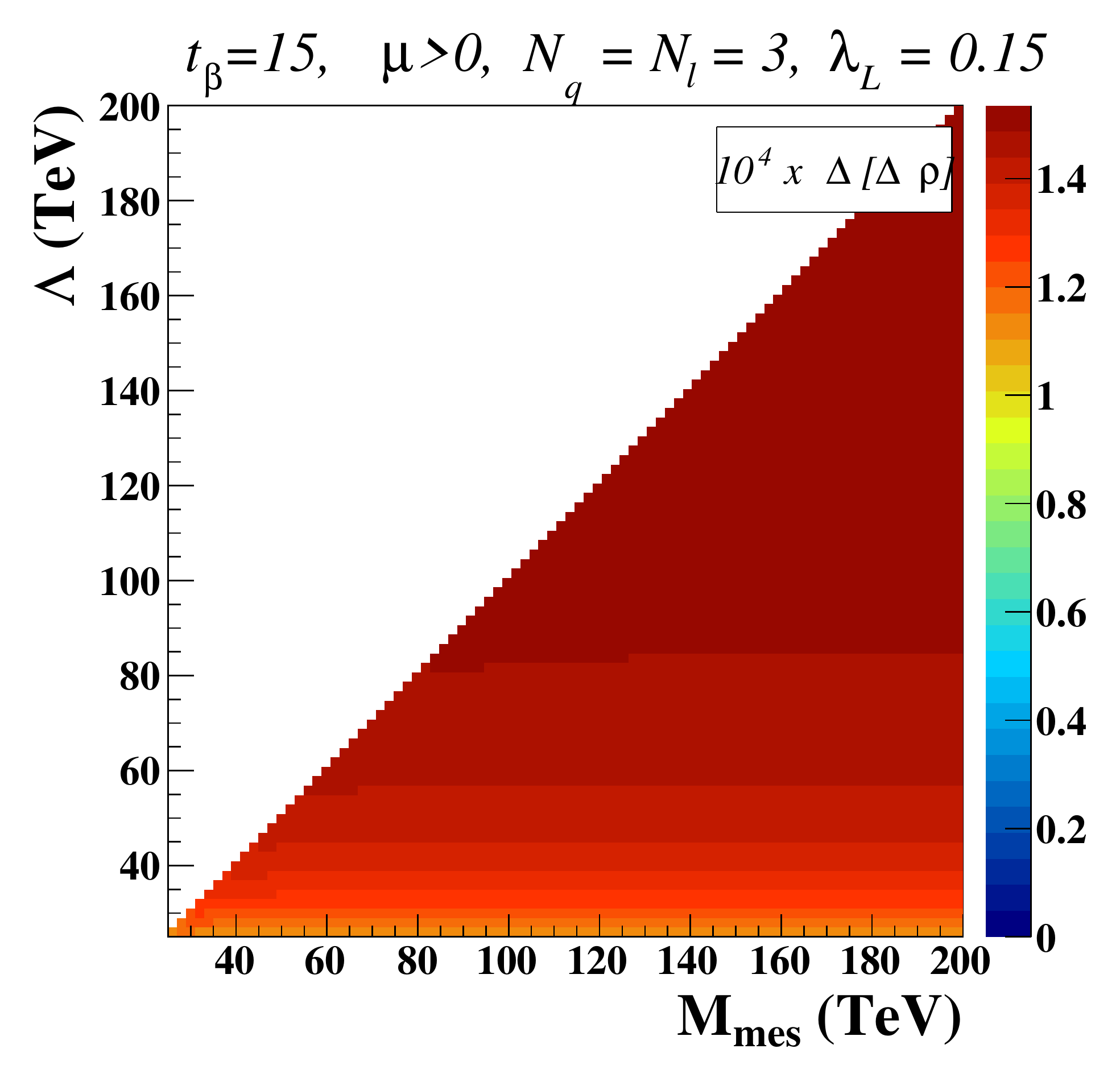}
 \includegraphics[width=.32\columnwidth]{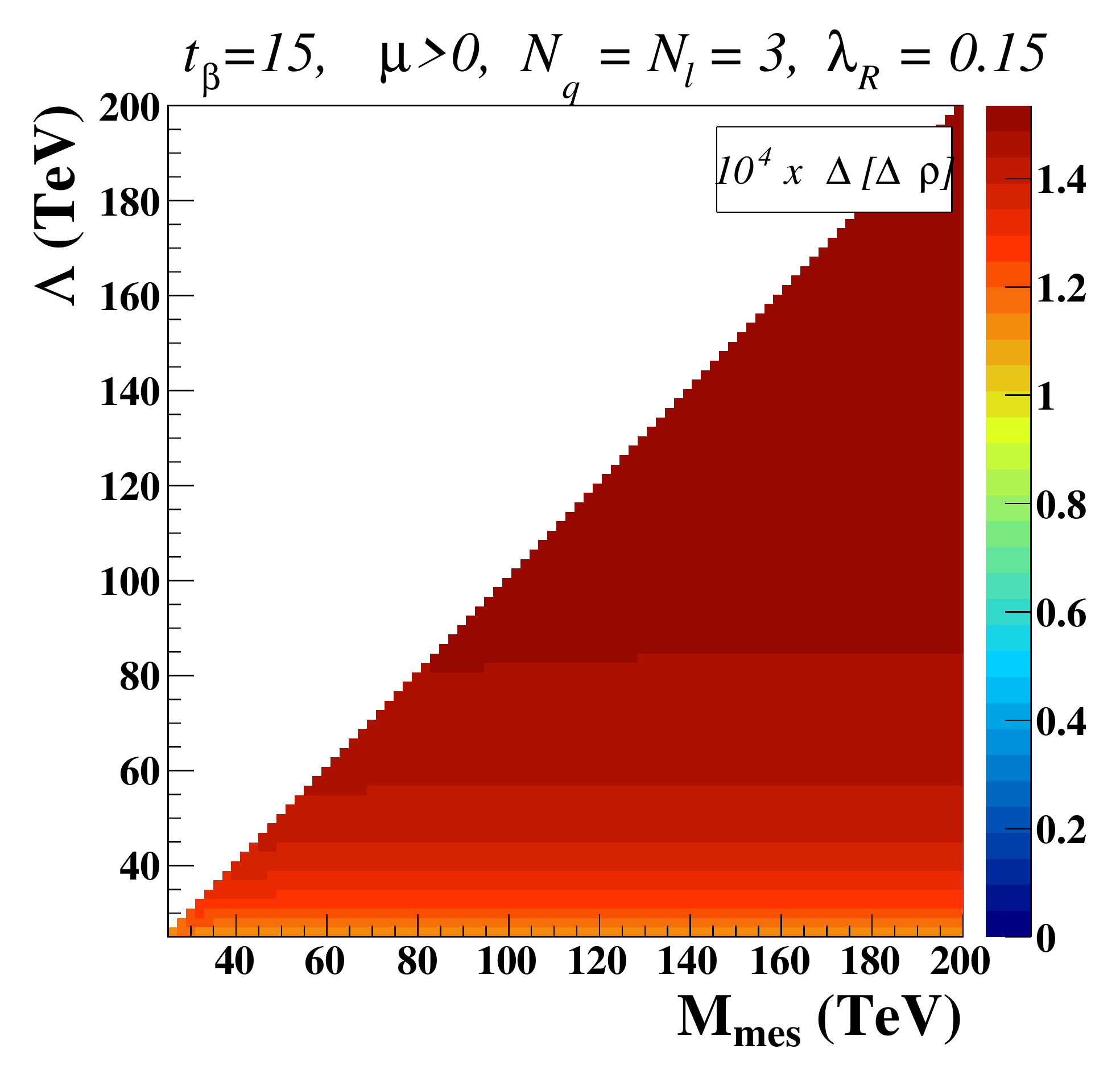}
 \caption{\label{fig:gmsb3_dr} Same as in Figure \ref{fig:cmssm10_dr} but
for the MSSM with gauge-mediated supersymmetry-breaking. We present $(M_{\rm
mes},\Lambda)$ planes with  $\tan\beta=15$, three
series of messenger fields $N_q = N_\ell =3$ and a positive Higgs
mixings parameter $\mu>0$.}
\end{figure}
%

In the context of the Minimal Supersymmetric Standard Model, radiative
contributions to $\Delta\rho$ are known at the
one-loop level \cite{Drees:1990dx}. Furthermore, the leading two-loop
contributions involving gluonic loops as well as top and bottom Yukawa couplings
have also been recently calculated 
\cite{Djouadi:1996pa, Djouadi:1998sq, Heinemeyer:2002jq, Heinemeyer:2004by}.

We present, in Figure \ref{fig:cmssm10_dr} and Figure \ref{fig:cmssm40_dr}, 
theoretical predictions for the $\Delta\rho$ quantity of Eq.~\eqref{eq:drho} in
the framework of cMSSM scenarios as calculated by the \spheno\ program. 
We scan over the two considered $(m_0,m_{1/2})$
benchmark planes ($\tan\beta=10$, $A_0 = 0$ GeV and $\tan\beta=40$, $A_0 = -500$
GeV). By design, the
$\Delta\rho$ parameter is directly sensitive to the mass splitting among the
particles running into the loop-diagrams contributing to the self
energies of the weak gauge bosons $\Sigma_Z$ and $\Sigma_W$. 
As shown in the left panel of the figures, the $\Delta\rho$ observable
does not allow to extract constraints on the parameter space
of the minimal models with $\lambda_L=\lambda_R=0$. However, when non-minimal 
flavor violation is included, the
situation changes for the low $\tan\beta$ region (see the middle and right panels
of the figures), the low-mass regions turning out to be excluded.

%
\begin{figure}[t!]
 \centering
 \includegraphics[width=.32\columnwidth]{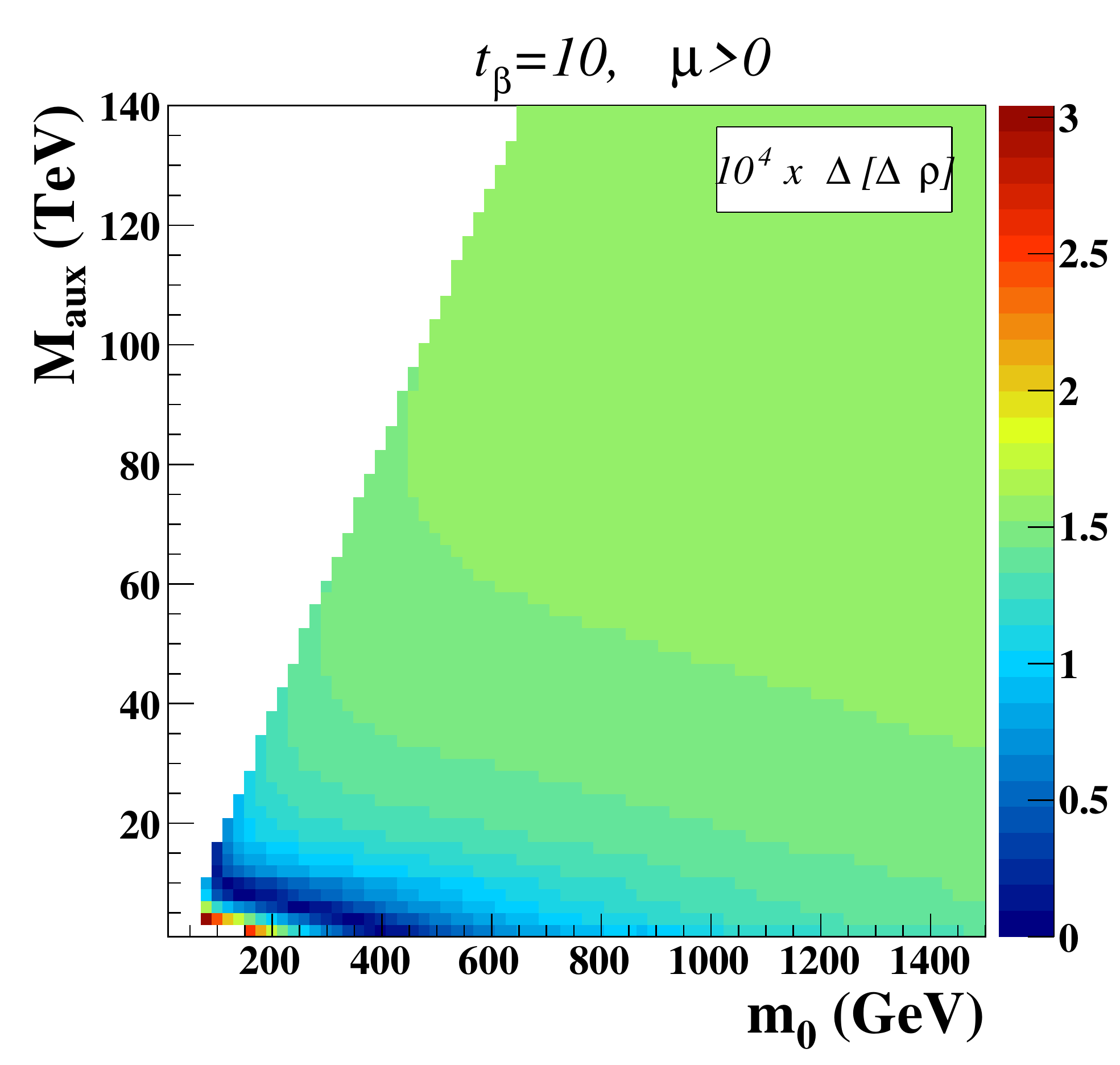}
 \includegraphics[width=.32\columnwidth]{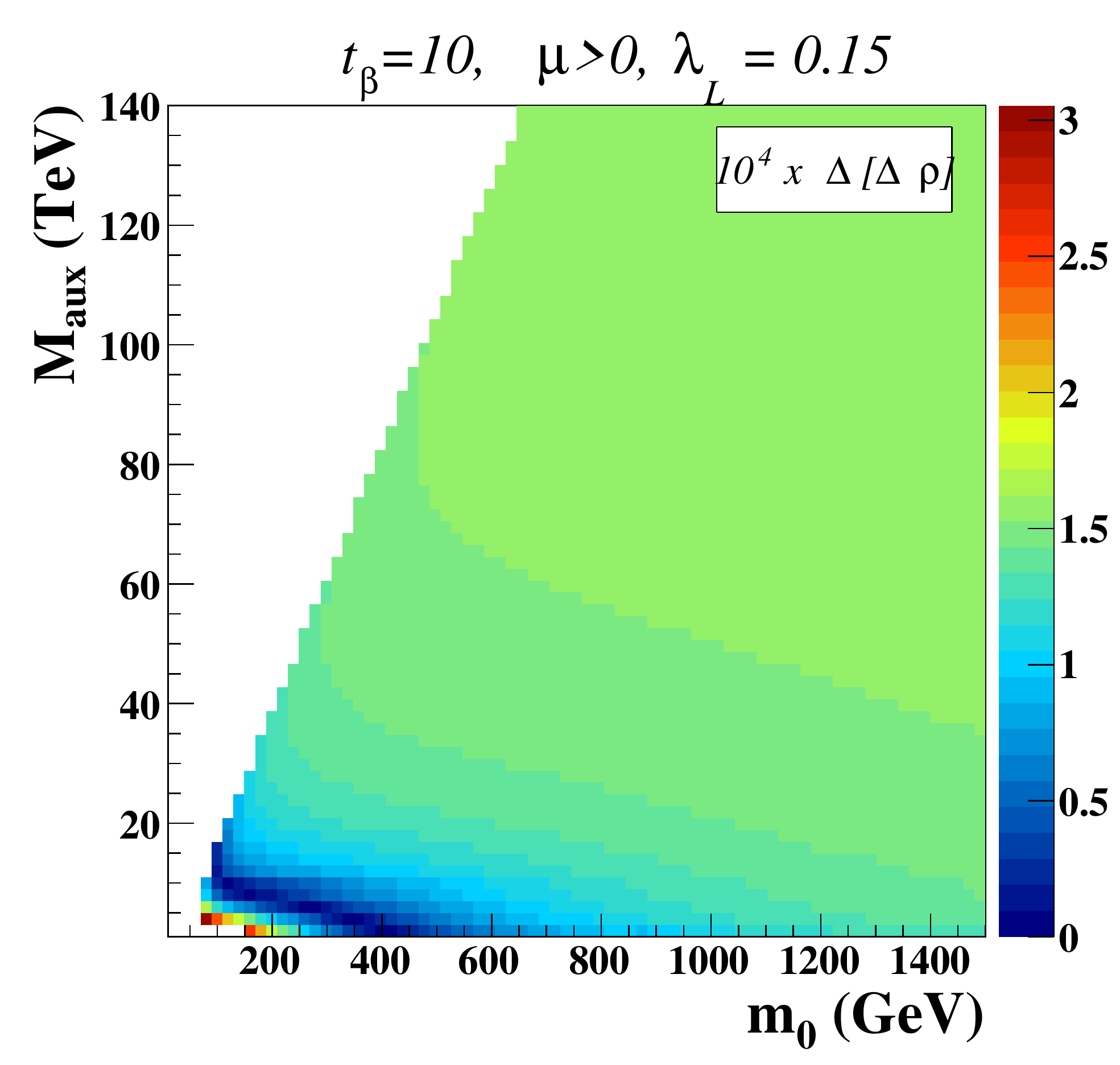}
 \includegraphics[width=.32\columnwidth]{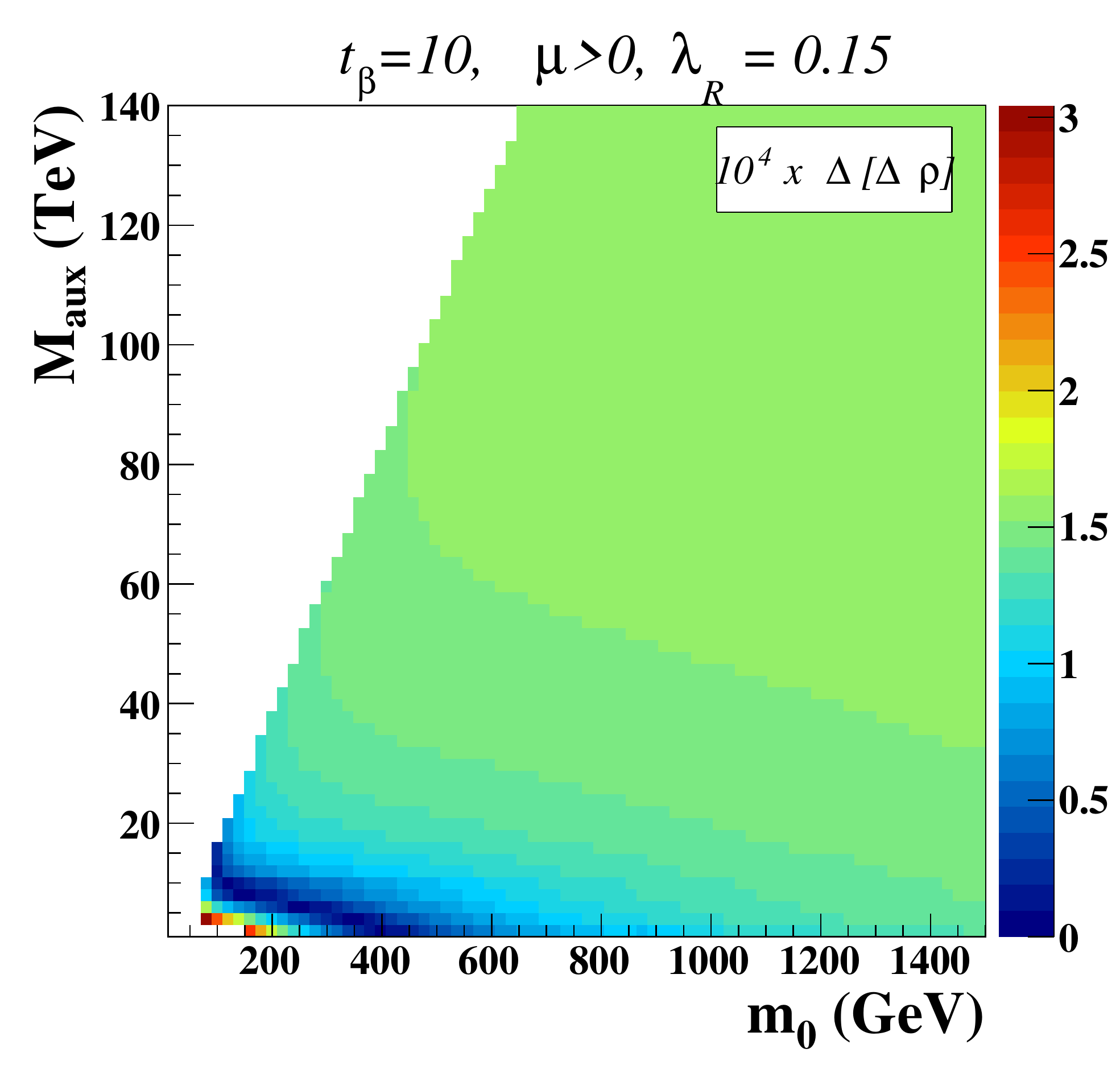}
 \caption{\label{fig:amsb_dr} Same as in Figure \ref{fig:cmssm10_dr} but
for the MSSM with anomaly-mediated supersymmetry-breaking. We present $(m_0,
M_{\rm aux})$ planes with
$\tan\beta=10$ and a positive Higgs mixing parameter $\mu>0$.}
\end{figure}
%

In Figure \ref{fig:gmsb1_dr} and Figure \ref{fig:gmsb3_dr}, we explore MSSM
scenarios with gauge-mediated super\-sym\-me\-try-breaking and show scans of the
$(M_{\rm mes}, \Lambda)$ planes with
$\tan\beta=15$ and respectively $N_q=N_\ell=1$ and  $N_q=N_\ell=3$, while we address
MSSM scenarios with
anomaly-mediated supersymmetry breaking in Figure
\ref{fig:amsb_dr} in which we present $(m_0, M_{\rm aux})$ planes for
$\tan\beta~=~10$. In all cases, almost all the scanned
parameter space regions are found compatible with the measurements, so that this observable 
does not play a major role in constraining the construction of
non-experimentally excluded benchmark scenarios.

\subsection{Summary}
In this section, we have investigated in details the MSSM effects on
several low-energy, flavor and electroweak precision observables.
We have considered, as a starting point for this study, two series of
scenarios designed in the context of the
cMSSM, two series of scenarios
featuring gauge mediated supersymmetry breaking and one series of scenarios where supersymmetry gets broken
via anomalies. Those scenarios have been chosen
after relying on considerations of the LHC Physics Center
at CERN and both the ATLAS and CMS collaborations after analyzing a fraction of
the 2011 LHC data and a large class of indirect constraints on new physics.

Computing for each point predictions for several rare $B$-meson decays and the frequency
of $B$-meson oscillations, it has been found that the low mass regions of the parameter
space are in general excluded at the $2\sigma$ level in the cMSSM and anomaly-mediated
supersymmetry-breaking cases, in contrast to gauge-mediated supersymmetry-breaking MSSM scenarios
where supersymmetric contributions to those observables are
reduced.

We have then started our exploration of
non-minimal supersymmetric models and included non-minimal flavor
violation in the squark sector for both left-left and right-right
chirality mixings. We have shown that, especially
for large flavor violation in the left-left sector, the constraints
are much stronger than in the flavor-conserving case
although large regions of the parameter space are still allowed by current flavor data.

We have next turned to electroweak constraints and
calculated predictions for the anomalous magnetic moment of the muon. We
have observed that there exist a substantial fraction
of the parameter space for which supersymmetric
loop diagrams allow to restore the agreement between
theory and data. Contrary, current bounds on the $\rho$-parameter are not sufficient
to induce any specific constraint at all, like when including non-minimal flavor
violation in the squark sector.

\mysection{Cosmological aspects}
\label{sec:mssm_cosmo}

\subsection{General features}

Among the most compelling evidences for physics beyond the Standard Model is the
presence of non-baryonic dark matter in the Universe. Its relic density is 
constrained to be, after combining seven-year data from WMAP with the latest
measurements related to supernov\ae, baryon acoustic oscillations and the
Hubble constant \cite{Komatsu:2010fb}\footnote{The recent results of
Planck~\cite{Ade:2013zuv} were not available at the time of writing and are thus ignored.},
\be
  \Omega_{\rm CDM} h^2 = 0.1126 \pm 0.0036 \ ,
\label{eq:omh2}\ee
where $h$ denotes the present Hubble expansion rate in units of 100 km
s$^{-1}$ Mpc$^{-1}$. This value is however derived from an interpretation
of cosmological data in the context of the
six-parameter vanilla concordance model of cosmology.

In order for new physics theories to be compatible with such an observation,
they must include a suitable dark matter candidate accounting for the quantity
of dark matter given in Eq.\ \eqref{eq:omh2}. In the context of supersymmetry,
we therefore require the lightest superpartner to be stable, electrically
neutral and singlet under the QCD gauge group  \cite{Goldberg:1983nd,
Ellis:1983ew}.
Consequently, this motivates $R$-parity conservation and leaves the MSSM with
three possible lightest supersymmetric particle, the lightest sneutrino, the
lightest neutralino and the gravitino. However, phenomenologically viable
scenarios with sneutrino dark matter are difficult to achieve by combining
cosmological and collider constraints. On the one hand, a correct dark
matter relic density can only be obtained if
the lightest sneutrino is very light or very heavy, preventing from
a too fast dark matter
annihilation into Standard Model particles via $Z$-boson exchange diagrams
\cite{Ibanez:1983kw, Hagelin:1984wv, Falk:1994es}. On the other hand, very light
sneutrinos are excluded by the invisible $Z$-boson width extracted at LEP
\cite{Beringer:2012zz} and very heavy sneutrinos are excluded by
dark matter direct detection searches~\cite{Falk:1994es}. We therefore focus in this
work on neutralino and gravitino dark matter scenarios.

\subsection{Neutralino dark matter in the constrained MSSM}
\label{sec:cosmocmssm}
Being inspired by gravity-mediated supersymmetry-breaking, cMSSM
scenarios include a gravitino field with a mass of the order of the TeV scale
(see the relation among the different superpartner masses in
Section \eqref{sec:grmsb}). The gravitino is therefore in general much
heavier than some of the other superpartners and thus not a viable dark
matter candidate. We consequently focus on neutralino dark matter.
The energy density of a neutralino of mass $M_{\tilde \chi_1^0}$ is directly
proportional to its present number density $n_0$,  
\be
  \Omega_{\rm CDM}h^2 = \frac{M_{\tilde \chi_1^0} n_0}{\rho_c} \ , 
\ee
where $\rho_c = 3H_0^2/(8\pi G_{\rm N})$ is the critical density of our
Universe, $G_{\rm N}$ being the gravitational constant and $H_0$ the present
value of the Hubble expansion parameter \cite{Bertone:2004pz}. The
neutralino relic abundance $\Omega_{\rm CDM} h^2$ 
can be evaluated after solving the Boltzmann equation  
\be
  \frac{\d n}{\d t} = -3 H n - \langle \sigma_{\rm eff} v \rangle
                    \big( n^2 -  n_{\rm eq}^2 \big) \ .
\label{eq:boltzman}\ee
The first contribution to the right-hand side of this equation is
proportional the time-dependent Hubble expansion parameter $H$ and describes a
dilution of the dark matter density with the expansion of the Universe. On
different footings, the second contribution is related to annihilations and
co-annihilations of dark matter particles into Standard Model fermions and
bosons. This term depends on the dark matter number density in thermal
equilibrium $n_{\rm eq}$ as well as on the annihilation and co-annihilation
effective cross section
$\sigma_{\rm eff}$ multiplied by the relative particle velocity $v$. This
product is then convoluted with the velocity distribution of the dark matter
candidate to obtain the thermally averaged cross section $\langle
\sigma_{\rm eff}v \rangle$. Taking into account a set of $N$ potentially
co-annihilating superparticles, heavier than the lightest neutralino and with
masses $M_1 \leq \dots \leq M_N$, the thermally averaged cross section is given
by \cite{Griest:1990kh, Edsjo:1997bg}
\be
  \langle \sigma_{\rm eff}v \rangle =
    \sum_{i,j=0}^N \langle\sigma_{ij}v_{ij}\rangle \frac{n_{\rm eq}^i n_{\rm
       eq}^j}{n_{\rm eq}^2}\ ,
\ee
where $n_{\rm eq}^i$ denotes the equilibrium number density of the particle $i$,
the index $i=0$ referring to the lightest neutralino. 
We have also introduced the cross sections $\sigma_{ij}$ associated with the
(co-)annihilation of the particles $i$ and $j$ and their relative velocity $v_{ij}$. 

This last equation can be entirely rewritten in terms of the particle masses,
the temperature $T$ and the number of internal degrees of freedom of each
particle species $g_i$, or equivalently their statistical weights accounting for
the number of possible combinations of related particle states,
\be
  \langle \sigma_{\rm eff}v \rangle =
    \sum_{i,j=0}^N \langle\sigma_{ij}v_{ij}\rangle \frac{g_i g_j}{g^2_{\rm eff}} 
      \Big[ \frac{M_i M_j}{M_{\tilde \chi_1^0}^2} \Big]^{3/2} \exp
      \Big[-\frac{(M_i+M_j-2M_{\tilde \chi_1^0})}{T}\Big]\ .
\label{eq:sigveff}\ee
This expression depends in addition on the effective number of degrees of
freedom $g_{\rm eff}$ that can be seen as an appropriate overall normalization
factor. The mass
differences between the superpartners hence play a crucial role in the
computation of the relic density, the exponential suppression factor of
Eq.\ \eqref{eq:sigveff} indicating that co-annihilations are only relevant for
almost mass-degenerate superparticles. 

In wide regions of the cMSSM parameter space, the annihilation of two
neutralino states into Standard Model particles contributes dominantly to the
computation of $\langle \sigma_{\rm eff}v \rangle$. However, other subprocesses
could be significant. For instance, it is clear from Eq.\ \eqref{eq:sigveff} that
non-trivial off-diagonal squark mixings that enhance splittings among the
squark mass-eigenstates directly affect the predictions for the lightest
neutralino
relic density. In addition, flavor-violating couplings also imply that new
channels could contribute to the effective annihilation and
co-annihilation cross section. For example, 
annihilations of neutralinos into a charm-antitop or a top-anticharm quark pair
become 
possible trough squark exchanges when the off-diagonal element of the squark
mass matrices are of the same order as the diagonal ones
\cite{Herrmann:2011xe}. 

%
\begin{figure}[t!]
 \centering
 \includegraphics[width=.32\columnwidth]{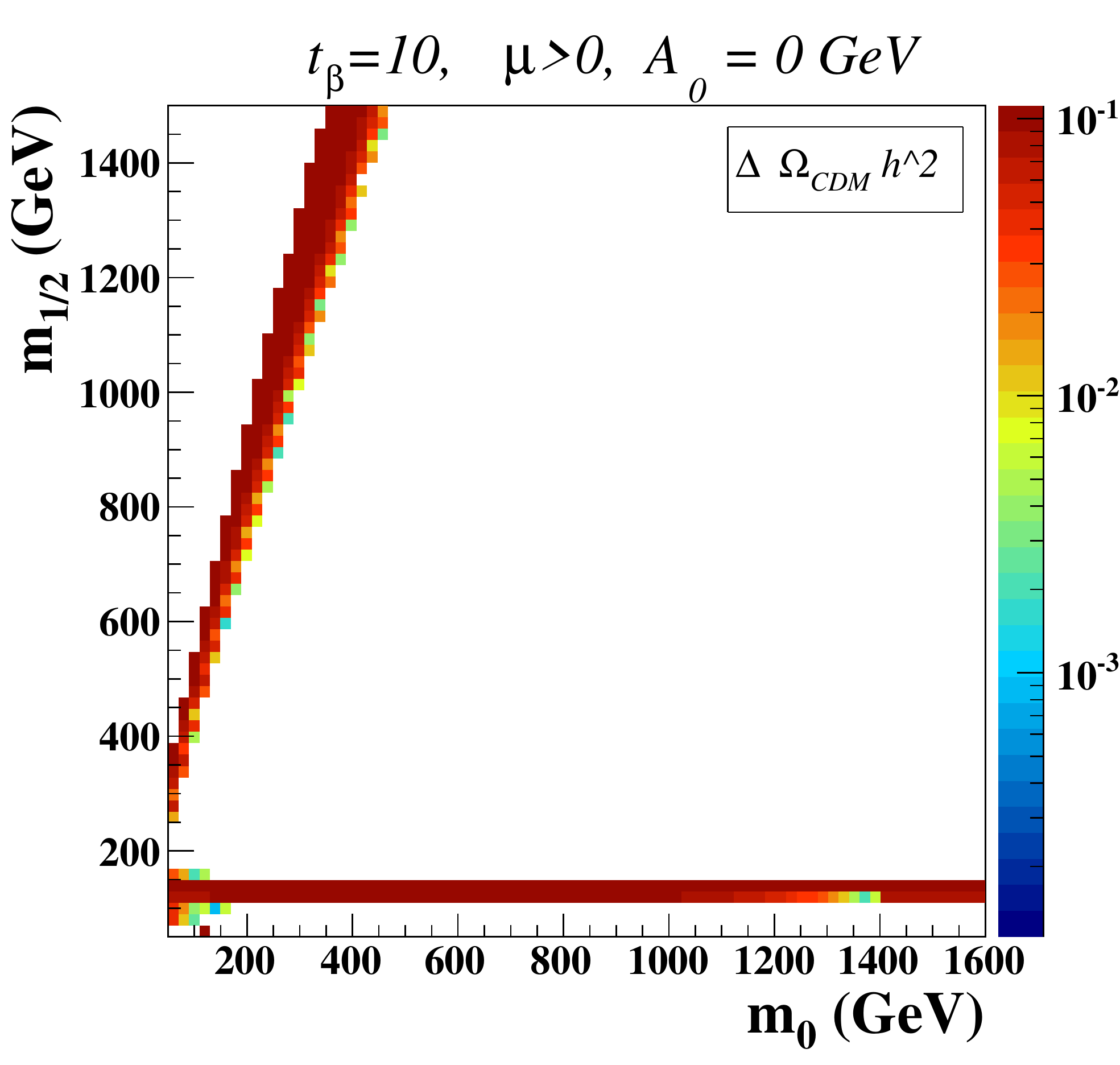}
 \includegraphics[width=.32\columnwidth]{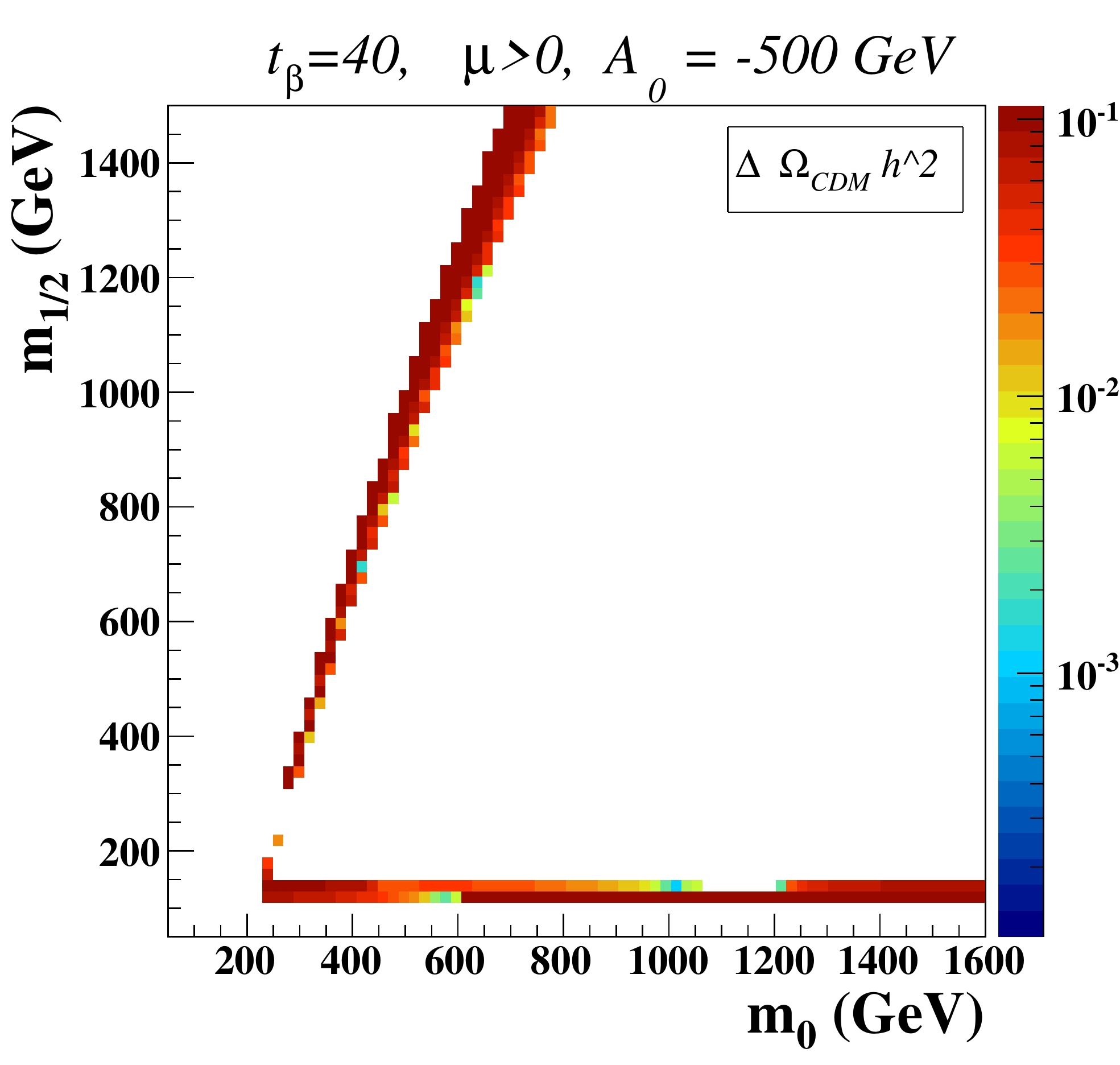}
 \caption{\label{fig:cmssm_omh} Theoretical predictions for the relic density
of the lightest neutralino represented as deviations from the
central value of Eq.~\eqref{eq:omh2}. We present the results in
$(m_0,m_{1/2})$-planes of the
cMSSM with $\tan\beta=10$, $A_0=0$ GeV (left) and
$\tan\beta=40$, $A_0=-500$ GeV (right) and a positive Higgs mixing parameter
$\mu>0$ in both case. The regions depicted in white
correspond either to excluded regions when applying the upper bound of Eq.\
\eqref{eq:omh2} at the $2\sigma$-level, to regions for which there is no
solution to the supersymmetric renormalization group equations or to regions
where the lightest neutralino is not the lightest supersymmetric particle (and 
therefore not a viable dark matter candidate).}
\end{figure}
%

In Figure \ref{fig:cmssm_omh}, we present scans of the two cMSSM
$(m_0,m_{1/2})$ benchmark planes already introduced in Section
\ref{sec:mssm_indirect}  and
show theoretical predictions for the neutralino dark matter relic density. 
After having firstly computed the particle masses and mixings by means of the
\spheno\ package version 3.2.1 \cite{Porod:2011nf},
the spectrum is in a second step exported
to the \darksusy\ code version 5.0.5 \cite{Gondolo:2004sc} which calculates the
associated neutralino relic density. The regions depicted in white correspond
either to regions for which there is no solution to the
supersymmetric renormalization group equations, to regions where the lightest
neutralino is not the lightest supersymmetric particle or to regions excluded
after applying the upper bound of Eq.\ \eqref{eq:omh2}. This conservative
assumption opens the possibility that either dark matter consists of several particle species,
the lightest neutralino being only one of them, or that an alternative
cosmological model is employed (see, \eg, Ref.~\cite{Kamionkowski:1990ni,Giudice:2000ex,%
Salati:2002md,Gelmini:2006pq,Arbey:2008kv,Arbey:2009gt}). Regions of
the parameter space compatible with the upper bound of Eq.\ \eqref{eq:omh2} are
indicated as colored areas, the color code being related to larger and larger deviations
from the central value of Eq.\ \eqref{eq:omh2}. The entire scanned
parameter space is found to be almost excluded by cosmological data. 

Furthermore, it is important to note that a strict
application of both the upper and lower limits of Eq.\ \eqref{eq:omh2}
leads to the exclusion
of all the scanned parameter space, with the exception of the frontiers of the
colored areas. Consequently, dark matter constraints turn out to be very
hard to accommodate. In this case,
allowing for non-minimal flavor violation does not help. 
Parameter space regions compliant with the
observations are known to
be rather insensitive to non-minimal flavor violation in the squark sector, the
dominant diagrams being sub-dominant for moderate values of the
$\lambda$-parameters of Eq.\ \eqref{eq:lambda} \cite{Bozzi:2007me}. 
One exception consists of fixing the $\lambda$-parameters to more extreme values,
close to $\lambda=1$,
as shown in Ref.~\cite{Herrmann:2011xe}.

The derivation of Eq.\ \eqref{eq:omh2}
has strongly relied on the assumption of the underlining cosmological model. The
properties of the early Universe are however relatively unknown so that
alternative cosmological 
models could be taken into account, such as models with a modified expansion
rate \cite{Arbey:2008kv, Arbey:2009gt} or with a modified entropy content
\cite{Arbey:2009gt, Arbey:2011gu}. Therefore, care must be taken when imposing
dark matter constraints on the building of phenomenological models that could be reasonably
evaded or modified.

\subsection{Gravitino dark matter in gauge-mediated supersymmetry breaking}
\label{sec:cosmogmsb}
For MSSM scenarios where supersymmetry is broken through gauge interactions,
the natural dark matter candidate is the gravitino. Depending on its
mass, it can account either for cold ($m_{3/2} \gtrsim 100$ keV), warm ($1\
{\rm keV} \lesssim m_{3/2} \lesssim 100$ keV), or hot ($m_{3/2} \lesssim 1$ keV)
dark matter. Two different sources contribute to the present gravitino abundance
in the Universe,
\be
  \Omega_{\rm CDM} h^2 = \Omega_{\rm CDM}^{\rm therm} h^2 +  \Omega_{\rm CDM}^{\rm
     non-therm} h^2 \ .
\label{eq:relicgrav}\ee
The first contribution shows that gravitinos can be thermally produced in the very early
Universe, the associated energy density reading \cite{Bolz:2000fu,
Pradler:2006qh, Rychkov:2007uq} 
\be\bsp
  \Omega_{\rm CDM}^{\rm therm} h^2 \approx 
    \Big[ \frac{m_{3/2}}{100\ \text{GeV}} \Big] 
    \Big[ \frac{T_R}{10^{10}\ \text{GeV}} \Big] 
    \sum_{i=1}^3 \bigg\{ \omega_i g_i^2 \big( 1 + \frac{M^2_i}{3 m^2_{3/2}} \big) 
          \log\frac{k_i}{g_i} \bigg\} \ .
\esp\label{eq:thgrav} \ee
The thermal contribution to the relic
density $\Omega_{\rm CDM}^{\rm therm}$ involves, in addition to the gravitino mass
$m_{3/2}$, a dependence on 
the reheating temperature $T_R$ which corresponds to the temperature
of the Universe 
after inflation. No stringent constraints on $T_R$ exist, but values of the order
${\cal O}(10^9)$ GeV and larger are preferred in scenarios that feature
leptogenesis to explain the cosmic baryon asymmetry
\cite{Buchmuller:2004nz}. The summation included in Eq.\ \eqref{eq:thgrav} runs
over the three gauge subgroups of
the MSSM, \ie, $U(1)_Y$, $SU(2)_L$, and $SU(3)_c$ for which the
couplings constants are denoted by $g_i$ and the soft mass parameters associated
with the three gaugino fields by $M_i$. The constants $\omega_i$ and $k_i$ are
taken as $\omega_i = 0.018$, $0.044$, $0.117$ and $k_i = 1.266$, $1.312$,
$1.271$ for $i=1,2,3$, respectively, as in Ref.\ \cite{Pradler:2006qh} where these
values are derived from a consistent gauge-invariant finite-temperature calculation
of all relevant squared matrix elements yielding gravitino production.

The second contribution to Eq.\ \eqref{eq:relicgrav} consists of
non-thermal production of gravitino through direct decays of the
next-to-lightest supersymmetric particle into its Standard Model partner,
together with a gravitino. The corresponding relic density $\Omega_{\rm
CDM}^{\rm non-therm}$ depends on both the
relic density of the next-to-lightest superparticle and its mass difference with
the gravitino,
\be
  \Omega_{\rm CDM}^{\rm non-therm} h^2 = \frac{m_{3/2}}{M_{\rm NLSP}} \Omega^{\rm
    therm}_{\rm NLSP}h^2 \ .
\label{eq:nthgrav} \ee
The mass of the next-to-lightest superpartner is denoted by $M_{\rm NLSP}$ and its 
thermal relic density that would have
been computed if it was stable by $\Omega^{\rm therm}_{\rm NLSP}h^2$. 
The would-be
thermal relic density of the next-to-lightest supersymmetric particle
$\Omega_{\rm NLSP}^{\rm therm}h^2$ is computed by solving the Boltzmann equation of
Eq.\ \eqref{eq:boltzman} by means of the 
\micromegas\ package \cite{Belanger:2001fz, Belanger:2004yn, Belanger:2006is,
Belanger:2008sj, Belanger:2010gh} since the 
\darksusy\ program is only adapted for neutralino dark matter. Furthermore,
we constrain the dark matter relic density as in
Ref.\ \cite{Fuks:2008ab}, 
\be
    0.094 \le \Omega_{\rm CDM} h^2 \le 0.136 
\label{eq:omh2old}\ee
at the $95\%$ confidence level. These numbers are extracted from three-year
WMAP data, instead of seven-year WMAP data, 
and a non-minimal, more general, cosmological scenario with eleven parameters is
assumed \cite{Hamann:2006pf}.

From Eq.\ \eqref{eq:thgrav} and Eq.\ \eqref{eq:nthgrav}, one observes that 
thermal gravitino production dominates for low values of the gravitino mass 
$m_{3/2}$ and/or a high reheating temperature $T_R$. In contrast, thermal
production is found negligible for heavier gravitinos.

\renewcommand{\arraystretch}{1.2}
\begin{table}[!t]
 \begin{center}
  \begin{tabular}{|l||ccccc|c|c|}
\hline
  & $\Lambda$ [TeV] & $M_{\rm mes}$ [TeV] & $N_q = N_\ell $ & $\tan\beta$ &
     sign$(\mu)$ & NLSP \\ 
  \hline 
  E & 65 & 90 & 1 & 15 & $>0$  &  $\tilde{\chi}_1^0$ \\
  F & 30 & 80 & 3 & 15 & $>0$  &  $\tilde{\tau}_1$  \\
  \hline
  \end{tabular}
  \end{center}
  \caption{\label{tab:benchgmsb}Benchmark MSSM scenarios featuring
gauge-mediated supersymmetry breaking. We indicate, in addition to the value of the 
parameters of Eq.\ \eqref{eq:gmsbprm}, the nature of the
next-to-lightest superpartner (NLSP).}
\end{table}
\renewcommand{\arraystretch}{1.}

Extra constraints on scenarios with gravitino dark matter arise from the
observed abundances of light elements in our Universe, 
which requires the next-to-lightest supersymmetric particle to decay fast enough
\cite{Pradler:2006qh}. Its 
lifetime $\tau_{\rm NLSP}$, given by the inverse of the total width, reads
\be\label{eq:lifetimenlsp}
  \tau_{\rm NLSP} \approx  6100\ {\text s}\  \Big[ \frac{1\
  \text{TeV}}{M_{\rm NLSP}} \Big]^5 \Big[\frac{m_{3/2}}{100 \text{GeV}} \Big]^2
   \ ,
\ee
after neglecting any source of flavor violation and chirality mixings among 
sfermions, the latter having been found to have only little impact
\cite{Bozzi:2007me}. In order to ensure correct predictions for the abundances
of the light elements as explained by primordial nucleosynthesis, the lifetime
of the next-to-lightest supersymmetric particle has to satisfy the constraint 
$\tau_{\rm NLSP} \lesssim 6000$ s \cite{Pospelov:2008ta}, favoring thus scenarios
with a light gravitino.

In order to confront predictions for the gravitino
relic density after imposing the constraint of Eq.\ \eqref{eq:omh2} at the 
$2\sigma$ level, we select two representative benchmark scenarios among those
included in the planes of Section \ref{sec:mssm_indirect}. Following
the conventions of Ref.\ \cite{Fuks:2008ab}, these scenarios are denoted by the
letters E and F, and defined in
Table \ref{tab:benchgmsb}. 

The benchmark point E has been chosen in the region both favored by 
electroweak precision and flavor data (see Section \ref{sec:mssm_indirect}).
The superparticles are here rather light, the next-to-lightest
supersymmetric particle being the lightest neutralino with a mass of 
$M_{\tilde{\chi}_1^0} = 95.4$ GeV and the three lighter charged sleptons have close
masses of about 100 GeV. The other sleptons, the sneutrinos, and
the gauginos have moderate masses below 300 GeV, while squarks and gluino are
quite heavy with masses above 700 GeV\footnote{This section is dedicated to the
illustration of several cosmological aspects associated with the three
supersymmetry-breaking scenarios investigated in this work. Therefore, we ignore
LHC constraints when designing benchmark scenarios and will address them
in the next section.}.

The point F is also compatible with low-energy and flavor constraints. 
In this case, the three lightest sleptons
have masses of about 100 GeV, the next-to-lightest supersymmetric particle being
the lightest stau with a mass $m_{\tilde{\tau}_1} = 90.7$ GeV. The other sleptons,
sneutrinos, and gauginos are a bit heavier but their masses are kept below 200
GeV. Finally, squarks and gluino have masses ranging up to 
700 GeV. 

%
\begin{figure}[t!]
  \centering
  \includegraphics[width=.42\columnwidth]{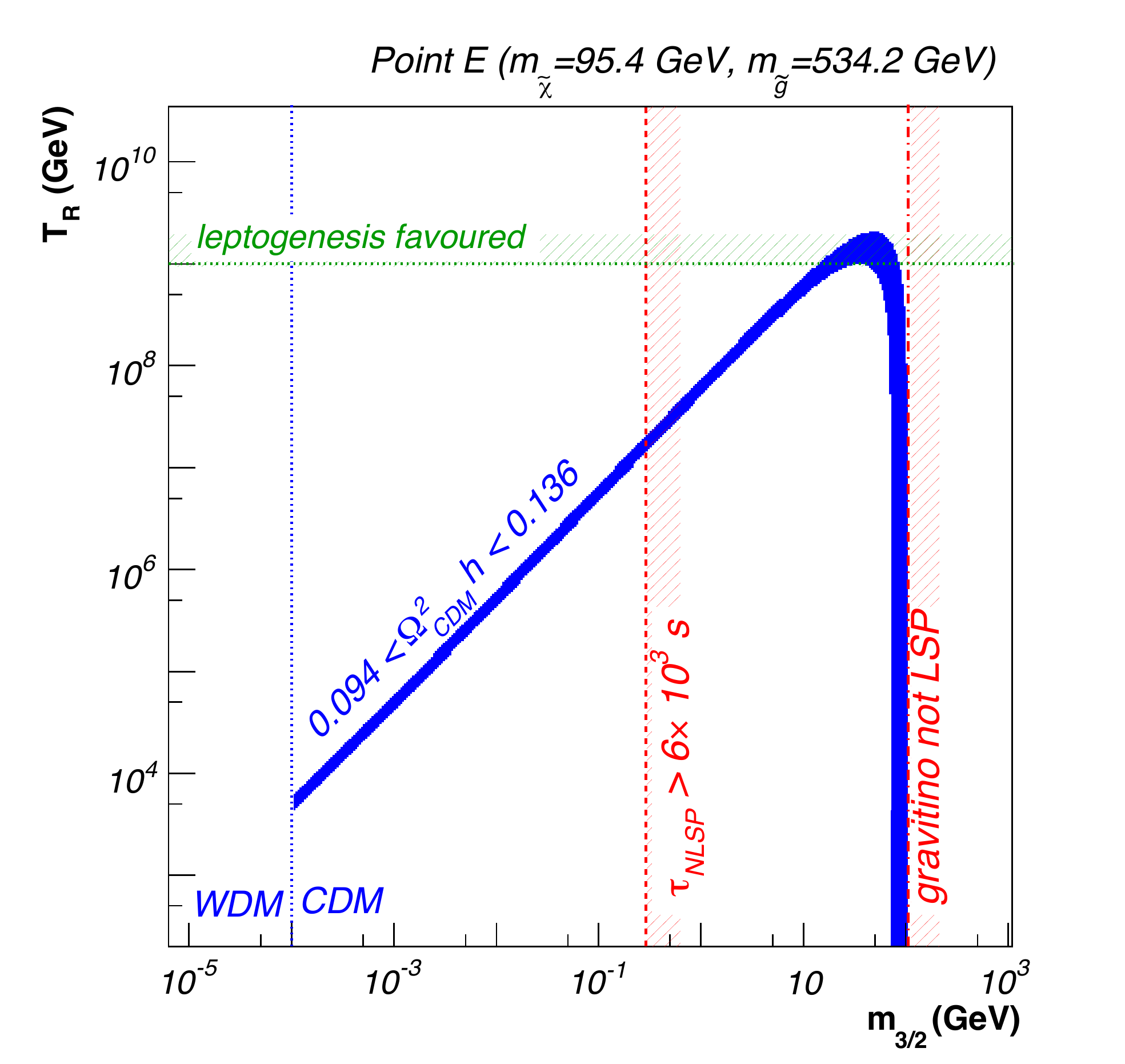}
  \includegraphics[width=.42\columnwidth]{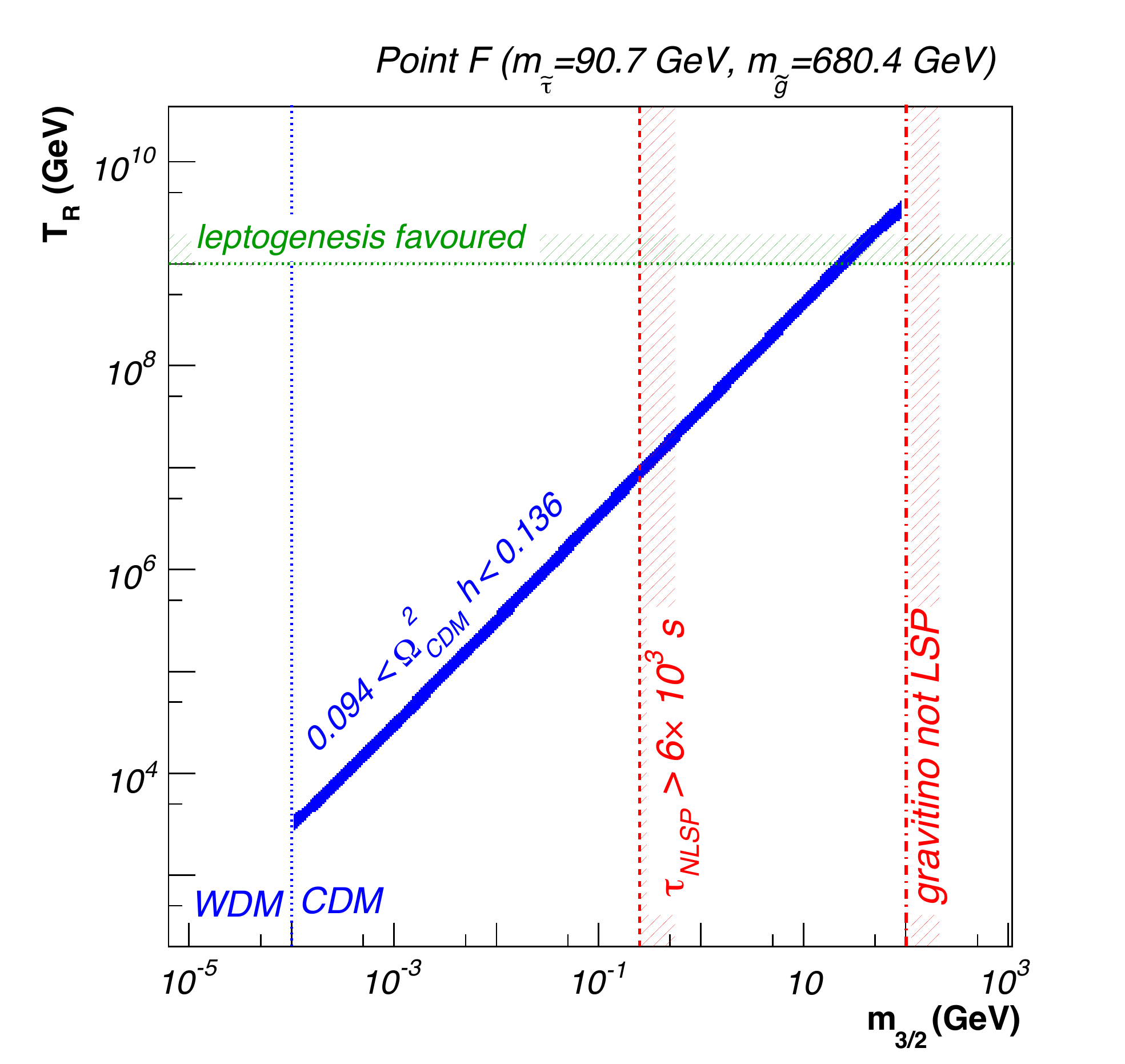}
  \caption{Cosmological constraints on the two MSSM scenarios with gauge-mediated
    supersymmetry breaking of Table~\ref{tab:benchgmsb} presented as
    $(m_{3/2},T_R)$ planes.
    Regions favored by WMAP data are shown in blue, 
    those predicting a correct lifetime for the next-to-lightest supersymmetric
    particle are indicated through a vertical red line and those favored
    by leptogenesis are shown by an horizontal green line. Regions where
    the gravitino consists of a warm (WDM) and cold (CDM) dark matter candidate
    are also pointed out, as well as those where the gravitino is not the lightest
    supersymmetric particle.}
  \label{fig:cosmogmsb}
\end{figure}
%

In Figure \ref{fig:cosmogmsb}, we confront the predictions for the gravitino
relic density after imposing the constraints of Eq.\ \eqref{eq:omh2} and present
the results in $(m_{3/2},T_R)$ planes for the two benchmark scenarios E (left
panel) and F (right panel). In both figures, we indicate the upper limit on the
gravitino mass deduced from the lifetime of the next-to-lightest superpartner. The
latter, computed as in Eq.\ \eqref{eq:lifetimenlsp}, is asked to be
shorter than 6000 seconds. In addition, the parameter space regions where the
gravitino is a warm dark matter
candidate, a cold dark matter candidate and not a suitable dark matter candidate 
are separated by means of vertical lines. Concerning the reheating temperature,
we indicate the regions favored by leptogenesis, for which $T_R \gtrsim 10^9$ GeV.

The contributions to the gravitino relic density induced by decays of the
next-to-lightest superparticle, $\Omega_{\rm CDM}^{\rm non-therm} h^2$, are only 
relevant for scenario E where the relic density of the lightest neutralino is
rather large, $\Omega_{\rm NLSP}^{\rm therm}h^2 = 0.1275$. This value lying well 
within
the interval favored by WMAP three-year data (see Eq.\ \eqref{eq:omh2old}), a 
band in the $(m_{3/2}, T_R)$ plane
around $m_{3/2} \approx M_{\tilde{\chi}_1^0} = 95.4$ GeV is found to be
favored by the constraints.
Concerning the benchmark point F,
the lightest stau is the next-to-lightest superpartner and its annihilation
cross section is large enough so that the corresponding relic density is negligible. 

From the results shown in Figure \ref{fig:cosmogmsb}, it is clear that all  
cosmological constraints cannot be fulfilled simultaneously.
For instance, constructing a scenario
featuring leptogenesis, \ie, where the reheating temperature is such that $T_R \gtrsim
10^9$ GeV, and predicting a correct value for the relic abundance 
leads to a too long predicted lifetime for the next-to-lightest
superpartner so that light element abundances are spoiled. This property still
holds for many other MSSM benchmark points with gauge-mediated supersymmetry
breaking, as shown in Ref.\ \cite{Fuks:2008ab}. However, the
reheating temperature constraint, linked to leptogenesis, can in general be
relaxed when building cosmologically viable MSSM scenarios. Mechanisms
alternative to thermal leptogenesis can be accounted for to
explain cosmic baryon asymmetry in the Universe, such as non-thermal
leptogenesis via inflaton decay \cite{Lazarides:1991wu, Murayama:1993em} or
Affleck-Dine leptogenesis \cite{Affleck:1984fy}. We deduce from our
results limits on an acceptable gravitino mass 
\be
    10^{-4} \text{ GeV } \lesssim m_{3/2} \lesssim 0.1 \text{ GeV }\ , 
\ee
which allows for gravitino cold dark matter whose relic density agrees with
WMAP data. In addition, the next-to-lightest superparticle lifetime is kept short
enough not to spoil light-element abundances.  

\subsection{Neutralino dark matter with anomaly-supersymmetry breaking}

In MSSM scenarios where supersymmetry is broken through anomalies, the lightest
of the four neutralinos is always the lightest superpartner and therefore a
viable dark matter candidate. In contrast to the cMSSM, renormalization group
evolution drives the wino mass parameter $M_2$ to a value smaller than the one 
of the two other gaugino mass parameters $M_1$ and $M_3$, dark matter
being thus mostly wino-like. However, 
chargino masses also depend on the soft parameter $M_2$ so that
the mass
difference between the lightest neutralino and the lightest chargino is often of
about a few GeV or less. From Eq.\ \eqref{eq:sigveff}, it turns out that
co-annihilations between these states are very
efficient and play a significant role in the computations of
predictions for the associated dark matter relic density.

\renewcommand{\arraystretch}{1.2}
\begin{table}
\begin{center}\begin{tabular}{|cccc||c c|}
  \hline
    $M_{\rm aux}$  [TeV] & $m_0$ [TeV] & $\tan\beta$ & sign$(\mu)$ &
    $M_{\tilde{\chi}^0_1}$ [GeV] & $M_{\tilde{\chi}^{\pm}_1}$ [GeV] \\ 
  \hline
  60   & 1    & 10  & $> 0$ &  174.5 & 174.7 \\
  \hline
\end{tabular}
\end{center}
\caption{MSSM benchmark scenario featuring anomaly-mediated supersymmetry
  breaking. We indicate the supersymmetry-breaking parameters at the high scale
  (see Eq.\ \eqref{eq:amsbprm}) and the masses, after
  renormalization group running, of the lightest neutralino and the lightest
  chargino.}
\label{tab:amsbbch}
\end{table}
\renewcommand{\arraystretch}{1.}

As a result of a large co-annihilation cross sections, the
neutralino relic density is usually, in anomaly-mediated supersymmetry-breaking
scenarios, one or two orders of magnitude below the range given in Eq.\
\eqref{eq:omh2} \cite{Chen:1996ap, Moroi:1999zb}. We illustrate this feature by
taking the specific example of one scenario of the
scans of Section~\ref{sec:mssm_indirect} that we define in
Table~\ref{tab:amsbbch}. This scenario is compatible with constraints
issued from flavor physics but do not include direct constraints on the
masses of the superpartners (as in Section~\ref{sec:cosmogmsb}).

As a generic feature of MSSM scenarios with anomaly-mediated supersymmetry
breaking, the lightest chargino and neutralino masses depend mainly on the
auxiliary mass $M_{\rm aux}$, as shown in Eq.\ \eqref{eq:amsbino}, and are roughly
independent of the
universal scalar mass $m_0$ introduced to solve the tachyonic slepton problem.
After renormalization group running, these masses 
are found
to be $M_{\tilde{\chi}^0_1} \sim
M_{\tilde{\chi}^{\pm}_1} \sim 175$ GeV. Moreover, the larger value of the $m_0$ parameter
leads masses above 1 TeV for the scalar superpartners and the gluino.
Using the public
code \darksusy\ \cite{Gondolo:2004sc}, we calculate the neutralino relic
density $\Omega_{\rm CDM} h^2 = 8.57 \times 10^{-4}$. As
expected, this value is far too low compared the measurements (see 
Eq.\ \eqref{eq:omh2}).

However, thermal production of neutralinos is not the only mechanism that has to
be considered since in the context of anomaly-mediated supersymmetry breaking,
several non-thermal production modes are possible. They include, for instance, 
the decay of heavy
fields such as moduli, gravitinos,  axions and axinos 
\cite{Moroi:1999zb, Covi:1999ty, Covi:2001nw, Baer:2010kd}. In this work, we
focus on two particular choices consisting first of decays of
neutral scalar fields, called moduli fields $\Phi$,
only coupling to matter via gravitational interactions and that are
necessary for the UV completion of many phenomenologically-based models for cosmology,
and secondly on gravitino fields.

\begin{figure}
\begin{center}
	\includegraphics[width=.42\columnwidth]{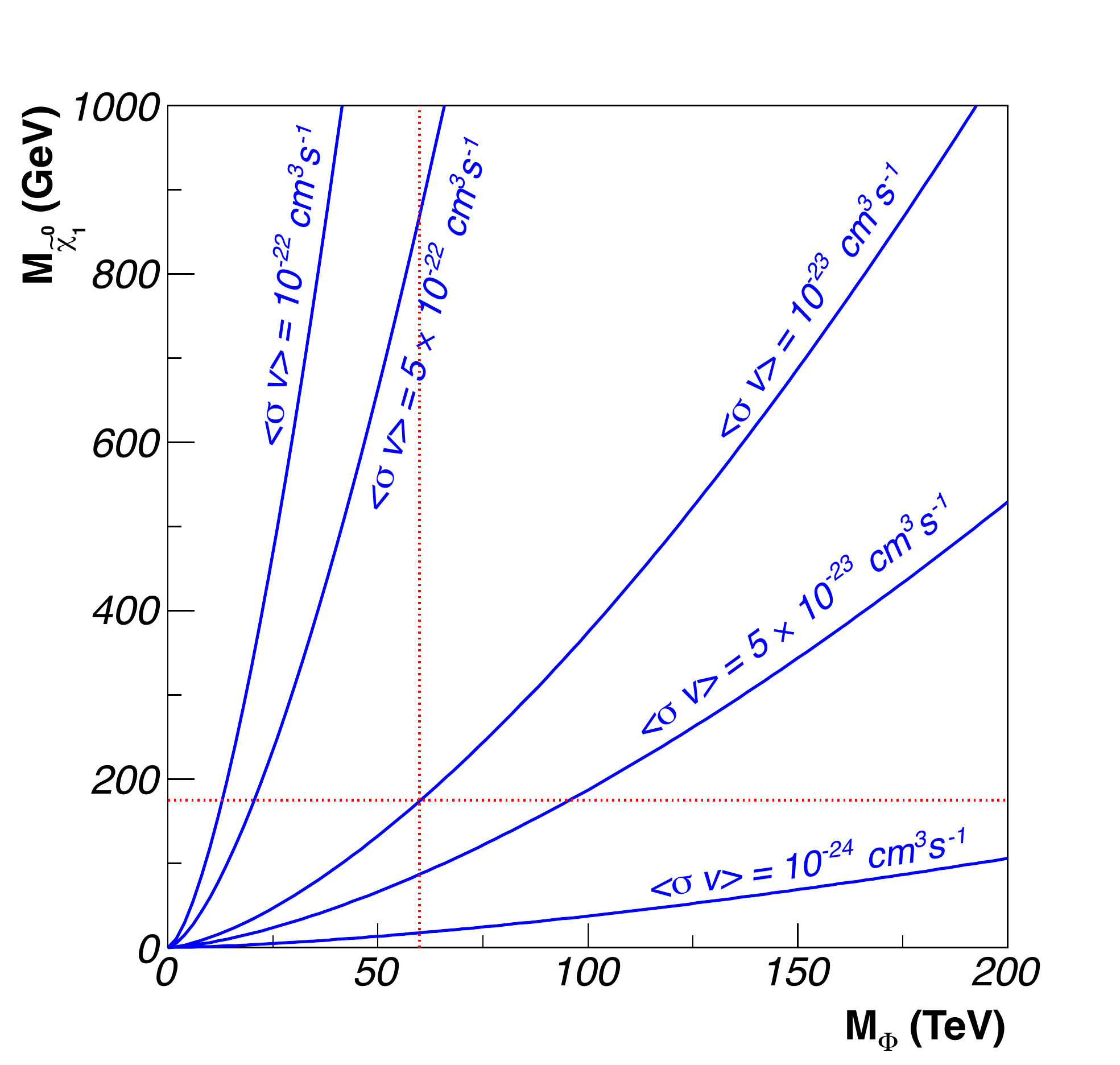}
	\includegraphics[width=.42\columnwidth]{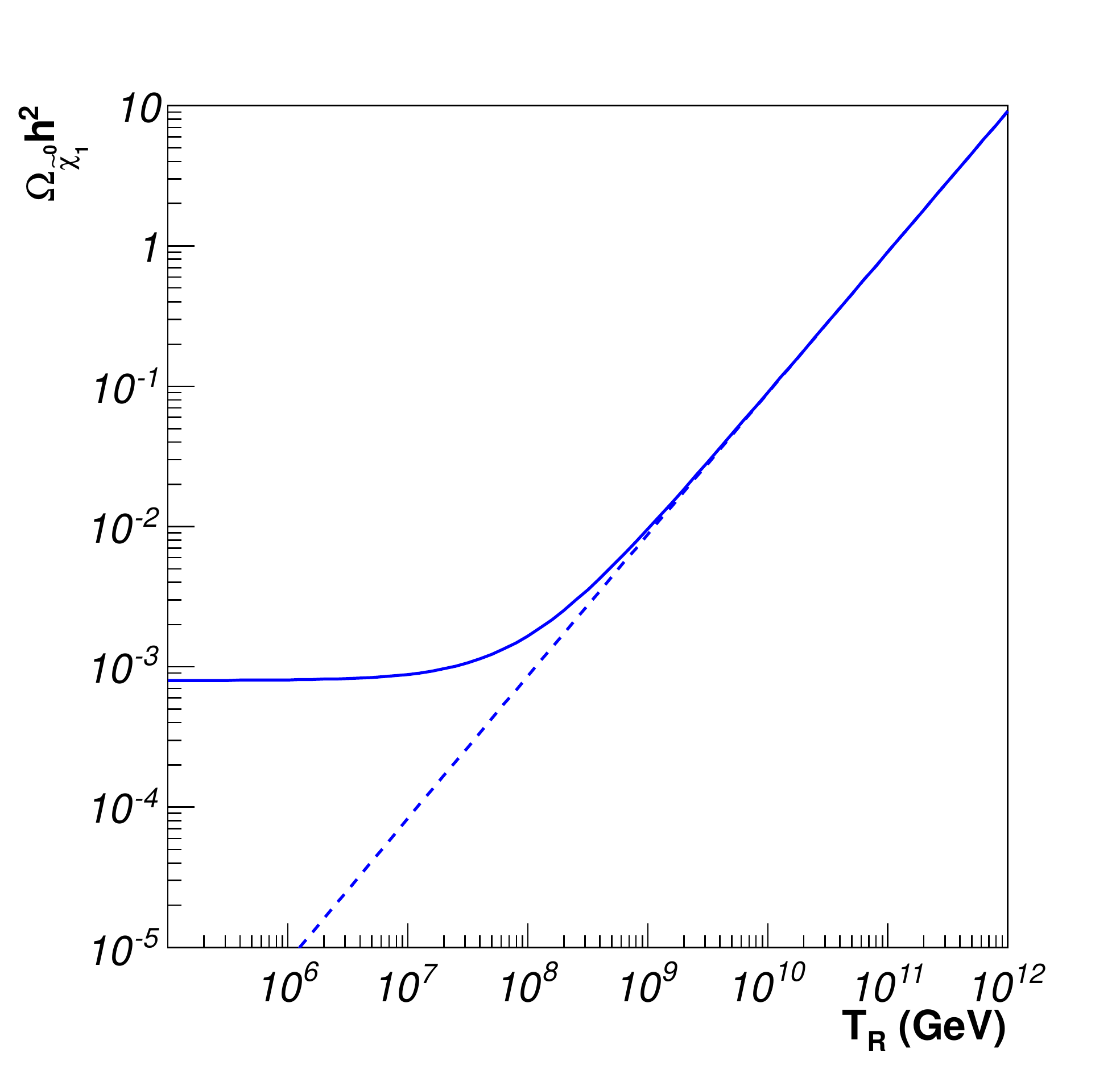}
\end{center}
\caption{We present, in the left panel, curves defined by $\Omega_{\rm
  CDM}^{\rm mod} h^2=0.1126$ in the $(M_{\Phi}, M_{\tilde{\chi}_1^0})$ plane
  for different values of the annihilation cross section $\langle\sigma
  v\rangle$. The dotted lines correspond to $M_{\tilde{\chi}^0_1}=175$ GeV and
  $M_{\Phi}=M_{\rm aux}=60$ TeV. In the right panel, we focus on the dependence
  on the reheating temperature $T_R$ of the neutralino
  relic density when both thermal and 
  gravitino-induced non-thermal contributions
  are included (solid line) for the benchmark scenario of Table
  \ref{tab:amsbbch}. In addition, the 
  pure non-thermal contribution is indicated as a dashed line.}
\label{fig:cosmoamsb}
\end{figure}

The moduli contributions $\Omega_{\rm CDM}^{\rm mod} h^2$ to the neutralino relic
density can be estimated as \cite{Acharya:2009zt}
\be
  \Omega_{\rm CDM}^{\rm mod} h^2 \approx 0.1 
    \Big[ \frac{M_{\tilde{\chi}^0_1}}{100 \text{ GeV}} \Big]\ 
    \Big[ \frac{10.75}{g_{\rm eff}} \Big]^{1/4} \
    \Big[ \frac{3 \cdot 10^{-24} \text{ cm}^3 \text{s}^{-1}}{\langle
        \sigma_{\rm eff} v \rangle} \Big] \
    \Big[ \frac{100 \text{  TeV}}{M_\Phi} \Big]^{3/2}  \ .
\ee
The quantity $\Omega_{\rm CDM}^{\rm mod}$ depends on, in addition to the mass
of the lightest neutralino $M_{\tilde{\chi}_1^0}$, the mass of the moduli
fields $M_\Phi$, the effective number of degrees of freedom $g_{\rm eff}$ and the
thermally averaged annihilation cross section $\langle \sigma_{\rm eff} v
\rangle$. We present, in the left panel of Figure \ref{fig:cosmoamsb}, isolines
in the $(M_{\tilde{\chi}_1^0}, M_{\Phi})$
plane where the neutralino relic density $\Omega_{\rm CDM} h^2 = 0.1126$ (as in
Eq.~\eqref{eq:omh2}).  These results assume an effective number of degrees 
of freedom equal to 
$g_{\rm eff} = 10.75$ \cite{Moroi:1999zb} and each curve is associated with a 
different value for the
neutralino annihilation cross section $\langle \sigma_{\rm eff} v \rangle$.
For the benchmark of Table \ref{tab:amsbbch}, the neutralino mass is 175 GeV and
the computed neutralino relic density reads $\Omega_{\rm CDM}h^2 = 8.57 \times
10^{-4}$. This corresponds to a neutralino annihilation cross section of $\langle
\sigma_{\rm eff} v \rangle \approx 10^{-23}$ cm$^3$ s$^{-1}$. Consequently,
constructing a scenario with moduli
masses of order of $M_{\rm aux} = 60$ TeV
allows to recover the measured dark matter abundance of Eq.\ \eqref{eq:omh2}.

A second way to increase the neutralino relic abundance is to include
contributions from gravitino decays, 
\be
  \Omega_{\rm CDM}^{\rm non-therm} h^2 = \frac{M_{\tilde{\chi}^0_1}}{m_{3/2}}
    \Omega_{3/2} h^2  \ , 
\ee
where the gravitino thermal abundance $\Omega_{3/2}h^2$ is given as in Eq.\
\eqref{eq:thgrav}. The mass ratio factor illustrates that 
each gravitino decays into one single stable neutralino. The corresponding effects
are described on the right panel of Figure \ref{fig:cosmoamsb} for 
the scenario defined in Table \ref{tab:amsbbch} where we present the total
neutralino relic density calculated as a function of the reheating temperature. For
low values of $T_R$, thermal neutralino production dominates so that the total
relic density, including both thermal and non-thermal contributions, has a
roughly constant value of $\Omega_{\rm CDM}h^2 \approx 8.57 \times 10^{-4}$. When the
reheating temperature reaches $10^7$ GeV, gravitino decays become dominant
and the relic density grows linearly with $T_R$. Consequently, agreement
with the observations corresponds to a reheating temperature of $T_R \approx
10^{10}$~GeV, a value in addition compatible with thermal leptogenesis
\cite{Buchmuller:2004nz}.

\mysection{Direct constraints}\label{sec:direct}
%
\begin{figure}[t!]
 \centering
 \includegraphics[width=.32\columnwidth]{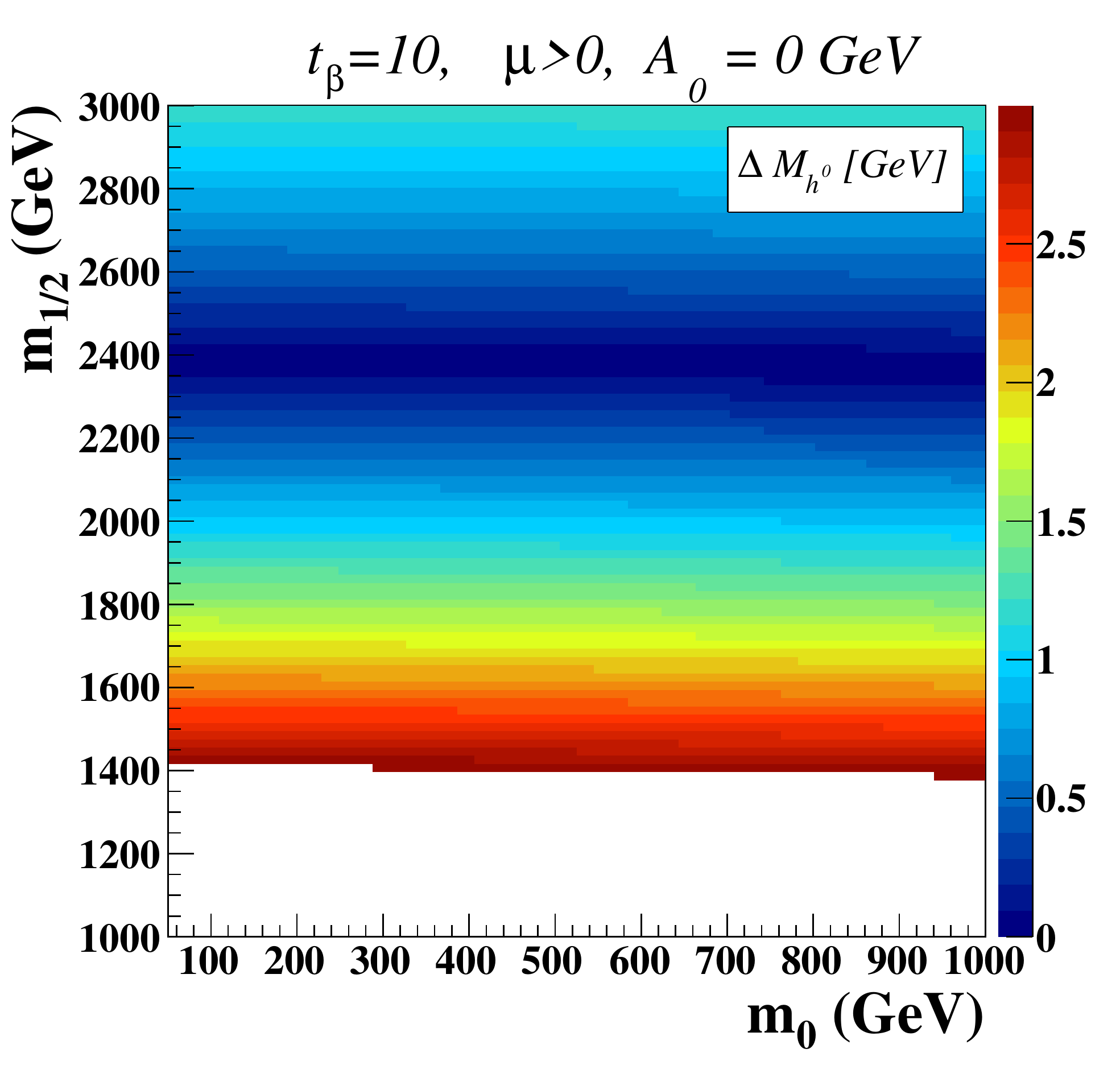}
 \includegraphics[width=.32\columnwidth]{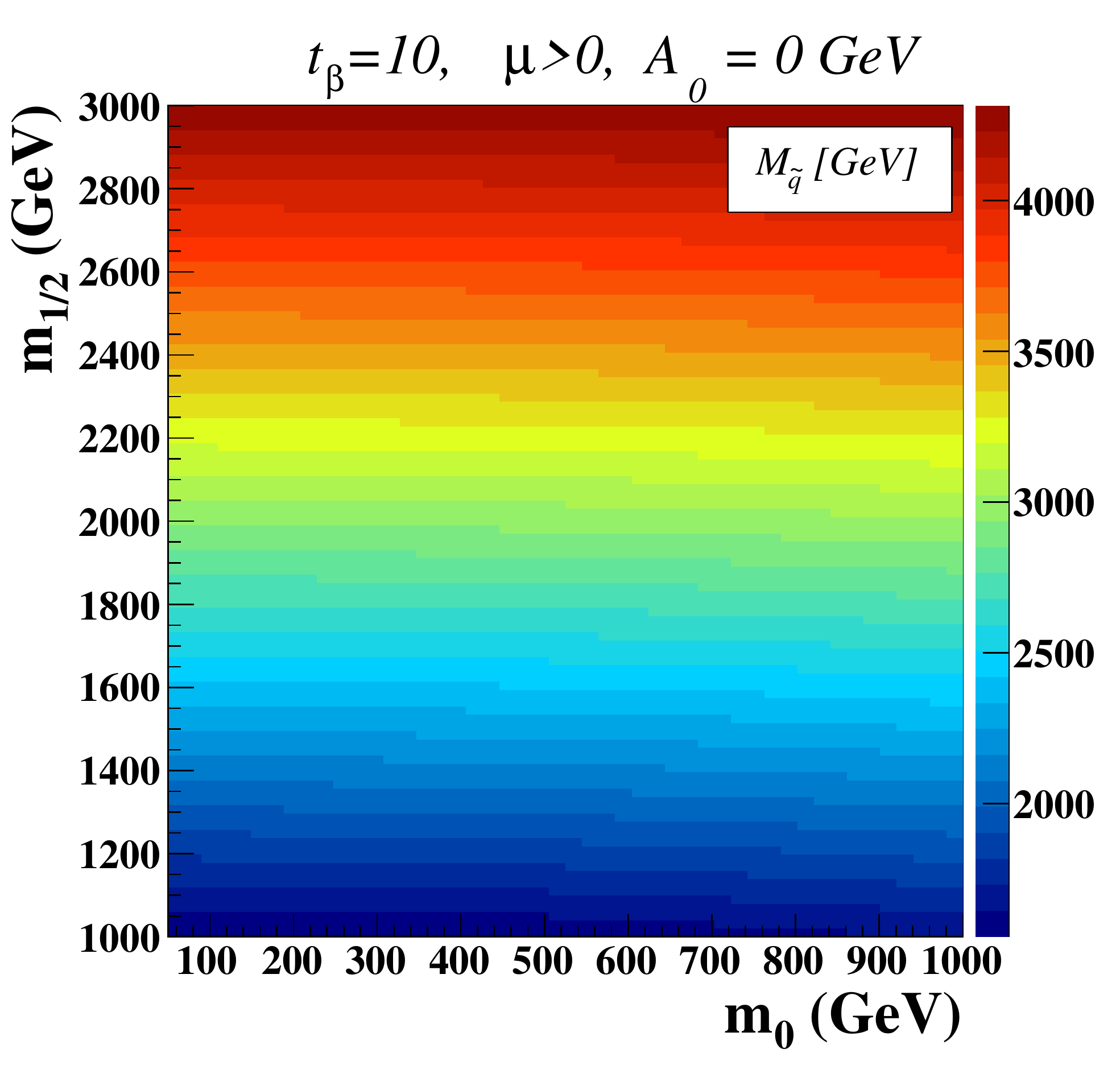}
 \includegraphics[width=.32\columnwidth]{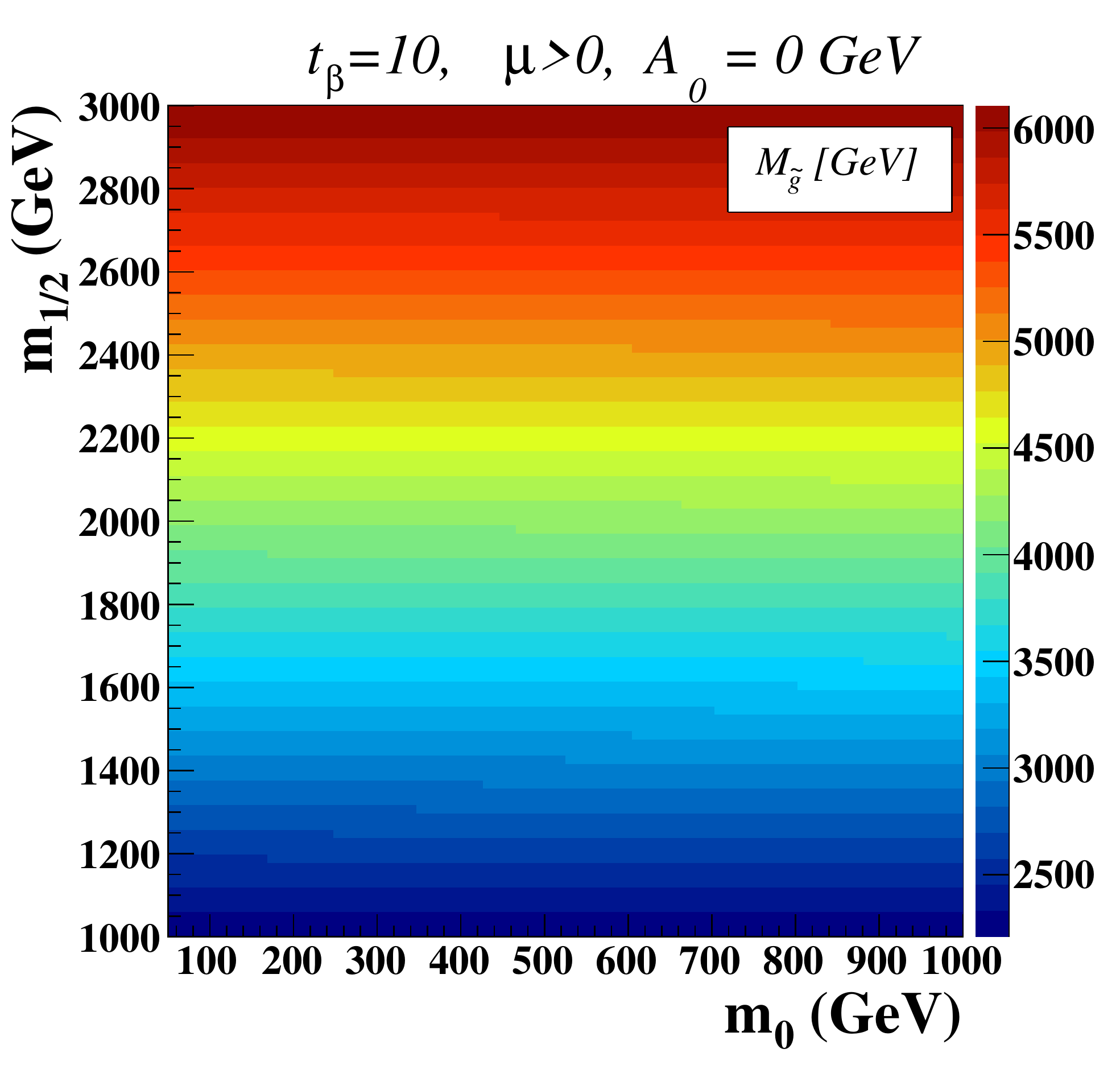}
\vspace{.3cm}
 \includegraphics[width=.32\columnwidth]{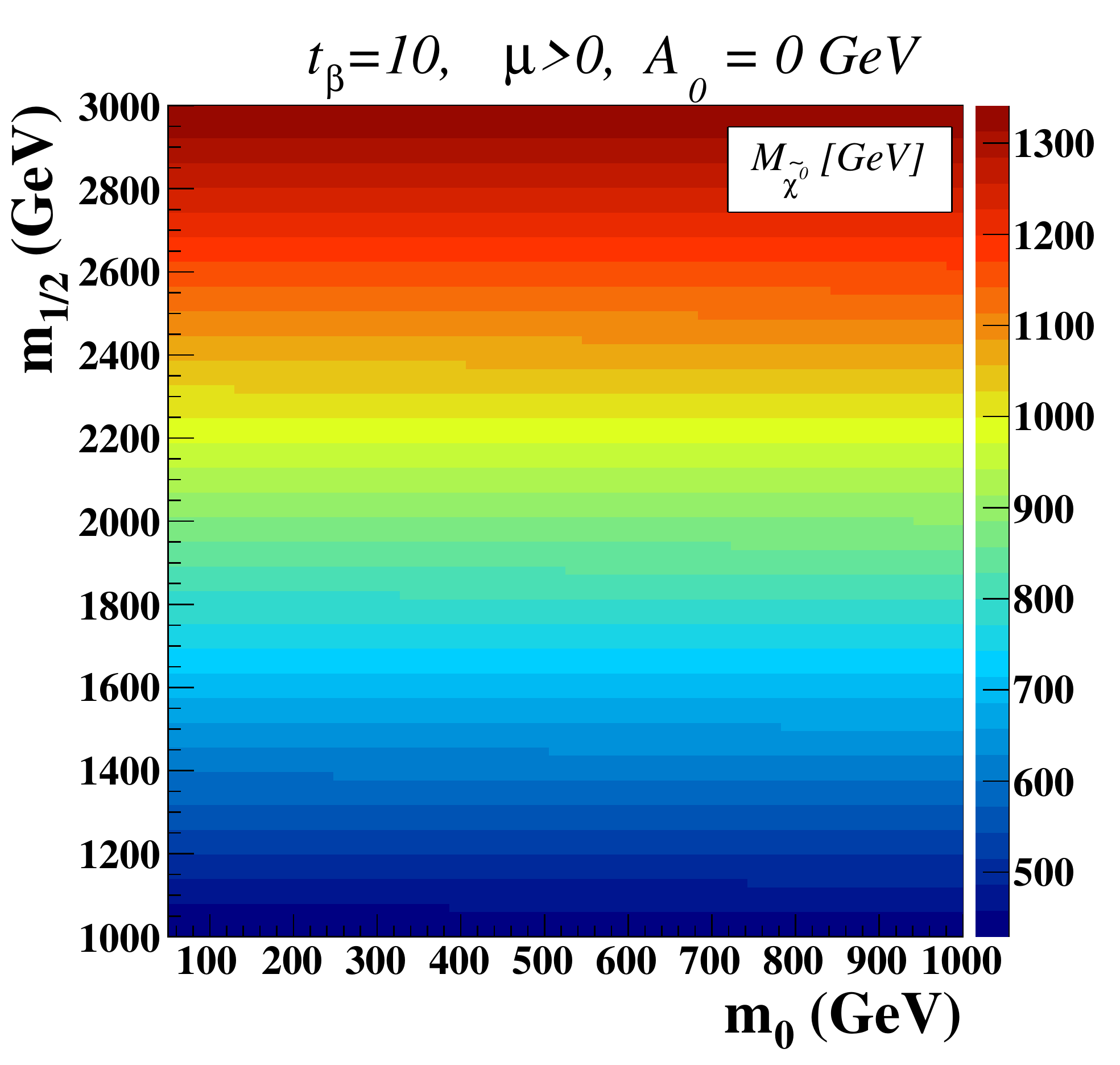}
 \includegraphics[width=.32\columnwidth]{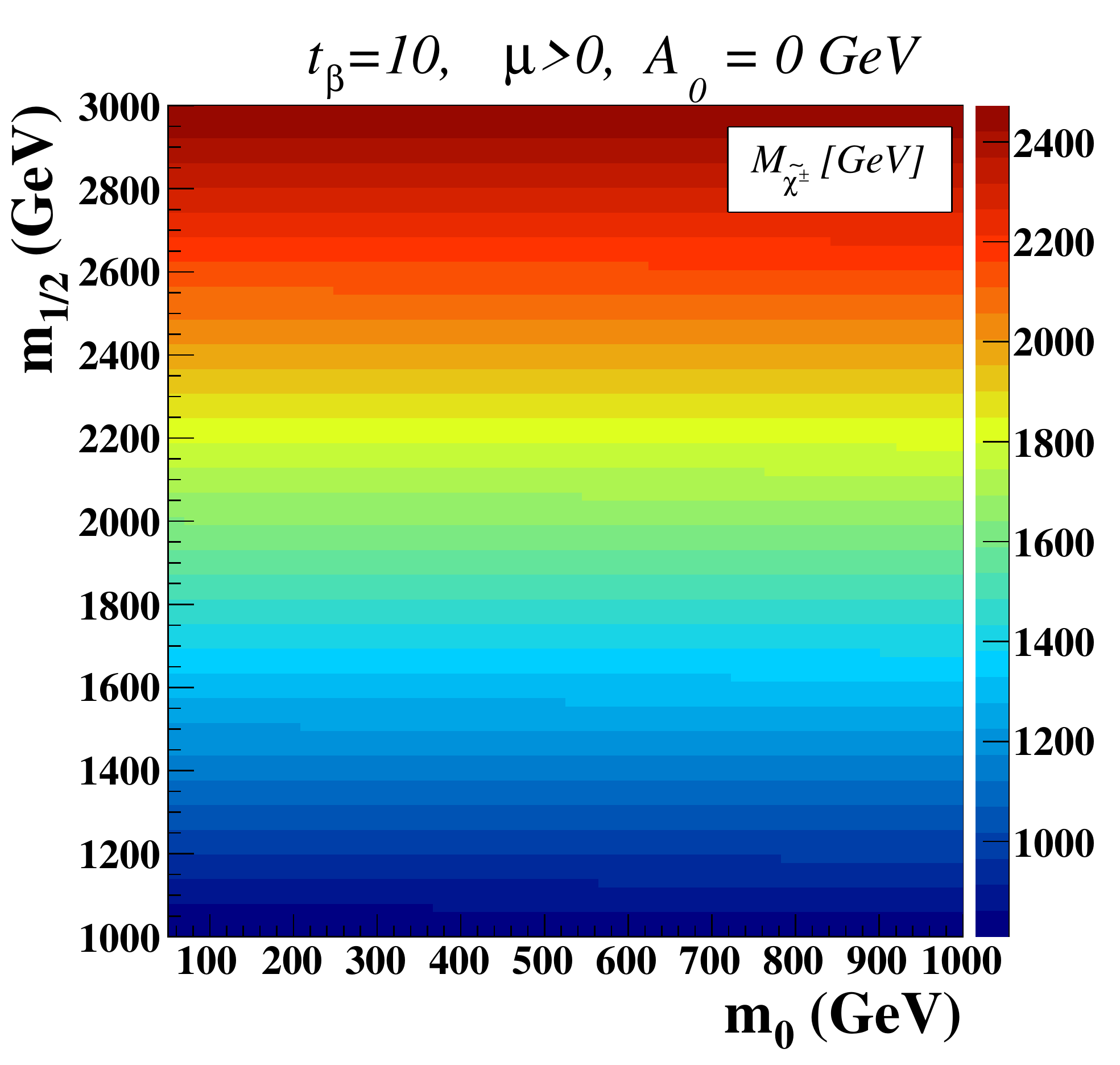}
 \includegraphics[width=.32\columnwidth]{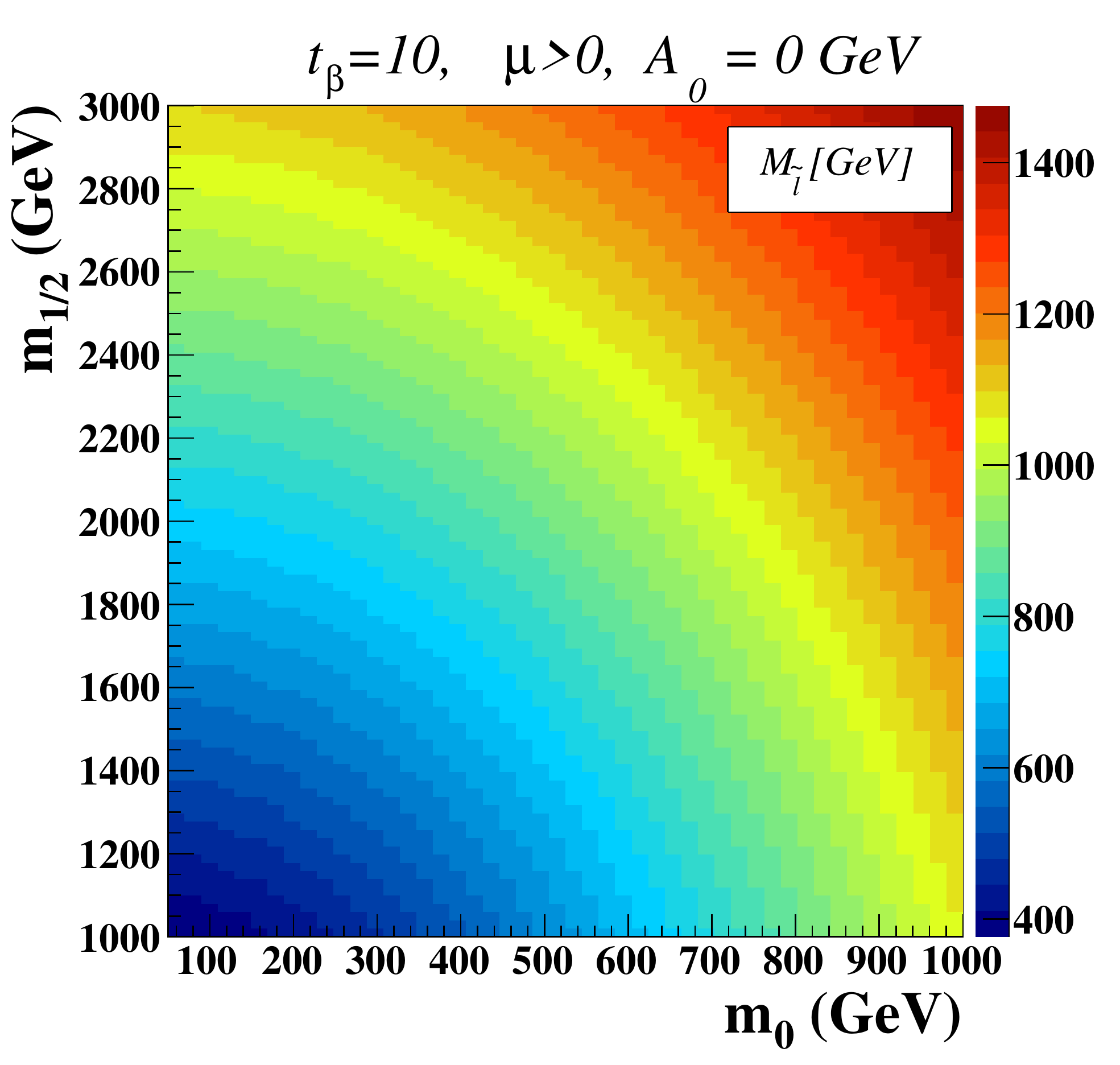}
 \caption{\label{fig:cmssm10_mass} Predictions for the lightest Higgs boson mass 
  (left panel, upper row) shown as deviations from the central measured value
  given in Eq.\ \eqref{eq:mhlhc}. We present the results in
  $(m_0,m_{1/2})$-planes of the cMSSM with $\tan\beta=10$, $A_0=0$ 
  GeV and a positive Higgs mixing parameter $\mu>0$. The predictions for the
  masses of the lightest
  squark (middle panel, upper row), the gluino (right panel, upper row), the lightest
  neutralino (left panel, lower row), the lightest chargino (middle panel, lower
  row) and the lightest slepton/sneutrino (right panel, lower row) are also
  depicted. In all figures, only the regions of the parameter space compatible
with the bounds of Eq.\ \eqref{eq:mhlhc} are indicated (1 TeV $< m_{1/2} <$ 3 TeV).}
\end{figure}
%
In the light of the latest experimental results, it is necessary to
account for results of the direct searches of the Higgs boson
\cite{Aad:2012gk, Chatrchyan:2012gu} when designing experimentally non-excluded 
scenarios. We therefore require 
\be
  M_{h^0} \approx 126\pm 3 \text{ GeV} \ ,
\label{eq:mhlhc}\ee
when scanning the MSSM parameter space. The central value is
obtained from the average of the values reported by the two experimental
collaborations and the bounds account for parametric uncertainties of the Standard Model
inputs \cite{Arbey:2012dq}. In Figure \ref{fig:cmssm10_mass} and Figure
\ref{fig:cmssm40_mass}, we scan over the two considered $(m_0, m_{1/2})$ planes 
with $\tan\beta=10$, $A_0=0$~GeV and $\tan\beta~=~40$,
$A_0=-500$ GeV, respectively, and present, in the left panel of the upper row of the
figures, predictions for the lightest Higgs boson mass shown as deviations from
the central experimental value of 126 GeV.  The theoretical
computations are performed by means of the \spheno\ package, version 3.1.2,
which includes one-loop and two-loop contributions to the  
Higgs mass matrices \cite{Chankowski:1992er, Dabelstein:1994hb,
Degrassi:2001yf, Brignole:2001jy, Brignole:2002bz, Dedes:2002dy, Dedes:2003km}.

Confronting the Higgs mass predictions to data, it turns out that viable 
cMSSM scenarios prefer large values of the universal gaugino mass $m_{1/2}
\gtrsim 1000$ GeV, while the universal scalar mass $m_0$ is 
left unconstrained. This has strong consequences on the superpartner
spectrum, the predictions for their masses being presented  
in the other panels of Figure~\ref{fig:cmssm10_mass} and
Figure~\ref{fig:cmssm40_mass} for the regions compatible with
Eq.\ \eqref{eq:mhlhc}.

%
\begin{figure}[t!]
 \centering
 \includegraphics[width=.32\columnwidth]{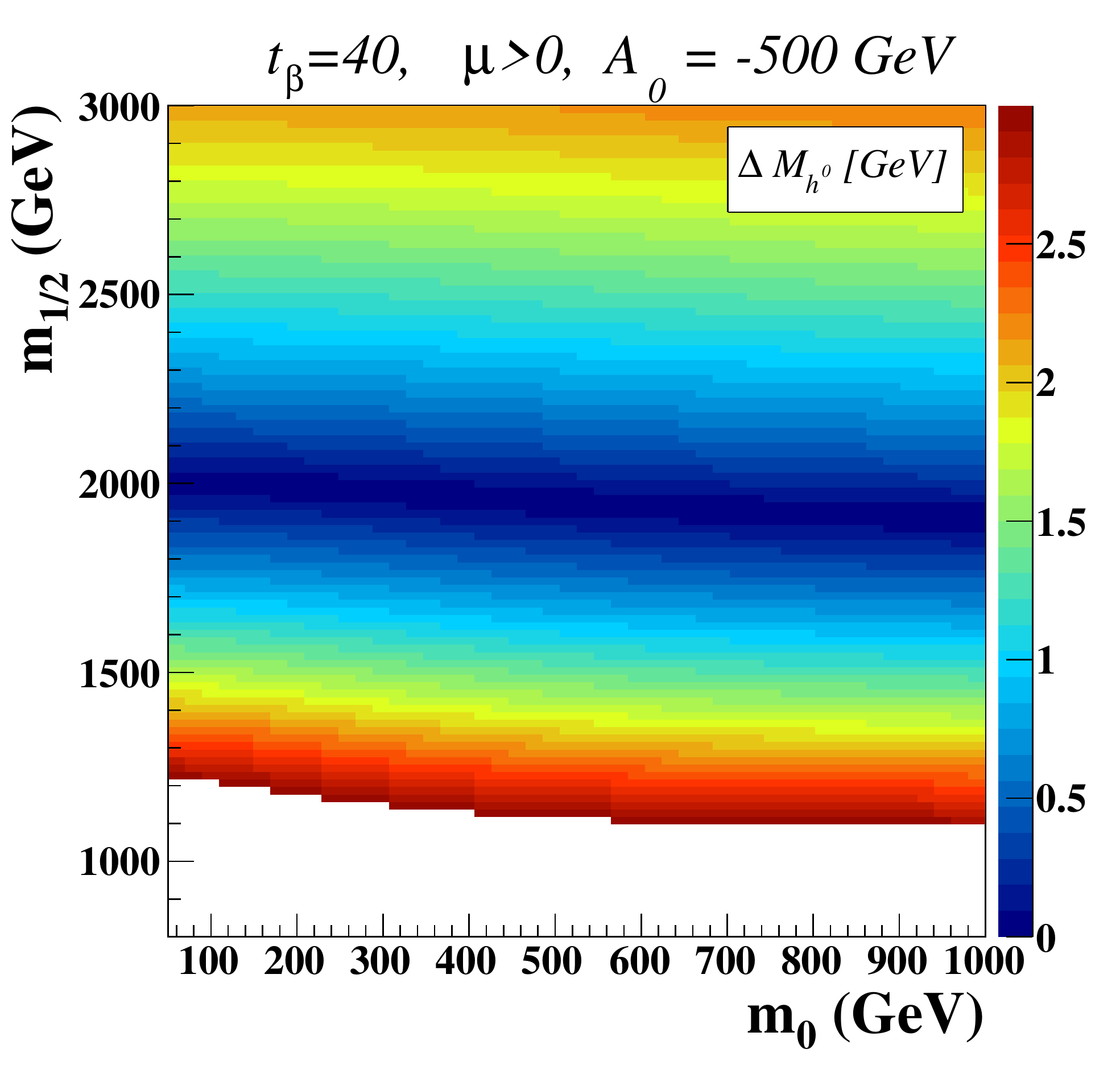}
 \includegraphics[width=.32\columnwidth]{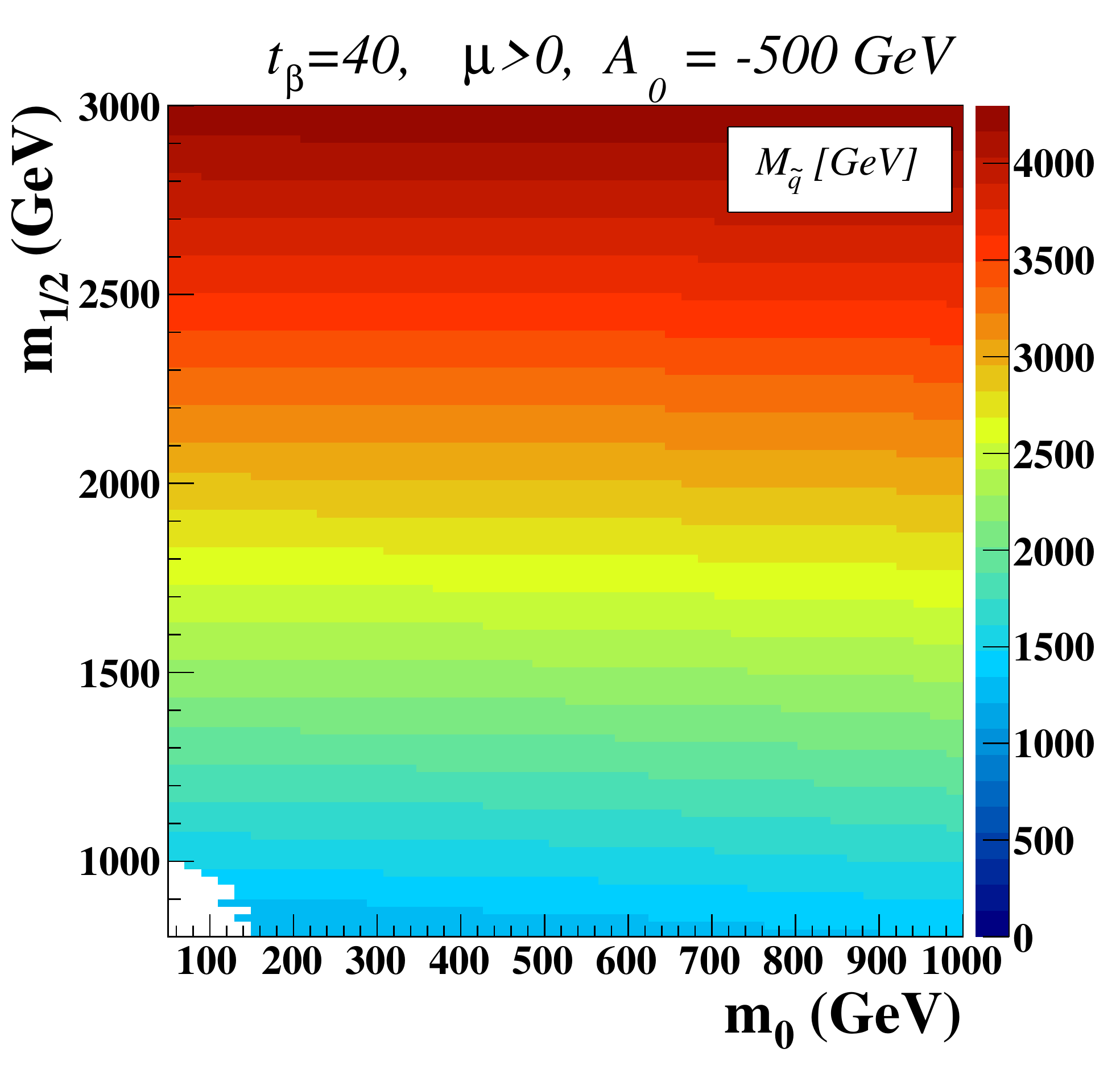}
 \includegraphics[width=.32\columnwidth]{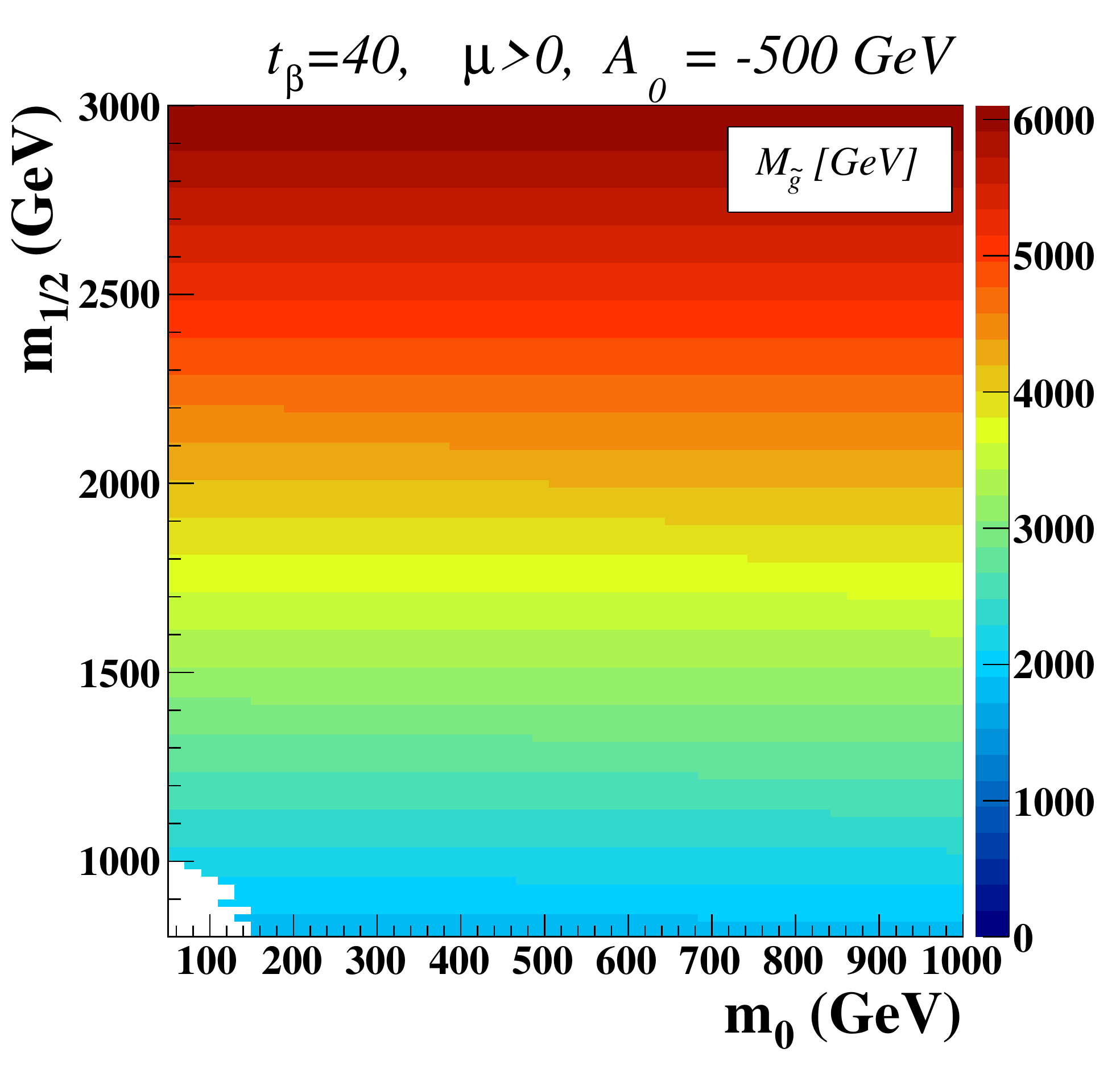}
\vspace{.3cm}
 \includegraphics[width=.32\columnwidth]{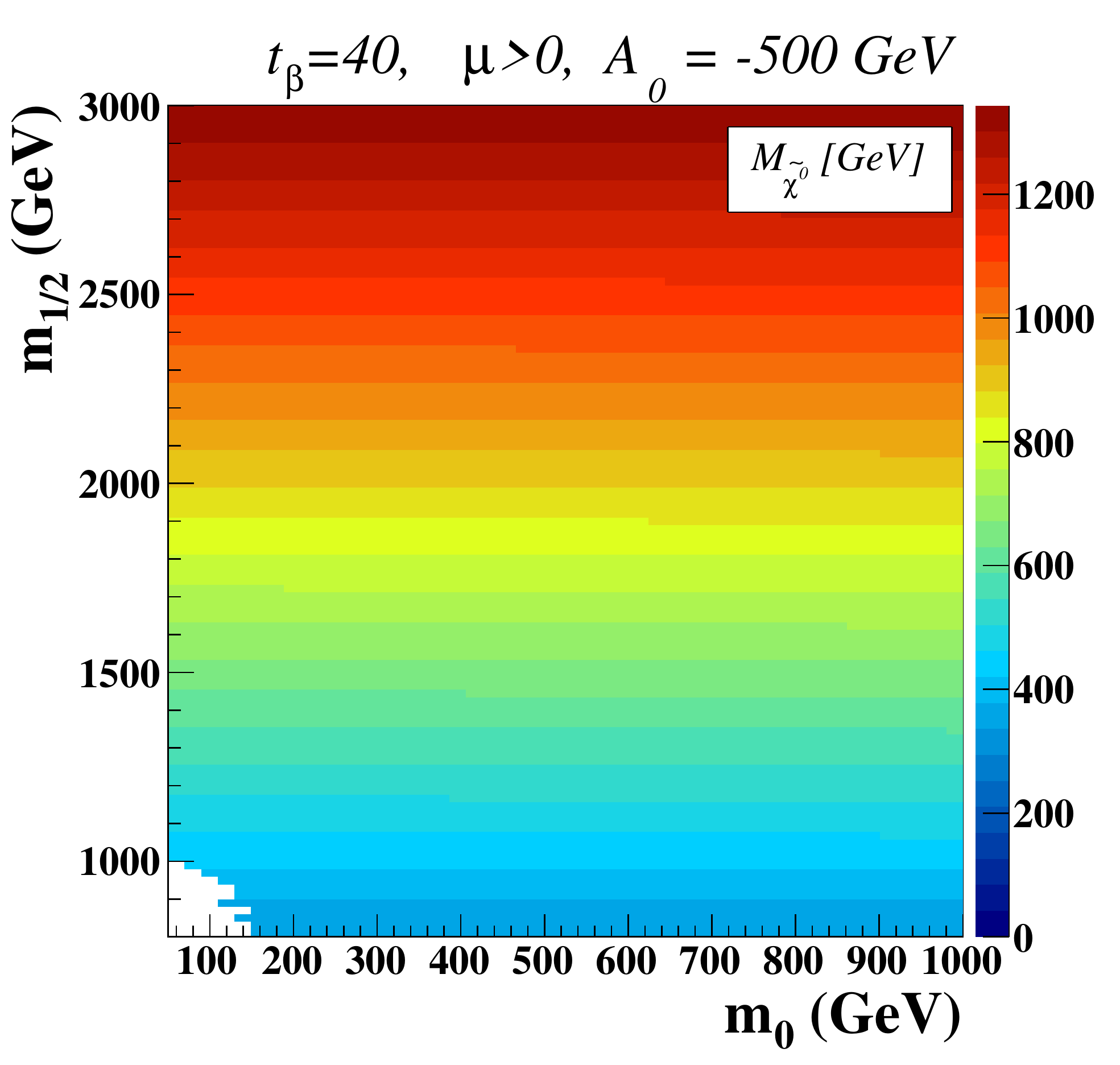}
 \includegraphics[width=.32\columnwidth]{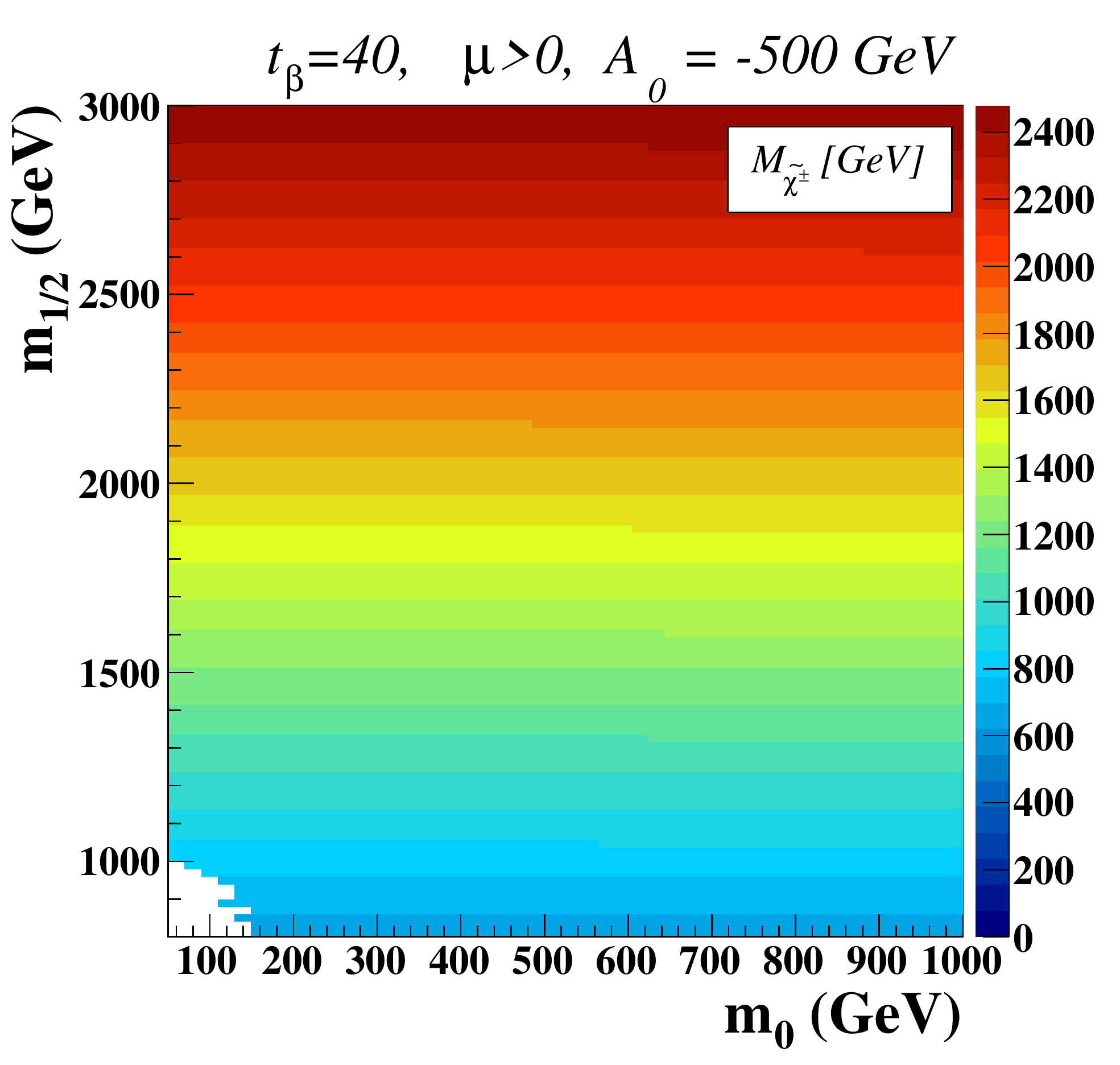}
 \includegraphics[width=.32\columnwidth]{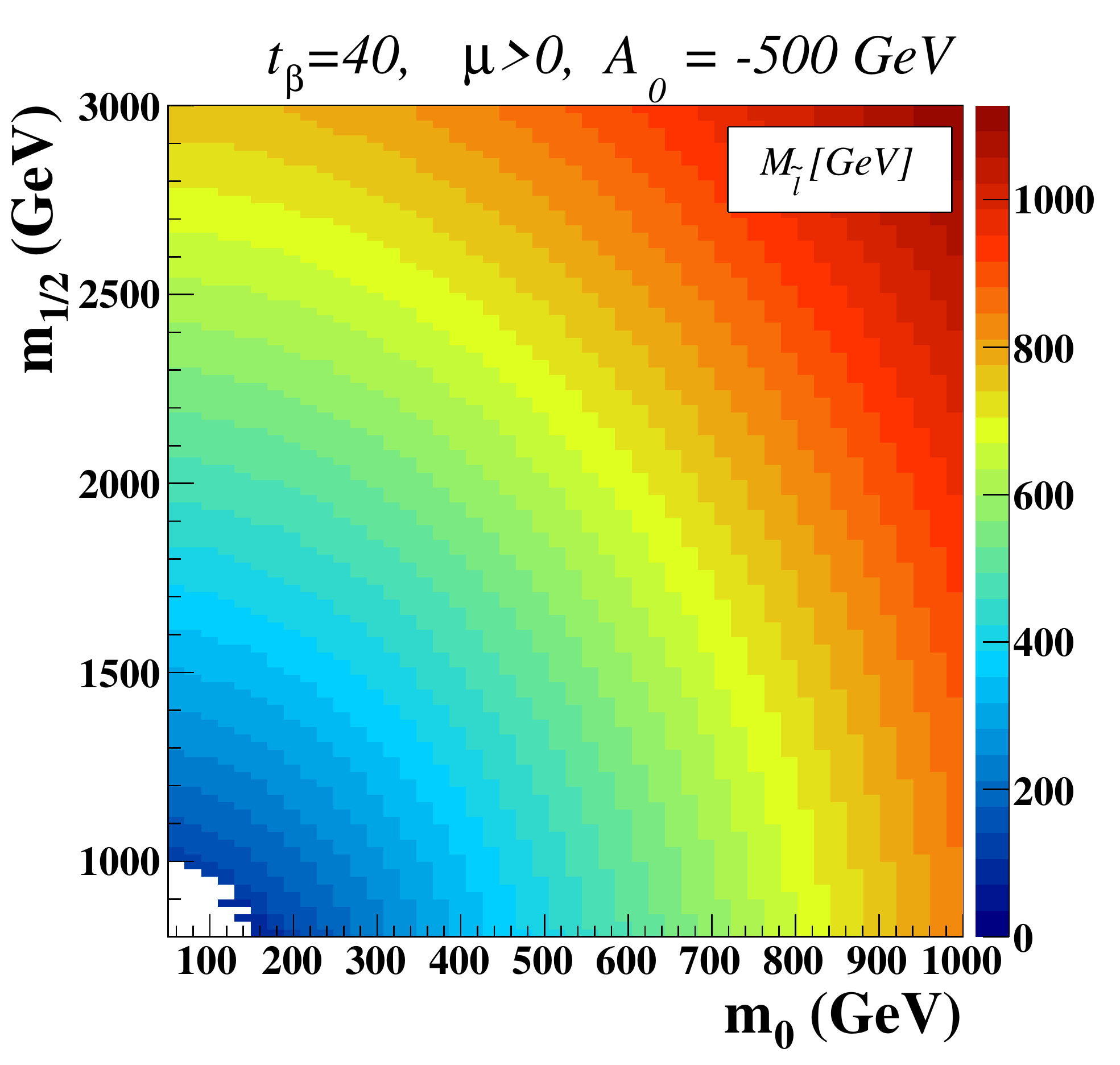}
 \caption{\label{fig:cmssm40_mass} Same as Figure \ref{fig:cmssm10_mass}, but
  for $\tan\beta=40$, $A_0=-500$ GeV and a positive Higgs mixing
  parameter $\mu>0$.}
\end{figure}
%

In the middle panel of the upper row of the figures, we show
the mass of the lightest of the squark
eigenstates, \ie, the lightest stop\footnote{First and second
generation squark masses lie above 1 TeV in the scanned regions
of interest. This agrees with LHC limits on these particles.}.
We observe that the regions of the
scanned parameter space which accommodate 
$M_{h^0} \approx 126$ GeV also include heavy squarks of masses larger than 
1 TeV. In this way, the tree-level Higgs mass of Eq.\ \eqref{eq:hcst} is 
sufficiently shifted so that predictions agree with data. This however in general
increases at the same time the amount of necessary fine-tuning as shown, \eg, by inspecting
the logarithmic terms of Eq.~\eqref{eq:scaltad}. In the right panel of the figures,
we show that these regions also exhibit heavy gluinos with masses above 2 TeV.
Moreover, including non-minimal flavor violation as 
in Eq.\ \eqref{eq:lambda} with $\lambda$-parameters equal to 0.15 does not
induce enough mass splitting to drastically change
the results. Although not considered in the examples studied in this work,
lighter stop masses can still be viable to accomodate a correct Higgs boson
mass. This specific setup requires an increased mixing among the two stop
superpartners, or equivalently larger values of the $A_0$ parameter,
and offers hence an alternative way for predicting a Standard Model-like
Higgs boson with a mass of about 126~GeV and with a reduced amount of fine-tuning.
We refer, \eg, to Refs.~\cite{Brummer:2012ns,Wymant:2012zp} for more information.

On the second row of the figures (left and middle panels), we present 
predictions for the masses of the lightest neutralino and chargino in the
regions accommodating a correct Higgs mass. These particles are 
expected to be lighter than squarks and gluino since the dominant
effects driving the evolution
of their masses with the energy are insensitive to the
$SU(3)_c$ gauge interactions. The predicted masses are found to be above several
hundreds of GeV for the lightest neutralino and above a TeV for the lightest 
chargino. Since the second lightest neutralino is mostly a wino state, as the lightest 
chargino, the range of its mass is can be inferred from the second panel of the 
second line of the figures. Finally,
sleptons are in general the only superpartner to be allowed to be light, with
masses of 100-200 GeV for large values of $\tan\beta$ being still possible.
However, their possible discovery at the LHC is much more complicated. Either 
they can be observed through the cascade decays of strongly produced superparticles,
or they are produced directly through the Drell-Yan mechanism. On the one hand,
the first option is unlikely due to the heavy squark and gluino masses
and the low cross sections~\cite{Beenakker:1996ch,Beenakker:1997ut,
Berger:1998kh,Berger:1999mc,Berger:2000iu,
Spira:2002rd,Jin:2002nu,Jin:2003ez,Hollik:2007wf,
Hollik:2008vm,Mirabella:2009ap, Kulesza:2008jb,Kulesza:2009kq,
Beenakker:2009ha, Beenakker:2010nq,Beenakker:2011fu, Kramer:2012bx}. On the
other hand, direct production based searches rely
on electroweak production processes~\cite{Baer:1997nh,
Beenakker:1999xh,Bozzi:2004qq, 
Bozzi:2006fw, Bozzi:2007qr, Bozzi:2007tea,Fuks:2012qx,Fuks:2013lya},
whose cross sections are often found to be
three or four orders of magnitude lower than those of the dominant $WW$ or 
$t\bar t$ backgrounds rendering the searches more challenging.

%
\begin{figure}[t!]
 \centering
 \includegraphics[width=.32\columnwidth]{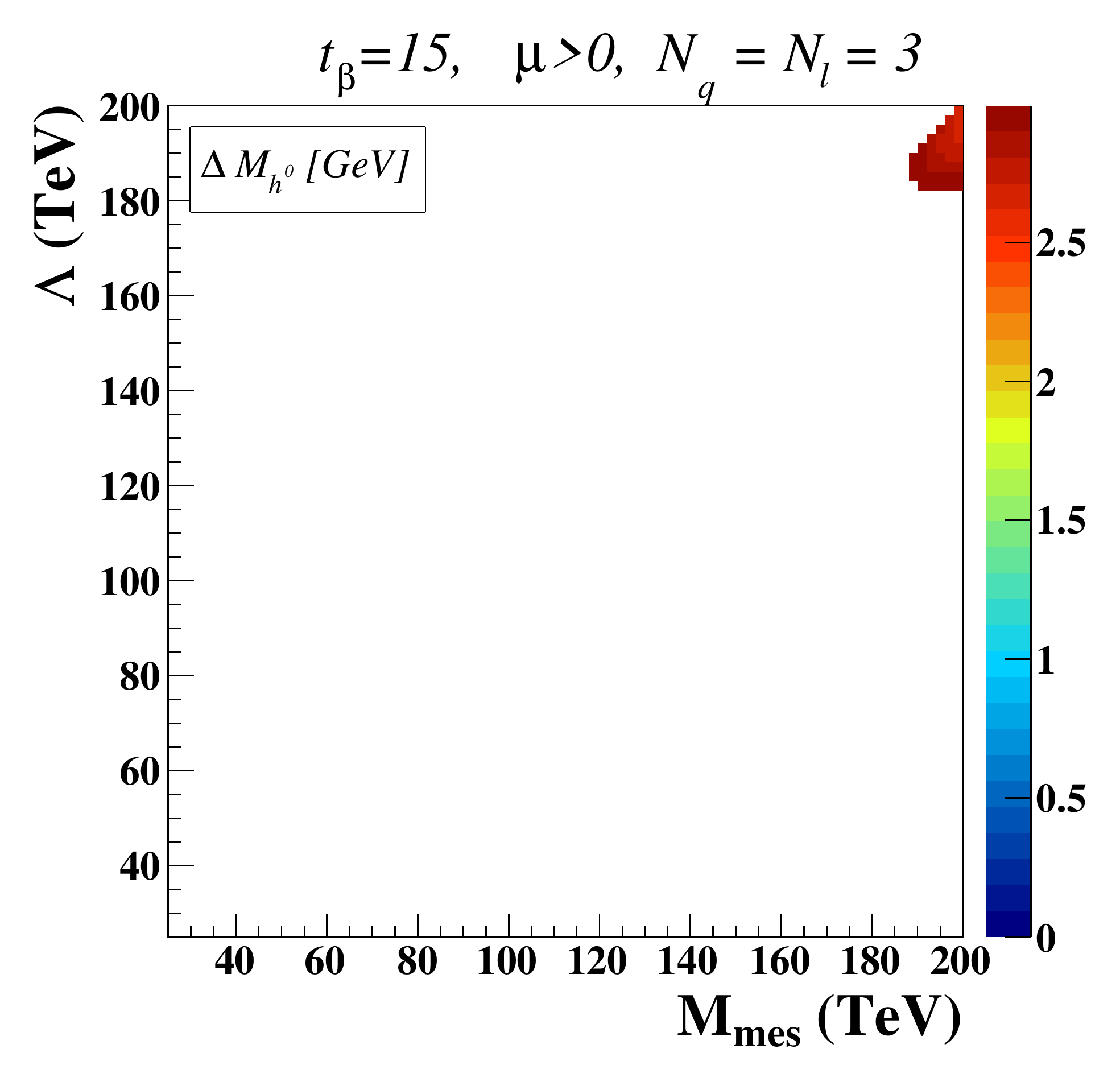}
 \includegraphics[width=.32\columnwidth]{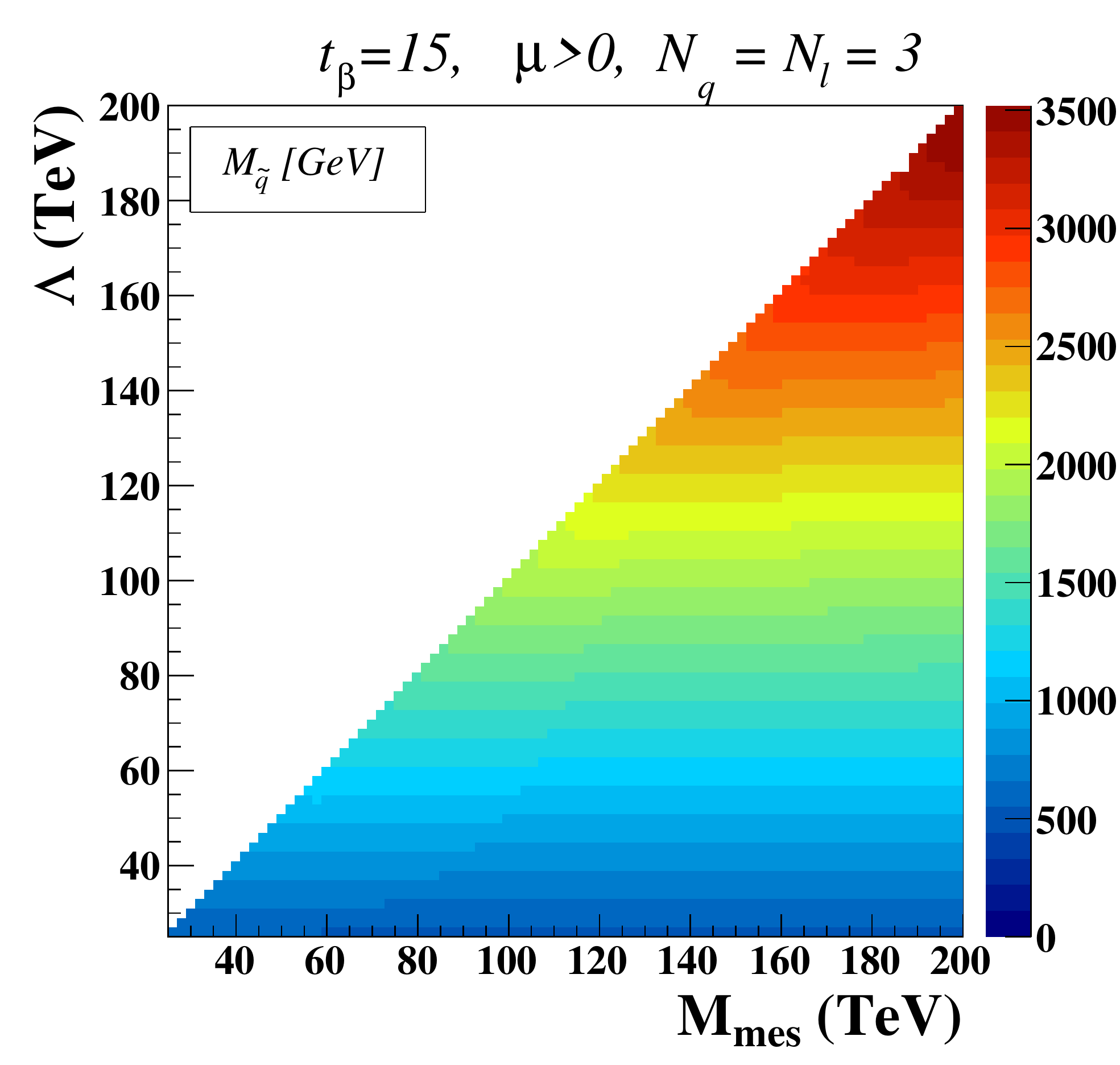}
 \includegraphics[width=.32\columnwidth]{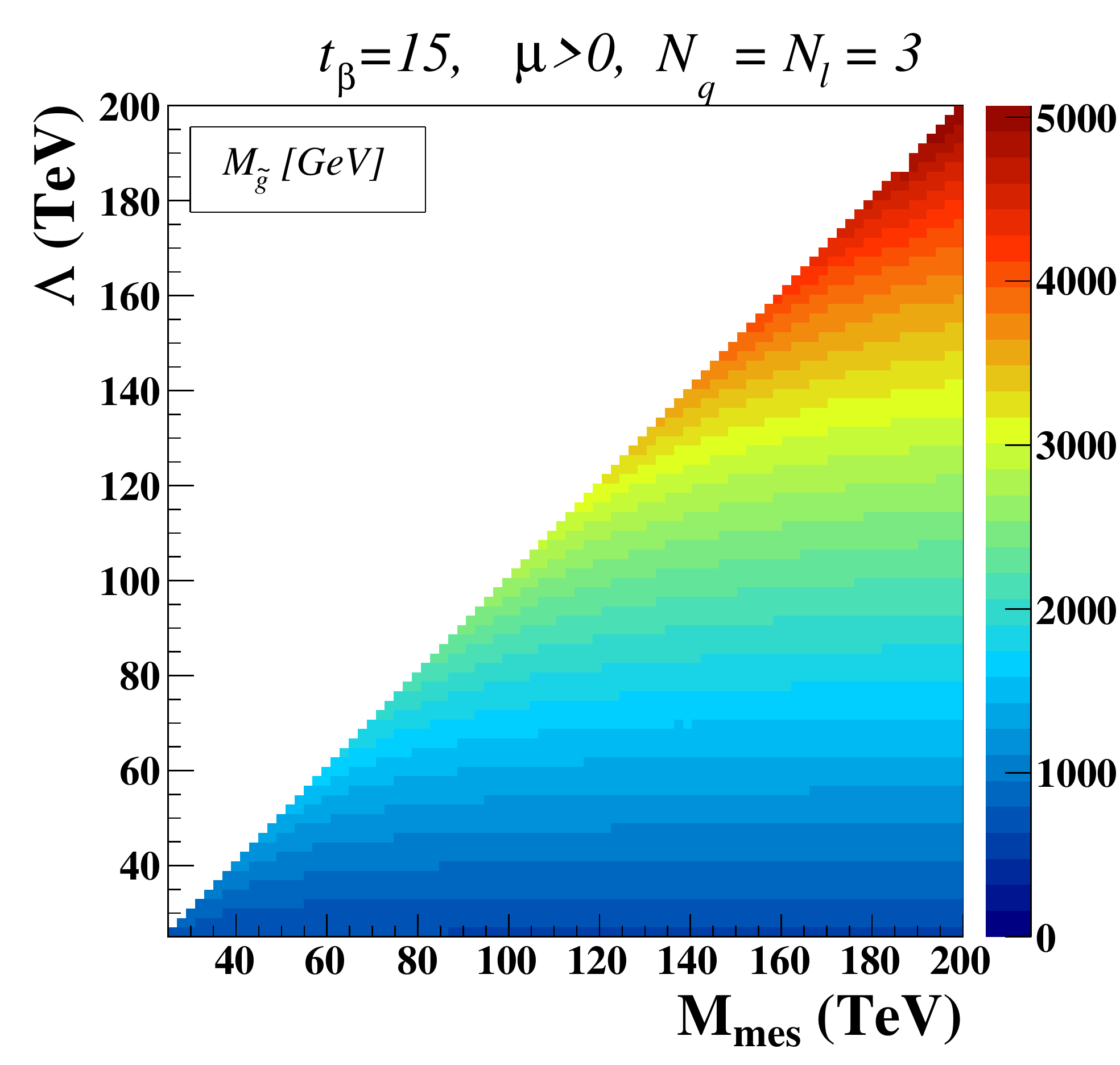}
\vspace{.3cm}
 \includegraphics[width=.32\columnwidth]{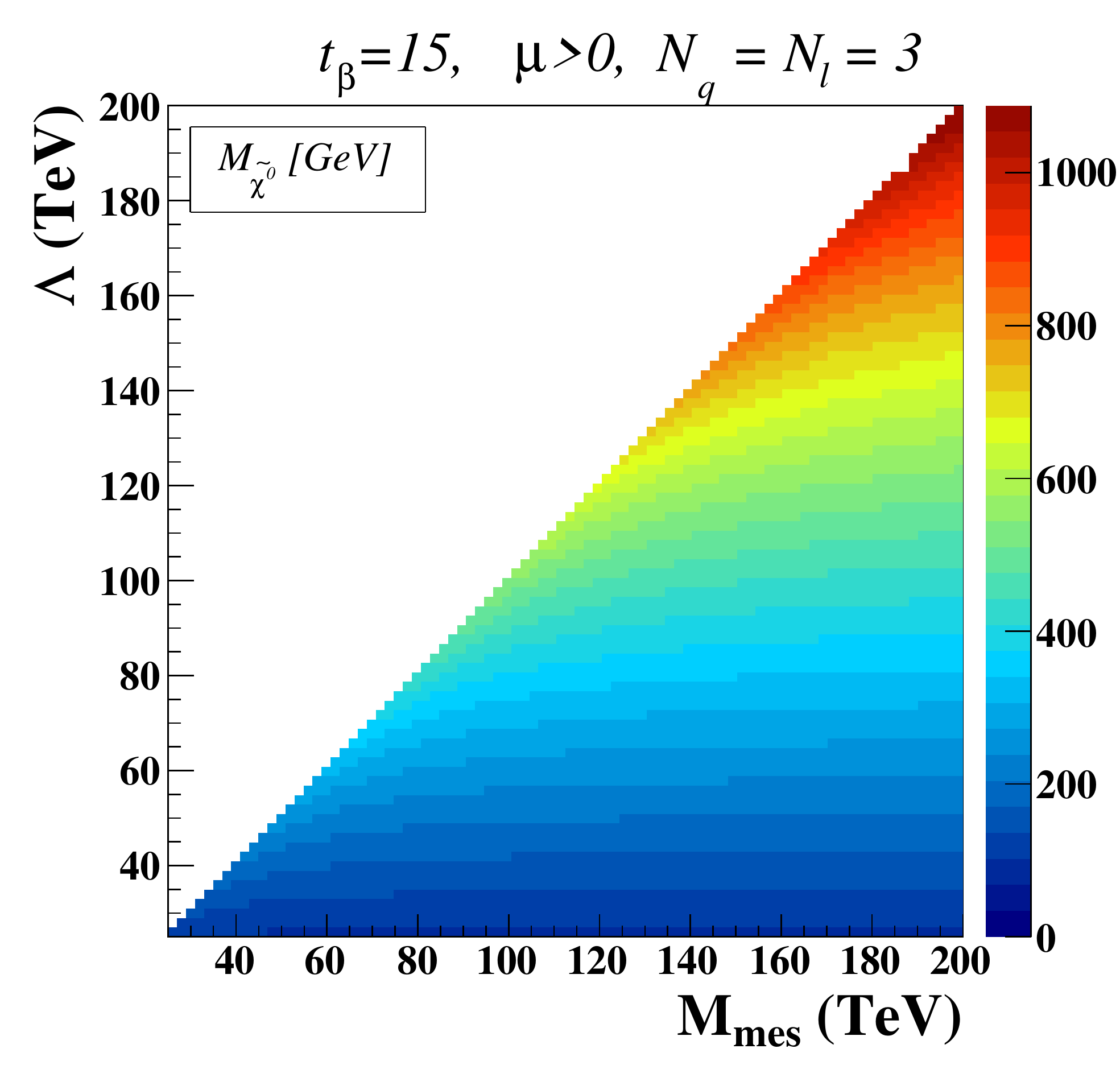}
 \includegraphics[width=.32\columnwidth]{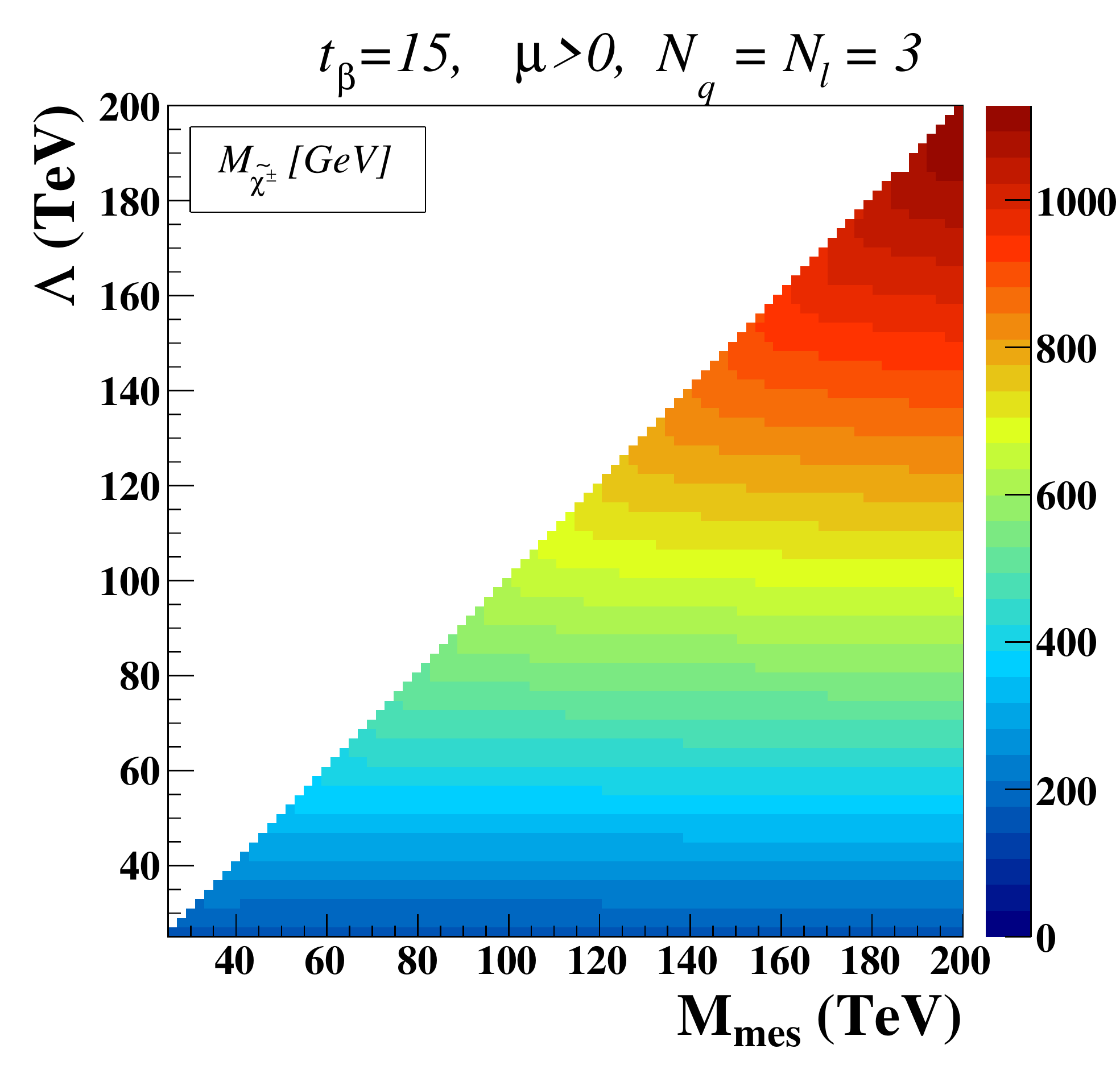}
 \includegraphics[width=.32\columnwidth]{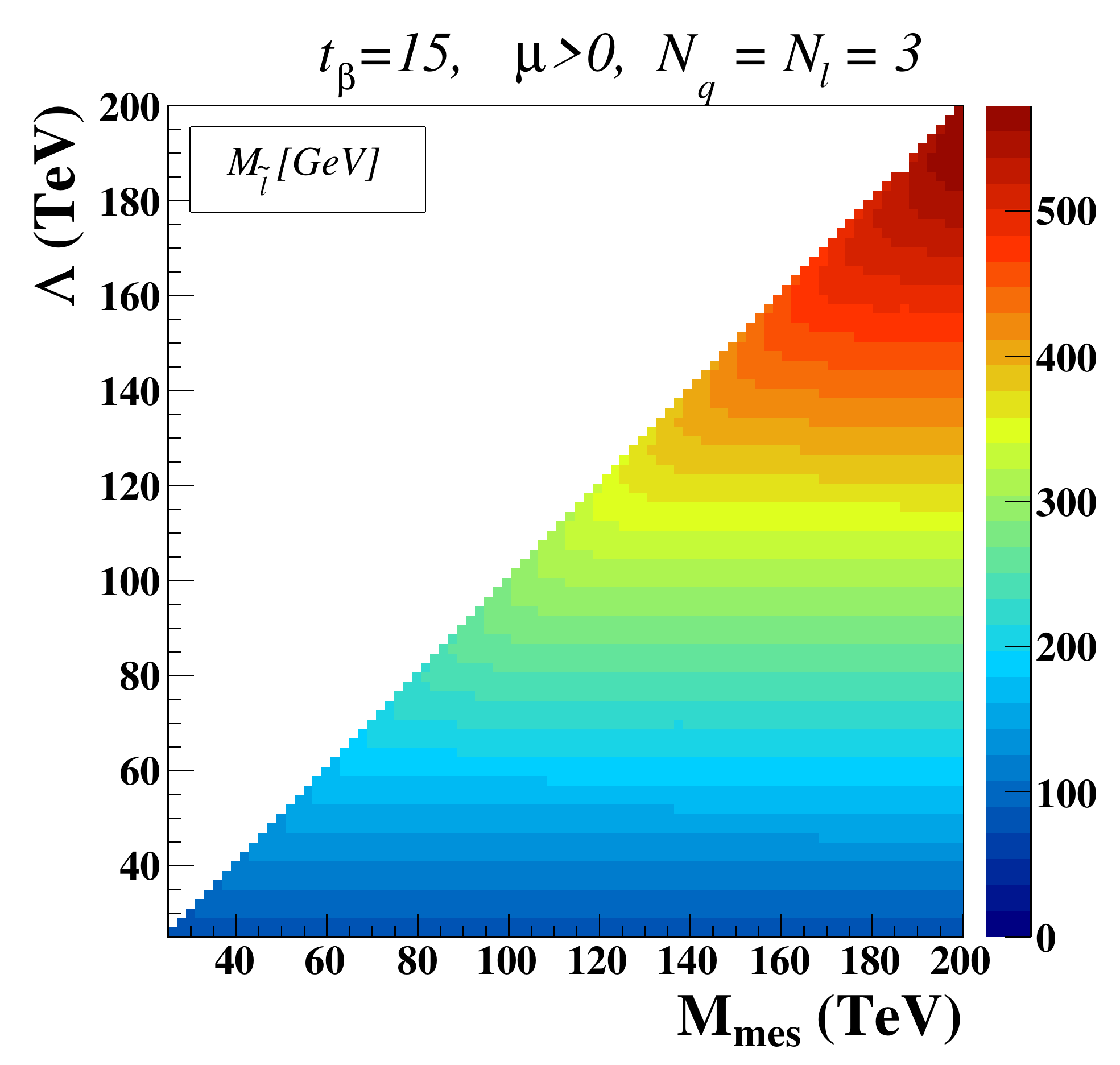}
 \caption{\label{fig:gmsb3_mass}  Same as in Figure \ref{fig:cmssm10_mass} but
for the MSSM with gauge-mediated supersymmetry-breaking. We present $(M_{\rm
mes},\Lambda)$ planes with $\tan\beta=15$,
$N_q = N_\ell =3$ messenger fields and a positive Higgs
mixings parameter $\mu>0$.}
\end{figure}
%

Two remarks are in order here. First, all these predictions are 
compatible with the current experimental bounds induced by supersymmetry searches
at the LHC experiments. 
With about 5 fb$^{-1}$ and 10 fb$^{-1}$ of data at a center-of-mass energy of 7
and 8 TeV, respectively, both the
ATLAS and CMS collaborations have been able to set strong limits on low-scale
supersymmetry. In particular, supersymmetry searches based on simplified
models (often inspired by the cMSSM) and on the cMSSM itself 
stringently constrain the colored
superpartner mass scale. Gluino and first and second generation squarks are
currently pushed above 1 or 2~TeV~\cite{Aad:2012hm,
Aad:2012pq, Aad:2012rz, Aad:2012cj, ATLAS145, ATLAS151, Chatrchyan:2012jx,
Chatrchyan:2012mfa, Chatrchyan:2012wa, Chatrchyan:2012pc,  CMSSUS028,
CMSSUS029}, while limits on third generation squark masses extend to about 500
GeV \cite{Chatrchyan:2012wa, CMSSUS028, CMSSUS029, Aad:2012gg, Aad:2012si, Aad:2012ar,
Aad:2012uu, ATLAS151, CMSSUS023}. These negative search results
have made the experimental attention shift
towards the weak production channels, so that limits on
chargino, neutralino and slepton masses have been improved 
to about 200-500 GeV \cite{Chatrchyan:2012mfa, Chatrchyan:2012pc, Aad:2012si, 
Chatrchyan:2012ewa, CMSSUS022, Aad:2012ku, ATLAS154}.
Those limits are however not general and hold in very specific
cases. Care should be taken when reinterpreted in different contexts such as a complete model,
for instance.

The direct measurements are compatible with results extracted from  
flavor physics and electroweak precision data (see Section \ref{sec:raredec}, 
Section \ref{sec:osc} and Section \ref{sec:drho}), heavy
superpartners being both favored by Higgs mass measurements and 
rare $B$-meson decays, neutral $B$-meson
oscillations and electroweak precision observable data. 
However, scenarios with too heavy superparticle masses forbid supersymmetry
to explain the gap between Standard Model predictions and measurements of the anomalous
magnetic moment of the muon (see Section~\ref{sec:gm2}).
Relaxing the
$a_\mu$ constraints, viable benchmark
scenarios can however be built, but are very collider-unfriendly
with superpartners above the TeV range.

%
\begin{figure}[t!]
 \centering
 \includegraphics[width=.32\columnwidth]{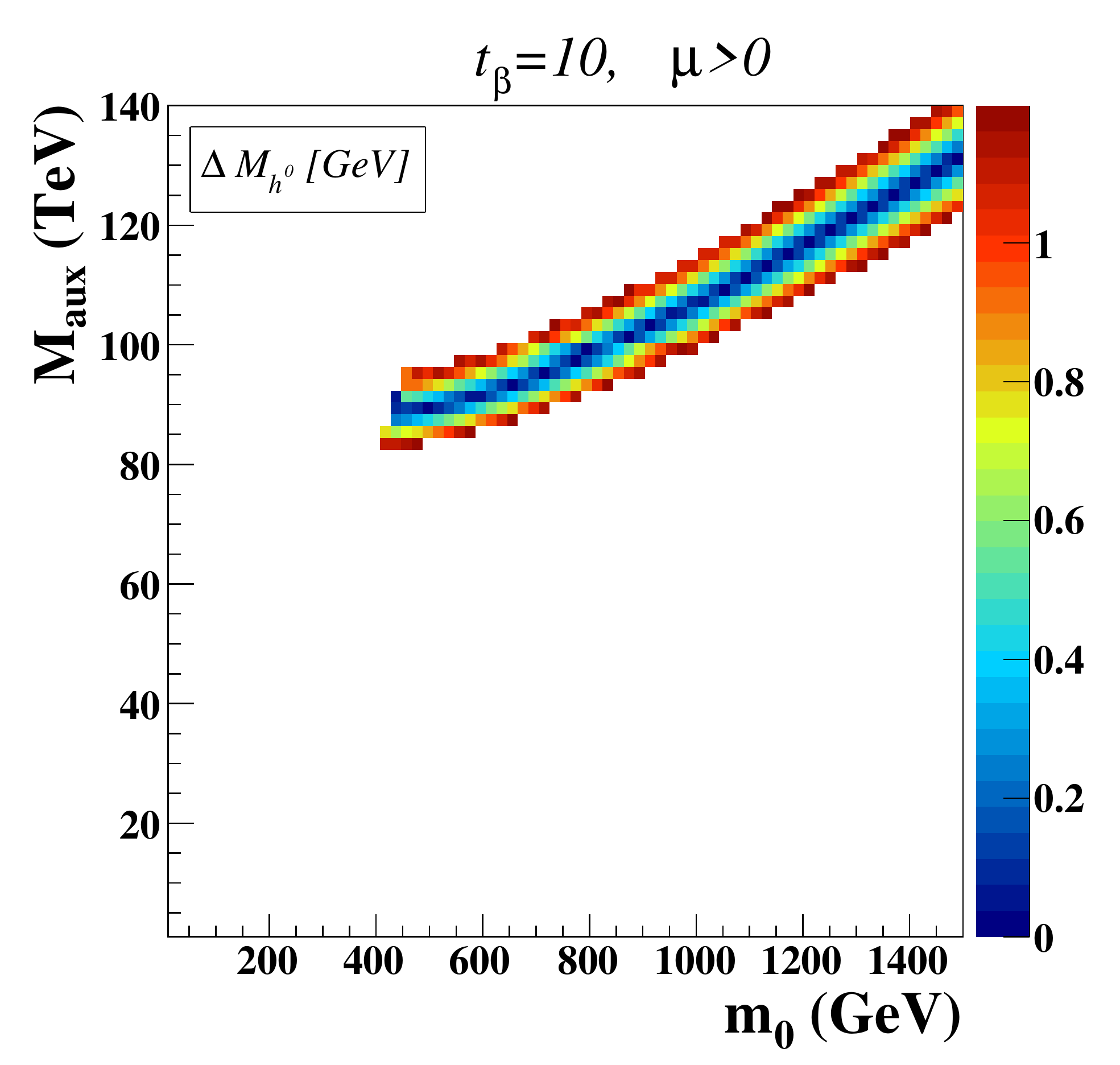}
 \includegraphics[width=.32\columnwidth]{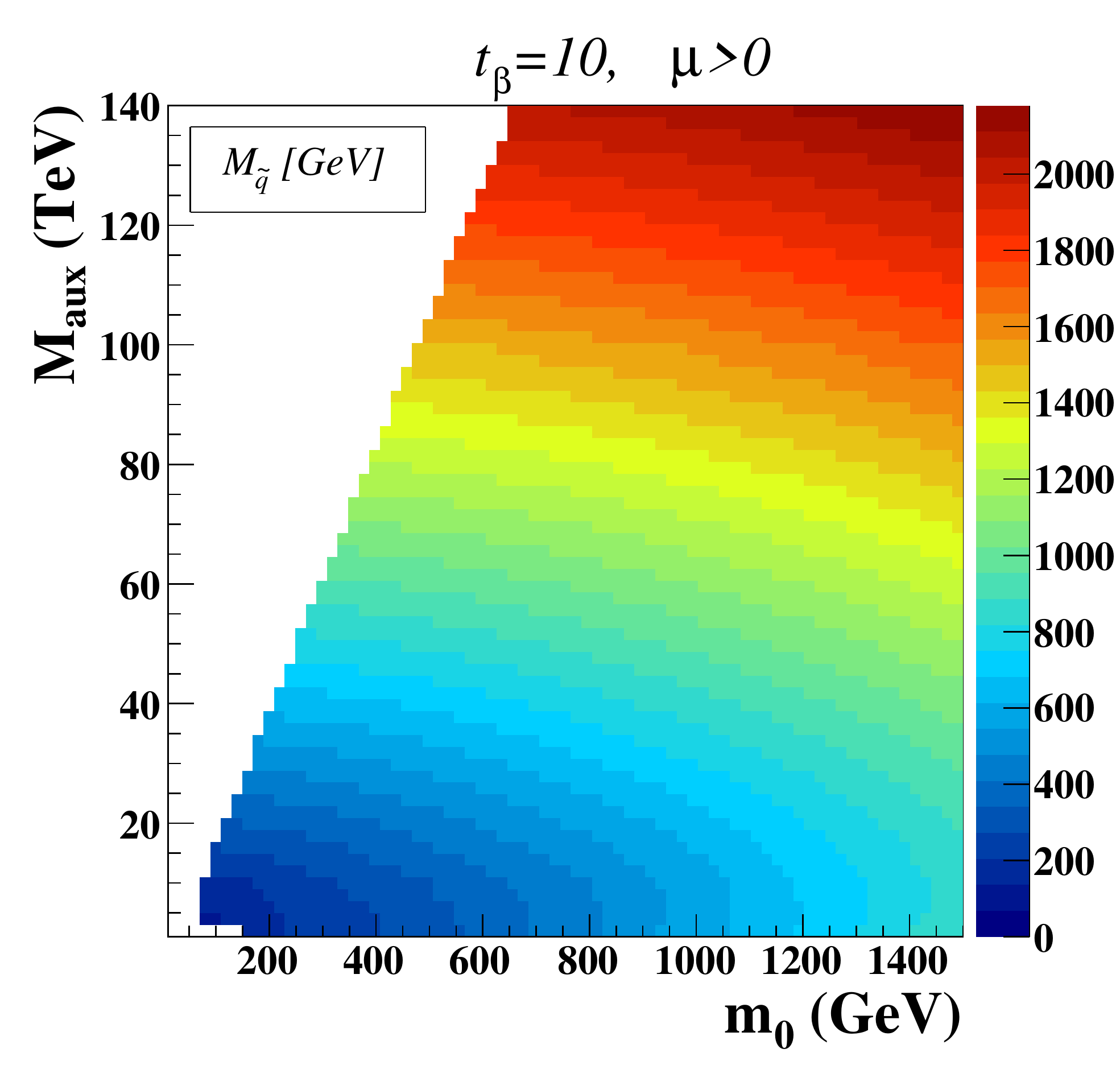}
 \includegraphics[width=.32\columnwidth]{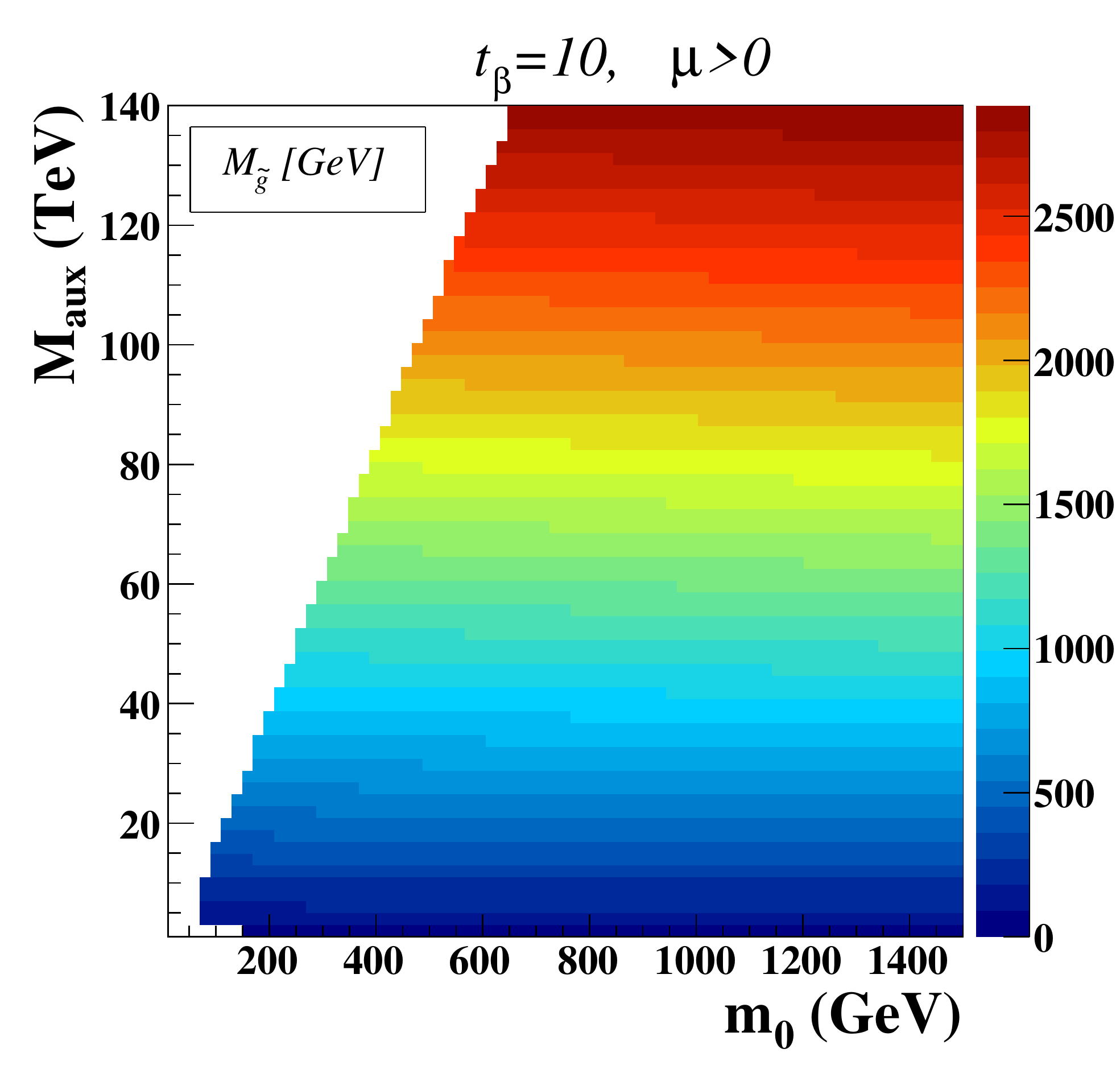}
\vspace{.3cm}
 \includegraphics[width=.32\columnwidth]{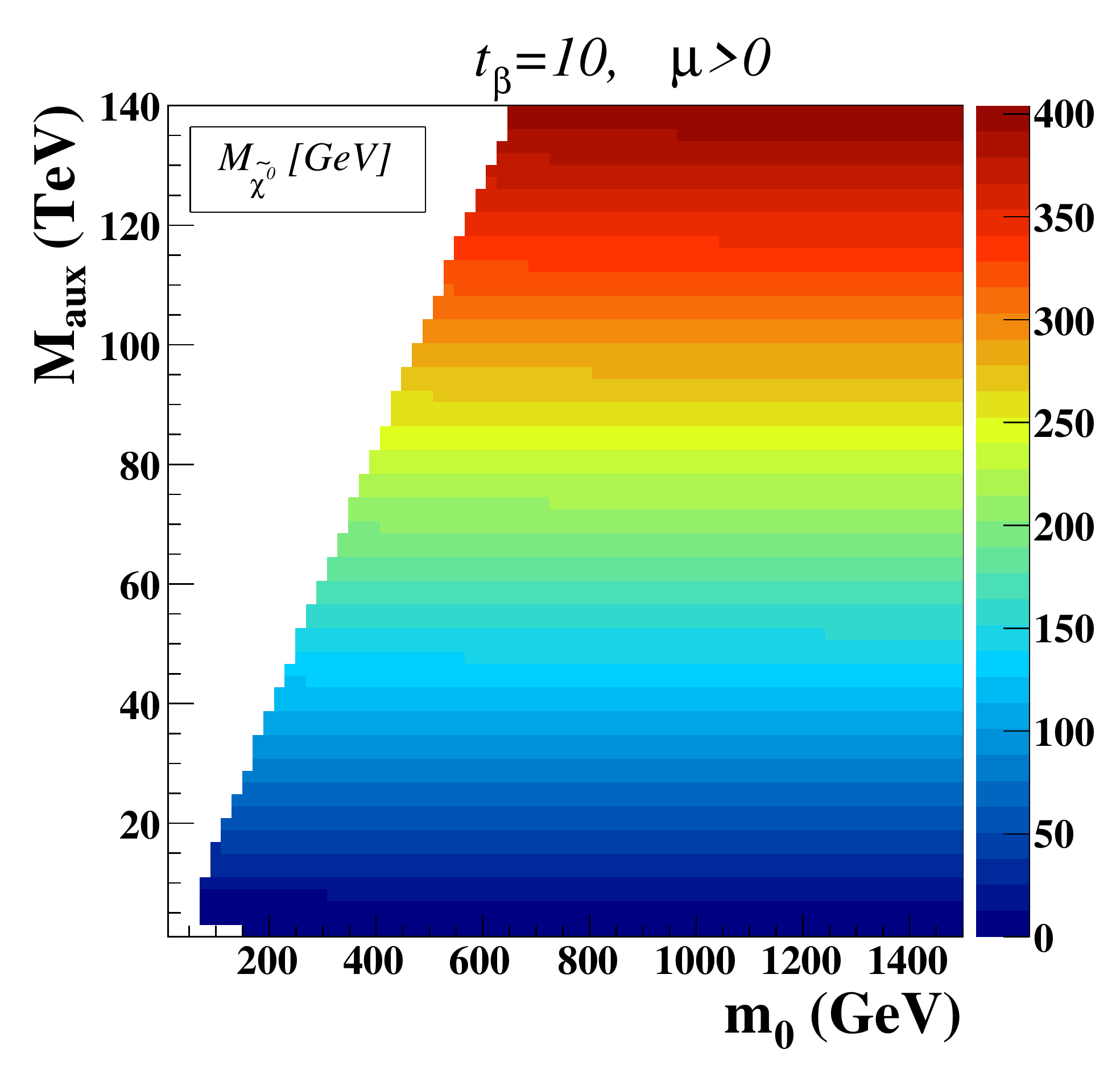}
 \includegraphics[width=.32\columnwidth]{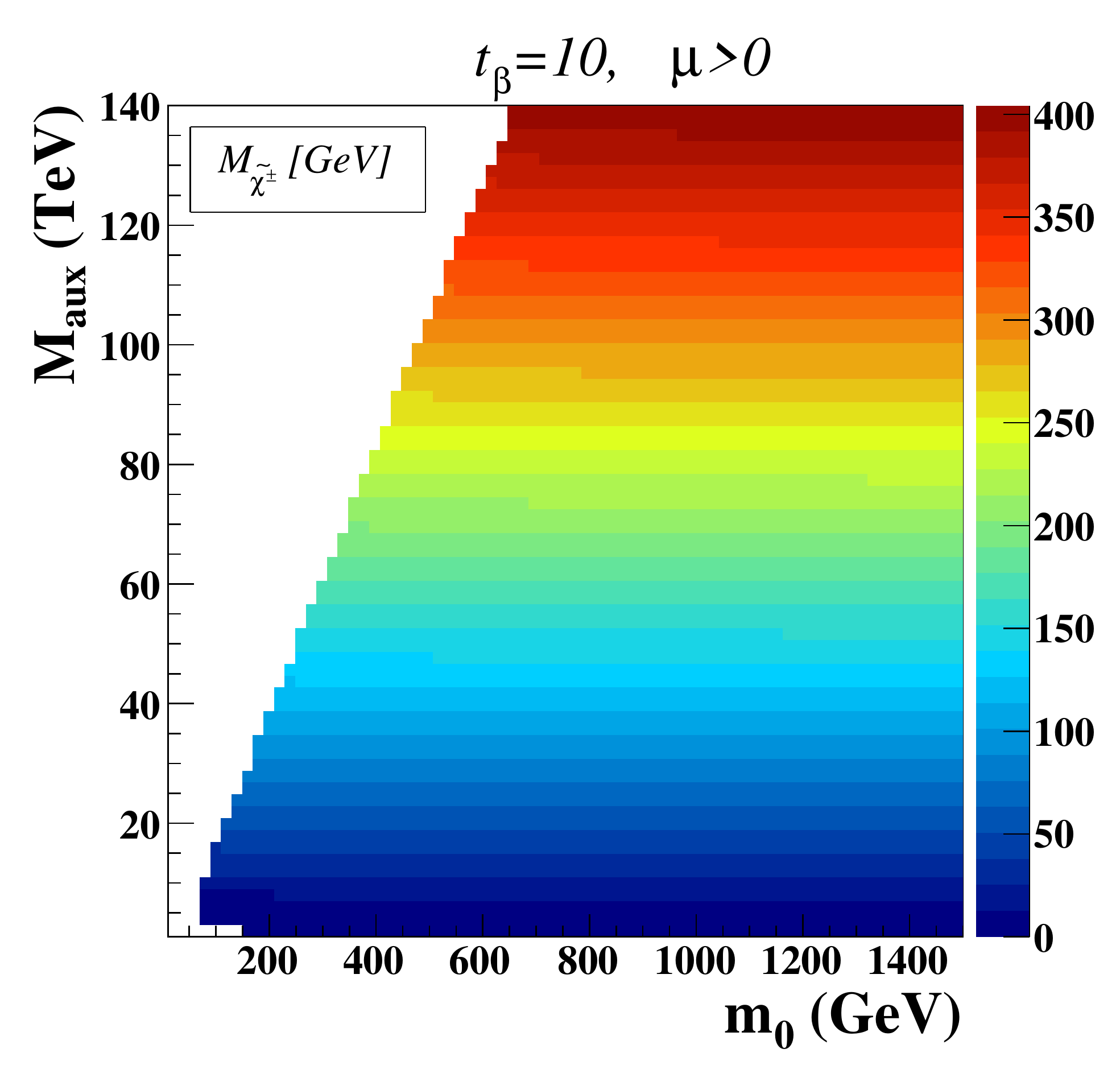}
 \includegraphics[width=.32\columnwidth]{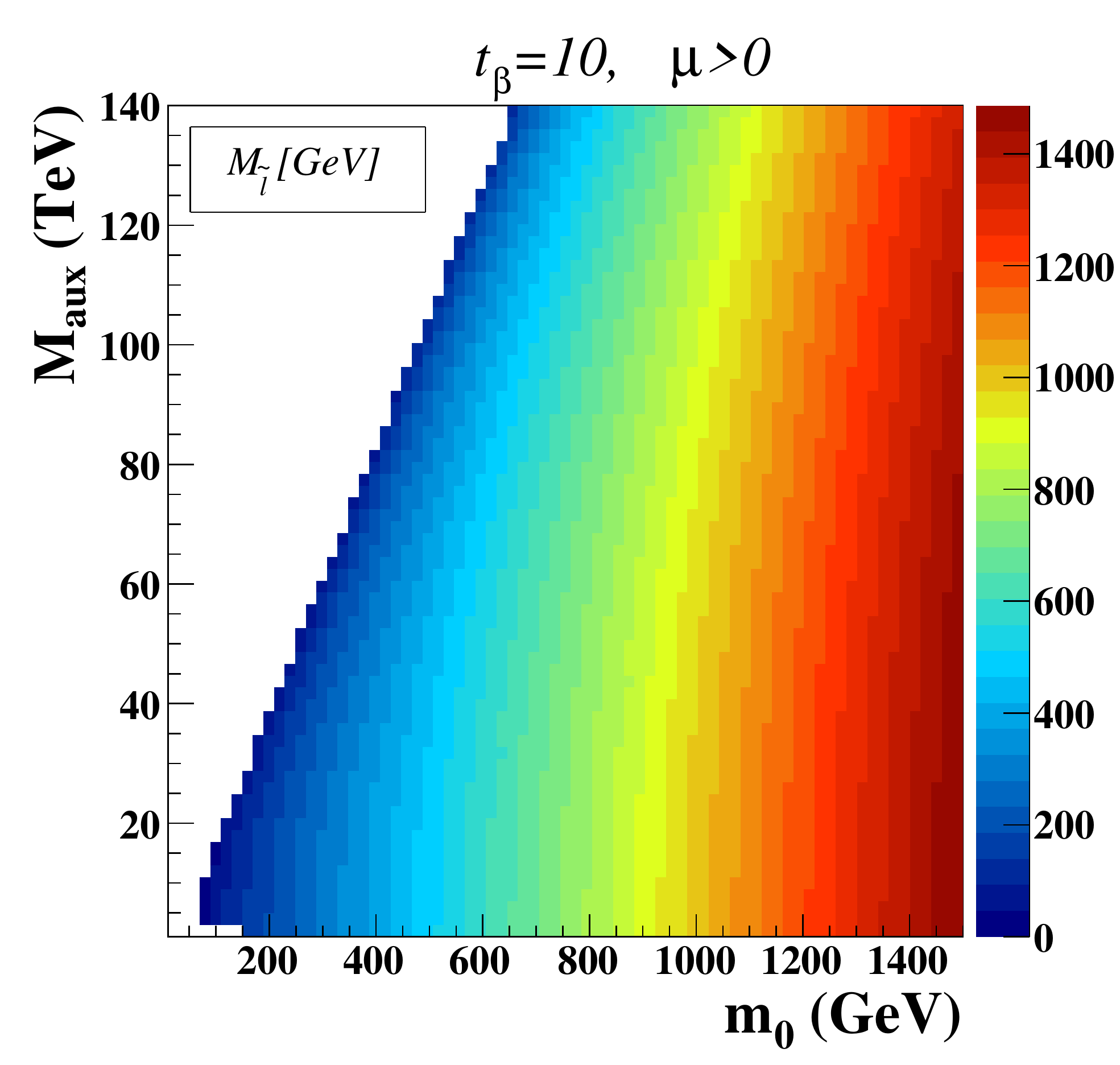}
 \caption{\label{fig:amsb_mass}  Same as in Figure \ref{fig:cmssm10_mass} but
for the MSSM with anomaly-mediated
supersymmetry-breaking. We present $(m_0, 
M_{\rm aux})$ planes with $\tan\beta=10$ and a positive Higgs mixing parameter $\mu>0$.}
\end{figure}
%

Turning to the MSSM with gauge-mediated supersymmetry-breaking, it is found 
that it is very difficult to accommodate the bounds on the Higgs mass of Eq.\
\eqref{eq:mhlhc} for a large fraction of the parameter space. For instance,
no scenario for which $\tan\beta=15$ and $N_q=N_\ell=1$ is
compliant with the measurements. Even for $N_q=N_\ell=3$, the only way
to restore agreement with data disfavors light supersymmetry
(see Figure~\ref{fig:gmsb3_mass}). The viable regions however
also satisfies the indirect constraints of Section 
\ref{sec:mssm_indirect}, with again the exception of the
anomalous magnetic moment of the muon. 
Signatures of
gauge-mediated super\-sym\-me\-try-breaking scenarios are different from those
expected in the context of the cMSSM. In particular, final states rich in photons
are expected due to the different
nature for the lightest superparticle. Dedicated experimental
analyses have led to excluded mass ranges very similar to those excluded in 
the cMSSM framework
\cite{Aad:2012afb, ATLAS:2012ht, ATLAS147, Chatrchyan:2012mx}. 

We finally present, in Figure \ref{fig:amsb_mass}, $(m_0, M_{\rm aux})$ planes
with $\tan\beta=10$ for MSSM scenarios with anomaly-mediated 
supersymmetry breaking. The Higgs mass predictions are shown in the left panel of
the first row of the figure and the masses of the (lightest) superpartners in the 
rest of
the figure. Only a narrow band of the scanned parameter space leads to an
agreement with 
the Higgs mass measurement, but also predicts
heavy superpartners of several TeV.

One important point to note is that in
this region, the numerical value of the lightest Higgs 
boson mass returned by \spheno\ 3 differs by several
GeV from those obtained by making use of \spheno\ version 2 \cite{Porod:2003um},
\suspect\ \cite{Djouadi:2002ze} or \softsusy\ \cite{Allanach:2001kg,
Allanach:2009bv, Allanach:2011de}, that all agree with each other. This
difference is explained by consistently including
in \spheno\ 3 gauge-invariant contributions to the Higgs
mixing parameters $b$ and $\mu$~\cite{Werner:PC}. As a consequence, \spheno\ 
predictions show that anomaly-mediated supersymmetry breaking might still be 
viable, contrary
to other spectrum generators predictions~\cite{Arbey:2012dq, Arbey:2011ab}. 
Experimental analyses exclusively dedicated to anomaly-mediated
supersymmetry-breaking scenarios exist, but only focus on the neutralino and chargino
sectors. Weak constraints have been extracted on the chargino and neutralino mass
required to be above ${\cal O}(100)$ GeV~\cite{ATLAS:2012jp}.

\mysection{Motivation for going beyond the MSSM}\label{sec:eqbyd} 

In the previous sections, we have presented a representative set of 
constraints that can be employed when constructing theoretically
motivated and not experimentally excluded benchmark scenarios. Direct
searches at colliders (and in particular at the LHC), indirect constraints
obtained from low-energy flavor and electroweak precision observables and 
cosmological data have been considered. 
We have concluded that it becomes very difficult to 
accommodate both light supersymmetry and 
all the constraints. Experimental results such as
the discovery of a Higgs boson or the absence of any supersymmetric
particle in current LHC data tend to show that if Nature is
supersymmetric, supersymmetry lies above the TeV range. Viable regions
are in general compatible with flavor and electroweak precision data, but are however
not able
to provide an explanation for the discrepancy between theory and experiment for the anomalous 
magnetic moment of the muon.

In order to reconcile supersymmetry, collider friendship and data, 
another option consists of leaving the
minimal picture. We have already shown that allowing for
non-minimal flavor violation in the MSSM can affect theoretical
predictions. Moreover, current experimental results may not be valid in
more general supersymmetric frameworks, in particular since most of all
the current bounds are derived from simplified models inspired by the cMSSM. 
Although typical supersymmetric
signatures, such as the presence of missing transverse energy in the
final state, are common for the MSSM and many 
non-minimal supersymmetric models, dedicated
phenomenological studies can be required to study new types of signatures.
In the next 
chapter of this work, several non-minimal models are investigated, namely the MSSM with
$R$-parity violation~\cite{Barbier:2004ez}, $N\!=\!1\!/\!N\!=\!2$ hybrid
supersymmetric theories~\cite{Fayet:1975yi, AlvarezGaume:1996mv, Plehn:2008ae,
Choi:2008pi, Choi:2008ub, Choi:2009jc, choi:2010gc, Choi:2010an,
Schumann:2011ji} and minimal $R$-symmetric supersymmetric
theories~\cite{Fayet:1974pd, Salam:1974xa, Kribs:2007ac}. To this aim, we will make use of the 
tools introduced in Chapter \ref{chap:FR} to perform LHC phenomenological studies based
on Monte Carlo simulations.

\cleardoublepage

%% file: nmsusy.tex
\label{chap:nonmin}

As stated in Section \ref{sec:eqbyd}, there are large classes of non-minimal 
supersymmetric theories valuable to be investigated. We choose to focus in this
work on two of these, 
namely the MSSM with $R$-parity violation \cite{Barbier:2004ez} 
and the minimal version of supersymmetric theories with an unbroken $R$-symmetry 
\cite{Fayet:1974pd, Salam:1974xa, Kribs:2007ac}. We first dedicate Section 
\ref{sec:bmssm2} to the description of the considered models and to the 
presentation of their 
main phenomenological implications. Next, Section \ref{sec:bmssmfr} and Section 
\ref{sec:frmg} are more technical and show how to efficiently 
perform phenomenological LHC analyses in the framework of these 
models by means of Monte Carlo simulations. 
For the sake of the example, 
two analyses
are finally addressed in Section \ref{sec:rpvmonotop} and Section
\ref{sec:mrssmsgluons}.

\mysection{Beyond minimal supersymmetry: two examples}
\label{sec:bmssm2}
\subsection{$R$-parity violation}\label{sec:rpv}
The conservation of $R$-parity is invoked to forbid the lepton-number-violating
and baryon-number-violating interactions and mass terms 
induced by the superpotential contributions 
\be\bsp
  W_{RPV} = 
      \frac12 \lambda_{ijk} L_L^i \!\cdot\! L_L^j E_R^k 
    + \lambda^\prime_{ijk} L_L^i \!\cdot Q_L^j D_R^k 
    + \frac12 \lambda^{\prime\prime}_{ijk} U_R^i D^j_R D_R^k
    - \kappa_i L_L^i \!\cdot\! H_U \ ,
\esp\ee
and the soft supersymmetry-breaking Lagrangian 
\be\bsp
  \lag_{{\rm soft},RPV} = &\ \Big[ 
      D_i \tilde \ell_L^i \!\cdot\! H_u 
    - ({\bf m_{LH}^2})^i \tilde \ell_{Li}^\dag H_d
    + \hc \Big]\\ 
   &\ - \Big[
      \frac12 T_{ijk} \tilde \ell_L^i \!\cdot\! \tilde \ell_L^j \tilde e_R^{k\dag}
    + T^\prime_{ijk} \tilde \ell_L^i \!\cdot\! \tilde q^j_L \tilde d_R^{k\dag}
    + \frac12 T^{\prime\prime}_{ijk}\,  \tilde u_R^{i\dag} \tilde d_R^{j\dag} 
        \tilde d_R^{k\dag} + \hc \Big] \ .
\esp\ee 
In those equations, all indices but the flavor ones are understood and the notations
for the superfields and their components are compliant with those  
introduced in Section \ref{sec:mssmfields}. The 
parameters $\kappa$, $D$ and ${\bf m^2_{LH}}$ are three-dimensional vectors 
and $\lambda$, $\lambda^\prime$, $\lpp$, $T$, 
$T^\prime$, $T^{\prime\prime}$ are $3 \times 3 \times 3$ tensors in 
generation space.

Switching to the super-CKM and super-PMNS basis, these parameters are redefined 
to prevent the rotation matrices of Eq.\ \eqref{eq:fermrot} and 
Eq.\ \eqref{eq:fermrot2} to explicitly
appear in the Lagrangian, having instead an explicit dependence on the CKM
and PMNS matrices. This leads to the $R$-parity violating
superpotential 
\be \boxed{\label{eq:spotrpv}\bsp
  W_{RPV} =&\ 
      (V_{\rm PMNS}^\dag)^i{}_\ip \hat\lambda_{ijk} 
         N_L^\ip E_L^j E_R^k 
    + (V_{\rm PMNS}^\dag)^i{}_\ip (V_{\rm CKM}^\dag)^j{}_\jp 
         \hat \lambda^\prime_{ijk} L_L^\ip \!\cdot Q_L^\jp D_R^k 
\\ &\
    + \frac12 \hat \lambda^{\prime\prime}_{ijk} U_R^i D^j_R D_R^k
    - \hat \kappa_i L_L^i \!\cdot\! H_U \ ,
\esp}\ee
and to the soft supersymmetry-breaking Lagrangian 
\be\label{eq:lsoftrpv}\boxed{\bsp
  \lag_{{\rm soft},RPV} = &\ \Big[ 
      \hat D_i \tilde \ell_L^i \!\cdot\! H_u 
    - ({\bf \hat m_{LH}^2})^i \tilde \ell_{Li}^\dag H_d
    + \hc \Big] - \Big[
      (V_{\rm PMNS}^\dag)^i{}_\ip 
         \hat T_{ijk} \tilde \nu_L^\ip \!\cdot\! \tilde e_L^j \tilde e_R^{k\dag}\\ 
   &\
    + (V_{\rm PMNS}^\dag)^i{}_\ip (V_{\rm CKM}^\dag)^j{}_\jp 
         \hat T^\prime_{ijk} \tilde \ell_L^\ip \!\cdot\! \tilde q^\jp_L 
          \tilde d_R^{k\dag}
    + \frac12 \hat T^{\prime\prime}_{ijk}\,  \tilde u_R^{i\dag} \tilde d_R^{j\dag} 
        \tilde d_R^{k\dag} + \hc \Big] \ ,
\esp}\ee 
the hatted parameters being used as input parameters according to the Supersymmetry 
Les Houches Accord conventions \cite{Allanach:2008qq}. We have also
introduced, in those expressions, 
the left-handed neutrino and left-handed charged lepton superfields $N_L$ and 
$E_L$, \ie, the components of the $SU(2)_L$ doublet $L_L$, 
and their scalar components $\tilde\nu_L$ and $\tilde e_L$.
Whereas there is not any theoretical motivation to forbid these $R$-parity 
violating couplings terms,
most of them are however highly constrained by experimental measurements.

The size of the lepton-number-violating operators is strongly restricted 
by the smallness 
of the neutrino masses \cite{Drees:1997id,Chun:1998gp,Joshipura:1999hr,
Cheung:1999az,Dey:2008ht}, the neutrino oscillations rate along their 
propagation within media \cite{Roulet:1991sm,Guzzo:1991hi} and the negative
results from
neutrinoless double beta decay searches \cite{Hirsch:1995ek,Babu:1995vh,
Faessler:1996ph,Faessler:1997db}. In addition, bounds on quadratic and 
quartic products of couplings are obtained from lepton rare decays 
\cite{Barger:1989rk,Dreiner:2001kc,Herz:2002gq}, 
strange or $B$-meson decays~\cite{Tahir:1998uk,Dreiner:2001kc,Grossman:1995gt,
Erler:1996ww,Saha:2002kt}, atomic parity 
violation data \cite{Rosner:2001ck, Ginges:2003qt} or 
magnetic and electric dipole moment measurements \cite{Kim:2001se,
Adhikari:2001ra,Godbole:1999ye,Abel:1999yz,Herczeg:1999me}. For superpartners 
with masses of the order of several hundreds of GeV, the lepton-number-violating 
couplings of Eq.~\eqref{eq:spotrpv} and Eq.~\eqref{eq:lsoftrpv} are
typically constrained to be smaller than about $0.01-0.1$
\cite{Barbier:2004ez,Kao:2009fg}.
Indirect limits on the baryon-number-violating operators arise either from single 
nucleon decay data, and in particular from the proton lifetime measurements, 
\cite{Chang:1996sw,Choi:1998ak,Bhattacharyya:1998dt,Hinchliffe:1992ad,
Carlson:1995ji,Smirnov:1996bg}, from $K$-meson or $B$-meson data 
\cite{Barbieri:1985ty,Abel:1996qj,Slavich:2000xm,Chakraverty:2000df,
BarShalom:2002sv} or from nucleon-antinucleon oscillations and 
double nucleon decays \cite{Hinchliffe:1992ad,Zwirner:1984is,Dimopoulos:1987rk}.
However, the most restrictive bounds are implied by cosmology, the  
observed flux of cosmic antiprotons inducing $\lpp < 10^{-19}-10^{-24}$
\cite{Baltz:1997ar} for most of the cases (see below).

Complementary to these indirect measurements, the ATLAS and CMS collaborations
are currently searching for $R$-parity violating supersymmetry within the LHC 
data. No hint of signal has been found so that limits on 
the superpartner masses up to 1-2 TeV and on the $R$-parity violating
coupling strengths of ${\cal O}(0.1)$ have been set 
\cite{Aad:2012zx,ATLAS:2012dp,ATLAS:2012kr,Aad:2012yw,Aad:2011qr,
ATLAS153,Chatrchyan:2011ff,Chatrchyan:2012mea,CMSSUS027},
extending older Tevatron  
\cite{Abazov:2006nw,Abulencia:2007mp} 
and H1 limits \cite{Aid:1996iw}.

All the above-mentioned bounds are nevertheless 
not applicable when considering the 
$\lambda^{\prime\prime}_{3jk}$ parameters, related to the (s)top sector, 
when the lightest supersymmetric particle is lighter than the 
top quark. In this framework, the lightest superpartner slowly decays 
to a four-body final state through both a virtual stop and a virtual top quark,
possibly outside the detector, so that 
typical supersymmetry search results are   
applicable. In addition, if we assume that the $T^{\prime \prime}$ parameters 
are non-zero, this class of scenarios also offers attractive solutions for baryogenesis 
and for the origin of the observed baryon asymmetry in the universe 
\cite{Dimopoulos:1987rk,Cline:1990bw,Scherrer:1991yu,Mollerach:1991mu}. 
We therefore choose to investigate in this work
a novel approach to probe the $\lambda^{\prime\prime}_{3jk}$ 
couplings at the LHC.

\subsection{The minimal $R$-symmetric supersymmetric model}\label{sec:mrssm}
As mentioned in Section \ref{sec:mssm_indirect}, there are many flavor 
observables allowing for constraining new physics. Data being compatible
with the Standard Model, it implies that supersymmetry breaking has to be,
in a good approximation, flavor blind such as in gauge-mediated or 
anomaly-mediated supersymmetry breaking. Another alternative consists of 
allowing for flavor violation, but screening it simultaneously so that new 
physics effects are hidden. An appealing option
for such scenarios lies in supersymmetric theories with an extended $R$-symmetry.

In Eq.\ \eqref{eq:poincaresalg}, we have shown that the Poincar\'e superalgebra 
automatically contains a continuous $R$-symmetry. Moreover, 
the hidden sector where supersymmetry is broken must necessarily be $R$-symmetric
to guarantee a \textit{spontaneous} breaking of supersymmetry 
\cite{Nelson:1993nf}. However, requiring 
an unbroken $R$-symmetry
in the visible sector of the model leads to various phenomenological issues. 
Firstly, both Majorana mass terms for the gaugino fields and superpotential 
$\mu$-terms
are forbidden. This challenges the current non-observation of massless 
gauginos and higgsinos as well as  
a proper electroweak symmetry breaking. Secondly, dynamical supersymmetry
breaking generally breaks at the same time the $R$-symmetry. This can 
however be avoided by means of metastable non-supersymmetric vacua naturally 
appearing
in supersymmetric gauge theories \cite{Intriligator:2006dd,Intriligator:2007cp,
Intriligator:2007py}. Finally, supersymmetry can only be embedded within 
the framework of supergravity if the cosmological constant is 
tunable to zero (by adding, \eg,
a constant term in the superpotential), which usually explicitly 
breaks the $R$-symmetry.
Consequently, $R$-symmetric supersymmetric theories have not received much
attention until quite recently where it has been shown 
that viable scenarios can be constructed by adding new matter 
fields in the visible sector \cite{Kribs:2007ac}.

We start from the MSSM field content of Section \ref{sec:mssmfields} and impose
an exact $R$-symmetry on the model. First, 
gauge spinorial superfields $W_\alpha$ have their usual $R$-charge
of $+1$ so that all gauge interaction and kinetic terms (including those
associated with the chiral content of the model) are 
$R$-symmetric and kept unchanged. 
Next, in order to avoid electroweak symmetry breaking to
spontaneously break the $R$-symmetry, the $R$-charges of the Higgs 
superfields are fixed to zero,
\be
  R(H_U) = R(H_D) = 0 \ .
\ee
In addition, including the usual Yukawa couplings leads to
\be
  R(Q_L) = R(U_R) = R(D_R) = R(L_L) = R(E_R) = 1 \ ,
\ee
since the superpotential has an $R$-charge of two units. 
Its most general renormalizable version satisfying the $R$-symmetry requirement
reads thus
\be
  W^{(1)}_{\rm MRSSM} =  
    ({\bf y^u})_{ij}  \  U_R^i\ Q_L^j \!\cdot\! H_U - 
    ({\bf y^d})_{ij}  \  D_R^i\ Q_L^j \!\cdot\! H_D  - 
    ({\bf y^e})_{ij}\   E_R^i\ L_L^j \!\cdot\! H_D \ . 
\label{eq:wmrssm1}\ee
The MSSM off-diagonal Higgs mixing term is absent and $R$-parity is 
automatically conserved\footnote{As in Chapter \ref{chap:mssm}, right-handed
neutrino superfields are not considered.}.

Soft supersymmetry breaking can also be achieved in an $R$-symmetric fashion.
To this aim, two hidden sector $F$-type and $D$-type spurion superfields 
are introduced,
\be
  \langle X \rangle = \theta\!\cdot\!\theta\ v_F
  \qquad\text{and}\qquad
  \langle W'_\alpha\rangle = \sqrt{2} \theta_\alpha\ v_D \ ,
\ee
where $X$ stands for a chiral superfield with an $R$-charge of $+2$ and
the spinorial superfield strength tensor $W'$, with its standard $R$-charge
of $+1$, is associated with a $U(1)'$ gauge 
symmetry of the hidden sector. 
The vacuum expectation values $v_F$ and $v_D$ are typically taken of the order of 
magnitude of the supersymmetric masses $M_{\rm susy}$ and the soft 
supersymmetry-breaking terms of the visible sector are constructed by coupling the
model superfields to
the spurions in an $R$-preserving manner.
In this way,
the $F$-type spurion allows us to write down mass terms for the
scalar fields of the theory and bilinear Higgs mixing terms, 
\be
  \lag_{\rm soft}^{(1)} =\int \d^2 \theta\d^2\thetabar\  
     \frac{X X^\dag}{M_{\rm susy}^2}\bigg[ 
        \Big( B H_u \!\cdot\! H_D + \hc \Big) + 
        \sum_{\Phi = \{H_U,H_D,Q_L,U_R,D_R,L_L,E_R\}} \Big(
           m_\Phi^2 \Phi^\dag \Phi \Big)
   \bigg]\ .
\label{eq:lrsoft1}\ee
The supersymmetry-breaking parameters $m_\Phi^2$ 
and $B$ are easily linked to their counterparts of Eq.\ \eqref{eq:lmssmbrk2}, 
and we recall that there is no $R$-symmetric way to generate 
trilinear scalar interactions. 
In addition, the $R$-symmetry also
forbids, on the one hand, 
the dangerous dimension-five operators $Q_LQ_LQ_LL_L$ and $U_RU_RD_RE_R$
yielding proton decay and allows, on the other hand,
for neutrino Majorana masses induced by the operator $H_UH_UL_LL_L$.

\renewcommand{\arraystretch}{1.3}
\begin{table}[!t]
 \begin{center}
  \begin{tabular}{|c||c|c|c|c|}
    \hline
    Supermultiplet & Fermion & Scalar & Representation \\ \hline
    $\tilde\Phi_B$ & $\widetilde B'$ & $\sigma_B$ & 
       $({\utilde{\bf 1}}, {\utilde{\bf 1}}, 0)$\\
    $\tilde\Phi_Y$ & $\widetilde W'$ & $\sigma_W$ & 
       $({\utilde{\bf 1}}, {\utilde{\bf 3}}, 0)$\\
    $\tilde\Phi_G$ & $\widetilde g'$ & $\sigma_G$ & 
       $({\utilde{\bf 8}}, {\utilde{\bf 1}}, 0)$\\
    \hline
       $R_U$ &$\widetilde R_u = \bpm \widetilde R_u^0 \\ \widetilde R_u^-\epm$&
       $R_u = \bpm R_u^0 \\ R_u^- \epm$ &         
        $({\utilde{\bf 1}}, {\utilde{\bf 2}}, -\frac12)$ \\
       $R_D$&$\widetilde R_d = \bpm \widetilde R_d^+ \\ \widetilde R_d^0\epm$&
       $R_d = \bpm R_d^+ \\ R_d^0 \epm$ &         
        $({\utilde{\bf 1}}, {\utilde{\bf 2}}, \frac12)$ \\
    \hline 
   \end{tabular}\end{center}
  \caption{\label{tab:rgauge}$R$-partners of the vector and Higgs superfields in
    the minimal $R$-symmetric supersymmetric model. They are given together with
    their scalar and fermionic components, as well as with their
    representations under the $SU(3)_c \times SU(2)_L \times
    U(1)_Y$ gauge group.}
\end{table}
\renewcommand{\arraystretch}{1}

The $R$-symmetric version of the MSSM introduced so far contains massless
gaugino and higgsino fields, which clearly contradicts experimental data.
Therefore, the field content of the model must be augmented appropriately.
Although Majorana gaugino masses are forbidden by the $R$-symmetry, 
Dirac masses are still allowed. Many phenomenologically viable models 
consequently include the pairing of each of the gaugino fields 
with the fermionic component of a new chiral superfield 
lying in the adjoint representation of the relevant gauge
subgroup~\cite{Polchinski:1982an,Dine:1992yw,Fox:2002bu}, 
\be
  \tilde\Phi_B = ({\utilde{\bf 1}}, {\utilde{\bf 1}}, 0) \ , \qquad  
  \tilde\Phi_W = ({\utilde{\bf 1}}, {\utilde{\bf 3}}, 0) \ , \qquad
  \tilde\Phi_G = ({\utilde{\bf 8}}, {\utilde{\bf 1}}, 0) \ .
\ee
Like the vector superfields of the theory, these new fields are 
uncharged under the $R$-symmetry.
The higgsino fields are rendered massive by allowing the two Higgs 
chiral superfields $H_D$ and $H_U$ to mix with two new chiral superfields 
$R_D$ and $R_U$, whose representations under the gauge group are given by 
\be
  R_D = ({\utilde{\bf 1}}, {\utilde{\bf 2}}, \frac12) \qquad\text{and}\qquad  
  R_U = ({\utilde{\bf 1}}, {\utilde{\bf 2}}, -\frac12) \  .
\ee
Their $R$-charge is fixed to two units so that superpotential mass terms
can be written. The notations for the component fields of these five 
new superfields are given in Table \ref{tab:rgauge}.

The new chiral superfields allow to extend the superpotential of Eq.\ 
\eqref{eq:wmrssm1} as
\be\boxed{
  W_{\rm MRSSM} =  W_{\rm MRSSM}^{(1)}+  
  \sum_{i=U,D} \bigg[ \lambda^B_i H_i \tilde\Phi_B Y R_i 
    + \lambda^W_i H_i \tilde\Phi_W^k \frac{\sigma_k}{2} R_i  
    + \mu_i H_i\!\cdot\!R_i \bigg] \ ,
}\label{eq:wmrssm2}\ee
so that it now contains higgsino masses as well as additional 
interactions among the Higgs superfields, their $R$-partners and the 
new chiral adjoint superfields $\tilde\Phi$. We have also explicitly 
indicated the $U(1)_Y$ operator $Y$ and the generators of $SU(2)_L$ in the 
fundamental representation $\sigma_k/2$. 
In addition, dimension-five and dimension-six operators softly breaking 
supersymmetry and involving $F$-type and $D$-type spurions,
\be\label{eq:mrssmsoft}\boxed{\bsp
  \lag_{\rm soft}^{\rm MRSSM} =&\
   \lag_{\rm soft}^{(1)} + 
     \sum_{k=B,W,G} \bigg[
       \frac{1}{2 g_k} m_k \int \d^2\theta\ \frac{W^{\prime\alpha}}{M_{\rm susy}}
       W_{k\alpha} \tilde\Phi_k + \hc \bigg] 
   \\&\ 
   +  \int \d^2 \theta\d^2\thetabar\ {\rm Tr} \big[\tilde\Phi_k \tilde\Phi_k \big]
      \sum_{k=B,W,G} \bigg[ (M_{\tilde\Phi_k})^2 \frac{X X^\dag}{M_{\rm susy}^2}
         +  ({\cal M}_{\tilde\Phi_k})^2 \frac{W'\!\cdot\!
      W'}{M_{\rm susy}^2} \bigg] \\
   &\   +\int \d^2 \theta\d^2\thetabar\  \frac{X X^\dag}{M_{\rm
    susy}^2} \bigg[ B_U H_U\!\cdot\!R_U + B_D H_D\!\cdot\!R_D +\hc \bigg] \ ,
\esp}\ee
allow for the generation of Dirac gaugino masses together with extra
multiscalar interactions. In this expression, 
$W_k$ denote the superfield strength tensor
associated with the vector superfield $V_k$ ($k = B,W,G$), 
$m_k$ being the corresponding mass parameter. Moreover, $M_{\tilde\Phi_k}$ and 
${\cal M}_{\tilde\Phi_k}$ are scalar mass parameters and 
$B_U$ and $B_D$ are bilinear mixing terms among Higgs and $R$-Higgs fields.

Gauge interaction and kinetic terms for the new chiral superfields are standard
and given by Eq.\ \eqref{eq:gensusyaction},
\be\label{eq:Lkin}\bsp
  \lag_K =&\ 
    \int \d^2\theta\d^2\thetabar\ \bigg[ \tilde\Phi_B^\dag \tilde\Phi_B +  
      \tilde\Phi_W^\dag e^{-2 g_w \tilde V_W} \tilde\Phi_W +  
      \tilde\Phi_G^\dag e^{-2 g_s \tilde V_G} \tilde\Phi_G \\
 &\ + 
    R_D^\dag \Big(e^{-g_y V_B} e^{-2 g_w V_W} \Big) R_D + 
    R_U^\dag \Big(e^{ g_y V_B} e^{-2 g_w V_W} \Big) R_U 
\bigg] \ ,
\esp\ee
where $\tilde V_W = V_W^k \tilde T_k$ and $\tilde V_G = V_G^a \tilde T_a$, 
the matrices $\tilde T_k$ and $\tilde T_a$ being taken as 
representation matrices of the $SU(2)$ and $SU(3)$ algebra in the adjoint
representation. We refer, for the rest of the notations, to Section 
\ref{sec:mssmsusylag}.  This Lagrangian
contains, in particular, couplings of the new scalar adjoint 
$\sigma$-fields to a single gauge boson and to pairs of 
gauge bosons through the usual gauge-covariant derivatives. However, these fields also
couple singly to up-type ($u$) and down-type ($d$)
quark pairs, as well as to gluon pairs, through loop-diagrams
involving squarks, gluinos, neutralinos and charginos. 
These interactions are described by the effective Lagrangian, expressed
in terms of four-component fermions,
\be\boxed{\bsp
 \lag_{\rm eff} =&\ 
   \sigma_G^a\ \bar d T_a \Big[ a^L_d P_L + a^R_d P_R \Big] d + 
   \sigma_G^a\ \bar u T_a \Big[ a^L_u P_L + a^R_u P_R \Big] u + 
   a_g\ d_a{}^{bc}\ \sigma_G^a\ G_{\mu\nu b} G^{\mu\nu}{}_c  \\ 
 &\ +  
   \sigma_W^k\ \bar d \frac{\sigma_k}{2} \Big[ b^L_d P_L + b^R_d P_R \Big] d + 
   \sigma_W^k\ \bar u \frac{\sigma_k}{2} \Big[ b^L_u P_L + b^R_u P_R \Big] u \\
   &\ + 
   \sigma_B\ \bar d Y  \Big[ c^L_d P_L + c^R_d P_R \Big] d + 
   \sigma_B\ \bar u Y  \Big[ c^L_u P_L + c^R_u P_R \Big] u + {\rm h.c.}  \ .
\esp}\label{eq:Leff}\ee 
The matrices $T_a$ and the tensor $d_a{}^{bc}$ are respectively the fundamental 
representation matrices and the symmetric structure constants of $SU(3)$, the
operators $P_L$ and $P_R$ the left-handed and right-handed chirality
projectors acting on four-component 
spin space and $G_{\mu\nu}{}^a$ the gluon field strength
tensor. We have also introduced the (internal) parameters $a_q^L$, $a_q^R$, 
$b_q^L$, $b_q^R$, $c_q^L$, $c_q^R$ (with
$q=u,d$) to model the strengths of the interactions among left-handed and
right-handed quarks and a single scalar adjoint field, as well as the 
parameter $a_g$ for the modeling of the
interactions among two gluons and the $\sigma_G$ field, also commonly dubbed sgluon.
The computation of the associated loop-diagrams lead to
effective couplings to quarks of the order of $M_q/M_{\rm susy}$, 
$M_q$ being the mass of the heaviest of the 
external quarks. These interactions are therefore 
non-suppressed only if at least one of the external quarks is a top quark.
Similar calculations lead to $a_g \sim 1/M_{\rm susy}$ \cite{Plehn:2008ae}.

On different footings, current experimental data implies that 
sgluon fields lighter than about 2 TeV
are excluded. These limits have however been obtained under the 
assumption that sgluons couple to light quarks and gluons by means of 
${\cal O}(1)$ effective interactions \cite{Aad:2011yh, Aad:2011fq, ATLAS:2012nna,
CMS:2012eza}. Since this setup 
does not apply to the framework presented in this section, 
no strong constrain exists on $R$-symmetric supersymmetric
sgluons that couple to top quarks or with realistic interaction strengths.
In the rest of this work, we choose to focus on these scalar $\sigma$-fields
lying in the adjoint representation of the QCD gauge group, assuming 
they dominantly couple to top quarks.
These fields also appear in the framework of hybrid 
$N\!=\!1\!/\!N\!=\!2$ supersymmetric theories \cite{Fayet:1975yi,
AlvarezGaume:1996mv, Plehn:2008ae,
Choi:2008pi, Choi:2008ub, Choi:2009jc, choi:2010gc, Choi:2010an,
Schumann:2011ji} where in this case, the three new chiral superfields 
$\tilde\Phi_B$, $\tilde\Phi_W$ and $\tilde\Phi_G$ are considered, 
together with the vector superfields $V_B$, $V_W$ and $V_G$
of Table \ref{tab:gauge}, as three complete vector representations of the 
$N=2$ supersymmetric algebra. 

The main difference between the MSSM and its $R$-symmetric version lies in 
the Dirac nature of the neutralino and gluino fields. Also, the neutral scalar adjoint 
fields $\sigma_B$ and $\sigma_W^3$ can obtain non-vanishing vacuum expectation values
at the minimum of the scalar potential so that the mechanism leading to the breaking
of the electroweak symmetry can be more involved. As a result, a 
more complicated particle mixing structure appears which we summarize below. 
 
Starting with the fermionic 
sector, the neutralino, chargino and gluino fields are now all Dirac fermions,%
\renewcommand{\arraystretch}{1.3}%
\be
  \psi_{\chi^0}^i = \bpm \chi^0_{Li} \\ \bar \chi^{0i}_R \epm \ , \qquad 
  \psi_{\chi^\pm}^i = \bpm \chi^\pm_i \\ \bar \chi^{\mp i} \epm \qquad\text{and}\qquad  
  \psi_{\tilde g} = \bpm i \tilde g \\ \overline {\tilde g'}\epm \ .
\label{eq:newdirac}\ee \renewcommand{\arraystretch}{1.}%
We have introduced in this expression 
the two-component neutralino and chargino states $\chi^0_L$, $\chi_R^0$ and $\chi^\pm$
which differ from their MSSM counterparts. 
The chargino $4\times 4$ mass matrix $M_{\tilde{\chi}^\pm}$ is diagonalized
by means of two unitary matrices $U$ and $V$
\be
  U^* M_{\tilde{\chi}^\pm} V^{-1} = {\rm diag} (M_{\tilde{\chi}^\pm_1},
    M_{\tilde{\chi}^\pm_2}, M_{\tilde{\chi}^\pm_3},
    M_{\tilde{\chi}^\pm_4}) \ ,
\ee
which relate the $\chi^+$ and $\chi^-$ mass eigenstates to the model 
gauge eigenstates as
\renewcommand{\arraystretch}{1.2}%
\be\boxed{ 
  \bpm \chi^+_1 \\ \chi^+_2 \\ \chi^+_3 \\ \chi^+_4 \epm = V 
  \bpm i \widetilde W^+ \\ \widetilde R_d^+ \\ \widetilde W^{\prime +} \\ 
     \widetilde H_u^+ \epm \qquad\text{and}\qquad
  \bpm \chi^-_1 \\ \chi^-_2 \\ \chi^-_3 \\ \chi^-_4 \epm = V 
  \bpm i \widetilde W^- \\ \widetilde R_u^- \\ \widetilde W^{\prime -} \\ 
     \widetilde H_d^- \epm \ . 
}\ee %
\renewcommand{\arraystretch}{1.}%
The charged wino states $\widetilde W^\pm$ and 
$\widetilde W^{\prime \pm}$ are defined as usual, from the  
diagonalization of the third generator of $SU(2)$ in the adjoint
representation, 
\be\label{eq:diagt3}
  \widetilde W^\pm = \frac{1}{\sqrt{2}} (\widetilde W^1 \mp i \widetilde
    W^2) \qquad\text{and}\qquad 
  \widetilde W^{\prime\pm} = \frac{1}{\sqrt{2}} (\widetilde W^{\prime 1} \mp i \widetilde
    W^{\prime 2}) \ . 
\ee

Similarly, the neutralino $4\times4$ mass matrix $M_{\tilde \chi^0}$
is diagonalized by two unitary  matrices $N$ and $N'$
\be
  N^{\prime*} M_{\tilde{\chi}^0} N^{-1} = {\rm diag} (M_{\tilde{\chi}^0_1},
    M_{\tilde{\chi}^0_2}, M_{\tilde{\chi}^0_3},
    M_{\tilde{\chi}^0_4}) \ ,
\ee
which relate the left-handed and right-handed neutralino mass-eigenstates $\chi^0_L$
and $\chi_R^0$ to the gauge eigenstate as
\renewcommand{\arraystretch}{1.2}%
\be\boxed{ 
  \bpm \chi^0_{L1} \\ \chi^0_{L2} \\ \chi^0_{L3} \\ \chi^0_{L4} \epm = N 
  \bpm i \widetilde B \\ i \widetilde W^3 \\ \widetilde R_u^0 \\ \widetilde R_d^0
     \epm \qquad\text{and}\qquad
  \bpm \chi^0_{R1} \\ \chi^0_{R2} \\ \chi^0_{R3} \\ \chi^0_{R4} \epm = N'
  \bpm i \widetilde B' \\ i \widetilde W^{\prime 3} \\ \widetilde H_d^0 \\ \widetilde 
     H_u^0 \epm \ .
}\ee %
\renewcommand{\arraystretch}{1.}%

Accounting for the scalar $R$-partners of the wino and bino states, the Higgs sector of 
the minimal $R$-symmetric supersymmetric model is enriched with
respect to the one of the MSSM. The components of the two Higgs doublets $h_u$ and 
$h_d$ and those of the scalar adjoint fields $\sigma_B$ and $\sigma_W$
mix and give
rise to four (three) neutral scalar (pseudoscalar) Higgs bosons $h^0$ ($A^0$), 
two charged Higgs bosons $H^\pm$ and three pseudo-Goldstone
bosons $G^0$ and $G^\pm$ 
eaten by the weak bosons. 
Introducing the associated mixing matrices $R^S$, $R^P$ and $R^\pm$, 
the physical eigenstates
are related to the gauge eigenstates by \renewcommand{\arraystretch}{1.2}%
\be\boxed{\bsp
 &\ 
 \bpm h^0_1 \\ h^0_2\\ h^0_3 \\ h^0_4 \epm = R^S \bpm \Re\{h_d^0\} \\ \Re\{h_u^0\} \\
   \Re\{\sigma_B\}\\ \Re\{\sigma_W^3\}  \epm 
 \qquad\text{and}\qquad
 \bpm G^0\\ A^0_1 \\ A^0_2\\ A^0_3 \epm = R^P \bpm \Im\{h_d^0\} \\ \Im\{h_u^0\} \\
   \Im\{\sigma_B\}\\ \Im\{\sigma_W^3\}  \epm 
 \ , \\
 &\
 \bpm (G^-)^\dag \\ G^+ \\ H^+_1 \\ H^+_2 \epm = R^\pm \bpm h_u^+ \\ (h_d^-)^\dag \\
   (\sigma_W^-)^\dag\\ \sigma_W^+ \epm \ , 
\esp}\ee\renewcommand{\arraystretch}{1.}%
where the charged $\sigma_W$ fields are defined as in Eq.\ \eqref{eq:diagt3},
\be
  \sigma_W^\pm = \frac{1}{\sqrt{2}} (\sigma_W^1 \mp i \sigma_W^2) \ . 
\ee
These three unitary matrices are derived from the diagonalization of the scalar,
pseudoscalar and charged Higgs squared mass matrices $M_S^2$, $M_P^2$ and $M^2_\pm$,
\be\bsp
 &\ R^S M_S^2 \R^{S\dag} = {\rm diag}(M^2_{h_1^0}, 
   M^2_{h_2^0}, M^2_{h_3^0}, M^2_{h_4^0}) \ , \quad
 R^P M_P^2 \R^{P\dag} = {\rm diag}(0, M^2_{A_1^0}, M^2_{A_2^0}, M^2_{A_3^0}) 
 \\&\ R^\pm M_\pm^2 \R^{\pm\dag} = {\rm diag}(0,0,M^2_{H_1^\pm}, M^2_{H_2^\pm}) \ .  
\esp\ee

We finally turn to the sector of the $R$-partners of the Higgs bosons,
the only scalar fields carrying non-vanishing 
$R$-charges. The neutral component
of the $SU(2)_L$ doublets $R_u^0$ and $R_d^0$ hence mix among themselves,%
\renewcommand{\arraystretch}{1.2}%
\be\boxed{
 \bpm R^0_1 \\ R^0_2 \epm = R^R \bpm R_d^0 \\ R_u^0  \epm  \ ,
}\ee\renewcommand{\arraystretch}{1.}%
while the charged fields do not undergo any additional mixing.
The unitary matrix $R^R$ is computed from the diagonalization of the 
neutral $R$-Higgs squared mass matrix,
\be
  R^R M_R^2 R^{R\dag} = {\rm diag}(M_{R^0_1}^2, M_{R^0_2}^2) \ .
\ee

\mysection{Implementation of supersymmetric models in \feynrules}
\label{sec:bmssmfr}
In this section, we start by describing the implementation of the MSSM in \feynrules, 
employing the superspace module of the package \cite{Duhr:2011se}. 
This serves as a first example on how to use
\feynrules\ for supersymmetric model implementation. Then, the cases of
the two models described in Section \ref{sec:rpv} and Section \ref{sec:mrssm} 
are respectively addressed in Section~\ref{sec:frrpv} and Section 
\ref{sec:frmrssm} \cite{Fuks:2012im}, all the model files being public 
and available on the \feynrules\ website~\cite{FRwebpage}. 

\subsection{The MSSM} \label{sec:FRMSSM}
\subsubsection{Gauge group}
The MSSM is based on the $SU(3)_c \times SU(2)_L \times U(1)_Y$ gauge group 
and one different vector superfield is associated with each gauge 
subgroup. The declaration 
in the \feynrules\ model file of each of these subgroups follows exactly 
the syntax introduced in Section \ref{sec:FRgroups}.
In this way, the abelian factor $U(1)_Y$ is implemented as
\begin{verbatim}
  U1Y  == { 
    Abelian          -> True,
    CouplingConstant -> gp,
    Superfield       -> BSF, 
    Charge           -> Y
  }
\end{verbatim}
where the declaration of the vector superfield {\tt BSF} has to be included in the
{\tt M\$Superfields} list (see below) and the one of the coupling constant
{\tt gp} in the {\tt M\$Parameters}
list.
The implementation of the non-abelian factors 
$SU(2)_L$ and $SU(3)_c$ contains a consistent definition of the
representation matrices relevant for the model field content 
(see Table \ref{tab:matter} and Table \ref{tab:higgs}), 
but the adjoint representation
which is internally handled by \feynrules\ \cite{Christensen:2008py,
Duhr:2011se}. Each matrix is in this manner linked to a specific index
to be carried by the (super)fields. 
The two series of \mathematica\ replacement rules
\begin{verbatim}
  SU2L == { 
    Abelian           -> False, 
    CouplingConstant  -> gw, 
    Superfield        -> WSF, 
    StructureConstant -> ep, 
    Representations   -> {Ta,SU2D}, 
    Definitions       -> { Ta[a__] -> PauliSigma[a]/2, ep -> Eps}
  }

  SU3C == 
  { 
    Abelian           -> False,
    CouplingConstant  -> gs, 
    SymmetricTensor   -> dSUN,
    Superfield        -> GSF,
    StructureConstant -> f,
    Representations   -> { {T,Colour}, {Tb,Colourb} }
  }
\end{verbatim}
allow to declare $SU(2)_L$ fundamental representation matrices, 
labeled by the symbol {\tt Ta} and related to the gauge 
index {\tt SU2D}, as well as $SU(3)_c$ fundamental and antifundamental 
representation matrices {\tt T} and {\tt Tb} mapped to the indices  
{\tt Colour} and {\tt Colourb}, respectively. The adjoint index label
is inferred from the indices carried by the vector superfield attached to each
gauge group. As for the declaration of $U(1)_Y$, the coupling constants must be 
declared in the parameter list. 

\subsubsection{Superfield and field content}
The Standard Model quarks and leptons are embedded, together with their 
squark and slepton partners, into the three
generations of chiral supermultiplets given in Table \ref{tab:matter}. 
The six associated chiral superfields, presented in Eq.\ \eqref{eq:SFnames}, 
are implemented following the instruction given in Section 
\ref{sec:FRSF}, while their
component fields are implemented as shown in Section~\ref{sec:FRfields}.
In order to be allowed to employ the automated function \texttt{CSFKineticTerms} to
generate kinetic and gauge interaction terms,
care must be taken when specifying the hypercharge quantum number  
and the attached gauge indices of each (super)field.
As examples, the weak isospin doublet of quarks $Q_L$ and the $SU(2)_L$ singlet
of charged leptons $E_R$ are implemented\footnote{We assume that the component
fields are properly declared in the {\tt M\$ClassesDescription} list.} as 
\begin{verbatim}
  CSF[1] == { 
    ClassName      -> QL, 
    Chirality      -> Left, 
    Weyl           -> QLw,  
    Scalar         -> QLs,  
    QuantumNumbers -> {Y-> 1/6}, 
    Indices        -> {Index[SU2D],Index[GEN],Index[Colour]}
  }

  CSF[2] == { 
    ClassName      -> ER, 
    Chirality      -> Left, 
    Weyl           -> ERw, 
    Scalar         -> ERs,
    QuantumNumbers -> {Y-> 1},
    Indices        -> {Index[GEN]}
  }
\end{verbatim}
In these two sets of \mathematica\ replacement rules, like in the full
model implementation, we follow a simple naming scheme for the component fields.
The symbols for the Weyl fermionic and for the scalar component of a chiral
superfield are respectively obtained by suffixing {\tt w} and {\tt s} to the
class name. The two chiral superfields $H_U$ and $H_D$ defining
the Higgs sector of the model (see Table~\ref{tab:higgs}) 
as well as the three vector superfields associated with
the gauge sector (see Table \ref{tab:gauge}) are implemented 
in a similar fashion. We recall that for the case of the vector superfields,
the {\tt Indices} attribute of the superfield class
has to refer to the name of the relevant adjoint 
index (labeled by {\tt SU2W} and {\tt Gluon} 
for $SU(2)_L$ and $SU(3)_c$, respectively). As an example, the $SU(2)_L$ vector
superfield implementation reads
\begin{verbatim}
  VSF[1] == { 
    ClassName  -> WSF, 
    GaugeBoson -> Wi,
    Gaugino    -> wow,
    Indices    -> { Index[SU2W] }
  }  
\end{verbatim}

\subsubsection{Lagrangian}
As stated in Section \ref{sec:susylag}, kinetic and gauge interaction
terms for all the chiral and vector superfields of the model are 
fixed by gauge invariance and supersymmetry. In the case of the MSSM, the
corresponding Lagrangian has already been presented in Section \ref{sec:mssmsusylag}
and the two Lagrangians of Eq.\ \eqref{eq:lagmssmvec} and Eq.\ \eqref{eq:lagmssmchir}
are implemented in the \feynrules\ model file 
as described in Section \ref{sec:frsusylag}, using the 
automated functions {\tt CSFKineticTerms} and {\tt VSFKineticTerms},
\begin{verbatim}
  LagKin = Theta2Thetabar2Component[CSFKineticTerms[]] + 
    Theta2Component[VSFKineticTerms[]] + Thetabar2Component[VSFKineticTerms[]]
\end{verbatim}
where all the terms are collected into the variable
{\tt LagKin}.

The superpotential is implemented by translating in terms of \mathematica\ notations
the content of Eq.\ \eqref{eq:wmssm3}, 
\begin{verbatim}
  SPot = yu[ff1,ff2] UR[ff1,cc1] (QL[1,ff2,cc1] HU[2] - QL[2,ff2,cc1] HU[1]) -
    yd[ff1,ff3] Conjugate[CKM[ff2,ff3]] DR[ff1,cc1] * 
     (QL[1,ff2,cc1] HD[2] - QL[2,ff2,cc1] HD[1]) -
    ye[ff1,ff2] ER[ff1] (LL[1,ff2] HD[2] - LL[2,ff2] HD[1]) +
    MUH (HU[1] HD[2] - HU[2] HD[1])]
\end{verbatim}
where the model superfields are denoted by the self-explained symbols
{\tt HU}, {\tt HU}, {\tt QL}, {\tt UR}, {\tt DR}, {\tt LL} and {\tt ER}. 
In these command lines, 
the Yukawa couplings are represented by the symbols {\tt yu},
{\tt yd}, {\tt ye}, the $\mu$-parameter of the superpotential
by the symbol {\tt MUH}, and the 
CKM matrix by the symbol \texttt{CKM}. Following the Supersymmetry Les Houches 
Accord conventions \cite{Skands:2003cj, Allanach:2008qq}, the Yukawa matrices are 
flavor-diagonal, their numerical values are related to the Les Houches
blocks {\tt YU}, {\tt YD} and {\tt YE}
and the CKM matrix explicitly appears in the superpotential (see 
Section \ref{sec:mssmewsb} for more details). In these conventions, the real
and imaginary parts of the CKM matrix are implemented as 
separate external parameters, within the Les Houches blocks {\tt
VCKM} and {\tt IMVCKM}, whereas the complete complex matrix is made an internal parameter.
Finally, the Higgs mixing parameter $\mu$ is stored in the block {\tt HMIX}.
The corresponding interaction Lagrangian is derived according to Eq.\ \eqref{eq:Sint},
which is converted to the \mathematica\ notations 
\begin{verbatim}
  LagW = Theta2Component[ SPot ] + Thetabar2Component[ HC[SPot] ]
\end{verbatim}
and the results are stored in the variable {\tt LagW}. 

Still following the Supersymmetry Les Houches Accord conventions, 
the supersymmetry-breaking Lagrangian is implemented 
from Eq.\ \eqref{eq:lmssmbrk2}. The gaugino mass terms are included 
in a variable denoted by {\tt ino}
\begin{verbatim}
  ino = Mx1*bow[s].bow[s] + Mx2*wow[s,k].wow[s,k] + Mx3*goww[s,a].goww[s,a]
\end{verbatim}
where {\tt Mx1}, {\tt Mx2} and {\tt Mx3} are the bino, wino and gluino mass
parameters, respectively. Their numerical value is stored in the 
Les Houches blocks
{\tt MSOFT} and {\tt IMSOFT}, after having split the parameters into their real
and imaginary parts. Finally, in the \mathematica\ line above, the symbols  
{\tt bow}, {\tt wow} and {\tt goww} are associated to the gaugino 
fields. 

The implementation of the scalar mass terms are assigned to a symbol {\tt sca} as
\begin{verbatim}
  sca = - mHu2*HC[hus[ii]]*hus[ii] - mHd2*HC[hds[ii]]*hds[ii] - 
   mL2[ff1,ff2]*HC[LLs[ii,ff1]]*LLs[ii,ff2] -
   mE2[ff1,ff2]*HC[ERs[ff1]]*ERs[ff2] - 
   CKM[ff1,ff2]*mQ2[ff2,ff3]*Conjugate[CKM[ff4,ff3]]*
     HC[QLs[ii,ff1,cc1]]*QLs[ii,ff4,cc1] -
   mU2[ff1,ff2]*HC[URs[ff1,cc1]]*URs[ff2,cc1] -  
   mD2[ff1,ff2]*HC[DRs[ff1,cc1]]*DRs[ff2,cc1] 
\end{verbatim}
where we have introduced two symbols \texttt{mHU2} and \texttt{mHd2}
representing the Higgs mass parameters, their numerical values being stored 
in the {\tt MSOFT} Les Houches block. The 
squark and slepton mass matrices are implemented within the variables 
\texttt{mQ2}, \texttt{mU2}, \texttt{mD2}, \texttt{mL2} and \texttt{mE2}, the numerical
values being stored in Les Houches blocks of the same name\footnote{The imaginary
parts of the matrices are stored into blocks of the same name, but with the prefix 
{\tt IM} appended.}. Following
the Les Houches conventions, the real and imaginary parts of these matrices are 
implemented, like the CKM matrix, as external parameters,  while the complete matrix
is internal. Finally, the bilinear and trilinear scalar soft interactions
deduced from the form of the superpotential are implemented into two variables
{\tt Tri} and {\tt Bil} as 
\begin{verbatim}
  Tri = -tu[ff1,ff2]*URs[ff1,cc1] * 
      (QLs[1,ff2,cc1] hus[2] - QLs[2,ff2,cc1] hus[1]) +
    Conjugate[CKM[ff3,ff2]]*td[ff1,ff2]*DRs[ff1,cc1] * 
      (QLs[1,ff3,cc1] hds[2] - QLs[2,ff3,cc1] hds[1]) +
    te[ff1,ff2]*ERs[ff1] (LLs[1,ff2] hds[2] - LLs[2,ff2] hds[1])

  Bil = -bb*(hus[1] hds[2] - hus[2] hds[1])
\end{verbatim}
In this expression, the $3\times 3$ tensors ${\bf \hat T^u}$, ${\bf \hat T^d}$ and 
${\bf \hat T^e}$ are represented by the symbols {\tt tu}, {\tt td} and {\tt te}, 
respectively. The information on their numerical value is again passed,
after having split their real and imaginary parts, into Les Houches blocks of the 
same name. Furthermore, the bilinear $b$-term is linked to the \mathematica\ symbol
{\tt bb}, an internal parameter whose the dependence on the other parameters is 
fixed by the Higgs potential minimization conditions.   
Collecting all the contributions above, the entire soft supersymmetry-breaking 
Lagrangian, represented by the symbol {\tt LS}, is implemented as 
\begin{verbatim}
  LS = (ino + HC[ino])/2 + sca + Tri + HC[Tri] + Bil + HC[Bil]
\end{verbatim} 

The complete model Lagrangian, stored in the variable {\tt Lag}, 
is eventually given by
\begin{verbatim}
  Lag = LagKin + LagW + LS
\end{verbatim}
The derivation of the Lagrangian is achieved by solving the equation of motions 
for the auxiliary fields so that they are eliminated as described in Section 
\ref{sec:frsusylag}. This step can be performed automatically with the help of the
{\tt SolveEqMotionD} and {\tt SolveEqMotionF} commands,
\begin{verbatim}
  Lag = SolveEqMotionD[ Lag ]

  Lag = SolveEqMotionF[ Lag ]
\end{verbatim}

\subsubsection{Particle mixings}
Rotations of the gauge eigenstates to the physical states of the model 
are implemented using the {\tt Definitions} attribute of the
particle class\footnote{An alternative implementation of the MSSM can be found on
the \feynrules\ website~\cite{FRwebpage}, using the new \feynrules\ module allowing
for automated mass diagonalization~\cite{Alloul:2013fw}.}.
As a first simple example, the $SU(2)_L$ gauge boson redefinitions are
implemented as 
\begin{verbatim}
  Definitions -> {
    Wi[mu_,1] -> (Wbar[mu]+W[mu])/Sqrt[2], 
    Wi[mu_,2] -> (Wbar[mu]-W[mu])/(I*Sqrt[2]), 
    Wi[mu_,3] -> cw Z[mu] + sw A[mu]
  } 
\end{verbatim}
where {\tt A}, {\tt Z} and {\tt W} correspond to the model file definitions of
the physical gauge bosons, as given, \eg, in Ref.\
\cite{Christensen:2008py}, and where {\tt sw} and {\tt cw} are the sine
and cosine of the weak mixing angle, implemented as internal parameters (see Section
\ref{sec:mssmewsb}). 

The redefinitions of the Higgs fields are a bit more involved since 
they include a dependence on the vacuum expectation value of the neutral
fields, in addition to the rotation angles $\alpha$ and $\beta$ 
presented in Section \ref{sec:mssmewsb}.
Following the Les Houches 
conventions \cite{Skands:2003cj, Allanach:2008qq}, the information
on the $\alpha$ angle is passed through a dedicated Les Houches block 
{\tt FRALPHA}\footnote{The Les Houches block name employed for the $\alpha$ angle
is different from the one included in the original Les Houches conventions, denoted by
{\tt ALPHA}. There reason is that we associate, in the \feynrules\ model file, 
a counter with the parameter that is absent in Ref.\ \cite{Skands:2003cj}.}
while
the numerical value of the tangent of the $\beta$ angle is included in the 
block {\tt HMIX}. The rotations are
then explicitly included in the Higgs field class declaration and read, for the 
example of the up-type Higgs doublet $H_u$ labeled by {\tt hus},
\begin{verbatim}
  Definitions -> { 
    hus[1] -> Cos[beta]*H + Sin[beta]*GP, 
    hus[2] -> (vu + Cos[alp]*h0 + Sin[alp]*H0 + 
       I*Cos[beta]*A0 + I*Sin[beta]*G0)/Sqrt[2]  
  } 
\end{verbatim}
In those replacement rules, 
{\tt H}, {\tt A0}, {\tt h0}, {\tt H0}, {\tt GP} and {\tt G0} are the labels
of the (properly declared) physical Higgs fields and unphysical Goldstone bosons.

In the fermionic sector, the real and imaginary parts of the neutralino and chargino 
mixing matrices $N$, $U$ and $V$ 
are considered as input parameters, following the Supersymmetry Les 
Houches conventions. Their numerical values are stored in the Les Houches blocks
({\tt IM}){\tt NMIX}, ({\tt IM}){\tt UMIX} and ({\tt IM}){\tt VMIX}, while 
the complex matrices are implemented as internal parameters. 
These matrices are subsequently employed in the rotation rules, which read, 
taking the example of the wino states 
\begin{verbatim}
  Definitions -> {
    wow[s_,1] :> Module[{i}, (Conjugate[UU[i,1]]*chmw[s,i] + 
        Conjugate[VV[i,1]]*chpw[s,i])/(I*Sqrt[2]) ], 
    wow[s_,2] :> Module[{i}, (Conjugate[UU[i,1]]*chmw[s,i] - 
        Conjugate[VV[i,1]]*chpw[s,i])/(-Sqrt[2]) ], 
    wow[s_,3] :> Module[{i}, -I*Conjugate[NN[i,2]]*neuw[s,i] ] 
  } 
\end{verbatim}
The symbols {\tt chmw}, {\tt chpw} and {\tt neuw} denote the labels of the physical
two-component Weyl fermions $\chi^-$, $\chi^+$ and $\chi^0$, respectively, 
whereas {\tt NN}, {\tt UU} and {\tt VV} are the mixing matrices.
Turning to the Standard Model quarks and leptons, the only rotations to be performed
are those introduced in Eq.\ \eqref{eq:rotsckm}. They are implemented in the \feynrules\
model file as 
\begin{verbatim}
  Definitions -> {
    QLw[s_, 1, ff_, cc_] -> uLw[s,ff,cc], 
    QLw[s_, 2, ff_, cc_] :> Module[{ff2}, CKM[ff,ff2] dLw[s,ff2,cc] ]
  }  
\end{verbatim}
and
\begin{verbatim}
  Definitions -> {
    LLw[s_, 1, ff_] :> Module[{ff2}, PMNS[ff,ff2]*vLw[s,ff2] ], 
    LLw[s_,2,ff_]   -> eLw[s,ff]
  } 
\end{verbatim}
where the Weyl fermion classes labeled by the symbols 
{\tt QLw} and \texttt{LLw} represent, according to our labeling scheme for the 
component fields, the 
fermionic component of the $SU(2)_L$ doublets of superfields 
$Q_L$ and $L_L$. 
Similarly, the symbols {\tt uLw}, {\tt dLw}, {\tt vLw} 
and {\tt eLw} that appear in the right-hand side of the rules 
denote the (two-component) quark and lepton mass eigenstates.
The PMNS matrix, labeled by the symbol {\tt PMNS}, 
is declared like the CKM matrix and is split
into its real and imaginary parts, whose numerical values are
respectively specified within the 
Les Houches blocks {\tt UPMNS} and {\tt IMUPMNS}.

Finally, the last rotations to be implemented concern the sfermion sector
(see Eq.\ \eqref{eq:sfmix}). The four associated rotation matrices $R^u$, $R^d$, $R^e$
and $R^\nu$ are again implemented after splitting them into their real and
imaginary parts, following the Supersymmetry Les Houches Accord conventions and
are stored 
respectively into the Les Houches blocks ({\tt IM}){\tt USQMIX}, ({\tt IM}){\tt DSQMIX},
({\tt IM}){\tt SELMIX} and ({\tt IM}){\tt SNUMIX}. 
As an example, the redefinition of the scalar component of the
superfield $U_R$
\begin{verbatim}
   Definitions -> { 
     URs[ff_, cc_] :> Module[{ff2}, subar[ff2,cc]*RuR[ff2,ff]]
   } 
\end{verbatim}
where {\tt RuR} refers to the three last columns of the mixing matrix $R^u$ and
{\tt su} is the symbol standing for the (properly declared) up-type
squarks.

\subsubsection{Dirac fermions}
Weyl fermions must eventually be reexpressed in terms of their four-component
counterparts (see Eq.\ \eqref{eq:weyltodirac}). 
In order to have \feynrules\ handling this automatically, the {\tt Weyl\-Com\-po\-nents} 
attribute of the particle class must be set appropriately \cite{Butterworth:2010ym}. 
As an example, the relations among the left-handed and right-handed Weyl components
of the Dirac charged lepton field are implemented as 
\begin{verbatim}
   F[1] == { 
     ClassName      -> l, 
     SelfConjugate  -> False, 
     Indices        -> {Index[GEN]}, 
     FlavorIndex    -> GEN, 
     WeylComponents -> {eLw,ERwbar}, 
     ...
   }
\end{verbatim}
where the dots stand for additional options such as those required by Monte
Carlo tools and {\tt eLw} and {\tt ERw} are the (left-handed) Weyl fermions,
{\tt ERwbar} representing thus a right-handed field.

The \feynrules\ function {\tt WeylToDirac} allows us to perform the replacement of
the Weyl fermions in terms of four-component fields at the Lagrangian level. 
However, we first need to
address an issue related to the antifundamental color representation which
the right-handed quark fields lie in, since only a single color representation, 
the fundamental one, is needed when declaring Dirac fermions.
Denoting $T$ and $\bar T$ the fundamental and antifundamental color
representation matrices and using the property $\bar T = - T^t$, the problem is solved
by implementing, in the model file, the instructions,
\begin{verbatim}
  Colourb = Colour  
  Lag = Lag /. { Tb[a_,i_,j_]->-T[a,j,i] } 
\end{verbatim}
Then, we start with an expansion of the $SU(2)_L$ multiplets in terms of 
their components,
\begin{verbatim}
  Lag = ExpandIndices[ Lag , FlavorExpand -> {SU2W, SU2D} ]  
\end{verbatim}
to remove fundamental and adjoint $SU(2)_L$ indices. 
This procedure enforces the $SU(2)_L$ field rotations from the
gauge basis to the mass basis, necessary for the function {\tt
WeylToDirac} to correctly perform the translation to Dirac and Majorana fermions
\cite{Duhr:2011se}. We finally eliminate all Weyl fermions from the Lagrangian
by means of the {\tt WeylToDirac} functions,
\begin{verbatim}
  Lag = WeylToDirac[ Lag ] 
\end{verbatim}

The Lagrangian obtained in this way is suitable either for the calculation of the 
Feynman rules
by means of the function {\tt FeynmanRules} or to be exported to the tools 
linked to \feynrules\ via the functions {\tt WriteCHOutput}, {\tt WriteFeynArtsOutput}, 
{\tt WriteMGOutput}, {\tt WriteSHOutput}, {\tt WriteUFO} or {\tt WriteWOOutput}.

\subsection{The MSSM with $R$-parity violation}
\label{sec:frrpv}
In $R$-parity violating MSSM scenarios described in 
Section \ref{sec:rpv}, 
the superfield content of the theory is by construction identical to
the one of the more standard $R$-parity conserving MSSM. Moreover, for
simplicity, we choose to neglect the bilinear terms included in 
Eq.~\eqref{eq:spotrpv} and Eq.~\eqref{eq:lsoftrpv}. All particle
mixings occurring after electroweak symmetry breaking 
are therefore left unchanged with respect to those of the MSSM. As a consequence, 
the associated \feynrules\ model implementation can be performed 
with minimal efforts, by loading simultaneously
into the \mathematica\ session several model files,
\begin{verbatim}
  LoadModel["mssm.fr", "rpv.fr"];
\end{verbatim}
The modifications specific to $R$-parity violation are implemented all
together 
in a file labeled by {\tt rpv.fr} whereas the file \texttt{mssm.fr} contains the MSSM
implementation described in Section~\ref{sec:FRMSSM}.

First of all, the file {\tt rpv.fr} includes
the declaration of the $R$-parity violating parameters $\hat\lambda$, $\hat \lambda'$, 
$\hlpp$, $\hat T$, $\hat T'$ and $\hat T''$ of Eq.\ \eqref{eq:spotrpv} and 
Eq.\ \eqref{eq:lsoftrpv}.
This follows the standard rules related to the implementation of parameters given
in Section \ref{sec:FRprm} and uses the self-explained Les Houches blocks
({\tt IM}){\tt RVLAMLLE}, ({\tt IM}){\tt RVLAMLQD}, ({\tt IM}){\tt RVLAMUDD}, 
{({\tt IM})\tt RVTLLE}, ({\tt IM}){\tt RVTLQD} 
and ({\tt IM}){\tt RVTUDD}\footnote{As in Section \ref{sec:FRMSSM}, the imaginary parts
of the parameters are stored into blocks whose name is appended with the prefix
{\tt IM}.} to store the numerical values of the 
$R$-parity violating couplings~\cite{Allanach:2008qq}. Secondly, this file also contains the
implementation of the model Lagrangian. The kinetic and gauge interaction terms
are similar to those of 
the MSSM and we therefore only show the superpotential interactions and 
the soft supersymmetry-breaking terms. 

Recalling that the $R$-parity conserving MSSM superpotential has been stored in 
the variable {\tt SPot} (see Section \ref{sec:FRMSSM}), the 
complete $R$-parity violating superpotential (both Eq.\ 
\eqref{eq:wmssm3} and  Eq.\ \eqref{eq:spotrpv}) is implemented as
\begin{verbatim}
  SupW = SPot + 
   LLLE[f1,f2,f3] Conjugate[PMNS[f4,f1]] * LL[1,f4] LL[2,f2] ER[f3] +
   LLQD[f4,f5,f3] Conjugate[CKM[f2,f5]] Conjugate[PMNS[f1,f4]] * 
      DR[f3,c1] (LL[1,f1] QL[2,f2,c1] - LL[2,f1] QL[1,f2,c1]) + 
   1/2 LUDD[f1,f2,f3] Eps[c1,c2,c3] UR[f1, c1] DR[f2,c2] DR[f3,c3]
\end{verbatim}
This drives interaction terms collected into a variable represented by the 
symbol {\tt LagW}, obtained by issuing 
\begin{verbatim}
  LagW = Theta2Component[ SupW ] + Thetabar2Component[ HC[SupW] ]
\end{verbatim}
In the definition of the superpotential above, the $\hat \lambda$, $\hat\lambda'$ and 
$\hat\lambda''$ couplings are represented by the symbols {\tt LLLE}, {\tt LLQD} and
{\tt LUDD}, respectively, while {\tt Eps} stands for
the fully antisymmetric tensor of rank three and {\tt LL}, {\tt ER}, {\tt QL},
{\tt UR} and {\tt DR} are, as in Section \ref{sec:FRMSSM}, the symbols 
respectively associated with the chiral superfields $L_L$, $E_R$, $Q_L$, $U_R$ and
$D_R$. 
 
In Section \ref{sec:FRMSSM}, we have implemented the $R$-parity conserving 
soft supersymmetry-breaking Lagrangian in the 
variable {\tt LS}. Therefore, it is enough to add the extra contributions 
of Eq.~\eqref{eq:lsoftrpv}, with the exception of the bilinear terms, 
\begin{verbatim}
  LSoft = LS + Tsoft + HC[Tsoft]
\end{verbatim}
into a new variable denoted by the symbol {\tt LSoft}, the quantity
{\tt Tsoft} being defined by 
\begin{verbatim}
  Tsoft = TLLE[f1,f2,f3] Conjugate[PMNS[f4,f1]] * LLs[1,f4] LLs[2,f2] ERs[f3] -
     TLQD[f4,f5,f3] Conjugate[CKM[f2,f5]] Conjugate[PMNS[f1,f4]] * 
         DRs[f3,c1] (LLs[1,f1] QLs[2,f2,c1]-LLs[2,f1] QLs[1,f2,c1]) - 
     1/2 TUDD[f1,f2,f3] Eps[c1,c2,c3] * URs[f1, c1] DRs[f2,c2] DRs[f3,c3]
\end{verbatim} 
The symbols {\tt TLLE}, {\tt TLQD} and {\tt TUDD} appearing in {\tt Tsoft} 
stand for the $\hat T$, $\hat T'$
and $\hat T''$ parameters of the soft-supersymmetry breaking Lagrangian of Eq.\
\eqref{eq:lsoftrpv} while {\tt
LLS}, {\tt ERs}, {\tt QLs}, {\tt URs} and {\tt DRs} denote the scalar component of
the $L_L$, $E_R$, $Q_L$, $U_R$ and $D_R$ superfields, respectively. 

The complete model Lagrangian, represented by the variable {\tt Lag}, 
is thus given by
\begin{verbatim}
  Lag = LagKin + LagW + LSoft
\end{verbatim}
In order to render this Lagrangian compliant with the requirements
of the Monte Carlo programs further linked to \feynrules, the auxiliary $F$- and
$D$-fields must be integrated out. In addition, all Weyl 
fermions are eventually replaced by their four-component
counterparts. These steps are achieved by issuing 
\begin{verbatim}
  Lag = SolveEqMotionD[ Lag ]
  Lag = SolveEqMotionF[ Lag ]
  Colourb = Colour
  Lag = Lag /. { Tb[a_,i_,j_]->-T[a,j,i] }
  Lag = ExpandIndices[ Lag , FlavorExpand -> {SU2W, SU2D} ]
  Lag = WeylToDirac[ Lag ]
\end{verbatim}
We refer to Section \ref{sec:FRMSSM} for more information about this set of commands.

\subsection{The minimal $R$-symmetric supersymmetric theory}
\label{sec:frmrssm}

Unlike the $R$-parity violating MSSM implementation presented in the previous section, 
the \feynrules\ implementation of the minimal $R$-symmetric supersymmetric
model described in Section \ref{sec:mrssm} cannot be performed by simply adjoining to 
the MSSM model file a new file with all the novelties. The mixing
relations, as well as the nature of the neutralino and gluino fields, have changed,
so that the {\tt mssm.fr} file must be deeply modified. Therefore, we start from a
copy of this file and update it accordingly. 

The implementation of the five new chiral superfields of Table \ref{tab:rgauge}
and the one of the new model free parameters strictly follow the rules presented in Chapter 
\ref{chap:FR}. For the sake of the example, we show and describe
the implementation of the $\tilde\Phi_G$ chiral superfield, 
relevant for the phenomenological
investigations performed in this work focusing on sgluon fields.
This superfield is declared in the \feynrules\ model file
as shown in Section \ref{sec:spacemod}, by including in the {\tt M\$Superfields} list 
the \mathematica\ equality
\begin{verbatim}
  CSF[100] == {
    ClassName -> SGL,
    Chirality -> Left,
    Scalar    -> sigG,
    Weyl      -> gopw,
    Indices   -> { Index[Gluon] } 
  }
\end{verbatim}
The replacement rules above allow us to assign the symbol {\tt SGL} to 
the $\tilde\Phi_G$ superfield and {\tt sigG} and {\tt gopw} to the sgluon
$\sigma_G$ and gluino $\tilde g'$ component fields, respectively. The latter 
can be declared together with the other fields of the model, within the {\tt 
M\$ClassesDescription} list (see Section \ref{sec:FRfields}), 
\begin{verbatim}
  S[100] == { 
    ClassName     -> sigG,
    Unphysical    -> True, 
    SelfConjugate -> False,
    Indices       -> { Index[Gluon] },
    Definitions   -> { sigG[aa_] -> (sig1[aa] + I sig2[aa])/Sqrt[2] } 
  }

  S[101] == { 
    ClassName     -> sig1, 
    SelfConjugate -> True,
    Indices       -> { Index[Gluon] },
    Mass          -> Msig1,
    Width         -> Wsig1
  }

  S[102] == { 
    ClassName     -> sig2, 
    SelfConjugate -> True,
    Indices       -> { Index[Gluon] },
    Mass          -> Msig2,
    Width         -> Wsig2
  }

  W[100]== { 
    ClassName     -> gopw,
    Unphysical    -> True, 
    Chirality     -> Left,
    SelfConjugate -> False,
    Indices       -> {Index[Gluon]}
  }
\end{verbatim}
Whilst {\tt sigG} represents the complex scalar field $\sigma_G$, \ie,
the component field of the chiral superfield $\tilde\Phi_G$, 
we have introduced the symbols {\tt sig1} and {\tt sig2} to respectively 
label its real scalar and pseudoscalar degrees of freedom since the different mass 
terms included in the soft supersymmetry-breaking Lagrangian of Eq.\ 
\eqref{eq:mrssmsoft} lead to their splitting.

In order to declare the physical four-component gluino field as a Dirac fermion,
it is enough to associate it with two different Weyl components by means of 
the \texttt{WeylComponents} attribute of the particle class.
Denoting by {\tt goww} the gaugino component of the vector superfield
$V_G$, after having absorbed a phase as shown in Eq.\ \eqref{eq:newdirac},
we include in the gluino declaration the rule 
\begin{verbatim}
  WeylComponents -> {goww, gopwbar}
\end{verbatim}

The complexity of the implementation of the mixing relations 
among the model gauge eigenstates being not that different as for the MSSM, 
they are therefore omitted from this manuscript and we refer to Section 
\ref{sec:FRMSSM} and the model implementation available from the \feynrules\
webpage~\cite{FRwebpage}
for technical details.

In addition, the implementation of the $R$-symmetric 
supersymmetric Lagrangian
is also similar to what has been performed for the two examples 
of Section \ref{sec:FRMSSM} and Section \ref{sec:frrpv}.
All kinetic and gauge interaction terms are implemented by issuing
 \begin{verbatim}
  LagKin = Theta2Thetabar2Component[ CSFKineticTerms[ ] ] + 
    Theta2Component[VSFKineticTerms[]] + Thetabar2Component[VSFKineticTerms[]]
\end{verbatim}
whereas the superpotential is included by translating Eq.\ \eqref{eq:wmrssm2} 
into the \mathematica\ declaration of a variable {\tt SuperW},
\begin{verbatim}
   SuperW =  ...
     -luB/2 (HU[1] PhiB RU[2] - HU[2] PhiB RU[1]) + 
      ldB/2 (HD[1] PhiB RD[2] - HD[2] PhiB RD[1]) + 
     luW PhiW[a] (HU[1] Ta[a,2,i] RU[i] - HU[2] Ta[a,1,i] RU[i]) + 
     ldW PhiW[a] (HD[1] Ta[a,2,i] RD[i] - HD[2] Ta[a,1,i] RD[i]) + 
     MUu (HU[1] RU[2] - HU[2] RU[1]) + 
     MUd (HD[1] RD[2] - HD[2] RD[1]) ]
\end{verbatim} 
In this expression, the dots stand for the trilinear Yukawa interactions 
of Eq.\ \eqref{eq:wmrssm1} that are identical to those of the MSSM (see Section
\ref{sec:FRMSSM}). In addition, 
the $SU(2)_L$ invariant contractions have been expanded and {\tt Ta} are the symbols
representing the fundamental representation matrices of $SU(2)_L$. 
The quantities {\tt HU},
{\tt HD}, {\tt RU} and {\tt RD} are the names of the classes associated with the
\mbox{($R$-)Higgs} superfields,
while {\tt PhiB} and {\tt PhiW} are those of the superfields $\tilde \Phi_B$ and 
$\tilde\Phi_W$. Finally, we denote the superpotential $\lambda$-parameters by {\tt
luB}, {\tt ldB}, {\tt luW} and {\tt ldW} whilst the $\mu$-parameters are
taken as {\tt MUu} and {\tt MUd}. As usual, we omit for brevity 
the description of the declaration of these parameters 
that is standard and refer, for more information, to Section \ref{sec:FRprm}. 
Like in Section \ref{sec:FRMSSM} and Section \ref{sec:frrpv}, the associated 
interaction Lagrangian is implemented as
\begin{verbatim}
  Lag = Theta2Component[SuperW] + Thetabar2Component[HC[SuperW]]
\end{verbatim}

Finally, the soft-supersymmetry breaking Lagrangian of Eq.\ \eqref{eq:mrssmsoft}
being given in terms of superfields, it can be directly implemented by
employing standard functions of the superspace extension of \feynrules\ (see
Section \ref{sec:spacemod}). As an example,  the gluino Dirac mass term 
could be implemented into a variable denoted by {\tt mgluino} by typing 
\begin{verbatim}  
    mgluino  = MG1/(2 gs) Ueps[be,al] * 
      nc[theta[al], SuperfieldStrengthL[GSF, be, a], PhiG[a]]
\end{verbatim}
where the symbol {\tt MG1} is associated with the product of the soft
supersymmetry-breaking mass by the vacuum expectation value of the
spurion superfield $W'$, \ie, it is equal to $m_1^G v_D$. The corresponding
Lagrangian, represented by the variable {\tt Lino}, is given by  
\begin{verbatim}  
    Lino = Theta2Component[mgluino] + Thetabar2Component[HC[mgluino]
\end{verbatim}
Al the other supersymmetry-breaking terms are implemented in a similar
fashion and details are left out of this document for brevity.

\mysection{From \feynrules\ to \madgraph\ 5}\label{sec:frmg} 

As already briefly mentioned in Section \ref{sec:bmssm2}, we consider 
in this work two phenomenological analyses performed in the framework of 
two different non-minimal supersymmetric theories.
First, we focus on the MSSM with $R$-parity violation and 
concentrate on the $\hat\lambda^{\prime\prime}_{3jk}$
superpotential interactions which are 
still almost unconstrained by experimental data in the case the lightest
neutralino is lighter than the top quark. Next, we dedicate our efforts 
to the study of the LHC sensitivity to the presence of sgluons 
in the context of the minimal $R$-symmetric version of the MSSM. This model, 
where sgluons dominantly couple to top quarks, is presently not addressed by
any of the current sgluon experimental LHC searches which assume ${\cal O}(1)$
couplings to light quarks and gluons. 

Among the whole set of existing automated Monte Carlo tools such as 
\comphep/\calchep~\cite{Pukhov:1999gg, Boos:2004kh,
Pukhov:2004ca, Belyaev:2012qa}, \mgme\ \cite{Stelzer:1994ta,
Maltoni:2002qb, Alwall:2007st, Alwall:2008pm, Alwall:2011uj}, \sherpa\
\cite{Gleisberg:2003xi, Gleisberg:2008ta} or
\whizard~\cite{Moretti:2001zz, Kilian:2007gr} that allow us to address
phenomenological studies at colliders, most all of them
contain restrictions on the set of supported color and Lorentz structures. 
While any structure that appears in the Standard Model or in the MSSM is 
in general allowed, vertices with non-standard color and/or Lorentz structures 
are most of the time not fulfilling the tools requirements 
and must therefore be discarded from the model implementations.

One possibility to overcome such a constrain is to 
compute the relevant squared matrix elements by hand and
implement the results into non-automated tools such as \herwig\
\cite{Corcella:2000bw,Corcella:2002jc,Bahr:2008pv,Arnold:2012fq} or \pythia\
\cite{Sjostrand:2000wi,Sjostrand:2006za,Sjostrand:2007gs}. Performing Monte Carlo
simulations of processes with a final states 
containing more than two particles is thus rather tedious.
There is however another option, that we adopt in this work, 
which relies on the flexibility of the UFO format~\cite{Degrande:2011ua}
employed by the \madgraph\ 5 generator \cite{Alwall:2011uj}.  
As stated in Section \ref{sec:ufo}, the UFO is by design agnostic of any 
restriction on the Lorentz and color structures allowed to appear in the interaction 
vertices. On the same lines, the \madgraph\ 5 program makes use of this strength
to compute automatically predictions for any new physics
theory, renormalizable or not and possibly containing non-standard 
structures\footnote{Although sextet and antisextet color representations are 
supported by \madgraph\ 5, in contrast to most of the other
publicly available tools, there is not any program capable so far to handle
color representations with a higher multiplicity.}. 

The UFO-\madgraph\ 5 setup is thus suitable for the two studies aimed to be 
performed in this work, both Lagrangians involving
non-standard color structures. The $R$-parity violating superpotential 
of Eq.\ \eqref{eq:spotrpv} includes a color structure
where three fields lying in the (anti)fundamental representation of $SU(3)_c$ 
are connected by means of a fully antisymmetric tensor, 
whereas the sgluon effective Lagrangian of Eq.\ 
\eqref{eq:Leff} contains interactions where three fields lying in
the adjoint representation of the QCD gauge group are connected through
the symmetric structure constants of $SU(3)_c$.

The Monte Carlo event generator \madgraph\ 5 allows for the automated
generation of tree-level matrix elements associated with any scattering 
processes, in particular occurring in proton-proton collisions as to be produced 
at the LHC, in a very efficient way. 
The main task left to the user consists of specifying the process of interest in
terms of initial and final state particles together with the considered particle 
physics model, the collision setup (including, \eg, the energy of the 
colliding beams) and a set of basic event selection criteria related to 
the analysis of interest. In addition, a UFO version of the 
model under investigation has to be provided by the user if not already 
included in the model library of \madgraph\ 5 which is 
built upon the \feynrules\ model database \cite{FRwebpage}. 
We recall that in order to convert a \feynrules\ model implementation into
its UFO version, an automated interface is 
included within the public version of \feynrules\ and can be called
by typing, in a \mathematica\ session, 
\begin{verbatim}
   WriteUFO[ Lag ]  
\end{verbatim}
where the variable {\tt Lag} contains the model Lagrangian,
expressed in terms of the usual fields of particle physics and
in the model mass eigenbasis. The {\tt WriteUFO}
function internally calls the \feynrules\ core method {\tt Feyn\-man\-Ru\-les} in
order to compute the interaction vertices of the model. They are subsequently 
expanded into a color $\otimes$ spin basis as in Eq.\ \eqref{eq:ufovert} and
exported, together with the rest of the model information, into the set of \python\ 
files described in Section \ref{sec:ufo}.
The output can eventually be loaded into \madgraph\ so that the user is able to
use the model, for event generation, as any other built-in model implementation. 

Once the user specifies a process, \madgraph\ 5 internally calls the 
\aloha\ package \cite{deAquino:2011ub}. This program generates from the UFO model
a series of subroutines, inspired by 
the \helas\ library \cite{Murayama:1992gi,Hagiwara:2008jb,Hagiwara:2010pi,%
Mawatari:2011jy}, allowing for the computation of helicity amplitudes
related to the process under consideration. These
amplitudes include helicity wave-functions associated with specific
substructures that can be 
further reused within different Feynman diagrams. This consequently 
leads to an efficient evaluation of the associated squared matrix elements. 

Supersymmetric theories contain, in their most general form, more than
several thousands of vertices. 
Taking the example of the MSSM, the $6\times6$ sfermion mixings of Eq.\ 
\eqref{eq:sfmix} lead to ${\cal O}(1000)$ four-scalar interactions, most of them
being flavor-violating. In the framework of the benchmark scenarios usually 
investigated in supersymmetry phenomenology (see, \eg, Section 
\ref{sec:mssm_indirect}), flavor violation in the sfermion sector is drastically 
restricted so that a large part of those ${\cal O}(1000)$ interaction vertices 
has to be vanishing or negligible.
At the level of the Monte Carlo generators, the explicit presence of 
such a large number of zero vertices considerably slows down event
generation, since they must be loaded into the computer memory on run time and 
diagrams with a vanishing contribution 
are generated. Therefore, it may be suitable to remove
these vertices from the model implementation.
This task can be done in an automatic way directly within \feynrules, 
with the help of the {\tt WriteRestrictionFile} and {\tt LoadRestriction} commands.

To this aim, the numerical values of all the model parameters must be 
firstly loaded by issuing, in the \mathematica\ session,
\begin{verbatim}
  ReadLHAFile[Input -> "susy.dat"]
\end{verbatim}
The file {\tt susy.dat} contains the model mass 
spectrum and particle mixings, together with the numerical values 
of the external parameters
provided in a form compatible with the Les Houches structure implemented
in the \feynrules\ model implementation. Then, the detection of the vanishing 
parameters is performed by issuing the commands 
\begin{verbatim}
  WriteRestrictionFile[ ]
 
  LoadRestriction["ZeroValues.rst"]
\end{verbatim}
The {\tt WriteRestrictionFile} method scans over the whole set of internal and 
external parameters of the model and generates a file, dubbed 
{\tt ZeroValues.rst}, with a list of \mathematica\ replacement rules mapping 
all the vanishing parameters to zero. The {\tt LoadRestriction} function reads 
this file and loads the list of rules
into the current \mathematica\ session. The \feynrules\ interfaces
subsequently apply the replacement rules to each vertex before  
writing it to the output files,
the zero contributions being in this way dropped before 
being translated to the considered Monte Carlo generator model format. 
In addition, vertices numerically evaluated 
to zero are ignored and not outputted.

After this optimization, hundreds of the remaining vertices still 
consist in four-scalar interactions which are, for tree-level computations, 
most of the time phenomenologically less relevant.
Therefore, the efficiency of the Monte Carlo tools can be highly 
improved by discarding these vertices from the output Monte Carlo model 
files. This task can be done automatically at the \feynrules\ level by means of 
the {\tt Exclude4Scalars} option of the interfaces.
In the UFO case, one would issue, in \mathematica, the command
\begin{verbatim}
 WriteUFO[ Lag, Exclude4Scalars -> True ]
\end{verbatim}

One must however keep in mind that the model files
including the two series of optimizations presented in this section
are not fully general. They instead depend on
the considered benchmark scenario (defined here in the file {\tt susy.dat}),
even though at the \feynrules\ level, the model implementation is as general as
possible.

\mysection{Monotop production in the MSSM with $R$-parity violation}
\label{sec:rpvmonotop}

\renewcommand{\arraystretch}{1.3}
\begin{table}[!t]
  \begin{center} \begin{tabular}{|cccccc|}
    \hline $M_t$ [GeV] & $M_b$ [GeV] &$M_Z$ [GeV]& $G_F$ [GeV$^{-2}$] &
      $\alpha_s(M_Z)$& $\alpha(M_Z)^{-1}$  \\
    \hline
    173.2 & 4.2 & 91.1876 & $1.16637 \times 10^{-5}$ & 0.1176 & 127.934 \\ 
   \hline\hline
    $m_0$ [GeV]  & $m_{1/2}$ [GeV] & $A_0$ [GeV] & 
      $\tan\beta$ & $\text{sign}(\mu)$ &
      $\hat\lambda''_{312}$ \\
    \hline
    100 & 400 & 0 & 10 & $>0$ & 0.2 \\ 
    \hline 
   \end{tabular}\end{center} 
  \caption{\label{tab:rpvbenchmark} 
    Input parameters associated with the chosen benchmark scenario for the
    $R$-parity violating supersymmetric explorations performed in this section.
    We recall that the supersymmetric parameters 
    are defined at the grand unification scale and that all other $\hlpp$-parameters,
    together with the $\hat\lambda$, $\hat\lambda'$, $\hat T$, $\hat T'$ and 
    $\hat T''$ couplings are taken equal to zero.}
\end{table}
\renewcommand{\arraystretch}{1.}

\subsection{Benchmark scenario and process of interest}
Specific MSSM benchmark scenarios have 
been recently proposed by the supersymmetry working groups of the ATLAS and 
CMS collaborations, together the LHC Physics Center at CERN \cite{AbdusSalam:2011fc}. 
We adopt one of those benchmark
scenarios, suitable for $R$-parity violating supersymmetry and lying 
along the so-called `RPV3-line' of the $R$-parity violating MSSM parameter
space. This point is derived from the cMSSM and 
defined by the usual four free parameters, 
one sign and the Standard Model inputs (see Section \ref{sec:mssmbrkex}). To these
inputs, one supplements the value
of one $\hlpp$ coupling that we choose to be $\hlpp_{3jk}$ in our case. 
This setup, where only a single $R$-parity violating parameter 
is non-vanishing, is not uncommon and inferred from the single coupling dominance
hypothesis \cite{Barger:1989rk, Dimopoulos:1988jw} often adopted when studying 
$R$-parity violation in supersymmetry. 

We fix the top quark pole mass to $M_t = 173.2$ GeV \cite{Lancaster:2011wr}
and the bottom quark mass to $M_b(M_b) = 4.2$ GeV. The electroweak sector
is defined by setting the $Z$-boson mass to $M_Z =
91.1876$ GeV, the Fermi constant to $G_F = 1.16637 \times
10^{-5}$ GeV$^{-2}$ and the electromagnetic coupling constant at the
$Z$-pole to $\alpha(M_Z)^{-1} = 127.924$, according to  
the 2010 Particle Data Group Review \cite{Nakamura:2010zzi}.
Finally, the strength of the strong interactions is determined from the $Z$-pole
value  
$\alpha_s(M_Z) = 0.1176$ \cite{Nakamura:2010zzi}. In the supersymmetric sector, 
we use a universal scalar mass of $m_0=100$ GeV, a universal gaugino mass of 
$m_{1/2}=400$ GeV and a universal
trilinear coupling set to $A_0=0$ GeV. The ratio of the vacuum expectation
values of the neutral component of the two Higgs doublets is taken as
$\tan\beta=10$, whilst the $\mu$-parameter is chosen positive and 
$\hat\lambda^{\prime\prime}_{312} = 0.2$ at the electroweak scale.

With this choice of parameters, summarized in Table \ref{tab:rpvbenchmark}, 
we employ the \spheno\ 3 package \cite{Porod:2011nf} to numerically evaluate
the model parameters at the electroweak scale by means of renormalization group 
running at the two-loop level (see Section \ref{sec:rge}).
In addition, \spheno\ 3 allows us to 
extract the particle spectrum and mixings at the two-loop level for the Higgs
sector and at the one-loop level for the other particles. 
Among the electroweak superpartners, the sleptons are found fairly light,
with masses of ${\cal O}(200-300)$ GeV, while the neutralino and chargino masses
range from 160 GeV for the lightest neutralino, being the lightest
supersymmetric particle, to 550 GeV for the heavier
states. The masses of the colored superparters are found larger,
ranging from 650 GeV to 900 GeV, the gluino being heavier than all squarks.

Since it is lighter than the top quark, the lightest neutralino can only 
decay to four-body final states and 
has a long lifetime. Therefore, when produced (directly or indirectly) 
at colliders such as the LHC, the neutralinos $\tilde\chi_1^0$ 
escape the detector invisibly \cite{Allanach:1999bf} so that 
bounds derived from standard supersymmetry searches (see Section 
\ref{sec:direct}) apply.
Consequently, by the time this manuscript was being completed,
the chosen benchmark scenario has been excluded by the most recent limits on 
the masses of the first and second generation squarks as well as by those on the 
gluino mass. There are
two obvious manners to restore agreement with data. Firstly, one can move along the 
`RPV3-line' and adopt a benchmark scenario with heavier squark and gluino 
masses, which is also attractive from the point of view of the experimental 
Higgs results. Secondly, one can leave the minimal picture and make the gluino and the 
down and strange squarks heavier without modifying the rest of the spectrum. 
However, the phenomenological results derived below
are mainly related to the mass of the lightest top squark, a quantity barely
affected by any of the two choices above for small modifications from
the original benchmark point. Therefore, we choose to keep the original
scenario and will consider that the results of Section \ref{sec:rpvpheno} 
will still be acceptable with a good approximation, even for similar but more realistic,
experimentally not excluded, benchmarks.

\begin{figure}\centering
  \vspace{-1.5cm}
  \includegraphics[width=\columnwidth]{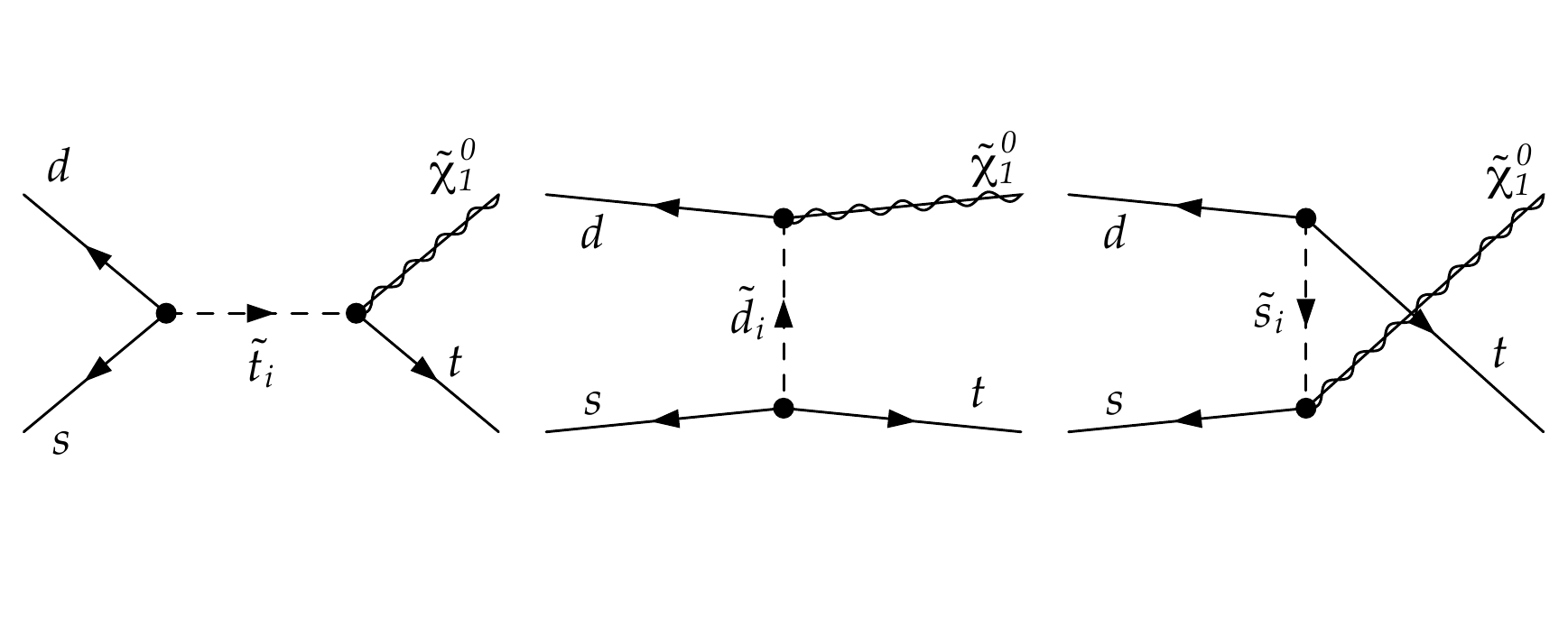}
  \vspace{-2.0cm}
  \caption{\label{fig:monotopdiag}
  Feynman diagrams associated with $R$-parity violating monotop
  production when the interactions related to the 
  $\hat\lambda^{\prime\prime}_{312}$ superpotential parameter are switched on. These
  diagrams have been generated by means of the program {\sc FeynArts} \cite{Hahn:2000kx}.}
\end{figure}

The $R$-parity violating MSSM scenario depicted above allows for the associated
production of a top quark with a neutralino, as shown in the Feynman diagrams of 
Figure \ref{fig:monotopdiag}. This signature, where a top quark is produced 
in association with missing energy, has been recently dubbed monotop
\cite{Andrea:2011ws}. It consists of a clear sign of new physics (although possibly
different from $R$-parity violating supersymmetry) since there is 
no process in the Standard Model that can lead to it at tree-level, the dominant
production mode being suppressed both by a loop factor and by the GIM mechanism.
In the benchmark scenario 
under consideration, the production cross section as computed by
\madgraph\ reaches about 300 fb when
employing the CTEQ6L1 set of parton densities \cite{Pumplin:2002vw} and for a 
center-of-mass energy of 7 TeV. 

In the next subsection, we will show that such a large cross section 
can lead to observable hints of new physics at the LHC, already for 
a center-of-mass energy of 7 TeV, a low luminosity of a few fb$^{-1}$, 
and after following 
a simple selection strategy. Then, instead of determining the LHC reach
to monotop production induced by $R$-parity violating supersymmetry, we will
make use of the designed search strategy to extend the analysis well beyond the framework of 
non-minimal supersymmetry and investigate
in Chapter \ref{chap:effective},  
monotop production in the context of an effective 
field theory to be interpreted within several beyond the Standard 
Model theories.

\subsection{Phenomenological investigations at 7 TeV}
\label{sec:rpvpheno}

In this chapter, we concentrate on event simulation for the LHC running
at a center-of-mass energy of 
$\sqrt{s} = 7$ TeV and for an integrated luminosity of 4 fb$^{-1}$.
Concerning both signal and background events, hard scattering matrix elements
are calculated with the automated Monte Carlo event generator 
\madgraph\ 5 \cite{Alwall:2011uj} and convoluted with the leading order set of 
the CTEQ6 parton density fit \cite{Pumplin:2002vw}. Moreover, both
renormalization and factorization scales are identified to the transverse 
mass of the produced (massive) particles. The events generated in this way 
are then matched to parton showering and hadronization as provided by the 
\pythia\ program. The version 6 of this code \cite{Sjostrand:2006za} is used 
for background events\footnote{In order to obtain more accurate predictions for 
the background, we merge matrix elements containing additional
hard jets according to the Mangano (MLM) merging procedure 
\cite{Mangano:2006rw,Alwall:2008qv}. We however concentrate, in this section,
on the physics results
and omit all technical details concerning background simulation. For the latter,
we refer to the 8 TeV analyses presented in Chapter \ref{chap:effective}. 
With the exception of the total rates, the energy of the beams
and the number of generated events,
the Monte Carlo setup is similar for both analyses.}, 
whilst the version~8~\cite{Sjostrand:2007gs} is employed for the $R$-parity
violating monotop signal due to the exotic color structure not compliant
with the requirements of \pythia\ 6. Fast detector simulation is eventually 
performed by means of the program \delphes\ \cite{Ovyn:2009tx}, using the publicly
available CMS detector card, and jet reconstruction is ensured by using the 
\fastjet\ package \cite{Cacciari:2005hq,Cacciari:2011ma} that contains
an anti-$k_{t}$
algorithm whose radius parameter is fixed to $R=0.5$~\cite{Cacciari:2008gp}. The
phenomenological analysis presented below is performed with the \madanalysis\ 5 package
\cite{Conte:2012fm}.

Monotop production can be classified according to the top quark decays,
\be
  p p \to t + \widetilde\chi^0_1 \to  b j j + \slashed{E}_T
    \qquad \text{or} \qquad 
  p p \to t + \widetilde\chi^0_1 \to  b \ell + \slashed{E}_T \ ,
\ee
where $j$ and $b$ denote light and $b$-jets, respectively, and $\ell$
a charged lepton. The missing transverse energy $\slashed{E}_T$ is associated with
the lightest neutralino $\widetilde \chi_1^0$ escaping the detector invisibly
due to its long lifetime, and also to the neutrino for
leptonically decaying top quarks. Since
leptonic monotops induced by $R$-parity violating
supersymmetry have been already investigated in the past 
\cite{Berger:1999zt,Berger:2000zk}, we focus instead on monotop 
events where the top quark decays hadronically.

\begin{figure}[!t]\centering
  \includegraphics[width=.48\columnwidth]{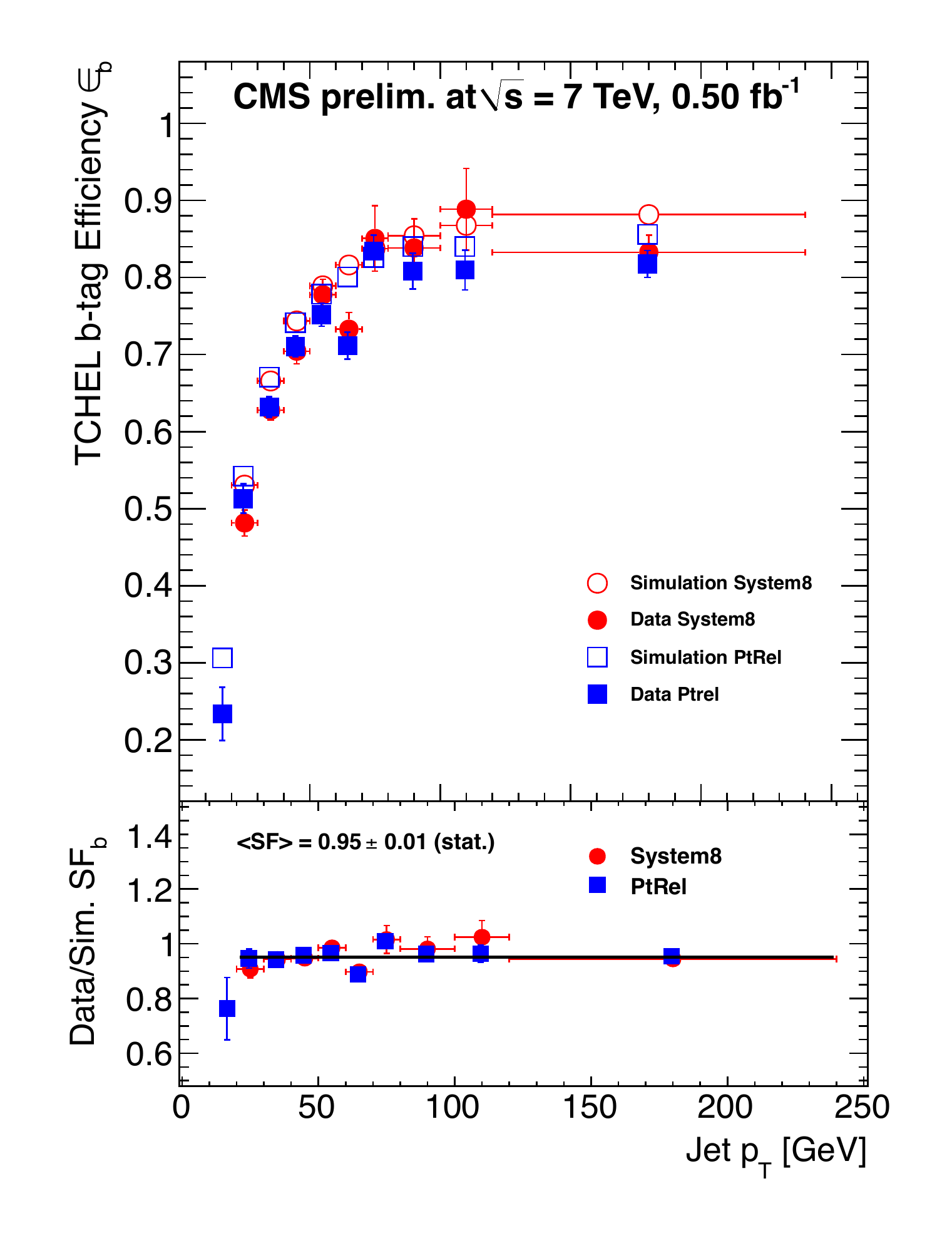}
  \includegraphics[width=.48\columnwidth]{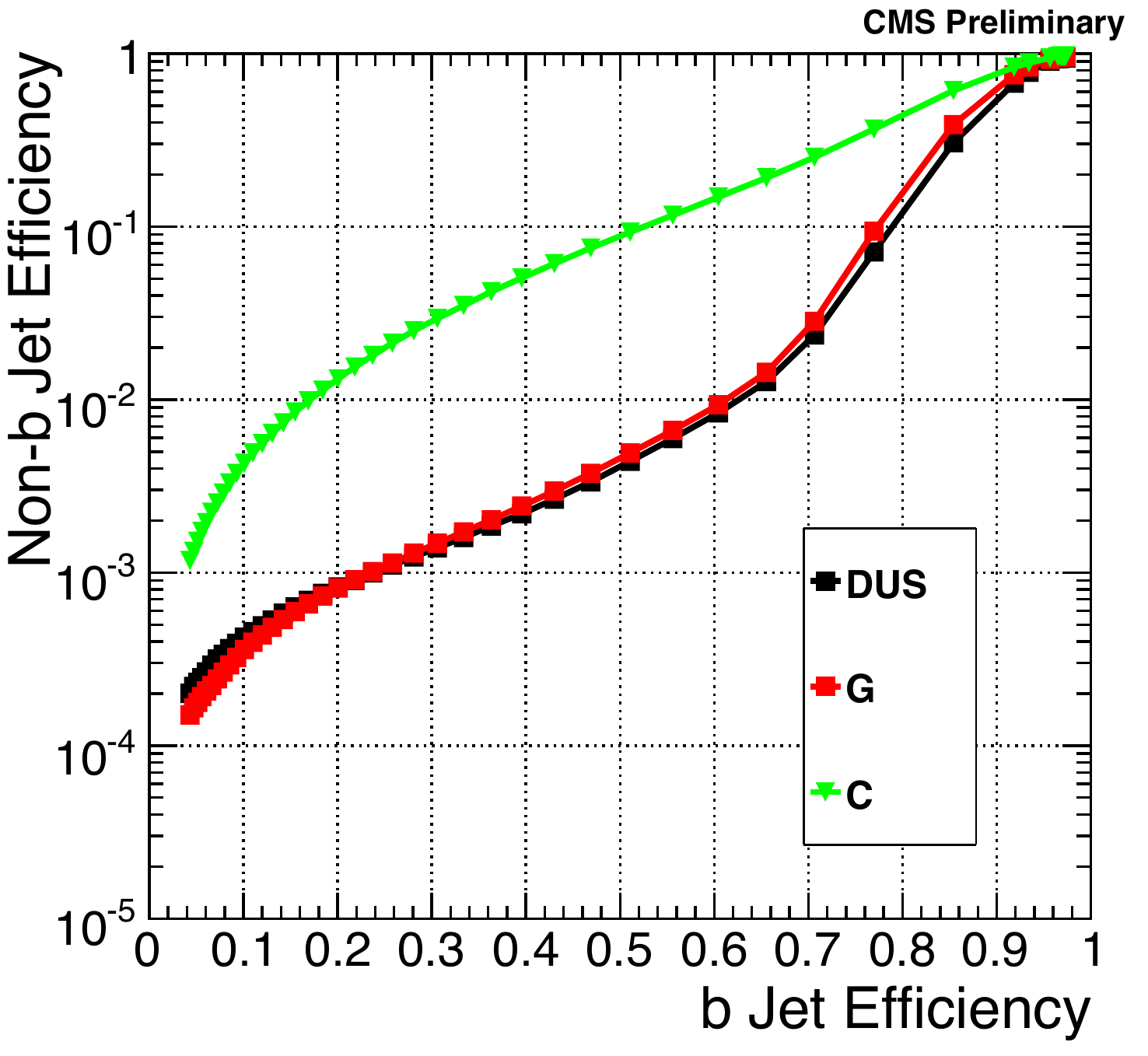}
  \caption{\label{fig:btag} Efficiency of tagging a jet originating
  from the fragmentation of a $b$-quark as a $b$-jet (left) and associated mistagging rates
  (right). They depend
  on the transverse momentum of the jet and follow the  
  `high efficiency $b$-tagging' algorithm of CMS (TCHEL). In the upper
  panel of the left figure (taken from Ref.\ \cite{CMS:2011cra}), the 
  measured and simulated $b$-tagging efficiencies from different methods are
  presented, while the ratio between data and simulation is shown on its lower panel.  
  In the right panel of the Figure (taken from Ref.\ \cite{CMS:2009gxa}),
  the associated mistagging rates are indicated, as a function of the $b$-tagging
  efficiency.} 
\end{figure}

The only source of irreducible Standard Model background 
consists of the production of an invisibly decaying $Z$-boson
together with at least three jets, one of them being originated from the fragmentation
of a $b$-quark.
However, many sources of instrumental background have also to be considered. 
On the one hand, multijet events with fake missing energy exactly mimic the signal.
On the other hand, events originating from the production of a $W$-boson, 
$t\bar{t}$ pair or a weak boson pair in association with jets 
contribute as well to the background when the leptons originating 
from the top quark and weak boson decays are
non-reconstructed. Finally, 
single top events including non-reconstructed or misrecontructed jets
must be considered too.

\begin{figure}\centering
  \includegraphics[width=.75\columnwidth]{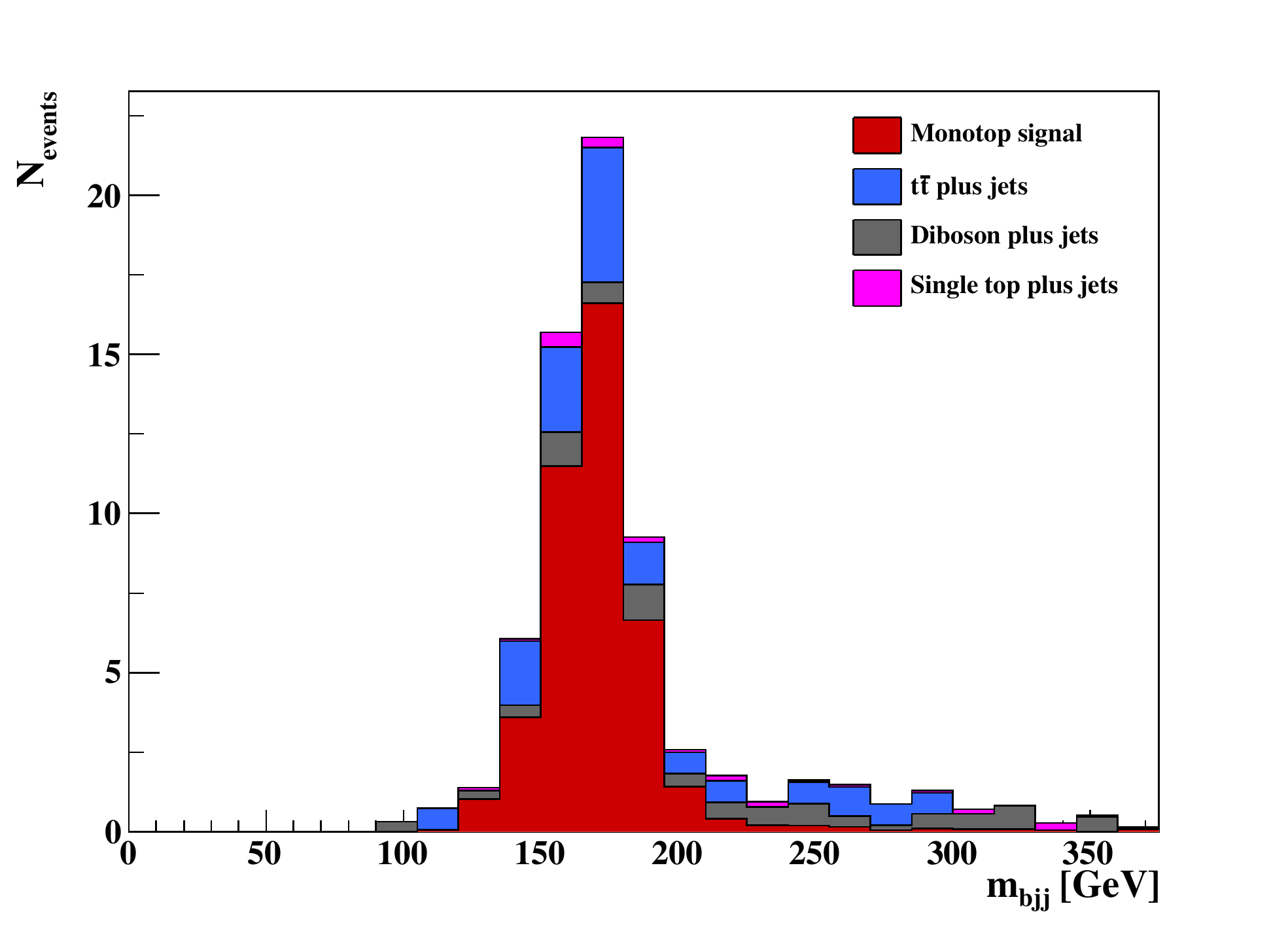}
  \caption{\label{fig:rpvmonotop}
  After applying the monotop selection strategy presented in the text, 
  we show the invariant-mass distribution 
  of the three jets $m_{bjj}$ both for the signal (red) and the dominant sources
  of background issued from single top (purple), $t \bar t$ (blue) and diboson (gray)
  processes.}
\end{figure}

It has been shown, in recent experimental analyses, that a simple 
selection strategy (as the one performed below)
allows to keep a good control over the
background~\cite{daCosta:2011qk,Collaboration:2011ida}. Inspired, in addition, 
by the parton-level results of Ref.\ \cite{Andrea:2011ws}, we preselect events
containing a large amount of missing transverse energy $\slashed{E}_T > 200$ GeV,
where
\be
  \slashed{E}_T = \bigg|\bigg| \sum_{\text{visible particles}} \vec p_T \bigg|
    \bigg| \ .
\label{eq:met}\ee
We then impose a veto on the presence of any charged lepton (electron or muon) 
with a transverse momentum $p_T \geq 10$ GeV and a pseudorapidity $|\eta|\leq 2.5$. 
These selections have been found not to affect the signal 
but sensibly reduce the contributions of the $t\bar t$, $Z$-boson, 
and $W$-boson events.
In a second stage, we exploit the presence of a hadronically decaying top quark
and demand exactly
one $b$-tagged jet with a transverse momentum $p_T \geq 50$ GeV and a
pseudorapidity $|\eta| \leq 2.5$, as well as exactly two light jets with a
transverse momentum $p_T \geq 30$ GeV and a pseudorapidity $|\eta| \leq 2.5$.
In our analysis, we estimate a $b$-tagging efficiency 
depending on the transverse momentum as on 
Figure \ref{fig:btag}  (left panel), together
with a charm and light jet mistagging rate as illustrated on the right panel
of the figure. This corresponds to an 
efficiency of correctly tagging a jet with a transverse momentum of 50 GeV 
as a $b$-jet of about 70\%, whilst the mistagging rate of a charm (light) jet
as a $b$-jet is of about 25 \% (2\%). 
Since the two selected 
light jets are issued from the decay of a $W$-boson, we constrain their 
invariant mass to be compatible with the $W$-mass, \ie,
lying in the $65$~GeV$-95$ GeV range.

The distribution of the invariant-mass of the three jets $m_{bjj}$ is
presented on Figure \ref{fig:rpvmonotop}. After applying all the selection
criteria described above, the remaining background contributions 
consists of $t \bar t$, diboson and single top events, while all other sources 
of background, such as $W$-boson or
$Z$-boson (or non-simulated multijet) events, are reduced to a(n expected to be)
barely visible level and
thus not presented in the figure. 
After further constraining  the system of the three selected jets
by requiring their invariant-mass to be compatible with the mass of the
top quark, lying in a 40 GeV mass window centered around the top mass, we 
obtain the number of events shown in Table
\ref{tab:monotop}, both for the signal and the dominant background contributions.

Defining the LHC
sensitivity to a monotop signal induced by $R$-parity violating supersymmetry
as the number of selected signal events over the total number of
selected events $S/\sqrt{S+B}$, the adopted benchmark scenario  
leads to a possible hint for 
monotops at the $4.95 \sigma$ level.
Conversely, a $3\sigma$-deviation from the Standard Model expectation can
already be 
observed for any value of the $R$-parity violating parameter 
$\hat\lambda^{\prime\prime}_{312} \geq 0.11$, assuming the supersymmetric spectrum 
to be unchanged. Since the
number of signal events is not expected to drastically change 
for moderate superpartner masses (below or around the TeV scale), the 
standard monotop search strategy presented above is expected to be sufficient 
to probe $R$-parity violating supersymmetric monotop signatures in
large regions of the cMSSM parameter space. This statement will be confirmed in the next
chapter.

\renewcommand{\arraystretch}{1.3}
\begin{table}[!t]
  \begin{center}\begin{tabular}{|c|c|}
    \hline  Event sample & Number of selected events\\
    \hline  \hline
    Top-antitop pair plus jets & $8.2 \pm 2.3$ \\ 
    Diboson plus jets           & $2.7 \pm 0.7$ \\
    Single top                  & $0.9 \pm 0.3$ \\
   \hline
    Total background & $11.8 \pm 2.4$ \\
    Monotop signal   & $33.2 \pm 1.0$ \\
   \hline
   \end{tabular} \end{center}
  \caption{\label{tab:monotop} 
    Number of selected events after applying the monotop search strategy
    described in the text for the different
    background contributions and for the signal. Since approximately no $W$-boson, 
    $Z$-boson and multijet event is passing the selection criteria, these channels 
    are not indicated in the table. These results correspond to 4 fb$^{-1}$ of LHC
    collisions at a center-of-mass energy of 7 TeV.}
\end{table}
\renewcommand{\arraystretch}{1.}

\mysection{Sgluon-induced multitop production in $R$-symmetric supersymmetry}
\label{sec:mrssmsgluons}

\subsection{Benchmark scenario and process of interest}
\label{sec:benchmrssm}

We now turn to an investigation of
some phenomenology related to sgluon fields dominantly coupling to top 
quarks as predicted in $R$-symmetric supersymmetric models.
Therefore, all superpartners and the numerous Higgs fields are irrelevant, which
motivates us to conceive
a practical benchmark scenario where all
mixing matrices related to the sfermion, neutralino, chargino and Higgs
sectors are set to zero so that the related particles do not play any role.
In the aim of using the event generator \madgraph~5,
the corresponding interaction vertices have been
removed from the UFO
model files generated by \feynrules\ 
by means of the optimization procedure described in Section
\ref{sec:frmg}. 

Our benchmark scenario is defined by fixing 
the Standard Model inputs,
together with the parameters related to the sgluon field, \ie, its mass $M_\sigma$ 
and its couplings to quarks $a_q^L$, $a_q^R$ (with $q=u$ and 
$d$) and gluons $a_g$ introduced in Eq.\ \eqref{eq:Leff}. Although in principle, 
the sgluon mass $M_\sigma$ depends on several soft parameters, \ie, $m_{\tilde\Phi_G}$,
$M_{\tilde\Phi_G}$ and ${\cal M}_{\tilde\Phi_G}$ (see Eq.\ \eqref{eq:lrsoft1}
and Eq.\ \eqref{eq:mrssmsoft}), and is different for scalar and pseudoscalar sgluons, 
we simplify the approach by decoupling the pseudoscalar degree of freedom 
and collecting the three contributions to the scalar mass 
into a single parameter $M_\sigma$\footnote{We recall that the
numerical value of the  mass of a particle is specified,
in a \feynrules\ model description, at the time of the
particle class declaration independently of the
Lagrangian mass terms.}. As already mentioned in Section
\ref{sec:mrssm}, recent ATLAS and CMS 
analyses have constrained the sgluon mass to be larger than 2 TeV 
\cite{Aad:2011yh, Aad:2011fq, ATLAS:2012nna, CMS:2012eza}. These limits 
however only hold when the sgluon field couple to light quarks and gluons 
with ${\cal O}(1)$
interaction strengths, and we therefore
evade this bound by setting all effective parameters to zero,
with the exception of $(a_u^L)^3{}_3 = (a_u^R)^3{}_3 = 3\times 10^{-3}$ and 
$a_g = 1.5\times 10^{-6}$ GeV$^{-1}$. These numbers are obtained by making use of the
explicit calculations of the relevant loop diagrams in Ref.~\cite{Plehn:2008ae},
and correspond to a benchmark scenario where squarks and gluinos have typical masses 
of about $1-2$~TeV and where non-minimal flavor violation in the squark sector is not
allowed.
Lower sgluon masses are thus viable and we set $M_\sigma= 500$ GeV.
The chosen numerical values for the model parameters are 
summarized in Table \ref{tab:sglbenchmark}, which also includes the relevant Standard
Model inputs taken from the 2010 Review of the Particle Data Group~\cite{Nakamura:2010zzi}.

\renewcommand{\arraystretch}{1.3}
\begin{table}[!t]
  \begin{center}\begin{tabular}{|ccc|cccc|}
    \hline  $M_t$ [GeV] &$M_Z$ [GeV] & $\alpha_s(M_Z)$ & 
       $M_\sigma$ [GeV] & $(a_u^L)^3{}_3$ &$(a_u^R)^3{}_3$ & $a_g$ [GeV$^{-1}$] \\
    \hline
    173.1 & 91.1876 & 0.1176 & 500 & $3\times 10^{-3}$ & $3\times 10^{-3}$ & $1.5\times 10^{-6}$ \\ 
   \hline
   \end{tabular}\end{center}
  \caption{\label{tab:sglbenchmark}
    Input parameters associated with the chosen benchmark scenario for the $R$-symmetric 
    supersymmetric exploration performed in this work.
    We recall that the superpartners are decoupled and
    irrelevant, and that all the omitted effective couplings are taken
    vanishing.}
\end{table}
\renewcommand{\arraystretch}{1.}

\begin{figure}\centering
  \vspace{-.3cm}
  \includegraphics[width=\columnwidth]{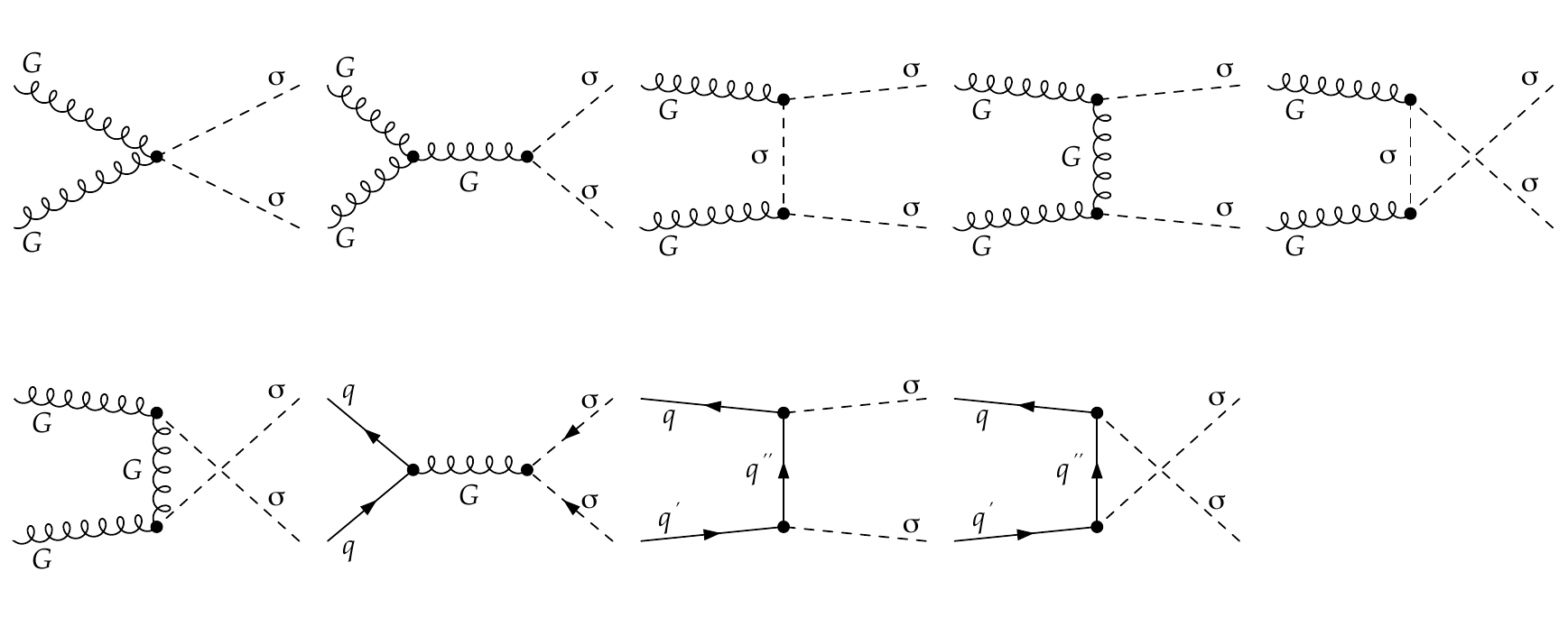}
  \vspace{-.9cm}
  \caption{\label{fig:sgluondiag}
    Tree-level Feynman diagrams associated with sgluon pair production
    at hadron colliders. These diagrams correspond to the interactions included
    in the Lagrangians of Eq.~\eqref{eq:Lkin} and Eq.~\eqref{eq:Leff} and have
    been generated by means of the program \feynarts~\cite{Hahn:2000kx}.} 
\end{figure}

In this scenario, the total sgluon-pair production cross section, as calculated
by \madgraph~5~\cite{Alwall:2011uj} 
after convoluting the hard scattering matrix-elements related to
the Feynman diagrams of Figure \ref{fig:sgluondiag} with the CTEQ6L1 set of
parton densities \cite{Pumplin:2002vw}, reach
the level of 0.20~pb. Including sgluon decays to a pair of top quarks, 
sgluon-induced four-top production occurs with a rate 
of about 42 fb, \ie, more than 140 times the (leading-order) Standard Model
predictions of 0.3 fb. Whereas the value of the effective
sgluon-gluon-gluon coupling of $1.5\times 10^{-6}$ GeV$^{-1}$ may seem
very small, the related diagrams contribute to the sgluon-pair 
production cross section up to about 15\%. 

\subsection{Phenomenological investigations at 7 TeV}
\label{sec:mrssmpheno}
Four-top production leads to
final states enriched in jets and leptons that originate from the top decays. 
Therefore, the main sources of Standard Model
background is expected to be related to rare processes with a high final state
multiplicity, such as the production of a top-antitop pair in association with one
or several gauge bosons or with jets. We generate both signal and background events
by employing the same setup as in Section \ref{sec:rpvpheno} and refer
to this section for more information.

We preselect events containing exactly two charged leptons with 
a transverse momentum $p_T \geq 20$ GeV and a pseudorapidity $|\eta| \leq 2.5$.
In addition, we impose them 
to be isolated so that electrons and muons 
at a relative distance $\Delta R  = \sqrt{\Delta\varphi^2 + \Delta\eta^2}
\leq 0.2$ from a jet are rejected, $\varphi$ standing for the azimuthal angle 
with respect to the beam direction. To ensure a good rejection of the background,
dominated by $t \bar t$ and Drell-Yan events, and to maintain
at the same time a important signal efficiency (of about 50\% in our case), we 
require that the two leptons carry the same electric charge\footnote{In our 
simplified detector simulation performed with \delphes, the charge of a
lepton is always correctly identified, contrary to simulation software employed
by the LHC experiments.}.
Moreover, leptonic top decays always imply
missing transverse energy $\slashed{E}_T$, so that we only keep events
with $\slashed{E}_T \geq 40$ GeV.

\begin{figure}\centering
  \includegraphics[width=.75\columnwidth]{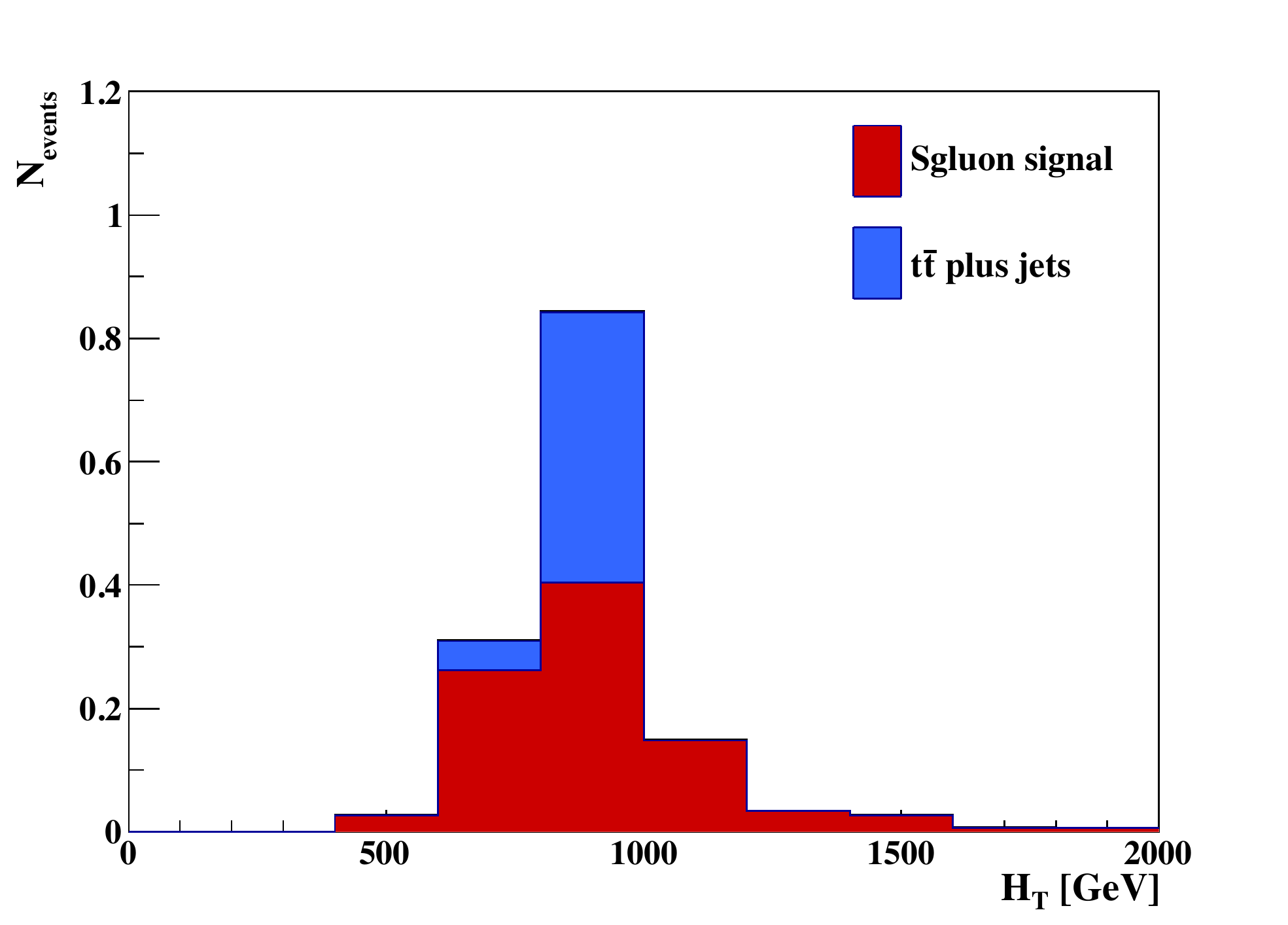}
\caption{\label{fig:4top} After applying the sgluon search strategy in 
  the dilepton channel presented in the text, we present the distribution of 
  the $H_T$ variable defined in Eq.\ \eqref{eq:htdef} for both the signal
  (red) and the dominant source of background consisting of $t \bar t$ events 
  (blue).}
\end{figure}

The rest of the proposed sgluon search strategy benefits from the important 
jet multiplicity specific to signal events. 
In particular, we
expect at least four $b$-tagged jets (one for each of the produced top quarks) 
and four additional light jets originating from the hadronically decaying top quarks.
Consequently, we demand the selected events to contain at least eight jets with
a transverse energy $E_{T} \geq 20$ GeV, and impose that at least three of them
are $b$-tagged, the $b$-tagging efficiency and the corresponding mistagging rates
being defined as in Section \ref{sec:rpvpheno}. 
The important hadronic activity in the final state suggests to
employ the $H_T$ variable, defined by 
\be\label{eq:htdef}
  H_T = \sum_{\text{jets, leptons, missing energy}} 
    \big|\big| \vec p_T \big| \big| \ ,
\ee
as a discriminating variable between signal and backgrounds. The results are
presented in Figure \ref{fig:4top} and in Table \ref{tab:sgluon}.
The dominant contributions to the background consist in $t\bar t$ events,
as well as, in a smaller extent (and therefore not shown), 
in events issued from the production and decay
of a $t \bar t$ pair accompanied by one or several gauge bosons.
Although the background rejection is efficient,
a 4 fb$^{-1}$ luminosity of 7 TeV collisions is unfortunately not sufficient 
to obtain a good sensitivity to the sgluon signal, at least for a 500~GeV sgluon mass.
This naive analysis has therefore to be improved, a task addressed in the next
chapter.

Two remarks are in order here. First, our event selection
criteria may seem very restrictive, in particular concerning the number 
of required jets and $b$-tags. However,
these selections are mandatory to ensure a good background rejection, as shown
in Ref.\ \cite{Fuks:2012im} where the effects or requiring 
different numbers of jets and $b$-tags have been investigated. Next, 
in our simulation setup, the multijet background, jets faking leptons and charge
misidentification have 
not been accounted for. On the basis of the analysis of Ref.\ 
\cite{ATLAS:2012hpa} where same sign dilepton events are investigated
after selection criteria similar to those applied in this analysis, these
sources of background have been found to be dominant. This
issue is addressed more into details in Chapter \ref{chap:effective}, as
this does not change the conclusions of the 7~TeV analysis of this section.

\renewcommand{\arraystretch}{1.3}
\begin{table}[!t]
  \begin{center}\begin{tabular}{|c|c|}
    \hline  Event sample & Number of selected events \\
    \hline  \hline
    Top-antitop pair plus jets & $0.5 \pm 0.3$ \\ 
   \hline
    Total background & $ 0.5 \pm 0.3$ \\
    Sgluon signal   & $0.9  \pm 0.1$ \\
   \hline
   \end{tabular} \end{center}
  \caption{Number of selected events after applying the dilepton sgluon
  search strategy presented in the text for the different
  background contributions and for the signal. Since approximately no 
  event related to all the other sources of background is passing the selection criteria,
  these channels are not indicated in the table. The results 
  correspond to an integrated luminosity of
  4 fb$^{-1}$ of proton-proton collisions at the LHC collider, running with a
  center-of-mass energy of 7 TeV.\label{tab:sgluon}} 
\end{table}
\renewcommand{\arraystretch}{1.}

\cleardoublepage

%% file: effective.tex
\label{chap:effective}

The search, in particular at the LHC, of tracks of new phenomena 
moves in several directions. The most beaten path, as illustrated in Chapter
\ref{chap:nonmin}, lies on a top-down approach. In this case, 
a theory extending the Standard Model of particle physics
is conceived on the basis of fundamental theoretical principles, such as an extended
symmetry group or additional spacetime dimensions. This theory is constructed in a way to reproduce
the Standard Model in the low energy limit and
to possibly address one or more of the Standard Model open issues, such as the hierarchy problem 
like in supersymmetry (see Section \ref{sec:hierarchy}).
Predictions of physical observables can then be made by making use
of perturbation theory. However, many new parameters, that cannot be fixed by 
experimental constraints, usually enter the calculations. For the sake of the examples,
we recall the large size of the parameter space of the models presented
in Chapter \ref{chap:mssm} and in Chapter \ref{chap:nonmin}, these models featuring
up to several hundreds of free parameters
in their general form. Benchmark scenarios must therefore be carefully designed in order to
be compatible with (most of) current data and they subsequently imply
typical signatures for the model that can be searched for in present and future experiments,
such as the missing energy requirement in supersymmetry.

While widely used, this theory-driven approach has notable limitations.
First, signatures are neither typical of a given benchmark nor
of a specific
model itself. For instance, universal extra dimensions and supersymmetry can have
very similar signatures involving cascade decays of heavy particles.
Next, the top-down approach can lead to strong biases 
in the experimental analyses. Therefore, it is important to pursue
a more pragmatic approach where beyond the Standard Model explorations based on a 
bottom-up construction of new physics models are supplemented to the more common 
top-down inspired analyses.

In the following, we employ the tool of effective 
low energy theories to explore different  scenarios built from the two experimental
signatures investigated in Section~\ref{sec:rpvpheno} and Section~\ref{sec:mrssmpheno}.
We analyze in this way several classes of models simultaneously for monotop searches
(Section~\ref{sec:monotopsEFT} and Section~\ref{sec:effmonotops}) and
sgluon-induced multitop final states (Section~\ref{sec:simpsgl} and Section~\ref{sec:effmultitops}).
Additionally, we detail in Section~\ref{sec:mc} our simulation setup for the Standard Model
background.

\mysection{Effective theories inspired by non-minimal supersymmetry} 
\subsection{An effective field theory for monotop production}
\label{sec:monotopsEFT}

\begin{figure}[t]\centering
  \includegraphics[width=.33\columnwidth]{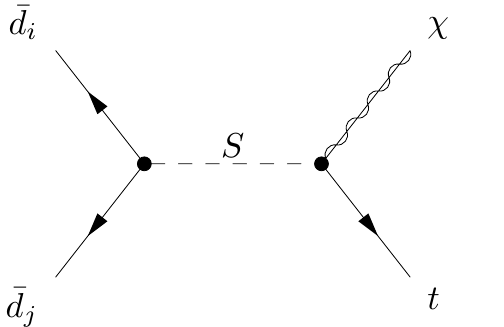}
  \includegraphics[width=.33\columnwidth]{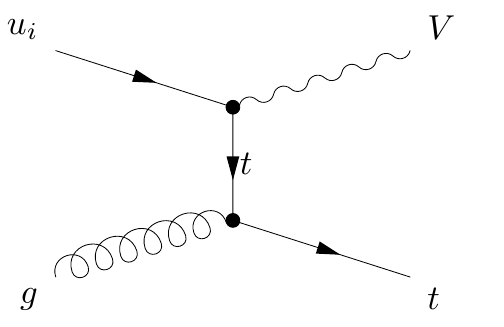}
  \caption{\label{fig:effmonotop} Representative Feynman diagrams leading to a monotop
    signature, either through the resonant exchange of a colored
    scalar field $S$ (left) or via a flavor-changing 
    interaction with a vector field $V$ (right). The $V$ and $\chi$ particles are
    here invisible and lead to missing energy.}
\end{figure}

Beyond supersymmetry and at tree-level, monotop production can occur 
via two main mechanisms. Either the top quark is produced, possibly resonantly, 
in association with
an invisible fermionic state (see, \eg, the representative Feynman
diagram shown on the left panel of Figure \ref{fig:effmonotop})
or through a flavor-changing interaction with an
invisible bosonic state (see, \eg, the representative Feynman diagram shown
on the right panel of Figure \ref{fig:effmonotop}). 

Within the first production mechanism,
the top quark is produced together
with an undetected fermion which we denote, in the following, by $\chi$. Possible diagrams
occur via $s$-channel (see Figure \ref{fig:effmonotop}), $t$-channel and $u$-channel
exchanges of a scalar ($S$) or vector ($V$) field lying in the (anti-)fundamental 
representation of $SU(3)_c$. As shown in Section \ref{sec:rpvmonotop}, 
such processes appear in $R$-parity-violating supersymmetry where,
similarly to the case discussed in Ref.\ \cite{Desai:2010sq}, the intermediate
particle is a (possibly on-shell) top squark and $\chi$ consists of the lightest neutralino,  
\be
  \bar d \bar s \to \tilde t_i \to t \tilde \chi_1^0 \ . 
\ee
In the limit of very heavy resonances, monotops can also be seen as produced through a baryon
number-violating four-fermion effective interaction \cite{Morrissey:2005uza, Dong:2011rh}. 
More exotic cases can involve invisible Rarita-Schwinger fields,
as in supersymmetric theories containing a spin-$3/2$ gravitino field, or a multiparticle
state with a global half-integer spin, as in hylogenesis scenarios for dark
matter \cite{Davoudiasl:2011fj}.

In the second class of models yielding a monotop signature,
the missing energy is carried by a neutral
bosonic state, either long-lived or decaying invisibly. Monotops are
arising from quark-gluon initial states undergoing a flavor-changing
interaction, as discussed, \eg, in Ref.~\cite{delAguila:1999ac}. In this case,
the missing energy can be either a spin-zero $S$, spin-one $V$ (see Figure \ref{fig:effmonotop})
or spin-two $G$ state,
\be
 u g \to t S \ , \quad u g \to t V \quad  \text{ or }\quad u g \to   t G \ ,
\ee
or can also be a continuous state containing an even number of fermions,
as in $R$-parity conserving supersymmetry with non-minimal
flavor violation (see Section \ref{sec:mssm_indirect}) \cite{Allanach:2010pp}, 
\be
 u g \to \tilde u_i \tilde\chi_1^0 \to t \tilde \chi_1^0 \tilde \chi_1^0 \ .
\ee

The properties of the produced top quark are induced by the features of the production
mechanism.
First, the partonic content of the
initial state and the nature (mass and spin) of the undetected recoiling object
play a key role. Next, the
possible presence of intermediate resonant states could alter
kinematical distributions such as the transverse-momentum spectrum of the final state
particles.
Following the spirit of
Ref.\ \cite{Alves:2011wf}, these considerations
suggest a model-independent approach including all the
cases within a single simplified theory.
Assuming the strong interactions to be flavor-conserving, as in the Standard Model, 
the flavor-changing neutral interactions are bound to the weak sector. 
For simplicity,  we also neglect
spin-two gravitons, since their flavor-changing interactions are
loop-suppressed \cite{Degrassi:2008mw}, as well as any of their
excitations which do not lead to a
missing energy signature.
Along the same lines, we do not consider spin-$3/2$ fields since their (flavor-violating
or not) couplings
are, at least in supersymmetric theories, suppressed by a high-energy scale. 
In addition,  four-fermion interactions are omitted as they are known not to
lead to a visible LHC signal \cite{Andrea:2011ws}. 

In our construction of an effective Lagrangian for monotop production, we denote by
$\phi$, $\chi$ and $V$ the possible scalar, fermionic and vector
particle leading to missing energy, respectively, and by  $\varphi$ and $X$ scalar and
vector fields lying in the fundamental representation of $SU(3)_c$ possibly inducing
resonant monotop production. The Lagrangian describing
the interactions of those fields is supplemented to the
Standard Model Lagrangian and reads, in the mass basis,
\be\label{eq:effmonotoplag}\bsp
 \lag &=\ \lag_{\rm kin} 
  + \bigg[\phi \bar u \Big[a^0_{FC}\!+\!b^0_{FC} \gamma_5 \Big] u \!+\!
     V_\mu \bar u \Big[a^1_{FC} \gmu \!+\! b^1_{FC} \gmu \gamma_5 \Big] u  \\
  &+\! \epsilon^{ijk} \varphi_i \bar d^c_j 
       \Big[a^q_{SR} \!+\! b^q_{SR} \gamma_5 \Big] d_k \!+\!
     \varphi_i \bar u^i \Big[a^{1/2}_{SR} \!+\! b^{1/2}_{SR} \gamma_5 \Big] \chi
  \\ & + \epsilon^{ijk} X_{\mu,i}\ \bar d^c_j 
       \Big[a^q_{VR}\gmu + b^q_{VR} \gmu\gamma_5 \Big] d_k
   + X_{\mu,i}\ \bar u^i 
       \Big[a^{1/2}_{VR} \gmu + b^{1/2}_{VR} \gmu \gamma_5 \Big] \chi + 
       {\rm h.c.} \bigg] \ ,
\esp\ee
where kinetic and gauge interaction terms for the new states are 
included in $\lag_{\rm kin}$ and where the indices $i$, $j$ and $k$ represent
color indices in the fundamental representation of $SU(3)_c$. Additionally,
flavor indices have been understood. The $3\times 3$ 
matrices (in flavor space) $a^{\{0,1\}}_{FC}$
and $b^{\{0,1\}}_{FC}$ contain quark interactions with the bosonic
invisible particles $\phi$ and $V$, while $a^{1/2}_{\{S,V\}R}$ and
$b^{1/2}_{\{S,V\}R}$ denote the couplings between up-type quarks,
the invisible fermion $\chi$ and the new colored states $\varphi$ and $X$. Moreover,
gauge invariance also allows the latter to couple to down-type 
quarks, the corresponding interaction strengths being given by 
the matrices $a^q_{\{S,V\}R}$ and $b^q_{\{S,V\}R}$. 

To illustrate the main features of monotop production, we consider
a series of simplified scenarios in which all axial couplings involving new particles vanish,
\be
  b = 0\ .
\ee
Furthermore, we only retain interactions that can be enhanced by parton densities
and set 
\be\bsp
  (a^0_{FC})_{13} = (a^0_{FC})_{31} = 
    (a^1_{FC})_{13} = (a^1_{FC})_{31} = a \ , \\
    (a^q_{SR})_{12} = \ -(a^q_{SR})_{21} = (a^{1/2}_{SR})_3 =
    (a^q_{VR})_{11} = (a^{1/2}_{VR})_3 = a \ ,
\esp \ee
the other couplings being fixed to zero.
Within the above settings, we define four scenarios, the first two, which we denote
by {\bf S.I} and {\bf S.II}, addressing resonant monotop
production and the last two, which we denote by {\bf S.III} and {\bf S.IV},
focusing on monotop production via baryon-number conserving but flavor-changing
interactions.
We now turn to the evaluation of monotop production cross section at the LHC,
running at a center-of-mass energy of 8~TeV and employ the QCD factorization theorem
to convolute the associated leading order squared matrix elements with the
leading order set of the CTEQ6 parton density fit
\cite{Pumplin:2002vw}, fixing both the renormalization and factorization scales
to the transverse mass of the monotop system. To this aim, we implement the Lagrangian
above into \feynrules,
and export the model into a UFO library which is then linked
to \madgraph~5.

\begin{figure}[t!]
  \centering
  \includegraphics[width=.49\columnwidth]{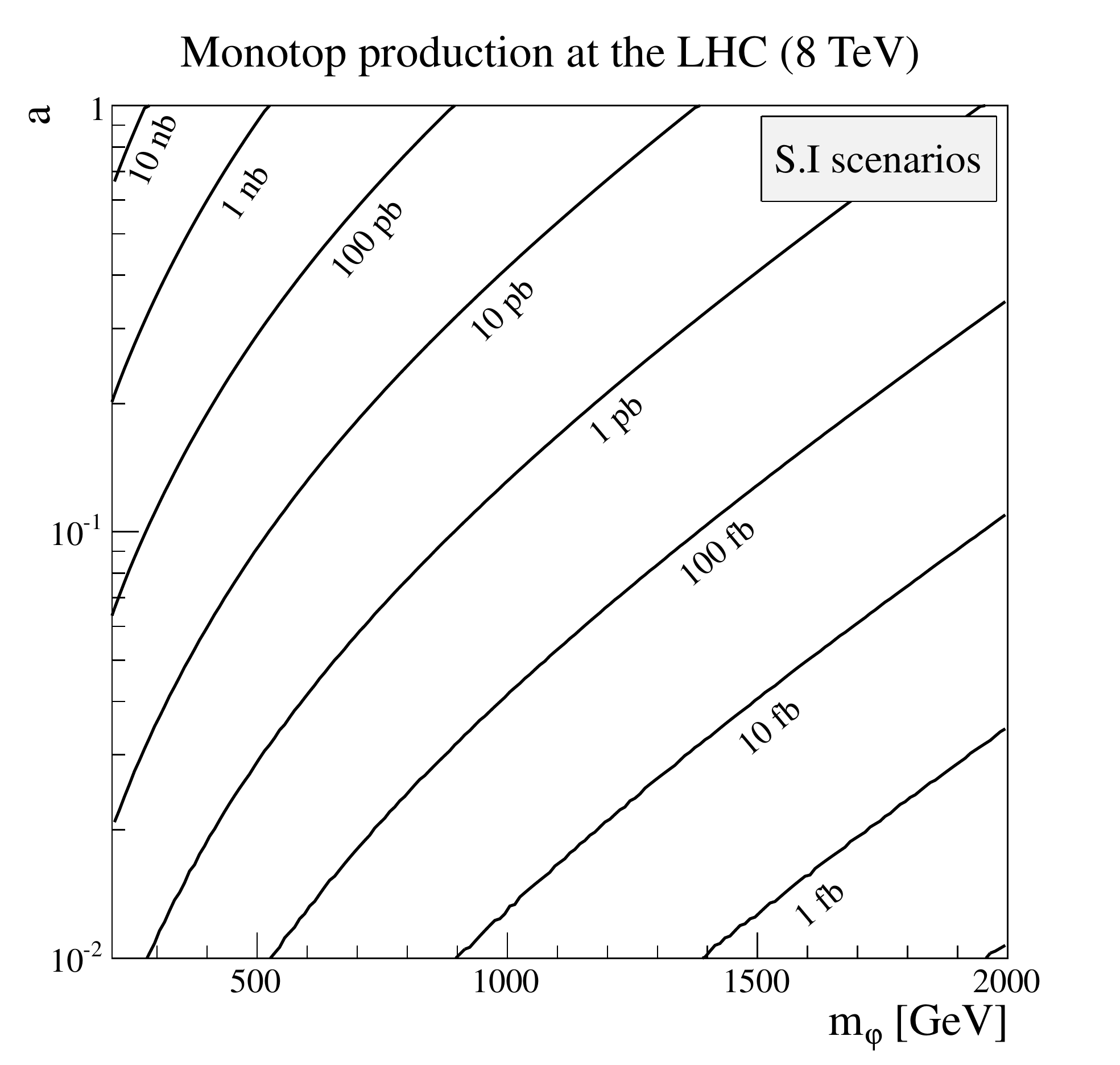}
  \includegraphics[width=.49\columnwidth]{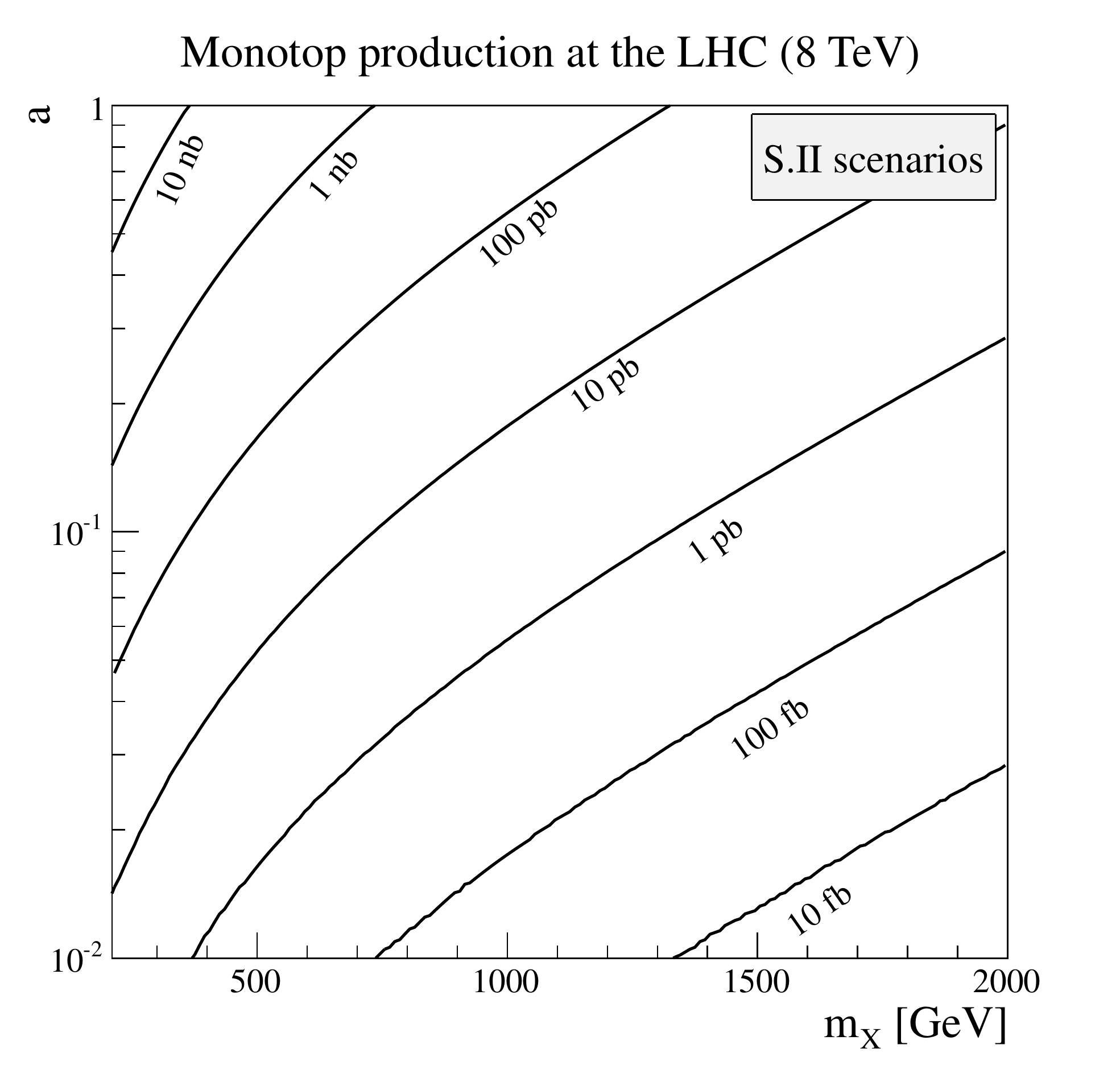}
  \caption{Total cross sections for monotop production at the LHC,
     running at a center-of-mass energy of 8~TeV, for scenarios of type
     {\bf S.I} (left panel) and {\bf S.II} (right panel) respectively
     featuring a scalar and vector colored resonance decaying with a branching
     fraction of one into a monotop state. The cross sections are presented as
     a function of the new physics coupling $a$ and of the resonance mass
     $m_\varphi$ (scalar resonance) and $m_X$ (vector resonance).}
  \label{fig:xsecS1-2}
\end{figure}

In scenarios of type {\bf S.I} and {\bf S.II}, we also assume that the
new colored scalar $\varphi$ and vector $X$
resonances decay into a top quark and an invisible particle $\chi$ with a
branching ratio equal to
one. Consequently, this renders our analysis insensitive to the parameters $(a^q_{SR})_3$ and 
$(a^q_{VR})_3$ and
the {\bf S.I} and {\bf S.II} scenarios are described by only three parameters, namely the couplings
of the down-type quarks to the new resonance $a$, the resonance mass $m_\varphi$ (for
scenarios of class {\bf S.I})
and $m_X$ (for scenarios of class {\bf S.II})
and the mass of the invisible particle $m_\chi$.
Monotop production cross sections are however
independent of the invisible particle mass as they
are equal to the colored new particle production
cross section, the subsequent branching ratio into a $t\chi$ pair being unity. The mass
difference between the resonance and the missing energy particle however alters
the selection efficiency of any monotop search strategy as it modifies the available
phase space for the decay (see Section~\ref{sec:effmonotops} for more details).
The dependence of the monotop production cross section on the resonance mass and
on the new physics coupling strength $a$ is illustrated on Figure
\ref{fig:xsecS1-2} for scenarios involving a scalar resonance {\bf S.I} (left panel
of the figure) and
a vector resonance {\bf S.II} (right panel of the figure).
Cross sections reaching the pb level are expected for a moderate coupling
strength of $a=0.1$ and resonance masses around (in the scalar case) or even
above (in the vector case) 1~TeV.
For given resonance mass and coupling strength,
it is also found that a
monotop signature induced by a vector state is produced with a larger rate
as when arising from the decay of a scalar state due to
the different Lorentz structure of the interactions
of the Lagrangian of Eq.~\eqref{eq:effmonotoplag} and to the larger number of
propagating degrees of freedom of a vector field.

\begin{figure}[t!]
  \centering
  \includegraphics[width=.49\columnwidth]{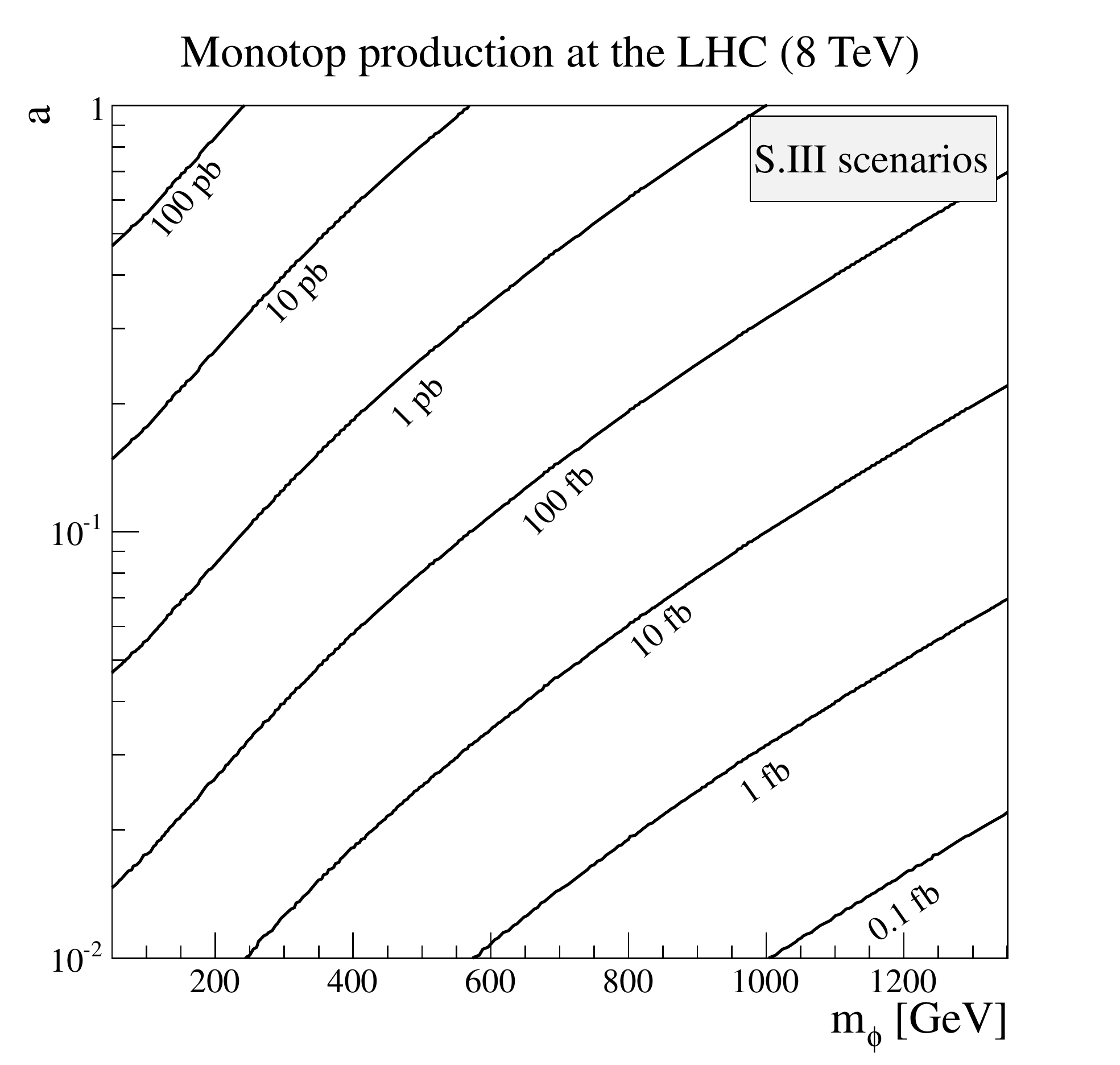}
  \includegraphics[width=.49\columnwidth]{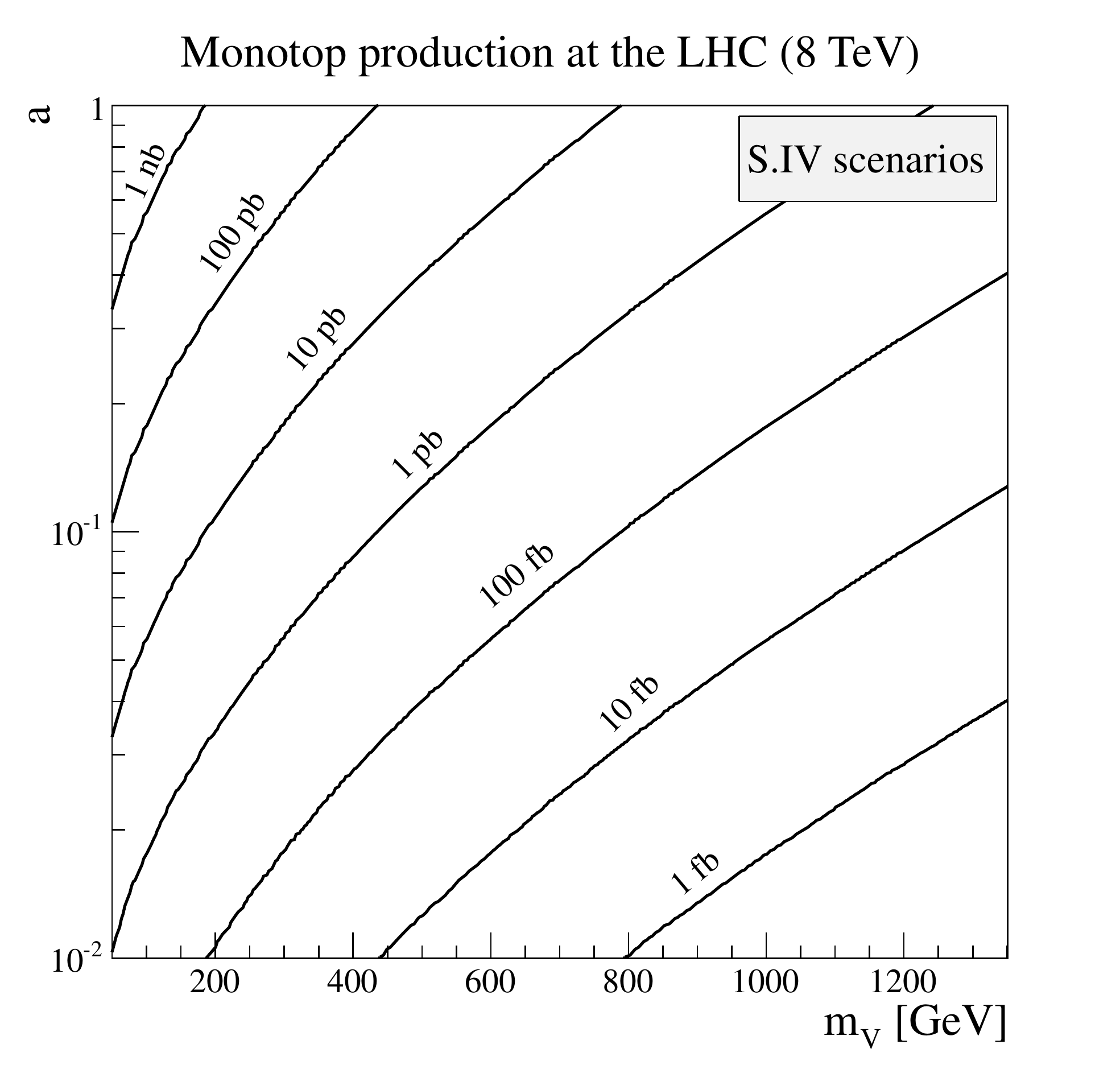}
  \caption{Total cross sections for monotop production at the LHC,
     running at a center-of-mass energy of 8~TeV, for scenarios of type
     {\bf S.III} (left panel) and {\bf S.IV} (right panel) respectively
     featuring a flavor-changing production of a top quark in association with a
     scalar and vector invisible state. The cross sections are presented as
     a function of the new physics coupling $a$ and of the invisible particle mass
     $m_\phi$ (scalar case) and $m_V$ (vector case).}
  \label{fig:xsecS3-4}
\end{figure}

In scenarios of class {\bf S.III} and {\bf S.IV}, the top quark is produced
in association with a bosonic particle through a flavor-changing interaction
in the weak sector. The bosonic state, being respectively
a scalar and vector particle in the two scenarios,
is further non-detected and gives rise to missing energy. The {\bf S.III} and {\bf S.IV}
scenarios are thus described by exactly two parameters, the mass of the
missing energy particle $m_\phi$ and $m_V$ in the scalar and vector cases, respectively,
and the strength of the flavor-changing interaction of this particle with
a pair of quarks comprised of one top quark and one up quark.
Cross sections for monotop production in
the flavor-changing mode are presented as a function of these two parameters
in Figure~\ref{fig:xsecS3-4} for a scalar (left panel of the figure) and
vector (right panel of the figure) invisible state. In contrast to
the first two scenarios where TeV-scale new physics can be 
associated with a cross section of 1~pb for a moderate coupling
strength of $a=0.1$, such a cross section value
corresponds to much lighter invisible particle masses of about 300~GeV and
500~GeV for scenarios of type {\bf S.III} and {\bf S.IV}, respectively, and an
identical coupling value of $a=0.1$.
Once again, larger cross sections are associated with the vector case
due to the more important number of polarization states and
the Lorentz structures of the possible interactions.

\subsection{A simplified model for sgluon production and decays at the LHC}
\label{sec:simpsgl}
In Section \ref{sec:mrssmpheno}, we have illustrated a phenomenological aspect of
sgluon fields in the framework of the minimal version of 
$R$-symmetric supersymmetric theories \cite{Fayet:1974pd, Salam:1974xa,
Kribs:2007ac}, considering their contributions to the production rate of
multitop final states. However, sgluon fields are also predicted
in other models, such as in $N\!=\!1\!/\!N\!=\!2$ hybrid
supersymmetric theories~\cite{Fayet:1975yi, AlvarezGaume:1996mv, Plehn:2008ae,
Choi:2008pi, Choi:2008ub, Choi:2009jc, choi:2010gc, Choi:2010an,
Schumann:2011ji}, in vector-like confining theories
\cite{Kilic:2008pm, Kilic:2008ub, Kilic:2009mi, Kilic:2010et,
Dicus:2010bm, Sayre:2011ed} or in extra-dimensional models \cite{Burdman:2006gy}.

All these new physics models contain a sgluon sector with similar properties
as in $R$-symmetric supersymmetry. This motivates us
to construct a simplified effective model describing a scalar
field lying in the octet representation of the QCD gauge group
and its interactions with the Standard Model
sector~\cite{Calvet:2012rk,Brooijmans:2012yi}. This leaves open the
possibility of reinterpreting the results in the context of 
any of the original models, or even in the framework of any other theory including
a sgluon field. In addition, it avoids the 
careful design of a theoretically motivated
and not experimentally excluded benchmark scenario for the complete model, 
the simplified model being instead 
described by a small number of couplings and masses.

We extend the Standard Model in a minimal way by supplementing to
its particle content one real scalar field $\sigma$ of mass $M_\sigma$ lying in the adjoint
representation of the QCD gauge group. Its kinetic, mass and gauge
interaction terms are standard,
\be
  \lag_{\rm kin} = \frac12 D_\mu \sigma^a D^\mu \sigma_a -
     \frac12 M_\sigma^2  \sigma^a \sigma_a \ ,
\label{eq:sglkin}\ee
and are expressed in terms of the QCD covariant derivative taken in the
adjoint representation
\be
  D_\mu \sigma^a = \partial_\mu \sigma^a + g_s\ f_{bc}{}^a\ g_\mu^b\ \sigma^c \
  ,
\ee
for which we recall that the strong coupling constant is denoted by $g_s$,
the antisymmetric structure constants of $SU(3)_c$ by $f_{bc}{}^a$ and that the
gluon field reads $g_\mu$.

In Eq.\ \eqref{eq:Leff}, we have introduced, in the context of
minimal $R$-symmetric supersymmetry,
loop-induced interactions of a
single sgluon to the Standard Model partons. This feature also holds
in other theories, where the presence of additional
particles in general implies loop diagrams yielding
dimension-four and dimension-five effective operators involving 
up-type quarks $u$, down-type quarks $d$ and gluons,
\be
 \lag_{\rm eff} = 
   \sigma^a \bar d T_a \Big[ a^L_d P_L + a^R_d P_R \Big] d + 
   \sigma^a \bar u T_a \Big[ a^L_u P_L + a^R_u P_R \Big] u + 
   a_g d_a{}^{bc} \sigma^a G_{\mu\nu b} G^{\mu\nu}{}_c + {\rm h.c.}  \ ,
\label{eq:sgleff}\ee
referring to Eq.\ \eqref{eq:Leff} for the notations. These interactions 
consequently open all possible sgluon decays into Standard Model colored particles. 
Inspecting the two Lagrangians of Eq.~\eqref{eq:sglkin} and Eq.~\eqref{eq:sgleff},
our simplified theory is described by one mass parameter, the
sgluon mass $M_\sigma$, and the effective couplings of sgluons to colored
partons described by a set of four complex $3\times 3$ matrices in flavor space
$a^L_d$, $a^R_d$, $a^L_u$ and $a^R_u$ and one real dimensionful number $a_g$.

As already mentioned in Section \ref{sec:benchmrssm}, sgluon masses up to about 2
TeV are excluded by dijet resonance searches \cite{Aad:2011fq} once we assume
${\cal O}(1)$ effective sgluon interactions to light quarks. We therefore focus,
motivated by $R$-symmetric supersymmetry, on 
scenarios where the sgluon field dominantly decays into final states containing 
at least one top quark and where its couplings to a pair of light quarks or to a
pair of gluons are reduced so that the experimental constraints can be evaded.
This brings us
to consider two series of benchmark scenarios.

\renewcommand{\arraystretch}{1.3}
\begin{table}[t]
\begin{center}
\begin{tabular}{| c || c  c |}
\hline
Parameters &  Scenarios of type {\bf S.I} & Scenarios of type {\bf S.II}\\
\hline
$a_g$ & $1.5 \times 10^{-6}$ GeV$^{-1}$ & $1.5 \times 10^{-6}$ GeV$^{-1}$ \\
$(a_u)^3{}_3$ & $3\times 10^{-3}$ & $3\times 10^{-3}$\\
$(a_u)^3{}_1 = (a_u)^1{}_3$ & $3\times 10^{-3}$ & 0 \\
$(a_u)^3{}_2 = (a_u)^2{}_3$ & $3\times 10^{-3}$ & 0 \\
$M_\sigma$ & [200-1000] GeV & [400-1000] GeV \\
$M_t$ & 172 GeV & 172 GeV\\
 \hline
\end{tabular}
\caption{\label{tab:params} Non-zero input parameters for benchmark scenarios of
class {\bf S.I} and {\bf S.II}. For all the other Standard Model parameters, we 
follow the conventions of Ref.\ \cite{Christensen:2009jx}.}
\end{center}
\end{table}
\renewcommand{\arraystretch}{1.0}

For the first set of scenarios, referred to as scenarios of class {\bf S.I}, sgluon particles are allowed to
decay in a universal way to any associated pair of up-type quarks containing
at least one top quark, so that 
\be\label{eq:Ia}
  (a^L_u)^3{}_i = (a^R_u)^3{}_i =  (a^L_u)^i{}_3 = (a^R_u)^i{}_3 = 3\times 10^{-3} \ ,
\ee
for $i=1,2,3$. We subsequently impose that any other interaction among quarks and a single sgluon
vanishes.
Concerning the parameter $a_g$, we choose the value 
\be\label{eq:Ib}
  a_g = 1.5\times 10^{-6}  \text{ GeV}^{-1}\ .
\ee
Both Eq.\ \eqref{eq:Ia} and Eq.\ \eqref{eq:Ib} correspond to a 
supersymmetry mass scale
of about 2~TeV (as shown by the analytical formulas presented in 
Ref.~\cite{Plehn:2008ae}), in agreement with the current experimental results
related to direct
searches for squarks and gluinos at the LHC (see Section~\ref{sec:direct}).

In our second class of scenarios, denoted as scenarios of class {\bf S.II}, 
we focus exclusively on sgluon-induced signatures
with four top quarks so that the only states into which a sgluon decays consist of
either a top-antitop or a gluon pair,
the only non-vanishing effective interactions being thus driven by the coupling parameters
\be
  (a^L_u)^3{}_3 =  (a^R_u)^3{}_3 = 3\times 10^{-3}
  \qquad\text{and}\qquad
  a_g = 1.5\times 10^{-6} \text{ GeV}^{-1} \ .
\ee

In addition, we fix for both classes of scenarios the mass of the top quark 
to $M_t=172$~GeV, all the other Standard Model parameters according to the conventions of Ref.\
\cite{Christensen:2009jx} and allow
the sgluon mass $M_\sigma$ to vary below 1~TeV.
We summarize in Table \ref{tab:params} the values of all
non-zero parameters of the Lagrangians 
of Eq.~\eqref{eq:Lkin} and Eq.~\eqref{eq:Leff}.

\renewcommand{\arraystretch}{1.3}
\begin{table}[!t]
\begin{center}
\begin{tabular}{|c|| c c || c c c || c c|}
\hline
  Scenario & $M_\sigma$ [MeV] & $\Gamma_\sigma$ [GeV] & BR$[t\bar t]$ &
  BR$[tj / \bar t j]$ & BR$[gg]$ & $\sigma_{\rm tot}$ [fb] & $K_{\rm
  NLO}$\\
\hline
  {\bf S.I} & 200 & 0.012& - & 80\% & 20\% & 98600 & 1.6\\
\hline
  {\bf S.I} & 300 & 0.105& - & 92.3\% & 7.7\% & 9802  & 1.6\\
\hline
  {\bf S.I} & \multirow{2}{*}{400} & 0.219 & 4.4 \% & 86.9\% & 8.7\% &
    \multirow{2}{*}{1625} & \multirow{2}{*}{1.7}\\
  {\bf S.II}&                      & 0.029& 33.3\% & -      & 66.7\% & & \\
\hline
  {\bf S.I} & \multirow{2}{*}{500} & 0.35 & 9.8 \% & 79.5\% & 10.1\% &
    \multirow{2}{*}{358.1} & \multirow{2}{*}{1.8}\\
  {\bf S.II}&                      & 0.072& 47.8\% & -      & 52.2\% & & \\
\hline
  {\bf S.I} & \multirow{2}{*}{600} & 0.485& 12 \% & 75\% & 13\% &
    \multirow{2}{*}{94.9} & \multirow{2}{*}{1.8}\\
  {\bf S.II}&                      &0.124& 48\% & -      & 52\% & & \\
\hline
  {\bf S.I} & \multirow{2}{*}{700} & 0.628& 13.2 \% & 70.5\% & 16.3\% &
    \multirow{2}{*}{28.4} & \multirow{2}{*}{1.9}\\
  {\bf S.II}&                      &0.185& 44.7\% & -      & 55.3\% & & \\
\hline
  {\bf S.I} & \multirow{2}{*}{800} &0.779 & 13.5 \% & 66.9\% & 19.6\% &
    \multirow{2}{*}{9.26} & \multirow{2}{*}{2.0}\\
  {\bf S.II}&                      &0.252& 41\% & -      & 59\% & & \\
\hline
  {\bf S.I}& \multirow{2}{*}{900} &0.943 & 13.5 \% & 63.4\% & 23.1\% &
    \multirow{2}{*}{3.22} & \multirow{2}{*}{2.1}\\
  {\bf S.II}&                      &0.345 & 36.9\% & -      & 63.1\% & & \\
\hline
  {\bf S.I} & \multirow{2}{*}{1000} & 1.12 & 13.2 \% & 60.2\% & 26.6\% &
    \multirow{2}{*}{1.17} & \multirow{2}{*}{2.2}\\
  {\bf S.II}&                      &0.447& 33.2\% & -      & 66.8\% & & \\
\hline
\end{tabular}
\caption{\label{tab:sigma} Dependence on the sgluon mass $M_\sigma$ of the
  sgluon total width ($\Gamma_\sigma$), of its branching fractions 
  to a top-antitop pair (BR$[t \bar t]$), to an associated pair of a top (anti)quark and a 
  light quark (BR$[tj/\bar tj]$) and to a gluon pair (BR$[gg]$), as well as of its total 
  pair-production cross section at leading order and at the LHC collider running at a 
  center-of-mass energy of 8 TeV ($\sigma_{\rm tot}$). The next-to-leading order $K$-factors 
  ($K_{\rm NLO}$) are also indicated.}
\end{center}
\end{table}
\renewcommand{\arraystretch}{1.0}

A key element in the multitop analysis of sgluon production and decay at the LHC
lies in the sgluon branching fraction to final states containing one or 
two top quarks. We investigate the evolution of these branching ratios with the
sgluon mass by implementing the Lagrangians of Eq.\ \eqref{eq:Lkin} and Eq.\ \eqref{eq:Leff}
into \feynrules,
following the syntax introduced in Chapter \ref{chap:FR}, and by subsequently 
exporting the model to the UFO format
so that it can be used within the \madgraph~5 framework.
This matrix-element generator is then employed to estimate the total sgluon width and the
different branching ratios into two gluons,
an associated pair of a top quark and a light quark and into two top quarks.
The results are shown in Table \ref{tab:sigma} for both classes of scenarios.

The branching of a light sgluon of a couple of hundreds of GeV 
into a $t \bar t$ pair is kinematically suppressed compared to
the other open decay channels for both types of scenarios.
This branching ratio then increases with the sgluon mass, although the
contributions of the dijet ($\sigma\to g g$) channel to the total width also become more
important. Therefore, the branching into a top-antitop
pair peaks for $M_\sigma\sim 800$ GeV and $M_\sigma\sim 600$ GeV for
scenarios of type {\bf S.I} and {\bf S.II}, respectively, and then decreases for heavier
sgluons.

Table \ref{tab:sigma} also contains the leading-order sgluon pair-production
cross sections as computed with the \madgraph~5 program. The numerical values are
calculated in the context of the LHC collider
running at a center-of-mass energy of 8 TeV and after convoluting 
the matrix elements related to the Feynman
diagrams of 
Figure \ref{fig:sgluondiag} with the leading order set of the CTEQ6 parton density fit
\cite{Pumplin:2002vw}, fixing both the renormalization and factorization scales 
to the transverse mass of the sgluon pair. Next-to-leading order corrections
to those cross sections have been recently computed within the 
\madgolem\ setup \cite{GoncalvesNetto:2012nt, LopezVal:2012ms}, 
and the corresponding $K$-factors are presented in the last column of the table.

\mysection{Monte Carlo simulations of the Standard Model background}
\label{sec:mc}
\subsection{Simulation setup}
\label{sec:mcsetup}

\renewcommand{\arraystretch}{1.3}
\begin{table}[!t]
\begin{center}
\begin{tabular}{| l || c c || c c || c c |}
\hline
Process & $k_T^{\rm min}$ [GeV] & $p_T^{\rm min}$ [GeV] & $n$ &
$Q^{\rm m}$ [GeV]& $\sigma$ [pb]& $N$\\
\hline
  $W(\to \ell\nu) + \text{jets}$ & 10 & 10  & 4 & 20 & 35678    & $2.56\cdot
    10^8$\\
  $\gamma^{*}/Z(\to 2 \ell/2 \nu) + \text{jets}$ & 10 & 10 & 4 & 20 & 10319 & $4
\cdot 10^7$\\
\hline 
  $t\bar{t} (\to 6 \text{jets}) \!+\! \text{jets}$& 20 & 20 & 2 & 30 & 116.2& $8\cdot 10^6$ \\
  $t\bar{t} (\to 4 \text{jets} \ 1\ell\ 1\nu) \!+\! \text{jets}$& 20 & 20 & 2 & 30 & 112.4& $9\cdot 10^6$ \\
  $t\bar{t} (\to 2\text{jets} \ 2 \ell\ 2\nu) \!+\! \text{jets}$& 20 & 20 & 2 & 30 & 27.2& $3\cdot 10^6$
    \\
\hline 
  $t/\bar t+ \text{jets}$ [$t$, incl.] & - & - & 0 & - & 87.2 & $6\cdot10^6$ \\
  $t/\bar t+ \text{jets}$ [$tW$, incl.]& - & - & 0 & - & 22.2 & $1\cdot10^6$ \\
  $t/\bar t+ \text{jets}$ [$s$, incl.] & - & - & 0 & - & 5.55 & $8\cdot10^5$ \\
\hline
  $WW(\to 1\ell\ 1\nu\ 2\text{jets}) \!+\! \text{jets}$ &10 & 10 & 2& 20 & 24.3 & $3\cdot10^6$\\
  $WW(\to 2\ell\ 2\nu) + \text{jets}$ &10 & 10 & 2& 20 & 5.87 & $8\cdot10^5$ \\
\hline
  $WZ(\to 1\ell\ 1\nu\ 2\text{jets}) + \text{jets}$ &10 & 10 & 2& 20 & 5.03 & $5\cdot10^5$ \\
  $WZ(\to 2\nu\ 2\text{jets}) + \text{jets}$ &10 & 10 & 2& 20 & 2.98 & $3\cdot10^5$ \\
  $WZ(\to 2\ell\ 2\text{jets}) + \text{jets}$ &10 & 10 & 2& 20 & 1.58 & $2\cdot10^5$ \\
  $WZ(\to 1\ell\ 3\nu) + \text{jets}$ &10 & 10 & 2& 20 & 1.44 & $2\cdot10^5$ \\
  $WZ(\to 3\ell\ 1\nu)+ \text{jets}$ &10 & 10 & 2& 20 & 0.76 & $2\cdot10^6$ \\
\hline
  $ZZ(\to 2\nu\ 2\text{jets})+ \text{jets}$ &10 & 10 & 2& 20 & 2.21 & $3\cdot10^5$ \\
  $ZZ(\to 2\ell\ 2\text{jets})+ \text{jets}$ &10 & 10 & 2& 20 & 1.18 & $1.5\cdot10^4$ \\
  $ZZ(\to 4\nu)+ \text{jets}$ &10 & 10 & 2& 20 & 0.63 & $1\cdot10^5$ \\
  $ZZ(\to 2\nu\ 2\ell)+ \text{jets}$ &10 & 10 & 2& 20 & 0.32 & $4\cdot10^4$ \\
  $ZZ(\to 4\ell)+ \text{jets}$ &10 & 10 & 2& 20 & 0.17 & $4\cdot10^4$ \\
\hline 
  $t\bar{t}W  + \text{jets}$ [incl.] & 10& 10 & 2 & 20  & 0.25   &
    $3\cdot10^4$\\
  $t\bar{t}Z  + \text{jets}$ [incl.] & 10& 10 & 2 & 20  & 0.21   & 
    $5\cdot10^4$\\
  $t/\bar{t} + Z + j + \text{jets}$ [incl.]& 6.5 & 20 & 1 & 10 & 0.046& $3\cdot10^5$ \\
  $t\bar{t}WW + \text{jets}$ [incl.] & 10& 10 & 2 & 20  & 0.013  & 
    $2\cdot10^3$\\
  $t\bar{t}t\bar{t} + \text{jets}$ [incl.]& - & - & 0 & - & $7 \cdot 10^{-4}$& $10^{3}$ \\
\hline
\end{tabular}
\caption{Simulated background processes given together with the 
applied parton-level selection criteria ($k_T^{\rm min}$ and $p_T^{\rm min}$),
the number of 
allowed extra hard emissions at the matrix-element level ($n$) and the merging
scale ($Q^{\rm m}$). The numerical values employed for the cross sections
($\sigma$) are
also shown, together with the number of generated events ($N$). We detail each of the
background contributions according to the final state signature, $\ell$ standing equivalently for 
electrons, muons, leptonic
and hadronic taus, $\nu$ for any neutrino, and jets or $j$ for any kind of jet. 
Moreover, the notation \textit{incl.} indicates 
that the produced samples are inclusive in the decays of the produced particles.
We refer to the rest of this section for more information, in particular 
on the adopted values for the cross sections.}
\label{tab:xsec}
\end{center}
\end{table}
\renewcommand{\arraystretch}{1.0}

This chapter aims to
estimate the LHC sensitivity to the presence of monotops and sgluons-induced multitop events
by means of Monte Carlo simulations. We consider
the LHC collider running at
a center-of-mass energy of $\sqrt{s}=8$ TeV and normalize our event samples
to an integrated luminosity of 20 fb$^{-1}$. The hard scattering processes related 
to the different sources of background are described with the
matrix-element generator \madgraph~5. Using the QCD factorization 
theorem, the matrix elements are convoluted with the leading order set of the CTEQ6 parton
density fit \cite{Pumplin:2002vw}, the renormalization and
factorization scales being fixed to the transverse mass of the produced heavy
particles and all quark masses but the top mass are neglected. 
Parton-level events are integrated into a full hadronic
environment by matching the hard scattering matrix elements with a parton 
showering and hadronization infrastructure as provided by the \pythia\ 6
package \cite{Sjostrand:2006za}. Moreover, since the generated parton-level
events are allowed to contain tau leptons, we make use of
the \tauola\ program \cite{Davidson:2010rw} 
to handle their decays.

Typical final states to be produced at the LHC contain 
in general abundant initial state QCD radiation that 
has important effects on the shapes of the kinematical distributions. In
particular, large logarithmic contributions arise in phase space
regions where these additional partons are neither widely 
separated nor hard. As a consequence, reliable theoretical predictions require a
consistent reorganization of the logarithmic terms which have to be 
resummed to all orders in the strong coupling, embedded in this way within the so-called
Sudakov form factor.

In phenomenological investigations relying on Monte
Carlo simulations, additional jet 
production is traditionally simulated using parton showering programs. These
tools describe QCD emissions as successive branchings of a mother parton into two
daughter partons, the associated probability laws 
being based on Markov chain techniques built upon
the Sudakov form factor. The latter must however be approximated in a way allowing for a 
description in terms of a Monte Carlo algorithm, which enforces to truncate it, in general,
to the leading logarithmic accuracy.
This description is formally only correct in the phase space regions where QCD radiation
is soft and collinear and fails when considering the
production of hard and widely separated additional partons. 
In this case, matrix elements describing the same final state 
together with an additional parton are required.

In order to obtain a good description over the whole kinematical range,
parton showering and matrix element methods have to be
eventually matched. We then allow for the matrix elements 
to contain up to $n$ additional hard jets and merge the $n+1$
different samples, after parton showering, following the
$k_T$-MLM merging scheme \cite{Mangano:2006rw} as implemented in the
\madevent\ generator \cite{Alwall:2008qv} interfaced to \pythia\ 6
\cite{Sjostrand:2006za}.
In this setup, two parton-level selection criteria are imposed. 
Firstly, parton-level final state jets are generated with a minimum jet measure
$k_T$ larger than a process-dependent threshold $k_T^{\rm min}$. The quantity $k_T$ is
defined, when considering two final state jets labeled by $i$ and $j$, by
\be
  k_T^2 = \min(p_{Ti}^2, p_{Tj}^2) R_{ij}\ ,
\ee
where $p_{Ti}$ and
$p_{Tj}$ are the transverse momenta of the jets and
$R_{ij}$ their angular distance in the $(\eta,\varphi)$ plane, $\eta$ denoting the
pseudorapidity and $\varphi$ the azimuthal angle with respect to the beam direction. Secondly,
in order to ensure better
QCD factorization properties with respect to initial-state collinear
singularities \cite{Catani:1993hr},
we define the jet measure related to an initial state splitting as
its transverse momentum, 
\be
  k_T = p_{Ti} \ ,
\ee
and ask it to be larger than a process-dependent value $p_T^{\rm min}$.

The events are then passed to
\pythia\ for parton showering and jets are reconstructed making use of a $k_T$-jet
algorithm~\cite{Cacciari:2008gp} with
a (process-dependent) cut-off scale $Q^{\rm m}$. A jet is said to be matched
to one of the original partons only if the jet measure between the jet and this parton 
is smaller than $Q^{\rm m}$.
In our procedure, only events where each jet is
matched to one parton and where each parton is related to one jet are retained, 
with the exception of events belonging to the sample with the highest 
jet multiplicity. In this case, 
extra jets are allowed, which maintains the full inclusiveness of the
merged sample. A correct choice for the merging parameters $k_T^{\rm min}$, $p_T^{\rm min}$
and $Q^{\rm m}$ is crucial since this allows for a
coherent splitting of the phase space among regions dominated by matrix-element-based
predictions and regions where parton showering correctly describes QCD emission.

In order to probe the smoothness of the transition between these regions, differential
jet rate spectra are traditionally investigated. This class of variables consists of the
distributions of
the scale at which a specific event switches from a $N$-jet configuration to a $N+1$-jet 
configuration, for various values of the integer number $N$. The 
merging procedure has been validated for all simulated background contributions. For brevity,
we only illustrate this validation in the context of $Z$-boson production (see Section 
\ref{sec:vjets}).

The values chosen for the parton-level selection thresholds $p_T^{\rm min}$ and 
$k_T^{\rm min}$, the
maximum number of included hard emissions $n$ and the merging scale $Q^{\rm m}$ are indicated in
Table \ref{tab:xsec} for the various background processes that have been simulated.
This table also contains the cross section values $\sigma$  employed for the
normalization of the different samples and the numbers of generated
events $N$.

\subsection{Single boson production in association with jets}\label{sec:vjets}

\begin{figure}[t!]
 \centering
 \includegraphics[width=.49\columnwidth]{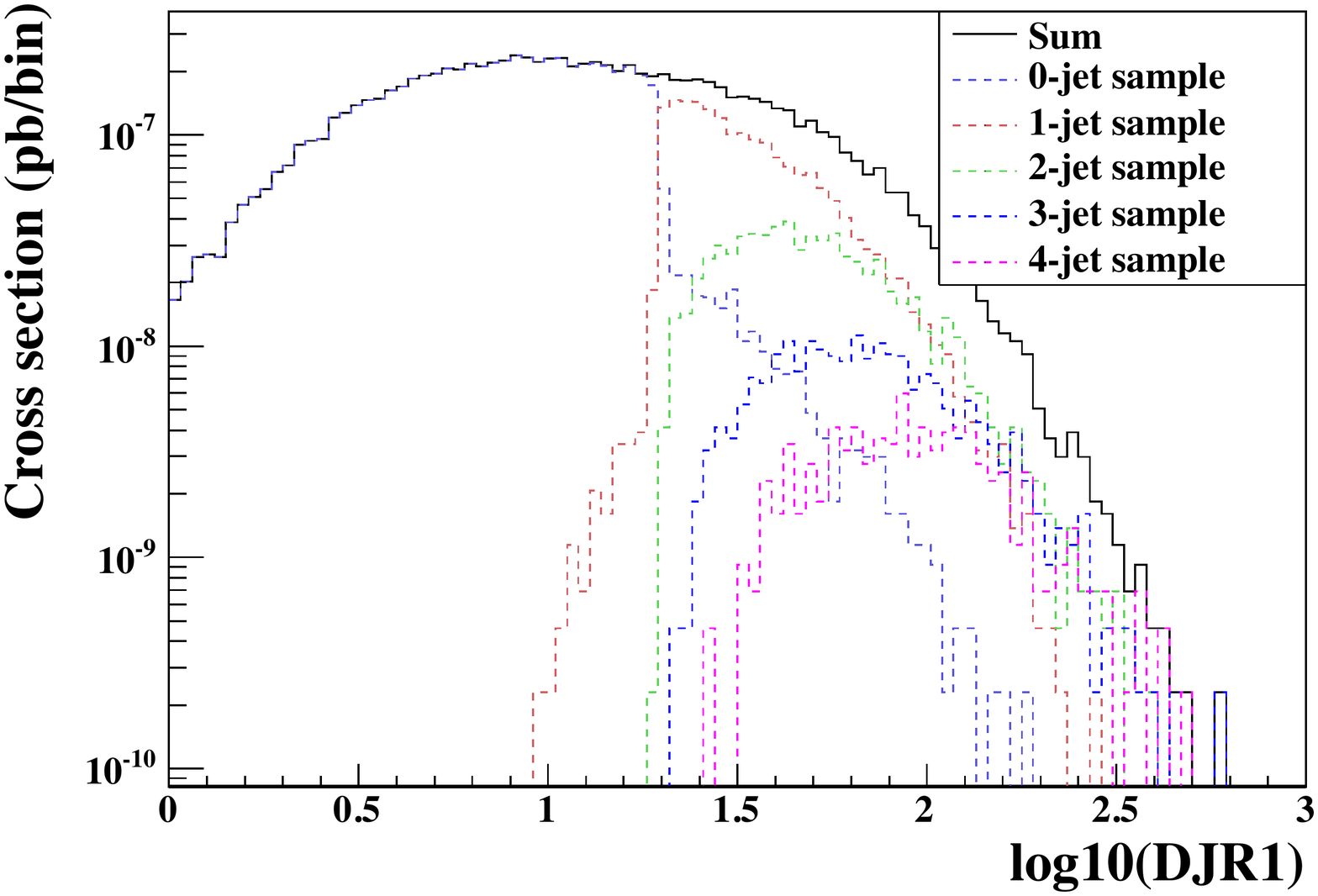}
 \includegraphics[width=.49\columnwidth]{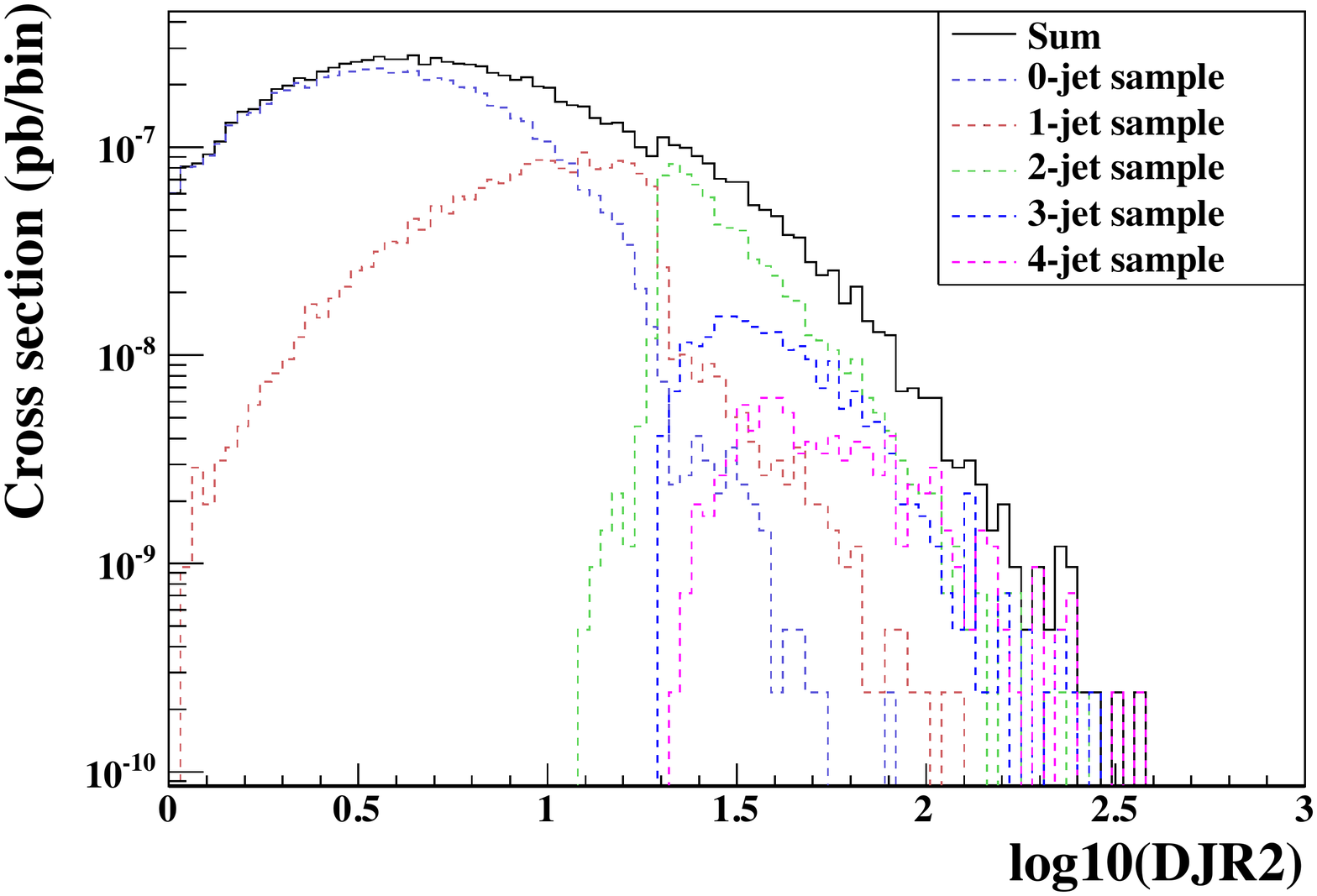}\\
 \includegraphics[width=.49\columnwidth]{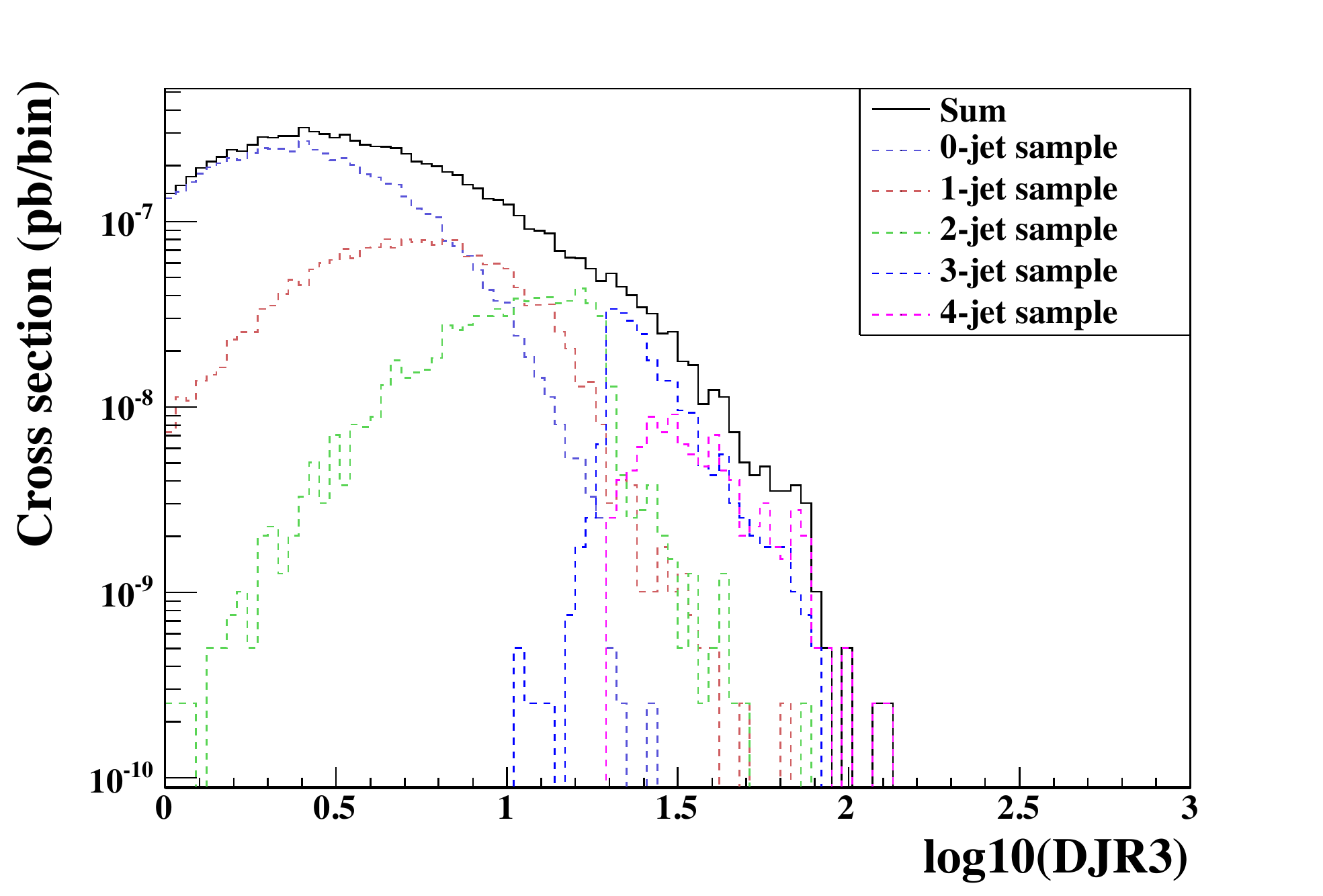}
 \includegraphics[width=.49\columnwidth]{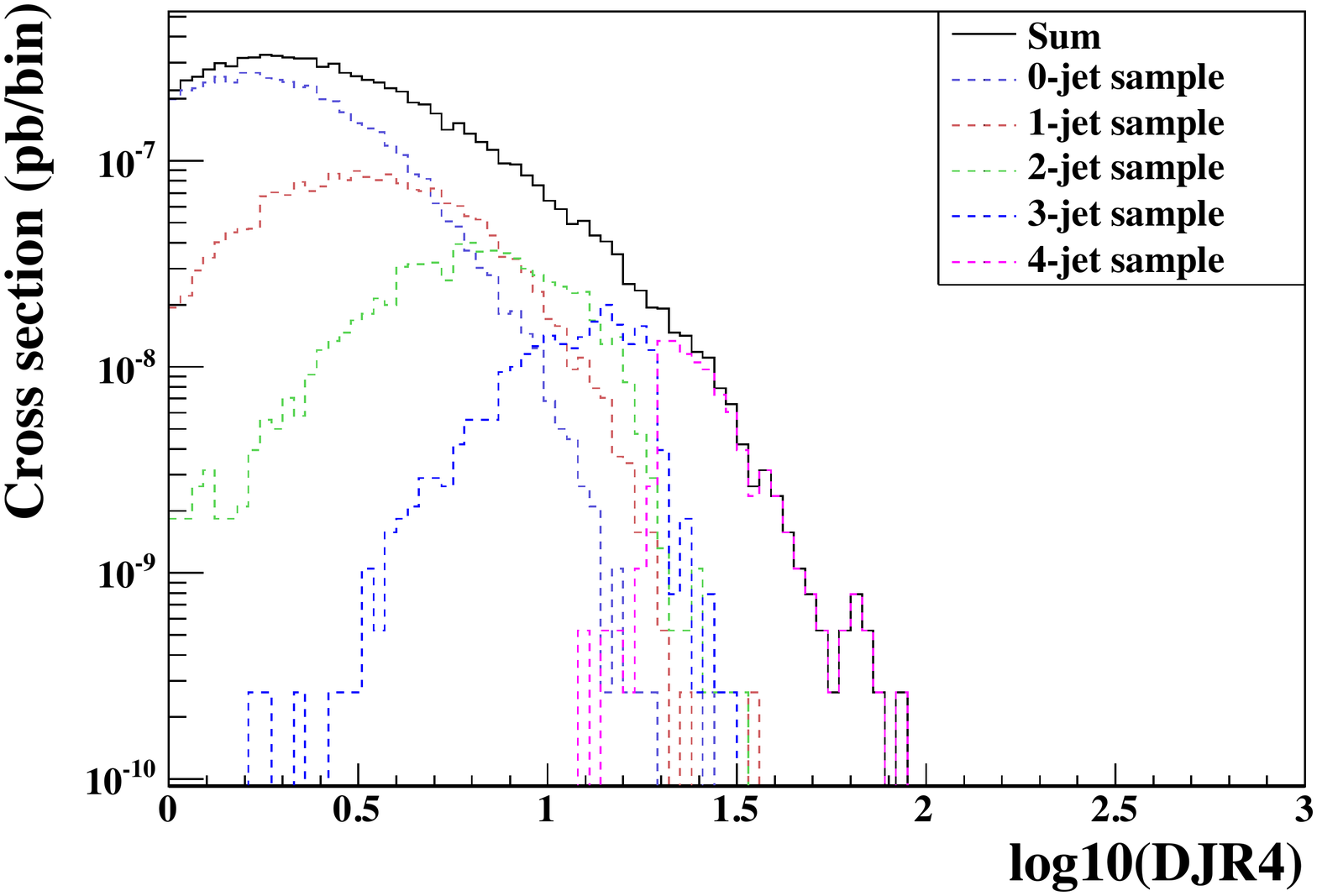}
 \caption{\label{fig:Zmatch}
 Differential jet rate distributions for the process $p p \to Z + $ 0, 1, 2, 3, and 4 jets,
  where multiparton matrix elements have been merged by employing the $k_T$-MLM scheme.}
\end{figure}

In terms of cross section, the main contributions to the Standard Model background originate
from single weak gauge boson production in association
with jets\footnote{In our setup,
QCD multijet production is not simulated since in order to investigate this
background contribution
properly, data-driven methods are more appropriate than Monte Carlo simulations.
We choose to resort instead on available experimental studies to ensure a good control
of the QCD background by designing an appropriate event selection strategy.}. 
After imposing the gauge bosons to decay either leptonically or invisibly,
\be\bsp
  p p \to W + \text{ jets } \to & \ \ \ell \nu + \text{ jets } \ , \\ 
  p p \to Z + \text{ jets } \to \ell \ell+ \text{ jets } \ , \qquad & 
  p p \to Z + \text{ jets } \to \nu  \nu + \text{ jets } \ , 
\esp \label{eq:singlebos}\ee
where $\ell$ and $\nu$ generically denote any charged lepton (including the tau) 
and neutrino, respectively,
we have merged event samples containing up to four additional hard
jets. We choose the jet measure thresholds to be $p_T^{\rm min} = k_T^{\rm min} = 10$ GeV
and fix the merging scale to $Q^{\rm m} = 20$~GeV. The suitability of these parameters
is checked on Figure \ref{fig:Zmatch} where we present differential jet rate 
distributions for $pp \to Z + 0,1,2,3$ and 4 jets events, the figures having been computed 
by making use of the \madanalysis\ 5 package \cite{Conte:2012fm}\footnote{As stated above,
we have checked the relevance of the adopted merging parameters for
all the simulated sources of background. Since all results are similar as in
Figure~\ref{fig:Zmatch}, the corresponding figures for the other background contributions
have been omitted for brevity.}.

%
%

On the left panel of the upper line of the figure, we show the $1\to 0$ 
differential jet rate distribution, or in other words, the distribution of the scale at which
an event passes from a configuration where no final state jet remains after reconstruction to
a configuration with one single final state jet does. It is found that
only events related to the production of a $Z$-boson without any extra hard jet 
contribute to the region to the left of the cut-off scale $Q^{\rm m}$ where parton
radiation is dominated by parton showering. All the other events (with one hard jet or more
at the matrix-element level) 
contribute to the region
located to the right of the cut-off scale, radiation being here correctly described by means
of matrix elements. From the figure, the cut-off scale is extracted and found
to be equal to 20 GeV, in agreement with the Monte Carlo setup.
Similar analyses can be performed for the $2\to 1$ 
(right panel of the upper line of the figure), $3\to 2$
(left panel of the lower line of the figure) and $4\to 3$
(right panel of the lower line of the figure) differential jet rate distributions.

We reweight the events according to the next-to-next-to-leading order
cross sections as obtained by making use of the \fewz\ package
\cite{Melnikov:2006kv,Gavin:2012sy,Gavin:2010az}. Employing the recent set of parton
densities CT10 provided by the CTEQ collaboration \cite{Lai:2010vv}, the results,
including the relevant weak boson decays of Eq.~\eqref{eq:singlebos}, read
\be
  \sigma_W = 35678 \text{ pb} \qquad \text{and}\qquad 
  \sigma_Z = 10319 \text{ pb} \ , 
\ee
where we omit any source of
theoretical uncertainties as neglected in  the prospective
studies performed in this work, although these are mandatory in
any experimental analysis aiming to a proper derivation of any
new physics limits.
Moreover, virtual photon contributions have been considered for
dileptonic final state production, after
imposing a necessary parton-level selection on the dilepton invariant mass in
the calculation of
the $\sigma_Z$ cross section, $m_{\ell\ell}~\geq~50$~GeV.

\subsection{Single and pair production of top quarks in association with jets}

We generate three distinct $t \bar t$ event samples according to the possible
decays of the $t \bar t$ pair, the latter being allowed to decay either into a
dileptonic final state, a semileptonic one or 
into a fully hadronic one. In our classification, tau leptons are considered as 
any other charged leptons, whatever they decay to. 
Parton-level events have been simulated based on 
matrix elements containing up to two additional jets after imposing the (parton-level)
jet measure to be larger than $p_T^{\rm min} = k_T^{\rm min} = 20$ GeV.
We then combine event samples related to the three processes 
$p p \to t \bar t$, $p p \to t \bar t$ plus one jet and
$p p \to t \bar t$ plus two
jets following to the MLM merging technique, setting the merging scale to 
$Q^{\rm m} = 30$~GeV. The results are reweighted according to the production cross section 
evaluated at the next-to-leading order accuracy, after including
genuine next-to-next-to-leading order contributions, as predicted by the \hathor\ program 
\cite{Aliev:2010zk,Baernreuther:2012ws}\footnote{
By the time of writing,
full next-to-next-to-leading order results for top-antitop production cross sections
have been made available~\cite{Czakon:2013goa}. While we do not include those
new results, the corresponding changes in the employed values for the total cross sections
are however small.},
\be
  \sigma_{t \bar t}(\text{hadronic})     \!=\! 116.2 \text{ pb,}\ \
  \sigma_{t \bar t}(\text{semileptonic}) \!=\! 112.4 \text{ pb}\ \ \text{and}\ \
  \sigma_{t \bar t}(\text{dileptonic})   \!=\! 27.2 \text{ pb.}
\ee
As for single gauge boson production,
those results have been obtained by convoluting partonic cross sections with the parton
density set CT10 of the CTEQ collaboration \cite{Lai:2010vv}.

Single top event generation has been split into the generation of three
different inclusive samples. We distinguish at the parton-level
$s$-channel diagrams where the top quark is produced in
association with a $b$ quark, $t$-channel diagrams where the top quark is
produced in association with a light jet, and $tW$ diagrams describing the
associated production of a top quark and a $W$-boson. In order to maintain this
distinction non-ambiguous, the MLM merging procedure has not been applied since it
possibly implies a double
counting over the three channels due to particular 
diagrams with extra radiation that can
belong to several of the categories, although the
kinematical regimes are different for all three channels. The events generated 
with the leading-order generator \madgraph\ 5 are 
reweighted according to the next-to-leading order precision, after including in the 
cross section calculations 
genuine next-to-next-to-leading order contributions \cite{Kidonakis:2010tc,
Kidonakis:2010ux, Kidonakis:2011wy, Kidonakis:2012db},
\be
  \sigma_t(s\text{-channel}) = 1.81 \text{ pb,}\quad
  \sigma_t(t\text{-channel}) = 28.4 \text{ pb}\quad\text{and}\quad
  \sigma_t(tW\text{-channel}) = 12.1 \text{ pb.}
\ee

\subsection{Diboson production in association with jets}

The simulation of events related to diboson production in association with jets has  been
split into several samples according to the weak boson decay
products (see Table \ref{tab:xsec}).
In the analyses performed in Section \ref{sec:effmonotops} and Section 
\ref{sec:effmultitops}, we always preselect events containing either at least one 
charged lepton or missing energy so that we do not consider pure multijet final states, 
for which the associated production
rate is also widely suppressed compared to the strong production channels. Moreover, we 
have included virtual photon contributions where relevant and consequently
imposed a parton-level selection on
the dilepton invariant-mass of $m_{\ell\ell}~\geq~50$~GeV when both leptons have the 
same flavor and an opposite electric charge.

We have merged hard matrix elements including up to two additional hard jets, 
making use of the MLM merging procedure as described in Section \ref{sec:mcsetup}. To this
aim, we impose that
parton-level jets satisfy the selection thresholds
$p_T^{\rm min} = k_T^{\rm min} = 10$ GeV
and we fix the merging scale to $Q^{\rm m} = 20$~GeV. The
cross sections have been normalized  to the next-to-leading order accuracy as provided by
the {\sc Mcfm} package \cite{Campbell:1999ah, Campbell:2011bn}, employing
the CT10 parton density sets~\cite{Lai:2010vv}. The rates related to 
$W$-boson pair-production are thus given, after including branching 
ratios relevant for the considered final states, by
\be
  \sigma_{W^+W^-}(1\ell\ 1\nu\ 2\text{jets}) = 24.3 \text{ pb} \qquad\text{and}\qquad
  \sigma_{W^+W^-}(2\ell\ 2\nu) = 5.87 \text{ pb,}
\ee
those related to the associated production of a $W$-boson with a $Z$-boson by
\be\bsp
   \sigma_{WZ}(1\ell\ 1\nu\ 2\text{jets}) = 5.03 \text{ pb,}\quad
   \sigma_{WZ}(2\nu\ 2\text{jets}) =&\ 2.98  \text{ pb,}\quad
   \sigma_{WZ}(2\ell\ 2\text{jets}) = 1.58  \text{ pb,}\\
   \sigma_{WZ}(1\ell\ 3\nu) = 1.44 \text{ pb}\quad\text{and}\quad
&  \sigma_{WZ}(3\ell\ 1\nu) = 0.762 \text{ pb,}
\esp\ee
and those related to $Z$-boson pair-production by
\be\bsp
 \sigma_{ZZ}(2\nu\ 2 \text{jets}) =  2.21 \text{ pb,} \quad
 \sigma_{ZZ}(2\ell\ 2 \text{jets}) =&\ 1.18 \text{ pb,} \quad
 \sigma_{ZZ}(4\nu) = 634 \text{ fb,} \\
 \sigma_{ZZ}(2\ell\ 2 \nu) = 319.9 \text{ fb} \quad\text{and}\quad
& \sigma_{ZZ}(4\ell) = 168.3 \text{ fb.} 
\esp\ee
In addition, we have also simulated specific samples describing the production of a pair of 
$W$-bosons with the same electric
charge. The total rate are normalized according to 
\be
  \sigma_{W^\pm W^\pm}(1\ell\ 1\nu\ 2\text{jets}) = 47.5 \text{ fb} \quad\text{and}\quad
  \sigma_{W^\pm W^\pm}(2\ell\ 2\nu) = 12.8 \text{ fb.} \\
\ee
as computed with the leading-order event generator \madgraph\ 5 and the leading order 
set of the CTEQ6 parton density fit \cite{Pumplin:2002vw}.

\subsection{Rare Standard Model processes}
Complementary to the main Standard Model background processes described 
in the previous sections, we have 
generated events related to several classes of rare Standard Model processes. 
First, we consider the production of a top-antitop pair in associated with one additional
(neutral or charged) 
weak boson and possibly extra jets. Our event simulation setup 
allows for the hard matrix elements to contain
up to two additional jets, the MLM-merging parameters 
being set to $p_T^{\rm min} = k_T^{\rm min} = 10$ GeV and $Q^{\rm m} = 20$~GeV. 
The produced samples are then normalized according to the next-to-leading order results
as provided by {\sc Mcfm} \cite{Campbell:2012dh},
\be
  \sigma_{ttW} = 254 \text{ fb} \qquad\text{and}\qquad
  \sigma_{ttZ} = 205 \text{ fb,} 
\ee
all the cross sections being inclusive in the top quark and weak boson decays.

Next, we also consider the production of a top-antitop pair in association with two
$W$-bosons. We normalize the events to the leading-order accuracy and employ the 
cross section as returned by \madgraph\ 5, 
\be
  \sigma_{ttWW} = 13.9 \text{ fb.}
\ee
As above, matrix elements are allowed to 
contain up to two additional jets and the merging parameters are taken 
similarly to the $t\bar t V$ production cases with $V=W,Z$.

We also consider the production of a single top or antitop quark in associated with a 
neutral $Z$-boson and a light or $b$-tagged jet.
Matrix elements containing up to one extra jet are merged, the MLM procedure parameters
being set to 
$p_T^{\rm min} = 20$~GeV, $k_T^{\rm min} = 6.5$ GeV and $Q^{\rm m} = 10$~GeV. We normalize the 
generated event sample to the leading order accuracy, making use of 
the total inclusive cross section as returned by \madgraph\ 5, 
\be
  \sigma_{tzj} = 45.6 \text{ fb,}
\ee
where the subscript $j$ equivalently denotes here light and $b$-tagged jets.

Finally, four-top quark production is simulated without applying
any merging procedure and the generated sample is normalized to the leading-order 
accuracy.  The total rate provided by \madgraph\ 5 reads
\be
  \sigma_{tttt} = 0.7 \text{ fb.}
\ee

\mysection{Monotop production with a CMS-like detector}
\label{sec:effmonotops}
In Section~\ref{sec:rpvpheno}, we have designed an event selection strategy
dedicated to the search for a single top squark produced
via $R$-parity violating supersymmetric interactions and decaying into
a monotop final state. In
this section, we show how this search strategy can serve as a basis for
a monotop event selection in a more general
context. We start from the Lagrangian of Eq.~\eqref{eq:effmonotoplag} and
investigate event samples associated with the four
scenarios {\bf S.I}, {\bf S.II}, {\bf S.III} and {\bf S.IV} of 
Section \ref{sec:monotopsEFT} in order to probe the LHC sensitivity
to monotops in many classes of new physics theories simultaneously\footnote{
The reinterpretation of the results in the framework of a given theory lies
however beyond the scope of this work.}.

In order
to evaluate the parameter space regions that can be probed with 20~fb$^{-1}$ of LHC
data recorded during the 2012 run, we simulate proton-proton collisions
at a center-of-mass energy of 8~TeV. Since the Standard Model background simulation
has already been introduced and detailed in Section~\ref{sec:mc}, we only
provide some details on the simulation of the signal. In this case,
event samples are produced by means of \madgraph~5,
after having implemented
the effective field theory presented in Section~\ref{sec:monotopsEFT} into
\feynrules\
in order to generate the necessary UFO model library for \madgraph~5.
Parton-level squared matrix elements have been convoluted with the
leading order set of the CTEQ6 parton density fit
\cite{Pumplin:2002vw}, fixing both the renormalization and factorization scales
to the transverse mass of the monotop system.
Detector simulation is finally performed for both signal and background
by means of the \delphes\
program~\cite{Ovyn:2009tx}. We have employed a detector setup based on the publicly
available CMS card which however includes a different modeling of the performances of the CMS
detector as described in Ref.~\cite{Ball:2007zza} and a more recent description of the
$b$-tagging efficiency and mistagging rates. The latter is
based, as in Section~\ref{sec:rpvpheno}, on the
TCHEL algorithm of CMS~\cite{CMS:2009gxa,CMS:2011cra}.
We eventually make use of 
the \madanalysis~5~framework~\cite{Conte:2012fm} to
analyze the generated events.

\subsection{Object definitions}
As shown in Section~\ref{sec:rpvpheno}, we can benefit from the
hadronically decaying top quark
plus missing energy signature of the final state to apply a monotop-dedicated
preselection aiming to already largely reduce
the Standard Model background. This selection employs several objects such as
jets that can be $b$-tagged or not, missing transverse energy and isolated charged
lepton (in order to reject any event whose final state features at least one
isolated charged lepton).

Isolated electrons and muons are first required to have their
transverse momentum satisfying
$p_T^\ell~\geq~10$~GeV and their pseudorapidity such that $|\eta^\ell| < 2.5$.
Concerning isolation,
we compute, for each candidate, a variable denoted by $I_{\rm rel}$ corresponding to
the amount of transverse energy, evaluated
relatively to the lepton $p_T^\ell$,
present in a cone of radius $R \!=\! \sqrt{\Delta\varphi^2 +
\Delta\eta^2} \!=\! 0.4$ centered on the lepton, $\varphi$ being the azimuthal angle with
respect to the beam direction. We then require this quantity to satisfy the constraint
$I_{\rm rel} \leq 20\%$.

Jets are reconstructed using
an anti-$k_T$ algorithm~\cite{Cacciari:2008gp} as implemented in the
\fastjet\ package \cite{Cacciari:2005hq,Cacciari:2011ma}, using a radius parameter 
set to $R=0.4$. Among all reconstructed jet candidates, we only
consider those lying within the detector geometrical acceptance, \ie,
those with a pseudorapidity satisfying $|\eta^j| \leq 2.5$. In addition,
we impose their transverse momentum $p_T^j$ to be greater than 30~GeV
and we demand that the ratio between the associated
hadronic and electromagnetic calorimeter deposits is larger than 30\%. 

Finally, the missing energy is calculated on the basis of Eq.~\eqref{eq:met},
\ie, as the scalar sum of the transverse momentum of all the visible objects.

\subsection{Seeking monotops at the LHC}
Hadronic monotop production leads to a final state containing missing
energy related to the undetected new state (denoted by $\chi$ in scenarios
of class {\bf S.I} and {\bf S.II} and respectively
by $\phi$ and $V$ in scenarios of class
{\bf S.III} and {\bf S.IV}) and the decay products of the top quark.
Focusing as in Section~\ref{sec:rpvpheno} on hadronic decays of the top quark,
these decay products consist of one single
$b$-tagged jet and two light (non $b$-tagged) jets.
We therefore preselect events whose final state is comprised of
one single $b$-tagged jet with a transverse momentum of at least 50~GeV
and two or three light jets. This selection allows us to
make use of monotop events containing initial or final state
radiation. This has been found to increase the sensitivity
defined as the significance $S/\sqrt{S+B}$,
$S$ and $B$ respectively being the number of signal and background events after
all selections described in the rest of this section.
In addition, any event with a least one identified charged lepton
is rejected.

After this preselection, about $8\cdot 10^5$ background events are expected,
although this number does not account for possible
QCD multijet contributions which have not been simulated.
As said in Section~\ref{sec:rpvpheno},
we base ourselves on existing experimental and pioneering
phenomenological analyses to ensure
that the full selection strategy presented below
is sufficient to have this source of background under good
control~\cite{Andrea:2011ws,daCosta:2011qk,Collaboration:2011ida}.
These $8\cdot 10^5$ Standard Model background events are comprised in
35\% of the cases of events issued from the production of a leptonically decaying
$W$-boson in association with jets, the charged lepton being either
too soft ($p_T^\ell < 10$~GeV), non-isolated, or
outside the detector acceptance ($|\eta^\ell| >
2.5$)\footnote{Within the \delphes\ simulation setup, leptons
cannot be mis-reconstructed as jets although this contribution has also
to be accounted for in any more realistic experimental analysis.}.
The next-to-leading background component (25\% of the Standard Model
background) is made up of top-antitop events. For half of the events,
both top quarks are found to decay
hadronically whereas for the other half, only one of the top quarks
decays hadronically, the other one decaying
to a non-reconstructed lepton.
The rest of the background finds its origin in
single top production (mainly in the $t$-channel mode) in
20\% of the cases and in the associated production of an invisibly-decaying
$Z$-boson with jets in
15\% of the cases. Any other contribution to the background,
such as events originating from diboson or rarer Standard Model processes,
is at this stage found to be subdominant.

\begin{figure}
  \begin{center}
    \includegraphics[width=0.80\columnwidth]{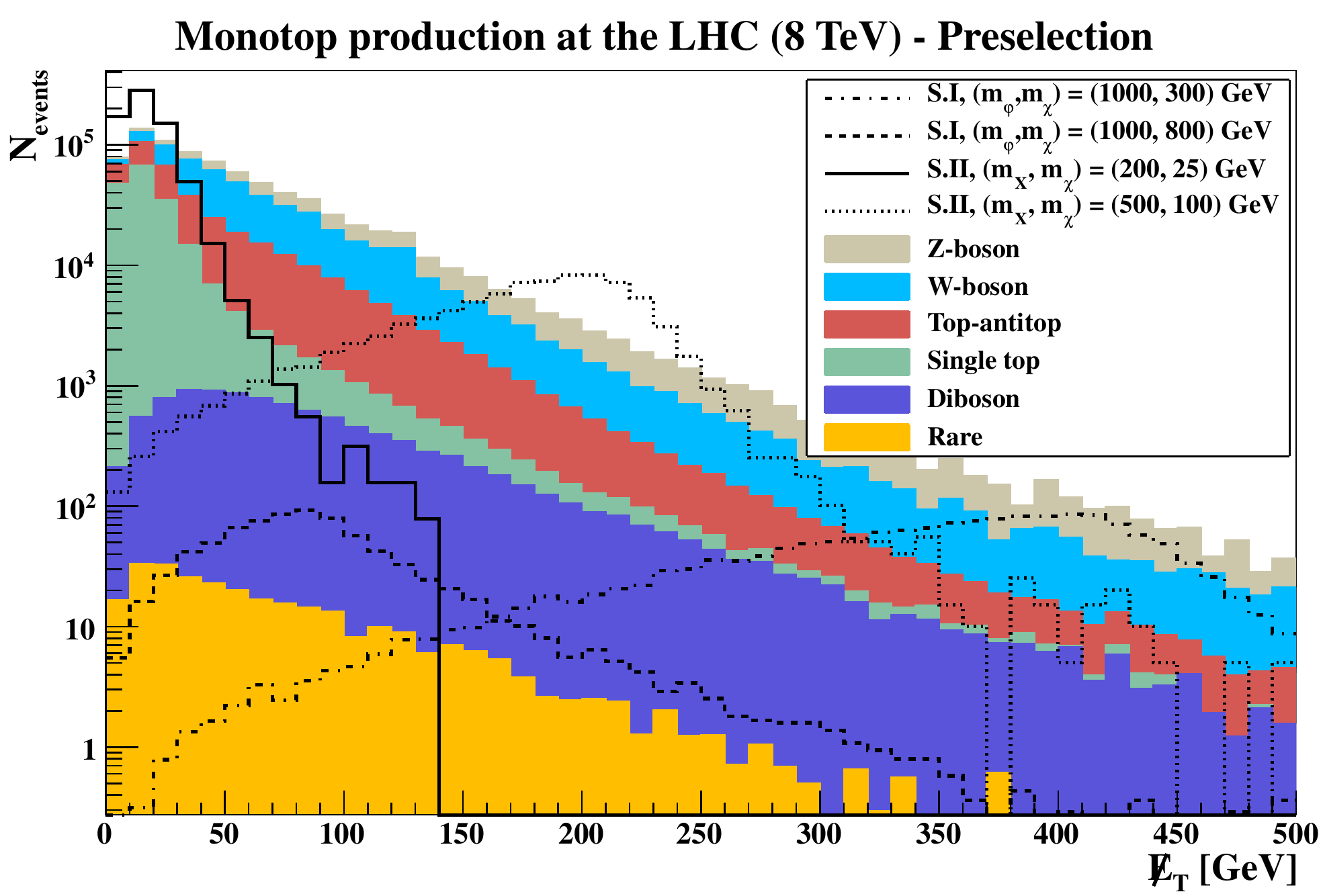}\\
\vspace{.4cm}
    \includegraphics[width=0.80\columnwidth]{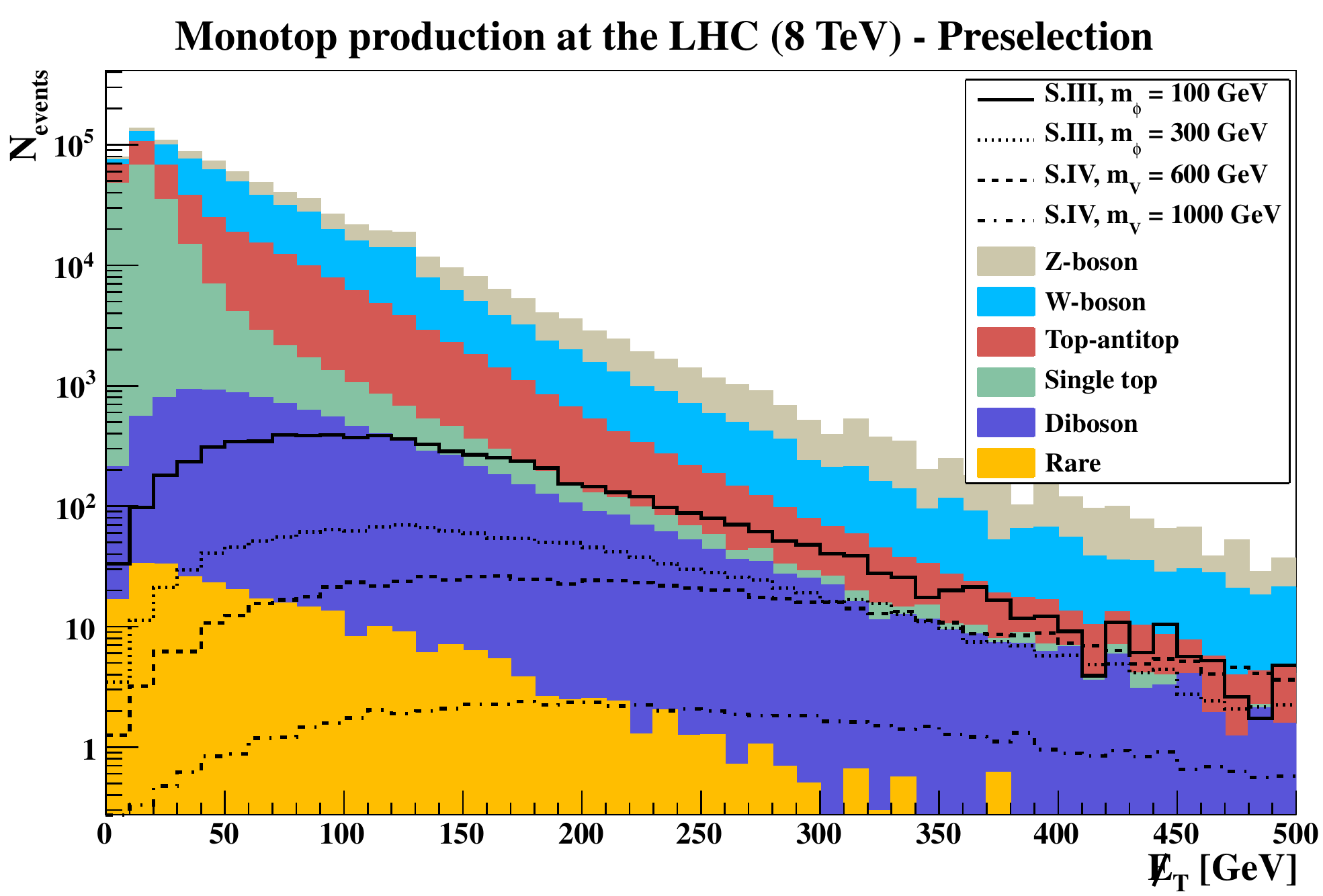}
    \caption{Missing transverse energy distributions after preselecting
    events containing exactly one single $b$-tagged jet, two or three light
    jets and no isolated charged leptons. We distinguish the various contributions
    to the Standard Model background and present results for the production of
    $Z$-boson (gray), $W$-boson (blue), top-antitop pairs (red), single top (green) and
    diboson (mauve) in association with jets, as well as those related to rarer
    Standard Model processes (orange). Predictions for four representative signal
    scenarios of class {\bf S.I} and {\bf S.II} 
    are superimposed to the Standard
    Model expectation for different choices of resonance and invisible particle masses
    on the upper panel of the figure while the lower panel addresses scenarios
    of class {\bf S.III} and {\bf S.IV}.
    The coupling strengths are taken as $a = 0.1$ in all cases.}
    \label{fig:monotopmet}
  \end{center}
\end{figure}

As already mentioned in Ref.~\cite{Andrea:2011ws}, a simple selection
on the missing energy is expected to be sufficient to suppress most of the
background. This is illustrated on Figure~\ref{fig:monotopmet} where
we present the missing transverse energy distribution for the
different background contributions as well as for a small set of
representative signal scenarios. On the upper panel of the figure,
we address scenarios of type {\bf S.I} and {\bf S.II} where the monotop
signature arises from the decay of a resonance. We choose four scenarios
depicting features associated with different regions of the 
parameter space. Two of the benchmark points are of class {\bf S.I}
and exhibit a heavy scalar resonant state
of mass $m_\varphi = 1000$~GeV. The invisible fermion is then either
moderately light ($m_\chi=300$~GeV)
or rather heavy ($m_\chi=800$~GeV). In the first case, the available
phase space for the decay into a monotop state is important while in the second
case, the monotop system has to be produced almost at threshold.
The two last illustrative scenarios are of type {\bf S.II} with a rather light vector
resonance of mass $m_X=200$~GeV and $m_X = 500$~GeV, respectively. For the first scenario,
the mass of the invisible particle must be chosen very light ($m_\chi=25$~GeV) in order
to allow the resonance to decay into a monotop state, while in the second one, its mass
is taken more moderately,
$m_\chi = 100$~GeV. Investigating the missing energy distributions
shown in the figure, we observe a typical resonant
behavior, the spectrum showing an edge (distorted
due to detector effects) at a value depending both
on the mass of the resonance and on the one of the invisible particle. For larger
mass differences, and in particular for very heavy resonances,
the missing energy spectrum extents to larger
values. This suggests a key event selection criterion requiring
an important quantity of missing energy which would yield the rejection
of most of the Standard Model background and allow
us to keep a large fraction of the
signal events. In contrast, when the resonance mass is close to the
sum of the top mass and the invisible particle mass, the position of the
edge of the spectrum lies at lower missing energy values, which renders
the observation of such a monotop state challenging due to the much larger
Standard Model background. We therefore design two different
monotop search strategies, one of them being dedicated to the low mass region
with a selection threshold on the missing energy taken as low as possible
(but reasonable enough to ensure a good control of the Standard Model background),
and another one with a harder selection on the missing energy, expected to be
more sensible to the high mass region.

On the lower panel of Figure~\ref{fig:monotopmet}, we superimpose
to the Standard Model predictions four representative
signal scenarios of class
{\bf S.III} and {\bf S.IV} where the monotop state arises from
a flavor-changing interaction. We consider lighter scalar invisible
states in the case of {\bf S.III} scenarios with
$m_\phi=100$~GeV and $m_\phi=300$~GeV and heavier vector invisible
state in the context of {\bf S.IV} scenarios with
$m_V = 600$~GeV and $m_V=1000$~GeV.
Compared to the resonant case, the
distributions are flatter and show a peak whose value depends
on the mass of the invisible particle. However, even for small masses,
the peak stands already at larger $\slashed{E}_T$ values compared to
the Standard Model background. Similarly, a key selection strategy
based on the missing energy would allow, in most of the cases, for a
good background rejection together with an important signal
selection efficiency.

From those considerations,
we define two options for a missing energy selection and
ask it to satisfy
\be
  \text{ either   } \slashed{E}_T \geq 150~\text{GeV}\qquad\text{or}\qquad
  \slashed{E}_T \geq 250~\text{GeV}\ . 
\ee
The first choice is driven by the missing
transverse energy value for which the ATLAS and CMS detectors can trig on with
an efficiency greater than 70\% \cite{atlasmet, cmsmet}, assuming the use
of missing energy only triggers\footnote{The use of missing energy plus jet triggers
is in principle also possible, but this has been found to decrease the
sensitivity~\cite{monoCMS}.}. Such
a low missing energy selection threshold is expected to increase the
sensitivity to monotop parameter space regions
where the invisible particle is light or,
for scenarios of type {\bf S.I} and {\bf S.II}, when the
intermediate resonance is not that heavy (see Figure~\ref{fig:monotopmet}).
In contrast, the second
choice on the missing transverse energy selection
is dedicated to parameter space regions with a heavier
invisible particle (but only for benchmark points where the production cross section
is large enough) or, in the context of
scenarios of class {\bf S.I} and {\bf S.II}
featuring a heavier resonance. A more stringent missing energy
selection is also associated with a
trigger efficiency closer to unity for both LHC experiments.
Additionally, both choices are also expected to lead to a good control of the non-simulated
multijet background, together with the current
preselection~\cite{Andrea:2011ws,daCosta:2011qk,Collaboration:2011ida}.

After imposing $\slashed{E}_T \geq 150$~GeV, about 45000 background events are found
to survive, most of them being related to the production of an invisibly
decaying $Z$-boson (43\%),  a $W$-boson (37\%) or a top-antitop pair
(15\%) with jets. With the harder selection on the missing energy
$\slashed{E}_T \geq 250$~GeV, only about 8000 background events remain.
In this case, the three main sources of Standard Model background consist of
events associated with the production of a $Z$-boson (53\%),
a $W$-boson (33\%) or of a top-antitop pair (8\%) in association with jets.

\begin{figure}
  \begin{center}
    \includegraphics[width=0.80\columnwidth]{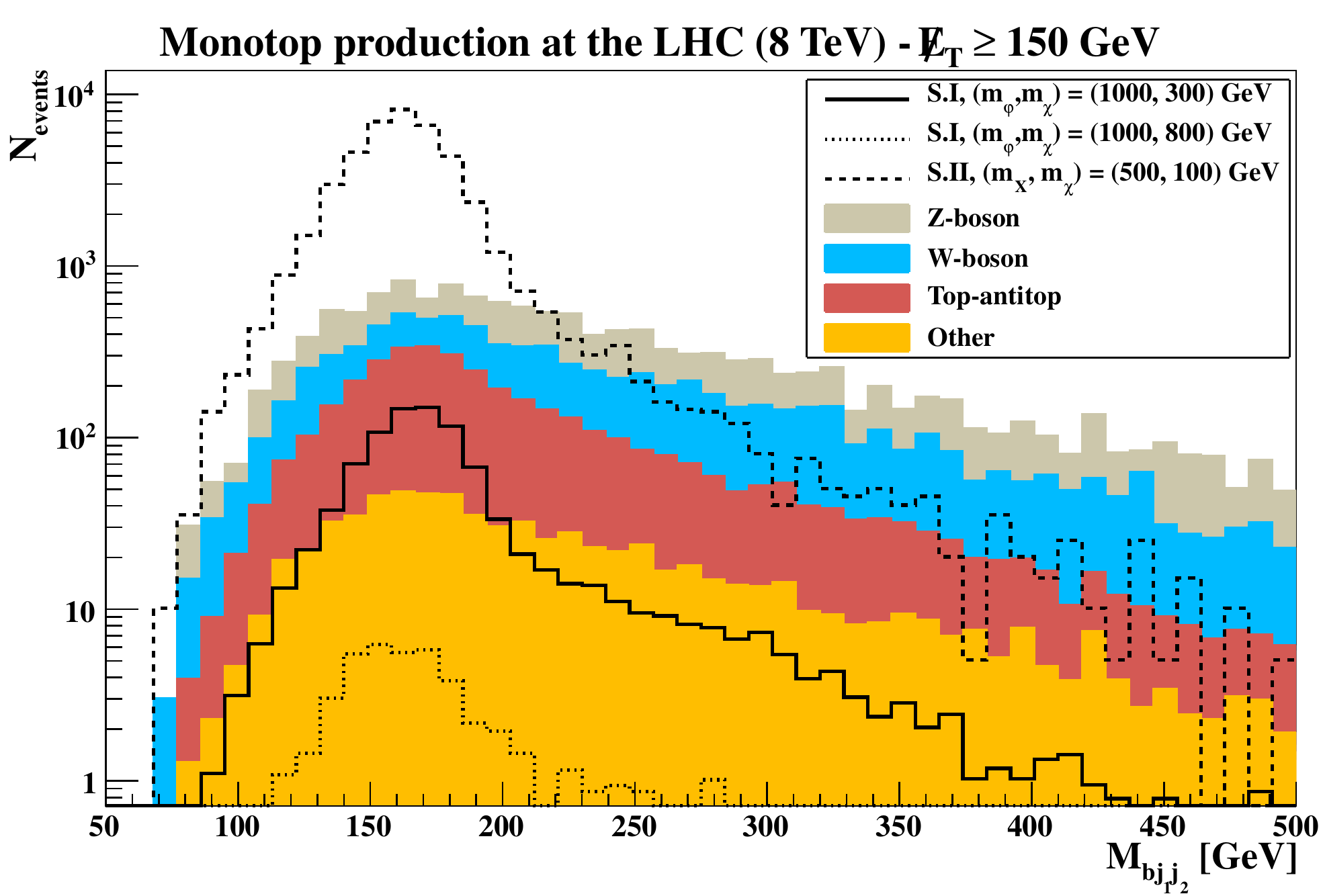}
    \vspace{.4cm}
    \includegraphics[width=0.80\columnwidth]{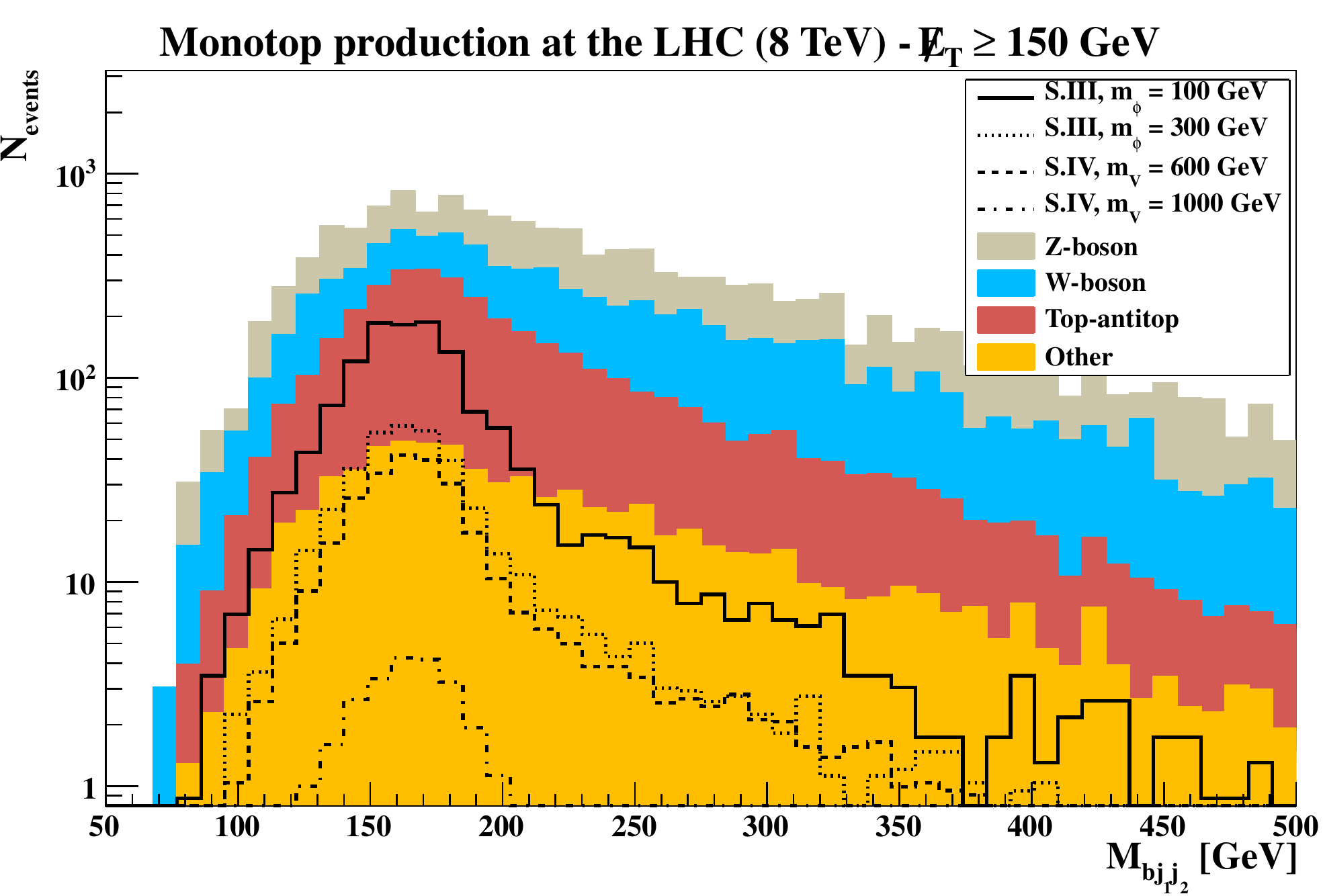}
    \caption{Invariant mass distribution of the three-jet system comprised of the $b$-tagged
    jet and the two light jets defined as originating from a $W$-boson decay,
    after applying the selection
    strategy described in the text with a missing transverse energy
    requirement of $\slashed{E}_T \geq 150$~GeV. We distinguish the various
    dominant contributions
    to the Standard Model background and present results for
    $Z$-boson (gray), $W$-boson (blue) and top-antitop pairs (red)
    production in association with jets, as well as those related to other
    Standard Model processes contributing in a smaller extent (orange).
    Predictions for four representative signal
    scenarios of class {\bf S.I} and {\bf S.II} are superimposed to the Standard
    Model expectation for different choices of resonance and invisible particle masses
    on the upper panel of the figure while the lower panel addresses scenarios
    of class {\bf S.III} and {\bf S.IV}.
    The coupling strengths are taken as $a = 0.1$ in all cases. There is no event
    surviving the selection strategy in the case of the fourth signal scenario
    included in the upper panel of Figure~\ref{fig:monotopmet}.}
    \label{fig:monotopmjjb_a}
  \end{center}
\end{figure}

\begin{figure}
  \begin{center}
    \includegraphics[width=0.80\columnwidth]{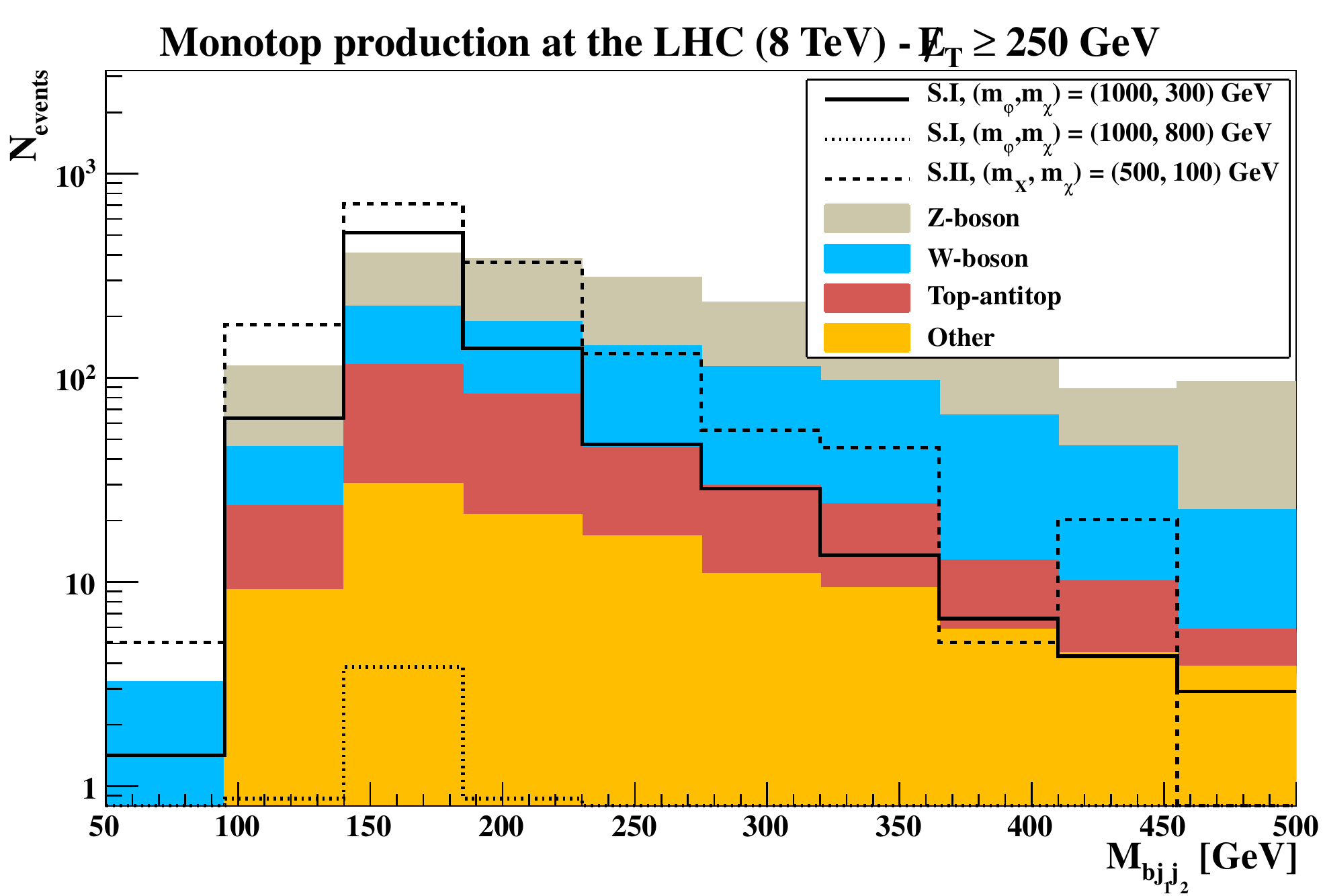}
    \vspace{.4cm}
    \includegraphics[width=0.80\columnwidth]{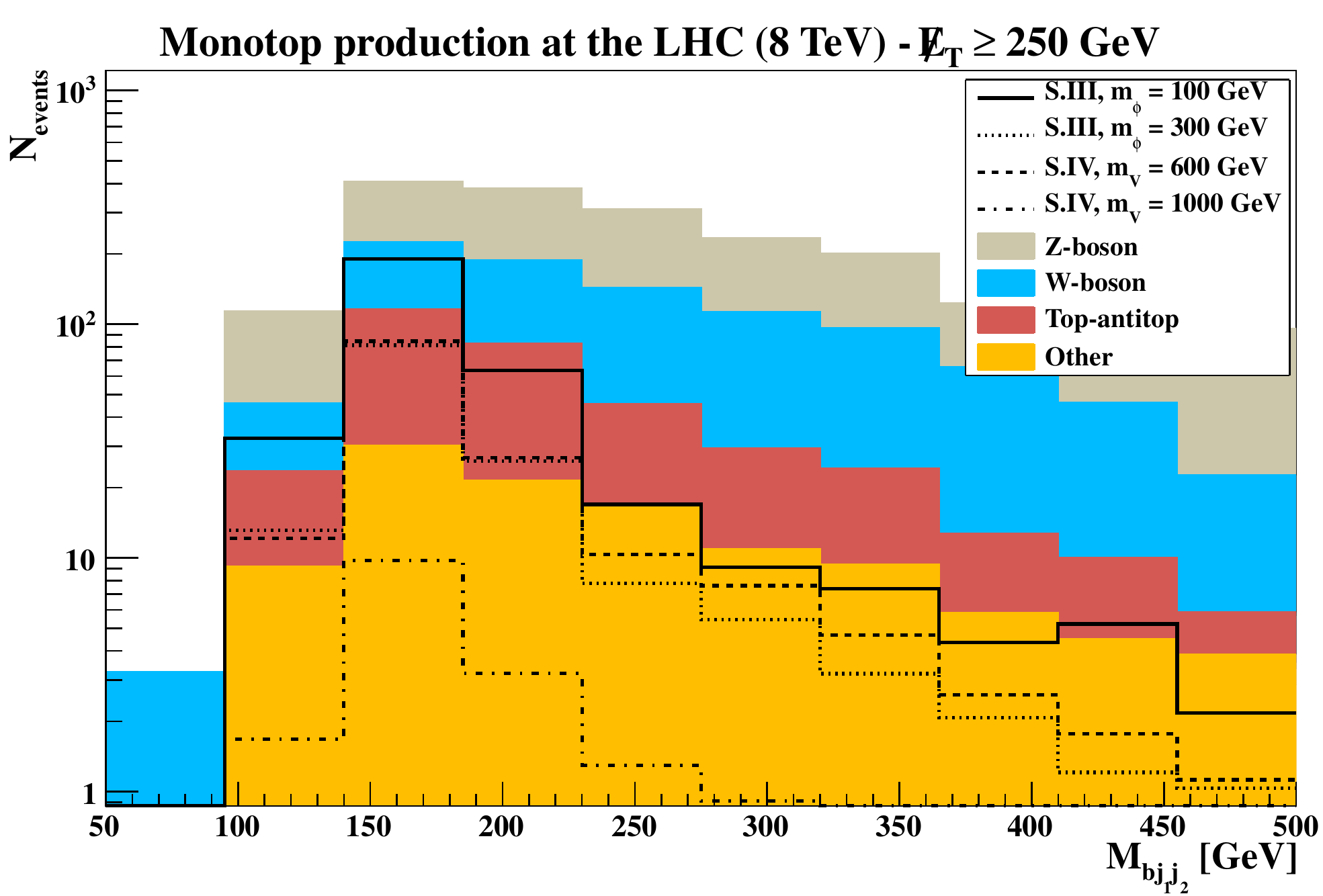}
    \caption{Same as in Figure~\ref{fig:monotopmjjb_a} but for
    a missing transverse energy
    requirement of $\slashed{E}_T~\geq~250$~GeV.}
    \label{fig:monotopmjjb_b}
  \end{center}
\end{figure}

The next steps of the selection take advantage of the configuration of the final
state for signal events. Two of the selected light jets $j_1$ and $j_2$
must have an invariant
mass $m_{j_1j_2}$ compatible with the mass of a $W$-boson so that we reject
events for which
\be
  m_{j_1j_2}\ \slashed{\in}\ [50, 105]~\text{GeV} \ . 
\ee
In the case of events containing
three light jets, we define as $m_{j_1 j_2}$ the invariant mass
of the dijet system whose invariant mass is the closest
to the $W$-boson mass. In addition, we require the leading jet momentum $\vec{p}(j_1)$
to be separated from the missing momentum $\vec{\slashed{p}}_T$
in the transverse plane,
\be
  \Delta\varphi \Big(\vec{\slashed{p}}_T, \vec{p} (j_1)\Big) \in [0.5, 5.75] \ ,
\ee
recalling that $\varphi$ stands for the azimuthal angle with respect
to the beam direction.
Taking into account the $b$-tagged jet $b$, we reconstruct the top quark as the system
comprised of the three jets $j_1$, $j_2$ and $b$. We first demand that the reconstructed
top is well separated from the missing momentum direction,
\be
  \Delta\varphi \Big(\vec{\slashed{p}}_T, \vec{p}(t)\Big) \in [1, 5] \ .
\ee

\begin{table}
  \begin{center}
    \begin{tabular}{|c||c|c|}
       \hline
         Background  & $N_{\rm events}$, $\slashed{E}_T \geq 150$~GeV & $N_{\rm events}$, $\slashed{E}_T \geq 250$~GeV\\
       \hline
           $Z$-boson plus jets & $1411 \pm 38$ & $210 \pm 15$ \\
           $W$-boson plus jets & $1064 \pm 33$ & $148 \pm 12$ \\
           Top-antitop pair plus jets & $1486 \pm 39$ & $105 \pm 10$ \\
           Other background sources & $262 \pm 15$ & $34.7 \pm 5.9$ \\
       \hline
       Total & $4223 \pm 65$ & $497 \pm 22$\\
       \hline
    \end{tabular}
     \caption{\label{tab:monotopfinalB}
  Number of expected monotop events ($N_{\rm events}$)
  for 20 fb$^{-1}$ of LHC collisions at a center-of-mass energy of 8~TeV,
  given together with the associated statistical uncertainties. These numbers have been derived
  after applying
  all the selections described in the text for two different requirements on the missing
  transverse energy. We present results for the different contributions to the
  Standard Model background.}\vspace{.5cm}
 \end{center}
\end{table}

At this stage of the analysis, the Standard Model background consists of about
15000 (2000) events when applying the  $\slashed{E}_T \geq 150$~GeV (250~GeV)
missing energy
requirement and is composed at 40\% (52\%), 33\% (31\%) and 22\% (11\%)
of events originating
from the production of a $Z$-boson, a $W$-boson and a top-antitop pair,
respectively, in association with jets. We illustrate the selection performed so far
on Figure~\ref{fig:monotopmjjb_a} and
Figure~\ref{fig:monotopmjjb_b} by presenting the trijet invariant-mass $m_{bj_1j_2}$
spectrum for the different
(dominant) background contributions and a few representative signal scenarios. The
first series of figures (Figure~\ref{fig:monotopmjjb_a})
is dedicated to the analysis strategy with the softer missing energy selection.
On the upper panel of the figure, we compare the background expectation to predictions
for signal scenarios of class {\bf S.I} and {\bf S.II}, whereas the lower panel
of the figure addresses scenarios of class {\bf S.III} and {\bf S.IV}. Similarly,
Figure~\ref{fig:monotopmjjb_b} concerns the analysis strategy with the hardest missing
energy requirement.

\begin{table}
  \begin{center}
    \begin{tabular}{|c||c|c|}
       \hline
       Signal scenario  & $N_{\rm events}$, $\slashed{E}_T \geq 150$~GeV & $N_{\rm events}$, $\slashed{E}_T \geq 250$~GeV\\
     \hline
     {\bf S.I}, $m_\varphi = 1000$~GeV, $m_\chi = 300$~GeV & $664  \pm  25$ & $581 \pm  23$ \\
     {\bf S.I}, $m_\varphi = 1000$~GeV, $m_\chi = 800$~GeV & $29.4 \pm 5.4$ & $4.1 \pm 2.0$\\
     {\bf S.II}, $m_X =  200$~GeV, $m_\chi =  25$~GeV & $\approx 0$ & $\approx 0$   \\
     {\bf S.II}, $m_X =  500$~GeV, $m_\chi = 100$~GeV & $31047 \pm 171$ & $334 \pm 18$ \\
     {\bf S.III}, $m_\phi =  100$~GeV & $885  \pm 29$  & $212 \pm 15$ \\
     {\bf S.III}, $m_\phi =  300$~GeV & $268 \pm 16$  & $92.0 \pm 9.5$\\
     {\bf S.IV}, $m_V =  600$~GeV     & $191  \pm 14 $ &  $95.6 \pm 9.7$ \\
     {\bf S.IV}, $m_V = 1000$~GeV     & $ 19.7 \pm 4.3$ & $11.0 \pm 3.3$ \\
     \hline
    \end{tabular}
     \caption{\label{tab:monotopfinalS}
  Same as above but for each of the eight
  representative signal scenarios introduced in this section. The free coupling parameter is
  chosen in each case as $a=0.1$.}
  \end{center}
\end{table}

As illustrated on the four subfigures, constraining the invariant mass of the
three jet system, or equivalently constraining the reconstructed top mass, can
help to reduce the background contamination. We hence enforce the $m_{bj_1j_2}$
quantity to lie close to the value of the top mass, since
contrary to the different signal spectra which present a clear
peak centered around
the mass of the top quark, the background expectation exhibits a continuum extending
to much larger values of the invariant mass $m_{bj_1j_2}$.
We therefore reject events for which
\be
  m_{bj_1j_2}\ \slashed{\in}\ [140, 195]~\text{GeV} \ ,
\ee
and obtain the number of events given in Table~\ref{tab:monotopfinalB}
for the different (dominant) contributions to the Standard Model background and in
Table~\ref{tab:monotopfinalS}
for the eight representative signal scenarios investigated in details in this section.
As already mentioned, resonant scenarios of class {\bf S.I} and {\bf S.II}
lead to a number of selected events
largely depending on both the resonant mass, whose production cross section depends on,
and on its difference with the sum of the top mass and the invisible
fermion mass which controls the position of the edge in the missing energy distribution.
In contrast, flavor-changing monotop production as featured in scenarios of class
{\bf S.III} and {\bf S.IV} predicts a number of events surviving the selection strategies
only depending on the invisible particle mass. In all cases, we have
chosen a given value
of the coupling strength $a$ fixed to $a= 0.1$ for the sake of the example.
Results for other values of $a$ can easily be deduced as the number of selected
signal events is proportional to $a^2$.

\begin{figure}
  \begin{center}
    \includegraphics[width=0.49\columnwidth]{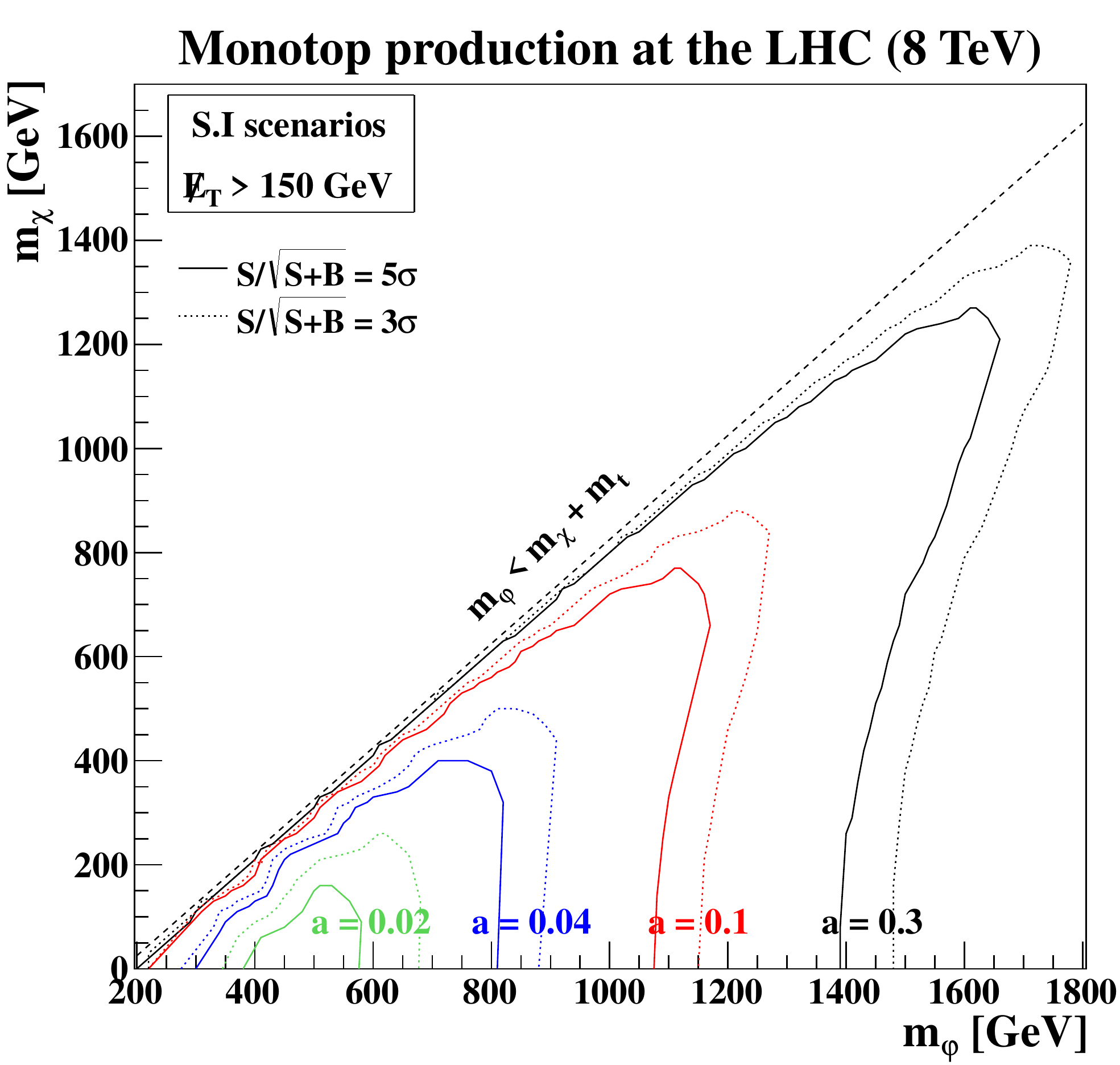}
    \includegraphics[width=0.49\columnwidth]{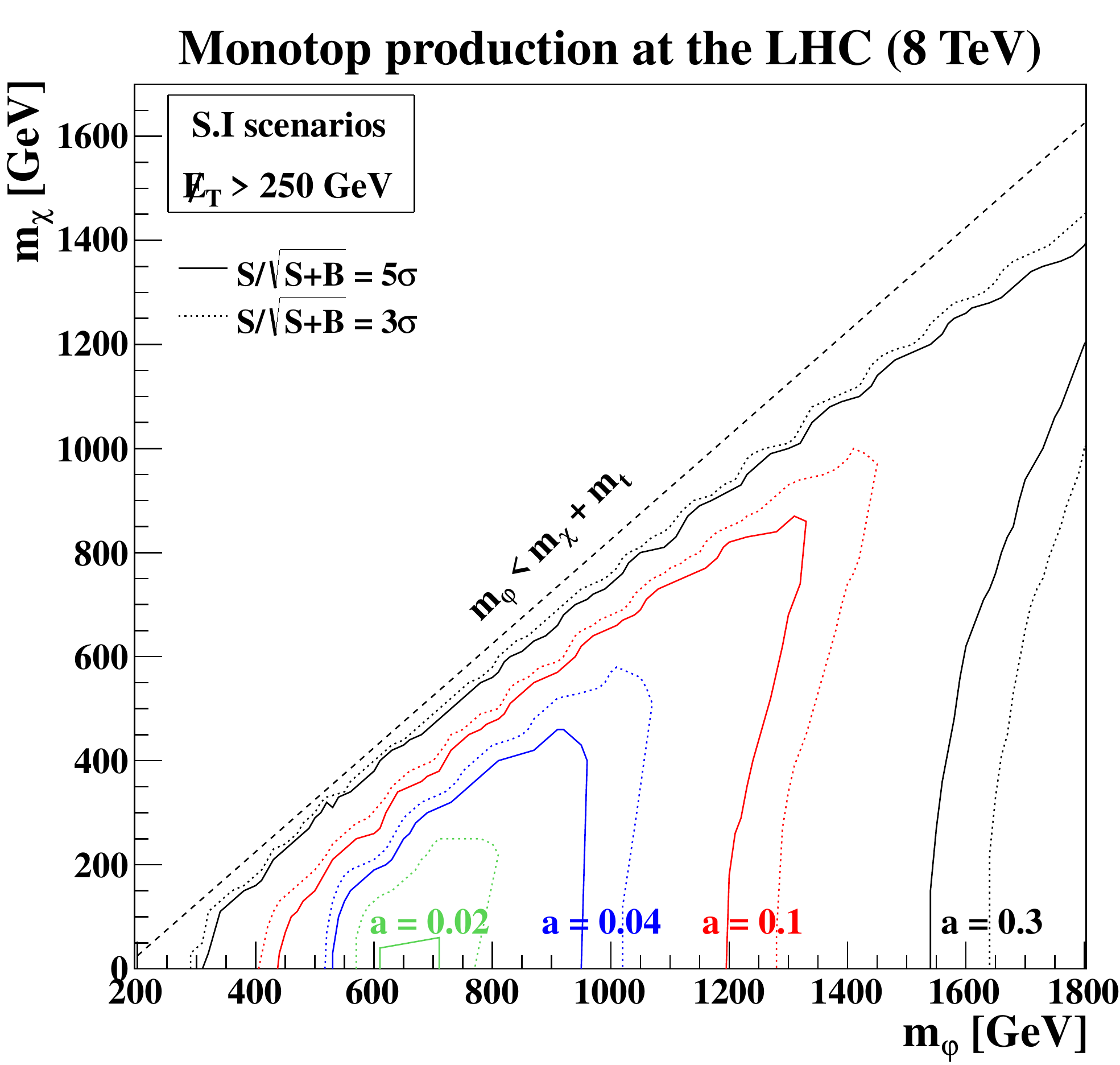}\\
    \includegraphics[width=0.49\columnwidth]{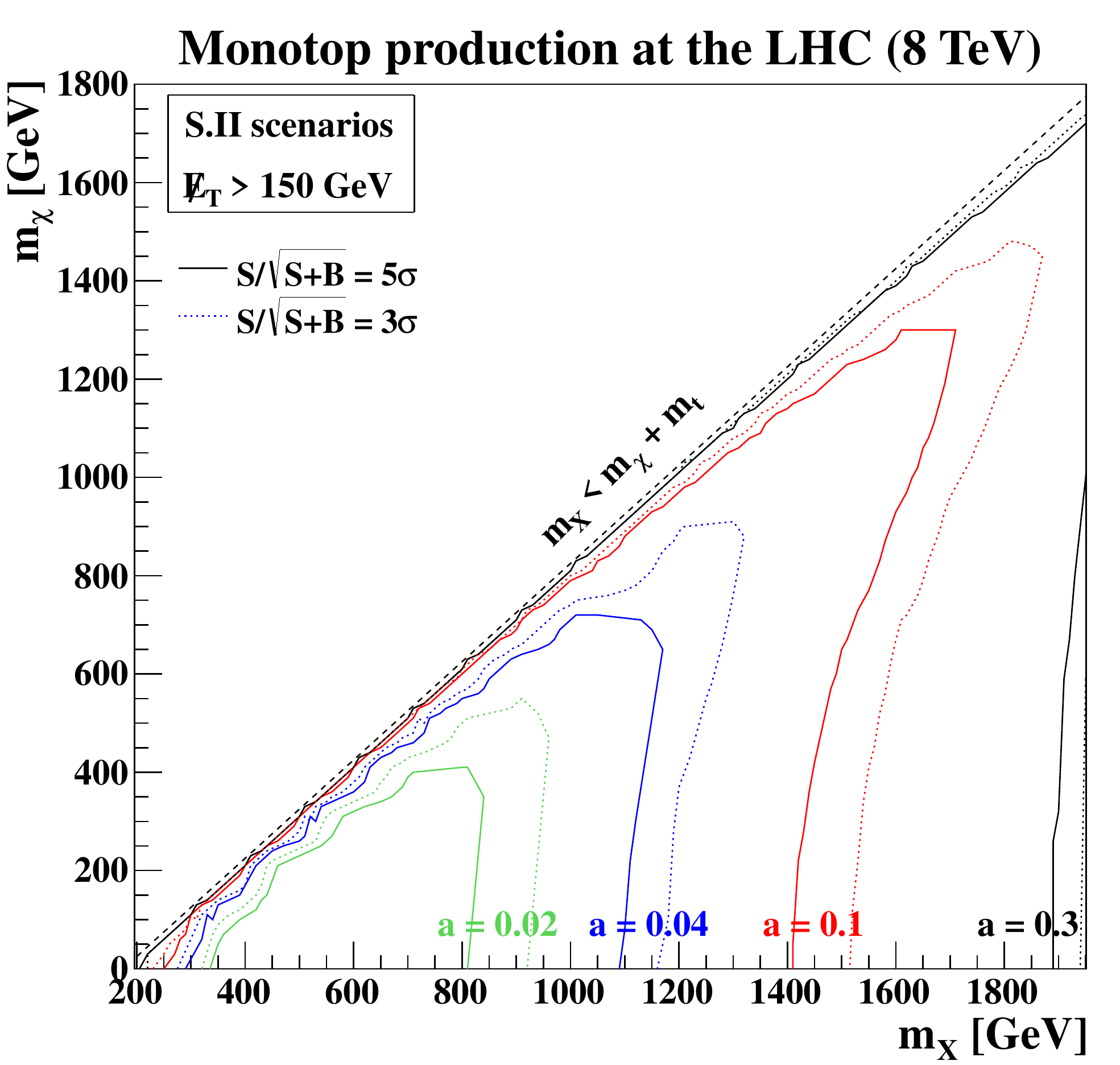}
    \includegraphics[width=0.49\columnwidth]{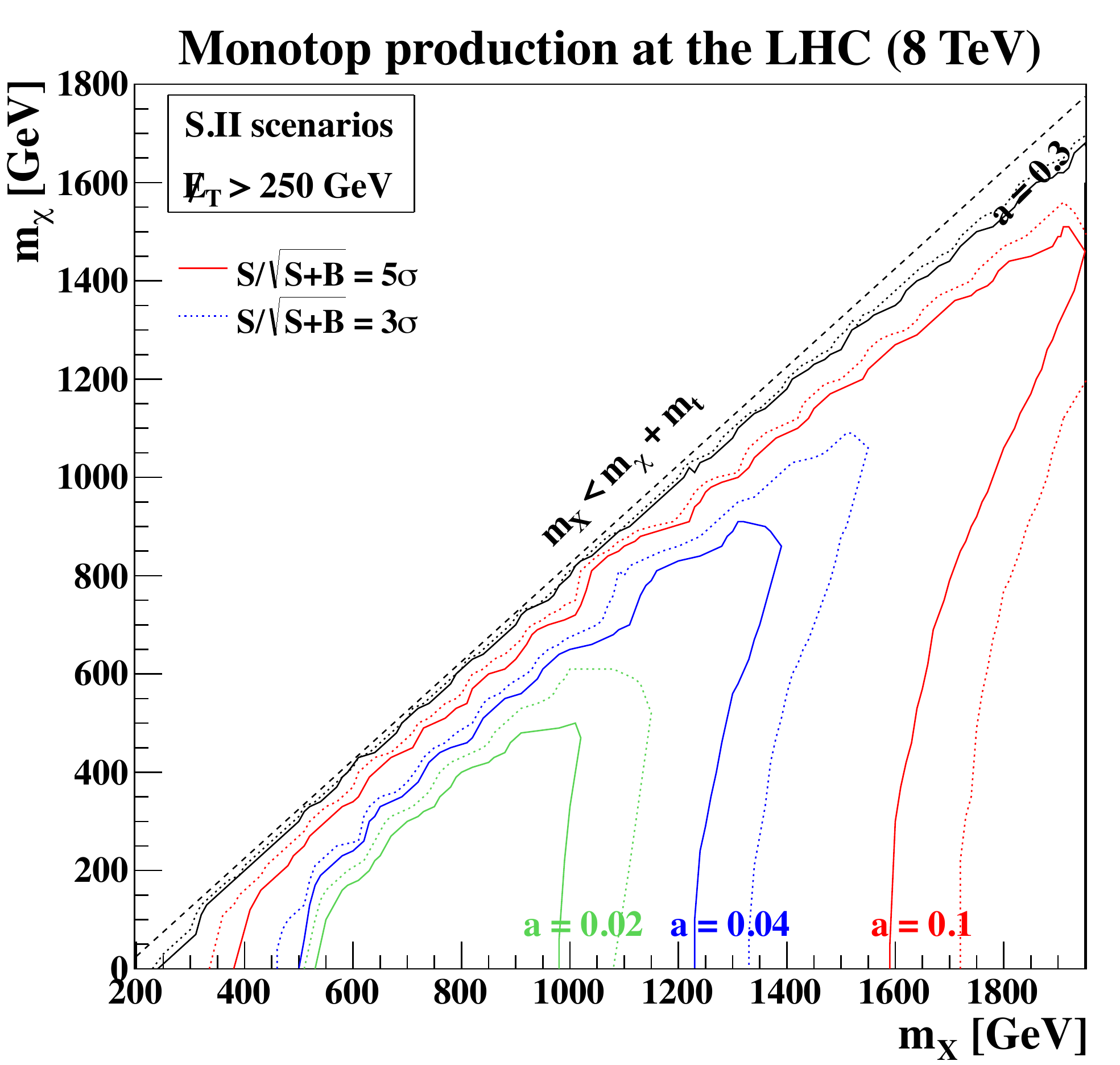}
    \caption{LHC sensitivity to monotop production in the context of scenarios of
    class {\bf S.I} (upper figures) and {\bf S.II} (lower figures) with 20~fb$^{-1}$ of
    proton-proton collisions at a center-of-mass energy of 8~TeV. The sensitivity
    is calculated as the ratio $S/\sqrt{S+B}$ where $S$ is the number of signal events
    surviving all selections presented in the text. The results, given in
    the $(m_\chi,m_\varphi)$ and $(m_\chi,m_X)$ planes for scenarios of class
    {\bf S.I} and {\bf S.II}, respectively, are presented for several values
    of the coupling parameter $a = 0.02$ (green), 0.04 (blue), 0.1 (red) and
    0.3 (black). Moreover, we focus on a search 
    strategy based on a missing energy requirement of $\slashed{E}_T \geq 150$~GeV
    in the left column of the figure, whereas those related to
    the $\slashed{E}_T \geq 250$~GeV selection are shown in the right column of
    the figure.}
    \label{fig:monotopsign_a}
  \end{center}
\end{figure}

\begin{figure}
  \begin{center}
    \includegraphics[width=0.49\columnwidth]{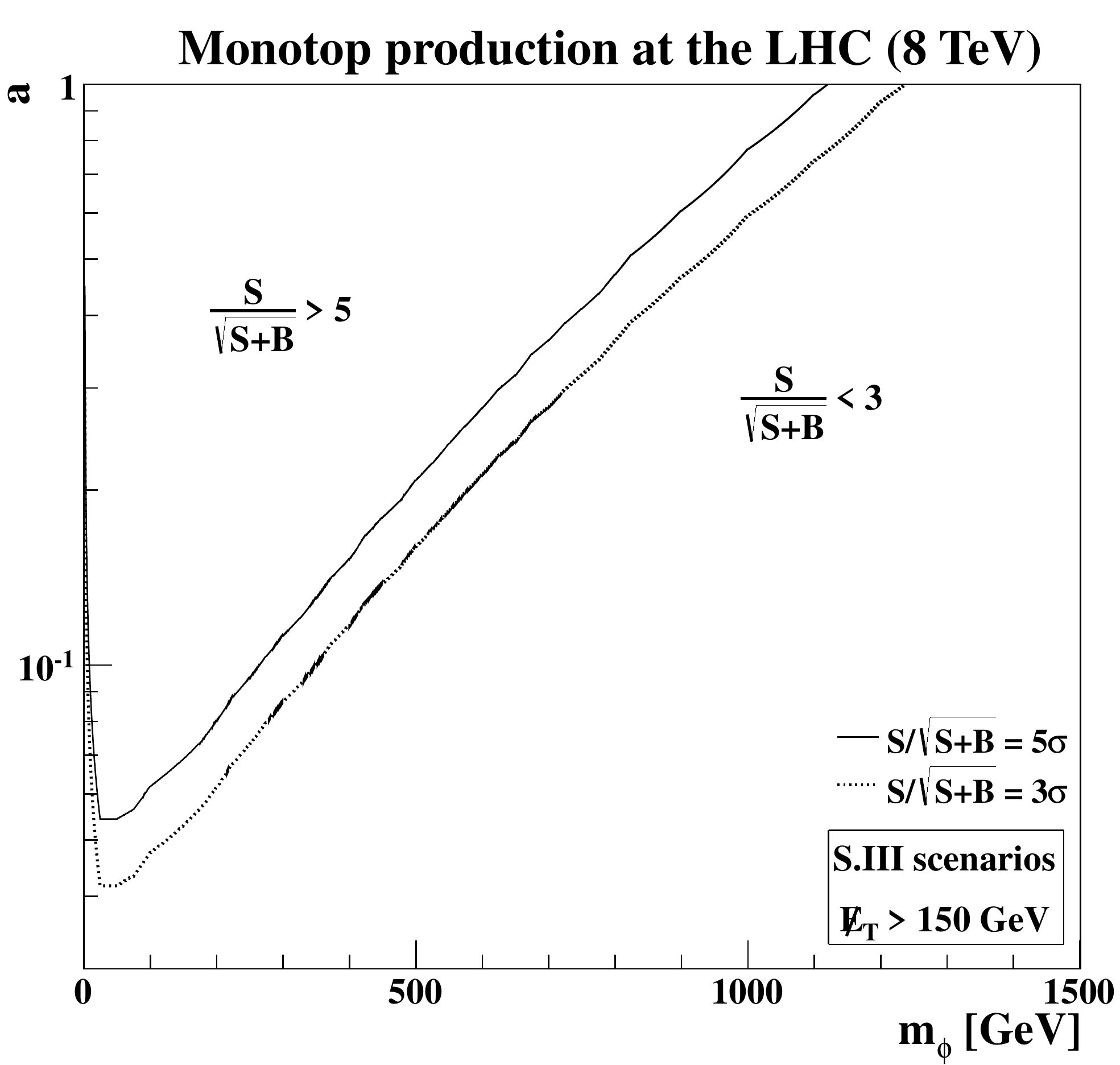}
    \includegraphics[width=0.49\columnwidth]{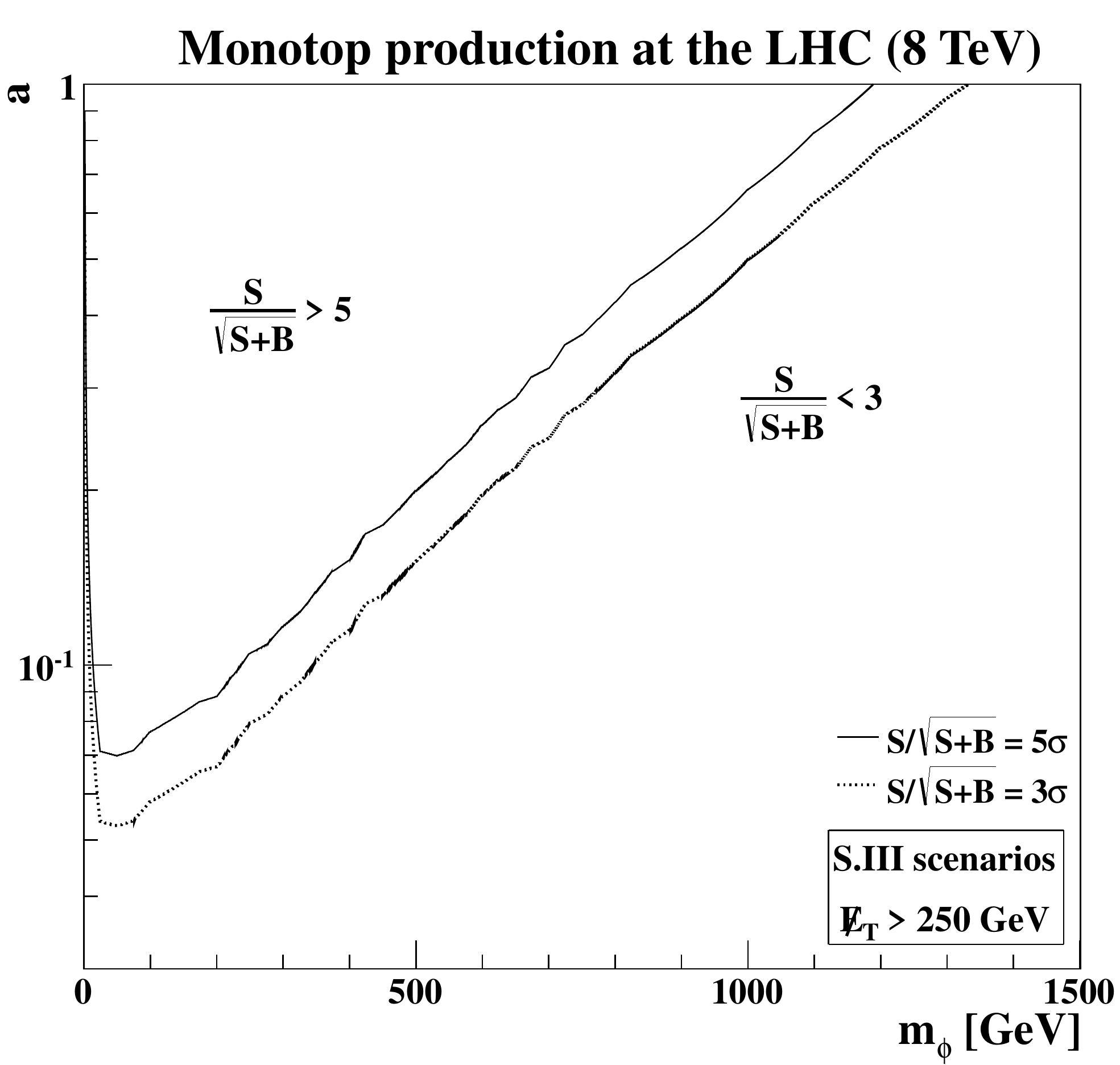}\\
    \includegraphics[width=0.49\columnwidth]{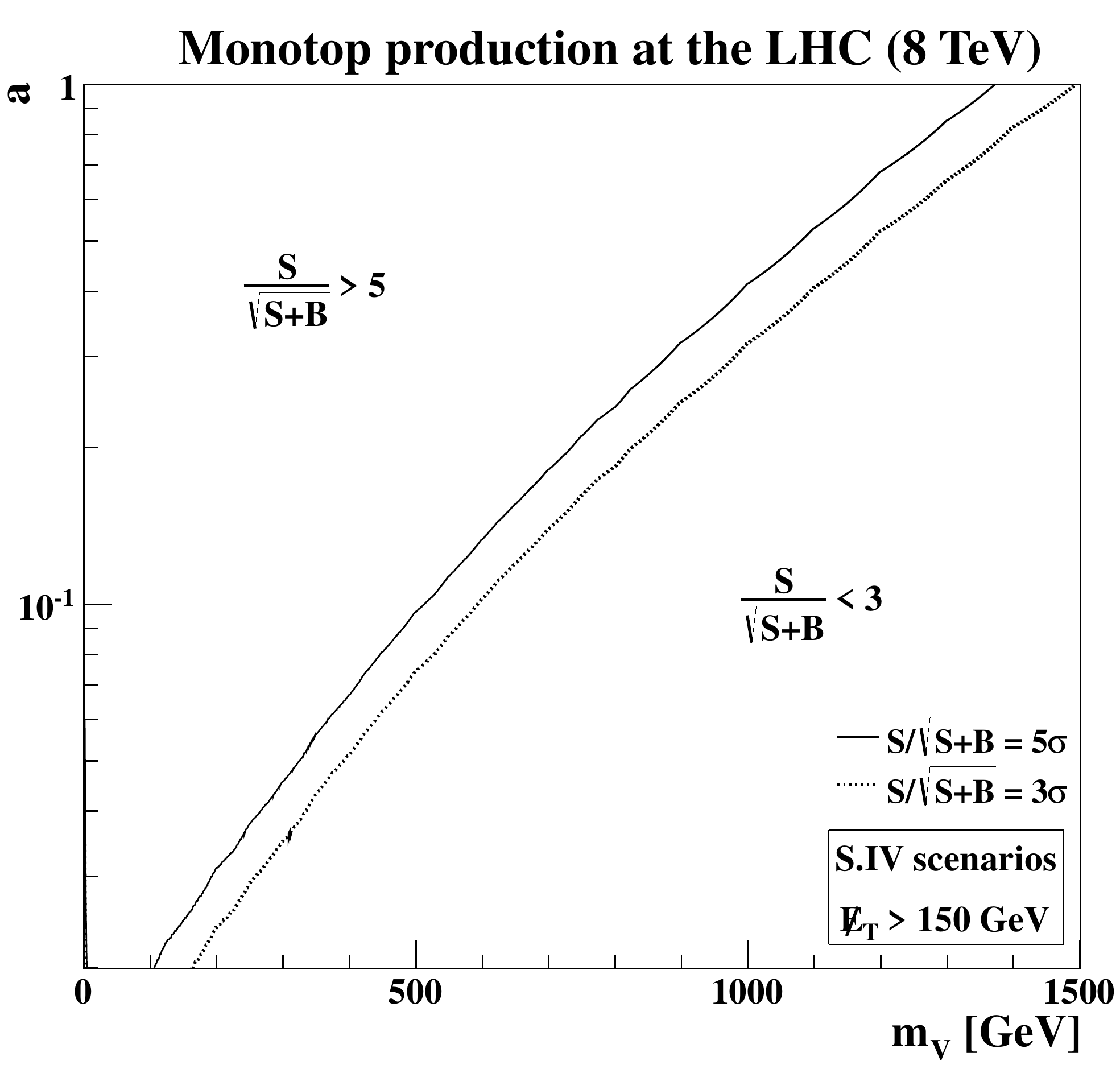}
    \includegraphics[width=0.49\columnwidth]{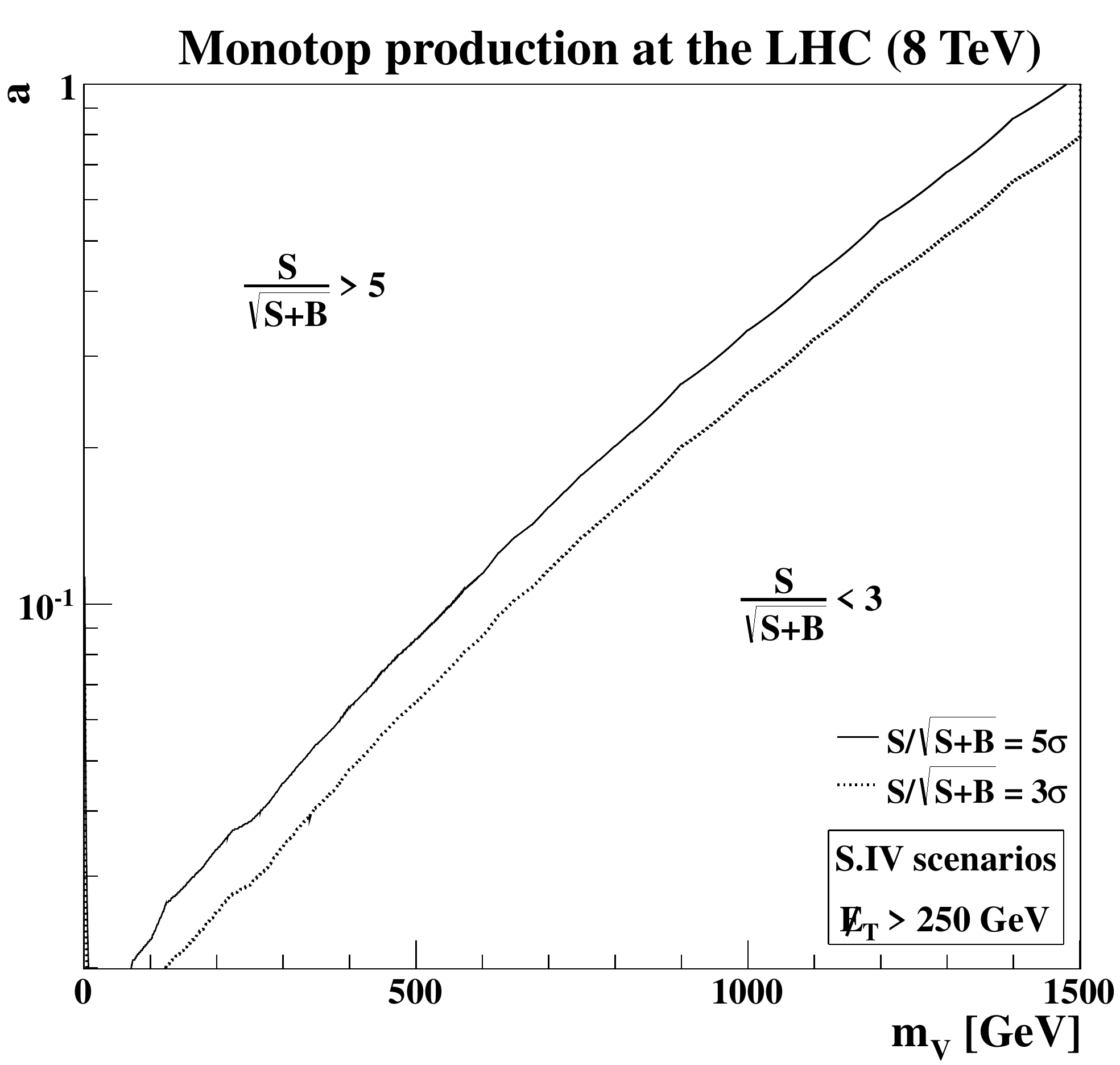}
    \caption{Same as in Figure~\ref{fig:monotopsign_a} but in the context of monotop
    scenarios of
    class {\bf S.III} (upper figures) and {\bf S.IV} (lower figures). The results are
    this time given in
    the $(m_\phi,a)$ and $(m_V,a)$ planes for scenarios of class
    {\bf S.III} and {\bf S.IV}, respectively.}
    \label{fig:monotopsign_b}
  \end{center}
\end{figure}

We now translate the number of signal ($S$) and background ($B$) events
passing all selections in terms of the LHC sensitivity,
with 20~fb$^{-1}$ of proton-proton collisions at a center-of-mass energy
of 8~TeV, to monotops as produced in the context of
scenarios of type {\bf S.I} (upper panel of Figure~\ref{fig:monotopsign_a}),
{\bf S.II} (lower panel of Figure~\ref{fig:monotopsign_a}),
{\bf S.III} (upper panel of Figure~\ref{fig:monotopsign_b}) and
{\bf S.IV} (lower panel of Figure~\ref{fig:monotopsign_b}).
In those figures, we define the sensitivity to each benchmark point
as the significance $s = S/\sqrt{S+B}$ and present the contour lines
where $s=3$ (dotted curves) and $s=5$ (plain curves).

On Figure~\ref{fig:monotopsign_a}, we extract the significance as a function
of both the resonant and the invisible fermion masses for different values
of the coupling strength $a$. We first observe that the LHC is more sensitive
to scenarios where the resonance decaying into a monotop state is a vector particle.
As already mentioned in Section~\ref{sec:monotopsEFT}, this effect is directly related
to the larger cross
section in scenarios of class {\bf S.II} for a given choice of masses. Next, it is found
that very large coupling values of $a \geq 0.1$ implies a very good coverage of
the parameter space by the LHC, both concerning a possible monotop discovery (the $5\sigma$
curves) or the observation of an important deviation (the $3\sigma$ curves) with
respect to the Standard Model expectation. In this way, resonance masses ranging up
to about 1-1.5~TeV (1.5-2~TeV) are accessible, for invisible particle masses of
800-1300~GeV (1200-1800~GeV) in scenarios of type {\bf S.I} ({\bf S.II}). For smaller
coupling strengths, the reaches are reduced, although the LHC remains a
promising machine for accessing the low mass regions of the parameter space. Finally,
comparing the left column of the figure to its right column, we again conclude that
a larger missing energy requirement increases the analysis sensitivity to the high
mass parameter space regions whereas it simultaneously decrease the sensitivity to the low
mass regions.

We recall that we have
neglected here the efficiencies of the missing energy only
triggers \cite{atlasmet, cmsmet}. Even if
requiring $\slashed{E}_T \geq 150$~GeV is above the thresholds for both the ATLAS
and CMS experiments,
the corresponding efficiencies are lower than in the case of a $\slashed{E}_T \geq 250$~GeV selection,
which could slightly changes the picture depicted in the figures.

Similar conclusions are found for scenarios of class {\bf S.III} and {\bf S.IV} in
Figure~\ref{fig:monotopsign_b}.
The results are here presented in two-dimensional planes
with the invisible particle mass on the $x$-axis ($m_\phi$ and $m_V$ for scenarios of
type {\bf S.III} and {\bf S.IV}, respectively) and the coupling strength $a$ on the
$y$-axis. In contrast to the resonant case for which there exists no
public results for monotop searches at a collider experiment, monotop production induced by a
flavor-changing interaction has been searched for by the CDF collaboration
in the case the invisible particle is a new vector state~\cite{Aaltonen:2012ek}
(see also Section~\ref{chap:monotopsCDF}).
Limits on the monotop production cross section
for masses of the invisible vector particle lying in the range 
[0, 150]~GeV have been extracted from data. The non-observation of any
signal event has implied that for a coupling strength of
$a=0.1$, benchmark scenarios for which $m_V\leq 140$~GeV are
excluded. From the curves shown on Figure~\ref{fig:monotopsign_b}, it is clear
that future results from monotop analyses at the LHC could greatly
improve the current constraints\footnote{Although both the ATLAS and CMS
collaborations are currently analyzing data for hints of a monotop signal,
no result is currently publicly available~\cite{monoATLAS, monoCMS} for comparison.}.

\mysection{Sgluon-induced multitop production with an ATLAS-like detector}
\label{sec:effmultitops}

In this section, we use the effective model describing sgluon pair production
and decay constructed in Section \ref{sec:simpsgl} to analyze  
the sensitivity of the LHC through two search strategies, the first one being based 
on a multilepton plus jets signature and the second one on 
a single lepton plus jets signature. The Standard Model background contributions 
are generated as in Section \ref{sec:mc} and we leave out any further detail from
this section. Concerning signal events, they have been simulated
by means of the \madgraph\ 5 program
and reweighted so that the production cross section matches the next-to-leading order
result~\cite{GoncalvesNetto:2012nt, LopezVal:2012ms}. The UFO
model files employed with
\madgraph~5 have been generated after implementing the model described in
Section~\ref{sec:simpsgl} into \feynrules.
While the leading order set of the CTEQ6 parton density fit
is again employed \cite{Pumplin:2002vw}, both the renormalization and factorization scales 
have been fixed to the transverse mass of the sgluon pair.
Parton showering, hadronization
and tau decays are then handled as in Section~\ref{sec:mc}, using
the \pythia~6 package and the \tauola\
program. Finally,
detector simulation is performed, both for the signal and the background,
by means of the \delphes\ program, using
the public ATLAS card.

\subsection{Object definitions}

\begin{figure}
  \begin{center}
    \includegraphics[width=0.65\columnwidth]{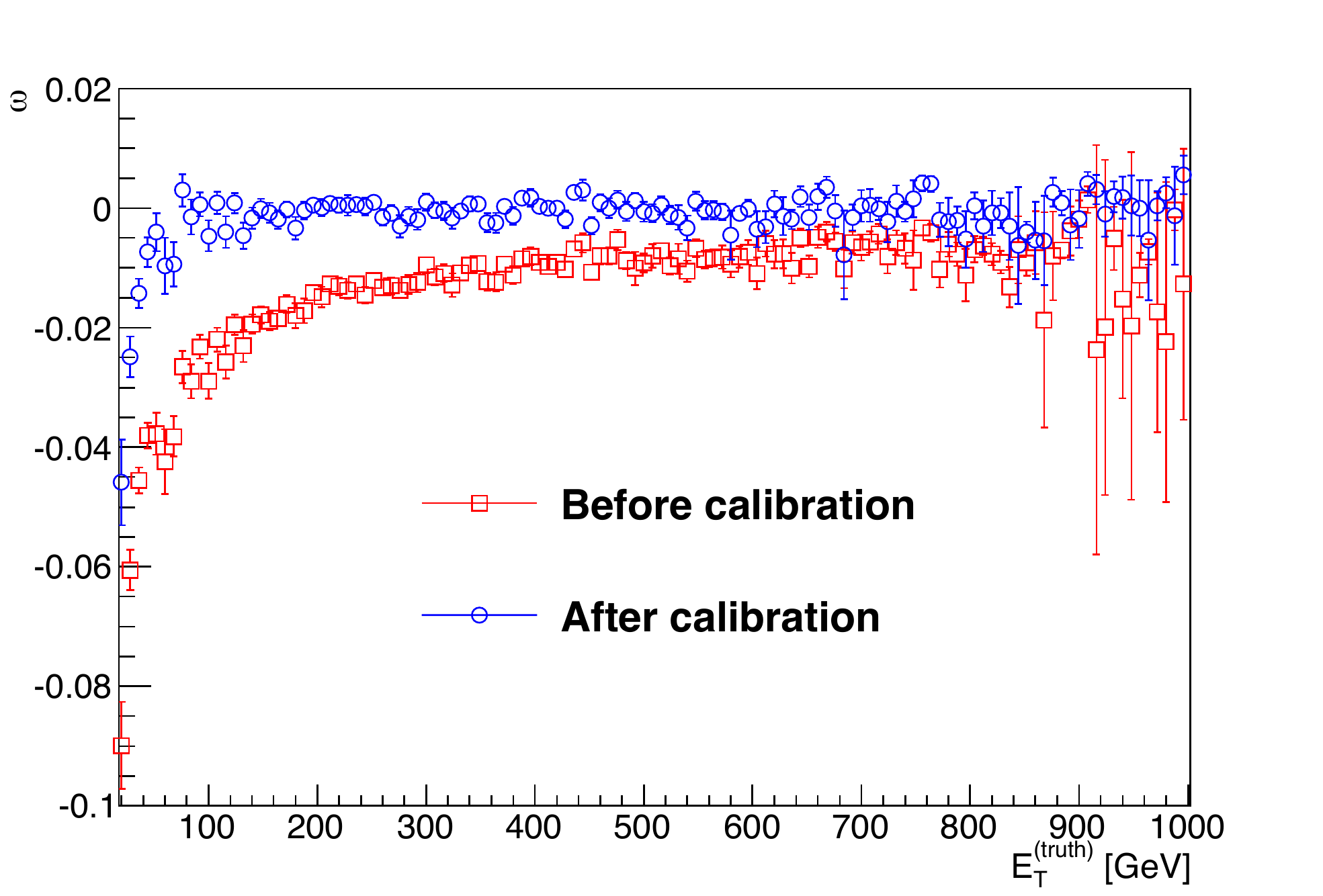}
    \caption{Evolution of the $\omega$-variable defined in Eq.\ \eqref{eq:omjes} 
       with respect to the true transverse energy of the reconstructed
       jet $E_T^{\text{(truth)}}$ before (red squares) and after (blue circles) 
       calibration. Figure taken from Ref.\ \cite{Calvet:2012rk}.}
    \label{fig:jes}
  \end{center}
\end{figure}

Jets are reconstructed using an anti-$k_T$ algorithm \cite{Cacciari:2008gp}
as provided by the 
\fastjet\ package~\cite{Cacciari:2005hq,Cacciari:2011ma},
the radius parameter being set to $R=0.4$. 
In addition, we apply a correction factor to the reconstructed jet transverse energy.
This allows us to account for magnetic field effects as simulated by \delphes\ which
are known to  
introduce large bias in energy reconstruction, in particular for jets with a low 
transverse momentum $p_T$ which get their energy spread out within the detector. 
Denoting by $E_T^{\text{(reco)}}$ the reconstructed jet transverse energy and
by $E_T^{\text{(truth)}}$ the jet transverse energy before detector simulation,
we model these effects through the variable
\be
  \omega = \frac{E_T^{\text{(reco)}} - E_T^\text{(truth)}}{E_T^\text{(truth)}} \ .
\label{eq:omjes}\ee
The evolution of this variable with the (true) jet energy is presented by red squares
on Figure~\ref{fig:jes}. This figure is based on dijet events originating from the
decay of a sequential $Z'$-boson, \ie, a massive vector boson with the same couplings to quarks
and leptons as the Standard Model $Z$-boson. In order to probe the entire
energy range, we have allowed the $Z'$ mass to vary in the range [200, 1000] GeV.
The energy loss reaches
about $5 \%$ for low-$p_T$ jets with $E_T^{\text{(truth)}} = 20$~GeV while it stabilizes
to about $1 \%$ for jets with a transverse energy $E_T^{\text{(truth)}}$ larger than 500~GeV.
Fitting the distribution, we account for magnetic field effects by an \textit{ad-hoc} 
energy calibration. This leads to the application on the reconstructed jet energy of 
the correction function 
\be
  E_T^{\rm (cal)} = \bigg[ 2.62 \cdot 10^{-3} - 
    \frac{ 0.451{\rm GeV}}{E_T^{\rm (reco)}}  \ln \frac{E_T^{\rm (reco)}}{1~{\rm
    GeV}} \bigg]  E_T^{\rm (reco)} \ ,  
\ee 
where $E_T^{\rm (cal)}$ is the jet transverse energy after calibration and where
all the energies are given in GeV. This
procedure allows us to recover a correct jet energy for transverse
energy as low as $E_T^{\text{(truth)}} \sim 40$ GeV, as shown by the blue circles 
on Figure \ref{fig:jes}.

In our analysis, only jets with a 
calibrated transverse energy $E_T^{({\rm cal})} \geq 20$ GeV and a pseudorapidity 
$|\eta| \leq 2.5$, are retained. In addition, we estimate a $b$-tagging efficiency of about 60\%, 
while the associated
charm and light flavor mistagging rate are assumed to be of about 10\% and 1\%, respectively.

Charged lepton candidates
are requested to have a transverse momentum $p_T \geq  20$ GeV and
a pseudorapidity $|\eta| \leq 2.47$ and $|\eta| \leq 2.5$ for electrons and muons,
respectively. In addition, we also impose
two isolation criteria. First, the closest jet to an electron is
removed from the event if their relative angular distance 
$\Delta R  = \sqrt{\Delta\varphi^2 + \Delta\eta^2}
\leq 0.1$, where $\varphi$ stands for the azimuthal angle with respect to the beam
direction. Secondly, in
the case at least one jet is present within a cone of radius $R=0.4$
centered on the lepton, the lepton is this time removed.

\subsection{Searching for sgluons via multitop events at the LHC}
In the context of the two classes of
scenarios introduced in Section \ref{sec:simpsgl} and summarized in Table \ref{tab:params},
sgluon pair production and decay lead to three topologies
comprised of two top quarks and two light jets ($tjtj$), three top quarks and one
light jet ($tjtt$) and four top quarks ($tttt$), denoting top and
antitop quarks by the common symbol $t$ and light jets by the symbol $j$.
In all channels, the final state is thus characterized by a large number of hard jets
(between four and twelve) with an important heavy-flavor content arising from
the top decays. 
We neglect full hadronic channels where each top quark is assumed
to decay into a pair of light jets and a single $b$-tagged jet. Although signal cross 
sections are larger than in the leptonic cases, the overwhelming multijet
background, whose a correct treatment requires data-driven methods, renders any signal
extraction from the background more challenging, so that we restrict ourselves to
leptonic final states, designing two analyses, a first one dedicated to events
containing exactly one lepton and a second one to events with at least two
leptons.

\subsubsection{Event selection strategy for a multilepton plus jets signature}

\begin{figure}
  \begin{center}
  \includegraphics[width=0.80\columnwidth]{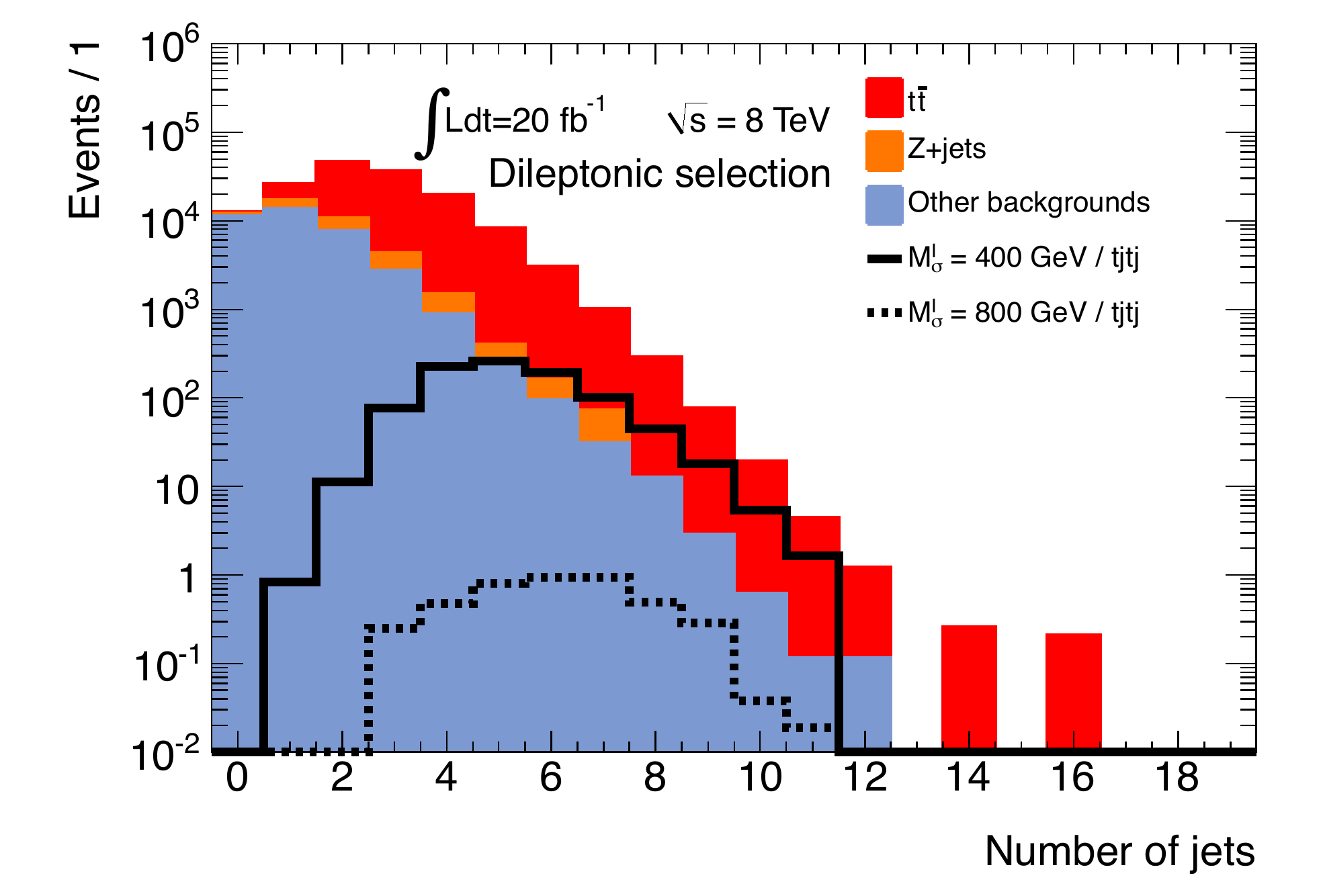}\\
  \includegraphics[width=0.80\columnwidth]{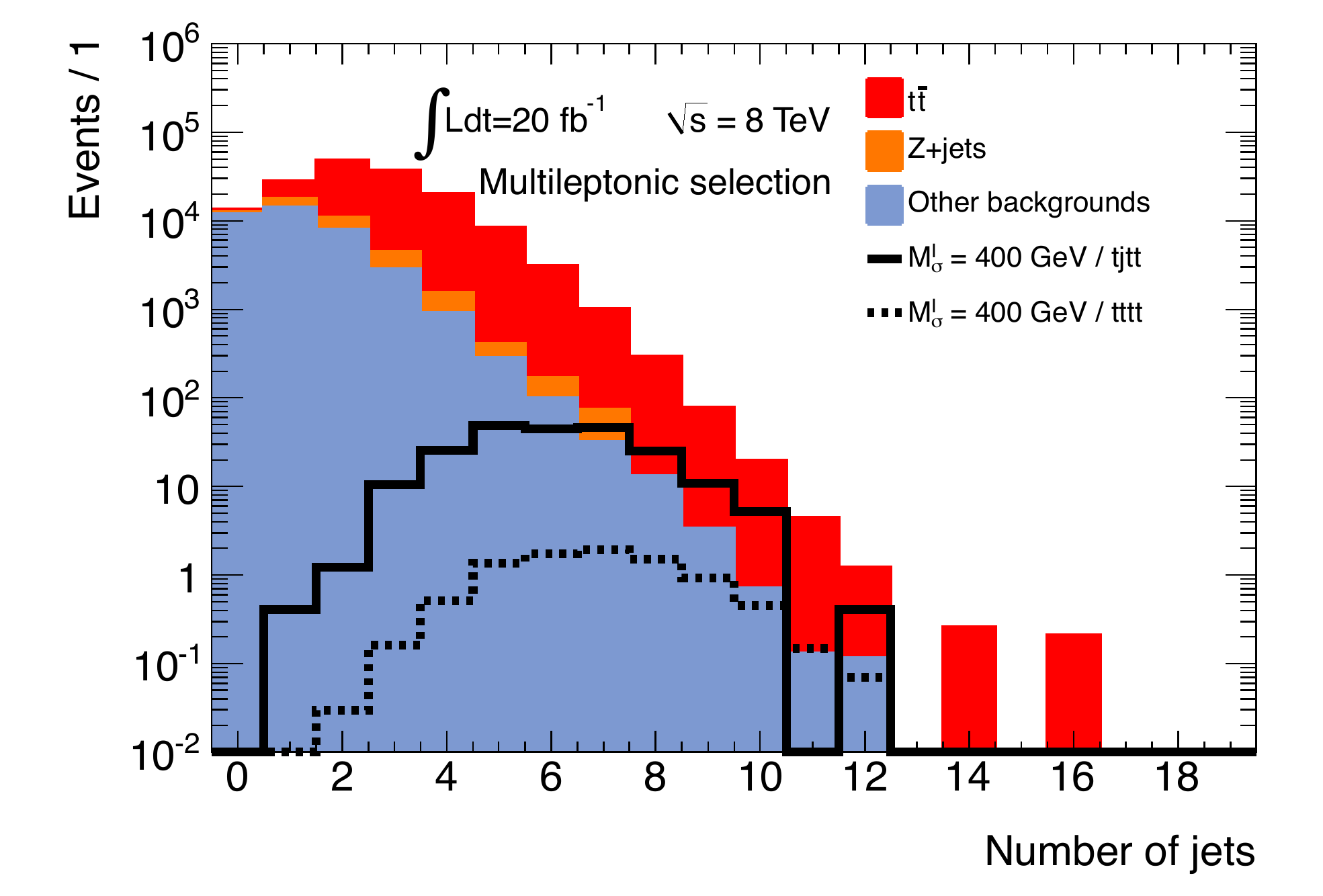}
  \caption{Jet multiplicity distribution after selecting events with exactly
    (upper panel) or at least (lower panel) two leptons, a certain amount of missing transverse
    energy $\slashed{E}_T \geq 40$ GeV and a dilepton invariant-mass $m_{\ell\ell} \geq 50$
    GeV. Contributions arising from top-antitop (red) and Drell-Yan
    (orange) events are factorized from the rest of the background (blue).
    Signal distributions for several representative benchmark scenarios are
    also indicated in the context of the $tjtj$ (upper panel) and the
    $tjtt$/$tttt$ (lower panel) by plain and dashed curves. Figures taken from Ref.\
    \cite{Calvet:2012rk}.}
    \label{fig:nj_2l}
  \end{center}
\end{figure}

Events are preselected with the requirement that they contain exactly (for the $tjtj$ topology)
or at least (for the other topologies) two charged 
leptons with a transverse momentum $p_T^\ell~\geq~20$~GeV. Moreover, the invariant mass
of a lepton pair is imposed to
be $m_{\ell \ell} \geq 50$~GeV to remove hadronic resonances decaying into a lepton pair.
After this preselection, the Standard Model
background contains a large fraction of Drell-Yan dileptonic events (98.7\% and 98.2\%
for the $tjtj$ and $tjtt$/$tttt$ topologies, respectively). To reduce this
background, a selection on the missing transverse energy $\slashed{E}_T$, defined
as in Eq.\ \eqref{eq:met}, is applied. By only considering events where $\slashed{E}_T \geq 
40$~GeV, we take advantage of the fact
that Drell-Yan events are characterized by a lack of missing energy, whereas
neutrinos arising from leptonic top decays ensure signal events to contain a
sensible quantity of missing energy.

\renewcommand{\arraystretch}{1.2}
\begin{table}[!t] 
\begin{center}
\begin{tabular}{ | c || c c c |}
\hline 
  \multirow{2}{*}{Selections}& \multicolumn{3}{c|}{ $tjtj$ channel}  \\ 
   & $M_\sigma^I = 400$ GeV & $M_\sigma^I = 800$ GeV &Backgrounds \\ 
\hline 
  $N_\ell = 2$ with $p^\ell_T \geq 20$ GeV  & $(1.26 \pm 0.02) \!\cdot\! 
     10^3$ & $4.86 \pm 0.30$ & $(1.721 \pm 0.002) \!\cdot\! 10^7$ \\ 

  $m_{\ell\ell} \geq 50$ GeV  & $(1.15 \pm 0.02)\!\cdot\!10^3$ & 
     $4.49 \pm 0.28$ & $(1.716 \pm 0.002) \!\cdot\! 10^7$ \\ 

  $\slashed{E}_T \geq 40$ GeV & $(9.38 \pm 0.20)\!\cdot\!10^2$ & $4.04 \pm
     0.27$ & $(1.549 \pm 0.004) \!\cdot\! 10^5$\\ 

  $N_j\geq 3$ with $p^j_T \geq 25$ GeV & $(9.18 \pm 0.19) \!\cdot\! 10^2$ &
     $4.04 \pm 0.27$ & $(5.693 \pm 0.020) \!\cdot\! 10^4$ \\ 

  $N_b\geq 1$ & $(6.05 \pm 0.16) \!\cdot\! 10^2$ & $2.80 \pm 0.22$ & $(4.089 \pm
     0.011) \!\cdot\! 10^4$ \\ 

  Same sign dilepton & $(2.81 \pm 0.11) \!\cdot\! 10^2$ & $1.06 \pm 0.14$ & 
     $(4.191 \pm 0.035) \!\cdot\! 10^3$ \\ 
\hline 
\end{tabular}\vspace{.25cm} 

\begin{tabular}{ | c || c c c |}
\hline 
  \multirow{2}{*}{Selections}& \multicolumn{3}{c|}{ $tjtt$ channel}  \\ 
   & $M_\sigma^I = 400$ GeV & $M_\sigma^I = 800$ GeV &Backgrounds \\ 
\hline 
  $N_\ell \geq 2$ with $p^\ell_T \geq 20$ GeV & $(2.89 \pm 0.11) \!\cdot\! 10^2$ &
    $4.71 \pm 0.17$ & $(1.722 \pm 0.002) \!\cdot\! 10^7$ \\ 
  $m_{\ell\ell} \geq 50$ GeV & $(2.63 \pm 0.10) \!\cdot\! 10^2$ & $4.44 \pm
    0.17$ & $(1.717 \pm 0.002) \!\cdot\! 10^7$ \\ 
  $\slashed{E}_T \geq 40$ GeV & $(2.17 \pm 0.09) \!\cdot\! 10^2$ & $4.12 \pm
    0.16$ & $(1.598 \pm 0.004) \!\cdot\! 10^5$ \\ 
  $N_j\geq 4$ with $p^j_T \geq 25$ GeV & $(1.97 \pm 0.09) \!\cdot\! 10^2$ & 
     $4.03 \pm 0.16$ & $(2.375 \pm 0.012) \!\cdot\! 10^4$ \\ 
  $N_b\geq 2$ & $83.0 \pm 6.0$ & $1.89 \pm 0.11$ & $(5.950 \pm 0.040) \!\cdot\!
    10^3$ \\ 
  Same sign dilepton & $36.0\pm 4.0$ & $0.77\pm 0.07$ & $(2.860\pm 0.080)
    \!\cdot\! 10^2$ \\ 
\hline 
\end{tabular} \vspace{.25cm}
 
\begin{tabular}{ | c || c c c |}
\hline 
  \multirow{2}{*}{Selections}& \multicolumn{3}{c|}{ $tttt$ channel}  \\ 
   & $M_\sigma^I = 400$ GeV & $M_\sigma^{II} = 800$ GeV &Backgrounds \\ 
\hline 
  $N_\ell \geq 2$ with $p^\ell_T \geq 20$ GeV & \phantom{$ \ \, $} 11.33 $\pm$ 0.33
\phantom{$\ \, $} & 7.90 $\pm$
    0.24 & $(1.722 \pm 0.002) \!\cdot\! 10^7$ \\ 
  $m_{\ell\ell} \geq 50$ GeV & 10.42 $\pm$ 0.32 &7.56 $\pm$ 0.22 & $(1.717 \pm
    0.002) \!\cdot\! 10^7$ \\ 
  $\slashed{E}_T \geq 40$ GeV &8.78 $\pm$ 0.30 &7.03 $\pm$ 0.21 & $(1.598 \pm
    0.004) \!\cdot\! 10^5$ \\ 
  $N_j\geq 5$ with $p^j_T \geq 25$ GeV & 7.50 $\pm$ 0.27 &6.60 $\pm$ 0.20 &
    $(8.11 \pm 0.06) \!\cdot\!10^3$ \\ 
  $N_b\geq 3$  &1.61 $\pm$ 0.13 &1.93 $\pm$ 0.11 & $(1.88 \pm 0.06) \!\cdot\! 10^2$\\
  Same sign dilepton  &0.69 $\pm$ 0.08 &0.82 $\pm$ 0.07 &10.3 $\pm$ 1.5 \\ 
\hline 
\end{tabular}
\caption{Flow charts of the number of events surviving the multilepton 
 selection strategy described in the text for the $tjtj$ (upper panel), 
 $tjtt$ (middle panel) and $tttt$ (lower panel) topologies, in the context of the LHC
 collider running at a center-of-mass energy of $\sqrt{s}=8$~TeV and 
 for an integrated luminosity of 20 fb$^{-1}$. We indicate,
 in addition, the associated statistical uncertainties.
 For the $tjtj$ and $tjtt$ channels, the signal event numbers correspond to scenarios 
 of class {\bf S.I} with a sgluon mass of $M_\sigma^I=400$~GeV and 800~GeV (second and 
 third column of the tables). Concerning the $tttt$ channel, we respectively present instead 
 the evolution of the number of signal events for a
 scenario of class {\bf S.I} with a sgluon mass of $M_\sigma^I = 400$ GeV (second column of the 
 last table) and for a scenario of class {\bf S.II} with a sgluon mass
 $M_\sigma^{II} = 800$ GeV (third column of the last table).
 The sum over all background contributions leads to the event
 numbers shown in the last column of the tables.}
 \label{tab:sel_multi} 
\end{center}
\end{table} 
\renewcommand{\arraystretch}{1.0}

On different footings, jets present in signal events originate mainly from the 
hadronization of the decay products of the top quarks. In contrast, the hadronic
activity in background events is mostly issued from initial-state radiation that
leads to a lower jet multiplicity. This is illustrated in Figure \ref{fig:nj_2l} where
we distinguish purely dileptonic final state arising from $tjtj$ events 
(upper panel of the figure) from
signatures possibly containing more than two leptons as induced by $tjtt$ and $tttt$ events
(lower panel of the figure). To simultaneously maintain a good sensitivity
to the signal and to discard a substantial part of the background, the
presence of at least three, four and five jets with $p_T^j \geq 25$ GeV is demanded
for the $tjtj$, $tjtt$ and $tttt$ final states, respectively. In addition, we
benefit from the presence of heavy-flavor jets arising from the fragmentation of long-lived
$b$-quarks issued from top decays, requiring respectively at least one, two and
three $b$-tagged jets for the $tjtj$, $tjtt$ and $tttt$ search channels.

After applying the above-mentioned requirements to the preselected events, the
signal selection efficiencies, computed from the information indicated in 
Table \ref{tab:sel_multi} containing the number of events surviving each step of the analysis, 
are found to range from 15\% to 50\% for a
sgluon mass of $M_\sigma = 400$ GeV (scenario of class {\bf S.I}) and from 25\% to 60\% for 
$M_\sigma = 800$ GeV (scenario of class {\bf S.I} for the $tjtj$ and $tjtt$ topologies and of 
class {\bf S.II} for the four-top channel). At this stage of the analysis, the 
Standard Model background has been divided by a factor of about 400, 3000 and 100000 for
the $tjtj$, $tjtt$ and $tttt$ search strategies, respectively, and is now largely
comprised of Drell-Yan and top-antitop (plus possibly one or 
two additional gauge bosons)
events. We therefore apply a
specific selection on the dileptonic events and retain only those
where the leptons have the same electric charge. The signal efficiency of such a
criterion is of about 50\% while only 10\% and 20\% of the background
events survive in the context of the $tjtj$ and $tjtt$/$tttt$ topologies. 
As a consequence, the background is eventually dominated by top-antitop 
events for all three search channels, as well as by events related to the associated 
production of a top-antitop pair with one or two additional gauge bosons in the case of 
the $tttt$ search strategy.

\begin{figure}[t!]
  \begin{center}
     \includegraphics[width=0.80\columnwidth]{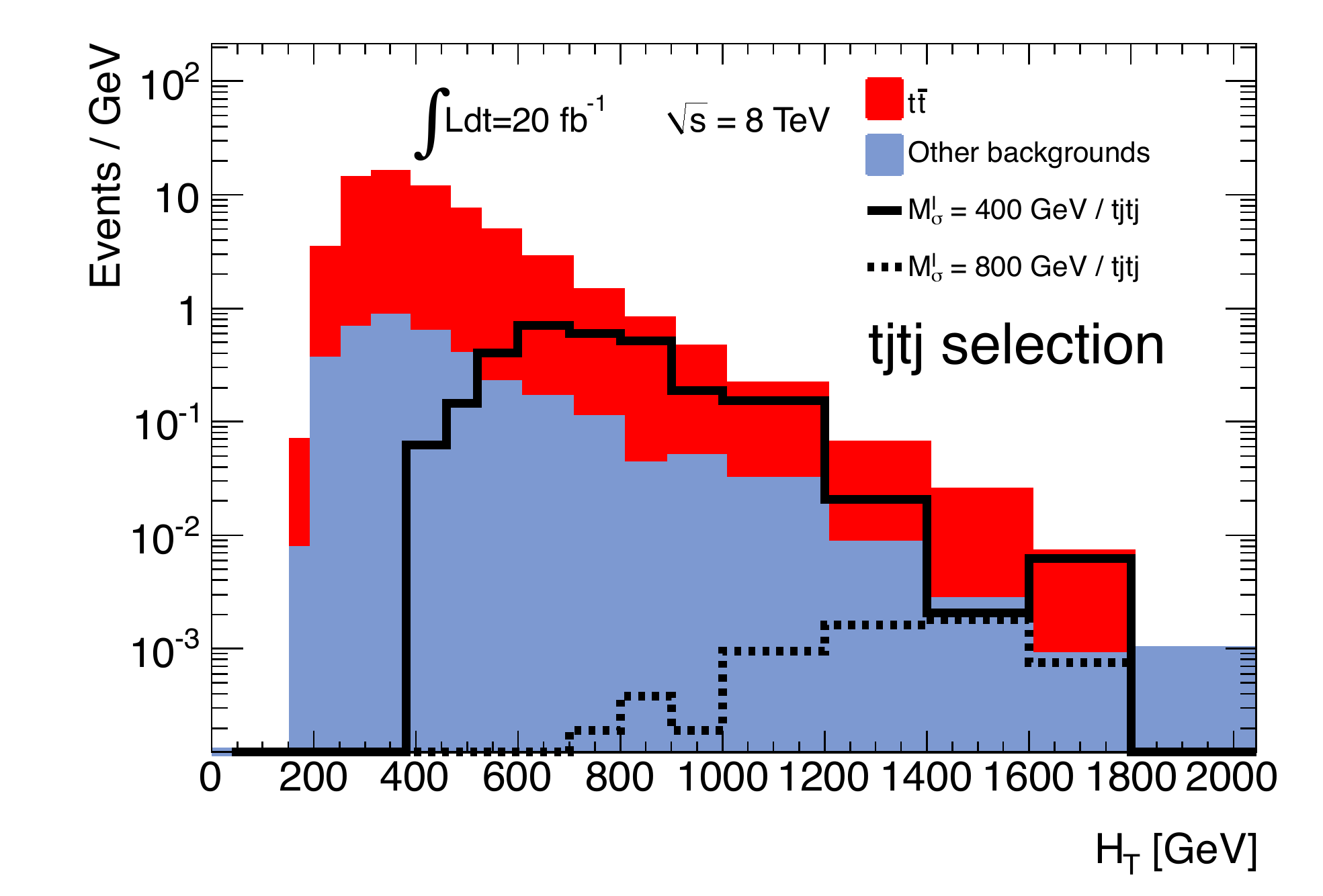}\\
     \includegraphics[width=0.80\columnwidth]{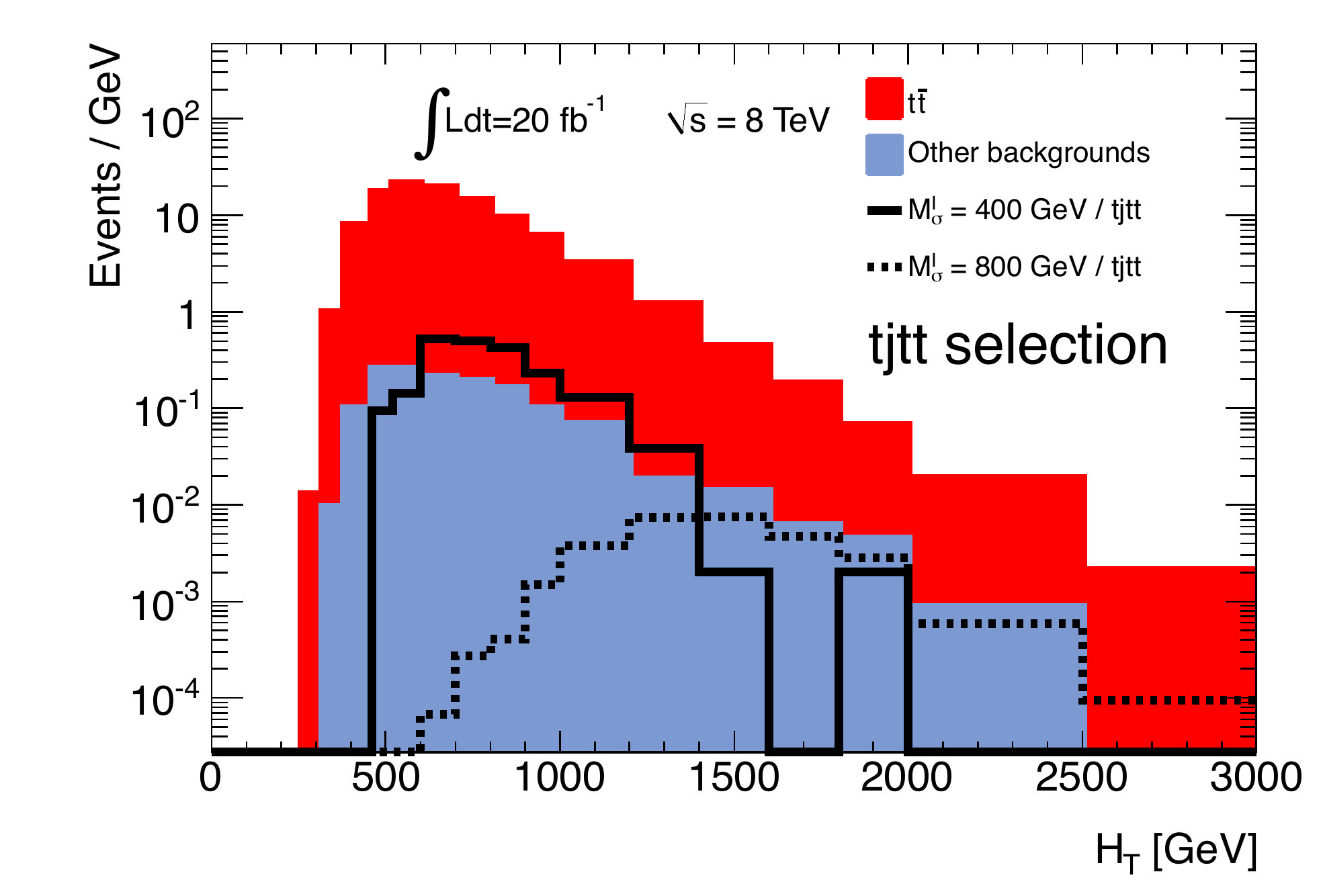}
    \caption{Distribution of the $H_T$ variable defined in Eq.\
  \eqref{eq:htdef} after applying the selection strategy presented in Table 
  \ref{tab:sel_multi} for the $tjtj$ (upper panel) and $tjtt$
  (lower panel) topologies. We distinguish the dominant source of background 
  related to the production of $t \bar t$ pairs in association with jets (red) from
  all the other contributions (blue). We superimpose the distributions obtained for two
  signal scenarios of class {\bf S.I} with respective sgluon
  mass of 400 GeV (plain) and 800 GeV (dashed). 
  Figures taken from Ref.\ \cite{Calvet:2012rk}.}
    \label{fig:HT_multi}
  \end{center}
\end{figure}

However, the multijet background, jets faking leptons and charge
misidentification have not been accounted for in our simulation setup, as already stated above.
Recently, the ATLAS collaboration has shown that after a selection strategy very similar to
the one performed in this work, neglecting these effects leads to an underestimation of 
the background contributions by a factor of ten \cite{ATLAS:2012hpa}. We
therefore adopt a conservative approach and derive below
two limits on sgluon-induced new physics in multitop events. First, limits
are extracted after omitting the non-simulated background contributions. Next, they are derived
after multiplying the number of background events by a factor of ten.

Since signal events are expected to contain more jets 
and leptons than background events, we consider the $H_T$ variable 
defined by Eq.\ \eqref{eq:htdef} to discriminate signal from background.
Omitting the $tttt$ channel as its statistical significance is very poor (see Table
\ref{tab:sel_multi} for two illustrative benchmark scenarios), we respectively present
$H_T$ distributions for the $tjtj$ and $tjtt$ topologies on the upper and lower panels of Figure
\ref{fig:HT_multi}. We depict curves associated with
signal scenarios of class {\bf S.I} where the sgluon mass is fixed to 400 GeV and 800
GeV, as in Table \ref{tab:sel_multi}. The distributions present a steep rise once the production
threshold is reached, followed by a large peak centered around twice the sgluon mass. 
We compare these curves to the background distributions, which are indicated after
having factorized the dominant $t\bar t$ contribution (in red) from the rest of
the background events (in blue). The large differences among the shapes of background and 
signal
distributions suggest us to probe the LHC sensitivity to the presence of sgluon fields coupling
dominantly to top quarks by means of this
$H_T$ observable, using the full spectrum rather than applying an (inefficient) selection
(see below).

\subsubsection{Event selection for a single lepton plus jets signature}
The branching ratios of the $tjtj$ and $tjtt/tttt$ topologies into states with one single
lepton are large and reach  36\% and about $41\%$, respectively. Therefore, events with a
single lepton
are expected to be copiously produced at the LHC from the production and decay of a sgluon
pair. To probe the sensitivity to such a signature, we preselect events by demanding
exactly one charged lepton with a transverse momentum $p_T^\ell \geq 25$ GeV. In this case,
the background from the Standard Model is dominated at $92\%$ by events
originating from the associated production of a $W$-boson with jets. We illustrate the effect
of such a selection on the signal by choosing representative scenarios with sgluon masses
of 400~GeV and 800~GeV, like in the previous subsection.
The results are presented in Table \ref{tab:sel_single} where it is shown that the
expected number of signal events ranges from 34.6 (23.2)
to 10600 (45.7) for a sgluon mass of 400~GeV (800~GeV).

\begin{figure}
  \begin{center}
    \includegraphics[width=0.80\columnwidth]{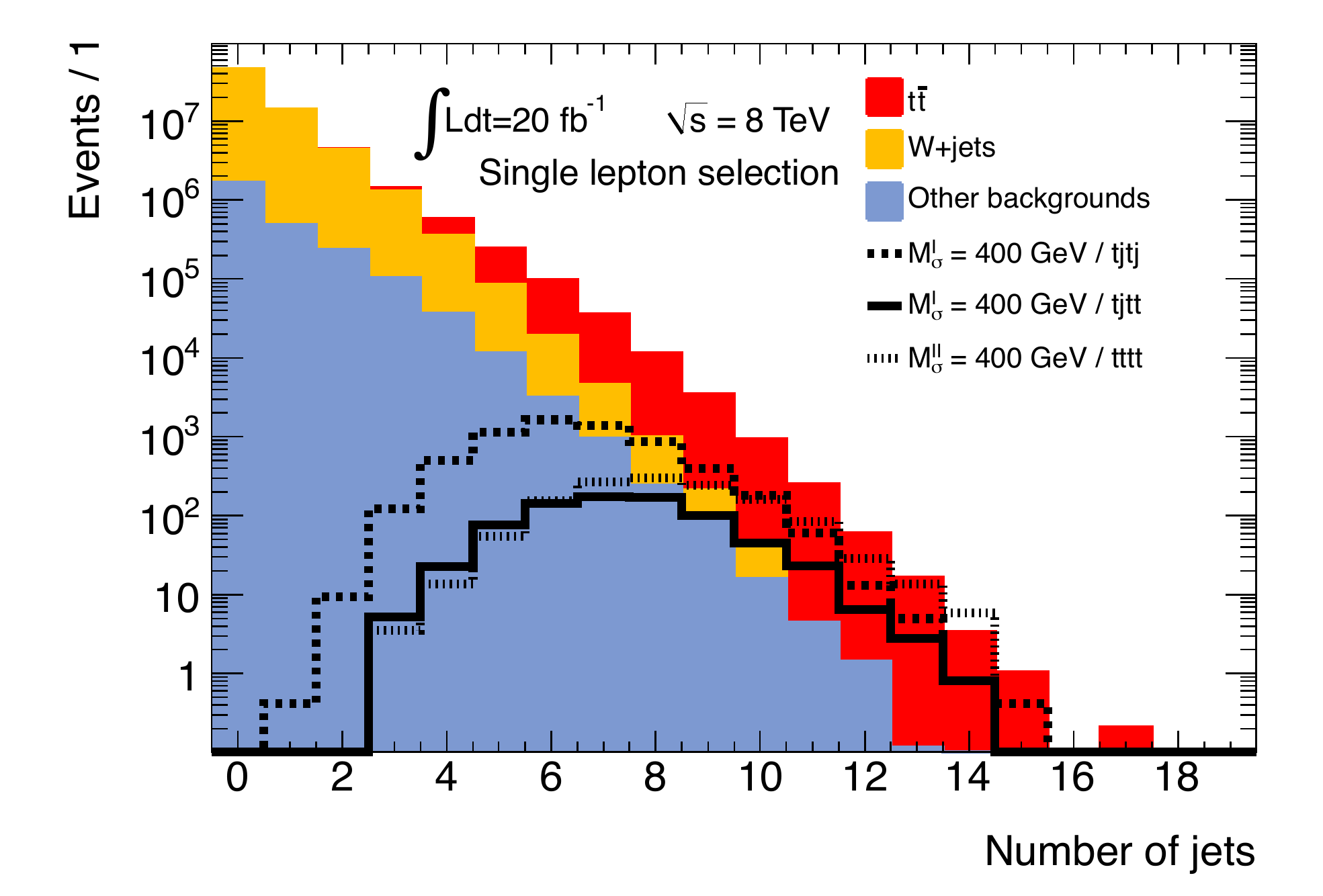}
    \caption{Jet multiplicity distribution after selecting events with exactly 
    one lepton, $\slashed{E}_T~\geq~40$ GeV and $M_T^W \geq 25$ GeV. We distinguish the 
   $t\bar t$
   (red) and $W$-boson plus jets (orange) contributions from the rest of the background
   (blue). We then superimpose signal distributions related to pair production of 400 GeV
   sgluons, the latter coupling
   universally to all up-type quarks (scenario of class {\bf S.I}) in the $tjtj$ (plain), $tjtt$
   (strong dashed) and $tttt$ (light dashed) channels. Figure taken from Ref.\
    \cite{Calvet:2012rk}.}
    \label{fig:nj_1l}
  \end{center}
\end{figure}

Multijet background contributions have not been accounted for in our simulation setup and we
need to remedy that lack. To this aim, we
rely on data-driven methods such as those introduced in the analysis of Ref.~\cite{Aad:2012wm}. 
The latter shows that selecting events with a missing transverse energy larger
than $\slashed{E}_T \geq 40$ GeV and requiring the reconstructed $W$-boson transverse mass,
\be
  M_T^W =  \sqrt{ 2 p_T^\ell \slashed{E}_T \Big[ 1 - \cos \Delta
    \varphi(\ell,\slashed{E}_T)\Big] } \ ,
\ee
to be larger than 25 GeV ensure a good control of that source of background. In the
equation above, we have introduced the quantity $\Delta\varphi(\ell,
\slashed{E}_T)$ standing for the angular distance, in the azimuthal direction with
respect to the beam, between the lepton and the missing energy. 

As in the multilepton analysis, signal events are expected to be rich in hard jets which arise
from the
hadronization of the decay products of the top quarks. This contrasts with the dominant
$W$-boson plus jets
background contributions where jets originate mainly from initial-state radiation (see
Figure~\ref{fig:nj_1l}).
Events are therefore selected with the requirement that they contain at least
six, seven and eight jets with a transverse-momentum $p_T^j \geq 25$ GeV in the context of
the searches in the
$tjtj$, $tjtt$ and $tttt$ topologies, respectively. For the same reasons, signal events are
also expected to include a high number of $b$-tagged jets. We therefore demand a minimal
number of one and two $b$-jets for the $tjtj$ and $tjtt$/$tttt$ search
channels, respectively. At this stage, the expected Standard Model background is
composed mainly of $t\bar t$ events, the top-antitop pair being possibly
produced in association with one or several gauge bosons.

\renewcommand{\arraystretch}{1.2}
\begin{table}[t] 
\begin{center}
\begin{tabular}{ | c || c c c |}
\hline 
  \multirow{2}{*}{Selections}& \multicolumn{3}{c|}{ $tjtj$ channel}  \\ 
   & $M_\sigma^I = 400$ GeV & $M_\sigma^I = 800$ GeV &Backgrounds \\ 
\hline 
  $N_\ell = 1$ with $p^\ell_T \geq 25$ GeV & 	$(1.06 \pm 0.01) \!\cdot\! 10^4$ &  $ 45.7 \pm 0.9 $ & $(2.376 \pm 0.003) \!\cdot\!10^8$ \\ 
  $\slashed{E}_T \geq 40$ GeV  & 		$(7.65 \pm 0.06) \!\cdot\! 10^3$ &  $ 37.9 \pm 0.8 $ & $(6.836 \pm 0.002) \!\cdot\! 10^7$ \\ 
  $M_T^W \geq 25$ GeV & 			$(6.43 \pm 0.05) \!\cdot\! 10^3$ &  $ 30.7 \pm 0.7 $ & $(6.722 \pm 0.002) \!\cdot\! 10^7$ \\ 
  $N_j \geq 6$ with$p_T^j \geq 25$ GeV & 	$(3.88 \pm 0.04) \!\cdot\! 10^3$ &  $ 24.9 \pm 0.7 $ & $(8.634 \pm 0.024) \!\cdot\! 10^4$ \\ 
  $N_b\geq 1$ & 				$(2.91 \pm 0.04)\!\cdot\! 10^3$  &  $ 19.3 \pm 0.6 $ & $(6.407 \pm 0.014) \!\cdot\! 10^4$ \\ 
\hline 
\end{tabular} \vspace{.25cm} 

\begin{tabular}{ | c || c c c |}
\hline 
  \multirow{2}{*}{Selections}& \multicolumn{3}{c|}{ $tjtt$ channel}  \\ 
   & $M_\sigma^I = 400$ GeV & $M_\sigma^I = 800$ GeV &Backgrounds \\ 
\hline 
  $N_\ell = 1$ with $p^\ell_T \geq 25$ GeV 	& $(1.21 \pm 0.22) \!\cdot\! 10^3$	&$21.3 \pm 0.4$ & $(2.376 \pm 0.001) \!\cdot\!10^8$ \\ 
  $\slashed{E}_T \geq 40$ GeV 			& $(8.81 \pm 0.19) \!\cdot\! 10^2$ 	&$18.1 \pm 0.3$ & $(6.836 \pm 0.002) \!\cdot\!10^7$ \\ 
  $M_T^W \geq 25$ GeV  				& $(7.66 \pm 0.18)\!\cdot\! 10^2$ 	&$15.4 \pm 0.3$ & $(6.722 \pm 0.002)\!\cdot\! 10^7$ \\ 
  $N_j \geq 7$ with $p_T^j \geq 25$ GeV 	& $(4.05 \pm 0.13)\!\cdot\! 10^2$ 	&$11.08 \pm 0.3$ & $(2.613 \pm 0.012)\!\cdot\! 10^4$ \\ 
  $N_b\geq 2$ 					& $(1.99 \pm 0.09) \!\cdot\! 10^2$ 	&$5.99 \pm 0.2$ & $(9.330\pm 0.050) \!\cdot\! 10^3$ \\ 
\hline 
\end{tabular} \vspace{.25cm} 

\begin{tabular}{ | c || c c c |}
\hline 
  \multirow{2}{*}{Selections}& \multicolumn{3}{c|}{ $tttt$ channel}  \\ 
   & $M^I_\sigma \ = \ 400$ GeV & $M^{II}_\sigma = 800$ GeV &Backgrounds \\ 
\hline 
  $N_\ell = 1$ with  $p^\ell_T \geq 25$ GeV 	& $34.6 \pm 0.6$ 	& $23.2 \pm 0.4$ & $(2.376 \pm 0.001)\!\cdot\! 10^8$ \\ 
  $\slashed{E}_T \geq 40$ GeV 			& $27.3 \pm 0.5$ 	& $20.2 \pm 0.4$ & $(6.836 \pm 0.002) \!\cdot\! 10^7$ \\ 
  $M_T^W \geq 25$ GeV 				& $23.6 \pm 0.5$ 	& $17.1 \pm 0.3$ & $(6.722 \pm 0.002) \!\cdot\! 10^7$ \\ 
  $N_j \geq 8$ with $p_T^j \geq 25$ GeV 	& $10.8 \pm 0.3$ 	& $12.3 \pm 0.3$ &$(7.020 \pm 0.060) \!\cdot\! 10^3$ \\ 
  $N_b\geq 2$ 					& $7.21 \pm 0.27$	& $8.47 \pm 0.23$ 	& $(2.658 \pm 0.026)  \!\cdot\! 10^3$ \\ 
\hline 
\end{tabular} 
\caption{Same as Table \ref{tab:sel_multi} but for a final state signature with one single
lepton. We refer to the text for a detailed description of the selection strategy.}
\label{tab:sel_single} 
\end{center}
\end{table} 
\renewcommand{\arraystretch}{1.0}

\begin{figure}
  \begin{center}
    \includegraphics[width=0.80\columnwidth]{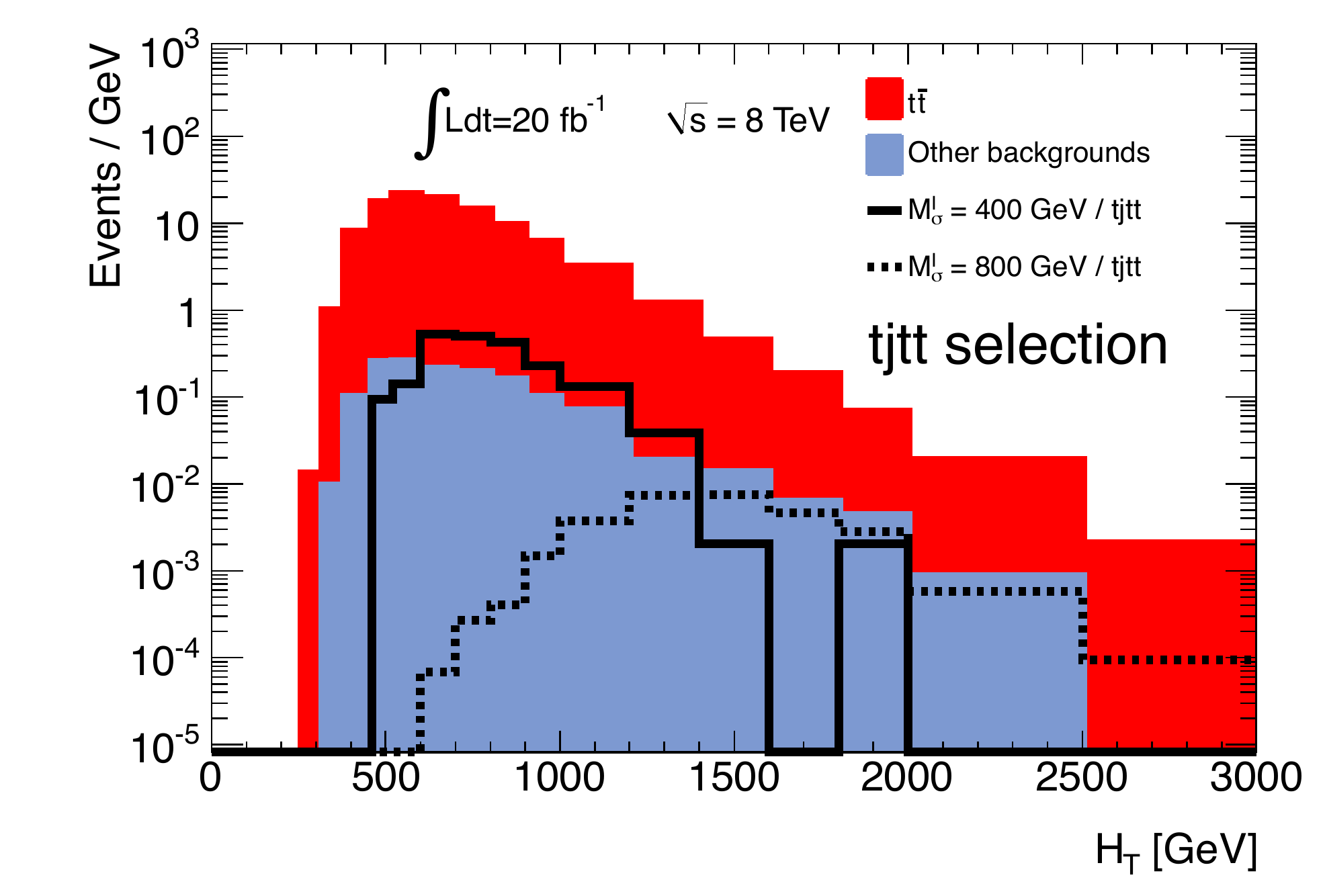}\\
    \includegraphics[width=0.80\columnwidth]{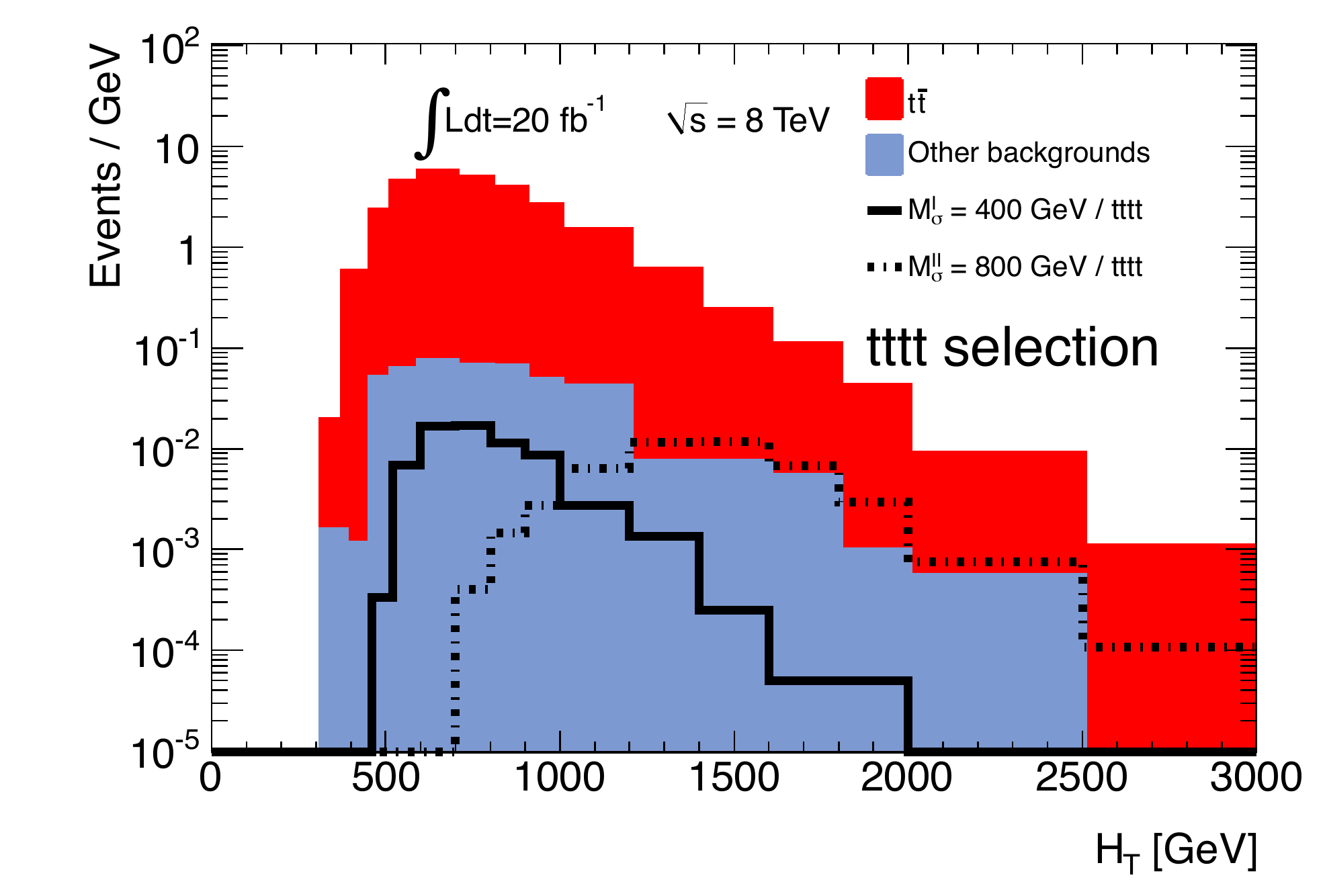}
    \caption{Distribution of the hadronic energy $H_T$,  defined in Eq.\
  \eqref{eq:htdef}, after the selection strategy presented in Table~\ref{tab:sel_single}.
  We distinguish background events issued from the production of a top-antitop pair 
  together with jets (red) from
  the other Standard Model contributions (blue). For the $tjtt$ topology (left panel), we
  superimpose the curves related to two signal scenarios of class {\bf S.I} with respective sgluon
  masses of 400 GeV (plain) and 800 GeV (dashed), while for the
  $tttt$ search strategy (right panel), we consider a scenario
  of class {\bf S.I} with a sgluon mass of 400 GeV (plain) and a scenario of
  class {\bf S.II} with a sgluon mass of 800 GeV (dashed).}
    \label{fig:HT_1l}
  \end{center}
\end{figure}

Details on the number of events surviving each of the selection criteria are given in Table
\ref{tab:sel_single} for two representative signal scenarios with $M_\sigma
= 400$ GeV and 800 GeV as well as for the sum of the background contributions. After all
selections, we predict 7.21 (8.47) to 2910 (19.3) signal events, depending on the search
channel and for a sgluon mass of 400 (800) GeV. In
contrast, the Standard Model expectation consists of  64070, 9330 and 2658
events for the $tjtj$, $tjtt$ and $tttt$ topologies, respectively.
The large hadronic activity proper to signal events motivates us to again consider
the $H_T$ variable as
discriminant between signal and background. This feature is depicted on
Figure \ref{fig:HT_1l} in the context of the $tjtt$ (upper panel) and $tttt$ (lower panel)
selection strategies. Signal distributions present a clear peaky behavior
centered around a $H_T$ value of about $1.5 M_\sigma$ and a tail which does not extend to 
very large hadronic energies, in contrast to the Standard Model results.
The latter, for which we distinguish events associated with
top-antitop production in association with jets (red) from the other contributions
(blue), show a steep rise once the
top-antitop production threshold is reached followed by a peak around
$H_T \approx 500$ GeV and a smooth fall with increasing energy.
This shape difference will be employed below to probe the sgluon mass possibly reachable at
the LHC.

While the $H_T$ variable is in principle a good discriminant between signal and background,
better limits on the sgluon mass can be extracted in the case of the $tjtj$ channel after a 
kinematical fit of the events, assuming that the missing energy is only
originating from a leptonic $W$-boson decay.
Assigning the labeling of the six jets according to 
\be
  p p \to \sigma \sigma \to (t j_5) (t j_6) \to (j_1 j_2 j_3 j_5) (j_4 \ell \nu
   j_6) \ ,
\label{eq:X2pattern}\ee
the true configuration of each event is defined as the jet permutation
minimizing the $\chi^2$-variable
\be\bsp
  \chi^2  =&\ 
  \left[ \frac{m_{j_1j_2}-m^{(r)}_W}{\sigma^{(r)}_W} \right]^{2} + \left[
    \frac{\big(m_{j_1j_2j_3}-m_{j_1j_2}\big) - m^{(r)}_{tW}}{\sigma^{(r)}_{tW}}
    \right]^{2} +
    \left[ \frac{m_{\ell\nu j_4}-m^{(r)}_{t\ell}}{\sigma^{(r)}_{t\ell}}
    \right]^{2} + \\
  &\  \left[\frac{(m_{\ell\nu j_4,j_6}-m_{\ell\nu
    j_4})-(m_{j_1j_2j_3,j_5}-m_{j_1j_2j_3})}
    {\sigma_{\sigma t}^{(r)} \big[ (m_{\ell\nu j_4,j_6}-m_{\ell\nu
    j_4})+(m_{j_1j_2j_3,j_5}-m_{j_1j_2j_3}) \big] } \right]^2 \ .
\esp\label{eq:chi2} \ee
The different contributions to this $\chi^2$-variable exactly mimic the decay chain
of Eq.~\eqref{eq:X2pattern}.
With the first two terms, we ensure that the three jets 
$j_1$, $j_2$ and $j_3$ consist of the decay products of 
a hadronically decaying top quark. More into details, they enforce the invariant mass of the
first two jets $m_{j_1 j_2}$ to be compatible with the 
$W$-boson mass and the three-jet system invariant mass $m_{j_1j_2j_3}$ to
be compatible with the top mass. Since these two observables are correlated,  we however
subtract from the reconstructed top mass $m_{j_1j_2j_3}$ the reconstructed dijet
invariant-mass $m_{j_1j_2}$ in the second term of Eq.~\eqref{eq:chi2}.
The values of the $\chi^2$ parameters are taken as $m^{(r)}_W =
80.7$ GeV, $\sigma^{(r)}_W = 8.9$ GeV,  $m^{(r)}_{tW} = 90.8$~GeV and
$\sigma^{(r)}_{tW} = 10.5$ GeV and have been extracted from
a fit based on the Monte Carlo truth where each reconstructed
object is correctly assigned according to the configuration of Eq.\
\eqref{eq:X2pattern}. The values of the widths 
of ${\cal O}(10\%)$  are compatible with the mass resolution inputted in our detector
simulation.

The third term of Eq.\ \eqref{eq:chi2} addresses the leptonically decaying
top quark and verifies that the invariant mass $m_{\ell\nu j_4}$ is compatible with the
top mass. The neutrino four-momentum is reconstructed after assuming that
the missing energy of the event as well as the identified charged lepton are both issued
from a $W$-boson decay. In this way, information on the reconstructed $W$-boson is
implicitly included in the $m_{\ell\nu j_4}$ term of the $\chi^2$
so that there is no need for a dedicated contribution.
From the Monte Carlo truth, we have extracted the parameters
 $m^{(r)}_{tl} = 167.8$ GeV and $\sigma^{(r)}_{tl} = 19.1$ GeV. 

Finally, both 
the $tj_5$ and $tj_6$ systems are the decay products of a sgluon field.
Therefore, the invariant masses $m_{j_1j_2j_3, j_5}$ and $m_{\ell\nu
j_4,j_6}$ must be compatible with each other, up to the detector
resolution. This leads to the last term of Eq.\ \eqref{eq:chi2}, after subtracting the
reconstructed top masses to avoid possible correlations among the different terms of the 
$\chi^2$ and after including an extra factor at the denominator to remove a too strong
dependence on the sgluon mass. From the Monte Carlo truth, we derive the
numerical value of the parameter $\sigma_{\sigma t}^{(r)}$, found equal to 0.098 GeV.

\begin{figure}
  \begin{center}
     \includegraphics[width=0.80\columnwidth]{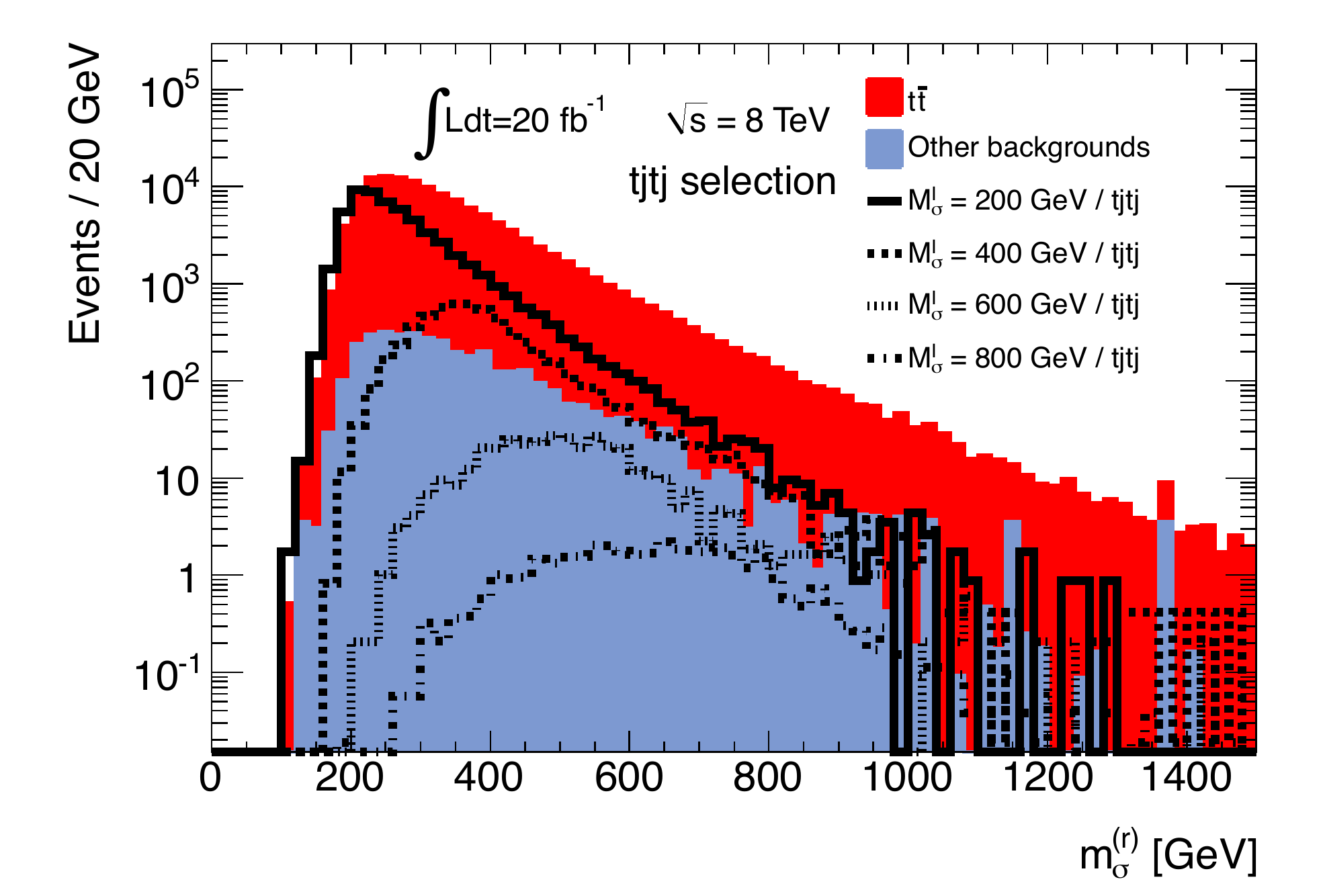}
    \caption{Reconstructed sgluon mass $M_\sigma^{(r)}$ following the procedure described in
    the text, after adopting the $tjtj$ search strategy
    (see the upper panel of Table \ref{tab:sel_single}). We distinguish the top-antitop
    contributions (red) from the rest of the background (blue) and superimpose
    the results associated with four signal scenarios of class {\bf S.I}, with respective sgluon mass 
    of 200~GeV (plain), 400 GeV (strong dotted), 600 GeV (light dotted) and 800 GeV
    (dash-dotted). Figure taken from Ref.\ \cite{Calvet:2012rk}.}
    \label{fig:Mtj}
  \end{center}
\end{figure}

After selecting the events as summarized in the upper panel of Table
\ref{tab:sel_single}, we reconstruct each surviving event according to the
pattern given in Eq.\ \eqref{eq:X2pattern} by minimizing the $\chi^2$-variable 
of Eq.\ \eqref{eq:chi2}. We then extract the 
sgluon mass $M_\sigma^{(r)}$ and present the related distributions on Figure
\ref{fig:Mtj} for both the background, distinguishing the top-antitop pair component
(in red) from the other contributions (in blue), and four signal scenarios
of class {\bf S.I} with a sgluon mass of
200 GeV, 400 GeV, 600 GeV and 800 GeV, respectively. The background
distribution presents a rising behavior once the
top-antitop production threshold is reached, followed by 
a tail extending up to rather large values of the reconstructed
sgluon mass $M_\sigma^{(r)}$. In contrast, signal distributions all show a
clear peak. For light sgluons ($M_\sigma=200$ GeV or 400
GeV), this peak is centered around the true sgluon mass while for
heavier sgluons ($M_\sigma=600$ GeV and 800 GeV), the detector resolution widens the peak
so that the central value is equal to about $70\%-80\%$ of the true sgluon mass.
The different shapes of the background and signal distributions can hence be used
to extract limits on the sgluon mass possibly reachable at the LHC (see below).

\subsubsection{LHC sensitivity to a sgluon field dominantly coupling to top quarks}

\begin{figure}
  \begin{center}
    \includegraphics[width=0.6\columnwidth]{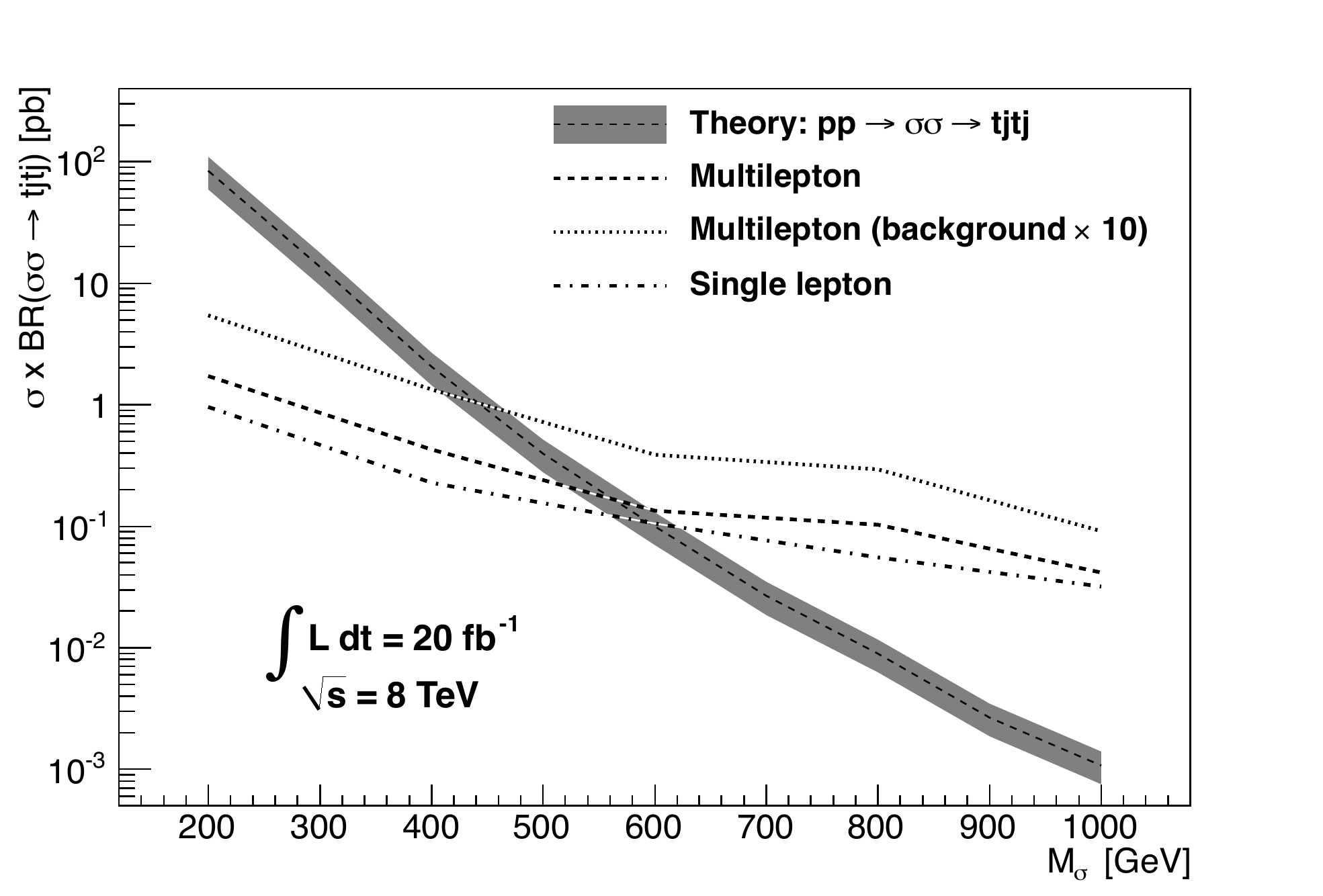} \\
    \includegraphics[width=0.6\columnwidth]{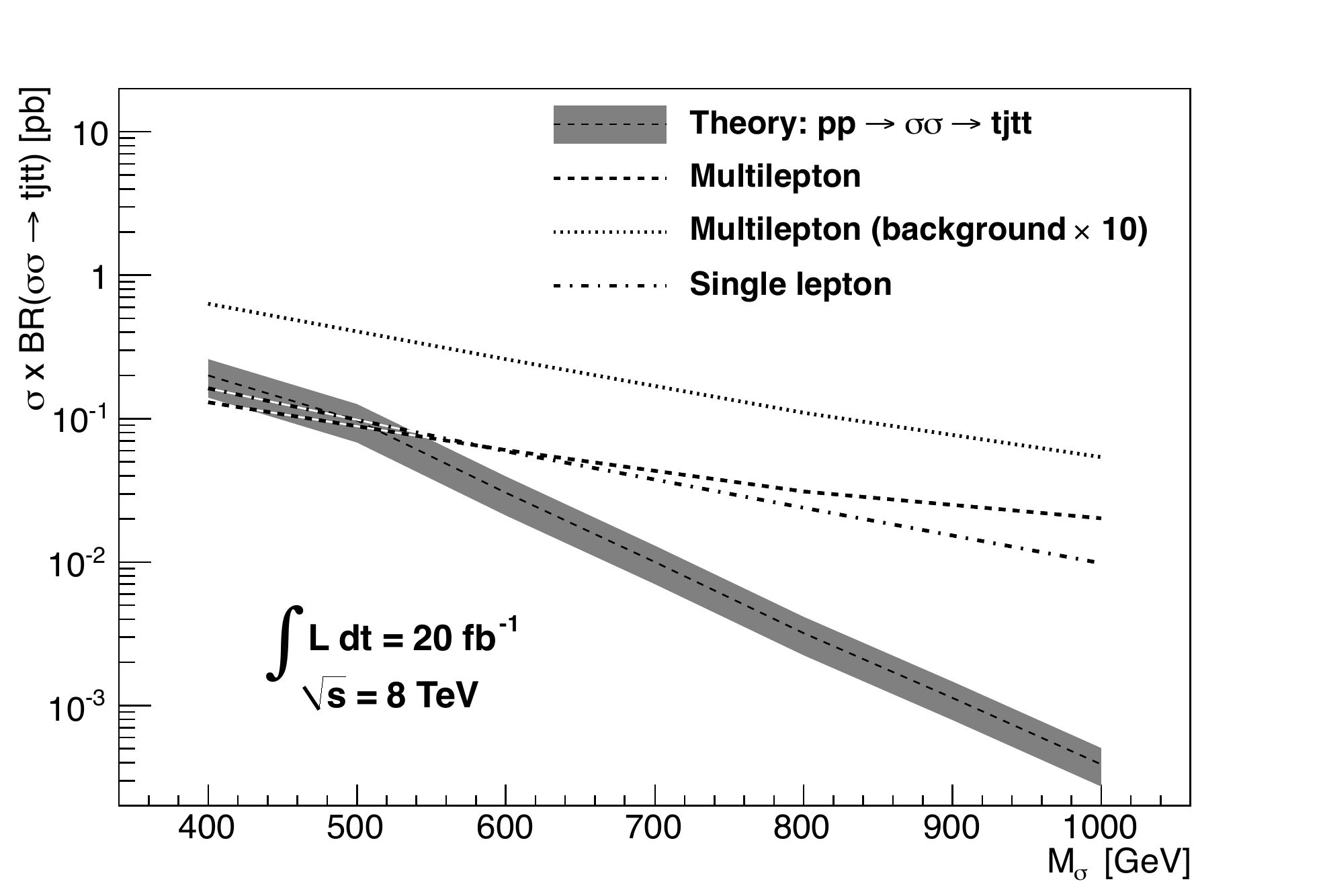} \\
    \includegraphics[width=0.6\columnwidth]{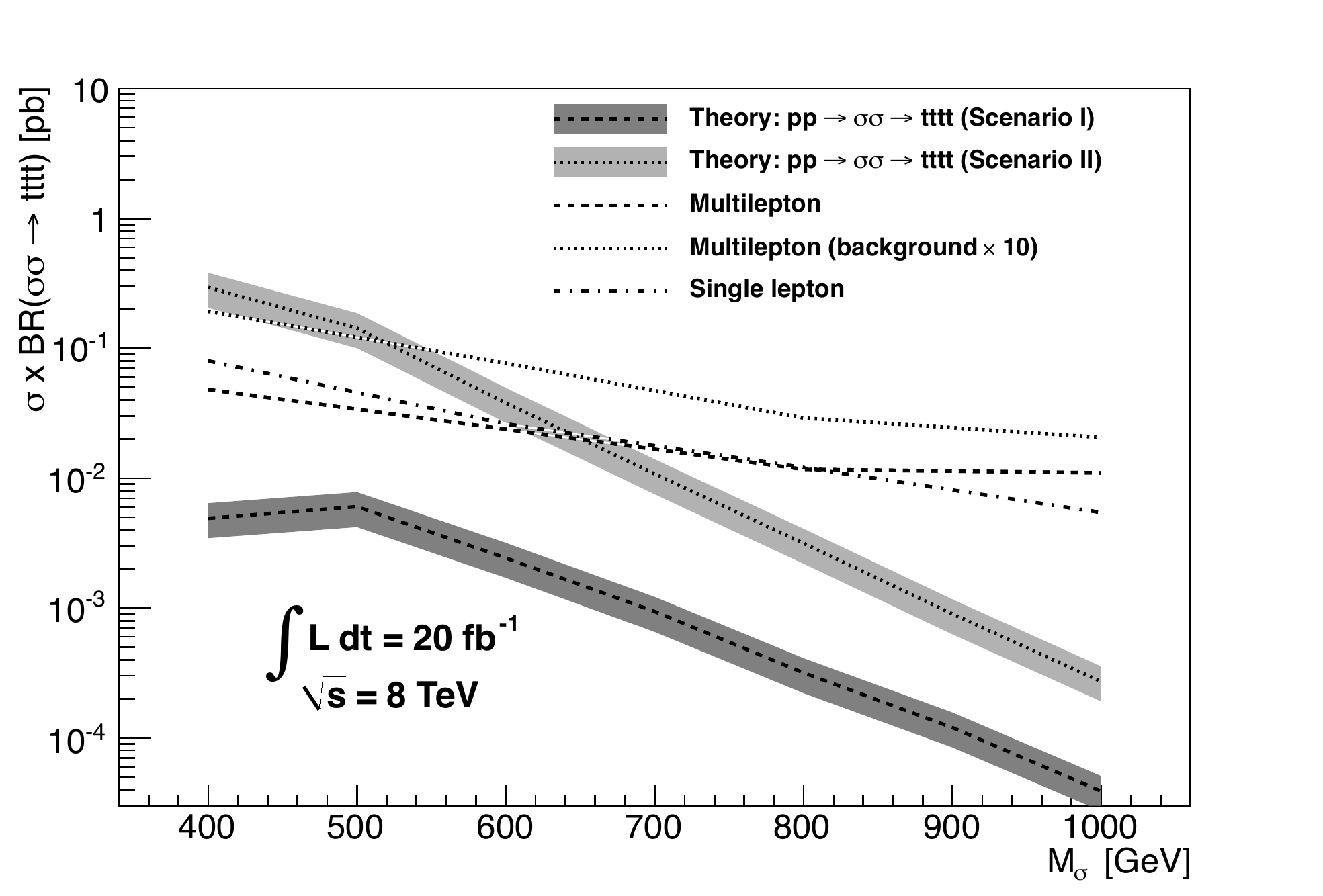} \\
    \caption{The 95\% confidence level expected signal cross sections as a
  function of the sgluon mass, for 20 fb$^{-1}$ of LHC collisions at $\sqrt{s}=8$
  TeV. The uncertainty bands associated with the next-to-leading order
  theoretical predictions correspond to
  30\% variations, as commonly driven by the dependence on the unphysical scales (see 
  Ref.~\cite{GoncalvesNetto:2012nt}), and are presented as light (dark) gray bands
  for scenarios of class {\bf S.I} ({\bf S.II}). Expected limits in the $tjtj$ (upper), $tjtt$
  (middle) and $tttt$ (lower) channels are given as dot-dashed (single lepton
  analysis), dashed (multilepton analysis) and dotted (multilepton analysis after enhancing
  the background by a factor of ten) curves. Figures taken from Ref.\ \cite{Calvet:2012rk}.}
    \label{fig:sgllim}
  \end{center}
\end{figure}

For each of the considered search channel and strategy, we combine the number of
expected signal and background events to calculate upper limits on the signal cross section
at the 95\% confidence level. More into details, we make use of the \texttt{CLs} technique
\cite{Read:2002hq} as
implemented in the {\sc MCLimit} software \cite{mclimit} and employ either the reconstructed
sgluon mass (for a $tjtj$ final state with an analysis based on a single lepton signature) or
the $H_T$ variable (in all the other cases) to discriminate
signal from background\footnote{Considering the reconstructed mass instead of the $H_T$
variable in the case of a single lepton analysis for the $tjtj$ topology
allows to improve the LHC sensitivity by about 15\%-20\% in the low mass region
without affecting the higher sgluon mass region.}.
The results are presented in Figure~\ref{fig:sgllim} for the $tjtj$ (upper panel), $tjtt$
(middle panel) and $tttt$ topology (lower panel) as dashed (multilepton analysis),
dotted (multilepton analysis after multiplying the background by ten to estimate the
effects of the non-simulated contributions) and
dot-dashed (single lepton analysis) curves. We also show
theoretical predictions for sgluon-induced
production of multitop final states as a function of the sgluon mass.
In addition to the central next-to-leading order results derived from
Table \ref{tab:sigma}, we include a $30\%$ uncertainty band corresponding to
typical effects induced by factorization and renormalization scale variations (light and 
dark gray for scenarios of class {\bf S.I} and {\bf S.II}, respectively)
\cite{GoncalvesNetto:2012nt}. 
In Table~\ref{tab:sgllim}, these results are translated in terms of the
sgluon mass that can be possibly excluded at the $95\%$ confidence level
for each final state and considered scenario.
In addition, $1\sigma$ variations have been derived after accounting for statistical
uncertainties.

\renewcommand{\arraystretch}{1.2}
\begin{table}[t] 
  \begin{center}
    \begin{tabular}{ | c || c || c || c |}
     \hline 
      & \multirow{2}{*}{Single lepton analysis} & 
        \multirow{2}{*}{Multilepton analysis} & Multilepton analysis \\
      &                                         &
        & (background $\times 10$) \\
     \hline
      $tjtj$ & 590 $^{+40}_{-30}$ GeV & 570 $^{+30}_{-50}$ GeV  & 440
       $^{+40}_{-15}$ GeV \\
      $tjtt$ & 480 $^{+70}_{-80}$ GeV & 520 $^{+35}_{-90}$ GeV  & -  \\
      $tttt$ ({\bf S.I})  & - & -  & -\\
      $tttt$ ({\bf S.II}) & 640 $^{+40}_{-30}$ GeV & 650 $^{+30}_{-40}$ GeV  &
         520 $^{+50}_{-110}$ GeV \\
     \hline 
    \end{tabular}
    \caption{Expected sensitivity of the LHC, running at a center-of-mass of 8 TeV and for an 
  integrated luminosity of 20 fb$^{-1}$, to a sgluon signal in multitop events. The results 
  are given, together with the associated $1\sigma$ statistical uncertainties,
  in terms of upper bounds on the sgluon mass possibly reachable, at the 95\%
  confidence level, for the different analyses under consideration.
  Expectations for scenarios of class {\bf S.I} are given for the
  $tjtj$, $tjtt$ and $tttt$ channels on the first three lines of the table, respectively.
  Scenarios of class {\bf S.II} are only investigated in the context of the $tttt$ search channel,
  the results being shown on the fourth line of the table.} 
    \label{tab:sgllim} 
  \end{center}
\end{table} 
\renewcommand{\arraystretch}{1.}

From the results of the multilepton analysis, we observe that sgluon masses ranging up to
570~GeV and 520 GeV can be excluded by an investigation of the $tjtj$ and $tjtt$ topologies,
respectively, in the context of scenarios of class {\bf S.I}. Equivalently, the LHC, when 
operating at $\sqrt{s}=8$ TeV,
is sensitive to sgluon-induced multitop production cross sections of ${\cal
O}(100)$ fb for both signatures. From Table~\ref{tab:sgllim}, it can be seen that the limits
obtained for the $tjtj$ search channel vary by less than $10\%$ when
accounting for $1\sigma$ statistical uncertainties. In contrast,
the bounds extracted from the analysis of the $tjtt$ topology are
found to be more sensitive to statistics since they could vary by up to 17\% with
respect to $1\sigma$ (un)lucky fluctuations. This feature is related to the
behavior of both cross sections (including the relevant branching ratios) in the 
500-600 GeV sgluon
mass range. In the $tjtj$ case, the cross section decreases with the mass, while in the
$tjtt$ case, it is rather constant. Due to the low sgluon branching ratio into a top-antitop
pair for scenarios of class {\bf S.I} (see Table \ref{tab:sigma}), the multilepton analysis is not
sensitive to final states with four top quarks, at least for an integrated luminosity of
20~fb$^{-1}$. In class {\bf S.II} scenarios, this branching ratio is $2.5-7.6$ times
more important so that sgluon masses ranging up 650 GeV, or cross sections of
${\cal O}(10)$ fb, can be excluded at the 95\% confidence level.
When accounting for non-simulated backgrounds by multiplying the 
number of expected background events by ten, we loose all sensitivity to the
$tjtt$ channel while the masses to be possibly excluded when analyzing $tjtj$ and $tttt$
(for scenarios of class {\bf S.II}) final states decrease from 570 GeV to 440 GeV and from 650 GeV to
520 GeV, respectively.

We now turn to single lepton analyses. First, considering $tjtj$ final states, we show that
the reconstructed sgluon mass can be used to exclude at the 95\% confidence
level sgluon as heavy as 590 GeV, these bounds 
being rather strong against statistical fluctuations. Next, considering $tjtt$ and $tttt$
topologies, the $H_T$ variable is employed as a
discriminant and it is found that sgluon masses up to 480 GeV and 640 GeV can be reached.
While in the $tjtt$ case, statistical fluctuations can lead to different expectations
by about $\pm 15\%$, the results related to a four top signature are
found only to change by about $5\%$ within the $1\sigma$ band.

\cleardoublepage

%% file: cdf.tex
\label{chap:monotopsCDF}

In the previous chapters of this work, we have investigated
two signatures
predicted by two non-minimal supersymmetric models. We have
addressed the production of
a monotop final state, consisting of a single top quark and missing energy,
originating from an $R$-parity-violating top squark decay as well as the
production of a multitop final state
arising from the pair-production and decay of a
new scalar state, dubbed sgluon, lying in the adjoint representation of the
QCD gauge group. We have then generalized the performed analyses by means of effective field
theories in order to facilitate their reinterpretation in the context of
various classes of new physics models,
supersymmetric or not, although the reinterpretation process in the framework
of a well-defined model by itself goes beyond the
scope of this work. All the predictions that have been made suggest that
the LHC is sensitive to large parts of the respective parameter spaces, which
has motivated dedicated ATLAS and CMS searches. These searches are more precisely
currently either already achieved~\cite{ATLAS-CONF-2013-051}
or on-going~\cite{monoATLAS, monoCMS}. Furthermore,
the obtained results at the phenomenology level have even triggered
a first monotop search at the Tevatron collider in 2012~\cite{Aaltonen:2012ek}.

The above-mentioned LHC and Tevatron experimental analyses
have followed the phenomenological studies performed
in this work. For obvious reasons, only analyses which are
achieved and publicly available are discussed, \ie, a monotop search
by the CDF collaboration~\cite{Aaltonen:2012ek} and an analysis of sgluons
decaying into multitop final states performed
by the ATLAS collaboration~\cite{ATLAS-CONF-2013-051}. In the following, the
CDF monotop analysis (which is a part of this work)
is detailed, from the point of view of a theorist,
in Section~\ref{sec:monotopcdf} whereas
only the conclusions of the ATLAS analysis
are presented in Section~\ref{sec:sgluonatlas}\footnote{
The rules of the LHC collaborations forbid anyone, including theorists,
to be member of several LHC collaborations simultaneously. This excludes
the author of this document to take part to any ATLAS analysis, as being a member of the
CMS collaboration for 2008. The main results of the
ATLAS analysis of Ref.~\cite{ATLAS-CONF-2013-051} are however included in this
document for completeness.}.

\mysection{Search for monotops with the CDF detector}
\label{sec:monotopcdf}

\subsection{Analysis strategy for monotop searches with the CDF detector}
\label{sec:cdfstrategy}
Among all the scenarios introduced in Section~\ref{sec:monotopsEFT},
benchmark points of class {\bf S.IV} have been considered for a confrontation to data
recorded by the CDF detector. In this case, we recall that the top quark is produced in
association with an invisible vector field through a flavor-changing interaction
of the new state with a top quark and an up quark. In the following, we adopt
the CDF notations and the new state,
that can in particular be compatible with a dark matter candidate, is denoted by the symbol~$D$.
The most relevant region of the parameter
space with respect to the Tevatron available center-of-mass energy and
integrated luminosity consists of the low-mass region. Therefore, the mass
of the invisible state $M_D$ is further assumed to satisfy
\be
  M_D \in [0-150]\text{~GeV.}
\ee
Such a light dark matter candidate is also motivated by new physics signals reported
by several experiments dedicated to direct
detection of dark matter. More into details, the DAMA project, based on a
detector made up of 250~kg of sodium iodide exploiting the annual modulation of the dark matter
signature due to Earth rotation, has first reported an effect
that could be explained by a light dark matter candidate~\cite{Bernabei:2008yi}.
Later, both the CoGeNT experiment, using a low-background germanium
detector looking for dark matter elastic interactions
with nuclei~\cite{Aalseth:2010vx,Aalseth:2011wp} and the CRESST-II collaboration, probing dark
matter detection via its interactions with nuclei inside CaWO$_4$
crystals~\cite{Angloher:2011uu}, have found discrepancies in several observables.
It has been found that a by light dark matter particle could explain
all the issues. In each of these cases, the light dark matter
candidate has to interact with the Standard Model sector in such a way that its detection
at collider experiments would be feasible.
The classes of scenarios {\bf S.III} and {\bf S.IV} designed in the context
of the monotop effective field theory of Section~\ref{sec:monotopsEFT}
are generic examples of models satisfying these constraints, so that
they can also be seen as extensions of
the Standard Model with a dark sector coupled to the top
quark in a flavor-violating fashion. Consequently, flavor-changing monotop
production has received more experimental attention than the resonant one so far.

In this section, we introduce the search strategy designed by the CDF collaboration
to probe possible signals of flavor-changing monotop production. This includes
7.7($\pm 0.55$)~fb$^{-1}$ of events recorded by the CDF-II detector,
one of the two general purpose detectors
employed for the study of $p\bar{p}$ collisions with a center-of-mass energy
of 1.96~TeV at the Tevatron.
The CDF-II detector contains a tracking system consisting of a cylindrical
open-cell drift chamber together with silicon microstrip detectors
immersed in a magnetic field of 1.4~T parallel to the beam axis.
This inner part of the detector is surrounded by a calorimetric system
comprised of both an electromagnetic and a hadronic piece in order to measure
particle energies. Additional drift chambers and muon scintillators are located
outside the calorimeters and are dedicated to muon identification.
A more detailed description of the detector and the Tevatron accelerating facility
can be found in Ref.~\cite{Acosta:2004yw} and references therein.

The results of the monotop analyses of Section~\ref{sec:rpvpheno}
and Section~\ref{sec:effmonotops} at the LHC were suggesting to employ a data acquisition
trigger based on the missing energy only. The CDF analysis presented in this section
investigates
instead events that have triggered the acquisition system by the presence in the final state
of two calorimeter clusters together with significant
missing transverse energy of at least 35~GeV~\cite{Aaltonen:2010fs}
and 30~GeV~\cite{5076030} for data recorded before and after 2007, respectively. In this case,
the missing transverse energy definition
relies on calorimetric deposits and is defined as
the magnitude of the vectorial sum of the transverse energy contained in
each calorimeter tower of the detector.

Jets are reconstructed by making use of the CDF {\sc JetClu} algorithm~\cite{Abe:1991ui}
using cones of radius $R=0.4$ to cluster calorimeter deposits
into jets. Reconstructed jet energies are then corrected
using standard techniques developed on the basis of
a deep comparison of Monte Carlo simulations of proton-antiproton
collisions in CDF and data for
several physics processes such as jet, single photon with jets, single gauge boson or $J/\psi$
production~\cite{Bhatti:2005ai}. In order to identify jets originating from the fragmentation
of a $b$-quark, a secondary-vertex-tagging algorithm is employed, accounting for the longer
lifetime of $B$-hadrons~\cite{Acosta:2004hw}. The efficiency of
such a $b$-tagging algorithm ranges from 30\% to 50\%
for jets with a transverse energy lower than 200~GeV,
for a mistagging rate of a few percents.

In order to achieve a full trigger system efficiency, event selection further requires
the amount of missing transverse energy to fulfill
\be
  \slashed{E}_T \geq 50\text{~GeV} \ .
\ee
In addition only those events whose final state contains exactly three jets are retained.
One of these jets is constrained to be identified as a $b$-tagged jet, whereas
the transverse energies of the three jets $j_1$, $j_2$ and $j_3$, ordered
by decreasing transverse energies, are required to
satisfy
\be
  E_T(j_1) \geq 35~\text{GeV}\ , \qquad E_T(j_2) \geq 25~\text{GeV} \qquad
  \text{and}\qquad E_T (j_3) \geq 15~\text{GeV}\ .
\ee
We further impose that either the leading or the next-to-leading jet
has a pseudorapidity such that  $|\eta| \leq 0.9$, and that all three jets
have their pseudorapidity satisfying $|\eta| < 2.4$. Finally, events containing
any identified charged lepton are vetoed in order to be compatible with the monotop
signature associated with a hadronically decaying top quark.

As a consequence of the low missing energy requirement, the event selection
gives at this stage a data sample dominated by QCD multijet events with fake
missing energy arising from the mismeasurement of jet energies.
The simulation of this type of background being prohibitive due to the
very large production rate and the important
theoretical uncertainties, a QCD multijet sample has instead been generated by using
a data-driven method~\cite{Aaltonen:2009jg}.
This method works in several steps. First, a control region enriched in multijet events is
defined by restricting the selection (without considering the $b$-tagging requirement)
to events with
a moderate amount of missing energy (below 70~GeV) and an angular separation
smaller than 0.3
between the next-to-leading jet and the missing momentum (multijet events featuring such a quantity
of missing energy have often their second jet aligned with the missing momentum).
Next, a probability for tagging a jet as a (real or fake) jet issued from a $b$-quark is derived as
a function of the hadronic activity of the event and several jet properties such as its transverse momentum
and its pseudorapidity. The derived probability is then applied as a per-event weight
to all the events satisfying
all the selection criteria introduced so far, but the $b$-tagging requirement.
The resulting weighted event sample consists of the desired data-driven multijet background once
the electroweak contributions, computed by means of Monte Carlo simulations, are subtracted.

All other sources of Standard Model background have been simulated by means of
Monte Carlo event generators. Events originating from diboson and top-antitop pair
production have been produced using \pythia~6~\cite{Sjostrand:2006za}
and reweighted according to the next-to-leading order
results as predicted by the {\sc Mcfm} program
for the diboson case~\cite{Campbell:1999ah,Campbell:2000bg}
and to the approximate next-to-next-to-leading order predictions
for the top-antitop case~\cite{Langenfeld:2009wd}.
Furthermore, events issued from single vector boson
production in association with (light and heavy flavored) jets have been
generated with {\sc AlpGen}~\cite{Mangano:2002ea}, parton showering and hadronization being
again performed by {\sc pythia}, and normalized on the basis of the next-to-leading
order cross section returned by {\sc Mcfm}.
Finally, single top events have been modeled with the \madgraph~5 package~\cite{Alwall:2011uj},
\pythia\ 6 taking again care of parton showering and hadronization. Each single top event
has then been reweighted
according to the next-to-leading order predictions~\cite{Harris:2002md,Sullivan:2004ie}.

On different footings, a non-negligible source of background consists of
events selected due to the mistagging of a light-flavor jet as a jet issued from a
$b$-quark. This contribution is modeled in two stages. As a first step, the mistagging
rate of a light jet as a $b$-jet is extracted from data~\cite{Acosta:2004hw}.
To this aim, one starts from the secondary vertex that has implied
the $b$-tag and calculates the projection onto the jet axis of the vector pointing
from the primary to the secondary vertex. This quantity, denoted $L_{2D}$, is dubbed
the two-dimensional decay length of the secondary vertex.
Its sign plays a critical
role in the estimation of the mistagging rate
as $b$-quark-initiated hadrons in general lead to large positive
values of  $L_{2D}$
whereas mistagged lighter jets exhibit smaller $|L_{2D}|$ values that occur with the same
rate for both signs. This feature finds its origin in the fact
that secondary vertices leading to fake $b$-jets in general arise
from tracks found displaced due to tracking errors. However, although
decays of $K_S$ mesons or $\Lambda$ hyperons contribute subdominantly,
their effect is non-negligible and has been considered. True $b$-tagged jet contributions
to the negative $L_{2D}$ region are then subtracted by comparing jet data to Monte Carlo simulations,
using the distribution of the pseudo-proper decay length $\hat L$
defined as
\be
  \hat L = \frac{L_{2D} M^{(v)}}{p_T^{(v)}}\ ,
\ee
where $M^{(v)}$ is the invariant mass of all tracks originating from the secondary vertex
and $p_T^{(v)}$ the transverse-momentum of the secondary vertex four-vector.
The quantity $\hat L$ turns out to be largely different for $b$-jets and lighter jets
and can thus be used as a discriminant.
Finally, the mistagging rate in the negative $L_{2D}$ region computed in this way
is extrapolated to the positive region.
As a second step, the calculated mistagging probability is applied to each light-flavor
jet included in the Monte Carlo generated events, the resulting sample being identified
as the so-called {\it mistag}
contributions to the background.

In order to probe the possible presence of monotop events in data, we have simulated
eleven signal samples in the framework of the {\bf S.IV}
class of scenarios. The benchmark points differ by the mass $M_D$ of the dark matter
candidate yielding missing energy. The values for this mass
have been chosen by sampling the [0,25]~GeV mass range in steps of 5~GeV
and the [25,150]~GeV range in steps of 25~GeV. Like in the previous sections,
signal simulation has been based on the UFO model extracted from the
\feynrules\
implementation of the effective model constructed in
Section~\ref{sec:monotopsEFT}, after fixing the mass of the top
quark to $M_t = 172.5$~GeV~\cite{Lancaster:2011wr,Galtieri:2011yd}.
Hard scattering events have then been generated
using the \madgraph~5 package and parton showering and hadronization
have been described by the \pythia~6 program.

Applying all the steps of the selection described above,
6471 data events survive. From the Standard Model expectation, it turns out that
most of these events (about 70\%) consists of QCD multijet events. In this case,
as already stated above, the missing momentum tends to be aligned with the momentum
of the second jet $j_2$. Therefore, a good fraction of
the multijet contribution to the background can be
rejected by imposing that the azimuthal distance between the missing transverse momentum
$\vec{\slashed{p}}_T$ and the momentum of the
second hardest jet $\vec{p}(j_2)$ satisfies
\be
  \Delta \varphi\Big(\vec{\slashed{p}}_T, \vec{p}(j_2)\Big) \geq 0.7 \ .
\ee

In order to further suppress the Standard Model background,
we follow the suggestions of the monotop selection strategy designed in the previous sections
and require that the
invariant mass of the three-jet system $m_{j_1j_2j_3}$ is compatible with the mass of the
top quark,
\be
   m_{j_1j_2j_3} \in [110,  200]\text{~GeV} \ .
\ee

We finally improve the background rejection (without too much affecting the signal
selection efficiency) by imposing a large missing transverse
energy significance,
\be
  \frac{\slashed{E}_T}{\sqrt{\sum E_T}} \geq 3.5~\sqrt{\text{GeV}} \ , 
\label{eq:metsign}\ee
and further constrain the transverse energy of the third jet,
\be
  E_T(j_3) \geq  25~\text{GeV}\ .
\ee
In Eq.~\eqref{eq:metsign}, the expression $\sum E_T$ refers to
the scalar sum of transverse energy deposited in both calorimeters.

\subsection{Results of the first monotop search at a hadron collider}

\renewcommand{\arraystretch}{1.2}
\begin{table}[t] 
 \begin{center}
  \begin{tabular}{|c|c|}
   \hline
     Processes &  Number of events\\ 
   \hline
     Signal scenario with $M_D = 20$ GeV & 2116.9 $\pm$ 121.4 \\
     Signal scenario with $M_D = 75$ GeV  & 232.3 $\pm$ 22.9 \\
     Signal scenario with $M_D = 100$ GeV & 129.8 $\pm$ 12.5 \\
     Signal scenario with $M_D = 125$ GeV & 94.5 $\pm$ 9.3 \\
  \hline
  \hline
  QCD multijet & 210.2 $\pm$ 54.5\\
  Top-antitop  & 182.8 $\pm$ 20.2 \\
  Single boson (plus jets) & 130.5 $\pm$ 33.8\\
  Mistag       & 96.9 $\pm$ 39.4\\
  Single top   & 24.3  $\pm$ 4.5\\
  Diboson      & 15.7  $\pm$ 2.7\\
  \hline
  Total background & 660.2 $\pm$ 78.1\\
  \hline
  \hline
  Data & 592 \\
  \hline
\end{tabular}
\caption{\label{tab:CDFmono} Number of expected signal (for different
  masses of the invisible vector particle $M_D$) and background events surviving
  the monotop selection strategy presented in the text, given together with the
  combined statistical and systematic uncertainties. We distinguish the different
  background contributions and compare the Standard Model expectation to data.}
\end{center}
\end{table}
\renewcommand{\arraystretch}{1.0}

Although the analysis strategy presented in Section~\ref{sec:cdfstrategy} has been
inspired by the phenomenological works of the previous sections,
it has more precisely been designed in the aim of optimizing
the significance $S/\sqrt{S+B}$, $S$ and $B$ being
the expected number of signal and background events, respectively. Out of the
6471 data events, 592 of them remain after the last selection steps. In order
to extract some conclusions on the possible presence of monotops in those data,
this number must be confronted to the Standard Model predictions.
This is achieved in Table \ref{tab:CDFmono}, where we distinguish the different contributions to
the Standard Model background and present the number of selected signal events
for several simulated scenarios of class {\bf S.IV}.

The uncertainty values quoted in the table encompass both statistic and systematic
uncertainties. For the latter, several contributions have been considered and are
listed below. The dominant
source of systematic uncertainties consist of the normalization of the
QCD multijet background, the mistagging rate of the $b$-tagging algorithm
(16.6\%) and the theoretical cross sections employed for the different other background
contributions ($6.5\% - 30\%$). In a smaller extent, uncertainties issued from
the jet energy scale ($2.8\% - 10.7\%$)~\cite{Bhatti:2005ai}, the measurement of the
luminosity (6\%)~\cite{Acosta:2002hx}, the efficiency of the $b$-tagging algorithm (5.2\%),
the initial-state and final-state parton radiation (4\%),
the parton densities (2\%), the veto on the presence of charged leptons in the
selection (2\%) and the trigger efficiency ($0.4\% - 0.9\%$) are included.
Finally, two additional sources of systematic uncertainties are also accounted for.
A first contribution is derived from
the shape variation of various kinematical distributions when a $\pm 1\sigma$
modification of the jet energy scale is performed. On a different line, a second
contribution consists of the uncertainties on the
efficiency of the data acquisition system.

\begin{figure}[t!]
  \begin{center}
  \includegraphics[width=.7\columnwidth]{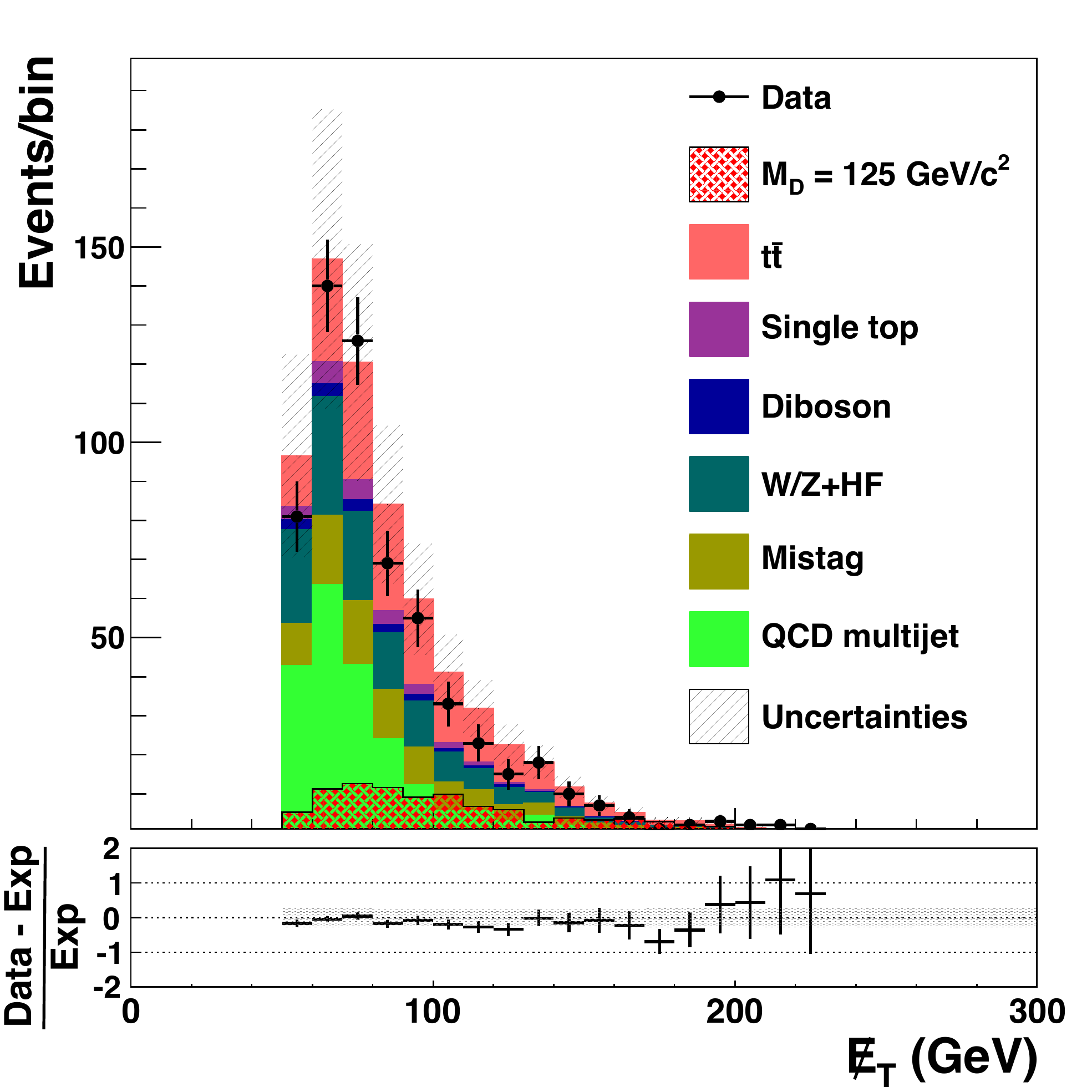}
  \caption{\label{fig:CDFmet}
  Missing transverse energy spectrum after applying the CDF monotop search strategy
  described in the Section~\ref{sec:cdfstrategy}.
  We compare the sum of the different contributions to the Standard
  Model background to data and superimpose the expected distribution of a monotop signal
  arising from a scenario of class {\bf S.IV} for an invisible particle mass of
  $M_D=125$~GeV. Figure taken from Ref.~\cite{Aaltonen:2012ek}. }
  \end{center}
\end{figure}

\begin{figure}[t!]
  \begin{center}
  \includegraphics[width=.7\columnwidth]{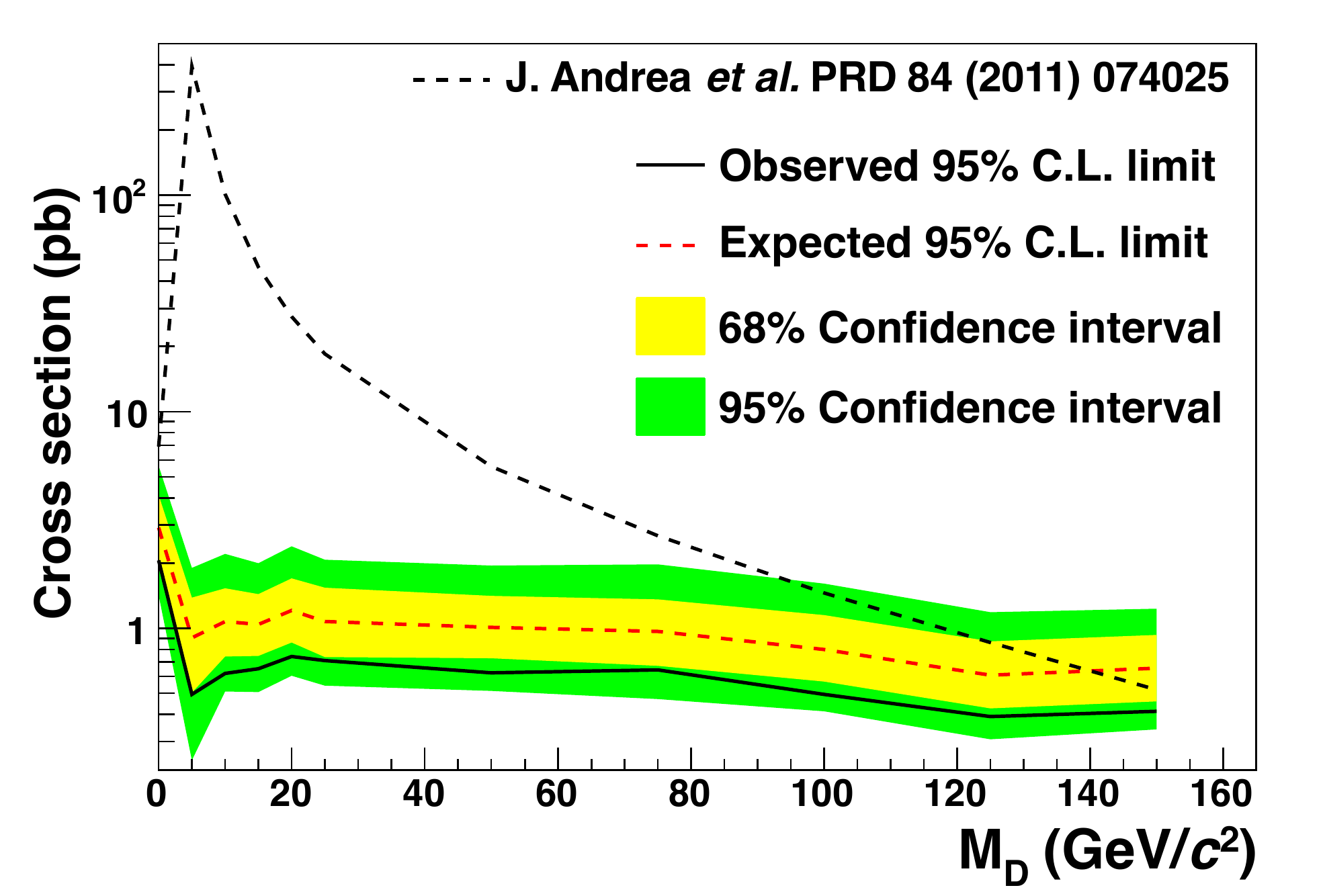}
  \caption{\label{fig:CDFcl}Exclusion curve for the monotop production cross section as a function
  of the mass of the invisible vector particle $M_D$. Theoretical cross section extracted from
  Ref.~\cite{Andrea:2011ws} have been superimposed to the curves. Figure taken from
  Ref.~\cite{Aaltonen:2012ek}.}
  \end{center}
\end{figure}

One of the key distribution allowing for a signal versus background discrimination
in the context of the described monotop search strategy at the Tevatron
consists of the missing energy spectrum. This holds in particular for kinematical
regions where $\slashed{E}_T$ is large, since in this case, a
significant signal contribution is expected together with a reduced background contamination.
This is illustrated on Figure~\ref{fig:CDFmet}
where we show the different contributions
to the Standard Model background and compare their sum to data. In addition,
we also present the spectrum induced by the presence of
a dark matter particle $D$ of mass $M_D = 125$~GeV after the selection.
The obtained results are compatible
with the Standard Model and no significant excess of signal-like events has been found
in the analyzed dataset.

Consequently, the results allow to extract 95\% confidence level (C.L.) upper limits on the
monotop production cross section by comparing the expected
shape of the transverse
missing energy distribution in the case there is a new physics contribution
with the one got from data. To this aim, a Bayesian maximum likelihood method
has been employed~\cite{Aaltonen:2010jr}.
The used likelihood function $L$ is defined as the product of Poisson probabilities
for each bin of the missing energy distribution. Since the Poisson probabilities are functions of
the number of observed data events $d_i$ and the predicted number of events $\mu_i$
in each bin, we have
\be
  L = \prod_{i=1}^{n_{\rm bins}}\frac{\mu_i^{d_i} e^{-\mu_i}}{d_i\!} \ ,
\ee
where the predictions $\mu_i$ have been computed as the sum over both signal and background
contributions. In the calculation of the $\mu_i$ quantities, a series of nuisance
parameters have been introduced, one of them being associated with each source of
systematic uncertainties. This allows to calculate predictions for the $\mu_i$
quantities encompassing
$\pm 1 \sigma$ variations possibly induced by the different sources of systematic uncertainties.
After assuming Gaussian priors centered on zero and with
a unit width for each of the nuisance parameters, the posterior
$L'$ is computed by marginalizing these parameters, so that it is sufficient to maximize
$L'$ to get the expected limit on the signal cross section.
The 68\% and 95\% confidence level uncertainty bands
are eventually obtained by integrating $L'$ over the signal cross section $\sigma_s$ varying
from zero to infinity when a uniform positive prior is associated with $\sigma_s$. The respective
bands are then defined by the
smallest cross section range containing 68\% and 95\% of the integral value.

The results are shown on Figure~\ref{fig:CDFcl}
as a function of the mass of the new vector state $M_D$. In addition, we have
indicated the theoretical predictions for the cross section as calculated
in the framework of the effective field theory developed in Section~\ref{sec:monotopsEFT},
for benchmark scenarios of type {\bf S.IV} and
for a coupling parameter set to $a=0.1$. From the figure, we observe that under the above-mentioned
hypotheses, new vector
states yielding a monotop signature via flavor-changing interactions are constrained to
be heavier than about 100~GeV, at the 95\% confidence level.

\mysection{Conclusions of a search for sgluons with the ATLAS detector}
\label{sec:sgluonatlas}

Using 14.3~fb$^{-1}$ of proton-proton collision at a center-of-mass energy
of 8~TeV, the ATLAS collaboration has investigated LHC data for hints of
sgluon fields dominantly coupling to the top quark,
following the work performed in Section~\ref{sec:effmultitops}~\cite{ATLAS-CONF-2013-051}.
In this case, only scenarios of class {\bf S.II} have been addressed, the sgluon field
only singly coupling to a pair of top quarks or gluons. As a result of the analysis,
sgluon masses of about 800~GeV have been excluded. Those
limits are slightly better than the predictions of Section~\ref{sec:effmultitops}
due to a more optimized analysis strategy. As stated at the
beginning of this section, we leave any detail of this experimental search
out of this work and refer to Ref.~\cite{ATLAS-CONF-2013-051} for more information.

\cleardoublepage

%% file: conclusions.tex
\label{chap:conclusions}

The Standard Model of particle physics provides a tremendously successful
description of most of currently observed high-energy physics data.
However, despite its success, it includes
several conceptual problems,
such as, \eg, the stabilization of the mass of the
recently observed Higgs boson with respect to quantum corrections.
This has triggered the construction of a plethora of new physics
theories over the last decades whose predict, for most of them,
phenomena expected to be observable at scales probed by current and future
experiments. In this work, we have focused of one class of such theories
dubbed supersymmetry
in the aim of investigating specific signatures predicted by
non-minimal supersymmetric models at the Large Hadron Collider.

We have started by showing how supersymmetric quantum field theories
naturally arise from the only knowledge of both N\oe ther and spin-statistics theorems
used jointly with a series of no-go theorems.
We have then described in details how the superspace formalism allows for the construction of
supersymmetric Lagrangians in a very efficient way. On different footings,
the collider phenomenology of any supersymmetric model can in general be probed
after having
implemented the Feynman rules associated with the underlying Lagrangian
in various high-energy physics tools such as, \eg, Monte Carlo event generators.
This has brought us to put some emphasis on the program \feynrules\ that
has been developed over the last
few years. It
allows, on the one hand, to derive supersymmetric Lagrangians in a very efficient
way, and, on the other hand, to extract the related Feynman rules and export them
into a programming language that can be understood by Monte Carlo codes.
In particular, translating the model into a UFO library offers a way
to pass all the information, regardless the use of non-usual Lorentz and/or color structures
that could be included in the vertices, to any program lying further in the simulation
chain.

Although supersymmetric theories are promising from the theoretical point of view,
not a single superpartner has been observed up to now. Therefore, supersymmetry
has to be broken so that the masses of the supersymmetric particles are shifted with respect to
those of their Standard Model counterparts.
We thus depict, in the last formal part of this manuscript,
some general properties of broken supersymmetric theories
and the most commonly employed mechanisms leading to
supersymmetry breaking.

Once these bases have been set,
we have moved on with more phenomenological topics.
We have detailed the construction of the
supersymmetric model resulting from the direct supersymmetrization of the
Standard Model, the so-called MSSM, together with its properties.
With the current status of the experimental data, designing phenomenologically viable
scenarios for the most constrained versions of the
MSSM becomes a more and more difficult task.
This feature has been illustrated by estimating the regions
of the parameter space (un)favored by several low-energy, flavor
and electroweak observables as well as by cosmological data and by results
of direct searches for supersymmetric particles at colliders. We have in a second stage
generalized the squark sector of the minimal model by allowing for non-minimal
flavor mixing among all up-type and down-type squark gauge-eigenstates and revisited
some of the above-mentioned constraints. We have found that important
flavor violation is still allowed by current data,
in particular for parameter space regions where the superpartners are heavy.

We have then turned to the building of two non-minimal supersymmetric
theories and the detailed investigation of two signatures
of such models at the LHC. First, we have illustrated how non-minimal
softly-broken supersymmetric theories can be efficiently and easily implemented
in \feynrules\ by using its superspace module and further exported to
Monte Carlo event generators for phenomenological investigations at colliders.
Next, we have focused on the study of monotop production in the framework of $R$-parity violating
supersymmetry and sgluon-induced multitop production within $R$-symmetric supersymmetry.
We have chosen two specific benchmark scenarios and show that the LHC is possibly capable
of unveiling the presence of new physics within the overwhelming background from the
Standard Model through simple search strategies.

However, observing a new physics signal such as those that can arise
in the framework of the two
investigated benchmark scenarios does neither imply that the relevant scenario consists
of the physics model chosen by Nature, nor that the underlying supersymmetric
theory is realized. A given signature can be predicted by many benchmarks of a given
theory and even by several different theories.
In order to allow for an easy recasting
of the results into any other
context (either another benchmark or another theory beyond the Standard Model),
we have followed a bottom-up path starting from the signature itself. In
this way, we have
constructed one effective field theory for each signature that
encompasses all its possible production modes. Consequently,
various new physics scenarios yielding
either the production of a monotop state or the one of a multitop state issued
from the decay of a sgluon pair have been explored simultaneously.
Using Monte Carlo simulations of 20 fb$^{-1}$ of collisions that have been produced
at the LHC in 2012, we have designed general search strategies which we have shown to be sensitive
to a large fraction of the effective field theory parameter spaces. It has been found
that new physics states lying within the TeV range can easily be observed
in most of the cases, for not too small interaction strengths.

Although the reinterpretation of
the general results in the framework of a very specific beyond the Standard Model theory
is beyond the scope of this work, this consists of its natural extension in a near future.
This future can also be seen as the recasting era in particle physics since both the ATLAS and CMS
collaborations will certainly not have enough manpower to dedicate (at least) one analysis
to each scenario of each new physics theory that
theorists can think of, even if we restrict ourselves to scenarios
valuable to be investigated. This is the reason why it is likely that both theorists
and experimentalists will share this task of reinterpreting the existing searches in the
context of large classes of models. In this work, we have shown that several existing tools,
such as \feynrules, \madgraph~5 or \madanalysis~5, are
ready to be used for rendering this step more straightforward from the point of
view of the user.

In the last part of this work, we have presented one additional example of possible
interactions among theorists and experimentalists and described an existing
analysis of Tevatron data that
has been motivated by our phenomenological results. It consists of a monotop
search with the CDF detector at the Tevatron, that we have presented with the eyes of a theorist.
We have observed that parameter space regions of the constructed monotop effective field theory,
relevant for providing an explanation of observed anomalies in dark matter direct detection experiments,
are already excluded by collider data.

\cleardoublepage

%% file: conventions.tex
\label{app:conv}
\mysection{Lorentz indices}
We employ, for the Minkowski metric $\eta_{\mu \nu}$ and its
inverse $\eta^{\mu\nu}$, the conventions 
\be
   \eta_{\mu \nu} = \text{diag}(1,-1,-1,-1)\qquad\text{and}\qquad
   \eta^{\mu \nu} = \text{diag}(1,-1,-1,-1)\ , 
\ee
so that the product of these two symmetric matrices is given by
\be
  \eta_{\mu \rho}\ \eta^{\rho \nu} = \delta_\mu{}^\nu \ . 
\ee 
These matrices allow for lowering and raising spacetime indices,
\be
   x^\mu = \eta^{\mu\nu} x_\nu \qquad\text{and}\qquad
   x_\mu = \eta_{\mu\nu} x^\nu \ ,   
\ee
where $x^\mu$ is an arbitrary four-vector. In this work, we use Greek letters
from the middle
of the alphabet for Lorentz indices ranging from zero to four, and Latin letters
from the middle of the alphabet for Euclidean indices ranging from one to three.
In our conventions, we define the rank-four antisymmetric tensor with lower
indices as
\be
  \e_{0123}=1\ .
\ee
Therefore, its counterpart with upper indices is derived as
\be
  \e^{\mu\nu\rho\sigma} = \e_{\tilde\mu\tilde\nu\tilde\rho\tilde\sigma}
     \ \eta^{\mu\tilde\mu}\ \eta^{\nu\tilde\nu}\ \eta^{\rho\tilde\rho}
     \ \eta^{\sigma\tilde\sigma} \ , 
\ee
or equivalently, we have $\e^{0123}=-1$.

\mysection{Spinor indices}
For Weyl spinors, we adopt the van der Waerden conventions for indices, so that
the tensorial structure of a left-handed Weyl spinor $\lambda$ reads
\be 
  \lambda \equiv \lambda_\alpha \ , 
\ee
and the one of a right-handed spinor $\chibar$ is given by
\be
  \chibar \equiv \chibar^\alphadot \ .
\ee
In both cases, we employ Greek letters from the beginning of the alphabet to
denote spin indices ranging from one to two.
Merging one two-component left-handed Weyl spinor $\lambda_\alpha$ and one right-handed
two-component Weyl spinor $\chibar^\alphadot$, one constructs 
four-component Dirac spinors which read, in the van der
Waerden notations
\be
  \psi_D=\bpm \lambda_\alpha\\ \chibar^\alphadot\epm\ .
\ee
In the case of a four-component Majorana spinor, the left-handed and
right-handed components are related by Hermitian conjugation, 
$(\lambda_\alpha)^\dag = \lambar_\alphadot$, so that the four-component object
reads 
\be
  \psi_M=\bpm \lambda_\alpha\\ \lambar^\alphadot\epm\ .
\ee

Getting back to Weyl spinors, their spin indices can be raised and lowered by means
of the rank-two antisymmetric tensors $\e_{\alpha \beta}$ and $\e^{\alpha
\beta}$ acting on left-handed spinors as well as
$\e_{\alphadot\betadot}$ and $\e^{\alphadot\betadot}$ acting on right-handed
spinors. We normalize these tensors as $\e_{12}=1$, $\e^{12}=-1$, $\e_{\dot 1
\dot 2}=1$ and $\e^{\dot 1 \dot
2}=-1$. Subsequently, the summation properties
\be
  \e_{\alpha \beta} \, \e^{\beta \gamma}=\delta_\alpha{}^\gamma
   \quad\text{and}\quad
  \e_{\alphadot\betadot}\, \e^{\betadot\gammadot}=\delta_\alphadot{}^\gammadot 
\ee
are verified and raising and lowering indices obey the rules
\be\label{eq:rank2eps}
  \lambda^\alpha= \e^{\alpha \beta} \lambda_\beta\ , \quad
  \lambda_\alpha= \e_{\alpha \beta} \lambda^\beta\ , \quad
  \chibar^\alphadot= \e^{\alphadot\betadot} \chibar_\betadot\quad\text{and}\quad
  \chibar_\alphadot= \e_{\alphadot\betadot} \chibar^\betadot \ .
\ee
Contracting spinors with upper and lower indices further allows to define scalar
products as 
\be\label{eq:scalprod}
  \lambda \cdot \lambda' = \lambda^\alpha \lambda'_\alpha \quad\text{and}\quad
  \chibar \cdot \chibar' = \chibar_\alphadot \chibar'{}^\alphadot \ .
\ee
where we have introduced a left-handed Weyl spinor $\lambda'$ and a right-handed
Weyl spinor $\chibar'$. 

\mysection{Pauli and Dirac matrices}
The three Pauli matrices are defined, in our conventions, by 
\be
\sigma^1 = \bpm 0&1\\1&0 \epm\ ,\quad
\sigma^2 = \bpm 0&-i\\i&0 \epm \quad\text{and}\quad
\sigma^3 = \bpm 1&0\\0&-1 \epm\ ,
\ee
and we remind that $\sigma_i=-\sigma^i$. Introducing $\sigma^0$ as the
dimension-two identity matrix, one can construct the two four-vectors 
\be 
  \sigma^\mu=(\sigma^0,\sigma^i) \quad\text{and}\quad
  \sibar^\mu=(\sigma^0,-\sigma^i) \ ,
\ee
whose the tensorial structure reads
\be
  \sigma^\mu \equiv \sigma^\mu{}_{\alpha\alphadot}\quad\text{and}\quad
  \sibar^\mu \equiv \sibar^\mu{}^{\alphadot\alpha} \ .
\ee
From the product of these four-vectors, one gets the generators of the
Lorentz algebra in the left-handed and right-handed spinorial representations,
\be\label{eq:simunu}
  \sigma^{\mu\nu} = \frac{i}{4} 
     \Big[\sigma^\mu\sibar^\nu-\sigma^\nu\sibar^\mu\Big]
  \quad\text{and}\quad
  \sibar^{\mu\nu} = \frac{i}{4} 
     \Big[\sibar^\mu\sigma^\nu-\sibar^\nu\sigma^\mu\Big] \ ,
\ee
respectively. 
The matrices $\sigma^\mu$ and $\sibar^\mu$ naturally appear in superspace
computations, together with products of these matrices. The latter can in
general be simplified by means of several useful identities. 
Firstly, the two four-vectors
$\sigma^\mu$ and $\sibar^\mu$ are related through metric tensors in spin
and Minkowski space, 
\be\label{eq:sisi}\bsp
  \sibar^\mu{}^{\betadot \beta} = \sigma^\mu{}_{\alpha \alphadot}
    \e^{\alpha\beta} \e^{\alphadot\betadot}\ ,\qquad &\
  \sigma^\mu{}_{\alpha \alphadot} \sibar_\mu{}^{\betadot \beta} = 2 
    \delta_\alpha{}^\beta \delta_\alphadot{}^\betadot \ , \\ 
  \sibar^{\mu\alphadot\alpha} \sibar_\mu{}^{\betadot \beta} = 2
    \e^{\alphadot\betadot} \e^{\alpha\beta} \ ,\qquad &\
  \sigma^\mu{}_{\alpha\alphadot} \sigma_\mu{}_{\beta\betadot} = 2
    \e_{\alpha\beta} \e_{\alphadot\betadot} \ , \\
  (\sibar^\mu\sigma^\nu)^\alphadot{}_\betadot = \eta^{\mu\nu}
  \delta^\alphadot{}_\betadot - 2 i (\sibar^{\mu\nu})^\alphadot{}_\betadot \ ,
  \qquad &\
  (\sigma^\mu\sibar^\nu)_\alpha{}^\beta = \eta^{\mu\nu} \delta_\alpha{}^\beta{}
    - 2 i (\sigma^{\mu\nu})_\alpha{}^\beta \ .
\esp\ee
Secondly, the tensors $\sigma^{\mu\nu}$ and $\sibar^{\mu\nu}$ fulfill
(anti-)self-duality properties, 
\be\label{eq:dual}
  \frac12 \e_{\mu \nu \rho \sigma} \sigma^{\rho \sigma} = -i
    \sigma_{\mu \nu} \qquad \text{and}\qquad
  \frac12 \e_{\mu \nu \rho \sigma} \sibar^{\rho \sigma} = i \sibar_{\mu \nu}\ .
\ee 
Finally, the product of three Pauli matrices can be simplified to
\be \label{eq:idsigma}\bsp
  \sibar^\mu \sigma^\nu \sibar^\rho =&\ i \e^{\mu\nu\rho\sigma} \sibar_\sigma 
    + \eta^{\mu\nu} \sibar^\rho + \eta^{\nu\rho} \sibar^\mu 
    - \eta^{\mu\rho} \sibar^\nu  \ , \\
  \sigma^\mu \sibar^\nu \sigma^\rho =&\ -i \e^{\mu\nu\rho\sigma} \sigma_\sigma 
    + \eta^{\mu\nu} \sigma^\rho + \eta^{\nu\rho} \sigma^\mu 
    - \eta^{\mu\rho} \sigma^\nu \ .
\esp\ee 

From Pauli matrices, one can construct the Dirac matrices in the Weyl
representation, 
\be
  \gamma^\mu=\bpm 0&\sigma^\mu\\ \sibar^\mu&0 \epm\ ,
\ee
while products of Dirac matrices define the four-component spinorial
representation of the Lorentz algebra
\be
  \gamma^{\mu\nu} = \frac{i}{4} \big[ \gamma^\mu,\gamma^\nu\big] \ .
\ee
This representation is reducible to the two-component representations defined
above so that one can also define a fifth Dirac matrix $\gamma^5$ as 
\be
  \gamma^5 = i\gamma^0 \gamma^1\gamma^2\gamma^3 = \bpm -1&0\\0&1 \epm \ ,
\ee
which commutes with all the generators of the algebra and which anticommutes 
with all the Dirac matrices.

\mysection{Grassmann variables}
A point in superspace is defined by adjoining to the spacetime coordinates
$x^\mu$ the Grassmann coordinates $\theta_\alpha$ and $\thetabar^\alphadot$
forming a Majorana spinor. The $\theta$-variables satisfy the Grassmann algebra
\be
  \big\{ \theta^\alpha, \theta^\beta \big\} = 
  \big\{ \thetabar_\alphadot, \thetabar_\betadot \big\} = 
  \big\{ \theta^\alpha, \thetabar_\alphadot \big\} = 0 \ .
\label{eq:grassmannalg}\ee
Since the square of an anticommuting object vanishes, it follows that 
\be\label{eq:thetarelations}
  \theta^\alpha \theta^\beta = -\frac12 \theta \!\cdot\! \theta \e^{\alpha \beta}\ ,
    \qquad 
  \thetabar^\alphadot \thetabar^\betadot = \frac12 \thetabar \!\cdot\! \thetabar
    \e^{\alphadot \betadot} \qquad\text{and}\qquad
  \theta^\alpha \thetabar^\alphadot = \frac12 \theta \sigma^\mu \thetabar
    \sibar_\mu{}^{\alphadot\alpha}\ , 
\ee
which is at the basis of any computation to be performed in superspace.
We define the operators $\del_\mu$, $\del_\alpha$ and $\delbar^\alphadot$
conjugate to the superspace coordinates $x^\mu$, $\theta^\alpha$ and
$\thetabar^\alphadot$. In other words, we define derivatives with respect to the
superspace coordinates that fulfill the properties
\be\label{eq:spaceconju}
 \big[ \del_\mu, x^\nu\big] = \delta_\mu{}^\nu \ , \qquad
 \big\{ \del_\alpha, \theta^\beta\big\} = \delta_\alpha{}^\beta
  \qquad\text{and}\qquad
 \big\{ \delbar_\alphadot, \thetabar^\betadot\big\} =
  \delta_\alphadot{}^\betadot \ .
\ee
Finally, integration upon Grassmann variables is defined by the relations
\be\label{eq:intgrass}
  \int \d^2\theta = 
  \int \d^2\theta\ \theta = 
  \int \d^2\thetabar = 
  \int \d^2\thetabar\ \thetabar =  0  \ , \qquad
  \int \d^2\theta\ \theta\!\cdot\!\theta = 
  \int \d^2\thetabar\ \thetabar\!\cdot\!\thetabar = 1 \ . 
\ee

\cleardoublepage

%% file: hdr.bbl
\providecommand{\href}[2]{#2}\begingroup\raggedright\begin{thebibliography}{100}

\bibitem{Glashow:1961tr}
S.~Glashow, ``{Partial Symmetries of Weak Interactions},''
\href{http://dx.doi.org/10.1016/0029-5582(61)90469-2}{{\em Nucl.Phys.} {\bf 22}
  (1961)  579--588}.

\bibitem{Salam:1964ry}
A.~Salam and J.~C. Ward, ``{Electromagnetic and Weak Interactions},''
\href{http://dx.doi.org/10.1016/0031-9163(64)90711-5}{{\em Phys.Lett.} {\bf 13}
  (1964)  168--171}.

\bibitem{Weinberg:1967tq}
S.~Weinberg, ``{A Model of Leptons},''
\href{http://dx.doi.org/10.1103/PhysRevLett.19.1264}{{\em Phys.Rev.Lett.} {\bf
  19} (1967)  1264--1266}.

\bibitem{salamsm}
A.~Salam, {\em Elementary Particle Theory}.
\newblock ed.~N.~Svartholm (Alm\-qvist and Forlag, Stockholm), 1968.

\bibitem{Glashow:1970gm}
S.~Glashow, J.~Iliopoulos, and L.~Maiani, ``{Weak Interactions with
  Lepton-Hadron Symmetry},''
\href{http://dx.doi.org/10.1103/PhysRevD.2.1285}{{\em Phys.Rev.} {\bf D2}
  (1970)  1285--1292}.

\bibitem{Weinberg:1971nd}
S.~Weinberg, ``{Mixing Angle in Renormalizable Theories of Weak and
  Electromagnetic Interactions},''
\href{http://dx.doi.org/10.1103/PhysRevD.5.1962}{{\em Phys.Rev.} {\bf D5}
  (1972)  1962--1967}.

\bibitem{Gross:1973ju}
D.~Gross and F.~Wilczek, ``{Asymptotically Free Gauge Theories. 1},''
\href{http://dx.doi.org/10.1103/PhysRevD.8.3633}{{\em Phys.Rev.} {\bf D8}
  (1973)  3633--3652}.

\bibitem{Kobayashi:1973fv}
M.~Kobayashi and T.~Maskawa, ``{CP Violation in the Renormalizable Theory of
  Weak Interaction},''
\href{http://dx.doi.org/10.1143/PTP.49.652}{{\em Prog.Theor.Phys.} {\bf 49}
  (1973)  652--657}.

\bibitem{Gross:1974cs}
D.~Gross and F.~Wilczek, ``{Asymptotically Free Gauge Theories. 2.},''
\href{http://dx.doi.org/10.1103/PhysRevD.9.980}{{\em Phys.Rev.} {\bf D9} (1974)
   980--993}.

\bibitem{Politzer:1974fr}
H.~D. Politzer, ``{Asymptotic Freedom: An Approach to Strong Interactions},''
\href{http://dx.doi.org/10.1016/0370-1573(74)90014-3}{{\em Phys.Rept.} {\bf 14}
  (1974)  129--180}.

\bibitem{Aad:2012gk}
{\bf ATLAS Collaboration}, G.~Aad {\em et al.}, ``{Observation of a New
  Particle in the Search for the Standard Model Higgs Boson with the ATLAS
  Detector at the LHC},''
\href{http://dx.doi.org/10.1016/j.physletb.2012.08.020}{{\em Phys.Lett} {\bf
  B716} (2012)  1--29}.

\bibitem{Chatrchyan:2012gu}
{\bf CMS Collaboration}, S.~Chatrchyan {\em et al.}, ``{Observation of a New
  Boson at a Mass of 125 GeV with the CMS Experiment at the LHC},''
\href{http://dx.doi.org/10.1016/j.physletb.2012.08.021}{{\em Phys.Lett} {\bf
  B716} (2012)  30--61}.

\bibitem{Georgi:1974sy}
H.~Georgi and S.~Glashow, ``{Unity of All Elementary Particle Forces},''
\href{http://dx.doi.org/10.1103/PhysRevLett.32.438}{{\em Phys.Rev.Lett.} {\bf
  32} (1974)  438--441}.

\bibitem{Georgi:1974yf}
H.~Georgi, H.~R. Quinn, and S.~Weinberg, ``{Hierarchy of Interactions in
  Unified Gauge Theories},''
\href{http://dx.doi.org/10.1103/PhysRevLett.33.451}{{\em Phys.Rev.Lett.} {\bf
  33} (1974)  451--454}.

\bibitem{Fritzsch:1974nn}
H.~Fritzsch and P.~Minkowski, ``{Unified Interactions of Leptons and
  Hadrons},''
\href{http://dx.doi.org/10.1016/0003-4916(75)90211-0}{{\em Annals Phys.} {\bf
  93} (1975)  193--266}.

\bibitem{Gursey:1975ki}
F.~Gursey, P.~Ramond, and P.~Sikivie, ``{A Universal Gauge Theory Model Based
  on E6},''
\href{http://dx.doi.org/10.1016/0370-2693(76)90417-2}{{\em Phys.Lett.} {\bf
  B60} (1976)  177}.

\bibitem{Chang:1984uy}
D.~Chang, R.~Mohapatra, and M.~Parida, ``{A New Approach to Left-Right Symmetry
  Breaking in Unified Gauge Theories},''
\href{http://dx.doi.org/10.1103/PhysRevD.30.1052}{{\em Phys.Rev.} {\bf D30}
  (1984)  1052}.

\bibitem{Witten:1981me}
E.~Witten, ``{Search for a Realistic Kaluza-Klein Theory},''
\href{http://dx.doi.org/10.1016/0550-3213(81)90021-3}{{\em Nucl.Phys.} {\bf
  B186} (1981)  412}.

\bibitem{ArkaniHamed:1998rs}
N.~Arkani-Hamed, S.~Dimopoulos, and G.~Dvali, ``{The Hierarchy Problem and New
  Dimensions at a Millimeter},''
\href{http://dx.doi.org/10.1016/S0370-2693(98)00466-3}{{\em Phys.Lett.} {\bf
  B429} (1998)  263--272}.

\bibitem{Randall:1999ee}
L.~Randall and R.~Sundrum, ``{A Large Mass Hierarchy from a Small Extra
  Dimension},''
\href{http://dx.doi.org/10.1103/PhysRevLett.83.3370}{{\em Phys.Rev.Lett.} {\bf
  83} (1999)  3370--3373}.

\bibitem{Appelquist:2000nn}
T.~Appelquist, H.-C. Cheng, and B.~A. Dobrescu, ``{Bounds on Universal Extra
  Dimensions},''
\href{http://dx.doi.org/10.1103/PhysRevD.64.035002}{{\em Phys.Rev.} {\bf D64}
  (2001)  035002}.

\bibitem{Golfand:1971iw}
Y.~Golfand and E.~Likhtman, ``{Extension of the Algebra of Poincare Group
  Generators and Violation of p Invariance},''
{\em JETP Lett.} {\bf 13} (1971)  323--326.

\bibitem{Volkov:1973ix}
D.~Volkov and V.~Akulov, ``{Is the Neutrino a Goldstone Particle?},''
\href{http://dx.doi.org/10.1016/0370-2693(73)90490-5}{{\em Phys.Lett.} {\bf
  B46} (1973)  109--110}.

\bibitem{Wess:1973kz}
J.~Wess and B.~Zumino, ``{A Lagrangian Model Invariant Under Supergauge
  Transformations},''
\href{http://dx.doi.org/10.1016/0370-2693(74)90578-4}{{\em Phys.Lett.} {\bf
  B49} (1974)  52}.

\bibitem{Wess:1974tw}
J.~Wess and B.~Zumino, ``{Supergauge Transformations in Four-Dimensions},''
\href{http://dx.doi.org/10.1016/0550-3213(74)90355-1}{{\em Nucl.Phys.} {\bf
  B70} (1974)  39--50}.

\bibitem{Wess:1974jb}
J.~Wess and B.~Zumino, ``{Supergauge Invariant Extension of Quantum
  Electrodynamics},''
\href{http://dx.doi.org/10.1016/0550-3213(74)90112-6}{{\em Nucl.Phys.} {\bf
  B78} (1974)  1}.

\bibitem{Salam:1974yz}
A.~Salam and J.~A. Strathdee, ``{Supergauge Transformations},''
\href{http://dx.doi.org/10.1016/0550-3213(74)90537-9}{{\em Nucl. Phys.} {\bf
  B76} (1974)  477--482}.

\bibitem{Salam:1974jj}
A.~Salam and J.~A. Strathdee, ``{On Superfields and Fermi-Bose Symmetry},''
\href{http://dx.doi.org/10.1103/PhysRevD.11.1521}{{\em Phys.Rev.} {\bf D11}
  (1975)  1521--1535}.

\bibitem{Ferrara:1974ac}
S.~Ferrara, J.~Wess, and B.~Zumino, ``{Supergauge Multiplets and
  Superfields},''
\href{http://dx.doi.org/10.1016/0370-2693(74)90283-4}{{\em Phys. Lett.} {\bf
  B51} (1974)  239}.

\bibitem{Ferrara:1974pu}
S.~Ferrara and B.~Zumino, ``{Supergauge Invariant Yang-Mills Theories},''
\href{http://dx.doi.org/10.1016/0550-3213(74)90559-8}{{\em Nucl.Phys.} {\bf
  B79} (1974)  413}.

\bibitem{Nilles:1983ge}
H.~P. Nilles, ``{Supersymmetry, Supergravity and Particle Physics},''
\href{http://dx.doi.org/10.1016/0370-1573(84)90008-5}{{\em Phys.Rept.} {\bf
  110} (1984)  1--162}.

\bibitem{Haber:1984rc}
H.~E. Haber and G.~L. Kane, ``{The Search for Supersymmetry: Probing Physics
  Beyond the Standard Model},''
\href{http://dx.doi.org/10.1016/0370-1573(85)90051-1}{{\em Phys.Rept.} {\bf
  117} (1985)  75--263}.

\bibitem{Ibanez:1981yh}
L.~E. Ibanez and G.~G. Ross, ``{Low-Energy Predictions in Supersymmetric Grand
  Unified Theories},''
\href{http://dx.doi.org/10.1016/0370-2693(81)91200-4}{{\em Phys.Lett.} {\bf
  B105} (1981)  439}.

\bibitem{Dimopoulos:1981yj}
S.~Dimopoulos, S.~Raby, and F.~Wilczek, ``{Supersymmetry and the Scale of
  Unification},''
\href{http://dx.doi.org/10.1103/PhysRevD.24.1681}{{\em Phys.Rev.} {\bf D24}
  (1981)  1681--1683}.

\bibitem{Ellis:1990wk}
J.~R. Ellis, S.~Kelley, and D.~V. Nanopoulos, ``{Probing the Desert Using Gauge
  Coupling Unification},''
\href{http://dx.doi.org/10.1016/0370-2693(91)90980-5}{{\em Phys.Lett.} {\bf
  B260} (1991)  131--137}.

\bibitem{Amaldi:1991cn}
U.~Amaldi, W.~de~Boer, and H.~Furstenau, ``{Comparison of Grand Unified
  Theories with Electroweak and Strong Coupling Constants Measured at LEP},''
\href{http://dx.doi.org/10.1016/0370-2693(91)91641-8}{{\em Phys.Lett.} {\bf
  B260} (1991)  447--455}.

\bibitem{Langacker:1991an}
P.~Langacker and M.-x. Luo, ``{Implications of Precision Electroweak
  Experiments for $M_t$, $\rho_{0}$, $\sin^2\theta_W$ and Grand Unification},''
\href{http://dx.doi.org/10.1103/PhysRevD.44.817}{{\em Phys.Rev.} {\bf D44}
  (1991)  817--822}.

\bibitem{Giunti:1991ta}
C.~Giunti, C.~Kim, and U.~Lee, ``{Running Coupling Constants and Grand
  Unification Models},''
\href{http://dx.doi.org/10.1142/S0217732391001883}{{\em Mod.Phys.Lett.} {\bf
  A6} (1991)  1745--1755}.

\bibitem{Witten:1981nf}
E.~Witten, ``{Dynamical Breaking of Supersymmetry},''
\href{http://dx.doi.org/10.1016/0550-3213(81)90006-7}{{\em Nucl.Phys.} {\bf
  B188} (1981)  513}.

\bibitem{Goldberg:1983nd}
H.~Goldberg, ``{Constraint on the Photino Mass from Cosmology},''
\href{http://dx.doi.org/10.1103/PhysRevLett.50.1419}{{\em Phys.Rev.Lett.} {\bf
  50} (1983)  1419}.

\bibitem{Ellis:1983ew}
J.~R. Ellis, J.~Hagelin, D.~V. Nanopoulos, K.~A. Olive, and M.~Srednicki,
  ``{Supersymmetric Relics from the Big Bang},''
\href{http://dx.doi.org/10.1016/0550-3213(84)90461-9}{{\em Nucl.Phys.} {\bf
  B238} (1984)  453--476}.

\bibitem{atlassusy}
 {\em
  {https://twiki.cern.ch/twiki/bin/view/AtlasPublic/SupersymmetryPublicResults}}
  .

\bibitem{cmssusy}
 {\em {https://twiki.cern.ch/twiki/bin/view/CMSPublic/PhysicsResultsSUS}}  .

\bibitem{Sjostrand:2000wi}
T.~Sjostrand, P.~Eden, C.~Friberg, L.~Lonnblad, G.~Miu, {\em et al.},
  ``{High-Energy Physics Event Generation with PYTHIA 6.1},''
\href{http://dx.doi.org/10.1016/S0010-4655(00)00236-8}{{\em
  Comput.Phys.Commun.} {\bf 135} (2001)  238--259}.

\bibitem{Sjostrand:2006za}
T.~Sjostrand, S.~Mrenna, and P.~Z. Skands, ``{Pythia 6.4 Physics and Manual},''
\href{http://dx.doi.org/10.1088/1126-6708/2006/05/026}{{\em JHEP} {\bf 0605}
  (2006)  026}.

\bibitem{Sjostrand:2007gs}
T.~Sjostrand, S.~Mrenna, and P.~Z. Skands, ``{A Brief Introduction to Pythia
  8.1},''
\href{http://dx.doi.org/10.1016/j.cpc.2008.01.036}{{\em Comput.Phys.Commun.}
  {\bf 178} (2008)  852--867}.

\bibitem{Gleisberg:2003xi}
T.~Gleisberg, S.~Hoeche, F.~Krauss, A.~Schalicke, S.~Schumann, {\em et al.},
  ``{SHERPA 1. Alpha: A Proof of Concept Version},''
\href{http://dx.doi.org/10.1088/1126-6708/2004/02/056}{{\em JHEP} {\bf 0402}
  (2004)  056}.

\bibitem{Gleisberg:2008ta}
T.~Gleisberg, S.~Hoeche, F.~Krauss, M.~Schonherr, S.~Schumann, {\em et al.},
  ``{Event Generation with SHERPA 1.1},''
\href{http://dx.doi.org/10.1088/1126-6708/2009/02/007}{{\em JHEP} {\bf 0902}
  (2009)  007}.

\bibitem{Corcella:2000bw}
G.~Corcella, I.~Knowles, G.~Marchesini, S.~Moretti, K.~Odagiri, {\em et al.},
  ``{Herwig 6: An Event generator for hadron emission reactions with
  interfering gluons (including supersymmetric processes)},''
{\em JHEP} {\bf 0101} (2001)  010.

\bibitem{Corcella:2002jc}
G.~Corcella, I.~Knowles, G.~Marchesini, S.~Moretti, K.~Odagiri, {\em et al.},
``{HERWIG 6.5 Release Note},''.

\bibitem{Bahr:2008pv}
M.~Bahr, S.~Gieseke, M.~Gigg, D.~Grellscheid, K.~Hamilton, {\em et al.},
  ``{Herwig++ Physics and Manual},''
\href{http://dx.doi.org/10.1140/epjc/s10052-008-0798-9}{{\em Eur.Phys.J.} {\bf
  C58} (2008)  639--707}.

\bibitem{Arnold:2012fq}
K.~Arnold, L.~d'Errico, S.~Gieseke, D.~Grellscheid, K.~Hamilton, {\em et al.},
``{Herwig++ 2.6 Release Note},''.

\bibitem{Mangano:2002ea}
M.~L. Mangano, M.~Moretti, F.~Piccinini, R.~Pittau, and A.~D. Polosa,
  ``{ALPGEN, a Generator for Hard Multiparton Processes in Hadronic
  Collisions},''
{\em JHEP} {\bf 0307} (2003)  001.

\bibitem{Pukhov:1999gg}
A.~Pukhov, E.~Boos, M.~Dubinin, V.~Edneral, V.~Ilyin, {\em et al.}, ``{CompHEP:
  A Package for Evaluation of Feynman Diagrams and Integration over
  Multiparticle Phase Space},''
\href{http://arxiv.org/abs/hep-ph/9908288}{{\tt arXiv:hep-ph/9908288
  [hep-ph]}}.

\bibitem{Boos:2004kh}
E.~Boos {\em et al.}, ``{CompHEP 4.4: Automatic Computations from Lagrangians
  to Events},''
\href{http://dx.doi.org/10.1016/j.nima.2004.07.096}{{\em Nucl.Instrum.Meth.}
  {\bf A534} (2004)  250--259}.

\bibitem{Pukhov:2004ca}
A.~Pukhov, ``{CalcHEP 2.3: MSSM, Structure Functions, Event Generation, Batchs,
  and Generation of Matrix Elements for Other Packages},''
\href{http://arxiv.org/abs/hep-ph/0412191}{{\tt arXiv:hep-ph/0412191
  [hep-ph]}}.

\bibitem{Belyaev:2012qa}
A.~Belyaev, N.~D. Christensen, and A.~Pukhov, ``{CalcHEP 3.4 for Collider
  Physics Within and Beyond the Standard Model},''
\href{http://arxiv.org/abs/1207.6082}{{\tt arXiv:1207.6082 [hep-ph]}}.

\bibitem{Kanaki:2000ey}
A.~Kanaki and C.~G. Papadopoulos, ``{HELAC: A Package to Compute Electroweak
  Helicity Amplitudes},''
\href{http://dx.doi.org/10.1016/S0010-4655(00)00151-X}{{\em
  Comput.Phys.Commun.} {\bf 132} (2000)  306--315}.

\bibitem{Cafarella:2007pc}
A.~Cafarella, C.~G. Papadopoulos, and M.~Worek, ``{Helac-Phegas: A Generator
  for all Parton Level Processes},''
\href{http://dx.doi.org/10.1016/j.cpc.2009.04.023}{{\em Comput.Phys.Commun.}
  {\bf 180} (2009)  1941--1955}.

\bibitem{Stelzer:1994ta}
T.~Stelzer and W.~Long, ``{Automatic Generation of Tree Level Helicity
  Amplitudes},''
\href{http://dx.doi.org/10.1016/0010-4655(94)90084-1}{{\em Comput.Phys.Commun.}
  {\bf 81} (1994)  357--371}.

\bibitem{Maltoni:2002qb}
F.~Maltoni and T.~Stelzer, ``{MadEvent: Automatic Event Generation with
  MadGraph},''
{\em JHEP} {\bf 0302} (2003)  027.

\bibitem{Alwall:2007st}
J.~Alwall, P.~Demin, S.~de~Visscher, R.~Frederix, M.~Herquet, {\em et al.},
  ``{MadGraph/MadEvent v4: The New Web Generation},''
\href{http://dx.doi.org/10.1088/1126-6708/2007/09/028}{{\em JHEP} {\bf 0709}
  (2007)  028}.

\bibitem{Alwall:2008pm}
J.~Alwall, P.~Artoisenet, S.~de~Visscher, C.~Duhr, R.~Frederix, {\em et al.},
  ``{New Developments in MadGraph/MadEvent},''
\href{http://dx.doi.org/10.1063/1.3052056}{{\em AIP Conf.Proc.} {\bf 1078}
  (2009)  84--89}.

\bibitem{Alwall:2011uj}
J.~Alwall, M.~Herquet, F.~Maltoni, O.~Mattelaer, and T.~Stelzer, ``{MadGraph 5
  : Going Beyond},''
\href{http://dx.doi.org/10.1007/JHEP06(2011)128}{{\em JHEP} {\bf 1106} (2011)
  128}.

\bibitem{Moretti:2001zz}
M.~Moretti, T.~Ohl, and J.~Reuter, ``{O'Mega: An Optimizing Matrix Element
  Generator},''
\href{http://arxiv.org/abs/hep-ph/0102195}{{\tt arXiv:hep-ph/0102195
  [hep-ph]}}.

\bibitem{Kilian:2007gr}
W.~Kilian, T.~Ohl, and J.~Reuter, ``{WHIZARD: Simulating Multi-Particle
  Processes at LHC and ILC},''
\href{http://dx.doi.org/10.1140/epjc/s10052-011-1742-y}{{\em Eur.Phys.J.} {\bf
  C71} (2011)  1742}.

\bibitem{Degrande:2011ua}
C.~Degrande, C.~Duhr, B.~Fuks, D.~Grellscheid, O.~Mattelaer, {\em et al.},
  ``{UFO - The Universal FeynRules Output},''
\href{http://dx.doi.org/10.1016/j.cpc.2012.01.022}{{\em Comput.Phys.Commun.}
  {\bf 183} (2012)  1201--1214}.

\bibitem{Semenov:1996es}
A.~Semenov, ``{LanHEP: A Package for Automatic Generation of Feynman Rules in
  Gauge Models},''
\href{http://arxiv.org/abs/hep-ph/9608488}{{\tt arXiv:hep-ph/9608488
  [hep-ph]}}.

\bibitem{Semenov:1998eb}
A.~Semenov, ``{LanHEP: A Package for Automatic gGneration of Feynman Rules from
  the Lagrangian},''
\href{http://dx.doi.org/10.1016/S0010-4655(98)00143-X}{{\em
  Comput.Phys.Commun.} {\bf 115} (1998)  124--139}.

\bibitem{Semenov:2002jw}
A.~Semenov, ``{LanHEP: A Package for Automatic Generation of Feynman Rules in
  Field Theory. Version 2.0},''
\href{http://arxiv.org/abs/hep-ph/0208011}{{\tt arXiv:hep-ph/0208011
  [hep-ph]}}.

\bibitem{Semenov:2008jy}
A.~Semenov, ``{LanHEP: A Package for the Automatic Generation of Feynman Rules
  in Field Theory. Version 3.0},''
\href{http://dx.doi.org/10.1016/j.cpc.2008.10.012}{{\em Comput.Phys.Commun.}
  {\bf 180} (2009)  431--454}.

\bibitem{Semenov:2010qt}
A.~Semenov, ``{LanHEP - a Package for Automatic Generation of Feynman Rules
  from the Lagrangian. Updated Version 3.1},''
\href{http://arxiv.org/abs/1005.1909}{{\tt arXiv:1005.1909 [hep-ph]}}.

\bibitem{Christensen:2008py}
N.~D. Christensen and C.~Duhr, ``{FeynRules - Feynman Rules Made Easy},''
\href{http://dx.doi.org/10.1016/j.cpc.2009.02.018}{{\em Comput.Phys.Commun.}
  {\bf 180} (2009)  1614--1641}.

\bibitem{Christensen:2009jx}
N.~D. Christensen, P.~de~Aquino, C.~Degrande, C.~Duhr, B.~Fuks, {\em et al.},
  ``{A Comprehensive Approach to New Physics Simulations},''
\href{http://dx.doi.org/10.1140/epjc/s10052-011-1541-5}{{\em Eur.Phys.J.} {\bf
  C71} (2011)  1541}.

\bibitem{Christensen:2010wz}
N.~D. Christensen, C.~Duhr, B.~Fuks, J.~Reuter, and C.~Speckner, ``{Introducing
  an Interface Between Whizard and FeynRules},''
\href{http://dx.doi.org/10.1140/epjc/s10052-012-1990-5}{{\em Eur.Phys.J.} {\bf
  C72} (2012)  1990}.

\bibitem{Duhr:2011se}
C.~Duhr and B.~Fuks, ``{A Superspace Module for the FeynRules Package},''
\href{http://dx.doi.org/10.1016/j.cpc.2011.06.009}{{\em Comput.Phys.Commun.}
  {\bf 182} (2011)  2404--2426}.

\bibitem{Fuks:2012im}
B.~Fuks, ``{Beyond the Minimal Supersymmetric Standard Model: from Theory to
  Phenomenology},''
\href{http://dx.doi.org/10.1142/S0217751X12300074}{{\em Int.J.Mod.Phys.} {\bf
  A27} (2012)  1230007}.

\bibitem{Alloul:2013fw}
A.~Alloul, J.~D'Hondt, K.~De~Causmaecker, B.~Fuks, and M.~Rausch~de
  Traubenberg, ``{Automated Mass Spectrum Generation for New Physics},''
\href{http://dx.doi.org/10.1140/epjc/s10052-013-2325-x}{{\em Eur.Phys.J.} {\bf
  C73} (2013)  2325}.

\bibitem{Christensen:2013aua}
N.~D. Christensen, P.~de~Aquino, N.~Deutschmann, C.~Duhr, B.~Fuks, {\em et
  al.}, ``{Simulating spin-$ \frac{3}{2}$ particles at colliders},''
\href{http://dx.doi.org/10.1140/epjc/s10052-013-2580-x}{{\em Eur.Phys.J.} {\bf
  C73} (2013)  2580}.

\bibitem{Alloul:2013bka}
A.~Alloul, N.~D. Christensen, C.~Degrande, C.~Duhr, and B.~Fuks, ``{FeynRules
  2.0 - A complete toolbox for tree-level phenomenology},''
\href{http://arxiv.org/abs/1310.1921}{{\tt arXiv:1310.1921 [hep-ph]}}.

\bibitem{Staub:2008uz}
F.~Staub, ``{SARAH},''
\href{http://arxiv.org/abs/0806.0538}{{\tt arXiv:0806.0538 [hep-ph]}}.

\bibitem{Staub:2009bi}
F.~Staub, ``{From Superpotential to Model Files for FeynArts and
  CalcHep/CompHep},''
\href{http://dx.doi.org/10.1016/j.cpc.2010.01.011}{{\em Comput.Phys.Commun.}
  {\bf 181} (2010)  1077--1086}.

\bibitem{Staub:2010jh}
F.~Staub, ``{Automatic Calculation of Supersymmetric Renormalization Group
  Equations and Self Energies},''
\href{http://dx.doi.org/10.1016/j.cpc.2010.11.030}{{\em Comput.Phys.Commun.}
  {\bf 182} (2011)  808--833}.

\bibitem{Staub:2012pb}
F.~Staub, ``{Linking SARAH and MadGraph Using the UFO Format},''
\href{http://arxiv.org/abs/1207.0906}{{\tt arXiv:1207.0906 [hep-ph]}}.

\bibitem{Conte:2012fm}
E.~Conte, B.~Fuks, and G.~Serret, ``{MadAnalysis 5, A User-Friendly Framework
  for Collider Phenomenology},''
{\em Comput.Phys.Commun.} {\bf 184} (2013)  222--256.

\bibitem{Mangano:2006rw}
M.~L. Mangano, M.~Moretti, F.~Piccinini, and M.~Treccani, ``{Matching Matrix
  Elements and Shower Evolution for Top-Quark Production in Hadronic
  Collisions},''
\href{http://dx.doi.org/10.1088/1126-6708/2007/01/013}{{\em JHEP} {\bf 0701}
  (2007)  013}.

\bibitem{Alwall:2008qv}
J.~Alwall, S.~de~Visscher, and F.~Maltoni, ``{QCD Radiation in the Production
  of Heavy Colored Particles at the LHC},''
\href{http://dx.doi.org/10.1088/1126-6708/2009/02/017}{{\em JHEP} {\bf 0902}
  (2009)  017}.

\bibitem{Ovyn:2009tx}
S.~Ovyn, X.~Rouby, and V.~Lemaitre, ``{Delphes, a Framework for Fast Simulation
  of a Generic Collider Experiment},''
\href{http://arxiv.org/abs/0903.2225}{{\tt arXiv:0903.2225 [hep-ph]}}.

\bibitem{Pauli:1940zz}
W.~Pauli, ``{The Connection Between Spin and Statistics},''
\href{http://dx.doi.org/10.1103/PhysRev.58.716}{{\em Phys.Rev.} {\bf 58} (1940)
   716--722}.

\bibitem{Noether:1918zz}
E.~Noether, ``{Invariant Variation Problems},''
\href{http://dx.doi.org/10.1080/00411457108231446}{{\em Gott.Nachr.} {\bf 1918}
  (1918)  235--257}.

\bibitem{Coleman:1967ad}
S.~R. Coleman and J.~Mandula, ``{All Possible Symmetries of the S Matrix},''
\href{http://dx.doi.org/10.1103/PhysRev.159.1251}{{\em Phys.Rev.} {\bf 159}
  (1967)  1251--1256}.

\bibitem{Haag:1974qh}
R.~Haag, J.~T. Lopuszanski, and M.~Sohnius, ``{All Possible Generators of
  Supersymmetries of the S Matrix},''
\href{http://dx.doi.org/10.1016/0550-3213(75)90279-5}{{\em Nucl.Phys.} {\bf
  B88} (1975)  257}.

\bibitem{livre}
B.~Fuks and M.~Rausch~de Traubenberg,
  \href{http://dx.doi.org/http://editions-ellipses.fr/supersymetrie-exercices-avec-solutions-p-7697.html}{{\em
  Supersym\'etrie~: exercices avec solutions}}.
\newblock Ellipses Editions, 2011.

\bibitem{Fayet:1974jb}
P.~Fayet and J.~Iliopoulos, ``{Spontaneously Broken Supergauge Symmetries and
  Goldstone Spinors},''
\href{http://dx.doi.org/10.1016/0370-2693(74)90310-4}{{\em Phys.Lett.} {\bf
  B51} (1974)  461--464}.

\bibitem{Wess:1978ns}
J.~Wess and B.~Zumino, ``{The Component Formalism Follows from the Superspace
  Formulation of Supergravity},''
\href{http://dx.doi.org/10.1016/0370-2693(78)90390-8}{{\em Phys.Lett.} {\bf
  B79} (1978)  394}.

\bibitem{Iliopoulos:1974zv}
J.~Iliopoulos and B.~Zumino, ``{Broken Supergauge Symmetry and
  Renormalization},''
\href{http://dx.doi.org/10.1016/0550-3213(74)90388-5}{{\em Nucl.Phys.} {\bf
  B76} (1974)  310}.

\bibitem{Fayet:1977vd}
P.~Fayet, ``{Mixing Between Gravitational and Weak Interactions Through the
  Massive Gravitino},''
\href{http://dx.doi.org/10.1016/0370-2693(77)90414-2}{{\em Phys.Lett.} {\bf
  B70} (1977)  461}.

\bibitem{Fayet:1979yb}
P.~Fayet, ``{Scattering Cross-Sections of the Photino and the Goldstino
  (Gravitino) on Matter},''
\href{http://dx.doi.org/10.1016/0370-2693(79)90836-0}{{\em Phys.Lett.} {\bf
  B86} (1979)  272}.

\bibitem{Ferrara:1979wa}
S.~Ferrara, L.~Girardello, and F.~Palumbo, ``{A General Mass Formula in Broken
  Supersymmetry},''
\href{http://dx.doi.org/10.1103/PhysRevD.20.403}{{\em Phys.Rev.} {\bf D20}
  (1979)  403}.

\bibitem{Witten:1982hu}
E.~Witten and J.~Bagger, ``{Quantization of Newton's Constant in Certain
  Supergravity Theories},''
\href{http://dx.doi.org/10.1016/0370-2693(82)90644-X}{{\em Phys.Lett.} {\bf
  B115} (1982)  202}.

\bibitem{Deser:1976eh}
S.~Deser and B.~Zumino, ``{Consistent Supergravity},''
\href{http://dx.doi.org/10.1016/0370-2693(76)90089-7}{{\em Phys.Lett.} {\bf
  B62} (1976)  335}.

\bibitem{Freedman:1976xh}
D.~Z. Freedman, P.~van Nieuwenhuizen, and S.~Ferrara, ``{Progress Toward a
  Theory of Supergravity},''
\href{http://dx.doi.org/10.1103/PhysRevD.13.3214}{{\em Phys.Rev.} {\bf D13}
  (1976)  3214--3218}.

\bibitem{Freedman:1976py}
D.~Z. Freedman and P.~van Nieuwenhuizen, ``{Properties of Supergravity
  Theory},''
\href{http://dx.doi.org/10.1103/PhysRevD.14.912}{{\em Phys.Rev.} {\bf D14}
  (1976)  912}.

\bibitem{Ferrara:1976um}
S.~Ferrara, J.~Scherk, and P.~van Nieuwenhuizen, ``{Locally Supersymmetric
  Maxwell-Einstein Theory},''
\href{http://dx.doi.org/10.1103/PhysRevLett.37.1035}{{\em Phys.Rev.Lett.} {\bf
  37} (1976)  1035}.

\bibitem{Cremmer:1978hn}
E.~Cremmer, B.~Julia, J.~Scherk, S.~Ferrara, L.~Girardello, {\em et al.},
  ``{Spontaneous Symmetry Breaking and Higgs Effect in Supergravity Without
  Cosmological Constant},''
\href{http://dx.doi.org/10.1016/0550-3213(79)90417-6}{{\em Nucl.Phys.} {\bf
  B147} (1979)  105}.

\bibitem{Cremmer:1978iv}
E.~Cremmer, B.~Julia, J.~Scherk, P.~van Nieuwenhuizen, S.~Ferrara, {\em et
  al.}, ``{SuperHiggs Effect in Supergravity with General Scalar
  Interactions},''
\href{http://dx.doi.org/10.1016/0370-2693(78)90230-7}{{\em Phys.Lett.} {\bf
  B79} (1978)  231}.

\bibitem{Cremmer:1982en}
E.~Cremmer, S.~Ferrara, L.~Girardello, and A.~Van~Proeyen, ``{Yang-Mills
  Theories with Local Supersymmetry: Lagrangian, Transformation Laws and
  SuperHiggs Effect},''
\href{http://dx.doi.org/10.1016/0550-3213(83)90679-X}{{\em Nucl.Phys.} {\bf
  B212} (1983)  413}.

\bibitem{Chamseddine:1982jx}
A.~H. Chamseddine, R.~L. Arnowitt, and P.~Nath, ``{Locally Supersymmetric Grand
  Unification},''
\href{http://dx.doi.org/10.1103/PhysRevLett.49.970}{{\em Phys.Rev.Lett.} {\bf
  49} (1982)  970}.

\bibitem{Barbieri:1982eh}
R.~Barbieri, S.~Ferrara, and C.~A. Savoy, ``{Gauge Models with Spontaneously
  Broken Local Supersymmetry},''
\href{http://dx.doi.org/10.1016/0370-2693(82)90685-2}{{\em Phys.Lett.} {\bf
  B119} (1982)  343}.

\bibitem{Ibanez:1982ee}
L.~E. Ibanez, ``{Locally Supersymmetric SU(5) Grand Unification},''
\href{http://dx.doi.org/10.1016/0370-2693(82)90604-9}{{\em Phys.Lett.} {\bf
  B118} (1982)  73}.

\bibitem{Ohta:1982wn}
N.~Ohta, ``{Grand Unified Theories Based on Local Supersymmetry},''
\href{http://dx.doi.org/10.1143/PTP.70.542}{{\em Prog.Theor.Phys.} {\bf 70}
  (1983)  542}.

\bibitem{Ellis:1982wr}
J.~R. Ellis, D.~V. Nanopoulos, and K.~Tamvakis, ``{Grand Unification in Simple
  Supergravity},''
\href{http://dx.doi.org/10.1016/0370-2693(83)90900-0}{{\em Phys.Lett.} {\bf
  B121} (1983)  123}.

\bibitem{AlvarezGaume:1983gj}
L.~Alvarez-Gaume, J.~Polchinski, and M.~B. Wise, ``{Minimal Low-Energy
  Supergravity},''
\href{http://dx.doi.org/10.1016/0550-3213(83)90591-6}{{\em Nucl.Phys.} {\bf
  B221} (1983)  495}.

\bibitem{Binetruy:2000zx}
P.~Binetruy, G.~Girardi, and R.~Grimm, ``{Supergravity couplings: A Geometric
  formulation},''
\href{http://dx.doi.org/10.1016/S0370-1573(00)00085-5}{{\em Phys.Rept.} {\bf
  343} (2001)  255--462}.

\bibitem{Dimopoulos:1981au}
S.~Dimopoulos and S.~Raby, ``{Supercolor},''
\href{http://dx.doi.org/10.1016/0550-3213(81)90430-2}{{\em Nucl.Phys.} {\bf
  B192} (1981)  353}.

\bibitem{Dine:1981za}
M.~Dine, W.~Fischler, and M.~Srednicki, ``{Supersymmetric Technicolor},''
\href{http://dx.doi.org/10.1016/0550-3213(81)90582-4}{{\em Nucl.Phys.} {\bf
  B189} (1981)  575--593}.

\bibitem{Derendinger:1982tq}
J.~Derendinger and C.~A. Savoy, ``{Gaugino Masses and a New Mechanism for
  Proton Decay in Supersymmetric Theories},''
\href{http://dx.doi.org/10.1016/0370-2693(82)90201-5}{{\em Phys.Lett.} {\bf
  B118} (1982)  347}.

\bibitem{Fayet:1978qc}
P.~Fayet, ``{Massive Gluinos},''
\href{http://dx.doi.org/10.1016/0370-2693(78)90474-4}{{\em Phys.Lett.} {\bf
  B78} (1978)  417}.

\bibitem{Dine:1981gu}
M.~Dine and W.~Fischler, ``{A Phenomenological Model of Particle Physics Based
  on Supersymmetry},''
\href{http://dx.doi.org/10.1016/0370-2693(82)91241-2}{{\em Phys.Lett.} {\bf
  B110} (1982)  227}.

\bibitem{Nappi:1982hm}
C.~R. Nappi and B.~A. Ovrut, ``{Supersymmetric Extension of the SU(3) x SU(2) x
  U(1) Model},''
\href{http://dx.doi.org/10.1016/0370-2693(82)90418-X}{{\em Phys.Lett.} {\bf
  B113} (1982)  175}.

\bibitem{AlvarezGaume:1981wy}
L.~Alvarez-Gaume, M.~Claudson, and M.~B. Wise, ``{Low-Energy Supersymmetry},''
\href{http://dx.doi.org/10.1016/0550-3213(82)90138-9}{{\em Nucl.Phys.} {\bf
  B207} (1982)  96}.

\bibitem{Dine:1993yw}
M.~Dine and A.~E. Nelson, ``{Dynamical Supersymmetry Breaking at
  Low-Energies},''
\href{http://dx.doi.org/10.1103/PhysRevD.48.1277}{{\em Phys.Rev.} {\bf D48}
  (1993)  1277--1287}.

\bibitem{Dine:1994vc}
M.~Dine, A.~E. Nelson, and Y.~Shirman, ``{Low-Energy Dynamical Supersymmetry
  Breaking Simplified},''
\href{http://dx.doi.org/10.1103/PhysRevD.51.1362}{{\em Phys.Rev.} {\bf D51}
  (1995)  1362--1370}.

\bibitem{Dine:1995ag}
M.~Dine, A.~E. Nelson, Y.~Nir, and Y.~Shirman, ``{New Tools for Low-Energy
  Dynamical Supersymmetry Breaking},''
\href{http://dx.doi.org/10.1103/PhysRevD.53.2658}{{\em Phys.Rev.} {\bf D53}
  (1996)  2658--2669}.

\bibitem{Giudice:1998bp}
G.~Giudice and R.~Rattazzi, ``{Theories with Gauge Mediated Supersymmetry
  Breaking},''
\href{http://dx.doi.org/10.1016/S0370-1573(99)00042-3}{{\em Phys.Rept.} {\bf
  322} (1999)  419--499}.

\bibitem{Randall:1998uk}
L.~Randall and R.~Sundrum, ``{Out of this World Supersymmetry Breaking},''
\href{http://dx.doi.org/10.1016/S0550-3213(99)00359-4}{{\em Nucl.Phys.} {\bf
  B557} (1999)  79--118}.

\bibitem{ArkaniHamed:1998kj}
N.~Arkani-Hamed, G.~F. Giudice, M.~A. Luty, and R.~Rattazzi, ``{Supersymmetry
  Breaking Loops from Analytic Continuation into Superspace},''
\href{http://dx.doi.org/10.1103/PhysRevD.58.115005}{{\em Phys.Rev.} {\bf D58}
  (1998)  115005}.

\bibitem{Bagger:1999rd}
J.~A. Bagger, T.~Moroi, and E.~Poppitz, ``{Anomaly Mediation in Supergravity
  Theories},''
{\em JHEP} {\bf 0004} (2000)  009.

\bibitem{Derendinger:1991kr}
J.-P. Derendinger, S.~Ferrara, C.~Kounnas, and F.~Zwirner, ``{All Loop Gauge
  Couplings from Anomaly Cancellation in String Effective Theories},''
\href{http://dx.doi.org/10.1016/0370-2693(91)90092-5}{{\em Phys.Lett.} {\bf
  B271} (1991)  307--313}.

\bibitem{Derendinger:1991hq}
J.~Derendinger, S.~Ferrara, C.~Kounnas, and F.~Zwirner, ``{On Loop Corrections
  to String Effective Field Theories: Field Dependent Gauge Couplings and Sigma
  Model Anomalies},''
\href{http://dx.doi.org/10.1016/0550-3213(92)90315-3}{{\em Nucl.Phys.} {\bf
  B372} (1992)  145--188}.

\bibitem{LopesCardoso:1991zt}
G.~Lopes~Cardoso and B.~A. Ovrut, ``{A Green-Schwarz Mechanism for D = 4, N=1
  Supergravity Anomalies},''
\href{http://dx.doi.org/10.1016/0550-3213(92)90390-W}{{\em Nucl.Phys.} {\bf
  B369} (1992)  351--372}.

\bibitem{LopesCardoso:1992yd}
G.~Lopes~Cardoso and B.~A. Ovrut, ``{Coordinate and Kahler Sigma Model
  Anomalies and their Cancellation in String Effective Field Theories},''
\href{http://dx.doi.org/10.1016/0550-3213(93)90675-F}{{\em Nucl.Phys.} {\bf
  B392} (1993)  315--344}.

\bibitem{Kaplunovsky:1994fg}
V.~Kaplunovsky and J.~Louis, ``{Field Dependent Gauge Couplings in Locally
  Supersymmetric Effective Quantum Field Theories},''
\href{http://dx.doi.org/10.1016/0550-3213(94)00150-2}{{\em Nucl.Phys.} {\bf
  B422} (1994)  57--124}.

\bibitem{sugra}
B.~Fuks and M.~Rausch~de Traubenberg, ``{A Supergravity Primer},'' {\em in
  preparation\!}  .

\bibitem{Bozzi:2007me}
G.~Bozzi, B.~Fuks, B.~Herrmann, and M.~Klasen, ``{Squark and Gaugino
  Hadroproduction and Decays in Non-Minimal Flavour Violating Supersymmetry},''
\href{http://dx.doi.org/10.1016/j.nuclphysb.2007.05.031}{{\em Nucl.Phys.} {\bf
  B787} (2007)  1--54}.

\bibitem{Fuks:2008ab}
B.~Fuks, B.~Herrmann, and M.~Klasen, ``{Flavour Violation in Gauge-Mediated
  Supersymmetry Breaking Models: Experimental Constraints and Phenomenology at
  the LHC},''
\href{http://dx.doi.org/10.1016/j.nuclphysb.2008.11.020}{{\em Nucl.Phys.} {\bf
  B810} (2009)  266--299}.

\bibitem{Fuks:2011dg}
B.~Fuks, B.~Herrmann, and M.~Klasen, ``{Phenomenology of Anomaly-Mediated
  Supersymmetry Breaking Scenarios with Non-Minimal Flavour Violation},''
\href{http://dx.doi.org/10.1103/PhysRevD.86.015002}{{\em Phys.Rev.} {\bf D86}
  (2012)  015002}.

\bibitem{Barbier:2004ez}
R.~Barbier, C.~Berat, M.~Besancon, M.~Chemtob, A.~Deandrea, {\em et al.},
  ``{R-parity Violating Supersymmetry},''
\href{http://dx.doi.org/10.1016/j.physrep.2005.08.006}{{\em Phys.Rept.} {\bf
  420} (2005)  1--202}.

\bibitem{Fayet:1974pd}
P.~Fayet, ``{Supergauge Invariant Extension of the Higgs Mechanism and a Model
  for the Electron and Its Neutrino},''
\href{http://dx.doi.org/10.1016/0550-3213(75)90636-7}{{\em Nucl.Phys.} {\bf
  B90} (1975)  104--124}.

\bibitem{Salam:1974xa}
A.~Salam and J.~Strathdee, ``{Supersymmetry and Fermion Number Conservation},''
\href{http://dx.doi.org/10.1016/0550-3213(75)90253-9}{{\em Nucl.Phys.} {\bf
  B87} (1975)  85}.

\bibitem{Kribs:2007ac}
G.~D. Kribs, E.~Poppitz, and N.~Weiner, ``{Flavor in Supersymmetry with an
  Extended R-Symmetry},''
\href{http://dx.doi.org/10.1103/PhysRevD.78.055010}{{\em Phys.Rev.} {\bf D78}
  (2008)  055010}.

\bibitem{Andrea:2011ws}
J.~Andrea, B.~Fuks, and F.~Maltoni, ``{Monotops at the LHC},''
\href{http://dx.doi.org/10.1103/PhysRevD.84.074025}{{\em Phys.Rev.} {\bf D84}
  (2011)  074025}.

\bibitem{Calvet:2012rk}
S.~Calvet, B.~Fuks, P.~Gris, and L.~Valery, ``{Searching for Sgluons in
  Multitop Events at a Center-of-Mass Energy of 8 TeV},''
\href{http://dx.doi.org/10.1007/JHEP04(2013)043}{{\em JHEP} {\bf 1304} (2012)
  043}.

\bibitem{Agram:2013wda}
J.-L. Agram, J.~Andrea, M.~Buttignol, E.~Conte, and B.~Fuks, ``{Monotop
  phenomenology at the Large Hadron Collider},''
\href{http://arxiv.org/abs/1311.6478}{{\tt arXiv:1311.6478 [hep-ph]}}.

\bibitem{Aaltonen:2012ek}
{\bf CDF Collaboration}, T.~Aaltonen {\em et al.}, ``{Search for a dark matter
  candidate produced in association with a single top quark in $p\bar{p}$
  collisions at $\sqrt{s} = 1.96$ TeV},''
\href{http://dx.doi.org/10.1103/PhysRevLett.108.201802}{{\em Phys.Rev.Lett.}
  {\bf 108} (2012)  201802}.

\bibitem{ATLAS-CONF-2013-051}
{\bf ATLAS collaboration}, ``Search for anomalous production of events with
  same-sign dileptons and $b$ jets in 14.3 fb$^{-1}$ of $pp$ collisions at
  $\sqrt{s} = 8$ tev with the atlas detector,''
  \href{http://arxiv.org/abs/ATLAS-CONF-2013-051}{{\tt ATLAS-CONF-2013-051}}.

\bibitem{monoATLAS}
{\bf ATLAS Collaboration,} {\em {Private Communication}}  .

\bibitem{monoCMS}
{\bf CMS Collaboration,} {\em {Private Communication}}  .

\bibitem{Salam:1976ib}
A.~Salam and J.~Strathdee, ``{Supersymmetry and Superfields},''
\href{http://dx.doi.org/10.1002/prop.19780260202}{{\em Fortsch.Phys.} {\bf 26}
  (1978)  57}.

\bibitem{Nahm:1977tg}
W.~Nahm, ``{Supersymmetries and their Representations},''
\href{http://dx.doi.org/10.1016/0550-3213(78)90218-3}{{\em Nucl.Phys.} {\bf
  B135} (1978)  149}.

\bibitem{Ferrara:1980ra}
S.~Ferrara, C.~A. Savoy, and B.~Zumino, ``{General Massive Multiplets in
  Extended Supersymmetry},''
\href{http://dx.doi.org/10.1016/0370-2693(81)90144-1}{{\em Phys.Lett.} {\bf
  B100} (1981)  393}.

\bibitem{weinbergqft}
S.~Weinberg, {\em The Quantum Theory of Fields, Volume 1: Foundations}.
\newblock Cambridge University Press, 2005.

\bibitem{Fayet:1976cr}
P.~Fayet and S.~Ferrara, ``{Supersymmetry},''
\href{http://dx.doi.org/10.1016/0370-1573(77)90066-7}{{\em Phys.Rept.} {\bf 32}
  (1977)  249--334}.

\bibitem{Zumino:1979et}
B.~Zumino, ``{Supersymmetry and Kahler Manifolds},''
\href{http://dx.doi.org/10.1016/0370-2693(79)90964-X}{{\em Phys.Lett.} {\bf
  B87} (1979)  203}.

\bibitem{AlvarezGaume:1981hm}
L.~Alvarez-Gaume and D.~Z. Freedman, ``{Geometrical Structure and Ultraviolet
  Finiteness in the Supersymmetric Sigma Model},''
\href{http://dx.doi.org/10.1007/BF01208280}{{\em Commun.Math.Phys.} {\bf 80}
  (1981)  443}.

\bibitem{Bagger:1983tt}
J.~Bagger and E.~Witten, ``{Matter Couplings in N=2 Supergravity},''
\href{http://dx.doi.org/10.1016/0550-3213(83)90605-3}{{\em Nucl.Phys.} {\bf
  B222} (1983)  1}.

\bibitem{Bagger:1982ab}
J.~A. Bagger, ``{Coupling the Gauge Invariant Supersymmetric Nonlinear Sigma
  Model to Supergravity},''
\href{http://dx.doi.org/10.1016/0550-3213(83)90411-X}{{\em Nucl.Phys.} {\bf
  B211} (1983)  302}.

\bibitem{Bordemann:1985xy}
M.~Bordemann, M.~Forger, and H.~Romer, ``{Homogeneous Kahler Manifolds: Paving
  the Way towards New Supersymmetric Sigma Models},''
\href{http://dx.doi.org/10.1007/BF01221650}{{\em Commun.Math.Phys.} {\bf 102}
  (1986)  605}.

\bibitem{Itoh:1985ha}
K.~Itoh, T.~Kugo, and H.~Kunitomo, ``{Supersymmetric Nonlinear Realization for
  Arbitrary Kahlerian Coset Space G/H},''
\href{http://dx.doi.org/10.1016/0550-3213(86)90118-5}{{\em Nucl.Phys.} {\bf
  B263} (1986)  295}.

\bibitem{Salam:1974ig}
A.~Salam and J.~Strathdee, ``{Supersymmetry and Nonabelian Gauges},''
\href{http://dx.doi.org/10.1016/0370-2693(74)90226-3}{{\em Phys.Lett.} {\bf
  B51} (1974)  353--355}.

\bibitem{Bagger:1982fn}
J.~Bagger and E.~Witten, ``{The Gauge Invariant Supersymmetric Nonlinear Sigma
  Model},''
\href{http://dx.doi.org/10.1016/0370-2693(82)90609-8}{{\em Phys.Lett.} {\bf
  B118} (1982)  103--106}.

\bibitem{Hull:1985pq}
C.~Hull, A.~Karlhede, U.~Lindstrom, and M.~Rocek, ``{Nonlinear Sigma Models and
  their Gauging in and out of Superspace},''
\href{http://dx.doi.org/10.1016/0550-3213(86)90175-6}{{\em Nucl.Phys.} {\bf
  B266} (1986)  1}.

\bibitem{wb}
J.~Wess and J.~Bagger, {\em Supersymmetry and Supergravity}.
\newblock Princeton University Press, second~ed., 1992.

\bibitem{Hahn:1998yk}
T.~Hahn and M.~Perez-Victoria, ``{Automatized One Loop Calculations in
  Four-Dimensions and D-Dimensions},''
\href{http://dx.doi.org/10.1016/S0010-4655(98)00173-8}{{\em
  Comput.Phys.Commun.} {\bf 118} (1999)  153--165}.

\bibitem{Hahn:2000kx}
T.~Hahn, ``{Generating Feynman Diagrams and Amplitudes with FeynArts 3},''
\href{http://dx.doi.org/10.1016/S0010-4655(01)00290-9}{{\em
  Comput.Phys.Commun.} {\bf 140} (2001)  418--431}.

\bibitem{Hahn:2009bf}
T.~Hahn, ``{FormCalc 6},''
{\em PoS} {\bf ACAT08} (2008)  121.

\bibitem{Agrawal:2011tm}
S.~Agrawal, T.~Hahn, and E.~Mirabella, ``{FormCalc 7},''
\href{http://arxiv.org/abs/1112.0124}{{\tt arXiv:1112.0124 [hep-ph]}}.

\bibitem{Cullen:2011ac}
G.~Cullen, N.~Greiner, G.~Heinrich, G.~Luisoni, P.~Mastrolia, {\em et al.},
  ``{Automated One-Loop Calculations with GoSam},''
\href{http://dx.doi.org/10.1140/epjc/s10052-012-1889-1}{{\em Eur.Phys.J.} {\bf
  C72} (2012)  1889}.

\bibitem{Cullen:2011xs}
G.~Cullen, N.~Greiner, G.~Heinrich, G.~Luisoni, P.~Mastrolia, {\em et al.},
  ``{GoSam: A Program for Automated One-Loop Calculations},''
\href{http://dx.doi.org/10.1088/1742-6596/368/1/012056}{{\em J.Phys.Conf.Ser.}
  {\bf 368} (2012)  012056}.

\bibitem{deAquino:2011ub}
P.~de~Aquino, W.~Link, F.~Maltoni, O.~Mattelaer, and T.~Stelzer, ``{ALOHA:
  Automatic Libraries Of Helicity Amplitudes for Feynman Diagram
  Computations},''
\href{http://dx.doi.org/10.1016/j.cpc.2012.05.004}{{\em Comput.Phys.Commun.}
  {\bf 183} (2012)  2254--2263}.

\bibitem{Butterworth:2010ym}
J.~Butterworth, F.~Maltoni, F.~Moortgat, P.~Richardson, S.~Schumann, {\em et
  al.}, ``{The Tools and Monte Carlo Working Group Summary Report},''
\href{http://arxiv.org/abs/1003.1643}{{\tt arXiv:1003.1643 [hep-ph]}}.

\bibitem{Alwall:xxx}
C.~Duhr, B.~Fuks, O.~Mattelaer, G.~\"Ozt\"urk, and C.-H. Shen, ``{Computing
  decay rates for new physics theories with FeynRules and MadGraph},'' {\em in
  preparation\!}  .

\bibitem{FRwebpage}
 {\em {http://feynrules.irmp.ucl.ac.be}}  .

\bibitem{Skands:2003cj}
P.~Z. Skands, B.~Allanach, H.~Baer, C.~Balazs, G.~Belanger, {\em et al.},
  ``{SUSY Les Houches Accord: Interfacing SUSY Spectrum Calculators, Decay
  Packages, and Event Generators},''
  \href{http://dx.doi.org/10.1088/1126-6708/2004/07/036}{{\em JHEP} {\bf 0407}
  (2004)  036}.

\bibitem{Allanach:2008qq}
B.~Allanach, C.~Balazs, G.~Belanger, M.~Bernhardt, F.~Boudjema, {\em et al.},
  ``{SUSY Les Houches Accord 2},''
  \href{http://dx.doi.org/10.1016/j.cpc.2008.08.004}{{\em Comput.Phys.Commun.}
  {\bf 180} (2009)  8--25}.

\bibitem{Beringer:2012zz}
{\bf Particle Data Group}, J.~Beringer {\em et al.}, ``{Review of Particle
  Physics (RPP)},''
\href{http://dx.doi.org/10.1103/PhysRevD.86.010001}{{\em Phys.Rev.} {\bf D86}
  (2012)  010001}.

\bibitem{Alwall:2007mw}
J.~Alwall, E.~Boos, L.~Dudko, M.~Gigg, M.~Herquet, {\em et al.}, ``{A Les
  Houches Interface for BSM Generators},''
\href{http://arxiv.org/abs/0712.3311}{{\tt arXiv:0712.3311 [hep-ph]}}.

\bibitem{Han:2009ya}
T.~Han, I.~Lewis, and T.~McElmurry, ``{QCD Corrections to Scalar Diquark
  Production at Hadron Colliders},''
\href{http://dx.doi.org/10.1007/JHEP01(2010)123}{{\em JHEP} {\bf 1001} (2010)
  123}.

\bibitem{Lowette:2012uh}
{\bf ATLAS and CMS collaborations}, S.~Lowette, ``{Supersymmetry Searches with
  ATLAS and CMS},''
\href{http://arxiv.org/abs/1205.4053}{{\tt arXiv:1205.4053 [hep-ex]}}.

\bibitem{O'Raifeartaigh:1975pr}
L.~O'Raifeartaigh, ``{Spontaneous Symmetry Breaking for Chiral Scalar
  Superfields},''
\href{http://dx.doi.org/10.1016/0550-3213(75)90585-4}{{\em Nucl.Phys.} {\bf
  B96} (1975)  331}.

\bibitem{west}
P.~C. West, {\em Introduction to Supersymmetry and Supergravity}.
\newblock World Scientific Pub Co Inc, 1986.

\bibitem{Wess:1977fn}
J.~Wess and B.~Zumino, ``{Superspace Formulation of Supergravity},''
\href{http://dx.doi.org/10.1016/0370-2693(77)90015-6}{{\em Phys.Lett.} {\bf
  B66} (1977)  361--364}.

\bibitem{Brans:1961sx}
C.~Brans and R.~Dicke, ``{Mach's Principle and a Relativistic Theory of
  Gravitation},''
\href{http://dx.doi.org/10.1103/PhysRev.124.925}{{\em Phys.Rev.} {\bf 124}
  (1961)  925--935}.

\bibitem{Howe:1978km}
P.~S. Howe and R.~Tucker, ``{Scale Invariance in Superspace},''
\href{http://dx.doi.org/10.1016/0370-2693(78)90327-1}{{\em Phys.Lett.} {\bf
  B80} (1978)  138}.

\bibitem{Wess:1978bu}
J.~Wess and B.~Zumino, ``{Superfield Lagrangian for Supergravity},''
\href{http://dx.doi.org/10.1016/0370-2693(78)90057-6}{{\em Phys.Lett.} {\bf
  B74} (1978)  51}.

\bibitem{Siegel:1979wr}
W.~Siegel, ``{Supergravity Supergraphs},''
\href{http://dx.doi.org/10.1016/0370-2693(79)90283-1}{{\em Phys.Lett.} {\bf
  B84} (1979)  197}.

\bibitem{Gates:1983nr}
S.~Gates, M.~T. Grisaru, M.~Rocek, and W.~Siegel, ``{Superspace Or One Thousand
  and One Lessons in Supersymmetry},''
{\em Front.Phys.} {\bf 58} (1983)  1--548.

\bibitem{Siegel:1978mj}
W.~Siegel and J.~Gates, S.~James, ``{Superfield Supergravity},''
\href{http://dx.doi.org/10.1016/0550-3213(79)90416-4}{{\em Nucl.Phys.} {\bf
  B147} (1979)  77}.

\bibitem{polo}
J.~Polonyi. Hungary Central Inst Res - KFKI-77-93 (unpublished).

\bibitem{Kaplunovsky:1993rd}
V.~S. Kaplunovsky and J.~Louis, ``{Model Independent Analysis of Soft Terms in
  Effective Supergravity and in String Theory},''
\href{http://dx.doi.org/10.1016/0370-2693(93)90078-V}{{\em Phys.Lett.} {\bf
  B306} (1993)  269--275}.

\bibitem{Barbieri:1993jk}
R.~Barbieri, J.~Louis, and M.~Moretti, ``{Phenomenological Implications of
  Supersymmetry Breaking by the Dilaton},''
\href{http://dx.doi.org/10.1016/0370-2693(93)90981-M}{{\em Phys.Lett.} {\bf
  B312} (1993)  451--460}.

\bibitem{Brignole:1993dj}
A.~Brignole, L.~E. Ibanez, and C.~Munoz, ``{Towards a Theory of Soft Terms for
  the Supersymmetric Standard Model},''
\href{http://dx.doi.org/10.1016/0550-3213(94)00068-9}{{\em Nucl.Phys.} {\bf
  B422} (1994)  125--171}.

\bibitem{Brignole:1997dp}
A.~Brignole, L.~E. Ibanez, and C.~Munoz, ``{Soft Supersymmetry Breaking Terms
  from Supergravity and Superstring Models},''
\href{http://arxiv.org/abs/hep-ph/9707209}{{\tt arXiv:hep-ph/9707209
  [hep-ph]}}.

\bibitem{Deser:1977uq}
S.~Deser and B.~Zumino, ``{Broken Supersymmetry and Supergravity},''
\href{http://dx.doi.org/10.1103/PhysRevLett.38.1433}{{\em Phys.Rev.Lett.} {\bf
  38} (1977)  1433}.

\bibitem{Passarino:1978jh}
G.~Passarino and M.~Veltman, ``{One Loop Corrections for e+ e- Annihilation
  Into mu+ mu- in the Weinberg Model},''
\href{http://dx.doi.org/10.1016/0550-3213(79)90234-7}{{\em Nucl.Phys.} {\bf
  B160} (1979)  151}.

\bibitem{Bagger:1995ay}
J.~Bagger, E.~Poppitz, and L.~Randall, ``{Destabilizing Divergences in
  Supergravity Theories at Two Loops},''
\href{http://dx.doi.org/10.1016/0550-3213(95)00463-3}{{\em Nucl.Phys.} {\bf
  B455} (1995)  59--82}.

\bibitem{Giudice:1988yz}
G.~Giudice and A.~Masiero, ``{A Natural Solution to the mu Problem in
  Supergravity Theories},''
\href{http://dx.doi.org/10.1016/0370-2693(88)91613-9}{{\em Phys.Lett.} {\bf
  B206} (1988)  480--484}.

\bibitem{Giudice:1998xp}
G.~F. Giudice, M.~A. Luty, H.~Murayama, and R.~Rattazzi, ``{Gaugino Mass
  Without Singlets},''
{\em JHEP} {\bf 9812} (1998)  027.

\bibitem{Pomarol:1999ie}
A.~Pomarol and R.~Rattazzi, ``{Sparticle Masses from the Superconformal
  Anomaly},''
{\em JHEP} {\bf 9905} (1999)  013.

\bibitem{Ferrara:1974fv}
S.~Ferrara, J.~Iliopoulos, and B.~Zumino, ``{Supergauge Invariance and the
  Gell-Mann - Low Eigenvalue},''
\href{http://dx.doi.org/10.1016/0550-3213(74)90372-1}{{\em Nucl.Phys.} {\bf
  B77} (1974)  413}.

\bibitem{Zumino:1974bg}
B.~Zumino, ``{Supersymmetry and the Vacuum},''
\href{http://dx.doi.org/10.1016/0550-3213(75)90194-7}{{\em Nucl.Phys.} {\bf
  B89} (1975)  535}.

\bibitem{Ferrara:1975ye}
S.~Ferrara and O.~Piguet, ``{Perturbation Theory and Renormalization of
  Supersymmetric Yang-Mills Theories},''
\href{http://dx.doi.org/10.1016/0550-3213(75)90573-8}{{\em Nucl.Phys.} {\bf
  B93} (1975)  261}.

\bibitem{Grisaru:1979wc}
M.~T. Grisaru, W.~Siegel, and M.~Rocek, ``{Improved Methods for Supergraphs},''
\href{http://dx.doi.org/10.1016/0550-3213(79)90344-4}{{\em Nucl.Phys.} {\bf
  B159} (1979)  429}.

\bibitem{Gaillard:1998bf}
M.~K. Gaillard, ``{One Loop Pauli-Villars Regularization of Supergravity 1.
  Canonical Gauge Kinetic Energy},''
\href{http://dx.doi.org/10.1103/PhysRevD.58.105027}{{\em Phys.Rev.} {\bf D58}
  (1998)  105027}.

\bibitem{Gherghetta:1999sw}
T.~Gherghetta, G.~F. Giudice, and J.~D. Wells, ``{Phenomenological Consequences
  of Supersymmetry with Anomaly Induced Masses},''
\href{http://dx.doi.org/10.1016/S0550-3213(99)00429-0}{{\em Nucl.Phys.} {\bf
  B559} (1999)  27--47}.

\bibitem{Feng:1999fu}
J.~L. Feng, T.~Moroi, L.~Randall, M.~Strassler, and S.-f. Su, ``{Discovering
  Supersymmetry at the Tevatron in Wino LSP Scenarios},''
\href{http://dx.doi.org/10.1103/PhysRevLett.83.1731}{{\em Phys.Rev.Lett.} {\bf
  83} (1999)  1731--1734}.

\bibitem{Feng:1999hg}
J.~L. Feng and T.~Moroi, ``{Supernatural Supersymmetry: Phenomenological
  Implications of Anomaly Mediated Supersymmetry Breaking},''
\href{http://dx.doi.org/10.1103/PhysRevD.61.095004}{{\em Phys.Rev.} {\bf D61}
  (2000)  095004}.

\bibitem{Rattazzi:1999qg}
R.~Rattazzi, A.~Strumia, and J.~D. Wells, ``{Phenomenology of Deflected Anomaly
  Mediation},''
\href{http://dx.doi.org/10.1016/S0550-3213(00)00130-9}{{\em Nucl.Phys.} {\bf
  B576} (2000)  3--28}.

\bibitem{Barr:2002ex}
A.~Barr, C.~Lester, M.~A. Parker, B.~Allanach, and P.~Richardson,
  ``{Discovering Anomaly Mediated Supersymmetry at the LHC},''
{\em JHEP} {\bf 0303} (2003)  045.

\bibitem{Derendinger:1983bz}
J.~Derendinger and C.~A. Savoy, ``{Quantum Effects and SU(2) x U(1) Breaking in
  Supergravity Gauge Theories},''
\href{http://dx.doi.org/10.1016/0550-3213(84)90162-7}{{\em Nucl.Phys.} {\bf
  B237} (1984)  307}.

\bibitem{Falck:1985aa}
N.~K. Falck, ``{Renormalization Group Equations for Softly Broken
  Supersymmetry: The Most General Case},''
\href{http://dx.doi.org/10.1007/BF01575432}{{\em Z.Phys.} {\bf C30} (1986)
  247}.

\bibitem{Martin:1993zk}
S.~P. Martin and M.~T. Vaughn, ``{Two Loop Renormalization Group Equations for
  Soft Supersymmetry Breaking Couplings},''
\href{http://dx.doi.org/10.1103/PhysRevD.50.2282,
  10.1103/PhysRevD.78.039903}{{\em Phys.Rev.} {\bf D50} (1994)  2282}.

\bibitem{Yamada:1993uh}
Y.~Yamada, ``{Two Loop Renormalization of Gaugino Mass in Supersymmetric Gauge
  model},''
\href{http://dx.doi.org/10.1016/0370-2693(93)90665-5}{{\em Phys.Lett.} {\bf
  B316} (1993)  109--111}.

\bibitem{Yamada:1993ga}
Y.~Yamada, ``{Two Loop Renormalization of Gaugino Masses in General
  Supersymmetric Gauge Models},''
\href{http://dx.doi.org/10.1103/PhysRevLett.72.25}{{\em Phys.Rev.Lett.} {\bf
  72} (1994)  25--27}.

\bibitem{Yamada:1994id}
Y.~Yamada, ``{Two loop Renormalization Group Equations for Soft SUSY Breaking
  Scalar Interactions: Supergraph Method},''
\href{http://dx.doi.org/10.1103/PhysRevD.50.3537}{{\em Phys.Rev.} {\bf D50}
  (1994)  3537--3545}.

\bibitem{Jack:1994rk}
I.~Jack, D.~T. Jones, S.~P. Martin, M.~T. Vaughn, and Y.~Yamada, ``{Decoupling
  of the Epsilon Scalar Mass in Softly Broken Supersymmetry},''
\href{http://dx.doi.org/10.1103/PhysRevD.50.R5481}{{\em Phys.Rev.} {\bf D50}
  (1994)  5481--5483}.

\bibitem{Farrar:1978xj}
G.~R. Farrar and P.~Fayet, ``{Phenomenology of the Production, Decay, and
  Detection of New Hadronic States Associated with Supersymmetry},''
\href{http://dx.doi.org/10.1016/0370-2693(78)90858-4}{{\em Phys.Lett.} {\bf
  B76} (1978)  575--579}.

\bibitem{Girardello:1981wz}
L.~Girardello and M.~T. Grisaru, ``{Soft Breaking of Supersymmetry},''
\href{http://dx.doi.org/10.1016/0550-3213(82)90512-0}{{\em Nucl.Phys.} {\bf
  B194} (1982)  65}.

\bibitem{Kim:1983dt}
J.~E. Kim and H.~P. Nilles, ``{The mu Problem and the Strong CP Problem},''
\href{http://dx.doi.org/10.1016/0370-2693(84)91890-2}{{\em Phys.Lett.} {\bf
  B138} (1984)  150}.

\bibitem{Fayet:1977yc}
P.~Fayet, ``{Spontaneously Broken Supersymmetric Theories of Weak,
  Electromagnetic and Strong Interactions},''
\href{http://dx.doi.org/10.1016/0370-2693(77)90852-8}{{\em Phys.Lett.} {\bf
  B69} (1977)  489}.

\bibitem{Ellis:1988er}
J.~R. Ellis, J.~Gunion, H.~E. Haber, L.~Roszkowski, and F.~Zwirner, ``{Higgs
  Bosons in a Nonminimal Supersymmetric Model},''
\href{http://dx.doi.org/10.1103/PhysRevD.39.844}{{\em Phys.Rev.} {\bf D39}
  (1989)  844}.

\bibitem{Drees:1988fc}
M.~Drees, ``{Supersymmetric Models with Extended Higgs Sector},''
\href{http://dx.doi.org/10.1142/S0217751X89001448}{{\em Int.J.Mod.Phys.} {\bf
  A4} (1989)  3635}.

\bibitem{Ellwanger:1993xa}
U.~Ellwanger, M.~Rausch~de Traubenberg, and C.~A. Savoy, ``{Particle Spectrum
  in Supersymmetric Models with a Gauge Singlet},''
\href{http://dx.doi.org/10.1016/0370-2693(93)91621-S}{{\em Phys.Lett.} {\bf
  B315} (1993)  331--337}.

\bibitem{Elliott:1994ht}
T.~Elliott, S.~King, and P.~White, ``{Unification Constraints in the
  Next-to-Minimal Supersymmetric Standard Model},''
\href{http://dx.doi.org/10.1016/0370-2693(95)00381-T}{{\em Phys.Lett.} {\bf
  B351} (1995)  213--219}.

\bibitem{Ellwanger:1995ru}
U.~Ellwanger, M.~Rausch~de Traubenberg, and C.~A. Savoy, ``{Higgs Phenomenology
  of the Supersymmetric Model with a Gauge Singlet},''
\href{http://dx.doi.org/10.1007/BF01553993}{{\em Z.Phys.} {\bf C67} (1995)
  665--670}.

\bibitem{Ellwanger:1996gw}
U.~Ellwanger, M.~Rausch~de Traubenberg, and C.~A. Savoy, ``{Phenomenology of
  Supersymmetric Models with a Singlet},''
\href{http://dx.doi.org/10.1016/S0550-3213(97)00128-4}{{\em Nucl.Phys.} {\bf
  B492} (1997)  21--50}.

\bibitem{Maniatis:2009re}
M.~Maniatis, ``{The Next-to-Minimal Supersymmetric Extension of the Standard
  Model Reviewed},''
\href{http://dx.doi.org/10.1142/S0217751X10049827}{{\em Int.J.Mod.Phys.} {\bf
  A25} (2010)  3505--3602}.

\bibitem{Ellwanger:2009dp}
U.~Ellwanger, C.~Hugonie, and A.~M. Teixeira, ``{The Next-to-Minimal
  Supersymmetric Standard Model},''
\href{http://dx.doi.org/10.1016/j.physrep.2010.07.001}{{\em Phys.Rept.} {\bf
  496} (2010)  1--77}.

\bibitem{ElKheishen:1992yv}
M.~El~Kheishen, A.~Aboshousha, and A.~Shafik, ``{Analytic Formulas for the
  Neutralino Masses and the Neutralino Mixing Matrix},''
\href{http://dx.doi.org/10.1103/PhysRevD.45.4345}{{\em Phys.Rev.} {\bf D45}
  (1992)  4345--4348}.

\bibitem{Gounaris:2001fx}
G.~Gounaris, C.~Le~Mouel, and P.~Porfyriadis, ``{A Description of the
  Neutralino Observables in Terms of Projectors},''
\href{http://dx.doi.org/10.1103/PhysRevD.65.035002}{{\em Phys.Rev.} {\bf D65}
  (2002)  035002}.

\bibitem{Ibrahim:2007fb}
T.~Ibrahim and P.~Nath, ``{CP Violation From Standard Model to Strings},''
\href{http://dx.doi.org/10.1103//RevModPhys.80.577}{{\em Rev.Mod.Phys.} {\bf
  80} (2008)  577--631}.

\bibitem{Hall:1985dx}
L.~J. Hall, V.~A. Kostelecky, and S.~Raby, ``{New Flavor Violations in
  Supergravity Models},''
\href{http://dx.doi.org/10.1016/0550-3213(86)90397-4}{{\em Nucl.Phys.} {\bf
  B267} (1986)  415}.

\bibitem{Chankowski:1992er}
P.~H. Chankowski, S.~Pokorski, and J.~Rosiek, ``{Complete On-Shell
  Renormalization Scheme for the Minimal Supersymmetric Higgs Sector},''
\href{http://dx.doi.org/10.1016/0550-3213(94)90141-4}{{\em Nucl.Phys.} {\bf
  B423} (1994)  437--496}.

\bibitem{Dabelstein:1994hb}
A.~Dabelstein, ``{The One Loop Renormalization of the MSSM Higgs Sector and its
  Application to the Neutral Scalar Higgs Masses},''
\href{http://dx.doi.org/10.1007/BF01624592}{{\em Z.Phys.} {\bf C67} (1995)
  495--512}.

\bibitem{Pierce:1996zz}
D.~M. Pierce, J.~A. Bagger, K.~T. Matchev, and R.-j. Zhang, ``{Precision
  Corrections in the Minimal Supersymmetric Standard Model},''
\href{http://dx.doi.org/10.1016/S0550-3213(96)00683-9}{{\em Nucl.Phys.} {\bf
  B491} (1997)  3--67}.

\bibitem{Dimopoulos:1995ju}
S.~Dimopoulos and D.~W. Sutter, ``{The Supersymmetric Flavor Problem},''
\href{http://dx.doi.org/10.1016/0550-3213(95)00421-N}{{\em Nucl.Phys.} {\bf
  B452} (1995)  496--512}.

\bibitem{Chacko:1999am}
Z.~Chacko, M.~A. Luty, I.~Maksymyk, and E.~Ponton, ``{Realistic Anomaly
  Mediated Supersymmetry Breaking},''
{\em JHEP} {\bf 0004} (2000)  001.

\bibitem{Katz:1999uw}
E.~Katz, Y.~Shadmi, and Y.~Shirman, ``{Heavy Thresholds, Slepton Masses and the
  Mu Term in Anomaly Mediated Supersymmetry Breaking},''
{\em JHEP} {\bf 9908} (1999)  015.

\bibitem{Jack:2000cd}
I.~Jack and D.~Jones, ``{Fayet-Iliopoulos D Terms and Anomaly Mediated
  Supersymmetry Breaking},''
\href{http://dx.doi.org/10.1016/S0370-2693(00)00501-3}{{\em Phys.Lett.} {\bf
  B482} (2000)  167--173}.

\bibitem{Jack:2003qg}
I.~Jack and D.~Jones, ``{Yukawa Textures and Anomaly Mediated Supersymmetry
  Breaking},''
\href{http://dx.doi.org/10.1016/S0550-3213(03)00310-9}{{\em Nucl.Phys.} {\bf
  B662} (2003)  63--88}.

\bibitem{Murakami:2003pb}
B.~Murakami and J.~D. Wells, ``{Abelian D Terms and the Superpartner Spectrum
  of Anomaly Mediated Supersymmetry Breaking},''
\href{http://dx.doi.org/10.1103/PhysRevD.68.035006}{{\em Phys.Rev.} {\bf D68}
  (2003)  035006}.

\bibitem{Kitano:2004zd}
R.~Kitano, G.~D. Kribs, and H.~Murayama, ``{Electroweak Symmetry Breaking via
  UV Insensitive Anomaly Mediation},''
\href{http://dx.doi.org/10.1103/PhysRevD.70.035001}{{\em Phys.Rev.} {\bf D70}
  (2004)  035001}.

\bibitem{Ibe:2004gh}
M.~Ibe, R.~Kitano, and H.~Murayama, ``{A Viable Supersymmetric Model with UV
  Insensitive Anomaly Mediation},''
\href{http://dx.doi.org/10.1103/PhysRevD.71.075003}{{\em Phys.Rev.} {\bf D71}
  (2005)  075003}.

\bibitem{Hodgson:2005en}
R.~Hodgson, I.~Jack, D.~Jones, and G.~Ross, ``{Anomaly Mediation,
  Fayet-Iliopoulos D-terms and Precision Sparticle Spectra},''
\href{http://dx.doi.org/10.1016/j.nuclphysb.2005.09.013}{{\em Nucl.Phys.} {\bf
  B728} (2005)  192--206}.

\bibitem{Jones:2006re}
D.~Jones and G.~Ross, ``{Anomaly Mediation and Dimensional Transmutation},''
\href{http://dx.doi.org/10.1016/j.physletb.2006.10.010}{{\em Phys.Lett.} {\bf
  B642} (2006)  540--545}.

\bibitem{Hodgson:2007kq}
R.~Hodgson, I.~Jack, and D.~Jones, ``{Anomaly mediation, Fayet-Iliopoulos
  D-terms and the Renormalisation Group},''
\href{http://dx.doi.org/10.1088/1126-6708/2007/10/070}{{\em JHEP} {\bf 0710}
  (2007)  070}.

\bibitem{Eliezer:1989cr}
D.~Eliezer and R.~Woodard, ``{The Problem of Nonlocality in String Theory},''
\href{http://dx.doi.org/10.1016/0550-3213(89)90461-6}{{\em Nucl.Phys.} {\bf
  B325} (1989)  389}.

\bibitem{Carena:1993ag}
M.~S. Carena, S.~Pokorski, and C.~Wagner, ``{On the Unification of Couplings in
  the Minimal Supersymmetric Standard Model},''
\href{http://dx.doi.org/10.1016/0550-3213(93)90161-H}{{\em Nucl.Phys.} {\bf
  B406} (1993)  59--89}.

\bibitem{Porod:2003um}
W.~Porod, ``{SPheno, a Program for Calculating Supersymmetric Spectra, SUSY
  Particle Decays and SUSY Particle Production at e+ e- Colliders},''
\href{http://dx.doi.org/10.1016/S0010-4655(03)00222-4}{{\em
  Comput.Phys.Commun.} {\bf 153} (2003)  275--315}.

\bibitem{Porod:2011nf}
W.~Porod and F.~Staub, ``{SPheno 3.1: Extensions Including Flavour, CP-Phases
  and Models Beyond the MSSM},''
\href{http://dx.doi.org/10.1016/j.cpc.2012.05.021}{{\em Comput.Phys.Commun.}
  {\bf 183} (2012)  2458--2469}.

\bibitem{Djouadi:2002ze}
A.~Djouadi, J.-L. Kneur, and G.~Moultaka, ``{SuSpect: A Fortran Code for the
  Supersymmetric and Higgs Particle Spectrum in the MSSM},''
\href{http://dx.doi.org/10.1016/j.cpc.2006.11.009}{{\em Comput.Phys.Commun.}
  {\bf 176} (2007)  426--455}.

\bibitem{Allanach:2001kg}
B.~Allanach, ``{SOFTSUSY: a Program for Calculating Supersymmetric Spectra},''
\href{http://dx.doi.org/10.1016/S0010-4655(01)00460-X}{{\em
  Comput.Phys.Commun.} {\bf 143} (2002)  305--331}.

\bibitem{Allanach:2009bv}
B.~Allanach and M.~Bernhardt, ``{Including R-parity Violation in the Numerical
  Computation of the Spectrum of the Minimal Supersymmetric Standard Model:
  SOFTSUSY},''
\href{http://dx.doi.org/10.1016/j.cpc.2009.09.015}{{\em Comput.Phys.Commun.}
  {\bf 181} (2010)  232--245}.

\bibitem{Allanach:2011de}
B.~Allanach, C.~Kom, and M.~Hanussek, ``{Computation of Neutrino Masses in
  R-parity Violating Supersymmetry: SOFTSUSY3.2},''
\href{http://dx.doi.org/10.1016/j.cpc.2011.11.024}{{\em Comput.Phys.Commun.}
  {\bf 183} (2012)  785--793}.

\bibitem{Baer:1985yd}
H.~Baer, J.~R. Ellis, G.~Gelmini, D.~V. Nanopoulos, and X.~Tata, ``{Squark
  Decays into Gauginos at the p anti-p Collider},''
\href{http://dx.doi.org/10.1016/0370-2693(85)90632-X}{{\em Phys.Lett.} {\bf
  B161} (1985)  175}.

\bibitem{Gamberini:1986eg}
G.~Gamberini, ``{Heavy Gluino and Squark Decays at p anti-p Collider},''
\href{http://dx.doi.org/10.1007/BF01571810}{{\em Z.Phys.} {\bf C30} (1986)
  605--613}.

\bibitem{Baer:1986au}
H.~Baer, V.~D. Barger, D.~Karatas, and X.~Tata, ``{Detecting Gluinos at Hadron
  Supercolliders},''
\href{http://dx.doi.org/10.1103/PhysRevD.36.96}{{\em Phys.Rev.} {\bf D36}
  (1987)  96}.

\bibitem{Barnett:1987kn}
R.~M. Barnett, J.~F. Gunion, and H.~E. Haber, ``{Gluino Decay Patterns and
  Signatures},''
\href{http://dx.doi.org/10.1103/PhysRevD.37.1892}{{\em Phys.Rev.} {\bf D37}
  (1988)  1892}.

\bibitem{Buchalla:1995vs}
G.~Buchalla, A.~J. Buras, and M.~E. Lautenbacher, ``{Weak Decays Beyond Leading
  Logarithms},''
\href{http://dx.doi.org/10.1103/RevModPhys.68.1125}{{\em Rev.Mod.Phys.} {\bf
  68} (1996)  1125--1144}.

\bibitem{Kagan:1998bh}
A.~L. Kagan and M.~Neubert, ``{Direct CP Violation in B to X(s) gamma Decays as
  a Signature of New Physics},''
\href{http://dx.doi.org/10.1103/PhysRevD.58.094012}{{\em Phys.Rev.} {\bf D58}
  (1998)  094012}.

\bibitem{Kagan:1998ym}
A.~L. Kagan and M.~Neubert, ``{QCD Anatomy of B to X(s gamma) Decays},''
\href{http://dx.doi.org/10.1007/s100529800959}{{\em Eur.Phys.J.} {\bf C7}
  (1999)  5--27}.

\bibitem{Hou:1992sy}
W.-S. Hou, ``{Enhanced Charged Higgs Boson Effects in B- to tau anti-neutrino,
  mu anti-neutrino and b to tau anti-neutrino + X},''
\href{http://dx.doi.org/10.1103/PhysRevD.48.2342}{{\em Phys.Rev.} {\bf D48}
  (1993)  2342--2344}.

\bibitem{Logan:2000iv}
H.~E. Logan and U.~Nierste, ``{$B_{s,d} \to \ell^+ \ell^-$ in a two Higgs
  Doublet Model},''
\href{http://dx.doi.org/10.1016/S0550-3213(00)00417-X}{{\em Nucl.Phys.} {\bf
  B586} (2000)  39--55}.

\bibitem{Baek:2001kh}
S.~Baek, T.~Goto, Y.~Okada, and K.-i. Okumura, ``{Muon Anomalous Magnetic
  Moment, Lepton Flavor Violation, and Flavor Changing Neutral Current
  Processes in SUSY GUT with Right-Handed Neutrino},''
\href{http://dx.doi.org/10.1103/PhysRevD.64.095001}{{\em Phys.Rev.} {\bf D64}
  (2001)  095001}.

\bibitem{Bobeth:2001jm}
C.~Bobeth, A.~J. Buras, F.~Kruger, and J.~Urban, ``{QCD corrections to $\bar{B}
  \to X_{d,s} \nu \bar{\nu}$, $\bar{B}_{d,s} \to \ell^{+} \ell^{-}$, $K \to \pi
  \nu \bar{\nu}$ and $K_{L} \to \mu^{+} \mu^{-}$ in the MSSM},''
\href{http://dx.doi.org/10.1016/S0550-3213(02)00141-4}{{\em Nucl.Phys.} {\bf
  B630} (2002)  87--131}.

\bibitem{Buras:2002vd}
A.~J. Buras, P.~H. Chankowski, J.~Rosiek, and L.~Slawianowska, ``{$\Delta
  M_{d,s}, B^0{d,s} \to \mu^{+} \mu^{-}$ and $B \to X_{s} \gamma$ in
  Supersymmetry at Large $\tan\beta$},''
\href{http://dx.doi.org/10.1016/S0550-3213(03)00190-1}{{\em Nucl.Phys.} {\bf
  B659} (2003)  3}.

\bibitem{Huber:2005ig}
T.~Huber, E.~Lunghi, M.~Misiak, and D.~Wyler, ``{Electromagnetic Logarithms in
  anti-B to X(s) l+ l-},''
\href{http://dx.doi.org/10.1016/j.nuclphysb.2006.01.037}{{\em Nucl.Phys.} {\bf
  B740} (2006)  105--137}.

\bibitem{Hahn:2005qi}
T.~Hahn, W.~Hollik, J.~Illana, and S.~Penaranda, ``{Interplay between H to b
  anti-s and b to s gamma in the MSSM with Non-Minimal Flavor Violation},''
\href{http://arxiv.org/abs/hep-ph/0512315}{{\tt arXiv:hep-ph/0512315
  [hep-ph]}}.

\bibitem{Lunghi:2006hc}
E.~Lunghi and J.~Matias, ``{Huge Right-Handed Current Effects in B to K*(K
  pi)l+l- in Supersymmetry},''
\href{http://dx.doi.org/10.1088/1126-6708/2007/04/058}{{\em JHEP} {\bf 0704}
  (2007)  058}.

\bibitem{Buras:1997bk}
A.~J. Buras, A.~Kwiatkowski, and N.~Pott, ``{On the Scale Uncertainties in the
  B to X(s) gamma Decay},''
\href{http://dx.doi.org/10.1016/S0370-2693(97)01142-8}{{\em Phys.Lett.} {\bf
  B414} (1997)  157--165}.

\bibitem{Chetyrkin:1996vx}
K.~G. Chetyrkin, M.~Misiak, and M.~Munz, ``{Weak Radiative B Meson Decay Beyond
  Leading Logarithms},''
\href{http://dx.doi.org/10.1016/S0370-2693(97)00324-9}{{\em Phys.Lett.} {\bf
  B400} (1997)  206--219}.

\bibitem{Amhis:2012bh}
{\bf Heavy Flavor Averaging Group}, Y.~Amhis {\em et al.}, ``{Averages of
  b-hadron, c-hadron, and tau-lepton properties as of early 2012},''
\href{http://arxiv.org/abs/1207.1158}{{\tt arXiv:1207.1158 [hep-ex]}}.

\bibitem{Hurth:2003dk}
T.~Hurth, E.~Lunghi, and W.~Porod, ``{Untagged Anti-B to X(s+d) Gamma CP
  Asymmetry as a Probe for New Physics},''
\href{http://dx.doi.org/10.1016/j.nuclphysb.2004.10.024}{{\em Nucl.Phys.} {\bf
  B704} (2005)  56--74}.

\bibitem{Misiak:2006zs}
M.~Misiak, H.~Asatrian, K.~Bieri, M.~Czakon, A.~Czarnecki, {\em et al.},
  ``{Estimate of B(anti-B) to X(s) gamma at O(alpha(s)**2)},''
\href{http://dx.doi.org/10.1103/PhysRevLett.98.022002}{{\em Phys.Rev.Lett.}
  {\bf 98} (2007)  022002}.

\bibitem{Huber:2007vv}
T.~Huber, T.~Hurth, and E.~Lunghi, ``{Logarithmically Enhanced Corrections to
  the Decay Rate and Forward Backward Asymmetry in $\bar{B} \to X_s \ell^+
  \ell^-$},''
\href{http://dx.doi.org/10.1016/j.nuclphysb.2008.04.028}{{\em Nucl.Phys.} {\bf
  B802} (2008)  40--62}.

\bibitem{Aaij:2012ct}
{\bf LHCb Collaboration}, R.~Aaij {\em et al.}, ``{First Evidence for the Decay
  Bs to mu+ mu-},''
\href{http://arxiv.org/abs/1211.2674}{{\tt arXiv:1211.2674 [hep-ex]}}.

\bibitem{Degrassi:2001yf}
G.~Degrassi, P.~Slavich, and F.~Zwirner, ``{On the Neutral Higgs Boson Masses
  in the MSSM for Arbitrary Stop Mixing},''
\href{http://dx.doi.org/10.1016/S0550-3213(01)00343-1}{{\em Nucl.Phys.} {\bf
  B611} (2001)  403--422}.

\bibitem{Brignole:2001jy}
A.~Brignole, G.~Degrassi, P.~Slavich, and F.~Zwirner, ``{On the O(alpha(t)**2)
  Two Loop Corrections to the Neutral Higgs Boson Masses in the MSSM},''
\href{http://dx.doi.org/10.1016/S0550-3213(02)00184-0}{{\em Nucl.Phys.} {\bf
  B631} (2002)  195--218}.

\bibitem{Brignole:2002bz}
A.~Brignole, G.~Degrassi, P.~Slavich, and F.~Zwirner, ``{On the Two Loop
  Sbottom Corrections to the Neutral Higgs Boson Masses in the MSSM},''
\href{http://dx.doi.org/10.1016/S0550-3213(02)00748-4}{{\em Nucl.Phys.} {\bf
  B643} (2002)  79--92}.

\bibitem{Dedes:2002dy}
A.~Dedes and P.~Slavich, ``{Two Loop Corrections to Radiative Electroweak
  Symmetry Breaking in the MSSM},''
\href{http://dx.doi.org/10.1016/S0550-3213(03)00173-1}{{\em Nucl.Phys.} {\bf
  B657} (2003)  333--354}.

\bibitem{Dedes:2003km}
A.~Dedes, G.~Degrassi, and P.~Slavich, ``{On the Two Loop Yukawa Corrections to
  the MSSM Higgs Boson Masses at Large tan beta},''
\href{http://dx.doi.org/10.1016/j.nuclphysb.2003.08.033}{{\em Nucl.Phys.} {\bf
  B672} (2003)  144--162}.

\bibitem{AbdusSalam:2011fc}
S.~AbdusSalam, B.~Allanach, H.~Dreiner, J.~Ellis, U.~Ellwanger, {\em et al.},
  ``{Benchmark Models, Planes, Lines and Points for Future SUSY Searches at the
  LHC},''
\href{http://dx.doi.org/10.1140/epjc/s10052-011-1835-7}{{\em Eur.Phys.J.} {\bf
  C71} (2011)  1835}.

\bibitem{Choudhury:1998ze}
S.~R. Choudhury and N.~Gaur, ``{Dileptonic Decay of B(s) Meson in SUSY Models
  with Large tan Beta},''
\href{http://dx.doi.org/10.1016/S0370-2693(99)00203-8}{{\em Phys.Lett.} {\bf
  B451} (1999)  86--92}.

\bibitem{Babu:1999hn}
K.~Babu and C.~F. Kolda, ``{Higgs Mediated $B^0 \to \mu^{+} \mu^{-}$ in Minimal
  Supersymmetry},''
\href{http://dx.doi.org/10.1103/PhysRevLett.84.228}{{\em Phys.Rev.Lett.} {\bf
  84} (2000)  228--231}.

\bibitem{Allanach:2002nj}
B.~Allanach, M.~Battaglia, G.~Blair, M.~S. Carena, A.~De~Roeck, {\em et al.},
  ``{The Snowmass Points and Slopes: Benchmarks for SUSY Searches},''
\href{http://dx.doi.org/10.1007/s10052-002-0949-3}{{\em Eur.Phys.J.} {\bf C25}
  (2002)  113--123}.

\bibitem{Gabbiani:1996hi}
F.~Gabbiani, E.~Gabrielli, A.~Masiero, and L.~Silvestrini, ``{A Complete
  Analysis of FCNC and CP Constraints in General SUSY Extensions of the
  Standard Model},''
\href{http://dx.doi.org/10.1016/0550-3213(96)00390-2}{{\em Nucl.Phys.} {\bf
  B477} (1996)  321--352}.

\bibitem{Hagelin:1992tc}
J.~S. Hagelin, S.~Kelley, and T.~Tanaka, ``{Supersymmetric Flavor Changing
  Neutral Currents: Exact Amplitudes and Phenomenological Analysis},''
\href{http://dx.doi.org/10.1016/0550-3213(94)90113-9}{{\em Nucl.Phys.} {\bf
  B415} (1994)  293--331}.

\bibitem{Brax:1995up}
P.~Brax and C.~A. Savoy, ``{Flavor Changing Neutral Current Effects from Flavor
  Dependent Supergravity Couplings},''
\href{http://dx.doi.org/10.1016/0550-3213(95)00216-F}{{\em Nucl.Phys.} {\bf
  B447} (1995)  227--251}.

\bibitem{Ciuchini:2007cw}
M.~Ciuchini, E.~Franco, D.~Guadagnoli, V.~Lubicz, M.~Pierini, {\em et al.},
  ``{$D$ - $\bar{D}$ Mixing and New Physics: General Considerations and
  Constraints on the MSSM},''
\href{http://dx.doi.org/10.1016/j.physletb.2007.08.055}{{\em Phys.Lett.} {\bf
  B655} (2007)  162--166}.

\bibitem{Tobe:2003nx}
K.~Tobe, J.~D. Wells, and T.~Yanagida, ``{Neutrino Induced Lepton Flavor
  Violation in Gauge Mediated Supersymmetry Breaking},''
\href{http://dx.doi.org/10.1103/PhysRevD.69.035010}{{\em Phys.Rev.} {\bf D69}
  (2004)  035010}.

\bibitem{Dubovsky:1998nr}
S.~Dubovsky and D.~Gorbunov, ``{Flavor Violation and tan Beta in Gauge Mediated
  Models with Messenger - Matter Mixing},''
\href{http://dx.doi.org/10.1016/S0550-3213(99)00346-6}{{\em Nucl.Phys.} {\bf
  B557} (1999)  119--145}.

\bibitem{Beneke:1998sy}
M.~Beneke, G.~Buchalla, C.~Greub, A.~Lenz, and U.~Nierste, ``{Next-to-leading
  Order QCD Corrections to the Lifetime Difference of B(s) Mesons},''
\href{http://dx.doi.org/10.1016/S0370-2693(99)00684-X}{{\em Phys.Lett.} {\bf
  B459} (1999)  631--640}.

\bibitem{Lenz:2006hd}
A.~Lenz and U.~Nierste, ``{Theoretical Update of $B_s - \bar{B}_s$ Mixing},''
\href{http://dx.doi.org/10.1088/1126-6708/2007/06/072}{{\em JHEP} {\bf 0706}
  (2007)  072}.

\bibitem{Lenz:2011ti}
A.~Lenz and U.~Nierste, ``{Numerical Updates of Lifetimes and Mixing Parameters
  of B Mesons},''
\href{http://arxiv.org/abs/1102.4274}{{\tt arXiv:1102.4274 [hep-ph]}}.

\bibitem{Abulencia:2006mq}
{\bf CDF Collaboration}, A.~Abulencia {\em et al.}, ``{Measurement of the
  $B^0_{s}-\bar{B}^0_s$ Oscillation Frequency},''
\href{http://dx.doi.org/10.1103/PhysRevLett.97.062003}{{\em Phys.Rev.Lett.}
  {\bf 97} (2006)  062003}.

\bibitem{Abazov:2006dm}
{\bf D0 Collaboration}, V.~Abazov {\em et al.}, ``{First Direct Two-Sided Bound
  on the $B^0_{s}$ Oscillation Frequency},''
\href{http://dx.doi.org/10.1103/PhysRevLett.97.021802}{{\em Phys.Rev.Lett.}
  {\bf 97} (2006)  021802}.

\bibitem{Aaij:2011qx}
{\bf LHCb Collaboration}, R.~Aaij {\em et al.}, ``{Measurement of the $B^0_s -
  \bar{B}^0_s$ Oscillation Frequency $\Delta m_s$ in $B^0_s \to D_s^-(3) \pi$
  Decays},''
\href{http://dx.doi.org/10.1016/j.physletb.2012.02.031}{{\em Phys.Lett.} {\bf
  B709} (2012)  177--184}.

\bibitem{Ball:2006xx}
P.~Ball and R.~Fleischer, ``{Probing New Physics through $B$ Mixing: Status,
  Benchmarks and Prospects},''
\href{http://dx.doi.org/10.1140/epjc/s10052-006-0034-4}{{\em Eur.Phys.J.} {\bf
  C48} (2006)  413--426}.

\bibitem{Bennett:2004pv}
{\bf Muon g-2 Collaboration}, G.~Bennett {\em et al.}, ``{Measurement of the
  Negative Muon Anomalous Magnetic Moment to 0.7 ppm},''
\href{http://dx.doi.org/10.1103/PhysRevLett.92.161802}{{\em Phys.Rev.Lett.}
  {\bf 92} (2004)  161802}.

\bibitem{Li:1992xf}
G.~Li, R.~Mendel, and M.~A. Samuel, ``{Precise Mass Ratio Dependence of Fourth
  Order Lepton Anomalous Magnetic Moments: The Effect of a New Measurement of
  m(tau)},''
\href{http://dx.doi.org/10.1103/PhysRevD.47.1723}{{\em Phys.Rev.} {\bf D47}
  (1993)  1723--1725}.

\bibitem{Laporta:1992pa}
S.~Laporta and E.~Remiddi, ``{The Analytical Value of the Electron Light-Light
  Graphs Contribution to the Muon (g-2) in QED},''
\href{http://dx.doi.org/10.1016/0370-2693(93)91176-N}{{\em Phys.Lett.} {\bf
  B301} (1993)  440--446}.

\bibitem{Laporta:1996mq}
S.~Laporta and E.~Remiddi, ``{The Analytical Value of the Electron (g-2) at
  Order alpha**3 in QED},''
\href{http://dx.doi.org/10.1016/0370-2693(96)00439-X}{{\em Phys.Lett.} {\bf
  B379} (1996)  283--291}.

\bibitem{Czarnecki:1998rc}
A.~Czarnecki and M.~Skrzypek, ``{The Muon Anomalous Magnetic Moment in QED:
  Three Loop Electron and Tau Contributions},''
\href{http://dx.doi.org/10.1016/S0370-2693(99)00076-3}{{\em Phys.Lett.} {\bf
  B449} (1999)  354--360}.

\bibitem{Erler:2000nx}
J.~Erler and M.-x. Luo, ``{Hadronic Loop Corrections to the Muon Anomalous
  Magnetic Moment},''
\href{http://dx.doi.org/10.1103/PhysRevLett.87.071804}{{\em Phys.Rev.Lett.}
  {\bf 87} (2001)  071804}.

\bibitem{Kinoshita:2004wi}
T.~Kinoshita and M.~Nio, ``{Improved alpha**4 term of the Muon Anomalous
  Magnetic Moment},''
\href{http://dx.doi.org/10.1103/PhysRevD.70.113001}{{\em Phys.Rev.} {\bf D70}
  (2004)  113001}.

\bibitem{Passera:2004bj}
M.~Passera, ``{The Standard Model Prediction of the Muon Anomalous Magnetic
  Moment},'' \href{http://dx.doi.org/10.1088/0954-3899/31/5/R01}{{\em J.Phys.}
  {\bf G31} (2005)  R75--R94},
\href{http://arxiv.org/abs/hep-ph/0411168}{{\tt arXiv:hep-ph/0411168
  [hep-ph]}}.

\bibitem{Kinoshita:2005ti}
T.~Kinoshita, ``{Theory of Lepton g-2: Improvement of QED Terms},''
\href{http://dx.doi.org/10.1016/j.nuclphysbps.2005.02.029}{{\em
  Nucl.Phys.Proc.Suppl.} {\bf 144} (2005)  206--213}.

\bibitem{Hughes:1999fp}
V.~Hughes and T.~Kinoshita, ``{Anomalous g Values of the Electron and Muon},''
\href{http://dx.doi.org/10.1103/RevModPhys.71.S133}{{\em Rev.Mod.Phys.} {\bf
  71} (1999)  S133--S139}.

\bibitem{Kinoshita:2001pn}
T.~Kinoshita, ``{Everyone Makes Mistakes: Including Feynman},''
\href{http://dx.doi.org/10.1088/0954-3899/29/1/302}{{\em J.Phys.} {\bf G29}
  (2003)  9--22}.

\bibitem{Czarnecki:2001pv}
A.~Czarnecki and W.~J. Marciano, ``{The Muon Anomalous Magnetic Moment: A
  Harbinger for 'New Physics'},''
\href{http://dx.doi.org/10.1103/PhysRevD.64.013014}{{\em Phys.Rev.} {\bf D64}
  (2001)  013014}.

\bibitem{Davier:2004gb}
M.~Davier and W.~Marciano, ``{The Theoretical Prediction for the Muon Anomalous
  Magnetic Moment},''
\href{http://dx.doi.org/10.1146/annurev.nucl.54.070103.181204}{{\em
  Ann.Rev.Nucl.Part.Sci.} {\bf 54} (2004)  115--140}.

\bibitem{Kataev:2005av}
A.~Kataev, ``{The Improved 10th Order QED Expression for a(mu): New Results and
  Related Estimates},''
\href{http://dx.doi.org/10.1016/j.nuclphysbps.2006.02.104}{{\em
  Nucl.Phys.Proc.Suppl.} {\bf 155} (2006)  369--371}.

\bibitem{Kinoshita:2005sm}
T.~Kinoshita and M.~Nio, ``{The Tenth-Order QED Contribution to the Lepton g-2:
  Evaluation of Dominant alpha**5 Terms of Muon g-2},''
\href{http://dx.doi.org/10.1103/PhysRevD.73.053007}{{\em Phys.Rev.} {\bf D73}
  (2006)  053007}.

\bibitem{Miller:2007kk}
J.~P. Miller, E.~de~Rafael, and B.~L. Roberts, ``{Muon (g-2): Experiment and
  Theory},''
\href{http://dx.doi.org/10.1088/0034-4885/70/5/R03}{{\em Rept.Prog.Phys.} {\bf
  70} (2007)  795}.

\bibitem{Jegerlehner:2007xe}
F.~Jegerlehner, ``{Essentials of the Muon g-2},''
{\em Acta Phys.Polon.} {\bf B38} (2007)  3021.

\bibitem{Brodsky:1966mv}
S.~J. Brodsky and J.~D. Sullivan, ``{W Boson Contribution to the Anomalous
  Magnetic Moment of the Muon},''
\href{http://dx.doi.org/10.1103/PhysRev.156.1644}{{\em Phys.Rev.} {\bf 156}
  (1967)  1644--1647}.

\bibitem{Burnett:1967467}
T.~Burnett and M.~Levine, ``Intermediate vector boson contribution to the
  muon's anomalous magnetic moment,''
  \href{http://dx.doi.org/10.1016/0370-2693(67)90274-2}{{\em Phys.Lett.} {\bf
  B24} (1967)  467--468}.

\bibitem{Jackiw:1972jz}
R.~Jackiw and S.~Weinberg, ``{Weak Interaction Corrections to the Muon Magnetic
  Moment and to Muonic Atom Energy Levels},''
\href{http://dx.doi.org/10.1103/PhysRevD.5.2396}{{\em Phys.Rev.} {\bf D5}
  (1972)  2396--2398}.

\bibitem{Bars:1972pe}
I.~Bars and M.~Yoshimura, ``{Muon Magnetic Moment in a Finite Theory of Weak
  and Electromagnetic Interaction},''
\href{http://dx.doi.org/10.1103/PhysRevD.6.374}{{\em Phys.Rev.} {\bf D6} (1972)
   374--376}.

\bibitem{Fujikawa:1972fe}
K.~Fujikawa, B.~Lee, and A.~Sanda, ``{Generalized Renormalizable Gauge
  Formulation of Spontaneously Broken Gauge Theories},''
\href{http://dx.doi.org/10.1103/PhysRevD.6.2923}{{\em Phys.Rev.} {\bf D6}
  (1972)  2923--2943}.

\bibitem{Altarelli:1972nc}
G.~Altarelli, N.~Cabibbo, and L.~Maiani, ``{The Drell-Hearn Sum Rule and the
  Lepton Magnetic Moment in the Weinberg Model of Weak and Electromagnetic
  Interactions},''
\href{http://dx.doi.org/10.1016/0370-2693(72)90833-7}{{\em Phys.Lett.} {\bf
  B40} (1972)  415}.

\bibitem{Bardeen:1972vi}
W.~A. Bardeen, R.~Gastmans, and B.~Lautrup, ``{Static Quantities in Weinberg's
  Model of Weak and Electromagnetic Interactions},''
\href{http://dx.doi.org/10.1016/0550-3213(72)90218-0}{{\em Nucl.Phys.} {\bf
  B46} (1972)  319--331}.

\bibitem{Kukhto:1992qv}
T.~Kukhto, E.~Kuraev, Z.~Silagadze, and A.~Schiller, ``{The Dominant Two Loop
  Electroweak Contributions to the Anomalous Magnetic Moment of the Muon},''
\href{http://dx.doi.org/10.1016/0550-3213(92)90687-7}{{\em Nucl.Phys.} {\bf
  B371} (1992)  567--596}.

\bibitem{Peris:1995bb}
S.~Peris, M.~Perrottet, and E.~de~Rafael, ``{Two Loop Electroweak Corrections
  to the Muon g-2: A New Class of Hadronic Contributions},''
\href{http://dx.doi.org/10.1016/0370-2693(95)00768-G}{{\em Phys.Lett.} {\bf
  B355} (1995)  523--530}.

\bibitem{Czarnecki:1995wq}
A.~Czarnecki, B.~Krause, and W.~J. Marciano, ``{Electroweak Fermion Loop
  Contributions to the Muon Anomalous Magnetic Moment},''
\href{http://dx.doi.org/10.1103/PhysRevD.52.R2619}{{\em Phys.Rev.} {\bf D52}
  (1995)  2619--2623}.

\bibitem{Czarnecki:1995sz}
A.~Czarnecki, B.~Krause, and W.~J. Marciano, ``{Electroweak Corrections to the
  Muon Anomalous Magnetic Moment},''
\href{http://dx.doi.org/10.1103/PhysRevLett.76.3267}{{\em Phys.Rev.Lett.} {\bf
  76} (1996)  3267--3270}.

\bibitem{Degrassi:1998es}
G.~Degrassi and G.~Giudice, ``{QED Logarithms in the Electroweak Corrections to
  the Muon Anomalous Magnetic Moment},''
\href{http://dx.doi.org/10.1103/PhysRevD.58.053007}{{\em Phys.Rev.} {\bf D58}
  (1998)  053007}.

\bibitem{Melnikov:2003xd}
K.~Melnikov and A.~Vainshtein, ``{Hadronic Light-by-Light Scattering
  Contribution to the Muon Anomalous Magnetic Moment Revisited},''
\href{http://dx.doi.org/10.1103/PhysRevD.70.113006}{{\em Phys.Rev.} {\bf D70}
  (2004)  113006}.

\bibitem{Erler:2006vu}
J.~Erler and G.~T. Sanchez, ``{An Upper Bound on the Hadronic Light-by-Light
  Contribution to the Muon g-2},''
\href{http://dx.doi.org/10.1103/PhysRevLett.97.161801}{{\em Phys.Rev.Lett.}
  {\bf 97} (2006)  161801}.

\bibitem{Moroi:1995yh}
T.~Moroi, ``{The Muon Anomalous Magnetic Dipole Moment in the Minimal
  Supersymmetric Standard Model},''
\href{http://dx.doi.org/10.1103/PhysRevD.53.6565,
  10.1103/PhysRevD.56.4424}{{\em Phys.Rev.} {\bf D53} (1996)  6565--6575}.

\bibitem{Ibrahim:1999hh}
T.~Ibrahim and P.~Nath, ``{CP Violation and the Muon Anomaly in N=1
  Supergravity},''
\href{http://dx.doi.org/10.1103/PhysRevD.61.095008}{{\em Phys.Rev.} {\bf D61}
  (2000)  095008}.

\bibitem{Heinemeyer:2003dq}
S.~Heinemeyer, D.~Stockinger, and G.~Weiglein, ``{Two Loop SUSY Corrections to
  the Anomalous Magnetic Moment of the Muon},''
\href{http://dx.doi.org/10.1016/j.nuclphysb.2004.04.017}{{\em Nucl.Phys.} {\bf
  B690} (2004)  62--80}.

\bibitem{Heinemeyer:2004yq}
S.~Heinemeyer, D.~Stockinger, and G.~Weiglein, ``{Electroweak and
  Supersymmetric Two-Loop Corrections to (g-2)(mu)},''
\href{http://dx.doi.org/10.1016/j.nuclphysb.2004.08.014}{{\em Nucl.Phys.} {\bf
  B699} (2004)  103--123}.

\bibitem{Martin:2001st}
S.~P. Martin and J.~D. Wells, ``{Muon Anomalous Magnetic Dipole Moment in
  Supersymmetric Theories},''
\href{http://dx.doi.org/10.1103/PhysRevD.64.035003}{{\em Phys.Rev.} {\bf D64}
  (2001)  035003}.

\bibitem{Allanach:2009ne}
B.~Allanach, G.~Hiller, D.~Jones, and P.~Slavich, ``{Flavour Violation in
  Anomaly Mediated Supersymmetry Breaking},''
\href{http://dx.doi.org/10.1088/1126-6708/2009/04/088}{{\em JHEP} {\bf 0904}
  (2009)  088}.

\bibitem{Veltman:1977kh}
M.~Veltman, ``{Limit on Mass Differences in the Weinberg Model},''
\href{http://dx.doi.org/10.1016/0550-3213(77)90342-X}{{\em Nucl.Phys.} {\bf
  B123} (1977)  89}.

\bibitem{Sirlin:1980nh}
A.~Sirlin, ``{Radiative Corrections in the SU(2)-L x U(1) Theory: A Simple
  Renormalization Framework},''
\href{http://dx.doi.org/10.1103/PhysRevD.22.971}{{\em Phys.Rev.} {\bf D22}
  (1980)  971--981}.

\bibitem{Kennedy:1988rt}
D.~Kennedy, B.~Lynn, C.~Im, and R.~Stuart, ``{Electroweak Cross-Sections and
  Asymmetries at the Z0},''
\href{http://dx.doi.org/10.1016/0550-3213(89)90243-5}{{\em Nucl.Phys.} {\bf
  B321} (1989)  83}.

\bibitem{Bardin:1989di}
D.~Y. Bardin, M.~S. Bilenky, G.~Mitselmakher, G.~Mitselmakher, T.~Riemann, {\em
  et al.}, ``{A Realistic Approach to the Standard Z Peak},''
\href{http://dx.doi.org/10.1007/BF01415565}{{\em Z.Phys.} {\bf C44} (1989)
  493}.

\bibitem{Hollik:1988ii}
W.~Hollik, ``{Radiative Corrections in the Standard Model and their Role for
  Precision Tests of the Electroweak Theory},''
\href{http://dx.doi.org/10.1002/prop.2190380302}{{\em Fortsch.Phys.} {\bf 38}
  (1990)  165--260}.

\bibitem{Drees:1990dx}
M.~Drees and K.~Hagiwara, ``{Supersymmetric Contribution to the Electroweak
  $\rho$ Parameter},''
\href{http://dx.doi.org/10.1103/PhysRevD.42.1709}{{\em Phys.Rev.} {\bf D42}
  (1990)  1709--1725}.

\bibitem{Djouadi:1996pa}
A.~Djouadi, P.~Gambino, S.~Heinemeyer, W.~Hollik, C.~Junger, {\em et al.},
  ``{Supersymmetric Contributions to Electroweak Precision Observables: QCD
  Corrections},''
\href{http://dx.doi.org/10.1103/PhysRevLett.78.3626}{{\em Phys.Rev.Lett.} {\bf
  78} (1997)  3626--3629}.

\bibitem{Djouadi:1998sq}
A.~Djouadi, P.~Gambino, S.~Heinemeyer, W.~Hollik, C.~Junger, {\em et al.},
  ``{Leading QCD Corrections to Scalar Quark Contributions to Electroweak
  Precision Observables},''
\href{http://dx.doi.org/10.1103/PhysRevD.57.4179}{{\em Phys.Rev.} {\bf D57}
  (1998)  4179--4196}.

\bibitem{Heinemeyer:2002jq}
S.~Heinemeyer and G.~Weiglein, ``{Leading Electroweak Two Loop Corrections to
  Precision Observables in the MSSM},''
{\em JHEP} {\bf 0210} (2002)  072.

\bibitem{Heinemeyer:2004by}
S.~Heinemeyer, W.~Hollik, F.~Merz, and S.~Penaranda, ``{Electroweak Precision
  Observables in the MSSM with Nonminimal Flavor Violation},''
\href{http://dx.doi.org/10.1140/epjc/s2004-02006-1}{{\em Eur.Phys.J.} {\bf C37}
  (2004)  481--493}.

\bibitem{Komatsu:2010fb}
{\bf WMAP Collaboration}, E.~Komatsu {\em et al.}, ``{Seven-Year Wilkinson
  Microwave Anisotropy Probe (WMAP) Observations: Cosmological
  Interpretation},''
\href{http://dx.doi.org/10.1088/0067-0049/192/2/18}{{\em Astrophys.J.Suppl.}
  {\bf 192} (2011)  18}.

\bibitem{Ade:2013zuv}
{\bf Planck Collaboration}, P.~Ade {\em et al.}, ``{Planck 2013 results. XVI.
  Cosmological parameters},''
\href{http://arxiv.org/abs/1303.5076}{{\tt arXiv:1303.5076 [astro-ph.CO]}}.

\bibitem{Ibanez:1983kw}
L.~E. Ibanez, ``{The Scalar Neutrinos as the Lightest Supersymmetric Particles
  and Cosmology},''
\href{http://dx.doi.org/10.1016/0370-2693(84)90221-1}{{\em Phys.Lett.} {\bf
  B137} (1984)  160}.

\bibitem{Hagelin:1984wv}
J.~S. Hagelin, G.~L. Kane, and S.~Raby, ``{Perhaps Scalar Neutrinos Are the
  Lightest Supersymmetric Partners},''
\href{http://dx.doi.org/10.1016/0550-3213(84)90064-6}{{\em Nucl.Phys.} {\bf
  B241} (1984)  638}.

\bibitem{Falk:1994es}
T.~Falk, K.~A. Olive, and M.~Srednicki, ``{Heavy Sneutrinos as Dark Matter},''
\href{http://dx.doi.org/10.1016/0370-2693(94)90639-4}{{\em Phys.Lett.} {\bf
  B339} (1994)  248--251}.

\bibitem{Bertone:2004pz}
G.~Bertone, D.~Hooper, and J.~Silk, ``{Particle Dark Matter: Evidence,
  Candidates and Constraints},''
\href{http://dx.doi.org/10.1016/j.physrep.2004.08.031}{{\em Phys.Rept.} {\bf
  405} (2005)  279--390}.

\bibitem{Griest:1990kh}
K.~Griest and D.~Seckel, ``{Three Exceptions in the Calculation of Relic
  Abundances},''
\href{http://dx.doi.org/10.1103/PhysRevD.43.3191}{{\em Phys.Rev.} {\bf D43}
  (1991)  3191--3203}.

\bibitem{Edsjo:1997bg}
J.~Edsjo and P.~Gondolo, ``{Neutralino Relic Density Including
  Coannihilations},''
\href{http://dx.doi.org/10.1103/PhysRevD.56.1879}{{\em Phys.Rev.} {\bf D56}
  (1997)  1879--1894}.

\bibitem{Herrmann:2011xe}
B.~Herrmann, M.~Klasen, and Q.~Le~Boulc'h, ``{Impact of Squark Flavour
  Violation on Neutralino Dark Matter},''
\href{http://dx.doi.org/10.1103/PhysRevD.84.095007}{{\em Phys.Rev.} {\bf D84}
  (2011)  095007}.

\bibitem{Gondolo:2004sc}
P.~Gondolo, J.~Edsjo, P.~Ullio, L.~Bergstrom, M.~Schelke, {\em et al.},
  ``{DarkSUSY: Computing Supersymmetric Dark Matter Properties Numerically},''
\href{http://dx.doi.org/10.1088/1475-7516/2004/07/008}{{\em JCAP} {\bf 0407}
  (2004)  008}.

\bibitem{Kamionkowski:1990ni}
M.~Kamionkowski and M.~S. Turner, ``{THERMAL RELICS: DO WE KNOW THEIR
  ABUNDANCES?},''
\href{http://dx.doi.org/10.1103/PhysRevD.42.3310}{{\em Phys.Rev.} {\bf D42}
  (1990)  3310--3320}.

\bibitem{Giudice:2000ex}
G.~F. Giudice, E.~W. Kolb, and A.~Riotto, ``{Largest temperature of the
  radiation era and its cosmological implications},''
\href{http://dx.doi.org/10.1103/PhysRevD.64.023508}{{\em Phys.Rev.} {\bf D64}
  (2001)  023508}.

\bibitem{Salati:2002md}
P.~Salati, ``{Quintessence and the relic density of neutralinos},''
\href{http://dx.doi.org/10.1016/j.physletb.2003.07.073}{{\em Phys.Lett.} {\bf
  B571} (2003)  121--131}.

\bibitem{Gelmini:2006pq}
G.~Gelmini, P.~Gondolo, A.~Soldatenko, and C.~E. Yaguna, ``{The Effect of a
  late decaying scalar on the neutralino relic density},''
\href{http://dx.doi.org/10.1103/PhysRevD.74.083514}{{\em Phys.Rev.} {\bf D74}
  (2006)  083514}.

\bibitem{Arbey:2008kv}
A.~Arbey and F.~Mahmoudi, ``{SUSY Constraints from Relic Density: High
  Sensitivity to Pre-BBN Expansion Rate},''
\href{http://dx.doi.org/10.1016/j.physletb.2008.09.032}{{\em Phys.Lett.} {\bf
  B669} (2008)  46--51}.

\bibitem{Arbey:2009gt}
A.~Arbey and F.~Mahmoudi, ``{SUSY Constraints, Relic Density, and Very Early
  Universe},''
\href{http://dx.doi.org/10.1007/JHEP05(2010)051}{{\em JHEP} {\bf 1005} (2010)
  051}.

\bibitem{Arbey:2011gu}
A.~Arbey, A.~Deandrea, and A.~Tarhini, ``{Anomaly Mediated SUSY Breaking
  Scenarios in the Light of Cosmology and in the Dark (Matter)},''
\href{http://dx.doi.org/10.1007/JHEP05(2011)078}{{\em JHEP} {\bf 1105} (2011)
  078}.

\bibitem{Bolz:2000fu}
M.~Bolz, A.~Brandenburg, and W.~Buchmuller, ``{Thermal Production of
  Gravitinos},''
\href{http://dx.doi.org/10.1016/S0550-3213(01)00132-8,
  10.1016/j.nuclphysb.2007.09.020}{{\em Nucl.Phys.} {\bf B606} (2001)
  518--544}.

\bibitem{Pradler:2006qh}
J.~Pradler and F.~D. Steffen, ``{Thermal Gravitino Production and Collider
  Tests of Leptogenesis},''
\href{http://dx.doi.org/10.1103/PhysRevD.75.023509}{{\em Phys.Rev.} {\bf D75}
  (2007)  023509}.

\bibitem{Rychkov:2007uq}
V.~S. Rychkov and A.~Strumia, ``{Thermal Production of Gravitinos},''
\href{http://dx.doi.org/10.1103/PhysRevD.75.075011}{{\em Phys.Rev.} {\bf D75}
  (2007)  075011}.

\bibitem{Buchmuller:2004nz}
W.~Buchmuller, P.~Di~Bari, and M.~Plumacher, ``{Leptogenesis for
  Pedestrians},''
\href{http://dx.doi.org/10.1016/j.aop.2004.02.003}{{\em Annals Phys.} {\bf 315}
  (2005)  305--351}.

\bibitem{Belanger:2001fz}
G.~Belanger, F.~Boudjema, A.~Pukhov, and A.~Semenov, ``{MicrOMEGAs: A Program
  for Calculating the Relic Density in the MSSM},''
\href{http://dx.doi.org/10.1016/S0010-4655(02)00596-9}{{\em
  Comput.Phys.Commun.} {\bf 149} (2002)  103--120}.

\bibitem{Belanger:2004yn}
G.~Belanger, F.~Boudjema, A.~Pukhov, and A.~Semenov, ``{micrOMEGAs: Version
  1.3},''
\href{http://dx.doi.org/10.1016/j.cpc.2005.12.005}{{\em Comput.Phys.Commun.}
  {\bf 174} (2006)  577--604}.

\bibitem{Belanger:2006is}
G.~Belanger, F.~Boudjema, A.~Pukhov, and A.~Semenov, ``{MicrOMEGAs 2.0: A
  Program to Calculate the Relic Density of Dark Matter in a Generic Model},''
\href{http://dx.doi.org/10.1016/j.cpc.2006.11.008}{{\em Comput.Phys.Commun.}
  {\bf 176} (2007)  367--382}.

\bibitem{Belanger:2008sj}
G.~Belanger, F.~Boudjema, A.~Pukhov, and A.~Semenov, ``{Dark Matter Direct
  Detection Rate in a Generic Model with MicrOMEGAs 2.2},''
\href{http://dx.doi.org/10.1016/j.cpc.2008.11.019}{{\em Comput.Phys.Commun.}
  {\bf 180} (2009)  747--767}.

\bibitem{Belanger:2010gh}
G.~Belanger, F.~Boudjema, P.~Brun, A.~Pukhov, S.~Rosier-Lees, {\em et al.},
  ``{Indirect Search for Dark Matter with MicrOMEGAs2.4},''
\href{http://dx.doi.org/10.1016/j.cpc.2010.11.033}{{\em Comput.Phys.Commun.}
  {\bf 182} (2011)  842--856}.

\bibitem{Hamann:2006pf}
J.~Hamann, S.~Hannestad, M.~S. Sloth, and Y.~Y. Wong, ``{How Robust are
  Inflation Model and Dark Matter Constraints from Cosmological Data?},''
\href{http://dx.doi.org/10.1103/PhysRevD.75.023522}{{\em Phys.Rev.} {\bf D75}
  (2007)  023522}.

\bibitem{Pospelov:2008ta}
M.~Pospelov, J.~Pradler, and F.~D. Steffen, ``{Constraints on Supersymmetric
  Models from Catalytic Primordial Nucleosynthesis of Beryllium},''
\href{http://dx.doi.org/10.1088/1475-7516/2008/11/020}{{\em JCAP} {\bf 0811}
  (2008)  020}.

\bibitem{Lazarides:1991wu}
G.~Lazarides and Q.~Shafi, ``{Origin of Matter in the Inflationary
  Cosmology},''
\href{http://dx.doi.org/10.1016/0370-2693(91)91090-I}{{\em Phys.Lett.} {\bf
  B258} (1991)  305--309}.

\bibitem{Murayama:1993em}
H.~Murayama and T.~Yanagida, ``{Leptogenesis in Supersymmetric Standard Model
  with Right-Handed Neutrino},''
\href{http://dx.doi.org/10.1016/0370-2693(94)91164-9}{{\em Phys.Lett.} {\bf
  B322} (1994)  349--354}.

\bibitem{Affleck:1984fy}
I.~Affleck and M.~Dine, ``{A New Mechanism for Baryogenesis},''
\href{http://dx.doi.org/10.1016/0550-3213(85)90021-5}{{\em Nucl.Phys.} {\bf
  B249} (1985)  361}.

\bibitem{Chen:1996ap}
C.~Chen, M.~Drees, and J.~Gunion, ``{A Nonstandard String / SUSY Scenario and
  its Phenomenological Implications},''
\href{http://dx.doi.org/10.1103/PhysRevD.60.039901,
  10.1103/PhysRevD.55.330}{{\em Phys.Rev.} {\bf D55} (1997)  330--347}.

\bibitem{Moroi:1999zb}
T.~Moroi and L.~Randall, ``{Wino Cold Dark Matter from Anomaly Mediated SUSY
  Breaking},''
\href{http://dx.doi.org/10.1016/S0550-3213(99)00748-8}{{\em Nucl.Phys.} {\bf
  B570} (2000)  455--472}.

\bibitem{Covi:1999ty}
L.~Covi, J.~E. Kim, and L.~Roszkowski, ``{Axinos as Cold Dark Matter},''
\href{http://dx.doi.org/10.1103/PhysRevLett.82.4180}{{\em Phys.Rev.Lett.} {\bf
  82} (1999)  4180--4183}.

\bibitem{Covi:2001nw}
L.~Covi, H.-B. Kim, J.~E. Kim, and L.~Roszkowski, ``{Axinos as Dark Matter},''
{\em JHEP} {\bf 0105} (2001)  033.

\bibitem{Baer:2010kd}
H.~Baer, R.~Dermisek, S.~Rajagopalan, and H.~Summy, ``{Neutralino, Axion and
  Axino Cold Dark Matter in Minimal, Hypercharged and Gaugino AMSB},''
  \href{http://dx.doi.org/10.1088/1475-7516/2010/07/014}{{\em JCAP} {\bf 1007}
  (2010)  014},
\href{http://arxiv.org/abs/1004.3297}{{\tt arXiv:1004.3297 [hep-ph]}}.

\bibitem{Acharya:2009zt}
B.~S. Acharya, G.~Kane, S.~Watson, and P.~Kumar, ``{A Non-Thermal WIMP
  Miracle},''
\href{http://dx.doi.org/10.1103/PhysRevD.80.083529}{{\em Phys.Rev.} {\bf D80}
  (2009)  083529}.

\bibitem{Arbey:2012dq}
A.~Arbey, M.~Battaglia, A.~Djouadi, and F.~Mahmoudi, ``{The Higgs sector of the
  phenomenological MSSM in the light of the Higgs boson discovery},''
\href{http://dx.doi.org/10.1007/JHEP09(2012)107}{{\em JHEP} {\bf 1209} (2012)
  107}.

\bibitem{Brummer:2012ns}
F.~Brummer, S.~Kraml, and S.~Kulkarni, ``{Anatomy of maximal stop mixing in the
  MSSM},''
\href{http://dx.doi.org/10.1007/JHEP08(2012)089}{{\em JHEP} {\bf 1208} (2012)
  089}.

\bibitem{Wymant:2012zp}
C.~Wymant, ``{Optimising Stop Naturalness},''
\href{http://dx.doi.org/10.1103/PhysRevD.86.115023}{{\em Phys.Rev.} {\bf D86}
  (2012)  115023}.

\bibitem{Beenakker:1996ch}
W.~Beenakker, R.~Hopker, M.~Spira, and P.~Zerwas, ``{Squark and Gluino
  Production at Hadron Colliders},''
\href{http://dx.doi.org/10.1016/S0550-3213(97)80027-2}{{\em Nucl.Phys.} {\bf
  B492} (1997)  51--103}.

\bibitem{Beenakker:1997ut}
W.~Beenakker, M.~Kramer, T.~Plehn, M.~Spira, and P.~Zerwas, ``{Stop Production
  at Hadron Colliders},''
\href{http://dx.doi.org/10.1016/S0550-3213(98)00014-5}{{\em Nucl.Phys.} {\bf
  B515} (1998)  3--14}.

\bibitem{Berger:1998kh}
E.~L. Berger, M.~Klasen, and T.~M. Tait, ``{Scale Dependence of Squark and
  Gluino Production Cross-Sections},''
\href{http://dx.doi.org/10.1103/PhysRevD.59.074024}{{\em Phys.Rev.} {\bf D59}
  (1999)  074024}.

\bibitem{Berger:1999mc}
E.~L. Berger, M.~Klasen, and T.~M. Tait, ``{Associated Production of Gauginos
  and Gluinos at Hadron Colliders in Next-to-Leading Order SUSY QCD},''
\href{http://dx.doi.org/10.1016/S0370-2693(99)00617-6}{{\em Phys.Lett.} {\bf
  B459} (1999)  165--170}.

\bibitem{Berger:2000iu}
E.~L. Berger, M.~Klasen, and T.~M. Tait, ``{Next-to-Leading Order SUSY QCD
  Predictions for Associated Production of Gauginos and Gluinos},''
\href{http://dx.doi.org/10.1103/PhysRevD.62.095014}{{\em Phys.Rev.} {\bf D62}
  (2000)  095014}.

\bibitem{Spira:2002rd}
M.~Spira, ``{Higgs and SUSY Particle Production at Hadron Colliders},''
\href{http://arxiv.org/abs/hep-ph/0211145}{{\tt arXiv:hep-ph/0211145
  [hep-ph]}}.

\bibitem{Jin:2002nu}
L.-G. Jin, C.-S. Li, and J.~J. Liu, ``{Next-to-Leading Order QCD Predictions
  for Associated Production of Top Squarks and Charginos},''
\href{http://dx.doi.org/10.1140/epjc/s2003-01261-x}{{\em Eur.Phys.J.} {\bf C30}
  (2003)  77}.

\bibitem{Jin:2003ez}
L.~G. Jin, C.~S. Li, and J.~J. Liu, ``{Associated Production of Top Squarks and
  Charginos at the CERN LHC in NLO SUSY QCD},''
\href{http://dx.doi.org/10.1016/S0370-2693(03)00422-2}{{\em Phys.Lett.} {\bf
  B561} (2003)  135--144}.

\bibitem{Hollik:2007wf}
W.~Hollik, M.~Kollar, and M.~K. Trenkel, ``{Hadronic Production of Top-Squark
  Pairs with Electroweak NLO Contributions},''
\href{http://dx.doi.org/10.1088/1126-6708/2008/02/018}{{\em JHEP} {\bf 0802}
  (2008)  018}.

\bibitem{Hollik:2008vm}
W.~Hollik, E.~Mirabella, and M.~K. Trenkel, ``{Electroweak Contributions to
  Squark-Gluino Production at the LHC},''
\href{http://dx.doi.org/10.1088/1126-6708/2009/02/002}{{\em JHEP} {\bf 0902}
  (2009)  002}.

\bibitem{Mirabella:2009ap}
E.~Mirabella, ``{NLO Electroweak Contributions to Gluino Pair Production at
  Hadron Colliders},''
\href{http://dx.doi.org/10.1088/1126-6708/2009/12/012}{{\em JHEP} {\bf 0912}
  (2009)  012}.

\bibitem{Kulesza:2008jb}
A.~Kulesza and L.~Motyka, ``{Threshold Resummation for Squark-Antisquark and
  Gluino-Pair Production at the LHC},''
\href{http://dx.doi.org/10.1103/PhysRevLett.102.111802}{{\em Phys.Rev.Lett.}
  {\bf 102} (2009)  111802}.

\bibitem{Kulesza:2009kq}
A.~Kulesza and L.~Motyka, ``{Soft Gluon Resummation for the Production of
  Gluino-Gluino and Squark-Antisquark Pairs at the LHC},''
\href{http://dx.doi.org/10.1103/PhysRevD.80.095004}{{\em Phys.Rev.} {\bf D80}
  (2009)  095004}.

\bibitem{Beenakker:2009ha}
W.~Beenakker, S.~Brensing, M.~Kramer, A.~Kulesza, E.~Laenen, {\em et al.},
  ``{Soft-Gluon Resummation for Squark and Gluino Hadroproduction},''
\href{http://dx.doi.org/10.1088/1126-6708/2009/12/041}{{\em JHEP} {\bf 0912}
  (2009)  041}.

\bibitem{Beenakker:2010nq}
W.~Beenakker, S.~Brensing, M.~Kramer, A.~Kulesza, E.~Laenen, {\em et al.},
  ``{Supersymmetric Top and Bottom Squark Production at Hadron Colliders},''
\href{http://dx.doi.org/10.1007/JHEP08(2010)098}{{\em JHEP} {\bf 1008} (2010)
  098}.

\bibitem{Beenakker:2011fu}
W.~Beenakker, S.~Brensing, M.~Kramer, A.~Kulesza, E.~Laenen, {\em et al.},
  ``{Squark and Gluino Hadroproduction},''
\href{http://dx.doi.org/10.1142/S0217751X11053560}{{\em Int.J.Mod.Phys.} {\bf
  A26} (2011)  2637--2664}.

\bibitem{Kramer:2012bx}
M.~Kramer, A.~Kulesza, R.~van~der Leeuw, M.~Mangano, S.~Padhi, {\em et al.},
  ``{Supersymmetry Production Cross Sections in $pp$ Collisions at $\sqrt{s}=7$
  TeV},''
\href{http://arxiv.org/abs/1206.2892}{{\tt arXiv:1206.2892 [hep-ph]}}.

\bibitem{Baer:1997nh}
H.~Baer, B.~Harris, and M.~H. Reno, ``{Next-to-Leading Order Slepton Pair
  Production at Hadron Colliders},''
\href{http://dx.doi.org/10.1103/PhysRevD.57.5871}{{\em Phys.Rev.} {\bf D57}
  (1998)  5871--5874}.

\bibitem{Beenakker:1999xh}
W.~Beenakker, M.~Klasen, M.~Kramer, T.~Plehn, M.~Spira, {\em et al.}, ``{The
  Production of Charginos / Neutralinos and Sleptons at Hadron Colliders},''
\href{http://dx.doi.org/10.1103/PhysRevLett.100.029901,
  10.1103/PhysRevLett.83.3780}{{\em Phys.Rev.Lett.} {\bf 83} (1999)
  3780--3783}.

\bibitem{Bozzi:2004qq}
G.~Bozzi, B.~Fuks, and M.~Klasen, ``{Slepton production in polarized hadron
  collisions},''
\href{http://dx.doi.org/10.1016/j.physletb.2005.01.060}{{\em Phys.Lett.} {\bf
  B609} (2005)  339--350}.

\bibitem{Bozzi:2006fw}
G.~Bozzi, B.~Fuks, and M.~Klasen, ``{Transverse-momentum resummation for
  slepton-pair production at the CERN LHC},''
\href{http://dx.doi.org/10.1103/PhysRevD.74.015001}{{\em Phys.Rev.} {\bf D74}
  (2006)  015001}.

\bibitem{Bozzi:2007qr}
G.~Bozzi, B.~Fuks, and M.~Klasen, ``{Threshold Resummation for Slepton-Pair
  Production at Hadron Colliders},''
\href{http://dx.doi.org/10.1016/j.nuclphysb.2007.03.052}{{\em Nucl.Phys.} {\bf
  B777} (2007)  157--181}.

\bibitem{Bozzi:2007tea}
G.~Bozzi, B.~Fuks, and M.~Klasen, ``{Joint resummation for slepton pair
  production at hadron colliders},''
\href{http://dx.doi.org/10.1016/j.nuclphysb.2007.10.021}{{\em Nucl.Phys.} {\bf
  B794} (2008)  46--60}.

\bibitem{Fuks:2012qx}
B.~Fuks, M.~Klasen, D.~R. Lamprea, and M.~Rothering, ``{Gaugino Production in
  Proton-Proton Collisions at a Center-of-Mass Energy of 8 TeV},''
\href{http://dx.doi.org/10.1007/JHEP10(2012)081}{{\em JHEP} {\bf 1210} (2012)
  081}.

\bibitem{Fuks:2013lya}
B.~Fuks, M.~Klasen, D.~R. Lamprea, and M.~Rothering, ``{Revisiting slepton pair
  production at the Large Hadron Collider},''
\href{http://arxiv.org/abs/1310.2621}{{\tt arXiv:1310.2621 [hep-ph]}}.

\bibitem{Aad:2012hm}
{\bf ATLAS Collaboration}, G.~Aad {\em et al.}, ``{Hunt for New Phenomena Using
  Large Jet Multiplicities and Missing Transverse Momentum with ATLAS in 4.7
  fb$^{-1}$ of $\sqrt{s}=7$ TeV Proton-Proton Collisions},''
\href{http://dx.doi.org/10.1007/JHEP07(2012)167}{{\em JHEP} {\bf 1207} (2012)
  167}.

\bibitem{Aad:2012pq}
{\bf ATLAS Collaboration}, G.~Aad {\em et al.}, ``{Search for Top and Bottom
  Squarks from Gluino Pair Production in Final States with Missing Transverse
  Energy and at Least Three b-jets with the ATLAS Detector},''
\href{http://dx.doi.org/10.1140/epjc/s10052-012-2174-z}{{\em Eur.Phys.J.} {\bf
  C72} (2012)  2174}.

\bibitem{Aad:2012rz}
{\bf ATLAS Collaboration}, G.~Aad {\em et al.}, ``{Search for Squarks and
  Gluinos with the ATLAS Detector in Final States with Jets and Missing
  Transverse Momentum Using 4.7 fb$^{-1}$ of $\sqrt{s}=7$ TeV Proton-Proton
  collision data},''
\href{http://arxiv.org/abs/1208.0949}{{\tt arXiv:1208.0949 [hep-ex]}}.

\bibitem{Aad:2012cj}
{\bf ATLAS Collaboration}, G.~Aad {\em et al.}, ``{Search for Supersymmetry in
  Events with Photons, Bottom Quarks, and Missing Transverse Momentum in
  Proton-Proton Collisions at a Centre-of-Mass Energy of 7 TeV with the ATLAS
  Detector},''
\href{http://arxiv.org/abs/1211.1167}{{\tt arXiv:1211.1167 [hep-ex]}}.

\bibitem{ATLAS145}
{\bf ATLAS Collaboration}, ``Search for gluino pair production in final states
  with missing transverse momentum and at least three b-jets using 12.8
  fb$^{-1}$ of pp collisions at sqrt(s) = 8 tev with the atlas detector.,''
  \href{http://arxiv.org/abs/ATLAS-CONF-2012-145}{{\tt ATLAS-CONF-2012-145}}.

\bibitem{ATLAS151}
{\bf ATLAS Collaboration}, ``Search for supersymmetry using events with three
  leptons, multiple jets, and missing transverse momentum in 13.0 fb$^{-1}$ of
  pp collisions with the atlas detector at $\sqrt{s}=8$ tev,''
  \href{http://arxiv.org/abs/ATLAS-CONF-2012-151}{{\tt ATLAS-CONF-2012-151}}.

\bibitem{Chatrchyan:2012jx}
{\bf CMS Collaboration}, S.~Chatrchyan {\em et al.}, ``{Search for
  Supersymmetry in Hadronic Final States Using MT2 in $pp$ Collisions at
  $\sqrt{s} = 7$ TeV},''
\href{http://dx.doi.org/10.1007/JHEP10(2012)018}{{\em JHEP} {\bf 1210} (2012)
  018}.

\bibitem{Chatrchyan:2012mfa}
{\bf CMS Collaboration}, S.~Chatrchyan {\em et al.}, ``{Search for New Physics
  in the Multijet and Missing Transverse Momentum Final State in Proton-Proton
  Collisions at $\sqrt{s} = 7$ TeV},''
\href{http://dx.doi.org/10.1103/PhysRevLett.109.171803}{{\em Phys.Rev.Lett.}
  {\bf 109} (2012)  171803}.

\bibitem{Chatrchyan:2012wa}
{\bf CMS Collaboration}, S.~Chatrchyan {\em et al.}, ``{Search for
  Supersymmetry in Final States with Missing Transverse Energy and 0, 1, 2, or
  at least 3 b-quark Jets in 7 TeV pp Collisions using the Variable alphaT},''
\href{http://arxiv.org/abs/1210.8115}{{\tt arXiv:1210.8115 [hep-ex]}}.

\bibitem{Chatrchyan:2012pc}
{\bf CMS Collaboration}, S.~Chatrchyan {\em et al.}, ``{Search for
  Supersymmetry in Final States with a Single Lepton, b-quark Jets, and Missing
  Transverse Energy in Proton-Proton Collisions at sqrt(s) = 7 TeV},''
\href{http://arxiv.org/abs/1211.3143}{{\tt arXiv:1211.3143 [hep-ex]}}.

\bibitem{CMSSUS028}
{\bf CMS collaboration}, ``Search for supersymmetry in final states with
  missing transverse energy and 0, 1, 2, 3, or at least 4 b-quark jets in 8 tev
  pp collisions using the variable alphat,'' \href{http://arxiv.org/abs/CMS PAS
  SUS-12-028}{{\tt CMS PAS SUS-12-028}}.

\bibitem{CMSSUS029}
{\bf CMS collaboration}, ``Search for supersymmetry in events with same-sign
  dileptons and b-tagged jets with 8 tev data,'' \href{http://arxiv.org/abs/CMS
  PAS SUS-12-029}{{\tt CMS PAS SUS-12-029}}.

\bibitem{Aad:2012gg}
{\bf ATLAS Collaboration}, G.~Aad {\em et al.}, ``{Search for Direct Slepton
  and Gaugino Production in Final States with Two Leptons and Missing
  Transverse Momentum with the ATLAS Detector in $pp$ Collisions at
  $\sqrt{s}=7$ TeV},''
\href{http://arxiv.org/abs/1208.2884}{{\tt arXiv:1208.2884 [hep-ex]}}.

\bibitem{Aad:2012si}
{\bf ATLAS Collaboration}, G.~Aad {\em et al.}, ``{Search for a Supersymmetric
  Partner to the Top Quark in Final States with Jets and Missing Transverse
  Momentum at $\sqrt{s}=7$ TeV with the ATLAS Detector},''
\href{http://arxiv.org/abs/1208.1447}{{\tt arXiv:1208.1447 [hep-ex]}}.

\bibitem{Aad:2012ar}
{\bf ATLAS Collaboration}, G.~Aad {\em et al.}, ``{Search for Direct Top Squark
  Pair Production in Final States with one Isolated Lepton, Jets, and Missing
  Transverse Momentum in $\sqrt{s}=7$ TeV $pp$ Collisions Using 4.7 $fb^{-1}$
  of ATLAS Data},''
\href{http://arxiv.org/abs/1208.2590}{{\tt arXiv:1208.2590 [hep-ex]}}.

\bibitem{Aad:2012uu}
{\bf ATLAS Collaboration}, G.~Aad {\em et al.}, ``{Search for a Heavy Top-Quark
  Partner in Final States with Two Leptons with the ATLAS Detector at the
  LHC},''
\href{http://dx.doi.org/10.1007/JHEP11(2012)094}{{\em JHEP} {\bf 1211} (2012)
  094}.

\bibitem{CMSSUS023}
{\bf CMS collaboration}, ``Search for direct top squark pair production in
  events with a single isolated lepton, jets and missing transverse energy at
  sqrt(s) = 8 tev,'' \href{http://arxiv.org/abs/CMS PAS SUS-12-023}{{\tt CMS
  PAS SUS-12-023}}.

\bibitem{Chatrchyan:2012ewa}
{\bf CMS Collaboration}, S.~Chatrchyan {\em et al.}, ``{Search for electroweak
  production of charginos and neutralinos using leptonic final states in $pp$
  collisions at $\sqrt{s}=7$ TeV},'' {\em JHEP} (2012)  ,
\href{http://arxiv.org/abs/1209.6620}{{\tt arXiv:1209.6620 [hep-ex]}}.

\bibitem{CMSSUS022}
{\bf CMS collaboration}, ``Search for direct ewk production of susy particles
  in multilepton modes with 8tev data,'' \href{http://arxiv.org/abs/CMS PAS
  SUS-12-022}{{\tt CMS PAS SUS-12-022}}.

\bibitem{Aad:2012ku}
{\bf ATLAS Collaboration}, G.~Aad {\em et al.}, ``{Search for Direct Production
  of Charginos and Neutralinos in Events with Three Leptons and Missing
  Transverse Momentum in $\sqrt{s}=7$ TeV $pp$ Collisions with the ATLAS
  Detector},''
\href{http://arxiv.org/abs/1208.3144}{{\tt arXiv:1208.3144 [hep-ex]}}.

\bibitem{ATLAS154}
{\bf ATLAS Collaboration}, ``Search for direct production of charginos and
  neutralinos in events with three leptons and missing transverse momentum in
  13.0 fb$^{-1}$ of pp collisions at $\sqrt{s}$=8 tev with the atlas
  detector,''.

\bibitem{Aad:2012afb}
{\bf ATLAS Collaboration}, G.~Aad {\em et al.}, ``{Search for Diphoton Events
  with Large Missing Transverse Momentum in 7 TeV Proton-Proton Collision Data
  with the ATLAS Detector},''
\href{http://arxiv.org/abs/1209.0753}{{\tt arXiv:1209.0753 [hep-ex]}}.

\bibitem{ATLAS:2012ht}
{\bf ATLAS Collaboration}, G.~Aad {\em et al.}, ``{Search for Supersymmetry in
  Events with Large Missing Transverse Momentum, Jets, and at Least One Tau
  Lepton in 7 TeV Proton-Proton Collision Data with the ATLAS Detector},''
\href{http://arxiv.org/abs/1210.1314}{{\tt arXiv:1210.1314 [hep-ex]}}.

\bibitem{ATLAS147}
{\bf ATLAS Collaboration}, ``Search for new phenomena in monojet plus missing
  transverse momentum final states using 10fb$^{-1}$ of pp collisions at
  $sqrt{s}=8$ tev with the atlas detector at the lhc,''
  \href{http://arxiv.org/abs/ATLAS-CONF-2012-147}{{\tt ATLAS-CONF-2012-147}}.

\bibitem{Chatrchyan:2012mx}
{\bf CMS Collaboration}, S.~Chatrchyan {\em et al.}, ``{Search for New Physics
  in Events with Photons, Jets, and Missing Transverse energy in pp Collisions
  at sqrt(s) = 7 TeV},''
\href{http://arxiv.org/abs/1211.4784}{{\tt arXiv:1211.4784 [hep-ex]}}.

\bibitem{Werner:PC}
W.~Porod, ``{Private communication},''.

\bibitem{Arbey:2011ab}
A.~Arbey, M.~Battaglia, A.~Djouadi, F.~Mahmoudi, and J.~Quevillon,
  ``{Implications of a 125 GeV Higgs for supersymmetric models},''
\href{http://dx.doi.org/10.1016/j.physletb.2012.01.053}{{\em Phys.Lett.} {\bf
  B708} (2012)  162--169}.

\bibitem{ATLAS:2012jp}
{\bf ATLAS Collaboration}, ``{Search for Direct Ghargino Production in
  Anomaly-Mediated Supersymmetry Breaking Models Based on a Disappearing-Track
  Signature in $pp$ Collisions at $\sqrt{s}=7$ TeV with the ATLAS Detector},''
\href{http://arxiv.org/abs/1210.2852}{{\tt arXiv:1210.2852 [hep-ex]}}.

\bibitem{Fayet:1975yi}
P.~Fayet, ``{Fermi-Bose Hypersymmetry},''
\href{http://dx.doi.org/10.1016/0550-3213(76)90458-2}{{\em Nucl.Phys.} {\bf
  B113} (1976)  135}.

\bibitem{AlvarezGaume:1996mv}
L.~Alvarez-Gaume and S.~Hassan, ``{Introduction to S Duality in N=2
  Supersymmetric Gauge Theories: A Pedagogical Review of the Work of Seiberg
  and Witten},''
{\em Fortsch.Phys.} {\bf 45} (1997)  159--236.

\bibitem{Plehn:2008ae}
T.~Plehn and T.~M. Tait, ``{Seeking Sgluons},''
\href{http://dx.doi.org/10.1088/0954-3899/36/7/075001}{{\em J.Phys.G} {\bf G36}
  (2009)  075001}.

\bibitem{Choi:2008pi}
S.~Choi, M.~Drees, A.~Freitas, and P.~Zerwas, ``{Testing the Majorana Nature of
  Gluinos and Neutralinos},''
\href{http://dx.doi.org/10.1103/PhysRevD.78.095007}{{\em Phys.Rev.} {\bf D78}
  (2008)  095007}.

\bibitem{Choi:2008ub}
S.~Choi, M.~Drees, J.~Kalinowski, J.~Kim, E.~Popenda, {\em et al.},
  ``{Color-Octet Scalars of N=2 Supersymmetry at the LHC},''
\href{http://dx.doi.org/10.1016/j.physletb.2009.01.040}{{\em Phys.Lett.} {\bf
  B672} (2009)  246--252}.

\bibitem{Choi:2009jc}
S.~Choi, M.~Drees, J.~Kalinowski, J.~Kim, E.~Popenda, {\em et al.},
  ``{Color-Octet Scalars at the LHC},''
{\em Acta Phys.Polon.} {\bf B40} (2009)  1947--1956.

\bibitem{choi:2010gc}
S.~Choi, D.~Choudhury, A.~Freitas, J.~Kalinowski, J.~Kim, {\em et al.},
  ``{Dirac Neutralinos and Electroweak Scalar Bosons of N=1/N=2 Hybrid
  Supersymmetry at Colliders},''
\href{http://dx.doi.org/10.1007/JHEP08(2010)025}{{\em JHEP} {\bf 1008} (2010)
  025}.

\bibitem{Choi:2010an}
S.~Choi, D.~Choudhury, A.~Freitas, J.~Kalinowski, and P.~Zerwas, ``{The
  Extended Higgs System in $R$-symmetric Supersymmetry Theories},''
\href{http://dx.doi.org/10.1016/j.physletb.2011.01.059,
  10.1016/j.physletb.2011.03.040}{{\em Phys.Lett.} {\bf B697} (2011)
  215--221}.

\bibitem{Schumann:2011ji}
S.~Schumann, A.~Renaud, and D.~Zerwas, ``{Hadronically Decaying Color-Adjoint
  Scalars at the LHC},''
\href{http://dx.doi.org/10.1007/JHEP09(2011)074}{{\em JHEP} {\bf 1109} (2011)
  074}.

\bibitem{Drees:1997id}
M.~Drees, S.~Pakvasa, X.~Tata, and T.~ter Veldhuis, ``{A Supersymmetric
  Resolution of Solar and Atmospheric Neutrino Puzzles},''
\href{http://dx.doi.org/10.1103/PhysRevD.57.5335}{{\em Phys.Rev.} {\bf D57}
  (1998)  5335--5339}.

\bibitem{Chun:1998gp}
E.~Chun, S.~Kang, C.~Kim, and U.~Lee, ``{Supersymmetric Neutrino Masses and
  Mixing with R-Parity Violation},''
\href{http://dx.doi.org/10.1016/S0550-3213(99)00034-6}{{\em Nucl.Phys.} {\bf
  B544} (1999)  89--103}.

\bibitem{Joshipura:1999hr}
A.~S. Joshipura and S.~K. Vempati, ``{Sneutrino Vacuum Expectation Values and
  Neutrino Anomalies through Trilinear R-parity Violation},''
\href{http://dx.doi.org/10.1103/PhysRevD.60.111303}{{\em Phys.Rev.} {\bf D60}
  (1999)  111303}.

\bibitem{Cheung:1999az}
K.-m. Cheung and O.~C. Kong, ``{Zee Neutrino Mass Model in SUSY Framework},''
\href{http://dx.doi.org/10.1103/PhysRevD.61.113012}{{\em Phys.Rev.} {\bf D61}
  (2000)  113012}.

\bibitem{Dey:2008ht}
P.~Dey, A.~Kundu, B.~Mukhopadhyaya, and S.~Nandi, ``{Two-Loop Neutrino Masses
  with Large R-Parity Violating Interactions in Supersymmetry},''
\href{http://dx.doi.org/10.1088/1126-6708/2008/12/100}{{\em JHEP} {\bf 0812}
  (2008)  100}.

\bibitem{Roulet:1991sm}
E.~Roulet, ``{MSW effect with Flavor Changing Neutrino Interactions},''
\href{http://dx.doi.org/10.1103/PhysRevD.44.R935}{{\em Phys.Rev.} {\bf D44}
  (1991)  935--938}.

\bibitem{Guzzo:1991hi}
M.~Guzzo, A.~Masiero, and S.~Petcov, ``{On the MSW Effect with Massless
  Neutrinos and no Mixing in the Vacuum},''
\href{http://dx.doi.org/10.1016/0370-2693(91)90984-X}{{\em Phys.Lett.} {\bf
  B260} (1991)  154--160}.

\bibitem{Hirsch:1995ek}
M.~Hirsch, H.~Klapdor-Kleingrothaus, and S.~Kovalenko, ``{Supersymmetry and
  Neutrinoless Double Beta Decay},''
\href{http://dx.doi.org/10.1103/PhysRevD.53.1329}{{\em Phys.Rev.} {\bf D53}
  (1996)  1329--1348}.

\bibitem{Babu:1995vh}
K.~Babu and R.~Mohapatra, ``{New vector - Scalar Contributions to Neutrinoless
  Double Beta Decay and Constraints on R-parity Violation},''
\href{http://dx.doi.org/10.1103/PhysRevLett.75.2276}{{\em Phys.Rev.Lett.} {\bf
  75} (1995)  2276--2279}.

\bibitem{Faessler:1996ph}
A.~Faessler, S.~Kovalenko, F.~Simkovic, and J.~Schwieger, ``{Dominance of Pion
  Exchange in R-parity Violating Supersymmetry Contributions to Neutrinoless
  Double Beta Decay},''
\href{http://dx.doi.org/10.1103/PhysRevLett.78.183}{{\em Phys.Rev.Lett.} {\bf
  78} (1997)  183--186}.

\bibitem{Faessler:1997db}
A.~Faessler, S.~Kovalenko, and F.~Simkovic, ``{Bilinear R-parity Violation in
  Neutrinoless Double Beta Decay},''
\href{http://dx.doi.org/10.1103/PhysRevD.58.055004}{{\em Phys.Rev.} {\bf D58}
  (1998)  055004}.

\bibitem{Barger:1989rk}
V.~D. Barger, G.~Giudice, and T.~Han, ``{Some New Aspects of Supersymmetry
  R-Parity Violating Interactions},''
\href{http://dx.doi.org/10.1103/PhysRevD.40.2987}{{\em Phys.Rev.} {\bf D40}
  (1989)  2987}.

\bibitem{Dreiner:2001kc}
H.~K. Dreiner, G.~Polesello, and M.~Thormeier, ``{Bounds on Broken R Parity
  from Leptonic Meson Decays},''
\href{http://dx.doi.org/10.1103/PhysRevD.65.115006}{{\em Phys.Rev.} {\bf D65}
  (2002)  115006}.

\bibitem{Herz:2002gq}
M.~Herz, ``{Bounds on Leptoquark and Supersymmetric, R Parity Violating
  Interactions from Meson Decays},''
\href{http://arxiv.org/abs/hep-ph/0301079}{{\tt arXiv:hep-ph/0301079
  [hep-ph]}}.

\bibitem{Tahir:1998uk}
F.~Tahir, M.~Sadiq, M.~Anwar~Mughal, and K.~Ahmed, ``{Bounds on R-Parity
  Violating SUSY Yukawa Couplings from Semileptonic Decays of Baryons},''
\href{http://dx.doi.org/10.1016/S0370-2693(98)01049-1}{{\em Phys.Lett.} {\bf
  B439} (1998)  316--318}.

\bibitem{Grossman:1995gt}
Y.~Grossman, Z.~Ligeti, and E.~Nardi, ``{New limit on Inclusive $B \to X_{s}$
  anti-neutrino neutrino Decay and Constraints on New Physics},''
\href{http://dx.doi.org/10.1016/0550-3213(96)00051-X}{{\em Nucl.Phys.} {\bf
  B465} (1996)  369--398}.

\bibitem{Erler:1996ww}
J.~Erler, J.~L. Feng, and N.~Polonsky, ``{A Wide Scalar Neutrino Resonance and
  b anti-b Production at LEP},''
\href{http://dx.doi.org/10.1103/PhysRevLett.78.3063}{{\em Phys.Rev.Lett.} {\bf
  78} (1997)  3063--3066}.

\bibitem{Saha:2002kt}
J.~P. Saha and A.~Kundu, ``{Constraints on R Parity Violating Supersymmetry
  from Leptonic and Semileptonic tau, B(d) and B(s) Decays},''
\href{http://dx.doi.org/10.1103/PhysRevD.66.054021}{{\em Phys.Rev.} {\bf D66}
  (2002)  054021}.

\bibitem{Rosner:2001ck}
J.~L. Rosner, ``{Role of Present and Future Atomic Parity Violation Experiments
  in Precision Electroweak Tests},''
\href{http://dx.doi.org/10.1103/PhysRevD.65.073026}{{\em Phys.Rev.} {\bf D65}
  (2002)  073026}.

\bibitem{Ginges:2003qt}
J.~Ginges and V.~Flambaum, ``{Violations of Fundamental Symmetries in Atoms and
  Tests of Unification Theories of Elementary Particles},''
\href{http://dx.doi.org/10.1016/j.physrep.2004.03.005}{{\em Phys.Rept.} {\bf
  397} (2004)  63--154}.

\bibitem{Kim:2001se}
J.~E. Kim, B.~Kyae, and H.~M. Lee, ``{Effective Supersymmetric Theory and
  (g-2)muon with R-Parity Violation},''
\href{http://dx.doi.org/10.1016/S0370-2693(01)01134-0}{{\em Phys.Lett.} {\bf
  B520} (2001)  298--306}.

\bibitem{Adhikari:2001ra}
R.~Adhikari and G.~Rajasekaran, ``{Anomalous Magnetic Moment of Muon and L
  Violating Supersymmetric Models},''
\href{http://arxiv.org/abs/hep-ph/0107279}{{\tt arXiv:hep-ph/0107279
  [hep-ph]}}.

\bibitem{Godbole:1999ye}
R.~Godbole, S.~Pakvasa, S.~Rindani, and X.~Tata, ``{Fermion Dipole Moments in
  Supersymmetric Models with Explicitly Broken R-Parity},''
\href{http://dx.doi.org/10.1103/PhysRevD.61.113003}{{\em Phys.Rev.} {\bf D61}
  (2000)  113003}.

\bibitem{Abel:1999yz}
S.~Abel, A.~Dedes, and H.~K. Dreiner, ``{Dipole Moments of the Electron,
  Neutrino and Neutron in the MSSM without R-Parity Symmetry},''
{\em JHEP} {\bf 0005} (2000)  013.

\bibitem{Herczeg:1999me}
P.~Herczeg, ``{P, T Violating Electron - Nucleon Interactions in the R-Parity
  Violating Minimal Supersymmetric Standard Model},''
\href{http://dx.doi.org/10.1103/PhysRevD.61.095010}{{\em Phys.Rev.} {\bf D61}
  (2000)  095010}.

\bibitem{Kao:2009fg}
Y.~Kao and T.~Takeuchi, ``{Single-Coupling Bounds on R-parity Violating
  Supersymmetry, an Update},''
\href{http://arxiv.org/abs/0910.4980}{{\tt arXiv:0910.4980 [hep-ph]}}.

\bibitem{Chang:1996sw}
D.~Chang and W.-Y. Keung, ``{New Limits on R-Parity Breakings in Supersymmetric
  Standard Models},''
\href{http://dx.doi.org/10.1016/S0370-2693(96)01271-3}{{\em Phys.Lett.} {\bf
  B389} (1996)  294--298}.

\bibitem{Choi:1998ak}
K.~Choi, K.~Hwang, and J.~S. Lee, ``{Constraints on R-Parity and B Violating
  Couplings in Gauge Mediated Supersymmetry Breaking Models},''
\href{http://dx.doi.org/10.1016/S0370-2693(98)00371-2}{{\em Phys.Lett.} {\bf
  B428} (1998)  129--135}.

\bibitem{Bhattacharyya:1998dt}
G.~Bhattacharyya and P.~B. Pal, ``{New Constraints on R-parity Violation from
  Proton Stability},''
\href{http://dx.doi.org/10.1016/S0370-2693(98)01012-0}{{\em Phys.Lett.} {\bf
  B439} (1998)  81--84}.

\bibitem{Hinchliffe:1992ad}
I.~Hinchliffe and T.~Kaeding, ``{B+L Violating Couplings in the Minimal
  Supersymmetric Standard Model},''
\href{http://dx.doi.org/10.1103/PhysRevD.47.279}{{\em Phys.Rev.} {\bf D47}
  (1993)  279--284}.

\bibitem{Carlson:1995ji}
C.~E. Carlson, P.~Roy, and M.~Sher, ``{New Bounds on R-Parity Violating
  Couplings},''
\href{http://dx.doi.org/10.1016/0370-2693(95)00895-R}{{\em Phys.Lett.} {\bf
  B357} (1995)  99--104}.

\bibitem{Smirnov:1996bg}
A.~Y. Smirnov and F.~Vissani, ``{Upper Bound on all Products of R-Parity
  Violating Couplings lambda-prime and lambda-prime-prime from Proton Decay},''
\href{http://dx.doi.org/10.1016/0370-2693(96)00495-9}{{\em Phys.Lett.} {\bf
  B380} (1996)  317--323}.

\bibitem{Barbieri:1985ty}
R.~Barbieri and A.~Masiero, ``{Supersymmetric Models with Low-Energy Baryon
  Number Violation},''
\href{http://dx.doi.org/10.1016/0550-3213(86)90136-7}{{\em Nucl.Phys.} {\bf
  B267} (1986)  679}.

\bibitem{Abel:1996qj}
S.~Abel, ``{CP Violation in R-Parity Violating Models},''
\href{http://dx.doi.org/10.1016/S0370-2693(97)01034-4}{{\em Phys.Lett.} {\bf
  B410} (1997)  173--180}.

\bibitem{Slavich:2000xm}
P.~Slavich, ``{Constraints on R-Parity Violating Stop Couplings from Flavor
  Physics},''
\href{http://dx.doi.org/10.1016/S0550-3213(00)00700-8}{{\em Nucl.Phys.} {\bf
  B595} (2001)  33--43}.

\bibitem{Chakraverty:2000df}
D.~Chakraverty and D.~Choudhury, ``{$B$ Physics Constraints on Baryon Number
  Violating Couplings: Grand Unification or R-Parity Violation},''
\href{http://dx.doi.org/10.1103/PhysRevD.63.112002}{{\em Phys.Rev.} {\bf D63}
  (2001)  112002}.

\bibitem{BarShalom:2002sv}
S.~Bar-Shalom, G.~Eilam, and Y.-D. Yang, ``{B $\to$ phi pi and B0 $\to$ phi phi
  in the Standard Model and New Bounds on R Parity Violation},''
\href{http://dx.doi.org/10.1103/PhysRevD.67.014007}{{\em Phys.Rev.} {\bf D67}
  (2003)  014007}.

\bibitem{Zwirner:1984is}
F.~Zwirner, ``{Observable Delta B=2 Transitions Without Nucleon Decay in a
  Minimal Supersymmetric Extension of the Standard Model},''
\href{http://dx.doi.org/10.1016/0370-2693(83)90230-7}{{\em Phys.Lett.} {\bf
  B132} (1983)  103--106}.

\bibitem{Dimopoulos:1987rk}
S.~Dimopoulos and L.~J. Hall, ``{Baryogenesis at the MeV Era},''
\href{http://dx.doi.org/10.1016/0370-2693(87)90593-4}{{\em Phys.Lett.} {\bf
  B196} (1987)  135}.

\bibitem{Baltz:1997ar}
E.~A. Baltz and P.~Gondolo, ``{Limits on R-Parity Violation from Cosmic Ray
  Anti-protons},''
\href{http://dx.doi.org/10.1103/PhysRevD.57.7601}{{\em Phys.Rev.} {\bf D57}
  (1998)  7601--7606}.

\bibitem{Aad:2012zx}
{\bf ATLAS Collaboration}, G.~Aad {\em et al.}, ``{Search for Long-Lived, Heavy
  Particles in Final States with a Muon and Multi-Track Displaced Vertex in
  Proton-Proton Collisions at $\sqrt{s}=7$ TeV with the ATLAS Detector},''
\href{http://arxiv.org/abs/1210.7451}{{\tt arXiv:1210.7451 [hep-ex]}}.

\bibitem{ATLAS:2012dp}
{\bf ATLAS Collaboration}, G.~Aad {\em et al.}, ``{Search for Pair Production
  of Massive Particles Decaying into Three Quarks with the ATLAS Detector in
  $\sqrt{s}=7$ TeV $pp$ Collisions at the LHC},''
\href{http://arxiv.org/abs/1210.4813}{{\tt arXiv:1210.4813 [hep-ex]}}.

\bibitem{ATLAS:2012kr}
{\bf ATLAS Collaboration}, G.~Aad {\em et al.}, ``{Search for
  R-Parity-Violating Supersymmetry in Events with Four or More Leptons in
  $\sqrt{s}=7$ TeV $pp$ Collisions with the ATLAS Detector},''
\href{http://arxiv.org/abs/1210.4457}{{\tt arXiv:1210.4457 [hep-ex]}}.

\bibitem{Aad:2012yw}
{\bf ATLAS Collaboration}, G.~Aad {\em et al.}, ``{Search for Lepton Flavour
  Violation in the emu Continuum with the ATLAS Detector in $\sqrt{s}=7$ TeV
  $pp$ Collisions at the LHC},''
\href{http://dx.doi.org/10.1140/epjc/s10052-012-2040-z}{{\em Eur.Phys.J.} {\bf
  C72} (2012)  2040}.

\bibitem{Aad:2011qr}
{\bf ATLAS Collaboration}, G.~Aad {\em et al.}, ``{Search for a Heavy weutral
  Particle Decaying into an Electron and a Muon Using 1 fb-1 of ATLAS Data},''
\href{http://dx.doi.org/10.1140/epjc/s10052-011-1809-9}{{\em Eur.Phys.J.} {\bf
  C71} (2011)  1809}.

\bibitem{ATLAS153}
{\bf ATLAS Collaboration}, ``Search for supersymmetry in events with four or
  more leptons in 13 fb-1 pp collisions at $\sqrt{s}$ = 8 tev with the atlas
  detector,'' \href{http://arxiv.org/abs/ATLAS-CONF-2012-153}{{\tt
  ATLAS-CONF-2012-153}}.

\bibitem{Chatrchyan:2011ff}
{\bf CMS Collaboration}, S.~Chatrchyan {\em et al.}, ``{Search for Physics
  Beyond the Standard Model Using Multilepton Signatures in $pp$ Collisions at
  $\sqrt{s}=7$ TeV},''
\href{http://dx.doi.org/10.1016/j.physletb.2011.09.047}{{\em Phys.Lett.} {\bf
  B704} (2011)  411--433}.

\bibitem{Chatrchyan:2012mea}
{\bf CMS Collaboration}, S.~Chatrchyan {\em et al.}, ``{Search for Anomalous
  Production of Multilepton Events in $pp$ Collisions at $\sqrt{s}=7$ TeV},''
\href{http://dx.doi.org/10.1007/JHEP06(2012)169}{{\em JHEP} {\bf 1206} (2012)
  169}.

\bibitem{CMSSUS027}
{\bf CMS collaboration}, ``Search for rpv supersymmetry with three or more
  leptons and b-tags,'' \href{http://arxiv.org/abs/CMS PAS SUS-12-027}{{\tt CMS
  PAS SUS-12-027}}.

\bibitem{Abazov:2006nw}
{\bf D0 Collaboration}, V.~Abazov {\em et al.}, ``{Search for R-parity
  Violating Supersymmetry via the LL anti-E couplings $\lambda_{121}$,
  $\lambda_{122}$ or $\lambda_{133}$ in $p \bar{p}$ Collisions at $\sqrt{s}$ =
  1.96-TeV},''
\href{http://dx.doi.org/10.1016/j.physletb.2006.05.077}{{\em Phys.Lett.} {\bf
  B638} (2006)  441--449}.

\bibitem{Abulencia:2007mp}
{\bf CDF Collaboration}, A.~Abulencia {\em et al.}, ``{Search for Anomalous
  Production of Multi-Lepton Events in $p \bar{p}$ Collisions at $\sqrt{s}$ =
  1.96-TeV},''
\href{http://dx.doi.org/10.1103/PhysRevLett.98.131804}{{\em Phys.Rev.Lett.}
  {\bf 98} (2007)  131804}.

\bibitem{Aid:1996iw}
{\bf H1 Collaboration}, S.~Aid {\em et al.}, ``{A Search for Squarks of Rp
  Violating SUSY at HERA},''
\href{http://dx.doi.org/10.1007/s002880050165}{{\em Z.Phys.} {\bf C71} (1996)
  211--226}.

\bibitem{Cline:1990bw}
J.~M. Cline and S.~Raby, ``{Gravitino Induced Baryogenesis: A Problem Made a
  Virtue},''
\href{http://dx.doi.org/10.1103/PhysRevD.43.1781}{{\em Phys.Rev.} {\bf D43}
  (1991)  1781--1787}.

\bibitem{Scherrer:1991yu}
R.~J. Scherrer, J.~M. Cline, S.~Raby, and D.~Seckel, ``{Gravitino Induced
  Baryogenesis, Primordial Nucleosynthesis, and the Tremaine-Gunn Limit},''
\href{http://dx.doi.org/10.1103/PhysRevD.44.3760}{{\em Phys.Rev.} {\bf D44}
  (1991)  3760--3766}.

\bibitem{Mollerach:1991mu}
S.~Mollerach and E.~Roulet, ``{Axino Induced Baryogenesis},''
\href{http://dx.doi.org/10.1016/0370-2693(92)91145-Y}{{\em Phys.Lett.} {\bf
  B281} (1992)  303--308}.

\bibitem{Nelson:1993nf}
A.~E. Nelson and N.~Seiberg, ``{R Symmetry Breaking versus Supersymmetry
  Breaking},''
\href{http://dx.doi.org/10.1016/0550-3213(94)90577-0}{{\em Nucl.Phys.} {\bf
  B416} (1994)  46--62}.

\bibitem{Intriligator:2006dd}
K.~A. Intriligator, N.~Seiberg, and D.~Shih, ``{Dynamical SUSY Breaking in
  Meta-Stable Vacua},''
\href{http://dx.doi.org/10.1088/1126-6708/2006/04/021}{{\em JHEP} {\bf 0604}
  (2006)  021}.

\bibitem{Intriligator:2007cp}
K.~A. Intriligator and N.~Seiberg, ``{Lectures on Supersymmetry Breaking},''
\href{http://dx.doi.org/10.1088/0264-9381/24/21/S02}{{\em Class.Quant.Grav.}
  {\bf 24} (2007)  S741--S772}.

\bibitem{Intriligator:2007py}
K.~A. Intriligator, N.~Seiberg, and D.~Shih, ``{Supersymmetry Breaking,
  R-Symmetry Breaking and Metastable Vacua},''
\href{http://dx.doi.org/10.1088/1126-6708/2007/07/017}{{\em JHEP} {\bf 0707}
  (2007)  017}.

\bibitem{Polchinski:1982an}
J.~Polchinski and L.~Susskind, ``{Breaking of Supersymmetry at
  Intermediate-Energy},''
\href{http://dx.doi.org/10.1103/PhysRevD.26.3661}{{\em Phys.Rev.} {\bf D26}
  (1982)  3661}.

\bibitem{Dine:1992yw}
M.~Dine and D.~MacIntire, ``{Supersymmetry, Naturalness, and Dynamical
  Supersymmetry Breaking},''
\href{http://dx.doi.org/10.1103/PhysRevD.46.2594}{{\em Phys.Rev.} {\bf D46}
  (1992)  2594--2601}.

\bibitem{Fox:2002bu}
P.~J. Fox, A.~E. Nelson, and N.~Weiner, ``{Dirac Gaugino Masses and Supersoft
  Supersymmetry Breaking},''
{\em JHEP} {\bf 0208} (2002)  035.

\bibitem{Aad:2011yh}
{\bf ATLAS Collaboration}, G.~Aad {\em et al.}, ``{Search for Massive Colored
  Scalars in Four-Jet Final States in $\sqrt{s}=7$ TeV Proton-Proton Collisions
  with the ATLAS Detector},''
\href{http://dx.doi.org/10.1140/epjc/s10052-011-1828-6}{{\em Eur.Phys.J.} {\bf
  C71} (2011)  1828}.

\bibitem{Aad:2011fq}
{\bf ATLAS Collaboration}, G.~Aad {\em et al.}, ``{Search for New Physics in
  the Dijet Mass Distribution using 1 fb$^{-1}$ of $pp$ Collision Data at
  $\sqrt{s}=7$ TeV collected by the ATLAS Detector},''
\href{http://dx.doi.org/10.1016/j.physletb.2012.01.035}{{\em Phys.Lett.} {\bf
  B708} (2012)  37--54}.

\bibitem{ATLAS:2012nna}
{\bf ATLAS Collaboration}, ``{Search for Massive Coloured Scalars with the
  ATLAS Detector in Four-Jet Final States Using 4.6 fb$^{-1}$ of $\sqrt{s} =
  7$~TeV Proton-Proton Collision Data},''
\href{http://arxiv.org/abs/ATLAS-CONF-2012-110, ATLAS-COM-CONF-2012-155}{{\tt
  ATLAS-CONF-2012-110, ATLAS-COM-CONF-2012-155}}.

\bibitem{CMS:2012eza}
{\bf CMS Collaboration}, ``{Search for Narrow Resonances Using the Dijet Mass
  Spectrum in pp Collisions at $\sqrt{s}$ of 8 TeV},''
\href{http://arxiv.org/abs/CMS-PAS-EXO-12-016}{{\tt CMS-PAS-EXO-12-016}}.

\bibitem{Murayama:1992gi}
H.~Murayama, I.~Watanabe, and K.~Hagiwara, ``{Helas: Helicity Amplitude
  Subroutines for Feynman Diagram Evaluations},''
\href{http://arxiv.org/abs/KEK-91-11}{{\tt KEK-91-11}}.

\bibitem{Hagiwara:2008jb}
K.~Hagiwara, J.~Kanzaki, Q.~Li, and K.~Mawatari, ``{Helas and MadGraph/MadEvent
  with Spin-2 Particles},''
\href{http://dx.doi.org/10.1140/epjc/s10052-008-0663-x}{{\em Eur.Phys.J.} {\bf
  C56} (2008)  435--447}.

\bibitem{Hagiwara:2010pi}
K.~Hagiwara, K.~Mawatari, and Y.~Takaesu, ``{Helas and MadGraph with Spin-3/2
  Particles},''
\href{http://dx.doi.org/10.1140/epjc/s10052-010-1529-6}{{\em Eur.Phys.J.} {\bf
  C71} (2011)  1529}.

\bibitem{Mawatari:2011jy}
K.~Mawatari and Y.~Takaesu, ``{Helas and MadGraph with Goldstinos},''
\href{http://dx.doi.org/10.1140/epjc/s10052-011-1640-3}{{\em Eur.Phys.J.} {\bf
  C71} (2011)  1640}.

\bibitem{Dimopoulos:1988jw}
S.~Dimopoulos and L.~J. Hall, ``{Lepton and Baryon Number Violating Collider
  Signatures from Supersymmetry},''
\href{http://dx.doi.org/10.1016/0370-2693(88)91418-9}{{\em Phys.Lett.} {\bf
  B207} (1988)  210}.

\bibitem{Lancaster:2011wr}
{\bf Tevatron Electroweak Working Group, CDF Collaboration, D0 Collaboration},
  ``{Combination of CDF and D0 Results on the Mass of the Top Quark Using up to
  5.8~fb-1 of Data},''
\href{http://arxiv.org/abs/1107.5255}{{\tt arXiv:1107.5255 [hep-ex]}}.

\bibitem{Nakamura:2010zzi}
{\bf Particle Data Group}, K.~Nakamura {\em et al.}, ``{Review of Particle
  Physics},''
\href{http://dx.doi.org/10.1088/0954-3899/37/7A/075021}{{\em J.Phys.} {\bf G37}
  (2010)  075021}.

\bibitem{Allanach:1999bf}
{\bf R parity Working Group Collaboration}, B.~Allanach {\em et al.},
  ``{Searching for R parity violation at Run II of the Tevatron},''
\href{http://arxiv.org/abs/hep-ph/9906224}{{\tt arXiv:hep-ph/9906224
  [hep-ph]}}.

\bibitem{Pumplin:2002vw}
J.~Pumplin, D.~Stump, J.~Huston, H.~Lai, P.~M. Nadolsky, {\em et al.}, ``{New
  Generation of Parton Distributions with Uncertainties from Global QCD
  Analysis},''
{\em JHEP} {\bf 0207} (2002)  012.

\bibitem{Cacciari:2005hq}
M.~Cacciari and G.~P. Salam, ``{Dispelling the $N^{3}$ Myth for the $k_t$
  Jet-Finder},''
\href{http://dx.doi.org/10.1016/j.physletb.2006.08.037}{{\em Phys.Lett.} {\bf
  B641} (2006)  57--61}.

\bibitem{Cacciari:2011ma}
M.~Cacciari, G.~P. Salam, and G.~Soyez, ``{FastJet User Manual},''
\href{http://dx.doi.org/10.1140/epjc/s10052-012-1896-2}{{\em Eur.Phys.J.} {\bf
  C72} (2012)  1896}.

\bibitem{Cacciari:2008gp}
M.~Cacciari, G.~P. Salam, and G.~Soyez, ``{The Anti-k(t) Jet Clustering
  Algorithm},''
\href{http://dx.doi.org/10.1088/1126-6708/2008/04/063}{{\em JHEP} {\bf 0804}
  (2008)  063}.

\bibitem{Berger:1999zt}
E.~L. Berger, B.~Harris, and Z.~Sullivan, ``{Single Top Squark Production via
  R-parity Violating Supersymmetric Couplings in Hadron Collisions},''
\href{http://dx.doi.org/10.1103/PhysRevLett.83.4472}{{\em Phys.Rev.Lett.} {\bf
  83} (1999)  4472--4475}.

\bibitem{Berger:2000zk}
E.~L. Berger, B.~Harris, and Z.~Sullivan, ``{Direct Probes of R-parity
  Violating Supersymmetric Couplings via Single Top Squark Production},''
\href{http://dx.doi.org/10.1103/PhysRevD.63.115001}{{\em Phys.Rev.} {\bf D63}
  (2001)  115001}.

\bibitem{CMS:2011cra}
{\bf CMS Collaboration}, ``{Performance of the b-jet identification in CMS},''
\href{http://arxiv.org/abs/CMS-PAS-BTV-11-001}{{\tt CMS-PAS-BTV-11-001}}.

\bibitem{CMS:2009gxa}
{\bf CMS Collaboration}, ``{Algorithms for b Jet identification in CMS},''
\href{http://arxiv.org/abs/CMS-PAS-BTV-09-001}{{\tt CMS-PAS-BTV-09-001}}.

\bibitem{daCosta:2011qk}
{\bf ATLAS Collaboration}, G.~Aad {\em et al.}, ``{Search for Squarks and
  Gluinos using Final States with Jets and Missing Transverse Momentum with the
  ATLAS Detector in $\sqrt{s}=7$ TeV Proton-Proton Collisions},''
\href{http://dx.doi.org/10.1016/j.physletb.2011.05.061}{{\em Phys.Lett.} {\bf
  B701} (2011)  186--203}.

\bibitem{Collaboration:2011ida}
{\bf CMS Collaboration}, S.~Chatrchyan {\em et al.}, ``{Search for New Physics
  with Jets and Missing Transverse Momentum in $pp$ collisions at $\sqrt{s}=7$
  TeV},''
\href{http://dx.doi.org/10.1007/JHEP08(2011)155}{{\em JHEP} {\bf 1108} (2011)
  155}.

\bibitem{ATLAS:2012hpa}
{\bf ATLAS Collaboration}, ``{Search for Exotic Same-Sign Dilepton Signatures
  ($b'$ Quark, $T_{5/3}$ and Four Top Quarks Production) in 4.7 fb$^{-1}$ of pp
  Collisions at $\sqrt{s}=7$ TeV with the ATLAS Detector},''
\href{http://arxiv.org/abs/ATLAS-CONF-2012-130, ATLAS-COM-CONF-2012-163}{{\tt
  ATLAS-CONF-2012-130, ATLAS-COM-CONF-2012-163}}.

\bibitem{Desai:2010sq}
N.~Desai and B.~Mukhopadhyaya, ``{R-Parity Violating Resonant Stop Production
  at the Large Hadron Collider},''
\href{http://dx.doi.org/10.1007/JHEP10(2010)060}{{\em JHEP} {\bf 1010} (2010)
  060}.

\bibitem{Morrissey:2005uza}
D.~Morrissey, T.~M. Tait, and C.~Wagner, ``{Proton Lifetime and Baryon Number
  Violating Signatures at the CERN LHC in Gauge Extended Models},''
\href{http://dx.doi.org/10.1103/PhysRevD.72.095003}{{\em Phys.Rev.} {\bf D72}
  (2005)  095003}.

\bibitem{Dong:2011rh}
Z.~Dong, G.~Durieux, J.-M. Gerard, T.~Han, and F.~Maltoni, ``{Baryon Number
  Violation at the LHC: the Top Option},''
\href{http://dx.doi.org/10.1103/PhysRevD.85.016006,
  10.1103/PhysRevD.85.039907}{{\em Phys.Rev.} {\bf D85} (2012)  016006}.

\bibitem{Davoudiasl:2011fj}
H.~Davoudiasl, D.~E. Morrissey, K.~Sigurdson, and S.~Tulin, ``{Baryon
  Destruction by Asymmetric Dark Matter},''
\href{http://dx.doi.org/10.1103/PhysRevD.84.096008}{{\em Phys.Rev.} {\bf D84}
  (2011)  096008}.

\bibitem{delAguila:1999ac}
F.~del Aguila, J.~Aguilar-Saavedra, and L.~Ametller, ``{Z t and gamma t
  Production via Top Flavor Changing Neutral Couplings at the Fermilab
  Tevatron},''
\href{http://dx.doi.org/10.1016/S0370-2693(99)00929-6}{{\em Phys.Lett.} {\bf
  B462} (1999)  310--318}.

\bibitem{Allanach:2010pp}
B.~C. Allanach, S.~Grab, and H.~E. Haber, ``{Supersymmetric Monojets at the
  Large Hadron Collider},''
\href{http://dx.doi.org/10.1007/JHEP07(2011)087, 10.1007/JHEP09(2011)027,
  10.1007/JHEP01(2011)138}{{\em JHEP} {\bf 1101} (2011)  138}.

\bibitem{Alves:2011wf}
{\bf LHC New Physics Working Group}, D.~Alves {\em et al.}, ``{Simplified
  Models for LHC New Physics Searches},''
\href{http://dx.doi.org/10.1088/0954-3899/39/10/105005}{{\em J.Phys.} {\bf G39}
  (2012)  105005}.

\bibitem{Degrassi:2008mw}
G.~Degrassi, E.~Gabrielli, and L.~Trentadue, ``{Flavor Changing
  Fermion-Graviton Vertices},''
\href{http://dx.doi.org/10.1103/PhysRevD.79.053004}{{\em Phys.Rev.} {\bf D79}
  (2009)  053004}.

\bibitem{Kilic:2008pm}
C.~Kilic, T.~Okui, and R.~Sundrum, ``{Colored Resonances at the Tevatron:
  Phenomenology and Discovery Potential in Multijets},''
\href{http://dx.doi.org/10.1088/1126-6708/2008/07/038}{{\em JHEP} {\bf 0807}
  (2008)  038}.

\bibitem{Kilic:2008ub}
C.~Kilic, S.~Schumann, and M.~Son, ``{Searching for Multijet Resonances at the
  LHC},''
\href{http://dx.doi.org/10.1088/1126-6708/2009/04/128}{{\em JHEP} {\bf 0904}
  (2009)  128}.

\bibitem{Kilic:2009mi}
C.~Kilic, T.~Okui, and R.~Sundrum, ``{Vectorlike Confinement at the LHC},''
\href{http://dx.doi.org/10.1007/JHEP02(2010)018}{{\em JHEP} {\bf 1002} (2010)
  018}.

\bibitem{Kilic:2010et}
C.~Kilic and T.~Okui, ``{The LHC Phenomenology of Vectorlike Confinement},''
\href{http://dx.doi.org/10.1007/JHEP04(2010)128}{{\em JHEP} {\bf 1004} (2010)
  128}.

\bibitem{Dicus:2010bm}
D.~A. Dicus, C.~Kao, S.~Nandi, and J.~Sayre, ``{Discovering Colorons at the
  Early Stage LHC},''
\href{http://dx.doi.org/10.1103/PhysRevD.83.091702}{{\em Phys.Rev.} {\bf D83}
  (2011)  091702}.

\bibitem{Sayre:2011ed}
J.~Sayre, D.~A. Dicus, C.~Kao, and S.~Nandi, ``{Searching for Colorons at the
  Large Hadron Collider},''
\href{http://dx.doi.org/10.1103/PhysRevD.84.015011}{{\em Phys.Rev.} {\bf D84}
  (2011)  015011}.

\bibitem{Burdman:2006gy}
G.~Burdman, B.~A. Dobrescu, and E.~Ponton, ``{Resonances from Two Universal
  Extra Dimensions},''
\href{http://dx.doi.org/10.1103/PhysRevD.74.075008}{{\em Phys.Rev.} {\bf D74}
  (2006)  075008}.

\bibitem{Brooijmans:2012yi}
G.~Brooijmans, B.~Gripaios, F.~Moortgat, J.~Santiago, P.~Skands, {\em et al.},
  ``{Les Houches 2011: Physics at TeV Colliders New Physics Working Group
  Report},''
\href{http://arxiv.org/abs/1203.1488}{{\tt arXiv:1203.1488 [hep-ph]}}.

\bibitem{GoncalvesNetto:2012nt}
D.~Goncalves-Netto, D.~Lopez-Val, K.~Mawatari, T.~Plehn, and I.~Wigmore,
  ``{Sgluon Pair Production to Next-to-Leading Order},''
\href{http://dx.doi.org/10.1103/PhysRevD.85.114024}{{\em Phys.Rev.} {\bf D85}
  (2012)  114024}.

\bibitem{LopezVal:2012ms}
D.~Lopez-Val, D.~Goncalves-Netto, K.~Mawatari, T.~Plehn, and I.~Wigmore,
  ``{MadGolem: automating NLO calculations for New Physics},''
{\em PoS} {\bf LL2012} (2012)  048.

\bibitem{Davidson:2010rw}
N.~Davidson, G.~Nanava, T.~Przedzinski, E.~Richter-Was, and Z.~Was,
  ``{Universal Interface of Tauola Technical and Physics Documentation},''
\href{http://dx.doi.org/10.1016/j.cpc.2011.12.009}{{\em Comput.Phys.Commun.}
  {\bf 183} (2012)  821--843}.

\bibitem{Catani:1993hr}
S.~Catani, Y.~L. Dokshitzer, M.~Seymour, and B.~Webber, ``{Longitudinally
  Invariant $K_t$ Clustering Algorithms for Hadron Hadron Collisions},''
\href{http://dx.doi.org/10.1016/0550-3213(93)90166-M}{{\em Nucl.Phys.} {\bf
  B406} (1993)  187--224}.

\bibitem{Melnikov:2006kv}
K.~Melnikov and F.~Petriello, ``{Electroweak Gauge Boson Production at Hadron
  Colliders through O(alpha(s)**2)},''
\href{http://dx.doi.org/10.1103/PhysRevD.74.114017}{{\em Phys.Rev.} {\bf D74}
  (2006)  114017}.

\bibitem{Gavin:2012sy}
R.~Gavin, Y.~Li, F.~Petriello, and S.~Quackenbush, ``{W Physics at the LHC with
  Fewz 2.1},''
\href{http://arxiv.org/abs/1201.5896}{{\tt arXiv:1201.5896 [hep-ph]}}.

\bibitem{Gavin:2010az}
R.~Gavin, Y.~Li, F.~Petriello, and S.~Quackenbush, ``{Fewz 2.0: A Code for
  Hadronic Z Production at Next-to-Next-to-Leading Order},''
\href{http://dx.doi.org/10.1016/j.cpc.2011.06.008}{{\em Comput.Phys.Commun.}
  {\bf 182} (2011)  2388--2403}.

\bibitem{Lai:2010vv}
H.-L. Lai, M.~Guzzi, J.~Huston, Z.~Li, P.~M. Nadolsky, {\em et al.}, ``{New
  Parton Distributions for Collider Physics},''
\href{http://dx.doi.org/10.1103/PhysRevD.82.074024}{{\em Phys.Rev.} {\bf D82}
  (2010)  074024}.

\bibitem{Aliev:2010zk}
M.~Aliev, H.~Lacker, U.~Langenfeld, S.~Moch, P.~Uwer, {\em et al.}, ``{HATHOR:
  HAdronic Top and Heavy quarks crOss section calculatoR},''
\href{http://dx.doi.org/10.1016/j.cpc.2010.12.040}{{\em Comput.Phys.Commun.}
  {\bf 182} (2011)  1034--1046}.

\bibitem{Baernreuther:2012ws}
P.~Baernreuther, M.~Czakon, and A.~Mitov, ``{Percent Level Precision Physics at
  the Tevatron: First Genuine NNLO QCD Corrections to $q \bar{q} \to t \bar{t}
  + X$},''
\href{http://dx.doi.org/10.1103/PhysRevLett.109.132001}{{\em Phys.Rev.Lett.}
  {\bf 109} (2012)  132001}.

\bibitem{Czakon:2013goa}
M.~Czakon, P.~Fiedler, and A.~Mitov, ``{The total top quark pair production
  cross-section at hadron colliders through ${\cal O}(alpha_S^4)$},''
\href{http://dx.doi.org/10.1103/PhysRevLett.110.252004}{{\em Phys.Rev.Lett.}
  {\bf 110} (2013)  252004}.

\bibitem{Kidonakis:2010tc}
N.~Kidonakis, ``{NNLL Resummation for s-channel Single Top Quark Production},''
\href{http://dx.doi.org/10.1103/PhysRevD.81.054028}{{\em Phys.Rev.} {\bf D81}
  (2010)  054028}.

\bibitem{Kidonakis:2010ux}
N.~Kidonakis, ``{Two-Loop Soft Anomalous Dimensions for Single Top Quark
  Associated Production with a W- or H-},''
\href{http://dx.doi.org/10.1103/PhysRevD.82.054018}{{\em Phys.Rev.} {\bf D82}
  (2010)  054018}.

\bibitem{Kidonakis:2011wy}
N.~Kidonakis, ``{Next-to-Next-to-Leading-Order Collinear and Soft Gluon
  Corrections for t-channel Single Top Quark Production},''
\href{http://dx.doi.org/10.1103/PhysRevD.83.091503}{{\em Phys.Rev.} {\bf D83}
  (2011)  091503}.

\bibitem{Kidonakis:2012db}
N.~Kidonakis, ``{Differential and Total Cross Sections for Top Pair and Single
  Top Production},''
\href{http://arxiv.org/abs/1205.3453}{{\tt arXiv:1205.3453 [hep-ph]}}.

\bibitem{Campbell:1999ah}
J.~M. Campbell and R.~K. Ellis, ``{An Update on Vector Boson Pair Production at
  Hadron Colliders},''
\href{http://dx.doi.org/10.1103/PhysRevD.60.113006}{{\em Phys.Rev.} {\bf D60}
  (1999)  113006}.

\bibitem{Campbell:2011bn}
J.~M. Campbell, R.~K. Ellis, and C.~Williams, ``{Vector Boson Pair Production
  at the LHC},''
\href{http://dx.doi.org/10.1007/JHEP07(2011)018}{{\em JHEP} {\bf 1107} (2011)
  018}.

\bibitem{Campbell:2012dh}
J.~M. Campbell and R.~K. Ellis, ``{$t \bar{t} W^{+-}$ Production and Decay at
  NLO},''
\href{http://dx.doi.org/10.1007/JHEP07(2012)052}{{\em JHEP} {\bf 1207} (2012)
  052}.

\bibitem{Ball:2007zza}
{\bf CMS Collaboration}, G.~Bayatian {\em et al.}, ``{CMS Technical Design
  Report, Volume II: Physics Performance},''
\href{http://dx.doi.org/10.1088/0954-3899/34/6/S01}{{\em J.Phys.} {\bf G34}
  (2007)  995--1579}.

\bibitem{atlasmet}
 {\em
  {https://twiki.cern.ch/twiki/bin/view/AtlasPublic/MissingEtTriggerPublicResults}}
  .

\bibitem{cmsmet}
{\bf CMS Collaboration,} {\em {Internal Note}}  .

\bibitem{Aad:2012wm}
{\bf ATLAS Collaboration}, G.~Aad {\em et al.}, ``{A Search for $t\bar{t}$
  Resonances with the ATLAS Detector in 2.05 fb$^{-1}$ of Proton-Proton
  Collisions at $\sqrt{s}=7$ TeV},''
\href{http://dx.doi.org/10.1140/epjc/s10052-012-2083-1}{{\em Eur.Phys.J.} {\bf
  C72} (2012)  2083}.

\bibitem{Read:2002hq}
A.~L. Read, ``{Presentation of Search Results: The CL(s) Technique},''
\href{http://dx.doi.org/10.1088/0954-3899/28/10/313}{{\em J.Phys.} {\bf G28}
  (2002)  2693--2704}.

\bibitem{mclimit}
 {\em {http://www-cdf.fnal.gov/$\sim$trj/mclimit/production/mclimit.html}}  .

\bibitem{Bernabei:2008yi}
{\bf DAMA Collaboration}, R.~Bernabei {\em et al.}, ``{First Results from
  DAMA/LIBRA and the Combined Results with DAMA/NaI},''
\href{http://dx.doi.org/10.1140/epjc/s10052-008-0662-y}{{\em Eur.Phys.J.} {\bf
  C56} (2008)  333--355}.

\bibitem{Aalseth:2010vx}
{\bf CoGeNT collaboration}, C.~Aalseth {\em et al.}, ``{Results from a Search
  for Light-Mass Dark Matter with a P-type Point Contact Germanium Detector},''
\href{http://dx.doi.org/10.1103/PhysRevLett.106.131301}{{\em Phys.Rev.Lett.}
  {\bf 106} (2011)  131301}.

\bibitem{Aalseth:2011wp}
C.~Aalseth, P.~Barbeau, J.~Colaresi, J.~Collar, J.~Diaz~Leon, {\em et al.},
  ``{Search for an Annual Modulation in a P-type Point Contact Germanium Dark
  Matter Detector},''
\href{http://dx.doi.org/10.1103/PhysRevLett.107.141301}{{\em Phys.Rev.Lett.}
  {\bf 107} (2011)  141301}.

\bibitem{Angloher:2011uu}
G.~Angloher, M.~Bauer, I.~Bavykina, A.~Bento, C.~Bucci, {\em et al.},
  ``{Results from 730 kg days of the CRESST-II Dark Matter Search},''
\href{http://dx.doi.org/10.1140/epjc/s10052-012-1971-8}{{\em Eur.Phys.J.} {\bf
  C72} (2012)  1971}.

\bibitem{Acosta:2004yw}
{\bf CDF Collaboration}, D.~Acosta {\em et al.}, ``{Measurement of the $J/\psi$
  Meson and $b-$Hadron Production Cross Sections in $p\bar{p}$ Collisions at
  $\sqrt{s} = 1960$ GeV},''
\href{http://dx.doi.org/10.1103/PhysRevD.71.032001}{{\em Phys.Rev.} {\bf D71}
  (2005)  032001}.

\bibitem{Aaltonen:2010fs}
{\bf CDF Collaboration}, T.~Aaltonen {\em et al.}, ``{Search for Single Top
  Quark Production in pbar $p$ Collisions at $\sqrt{s}=1.96$ TeV in the Missing
  Transverse Energy plus Jets Topology},''
\href{http://dx.doi.org/10.1103/PhysRevD.81.072003}{{\em Phys.Rev.} {\bf D81}
  (2010)  072003}.

\bibitem{5076030}
A.~Bhatti, A.~Canepa, M.~Casarsa, M.~Convery, G.~Cortiana, S.~Donati,
  G.~Flanagan, V.~Greco, H.~Frisch, T.~Fukun, P.~Giannetti, D.~Krop, T.~Liu,
  D.~Lucchesi, M.~Piendibene, L.~Ristori, L.~Rogondino, V.~Rusu, L.~Sartori,
  M.~Vidal, and L.~Zhou, ``{Level-2 Calorimeter Trigger Upgrade at CDF},''
  \href{http://dx.doi.org/10.1109/TNS.2009.2016420}{{\em Nuclear Science, IEEE
  Transactions on} {\bf 56} (2009) no.~3, 1685--1689}.

\bibitem{Abe:1991ui}
{\bf CDF Collaboration}, F.~Abe {\em et al.}, ``{The Topology of Three Jet
  Events in $\bar{p}p$ Collisions at $\sqrt{s} = 1.8$ TeV},''
\href{http://dx.doi.org/10.1103/PhysRevD.45.1448}{{\em Phys.Rev.} {\bf D45}
  (1992)  1448--1458}.

\bibitem{Bhatti:2005ai}
{\bf CDF Collaboration}, A.~Bhatti {\em et al.}, ``{Determination of the Jet
  Energy Scale at the Collider Detector at Fermilab},''
\href{http://dx.doi.org/10.1016/j.nima.2006.05.269}{{\em Nucl.Instrum.Meth.}
  {\bf A566} (2006)  375--412}.

\bibitem{Acosta:2004hw}
{\bf CDF Collaboration}, D.~Acosta {\em et al.}, ``{Measurement of the
  $t\bar{t}$ Production Cross Section in $p\bar{p}$ Collisions at $\sqrt{s} =
  1.96$ TeV Using Lepton + Jets Events with Secondary Vertex $b-$Tagging},''
\href{http://dx.doi.org/10.1103/PhysRevD.71.052003}{{\em Phys.Rev.} {\bf D71}
  (2005)  052003}.

\bibitem{Aaltonen:2009jg}
{\bf CDF Collaboration}, T.~Aaltonen {\em et al.}, ``{A Search for the Higgs
  Boson Using Neural Networks in Events with Missing Energy and b-quark Jets in
  p anti-p Collisions at s**(1/2) = 1.96-TeV},''
\href{http://dx.doi.org/10.1103/PhysRevLett.104.141801}{{\em Phys.Rev.Lett.}
  {\bf 104} (2010)  141801}.

\bibitem{Campbell:2000bg}
J.~M. Campbell and R.~K. Ellis, ``{Radiative Corrections to Z b anti-b
  Production},''
\href{http://dx.doi.org/10.1103/PhysRevD.62.114012}{{\em Phys.Rev.} {\bf D62}
  (2000)  114012}.

\bibitem{Langenfeld:2009wd}
U.~Langenfeld, S.~Moch, and P.~Uwer, ``{Measuring the Running Top-Quark
  Mass},''
\href{http://dx.doi.org/10.1103/PhysRevD.80.054009}{{\em Phys.Rev.} {\bf D80}
  (2009)  054009}.

\bibitem{Harris:2002md}
B.~Harris, E.~Laenen, L.~Phaf, Z.~Sullivan, and S.~Weinzierl, ``{The Fully
  Differential Single Top Quark Cross-Section in Next to Leading Order QCD},''
\href{http://dx.doi.org/10.1103/PhysRevD.66.054024}{{\em Phys.Rev.} {\bf D66}
  (2002)  054024}.

\bibitem{Sullivan:2004ie}
Z.~Sullivan, ``{Understanding Single-Top-Quark Production and Jets at Hadron
  Colliders},''
\href{http://dx.doi.org/10.1103/PhysRevD.70.114012}{{\em Phys.Rev.} {\bf D70}
  (2004)  114012}.

\bibitem{Galtieri:2011yd}
A.~B. Galtieri, F.~Margaroli, and I.~Volobouev, ``{Precision Measurements of
  the Top Quark Mass from the Tevatron in the pre-LHC Era},''
\href{http://dx.doi.org/10.1088/0034-4885/75/5/056201}{{\em Rept.Prog.Phys.}
  {\bf 75} (2012)  056201}.

\bibitem{Acosta:2002hx}
D.~Acosta, S.~Klimenko, J.~Konigsberg, A.~Korytov, G.~Mitselmakher, {\em et
  al.}, ``{The Performance of the CDF Luminosity Monitor},''
\href{http://dx.doi.org/10.1016/S0168-9002(02)01445-6}{{\em Nucl.Instrum.Meth.}
  {\bf A494} (2002)  57--62}.

\bibitem{Aaltonen:2010jr}
{\bf CDF Collaboration}, T.~Aaltonen {\em et al.}, ``{Observation of Single Top
  Quark Production and Measurement of |Vtb| with CDF},''
\href{http://dx.doi.org/10.1103/PhysRevD.82.112005}{{\em Phys.Rev.} {\bf D82}
  (2010)  112005}.

\end{thebibliography}\endgroup\newpage
